\documentclass[onecolumn,notitlepage,floats,floatfix,amssymb,aps,prd,showpacs,superscriptaddress,groupedaddress,nofootinbib]{revtex4-2}  

\usepackage[T1]{fontenc}
\usepackage{lmodern} 
\usepackage{graphicx}  
\usepackage{dcolumn}   
\usepackage{bm}        
\usepackage{amssymb}   
\usepackage{amsmath}
\usepackage{bbold}
\usepackage{array}
\usepackage{makecell}
\usepackage{braket}
\usepackage{longtable}
\usepackage{supertabular,booktabs}
\usepackage[toc,page]{appendix}
\usepackage{moresize}

\usepackage{titlesec} 
\usepackage[colorlinks,citecolor=blue,urlcolor=blue,hypertexnames=true]{hyperref}
\usepackage{subfigure}

\usepackage[usenames,dvipsnames,svgnames,table]{xcolor} 

\renewcommand{\arraystretch}{2}

\newcommand{\be}{\begin{equation}}
\newcommand{\ee}{\end{equation}}
\newcommand{\bea}{\begin{eqnarray}}
\newcommand{\eea}{\end{eqnarray}}
\newcommand{\bml}{\begin{subequations}}
\newcommand{\eml}{\end{subequations}}
\newcommand{\bfig}{\begin{figure}}
\newcommand{\efig}{\end{figure}}

\newcommand{\del}{\delta}

\newcommand{\Del}{\Delta}

\newcommand{\mbf}{\mathbf}

\newcommand{\slashed}{\hspace{-1.1ex}/}

\newcommand{\bmat}{\begin{pmatrix}}
\newcommand{\emat}{\end{pmatrix}}
\usepackage{graphicx,slashed,booktabs,xcolor,multirow,float,
amsfonts,bbold,mathtools,sidecap,tikz,bm,enumitem}
\usepackage{multirow}
\usepackage{bbding}
\usepackage{titlesec}
\usepackage{hyperref}
\usepackage{wasysym}
\usepackage{amssymb}
\usepackage{pifont}
\newcommand{\grad}{\nabla}

\renewcommand{\d}{\mathrm{d}}

\renewcommand{\leq}{\leqslant}
\renewcommand{\geq}{\geqslant}

\usepackage[dvipsnames, usenames]{xcolor}

\definecolor{linkcolor}{rgb}{0.55, 0.13, .32}

\definecolor{oucrimsonred}{rgb}{0.6, 0.0, 0.0}
\definecolor{persianblue}{rgb}{0.11, 0.22, 0.73}
\definecolor{forestgreen}{rgb}{0.13,0.35,0.13}
\definecolor{lightgray}{rgb}{0.83, 0.83, 0.83}
 \hypersetup{colorlinks, citecolor=oucrimsonred, linkcolor=persianblue, urlcolor=oucrimsonred}
\definecolor{cornellred}{rgb}{0.7, 0.11, 0.11}
\definecolor{navyblue}{rgb}{0.0, 0.0, 0.5}
\definecolor{amethyst}{rgb}{0.6, 0.4, 0.8}
\definecolor{yellow}{rgb}{1.0, 1.0, 0.0}
\definecolor{firebrick}{rgb}{0.7, 0.13, 0.13}
\definecolor{tangerineyellow}{rgb}{1.0, 0.8, 0.0}
\definecolor{deepfuchsia}{rgb}{0.76, 0.33, 0.76}
\definecolor{amber}{rgb}{1.0, 0.75, 0.0}
\definecolor{VioletRed4}{rgb}{0.55, 0.13, .32}
\definecolor{indiagreen}{rgb}{0.07, 0.53, 0.03}
\definecolor{VioletRed4}{rgb}{0.55, 0.13, .32}

\usepackage{hyperref}

\usepackage{graphics, appendix,afterpage,makecell}

\usepackage{bbold}
\usepackage{tikz}
\usepackage{adjustbox}
\usepackage{tikz-feynman}
\usepackage{tcolorbox}

\def\abs#1{\left| #1\right|}

\definecolor{oucrimsonred}{rgb}{0.6, 0.0, 0.0}
\definecolor{persianblue}{rgb}{0.11, 0.22, 0.73}
\definecolor{forestgreen}{rgb}{0.13,0.35,0.13}
\definecolor{lightgray}{rgb}{0.83, 0.83, 0.83}
 \hypersetup{colorlinks, citecolor=oucrimsonred, linkcolor=persianblue, urlcolor=oucrimsonred}
\definecolor{cornellred}{rgb}{0.7, 0.11, 0.11}
\definecolor{navyblue}{rgb}{0.0, 0.0, 0.5}
\definecolor{amethyst}{rgb}{0.6, 0.4, 0.8}
\definecolor{yellow}{rgb}{1.0, 1.0, 0.0}
\definecolor{firebrick}{rgb}{0.7, 0.13, 0.13}
\definecolor{tangerineyellow}{rgb}{1.0, 0.8, 0.0}
\definecolor{deepfuchsia}{rgb}{0.76, 0.33, 0.76}
\definecolor{amber}{rgb}{1.0, 0.75, 0.0}
\definecolor{VioletRed4}{rgb}{0.55, 0.13, .32}
\definecolor{indiagreen}{rgb}{0.07, 0.53, 0.03}
\definecolor{VioletRed4}{rgb}{0.55, 0.13, .32}

\definecolor{oucrimsonred}{rgb}{0.6, 0.0, 0.0}
\newcommand\vertarrowbox[3][6ex]{%
  \begin{array}[t]{@{}c@{}} #2 \\
  \left\uparrow\vcenter{\hrule height #1}\right.\kern-\nulldelimiterspace\\
  \makebox[0pt]{\scriptsize#3}
  \end{array}%
}

\definecolor{mtcolor}{rgb}{.8,.3,.1}

\definecolor{violachiaro}{rgb}{1,0.6,1}

\definecolor{gbcolor}{rgb}{.43,.22,.12}
 
\definecolor{gbcolor2}{rgb}{.9,.2,.6}
\definecolor{gbcolor3}{rgb}{.3,.2,.6}

\definecolor{verdechiaro}{rgb}{0.6,1,0.6}
\definecolor{giallochiaro}{rgb}{1,1,0.6}
\definecolor{bluscuro}{rgb}{0.15, 0.2, 0.9}
\definecolor{verdes}{rgb}{0.1, 0.5, 0.1}%
\definecolor{tangerineyellow}{rgb}{1.0, 0.8, 0.0}
\definecolor{smokyblack}{rgb}{0.06, 0.05, 0.03}

\definecolor{americanrose}{rgb}{1.0, 0.01, 0.24}
\definecolor{cobalt}{rgb}{0.0, 0.28, 0.67}
\definecolor{brandeisblue}{rgb}{0.0, 0.44, 1.0}
\definecolor{mycolor}{rgb}{0.0, 0.0, 0.5}
\definecolor{oxfordblue}{rgb}{0.0, 0.13, 0.28}
\definecolor{azure}{rgb}{0.0, 0.5, 1.0}
\definecolor{turquoiseblue}{rgb}{0.0, 1.0, 0.94}
\newtcolorbox{mynewbox}[1]{colback=white!5!white,colframe=azure!75!black,fonttitle=\bfseries,title=#1}
\newtcolorbox{mybox}{colback=mycolor!5!white,colframe=azure!75!black}
\newtcolorbox{mynamedbox}[1]{colback=mycolor!5!white,colframe=azure!75!black,title=#1}
\definecolor{venetianred}{rgb}{0.78, 0.03, 0.08}
\newtcolorbox{mynamedbox1}[1]{colback=venetianred!5!white,colframe=venetianred!80!black,title=#1}
\newtcolorbox{mynamedbox2}[1]{colback=azure!5!white,colframe=azure!80!black,title=#1}

\definecolor{rossocorsa}{rgb}{0.83, 0.0, 0.0}

\tikzset{->-/.style={decoration={
  markings,
  mark=at position #1 with {\arrow{>}}},postaction={decorate}}}
\tikzset{-<-/.style={decoration={
  markings,
  mark=at position #1 with {\arrow{<}}},postaction={decorate}}} 

\def\be{\begin{equation}}
\def\ee{\end{equation}}
\def\ba{\begin{eqnarray}}
\def\ea{\end{eqnarray}}

\def\d{\mathrm{d}}
\def\p{{\cal P}}

\def\L*{{\cal L}_*}
\def\L{\mathcal{L}}
\def\({\left(}
\def\){\right)}

\def\p{\partial}

\def\p{\partial}

\def\<{\langle}
\def\>{\rangle}

\def\cs2{c_{s}^{2}}

 \def\p{\partial}

 \def\be   {\begin{equation}}   \def\ee   {\end{equation}}
 \def\ba   {\begin{array}}      \def\ea   {\end{array}}
 \def\bea  {\begin{eqnarray}}   \def\eea  {\end{eqnarray}}
 \def\bean {\begin{eqnarray*}}  \def\eean {\end{eqnarray*}}

\newcommand{\fnl}{f_{\mathrm{NL}}}

\titleclass{\subsubsubsection}{straight}[\subsection]

\newcounter{subsubsubsection}[subsubsection]
\renewcommand\thesubsubsubsection{\thesubsubsection.\arabic{subsubsubsection}}

\titleformat{\subsubsubsection}
  {\normalfont\normalsize\bfseries}{\thesubsubsubsection}{1em}{}
\titlespacing*{\subsubsubsection}
{0pt}{3.25ex plus 1ex minus .2ex}{1.5ex plus .2ex}

\makeatletter
\renewcommand\paragraph{\@startsection{paragraph}{5}{\z@}%
  {3.25ex \@plus1ex \@minus.2ex}%
  {-1em}%
  {\normalfont\normalsize\bfseries}}
\renewcommand\subparagraph{\@startsection{subparagraph}{6}{\parindent}%
  {3.25ex \@plus1ex \@minus .2ex}%
  {-1em}%
  {\normalfont\normalsize\bfseries}}
\def\toclevel@subsubsubsection{4}
\def\toclevel@paragraph{5}
\def\toclevel@paragraph{6}
\def\l@subsubsubsection{\@dottedtocline{4}{7em}{4em}}
\def\l@paragraph{\@dottedtocline{5}{10em}{5em}}
\def\l@subparagraph{\@dottedtocline{6}{14em}{6em}}
\makeatother

\begin{document}


\definecolor{lime}{HTML}{A6CE39}
\DeclareRobustCommand{\orcidicon}{\hspace{-2.1mm}
\begin{tikzpicture}
\draw[lime,fill=lime] (0,0.0) circle [radius=0.13] node[white] {{\fontfamily{qag}\selectfont \tiny \,ID}}; \draw[white, fill=white] (-0.0525,0.095) circle [radius=0.007]; 
\end{tikzpicture} \hspace{-3.7mm} }
\foreach \x in {A, ..., Z}{\expandafter\xdef\csname orcid\x\endcsname{\noexpand\href{https://orcid.org/\csname orcidauthor\x\endcsname} {\noexpand\orcidicon}}}
\newcommand{\orcidauthorA}{0000-0002-0459-3873}
\newcommand{\orcidauthorB}{0000-0003-1081-0632}


\title{\textcolor{Sepia}{\textbf \Huge\Large\LARGE  
Large fluctuations and Primordial Black Holes } 
}


\author{{\large  Sayantan Choudhury\orcidA{}${}^{1}$}}
\email{sayantan\_ccsp@sgtuniversity.org,  sayanphysicsisi@gmail.com } 

\author{ \large M.~Sami\orcidB{}${}^{1,2,3}$}
\email{ sami\_ccsp@sgtuniversity.org,  samijamia@gmail.com}

\affiliation{ ${}^{1}$Centre For Cosmology and Science Popularization (CCSP),\\
        SGT University, Gurugram, Delhi- NCR, Haryana- 122505, India.}
\affiliation{${}^{2}$Center for Theoretical Physics, Eurasian National University, Astana 010008, Kazakhstan.}
	\affiliation{${}^{3}$Chinese Academy of Sciences,52 Sanlihe Rd, Xicheng District, Beijing.}

\begin{abstract}

In this paper, we review in detail different mechanisms of generation of large primordial fluctuations and their implications for the production of primordial black holes (PBHs) and scalar-induced secondary gravity waves (SIGW), with the ultimate aim of understanding the impact of loop correction on quantum correlations and the power spectrum. To accomplish the goal, we provide a concise, comprehensive, but in depth review of conceptual and technical details of the standard model of the universe, namely, causal structure and inflation, quantization of primordial perturbations and field theoretic techniques such as "in-in" formalism needed for the estimation of loop correction to the power spectrum. We discuss at length the severe constraints (no-go) on PBH production in single-field inflation imposed by appropriately renormalized quantum loop corrections, computed while maintaining the validity of the perturbation framework and assuming sufficient inflation to address the causality problem. Thereafter, we discuss in detail the efforts to circumvent the no-go result in Galileon inflation, multiple sharp transition (MST)-induced inflation, and stochastic single field inflation using an effective field theoretic (EFT) framework applicable to a variety of models. We provide a thorough analysis of the Dynamical Renormalization Group (DRG) resummation approach, adiabatic and late-time renormalization schemes, and their use in producing solar and sub-solar mass PBHs. Additionally, we give a summary of how scalar-induced gravitational waves (SIGWs) are produced in MST setups and Galileon inflation.
Finally, the PBH overproduction issue is thoroughly discussed. 



\end{abstract}

\maketitle
\begin{center}
   \textcolor{Sepia}{\large\it Dedicated to the memory of Alexei A. Starobinsky.}
\end{center}

\tableofcontents
\newpage


\section{Introduction}

Cosmological inflation \cite{Borde:2001nh,Sato:1980yn,Guth:1982pn,Guth:1982ec,Guth:1980zm,Starobinsky:1980te,Starobinsky:1982ee,Linde:1983gd,Linde:1981mu,La:1989za,Steinhardt:1990zx,La:1989pn,Sato:2015dga} is a promising theoretical framework that led to a paradigm shift in the history of theoretical physics, namely, after its invention in 1981–82, cosmology not only got integrated with high-energy physics, but it also became a laboratory for testing and constraining new models. Inflation beautifully resolves the inconsistencies of the hot Big Bang, in particular, the causality problem, the model is plagued with. Perhaps what is most remarkable about inflation is the discovery that structures in our universe can be a reflection of the tiny fluctuations created quantum mechanically during inflation.
Furthermore, it was found that large fluctuations generated by inflation could trigger gravitational collapse during the radiation era, leading to the formation of astrophysical objects that behave like black holes and are named Primordial Black Holes (PBHs)\cite{Hawking:1974rv,Carr:1974nx,Carr:1975qj,Chapline:1975ojl,Carr:1993aq,Ivanov:1994pa,Yokoyama:1995ex,Kawasaki:1997ju,Yokoyama:1998pt,Kawasaki:1998vx,Rubin:2001yw,Khlopov:2002yi,Khlopov:2004sc,Saito:2008em,Khlopov:2008qy,Carr:2009jm,Choudhury:2011jt,Lyth:2011kj,Drees:2011yz,Drees:2011hb,Ezquiaga:2017fvi,Kannike:2017bxn,Hertzberg:2017dkh,Pi:2017gih,Gao:2018pvq,Dalianis:2018frf,Cicoli:2018asa,Ozsoy:2018flq,Byrnes:2018txb,Ballesteros:2018wlw,Belotsky:2018wph,Martin:2019nuw,Ezquiaga:2019ftu,Motohashi:2019rhu,Fu:2019ttf,Ashoorioon:2019xqc,Auclair:2020csm,Vennin:2020kng,Nanopoulos:2020nnh,Inomata:2021uqj,Stamou:2021qdk,Ng:2021hll,Wang:2021kbh,Kawai:2021edk,Solbi:2021rse,Ballesteros:2021fsp,Rigopoulos:2021nhv,Animali:2022otk,Correa:2022ngq,Frolovsky:2022ewg,Escriva:2022duf,Kristiano:2022maq,Riotto:2023hoz,Kristiano:2023scm,Riotto:2023gpm,Karam:2022nym,Ozsoy:2023ryl,Choudhury:2023vuj,Choudhury:2023jlt,Choudhury:2023rks,Choudhury:2023hvf,Choudhury:2023kdb,Choudhury:2023hfm,Bhattacharya:2023ysp,Choudhury:2023fwk,Choudhury:2023fjs,Choudhury:2024one,Choudhury:2024ybk,Choudhury:2024jlz,Choudhury:2024dei} to differentiate them from stellar black holes. 
One of the distinguished issues of cosmology and high energy physics is associated with the missing understanding about dark matter whose presence is necessary for structure formation; PBH might unveil the secret about the identity of the major component of matter in the universe
\cite{Ivanov:1994pa,Afshordi:2003zb,Frampton:2010sw,Carr:2016drx,Kawasaki:2016pql,Inomata:2017okj,Espinosa:2017sgp,Ballesteros:2017fsr,Sasaki:2018dmp,Ballesteros:2019hus,Dalianis:2019asr,Cheong:2019vzl,Green:2020jor,Carr:2020xqk,Ballesteros:2020qam,Carr:2020gox,Ozsoy:2020kat,Dolgov:2008wu,Dolgov:2000ht,Sami:2021ufn}. 
Another outstanding problem  related to baryon asymmetry could also be addressed through asymmetric Hawking evaporation of PBHs.

In addition to being a possible answer to the origin of dark matter and  baryon asymmetry, PBHs have attracted a lot of attention since the Laser Interferometer Gravitational-Wave Observatory (LIGO) \cite{LIGOScientific:2016aoc} recently observed gravitational waves (GWs)\cite{Baumann:2007zm,Saito:2008jc,Saito:2009jt,Choudhury:2013woa,Sasaki:2016jop,Raidal:2017mfl,Ali-Haimoud:2017rtz,Di:2017ndc,Raidal:2018bbj,Cheng:2018yyr,Vaskonen:2019jpv,Drees:2019xpp,Hall:2020daa,Ballesteros:2020qam,Ragavendra:2020sop,Carr:2020gox,Ozsoy:2020kat,Ashoorioon:2020hln,Ragavendra:2020vud,Papanikolaou:2020qtd,Ragavendra:2021qdu,Wu:2021zta,Kimura:2021sqz,Solbi:2021wbo,Teimoori:2021pte,Cicoli:2022sih,Ashoorioon:2022raz,Papanikolaou:2022chm,Wang:2022nml} from merging black holes. 
Needless to mention that PBHs related investigations have attracted enormous attention of researchers world wide in the recent years\cite{Ozsoy:2023ryl}. 


Formation of PBHs  in the early universe is mostly ascribed to the evolution of the inflaton field on its flat potential that might enhance fluctuation at a particular scale. 
In an inflationary setting, two primary avenues are investigated. Scenarios where the scalar field while evolving along its  potential  encounters one or more tiny, transient bumps/dips \cite{Mishra:2019pzq,ZhengRuiFeng:2021zoz}. This results in large field fluctuations that leave an impression on the corresponding scales. Interestingly, the said enhancement may be achieved in models in which the potential experiences a transition at a point of inflection from slow roll (SR) to the ultra-slow roll (USR) regime. In this review, our primary focus is on the generation of PBHs in a single-field inflationary framework. Although phenomenologically possible, adding a bump to an otherwise flat potential is a liability on theoretical grounds. However, a transition from SR to USR is an attractive avenue that could cause volatility to rise, eventually resulting in PBH production. This approach allows for a model-independent study of quantum phenomena without requiring knowledge of the inflaton potential.

Recent investigations reveal that generation of PBHs in scenarios based upon single-field inflation is faced with difficult issues of field theoretic nature, \cite{Kristiano:2022maq,Kristiano:2023scm,Kristiano:2024ngc,Kristiano:2024vst}. The one-loop corrections to the tree-level power spectrum is argued to be large on a wide scale, which forms the basis of the related conclusions. The compelling case is primarily supported by the observation that, in addition to logarithmic divergent effects, there are quadratic divergent contributions at the one-loop level. 
The slow roll  regime produced in the sub-horizon regime was originally claimed to  support only logarithmic divergences at the level of one-loop corrections\cite{Sloth:2006az,Seery:2007we,Seery:2007wf,Bartolo:2007ti,Senatore:2009cf,Seery:2010kh,Bartolo:2010bu,Senatore:2012ya,Senatore:2012nq,Pimentel:2012tw,Chen:2016nrs,Markkanen:2017rvi,Higuchi:2017sgj,Syu:2019uwx,Rendell:2019jnn,Cohen:2020php,Green:2022ovz,Premkumar:2022bkm}. This behavior is maintained up to the super-horizon scales.
In Ref.\cite{Chen:2016nrs}, it was demonstrated through comprehensive computation that while the quadratic and other power law divergent effects appear in the sub-Hubble regime of the one-loop calculation, they do not  survive on the super-Hubble scales in the  limit, $\tau\to 0$, applied to dimensionally regularized and appropriately renormalized version of the  one-loop corrected two-point correlation function. It has been  demonstrated clearly that, if the late time restriction is applied appropriately, only logarithmic effects remain in the final form of the two-point correlator, with no quadratic or other power law divergence effects present. A similar reasoning holds true for the model-independent framework, where the generation of PBHs occurs during the SR-USR transition followed by the end of inflation.
Recently, there has been a very interesting and intensive debate on the role/no-role of quadratic divergence on PBH formation in the single field framework of inflation and we refer the reader to 
this development\cite{Kristiano:2022maq,Kristiano:2023scm,Kristiano:2024ngc,Kristiano:2024vst,Riotto:2023hoz,Riotto:2023gpm,Riotto:2023hoz,Riotto:2023gpm,Firouzjahi:2023aum,Firouzjahi:2023ahg,Firouzjahi:2023bkt,Choudhury:2023vuj,Choudhury:2023jlt,Choudhury:2023rks,Choudhury:2024ybk,Choudhury:2024dei,Choudhury:2024jlz,Choudhury:2023fjs,Bhattacharya:2023ysp,Motohashi:2023syh,Franciolini:2023lgy,Cheng:2023ikq,Tasinato:2023ukp,Tasinato:2023ioq, Iacconi:2023ggt,Davies:2023hhn,Kristiano:2022maq,Kristiano:2023scm,Kristiano:2024ngc,Kristiano:2024vst}.

At present, enormous efforts are being made to understand the cosmological implications of large
quantum fluctuations within the framework of cosmological perturbation theory, Effective Field Theory and
Quantum Field Theory of de Sitter space. 
In  Refs.\cite{Choudhury:2023vuj,Choudhury:2023jlt,Choudhury:2023rks},  
it was clarified 
as how to deal with large
quantum fluctuations within the framework of Primordial Cosmology by maintaining the validity of perturbation theory. To this effect, one needs to deal with
Ultra-Violet (UV) and Infra-Red (IR) divergences. In most cosmological
scenarios, it is expected to have logarithmic type of IR divergences, however, in the presence of large quantum
fluctuation, the fate of the quadratic UV divergences  
 is extensively debated in the literature 
\cite{Kristiano:2022maq,Kristiano:2023scm,Kristiano:2024ngc,Kristiano:2024vst,Riotto:2023hoz,Riotto:2023gpm}. In this case,
schemes of regularization, renormalization, and resummation of the
obtained quantum correlations within the framework of primordial cosmological
perturbation theory become important. These techniques are designed in a gauge invariant fashion,
and can completely remove the quadratic UV divergent contributions, smoothing out and shifting the IR divergent effects
in the cosmological correlators  next to leading order of the perturbation theory in a trustworthy fashion. Importantly, the DRG resummed result  mimics the findings of $\delta {\cal N}$ formalism \cite{Sugiyama:2012tj}. Such identifications help us to connect the obtained result from the DRG resummation technique to the non-perturbative  result obtained from the well-known $\delta {\cal N}$ formalism derived using the separate Universe approach\cite{Burgess:2015ajz,Burgess:2009bs,Dias:2012qy}. The said issues are under active consideration at present. In this review, we exhibit the mentioned connections explicitly in great detail in the framework of  Effective Field Theory (EFT) of single inflation, { higher derivative ghost-free
Galileon single-field inflation} and stochastic theory of single-field inflation developed in the background of an open EFT setup. Needless to say that, in particular, a comprehensive review of the mentioned techniques is asked for by the researchers in this field. It should be noted that the issue related to the implications of sharp/smooth transitions between SR and USR
\cite{Riotto:2023hoz,Riotto:2023gpm,Firouzjahi:2023ahg,Firouzjahi:2023bkt,Firouzjahi:2023aum} also needs clarification~\footnote{It is important to note that, smooth transition discussed in the
literature are a very specific transition that satisfies Wands
duality condition \cite{Wands:1998yp}, which we discuss in great detail in ref \cite{Kristiano:2024vst}. }. To better comprehend the assertions stated in Ref.\cite{Kristiano:2022maq,Kristiano:2023scm,Kristiano:2024ngc,Kristiano:2024vst,Riotto:2023hoz,Riotto:2023gpm}, we re-examined in great depth the one-loop corrections, renormalization, and Dynamical Renormlization Group (DRG) resummation approach \cite{Kristiano:2022maq,Kristiano:2023scm,Kristiano:2024ngc,Kristiano:2024vst,Riotto:2023hoz,Riotto:2023gpm} with the help of the available tools and techniques of effective field theory (EFT) of single-field inflation \cite{Weinberg:2008hq,Cheung:2007st,Choudhury:2017glj,Delacretaz:2016nhw} in this review. Also it is important to note that
PBH formation in single field inflation is heavily constrained by one loop corrections to primordial power spectrum
\cite{Choudhury:2023vuj,Choudhury:2023jlt,Choudhury:2023rks,Choudhury:2024ybk}. This would be demonstrated by correctly implementing regularization, renormalization, and resummation techniques. We shall also present a novel {\it No-go theorem} on the PBH mass to support this reasoning, which restricts the possibility of creating big mass (solar or sub-solar mass) PBHs within the current framework to only very tiny PBHs with mass, ${\cal O}(10^{2}{\rm gm})$. Furthermore, we will present the most effective workarounds for the proposed {\it No-go theorem} on the PBH mass \cite{Choudhury:2023hvf,Choudhury:2023kdb,Choudhury:2023hfm,Choudhury:2023fwk,Choudhury:2024one,Bhattacharya:2023ysp,Choudhury:2023fjs,Choudhury:2024ybk,Choudhury:2024jlz}. 

It should be emphasized that the presence of an ultra slow-roll-phase has  direct impact on the production of scalar-induced gravitational waves (SIGWs) in second-order cosmological perturbation theory which has immediate observational consequences\cite{Kohri:2018awv,Baumann:2007zm}.{  It is expected that the theoretically consistent enhancement of amplitude of the primordial power spectrum due to large fluctuaion for the scalar modes will be automatically reflected in the SIGW spectrum . }
In this case, a larger amplitude of the GW will be produced from the previously mentioned GW framework. From various theoretical setups of PBH formation, which can  generate the PBHs mass, ${\cal O}(10^5M_{\odot}-10^{-36}M_{\odot})$ ( $M_{\odot}\sim 2\times 10^{30}{\rm kg}$, ), would cover a large frequency range for GW, lying within ${\cal O}(10^{-9}{\rm Hz}-10^2{\rm Hz})$. As an immediate consequence, such SIGWs can produce amplitude consistent with  NANOGrav 15-year data, and the sensitivities
outlined by other terrestrial and space-based experiments (e.g.: LISA, HLVK, BBO, HLV(O3), etc.). 
At present, a hardcore analysis is being pursued  to explore the generation of SIGWs in different frameworks of inflation for PBH formation. It is interesting to extend the mentioned analysis to first-order phase transitions, cosmic strings  and domain walls where the statistical evidence of producing large SIGW amplitude becomes significant\cite{NANOGrav:2023hvm}. It should  be pointed out that large negative primordial non-Gaussianities can be treated as the saviour of the long-standing PBH overproduction issue in the light of the recently observed enhancement in the NANGRav 15 data \cite{NANOGrav:2023hvm} in SIGW spectrum. Presently, the PBH overproduction issue \cite{DeLuca:2023tun} is receiving a lot of attention in the literature. However, the NANOGrav15 collaboration \cite{NANOGrav:2023hvm} { pointed out that to get the proper enhanced peak amplitude of the SIGW spectrum , the theoretical framework, in addition to inflation, might include the mentioned sources also}. Additionally, it is also pointed out that in the non-attractor models of single field inflation, it is almost  impossible to generate the required amount of large negative non-Gaussinaity to address the issue of PBH overproduction. To date, this issue is addressed with curvaton models \cite{Franciolini:2023pbf} and spectator scenarios \cite{Gorji:2023sil}. It is important to examine mechanisms and theoretical constructions within the framework of non-attractor single field models of inflation 
to consistently generate a large peak amplitude of SIGW-compatible with NANOGrav15 data \cite{NANOGrav:2023hvm}.
Apart from introducing large local non-Gaussianity in the presence of a sufficient controllable amount of non-linearities in the compaction function technique-based peak statistics \cite{DeLuca:2023tun,Franciolini:2023pbf}, one can address the PBH overproduction issues considering the general equation of state parameter ($w$), which is one of the key ingredients in cosmology. Earlier, the computation was performed mostly for the radiation-dominated era.
However, this framework could not resolve the previously mentioned PBH overproduction issue which we believe could be addressed by including non-Gaussinities and non-linearities in this computation. 
Analysis is  extended to scenario with general equation of state parameter such as $w=0$ (Matter), $w=1$ (Kinetion), and $w=2/3$ and many more cases, which we believe can be treated as one of the possible saviours from the PBH overproduction issue and can be compatible with the observed enhancement of the amplitude of GW spectrum in NANOGrav15. See refs. \cite{Domenech:2019quo,Domenech:2021ztg,Balaji:2023ehk} for more details. 
Further extending the discussion regarding the production of GWs, it is important to point out that in this specific case, in the first order of cosmological perturbation, the produced amplitude of the GW is excessively small. On the contrary, in the second order, since the tensor modes are induced by the scalar modes, there is a power enhancement in the scalar sector that automatically propagates to the SIGW spectrum. It might be important to investigate the production of the GWs in the third order to check the
validity of the perturbative approximation.
We hope to present a coherent description of the
aforesaid themes in this review.

 The review is organized as follows: In Section \ref{s1}, we elaborately discuss the underlying conceptual and technical details of the standard model of the universe to be used in the review. Fresh and in-depth discussions on causal structure, the need for an inflationary paradigm and quantization of perturbations have been included keeping in mind the often encountered misconceptions. A comprehensive and concise discussion of "in-in" formalism used for the estimation of quantum correlations and power spectrum is also presented in this section.
 We hope that the aforesaid would benefit the young researchers interested in the related theme. We hope that the young reader, equipped with presented conceptual and technical details  of the theoretical framework of cosmology and related quantum field theoretic framework,
 would be able to easily work through the review.
Further, in section \ref{s2}, we discuss the possible methods of generating large fluctuations within the primordial cosmological setup, which are necessarily required to understand the process of generation of PBHs. In section \ref{s3}, we provide the details of the PBH formation mechanisms. In section \ref{s4}, we discuss the underlying theory and the computational details of Scalar Induced Gravity Waves (SIGWs). Section \ref{s5} is devoted to Effective Field Theory (EFT) methods applied to large fluctuations and PBH formation. In section \ref{s6}, we elaborately discuss Galileon inflation in the light of large fluctuations and its implications for generation of PBHs  and  SIGWs.  In section \ref{s7}, we provide details of Stochastic Single Field Inflation in the light of large fluctuations and its implications for PBH production. We conclude the review in section \ref{s8}. Finally, in Appendix \ref{Ap},\ref{A1a},\ref{A2a},\ref{A3a},\ref{A4a},\ref{A5a},\ref{A6a},\ref{A7a},\ref{A8a} and \ref{A9a}, we provide various computational details which will be helpful to understand the content of review presented in different sections. 

{\bf Notation:} In this review, we have used the metric signature $(-,+,+,+)$, which is frequently used in the literature. Also, we have used reduced Planck mass $M_{pl}=1/\sqrt{8\pi G}=1$ (where  $G$ is the gravitational coupling constant in $1+3$ dimensional space-time) in our discussions. All the vectors are denoted by bold symbol ${\bf x}$ instead of using arrow $\vec{x}$ in our discussions. Also, the magnitude of the vector is denoted by the notation $x=|{\bf x}|$. Conformal time coordinate is represented by the symbol $\tau$ and the second-slow roll parameter is denoted by the symbol $\eta$. In our discussions, a dot represents the ordinary time derivative with respect to cosmic time  e.g. $\dot{a}=da/dt$. Similarly, a prime represents time derivative with respect to the conformal time coordinate e.g. $a^{'}=da/d\tau$. We use the natural unite system, $\hbar=c=1$.

\section{Glimpses of the standard model of Universe}
\label{s1}
What is known as the standard model of the universe is the hot big-bang model sandwiched between two phases of accelerated expansion: inflation at early epochs and late-time acceleration around the present epoch. The hot big bang has several important successes to its credit, including the predictions of expanding universe, micro-wave background, and the primordial synthesis of light elements. On top of that, it has an inbuilt mechanism of gravitational instability that would amplify the assumed primordial density perturbation into the structure we see today in the universe. COBE in 1992 confirmed the existence of these primordial perturbations, though their origin is beyond the big bang model. Despite the listed successes, the model is plagued with serious inconsistencies related to both the early and late epochs. It may be noted that most of the contribution to the age of the universe comes from matter dominated regime; the universe was just about $10^{5}$ years old at radiation-matter equality. The age computed in the model falls short of the age of known objects in the universe \cite{Zeldovich:1983cr}. The only way to increase it to the observed values is through the introduction of late-time acceleration, caused by cosmological constants, quintessences, or large-scale modifications of gravity, which slows down the Hubble expansion rate and improves the age of the universe \cite{Copeland:2006wr}. As for early evolution, there are serious issues related to causality, dubbed the {\it horizon problem}, the flatness problem, the entropy problem, and others. In our opinion, the problem of causality is a serious issue that extends even beyond science; this is a law of nature; if it is solved, other issues will get resolved. Inflation \cite{Borde:2001nh,Guth:1982pn,Guth:1982ec,Guth:1980zm,Starobinsky:1980te,Starobinsky:1982ee,Linde:1983gd,Linde:1981mu,La:1989za,Steinhardt:1990zx,La:1989pn} successfully addresses the horizon problem, and the flat universe, in particular, might become one of its predictions \cite{Lyth:1998xn,Riotto:2002yw}. As a bonus, it turns out that primordial density perturbations that seed the structure in the universe could originate quantum-mechanically from inflation. It is remarkable that inflation addresses a long-standing problem about our origin: we originated from quantum fluctuations!
At what scale inflation happened is an open question: it could have taken place anywhere between the Planck scale and one $TeV$; we hardly know any physics beyond the standard model of particle physics. However, it is not advisable to have inflation at low energy scales. First, there should definitely be a lot of new physics beyond the standard model of particle physics at high energy scales—the origin of baryon asymmetry and dark matter, in particular \cite{Rubakov:2017xzr,Gorbunov:2011zzc}. Maybe the {\it primordial black holes} that formed due to large fluctuations generated during inflation are hiding these secrets \cite{Baumann:2007np}. On the other hand,it is difficult to realise inflation at low energy scales. The equation of state of the scalar field used to realise inflation needs to be heavily fine-tuned in this case \cite{Steinhardt:2004rf}. It is therefore reasonable to introduce this phase around the GUT scale. 

In what follows, we present a brief and  concise discussion on the conceptual and technical issues of the standard model of universe and field theoretic framework necessary for the estimation of quantum correlations in cosmology. This discussion is based upon lectures delivered by one of us at various fora keeping in mind the misconceptions that often encountered by the young researchers. 
\subsection{Friedmann–Lemaître–Robertson–Walker (FLRW) space time}
Observations reveal that the universe is spatially flat to a good accuracy and is approximately homogeneous and isotropic at large scales. The small deviations from the smooth background dubbed the FLRW universe present in the early universe are believed to have grown into the structure present in the universe. Owing to homogeneity, isotropy, and flatness, the metric assumes a simple form \cite{Sami:2007zz},
\begin{equation}
\label{metric}
 ds^2=-dt^2+a^2(t)(dx^2+dy^2+dz^2 
 )  
\end{equation}
where all the information of dynamics is contained in the scale factor $a(t)$;
coordinates ($x,y,z$) are known as co-moving coordinates, a freely moving particles comes to rest in these coordinates. Einstein equations,
\begin{equation}
 G^\mu_\nu\equiv R^\mu_\nu-\frac{1}{2}\delta _\nu ^\mu R=8\pi G T^\mu_\nu   
\end{equation}
where symbols have usual meaning,
are simplified thanks to presence of symmetry under consideration and give rise evolution equations that determine the scale factor provided matter content present in the universe is specified.
 The energy momentum tensor, in this case, takes the simple form,
 \begin{equation}
 T^\mu_\nu =diag(-\rho(t),p(t),p(t),p(t)) 
 \end{equation}
 reminiscent of ideal perfect fluid, where symbols have their usual meaning.
In the present background, the $G_0^0~ \&~G_{i}^j$  are non-vanishing and give rise to following evolution equations,
\begin{eqnarray}
\label{friedmann}
&& H^2\equiv \frac{\dot{a}^2}{a^2}    =\frac{8\pi G}{3}\rho\equiv \frac{1}{3 M^2_{pl}}\rho\\
&& \frac{\ddot{a}}{a}=-\frac{1}{6M^2_{pl}}(\rho+3 p)
\label{acceleration}
\end{eqnarray}
Conservation of energy momentum tensor due to Bianchi identity gives rise to continuity equation,
\begin{equation}
\label{cons}
\dot{\rho}+3 H(\rho+p)    =0
\end{equation}
The redundant set of equations (\ref{friedmann}),(\ref{acceleration}) $\&$ (\ref{cons}) supplemented by the equation of state, namely, $p=p(\rho)$
uniquely determine, $a(t), p(t) ~\& ~\rho(t)$. Assuming further that universe is filled with perfect barotropic fluid, $\omega=p/\rho$, gives rise to following solution of evolution equations,
\begin{eqnarray}
&&a(t)\propto t^n   \\
&&\rho\propto a^{-2/n};~~ n=\frac{2}{3(1+\omega)}
\end{eqnarray}
where $n<1$ corresponds to standard matter($n=1/2$-radiation, $n=2/3$-cold matter); $n>1$ corresponds to an exotic matter with large negative pressure,
\begin{eqnarray}
&& \ddot{a}>0 ~~inflation~~~~~~n>1\\
&& \ddot{a}<0~~deceleration~~~n<1
\end{eqnarray}
 
 For $n\to \infty$($\omega \to -1)$ and $\rho\to const$ giving rise to exponential expansion,
 \begin{equation}
a(t)=a_i e^{H(t-t_i)}   
 \end{equation}
 as expected.
Let us note that the evolution equations for the spatially flat universe under consideration remain unchanged if the scale factor is multiplied by a constant,
using which the scale factor can be chosen to have a convenient value at a given epoch, say $a(t)=a_0=1$ at the present epoch. Thus, the scale factor has no physical meaning in the present setting, but its ratio at two given epochs, say, $t~ \&~t_0$,
namely, $N\equiv a_0/a(t)=(1+z)$, number of e-folds or red- shifts has a well defined meaning.

\begin{figure*}[htb!]
    	\centering
    {
       \includegraphics[width=20cm,height=10.5cm]{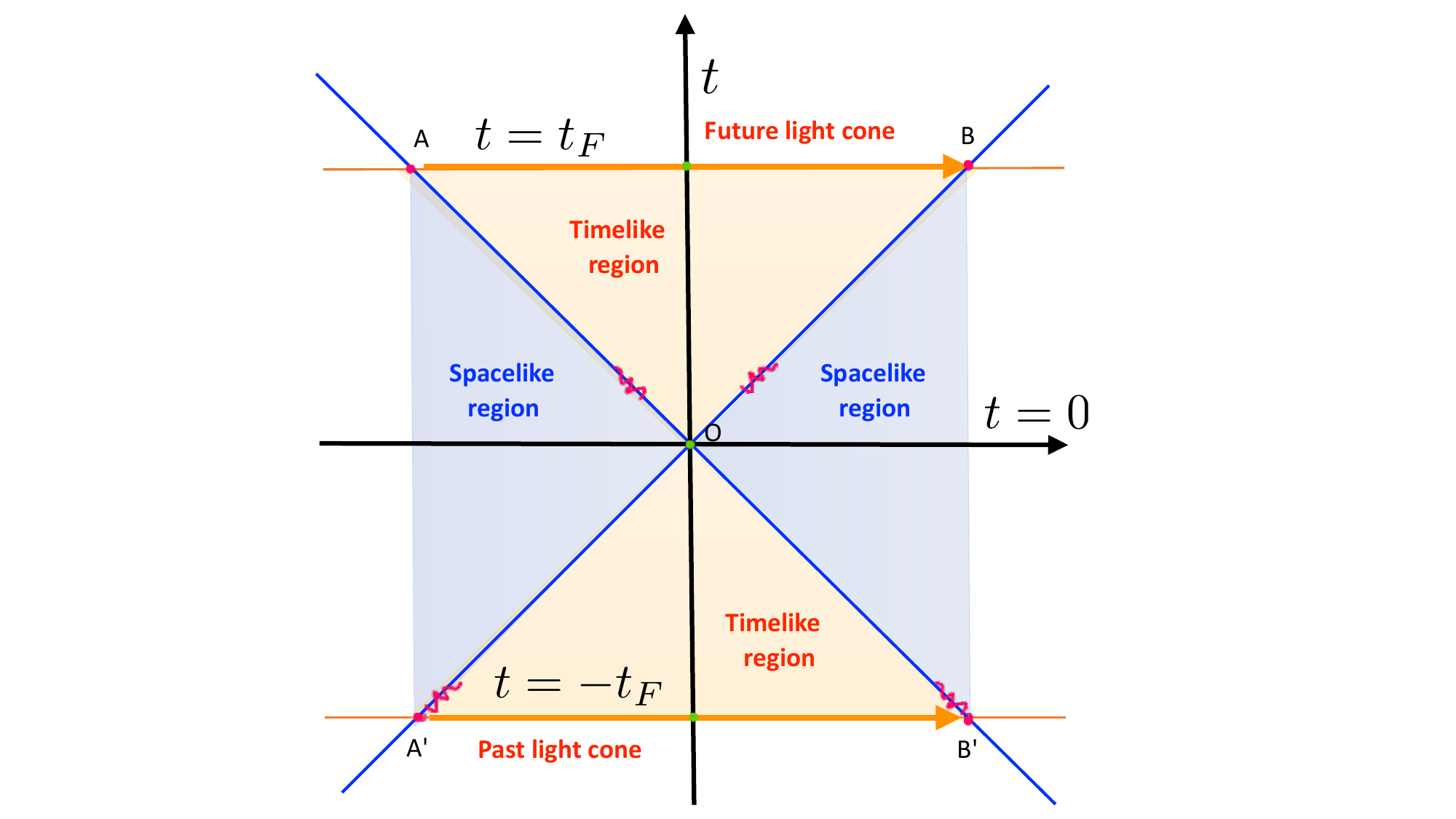}
    } 
    \caption[Optional caption for list of figures]{Diagram featuring causal structure of Minkowski space -time. The future light cone defines the maximum distance of  causal contact between points $A$ and $B$ on the hyper surface, $t=t_F$. The past light defines the maximum distance that could be visible to the observer at the hyper surface, $t=-t_F $  at "O" at  the present time. In Minkowski space time, $t_F$ could be arbitrarily large$-$ future (past) light cone is not bounded from below(above). }
\label{MI}
    \end{figure*}

 We may now discuss causal structure of FLRW space time which is dictated by its dynamics. 
It is, however, instructive to first revisit the causal structure of Minkovsky space-time,
\begin{equation}
ds^2=-dt^2+dx^2+dy^2+dz^2\equiv \eta_{\mu \nu}dx^\mu dx^\nu=-dt^2+dr^2+r^2 d\theta^2+r^2\sin^2\theta d\phi^2    
\end{equation}
which is static.
Considering the radial flow of light, one finds,
\begin{equation}
0=-dt^2+dr^2 \to r=\pm t   
\end{equation}
which is the distance travelled by the photon on the hyper surface, $t=const$. Let us now consider the {\it future light cone}, see Fig.\ref{MI}. Photons emitted from the origin "O" reach the hyper-surface $t=t_F$ in time "$t_F$". During this time a distance is travelled by photons on this hyper surface,
\begin{eqnarray}
 D_{AB}=2 t_F  
\end{eqnarray}
which is the maximum distance on the hyper surface, under consideration, where the emitter could influence events . Imagine an emitter situated at "O" and emitting photons in all directions. At time "$t_F$" the photons reach the surface of the sphere of diameter "2$t_F$" which is the surface of sphere of maximum diameter that could be influenced at time $t_F$ by the emitter at "O".
As for the {\it past light cone}, consider,  photons emitted from diametrically opposite points $A'$ and $B'$ on the hyper surface, $t=-t_F$ in the past. By the time, the photons reach the observer at "O" , a distance,
\begin{equation}
D_{A'B'}=2t_F    
\end{equation}
is travelled on the hyper surface under consideration. This is surface of the sphere of maximum diameter  which is visible to the observer at "O" at the present time.
Minkowski space time is static and the future and past light cones are not bounded either from above or below; $t_F$ could be as large as we wish. Thus, there is no issue of causality in Minkowski space time where we have glorious past and bright future!. The FLRW space time, on the other hand, has dynamics which could impose  restriction on the future and past light cones leading to serious causality issues \cite{Riotto:2002yw,Riotto:2018pcx,Baumann:2022mni,Baumann:2009ds,Baumann:2018muz,Senatore:2016aui,Trodden:2004st}. For instance, we can not extend a light cone  below the big-bang hyper surface.
\subsection{Causal structure of FLRW space time and Inflation}

In relativity, there is no preferred time and it is legitimate to use the convenient one. 
In order to investigate the causal structure of FLRW, we transform from cosmic time to conformal time,
\begin{equation}
t\equiv a(t) \tau \to \tau =\int {\frac{1}{a(t)}dt}    
\end{equation}
such that FRW metric assumes the following form,
 \begin{eqnarray} 
 && ds^2=a^2(\tau)(-d\tau^2+dx^2+dy^2+dz^2)=a^2(\tau)\eta_{\mu \nu}dx^\mu dx^\nu\\
&& ds^2=a^2(\tau)(-d\tau^2+dr^2+r^2 d\theta^2+r^2\sin^2\theta d\phi^2  ) 
\end{eqnarray}
which is conformally equivalent to Minkowski metric.
 The causal structure of FRW in conformal time is formally identical to that of Minkowski space-time. However, dynamics might impose bounds on past and future light cones; for instance, a past light cone is liable to terminate on the Big Bang singularity surface.
  Indeed, for light travelling radially,
  \begin{equation}
0=-d\tau^2+dr^2\to r(t)=\tau(t)=  \int_0^t {\frac{dt}{a(t)}} =  \int_0^{a(t)} {\frac{da}{ a^2 H}}  
\label{ch}
  \end{equation}
\begin{figure*}[htb!]
    	\centering
    {
       \includegraphics[width=17cm,height=9.5cm]{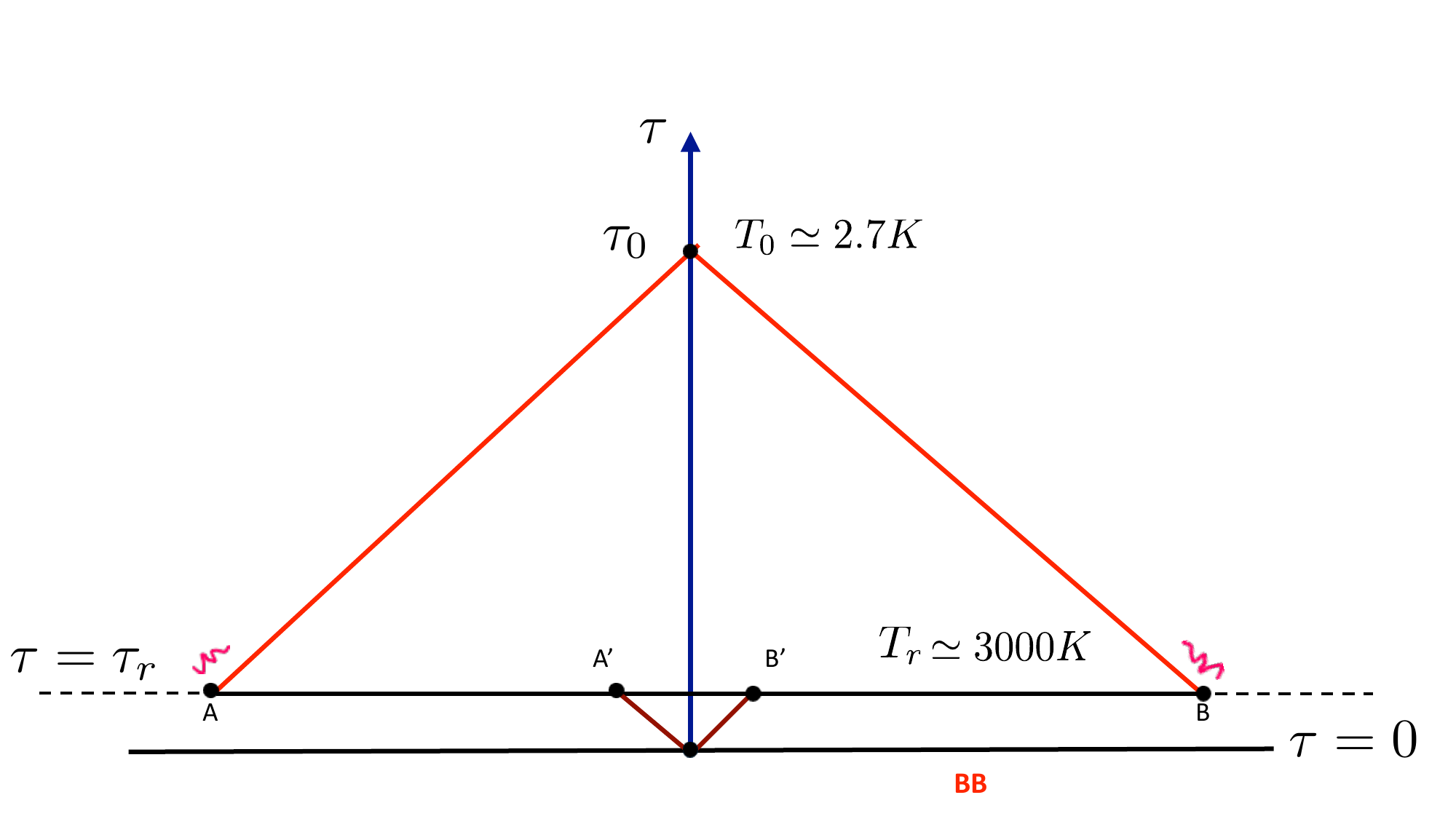}
    } 
    \caption[Optional caption for list of figures]{\small{Schematic diagram featuring the causal structure of the big bang, where "BB" and $\tau=\tau_r$ designate the big bang and last scattering hyper surfaces (with temperatures $T_0$ and $T_r$), respectively. The future light cone drawn from "BB" to the last scattering surface defines the distance of causal contact on this surface. The past light cone drawn from the last scattering surface defines the distance between points A and B from where the observer receives photons today : diameter of the sphere whose surface is visible to us today. This distance of causal contact, $D_{A'B'}$ on the surface $\tau=\tau_r$ is much smaller than the distance $D_{AB}$ visible to the observer at the present epoch corresponding to $\tau=\tau_0$
    .} }
\label{f1}
    \end{figure*}
 which gives the coordinate or co-moving distance travelled by photon along the hyper surface $t=const$ as the photon emitted from the big bang hyper surface lands there.
  The corresponding physical distance designated by $R_{H}(t)$ is given by,
 \begin{equation}
R_H(t)=a(t) \tau(t) 
\label{ph}
 \end{equation}
which defines the horizon distance at time "t".  
   The FLRW space time has dynamics embodied  in the scale factor  which  is reflected in the conformal time and which results into restrictions on the light cone. Indeed, the structure of integral in (\ref{ch}) or (\ref{ph}) crucially depends upon the nature of matter filling the universe which is reflected in the conformal time. Indeed, let us consider the coordinate distance travelled by the photon from $t_i$ to $t_{f}$,
   \begin{eqnarray}
   a(t)\sim t^n:~~\tau= \int_{a_i}^{a_{f}} {\frac{1}{ a^2 H }da}  \propto  a^{(1-n/n)} |_{a_i}^{a_{f} }\propto \left(\frac{1}{H a}  \right)^{a_{f}}_{a_i}
   \label{ns}
   \end{eqnarray}
In case of $n<1$, we have positive power of scale factor in the numerator, thereby, the important contribution to the integral in (\ref{ns}) comes from upper limit or the {\it integral  saturates at the upper limit},
\begin{equation}
\tau \propto  \frac{1}{a_{f}H_{f}}    
\end{equation}
where the proportionality constant would be reinstated when needed.
 In this case, it is reasonable to fix the zero of conformal time at the lower limit pushing it to zero\footnote{We keep in mind that $\tau\sim t^{1/2}(\tau\sim t^{1/3})$ during radiation(matter) dominated era. }, namely, $a(t_i=0)=0=\tau(t_i=0)$. 

Let us
note that the horizon distance in case of the standard matter($n<1$) is given by the Hubble radius,
\begin{equation}
 R_{H}(t)\equiv a(t)\tau(t)=a(t)\int_0^1 {\frac{da}{a^2 H } }\propto   H^{-1}(t)
\end{equation}
such that $R_{H_0}=\tau_0\sim H_0^{-1}$ defines the present horizon of universe:{ a distance travelled along the hyper surface} $t_0=const$ { by a hypothetical photon emitted from the big bang hyper surface\footnote{We imagine in this case a light cone drawn from the big bang singularity to $t=t_0$ hyper surface.}}.
 Situation, however, gets reversed for $n>1$; in this case, {\it integral } in (\ref{ch}) or in (\ref{ph}) {\it saturates on the lower limit},
\begin{eqnarray}
 n>1:~~\tau \propto -\frac{1}{a_i H_i};~~\tau_{end}\propto \frac{1}{a_{end}H_{end}} \simeq 0  ;~~~~
 \tau \to -\infty,~a_i \to 0
\label{etainf}
\end{eqnarray}
and zero of conformal time can be chosen at the upper limit by taking $a_{end}>>a_i$. We have in mind the quasi exponential expansion corresponding to large numerical value of $n$ where Hubble parameter remains nearly constant.

Let us now consider the conformal diagram in Fig.\ref{f1} that expresses the causal structure of FLRW spacetime. Photons emitted from diametrically opposite points $A~\&~ B$ at the last scattering surface, $\tau=\tau_r$, reach the observer today. By the time a photon either from $A$ or $B$ reaches the observer today, a coordinate distance is travelled along the last scattering surface, which can be estimated assuming matter domination,
\begin{equation}
\label{tau0}
 \int_{a_r}^ 1{\frac{da}{a^2 H }} \propto \frac{1}{a_0 H_0} \sim \tau_0 
\end{equation}

which is approximately same as the present day horizon of universe(see (\ref{hp1}) below), however, the physical distance travelled along the last scattering surface in this case is different, namely, $a_r \tau_0$\footnote{We are dealing here with the past light cone.}.
Drawing the light cone from big bang hyper surface to the surface of last scattering, we find the coordinate distance,
\begin{equation}
 \tau_r=\int_{0}^ {a_r}{\frac{da}{a^2 H }} \propto \frac{1}{a_r H_r}  
 \label{etar}
\end{equation}
where we again assumed matter domination which is justified around the upper limit $(a=a_r$) where the integral saturates. This is the coordinate distance travelled by the photon on the last scattering surface.
The ratio of physical distances $D_{AB}$ and $D_{A'B'}$ travelled along the last scattering surface is given by,
\begin{equation}
 \frac{D_{A'B'}} {D_{AB}}=\frac{2 a_r \tau_r}{2 a_r \tau_0}=\frac{a_0 H_0}{a_r H_r}= \left(\frac{T_0}{T_r}\right)^{1/2}=\left(\frac{1}{1+z}\right)^{1/2}\simeq \frac{1}{33}
 \label{hp}
\end{equation}
where we assumed matter domination around the present epoch as well as on the last scattering surface such that
, $H\sim a^{-2/3}$.
The estimate of (\ref{hp}) can further be corrected remembering that dark energy dominates around the present epoch where the  integral in  (\ref{tau0}) or (\ref{etar})  saturates. Secondly, in order to get the coefficient correct, one should also include contribution of radiation  in (\ref{etar}) as the last scattering surface is not far from radiation-matter equality. These corrections then yield the
 improved estimate\footnote{Indeed,
$\tau_0=\int_0^1{\frac{da}{a^2 H}}=
H_0^{-1}\int_0^\infty{\frac{dz}{\sqrt{\Omega_m^0(1+z)^3+\Omega_\Lambda}}}\simeq 3.5 H_0^{-1}$~~~~~~~~~($\Omega^0_m\simeq 0.3;~~\Omega_\Lambda\simeq 0.7$);\\
$\tau_r=\int_0^{a_r}{\frac{da}{a^2 H}}=
H_0^{-1}\int_{z_r}^\infty{\frac{dz}{\sqrt{\Omega_m^0(1+z)^3+\Omega^0_R(1+z)^4}}}\simeq 0.065 H_0^{-1}$
~~~~($\Omega^0_R\simeq 8\times 10^{-5}$),~~ which then yields the estimate,\\
$\frac{\tau_r}{\tau_0}\simeq \frac{1}{55}$

}
\begin{equation}
\label{hp1}
  \frac{D_{A'B'}} {D_{AB}}=\frac{\tau_r}{ \tau_0}  \simeq \frac{1}{55}
\end{equation}
Let us note that,
\begin{equation}
\frac{\lambda^{(0)}_r}{\lambda_{H_0}}=\frac{\tau_r}{\tau_0}\simeq \frac{1}{55}
\end{equation}
where $\lambda_{H_0}\equiv H_0^{-1}$ and $\lambda^{(0)}_r\equiv H^{-1}_r a_0/a_r$ represents the horizon length at last scattering stretched  to the present epoch. Equivalently, the present horizon length squeezed to the last scattering surface is larger by the same factor than $H^{-1}_r$,
\begin{equation}
\frac{\frac{a_r}{a_0}H^{-1}_0  }  {H^{-1}_r } \simeq 55
\end{equation}
In fact, any length scale today was out side the Hubble radius in the past which stems from the fact that the universe, in hot big bang framework, is dominated by the standard fluid $(a(t)\sim t^n;~n<1) $ such that the Hubble radius ($H^{-1} \sim a^{1/n}$ ) grows (diminishes) faster than any physical length which scales with the scale factor in FRW. The trend is reversed in case of an exotic fluid with $n>1 $ or $\omega<-1/3$ when deceleration ($\ddot{a}<0$)  changes to acceleration ($\ddot{a}>0$).

Relation (\ref{hp1}) poses a serious causality problem, namely, photons are received today from the last scattering surface from regions which were not in causal contact but the received photons display identical characteristic. Infect, the photons come to us from the  the surface of last scattering sphere of diameter $D_{AB}$. We can easily estimate the number of causally disconnected patches on surface of last scattering sphere as,
\begin{equation}
\frac{D^2_{AB}}   {D_{A'B'}^2} \simeq 3000
\end{equation}
which should have resulted in large patchiness in the sky, which is not observed. On the contrary, CMB is smooth to great accuracy; the small observed temperature anisotropy exhibits statistical isotropy, which cannot be understood as photons were emitted from patches that were not in causal contact. This is a serious inconsistency of hot the big bang model. The underlying reason for the problem is attributed to the fact that the last scattering surface is pretty close to the big bang hyper surface, and one cannot go below the big bang hyper surface to open up the light cone wide enough to increase the distance of causal contact on the last scattering surface. And the last scattering surface is fixed by well-known physical processes taking place in the expanding universe and cannot be pushed up. On the other hand, the big bang hypersurface cannot be pushed down within the framework under consideration, where evolution is dictated by radiation around the big bang hypersurface. It is interesting to note that in Minkovsky space time, the future light cone is not bound from below and is not forced to terminate on any hypersurface. In this case, one can draw a future light cone to a given hypersurface, say around $t=t_0$, as wide as one wishes by going down to a desired epoch in the past.

Inflation seems to provide natural way to push down the big bang hyper surface if introduced at early epochs (see Eq.(\ref{etainf})). Indeed, assuming exponential expansion in a small slice around the big bang hyper surface, with further assumption that it commenced in the post Planck epoch and ended on the top of the chosen time slice with $\tau=\tau_{end}\simeq 0$, we have(see Fig.\ref{f2}), 
\begin{eqnarray}
 \tau = -\frac{1}{a_i H_I};~~ 
\tau \to -\infty,~a_i \to 0 
\end{eqnarray}

 Thus, inflation opens up a see of negative values of conformal time $\tau$ below $\tau_{end}\simeq 0$ such that we have standard big-bang evolution for $\tau_{end}> 0$. Similar to Minkovsky space time, we can now draw a light cone starting from a large negative value of $\tau$, increasing the distance of causal contact on the last scattering surface to a desired value.
 The amount of inflation
and its duration depend upon the energy scale where inflation commenced. Indeed,
in radiation domination era around big bang,
\begin{equation}
H \simeq \frac{T^2}{M_{pl}}
\label{IS0}
\end{equation}
We should first fix the temperature $T=T_{end}$ where inflation should commence which fixes energy scale of inflation and time when inflation happened. After the end of inflation, universe is expected to reheat to $T_{end}$ dubbed reheating temperature. Assuming exponential expansion, one finds,
\begin{eqnarray}
 N=\int_{a_i}^{a_{end}}{d\ln a}=\int_{t_i}^{t_{end}}{H_Idt}=
 H_I \Delta t  
\end{eqnarray}
which then using (\ref{IS0}) allows to estimate the duration of inflation,
\begin{equation}
 \Delta t=\frac{{N}}
{T^2_{end}} M_{pl}   
\end{equation}
if the number of e-folds is known; the latter is fixed from the requirement of addressing the horizon problem. 
As pointed out in the preceding discussion, it is not advisable to have inflation at low energy scales, as it would close the window on the early universe physics expected to reveal the secret of beyond-standard model physics. For instance, 
the GUT scale physics could reveal the secret of baryon asymmetry and dark matter; the recent NANOGrav findings are believed to be linked to the melting of cosmic strings, domain walls, or phase transitions reminiscent of high energy scales.
On the other hand, the low-energy scale inflation is heavily fine-tuned if we adhere to the observed value of primordial density perturbation. In the discussion to follow, we shall bring out this argument. 
\begin{figure*}[htb!]
    	\centering
    {
       \includegraphics[width=17cm,height=14.5cm]{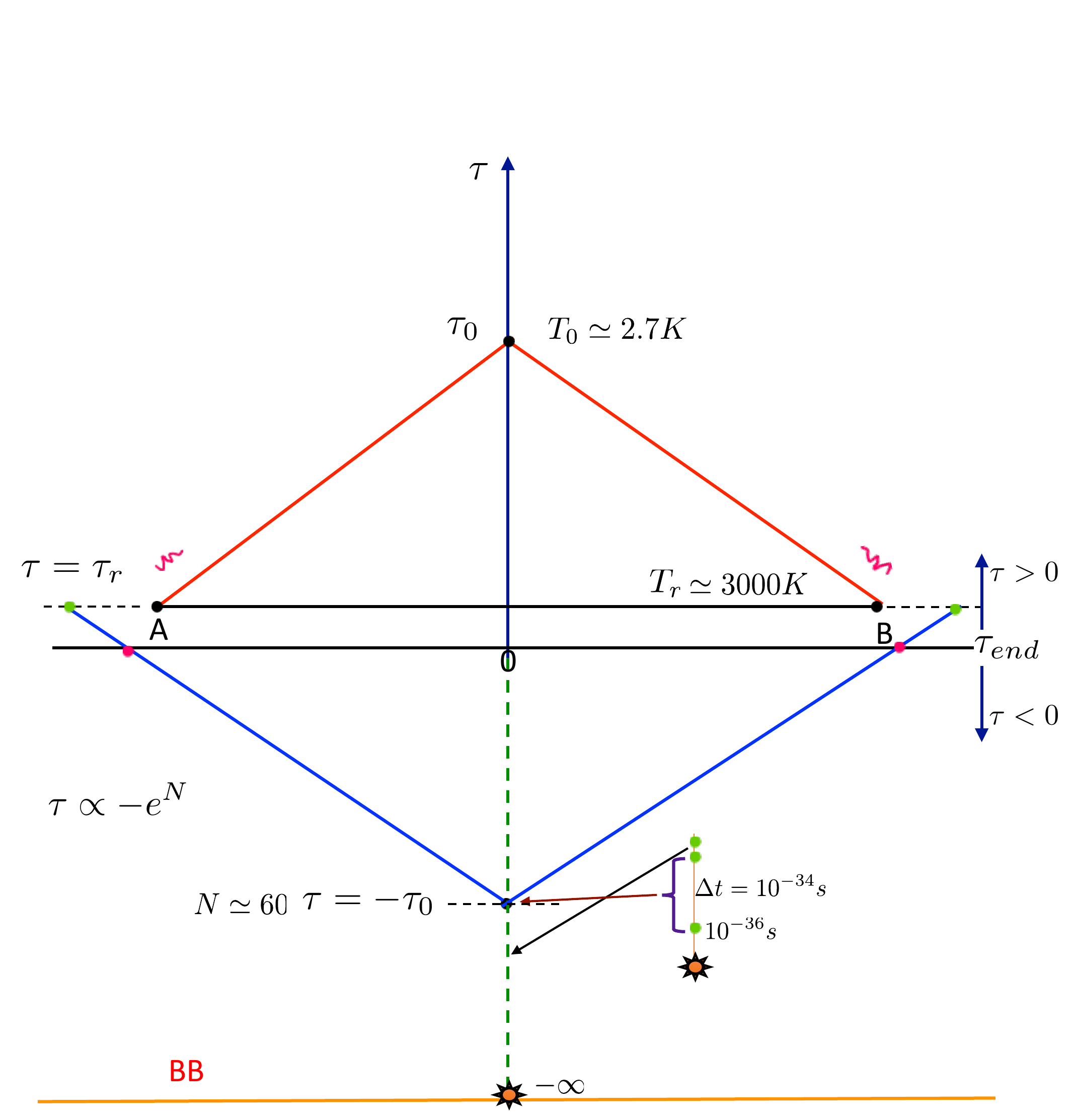}
    } 
    \caption[Optional caption for list of figures]{Schematic diagram which displays the causal structure of big bang after being complimented by inflation in a small time slice around the big bang hyper surface "BB"  such that one has standard big bang evolution for $\tau>\tau_{ends}$. Inflation commences at $10^{-36} s$ post big bang and continues for an interval of cosmic time, $\Delta t\simeq 10^{-34} s$; it ends at the top of the time slice $\tau=\tau_{end}\simeq 0$. In this process, one obtains, $\tau=-\tau_0$ corresponding to $N\simeq 60$ and the part of last scattering surface visible to the observer today comes with the causal contact. Increasing the duration of inflation, slightly without significantly affecting the corresponding value of time at the top of the slice($\tau_{end}\simeq 0$) , exponentially increases the negative value of conformal time, pushing down the big bang singularity towards minus infinity. As one moves down $\tau_{end}=0$, the distance of causal contact on the $\tau=\tau_r$ hypersurface increases, bringing its visible part (today) within the causal contact. }
\label{f4}
    \end{figure*}

In view of the aforesaid, one should be comfortable introducing the phase of inflation around the GUT scale, say, $T_{end}\simeq 10^{15} GeV$, corresponding to $10^{-36}~s$  after the big bang. 
 To this effect, let us consider a small strip of time $\Delta t\ll 1$ 
 around the big bang hyper surface such that $\tau\simeq 0$ on the top of the strip, and introduce inflation or exponential expansion in the said interval such that it commences around the GUT scale and ends on the top of the strip at $\tau\simeq 0$. In this process, we earn a certain number of e-folds, giving rise to an exponentially large negative value of $\tau$.
\begin{equation}
 \tau \propto -e^N  ; ~~N=H_I \Delta t~~(\Delta t= t_{end}-t_i )
\end{equation}

\begin{figure*}[htb!]
    	\centering
    {
       \includegraphics[width=17cm,height=9.5cm]{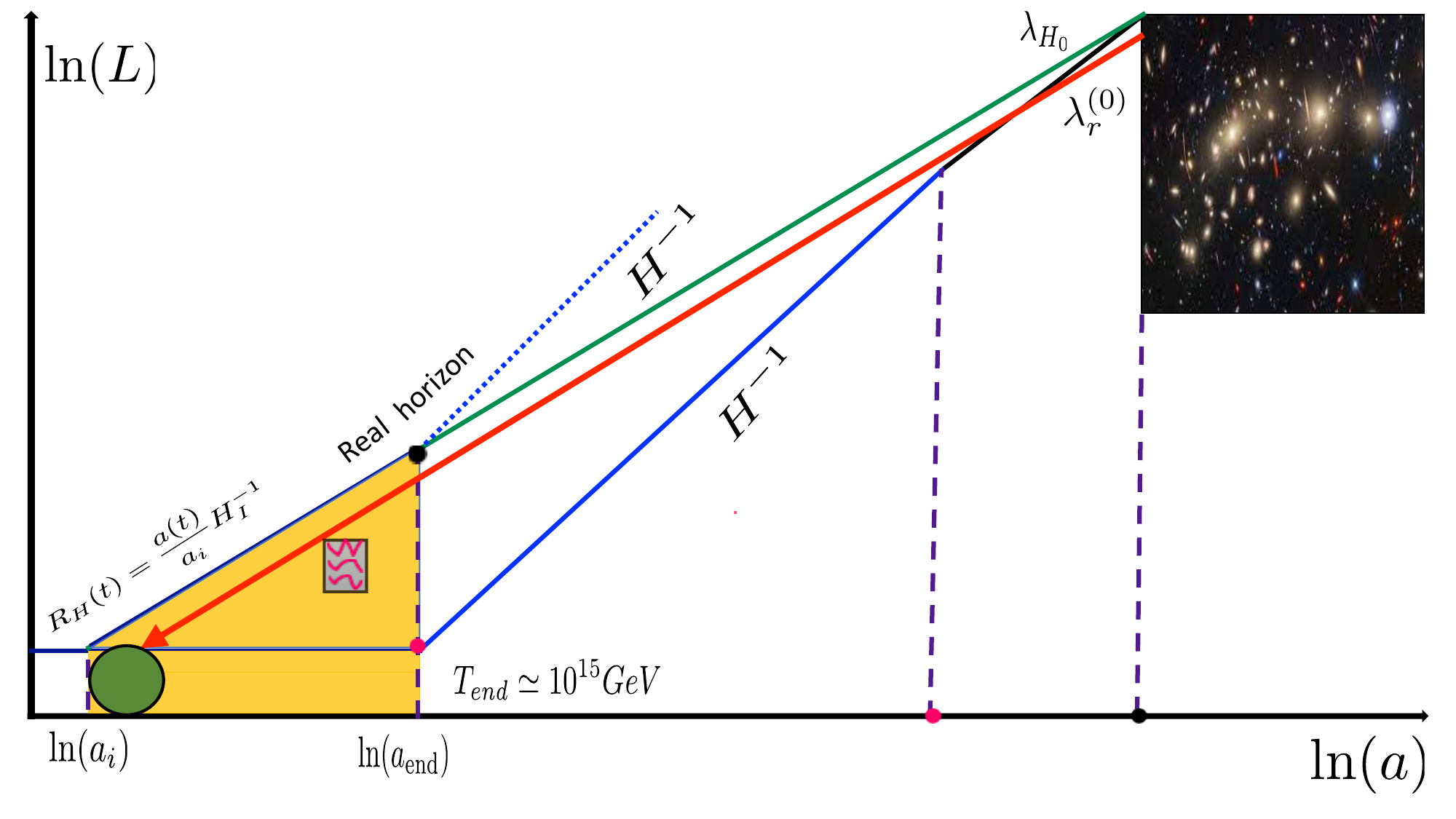}
    } 
    \caption[Optional caption for list of figures]{Schematic diagram which presents evolution of relevant length scales in the universe versus the scale factor on Log scale from the present epoch back to inflationary era  corresponding to about $60$ e-folds. The yellow shaded region  is  inside the inflationary horizon ($R_H=a(t) H_I^{-1}/a_i; a_i\leq a(t)\leq a_{end}$  ) depicted by the thick blue line; the inflationary horizon scales with the scale factor from the commencement of inflation to its end. $R_H$ is continued by $H^{-1}$(dashed blue line) from the end of inflation.
    The latter presents the real horizon compared naive horizon
    given by $H^{-1}$.
    The length scale $\lambda_{H_0}$, the present horizon of the universe which is larger than the Hubble radius when extrapolated back to the surface of last scattering, enters the inflationary horizon when squeezed back to the end of inflation; it crosses below $H_I^{-1}$ when further squeezed to the epoch, $a=a_i$ where inflation commenced. The length scale $\lambda^{(0)}_r$(Hubble radius at last scattering stretched to the present epoch) relevant to last scattering had entered the Hubble radius during inflation few e-folds after the commencement of inflation. Obviously, all the length scales today in the universe that had gone out of the Hubble radius $H^{-1}$, in the past, were inside the inflationary horizon $R_H(t=t_{end})$ at the end of inflation. Tiny perturbations born deep inside the Hubble radius during inflation are shown to have grown into the structure we see in the universe today.}
\label{f5}
    \end{figure*}

The conformal diagram dictates the condition for resolution of horizon problem(see, Fig.\ref{f4}),
\begin{equation}
\label{hconf}
 \tau\gtrsim \tau_0 \Longrightarrow   \frac{1}{a_i H_I}  \gtrsim \frac{1}{a_0 H_0}
\end{equation}
 Clearly, the minimum number of e-folds is the one that corresponds to $\tau=-\tau_0$. We then draw light cone from this point to the $\tau=\tau_{end}=0$ hyper surface where inflation ends and universe  reheats  to the temperature, $T=T_{end}$. Clearly, the part of the last scattering surface which is visible to us today is automatically in the causal contact and the horizon problem gets resolved.

Multiplying (\ref{hconf}) left right by $a_{end}$, we 
 estimate the required number of e-folds,
\begin{eqnarray}
 &&\frac{a_{end}}{a_i H_I}  \gtrsim \frac{a_{end}}{a_0 H_0} \Longrightarrow e^N\gtrsim \left(\frac{T_0}{H_0}\right)\left(\frac{H_I}{T_{end}} \right) =\left(\frac{T_0}{H_0}\right)\left(\frac{T_{end}}{M_{pl}} \right)\\
 && N\gtrsim \ln\left(\frac{T_0}{H_0}\right)+\ln \left(\frac{T_{end}}{M_{pl}} \right)=67+\ln \left(\frac{T_{end}}{M_{pl}} \right)
 \label{NIF}
\end{eqnarray}
which is attributed to a definite value of $a_i$ as $a_{end}$ is held fixed.
For the chosen value of $T_{end}$, we find the minimum value of the number of e-folds, $N\simeq  60$,  which then allows us to estimate the minimum duration of inflation, $\Delta t \simeq 0.8 \times 10^{-34} s $ such that $\tau=\tau_{end}\simeq 0$. Enhancing the duration of inflation by a factor of few,  correspondingly increases $N$,  giving rise to an exponential increase in the conformal time $\tau$, which pushes the big bang singularity down towards minus infinity, see Fig.\ref{f4}.

 In a nutshell, we cut out a small strip of time, $\Delta t\simeq 10^{-34} s$ around the big bang hyper surface, above which we have standard big bang evolution with reheating temperature $T_{end}$ on the top of the strip corresponding to $\tau\simeq 0$.
We believe that nothing important happened in this interval of time around the big bang hyper surface. We then introduce inflation in this strip, which opens up a sea of negative conformal time below $\tau=\tau_{end}\simeq 0$. A slight increase in the duration of inflation pushes the big bang singularity down to minus infinity. Though inflation does not address the singularity problem and it is not meant  for that  but it pushes the big bang singularity away from our consideration so that it no longer bothers us!

In order to further decode the meaning of (\ref{hconf}), let define inflationary horizon,
\begin{eqnarray}
&& a(t) =a_ie^{H_I(t-t_i)}\\
&&R_{H}(t)=a(t) \int_{a_i}^{a(t)}{\frac{da}{a^2 H_I}} =\frac{a(t)}{a_i H_I} ~~  ~~~(a_i<a(t)<a_{end})
\label{infh}
\end{eqnarray}
where, for simplicity, we assumed the Hubble parameter to be constant during inflation.
 It should be noted that $R_{H}(t)$ scales with the scale factor similar to a length scale which happens only in de-Sitter inflation. (\ref{hconf}) then reads for minimum number number of e-folds,
 \begin{equation}
 \label{hubblecr}
\frac{a_{end}}{a_0} H^{-1}_0= R^{end}_{H}=\frac{a_{end}}{a_i H_{I}} 
 \end{equation}
 which implies that the largest length scale in the universe (corresponding present horizon), when squeezed to the end of inflation, just enters the inflationary horizon($R_H$) and crosses the Hubble radius when further squeezed to the epoch where inflation commences at $a(t)=a_i$(see, Fig.\ref{f5}) which corresponds to approximately $60$ e-folds\footnote{$H^{-1}_0$ squeezed to the end of inflation matches there with the inflationary horizon and touches the Hubble radius $H_I^{-1} $ when further squeezed to $a_i$ as the inflationary horizon scales similar to the length scale. This is obviously consistent with (\ref{hubblecr}) which expresses the fact that the present horizon size squeezed to $a_i$ is $H_I^{-1}$.}($R_{H}(t=t_i)=H^{-1}_I$).
 It is important to note that any other length scale in the universe was inside the inflationary horizon at the end of inflation which crossed the Hubble radius at $a'_i>a_i$ \footnote{One is thinking in the backward direction in $a$.}such that the number of e-folds from $a'_i$ to $a_{end}$ is little less than the number of e-folds earned during inflation from $a_i$ to $a_{end}$. To this effect, let us consider the following ratio,
 \begin{equation}
 \frac{\tau(t)}{\tau_0}    =\frac{a_0 H_0}{a(t) H(t)}=\frac{\lambda^{(0)}(t)}{\lambda_{H_0}}
 \end{equation}
 where $\lambda^{(0}(t)$ represents the length scale $H^{-1}(t)$ stretched to today such that $\lambda^{(0)}(t)<\lambda_{H_0}$. This wavelength would be inside the inflationary horizon if squeezed to the end of inflation,
 \begin{equation}
  \left(\frac{a_{end}}{a_0 } \right)  \lambda^{(0)}(t) =\frac{a_{end}}{a'_i H_I}\Longrightarrow \frac{\lambda^{(0)}(t)}{\lambda_{H_0}}=\left(\frac{T_{end}}{T_0}\right)\left(\frac{H_0}{H_I}\right)e^{N'}=e^{(N'-N)}
 \end{equation}
 where we used (\ref{NIF});
   $N'$ is the number of e-folds earned during inflation from $a'_i$ to $a_{end}$. We then arrive at an interesting result,
  \begin{equation}
 N'=N+\ln \left( \frac{\lambda^{(0)}(t)}{\lambda_{H_0}} \right)   
  \end{equation}
 which allows us to estimate, $N'$ in case of length scale corresponding to last scattering ($H^{-1}_r$ ),
 \begin{eqnarray}
 \label{cmbl}
     N'=N+\ln \left( \frac{\lambda^{(0)}_r}{\lambda^{(0)}} \right)=N+\ln\left( \frac{\tau_r}{\tau_0}\right)\simeq 56;~~~
  \lambda^{(0)}_r\equiv \lambda^{(0)}(t_r);~~\lambda^{(0)}\equiv \lambda_{H_0}
 \end{eqnarray}
 where $\lambda^{(0)}_r$ is the length scale $H^{-1}_r$ stretched to today.
 Relation(\ref{cmbl}) tells that that the length scale relevant to CMB went out of Hubble radius during inflation few e-foldings after the commencement of inflation\footnote{If one thinks going in the forward direction in $a$.}. 
 
 It should be emphasized that there is nothing super-horizon if required amount of  inflation has taken place. In this case, we should clearly distinguish between the Hubble radius and the Horizon distance. During inflation, horizon  $R_H$ scales  with the scale factor  from the commencement to the end and thereafter it is continued by the Hubble radius, $H^{-1}$. All the length scales including $H_0^{-1}$
 are inside the horizon. All of them were outside the Hubble radius in the post-inflationary era but re-entered later, Fig.\ref{f5}. It is amazing to think how much information is imbibed in a simple relation (\ref{hconf})!

It is interesting to note that
by taking the number of e-folds larger than the minimum one, flat universe become a prediction of inflation: universe inflates to the level that its visible part we habitat is flat to great accuracy irrespective of the underlying geometry.
inflation seems to be a natural habitat for addressing the inconsistencies of the hot big bang model.
 However, one might be puzzled due to exponential sensitivity
of conformal time with regard to the number of e-foldings used to resolve the causality issue. To this effect, inflation involves the mapping of the smallest to the largest, namely, the GUT scale, $T_{end}\sim 10^{15} GeV$, is mapped to the temperature $T_0$ today! Infact, assuming radiation domination from the GUT scale till today, approximately retains the estimate (obtained in (\ref{hconf}) ) for the minimum number of e-folds, namely, $ T_{end}\simeq T_0~\exp(N)$.

Let us reiterate that  issues related to entropy, flatness, monopoles  and others get resolved once the horizon problem is addressed.  No doubt, inflation is a beautiful paradigm that successfully addresses the initial conditions of the big bang model. The prediction of primordial perturbations generated quantum mechanically during inflation adds to the strength of the paradigm. However, inflation might be plagued by its own initial condition problem. It has been pointed out that the probability measure of "generic" initial conditions, that could give rise to inflation , might be very thin. Bringing in quantum effects and issues related to eternal inflation might further  complicate the situation. Inflation is believed to have happened in the classical regime, below the Planck scale. It is plausible  that quantum gravity effects around the Planck scale would prepare a conducive background for the initiation of inflation.

  We close the present discussion with a comment on the scale of inflation $T_{end}$ in our notation \cite{Steinhardt:2004rf}. Inflation is often implemented by a slowly rolling scalar field $\phi$ with an equation of state parameter $\omega_\phi\equiv p_\phi/\rho_\phi$ close to minus one, where $\rho_\phi(p_\phi)$ designates the energy density (pressure) of the field. The observed value of the primordial density contrast $\delta$ is linked to the scale of inflation and equation of state parameter $\omega_\phi$ of the inflaton,
  \begin{equation}
   \label{deltaf}
  \delta =\frac{H^2}
  {\dot{\phi}}\sim \frac{1}{M^2_{pl}}\frac{\rho_\phi}{\sqrt{\rho_\phi+p_\phi}} 
  \sim \frac{1}{M^2_{pl}} \sqrt{\frac{\rho_\phi}{1+\omega_\phi}}
 \end{equation}
 where the inflaton energy density $\rho_\phi$ is nearly constant during inflation and is expected to decay into relativistic species  at the end of inflation such that
$\rho_\phi^{1/4}\sim T_{end}/M_{pl} $ which defines the scale of inflation. Using (\ref{deltaf}), we then find the estimate,
\begin{equation}
(1+\omega_\phi) \simeq 10^{10}\times \left(\frac{T_{end}}{M_{pl}}\right)^4 
\end{equation}
which reveals that the equation of the state parameter should cancel  minus one to an accuracy of $10^{-14}$, an incredible fine tuning,  for $T_{end}\simeq 10^{12} GeV$. In the case of $T_{end}\simeq 10^{15} GeV$, one has a reasonable adjustment, namely, $1+\omega_\phi\simeq 10^{-2}$. Obviously, it is not easy to implement inflation at low energy scales.

\subsection{Quantum effects in non-trivial background}
As mentioned before, it is common belief that the primordial density perturbations generated quantum mechanically from the vacuum during inflation seeded the structure in our universe. The study of the generation and evolution of these perturbations is a central theme of modern cosmology. Unlike the Minkowski space-time, the FRW background is dynamical and space-time is curved, which throws challenges to the field theoretic framework in this background. Indeed, in particular, the definition of vacuum, in this case, is ambiguous contrary to Minkowski space time, where the vacuum state is Poincare invariant. 
There are other difficult issues related to renormalization of field theory in non-trivial background.
There are also  gauge issues that can be addressed by either opting to work in a given gauge or using a gauge-invariant formalism. Interestingly, the inflaton and metric perturbations can be combined into a gauge invariant variable dubbed curvature perturbation, whose action resembles that of a free scalar field in a (quasi) Sitter background. As for the quantum nature of perturbation, it is an unsettled issue and we shall come back to it later in this section.

 Since cosmological perturbations  mimic a free scalar field, our discussion confines to scalar field quantization.
Let us initiate the discussion considering a free massive scalar field in Minkowski background,
\begin{equation}
\label{LSF}
 \mathcal{L}  =-\frac{1}{2}\partial^\mu\phi\partial_\mu \phi-\frac{1} 
 {2}m^2 \phi^2
\end{equation}
with the conjugate momentum,
\begin{equation}
\label{PiSF}
    \pi(x)=\frac{\partial \mathcal{L}}{\partial \dot{\phi}}=\dot{\phi}(x)
\end{equation}
The Hamiltonian is then given by,
\begin{equation}
\label{HSF}
H=\int{\mathcal{H}d^3x} = \int{\left( \frac{1}{2}\pi^2(x)+\frac{1}{2}{\bf \nabla}\phi.{\bf\nabla}\phi+\frac{1}{2}m^2\phi^2(x)\right)}d^3x 
\end{equation}
The Euler-Lagrange equation that follows from (\ref{LSF}) dubbed Klein-Gordon equation is,
\begin{equation}
\label{SFEQ}
 (\partial_\mu  \partial^\mu +m^2)\phi(x)=0 
\end{equation}

Expanding the field into Fourier integral,
\begin{equation}
\label{FE}
    \phi({\bf x},t)=\frac{1}{({2\pi})^{3}}\int\phi_{\bf k}(t) e^{{i\bf k}.{\bf x}}{d^3{\bf k}}
\end{equation}
transforms the field equation into the equations of harmonic oscillators for the Fourier coefficients(modes),
\begin{equation}
\label{OSE}
\ddot{\phi}_{\bf k}(t)+\omega_k^2 \phi_{\bf k}=0;~~\omega_k=k^2+m^2    
\end{equation}
which shows that field is represented by non-interacting harmonic oscillators labelled by continuous  variables, ${\bf k}=(k_x,k_y, k_z)$.
Equation (\ref{OSE}) is readily solved,
\begin{equation}
\label{OSS}
\phi_{\bf k}(t) =\frac{1}{\sqrt{2\omega_k}} \left(a_{\bf k}e^{-i\omega_k t}
+a^*_{\bf -k} e^{+i\omega_k t}\right)
\end{equation}



where we have taken into account that the field is real ($\phi^*=\phi\to \phi_{\bf k}=\phi^*_{-\bf k}$).
 Plugging (\ref{OSS}) into (\ref{FE})
 one then finds,
\begin{equation}
\label{SFEF}
 \phi({\bf x},t)=\frac{1}{ (2\pi)^3}\int{\frac{1}{\sqrt{ 2\omega_k}}[a_{\bf k}e^{+i k_\mu x^\mu}+a^*_{\bf k}e^{-i k_\mu x^\mu}]d^3k}   
\end{equation}
where we have replaced in second term, ${\bf k}$ by ${\bf- k}$ which allows us to write the arguments of the exponential in covariant form.
The canonical conjugate momentum has the expansion,
\begin{equation}
\label{piEF}
 \pi({\bf x},t)=-\frac{i}{(2\pi)^3}\int{\sqrt{\frac{\omega_k}{2} }\left[a_{\bf k}e^{+i k_\mu x^\mu}-a^*_{\bf k}e^{-i k_\mu x^\mu}\right]d^3k}
\end{equation}
Making use of (\ref{SFEF}) and (\ref{piEF}), one finds the expression of the Hamiltonian density,
\begin{equation}
\label{HAM}
H=
\int{\left(\frac{1}{2}|\dot{\phi}_{\bf k} |^2+\frac{1}{2}\omega^2_k|\phi_{\bf k}|^2\right)}
=\int{\frac{\omega_k}{2}\left[a^*_{\bf k} a_{\bf k}+a_{\bf k} a^*_{\bf k}\right]d^3k}  =\int{\frac{\omega_k }{2}(a^*_{\bf k} a_{\bf k}+a^*_{\bf k} a_{\bf k})d^3k}  \equiv \int{\omega_k a^*_{\bf k} a_{\bf k}d^3k} 
\end{equation}
In classical framework, order of expansion coefficients is not important. However, order becomes crucial when we turn to quantization.
Canonical quantization is carried out by treating $\phi$ and $\pi$ as Hermitian operators(designated by hates),
\begin{eqnarray}
\label{FO}
&& \hat{\phi}({\bf x},t)=\frac{1}{(2\pi)^3}\int{\frac{1}{\sqrt{ 2\omega_k}}[\hat{a}_{\bf k}e^{+i k_\mu x^\mu}+\hat{a}^+_{\bf k}e^{-i k_\mu x^\mu}]d^3k} \equiv\frac{1}{(2\pi)^3}\int{[v_k(t)\hat{a}_{\bf k}+v^*_k(t)\hat{a}^+_{-\bf k}]e^{+i {\bf k} {\bf x}}d^3k} \\
&&  \hat{\pi}({\bf x},t)=-\frac{i}{(2\pi)^3}\int{\sqrt{\frac{\omega_k}{2} }\left[\hat{a}_{\bf k}e^{+i k_\mu x^\mu}-\hat{a}^+_{\bf k}e^{-i k_\mu x^\mu}\right]d^3k}
\end{eqnarray}
which satisfy the equal time commutation relations,
\begin{eqnarray}
&& [\hat{\phi}({\bf x},t),\hat{\pi}({\bf y},t)]_{x^0=y^0}  =[\hat{\pi}({\bf x},t),\hat{\pi}({\bf y},t)]_{x^0=y^0}=0 \\
&&[\hat{\phi}({\bf x},t),\hat{\pi}({\bf y},t)]_{x^0=y^0}=i\delta^3({\bf x}-{\bf y})
\end{eqnarray}
that in turn imply commutation relations for coefficients in the  field expansion in (\ref{SFEF}),
\begin{eqnarray}
\label{COMR}
 && [\hat{a}_{\bf k},\hat{a}_{\bf k'}]  = [\hat{a}^+_{\bf k},\hat{a}^+_{\bf k'}]=0\\
 && [\hat{a}_{\bf k},\hat{a}^+_{\bf k'}]  =(2\pi)^3 \delta^3({\bf k}-{\bf k'})
\end{eqnarray}
This shows that coefficients of field operator expansion resemble annihilation and creation operators; their number is infinite in this case.  Fourier modes corresponding to different
${\bf k}'s$ do not couple by virtue of field equation being linear. A free field can be thought as an ideal gas of harmonic oscillators.

 While computing the physical quantities, in particular, the Hamiltonian operator, there is always an ambiguity   related to ordering of operators which occurs when we replace the classical quantities by corresponding operators that gives rise to zero point energy which is divergent (see Eq.(\ref{HAM}). The latter is removed by choosing the so called normal ordering such that the creation operators are arranged on the left of annihilation operators,
 \begin{equation}
 \label{HAMN}
 \hat{H}:=\int{\omega_k\hat{a}^+_{\bf k}\hat{a}_{\bf k}d^3{\bf k}}   
 \end{equation}
such that, $\hat{H}\ket{0}=0$\footnote{Vacuum state is defined as,~$\hat{a}_{\bf k}|0\rangle =0$ and the vacuum energy is chosen to be zero in the present setting. }.
Let us consider a general expression for the field operator: a deformation of (\ref{FO}),
\begin{equation}
\label{GFE}
 \hat{\phi}(x)=\frac{1}{{(2\pi)^3}} \int{\left(v_{ k}(t)e^{+i {\bf  k}.{ \bf x}}\hat{a_{\bf k}}+v^*_{ k}(t)e^{-i {\bf  k}.{ \bf x}}\hat{a}^+_{\bf k}\right)d^3{\bf k}}  
\end{equation}
assuming that the commutation relations (\ref{COMR}) hold. 
Let us  note that the mode function,
\begin{equation}
\label{BD}
v_{k }(t)=\frac{e^{-i\omega_k t}} {\sqrt{2\omega_k}}   
\end{equation}
is special amongst all the mode functions, it is specific to free field in Minkowski space time. Indeed, let us compute the Hamiltonian operator using (\ref{GFE}) and commutation relations (\ref{COMR}),
\begin{eqnarray}
\label{DH}
 \hat{H}=\frac{1}{2}\int{\frac{d^3{\bf k}}{(2\pi)^3}\left[ ( v^{*'}_{ k} v^{*'}_{k}+\omega^2_k {v}^*_{ k} {v}^*_{ k})a^+_{\bf k}a^+_{\bf- k}+({v'}_{k}{v'}_{ k}+\omega^2_kv_{ k}v_{k})a_{\bf k}  a_{\bf -k} +2(|{v'}_{ k}|^2 +\omega^2_k|v_{ k} |^2 )a^+_{\bf k} a_{\bf k}  \right] }    
\end{eqnarray}
Here $'$ indicates the derivative with respect to time in Minkowski space. 
Vacuum state is Poincare invariant in the Minkowski background, such that $\hat{H}|0\rangle=0$ implies that the coefficient of the second term in (\ref{DH}) is identically zero (the coefficient of the first term, which is the conjugate of the second, also vanishes). In case of (\ref{BD}), the second term obviously vanishes and the third reduces to $\omega_k$.
However, 
 if the  second term  in (\ref{DH})  non-vanishing, it gives rise to pair production with momenta $\bf{k}~\&-\bf{k}$ in accordance with momentum conservation. Even if we choose an instantaneous vacuum state,
 due to particle production, it will not remains to be a vacuum state, it will have particles in it. In this case, there is no unambiguous definition of vacuum state. 
The generalization or deformation in (\ref{GFE}) could be accomplished by introducing time dependence in the frequency which is induced by interaction of free field to a classical source  that gives rise to particle production $\hat{\rm a}$  {\it  la} {\it Schwinger process}. 
The latter, in cosmology, is a manifestation of considering quantum field in a non-trivial background, say, de Sitter. 
In quantum field theory, in Minkowski space time, particle creation is caused by interaction term higher than the quadratic one.
In the event that frequency becomes constant in a region, say, $t<T$, and deformation switches off giving rise  Minkowski background,  allows us to unambiguously define the vacuum state in this region. 
Actually, we have similar situation during inflation where by virtue of non-trivial (quasi de Sitter) background, we deal with parametric oscillator with time dependent frequency to be discussed later in this section.

As noted before, a free field, in Minkowski background, is composed of infinite number of modes(harmonic oscillators)
which evolve  independently of each other. The wave function of each mode is a Gaussian function
such that the wave function(al) of ground state of the system is given by the product of individual wave functions\footnote{$\langle\Psi(\phi)|0\rangle\equiv \Psi_0(\phi)$.},
\begin{equation}
\label{GBD}
 |0\rangle:=\Psi_0(\phi)=A  \prod_{\bf k} e^{-\frac{1}{2}\omega_k |\phi_{\bf k}|^2} =A e^{{-\frac{1}{2}\int{\omega_k |\phi_{\bf k}|^2d^3{\bf k}  } }}
\end{equation}
where $A$ is a normalization constant. An important comment is in order:{\it A free quantum field in vacuum is a Gaussian random field which naturally obeys the Wick's theorem.} This aspect plays important role while discussing quantization of cosmological perturbations.

Summarising the aforesaid, it can be stated that in a regime where the deformation in (\ref{GFE}) is absent, the mode function takes the form (\ref{BD})  $\hat{\rm a}$  {\it  la} Minkowski allowing an unambiguous definition of vacuum state. In this case, 
the ground state of the system takes the form
of a Gaussian wave functional (\ref{GBD}) referred to as {\it Bunch Davies} vacuum with mode function given by (\ref{BD}).

The central theme in quantum field theory is associated with the computations of correlations, particularly the two-point correlation. Before embarking upon the non-trivial background, say de Sitter, it is instructive to quote the result in Minkowski space time, 
\bea
\langle 0|\hat{\phi}(x) \hat{\phi}(y)|0\rangle &=&\frac{1}{(2\pi)^6}\int\int \frac{d^{3}{\bf k}}{\sqrt{2\omega_{ k}}}\frac{d^{3}{\bf k}^{'}}{\sqrt{2\omega_{{ k}^{'}}}}\langle 0|\left(\hat{a}_{\bf k}\hat{a}_{{\bf k}^{'}}~e^{i(kx+k^{'}y)}+\hat{a}_{\bf k}\hat{a}^{\dagger}_{{\bf k}^{'}}~e^{i(kx-k^{'}y)}\right.\nonumber\\
&&\left.\quad\quad\quad\quad\quad\quad\quad\quad\quad\quad\quad\quad\quad\quad\quad +\hat{a}^{\dagger}_{\bf k}\hat{a}_{{\bf k}^{'}}~e^{-i(kx-k^{'}y)}+\hat{a}^{\dagger}_{\bf k}\hat{a}^{\dagger}_{{\bf k}^{'}}~e^{-i(kx+k^{'}y)}\right)|0\rangle\nonumber\\
&=&\frac{1}{(2\pi)^6}\int\int \frac{d^{3}{\bf k}}{\sqrt{2\omega_{ k}}}\frac{d^{3}{\bf k}^{'}}{\sqrt{2\omega_{{ k}^{'}}}}\underbrace{\langle 0|\left[\hat{a}_{\bf k},\hat{a}^{\dagger}_{{\bf k}^{'}}\right]|0\rangle}_{=(2\pi)^3\delta^3({\bf k}-{\bf k}^{'})}~e^{i(kx-k^{'}y)}\nonumber\\
&=& \frac{1}{(2\pi)^{3}} \int \frac{d^3 {\bf k}}{2\omega_k}  e^{i k (x-y)}
\label{CORRF}
\eea
where we used (\ref{FO}) and commutation relations (\ref{COMR}).  
It may be noted that the correlation function  (\ref{CORRF}) has oscillatory character in time and in order to give an unambiguous meaning to this quantity, one considers, the equal time correlation,
\bea
&& \langle 0|\hat{\phi}({\bf x},t) \hat{\phi}({\bf y},t)|0\rangle=\frac{1}{(2\pi)^3}  \int\frac{d^3{\bf k}}{2\omega_{ k}}~e^{i {\bf k}. ({\bf x}-{\bf y})}\nonumber\\
&&= \frac{1}{(2\pi)^6}\int\int d^3{\bf k}  d^3{\bf k'} \langle 0|\hat{\phi}_{\bf k}(t)  \hat{\phi}_{{\bf k'}} (t) |0\rangle~e^{i {\bf k}. {\bf x} } e^{i{\bf k'}.{\bf  y}} ~\left(\langle 0|\hat{\phi}_{\bf k}(t)  \hat{\phi}_{{\bf k'}} (t) |0\rangle\equiv \frac{1}{\sqrt{4\omega_{ k}\omega_{ k'}}}  (2\pi)^3 \delta({\bf k}+\bf k') \right)           \eea
where $\langle 0|\hat{\phi}_{\bf k}(t)  \hat{\phi}_{\bf k'} (t) |0\rangle$ denotes the correlation function in Fourier space. Let us estimate the  correlation function to check how it varies with the distance,
\bea
\langle 0|\hat{\phi}({\bf x},t) \hat{\phi}({\bf y},t)|0\rangle& =&\frac{1}{(2\pi)^3}  \int\frac{d^3{\bf k}}{2\omega_{ k}}~e^{i {\bf k}. ({\bf x}-{\bf y})}=\frac{1}{(2\pi)^3}  \int\frac{1}{2\omega_{ k}}~k^2dk\sin\theta d\theta d\phi~ e^{ik |{\bf x}-{\bf y}|\cos\theta}\nonumber\\
&=&\frac{1}{(2\pi)^2}  \int dk~\frac{k^2}{\omega_{ k}}~\frac{\sin(k|{\bf x}-{\bf y}|)}{k|{\bf x}-{\bf y}|}\nonumber\\
&=&\frac{1}{2\pi^2|{\bf x}-{\bf y}|}  \int^{L_{\bf UV}}_{0} dk~\sin(k|{\bf x}-{\bf y}|)\nonumber\\
&=&\frac{1}{2\pi^2|{\bf x}-{\bf y}|^2}\left\{1-\cos (L_{\bf UV}|{\bf x}-{\bf y}|) \right\},
\eea
where, for the sake of simplicity, we considered the mass-less field\footnote{Including mass term does not alter the conclusion.};
$L_{\bf UV}$ denotes UV cutoff here. This result sounds intuitive\footnote{In the sense that correlation decreases with distance.}, namely, the quantum correlation at two given points  in the coordinate space diminishes as inverse  square of the distance between the points. In the discussion to follow, this result would be confronted to its counter part in the de Sitter space where it is counter intuitive; cosmological correlations throws a surprise to us.


However, we shall first extend the aforesaid to the case of  interaction in Minkowski space time, which is a problem of incredible complexity as we do not know the interacting vacuum state and the fields in this case. The problem is handled in a perturbation framework where the interacting vacuum state can be expressed through the free vacuum $\ket{0}$ and interaction is taken into account perturbatively, order by order. As a final step, in what follows, the framework would be promoted to de Sitter background relevant to cosmology. To this effect, we briefly review the "in-out formalism to demonstrate its equivalence to the "in-in" framework in the Minkowski space, which is then carried over to de Sitter background\footnote{The effect of the non-trivial background manifests as the interaction of the quantum field with a classical source, which leads to a parametric oscillator. This interaction is qualitatively different from $H_{int}$ which is treated perturbatively in quantum field theory order by order. 
 }. However, in that case, naturally\footnote{In this case, an extra construction might allow pasting a Minkowsky patch on to de Sitter where one can define an "out" state \cite{Donath:2024utn}.}, there is no "out" state and one has no option then other than to opt for "in-in" formalism. The formalism will be used for the estimation of loop corrections to the power spectrum which plays important role in case of large primordial fluctuations and PBHs.
In our opinion, a coherent discussion on this theme is needed for a wider audience in cosmology.

\subsection{In-out, In-in and all that}
 The interacting quantum field theory framework is best understood in terms of
perturbation expansion.
The interaction Hamiltonian that contains a small parameter is added to 
the free field Hamiltonian and  perturbation expansion is then developed in the coupling. 
To this effect,
let us split the  field Hamiltonian into Klein-Gordon and self-interaction parts\footnote{Hereafter, we drop hats on operators if not specifically required. }, 
\begin{equation}
 H=H_0+H_{int}
\end{equation}
where $H_{int}$ includes a small coupling(for instance $\lambda \phi^4$ theory) that allows us to treat the problem perturbatively. In Schrodinger's picture, the state  evolves with time, and field operators are time-independent,
\begin{equation}
 i\frac{\partial}{\partial t}|\psi(t)\rangle=H   |\psi(t)\rangle \Longrightarrow |\psi(t)\rangle=U(t,t_0)|\psi(t_0)\rangle \equiv e^{-i H(t-t_0)}|\psi(t_0)\rangle
 \label{SFE}
\end{equation}
where $t_0$ is a reference time.
In Heisenberg's picture, time dependence is transferred to field operators,
\begin{equation}
\mathcal {O}_H(t)  = U^{-1}(t,t_0)\mathcal{O} U(t,t_0) ;~~|\psi\rangle_H\equiv |\psi(t_0)\rangle
\end{equation}
where the quantities specific to Schrodinger picture do not carry sub(super) script.  The evolution operator, $U(t,t_0)$ in Eq.(\ref{SFE}) should be computed perturbatively, the best way of which is provided by going to
an intermediate representation dubbed "Interaction Picture",
\begin{equation}
|\psi(t)\rangle_I\equiv U_0^{-1} (t, t_0) |\psi(t)\rangle  
\end{equation}
which implies that,
\begin{eqnarray}
 \mathcal{O}_I(t) =U_0^{-1}(t,t_0) \mathcal{O}U(t,t_0)\Longrightarrow \mathcal{O}_H(t)   =\mathcal{U}^{-1}_I(t,t_0)\mathcal{O}_I(t)\mathcal{U}_I(t,t_0);~~\mathcal{U}_I(t,t_0)=U_0^{-1}(t,t_0)U(t,t_0)
\end{eqnarray}
Interestingly, the dynamics of evolution operator in this picture is governed by the interaction Hamiltonian computed in the interaction picture,
\begin{equation}
i\frac{\partial}{\partial t}\mathcal{U}_I(t,t_0)=H^I_{\rm int} \mathcal{ U}_I(t,t_0) ;~~~H^I_{\rm int}=U_0^{-1}(t,t_0) H_{\rm int}U_0(t,t_0)
\label{EVOLOPEI}
\end{equation}
where $\mathcal{U}_I$ designates evolution operator in the interaction picture. {Evolution operator has the following defining properties,
\begin{equation}
\label{UG}
\mathcal{U}_I(t,t_1)\mathcal{U}_I(t_1,t_0)=\mathcal{U}_I(t,t_0);~~\mathcal{U}_I(t,t)=I;~\mathcal{U}_I(t_0,t)=\mathcal{U}_I^{-1}(t,t_0)=\mathcal{U}_I^{\dag}(t,t_0).
\end{equation}
Eq.(\ref{EVOLOPEI}) can be solved iteratively and
has a following formal solution in terms of a time ordererd exponential,
\begin{equation}
\mathcal{U}_I(t,t_0)=Te^{-i \int^t_{t_0}{H^I_{\rm int}(t') dt'}}   
\label{EVOLOPI}
\end{equation}
which should be understood in terms of series expansion.
Let us note that the states at reference point $t=t_0$  coincide in all the three pictures,
\begin{equation}
|\psi\rangle_H=  |\psi(t_0)\rangle_I  =|\psi(t_0)\rangle
\end{equation}
One of the important concepts in field theory is related to the construction of asymptotic states known as "in" and "out" states defined as eigen-states of free Hamiltonian in the far past, $t\to -\infty$, and distant future, $t \to +\infty$, namely, $|in\rangle\equiv |\psi(-\infty)\rangle_I$ and  $|out\rangle\equiv |\psi(+\infty)\rangle_I$. These states can be evolved to the uniquely defined Heisenberg state using (\ref{EVOLOPI}),
\begin{eqnarray}
&& |\psi\rangle_H= \mathcal{U}_I(t_0,-\infty) |\psi(-\infty)\rangle_I=\mathcal{U}_I(t_0,-\infty)|in\rangle\\
&& |\psi\rangle_H=\mathcal{U}_I(t_0,+\infty)|\psi(+\infty)\rangle_I=\mathcal{U}_I(t_0,+\infty)|out\rangle
\end{eqnarray}
Let us denote the ground state of the interacting system, by, $|\Omega\rangle_H$,
\begin{equation}
|\Omega(t)\rangle_I=\mathcal{U}_I(t,t_0)|\Omega(t_0)\rangle_I=\mathcal{U}_I(t,t_0)|\Omega_H\rangle    
\end{equation}
It should be emphasized that identifying the "in" and "out" states will be misleading. Indeed, the evolution operator $\mathcal{U}_I(t,t_0)$ is ill-defined at $t= \pm \infty$ and requires adiabatic regularization, asking for different regulators in both asymptotic regimes.
Indeed, let us consider, the action of the evolution operator on an "in" state at linear order,
\begin{equation}
 U(0,-\infty)|in\rangle-I\simeq 
 -i\int^{0}_{-\infty}{ dt ~e^{iH_0t}H_{int} e^{-iH_0t}}|in\rangle=-i\int^{0}_{-\infty}{dt~ e^{-i(E_i-H_0)t}H_{int}}|in\rangle;~~H_0|in\rangle=E_i=|in\rangle
\end{equation}
which shows that the evolution operator is ill defined at the lower limit. 
One therefore needs a kind of regularization
known as "adiabatic switching" of the interaction in far past which cab be accomplished by the following prescription,
\begin{equation}
  H^I_{int}  \to \lim_{\epsilon\to 0^+}e^{-\epsilon|t|}H^I_{int}
\end{equation}
such that,
\begin{equation}
 \mathcal{U}_I(0,-\infty)|in\rangle-I\simeq \lim_{\epsilon\to 0^+} \lim_{t_0\to -\infty}
 -i\int^{0}_{t_0}{dt e^{-i(E_i-H_0)t}e^{ \epsilon t}H_{int}}|in\rangle=\lim_{\epsilon\to 0^+} \frac{1}{E_i-H_0+i\epsilon}H_{int}|in\rangle
\end{equation}
 transforming the fast oscillating exponential into a damped one.
One is faced with similar problem while considering the action of evolution operator, $\mathcal{U}_I(\infty,0)$ on the "out" state; situation is again remedied by the same prescription. 
It should be noticed that one needs different regulators, namely, $e^{+\epsilon t}~\&~e^{-\epsilon t} $ in far past and distant future. 

The regularization is essential for giving meaning to the evolution operators in the asymptotic regions.
It is clear that the concept of  asymptotic "in" and "out" states is intrinsically linked to adiabatic regularization.
The need of regularization is also obvious at the level of full evolution operator,
\begin{equation}
 \mathcal{U}_I(0,t_0)|\Omega(t_0)\rangle=e^{-iH_0 t_0}   e^{iHt_0}|\Omega_H\rangle=e^{-iH_0 t_0}   e^{iEt_0}|\Omega_H\rangle
\end{equation}
where $|\Omega_H\rangle$ is the ground state of the interacting system, $H|\Omega_H\rangle=E |\Omega_H\rangle$.

Since we treat the interaction part as a perturbation to the free system Hamiltonian, $|\Omega_H\rangle$ bears a definite relation to the ground state of free Hamiltonian, $H_0|0\rangle=E_0|0\rangle$\footnote{$E_0=0$ in our setting.}, namely, there is a  energy shift due to interaction such that the free Hamiltonian ground state has non-vanishing overlap with the interacting vacuum,
\begin{equation}
\langle 0|\Omega_H\rangle\neq 0
\end{equation}
where we are referring to Gell-Mann and Low theorem.
We next expand the interacting vacuum state into the eigen states of the free Hamiltonian,
\begin{equation}
\mathcal{U}_I(0,t_0)|\Omega_H\rangle=e^{i(E-E_0)t_0}  \sum_n{\langle n|\Omega_H\rangle|n\rangle}  
\end{equation}
which for $t_0\to -\infty$ has unwanted behaviour (exponential oscillates infinitely fast) that needs correction. The correction or regularization can be implemented by slightly rotating the time axis into complex plane.
We can then turn to the computation of the correlation,
\begin{equation}
\langle\Omega_H|\mathcal{O}_H(t)|\Omega_H\rangle =  
\langle\Omega_H|\mathcal{U}^{-1}_I(t,t_0)\mathcal{O}_I (t)\mathcal{U}_I(t,t_0)|\Omega_H\rangle
\end{equation}
where $|\Omega_H\rangle$ designates the ground state of the interacting system  corresponding
to the reference time $t_0$
. 

In what follows, we shall express the ground state $\Omega_H$ in terms of the eigen states of free Hamiltonian. Regularization would than allow us to isolate the ground state by sending $t_0\to- \infty $ \cite{Senatore:2016aui}. 
Since the evolution operator exhibits the integrating effect, we can confine the effect of regularization around $t=t_0$ by using the group  structure of the evolution operator\footnote{$\mathcal{U}_I(t,t_0(1+i\epsilon))=\mathcal{U}^\dag_I(t_0(1-i\epsilon),t)$ which is consistent with our earlier prescription.},
\begin{equation}
  \mathcal{U}_I(t,t_0) =   \mathcal{U}_I(t,t_0(1-i\epsilon )) \mathcal{U}_I(t_0(1-i\epsilon ),t_0)
\end{equation}
which allows us to project out the the effect of exited states and express the ground state of the interacting system through  the vacuum state of the free Hamiltonian, $H_0$. Indeed,  
\begin{eqnarray}
&&  \mathcal{U}_I(t,t_0) |\Omega(t_0)\rangle \equiv  \mathcal{U}_I(t_0(1-i\epsilon ),t_0) |\Omega(t_0)\rangle=  e^{iH_0(t-t_0)}e^{-iE(t-t_0)} |\Omega_H\rangle=\sum_{n=0}{e^{i(E_n-E)(t-t_0)} \langle n|\Omega_H\rangle|n\rangle} =\nonumber\\
 &&= e^{\epsilon t_0(E_0-E)}\langle 0|\Omega_H\rangle|0\rangle+ \sum_{n\neq 0}{e^{+(E_n-E)\epsilon t_0} \langle n|\Omega_H\rangle|n\rangle} \rightarrow \lim_{\epsilon \to 0}\lim_{t_0\to -\infty} e^{+\epsilon t_0(E_0-E)}\langle 0|\Omega_H\rangle|0\rangle\equiv \mathcal{N} |0\rangle
 \label{OUTR}
\end{eqnarray}
where the  first term contributes most in the limit under consideration. 
 Regularization operates between $t_0~ \&~ t_0(1-i\epsilon)$($t_0\to -\infty)$ that transforms the previously fast oscillating exponential in (\ref{OUTR}) 
 into a damped one and raises the interacting vacuum state at minus infinity, in the form of free Hamiltonian vacuum state , to a slightly tilted complex time axis as an evolutionary track. 
The regularization sets the correct track of evolution between the free states in the asymptotic regions.
Evolution of interacting vacuum is given by,
\begin{equation}
\label{OMT}
|\Omega_I(t)\rangle=\mathcal{U}_I(t,t_0)|\Omega(t_0)\rangle= \mathcal{U}_I(t,t_0(1-i\epsilon ))   \mathcal{U}_I(t_0(1-i\epsilon ),t_0)|\Omega(t_0)\rangle\to \mathcal{N}\mathcal{U}_I(t,t_0(1-i\epsilon ))|0\rangle 
\end{equation}
in the, $ t_0\to -\infty$ limit.
Thus the interacting vacuum in the far past gets linked to free vacuum $|0\rangle$,
\begin{equation}
 \mathcal{U}_I(-\infty(1-i\epsilon ),-\infty)|\Omega(-\infty)\rangle=|\Omega(-\infty(1-i\epsilon)\rangle=\mathcal{N}  |0\rangle  
\end{equation}
which implies that interaction switches off in the far past and interacting vacuum reduces to free vacuum state (multiplying a state by a constant does not change the state) which is in the spirit of adiabatic regularization. 
Thus adiabatic regularization successfully switched of the interaction in fat past transforming the interacting vacuum state into the free Hamiltonian vacuum at $t_0=-\infty(1-i\epsilon)$ which then evolves, along time axis slightly rotated into the complex plane,  to $\Omega_I(t)$, see Eq.(\ref{OMT}). 
On the similar note, one has,
\begin{eqnarray}
 &&  \mathcal{U}^{-1}_I (t,t_0)= \mathcal{U}^{\dag}_I (t,t_0)=
 \mathcal{U}_I (t_0,t) =\mathcal{U}_I (t_0,t_0(1+i\epsilon))\mathcal{U}_I (t_0(1+i\epsilon),t))\nonumber\\
 &&\langle\Omega(t_0)|\mathcal{U}_I (t_0,t_0(1+i\epsilon))=\langle\Omega(t_0)|\mathcal{U}^\dag_I (t_0(1-i\epsilon),t_0)=\nonumber\\
 &&\langle\Omega(t_0)|e^{iH(t_0(1+i\epsilon)-t_0)}e^{-iH_0(t_0(1+i\epsilon)-t_0)}\rightarrow \lim_{\epsilon \to 0}\lim_{t_0\to -\infty} e^{\epsilon t_0(E_0-E)}\langle 0|\Omega_H\rangle |0\rangle\equiv \mathcal{N}^* |0\rangle
\end{eqnarray}
Let us note,
\begin{equation}
\label{normN}
1=\langle\Omega_I(t)|\Omega_I(t)\rangle = \langle0|\mathcal{U}_I (t_0(1+i\epsilon),t_0(1-i\epsilon))|0\rangle |\mathcal{N}|^2
\end{equation}
We can now compute the correlation that we had started from,
\begin{equation}
\langle\Omega_H|\mathcal{O}_H(t)|\Omega_H\rangle=|\mathcal{N} |^2\langle 0| \mathcal{U}_I (t_0(1+i\epsilon),t)\mathcal{O}(t) \mathcal{U}_I (t, t_0(1-i\epsilon))|0\rangle
\end{equation}
where we can get rid of  pre-factor using (\ref{normN}),
\begin{eqnarray}
\label{O1}
\langle\Omega_H|\mathcal{O}_H(t)|\Omega_H\rangle=\langle in|\mathcal{O}_H(t)|in\rangle
&=&\lim_{\epsilon\to 0}\lim_{t_0\to -\infty}\frac{\langle 0| \mathcal{U}_I (t_0(1+i\epsilon),t)\mathcal{O}_I(t) \mathcal{U}_I (t, t_0(1-i\epsilon))|0\rangle}{\langle0|\mathcal{U}_I (t_0(1+i\epsilon),t_0(1-i\epsilon))|0\rangle } \nonumber\\
&=&\frac{\langle 0| \mathcal{U}_I (-\infty(1+i\epsilon),t)\mathcal{O}_I(t) \mathcal{U}_I (t, -\infty(1-i\epsilon))|0\rangle}{\langle0|\mathcal{U}_I (-\infty(1+i\epsilon),t_0(1-i\epsilon))|0\rangle }
\end{eqnarray}
where $|\Omega_H\rangle=|\Omega(t_0\to -\infty(1-i\epsilon))\rangle; \langle\Omega(t_0\to -\infty(1+i\epsilon))|$
represent "in" vacuum states of the interacting system in far past. The evolution operators of interest, that evolve the system from far past to the epoch at time "t",
are given by,
\begin{eqnarray}
&&\mathcal{U}^{-1}_I (t, -\infty(1+i\epsilon))   = \bar{T}\exp\left(i\int^{t}_{-\infty(1+i\epsilon)}dt^{'}\;H^I_{\rm int}(t^{'})\right)\\
&&\mathcal{U}_I (t, -\infty(1-i\epsilon))   = {T}\exp\left(i\int^{t}_{-\infty(1-i\epsilon)}dt^{'}\;H^I_{\rm int}(t^{'})\right)
\end{eqnarray}
where $\bar{T}$ designates anti-time ordering operator.
\begin{figure*}[htb!]
    	\centering
   {
      \includegraphics[width=17cm,height=9cm]{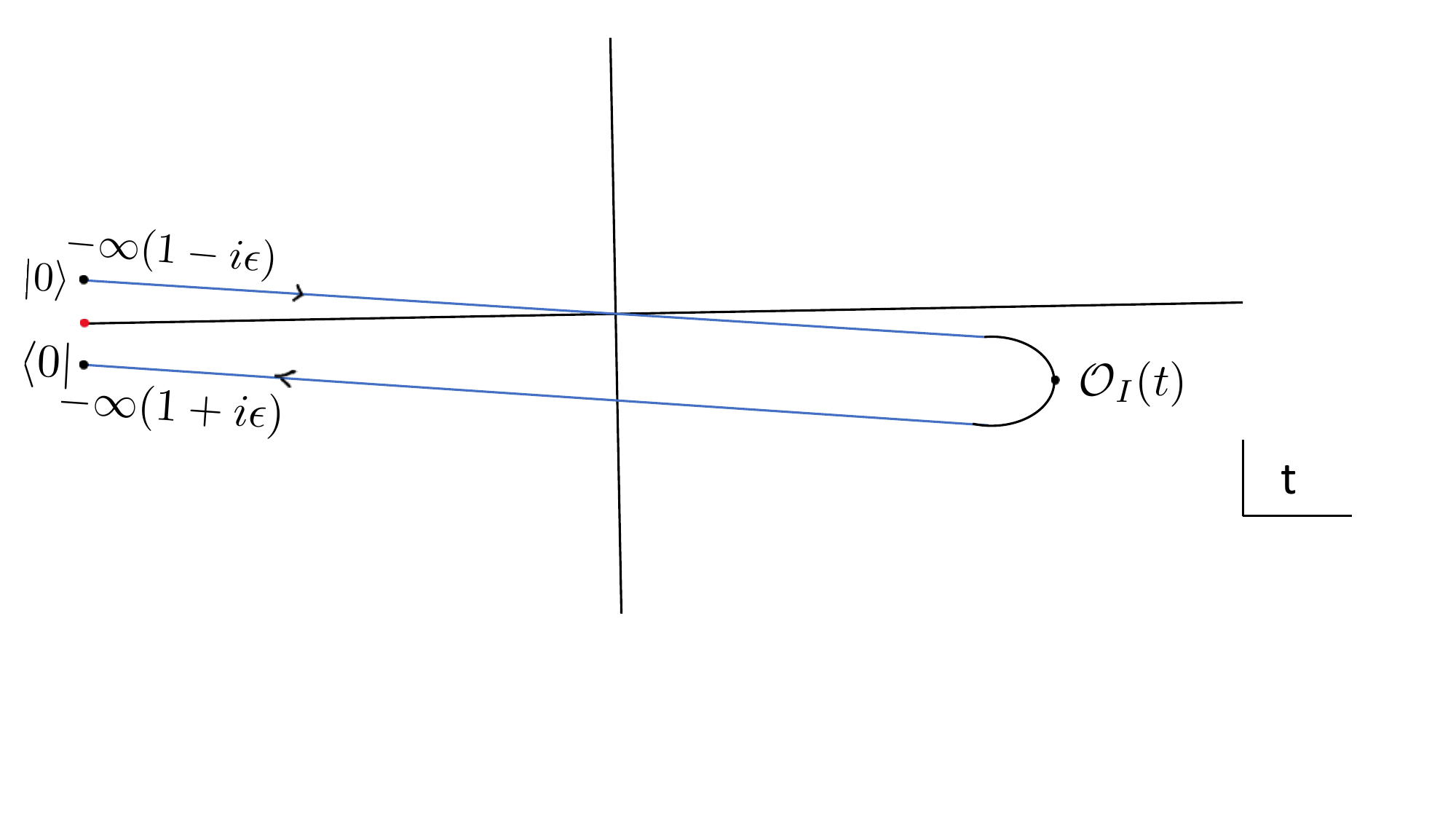}
   } 
    \caption[Optional caption for list of figures]{Schematic diagram of "in-in" correlator. Free vacuum states at $t_0=-\infty(1-i\epsilon)~\&~t_0=-\infty(1+i\epsilon)$ in the asymptotic region in far past are connected by evolution to the interacting vacuum at time $t$ where correlation is evaluated. In the process of regularization, the interacting vacuum state at $t_0=-\infty$, gets lifted to $t_0(1-i\epsilon)$ at the time axis which is slightly rotated towards the imaginary axis. At the new location, the interacting vacuum state reduces to $\ket{0}$. The tilted time axis is the right track for evolution that adheres to adiabatic regularization. The vacuum state $\ket{0}$ evolves along the said track from $-\infty(1-i\epsilon)$  to time "t" and then back to free Hamiltonian vacuum state at $-\infty(1+i\epsilon)$. }
\label{ininsk}
   \end{figure*}
   
Looking at the right-hand side of the expression (\ref{O1}), evolution can be thought to have commenced in the far past ($t_0=-\infty(1-i\epsilon))$ from the free vacuum $|0\rangle_{in}$ and proceeds to the future along a time axis, which is slightly rotated on the complex plane but turns back at time "t" to the past to end in the "in" vacuum  in the far past ($t_0=-\infty(1+i\epsilon))$ , see Fig.\ref{ininsk}. Evolution along the tilted axis is the requirement of adiabatic regularization. It may  be noticed that the evolution operator requires different regulators in both cases, and the contour does not close. Let us also note that the denominator in (\ref{O1}) describes vacuum-to-vacuum processes represented by bubble diagrams. As mentioned before, the evolution operator is understood in terms of the series expansion. In each order of perturbation expansion, the numerator can be expressed as the product of connected and disconnected diagrams; the disconnected part (due to bubble diagrams) is cancelled by the denominator order by order, leaving behind contributions from connected diagrams.
Though we formally focus on Minkowski space-time, our conclusions are valid for quantum fields in non-trivial backgrounds due to gravity. In Minkowski's background, one normally deals with the "in-out" correlation $\langle in|\mathcal{O}_H(t)|out\rangle$ rather than the "in-in" construct. Let us check for their interrelationship. To this effect, let us multiply (\ref{O1}) upside down by, $\langle 0|\mathcal{U}_I(+\infty(1-i\epsilon),-\infty(1+i\epsilon)|0\rangle$,

\begin{eqnarray}
\label{O2}
\langle in|\mathcal{O}_H(t)|in\rangle
&=&\frac{\langle 0|\mathcal{U}_I(+\infty(1-i\epsilon),-\infty(1+i\epsilon))|0>\langle 0| \mathcal{U}_I (-\infty(1+i\epsilon),t)\mathcal{O}(t) \mathcal{U}_I (t, -\infty(1-i\epsilon))|0\rangle}{\langle 0|\mathcal{U}_I(+\infty(1-i\epsilon),-\infty(1+i\epsilon))|0>\langle0|\mathcal{U}_I (-\infty(1+i\epsilon),-\infty(1-i\epsilon))|0\rangle }\nonumber \\
&=&\frac{\langle 0|\mathcal{U}_I(+\infty(1-i\epsilon),-\infty(1+i\epsilon)) \mathcal{U}_I (-\infty(1+i\epsilon),t)\mathcal{O}(t) \mathcal{U}_I (t, -\infty(1-i\epsilon))|0\rangle}{\langle 0|\mathcal{U}_I(+\infty(1-i\epsilon),-\infty(1+i\epsilon))\mathcal{U}_I (-\infty(1+i\epsilon),-\infty(1-i\epsilon))|0\rangle }.
\end{eqnarray}
where we used the fact that the contribution of exited state of free Hamiltonian in the projection operator, $\sum_{n}{|n\rangle\langle n|}=I$ is suppressed leaving behind, $|0\rangle\langle0|\to I$. We can then further simplify (\ref{O2}) using the property of evolution operator given in expression (\ref{UG}) \cite{Donath:2024utn},
\begin{eqnarray}
\langle in|\mathcal{O}_H(t)|in\rangle
&=&\frac{\langle 0|\mathcal{U}_I (+\infty(1-i\epsilon),t)\mathcal{O}_I(t) \mathcal{U}_I (t, -\infty(1-i\epsilon))|0\rangle}{\langle 0|\mathcal{U}_I (+\infty(1-i\epsilon),-\infty(1-i\epsilon))|0\rangle }\nonumber \\
&=&\frac{\langle 0|T \mathcal{O}_I(t)\exp\left(-i\int^{+\infty(1-i\epsilon)}_{-\infty(1-i\epsilon)}dt^{''}\;H^I_{\rm int}(t^{''})\right)|0\rangle}{\langle 0|T\exp\left(-i\int^{+\infty(1-i\epsilon)}_{-\infty(1-i\epsilon)}dt^{''}\;H^I_{\rm int}(t^{''})\right)|0\rangle}=\langle out|\mathcal{O}_H(t)|in\rangle
\end{eqnarray}
where the "out" vacuum of the interacting system is symbolized by, $|\Omega(+\infty(1-i\epsilon))\rangle$.
Evolution in this case commences in the vacuum state $|0\rangle$ (the "in vacuum") proceeds to time $t$ (where correlation is computed) along time axis in the complex plane and ends in the out vacuum in distant future.
\begin{figure*}[htb!]
    	\centering
   {
      \includegraphics[width=17cm,height=8cm]{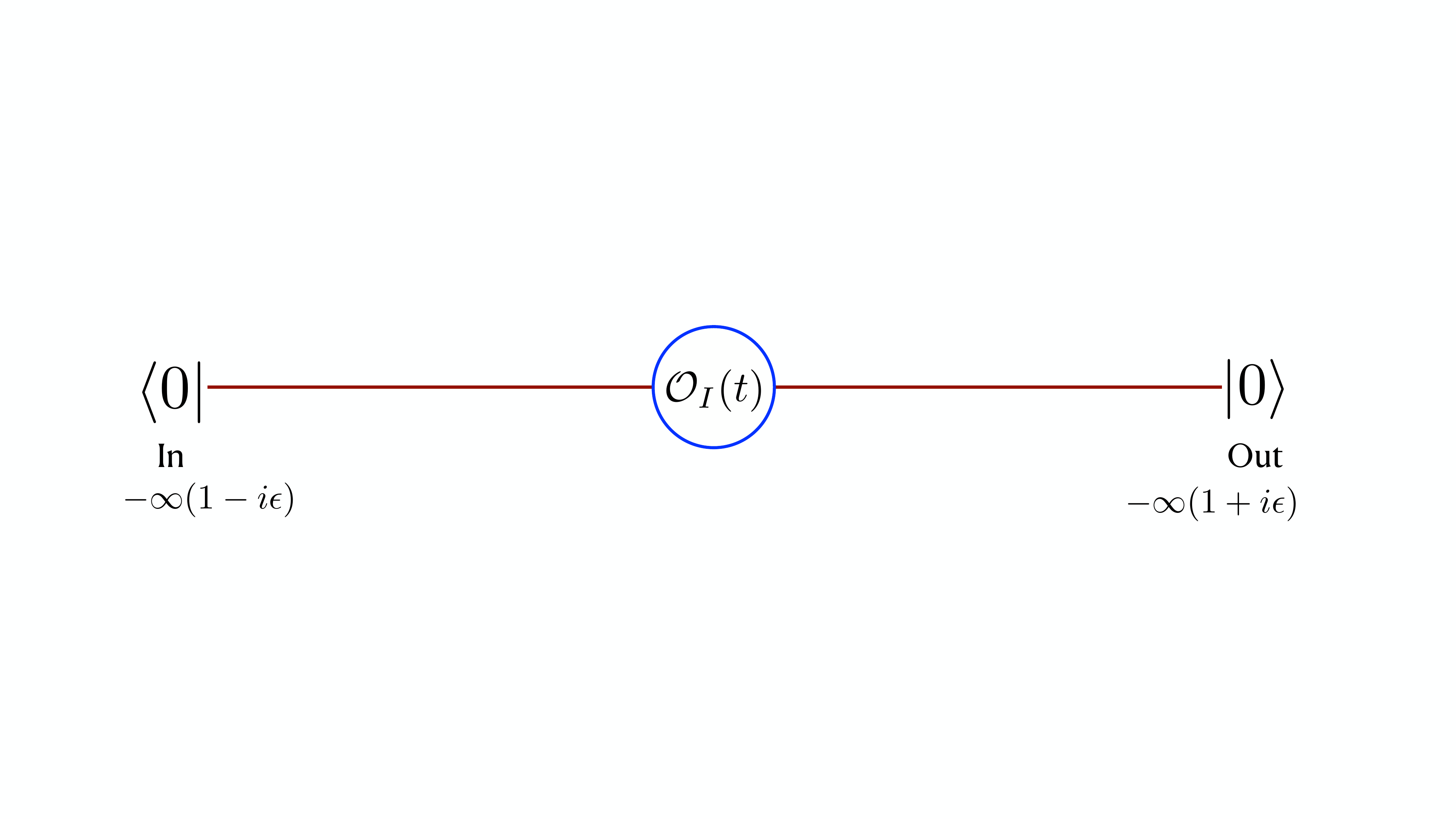}
   } 
    \caption[Optional caption for list of figures]{Diagrammatic  representation of correlation in the "in-out" formalism. Evolution begins from free vacuum in the far past where interaction is switched off due to adiabatic boundary condition, enters the region of interaction and proceeds to distant future. Asymptotically free vacuum states from past and future are connected to interacting vacuum state at time $t$ where correlation is evaluated.  }
\label{sk}
   \end{figure*}
The aforesaid demonstrates the equivalence of the "in-in" and "in-out" formalisms, and we can use either.
The set-up discussed here also applies to the quantum framework in the non-trivial background used in cosmology. In that case, the vacuum state dubbed {\it Bunch-Davies} can be chosen in the deep sub-Hubble region where the background reduces to Minkowski space time
which infect  is a unique possibility in cosmology. 
The dynamical coupling due to gravity becomes important around the Hubble radius and thereafter, such that there is no "out" state in this case. And this leaves no option other than to turn back to the Bunch-Davies vacuum, allowing us to define the "in-in" correlation. 
Thus, we shall focus on the "in-in" formalism to be used in cosmology. We shall be interested in higher-order corrections to the correlator that are important for the calculation of the power spectrum of primordial perturbations.
To that effect, by expanding the in-in correlator in the Taylor series, we have,

\bea \label{inin}
\langle \mathcal{O}_{H}(t)\rangle&=&\langle in|\mathcal{O}_{H}(t)|in\rangle\nonumber\\
&=&\langle 0|\left(\bar{T}\exp\left(i\int^{t}_{-\infty(1+i\epsilon)}dt^{'}\;H^{I}_{\rm int}(t^{'})\right)\right)\mathcal{O}_I(t)\left(T\exp\left(-i\int^{t}_{-\infty(1-i\epsilon)}dt^{''}\;H^{I}_{\rm int}(t^{''})\right)\right)|0\rangle\nonumber\\
&=&\langle 0|\left(1+i\int^{t}_{-\infty(1+i\epsilon)}dt^{'}\;H^{I}_{\rm int}(t^{'})-\frac{1}{2}\int^{t}_{-\infty(1+i\epsilon)}dt^{'}\int^{t}_{-\infty(1+i\epsilon)}dt^{'''}\;H^{I}_{\rm int}(t^{'})\;H^{I}_{\rm int}(t^{'''})+\cdots\right)\mathcal{O}_I(t)\nonumber\\
&&\quad\quad\left(1-i\int^{t}_{-\infty(1-i\epsilon)}dt^{''}\;H^{I}_{\rm int}(t^{''})-\frac{1}{2}\int^{t}_{-\infty(1-i\epsilon)}dt^{''}\int^{t}_{-\infty(1-i\epsilon)}dt^{''''}\;H^{I}_{\rm int}(t^{''})\;H^{I}_{\rm int}(t^{''''})+\cdots\right)|0\rangle\nonumber\\
&=&\langle \mathcal{O}_H(t)\rangle_{(0,0)}+\langle \mathcal{O}_H(t)\rangle_{(0,1)}+\langle \mathcal{O}_H(t)\rangle^{\dagger}_{(0,1)}+\langle \mathcal{O}_H(t)\rangle_{(0,2)}+\langle \mathcal{O}_H(t)\rangle^{\dagger}_{(0,2)}+\langle \mathcal{O}_H(t)\rangle_{(1,1)}+\cdots
\label{PEO}
\eea
where we symbolically define the zero-th, first and second order contributions by the following expressions:
\begin{itemize}
    \item {\bf Zero-th~order expression:}
    \bea &&\langle \mathcal{O}_H(t)\rangle_{(0,0)}=\langle 0|\mathcal{O}_H(t)|0\rangle,\eea

    \item  {\bf First~order expression:}
    \bea &&\langle \mathcal{O}_H(t)\rangle_{(0,1)}+\langle \mathcal{O}_H(t)\rangle^{\dagger}_{(0,1)}=2\;{\rm Re}\bigg[-i\int^{t}_{-\infty(1+i\epsilon)}dt^{'}\;\langle 0|\mathcal{O}_I(t)H^I_{\rm int}(t^{'})|0\rangle\bigg],\eea

    \item {\bf Second~order expression:}
   \bea  &&\langle \mathcal{O}_H(t)\rangle_{(0,2)}+\langle \mathcal{O}_H(t)\rangle^{\dagger}_{(0,2)}=-2\;{\rm Re}\bigg[\int^{t}_{-\infty(1+i\epsilon)}dt^{'}\int^{t}_{-\infty(1+i\epsilon)}dt^{''}\;\langle 0|\mathcal{O}_I(t)H^I_{\rm int}(t^{'})H^I_{\rm int}(t^{''})|0\rangle\bigg],\quad\quad\quad\\
 &&\langle \mathcal{O}_H(t)\rangle_{(1,1)}=\int^{t}_{-\infty(1+i\epsilon)}dt^{'}\int^{t}_{-\infty(1-i\epsilon)}dt^{''}\;\langle 0|H^I_{\rm int}(t^{'})\mathcal{O}_I(t)H^I_{\rm int}(t^{''})|0\rangle,\eea
\end{itemize}
The formalism developed here woulds be carried forward to de Sitter background
to be used in the forthcoming discussion.

\begin{figure*}[htb!]
    	\centering
   {
      \includegraphics[width=14cm,height=7cm]{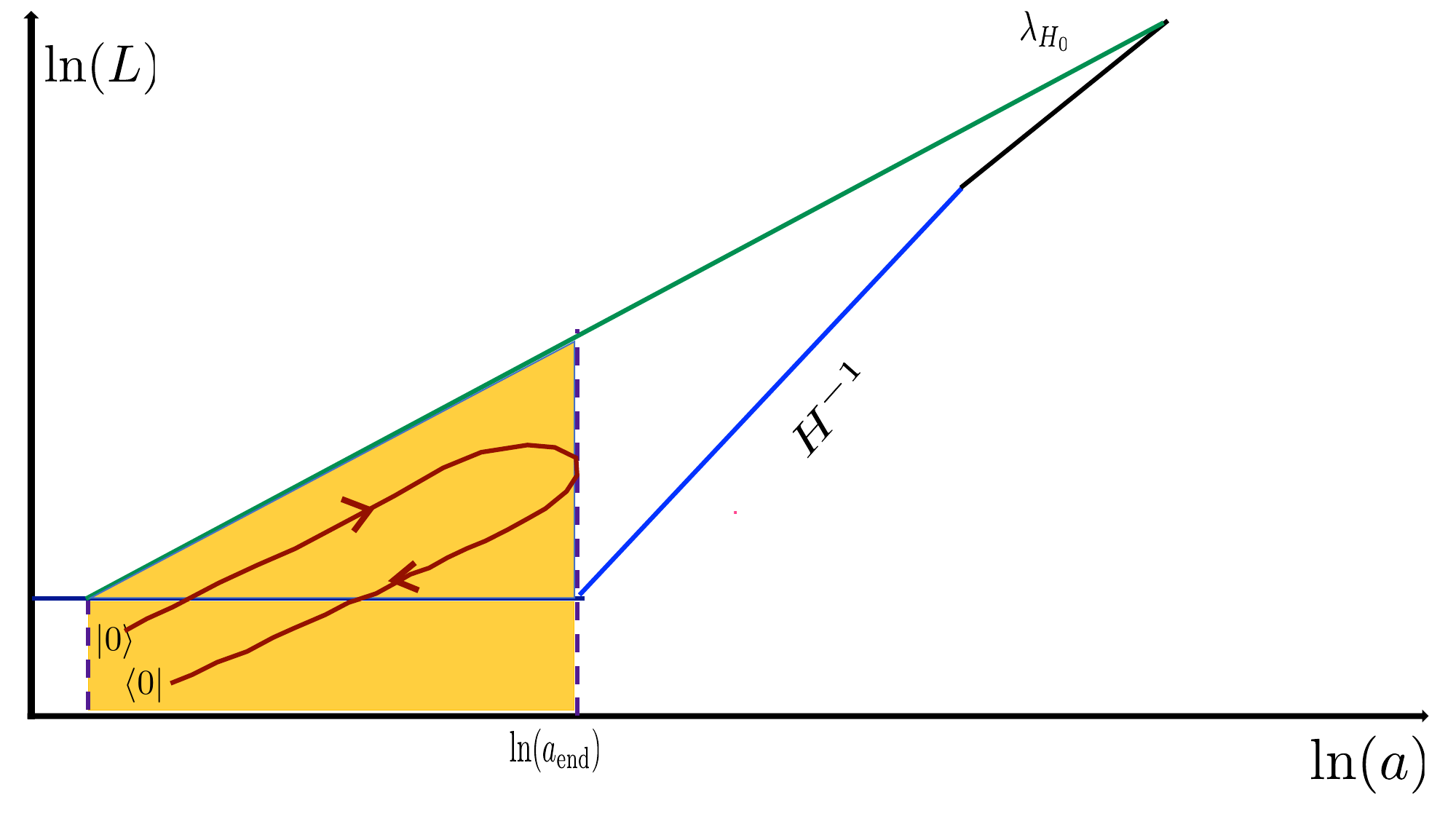}
   } 
   \caption[Optional caption for list of figures]{Schematic diagram representation of the contour used to evaluate the "in-in" correlation. Free vacuum state is evolved from far past ($ {\tau_0=(-\infty(1-i\epsilon}$) to time "$\tau$" where correlation is evaluated and then back to far past to vacuum state($\tau_0=-\infty(1-i\epsilon)$) such that contour does not close.
    It is reasonable to choose, $\tau=\tau_{end}=0$ in the super Hubble regime where perturbation turns classical and freezes. }
\label{sk}
   \end{figure*}



   \subsection{Cosmological perturbations in a nutshell}
The observed CMB anisotropy reveals the presence of primordial density perturbations, $\delta\rho/\rho\sim 10^{-5}$, that grew due to gravitational instability after recombination into the observed structure in the universe. The latter allows us to treat the density mismatch from a smooth FRW background perturbatively, confining most of the time to the first order. These perturbations can be classified into scalar, vector, and tensor perturbations, which evolve independently in the leading order of perturbation theory. The redundancy in this description can be removed by fixing the gauge degrees of freedom, allowing us to work in a particular gauge. One also constructs the gauge invariant variables, for instance, the curvature perturbation $\zeta$ that possesses the remarkable properties: It conserves on supper-Hubble scales, and it mimics density perturbation. Indeed, let us consider the small perturbations of the metric around the FRW background \cite{Mukhanov:2005sc,Riotto:2018pcx,Riotto:2002yw,Lyth:1998xn,Baumann:2022mni,Baumann:2009ds,Baumann:2018muz,Senatore:2016aui},
\begin{equation*}
g_{\mu\nu}=g^{(0)}_{\mu\nu}+\delta g_ {\mu\nu};~~ds^2=a^2(\tau)(\eta_{\mu\nu}+h_{\mu\nu})   
\end{equation*}
Similarly, the energy momentum tensor is perturbed,
\begin{equation*}
T_{\mu\nu}=T_{\mu\nu}^{(0)}+\delta T_{\mu\nu}    
\end{equation*}
 where, $T^{(0)}_{\mu\nu}$ obeys Einstein equations and covariant conservation such that,
 \begin{equation}
 \delta G_{\mu\nu}=8\pi G \delta T_{\mu\nu}  
 \label{LEE}
 \end{equation}
 Since we consider small perturbations $h_{\mu\nu}$ around the Minkowski background $\eta_{\mu\nu}$, the raising (lowering) of indices is accomplished with $\eta^{\mu\nu}$ ($\eta_{\mu\nu}$). The general coordinate transformation in Einstein gravity at linear order, $\tilde{x}^\mu=x^\mu+\xi^\mu(x)$ gives rise to following  gauge transformation for perturbations,
 \begin{equation}
 \label{GT}
     \tilde{h}_{\mu\nu}={h}_{\mu\nu}+\partial_\mu \xi_\nu-\partial_\nu \xi_\mu-\frac{2}{a}\eta_{\mu\nu}\xi^\alpha \partial_\alpha a
 \end{equation}
 where $\xi^\mu(x)$ are arbitrary functions and $h_{\mu\nu}$ and $\xi_{\mu}$
 are quantities of the same order of smallness. 
 
 Similar to a vector which can be decomposed into divergence free and curl free parts,  $h_{\mu \nu}$ can be decomposed into a scalar part (4 scalars, 2 gauge degrees) plus a vector part (two transverse vectors, 2 gauge degrees) plus the transverse trace less part of the perturbed metric $h^T_{\mu\nu} $(2 degrees $-$ gravity waves). On the similar note, one can classify the $T_{\mu \nu}$. 
  In this case, 5 degrees of freedom are attributed to perfect fluid whereas the other 5 link to imperfect fluid. 
  The classification of cosmological perturbations was carried out by Evgeny Lifshitz as early as 1946!
  The vector part of the perturbations decay with evolution and can be ignored during inflation.  The transverse trace less part $h^T_{\mu\nu}$ has two physical degrees of freedom and can be dealt separately\footnote{While working to the accuracy of first order in perturbation.}. The scalar part of the metric has the following form,
 \begin{equation}
ds^2=a^2(\tau)\left[-(1-2\phi)d\tau^2+2B_{,i}dx^id\tau((1-2\psi)\delta_{ij}+E_{,ij})dx^i dx^j\right]  ;~~B_{,i}\equiv \partial_i B ,~E_{,ij}\equiv \partial_i  \partial_j E
\label{SM}
 \end{equation}
The gauge transformation that  preserves the form of the metric (\ref{SM}) transforms the scalar quantities, $\phi,\psi,B~\&~E$ in a specific way and one can use this freedom to get rid of two extra degrees of freedom; for instance, we can turn $E$ and $B$ to zero $\hat{\rm a}$  {\it  la}   Newtonian gauge. However, it is more useful to identify the gauge invariant combinations made out of these functions. Secondly, we should also do the similar procedure for energy momentum tensor and since we are dealing here with perturbations during inflation, $T_{\mu\nu}$ will be contributed by scalar field(s)($\varphi$ in case of single field inflation). Simplification  occurs in case of the perfect fluid such as the scalar field for which $\phi=\psi$. In that case\footnote{Since  the gravitational potentials are denoted by $\phi$ and $\psi$, we denote here the scalar field by $\varphi$, later we shall return to original notation.},
\begin{eqnarray}
&& \Phi(\tau,{\bf x})=\phi+\frac{1}{a}[a(B-E')]'   \\
&&\tilde{\varphi}(\tau,{\bf x})=\delta\varphi+\varphi'_0(B-E');~~'=\frac{d}{d\tau};~~\varphi(\tau,{\bf x})=\varphi_0(\tau)+\delta \varphi(\tau,{\bf x})
\end{eqnarray}
are gauge invariant and contain all information about scalar perturbations. Since matter and metric perturbations are related through Einstein equation (\ref{LEE}), we can form a specific combination of these quantities known as curvature perturbation,
\begin{eqnarray}
 &&\zeta(\tau,{\bf x})\equiv \frac{\mathcal{H}}  {\varphi'_0} \tilde{\varphi}+\Phi=\frac{a}{z}\left( \tilde{\varphi}+\frac{\varphi'_0}{\mathcal{H}} \Phi  \right) \nonumber\\
 && z=a M_{pl} \sqrt{2\epsilon};~~\epsilon=-\frac{\dot{H}}{H^2}=1-\frac{\mathcal{H}'}{\mathcal{H}^2}=\frac{1}{2 M^2_{pl}}\frac{\varphi^{'2}_0}{\mathcal{H}^2},
 \label{CURP}
\end{eqnarray}
where $\epsilon$ is the slow roll parameter which can be treated as constant in the leading order;
 we used the Friedmann and acceleration equations\footnote{$d \tau=\frac{dt}{a};~~\mathcal{H}=\frac{a'}{a};~~H=\frac{a'}{a^2}=\frac{\mathcal{H}}{a};\\
\ddot{a}=\frac{a''}{a^2}-\frac{\mathcal{H}^2}{a};~~\dot{H}=\frac{\mathcal{H}'}{a^2}-\frac{\mathcal{H}^2}{a^2}$}.
We then expand the Einstein-Hilbert action using the perturbed metric,
\begin{equation}
S=S_0+S_2    
\end{equation}
where $S_0$ is the background action and $S_2$ is quadratic in perturbation\footnote{Linear term is absent as we carry out expansion about the solution of equations of motion. }. Interestingly, the perturbed action is expressed through the curvature perturbation, namely, $v(\tau,{\bf x})\equiv z \zeta$,
\begin{equation}
\label{ACTIONV}
 S_2= \frac{1}{2}\int{\left(v'^2-(\nabla v)^2+\frac{z''}{z} v^2\right)}d^4x  
\end{equation}
which is the action of a mass-less scalar field in the Minkowski background($z''=0$). The last term in the action represents the coupling of the field to a classical source due to non-trivial background. The equation of motion known as {\it Mukhanov-Sassaki} equation follows from the action
(\ref{ACTIONV}),
\begin{equation}
 v''-\nabla^2v-\frac{z''}{z}  v=0 
 \label{veq1}
\end{equation}
As in case of a free field in Minkowski space time,  $v$ is expanded in Fourier integral,
\begin{equation}
 v(\tau,{\bf x})=\frac{1}{(2\pi)^{3}}\int{\left(  v_k(\tau)a_{\bf k}e^{i{\bf k}.{\bf x}}+v^*_k(\tau) a^*_{\bf k}e^{-i{\bf k}.{\bf x}} 
 \right)d^3k}   
 \label{vF}
\end{equation}
transforming  (\ref{veq1}) into the equation for modes,
\begin{equation}
v''_k(\tau)+\left(k^2-\frac{z''(\tau)}{z(\tau)} \right)v_k(\tau)=0;~~\omega_k=k   
\label{veq2}
\end{equation}
which is an equation of parametric oscillator
of time dependent frequency which is determined by the geometry of background space time. For the sake of simplicity, we assume background space time to be de Sitter during inflation\footnote{ Conclusions remain  valid for the case of quasi de Sitter.} ($a(\tau)\sim \tau^{-1}$) such that $z''/z=-2/\tau^2\sim a^2$. Eq.(\ref{veq2}) has an exact solution, it is however, instructive to examine the solutions in different regimes of the parametric oscillator. (1) Sub Hubble regime: $k\gg aH$ or $k|\tau|\gg 1$ where (\ref{veq2}) reduces to an equation of simple harmonic oscillator,
\begin{equation}
 v_k(\tau)=\alpha_k\frac{e^{-ik\tau}}{\sqrt{2k}}+\beta_k    \frac{e^{ik\tau}}{\sqrt{2k}};~~k|\tau|\gg 1
\label{SUBH}
\end{equation}
where $\alpha_k~\&~ \beta_k$ are two integration constants.
In this limit, the length scales are arbitrarily small and space time mimics a  flat background. However, with evolution, length scales fast grow to the Hubble radius and space time curvature becomes important. (2) Supper Hubble regime : $k\ll aH$ or $k|\tau|\ll 1$ where (\ref{veq2}) goes over to, $v_k''/v_k=z''/z$ and has a following solution,
\begin{equation}
v_k(\tau)=A_k z+B_k z\int^\tau{\frac{d\tau_1}{z^2}};
~~k|\tau|\ll 1
\label{SH}
\end{equation}
in which the first term grows with evolution whereas the second has a decaying character\footnote{Indeed, $z\sim a\sim\tau^{-1}$ and conformal time diminishes  as the end of inflation is approached.}. The latter has an important implication, namely, $\zeta_k=v_k/z$ is constant on supper Hubble scales. One can match the two solutions (\ref{SUBH}) and (\ref{SH}) if $\alpha_k$ and $\beta_k$ are known. 
The limiting solutions of (\ref{vF}) conform to its exact solution,
\begin{equation}
 v_k(\tau)=\alpha_k \frac{e^{-ik\tau}}{2k} \left ( 1-\frac{i}  {k\tau}   \right)  +\beta_k\frac{e^{ik\tau}}{2k}\left ( 1+\frac{i}  {k\tau}   \right) 
\end{equation}
\begin{figure*}[htb!]
    	\centering
    {
       \includegraphics[width=12cm,height=9cm]{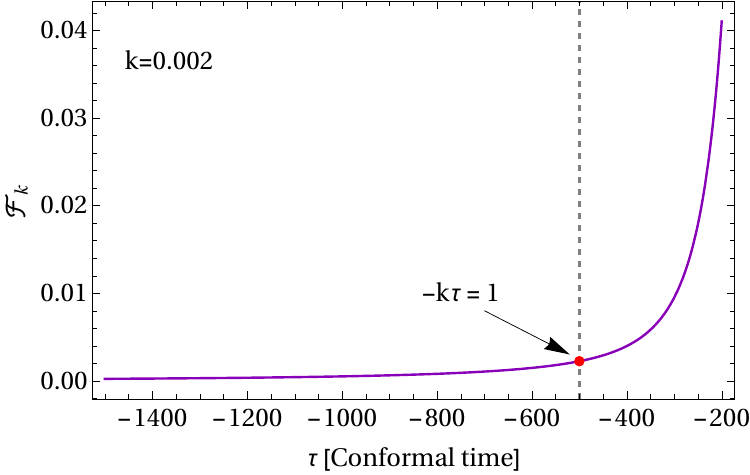}
    } 
    \caption[Optional caption for list of figures]{ Plot of $|\mathcal{F}_k|$ for a given $k$ ($k=0.002 Mpc^{-1}$) versus conformal time from large negative values of $\tau$ to $\tau\to 0$ where inflation ends. Plot shows that $|\mathcal{F}_k|$ remains close to zero in deep sub-Hubble regime ($-k\tau\ll 1$) but starts increasing  as $\tau$ moves towards Hubble crossing ($-k\tau=1$ ) and keeps growing thereafter.}
\label{ftau}
    \end{figure*}
There is no clue as to how to fix these two arbitrary constants $\alpha_k~\&~\beta_k$; no physics is known to accomplish this at the classical level. However, the quantization of perturbation with the assumption that it emerges from the ground state during inflation allows us to uniquely fix the integration constants. Indeed, quantization of $v(\tau,{\bf x})$ field implies that expansion coefficients in (\ref{vF}) are replaced by the annihilation and creation operators, $a_{\bf k}\to \hat{a}_{\bf k}$ and $a^*_{\bf k}\to \hat{a}^+_{\bf k}$ which obey the prescribed commutation relations and imply a condition for the mode function,
 \be 
 W[v_{ k},v^{'}_{ k}]:=\begin{vmatrix}
     v_{ k} & v^{*}_{ k} \\ 
     v^{'}_{ k} & v^{'*}_{ k} 
\end{vmatrix}=v^{'*}_{ k}v_{ k}-v^{'}_{ k}v^{*}_{k}=i.
\label{WOR}
\ee
The corresponding Hamiltonian operator is identical to (\ref{DH}) ,
\begin{eqnarray}
\label{DHQ}
 \hat{H}=\frac{1}{2}\int{\frac{d^3{\bf k}}{(2\pi)^3}\left  [\mathcal{F}^*_k \hat{a}_{\bf k}  \hat{a}_{\bf -k}+\mathcal{F}_k\hat{a}^+_{\bf k} \hat{a}^+_{-\bf k}+\mathcal{G}_k \hat{a}^+_{\bf k} \hat{a}_{\bf k}  \right] } 
\end{eqnarray}
The system under consideration reduces to simple harmonic oscillator for $\tau\to-\infty$ (see Eq.(\ref{SUBH})) where (\ref{WOR}) imply a condition on integration constants, 
\begin{equation}
|\alpha_k|^2-|\beta_k|^2  =1  
\label{CRC}
\end{equation}
We further assume that the system (field $v({\bf x},\tau)$) was in the ground state $|0\rangle$ of the simple harmonic oscillator (known as the Bunch-Davies vacuum) in the deep sub-Hubble regime or the perturbation originates from the vacuum state such that $\hat{H}|0\rangle=0$ in our setting. 
The first term in (\ref{DHQ}) annihilates the vacuum but the second term in general does not  thereby the coefficient of $a^+_{\bf k}a^+_{-\bf k}$ should vanish which yields a condition, 
\begin{equation}
\mathcal{F}_k\equiv {v'^2_ k+k^2 v^2_k}=0\Longrightarrow 4 \alpha_k \beta_k=0   
\end{equation}
We then demand that the Hamiltonian should take the form akin to a simple harmonic oscillator,
\begin{equation}
 \mathcal{G}_k\equiv 2( |{v'}_{ k}|^2 +k^2|v_{ k} |^2  ) =2k(|\alpha_k|^2+|\beta_k|^2)=2 k\Longrightarrow |\alpha_k|^2+|\beta_k|^2=1
\end{equation}
allowing us to finally fix the integration constants as, $\alpha_k=1$ and $\beta_k=0$, where we ignore
the constant phase factor in $\alpha_k$ which can always be absorbed in the redefinition of conformal time.

After fixing the integration constants, we finally have the unique solution for the mode function,
\begin{equation}
 v_k(\tau)=\frac{e^{-ik\tau}}{\sqrt{2k} } \left(1-\frac{i}{k\tau} \right)=-\frac{i}{k\tau}\frac{e^{-ik\tau}}{\sqrt{2k} }(1+ik\tau)
 \label{MODF}
\end{equation}
which in the sub-Hubble regime,
gives mode function specific to Bunch-Davies vacuum,
\begin{equation}
v_k(\tau) =\frac{e^{-ik\tau}}{\sqrt{2k} },~~-k\tau\to \infty  
\end{equation}
for which $\mathcal{F}_k(\tau)$ vanishes. 
 In the supper-Hubble regime (\ref{MODF}) yields,
\begin{equation}
    v_k(\tau)=i\frac{aH}{\sqrt{2k^3}},~~\zeta_k=\frac{v_k}{z}=\text{Const}~(z= a M_{pl} \sqrt{2\epsilon}),~~~-k\tau\to 0
    \label{SUPH}
\end{equation}
where we used the de Sitter value for the conformal time and the fact that $z\sim a$; in this case, $\mathcal{F}_k\neq 0$.} 

Since the background reduces to Minkowski space-time in the deep sub-Hubble regime, it allows us to define the Bunch-Davies vacuum there. In this case, $\mathcal{F}_k=0$ ($\beta_k=0$), and consequently, there is no particle production. However, with evolution, the system fast approaches the Hubble radius, making $\mathcal{F}_k\neq 0$ and giving rise to pair production\footnote{One can estimate density of these particles using the Bogoliubov transformation.}. Indeed using (\ref{MODF}), we have the expression of $\mathcal{F}_k$,
\begin{equation}
    \mathcal{F}_k = \frac{e^{-2 i k \tau } \left(\tau ^2 k^2 (k \tau -i)^2-(-1+k \tau  (k \tau -i))^2\right)}{2 k^3 \tau ^4};~~|\mathcal{F}_k |= k \frac{\sqrt{4 k^4 \tau ^4+1}}{2 k^4 \tau ^4}
\end{equation}
 We have plotted $|\mathcal{F}_k|$ versus $\tau$ for a given mode in Fig.\ref{ftau}    which clearly shows that this quantity vanishes in the deep sub-Hubble region and starts growing as $\tau$ moves towards the Hubble crossing; thereafter, $|\mathcal{F}_k(\tau)|\sim \tau^{-2}$ in the supper-Hubble regime($\tau\to 0$). This behaviour is consistent with solutions (\ref{SUBH}) and (\ref{SUPH}) in the asymptotic regions.

In view of the aforesaid, we are given a unique habitat in terms of the sub-Hubble region where vacuum is defined. This aspect is important for the computation of correlations; namely, in this case, we can define "in-in" correlation only. We previously demonstrated that the "in-in" correlator is equivalent to its "in-out" counterpart which is often used in the Minkowki space time. The framework is carried forward to the de Sitter background.
However, in that case, there is no natural "out" state, and we have no option other than to turning back to the "in" state (see Fig.\ref{sk}) and opt for the "in-in" formalism.

To conclude, the integration constants and the accompanying mode function are uniquely fixed by the assumption that perturbation has a quantum origin and emerges from the free vacuum state, which is not possible at the classical level.
  What a compelling case for quantization, in fact!

 Even if we believe in the quantum nature of perturbation, there is no  
{\it a priory }reason to assume that it originates from the ground state, it could well take off from an exited state of the simple harmonic oscillator with large number of quanta in it. In that case, perturbation is large and this might upset the inflation itself. Needless to mention that the number of such quantum state is infinite in number.
 Secondly, the large primordial perturbation is unwanted from observational point of view also. Indeed, the curvature perturbation responsible for density contrast can grow only after recombination where it is observed to be small $\hat{\rm a}$  {\it  la}  {\it Anthropic Principle}. 
 Another argument in support of the quantum nature of primordial perturbation is the following: Since perturbation is a random field, it should be accompanied by a probability distribution. On a simple note, one could assume it to be Gaussian, which agrees extremely well with observation. One could further assume an underlying stochastic process  (a classical stochastic variable) to produce the Gaussian distribution (or a deformed one to capture the additional non-Gaussian effects). Alternatively, one could think of perturbation as a quantum field as well, namely, { \it a free quantum field in vacuum, which is known to be a Gaussian random field and obeys Wick's theorem }.

In our opinion, the aforesaid is an unsettled issue and deserves attention.
To be rigorous, one should employ the method of the Wigner distribution function to conclude the classical/quantum nature of perturbations \cite{Martin:2022kph,Martin:2012ua,Martin:2019wta,Martin:2007bw,Martin:2015qta,Martin:2016tbd,Martin:2021znx,Polarski:1995jg,Lesgourgues:1996jc,Kiefer:1998qe,Choudhury:2016cso,Choudhury:2016pfr,Choudhury:2017bou,Choudhury:2017qyl,Choudhury:2018ppd,Choudhury:2022mch}. In this case, however, so far, one does not reach a definite conclusion using the positivity/negativity of the Wigner distribution function. The use of a classical stochastic variable might suffice, but observation cannot distinguish between the two approaches as the difference between their outcomes is incredibly small. It is not unfair to say that the { \it quantization of cosmological perturbations, as of now, is nothing but an efficient device to fix the integration constants, which are otherwise unknown, or a way to produce a desired probability distribution.}

\subsection{Quantum  correlations and power spectrum}

In this subsection, our objective is to compute the quantum correlation function of curvature perturbation and its associated power spectrum, which helps in making  connection between the underlying theory and observation.
To this effect, let us begin with the computation of the zero-point fluctuations by promoting $v(\tau,{\bf x})$ to  operator,
\bea \hat{v}(\tau,{\bf x})=\int \frac{d^3{\bf k}}{(2\pi)^3}\left[v_{ k}(\tau)\hat{a}_{\bf k}~e^{i{\bf k}.{\bf x}}+v^{*}_{ k}(\tau)a^{\dagger}_{{\bf k}}~e^{-i{\bf k}.{\bf x}}\right].\eea
Next we compute the expectation value of the operator $\hat{v}$ which turns out be:
\bea \langle \hat{v}\rangle&=&\langle 0| \hat{v}(\tau,{\bf x})|0\rangle
=\int \frac{d^3{\bf k}}{(2\pi)^3}\langle 0|\left[v_{ k}(\tau)\hat{a}_{\bf k}~e^{i{\bf k}.{\bf x}}+v^{*}_{ k}(\tau)a^{\dagger}_{{\bf k}}~e^{-i{\bf k}.{\bf x}}\right]|0\rangle=0
\eea
Nonetheless, non-zero quantum fluctuations are received by the variance of inflaton fluctuations:
\bea \langle \hat{v}\hat{v}\rangle&=&\langle 0| \hat{v}(\tau,{\bf x})\hat{v}(\tau,{\bf y})|0\rangle\nonumber\\
&=&\int \frac{d^3{\bf k}}{(2\pi)^3}\int \frac{d^3{\bf k}^{'}}{(2\pi)^3}\langle 0|\left[v_{ k}(\tau)\hat{a}_{\bf k}~e^{i{\bf k}.{\bf x}}+v^{*}_{ k}(\tau)a^{\dagger}_{{\bf k}}~e^{-i{\bf k}.{\bf x}}\right]\left[v_{{ k}^{'}}(\tau)\hat{a}_{{\bf k}^{'}}~e^{i{\bf k}^{'}.{\bf y}}+v^{*}_{{ k}^{'}}(\tau)a^{\dagger}_{{\bf k}^{'}}~e^{-i{\bf k}^{'}.{\bf y}}\right]|0\rangle 
\nonumber\\
&=&\int \frac{d^3{\bf k}}{(2\pi)^3}\int \frac{d^3{\bf k}^{'}}{(2\pi)^3}\langle 0|\left[v_{ k}(\tau)v_{{ k}^{'}}(\tau)\hat{a}_{\bf k}\hat{a}_{{\bf k}^{'}}~e^{i({\bf k}.{\bf x}+{\bf k}^{'}.{\bf y})}+v_{ k}(\tau)v^{*}_{{ k}^{'}}(\tau)a_{\bf k}a^{\dagger}_{{\bf k}^{'}}~e^{i({\bf k}.{\bf x}-{\bf k}^{'}.{\bf y})}\right.\nonumber\\
&&\left. \quad\quad\quad\quad\quad\quad\quad\quad\quad+v^{*}_{ k}(\tau)v_{{ k}^{'}}(\tau)a^{\dagger}_{{\bf k}}\hat{a}_{{\bf k}^{'}}~e^{-i({\bf k}.{\bf x}-{\bf k}^{'}.{\bf y})}+v^{*}_{ k}(\tau)v^{*}_{{ k}^{'}}(\tau)\hat{a}^{\dagger}_{\bf k}a^{\dagger}_{{\bf k}^{'}}~e^{-i({\bf k}.{\bf x}+{\bf k}^{'}.{\bf y})}\right]|0\rangle 
\nonumber\\
&=& \int \frac{d^3{\bf k}}{(2\pi)^3}\int \frac{d^3{\bf k}^{'}}{(2\pi)^3}v_{ k}(\tau)v^{*}_{{ k}^{'}}(\tau)\underbrace{\langle 0|\left[a_{\bf k},a^{\dagger}_{{\bf k}^{'}}\right]|0\rangle}_{=(2\pi)^3\delta^3({\bf k}-{\bf k}^{'})}~e^{i({\bf k}.{\bf x}-{\bf k}^{'}.{\bf y})}
\nonumber\\
&=& \int \frac{d^3{\bf k}}{(2\pi)^3}~|v_{ k}(\tau)|^2~e^{i{\bf k}.({\bf x}-{\bf y})}\nonumber\\
&=&\frac{1}{2}\int \int d\ln k~ sin\theta~d\theta~\frac{k^3}{2\pi^2}|v_{ k}(\tau)|^2~e^{ik|{\bf x}-{\bf y}|\cos\theta}\nonumber\\
&\approx&\frac{1}{8\pi^2\tau^2}\int \int d\ln k~ sin\theta~d\theta~e^{ik|{\bf x}-{\bf y}|\cos\theta}\nonumber\\
&\approx& \frac{1}{8\pi^2\tau^2}\ln\left(\frac{L_{\bf IR}}{|{\bf x}-{\bf y}|}\right),\eea
where $L_{\bf IR}$ is the IR cut-off of the momentum integral. Consequently, translating the entire calculation in terms of comoving curvature perturbation, the corresponding two-point function in coordinate and the Fourier space can be computed as:
\bea &&\langle \zeta(\tau,{\bf x})\zeta(\tau,{\bf y})\rangle=\left(\frac{H^2}{8\pi^2M^2_{pl}\epsilon}\right)_{*}\ln\left(\frac{L_{\bf IR}}{|{\bf x}-{\bf y}|}\right)\sim \left(\frac{H^2}{8\pi^2M^2_{pl}\epsilon}\right)_{*},\\
&& \langle \zeta_{\bf k}\zeta_{{\bf k}^{'}}\rangle=\langle \zeta_{\bf k}(\tau)\zeta_{{\bf k}^{'}}(\tau)\rangle=(2\pi)^3\delta^3({\bf k}+{\bf k}^{'})\frac{2\pi^2}{k^3}\Delta^2_{\zeta}(k),\eea
where we define the dimensionless power spectrum by the following expression:
\bea \Delta^2_{\zeta}(k)=\frac{k^3}{2\pi^2}|\zeta_{\bf k}|^2=\frac{k^3}{2\pi^2}\frac{|v_{\bf k}|^2}{z^2}=\frac{k^3}{2\pi^2}\frac{|v_{\bf k}|^2}{2a^2M^2_{pl}\epsilon}=\frac{H^2}{8\pi^2M^2_{pl}\epsilon}(1+k^2\tau^2)\xrightarrow{-k\tau\rightarrow 0} \left(\frac{H^2}{8\pi^2M^2_{pl}\epsilon}\right).\eea
 Scale invariance of the power spectrum is indicated by the fact that the correlation of curvature perturbation in coordinate spaceFor further information on this topic, 
is constant (up to a logarithmic  sensitivity to distance scale).
An important comment about the origin of perturbation is in order.
The assumption that perturbation originates from the ground state of harmonic  $|0\rangle$ dubbed free vacuum state allows to uniquely fix the integration constants. We could also think of an exited state with a number of quanta in it as an initial state. This situation can be captured by parametrization of $\alpha_k$ and $\beta_k$,
\bea
&& \alpha_k=\cosh \alpha\\
&& \beta_k=e^{i\delta} \sinh\alpha;~~\delta\in(-\pi,+\pi),~\alpha \in(0,\infty)
\eea
 consistent with (\ref{CRC}) and de Sitter symmetry assumed  during inflation; the Bunch-Davies vacuum corresponds to $\alpha=0$. The state under consideration is given by,
\bea
|\alpha,\delta\rangle &=&\prod_k{\frac{1}{\sqrt{|\alpha_k|}}\exp\left[-\frac{i}{2}\frac{\beta^*_{\bf k}}{\alpha^*_k}\hat{a}_{\bf k}^\dag \hat{a}_{-\bf k}^\dag \right] }|0\rangle\nonumber\\
&=&{\frac{1}{\sqrt{\cosh \alpha}}\prod_k\exp\left[-\frac{i}{2}e^{-i\delta}{\rm tanh}\alpha~\hat{a}_{\bf k}^\dag \hat{a}_{-\bf k}^\dag \right] }|0\rangle\nonumber\\
&=&{\frac{1}{\sqrt{\cosh \alpha}}\exp\left[-\frac{i}{2}e^{-i\delta}{\rm tanh}\alpha~\int\frac{d^3{\bf k}}{(2\pi)^3}~\hat{a}_{\bf k}^\dag \hat{a}_{-\bf k}^\dag \right] }|0\rangle
\eea
known as $\alpha$ vacuum\footnote{Actually, the state, $|\alpha,\delta\rangle$ is called in the literature the Allan-Motta vacuum and referred to as the $\alpha$ vacuum if $\delta=0$. For brevity, we call $|\alpha,\delta\rangle$ as  $\alpha$ vacuum, see Refs.                                                                                                                                                                                                                                                                                                                                                                                                                                                                                                                                                                                                                                                                                                                                                                                                                                                                                                                                                                                                                                                                                                                                                                                                                                                                                                                                                                                                                                                                                                                                                                                                                                                                                                                                                                                                                                                                                                                                                                                                                                                                                                                                                                                                                                                                                                                                                                                                                                                                                                                                                                                                                                                                                                                                                                                                                                                                                                                                                                                                                                                                                                                                                                                                                                                                                                                                                                                                                                                                                                                                                                                                                                                                                                                                                                                                                                                                                                                                                                                                                                                                                                                                                                                                                                                                                                                                                                                                                                                                                                                                                                                                                                                                                                                                                                                                                                                                                                                                                                                                                                                                                                                                                                                                                                                                                                                                                                                                                                                                                                                                                                                                                                                                                                                                                                                                                                                                                                                                                                                                                                                                                                                                                                                                                                                                                                                                                                                                                                                                                                                                                                                                                                                                                                                                                                                                                                                                                                                                                                                                                                                                                                                                                                                                                                                                                                                                                                                                                                                                                                                                                                                                                                                                                                                                                                                                                                                                                                                                                                                                                                                                                                                                                                                                                                                                                                                                                                                                                                                                                                                                                                                                                                                                                                                                                                                                                                                                                                                                                                                                                                                                                                                                                                                                                                                                                                                                                                                                                                                                                                                                                                                                                                                                                                                                                                                                                                                                                                                                                                                                                                                                                                                                                                                                                                                                                                                                                                                                                                                                                                                                                                                                                                                                                                                                                                                                                                                                                                                                                                                                                                                                                                                                                                                                                                                                                                                                                                                                                                                                                                                                                                                                                                                                                                                                                                                                                                                                                                                                                                                                                                                                                                                                                                                                                                                                                                                                                                                                                                                                                                                                                                                                                                                                                                                                                                                                                                                                                                                                                                                                                                                                                                                                                                                                                                                                                                                                                                                                                                                                                                                                                                                                                                                                                                                                                                                                                                                                                                                                                                                                                                                                                                                                                                                                                                                                                                                                                                                                                                                                                                                                                                                                                                                                                                                                                                                                                                                                                                                                                                                                                                                                                                                                                                                                                                                                                                                                                                                                                                                                                                                                                                                                                                                                                                                                                                                                                                                                                                                                                                                                                                                                                                                                                                                                                                                                                                                                                                                                                                                                                                                                                                                                                                                                                                                                                                                                                                                                                                                                                                                                                                                                                                                                                                                                                                                                                                                                                                                                                                                                                                                                                                                                                                                                                                                                                                                                                                                                                                                                                                                                                                                                                                                                                                                                                                                                                                                                                                                                                                                                                                                                                                                                                                                                                                                                                                                                                                                                                                                                                                                                                                                                                                                                                                                                                                                                                                                                                                                                                                                                                                                                                                                                                                                                                                                                                                                                                                                                                                                                                                                                                                                                                                                                                                                                                                                                                                                                                                                                                                                                                                                                                                                                                                                                                                                                                                                                                                                                                                                                                                                                                                                                                                                                                                                                                                                                                                                                                                                                                                                                                                                                                                                                                                                                                                                                                                                                                                                                                                                                                                                                                                                                                                                                                                                                                                                                                                                                                                                                                                                                                                                                                                                                                                                                                                                                                                                                                                                                                                                                                                                                                                                                                                                                                                                                                                                                                                                                                                                                                                                                                                                                                                                                                                                                                                                                                                                                                                                                                                                                                                                                                                                                                                                                                                                                                                                                                                                                                                                                                                                                                                                                                                                                                                                                                                                                                                                                                                                                 \cite{Allen:1985ux,Shukla:2016bnu,Choudhury:2017glj,Adhikari:2021ked,Choudhury:2018ppd,Choudhury:2017qyl} for more details. } which has non-vanishing particle number density in it\footnote{One may check that ${\mathcal{F}_k}$ which vanishes for Bunch-Davies, becomes large proportional to number density of particles in case of $\alpha$ vacuum.}. In the presence of $\alpha$ vacua, non-zero quantum fluctuations are received by the variance of inflaton fluctuations:
\bea \langle \hat{v}\hat{v}\rangle&=&\langle 0| \hat{v}(\tau,{\bf x})\hat{v}(\tau,{\bf y})|0\rangle\nonumber\\
&=&\langle 0| \hat{v}(\tau,{\bf x})\hat{v}(\tau,{\bf y})|0\rangle_{\alpha}-\langle 0| \hat{v}(\tau,{\bf x})\hat{v}(\tau,{\bf y})|0\rangle_{\rm BD}\nonumber\\\nonumber\\
&=&\int \frac{d^3{\bf k}}{(2\pi)^3}~\left(|v^{\alpha}_{ k}(\tau)|^2-|v^{\rm BD}_{ k}(\tau)|^2\right)~e^{i{\bf k}.({\bf x}-{\bf y})}\nonumber\\
&=&\frac{1}{8\pi^2\tau^2}\int \int d\ln k~ sin\theta~d\theta~\left(2|\beta_k|^2-\alpha^{*}_k\beta_k-\alpha_k\beta^{*}_k\right)~e^{ik|{\bf x}-{\bf y}|\cos\theta}\nonumber\\
&\approx&\frac{1}{8\pi^2\tau^2}\int \int d\ln k~\left(n_k-\sqrt{n_k(n_k+1)}\cos\delta\right) sin\theta~d\theta~e^{ik|{\bf x}-{\bf y}|\cos\theta}\nonumber\\
&\approx& \frac{1}{8\pi^2\tau^2}\left(n_k-\sqrt{n_k(n_k+1)}\cos\delta\right)\nonumber\\
&\approx& \frac{1}{8\pi^2\tau^2}\times2 n_k \sin^2 (\delta/2)\eea
where we have considered the large $\alpha$ limit and ignored the logarithmic term; $n_k\equiv|\beta_k|^2=\sinh^2 \alpha$
designates the particle number density in the $\alpha$ vacuum\footnote{$n_k\equiv<0| \hat{a}_k^{\dag (\alpha)}\hat{a}_k^{(\alpha)}|0>=|\beta_k|^2$}. 
Translating the entire calculation in terms of co-moving curvature perturbation, the corresponding two-point function in coordinate and the Fourier space are given by,
\bea &&\langle \zeta(\tau,{\bf x})\zeta(\tau,{\bf y})\rangle=\left(\frac{H^2}{8\pi^2M^2_{pl}\epsilon}\right)\times2 n_k \sin^2 (\delta/2),\\
&& \langle \zeta_{\bf k}\zeta_{{\bf k}^{'}}\rangle=\langle \zeta_{\bf k}(\tau)\zeta_{{\bf k}^{'}}(\tau)\rangle=(2\pi)^3\delta^3({\bf k}+{\bf k}^{'})\frac{2\pi^2}{k^3}\Delta^2_{\zeta}(k),\eea
The latter has important  implication for inflation. Indeed, if the perturbation originates from 
$\alpha$ vacuum,
\bea
\Delta^2_\zeta(k)=\left(\frac{H^2}{8\pi^2M^2_{pl}\epsilon}\right)\times|\cosh\alpha -e^{i\delta}\sinh\alpha|^2\simeq \left(\frac{H^2}{8\pi^2M^2_{pl}\epsilon}\right)\times 
2 n_k \sin^2 (\delta/2)
\label{POWERAL}
\eea
for large $\alpha$ (large $n_k$). The extra factor due to $\alpha$ vacuum in (\ref{POWERAL}), in general, is large if $n_k$ is large which has serious implication for inflation. In particular, the scale of inflation would be pushed down by COBE normalization
 which in turn  would severely constrain the equation of state of the inflaton. Needless to say that large perturbation might upset the inflation itself.

For the sake of completeness, let us also note that in the slow-roll limit,  
\bea H^2\approx\frac{V}{3M^2_{pl}}\quad\quad\quad {\rm where}\quad V\gg \frac{\dot{\phi}^2_0}{2},\eea
 the power spectrum can be expressed in terms of inflaton potential,
\bea \Delta^2_{\zeta}(k)=\left(\frac{1}{8\pi^2M^2_{pl}\epsilon}\frac{V}{3M^2_{pl}}\right)=\frac{V}{24\pi^2M^4_{pl}\epsilon}.\eea
and allows to compute the  spectral tilt for the scalar perturbation, 
\bea n_{\zeta}-1=\left(\frac{d\ln \Delta^2_{\zeta}(k)}{d\ln k}\right)=-4\epsilon +2\eta\approx -6\epsilon_{V}+2\eta_{V},\eea
where $\epsilon_V$ and $\eta_V$ are potential dependent slow-roll parameters, which are defined by the following expressions:
\bea \epsilon_V=\frac{M^2_{pl}}{2}\left(\frac{V^{'}}{V}\right)^2, \quad\quad {\rm and}\quad\quad \eta_V=M^2_{pl}\left(\frac{V^{''}}{V}\right).\eea
Additionally, the following relation connect Hubble slow-roll parameters $(\epsilon,\eta)$ with the potential dependent slow-roll parameters, $(\epsilon_V,\eta_V)$:
\bea \epsilon\approx \epsilon_V,\quad\quad \eta\approx \eta_V- \epsilon_V.\eea
Further, it is important to note that, from Planck data \cite{Planck:2018jri} we have the following CMB constraints available for the amplitude and spectral tilt of the power spectrum:
\bea \ln(10^{10}\Delta^2_{\zeta})=3.045 \pm 0.016, \quad\quad n_{\zeta}=0.9649\pm 0.0042~~  {\rm (both~at~ 68\%~ CL)},\eea
which needs to be satisfied by the inflationary models under consideration. 

Let us note that until now the power spectrum has been derived from the second-order perturbed action from the gauge-invariant comoving scalar curvature perturbation, and most importantly, it describes the tree-level power spectrum, which we frequently used in the later parts of this review. Further, 
the estimation
of one-loop effects and primordial non-Gaussian effects can be accomplished
perturbatively using the third-order  action by employing (\ref{PEO}),
\bea
\langle \mathcal{O}_{H}(\tau)\rangle=\langle \mathcal{O}_{H}(\tau)\rangle_{(0,0)}+\langle \mathcal{O}_{H}(\tau)\rangle_{(0,1)}+\langle \mathcal{O}_{H}(\tau)\rangle^{\dagger}_{(0,1)}+\langle \mathcal{O}_{H}(\tau)\rangle_{(0,2)}+\langle \mathcal{O}_{H}(\tau)\rangle^{\dagger}_{(0,2)}+\langle \mathcal{O}_{H}(\tau)\rangle_{(1,1)}+\cdots
\eea
where $\mathcal{O}_{H}(\tau)$ is Heisenberg operator. The presented estimation of tree-level power spectrum corresponds to 
$\langle \mathcal{O}_{H}(\tau)\rangle_{(0,0)}$ which is the only term term for quadratic action.
When we consider the impacts of the large fluctuations, which we introduce in the next section of this review, the third-order contributions play a very crucial role in enhancing the amplitude of the perturbations and explaining the formation of PBHs in the present context of the discussion. In the latter half of this review, we shall  explicitly discuss the physical implications of such higher-order terms in great detail. 

\section{Generation of large fluctuations} 
\label{s2}

In this section, our prime objective is to briefly indicate the mechanisms that generate large fluctuations in the cosmological scales, which are necessarily required to generate Primordial Black Holes (PBHs), primarily focused on in this review. Additionally, it is important to note that the generation of such large fluctuations produces an enhanced amplitude of scalar-induced gravitational waves (SIGWs), which has direct observational consequences for various gravitational wave detectors. We discuss such possibilities in detail in the later half of this review.

\subsection{Method-I: Using Sharp/Smooth transition}

Using sharp or smooth transitions, it is possible to generate large  fluctuations on the cosmological scale, which is the necessary ingredient for the formation of PBHs. Implementation of the sharp or smooth transitions within the framework of the cosmological perturbation theory setup comes with the insertion of an ultra-slow roll (USR) phase in the usual well-known single-field slow-roll (SR) inflationary paradigm. Such a transition is theoretically motivated by the inflection point inflationary paradigm\cite{Choudhury:2011jt,Choudhury:2013jya,Choudhury:2013iaa,Choudhury:2013woa,Choudhury:2014sxa,Choudhury:2014uxa} \footnote{In case of an effective potential $V(\phi)$, let us consider the inflection point, $\phi=\phi_0$ around which we  expand the effective potential in Taylor series, 
\bea V(\phi)=A+B\left(\phi-\phi_0\right)^3+C\left(\phi-\phi_0\right)^4+\cdots\eea
where the co-efficients $A$, $B$ and $C$ are given by the following expressions:
\bea A=V(\phi_0),\quad B=\frac{1}{3!}\left(\frac{d^3V(\phi)}{d\phi^3}\right)_{\phi=\phi_0}\quad C=\frac{1}{4!}\left(\frac{d^4V(\phi)}{d\phi^4}\right)_{\phi=\phi_0}.\eea
To implement the inflection point, $\phi=\phi_0$, in the effective potential, the following constraints need to be maintained:
\bea \left(\frac{dV(\phi)}{d\phi}\right)_{\phi=\phi_0}=0\quad\quad{\rm and}\quad\quad \left(\frac{d^2V(\phi)}{d\phi^2}\right)_{\phi=\phi_0}=0.\eea
For a given explicit structure of the effective potential $V(\phi)$ one can be able to explicitly compute the expression for the inflection field value $\phi_0$ and the coefficients $A$, $B$, $C$ in terms of the model parameters.} which can be engineered with the help of a single field effective potential. 
During inflation, the background energy density reduces very slowly which is quantified by the first slow roll parameter,
\bea
\epsilon\equiv -\frac{\dot{H}}{H^2}\ll 1\to V\gg\frac{\dot{\phi}^2}{2}
\eea
The requirement that inflation be sustained for stipulated duration in-order to address the causality problem puts a restriction on the second slow roll parameter,
\bea
\eta\equiv \frac{\dot{\epsilon}}{\epsilon H}
\eea
which crucially depends whether the field potential is flat or very flat, namely, $V_\phi=0$. 
In what follows, we give a broad description of inflationary phases, intending to define a prescribed background setting where the dynamics of perturbation will be studied in the forthcoming sections.

    \begin{figure*}[htb!]
    	\centering
    	\subfigure[Potential with point of inflection.]{
      	\includegraphics[width=8cm,height=7cm] {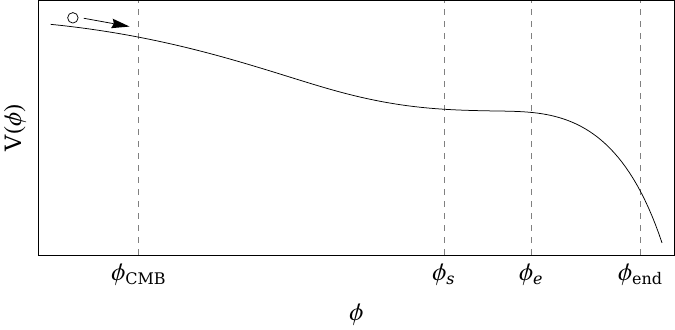}
        \label{Gc1}
    }
    \subfigure[Potential with bump.]{
       \includegraphics[width=8cm,height=7cm] {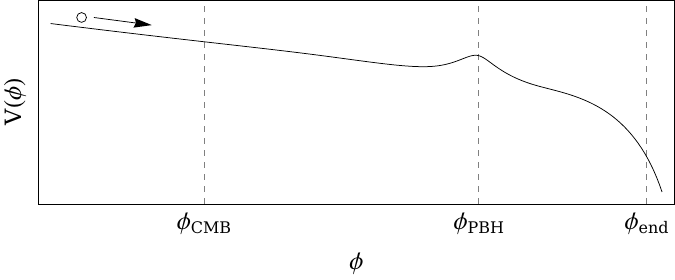}
        \label{Gc2}
       }
        \subfigure[Scalar power spectrum with USR and bump.]{
       \includegraphics[width=12cm,height=7cm] {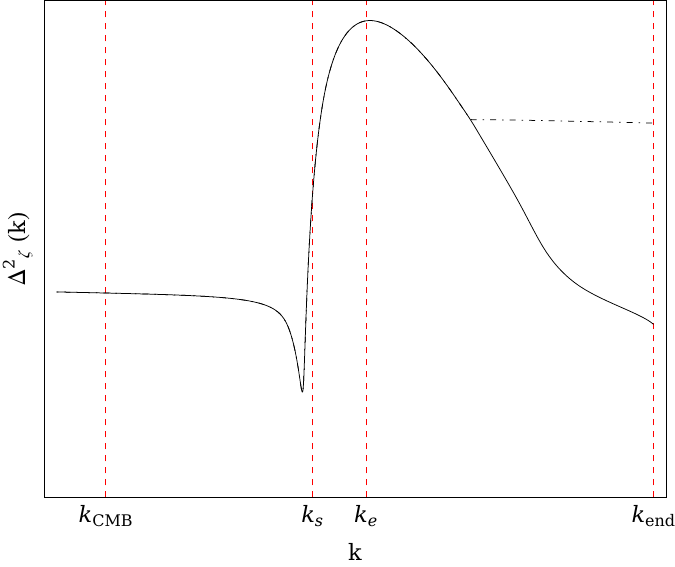}
        \label{Gc3}
       }    
    	\caption[Optional caption for list of figures]{Schematic plot of the effective potential in (\ref{Gc1}) with point of inflection which has a USR region between $\phi_s$ and $\phi_e$ where the potential is very flat, $V_\phi=0$. Field rolls down its flat potential (SRI) from a field location little before the scale relevant to 
     CMB exits the Hubble radius at $\phi=\phi_{CMB}$. The USR is followed by SRII which extends to $\phi=\phi_{end}$ where the inflation terminates.
      Fig.(\ref{Gc2}) shows the effective potential with a bump at a chosen location where $\phi=\phi_{PBH}$ corresponds to the top of the bump.  
     Fig.\ref{Gc3} is a cartoon plot that displays general features of the scalar power spectrum versus the co-moving wave number $k$. The power spectrum remains constant during SRI, enhances during USR ($k_s\leq k\leq k_e$),
     drops thereafter in SRII and approaches a constant value shown by dotted line whereas it drops to the base value it had started with in SRI in the case of a bump.
     } 
    	\label{Pow}
    \end{figure*}
\begin{itemize}[label=\ding{212}]

\item \underline{\bf First slow-roll phase (SRI):}This phase is characterized 
by $|\epsilon|, |\eta|<<1$. 
To determine the explicit behaviour of the mode functions, we choose Bunch Davies vacuum state as an initial condition. Then, the {\it Mukhanov-Sasaki} equation, 
has a unique solution that yields,
\bea
 \zeta_{ k}(\tau)&=&\frac{v_{ k}(\tau)}{zM_{ pl}}=\left(\frac{iH}{2M_{\rm pl}\sqrt{\epsilon}}\right)\frac{1}{k^{3/2}}\left(1+ik\tau\right)\; e^{-ik\tau}.\quad
 \eea
 Let us reiterate that  perturbation freezes at supper-Hubble scales($-k\tau\to 0$) whose amplitude is fixed by COBE normalization.
 We further note that, to write down the above-mentioned expression, we assumed  canonical single field models of inflation where the scalar field is minimally coupled to gravitational sector at the background level. Hence the above-mentioned expression and henceforth the other derived expressions in this section do not include effective sound speed parameter $c_s$. In the later part of the review, we have further generalized this framework to Effective Field Theory (EFT) of single field inflation which captures both canonical as well as non-canonical scenarios, commonly designated as $P(X,\phi)$ models of inflation where all the derived expressions contain the effective sound speed parameter $c_s$\cite{Choudhury:2023vuj,Kristiano:2022maq,Riotto:2023hoz,Kristiano:2023scm,Riotto:2023gpm}. 
 
\item \underline{\bf Ultra slow-roll phase (USR):} From the dynamical point of view, this phase is characterized by the first slow-roll parameter, $\epsilon$, being small, and the second slow-roll parameter, \bea 
\label{etusr}
\eta=\frac{\d,,,,,,,,,,,,,,,,,,,,,,,,,,,,,,,,,,,,,,,,,,,,,,,,,,,,,,,,,,,,,,,,,,,,,,,,,,,,,,,,,,,,,,,,,,,,,,,,,,,,,,,,,,,,,,,,,,,,,,,,,,,,,,,,,,,,,,,,,,,,,,,,,,,,,,,,,,,,,,,,,,,,,,,,,,,,,,,,,,,,,,,,,,,,,,,,,,,,,,,,,,,,,,,,,,,,,,,,,,,,,,,,,,,,,,,,,,,,,,,,,,,,,,,,,,,,,,,,,,,,,,,,,,,,,,,,,,,,,,,,,,,,,,,,,,,,,,,,,,,,,,,,,,,,,,,,,,,,,,,,,,,,,,,,,,,,,,,,,,,,,,,,,,,,,,,,,,,,,,,,,,,,,,,,,,,,,,,,,,,,,,,,,,,,,,,,,,,,,,,,,,,,,,,,,,,,,,,,,,,,,,,,,,,,,ot{\epsilon}}{H\epsilon}=2\left(\frac{\ddot{\phi}}{H\dot{\phi}}+\epsilon\right)=-6-\frac{2V_\phi}{H\dot{\phi}}+2\epsilon\sim {\cal O}(-6),\eea
where the potential is very flat, $V_\phi=0$ and acceleration 
term can not be neglected. An example of this is provided by a potential  with point of inflection, $V_\phi=V_{\phi\phi}=0$ where, $V_\phi$ and $V_{\phi\phi}$ denote the first and second derivative of the potential with respect to the field, see Fig.\ref{Gc1}.

In the simple setting to be followed, a short USR phase shall be inserted in SR regime
such that there are  three phases, SRI, USR and SRII followed by the end of inflation. 
The USR phase commences at $\tau=\tau_s$ where  one implements a sharp or smooth transition at a given scale. The USR phase persists within the domain, $\tau_s\leq \tau\leq \tau_e$, where $\tau_e$ represents the conformal time scale corresponding to the end of the USR phase which is followed by SRII. It follows from (\ref{etusr}) that the first slow roll parameter is not constant in the USR phase, 
\bea
\epsilon(\tau)=\epsilon \left(\frac{\tau}{\tau_s} \right)^6\propto a^{-6}
\eea
 where $\epsilon$ is the slow-roll parameter in SRI. Using this information, the expression for the gauge invariant scalar modes can be obtained by solving the {\it Mukhanov-Sasaki} equation as:
\bea
\label{zeetasr2}
 \zeta_{ k}(\tau)&=&\left(\frac{iH}{2M_{ pl}\sqrt{\epsilon}}\right)\left(\frac{\tau_s}{\tau}\right)^{3}\frac{1}{k^{3/2}}\times\bigg[\alpha^{(2)}_{ k}\left(1+ik\tau\right)\; e^{-ik\tau}-\beta^{(2)}_{ k}\left(1-ik\tau\right)\; e^{ik\tau}\bigg],
 \eea
 where Bogoliubov coefficients $\alpha^{(2)}_{ k}$ and  $\beta^{(2)}_{ k}$ characterize the new structure of the corresponding vacuum, which is shifted from Bunch Davies vacuum due to having the sharp or smooth transition. Using the continuity and differentiability of the corresponding modes at the transition points~\footnote{We use, $[\zeta(\tau=\tau_s)]_{\bf SRI}=[\zeta(\tau=\tau_s)]_{\bf USR}$ and $[\zeta^{'}(\tau=\tau_s)]_{\bf SRI}=[\zeta^{'}(\tau=\tau_s)]_{\bf USR}$ at the transition scale $\tau_s$.}, the Bogoliubov coefficients $\alpha^{(2)}_{k}$ and  $\beta^{(2)}_{ k}$ can be computed as:
 \bea \label{b2aq}\alpha^{(2)}_{ k}&=&1-\frac{3}{2ik^{3}\tau^{3}_s}\left(1+k^{2}\tau^{2}_s\right),\\
\label{b2bq}\beta^{(2)}_{ k}&=&-\frac{3}{2ik^{3}\tau^{3}_s}\left(1+ik\tau_s\right)^{2}\; e^{-2ik\tau_s}.\eea

It follows from (\ref{zeetasr2}) that perturbation does not remain frozen in the regime under consideration, it enhances with evolution, $\zeta_k\sim a^3$. Clearly, this regime should be very short otherwise perturbation could grow beyond the validity of perturbation theory. , if the initial velocity of the field is insufficient to overcome the ultra-slow roll region ($\dot{\phi}\sim a^{-3}$), the field may become trapped there, leading to an eternal inflation.

\item \underline{\bf Second slow-roll phase (SRII):} In this phase around the transition from USR, the second slow roll parameter increases to a small negative value and settles there and the first slow roll parameter gradually increases in this phase \cite{Franciolini:2023agm}. 
  In the conformal time scale, SRII phase starts at $\tau=\tau_e$ and the duration of this phase is $\tau_e\leq \tau\leq \tau_{\rm end}$, where $\tau_{\rm end}$ describes the scale corresponding to the end of inflation and SRII phase. In the SRII phase the conformal time dependence of the first slow-roll parameter is known. Using this information the expression for the gauge invariant scalar modes can be obtained by solving the {\it Mukhanov-Sasaki} equation as:
\bea
 \zeta_{ k}(\tau)&=&\left(\frac{iH}{2M_{ pl}\sqrt{\epsilon}}\right)\left(\frac{\tau_s}{\tau_e}\right)^{3}\frac{1}{k^{3/2}}\times\bigg[\alpha^{(3)}_{ k}\left(1+ik\tau\right)\; e^{-ik\tau}-\beta^{(3)}_{ k}\left(1-ik\tau\right)\; e^{ik\tau}\bigg],
 \eea
 where Bogoliubov coefficients $\alpha^{(3)}_{ k}$ and  $\beta^{(3)}_{ k}$ characterize the new structure of the corresponding vacuum, which is shifted from the previous vacuum due to having the sharp or smooth transition from USR to SRII phase. Using the continuity and differentiability of the corresponding modes~\footnote{ We use the conditions: $[\zeta(\tau=\tau_e)]_{\bf SRII}=[\zeta(\tau=\tau_e)]_{\bf USR}$ and $[\zeta^{'}(\tau=\tau_e)]_{\bf SRII}=[\zeta^{'}(\tau=\tau_e)]_{\bf USR}$ at the scale $\tau_e$ where the USR ends and SRII phase starts.}, the Bogoliubov coefficients $\alpha^{(3)}_{ k}$ and  $\beta^{(3)}_{ k}$ can be computed as:
 \bea \label{b3aQ}\alpha^{(3)}_{ k}&=&-\frac{1}{4k^6\tau^3_s\tau^3_e}\Bigg[9\left(k\tau_s-i\right)^2\left(k\tau_e+i\right)^2 e^{2ik(\tau_e-\tau_s)}\nonumber\\
&&\quad\quad\quad\quad\quad\quad\quad\quad\quad\quad\quad\quad\quad\quad\quad\quad-
\left\{k^2\tau^2_e\left(2k\tau_e-3i\right)-3i\right\}\left\{k^2\tau^2_s\left(2k\tau_s+3i\right)+3i\right\}\Bigg],\\
\label{b3bQ}\beta^{(3)}_{ k}&=&\frac{3}{4k^6\tau^3_s\tau^3_e}\Bigg[\left(k\tau_s-i\right)^2\left\{k^2\tau^2_e\left(3-2ik\tau_e\right)+3\right\}e^{-2ik\tau_s}\nonumber\\
&&\quad\quad\quad\quad\quad\quad\quad\quad\quad\quad\quad\quad\quad\quad\quad\quad+i\left(k\tau_e-i\right)^2\left\{3i+k^2\tau^2_s\left(2k\tau_s+3i\right)\right\}e^{-2ik\tau_e}\Bigg].\eea
\end{itemize}
We illustrate the benefits and ramifications of integrating a sudden change in direction in our arrangement with the USR. Heaviside Theta functions $\Theta(\tau-\tau_{s})$ and $\Theta(\tau-\tau_{e})$, which are used at the exact point of transition between $\tau_{s}$ and $\tau_{e}$, allow to execute the sharp transition throughout. The purpose of this function is to show the characteristics of the $\eta$ parameter, which is subsequently utilised to produce the features that are evident for the $\epsilon$ parameter. We select a certain parameterization for the $\eta$, 
 \bea
\eta(\tau)  
&=& \eta_{\rm SRI}(\tau_{*} < \tau < \tau_{s}) + \eta_{\rm USR}(\tau_{s} < \tau < \tau_{e}) - \Delta\eta(\tau)\{\Theta(\tau-\tau_{s}) - \Theta(\tau-\tau_{e})\} + \eta_{\rm SRII}(\tau_{e} < \tau < \tau_{\rm end}).
\eea
Upon taking the conformal time derivative of this $\eta$ parameterization at the transition moments, we may derive the following:
\bea
\bigg(\eta(\tau)\bigg)' &=& \underbrace{\bigg(\eta_{\rm SRI}(\tau_{*} < \tau < \tau_{s})\bigg)'}_{=0} + \underbrace{\bigg(\eta_{\rm USR}(\tau_{s} < \tau < \tau_{e})\bigg)'}_{=0} \nonumber\\
&& \quad\quad\quad\quad\quad\quad\quad\quad\quad\quad - \Delta\eta(\tau)\{\delta(\tau-\tau_{s}) - \delta(\tau-\tau_{e})\} + \underbrace{\bigg(\eta_{\rm SRII}(\tau_{e} < \tau < \tau_{\rm end})\bigg)'}_{=0},
\eea
where a conformal time derivative is indicated by the prime notation. In the case of a smooth transition, each of the Heaviside functions is replaced by the tan hyperbolic functions, and modified expressions for the second slow roll parameter and its conformal time derivative are given by:
\bea
\eta(\tau)
&=& \eta_{\rm SRI}(\tau_{*} < \tau < \tau_{s}) + \eta_{\rm USR}(\tau_{s} < \tau < \tau_{e}) - \Delta\eta(\tau)\left\{{\rm tanh}\left(\frac{\tau-\tau_{s}}{\Delta\tau}\right)- {\rm tanh}\left(\frac{\tau-\tau_{e}}{\Delta\tau}\right)\right\}\nonumber\\
&&\quad\quad\quad\quad\quad\quad\quad\quad\quad\quad\quad\quad\quad\quad\quad\quad\quad\quad\quad\quad\quad+ \eta_{\rm SRII}(\tau_{e} < \tau < \tau_{\rm end}),\\
\bigg(\eta(\tau)\bigg)' &=& - \frac{\Delta\eta(\tau)}{\Delta\tau}\left\{{\rm sech}^2\left(\frac{\tau-\tau_{s}}{\Delta\tau}\right)- {\rm sech}^2\left(\frac{\tau-\tau_{e}}{\Delta\tau}\right)\right\}.
\eea
where, $\Delta\tau\equiv|\tau_e-\tau_s|$. In the above two expressions once we take the limit $\Delta\tau\rightarrow 0$, then these above-mentioned two expressions computed for smooth transition will approach the results obtained using the sharp transition. 
In both scenarios, with smooth or sharp transitions, the conformal time derivative of the second slow-roll parameter plays a very crucial role in determining the one-loop correction to the two-point and three-point correlation functions. In the later half of this review, we discuss such implications in detail. Meanwhile, we summarise the implication of the insertion of a short ultra-slow roll phase in the standard slow-roll regime. The power spectrum remains constant in SRI ($\epsilon(\tau)$=const) but enhances during USR phase where $\epsilon(\tau)\sim a^{-6}$. It drops as it enters the SRII and then becomes constant, see Fig.\ref{Gc3}. Using the prescribed prescription, one can produce a large fluctuation that would collapse to PBH provided the size of the fluctuation exceeds a critical value.
The aforementioned lays out the inflationary background, where detailed investigations of the dynamics of perturbations taking into account the loop corrections will be discussed in the forthcoming sections.

Before proceeding ahead, let us note that the inflaton with the point of inflection is not the only way to generate large perturbations; we could mimic the same effect by introducing a bump (dip) in the inflationary potential(Fig.\ref{Gc2}), which depresses the first slow roll parameter during the field journey from the bottom to the top of the bump. In this process, the power spectrum would first enhance and then drop to its base value it had started with, see Fig.\ref{Gc3}. Needless to say that process of PBH formation is constrained by various  considerations, in particular, avoidance of eternal inflation and validity of perterbative framework.

\subsection{Method-II: Using dip/bump in inflationary potential}

The generic structure of the inflationary effective potential in the presence of dip/bump is described by the following expression \cite{Mishra:2019pzq}:
\bea V(\phi)=V_{b}(\phi)\left[1+g(\phi)\right],\eea
where $V_b(\phi)$ is the fundamental inflationary potential that produces quantum fluctuations that are consistent with the restrictions imposed by the CMB. Additionally, it is important to note that, the phenomenological term $g(\phi)$ signifies the tiny dip/bump which has a width $\sigma$. To insert such a tiny dip/bump one needs to also have $g(\phi)\ll 1$ in the present context of the discussion. Depending on the underlying structure of the base effective potential the mathematical form of the function $g(\phi)$ is chosen in this scenario. The accurate position of such a tiny dip/bump is described by the field value $\phi_0$.

To keep things simple, we will use a concave base potential that is asymptotically flat and locally modified by a Gaussian bump. In the refs. \cite{Mishra:2019pzq,Atal:2019cdz}, such Gaussian bump is identified with a speed breaker, which is parametrized by the following functional form:
\bea\label{pwe1} g(\phi)=A\exp\left[-\frac{1}{2}\frac{(\phi-\phi_0)^2}{\sigma^2}\right].\eea
In the above-mentioned expression height, width, and position of the Gaussian bump are characterized by the symbols, $A$, $\sigma$, and $\phi_0$ respectively.  In ref \cite{Mishra:2019pzq}, the authors have also mentioned the possibility of having another simplified profile described by the following phenomenological functional form:
\bea\label{pwe2} g(\phi)=A {\rm sech}^2\left(\frac{(\phi-\phi_0)}{\sigma}\right).\eea
Both of the profiles mentioned in equations (\ref{pwe1}) and (\ref{pwe2}), technically mimics the role of smooth transition as mentioned in the previous method. Once we take the limit, $\sigma\rightarrow 0$, the result converges to the case of sharp transition as demonstrated before. Comparing with the previously mentioned method one can easily infer that position of the bump $\phi_0$ exactly mimics the role of inflection point where smooth/sharp transition is implemented.








\section{PBH formation}
\label{s3}
 \subsection{Press Sechter formalism}

 To explain PBH generation in an age of constant EoS $w$, we opt to deal with the traditional technique of threshold statistics. Using this approach, the primordial density fluctuations must meet a threshold requirement on the perturbation over-density for them to gravitationally collapse and produce PBHs. This is where the amplitude of scalar power  comes in, and this section will go into more detail about it. Working with the Press-Schechter formalism, which has been altered to include the constant EoS $w$, will be our main emphasis. 

 The mass that was present in the Horizon at its development is still proportionate to the mass of the produced PBH. Nevertheless, the perturbation over-densities must meet the threshold requirement $\delta\rho/\rho \equiv \delta>\delta_{\rm th}$ in order to start the formation. We decide to utilise $c_{s}^{2}=1$ as Carr's criterion. We consider ref. \cite{1975ApJ...201....1C}, which provides the threshold connection as follows:
\bea
\label{deltathX}
\delta_{\rm th} = \frac{3(1+w)}{5+3w}.
\eea
Furthermore, we assume that the density contrast in the super-Horizon regime and the co-moving curvature perturbation are approximately linear:
\bea \label{deltalinearX}
\delta(t,\mathbf{x}) \cong \frac{2(1+w)}{5+3w}\left(\frac{1}{aH}\right)^{2}\nabla^{2}\zeta(\mathbf{x}).
\eea
In refs. \cite{Ferrante:2022mui,Franciolini:2023pbf,Franciolini:2023wun}, PBH production in the presence of non-linearities in the aforementioned connection is analysed. In the way illustrated in \cite{Alabidi:2013lya}, $w$ modifies the produced PBHs final mass.
\bea \label{mpbhX}
M_{\rm PBH} = 1.13 \times 10^{15} \times \bigg(\frac{\gamma}{0.2}\bigg)\bigg(\frac{g_{*}}{106.75}\bigg)^{-1/6}\bigg(\frac{k_{*}}{k_{\rm s}}\bigg)^{\frac{3(1+w)}{1+3w}} M_{\odot}, \eea
where $k_{*}=0.02{\rm Mpc^{-1}}$ denotes the pivot scale value, $M_{\odot}$ denotes the solar mass, and $\gamma \sim 0.2$ represents the efficiency factor of collapse. We need to estimate the variation in the primordial overdensity distribution in order to get the PBH abundance. One way to compute this variance is as follows: 
\bea
\sigma_{\rm M_{\rm PBH}} ^2 = \bigg(\frac{2(1+w)}{5+3w}\bigg)^2 \int \frac{dk}{k} \; (k
R)^4 \; W^2(kR) \;\Delta^{2} _{\zeta }(k).\eea
the importance of the amplitude $A$ of the scalar power spectrum is evident. Our numerical results for abundance detailed in the next sections show that the change in the amplitude $A$ is extremely sensitive to the variance estimations.

Here, $W(kR)=\exp{(-k^2 R^2 /4)}$ represents the Gaussian smoothing function across the PBH formation scales, $R=1/k_{s}$. The acceptable threshold regime where the collapse of perturbations, creating a significant abundance of PBH, is attained based on the initial form of the power spectrum is limited by the assumption of working with the linear relation in equation (\ref{deltalinearX}). Numerous numerical studies have been conducted on this regime, and the results show that $2/5 \leq \delta_{\rm th} \leq 2/3$ \cite{Musco:2020jjb}. We examine the interval $-5/9 \leq w \leq 1/3$ in terms of $w$. We want to utilise this approximation to identify the feasible regime that contributes to the production of the targeted PBH abundance and the induced gravitational wave signature consistent with the latest NANOGrav15 output. According to \cite{Sasaki:2018dmp}, the mass fraction of PBHs is now as follows:
\bea
\beta(M_{\rm PBH})= \gamma \frac{\sigma_{\rm M_{\rm PBH}}}{\sqrt{2\pi}\delta_{\rm th}}\exp{\bigg(-\frac{\delta_{\rm th}^2}{2\sigma_{\rm M_{\rm PBH}}^2}\bigg)},
\eea
Additionally, the mass fraction now includes the mass and $w$ dependency resulting from the variance. Since the mass fraction is a function of Gaussian statistics for $\delta$, our decision to ignore any non-linear contributions in the density contrast is reflected in it. After that, the phrase is used to write the PBHs' current abundance:
\bea
f_{\rm PBH} \equiv \frac{\Omega_{\rm PBH}}{\Omega_{\rm CDM}}= 1.68\times 10^{8} \bigg(\frac{\gamma}{0.2}\bigg)^{1/2} \bigg(\frac{g_{*}}{106.75}\bigg)^{-1/4} \left(M_{\rm PBH}\right)^{-\frac{6w}{3(1+w)}}\times \beta(M_{\rm PBH}),\eea
where the relativistic degrees of freedom are denoted by $g_{*}=106.75$. It should be noted that $f \simeq 1.6\times 10^{-15}(k/{\rm Mpc}^{-1})$ connects the frequency and wavenumber. To calculate the abundance estimate and the permissible range for the EoS $w$ that can still give a sizable abundance after staying within the numerically allowed range of the threshold, we employ the scalar power spectrum.

 \subsection{Compaction function approach}

 When using the threshold statistics with the compaction function, the computations that must be done are explained in this section. Double the excess mass over the areal radius is the definition of this function, which was first presented by Shibata and Sasaki in \cite{Shibata:1999zs}:
\bea \label{compaction}
{\cal C}(r,t)&=& 2 \times\bigg(\frac{M(r,t)-M_{o}(r,t)}{R(r,t)} \bigg)=\frac{2}{R(r,t)}\int_{S_{R}^{2}} d^{3}\Vec{x} \rho_o(t)\delta(\Vec{x},t),
\eea
where $M_{o}(r,t)=\rho_{0}(t)V_{o}(r,t)$, where $V_{o}(r,t)=(4\pi/3)R(r,t)^{3}$, represents the background mass relative to the energy density $\rho_{o}(t)$, and $M(r,t)$ represents the Misner-Sharp mass inside the areal radius sphere $S_{R}^{2}$ of areal radius, $R(r,t)\equiv a(t)r\exp{(\zeta(r))}$, expressed in comoving radial coordinates. Using the long-wavelength approximation (also called the gradient expansion) on such super-horizon scales, one can now accommodate the non-linearities from the conserved curvature perturbation, $\zeta(r)$, into the density contrast field $\delta(\vec{x},t) = \delta\rho(\vec{x},t)/\rho_o(t)$ after analysing the perturbations well beyond the cosmological horizon. The aforementioned non-linearities in the radial coordinates for a generic backdrop with the equation of state $w$ originate from the formula \cite{Musco:2018rwt,Harada:2015yda}:
\bea \label{NLdensity}
\delta(r,t) &=& -\frac{2u}{3}\left(\frac{1}{aH}\right)^{2}e^{-5\zeta(r)/2}\nabla^{2}e^{\zeta(r)/2}\approx -\frac{2u}{3}\frac{1}{(aH)^2}e^{-2\zeta(r)} \bigg[\zeta^{''}(r) + \frac{2}{r}\zeta^{'}(r) + \frac{1}{2}\zeta^{'}(r)^{2}\bigg],
\eea
where a prime denotes a radial derivative and $u=3(1+w)/5+3w$. The analysis carried out in this letter is not specific to any one component; rather, it may be used to any equation of state, including Galileon, throughout the range $0 \leq w \leq 1$, which includes $w = \{0,1/3,1\}$, which describes the kination, radiation, and matter-dominated eras, respectively. The following formula clearly shows that NGs will always be present, independent of the statistics of the curvature perturbation $\zeta(r)$ \cite{Young:2019yug, DeLuca:2019qsy}. The explicit non-linear effects are introduced by the Laplacian in radial coordinates and the exponential terms. Using equation (\ref{compaction}) compaction function obtains the new time-independent definition by the use of the non-linear equation (\ref{NLdensity}):
\bea \label{compaction2}
{\cal C}(r)=-2ur\zeta'(r)\bigg(1+\frac{r}{2}\zeta'(r)\bigg).
\eea
The following is a breakdown of the linear and non-linear parts of the aforementioned equation:
\bea
{\cal C}(r)={\cal C}_{\rm L}(r)-{\cal C}_{\rm NL}(r)\quad\quad\quad{\rm where}\quad\quad{\cal C}_{\rm L}(r)=-2ur\zeta'(r),\quad\quad{\rm and}\quad\ {\cal C}_{\rm NL}(r)={\cal C}_{\rm L}(r)^2/4u.
\eea
According to equation (\ref{compaction2}), the criteria for PBH formation now arises from maximising the compaction function at the scale $r_{m}$. This allows for the fulfilment of the following condition: 
\bea \label{maxcond}
\zeta'(r_{m})+r_{m}\zeta''(r_{m})=0.
\eea
Through the conventional relation in terms of the related wavenumbers, $\tilde{c_{s}}k_{H} = r_{m}^{-1}$, that source the PBH producing perturbation during horizon re-entry (crossing), the same comoving scale $r_{m}$ may also be connected to the Horizon mass $M_{H}$:
\bea
M_H \approx 17M_{\odot}\bigg[\frac{g_{*}}{10.75}\bigg]^{-1/6}\bigg[\frac{\tilde{c_{s}}k_{H}}{10^6 \rm Mpc^{-1}}\bigg]^{-2}.
\eea
Hence, in the current situation, $k_{H}$ corresponds to the transition scale $k_{s}$ into the USR area, and $g^{*}=106.75$ represents the relativistic d.o.f in the radiation-dominated period. In order to ensure that $\delta \ll 1$, we use $\tilde{c_{s}} \approx 1\pm\delta$ to add the effective sound speed at the transition scale. Let us now additionally look at the surplus mass volume-averaged across an area of radius $R$. At the scale $r_{m}$ and the accompanying re-entry time $t_{H}$, this quantity has a unique relationship with the compaction function at that scale:
\bea \label{reentryeql} 
\delta_{m} = \frac{\delta M(r,t)}{M_{o}(r,t)} \equiv \frac{3}{R_{m}}\int^{R_{m}}_{0}\frac{\delta\rho}{\rho_{o}}R^{2}dR = {\cal C}(r_{m}) = 3\delta(r_{m},t_{H}),
\eea
Here, the physical length scale of the perturbation during re-entry is denoted by $R_{m} = a(t_{H})r_{m}e^{\zeta(r_{m})}$. When equation (\ref{NLdensity}), equation (\ref{maxcond}), and the Horizon crossing condition $R_{m}H(t_{H})=1$ are used, the last two equalities can only be realised while taking equation (\ref{compaction2}) into consideration.
A noteworthy conclusion is provided by equation (\ref{reentryeql}), which makes use of the compaction function upon horizon re-entry to directly introduce the volume-averaged density contrast, $\delta_{m}$.

At the associated scale $r_{m}$ and perturbation mode $k_{H}$, current estimates for the volume-averaged density contrast on superhorizon scales, obtained through numerical analysis, result in the interval, $2/5 \leq \delta_{\rm th} \leq 2/3$, where under the first approximation, that is ignoring higher-order effects in the gradient expansion approach, is said to provide a reasonable outcome for the perturbation amplitude during horizon re-entry \cite{Musco:2018rwt, Musco:2020jjb}. To get the above interval for $\delta_{\rm th}$, it is crucial to keep in mind that in the aforementioned studies, only the NG effects resulting from the non-linearities from equation (\ref{NLdensity}) are considered. Additionally, it is assumed that the curvature perturbation only follows Gaussian statistics. The impacts of quadratic primordial NGs on the threshold were examined in a recent research \cite{Kehagias:2019eil}, and very slight alterations were seen. We can therefore ascertain the range of the threshold values ${\cal C}_{\rm th}$ directly using the equality in equation (\ref{reentryeql}).

This also occurs inside equation (\ref{compaction2}), which we now state as follows as we are contemplating a generic formula for the curvature perturbation $\zeta(r)$ that contains inside the primordial NG information:
\bea
{\cal C}(r) = {\cal C}_{G}(r)\frac{d\zeta}{d\zeta_{G}} - \frac{1}{4u}\left({\cal C}_{G}(r)\frac{d\zeta}{d\zeta_{G}}\right)^{2},
\eea 
using the formula ${\cal C}_{G}(r)= -2ur\zeta'_{G}(r)$.
First, based on the threshold of the compaction function ${\cal C}_{\rm th}$, one must require the condition ${\cal C} \geq {\cal C}_{\rm th}$ in order to determine the conditions under which a cosmological perturbation undergoes collapse to create PBH. Although this is the case, maximising the compaction function establishes an additional boundary condition that gives us the appropriate range of values: 
\bea
{\cal C}'(r)&=& {\cal C}_{\rm L}'(r)-{\cal C}_{\rm NL}'(r)=
 {\cal C}_{\rm L}'(r)-\frac{d}{dr}\bigg(\frac{{\cal C}_{\rm L}(r)^2}{4u}\bigg)=
 {\cal C}_{\rm L}'(r) - \frac{1}{2u}\bigg({\cal C}_{\rm L}(r)\;{\cal C}_{\rm L}'(r)\bigg)=0\implies {\cal C}_{\rm L}(r) = 2u.
\eea
The second boundary condition may be satisfied by finding a double derivative, which yields an affirmative result of ${\cal C}_{\rm L}(r) \le 2u$ or $ {\cal C}_{G}(d\zeta/d\zeta_{G}) \leq 2u$.
Consequently, following the application of threshold statistics to the compaction function, the whole domain for integrating the PBH mass fraction is as follows: ${\cal D} = \{{\cal C}(r) \geq {\cal C}_{th} \wedge {\cal C}_{G}(d\zeta/d\zeta_{G}) \leq 2u\}$. 

Currently, the relation \cite{Ferrante:2022mui, Franciolini:2023pbf} provides the overall percentage of dark matter contained in PBHs after integrating across a range of horizon masses ($M_{H}$): 
\bea \label{fpbh} f_{\rm PBH} &=& \frac{1}{\Omega_{\rm DM}}\int d\;\ln{M_{H}}\left(\frac{M_{H}}{M_{\odot}}\right)^{-\frac{1}{2}} \times\left(\frac{g_{*}}{106.75}\right)^{\frac{3}{4}}\left(\frac{g_{*s}}{106.75}\right)^{-1}\left(\frac{\beta_{\rm NG}(M_{H})}{7.9 \times 10^{-10}}\right),
\eea
where $g_{*},\;g_{*s}$ denote the effective energy and entropy degrees of freedom, and $\Omega_{\rm DM} \simeq 0.264$ denotes the dark matter density of the universe. The creation of PBH is recognised to exhibit exponential sensitivity to the tail of the density fluctuations' Probability Distribution Function (PDF), where non-Gaussian effects are significant \cite{DeLuca:2022rfz, Taoso:2021uvl,Atal:2018neu, Young:2013oia,Byrnes:2012yx,Bullock:1996at}. Currently, the two Gaussian random variables, $\cal{C}_G$ and $\zeta_G$, have non-zero auto and cross-correlations, which result in the two-dimensional joint PDF:
\bea
   \label{PDF1}
 P_G({\cal C}_G,\zeta_G) &=& \frac{1}{2\pi\sigma_c \sigma_{r}\sqrt{1-\gamma_{\rm cr}^2}}\exp{\left(-\frac{\zeta_G ^2}{2\sigma_r ^2}\right)}\times \exp{\bigg[-\frac{1}{2(1-\gamma_{\rm cr}^2)}\bigg(\frac{{\cal C}_G}{\sigma_c}-\frac{\gamma_{\rm cr}\zeta_G}{\sigma_r}\bigg)^2\bigg]},
   \eea
   Through the aforementioned non-linear connection, the further incorporation of primordial NGs in the conserved curvature perturbation random field $\zeta(r)$ would subsequently convert into the density contrast field. In this study, we concentrate on the quadratic NG model, denoted by the well-known ansatz:
\bea \zeta = \zeta_{G} + \frac{3}{5}f_{\rm NL}\zeta_{G}^{2}, \eea
where the quantity of NG in the theory is measured by $f_{\rm NL}$, and Gaussian statistics are followed by $\zeta_{G} \equiv \zeta_{G}(r)$. The correlation coefficient is represented as $\gamma_{cr}=\sigma^{2}_{cr}/(\sigma_{c}\sigma_{r})$. Using the threshold statistics on the compaction function as a guide, this PDF computes the necessary mass fraction of PBHs \cite{Ferrante:2022mui}:
\bea   
\label{Beta}
\beta_{\rm NG}(M_{H}) = \int_{\cal D}{\cal K}({\cal C}-{\cal C}_{\rm th})^{\gamma}P_{G}({\cal C}_G, \zeta_G) d{\cal C}_G\;d\zeta_{G}, \eea
where for the PBH mass generated during horizon re-entry, the \textit{critical scaling relation} is included through ${\cal K}({\cal C}-{\cal C}_{\rm th})^{\gamma}$ \cite{Choptuik:1992jv, Evans:1994pj}. The simulations yielded $\gamma \sim 0.36$ for the RD era and values for the constant ${\cal K}\sim {\cal O}(1-10)$ and the threshold ${\cal C}_{\rm th}$, which are based on \cite{Musco:2020jjb}. 

We now present the definitions for the different correlations found in the 2D joint PDF \cite{Franciolini:2023pbf,Young:2022phe}:

\bea \label{sigcr}
\sigma_{cr}^{2} &=& \frac{2u}{3} \int_{0}^{\infty}\frac{dk}{k}(k r_{m})^{2}W_{g}(k,r_{m})W_{s}(k,r_{m})\Delta^{2}_{\zeta}(k),\quad
\\
\label{sigc} \sigma_{c}^{2} &=& \left(\frac{2u}{3}\right)^{2}\int_{0}^{\infty}\frac{dk}{k}(k r_{m})^{4}W_{g}^{2}(k,r_{m})\Delta^{2}_{\zeta}(k),
\\
\label{sigr} \sigma_{r}^{2} &=& \int_{0}^{\infty}\frac{dk}{k}W^{2}_{s}(k,r_{m})\Delta^{2}_{\zeta}(k),
\eea
where it is decided to use the Gaussian smoothing functions $W_{g}(k,r)$ and $W_{s}(k,r)$, i.e., $\exp{(-k^{2}r^{2}/2)}$. This is often not done, particularly when using the so-called spherical-shell window function $W_{s}(k,r)$. This is a good decision for our theory since using the top-hat or other sinusoidal functions creates huge oscillations that are ineffective at adequately smoothing out the small-scale disturbances. Utilising the smoothing characteristic of the radiation transfer function $T(k,\tau) = 3(\sin{l}-l\cos{l})/l^{3}$, with $l=k\tau/\sqrt{3}$ and $\tau=1/aH$, we established the new power spectrum form as $\tilde{\Delta}^{2}_{\zeta}(k) = T^{2}(k,r_{m})\Delta^{2}_{\zeta}(k)$.
Note that the horizon re-entry scale, $r_{m} = (\tilde{c_{s}}k_{H})^{-1}$, corresponds to the wavenumber during PBH formation and the scale where the compaction function maximises. This is a crucial point to remember when evaluating equations (\ref{sigcr}-\ref{sigr}) discussed in the later part of this review when applied for a specific model.

The occurrence of a USR phase during inflation makes NG a key component of the comoving curvature perturbation. Currently, the tiny values of the slow-roll parameters inhibit the NG corrections at the CMB scales. But one major contribution to the NGs is the creation of a USR phase. Here, we go into further detail on the implications of departing from the Gaussian limit when taking the local type of primordial NGs into account in the curvature perturbations. We are specifically thinking about the quadratic parameterization of the perturbation in terms of the Gaussian random variable $\zeta_{G}$, which is $\zeta = \zeta_{G} + 3/5 {f_{\rm NL}}\zeta_{G}^{2}$. 

Commencing with the previously addressed 2D joint PDF in equation (\ref{PDF1}), we inquire as to what modifications the departure from a completely correlated situation (basically, using a Dirac delta type of distribution) brings to the outcomes of the primordial abundance of black holes. After Taylor expanding the current PDF around the value of $\gamma_{cr}=1$, between the $2$ Gaussian variables, ${\cal C}_{G},\zeta_{G}$, we obtain the following expression:
\bea \label{taylorPg}
P_{G}({\cal C}_{G},\zeta_{G}) &=& \frac{1}{2 \sqrt{2} \pi \sigma_{c}\sigma_{r} \sqrt{1-\gamma_{cr}}}\exp \bigg(\displaystyle{\frac{1}{4 (\gamma_{cr}
   -1)}\left(\frac{{\cal C}_{G}}{\sigma_{c}}-\frac{\zeta_{G}}{\sigma_{r}}\right)^{2}+\sum_{n=3}^{\infty}\left(\frac{-1}{2}\right)^{n}\left(\frac{{\cal C}_{G}}{\sigma_{c}}+\frac{\zeta_{G}}{\sigma_{r}}\right)^{2}(\gamma_{cr}-1)^{n-3}}\bigg)\nonumber\\
   && \times\left(\displaystyle{1+\frac{1-\gamma_{cr}}{4}+\frac{3 (1-\gamma_{cr})^{2}}{32} +\frac{5 (1-\gamma_{cr})^{3}}{128}+\frac{35 (1-\gamma_{cr})^{4}}{2048}+\frac{63 (1-\gamma_{cr})^{5}}{8192}+\frac{231 (1-\gamma_{cr})^{6}}{65536}+\cdots}\right), \quad\quad\quad\eea
where the higher-order elements in the Taylor expansion are represented by the $\cdots$. The distribution above simplifies to the following under the specific limit $\gamma_{cr} \rightarrow 1$:
\bea \label{peakpdf}
\lim_{\gamma_{cr}\rightarrow 1}P_{G}({\cal C}_{G},\zeta_{G}) &=& \frac{1}{\sqrt{2\pi}\sigma_{c}\sigma_{r}}\exp{\left(-\frac{1}{8}\left(\frac{{\cal C}_{G}}{\sigma_{c}}+\frac{\zeta_{G}}{\sigma_{r}}\right)^{2}\right)}\lim_{\gamma_{cr}\rightarrow 1}\frac{1}{\sqrt{2\pi(2(1-\gamma_{cr}))}}\exp{\left(\frac{1}{4(\gamma_{cr}-1)}\left(\frac{{\cal C}_{G}}{\sigma_{c}}-\frac{\zeta_{G}}{\sigma_{r}}\right)^{2}\right)},\nonumber\\
&=&\frac{1}{\sqrt{2\pi}\sigma_{c}\sigma_{r}}\exp{\left(-\frac{1}{8}\left(\frac{{\cal C}_{G}}{\sigma_{c}}+\frac{\zeta_{G}}{\sigma_{r}}\right)^{2}\right)}\delta\left(\frac{{\cal C}_{G}}{\sigma_{c}}-\frac{\zeta_{G}}{\sigma_{r}}\right).
\eea
Here, the remaining quadratic expression is reduced to a sharp or highly correlated distribution function by the Dirac delta function, which enforces the condition ${\cal C}_{G}/\sigma_{c}=\zeta_{G}/\sigma_{r}$. This must be the case since the correlation coefficient $\gamma_{cr}$ reaches its maximum value of $\gamma_{cr}=1$. The series expansion fails at smaller values of $\gamma_{cr}$. The expansion in equation (\ref{taylorPg}) is only viable when the distribution is near being strongly correlated and is indefinitely differentiable, which appears to be the case here.

In light of the peaked distribution found in equation (\ref{peakpdf}), its share of the overall mass fraction may be expressed as follows: 
\bea
\beta_{\rm peak}(M_{H}) &=& \frac{1}{\sqrt{2\pi}\sigma_{c}\sigma_{r}}\int_{\cal D}{\cal K}({\cal C}-{\cal C}_{\rm th})^{\gamma}\exp{\left(-\frac{1}{8}\left(\frac{{\cal C}_{G}}{\sigma_{c}}+\frac{\zeta_{G}}{\sigma_{r}}\right)^{2}\right)}\delta\left(\frac{{\cal C}_{G}}{\sigma_{c}}-\frac{\zeta_{G}}{\sigma_{r}}\right) d{\cal C}_G\;d\zeta_{G},\nonumber\\
&=& \frac{1}{\sqrt{2\pi}\sigma_{r}}\int_{\cal D}{\cal K}\left\{\left[g(\zeta_{G})-\frac{1}{4u}g(\zeta_{G})^{2}\right]-{\cal C}_{\rm th}\right\}^{\gamma}\exp{\left(-\frac{\zeta_{G}^{2}}{2\sigma_{r}^{2}}\right)}d\zeta_{G}.
\eea
The Dirac delta function allows one to get an integral over the variable $\zeta_{G}$. For convenience, we define the function $g(\zeta_{G})=\displaystyle{\frac{\sigma_{c}}{\sigma_{r}}\zeta_{G}\frac{d\zeta}{d\zeta_{G}}}$. The remaining constants, $\{{\cal K}, u, {\cal C}_{\rm th}, \gamma\}$, are the same as they were when the joint PDF was initially established. The PBH mass fraction further permits the calculation of the associated fractional abundance after integration over various horizon masses, therefore designating the potential epochs of black hole creation.
The following equation is used to determine the abundance that corresponds to the peak distribution:
\bea
f^{\rm peak}_{\rm PBH} \equiv \frac{\Omega^{\rm peak}_{\rm PBH}}{\Omega_{\rm DM}}=\frac{1}{\Omega_{\rm DM}}\int d\ln{M_{H}}\left(\frac{M_{H}}{M_{\odot}}\right)^{-1/2}\left(\frac{g_{*}}{106.75}\right)^{3/4}\left(\frac{g_{*s}}{106.75}\right)^{-1}\left(\frac{\beta_{\rm peak}(M_{H})}{7.9 \times 10^{-10}}\right).
\eea
In the above, $M_{H} \equiv M_{H}(t)= (4\pi/3)\rho(t)/H^{3}(t)$ is the definition of the time-dependent horizon mass contained within the Universe of radius $1/H(t)$ at cosmic time $t$. The dark matter density of the Universe is represented by $\Omega_{\rm DM}\simeq 0.264$, while $\rho(t)$ is the total energy density of the Universe at that moment. Getting the largest contribution to the overall abundance in the computation above is the responsibility of the Gaussian PDF. The remaining terms in the PDF expansion that deviate from Gaussian nature contribute less to the total mass fraction and, as a result, yield an estimate of the fractional abundance that varies less dramatically, particularly for the correlation coefficient values, $\gamma_{cr}$, in our instance. Compared with the primordial NGs, which determine the domain and the scalar power spectrum amplitude, which is present inside the variances and leads to an exponential sensitivity to the fractional abundance, this analysis indicates that the fractional abundance is not highly sensitive to other terms in the PDF expansion.

 \section{Scalar Induced Gravity Waves (SIGWs)}
\label{s4}
The theory of scalar induced gravitational waves (SIGW) produced in the presence of a generic cosmic background with a constant EoS $w$ is covered in this section. After establishing the appropriate mathematical framework and examining the underlying theoretical setting, we apply the Galileon theory scalar power spectrum and analyse the resultant GW spectrum. Since we are investigating this particular theory for the first time and will be incorporating its findings towards the main objective of this study, our purpose in this section is to convey the theoretical background in a self-consistent manner.

 \subsection{The underlying framework}

Due to their capacity to explain primordial universe events that are beyond the reach of current observational methods like the BBN and CMB, gravitational waves (GW) have garnered significant interest in recent research. As an example, the CMB anisotropies bear the marks of the primordial oscillations. That data is available on a broader scale, though, and it provides inadequate details on the later phases of inflation. This is where the physics of GW becomes important. It can provide information about the latter phases of inflation and allow for investigation into the early universe, even before the Big Bang nucleosynthesis. These induced GWs may now be examined from several theoretical perspectives, including inflationary scenarios, domain barriers, cosmic strings, and first-order phase transitions, to mention a few \cite{Choudhury:2023hfm,Bhattacharya:2023ysp,Franciolini:2023pbf,Inomata:2023zup,Wang:2023ost,Balaji:2023ehk,HosseiniMansoori:2023mqh,Gorji:2023sil,DeLuca:2023tun,Choudhury:2023kam,Yi:2023mbm,Cai:2023dls,Cai:2023uhc,Huang:2023chx,Vagnozzi:2023lwo,Frosina:2023nxu,Zhu:2023faa,Jiang:2023gfe,Cheung:2023ihl,Oikonomou:2023qfz,Liu:2023pau,Liu:2023ymk,Wang:2023len,Zu:2023olm, Abe:2023yrw, Gouttenoire:2023bqy,Salvio:2023ynn, Xue:2021gyq, Nakai:2020oit, Athron:2023mer,Ben-Dayan:2023lwd, Madge:2023cak,Kitajima:2023cek, Babichev:2023pbf, Zhang:2023nrs, Zeng:2023jut, Ferreira:2022zzo, An:2023idh, Li:2023tdx,Blanco-Pillado:2021ygr,Buchmuller:2021mbb,Ellis:2020ena,Buchmuller:2020lbh,Blasi:2020mfx, Madge:2023cak, Liu:2023pau, Yi:2023npi,Gangopadhyay:2023qjr,Vagnozzi:2020gtf,Benetti:2021uea,Inomata:2023drn,Lozanov:2023rcd,Basilakos:2023jvp,Basilakos:2023xof,Li:2023xtl,Domenech:2021ztg,Yuan:2021qgz,Chen:2019xse,Cang:2023ysz,Cang:2022jyc,Konoplya:2023fmh,Huang:2023chx}. We pay particular attention to GWs that are called Scalar Induced Gravitational Waves (SIGWs) that are caused by the mode coupling between scalar perturbations. 
The scalar perturbations must be greatly amplified at the smaller scales compared to the CMB scale in order to generate a sizable abundance of the SIGWs at observationally relevant scales. Most of the research has been done with the assumption that induced GWs emerge during the RD period. Still, there may be opportunities for them to be produced in other eras as well. Therefore, it is worthwhile to investigate, as we have done in the context of this work, the generation of SIGWs for a broad EoS parameter $w$.

In the next sections, we calculate the tensor power spectrum amplitude formula for second-order cosmological perturbation theory SIGWs. The spatially flat FLRW metric expressed in the transverse-traceless gauge will be our starting point.
\bea
ds^2 = a^2(\tau)[-(1+2\Phi)d\tau^2 + (\del_{ij}\left(1-2\Psi\right)+2h_{ij})dx^i dx^j],
\eea
where the scalar potentials are $\Phi$ and $\Psi$, and the scale factor is represented by $a(\tau)$. The tensor modes in linear order, denoted by $h_{ij}$ in the equation above, are understood to be the primordial gravitational waves. 
Assuming that the universe's matter is represented by a perfect fluid, we may express its energy-momentum tensor as follows:
\bea
T_{\mu \nu} =(\rho+P)u_{\mu}u_{\nu}+Pg_{\mu \nu},
\eea
where $g_{\mu \nu}$ is our spacetime metric tensor and $u^{\nu}$ is the four-velocity of the fluid under consideration. The characteristic equation of state for this fluid, $w=P/\rho$, is the main topic of study in this work. The scalar and tensor modes in perturbation theory exhibit mixing at the second order. To get the induced GWs in the second order, we must solve the first-order equations of motion. Therefore, SIGWs are the solutions that are found and serve as a source for the GW.

The analysis of gravitational scalar potentials and GWs when studied in the leading order is briefly mentioned below. When there are no anisotropies, it is demonstrated that both values fulfil the following equations:
\bea
\Phi'' + \frac{6(1+w)}{\tau(1+3w)}\Phi' + wk^{2}\Phi = 0, \quad\quad h''_{\lambda} + \frac{4}{\tau(1+3w)}h'_{\lambda} + k^{2}h_{\lambda} = 0.
\eea
for the two polarisation modes $\lambda=+,\times$ to occur. The following are the answers to the aforementioned equations \cite{Domenech:2021ztg,Domenech:2019quo}:
\bea \label{O1scalartensor}
\Phi(x) = (\sqrt{w}x)^{-s}\left[C_{1}(k)J_{s}(\sqrt{w}x) + C_{2}(k)Y_{s}(\sqrt{w}x)\right], \quad\quad 
h_{\lambda}(x) = x^{-t}\left[C_{1}(k)J_{t}(x) + C_{2}(k)Y_{t}(x)\right]
\eea
where the first and second class Bessel functions are denoted by $x\equiv k\tau$ and $\{J_{\alpha},Y_{\alpha}\}$, respectively. The $w$-dependent form of the indexes $s,t$ is as follows:
\bea
s = \frac{5+3w}{2(1+3w)}, \quad\quad\quad t = \frac{3(1-w)}{2(1+3w)},
\eea
such that, when modes reach sub-Horizon, the solutions stated contain the effects of damped oscillations. Later on, the equivalence $1+t=s$ will be useful, and $t$ is not to be mistaken with any time variable. In order to comprehend the general solution for the tensor modes at the second order, these solutions will thereafter be crucial. We now go over the scenario of caused GW.  

We begin with the equation of motion for the tensor modes derived at second order in perturbation theory in Fourier space:
\bea \label{tensoreqn}
h''_{\lambda} + 2 {\cal H}h'_{\lambda}+k^2 h_{\lambda} = S_{\lambda}(\mbf{k}),
\eea
where $S_{\lambda}(\mbf{k})$ signifies the source term contribution which is denoted by:
\bea \label{source}
S_{\lambda} = 4 \int \frac{d^3q}{(2\pi)^3}e_{\lambda}^{ij}(k)q_{i}q_{j}\left\{\Phi_{\mbf{q}}\Phi_{\mbf{k-q}} + \frac{1+b}{2+b}\bigg[\Phi_{\mbf{q}}+\frac{\Phi'_{\mbf{q}}}{{\cal H}}\bigg]\;\bigg[\Phi_{\mbf{k-q}}+\frac{\Phi'_{\mbf{k-q}}}{{\cal H}}\bigg]\right\},
\eea
where $b$ is a $w$-dependent combination determined by:
\bea 
b= \frac{1-3w}{1+3w},
\eea
and $\Phi_{\mbf{q}}$ are the Fourier components of the scalar potential.

The polarisation tensors of GWs provided by are also available:
\bea
e_{ij}^{+}(\mbf{k})= \frac{1}{\sqrt{2}}\bigg(e_{i}(\mbf{k})e_{j}(\mbf{k})-\bar{e}_{i}(\mbf{k})\bar{e}_{j}(\mbf{k})\bigg),\quad\quad
e_{ij}^{\times}(\mbf{k})= \frac{1}{\sqrt{2}}\bigg(e_{i}(\mbf{k})\bar{e}_{j}(\mbf{k})+\bar{e}_{i}(\mbf{k})e_{j}(\mbf{k})\bigg),
\eea
for the two polarisation modes.  The tensor modes $h_{\lambda}(\tau)$ may be solved using the Green's function approach, which can be obtained from equation (\ref{tensoreqn}) as follows:
\bea
h_{\lambda}(\tau) = \int_{\tau_i} ^{\tau} d\Tilde{\tau}\;{\cal G}(\tau,\Tilde{\tau})S_{\lambda}(\Tilde{\tau}).
\eea

The initial conditions at time $\tau_i$ are given along with the solution, which follows the formula $h_{\lambda}(\tau_i)= h_{\lambda}^{'}(\tau_i)=0$. 
The last objective is to determine the SIGW power spectrum expression, which requires the two-point correlation function of the tensor modes, which is given as follows:
\bea
\big<h_{\mbf{k}}(\tau)h_{\mbf{k'}}(\tau)\big> = \int_{0}^{\tau}\;d\tau_1\;\int_{0}^{\tau}\;d\tau_2 \;{\cal G}(\tau,\tau_1)G(\tau,\tau_2)\big<S_{\lambda}(k,\tau_1)S_{\lambda}(k',\tau_{2})\big>.
\eea
This necessitates further understanding of the source term's two-point correlation function, which may be obtained by computing (with the exception of any non-Gaussian characteristics in the primordial power spectrum):
\bea
\langle S_{\lambda}(\mbf{k},\tau_{1})S_{\lambda}(\mbf{k'},\tau_{2})\rangle &=& 16 \int \frac{d^3 q}{(2\pi)^3} \int \frac{d^{3}q'}{(2\pi)^3} e_{\lambda}^{ij}(\mbf{k})q_{i}q_{j}e_{\lambda}^{ij}(\mbf{k'})q'_{i}q'_{j} \times f(\tau_{1},q,k)\; f(\tau_{2}, q', k')\langle\phi_{\mbf{q}}\phi_{\mbf{k-q}}\phi_{\mbf{q'}}\phi_{\mbf{k'-q'}}\rangle,
\quad\quad\quad\eea
where the splitting of the Fourier modes of scalar potential is implemented into the transfer function $\Phi(q\tau)$ and the primordial fluctuations $\phi_{\mbf{q}}$. Utilising the primordial scalar power spectrum and learning about the scalar potential's development are made possible by such a divide. The function $f(\tau,q,k)$ denotes the source function found in equation (\ref{source}), expressed in terms of the gravitational potential's transfer function that was previously discussed. The appropriate Wick contractions between the scalar fluctuations in the above RHS must be taken in order to fully calculate the tensor power spectrum. The dimensionless tensor power spectrum that we employ in this context is defined as follows: 
\bea
\langle h_{k}(\tau)h_{k'}(\tau)\rangle = \frac{2\pi^{2}}{k^3}\Delta^{2}_{h}(k,\tau)\;\delta^3\left(\mbf{k+k'}\right).
\eea
The primordial power spectrum of the induced GW may now be expressed as follows by using the two relations mentioned above:
\bea \label{tensorpspec}
\Delta^{2}_{h}(k,\tau)&=& 4\int_{0}^{\infty}\;dv\int_{|1-v|}^{1+v}\;du\;\bigg(\frac{4v^2-(1+v^2-u^2)^2}{4uv}\bigg)^2\; {\cal I}^2 (u,v,x)\;\Delta^{2}_{\zeta}(ku)\;\Delta^{2}_{\zeta}(kv), \eea
where the following expressions define the new variables $u$ and $v$ that we have introduced:
\bea v\equiv \frac{q}{k},  \quad \quad \quad u \equiv \frac{|\mbf{k-q}|}{k}.
\eea
Here, the kernel is represented by the function ${\cal I}$, which we will address in more depth in the next section. Here, we give the formulation of the kernel in terms of the source and Green's functions:
\bea \label{kernel}
{\cal I}(u,v,x)= 2\bigg(\frac{3+3w}{5+3w}\bigg)^2 \int_{0}^{x} d\Tilde{x}\;{\cal G}(x,\Tilde{x})f(\Tilde{x},u,v),
\eea
using the equation (\ref{O1scalartensor}) to find the Green's function ${\cal G}$ and the source function $f$, which reduces to:
\bea \label{simplG}
{\cal G}(x,\Tilde{x}) &=& \frac{\pi \Tilde{x}^{s}}{2x^{t}}[J_{t}(\Tilde{x})Y_{t}(x)-Y_{t}(\Tilde{x})J_{t}(x)], \\
\label{sourcef}
f(\Tilde{x},u,v) &=& \frac{4^{s}}{6s}\Gamma^2
[1+s]\frac{1+3w}{1+w}(uvc^{2}_{s}\Tilde{x}^2)^{1-s} \nonumber \\
&&\quad \quad \quad \quad \quad \quad \quad \quad \times \Bigg[J_{s-1}(uc_{s}\Tilde{x})J_{s-1}(vc_{s}\Tilde{x})+ \frac{3(1+w)}{2}J_{s+1}(uc_{s}\Tilde{x})J_{s+1}(vc_{s}\Tilde{x})\bigg].\eea
Here, the Bessel functions are related using the following relation: $J_{s-1}(x)+J_{s+1}(x) = (2s/x)J_{s}(x)$ to express $J_{s}$ as $J_{s\pm 1}$. The next section presents the analytical formulation of the kernel for a generic $w$, which is derived from this reduced source term.

 \subsection{Semi-Analytical computation of the transfer function}

 Although we may evaluate the kernel from equation (\ref{kernel}) numerically, we can simplify the computation by getting an analytic expression for the kernel (or transfer) function contained in the power spectrum equation (\ref{tensorpspec}). We have adhered to the methodology employed in \cite{Domenech:2021ztg}. The kernel may now be expressed more simply by inserting the relations from equations (\ref{simplG},\ref{sourcef}) into eqn.(\ref{kernel}):
 \bea \label{simplekern}
{\cal I}(u,v,x)= 4^{t}\frac{3\pi}{2s^{3}}\frac{1+w}{1+3w}\Gamma^{2}(t+2)(c^{2}_{s}uvx)^{-t}\{Y_{t}(x)I_{J}^{x}(u,v,w)-J_{t}(x)I_{Y}^{x}(u,v,w)\},
\eea
And the following definitions apply to the other new integrals in the RHS:
\bea \label{besselprod}
I^{x}_{B}(u,v,w) = \int^{x}_{0}d\tilde{x}\;\tilde{x}^{1-t}B_{t}(\tilde{x})\left\{J_{t}(uc_{s}\tilde{x})J_{t}(vc_{s}\tilde{x}) + \frac{3(1+w)}{2}J_{t+2}(uc_{s}\tilde{x})J_{t+2}(vc_{s}\tilde{x}) \right\} 
\eea
where the two Bessel functions, $B=J,\;Y$, both of order $t$ and $t+2$, are labelled by $B$. We operate inside this regime for our future study because the above integral cannot provide analytical findings for random values of $x$. However, in the limit $x \gg 1$, this integral can still be solved analytically, which also corresponds to the scales deep inside the horizon. Pushing the upper limit for the integral to $x\rightarrow \infty$ will allow us to collect the leading order effects that are necessary for our investigation of the induced GWs. By Gervois and Navelet \cite{gervois1985integrals}, such integrals have already been studied to provide analytical findings in terms of the Legendre and related Legendre polynomials. The form of the two integrals stated in the sub-Horizon regime ($x\gg 1$) is the resultant form of the kernel that we will now discuss:
\bea
I_{J}(u,v,w,c_{s}) &=& \cos{\left(x-\frac{b\pi}{2}\right)}\bigg(P_{t-1/2}^{-t+1/2}(y) + \frac{3(1+w)}{2}P_{t+3/2}^{-t+1/2}(y)\bigg)\Theta(c_{s}(u+v)-1), \nonumber\\
I_{Y}(u,v,w,c_{s}) &=& \frac{2}{\pi}\sin{\left(x-\frac{b\pi}{2}\right)}\bigg[\bigg\{Q_{t-1/2}^{-t+1/2}(y) + \frac{3(1+w)}{2}Q_{t+3/2}^{-t+1/2}(y)\bigg\}\Theta(c_{s}(u+v)-1)\\
&& \quad\quad\quad\quad\quad\quad -\bigg\{ {\cal Q}_{t-1/2}^{-t+1/2}(-y) + 3(1+w){\cal Q}_{t+3/2}^{-t+1/2}(-y)\bigg\}\Theta(1-c_{s}(u+v)) \bigg],\\
{\cal I}(u,v,w,c_{s},x\gg 1) &=& x^{-(t+1/2)}\frac{4^{t}s}{3uvc^{2}_{s}}\frac{1+3w}{1+w}\Gamma^{2}(t+1)\bigg(\frac{Z}{2uv}\bigg)^{t-1/2}\times\bigg(I_{J}(u,v,w,c_{s})+I_{Y}(u,v,w,c_{s})\bigg).
\eea
where $2t = 2b+1$. Additionally, two additional variables, $Z$ and $y$, have been added. These variables are defined by the following expressions:
\bea 
Z^{2}=4u^{2}v^{2}(1-y^{2}), \quad\quad y = -1+\frac{c_{s}^{2}(u+v)^{2}-1}{2c_{s}^{2}uv},
\eea 
and the Bessel functions are expanded for big parameters. We may now talk about some important characteristics of the given results. For such sub-Horizon integrals, the associated Legendre polynomial of the second kind, ${\cal Q}_{\alpha}^{\beta}$, is the Olver's function; it has solutions for $|y|>1$. The Ferrer's functions, $P_{\alpha}^{\beta}(y)$ and $Q_{\alpha}^{\beta}(y)$, are valid for specific values of $|y|<1$. Resonant conditions occur when $c_{s}(u+v)=1$, as shown by the Heaviside theta argument. This condition also matches the situation in which the tensor mode's wavenumbers equal the sum of two scalar modes. The Heaviside Theta, $\Theta(c_{s}(u+v)-1)$, aids in the differentiation of situations that deviate from resonance.
Taking the square and oscillation averaged value from the integrals and ignoring any non-Gaussian contributions for the time being, we present the kernel's final result directly, assuming only Gaussian fluctuations. This gives us:
\bea \label{kernelavg}
\overline{{\cal I}^{2}(u,v,w,c_{s},x)} &=& x^{-(2t+1)}\frac{4^{2t}s^{2}}{9u^{2}v^{2}c_{s}^{4}}\frac{1+3w}{1+w}\Gamma^{4}(t+1)\bigg(\frac{Z}{2uv}\bigg)^{2t-1}\nonumber\\
&& \quad\quad\quad\quad \times\bigg\{\bigg(P_{t-1/2}^{-t+1/2}(y) + \frac{3(1+w)}{2}P_{t+3/2}^{-t+1/2}(y)\bigg)^{2}\Theta(c_{s}(u+v)-1) \nonumber\\
&& \quad\quad\quad\quad + \frac{4}{\pi^{2}}\bigg(Q_{t-1/2}^{-t+1/2}(y) + \frac{3(1+w)}{2}Q_{t+3/2}^{-t+1/2}(y)\bigg)^{2}\Theta(c_{s}(u+v)-1) \nonumber\\
&& \quad\quad\quad\quad + \frac{4}{\pi^{2}}\bigg({\cal Q}_{t-1/2}^{-t+1/2}(-y) + 3(1+w){\cal Q}_{t+3/2}^{-t+1/2}(-y)\bigg)^{2}\Theta(1-c_{s}(u+v)) \bigg\}
\eea
For our purposes of calculating the induced GW spectrum for a broad EoS background, this formula is crucial. We draw attention to the fact that the notation $c_{s}$, which is used to indicate the speed of propagation in the hypothetical cosmic fluid with EoS $w$, where the metric fluctuations propagate and eventually participate in the development of the tensor modes, is important to understand when deriving the transfer function.

 \subsubsection{Special Case I: Radiation Domination}

We have $w=1/3$ for the radiation-dominated period, which suggests that $b=0$. After plugging this into eqn. (\ref{kernelavg}) and making the necessary approximations, we get the radiation epoch transfer function that looks like this:
\bea \label{rdtransfer}
{\cal T}_{\rm RD} (u,v, w=1/3,c_{s})&=&
\bigg[\frac{4v^2 - (1-u^2 +v^2)^2}{4u^2v^2}\bigg]^{2}\times\overline{{\cal I}^{2}_{\rm RD} (u,v, w=1/3,c_{s})} \nonumber\\
&=& \frac{y^2}{3c_s ^4}\bigg[\frac{4v^2 - (1-u^2 +v^2)^2}{4u^2v^2}\bigg]^2 \times \bigg[\frac{\pi ^2 y^2}{4}\Theta[c_s(u+v)-1]+\bigg(1-\frac{1}{2}y \ln \bigg|\frac{1+y}{1-y}\bigg|\bigg)^2\bigg].
\eea
This outcome provides the averaged tensor power spectrum and, hence, the density of GWs in eqn. (\ref{tensorpspec}). The obtained kernel in \cite{Kohri:2018awv} fits this result correctly. 

 \subsubsection{Special Case II: Matter Domination}

 We have $w=0$ for the matter-dominated period, implying $b=1$. This condition may also be used to describe a pressure-free fluid backdrop and a scalar field oscillating coherently around the bottom of a potential. Using the given values for $b,w$ in eqn.(\ref{kernelavg}), the kernel for this example simplifies to:
\bea
{\cal T}_{\rm MD}(u,v,w=0,c_{s})&=& 
\bigg[\frac{4v^2 - (1-u^2 +v^2)^2}{4u^2v^2}\bigg]^{2}\times\overline{{\cal I}^{2}_{\rm MD} (u,v, w=0,c_{s})} \nonumber\\
&=&\frac{3^3 5^2}{2^{14}c_s ^4}\bigg[\frac{4v^2-(1-u^2+v^2)^2}{4u^2v^2}\bigg]^2  \times \bigg[\frac{\pi ^2}{4}(1-y^2)^2 (1+3y^2)^2 \Theta[c_s (u+v)-1]\nonumber \\
&& \quad \quad \quad \quad +\bigg(y(1-3y^2)-1/2(1+2y^2-3y^4)\ln{\abs{\frac{1+y}{1-y}}}\bigg)^2 \bigg].
\eea

 \subsubsection{Special Case III: Kinetic Domination}

 We know $w=1$ for the kinetic dominance epoch, which therefore $b=0$. This describes a situation in which the kinetic energy dominates the potential energy and the scalar field follows a steep potential. With the values from eqn.(\ref{kernelavg}), the kernel's final form is as follows:
\bea
{\cal T}_{\rm KD}(u,v,w=1,c_s)&=&
\bigg[\frac{4v^2 - (1-u^2 +v^2)^2}{4u^2v^2}\bigg]^{2}\times\overline{{\cal I}^{2}_{\rm KD} (u,v, w=1,c_{s})} \nonumber\\
&=&\frac{4}{3 \pi c_s^4 \abs{1-y^2}}\bigg[\frac{4v^2 - (1-u^2 +v^2 )^2}{4u^2v^2}\bigg]^2 \nonumber \\
&& \times \bigg[(1+3y^2)\; \Theta[c_s(u+v)-1]+\big(1-3y^2+3y\sqrt{\abs{1-y^2}}\big)^2\; \Theta[1-c_s(u+v)]\bigg].
\eea

  \subsubsection{Special Case IV: Soft Fluid Domination}

  We know $w=1/9$ for the soft fluid situation, which means $b=1/2$. From eqn.(\ref{kernelavg}), the kernel provides the simplified form:
\bea
{\cal T}_{\rm SFD}(u,v,w=1/9,c_{s})&=&
\bigg[\frac{4v^2 - (1-u^2 +v^2)^2}{4u^2v^2}\bigg]^{2}\times\overline{{\cal I}^{2}_{\rm SFD} (u,v, w=1/9,c_{s})} \nonumber\\
&=&\frac{2^8}{3^8 \pi c_s^4}\bigg[\frac{4v^2 - (1-u^2+v^2)^2}{4u^2v^2}\bigg] \times \bigg[(4+45y^2)\; \Theta [c_s(u+v)-1]\nonumber \\
&& \quad \quad \quad \quad +\big(y(3-10y^2)+(2+10y^2)\sqrt{\abs{1-y^2}}\big)^2 \Theta[1-c_s(u+v)]\bigg].
\eea

    \subsubsection{Special Case V: Negative EoS Fluid Domination}

    In light of the induced GW generation, another situation to take into account is the negative EoS, $w<0$, condition. We provide a particular example pertaining to this scenario, in which $w=-1/9$ yields $b=2$, and the kernel is reduced to the readable form by these values in eqn.(\ref{kernelavg}):
\bea
{\cal T}_{\rm NEoS}(u,v,w=-1/9,c_{s}) &=&
\bigg[\frac{4v^2 - (1-u^2 +v^2)^2}{4u^2v^2}\bigg]^{2}\times\overline{{\cal I}^{2}_{\rm NEoS} (u,v, w=-1/9,c_{s})} \nonumber\\
&=&\frac{5^2 7^2}{2^8 3^9 c_s ^4}\bigg(\frac{4v^2 -(1-u^2+v^2)^2}{4u^2 v^2}\bigg)^2 \times \bigg[\frac{225 \pi^2}{4}(1-y^2)^2 (1+y^2-2y^4)^2 \;\Theta [c_s(u+v)-1] \nonumber \\ 
&& \quad \quad \quad \quad \quad \quad + \bigg(y(9+35y^2 - 30y^4)+\frac{15}{2}(1-3y^4+2y^6)\ln{\abs{\frac{1+y}{1-y}}\bigg)^2}\bigg].
\eea
We would want to make it clear that we will not be discussing the case outcomes in detail since we discovered that they did not yield any useful information when the NANOGrav15 signal was present. As a result, we decide to disregard these findings and merely provide them here for the sake of being thorough.

\subsection{Computing SIGW spectrum}

The generation of SIGWs is studied in this part by looking at their resultant spectrum, using the generic equations for the kernel (or transfer) function discussed in the preceding sections. When the Universe eventually recovers the hot Big Bang scenario, the $w$-SIGW phenomenon corresponds to an arbitrary background EoS $w$, which is postulated during the last phases of inflation. We take use of this circumstance to show the induced GW spectrum under such a broad cosmic backdrop by means of sourcing from the scalar modes sub-Horizon. Thus, with a generic $w$ backdrop, the induced GW spectrum is expressed as follows:
\bea
\label{GWdensityQ}
\Omega_{\rm{GW},0}h^2 = 1.62 \times 10^{-5}\;\bigg(\frac{\Omega_{r,0}h^2}{4.18 \times 10^{-5}}\bigg) \bigg(\frac{g_{*}(T_c)}{106.75}\bigg)\bigg(\frac{g_{*,s}(T_c)}{106.75}\bigg)^{-4/3}\Omega_{\rm GW,c},
\eea
where the radiation energy density as measured today is denoted by $\Omega_{r,0}h^2$, and the energy and entropy effective degrees of freedom are represented by $g_{*},g_{*,s}$. When such generated GWs behave as freely propagating GWs throughout the radiation-dominated period, indicated by the instant ``c'', the number $\Omega_{\rm GW,c}$ reflects the GW energy density fraction. 

Using the kernel functions for the modes and scales that fulfill $k \geq k_{*}$, we further describe the energy density $\Omega_{\rm GW,c}$ as follows:
\bea
\label{omegacQ}
\Omega_{\rm {GW},c}&=& \frac{k^{2}}{12a^{2}H^{2}}\times\Delta^{2}_{h}(k,\tau) = \bigg(\frac{k}{k_{*}}\bigg)^{-2b}\int_{0}^{\infty}dv \int_{|{1-v}|}^{1+v} du \; {\cal T}(u,v,w,c_s) \;\;\Delta^{2}_{\zeta}(ku) \times\Delta^{2}_{\zeta}(kv),
\eea
having a constant propagation speed $c_{s}$ and the transfer function with a constant EoS background. The pivot scale in our configuration, which we will use to do our next analysis of the GW spectrum, is denoted by the value $k_{*}$. We refer to the final version of the transfer function used above, ${\cal T}(u,v,w,c_{s})$, for clarity's sake. This version is the result of applying eqn.(\ref{kernelavg}) and correcting for additional multiplicative factors found in eqn.(\ref{tensorpspec}):
\bea \label{transferQ}
{\cal T}(u,v,w,c_{s}) &=& (b+1)^{-2(b+1)}\frac{4^{2b}}{3c^{4}_{s}}\bigg[\frac{3(1+w)}{1+3w}\bigg]^{2}\Gamma^{4}(b+3/2)\bigg[\frac{4v^2 - (1-u^2 +v^2 )^2}{4u^2v^2}\bigg]^{2}\bigg(\frac{Z}{2uv}\bigg)^{2b} \nonumber\\
&& \quad\quad\quad\quad \times\bigg\{\bigg(P_{-b}^{b}(y) + \frac{3(1+w)}{2}P_{b+2}^{-b}(y)\bigg)^{2}\Theta(c_{s}(u+v)-1) \nonumber\\
&& \quad\quad\quad\quad + \frac{4}{\pi^{2}}\bigg(Q_{-b}^{b}(y) + \frac{3(1+w)}{2}Q_{-b}^{b+2}(y)\bigg)^{2}\Theta(c_{s}(u+v)-1) \nonumber\\
&& \quad\quad\quad\quad + \frac{4}{\pi^{2}}\bigg({\cal Q}^{-b}_{b}(-y) + 3(1+w){\cal Q}^{-b}_{b+2}(-y)\bigg)^{2}\Theta(1-c_{s}(u+v)) \bigg\}.
\eea
We assess the GW density and examine the spectrum's behavior for different EoS situations using the transfer function mentioned above, with different cases of $b$ or $w$ values. Prior to delving into more findings, we clarify a crucial aspect concerning the suitability of our theory for producing induced gravitational waves from a generic $w$ background. Look at the Appendix \ref{A7a}, \ref{A8a} and \ref{A9a} for more details on this computation.

\section{Effective Field Theory methods applied to large fluctuations}
\label{s5}

For the past several years, there has been intense research into primordial black holes (PBHs) \cite{Zeldovich:1967lct,Hawking:1974rv,Carr:1974nx,Carr:1975qj,Chapline:1975ojl,Carr:1993aq,Kawasaki:1997ju,Yokoyama:1998pt,Kawasaki:1998vx,Rubin:2001yw,Khlopov:2002yi,Khlopov:2004sc,Saito:2008em,Khlopov:2008qy,Carr:2009jm,Choudhury:2011jt,Lyth:2011kj,Drees:2011yz,Drees:2011hb,Ezquiaga:2017fvi,Kannike:2017bxn,Hertzberg:2017dkh,Pi:2017gih,Gao:2018pvq,Dalianis:2018frf,Cicoli:2018asa,Ozsoy:2018flq,Byrnes:2018txb,Ballesteros:2018wlw,Belotsky:2018wph,Martin:2019nuw,Ezquiaga:2019ftu,Motohashi:2019rhu,Fu:2019ttf,Ashoorioon:2019xqc,Auclair:2020csm,Vennin:2020kng,Nanopoulos:2020nnh,Gangopadhyay:2021kmf,Inomata:2021uqj,Stamou:2021qdk,Ng:2021hll,Wang:2021kbh,Kawai:2021edk,Solbi:2021rse,Ballesteros:2021fsp,Rigopoulos:2021nhv,Animali:2022otk,Correa:2022ngq,Frolovsky:2022ewg,Escriva:2022duf,Kristiano:2022maq,Karam:2022nym,Riotto:2023hoz,Kristiano:2023scm,Riotto:2023gpm,Ozsoy:2023ryl,Ivanov:1994pa,Afshordi:2003zb,Frampton:2010sw,Carr:2016drx,Kawasaki:2016pql,Inomata:2017okj,Espinosa:2017sgp,Ballesteros:2017fsr,Sasaki:2018dmp,Ballesteros:2019hus,Dalianis:2019asr,Cheong:2019vzl,Green:2020jor,Carr:2020xqk,Ballesteros:2020qam,Carr:2020gox,Ozsoy:2020kat,Baumann:2007zm,Saito:2008jc,Saito:2009jt,Choudhury:2013woa,Sasaki:2016jop,Raidal:2017mfl,Papanikolaou:2020qtd,Ali-Haimoud:2017rtz,Di:2017ndc,Raidal:2018bbj,Cheng:2018yyr,Vaskonen:2019jpv,Drees:2019xpp,Hall:2020daa,Ballesteros:2020qam,Ragavendra:2020sop,Carr:2020gox,Ozsoy:2020kat,Ashoorioon:2020hln,Ragavendra:2020vud,Papanikolaou:2020qtd,Ragavendra:2021qdu,Wu:2021zta,Kimura:2021sqz,Solbi:2021wbo,Teimoori:2021pte,Cicoli:2022sih,Ashoorioon:2022raz,Papanikolaou:2022chm,Wang:2022nml,Mishra:2019pzq,ZhengRuiFeng:2021zoz,Cohen:2022clv,Arya:2019wck,Bastero-Gil:2021fac,Correa:2022ngq,Gangopadhyay:2021kmf,Cicoli:2022sih,Brown:2017osf,Palma:2020ejf,Geller:2022nkr,Braglia:2022phb,Kawai:2022emp,Frolovsky:2023xid,Aldabergenov:2023yrk,Aoki:2022bvj,Frolovsky:2022qpg,Aldabergenov:2022rfc,Ishikawa:2021xya,Gundhi:2020kzm,Aldabergenov:2020bpt,Cai:2018dig,Fumagalli:2020adf,Cheng:2021lif,Balaji:2022rsy,Qin:2023lgo,Riotto:2023hoz}. The first reason for studying their genesis in the early cosmos was to comprehend the origin of supermassive black holes (PHBs), which may have started the formation of the one at the center of our galaxy.
It is plausible that their origin resulted from the collapse of large fluctuations in the early cosmos. Shortly after its creation, inflation emerged as a natural setting where such massive fluctuations appropriate for PBH synthesis might be produced in a quantum mechanical manner. These strange items could be concealing true natural mysteries. To this end, let us observe that despite the great achievements of contemporary cosmology, there remain notable open questions associated with each of the four phases of the universe's evolution; for example, the nature and origin of dark matter in the matter-dominated regime and baryon asymmetry (radiation era) remain to be clarified. Surprisingly, these things may be able to solve the aforementioned unresolved problems with contemporary cosmology. In fact, PBHs experience Hawking evaporation following creation, and the byproducts of this process may contribute to the observed baryon imbalance in the cosmos as well as dark matter.

Note that the process of PBH creation during inflation is currently the subject of extensive research in the literature \cite{Kristiano:2022maq,Riotto:2023hoz,Kristiano:2023scm,Riotto:2023gpm}. One-loop corrections \footnote{To know more about one-loop effects on inflation and de Sitter space see the refs. \cite{Adshead:2008gk,Senatore:2009cf,Senatore:2012nq,Pimentel:2012tw,Sloth:2006az,Seery:2007we,Seery:2007wf,Bartolo:2007ti,Seery:2010kh,Bartolo:2010bu,Senatore:2012ya,Chen:2016nrs,Markkanen:2017rvi,Higuchi:2017sgj,Syu:2019uwx,Rendell:2019jnn,Cohen:2020php,Green:2022ovz,Premkumar:2022bkm}.} have been shown to significantly restrict their creation during abrupt transitions from Slow Roll (SR) to Ultra Slow Roll (USR), according to an argument made using the classic single field inflation framework. In ref.\cite{Choudhury:2023jlt}, it was also mentioned that adding non-canonical features will make the result much worse. To this purpose, a generic framework of single-field inflation that is applicable to both the canonical and non-canonical scenarios and is model-independent is highly desirable. For further information on this topic, where the authors have examined a variety of options along the same subject line with smooth and sharp transitions, see refs. \cite{Choudhury:2023hvf,Choudhury:2023kdb,Choudhury:2023hfm,Bhattacharya:2023ysp,Motohashi:2023syh,Firouzjahi:2023ahg,Franciolini:2023lgy,Firouzjahi:2023aum,Cheng:2023ikq,Tasinato:2023ukp}. An effective action that is valid below the UV cut-off scale is the basis for the appropriate scenario known as effective field theory (EFT) of inflation \cite{Weinberg:2008hq,Cheung:2007st,Choudhury:2017glj,Delacretaz:2016nhw}.  Furthermore, there is an underlying symmetry that constrains the shape of the EFT action.  The main goal is developing a UV full theory that the underlying symmetries permit. The full-proof version of the UV complete theory, which can be built at very high energy scales when the UV cut-off scale is fixed, is, however, still not fully understood.  The scale $\Lambda_{UV}=M_{pl}\sim 10^{19}{\rm Gev}$ can theoretically be fixed at this cut-off. In this kind of analysis, the theory above the scale $\Lambda_{UV}$ is considered as a black box, since it is entirely unknown. Technically speaking, in the corresponding EFT literature, this is sometimes referred to as the hidden sector of the theory. As all of these degrees of freedom should ultimately be path integrated out to create a workable description of the EFT composed of only observable sector field degrees of freedom, we actually don't need to consider this sector in theory. According to $\Lambda_{EFT}\leq\Lambda_{UV}$, the maximum can run up to the specified scale, however this visible sector description of the EFT setup should be valid below the indicated UV cut-off scale. Thus, as each coupling parameter is in essence a function of the running scale of the underlying theory, the coupling parameters that appear in front of each EFT operator and characterize the visible sector of the theory can also run from the lower scale to the specified cut-off scale. If we possess a thorough understanding of how a UV complete theory is constructed from basic physical principles, we can solve Renormalization Group (RG) flow equations to compute the behavior, pole structure, and dependence of the mentioned coupling parameters with respect to the scale of the underlying theory. See refs. \cite{Assassi:2013gxa,Baumann:2019ghk,Green:2022ovz,Cohen:2020php,Green:2020txs,Gorbenko:2019rza,Burgess:2022rdo,Burgess:2021luo,Burgess:2020tbq,Burgess:2020qsc,Burgess:2020fmg,Burgess:2017ytm,Burgess:2015ajz,Burgess:2014eoa,Collins:2012nq,Boyanovsky:2011xn,Burgess:2010dd} for more details.

Three options are available to build a basic EFT set-up; these are listed below, point-by-point:
\begin{enumerate}
    \item \underline{\bf Top-down approach:}
The UV complete theory is a necessary starting point for this description, and it is true at $\Lambda_{EFT}\leq\Lambda_{UV}$. It is quite challenging to begin building an EFT theory using this specific method because the correct understanding of how the UV complete theory is constructed is still incomplete. In spite of the technical challenges, much work has gone toward building an EFT from string theory as, among alternative options, this is the most effective one that can be used to describe a UV full version of quantum field theory. But one can have several kinds of EFT theories that can handle diverse cosmological frameworks, from inflation to the late time acceleration of our universe, because of the vast amount of compactification schemes in the four-dimensional reduced version. We have not taken advantage of this opportunity to build the EFT for primordial cosmology in this review.

\item \underline{\bf Intermediate approach:} 
We employ the notion of hidden and visible sectors, which we discussed at the beginning of this section, in place of beginning with a UV full theory, as in String Theory. We examine a simplified variant of this description in which a low-mass scalar field interacts with a hefty scalar field. In this case, the visible sector is represented by the low mass field, and the hidden sector by the heavy field. The heavy field must be integrated out of the UV complete two-field interaction theory in order to build an EFT of the visible sector field in four dimensions for primordial cosmology. Upon completion of this process, one should possess both non-renormalizable (irrelevant) and renormalizable (relevant and marginal) higher mass dimensional operators. These are all suppressed by the underlying EFT scale, which is $\Lambda_{EFT}\leq\Lambda_{UV}$. In order for the Wilsonian point of view of EFT to represent an underlying cosmic setting, all the coefficients that appear in front of these operators must be fixed. This method, however, is somewhat phenomenological at its beginning, highly laborious, and exclusively model reliant. That's why we have rejected the idea of building the EFT for primordial cosmology in this study.

 \item \underline{\bf Bottom-up approach:}
Ultimately, applying the underlying symmetry principle, we introduce a pure gravitational sector characterized by Einstein theory written in a quasi de Sitter background. This is in contrast to the intermediate approach, which starts with a particular version of UV complete theory or uses a minimalistic phenomenological version of this type of theory. To further replicate the function of scalar field action in the conventional generic $P(X,\phi)$ theory, two more time-dependent operators are included \cite{Chen:2010xka,Chen:2006nt}. Besides the previously described contributions, we also present several combinations of gravitational fluctuations in terms of the extrinsic curvature and the perturbations on the temporal component of the metric tensor. These operators are all akin to Wilsonian operators, which can encompass both the non-renormalizable (irrelevant) larger mass dimensional operators and the renormalizable (relevant and marginal) operators. The advantage of this strategy is that it is not dependent on the existence of any scalar field; rather, it is based on the particular structure of the kinetic contributions and the effective potential of the specified scalar field. This enables us to take this suggested theoretical setting and turn it into a fully model-independent forecast. Here, we fix the Wilson coefficients, which constrain both the two-point primordial cosmic correlation function estimated from this particular EFT configuration employed in this work, and the effective sound speed $c_s$. As we have not utilized any particular structure of the $P(X,\phi)$ model action, one may naturally wonder where the scalar curvature perturbation originates, and whose quantum loop corrected two-point correlation is the most desired result of our work. Goldstone modes are produced as a result of temporal diffeomorphism symmetry breaking in the unitary gauge, and the St$\ddot{u}$ckelberg approach provides the answer to this pivotal question. This bears a striking resemblance to spontaneous symmetry breaking as it occurs in $SU(N)$ gauge theory. One may directly relate the scalar curvature perturbation mode to the previously discussed Goldstone mode if we limit our research to the linear regime. After confirming this identification, the remaining study can be performed using the terminology of scalar curvature perturbation in the context of primordial cosmology. In this study, we have adhered completely to this strategy, which will be covered in more detail in the latter part of this review. Finally, one might further comment on how precisely the limitation on the effective sound speed $c_s$ would further fix the couplings of these models, to establish a fruitful relationship with the known models of $P(X,\phi)$ theories. At the conclusion of this chapter of the review, we go into further depth about this option.
\end{enumerate}

The framework belongs to the following class of theories based on the value of the sound speed parameter $c_s$:
\begin{enumerate}
     \item \underline{\bf Ghost EFT:}
            Specifically explored in the context of inflation, it is described as $c_s=0$, which was proposed by Arkani-Hameda, Creminelli, Mukohyama, and Zaldarriagaa in ref.\cite{Arkani-Hamed:2003juy}. The scope of our study in this paper does not extend to this specific model. Nonetheless, given the current situation, this might be a wise way to search for the results.

      \item \underline{\bf Non-canonical and causal EFT:} This is characterized by $0<c_s<1$, which is also a domain that has been extensively investigated in several situations in cosmology. It is noteworthy that in the UV complete description, this particular EFT setup, where causality ($c_s<1$) and ghost-free criteria are strictly maintained, can describe all classes of $P(X,\phi)$ \footnote{The kinetic term $X=-1/2\left(\partial_{\mu}\phi\right)^2$ and the scalar field $\phi$ can be represented as a general functional here, denoted by $P(X,\phi)$. To illustrate, consider the following functions for Tachyon and $K$ inflation: $P(X,\phi)=-V(\phi)\sqrt{1-2\alpha^{'}X}$ and $P(X,\phi)=K(X)-V(\phi)$, where $V(\phi)$ denotes the effective potential in this description.} theories minimally coupled to gravity having the general structure of $P(X,\phi)$ as well as different kinds of modified gravity theories. Dirac-Born-Infeld (DBI), Tachyon, $K$ inflation, and other instances are a handful that are currently known. For further information, see refs. \cite{Alishahiha:2004eh,Mazumdar:2001mm,Choudhury:2002xu,Panda:2005sg,Chingangbam:2004ng,Armendariz-Picon:1999hyi,Garriga:1999vw}.

      \item \underline{\bf Canonical and causal EFT:}  It is represented by $c_s=1$, which is essentially a minimally coupled scalar field theory to gravity with the structure $P(X,\phi)=\left(X-V(\phi)\right)$, where $V(\phi)$ is the effective potential of a single scalar field $\phi$ and $X$ is its kinetic term. This kind of theory has been investigated in different cosmological contexts and strictly respects the causality requirement, $c_s<1$ in the EFT definition. For further information on these aspects, see refs. \cite{Choudhury:2017glj,Naskar:2017ekm,Choudhury:2015pqa,Choudhury:2014sua,Choudhury:2014kma,Choudhury:2013iaa,Baumann:2022mni,Baumann:2018muz,Baumann:2015nta,Baumann:2014nda,Baumann:2009ds}.

        \item \underline{\bf Non-canonical and a-causal EFT:} The corresponding literature has a fairly limited class of EFTs that satisfy $c_s>1$, which characterizes this. The causality constraint is broken by this kind of theory, which also needs to be non-canonical. This is specifically why it is referred to as a-causal EFT, or prominently superluminal EFT. In GTachyon \footnote{In this context, GTachyon is considered a particular class of $P(X,\phi)$ theory, which is essentially a generalized Tachyon as it appears in String Theory. The effective sound speed in such a model shifts from a causal to an acausal regime over a wide range.}, one of the most promising instances of this kind of theory, the superluminal effect is well understood within a certain portion of the theory's parameter space. Perfectly matching the expectations from the standard Tachyon, the other part of the parameter space validates causality. Refer to \cite{Choudhury:2015hvr} for further information.

\end{enumerate}
 
In this study, we follow the well-known St$\ddot{u}$ckelberg approach in the unitary gauge where the Goldstone modes are generated, which in turn breaks the temporal diffeomorphism symmetry while generating the scalar comoving cosmological perturbation. All classes of $P(X,\phi)$ theories that result in a single-field inflationary paradigm are limited to using this methodology. However, an extension of this configuration to multiple scalar field theories is possible \cite{Senatore:2010wk,Khosravi:2012qg,Shiu:2011qw}. Here we report in detail an investigation of one-loop corrections to the power spectrum in the context of EFT of inflation with a single field. This review goes into considerable length about the latter, which leads to significant limits on the creation process of PBHs in the single-field inflation paradigm.


\subsection{The Effective Field Theory (EFT) of Single Field Inflation in a nutshell}

It is noteworthy that, under the following full diffeomorphism symmetry, such contribution becomes a scalar, if we choose to express the formulation in terms of scalar field degrees of freedom:
\be x^{\nu}\longrightarrow x^{\nu}+\xi^{\nu}(t,{\bf x})~~~\forall~ \nu=0,1,2,3~.\ee  
Within the present discourse, $\xi^{\nu}(t,{\bf x})$ represents the diffeomorphism parameter. As an alternative to using the full symmetry, the perturbation on the field $\delta \phi$ changes in the following two-fold parallel scenarios:
\begin{enumerate}
    \item Under the spatial component of the space-time diffeomorphism symmetry, only as a scalar, and

    \item Solely under the temporal component of the space-time diffeomorphism symmetry in a non-linear manner. 
\end{enumerate}
The following is one way to express these particular changes:
\bea
{\textcolor{Sepia}{\bf Broken~ spatial~ diffeo:}}~~	t&&\longrightarrow t,~x^{i}\longrightarrow   x^{i}+\xi^{i}(t,{\bf x})~~~\forall~ i=1,2,3
~~~\delta\phi\longrightarrow \delta\phi,\\	
{\textcolor{Sepia}{\bf Broken~ temporal~ diffeo:}}~~	t&&\longrightarrow t+ \xi^{0}(t,{\bf x}),~x^{i}\longrightarrow x^{i}~~~\forall~ i=1,2,3
	~~~\delta\phi\longrightarrow \delta\phi +\dot{\phi}_{0}(t)\xi^{0}(t,{\bf x}).\quad\quad
\eea
Under this setting, $\xi^{0}(t,{\bf x})$ and $\xi^{i}(t,{\bf x})\forall i=1,2,3$, respectively, characterize the spatial and temporal diffeomorphism parameters. We use the gravitational gauge $\phi(t,{\bf x})=\phi_{0}(t),$ in this case. The background time-dependent scalar field is embedded in homogeneous isotropic spatially flat FLRW space-time, and its $\phi_{0}(t)$ reflects this background. Furthermore, this means that in this gauge selection, $\delta \phi(t,{\bf x})=0$.

The essential components needed to assemble this EFT setup are as follows:
\begin{itemize}
    \item For EFT operators, a function of the gravitational space-time metric $g_{\mu\nu}$ must be needed.  It is now possible to compute the Ricci tensor $R_{\mu\nu},$ the Riemann tensor $R_{\mu\nu\alpha\beta}$, and the Ricci scalar $R$, which are also components of the EFT action, using the derivatives of the space-time metric.

    \item To generate the EFT action, one must first obtain the polynomial powers of the temporally perturbed component of the metric ($\delta g^{00}$), which is represented as $\delta g^{00}= \left(g^{00}+1\right)$. In this case, $g^{00}$ represents the time element of the background metric.  Under the spatial diffeomorphism symmetry transformation, this operator must be invariant.

    \item Within this setting, the quasi de Sitter solution is represented by the spatially flat FLRW space-time metric, which describes the background geometry, $ds^2=a^2(\tau)\left(-d\tau^2+d{\bf x}^2\right),$ where the scale factor is given by, $a=-1/H\tau$ with $-\infty<\tau<0$. Here $H$ is the Hubble parameter in the quasi de Sitter space.

    \item Moreover, the powers of the perturbed extrinsic curvature component at constant time slice ($\delta K_{\mu\nu}$) are required to generate the EFT action which is described in terms of, $\delta K_{\mu\nu}=\left(K_{\mu\nu}-a^2Hh_{\mu\nu}\right),$  where the extrinsic curvature ($K_{\mu\nu}$), unit normal ($n_{\mu}$),  and induced metric ($h _{\mu\nu}$) are defined in this context as:
\bea K_{\mu \nu}&=&h^{\sigma}_{\mu}\nabla_{\sigma} n_{\nu}\nonumber\\
&=&\left[\frac{\delta^{0}_{\mu}\partial_{\nu}g^{00}+\delta^{0}_{\nu}\partial_{\mu}g^{00}}{2(-g^{00})^{3/2}}
+\frac{\delta^{0}_{\mu}\delta^{0}_{\nu}g^{0\sigma}\partial_{\sigma}g^{00}}{2(-g^{00})^{5/2}}-\frac{g^{0\rho}\left(\partial_{\mu}g_{\rho\nu}+\partial_{\nu}g_{\rho\mu}-\partial_{\rho}g_{\mu\nu}\right)}{2(-g^{00})^{1/2}}\right],\\
h_{\mu \nu}&=&g_{\mu \nu}+n_{\mu} n_{\nu},\\
n_{\mu}&=&\frac{\partial_{\mu}t}{\sqrt{-g^{\mu \nu}\partial_{\mu}t \partial_{\nu}t}}
=\frac{\delta_{\mu}^0}{\sqrt{-g^{00}}},\\
\left[\delta K\right]^{m+2} &=&\delta K^{\mu_{1}}_{\mu_{2}}\delta K^{\mu_{2}}_{\mu_{3}}\delta K^{\mu_{3}}_{\mu_{4}}\cdots\delta K^{\mu_{m+1}}_{\mu_{m+2}}\delta K^{\mu_{m+2}}_{\mu_{1}}. \eea

\end{itemize}
Therefore, the EFT action can be described as:
\bea
	S&=&\displaystyle\int d^{4}x \sqrt{-g}\left[\frac{M^2_{pl}}{2}R-c(t)g^{00}-\Lambda(t)+\sum^{\infty}_{n=2}\frac{M^4_n(t)}{n!}
	(\delta g^{00})^n-\sum^{\infty}_{q=0}\frac{\bar{M}^{3-q}_1(t)}{(q+2)!}\delta g^{00}
	\left(\delta K_{\mu}^{\mu}\right)^{q+1}~~~~~~~~\nonumber\right.\\&&
	\left.\displaystyle~~~~~~~~~~~~~~~~~~~~~~~~~~~~~~~~~~~~~~~~~~-\sum^{\infty}_{m=0}\frac{\bar{M}^{2-m}_2(t)}{(m+2)!}
	\left(\delta K_{\mu}^{\mu}\right)^{m+2}-\sum^{\infty}_{p=0}\frac{\bar{M}^{2-p}_3(t)}{(m+2)!}
	\left[\delta K\right]^{m+2}+\cdots \right].
	\eea
We have now restricted it to the following shortened EFT operation since we find it important to compute the correlation functions:
\bea
 S&=&\displaystyle\int d^{4}x \sqrt{-g}\left[\frac{M^2_{pl}}{2}R-c(t)g^{00}-\Lambda(t)+\frac{M^{4}_{2}(t)}{2!}\left(g^{00}+1\right)^2+\frac{M^{4}_{3}(t)}{3!}\left(g^{00}+1\right)^3~~~~~~~~\nonumber\right.\\&&
	\left.\displaystyle~~~~~~~~~~~~~~~~~~~~~~~~~~~~~~~~~~~~~~~~~~~~~~~~-\frac{\bar{M}^{3}_{1}(t)}{2}\left(g^{00}+1\right)\delta K^{\mu}_{\mu}-\frac{\bar{M}^{2}_{2}(t)}{2}(\delta K^{\mu}_{\mu})^2-\frac{\bar{M}^{2}_{3}(t)}{2}\delta K^{\mu}_{\nu}\delta K^{\nu}_{\mu}\right].
	\eea
 Here
the time-dependent coefficients $M_1(t)$, $M_3(t)$,  $\bar{M}_1(t)$,  $\bar{M}_2(t)$ and $\bar{M}_3(t)$ are mimicking the role of Wilson coefficients which we need to be fixed from our analysis.

The associated Friedmann equations can be represented as follows if we further examine solely the background contributions in the aforementioned EFT action:
\bea
\left(\frac{\dot{a}}{{a}}\right)^2=H^2&=&\frac{1}{3M^2_{ pl}}\Bigg(c(t)+\Lambda(t)\Bigg)=\frac{{\cal H}^2}{a^2},\\
\frac{\ddot{a}}{{a}}=\dot{H}+H^2&=&-\frac{1}{3M^2_{ pl}}\Bigg(2c(t)-\Lambda(t)\Bigg)
=\frac{{\cal H}^{'}}{a^2},
\eea
where $'$ represents the derivative with respect to the conformal time $\tau$. Additionally,  it is important to note that, $\displaystyle {\cal H}=\frac{a^{'}}{a}=aH.$ After solving these equations, we obtained the following formulas for the background-level time-dependent parameters, $\Lambda(t)$ and $c(t)$:
\bea c(t)&=&-M^2_{pl} \dot{H}=-\frac{M^2_{pl}}{a^2}\Bigg({\cal H}^{'}-{\cal H}^2\Bigg),\\
\Lambda(t)&=&M^2_{pl} \left(3H^2+\dot{H}\right)=\frac{M^2_{pl}}{a^2}\Bigg(2{\cal H}^2+{\cal H}^{'}\Bigg).\eea
Upon replacing the time-dependent parameters $c(t)$ and $\Lambda(t)$ with their respective expressions, the EFT action can be expressed as follows:
\bea
 S&=&\displaystyle\int d^{4}x \sqrt{-g}\left[\frac{M^2_{pl}}{2}R+M^2_{pl} \dot{H} g^{00}-M^2_{pl} \left(3H^2+\dot{H}\right)+\frac{M^{4}_{2}(t)}{2!}\left(g^{00}+1\right)^2+\frac{M^{4}_{3}(t)}{3!}\left(g^{00}+1\right)^3~~~~~~~~\nonumber\right.\\&&
	\left.\displaystyle~~~~~~~~~~~~~~~~~~~~~~~~~~~~~~~~~~~~~~~~~~-\frac{\bar{M}^{3}_{1}(t)}{2}\left(g^{00}+1\right)\delta K^{\mu}_{\mu}-\frac{\bar{M}^{2}_{2}(t)}{2}(\delta K^{\mu}_{\mu})^2-\frac{\bar{M}^{2}_{3}(t)}{2}\delta K^{\mu}_{\nu}\delta K^{\nu}_{\mu}\right].
	\eea
 It is also necessary to introduce the subsequent deviation parameters, which are also referred to as the slow-roll parameters:
\bea
\epsilon &=&\bigg(1-\frac{{\cal H}^{'}}{{\cal H}^2}\bigg),\quad\quad
\eta =\frac{\epsilon^{'}}{\epsilon {\cal H}}.
\eea
The SR region cannot experience inflation unless the following circumstances are satisfied: $ |\eta|\ll 1.$, $\epsilon\ll 1.$

\subsection{Goldstone modes from EFT and decoupling limit}

Under the temporal diffeomorphism symmetry, the Goldstone mode ($\pi(t, {\bf x})$) transforms as follows:
\bea
\pi(t, {\bf x})\rightarrow\tilde{\pi}(t, {\bf x})=\pi(t, {\bf x})-\xi^{0}(t,{\bf x}).
\eea
where the local parameter is $\xi^{0}(t,{\bf x})$. In this study, the role of these Goldstone modes is analogous to that of the scalar modes in cosmic perturbation. The fixing requirement for the relevant unitary gauge is then given by, $\pi(t,{\bf x})=0$ which implies, $\tilde{\pi}(t,{\bf x})=-\xi^{0}(t,{\bf x})~.$

It is now critical to talk about the changes in the space-time metric, the Ricci tensor, the Ricci scalar, the perturbation on the extrinsic curvature, the time-dependent coefficients, and the slowly evolving Hubble parameter that result from the broken time diffeomorphism symmetry:
\begin{itemize}
     \item \underline{{\bf Space-time metric:}} The breaking of the temporal diffeomorphism symmetry results in the following transformations of the contravariant and covariant metrics:
\bea
		&&{g}^{00}\longrightarrow
	(1+\dot{\pi}(t,{\bf x}))^2 {g}^{00}+2(1+\dot{\pi}(t,{\bf x})){g}^{0 i}\partial_{i}\pi(t,{\bf x})+{g}^{ij}\partial_{i}\pi(t,{\bf x})\partial_{j}\pi(t,{\bf x}),\quad\quad\quad\\ 
		&&{g}^{0i}\longrightarrow
		(1+\dot{\pi}(t,{\bf x})){g}^{0i}+{g}^{ij}\partial_j \pi(t,{\bf x}),\\
		&&{g}^{ij}\longrightarrow{g}^{ij}.
	\\
			&&g_{00}\longrightarrow  (1+\dot{\pi}(t,{\bf x}))^2 g_{00},\\
		&&g_{0i}\longrightarrow (1+\dot{\pi}(t,{\bf x})){g}_{0i}+{g}_{00}\dot{\pi}(t,{\bf x})\partial_{i}\pi(t,{\bf x}),\\ && g_{ij}\longrightarrow{g}_{ij}+{g}_{0j}\partial_{i}\pi(t,{\bf x})+{g}_{i0}\partial_{j}\pi(t,{\bf x}).
	\eea

    \item \underline{{\bf Ricci scalar and Ricci tensor:}} Under the broken time diffeomorphism symmetry, the spatial components of the Ricci tensor and the Ricci scalar on a 3-hypersurface transform as follows:
\bea
			 &&{}^{(3)}R\longrightarrow\displaystyle {}^{(3)}R+\frac{4}{a^2}H(\partial^2\pi(t,{\bf x})),\\
			 &&{}^{(3)}R_{ij}\longrightarrow{}^{(3)}R_{ij}+H(\partial_{i}\partial_{j}\pi(t,{\bf x})+\delta_{ij}\partial^2\pi(t,{\bf x})).\quad
			\quad\eea

    \item \underline{{\bf Extrinsic curvature:}} Under the broken time diffeomorphism symmetry, the trace and the spatial, time, and mixed components of the extrinsic curvature transform as follows:
\bea
	 &&\delta K\longrightarrow \displaystyle \delta K-3\pi\dot{H}-\frac{1}{a^2}(\partial^2\pi(t,{\bf x})),\\
		&&\delta K_{ij}\longrightarrow\delta K_{ij}-\pi(t,{\bf x})\dot{H}h_{ij}-\partial_{i}\partial_{j}\pi(t,{\bf x}),\\
		&&\delta K^{0}_{0}\longrightarrow\delta K^{0}_{0},\\
		&&\delta K^{0}_{i}\longrightarrow\delta K^{0}_{i},\\
		&&\delta K^{i}_{0}\longrightarrow\delta K^{i}_{0}+2Hg^{ij}\partial_j\pi(t,{\bf x}).
	\eea

    \item \underline{{\bf Time-dependent Wilson coefficients:}} Under the broken time diffeomorphism symmetry, the time-dependent EFT coefficients convert as follows after canonical normalisation $\pi_c(t,{\bf x})=Q^2(t)\pi(t,{\bf x})$:
\bea		
		&&Q(t)\longrightarrow Q(t+\pi(t,{\bf x}))=\displaystyle \sum^{\infty}_{n=0}\underbrace{\frac{\pi^{n}_c(t,{\bf x})}{n!Q^{2n}(t)}}_{{\bf \ll 1}}\frac{d^{n}Q(t)}{dt^n}\approx Q(t)~.\quad\quad\eea
in this description the time-dependent Wilson coefficients are represented by $Q(t)$.

    \item \underline{{\bf Hubble parameter:}} Upon breaking the symmetry of temporal diffeomorphism, the Hubble parameter becomes:
\bea
	H(t)\longrightarrow H(t+\pi(t,{\bf x}))&=&\displaystyle \sum^{\infty}_{n=0}\frac{\pi^{n}}{n!}\frac{d^{n}H(t)}{dt^n}=\displaystyle\left[1-\underbrace{\pi(t,{\bf x}) H(t) \epsilon-\frac{\pi^2(t,{\bf x})H(t)}{2}\left(\dot{\epsilon}-2\epsilon^2\right)+\cdots}_{{\bf sub-leading~contributions}}\right]H(t)~.\quad\quad\quad\eea
    
\end{itemize}
To construct the EFT action, we now need to have a deeper understanding of the decoupling limit. In this limit, it is simple to ignore the mixing contributions of the Goldstone modes and gravity. To prove this statement is true, let's begin with the EFT operator $-\dot{H}M_{pl}^2g^{00}$, which is required for additional computation.

Under the broken time diffeomorphism symmetry, this operator takes on the subsequent transformation:
\bea &&-\dot{H}M_{pl}^2g^{00}\longrightarrow -\dot{H}M_{pl}^2\bigg[ (1+\dot{\pi}(t,{\bf x}))^2g^{00}
+\left(2(1+\dot{\pi}(t,{\bf x}))\partial_i \pi(t,{\bf x}) g^{0i}+g^{ij}\partial_i\pi(t,{\bf x}) \partial_j \pi(t,{\bf x})\right)\bigg].\eea
The temporal part of the metric after perturbation can be expressed as, $g^{00}=\bar{g}^{00}+\delta g^{00},$ where the perturbation is indicated by $\delta g^{00}$, and the temporal component of the baseline quasi-de Sitter metric is shown by $\bar{g}^{00}=-1$. A mixing contribution $M_{pl}^2\dot{H}\dot{\pi}\delta g^{00}$ and a kinetic contribution $M_{pl}^2\dot{H}\dot{\pi}^2\bar{g^{00}}$ make up the remaining contributions.
Additionally, $\delta g^{00}_c=M_{pl}\delta g^{00},$ is employed as a standard normalized metric perturbation mixing contribution after canonical normalization is given by, $M_{pl}^2\dot{H}\dot{\pi}\delta g^{00}= \sqrt{\dot{H}}\dot{\pi}_c\delta g^{00}_c.$ This mixing term above the energy scale, $E_{mix}=\sqrt{\dot{H}},$ can be conveniently ignored in the decoupling limit. Alternatively, one can combine the two essential contributions:
\begin{enumerate}
    \item Term contains the operator $M_{pl}^2\dot{H}\dot{\pi}^2\delta{g^{00}}$,

    \item  Term contains the operator $\pi M_{pl}^2\ddot{H}\dot{\pi}\bar{g}^{00}$,
\end{enumerate}
which are expressed after canonical normalization as, $M_{pl}^2\dot{H}\dot{\pi}^2\delta{g^{00}}=\dot{\pi}_c^2\delta{g^{00}_c}/M_{pl},$ and 
$\pi M_{pl}^2\ddot{H}\dot{\pi}\bar{g}^{00}=\ddot{H}\pi_c\dot{\pi}_c\bar{g}^{00}/\dot{H}.$
It is possible to ignore the contribution of the $M_{pl}^2\dot{H}\dot{\pi}\delta{g^{00}}$ term when $E>E_{mix}$. Regarding the decoupling limit, the following simplification holds true:
\bea -\dot{H}M_{pl}^2g^{00}\rightarrow -\dot{H}M_{pl}^2g^{00}\left[\dot{\pi}^2-\frac{1}{a^2}(\partial_i\pi)^2\right].\eea

\subsection{Second order perturbation from scalar mode: Connecting with Goldstone EFT}

Consider the decoupling limit $E>E_{mix}=\sqrt{\dot{H}}$, where all mixing contributions are readily discarded. In this case, the second-order Goldstone action ($S^{(2)}_{\pi}$) is expressed as follows:
 \bea 
  	S^{(2)}_{\pi}&=&\displaystyle \int d^{4}x ~a^3\left[-M^2_{p}\dot{H}\left(\dot{\pi}^2-\frac{1}{a^2}(\partial_{i}\pi)^2\right)
   	+2M^4_2 \dot{\pi}^2\nonumber\right.\\ &&\left. ~~~~~~~~\displaystyle
   	+\frac{1}{2}\left(\bar{M}^2_3+3\bar{M}^2_2\right)H^2(1-\epsilon)\frac{\left(\partial_{i}\pi\right)^2}{a^2}-\left(\bar{M}^2_3+3\bar{M}^2_2\right)H^2\frac{\left(\partial_{i}\pi\right)^2}{a^2}
   	-\frac{\bar{M}^3_1}{2}~H~\frac{\left(\partial_i\pi\right)^2}{a^2}\right]\nonumber\\
    &=&\int d^{4}x ~a^3~\left(-\frac{M^2_p\dot{H}}{c^2_s}\right)\left[\dot{\pi}^2
   -c^2_s\left(1-\frac{\bar{M}^3_1 H}{M^2_p \dot{H}}-\left(\bar{M}^2_3+3\bar{M}^2_2\right)\frac{H^2(1+\epsilon)}{2M^2_p\dot{H}}\right)\frac{1}{a^2}(\partial_{i}\pi)^2
   \right].~~~~~\quad\quad\eea 
Since the contributions from all the other coefficients are strongly suppressed except $M^4_2$ at the level of the second-order perturbed action for the Goldstone modes we have neglected such contributions henceforth from our analysis. For this reason, we get the following simplified version of the action, which is given by:
\bea 
  	S^{(2)}_{\pi}&\approx&\displaystyle \int d^{4}x ~a^3\left[-M^2_{pl}\dot{H}\left(\dot{\pi}^2-\frac{1}{a^2}(\partial_{i}\pi)^2\right)
   	+2M^4_2 \dot{\pi}^2\right]=\displaystyle \int d^{4}x ~a^3\left(\frac{-M^2_{pl}\dot{H}}{c^2_s}\right)\Bigg(\dot{\pi}^2-c^2_s\frac{\left(\partial_{i}\pi\right)^2}{a^2}\Bigg).~~~~~\quad\quad\eea 
   The effective sound speed is defined in terms of the EFT Wilson coefficient, which is provided by:
 \bea c_{s}\equiv \frac{1}{\displaystyle \sqrt{1-\frac{2M^4_2}{\dot{H}M^2_{pl}}}},\eea
 The effective sound speed is readily connected with a broad class of $P(X, \phi)$ theories. In typical $P(X,\phi)$ theories, the effective sound speed may be calculated as follows \cite{Chen:2010xka,Chen:2006nt}:
\be c_s=\sqrt{\frac{P_{,\bar{X}}}{P_{,\bar{X}}+2\bar{X}P_{,\bar{X}\bar{X}}}}.\ee
 Here $\bar{X}$ is the background value of $X$.  To connect realistic $P(X,\phi)$ models with the model-independent version of EFT, we can use the dimensionless coupling parameter $\displaystyle \frac{M^4_2}{\dot{H}M^2_{pl}}$, which can be written as:
\be \frac{M^4_2}{\dot{H}M^2_{pl}}=\frac{1}{2}\Bigg(1-\frac{P_{,\bar{X}}}{P_{,\bar{X}}+2\bar{X}P_{,\bar{X}\bar{X}}}\Bigg)=\Bigg(\frac{\bar{X}P_{,\bar{X}\bar{X}}}{P_{,\bar{X}}+2\bar{X}P_{,\bar{X}\bar{X}}}\Bigg).\ee
It is vital to remember that the spatial component of the metric fluctuation is defined by:
\bea g_{ij}\sim a^{2}(t)\left[\left(1+2\zeta(t,{\bf x})\right)\delta_{ij}\right]~~\forall~~~i=1,2,3,\eea
The scale factor in the FLRW quasi-de Sitter background space-time is $a(t)=\exp(Ht)$. In addition, the scalar comoving curvature perturbation is denoted by the notation $\zeta(t,{\bf x})$. In this scenario, the scale factor $a(t)$ changes as follows under the broken time diffeomorphism:
\bea a(t)\Longrightarrow  a(t-\pi(t,{\bf x}))&=& a(t)-H\pi(t,{\bf x})a(t)+\cdots\approx a(t)\left(1-H\pi(t,{\bf x})\right)~,\eea
    using which we get:
    \bea a^2(t)\left(1-H\pi(t,{\bf x})\right)^2&\approx & a^2(t)\left(1-2H\pi(t,{\bf x})\right)=a^{2}(t)\left(1+2\zeta(t,{\bf x})\right).\eea 
This implies that the scalar curvature perturbation can be expressed using the linking relationship that follows. In terms of Goldstone modes $\pi(t,{\bf x})$, $\zeta(t,{\bf x})$ can be written as, $\zeta(t,{\bf x})\approx-H\pi(t,{\bf x})$. Here, we have limited our analysis to the linear relationship between the Goldstone mode and the comoving curvature perturbation. If we do not confine ourselves to the linear regime, then one can, in theory, take into consideration the following relationship:
\be \zeta(t,{\bf x})=-H\pi(t,{\bf x})-\frac{1}{2}\left(\epsilon-\eta\right)H^{2}\pi^{2}(t,{\bf x})+\cdots\ee
where the slow-roll parameters, $\epsilon$ and $\eta$, are explicitly expressed as they were previously discussed. Once we assume the limit $\tau_0\rightarrow 0$, this additional non-linear factor in the SR regime is minuscule and will be considered as a little correction. Since the linear relationship is all that matters in this situation, we have disregarded the non-linear suppressed contributions from our analysis.

We wrote the entire description in terms of the Goldstone mode in this paper, which we then translated into the comoving curvature perturbation language. That being said, the comoving curvature perturbation can be represented in terms of the scalar field fluctuation if we attempt to connect with any universal $P(X,\phi)$ single scalar field theory. This is one way to express $\delta\phi(t,{\bf x})$:
\be \zeta(t,{\bf x})=-\frac{H}{\dot{\phi}_0(t)}\delta\phi(t,{\bf x}).\ee
In this explanation, the background field is $\phi_0(t)$. This also suggests that the Goldstone mode and the scalar field fluctuation are related in the $P(X,\phi)$ theoretical description by the following expression:
\be \pi(t,{\bf x})=\frac{1}{\dot{\phi}_0(t)}\delta\phi(t,{\bf x}).\ee
Upon establishing this link, the expression for the co-moving curvature perturbation $\zeta(t,{\bf x})$ can be utilized to define the second-order perturbed action ($S^{(2)}_{\zeta}$) as follows:
 \bea 
  	S^{(2)}_{\zeta}&=&\displaystyle \int d^{4}x ~a^3\left(\frac{M^2_{pl}\epsilon}{c^2_s}\right)\Bigg(\dot{\zeta}^2-c^2_s\frac{\left(\partial_{i}\zeta\right)^2}{a^2}\Bigg).~~~~~\quad\quad\eea
   Furthermore, we will now utilize the conformal time coordinate to simplify the computation rather than the physical time coordinate for the remaining calculations. This allows us to write the second-order perturbed action as follows:
 \bea S^{(2)}_{\zeta}&=&M^2_{pl}\displaystyle \int d\tau\;  d^3x\;  a^2\;\left(\frac{\epsilon}{c^2_s}\right)\Bigg(\zeta^{'2}-c^2_s\left(\partial_i\zeta\right)^2\Bigg).~~~~\quad\eea  

    \subsection{Constructing Mukhanov Sasaki equation from EFT}

In order to obtain the field equation from the second-order perturbed action, the comoving curvature perturbation field redefinition requires the definition of a new variable, which is provided by:
   \bea v=zM_{ pl}\zeta \quad{\rm with}\quad z=\displaystyle \frac{a\sqrt{2\epsilon}}{c_s},\eea
In the context of this discussion, it is often referred to as the Mukhanov Sasaki (MS) variable \cite{Sasaki:1986hm,Garriga:1999vw}. Speaking in terms of the MS variable, the second order perturbed action that was previously addressed has the following canonically normalized form:
\bea
S^{(2)}_{\zeta}=\frac{1}{2}\int d\tau\;  d^3x\;  \bigg(v^{'2}-c^2_s\left(\partial_iv\right)^2+\frac{z^{''}}{z}v^{2}\bigg).
   \eea
Next, we will write the previously indicated action in the Fourier space using the following ansatz for the Fourier transformation:
\bea
   v(\tau,{\bf x})=\int\frac{d^3{\bf k}}{(2\pi)^3}\; e^{i{\bf k}.{\bf x}}\;v_{\bf k}(\tau).
   \eea
Further recasting of the previously indicated action results in the following Fourier transformed scalar modes:
  \bea
S^{(2)}_{\zeta}&=&\frac{1}{2}\int \frac{d^3{\bf k}}{(2\pi)^3}\;d\tau\;\bigg(|v^{'}_{ k}|^{2}-\omega^2(k,c_s,\tau)|v_{ k}|^{2}\bigg).
   \eea
   Subsequently, the MS equation for the scalar perturbed modes can be stated as follows after varying the described action:
 \bea
 v^{''}_{ k}+\omega^2(k,c_s,\tau)v_{ k}=0\,,
   \eea
where the following expression provides the expression for the effective time-dependent frequency: 
 \bea \omega^2(k,c_s,\tau):=\left(c^2_sk^2-\frac{z^{''}}{z}\right)\quad\displaystyle{\rm where}\quad\frac{z^{''}}{z}\approx\frac{2}{\tau^2}.\eea
The normalisation condition described before, which is given in terms of the Klein Gordon product for the scalar perturbed modes, is now used to fix the mathematical structure of the general solution:
   \be W[v_{ k},v^{'}_{ k}]:=\begin{vmatrix}
     v_{ k} & v^{*}_{ k} \\ 
     v^{'}_{ k} & v^{'*}_{ k} 
\end{vmatrix}=v^{'*}_{ k}v_{ k}-v^{'}_{ k}v^{*}_{k}=i.\ee
We will solve the Mukhanov Sasaki equation in different stages in the next two subsections, which will be very helpful for the remainder of this review's topic.

\subsection{Semi-Classical modes from Goldstone EFT with single sharp transition} 

The physical framework under consideration in this review consists of the three phases listed below, which are discussed in chronological order point-wise:
 \begin{enumerate}
       \item \underline{\bf Region I:} We first examine an area known as the Slow Roll (SRI) that endures over the conformal time scale $\tau<\tau_s$. The SRI transits to an Ultra Slow Roll (USR) zone at $\tau=\tau_s$. This suggests that in this arrangement, SRI terminates at $\tau=\tau_s$.

       \item \underline{\bf Region II:} Next, we take into consideration a region known as the Ultra Slow Roll (USR), which terminates at the scale $\tau=\tau_e$ and begins at the conformal time scale $\tau=\tau_s$. Here, $\tau=\tau_e$ is considered the second sharp transition scale from USR to the second Slow Roll (SRII) sharp transition.

       \item \underline{\bf Region III:} Lastly, we take into consideration the second Slow Roll (SRII) area, where the inflation terminates at $\tau=\tau_{\rm end}$ after a brief period of time, starting at $\tau=\tau_e$.

\end{enumerate}
In the ensuing subsections of this review, our task is to explicitly investigate the classical solution and its quantum consequences from these three locations independently.

\subsubsection{Region I: First Slow Roll (SRI) region}

The MS equation for the scalar perturbed mode can be generally solved as follows, and it is explicitly valid for the first SR (SRI) period ($\tau\leq\tau_s$):
\bea
   v_{ k}(\tau)&=&\frac{\alpha^{(1)}_{ k}}{\sqrt{2c_sk}}\left(1-\frac{i}{kc_s\tau}\right)\; e^{-ikc_s\tau}+\frac{\beta^{(1)}_{ k}}{\sqrt{2c_sk}}\left(1+\frac{i}{kc_s\tau}\right)\; e^{ikc_s\tau},
   \eea
where the coefficients $\beta^{(1)}_{ k}$ and $\alpha^{(1)}_{ k}$, commonly known as Bogoliubov coefficients which appear in the above equation, fix the solution in terms of the initial condition that is correctly selected. In terms of comoving curvature perturbation in the SRI period the solution can be further expressed in the following form:
 \bea
 \zeta_{ k}(\tau)&=&\frac{v_{ k}(\tau)}{zM_{ pl}}=\left(\frac{ic_sH}{2M_{\rm pl}\sqrt{\epsilon}}\right)\frac{1}{(c_sk)^{3/2}}\left[\alpha^{(1)}_{ k}\left(1+ikc_s\tau\right)\; e^{-ikc_s\tau}-\beta^{(1)}_{ k}\left(1-ikc_s\tau\right)\; e^{ikc_s\tau}\right].\quad
 \eea
As long as the observable restrictions for inflation from the CMB are met within the SR time, one can, in theory, select any beginning quantum vacuum state. Nonetheless, we opt for the most widely recognized one, the renowned Bunch Davies quantum vacuum state, in order to align with the Mikowski vacuum solution. In our case, the Bunch Davies vacuum, which is defined by the following equation, is essentially a Euclidean vacuum that fixes the initial condition in the first SR (SRI) period:
\bea \label{b1a}&&\alpha^{(1)}_{ k}=1, \\
 \label{b1b}&&\beta^{(1)}_{ k}=0.\eea
Making this decision will be very helpful for the additional calculations made throughout the remainder of the review. The following basic abbreviation can be used to define the scalar mode function in terms of the previously mentioned initial condition in the first SR (SRI) region:
\bea
 v_{ k}(\tau)=\frac{1}{\sqrt{2c_sk}}\left(1-\frac{i}{kc_s\tau}\right)\; e^{-ikc_s\tau}.
 \eea
 Utilizing the aforementioned MS equation solution for the scalar mode, which is fixed in terms of the Bunch Davies initial condition, in the first SR regime, $\tau\leq \tau_s$, one can additionally formulate the comoving curvature perturbation formula in terms of the momentum modes, conformal time, and effective sound speed parameter $c_s$:
  \bea
 \zeta_{ k}(\tau)&=&\frac{v_{ k}(\tau)}{zM_{ pl}}=\left(\frac{ic_sH}{2M_{\rm pl}\sqrt{\epsilon}}\right)\frac{1}{(c_sk)^{3/2}}\left(1+ikc_s\tau\right)\; e^{-ikc_s\tau}.\quad
 \eea
Although $\epsilon$, the first-slow roll parameter, varies extremely slowly with time scale, it is roughly a constant and small number in the SRI region. It is noteworthy that the sub-horizon area, denoted by the limit $-kc_s\tau\gg 1$, is where the loop corrections become substantial. Quite the contrary, the associated scalar modes become fully frozen and exhibit classical behavior in the super-horizon scale, denoted by $-kc_s\tau\ll 1$. This further suggests that we must employ the semi-classical approximations, which are the mixing of quantum and classical effects, at the horizon crossing scale, which is governed by $-kc_s\tau= 1$.

\subsubsection{Region II: Ultra Slow Roll (USR) region}

We now turn our attention to the Ultra Slow Roll (USR) region, which corresponds to the valid scalar modes and may be seen in the conformal time scale window $\tau_s\leq \tau\leq \tau_e$. The steep transition scale $\tau_s$, which marks the change from the first SR (SRI) to USR, is identified in this description. The temporal scale, $\tau_e$, on the other hand, describes the end of the USR period and the inflationary paradigm. Time dependence of the first slow-roll parameter in the USR regime can be explicitly recorded and is translated into the following expression in terms of the first SR contribution:
\be \epsilon(\tau)=\epsilon \;\left(\frac{a(\tau_s)}{a(\tau)}\right)^{6}=\epsilon  \;\left(\frac{\tau}{\tau_s}\right)^{6}\quad\quad\quad{\rm where}\quad\quad\tau_s\leq\tau\leq \tau_e.\ee
In this case, the initial slow-roll parameter in the SR zone is $\epsilon$, which we clearly established in the first section of the discussion. The mathematical form described earlier indicates that this parameter approaches a constant value at the point when the sharp transition from the first SR (SRI) to USR occurs, i.e., at $\tau=\tau_s$, where $\epsilon(\tau_s)=\epsilon$, which is really at the first SR (SRI) regime. Following the initial rapid transition from SR (SRI) to USR at the scale $\tau>\tau_s$, the departure from the constant behavior manifests itself. It is significant to note that, for the purposes of this computation, we have especially taken into account a sharp transition from the first SR (SRI) to the USR area, which will provide incredibly helpful information for the remainder of the study. With respect to the particular time-dependent behavior and non-constancy of the first slow-roll parameter, the following simplified expression can be used to represent the solution of the MS equation for the comoving curvature perturbation in the USR period:
\bea
 \zeta_{ k}(\tau)&=&\left(\frac{ic_sH}{2M_{ pl}\sqrt{\epsilon}}\right)\left(\frac{\tau_s}{\tau}\right)^{3}\frac{1}{(c_sk)^{3/2}}\times\bigg[\alpha^{(2)}_{ k}\left(1+ikc_s\tau\right)\; e^{-ikc_s\tau}-\beta^{(2)}_{ k}\left(1-ikc_s\tau\right)\; e^{ikc_s\tau}\bigg],
 \eea
It is noteworthy to mention that the explicit structure of the momentum, sharp transition scale time $\tau_s$, and the effective sound speed dependent coefficients $\alpha^{(2)}_{ k}$ and $\beta^{(2)}_{ k}$ in the aforementioned solution obtained in the USR region can be expressed in terms of the initial condition fixed in terms of Bunch Davies vacuum in the first SR (SRI) region via Bogoliubov transformations. This further suggests that, in contrast to the Bunch Davies initial state, the underlying structure of the vacuum state is altered in the USR region. Our goal in the USR region is to find these Bogoliubov coefficients $\beta^{(2)}_{ k}$ and $\alpha^{(2)}_{ k}$. 

The two boundary conditions that follow can be used to accomplish this. These are provided by the following formulas and can be regarded as Israel junction conditions that we must apply at the first sharp transition scale (SRI) to the USR, $\tau=\tau_s$:

\begin{itemize}

    \item \underline{\bf Condition I:} The scalar modes derived from the USR region and the first SR (SRI) are implied to become continuous at the steep transition point $\tau=\tau_s$ i.e. $\left[\zeta_{ k}(\tau)\right]_{\rm SR, \tau=\tau_s}= \left[\zeta_{ k}(\tau)\right]_{\rm USR, \tau=\tau_s}$.

    \item \underline{\bf Condition II:} It suggests that at the sharp transition point $\tau=\tau_s$, the momentum modes for the scalar perturbation derived from the first SR (SRI) and USR region become continuous i.e. $\left[\zeta^{'}_{ k}(\tau)\right]_{\rm SR, \tau=\tau_s}= \left[\zeta^{'}_{ k}(\tau)\right]_{\rm USR, \tau=\tau_s}$.

\end{itemize}
The Bogoliubov coefficients in the USR region, $\alpha^{(2)}_{\bf k}$ and $\beta^{(2)}_{\bf k}$, can be expressed in the following closed form when the aforementioned junction requirements are imposed:
\bea \label{b2a}\alpha^{(2)}_{ k}&=&1-\frac{3}{2ik^{3}c^{3}_s\tau^{3}_s}\left(1+k^{2}c^{2}_s\tau^{2}_s\right),\\
\label{b2b}\beta^{(2)}_{ k}&=&-\frac{3}{2ik^{3}c^{3}_s\tau^{3}_s}\left(1+ikc_s\tau_s\right)^{2}\; e^{-2ikc_s\tau_s}.\eea
Now that the structure of the Bogoliubov coefficients has been fixed with respect to the momentum mode, effective sound speed, and conformal time scale at the first sharp transition point $\tau=\tau_s$ from SR to USR, the corresponding modified structure of the quantum vacuum state has also been fixed. The analysis carried out in the remaining portion of the study will greatly benefit from this information.

\subsubsection{ Region III: Second Slow Roll (SRII) region}

Our attention will now be directed towards the second Slow Roll (SRII) area, which can be seen as $\tau_e\leq \tau\leq \tau_{\rm end}$ in the conformal time scale window. The time scale, $\tau_e$, in this definition represents the sharp transition scale from the USR to SRII. On the other hand, the time scales $\tau_e$ and $\tau_{\rm end}$, respectively, characterize the end of the USR period and the inflationary paradigm. The first slow-roll parameter's time dependence in the SRII regime can be explicitly recorded and is represented as follows in terms of the first SR (SRI) contribution:
\be \epsilon(\tau)=\epsilon \;\left(\frac{a(\tau_s)}{a(\tau_e)}\right)^{6}=\epsilon  \;\left(\frac{\tau_e}{\tau_s}\right)^{6}\quad\quad\quad{\rm where}\quad\quad\tau_e\leq\tau\leq \tau_{\rm end}.\ee
In this case, the initial slow-roll parameter in the SRI region is $\epsilon$, which we clearly established in the first section of the discussion. According to the previously given mathematical form, this parameter is approximately a non-constant amount at the moment of the sudden transition from the USR to the SRII, or at $\tau=\tau_e$, which is truly the end of the USR regime. This number will not change in the interval between $\tau_e<\tau<\tau_{\rm end}$ and the time scale corresponding to the end of inflation, but the departure from the constant behavior persists up until that point. At the temporal scale, $\tau=\tau_e$, we are now examining a sharp transition. Under the current circumstances, the solution of the MS equation varies based on the particular time-dependent behavior and non-constancy of the first slow-roll parameter.
The second SR (SRII) region, which is expected to occur at the sharp transition point $\tau=\tau_e$, requires consideration, as we have already mentioned. In the region $\tau>\tau_e$, the SR features continue to exist with the non-constant value of the first slow-roll parameter. Accordingly, the MS equation in terms of the scalar modes in the $\tau_e\leq\tau\leq \tau_{\rm end}$ region is given by:
\bea
 \zeta_{ k}(\tau)&=&\left(\frac{ic_sH}{2M_{ pl}\sqrt{\epsilon}}\right)\left(\frac{\tau_s}{\tau_e}\right)^{3}\frac{1}{(c_sk)^{3/2}}\times\bigg[\alpha^{(3)}_{\bf k}\left(1+ikc_s\tau\right)\; e^{-ikc_s\tau}-\beta^{(3)}_{\bf k}\left(1-ikc_s\tau\right)\; e^{ikc_s\tau}\bigg],
 \eea
In this case, $\tau_{\rm end}$ denotes the conformal time scale at the end of inflation, which occurs, strictly speaking, at the end of the SRII phase. Additionally, $k_{\rm end}$ provides the momentum scale linked with the conformal time scale $\tau_{\rm end}$, which will be indispensable for the remaining discussions. Notably, the explicit structure of the momentum, the sharp transition scale time $\tau_e$, and the effective sound speed dependent coefficients $\alpha^{(3)}_{\bf k}$ and $\beta^{(3)}_{\bf k}$ in the aforementioned solution obtained in the second SR region can all be expressed in terms of the boundary condition fixed in terms of new modified vacuum in the USR region via Bogoliubov transformations. This further suggests that, in contrast to the previously computed vacuum state in the USR region, the underlying structure of the vacuum state changes in the second SR (SRII) region. Our goal in the second SR (SRII) region is to explicitly find these Bogoliubov coefficients $\beta^{(3)}_{\bf k}$ and $\alpha^{(3)}_{\bf k}$.

This can be accomplished by applying the next two boundary conditions, which are provided by the following formulas and which, in theory, can be understood as Israel junction conditions that we must apply at the USR to the second SR (SRII) sharp transition scale, $\tau=\tau_e$:

\begin{itemize}
    \item \underline{\bf Condition I:} This suggests that at the sharp transition point, $\tau = \tau_e$ the scalar modes from the USR and the second SR (SRII) area become continuous i.e. $\left[\zeta_{ k}(\tau)\right]_{\rm USR, \tau=\tau_e}= \left[\zeta_{ k}(\tau)\right]_{\rm SR, \tau=\tau_e}$.

 \item \underline{\bf Condition II:} It suggests that at the sharp transition point $\tau=\tau_e$, the momentum modes for the scalar perturbation derived from USR and the second SR region (SRII) become continuous i.e. $\left[\zeta^{'}_{ k}(\tau)\right]_{\rm USR, \tau=\tau_e}= \left[\zeta^{'}_{ k}(\tau)\right]_{\rm SR, \tau=\tau_e}$.

\end{itemize}
Following the imposition of the aforementioned junction conditions, two constraint equations arise, which when solved yield the closed forms of the Bogoliubov coefficients in the second SR (SRII) region, $\beta^{(3)}_{ k}$ and $\alpha^{(3)}_{ k}$, which are given by:
\bea \label{b3a}\alpha^{(3)}_{ k}&=&-\frac{1}{4k^6c^6_s\tau^3_s\tau^3_e}\Bigg[9\left(kc_s\tau_s-i\right)^2\left(kc_s\tau_e+i\right)^2 e^{2ikc_s(\tau_e-\tau_s)}\nonumber\\
&&\quad\quad\quad\quad\quad\quad\quad\quad\quad\quad\quad\quad\quad\quad\quad\quad-
\left\{k^2c^2_s\tau^2_e\left(2kc_s\tau_e-3i\right)-3i\right\}\left\{k^2c^2_s\tau^2_s\left(2kc_s\tau_s+3i\right)+3i\right\}\Bigg],\\
\label{b3b}\beta^{(3)}_{ k}&=&\frac{3}{4k^6c^6_s\tau^3_s\tau^3_e}\Bigg[\left(kc_s\tau_s-i\right)^2\left\{k^2c^2_s\tau^2_e\left(3-2ikc_s\tau_e\right)+3\right\}e^{-2ikc_s\tau_s}\nonumber\\
&&\quad\quad\quad\quad\quad\quad\quad\quad\quad\quad\quad\quad\quad\quad\quad\quad+i\left(kc_s\tau_e-i\right)^2\left\{3i+k^2c^2_s\tau^2_s\left(2kc_s\tau_s+3i\right)\right\}e^{-2ikc_s\tau_e}\Bigg].\eea

\subsection{Semi-Classical modes from Goldstone EFT generalized for multiple sharp transitions}

Within this arrangement of multiple (here we fix at six) successive sharp transitions, the general $n$th USR phase is designated as USR$_{n}$, and in a similar manner, the $(n+1)$th SR phase is designated as SR$_{n+1}$. The index $n$ ranges from $1\to 6$, and the phase SR$_{1}$ represents the first slow-roll, SRI, phase that satisfies $\Delta{\cal N}_{\rm SRI}=\ln{(k_{s_{1}}/k_{*})}\sim{\cal O}(12)$ for the fixed value of the pivot scale ($k_{*}=0.02{\rm Mpc^{-1}}$). As a result, after resolving the MS equation, the appropriate mode solutions for these SR and the USR phases are as follows: 
\bea
\zeta_{k, \rm SR_{1}}(\tau) &=& \bigg[\frac{ic_s H}{2 M_p \sqrt{\epsilon}}\bigg]_{*}\times \frac{1}{(c_s k)^{3/2}} \bigg[\alpha_{k}^{(1)}(1+ikc_s \tau)\exp{(-ikc_s \tau)}-\beta_{k}^{(1)}(1-ikc_s \tau)\exp{(ikc_s\tau)}\bigg], \nonumber \\
\zeta_{k, \rm USR_n}(\tau) &=& \bigg[\frac{ic_s H}{2 M_p \sqrt{\epsilon}}\bigg]_{*} \times \bigg(\frac{\tau_{s_{n}}}{\tau}\bigg)^3 \bigg[\alpha_{k}^{(2n)}(1+ikc_s \tau)\exp{(-ikc_s \tau)}-\beta_{k}^{(2n)}(1-ikc_s \tau)\exp{(ikc_s\tau)}\bigg], \nonumber \\
\zeta_{k, \rm SR_{n+1}}(\tau) &=& \bigg[\frac{ic_s H}{2 M_p \sqrt{\epsilon}}\bigg]_{*} \times \bigg(\frac{\tau_{s_{n}}}{\tau_{e_{n}}}\bigg)^3 \bigg[\alpha_{k} ^{(2n+1)}(1+ikc_s \tau)\exp{(-ikc_s \tau)}-\beta_{k}^{(2n+1)}(1-ikc_s \tau)\exp{(ikc_s\tau)}\bigg].
\eea
The pivot scale is used to assess the expression enclosed in parenthesis. The values of $\beta_{k} ^{(i)}$ and $\alpha_{k} ^{(i)}$The Bogoliubov coefficients that follow from applying the Israel junction conditions at the boundaries across the conformal times $\tau=\tau_{s_{n}}$ and $\tau=\tau_{e_{n}}$ are denoted as  (where $i\in 1,2,3,\cdots$). We will now talk about how the mode solutions mentioned above give us the ability to apply the MST theory. To do this, we start with an initial SR$_{1}$ phase that incorporates the pivot scale $k_{*}$. In this phase, the second SR parameter, $\eta = \epsilon'/\epsilon{\cal H}$, is a very small number that is treated as constant, whereas the first SR parameter, $\epsilon$, is almost a constant. This phase continues until the conformal time $\tau=\tau_{s_{1}}$, at which an abrupt transition is encountered. The first USR phase, USR$_{1}$, now begins. It corresponds to the conformal time period $\tau_{s_{1}} \leq \tau \leq \tau_{e_{1}}$. Notably, the $\eta$ parameter abruptly jumps from $\eta \rightarrow 0$ to $\eta \rightarrow -6$ during the transition into this phase. As we shall shortly demonstrate, this sharp increase in value has a cascading effect that leads to substantial one-loop contributions. The required bound on the e-foldings is respected by this phase: $\Delta{\cal N}_{\rm USR_{1}} \sim {\cal O}(2)$. Another abrupt change into the new SR$_{2}$ phase occurs from USR$_{1}$ at $\tau_{e_{1}}$, and it lasts for the interval $\tau_{e_{1}} \leq \tau \leq \tau_{s_{2}}$. The $\epsilon$ parameter returns to being virtually constant in this new SR phase, with a value of $\epsilon(\tau) = \epsilon(\tau_{e_{1}}/\tau_{s_{1}})^{6}$, where $\epsilon$ is the same as in the SR$_{1}$. The $\eta$ parameter also decreases to its beginning value of $\eta \rightarrow 0$. Subsequently, until $\tau = \tau_{e_{2}}$, we see another abrupt transition at $\tau = \tau_{s_{2}}$ into a new USR$_{2}$ phase, which respects the same constraint on its interval of $\Delta{\cal N}_{\rm USR_{2}} \sim {\cal O}(2)$. This pattern is maintained in a similar manner until the requirement of having all e-foldings satisfy $\Delta{\cal N}_{\rm Total} \sim {\cal O}(60-70)$ is eventually met.  This is mainly the reason why we chose to consider six sharp transitions in our analysis. Keep in mind that six abrupt transitions are the bare minimum needed to meet the requirements for solving the Horizon Problem. In actuality, you can solve the Horizon problem automatically and make more abrupt transitions that nevertheless adhere to the number of efoldings required for inflation. Next, we will talk about and give the formulae for the Bogoliubov coefficients ($\alpha_{k}$, $\beta_{k}$) for the various phases. For any given SR and USR phase, the Bogoliubov coefficients are as follows: 
\bea
\alpha^{(2n)}_{k}&=&-\frac{k_{e_{n-1}}^{3}k_{s_{n}}^{3}}{2k^{3}k_{s_{n-1}}^{3}}
\bigg[\bigg(3i-\frac{2k^{3}}{k_{s_{n}}^{3}}+\frac{3ik^{2}}{k_{s_{n}}^{2}}\bigg)\alpha^{(2n-1)}_{k}+3i\exp{\left(-\frac{2ik}{k_{s_{n}}}\right)}\bigg(i-\frac{k}{k_{s_{n}}}\bigg)^{2}\beta^{(2n-1)}_{k}\bigg],\\
\beta^{(2n)}_{k}&=&
\exp{\left(\frac{2ik}{k_{s_{n}}}\right)}\frac{k_{e_{n-1}}^{3}k_{s_{n}}^{3}}{2k^{3}k_{s_{n-1}}^{3}}
\bigg[3i\bigg(i+\frac{k}{k_{s_{n}}}\bigg)^{2}\alpha^{(2n-1)}_{k}+\exp{\left(-\frac{2ik}{k_{s_{n}}}\right)}\bigg(3i+\frac{2k^{3}}{k_{s_{n}}^{3}}+\frac{3ik^{2}}{k_{s_{n}}^{2}}\bigg)
\beta^{(2n-1)}_{k}\bigg],\\
\alpha^{(2n+1)}_{k}&=&\frac{k_{e_{n}}^{3}}{2k^{3}}
\bigg[\bigg(3i+\frac{2k^{3}}{k_{e_{n}}^{3}}+\frac{3ik^{2}}{k_{e_{n}}^{2}}\bigg)\alpha^{(2n)}_{k}+3i\exp{\left(-\frac{2ik}{k{e_{n}}}\right)}\bigg(i-\frac{k}{k_{e_{n}}}\bigg)^{2}\beta^{(2n)}_{k}\bigg],\\
\beta^{(2n+1)}_{k} &=& -\exp{\left(\frac{2ik}{k_{e_{n}}}\right)}\frac{k_{e_{n}}^{3}}{2k^{3}}
\bigg[3i\bigg(i+\frac{k}{k_{e_{n}}}\bigg)^{2}\alpha^{(2n)}_{k}+\exp{\left(-\frac{2ik}{k_{e_{n}}}\right)}\bigg(3i-\frac{2k^{3}}{k_{e_{n}}^{3}}+\frac{3ik^{2}}{k_{e_{n}}^{2}}\bigg)
\beta^{(2n)}_{k}\bigg].
\eea
When the Horizon crossing requirements are used, these expressions become simpler: $-k_{e_{n}}c_{s}\tau_{e_{n}} = 1$ is the result of $-k_{s_{n}}c_{s}\tau_{s_{n}}$. The speed at which sound is produced effectively $c_s$ is essential to the current arrangement of abrupt transitions. $c_{s}(\tau_{*})=c_{s}$ is the value that was previously determined at the pivot scale. However, its value changes to $c_{s}(\tau_{s_{n}}) = c_{s}(\tau_{e_{n}}) = \tilde{c_{s}} = 1 \pm \delta$ at the moments when the sharp transition occurs, where $\delta \ll 1$. The value of $\tilde{c_{s}}$ is constant at every subsequent conformal time of the transition $(\tau_{s_{n}},\tau_{e_{n}})$ for $n=1\rightarrow 6$. Here, $\tilde{c_{s}} \ne c_{s}$. This sound speed behavior design facilitates the development of PBH from our setup by allowing for the required amplification to the scalar perturbations; it will also prove to be significant in the long run from the one-loop contributions perspective.

\subsection{Tree level primordial power spectrum from scalar modes}

\subsubsection{Quantization of classical scalar mode}

In order to obtain the expression for the two-point correlation function and the corresponding power spectrum in Fourier space—which are required to calculate the cosmic correlations—we must explicitly quantize the scalar modes suitably. To do this, we first need to build the creation operator, $\hat{a}^{\dagger}_{\bf k}$, and the annihilation operator, $\hat{a}_{\bf k}$, which will create and destroy an excited state from the original Bunch Davies state, respectively. For the remainder of the purposes, we now identify $|0\rangle$ as the initial state of Bunch Davies, which has to follow the following constraint: 
\be \hat{a}_{\bf k}|0\rangle=0\quad\forall {\bf k}.\ee
Now, the scalar perturbed mode and its accompanying canonically conjugate momenta have to satisfy the equal time commutation relations:
\bea &&\left[\hat{\zeta}_{\bf k}(\tau),\hat{\zeta}^{'}_{{\bf k}^{'}}(\tau)\right]_{\rm ETCR}=i\;\delta^{3}\left({\bf k}+{\bf k}^{'}\right),~~~~~
\left[\hat{\zeta}_{\bf k}(\tau),\hat{\zeta}_{{\bf k}^{'}}(\tau)\right]_{\rm ETCR}=0,~~~~~
\left[\hat{\zeta}^{'}_{\bf k}(\tau),\hat{\zeta}^{'}_{{\bf k}^{'}}(\tau)\right]_{\rm ETCR}=0.\eea
Regarding the scalar mode and its conjugate momenta, the following formulas denote the proper quantum mechanical operators:
\bea \hat{\zeta}_{\bf k}(\tau)&=&\bigg[{\zeta}_{\bf k}(\tau)\hat{a}_{\bf k}+{\zeta}^{*}_{\bf k}(\tau)\hat{a}^{\dagger}_{-{\bf k}}\bigg]=\frac{c_s}{a\sqrt{2\epsilon} M_{pl}}\bigg[v_{ k}(\tau)\hat{a}_{\bf k}+v^{*}_{ k}(\tau)\hat{a}^{\dagger}_{-{\bf k}}\bigg],\\
\hat{\zeta}^{'}_{\bf k}(\tau)&=&\bigg[{\zeta}^{'}_{\bf k}(\tau)\hat{a}_{\bf k}+{\zeta}^{*'}_{\bf k}(\tau)\hat{a}^{\dagger}_{-{\bf k}}\bigg]=\frac{c_s}{a\sqrt{2\epsilon} M_{pl}}\bigg[v^{'}_{ k}(\tau)\hat{a}_{\bf k}+v^{*'}_{ k}(\tau)\hat{a}^{\dagger}_{-{\bf k}}\bigg]-\frac{c^{2}_s}{2\epsilon a^{2} M_{pl}}\bigg(\frac{a\sqrt{2\epsilon}}{c_s}\bigg)^{'}\bigg[v_{ k}(\tau)\hat{a}_{\bf k}+v^{*}_{ k}(\tau)\hat{a}^{\dagger}_{-{\bf k}}\bigg].\quad\quad \eea
which, in the next part, when we employ the in-in formalism to accomplish the one-loop computation, will be quite useful.

As previously mentioned, this can also be represented in terms of every possible commutation connection between the creation and annihilation operators:
\bea \left[\hat{a}_{\bf k},\hat{a}^{\dagger}_{{\bf k}^{'}}\right]&=&\delta^{3}\left({\bf k}-{\bf k}^{'}\right),~~~~
 \left[\hat{a}_{\bf k},\hat{a}_{{\bf k}^{'}}\right]=0=\left[\hat{a}^{\dagger}_{\bf k},\hat{a}^{\dagger}_{{\bf k}^{'}}\right].\eea

\subsubsection{Computing tree level primordial power spectrum}



The comoving curvature perturbation is known to occur at the late time scale, $\tau\rightarrow 0$, using which  the following represents the relevant tree-level contribution to the two-point cosmic correlation function for the scalar comoving curvature perturbation:
\bea \langle \hat{\zeta}_{\bf k}\hat{\zeta}_{{\bf k}^{'}}\rangle_{{\bf Tree}} &=&(2\pi)^{3}\;\delta^{3}\left({\bf k}+{\bf k}^{'}\right)P^{\bf Tree}_{\zeta}(k),\quad\eea
In the Fourier space, the dimensionful power spectrum is represented as $P^{\bf Tree}_{\zeta}(k)$, which can be expressed in a simplified manner as follows:
\bea P^{\bf Tree}_{\zeta}(k)=\langle \hat{\zeta}_{\bf k}\hat{\zeta}_{-{\bf k}}\rangle_{(0,0)}=\left[{\zeta}_{\bf k}(\tau){\zeta}_{-{\bf k}}(\tau)\right]=|{\zeta}_{\bf k}(\tau)|^{2}.\quad\eea
Nonetheless, it is always pertinent to deal with a dimensionless form of the power spectrum in Fourier space, which is represented by the following expression, for practical purposes and to make a connection with cosmological observation:
\bea \Delta^{2}_{\zeta,{\bf Tree}}(k)=\frac{k^{3}}{2\pi^{2}}P^{\bf Tree}_{\zeta}(k)=\frac{k^{3}}{2\pi^{2}}|{\zeta}_{\bf k}(\tau)|^{2}_{\tau\rightarrow 0}.\eea

Now using both scenarios, with single and multiple sharp transitions, we now mention the possible outcomes at the tree level.
\subsubsubsection{For single sharp transition}
It is evident from the current study that the modes of solution for the scalar cosmological perturbations differ in the first SR (SRI), USR, and second SR (SRII) regions, as we have specifically computed in this review. Here, at the tree level, the computed scalar modes from the first SR (SRI), USR, and in the second SR (SRII) region can be used to calculate the dimensionless power spectrum as follows:
\bea   \bigg[\Delta^{2}_{\zeta,{\bf Tree}}(k)\bigg]
&=& \left(\frac{H^{2}}{8\pi^{2}M^{2}_{ pl}\epsilon c_s}\right)_* \nonumber\\
&&\times
\left\{
	\begin{array}{ll}
		\displaystyle\left(1+k^{2}c^{2}_s\tau^{2}\right)& \mbox{when}\quad  k\ll k_s  \;(\rm SRI)  \\  
			\displaystyle 
			\displaystyle\left(\frac{k_e}{k_s }\right)^{6}\times\left|\alpha^{(2)}_{\bf k}\left(1+ikc_s\tau\right)\; e^{-ikc_s\tau}-\beta^{(2)}_{\bf k}\left(1-ikc_s\tau\right)\; e^{ikc_s\tau}\right|^{2} & \mbox{when }  k_s\leq k\leq k_e  \;(\rm USR)\\ 
   \displaystyle 
			\displaystyle\left(\frac{k_e}{k_s }\right)^{6}\times\left|\alpha^{(3)}_{\bf k}\left(1+ikc_s\tau\right)\; e^{-ikc_s\tau}-\beta^{(3)}_{\bf k}\left(1-ikc_s\tau\right)\; e^{ikc_s\tau}\right|^{2} & \mbox{when }  k_e\leq k\leq k_{\rm end}  \;(\rm SRII) 
	\end{array}
\right. .\eea
It is noteworthy that the statement given above holds true in the following scales: sub-horizon ($-kc_s\tau\gg 1$), super-horizon ($-kc_s\tau\ll 1$), and horizon re-entry ($-kc_s\tau= 1$). Here, the explicit derivations of the Bogoliubov coefficients for the USR region, $(\alpha^{(2)}_{\bf k},\beta^{(2)}_{\bf k})$, and the second SR region (SRII), $(\alpha^{(3)}_{\bf k},\beta^{(3)}_{\bf k})$, can be found in equations (\ref{b2a}), (\ref{b2b}), (\ref{b3a}), and (\ref{b3b}), separately. The Bunch Davies starting condition, which can be found in equations (\ref{b1a} and (\ref{b1b}), fixes the Bogoliubov coefficients for the first SR region (SRI), $(\alpha^{(1)}_{\bf k},\beta^{(1)}_{\bf k})$.  For the current computation, it is also necessary to take into account the wave numbers $k_e$ and $k_s$ that correspond to the time scales, $\tau_e$ and $\tau_s$. The pivot scale that is pertinent to CMB is represented by the symbol $*$. In the current context of the discussion, it is possible to discern the specific contributions originating from the first SR region (SRI), USR region, and second SR region (SRII) based on the structure mentioned above. 

For the three successive phases, SRI, USR, and SRII, the individual contribution of the tree-level scalar power spectrum can be further simplified as follows in the limit, $\tau\to 0$ of interest, corresponding to super-hubble scales, $-kc_s\tau\ll 1$:
\bea   \bigg[\Delta^{2}_{\zeta,{\bf Tree}}(k)\bigg]
&=& \left(\frac{H^{2}}{8\pi^{2}M^{2}_{ pl}\epsilon c_s}\right)_* \times
\left\{
	\begin{array}{ll}
		\displaystyle 1& \mbox{when}\quad  k\ll k_s  \;(\rm SRI)  \\  
			\displaystyle 
			\displaystyle\left(\frac{k_e}{k_s }\right)^{6}\times\left|\alpha^{(2)}_{\bf k}-\beta^{(2)}_{\bf k}\right|^{2} & \mbox{when }  k_s\leq k\leq k_e  \;(\rm USR)\\ 
   \displaystyle 
			\displaystyle\left(\frac{k_e}{k_s }\right)^{6}\times\left|\alpha^{(3)}_{\bf k}-\beta^{(3)}_{\bf k}\right|^{2} & \mbox{when }  k_e\leq k\leq k_{\rm end}  \;(\rm SRII) 
	\end{array}
\right. \eea  
In the limit $\tau\to 0$, it can be observed that the scalar modes first exit the horizon and reach the super-hubble regime, then eventually return to the horizon. The amplitude that corresponds to the modes' departure from the horizon freezes and they turn classical. Consequently, the super-horizon scale contributions are utilized to provide an approximation of the scalar power spectrum amplitude at the horizon re-entry.

Accordingly, the resulting formula for the tree-level amplitude of the primordial scalar power spectrum can be characterized by the following expression in the super-horizon scale ($-kc_s\tau\ll 1$) when the contributions from SRI, USR, and SRII phases are added together in the presence of sharp transition \footnote{It is noteworthy to mention that the first term in the equation (\ref{totree}) represents the contribution originating from the SRI phase. This specific contribution won't change from the small to the large size if Framework I's other stages are missing. But for the purposes of this discussion, we're primarily interested in discussing how PBHs are produced, which can be achieved by inserting a USR phase followed by another SRII phase. To produce PBHs with the proper amplitude in both of these indicated phases, the scalar power spectrum needs to be enhanced. Because of this, there are two more terms in the equation (\ref{totree}) besides the contribution from the SRI phase.
}:
\bea \label{totree}\bigg[\Delta^{2}_{\zeta,{\bf Tree}}(k)\bigg]_{\bf Total}=\bigg[\Delta^{2}_{\zeta,{\bf Tree}}(k)\bigg]_{\bf SR}\times\Bigg\{1+\left(\frac{k_e}{k_s }\right)^{6}\times\bigg[\Theta(k-k_s)\left|\alpha^{(2)}_{\bf k}-\beta^{(2)}_{\bf k}\right|^{2}+\Theta(k-k_e)\left|\alpha^{(3)}_{\bf k}-\beta^{(3)}_{\bf k}\right|^{2}\bigg]\Bigg\},\quad\quad\quad\eea
where the amplitude of the scalar power spectrum in the SRI phase is represented by the subscript {\bf SR}. This appears as an overall common factor, and in the super-horizon scale, it may be found using the following expression:
\bea \label{sramp}\bigg[\Delta^{2}_{\zeta,{\bf Tree}}(k)\bigg]_{\bf SR}:=\left(\frac{H^{2}}{8\pi^{2}M^{2}_{ pl}\epsilon c_s}\right)_*.\eea
The first term in the equation (\ref{totree}) represents the contribution from the SRI phase, which is crucial to note in this context. The specific contribution won't change from the small to the large size if Framework I doesn't include any more phases. The creation of PBHs, which can be accomplished by inserting a USR phase followed by another SRII phase, is what interests us in the current situation. The production of PBHs with the appropriate amplitude in both of these phases necessitates amplification of the scalar power spectrum. This explains why, in addition to the contribution from the SRI phase, the equation (\ref{totree}) has two more terms. To appropriately fulfill the need of sharp transitions at the scales, $k_s$ and $k_e$, respectively, two Heaviside theta functions are placed here. Instead of sharp transition if we implement the smooth transition in the present context of discussion then the expression in that case is modified by the following expression:
\bea \label{totreex}\Delta^{2}_{\zeta,{\bf Tree,Total}}(k)&=&\bigg[\Delta^{2}_{\zeta,{\bf Tree}}(k)\bigg]_{\bf SR}\nonumber\\
&&\quad\quad\quad\times\Bigg\{1+\left(\frac{k_e}{k_s }\right)^{6}\times\bigg[{\rm tanh}\left(\frac{k-k_s}{\Delta k}\right)\left|\alpha^{(2)}_{\bf k}-\beta^{(2)}_{\bf k}\right|^{2}+{\rm tanh}\left(\frac{k-k_e}{\Delta k}\right)\left|\alpha^{(3)}_{\bf k}-\beta^{(3)}_{\bf k}\right|^{2}\bigg]\Bigg\},\quad\quad\quad\eea
where $\Delta k=k_e-k_s$ represents the width of the USR phase for the case when smooth transition is considered. When we take the limit $\Delta k$ is very small then one can get back the same result as written for the sharp transition. However, it is important to note that, apart from smoothening the behaviour of the joining functions at the transitions points in both the cases for tree level power spectrum we get approximately the same physical outcomes. This further implies that, the behaviour of the transitions will not change the final outcomes at least at the tree-level result of the primordial power spectrum for the scalar modes. The overall contribution appearing in the SRI phase is consistent with the canonical ($c_s=1$) \cite{Baumann:2009ds,Chen:2010xka,Baumann:2018muz} and non-canonical ($c_s\neq 1$) \cite{Chen:2006nt,Chen:2010xka} models of single-field inflation. The authors have taken advantage of the fact that $M_{pl}=1$ in the aforementioned refs. \cite{Chen:2006nt,Chen:2010xka}. The power spectrum in this Planckian unit is dimensionless in and of itself and can be expressed as follows: $\bigg[\Delta^{2}_{\zeta,{\bf Tree}}(k)\bigg]_{\bf SR}:=\displaystyle\left(\frac{H^{2}}{8\pi^{2}\epsilon c_s}\right)_*$. Nevertheless, we have maintained the factor $M_{pl}$ throughout our computation, and in this specific instance, $H/M_{pl}$ is a dimensionless factor. As a result, the numerical estimate of the scalar power spectrum is dimensionless and consistent with the observed value, $\bigg[\Delta^{2}_{\zeta,{\bf Tree}}(k)\bigg]_{\bf SR}\sim 2.2\times 10^{-9}$. We have chosen to offer this clarification in order to prevent any more misunderstandings and because we think it will be useful in examining the physics behind this section. They can be combined and expressed in terms of a single formula within the single field EFT framework. It is important to note that the results produced in this section are fully consistent with the previously investigated numerous related topics, as mentioned in the references, which serves as additional justification for the validity of the tree-level result \cite{Kristiano:2022maq,Riotto:2023hoz,Kristiano:2023scm,Riotto:2023gpm,Cheng:2023ikq, Tasinato:2023ukp,Bhattacharya:2023ysp,Motohashi:2023syh,Firouzjahi:2023ahg,Franciolini:2023lgy,Firouzjahi:2023aum,Cheng:2023ikq,Tasinato:2023ukp}.

  \subsubsubsection{For multiple sharp transitions}
  The dimensionless scalar power spectrum at each region of SR$_1$, USR$_n$, and SR$_{n+1}$ will differ in value since the mode solutions are different for each of these regions in presence of multiple sharp transitions. But there will be a single general phrase that is applicable to all of the scales—subhorizon, superhorizon, and at horizon re-entry—and it is as follows:

\bea
\bigg[\Delta^{2}_{\zeta,{\bf Tree}}(k)\bigg] &=& \bigg[\frac{H^2}{8\pi ^2 M_{pl}^2 \epsilon c_s}\bigg]_{*} \nonumber \\ 
\quad\quad\quad  &\times&  \begin{cases} (1+k^2c_s^2\tau^2)
& \mbox{if } k\ll k_{s_1} (\rm SR_{1})\\
    \bigg(\displaystyle{\frac{k_{e_n}}{k_{s_n}}}\bigg)^{6} \bigg|\alpha_{\bf k}^{(2n)}(1+ikc_s\tau)\exp{(-ikc_s\tau)} \\              \quad\quad\quad\quad\quad\quad\quad\quad\quad\quad\quad - \beta_{\bf k}^{(2n)}(1-ikc_s\tau)\exp{(ikc_s\tau)}\bigg|^{2} &\mbox{if } k_{s_n} \le k\le k_{e_n} (\rm USR_{n}) \\
      \bigg(\displaystyle{\frac{k_{e_n}}{k_{s_n}}}\bigg)^{6} \bigg|\alpha_{\bf k}^{(2n+1)}(1+ikc_s\tau) \exp{(-ikc_s\tau)}  \\ \quad\quad\quad\quad\quad\quad\quad\quad\quad\quad\quad - \beta_{\bf k}^{(2n+1)}(1-ikc_s\tau) \exp{(ikc_s\tau)} \bigg|^{2} & \mbox{if } k_{e_n}\le k \le k_{s_{n+1}} (\rm SR_{n+1})
    
    \end{cases}
\quad \eea
In the pivot scale $k_{*}$, the expression included in parenthesis outside of every region is evaluated. Using the late time scale $\tau \to 0$, we can now write the scalar power spectrum as follows:
\begin{equation}
    \bigg[\Delta^{2}_{\zeta,{\bf Tree}}(k)\bigg] = \bigg[\frac{H^2}{8\pi ^2 M_{pl}^2 \epsilon c_s}\bigg]_{*} \times \begin{cases} 1  &\mbox{if } k\ll k_{s_1} (\rm SR_{1})\\
    \bigg(\displaystyle{\frac{k_{e_n}}{k_{s_n}}}\bigg)^{6}\times \bigg|\alpha_{\bf k}^{(2n)}-\beta_{\bf k}^{(2n)}\bigg|^{2} & \mbox{if } k_{s_n} \le k\le k_{e_n} (\rm USR_{n}) \\
      \bigg(\displaystyle{\frac{k_{e_n}}{k_{s_n}}}\bigg)^{6} \times \bigg|\alpha_{\bf k}^{(2n+1)} - \beta_{\bf k}^{(2n+1)}\bigg|^{2} & \mbox{if } k_{e_n}\le k \le k_{s_{n+1}} (\rm SR_{n+1})
    
    \end{cases}
\end{equation}
By summing up the power spectrums for each region, we can calculate the overall tree-level scalar power spectrum at the superhorizon scale ($-kc_s\tau \ll 1$). This may be expressed in terms of the previously determined Bogoliubov coefficients. 
\bea
   \label{treeps}
\bigg[\Delta^{2}_{\zeta,{\bf Tree}}(k)\bigg]_{\bf Total} &=&\bigg[\Delta^{2}_{\zeta,{\bf Tree}}(k)\bigg]_{\textbf{SR}_{1}} \times \bigg[1+\sum^{N}_{n=1}\Theta(k-k_{s_{n}}) \left(\frac{k_{e_{n}}}{k_{s_{n}}}\right)^{6}\left|\alpha^{(2n)}_{\bf k}-\beta^{(2n)}_{\bf k}\right|^2\nonumber\\
&&\quad\quad\quad\quad\quad\quad\quad\quad\quad\quad\quad\quad\quad\quad\quad+\sum^{N}_{n=1}\Theta(k-k_{e_{n}})\left(\frac{k_{e_{n}}}{k_{s_{n}}}\right)^{6}\left|\alpha^{(2n+1)}_{\bf k}-\beta^{(2n+1)}_{\bf k}\right|^2\bigg].\quad\quad
\eea
In our computation we need at least $N=6$ USR periods which are inserted along with multiple sharp transitions. The SR$_1$ contribution is represented by the common factor for the whole power spectrum, which is valid for all scales (subhorizon, superhorizon, and horizon re-entry) and is as follows:
\bea \bigg[\Delta^{2}_{\zeta,{\bf Tree}}(k)\bigg]_{\textbf{SR}_{1}} = \displaystyle{ \bigg[\frac{H^2}{8\pi ^2 M_{p}^2 \epsilon c_s}\bigg]_{*}}\left[1+(k/k_{s_{1}})^{2}\right] \xrightarrow{\rm Super-horizon \; scale \; k\ll k_{s_1}} \displaystyle{ \bigg[\frac{H^2}{8\pi ^2 M_{p}^2 \epsilon c_s}\bigg]_{*}}. \eea
It is also important to note that the Heaviside theta function in eqn.(\ref{treeps}) is used to emphasize the need for an abrupt transition at the scales where PBH creation takes place.  

Instead of sharp transition if we implement the smooth multiple transition in the present context of discussion then the expression in that case is modified by the following expression:
\bea \bigg[\Delta^{2}_{\zeta,{\bf Tree}}(k)\bigg]_{\bf Total} &=&\bigg[\Delta^{2}_{\zeta,{\bf Tree}}(k)\bigg]_{\textbf{SR}_{1}} \times \bigg[1+\sum^{N}_{n=1}\bigg[{\rm tanh}\left(\frac{k-k_{s_{n}}}{\Delta k_n}\right)\left(\frac{k_{e_{n}}}{k_{s_{n}}}\right)^{6}\left|\alpha^{(2n)}_{\bf k}-\beta^{(2n)}_{\bf k}\right|^2\nonumber\\
&&\quad\quad\quad\quad\quad\quad\quad\quad\quad\quad\quad\quad\quad\quad\quad+\sum^{N}_{n=1}{\rm tanh}\left(\frac{k-k_{e_{n}}}{\Delta k_n}\right)\left(\frac{k_{e_{n}}}{k_{s_{n}}}\right)^{6}\left|\alpha^{(2n+1)}_{\bf k}-\beta^{(2n+1)}_{\bf k}\right|^2\bigg],\quad\quad\eea
where $\Delta k_n=k_{e_{n}}-k_{s_{n}}$ represents the width of the all USR phases for the case when smooth transition is considered. When we take the limit $\Delta k_n$ is very small then one can get back the same result as written for the multiple sharp transition. However, it is important to note that, apart from smoothening the behaviour of the joining functions at the transition points in both the cases for tree level power spectrum we get approximately the same physical outcomes. This further implies that the behaviour of the transitions will not change the final outcomes at least at the tree-level result of the primordial power spectrum for the scalar modes.

\subsection{Cut-off regulated one-loop scalar power spectrum from EFT
}

\subsubsection{Third order perturbation from scalar mode}
The effect of the one-loop correction to the power spectrum from the scalar modes of the perturbation will be directly calculated as the next step in our research. Assuming that the curvature perturbation expands the normal EFT action in third order, the following computation will be carried out \footnote{We would like to give credit to \cite{Kristiano:2022maq} for realizing
that cubic action with $\eta'$ is the most important for
one-loop correction. } \footnote{Here it is important to note that the cubic interaction
proportional to $\eta'$ generates bispectrum that satisfies
Maldacena’s theorem \cite{Kristiano:2023scm}.}:
\bea &&S^{(3)}_{\zeta}=\int d\tau\;  d^3x\;  M^2_{ pl}a^2\; \bigg[\left(3\left(c^2_s-1\right)\epsilon+\epsilon^2-\frac{1}{2}\epsilon^3\right)\zeta^{'2}\zeta+\frac{\epsilon}{c^2_s}\bigg(\epsilon-2s+1-c^2_s\bigg)\left(\partial_i\zeta\right)^2\zeta\nonumber\\ 
&&\quad\quad\quad\quad\quad\quad\quad\quad\quad\quad\quad-\frac{2\epsilon}{c^2_s}\zeta^{'}\left(\partial_i\zeta\right)\left(\partial_i\partial^{-2}\left(\frac{\epsilon\zeta^{'}}{c^2_s}\right)\right)-\frac{1}{aH}\left(1-\frac{1}{c^2_{s}}\right)\epsilon \bigg(\zeta^{'3}+\zeta^{'}(\partial_{i}\zeta)^2\bigg)
     \nonumber\\
&& \quad\quad\quad\quad\quad\quad\quad\quad\quad\quad\quad+\frac{1}{2}\epsilon\zeta\left(\partial_i\partial_j\partial^{-2}\left(\frac{\epsilon\zeta^{'}}{c^2_s}\right)\right)^2
+\boxed{\underbrace{\frac{1}{2c^2_s}\epsilon\partial_{\tau}\left(\frac{\eta}{c^2_s}\right)\zeta^{'}\zeta^{2}}_{\bf Most~dominant ~term~in~USR}}\nonumber\\
&&+\underbrace{\frac{3}{2}\frac{1}{aH}\frac{\bar{M}^3_1}{ HM^2_{ pl}}
	  \zeta^{'}(\partial_{i}\zeta)^2+\frac{9}{2}\frac{\bar{M}^3_1}{ HM^2_{ pl}}\zeta \zeta^{'2}+\bigg(\frac{3}{2}\frac{\bar{M}^3_1}{ HM^2_{ pl}}-\frac{4}{3}\frac{M^4_3}{H^2M^2_{ pl}}\bigg)\zeta^{'3}-\frac{3}{2}\frac{1}{aH}\frac{\bar{M}^3_1}{ HM^2_{ pl}} \zeta\partial_{\tau}\left(\partial_{i}\zeta\right)^2}_{\bf Suppressed~contributions
	  ~in~the~one-loop~corrected~power~spectrum~from~SR~region~only}
  \bigg],\quad\quad\eea
It is significant to remember that in this case, $c_s=1$ results from setting $M_2=0$. Furthermore, the third order action for the standard single field slow-roll model is returned if we set $M_3=0$, $\bar{M}_1=0$. The contributions with EFT coefficients other than $M_2$ are strongly suppressed in the SR area of the one-loop contribution, thus they won't significantly alter the one-loop corrected equation for the scalar power spectrum in the end. These specified EFT coefficients may not be precisely zero for the various kinds of non-canonical or non-minimal $P(X,\phi)$ models of single-field inflation. That being said, we must limit our calculations to modest values of these parameters in order to maintain the perturbation theory for the scalar modes precisely during the one-loop computation in the current scenario. The latter portion of this examination will cover this topic with appropriate numerical examples.

In the current EFT framework of cosmic perturbation, we additionally include another slow roll parameter, $s$, which appears as a result of having an effective sound speed $c_s$ and its slow change with respect to the underlying conformal time scale. It is given by the following expression:
\bea s=\frac{\dot{c}_s}{Hc_s}=\frac{1}{aH}\frac{c^{'}_s}{c_s}=\frac{1}{{\cal H}}\frac{c^{'}_s}{c_s}.\eea
To extract the correction from the one-loop quantum effect or to see the large enhancement in the primordial non-Gaussian amplitudes, we shall make explicit use of all the contributions indicated previously in the presence of both single and multiple sharp transitions. Here it is additionally important to note that the conclusions that we are deriving in the rest of the computations are not going to change for a single or multiple smooth transitions, only the cumbersome mathematical expressions we need to tackle throughout the computations in those situations. The highlighted box represents the largest contribution, which goes into the first SR (SRI) region, second SR (SRII) region, and USR region as ${\cal O}(\epsilon^{3})$, ${\cal O}(\epsilon^{3})$, and ${\cal O}(\epsilon)$, respectively for the case of single sharp transition. However, the highlighted term contributes as ${\cal O}(\epsilon^{3})$ and ${\cal O}(\epsilon)$ in the USR and SR regions, respectively. The main source of the significant shift in the overall contribution for this specifically highlighted term from the SRI and SRII regimes to the USR regime is the second slow roll parameter, which shifts during the aforementioned rapid transition from $\eta\sim 0$ (SRI) to $\eta\sim -6$ (USR). The USR to SRII sharp transition then occurs in the second sharp transition, which is $\eta\sim 0$ (SRII) and lasts until the end of the inflation. For the current computation, the only thing that is particularly relevant to us is the highlighted term, which corresponds to the leading cubic self-interaction. The one-loop corrected equation for the scalar power spectrum only in the SR region is set to benefit from the five contributions that are shown in the final line of the preceding expression. These terms exhibit greater suppression in the USR region in the one-loop correction stated than in the SR region. Because of this, these terms in the final equation will not alter the improved one-loop corrected scalar power spectrum. Once more, in the equivalent one-loop result, the remaining contributions show higher suppression in the USR region relative to the SR region. For the sake of completeness, it is also crucial to mention that, if we compare the strengths of each term—including the boxed one—we can see with ease that, in the final line of the third-order action, each of the five contributions makes a minuscule adjustment to the one-loop correction necessary to maintain the perturbative approximations in the given situation. This indicates that when we significantly increase the strength of these five contributions, the perturbative approximation fails in the SRI, SRII, and USR regions. As a result, the scalar power spectrum during PBH generation will have the incorrect augmented amplitude. In the case of multiple sharp transitions, in this action contributions appear as ${\cal O}(\epsilon^3)$ in the SR$_1$ and SR$_{n+1}$ phase and as ${\cal O}(\epsilon)$ in the USR$_n$ phase. The rest of the discussions that we have provided will all hold good for multiple sharp transitions as well. It will be going to be more clear as we proceed with our computations and related discussions.

\subsubsection{One-loop computation of primordial power spectrum for scalar mode}

We will explicitly investigate every term in the next part of the evaluation, with special attention paid to the most important boxed phrase that appears as a byproduct of the EFT framework that we have chosen for our investigation. To this end, we utilize the well-known in-in formalism.  This means that the two-point correlation function of the subsequent quantum operator at $\tau\rightarrow 0$, the late time scale (end of inflation), can be expressed as follows:
  \bea \label{Hamx}\langle\hat{\zeta}_{\bf p}\hat{\zeta}_{-{\bf p}}\rangle:&=&\langle\hat{\zeta}_{\bf p}(\tau)\hat{\zeta}_{-{\bf p}}(\tau)\rangle_{\tau\rightarrow 0}\nonumber\\
    &=&\left\langle\bigg[\overline{T}\exp\bigg(i\int^{\tau}_{-\infty(1-i\epsilon)}d\tau^{'}\;H_{\rm int}(\tau^{'})\bigg)\bigg]\;\;\hat{\zeta}_{\bf p}(\tau)\hat{\zeta}_{-{\bf p}}(\tau)
\;\;\bigg[{T}\exp\bigg(-i\int^{\tau}_{-\infty(1+i\epsilon)}d\tau^{''}\;H_{\rm int}(\tau^{''})\bigg)\bigg]\right\rangle_{\tau\rightarrow 0},\quad\quad \eea
where $\overline{T}$ and $T$ denote the time ordering and anti-time operations, respectively. The interaction Hamiltonian in the current context is represented by $H_{\rm int}(\tau)$, which may be derived from the third order EFT action as follows:
\bea && H_{\rm int}(\tau)=-\int d^3x\;  M^2_{ pl}a^2\; \bigg[\left(3\left(c^2_s-1\right)\epsilon+\epsilon^2-\frac{1}{2}\epsilon^3\right)\zeta^{'2}\zeta+\frac{\epsilon}{c^2_s}\bigg(\epsilon-2s+1-c^2_s\bigg)\left(\partial_i\zeta\right)^2\zeta\nonumber\\ 
&&\quad\quad\quad\quad\quad\quad\quad\quad\quad\quad\quad-\frac{2\epsilon}{c^2_s}\zeta^{'}\left(\partial_i\zeta\right)\left(\partial_i\partial^{-2}\left(\frac{\epsilon\zeta^{'}}{c^2_s}\right)\right)-\frac{1}{aH}\left(1-\frac{1}{c^2_{s}}\right)\epsilon \bigg(\zeta^{'3}+\zeta^{'}(\partial_{i}\zeta)^2\bigg)
     \nonumber\\
&& \quad\quad\quad\quad\quad\quad\quad\quad\quad\quad\quad+\frac{1}{2}\epsilon\zeta\left(\partial_i\partial_j\partial^{-2}\left(\frac{\epsilon\zeta^{'}}{c^2_s}\right)\right)^2
+\boxed{\underbrace{\frac{1}{2c^2_s}\epsilon\partial_{\tau}\left(\frac{\eta}{c^2_s}\right)\zeta^{'}\zeta^{2}}_{\bf Most~dominant ~term~in~USR}}\nonumber\\
&&+\underbrace{\frac{3}{2}\frac{1}{aH}\frac{\bar{M}^3_1}{ HM^2_{ pl}}
	  \zeta^{'}(\partial_{i}\zeta)^2+\frac{9}{2}\frac{\bar{M}^3_1}{ HM^2_{ pl}}\zeta \zeta^{'2}+\bigg(\frac{3}{2}\frac{\bar{M}^3_1}{ HM^2_{ pl}}-\frac{4}{3}\frac{M^4_3}{H^2M^2_{ pl}}\bigg)\zeta^{'3}-\frac{3}{2}\frac{1}{aH}\frac{\bar{M}^3_1}{ HM^2_{ pl}} \zeta\partial_{\tau}\left(\partial_{i}\zeta\right)^2}_{\bf Suppressed~contributions
	  ~in~the~one-loop~corrected~power~spectrum~from~SR~region~only}
  \bigg],\quad\quad\eea
  which is simply ${\cal H} = -{\cal L}^{(3)}_{\rm int}$. For the two-point correlation function for the scalar modes, we obtain the following expression by using equation (\ref{Hamx}) and taking into account the contribution up to the one-loop correction in the Dyson Swinger series:
  \bea  &&\label{g}\langle\hat{\zeta}_{\bf p}\hat{\zeta}_{-{\bf p}}\rangle= \underbrace{\langle\hat{\zeta}_{\bf p}\hat{\zeta}_{-{\bf p}}\rangle_{(0,0)}}_{\bf Tree\;level\;result}+\underbrace{\langle\hat{\zeta}_{\bf p}\hat{\zeta}_{-{\bf p}}\rangle_{(0,1)}+\langle\hat{\zeta}_{\bf p}\hat{\zeta}_{-{\bf p}}\rangle^{\dagger}_{(0,1)}+\langle\hat{\zeta}_{\bf p}\hat{\zeta}_{-{\bf p}}\rangle_{(0,2)}+\langle\hat{\zeta}_{\bf p}\hat{\zeta}_{-{\bf p}}\rangle^{\dagger}_{(0,2)}+\langle\hat{\zeta}_{\bf p}\hat{\zeta}_{-{\bf p}}\rangle_{(1,1)}}_{\bf One-loop\;level\;result},
\eea
where the following can be written for each contribution:
\bea
     &&\label{g0}\langle\hat{\zeta}_{\bf p}\hat{\zeta}_{-{\bf p}}\rangle_{(0,0)}=\left[\langle \hat{\zeta}_{\bf p}(\tau)\hat{\zeta}_{-{\bf p}}(\tau)\rangle\right]_{\tau\rightarrow 0},\\
    &&\label{g1}\langle\hat{\zeta}_{\bf p}\hat{\zeta}_{-{\bf p}}\rangle_{(0,1)}=\left[-i\int^{\tau}_{-\infty}d\tau_1\;\langle \hat{\zeta}_{\bf p}(\tau)\hat{\zeta}_{-{\bf p}}(\tau)H_{\rm int}(\tau_1)\rangle\right]_{\tau\rightarrow 0},\\
 &&\label{g2}\langle\hat{\zeta}_{\bf p}\hat{\zeta}_{-{\bf p}}\rangle^{\dagger}_{(0,1)}=\left[-i\int^{\tau}_{-\infty}d\tau_1\;\langle \hat{\zeta}_{\bf p}(\tau)\hat{\zeta}_{-{\bf p}}(\tau)H_{\rm int}(\tau_1)\rangle^{\dagger}\right]_{\tau\rightarrow 0},\\
 &&\label{g3}\langle\hat{\zeta}_{\bf p}\hat{\zeta}_{-{\bf p}}\rangle_{(0,2)}=\left[\int^{\tau}_{-\infty}d\tau_1\;\int^{\tau}_{-\infty}d\tau_2\;\langle \hat{\zeta}_{\bf p}(\tau)\hat{\zeta}_{-{\bf p}}(\tau)H_{\rm int}(\tau_1)H_{\rm int}(\tau_2)\rangle\right]_{\tau\rightarrow 0},\\
 &&\label{g4}\langle\hat{\zeta}_{\bf p}\hat{\zeta}_{-{\bf p}}\rangle^{\dagger}_{(0,2)}=\left[\int^{\tau}_{-\infty}d\tau_1\;\int^{\tau}_{-\infty}d\tau_2\;\langle \hat{\zeta}_{\bf p}(\tau)\hat{\zeta}_{-{\bf p}}(\tau)H_{\rm int}(\tau_1)H_{\rm int}(\tau_2)\rangle^{\dagger}\right]_{\tau\rightarrow 0},\\
  &&\label{g5}\langle\hat{\zeta}_{\bf p}\hat{\zeta}_{-{\bf p}}\rangle^{\dagger}_{(1,1)}=\left[\int^{\tau}_{-\infty}d\tau_1\;\int^{\tau}_{-\infty}d\tau_2\;\langle H_{\rm int}(\tau_1)\hat{\zeta}_{\bf p}(\tau)\hat{\zeta}_{-{\bf p}}(\tau)H_{\rm int}(\tau_2)\rangle^{\dagger}\right]_{\tau\rightarrow 0}.\eea
  See equation (\ref{inin}) and related section \ref{s1} for more details on this aspect.
\subsubsubsection{For single sharp transition}
The next thing we will examine is the Hamiltonian's boxed highlighted cubic self-interaction, which will also have an impact on the two-point correlation function of the scalar modes at the one-loop level during the insertion of a single USR period with the sharp transition. This function may be expressed as follows:
 \bea   \langle\hat{\zeta}_{\bf p}\hat{\zeta}_{-{\bf p}}\rangle_{(0,1)}&=& -\frac{iM^2_{ pl}}{2}\int^{0}_{-\infty}d\tau\frac{a^2(\tau)}{c^2_s(\tau)}\epsilon(\tau)\partial_{\tau}\left(\frac{\eta(\tau)}{c^2_s(\tau)}\right)\nonumber\\
  &&\times\int \frac{d^{3}{\bf k}_1}{(2\pi)^3} \int \frac{d^{3}{\bf k}_2}{(2\pi)^3} \int \frac{d^{3}{\bf k}_3}{(2\pi)^3} \nonumber\\
  && \times\delta^3\bigg({\bf k}_1+{\bf k}_2+{\bf k}_3\bigg) \times \langle \hat{\zeta}_{\bf p}\hat{\zeta}_{-{\bf p}}\hat{\zeta}^{'}_{{\bf k}_1}(\tau)\hat{\zeta}_{{\bf k}_2}(\tau)\hat{\zeta}_{{\bf k}_3}(\tau)\rangle,\\
   \langle\hat{\zeta}_{\bf p}\hat{\zeta}_{-{\bf p}}\rangle_{(0,1)}&=& -\frac{iM^2_{ pl}}{2}\int^{0}_{-\infty}d\tau\frac{a^2(\tau)}{c^2_s(\tau)}\epsilon(\tau)\partial_{\tau}\left(\frac{\eta(\tau)}{c^2_s(\tau)}\right)\nonumber\\
  &&\times\int \frac{d^{3}{\bf k}_1}{(2\pi)^3} \int \frac{d^{3}{\bf k}_2}{(2\pi)^3} \int \frac{d^{3}{\bf k}_3}{(2\pi)^3} \nonumber\\
  && \times\delta^3\bigg({\bf k}_1+{\bf k}_2+{\bf k}_3\bigg) \times \langle \hat{\zeta}_{\bf p}\hat{\zeta}_{-{\bf p}}\hat{\zeta}^{'}_{{\bf k}_1}(\tau)\hat{\zeta}_{{\bf k}_2}(\tau)\hat{\zeta}_{{\bf k}_3}(\tau)\rangle^{\dagger},\\
     \langle\hat{\zeta}_{\bf p}\hat{\zeta}_{-{\bf p}}\rangle_{(0,2)}&=& -\frac{M^4_{ pl}}{4}\int^{0}_{-\infty}d\tau_1\frac{a^2(\tau_1)}{c^2_s(\tau_1)}\epsilon(\tau_1)\partial_{\tau_1}\left(\frac{\eta(\tau_1)}{c^2_s(\tau_1)}\right)\;\int^{0}_{-\infty}d\tau_2\;\frac{a^2(\tau_2)}{c^2_s(\tau_2)}\epsilon(\tau_2)\partial_{\tau_2}\left(\frac{\eta(\tau_2)}{c^2_s(\tau_2)}\right)\nonumber\\
  &&\times\int \frac{d^{3}{\bf k}_1}{(2\pi)^3} \int \frac{d^{3}{\bf k}_2}{(2\pi)^3} \int \frac{d^{3}{\bf k}_3}{(2\pi)^3} \int \frac{d^{3}{\bf k}_4}{(2\pi)^3} \int \frac{d^{3}{\bf k}_5}{(2\pi)^3} \int \frac{d^{3}{\bf k}_6}{(2\pi)^3}\nonumber\\
  &&\times \delta^3\bigg({\bf k}_1+{\bf k}_2+{\bf k}_3\bigg) \delta^3\bigg({\bf k}_4+{\bf k}_5+{\bf k}_6\bigg)\nonumber\\
  &&\times \langle \hat{\zeta}_{\bf p}\hat{\zeta}_{-{\bf p}}\hat{\zeta}^{'}_{{\bf k}_1}(\tau_1)\hat{\zeta}_{{\bf k}_2}(\tau_1)\hat{\zeta}_{{\bf k}_3}(\tau_1)\hat{\zeta}^{'}_{{\bf k}_4}(\tau_2)\hat{\zeta}_{{\bf k}_5}(\tau_2)\hat{\zeta}_{{\bf k}_6}(\tau_2)\rangle,\eea\bea
  \langle\hat{\zeta}_{\bf p}\hat{\zeta}_{-{\bf p}}\rangle^{\dagger}_{(0,2)}&=& -\frac{M^4_{ pl}}{4}\int^{0}_{-\infty}d\tau_1\frac{a^2(\tau_1)}{c^2_s(\tau_1)}\epsilon(\tau_1)\partial_{\tau_1}\left(\frac{\eta(\tau_1)}{c^2_s(\tau_1)}\right)\;\int^{0}_{-\infty}d\tau_2\;\frac{a^2(\tau_2)}{c^2_s(\tau_2)}\epsilon(\tau_2)\partial_{\tau_2}\left(\frac{\eta(\tau_2)}{c^2_s(\tau_2)}\right)\nonumber\\
  &&\times\int \frac{d^{3}{\bf k}_1}{(2\pi)^3} \int \frac{d^{3}{\bf k}_2}{(2\pi)^3} \int \frac{d^{3}{\bf k}_3}{(2\pi)^3} \int \frac{d^{3}{\bf k}_4}{(2\pi)^3} \int \frac{d^{3}{\bf k}_5}{(2\pi)^3} \int \frac{d^{3}{\bf k}_6}{(2\pi)^3}\nonumber\\
  &&\times \delta^3\bigg({\bf k}_1+{\bf k}_2+{\bf k}_3\bigg) \delta^3\bigg({\bf k}_4+{\bf k}_5+{\bf k}_6\bigg)\nonumber\\
  &&\times \langle \hat{\zeta}_{\bf p}\hat{\zeta}_{-{\bf p}}\hat{\zeta}^{'}_{{\bf k}_1}(\tau_1)\hat{\zeta}_{{\bf k}_2}(\tau_1)\hat{\zeta}_{{\bf k}_3}(\tau_1)\hat{\zeta}^{'}_{{\bf k}_4}(\tau_2)\hat{\zeta}_{{\bf k}_5}(\tau_2)\hat{\zeta}_{{\bf k}_6}(\tau_2)\rangle^{\dagger},\\
 \langle\hat{\zeta}_{\bf p}\hat{\zeta}_{-{\bf p}}\rangle_{(1,1)}&=& \frac{M^4_{ pl}}{4}\int^{0}_{-\infty}d\tau_1\frac{a^2(\tau_1)}{c^2_s(\tau_1)}\epsilon(\tau_1)\partial_{\tau_1}\left(\frac{\eta(\tau_1)}{c^2_s(\tau_1)}\right)\;\int^{0}_{-\infty}d\tau_2\;\frac{a^2(\tau_2)}{c^2_s(\tau_2)}\epsilon(\tau_2)\partial_{\tau_2}\left(\frac{\eta(\tau_2)}{c^2_s(\tau_2)}\right)\nonumber\\
  &&\times\int \frac{d^{3}{\bf k}_1}{(2\pi)^3} \int \frac{d^{3}{\bf k}_2}{(2\pi)^3} \int \frac{d^{3}{\bf k}_3}{(2\pi)^3} \int \frac{d^{3}{\bf k}_4}{(2\pi)^3} \int \frac{d^{3}{\bf k}_5}{(2\pi)^3} \int \frac{d^{3}{\bf k}_6}{(2\pi)^3}\nonumber\\
  &&\times \delta^3\bigg({\bf k}_1+{\bf k}_2+{\bf k}_3\bigg) \delta^3\bigg({\bf k}_4+{\bf k}_5+{\bf k}_6\bigg)\nonumber\\
  &&\times \langle \hat{\zeta}^{'}_{{\bf k}_1}(\tau_1)\hat{\zeta}_{{\bf k}_2}(\tau_1)\hat{\zeta}_{{\bf k}_3}(\tau_1)\hat{\zeta}_{\bf p}\hat{\zeta}_{-{\bf p}}\hat{\zeta}^{'}_{{\bf k}_4}(\tau_2)\hat{\zeta}_{{\bf k}_5}(\tau_2)\hat{\zeta}_{{\bf k}_6}(\tau_2)\rangle. \eea
  It is noteworthy to mention that in order to compute the aforementioned correlators explicitly, we must make use of the fact that, for the EFT $c_s$ and the second slow-roll parameter, the effective sound speed parameter is assumed to be nearly constant throughout the SRI, SRII, and USR periods, with the exception of $\tau=\tau_s$ and $\tau=\tau_e$, where we have taken into account abrupt transitions from SRI to USR and, respectively, from USR to SRII. In the process of computing, we have taken into account abrupt changes at the conformal time scales $\tau=\tau_s$ (from SRI to USR) and $\tau=\tau_e$ (from USR to SRII), where we establish the subsequent formula \footnote{We would like to give credit to \cite{Kristiano:2022maq} for realizing
that cubic action with $\eta'$ is the most important for
one-loop correction. }:
\bea \partial_{\tau}\left(\frac{\eta(\tau)}{c^2_s(\tau)}\right)\approx\frac{\Delta \eta(\tau)}{c^2_s(\tau)}\bigg(\underbrace{\delta(\tau-\tau_e)}_{\bf USR\rightarrow SRII}-\underbrace{\delta(\tau-\tau_s)}_{\bf SRI\rightarrow USR}\bigg).\eea
As a direct result, one can thus construct the following equation right away:
\be \displaystyle \partial_{\tau}\left(\frac{\eta(\tau)}{c^2_s(\tau)}\right)\approx 0,\ee 
The aforementioned findings, therefore, are entirely valid in the following regions: $\tau>\tau_e$, $\tau_s<\tau<\tau_e$, and the area $\tau<\tau_s$. Outside of the aforementioned areas, the above approximation is not applicable at the SRI to USR sharp transition point, $\tau=\tau_s$, and at the USR to SRII sharp transition point, $\tau=\tau_e$. It is therefore necessary to collect the contributions only from the boundaries of the USR region that are attached to the SRI region and SRII region, and appear at the scale $\tau=\tau_s$ and $\tau=\tau_e$, as previously mentioned, rather than computing the full integral over the conformal time scale within the range $-\infty<\tau<0$. With the help of the previously mentioned definition of sharp transition and utilizing the behaviour of the factor $\eta/c^2_s$ at the transition point one can compute the temporal part of the one-loop integrals using the following expressions:
\bea &&\int^{0}_{-\infty}d\tau\;\frac{1}{c^2_s(\tau)}\partial_{\tau}\left(\frac{\eta(\tau)}{c^2_s(\tau)}\right)\; {\cal K}_{1}(\tau)
=\bigg(\frac{\Delta \eta(\tau_{e})}{c^2_s}\; {\cal K}_{1}(\tau=\tau_{e})-\frac{\Delta \eta(\tau_{s})}{c^2_s}\; {\cal K}_{1}(\tau=\tau_{s})\bigg)-\underbrace{\int^{0}_{-\infty}d\tau\;\left(\frac{\eta(\tau)}{c^2_s(\tau)}\right)\; {\cal K}^{'}_{1}(\tau)}_{\approx 0}\nonumber\\&&\quad\quad\quad\quad\quad\quad\quad\quad\quad\quad\quad\quad\quad\quad\approx\bigg(\frac{\Delta \eta(\tau_{e})}{c^2_s}\; {\cal K}_{1}(\tau_{e})-\frac{\Delta \eta(\tau_{s})}{c^2_s}\; {\cal K}_{1}(\tau_{s})\bigg),
\eea
where the residual momentum integrals are included in the kernel ${\cal K}_{1}(\tau)$. Likewise, we may represent the product of two such temporal derivatives as an integral:
\bea
&&\int^{0}_{-\infty}d\tau_1\;\int^{0}_{-\infty}d\tau_2\;\frac{1}{c^2_s(\tau_1)}\frac{1}{c^2_s(\tau_2)}\partial_{\tau_1}\left(\frac{\eta(\tau_1)}{c^2_s(\tau_1)}\right)\;\partial_{\tau_2}\left(\frac{\eta(\tau_2)}{c^2_s(\tau_2)}\right)\; {\cal K}_{2}(\tau_1,\tau_2)\nonumber\\
&&=\int^{0}_{-\infty}d\tau_2\frac{1}{c^2_s(\tau_2)}\partial_{\tau_2}\left(\frac{\eta(\tau_2)}{c^2_s(\tau_2)}\right)\;\bigg(\frac{\Delta \eta(\tau_e)}{c^4_s}\; {\cal K}_{2}(\tau_1=\tau_{e},\tau_2)-\frac{\Delta \eta(\tau_{s})}{c^4_s}\; {\cal K}_{2}(\tau_1=\tau_{s},\tau_2)\bigg)\nonumber\\
&&\quad\quad-\underbrace{\int^{0}_{-\infty}d\tau_1\;\int^{0}_{-\infty}d\tau_2\;\frac{1}{c^2_s(\tau_1)}\frac{1}{c^2_s(\tau_2)}\left(\frac{\eta(\tau_1)}{c^2_s(\tau_1)}\right)\;\partial_{\tau_2}\left(\frac{\eta(\tau_2)}{c^2_s(\tau_2)}\right)\; \partial_{\tau_1}{\cal K}_{2}(\tau_1,\tau_2)}_{\approx 0}\\&&\approx \bigg(\frac{\Delta \eta(\tau_{e})}{c^8_s}\; {\cal K}_{2}(\tau_1=\tau_{e},\tau_2=\tau_{e})-\frac{\Delta \eta(\tau_s)}{c^8_s}\; {\cal K}_{2}(\tau_1=\tau_{s},\tau_2=\tau_{s})\bigg)\nonumber\\&&-\underbrace{\int^{0}_{-\infty}d\tau_2\left(\frac{\eta(\tau_2)}{c^4_s(\tau_2)}\right)\;\bigg(\frac{\Delta \eta(\tau_{e})}{c^4_s}\; \partial_{\tau_2}{\cal K}_{2}(\tau_1=\tau_e,\tau_2)-\frac{\Delta \eta(\tau_{s_{n}})}{c^4_s}\; \partial_{\tau_2}{\cal K}_{2}(\tau_1=\tau_{s},\tau_2)\bigg)}_{\approx 0}\nonumber\\
&&\approx \bigg(\frac{\Delta \eta(\tau_{e})}{c^8_s}\; {\cal K}_{2}(\tau_{e})-\frac{\Delta \eta(\tau_{s})}{c^8_s}\; {\cal K}_{2}(\tau_{s})\bigg)\nonumber.
\eea 
This furthermore has a distinct kernel denoted as ${\cal K}_{2}(\tau_{1},\tau_{2})$. We have also taken use of the fact that under the SR and USR regime, the initial slow-roll parameter $\epsilon$ is also constant in the current calculation. As a result, the restriction $\epsilon^{'}(\tau)\approx 0$ may be obtained right away, implying ${\cal H}^{''}(\tau)\approx 2{\cal H}^{'}(\tau)$. Thus, the following vital current data are available to us: 
$\partial_{\tau}{\cal J}_1(\tau)\approx 0$, $\partial_{\tau_1}{\cal J}_2(\tau_1,\tau_2)\approx 0$,
In particular, $\partial_{\tau_2}{\cal J}_2(\tau_1=\tau_e,\tau_2)\approx 0$ and $\partial_{\tau_2}{\cal J}_2(\tau_1=\tau_s,\tau_2)\approx 0$.
For the purpose of streamlining the one-loop contribution to the two-point scalar correlation in the USR area, each of the implications is quite beneficial.

In the current frame of discussion, let us also explore the time-dependent parametrization of the effective sound speed parameter for the goal of better comprehension. The effective sound speed parameter is fixed at the CMB pivot scale value, $c_s(\tau_*)=c_{s}$, during the time scale evolution. It is parameterized as follows, but, at the abrupt transition points: $c_s(\tau_e)=c_s(\tau_s)=\tilde{c}_s=1\pm \delta$. Here, $\delta$ denotes a fine-tuning parameter, which is kept constant by keeping $\delta\ll 1$. When computing the normalized and resumed contribution of the one-loop contribution to the momentum loop integrals computed in the various phases, this extra information will be of great use.

The one-loop contribution to the two-point correlation function of the scalar perturbation, after employing all possible Wick contractions, may be further expressed as follows:
 \bea \label{ff1}\langle\langle\hat{\zeta}_{\bf p}\hat{\zeta}_{-{\bf p}}\rangle\rangle_{\bf One-loop}&=&\langle\langle\hat{\zeta}_{\bf p}\hat{\zeta}_{-{\bf p}}\rangle\rangle_{(1,1)}+2{\rm Re}\bigg[\langle\langle\hat{\zeta}_{\bf p}\hat{\zeta}_{-{\bf p}}\rangle\rangle_{(0,2)}\bigg]\nonumber\\
 &\approx & \frac{M^4_{ pl}}{4}\Bigg(a^4(\tau_e)\epsilon^2(\tau_e)\frac{\left(\Delta\eta(\tau_e)\right)^2}{c^8_s} {\cal D}({\bf p},\tau_e)-a^4(\tau_s)\epsilon^2(\tau_s)\frac{\left(\Delta\eta(\tau_s)\right)^2}{c^8_s}{\cal D}({\bf p},\tau_s)\Bigg)\nonumber\\
 &&\quad\quad\quad\quad+\frac{M^2_{ pl}}{2} \Bigg(a^2(\tau_e)\epsilon(\tau_e)\frac{\left(\Delta\eta(\tau_e)\right)}{c^4_s} {\cal E}({\bf p},\tau_e)-a^2(\tau_s)\epsilon(\tau_s)\frac{\left(\Delta\eta(\tau_s)\right)}{c^4_s}{\cal E}({\bf p},\tau_s)\Bigg).\eea
 Here it is important to note that, 
\be \langle\hat{\zeta}_{\bf p}\hat{\zeta}_{-{\bf p}}\rangle_{(0,1)}+\langle\hat{\zeta}_{\bf p}\hat{\zeta}_{-{\bf p}}\rangle^{\dagger}_{(0,1)}=2{\rm Re}\left[\langle\langle\hat{\zeta}_{\bf p}\hat{\zeta}_{-{\bf p}}\rangle\rangle_{(0,1)}\right]=0,\ee
by applying the Wick contraction. For this reason equation (\ref{g0}) and equation (\ref{g1}) will not finally contribute to the one-loop correction of the primordial power spectrum for the scalar perturbation.
 It is also noteworthy that, in this instance, we establish a new function, denoted by ${\cal E}({\bf p},\tau)$ and ${\cal D}({\bf p},\tau)$, which are conformal time dependent functions \footnote{Also, we would like to
give credit to \cite{Kristiano:2022maq} for first pointing that the form of one-loop correction that is
proportional to ${\rm Im}(\zeta^{'}_p \zeta_p^*)$. }:
\bea {\cal D}({\bf p},\tau)&=&16\int \frac{d^{3}{\bf k}}{(2\pi)^3}\bigg[|\zeta_{\bf p}|^{2}|\zeta_{{\bf k}-{\bf p}}|^{2}{\rm Im}\bigg(\zeta^{'}_{{\bf p}}\zeta^{*}_{{\bf p}}\bigg){\rm Im}\bigg(\zeta^{'}_{{\bf k}}\zeta^{*}_{{\bf k}}\bigg)\bigg],\\
 {\cal E}({\bf p},\tau)&=&\int \frac{d^{3}{\bf k}}{(2\pi)^3}\bigg[|\zeta_{\bf p}|^{2}|\zeta_{{\bf k}-{\bf p}}|^{2}\frac{d\ln |\zeta_{{\bf k}-{\bf p}}|^{2} }{d\ln k}\bigg].\eea
 The loop momenta, $k$, is significantly larger than the external momenta, $p$, or $k\gg p$, in this case. It is possible to overlook the contribution of momenta $p$ while conducting the loop integral because of this. Thus, the integrals ${\cal D}({\bf p},\tau)$ and ${\cal E}({\bf p},\tau)$ may be further approximated using the subsequent formulas:
\bea {\cal D}({\bf p},\tau)&=&16|\zeta_{\bf p}|^{2}\int \frac{d^{3}{\bf k}}{(2\pi)^3}\bigg[|\zeta_{{\bf k}}|^{2}{\rm Im}\bigg(\zeta^{'}_{{\bf p}}\zeta^{*}_{{\bf p}}\bigg){\rm Im}\bigg(\zeta^{'}_{{\bf k}}\zeta^{*}_{{\bf k}}\bigg)\bigg],\\
 {\cal E}({\bf p},\tau)&=&|\zeta_{\bf p}|^{2}\int \frac{d^{3}{\bf k}}{(2\pi)^3}\bigg[|\zeta_{{\bf k}}|^{2}\frac{d\ln |\zeta_{{\bf k}}|^{2} }{d\ln k}\bigg].\eea
 Assessing this integral inside the USR area is now our main goal, as is attempting to determine the precise role that this cubic self-interaction component plays in the one-loop corrected primordial power spectrum for scalar modes.

We must utilize the following information to assess these integrals:
\bea  &&\bigg[{\rm Im}\bigg(\zeta^{'}_{{\bf k}}\zeta^{*}_{{\bf k}}\bigg)\bigg]_{\tau=\tau_e}=-\frac{1}{3\left(k_e\right)^3}\left(\frac{k_e}{k_s}\right)^{6}\left(\frac{H^{2}}{4\pi^{2}M^{2}_{ pl}\epsilon c_s}\right)_*,\\
&&\bigg[{\rm Im}\bigg(\zeta^{'}_{{\bf k}}\zeta^{*}_{{\bf k}}\bigg)\bigg]_{\tau=\tau_s}=-\frac{1}{3\left(k_s\right)^3}\left(\frac{H^{2}}{4\pi^{2}M^{2}_{pl}\epsilon c_s}\right)_*.\quad\quad\eea
These outcomes allow us to arrive at the following simplified conclusion: 
\bea &&\bigg[\Delta^{2}_{\zeta, {\bf One-loop}}(p)\bigg]_{\bf USR\;on\;SR}=\bigg\{\bigg[\frac{1}{4}\bigg[\Delta^{2}_{\zeta,{\bf Tree}}(p)\bigg]^2_{\bf SR}\times\bigg(\frac{\left(\Delta\eta(\tau_e)\right)^2}{c^8_s} \left(\frac{k_e}{k_s}\right)^{6}{\bf I}(\tau_e)- \frac{\left(\Delta\eta(\tau_s)\right)^2}{c^8_s}{\bf I}(\tau_s)\bigg)\nonumber\\
&&\quad\quad\quad\quad\quad\quad\quad\quad\quad\quad\quad\quad+\frac{1}{2}\bigg[\Delta^{2}_{\zeta,{\bf Tree}}(p)\bigg]^2_{\bf SR}\times\bigg(\frac{\left(\Delta\eta(\tau_e)\right)}{c^4_s}\left(\frac{k_e}{k_s}\right)^{6}{\bf F}(\tau_e)-\frac{\left(\Delta\eta(\tau_s)\right)}{c^4_s}{\bf F}(\tau_s)\bigg)\bigg]-c_{\bf USR}\bigg\},\quad\eea
Here we now explicitly compute the momentum integrals in the following:
\begin{enumerate}
    \item \underline{\bf First USR integral ${\bf I}$:}\\
    Initially, we will express the contribution using a momentum-dependent integral that is present in the USR region:
\bea  \label{gk1} &&{\bf I}(\tau):=\int^{k_e}_{k_s}\frac{dk}{k}\;\left|{\cal G}_{ k}(\tau)\right|^{2},\eea
In this instance, a new function ${\cal G}_{ k}(\tau)$ is defined as follows:
\bea \label{hhgx} {\cal G}_{ k}(\tau)&=&\bigg[\alpha^{(2)}_{ k}\left(1+ikc_s\tau\right)\; e^{-ikc_s\tau}-\beta^{(2)}_{ k}\left(1-ikc_s\tau\right)\; e^{ikc_s\tau}\bigg].\eea
In the USR phase, the Bogoliubov coefficients $\alpha^{(2)}_{ k}$ and $\beta^{(2)}_{ k}$ are denoted by the following expressions:
\bea \alpha^{(2)}_{ k}&=&1-\frac{3}{2ik^{3}c^{3}_s\tau^{3}_s}\left(1+k^{2}c^{2}_s\tau^{2}_s\right),
  \quad\quad\quad
\beta^{(2)}_{ k}=-\frac{3}{2ik^{3}c^{3}_s\tau^{3}_s}\left(1+ikc_s\tau_s\right)^{2}\; e^{-2ikc_s\tau_s}.\eea
Following the substitution of the function ${\cal G}_{\bf k}(\tau)$ in its particular form in equation (\ref{gk1}), we obtained the following simplified relation \footnote{We would like to give credit to ref \cite{Kristiano:2022maq} for realizing logarithmic and quadratic divergences associated with short modes at $\tau = \tau_e$.}:
\bea \label{c11}{\bf I}(\tau)&=&\Bigg\{\frac{1}{2}\left(k^2_e-k^2_s\right)c^2_s\tau^2+\left(1+\frac{9}{2}\left(\frac{\tau}{\tau_s}\right)^2\right) \ln\left(\frac{k_e}{k_s}\right)+\cdots\Bigg\}.\eea
Here the $\cdots$ represents the suppressed contribution in which we are not interested in the present discussion. To know more about the full contribution of the integral ${\bf I}$ look at the appendix where we have presented the detailed calculation.

 \item \underline{\bf Second USR integral ${\bf F}$:}\\
 The sub-leading contribution may be expressed as follows in terms of a momentum-dependent integral that appears in the USR region:
\bea  \label{gslk1} &&{\bf F}(\tau):=\int^{k_e}_{k_s}\frac{dk}{k}\;|{\cal G}_{ k}(\tau)|^2\bigg(\frac{d\ln|{\cal G}_{ k}(\tau)|^2 }{d\ln k}\bigg)=\int^{k_e}_{k_s}d\ln k\;\bigg(\frac{d|{\cal G}_{ k}(\tau)|^2 }{d\ln k}\bigg)=\Bigg[|{\cal G}_{ k}(\tau)|^2\Bigg]^{k_e}_{k_s},\eea
The function ${\cal G}_{\bf k}$ has previously been defined in equation (\ref{hhgx}). 
\end{enumerate}
Look at appendix which has the computation specifics. Moreover, it is noteworthy that the term $c_{\bf USR}$ emulates the function of a counterterm throughout the renormalization process inside a designated scheme. Various comparable schemes—late-time renormalization and adiabatic/wave function renormalization—will be covered in the second part of the relevant topic. These schemes have the capability to fix the explicit form of the counter term of a particular scheme. 

 Finally, the following abbreviated form may be used to express, on the super-horizon scale, the one-loop contribution to the power spectrum of the scalar perturbation arising from the USR period on SR contribution:
\bea \bigg[\Delta^{2}_{\zeta, {\bf One-loop}}(p)\bigg]_{\bf USR\;on\;SR}=\bigg[\Delta^{2}_{\zeta,{\bf Tree}}(p)\bigg]_{\bf SR}\times V.\quad\quad\eea
Here the quantity $V$ is defined as:
\bea V&=&\bigg({\bf Z}_1+{\bf Z}_2-c_{\bf USR}\bigg),\eea
in which the following formulas describe the elements ${\bf Z}_1$ and ${\bf Z}_2$:
\bea {\bf Z}_1:&=&\frac{1}{4}\bigg[\Delta^{2}_{\zeta,{\bf Tree}}(p)\bigg]_{\bf SR}\times \bigg(\frac{\left(\Delta\eta(\tau_e)\right)^2}{c^8_s} \left(\frac{k_e}{k_s}\right)^{6}{\bf I}(\tau_e)- \frac{\left(\Delta\eta(\tau_s)\right)^2}{c^8_s}{\bf I}(\tau_s)\bigg),\\
{\bf Z}_2:&=&\frac{1}{2}\bigg[\Delta^{2}_{\zeta,{\bf Tree}}(p)\bigg]_{\bf SR}\times\bigg(\frac{\left(\Delta\eta(\tau_e)\right)}{c^4_s}\left(\frac{k_e}{k_s}\right)^{6}{\bf F}(\tau_e)-\frac{\left(\Delta\eta(\tau_s)\right)}{c^4_s}{\bf F}(\tau_s)\bigg).\eea
More discussions on the relative amplitudes of ${\bf Z}_1$ and ${\bf Z}_2$ are discussed in the appendix. Just to note that, ${\bf Z}_2\ll {\bf Z}_1$ for which one can approximate, ${\bf Z}_1+{\bf Z}_2={\bf Z}_1\left(1+{\bf Z}_2/{\bf Z}_1\right)\approx {\bf Z}_1$. Consequently, we can write $V\approx \left({\bf Z}_1-c_{\bf USR}\right)$.

Likewise, the combined one-loop contribution for the SRI and SRII periods is provided by:
  \bea \bigg[\Delta^{2}_{\zeta, {\bf One-loop}}(p)\bigg]_{\bf SR}=\bigg[\Delta^{2}_{\zeta,{\bf Tree}}(p)\bigg]_{\bf SR}\times U,\eea
where the definition of quantity $U$ is as follows:
\bea U=U_{\bf SRI}+U_{\bf SRII},\eea
The following terms indicate each of the contributions made by SRI and SRII:
\bea U_{\bf SRI}&=&\bigg[\Delta^{2}_{\zeta,{\bf Tree}}(p)\bigg]_{\bf SR}\times\Bigg(1+\frac{2}{15\pi^2}\frac{1}{c^2_{s}p^2_*}\bigg(-\left(1-\frac{1}{c^2_{s}}\right)\epsilon+6\frac{\bar{M}^3_1}{ HM^2_{ pl}}-\frac{4}{3}\frac{M^4_3}{H^2M^2_{ pl}}\bigg)\Bigg)\times\Bigg(c_{\bf SR_{I}}-\frac{4}{3}{\bf K}\Bigg),\\ 
U_{\bf SRII}&=&\bigg[\Delta^{2}_{\zeta,{\bf Tree}}(p)\bigg]_{\bf SR}\times\Bigg(1+\frac{2}{15\pi^2}\frac{1}{c^2_{s}p^2_*}\bigg(-\left(1-\frac{1}{c^2_{s}}\right)\epsilon+6\frac{\bar{M}^3_1}{ HM^2_{ pl}}-\frac{4}{3}\frac{M^4_3}{H^2M^2_{ pl}}\bigg)\Bigg)\times\Bigg(c_{\bf SR_{II}}+{\bf O}\Bigg),\quad\quad\eea
Now we comment on the explicit structure of the momentum integrals that contribute to the region SRI and SRII, which are given by:
\begin{enumerate}
    \item \underline{\bf SRI integral ${\bf K}$:}\\
    We shall discuss the following integral here, which shows up in the computation of the primordial power spectrum of scalar modes in the SRI area after a one-loop correction. Now let's assess the integral that follows:
\bea\label{intSR} {\bf K}(\tau):=\int^{k_e}_{p_*}\frac{dk}{k}\;\left|{\cal M}_{ k}(\tau)\right|^{2},\eea
in which a new function ${\cal M}_{ k}(\tau)$ is defined in the following way:
\bea {\cal M}_{ k}(\tau)&=&\left(1+ikc_s\tau\right)\; e^{-ikc_s\tau}.\eea
Upon replacing the explicit form of the previously mentioned function ${\cal M}_{\bf k}(\tau)$ in equation (\ref{intSR}), we obtain the subsequent outcome \footnote{We would like to give credit to ref \cite{Kristiano:2022maq} for realizing logarithmic and quadratic divergences associated with short modes at $\tau = \tau_e$.}:
\bea{\bf K}(\tau)=\int^{k_e}_{p_*}\frac{dk}{k}\;\left(1+k^2c^2_s\tau^2\right)=\bigg[\ln\left(\frac{k_e}{p_*}\right)+\frac{1}{2}\left(k^2_e-p^2_*\right)c^2_s\tau^2\bigg]=\bigg[\ln\left(\frac{k_e}{p_*}\right)+\frac{1}{2}\left(\frac{k_e}{p_*}\right)^2-\frac{1}{2}\bigg].\eea
    \item \underline{\bf SRII integral ${\bf O}$:}\\
    To begin with, let us express the contribution as a momentum-dependent integral that appears in the SRII region:
\bea  \label{gkk1} &&{\bf O}(\tau):=\left(\frac{\tau_s}{\tau_e}\right)^6\int^{k_{\rm end}}_{k_e}\frac{dk}{k}\;\left|{\cal X}_{ k}(\tau)\right|^{2},\eea
where the function ${\cal X}_{ k}(\tau)$ is defined as:
\bea {\cal X}_{ k}(\tau)&=&\bigg[\alpha^{(3)}_{ k}\left(1+ikc_s\tau\right)\; e^{-ikc_s\tau}-\beta^{(3)}_{ k}\left(1-ikc_s\tau\right)\; e^{ikc_s\tau}\bigg].\eea
In this case, the Bogoliubov coefficients in the SRII phase, $\beta^{(3)}_{ k}$ and $\alpha^{(3)}_{ k}$, are as follows:
\bea \alpha^{(3)}_{ k}&=&-\frac{1}{4k^6c^6_s\tau^3_s\tau^3_e}\Bigg[9\left(kc_s\tau_s-i\right)^2\left(kc_s\tau_e+i\right)^2 e^{2ikc_s(\tau_e-\tau_s)}\nonumber\\
&&\quad\quad\quad\quad\quad\quad\quad\quad\quad\quad\quad\quad\quad\quad\quad\quad-
\left\{k^2c^2_s\tau^2_e\left(2kc_s\tau_e-3i\right)-3i\right\}\left\{k^2c^2_s\tau^2_s\left(2kc_s\tau_s+3i\right)+3i\right\}\Bigg],\quad\quad\quad\\
\beta^{(3)}_{ k}&=&\frac{3}{4k^6c^6_s\tau^3_s\tau^3_e}\Bigg[\left(kc_s\tau_s-i\right)^2\left\{k^2c^2_s\tau^2_e\left(3-2ikc_s\tau_e\right)+3\right\}e^{-2ikc_s\tau_s}\nonumber\\
&&\quad\quad\quad\quad\quad\quad\quad\quad\quad\quad\quad\quad\quad\quad\quad\quad+i\left(kc_s\tau_e-i\right)^2\left\{3i+k^2c^2_s\tau^2_s\left(2kc_s\tau_s+3i\right)\right\}e^{-2ikc_s\tau_e}\Bigg].\eea
The aforementioned integral makes it simple to compute explicit results. Unfortunately, we will not be quoting the specific results in full here owing to the lengthy language. The following result, which will go toward the one-loop integral, has now been discovered in the super-horizon limiting scale:
\bea {\bf O}(\tau_e)={\bf O}(\tau_{\rm end})\approx -\ln \left(\frac{k_e}{k_{\rm end}}\right)-\frac{27}{32}\Bigg\{1-\left(\frac{k_e}{k_{\rm end}}\right)^{12}\Bigg\}\approx -\ln \left(\frac{k_e}{k_{\rm end}}\right).\eea

\end{enumerate}
Then the total one-loop regularized but unrenormalized power spectrum for the scalar modes is described by the following simplified expression:
\bea \bigg[\Delta^{2}_{\zeta, {\bf EFT}}(p)\bigg]&=&\bigg[\Delta^{2}_{\zeta,{\bf Tree}}(p)\bigg]_{\bf SR}+\bigg[\Delta^{2}_{\zeta, {\bf One-loop}}(p)\bigg]_{\bf SR}+\bigg[\Delta^{2}_{\zeta, {\bf One-loop}}(p)\bigg]_{\bf USR\;on\;SR}\nonumber\\
&=&\bigg[\Delta^{2}_{\zeta,{\bf Tree}}(p)\bigg]_{\bf SR}\times \left(1+U+V\right).\eea
Look at the Appendix \ref{A3a} for more details on this computation.
\subsubsubsection{For multiple sharp transitions}
 The next thing we will examine is the Hamiltonian's boxed highlighted cubic self-interaction, which will also have an impact on the two-point correlation function of the scalar modes at the one-loop level during the insertion of multiple USR periods with sharp transitions. In case of multiple sharp transitions (MSTs) we need to follow the similar steps that we have performed for the single sharp transition. We utilize an important fact about both the effective sound speed $c_{s}$ and the second slow-roll parameter $\eta$ to explicitly compute the above set of integrals, with an emphasis on the temporal integrals. Except during the abrupt transitions $\tau_{s_{n}}$ and $\tau_{e_{n}}$, both of these parameters are constant throughout the SRI, USR, and SRII. We express the temporal derivative term as follows in order to appropriately evaluate the integral at the boundaries of USR phases:
\bea \partial_{\tau}\left(\frac{\eta(\tau)}{c^2_s(\tau)}\right)\approx\frac{\Delta \eta(\tau)}{c^2_s(\tau)}\bigg(\delta(\tau-\tau_{e_{n}})-\delta(\tau-\tau_{s_{n}})\bigg)\quad\quad \forall\quad n.\eea
At the conformal time limits $\tau_{s_{n}}$ and $\tau_{e_{n}}$, the Dirac delta makes a contribution.
Using the aforementioned formula, the two-point correlators in the USR may be expressed as follows.
\bea &&\int^{0}_{-\infty}d\tau\;\frac{1}{c^2_s(\tau)}\partial_{\tau}\left(\frac{\eta(\tau)}{c^2_s(\tau)}\right)\; {\cal K}_{1}(\tau)
=\bigg(\frac{\Delta \eta(\tau_{e_{n}})}{c^2_s}\; {\cal K}_{1}(\tau=\tau_{e_{n}})-\frac{\Delta \eta(\tau_{s_{n}})}{c^2_s}\; {\cal K}_{1}(\tau=\tau_{s_{n}})\bigg)-\underbrace{\int^{0}_{-\infty}d\tau\;\left(\frac{\eta(\tau)}{c^2_s(\tau)}\right)\; {\cal K}^{'}_{1}(\tau)}_{\approx 0}\nonumber\\&&\quad\quad\quad\quad\quad\quad\quad\quad\quad\quad\quad\quad\quad\quad\approx\bigg(\frac{\Delta \eta(\tau_{e_{n}})}{c^2_s}\; {\cal K}_{1}(\tau_{e_{n}})-\frac{\Delta \eta(\tau_{s_{n}})}{c^2_s}\; {\cal K}_{1}(\tau_{s_{n}})\bigg),
\eea
where the residual momentum integrals are included in the kernel ${\cal K}_{1}(\tau)$. Likewise, we may represent the product of two such temporal derivatives as an integral.
\bea
&&\int^{0}_{-\infty}d\tau_1\;\int^{0}_{-\infty}d\tau_2\;\frac{1}{c^2_s(\tau_1)}\frac{1}{c^2_s(\tau_2)}\partial_{\tau_1}\left(\frac{\eta(\tau_1)}{c^2_s(\tau_1)}\right)\;\partial_{\tau_2}\left(\frac{\eta(\tau_2)}{c^2_s(\tau_2)}\right)\; {\cal K}_{2}(\tau_1,\tau_2)\nonumber\\
&&=\int^{0}_{-\infty}d\tau_2\frac{1}{c^2_s(\tau_2)}\partial_{\tau_2}\left(\frac{\eta(\tau_2)}{c^2_s(\tau_2)}\right)\;\bigg(\frac{\Delta \eta(\tau_e)}{c^4_s}\; {\cal K}_{2}(\tau_1=\tau_{e_{n}},\tau_2)-\frac{\Delta \eta(\tau_{s_{n}})}{c^4_s}\; {\cal K}_{2}(\tau_1=\tau_{s_{n}},\tau_2)\bigg)\nonumber\\
&&\quad\quad-\underbrace{\int^{0}_{-\infty}d\tau_1\;\int^{0}_{-\infty}d\tau_2\;\frac{1}{c^2_s(\tau_1)}\frac{1}{c^2_s(\tau_2)}\left(\frac{\eta(\tau_1)}{c^2_s(\tau_1)}\right)\;\partial_{\tau_2}\left(\frac{\eta(\tau_2)}{c^2_s(\tau_2)}\right)\; \partial_{\tau_1}{\cal K}_{2}(\tau_1,\tau_2)}_{\approx 0}\\&&\approx \bigg(\frac{\Delta \eta(\tau_{e_{n}})}{c^8_s}\; {\cal K}_{2}(\tau_1=\tau_{e_{n}},\tau_2=\tau_{e_{n}})-\frac{\Delta \eta(\tau_s)}{c^8_s}\; {\cal K}_{2}(\tau_1=\tau_{s_{n}},\tau_2=\tau_{s_{n}})\bigg)\nonumber\\&&-\underbrace{\int^{0}_{-\infty}d\tau_2\left(\frac{\eta(\tau_2)}{c^4_s(\tau_2)}\right)\;\bigg(\frac{\Delta \eta(\tau_{e_{n}})}{c^4_s}\; \partial_{\tau_2}{\cal K}_{2}(\tau_1=\tau_e,\tau_2)-\frac{\Delta \eta(\tau_{s_{n}})}{c^4_s}\; \partial_{\tau_2}{\cal K}_{2}(\tau_1=\tau_{s_{n}},\tau_2)\bigg)}_{\approx 0}\nonumber\\
&&\approx \bigg(\frac{\Delta \eta(\tau_{e_{n}})}{c^8_s}\; {\cal K}_{2}(\tau_{e_{n}})-\frac{\Delta \eta(\tau_{s_{n}})}{c^8_s}\; {\cal K}_{2}(\tau_{s_{n}})\bigg)\nonumber,
\eea 
This furthermore has a distinct kernel denoted as ${\cal K}_{2}(\tau_{1},\tau_{2})$.

These outcomes allow us to arrive at the following simplified conclusion: 
\bea \bigg[\Delta^{2}_{\zeta,\textbf{One-loop}}(k)\bigg]_{\textbf{USR}_{n}} &=&  \frac{1}{4}\bigg[\Delta^{2}_{\zeta,\textbf{Tree}}(k)\bigg]_{\textbf{SR}_{1}}^{2}
    \times \bigg\{\bigg[\bigg(\frac{\Delta\eta(\tau_{e_{n}})}{\tilde{c}^{4}_{s}}\bigg)^{2}{\cal J}_{\textbf{USR}_{n}}^{(1)}(\tau_{e_{n}}) - \left(\frac{\Delta\eta(\tau_{s_{n}})}{\tilde{c}^{4}_{s}}\right)^{2}{\cal J}_{\textbf{USR}_{n}}^{(1)}(\tau_{s_{n}})\bigg] \nonumber \\ 
    &\quad \quad +& 2 \bigg[\bigg(\frac{\Delta\eta(\tau_{e_{n}})}{\tilde{c}^{4}_{s}}\bigg) {\cal J}_{\textbf{USR}_{n}}^{(2)}(\tau_{e_{n}}) - \bigg(\frac{\Delta\eta(\tau_{e_{n}})}{\tilde{c}^{4}_{s}}\bigg) {\cal J}_{\textbf{USR}_{n}}^{(2)}(\tau_{s_{n}})\bigg]-c_{\textbf{USR}_{n}} \bigg\}\eea
  Here we now explicitly compute the momentum integrals in the following:
\begin{enumerate}
    \item \underline{\bf First USR${}_n$ integral ${\cal J}_{{\bf USR}_{n}}^{(1)}$:}\\
    We have explicitly written the expressions for the associated momentum integrals which directly contribute to the final result of the regularized one-loop corrected power spectrum. These integrals are appended below:
    \bea {\cal J}_{\textbf{USR}_{n}}^{(1)}(\tau_{s_{n}}) &=& \int_{k_{s_{n}}}^{k_{e_{n}}}\frac{dk}{k}\bigg|\alpha^{(2n)}_{ k}\left(1+ikc_s\tau\right)e^{-ikc_s\tau}-\beta^{(2n)}_{ k}\left(1-ikc_s\tau\right)e^{ikc_s\tau}\bigg|^2\nonumber\\
&\approx &\ln\left(\frac{k_{e_{n}}}{k_{s_{n}}}\right) + \frac{1}{2} \bigg(\frac{k_{e_n}}{k_{s_n}}\bigg)^2 - \frac{1}{2} + \cdots,\quad\quad \eea
and \bea
{\cal J}_{\textbf{USR}_{n}}^{(1)} (\tau_{e_{n}}) &=& \left(\frac{k_{e_{n}}}{k_{s_{n}}}\right)^{6} \int_{k_{s_{n}}}^{k_{e_{n}}}\frac{dk}{k}\bigg|\alpha^{(2n)}_{ k}\left(1+ikc_s\tau\right)e^{-ikc_s\tau}-\beta^{(2n)}_{ k}\left(1-ikc_s\tau\right)e^{ikc_s\tau}\bigg|^2\nonumber\\
&\approx&\left(\frac{k_{e_{n}}}{k_{s_{n}}}\right)^{6}{\cal J}_{\textbf{USR}_{n}}^{(1)}(\tau_{s_{n}}),\quad\quad\eea
    where the Bogoliubov coefficients, $\alpha^{(2n)}_{\bf k}$ and $\beta^{(2n)}_{\bf k}$ are explicitly mentioned before. Also the $\cdots$ represent contributions which is suppressed in the context of the present computation. 
      \item \underline{\bf First USR${}_n$ integral ${\cal J}_{{\bf USR}_{n}}^{(2)}$:}
      The second subdominant contribution in the one-loop momentum integrals are given by the following expressions:
      \bea {\cal J}_{\textbf{USR}_{n}}^{(2)}(\tau_{s_{n}}) &=& \int_{k_{s_n}}^{k_{e_n}} d\ln{k} \bigg(\frac{d}{d\ln{k}} \bigg|\alpha^{(2n)}_{ k}\left(1+ikc_s\tau\right)e^{-ikc_s\tau}-\beta^{(2n)}_{ k}\left(1-ikc_s\tau\right)e^{ikc_s\tau}\bigg|^2 \bigg) \nonumber \\
&=& \bigg[\bigg|\alpha^{(2n)}_{ k}\left(1+ikc_s\tau\right)e^{-ikc_s\tau}-\beta^{(2n)}_{ k}\left(1-ikc_s\tau\right)e^{ikc_s\tau}\bigg|^2 \bigg]_{k_{s_n}}^{k_{e_n}}, \eea
and 
\bea
{\cal J}_{\textbf{USR}_{n}}^{(2)}(\tau_{e_{n}}) &=& \left(\frac{k_{e_{n}}}{k_{s_{n}}}\right)^{6} 
{\cal J}_{\textbf{USR}_{n}}^{(2)}(\tau_{e_{n}}), \eea
\end{enumerate}
Finally, for the $n$-th USR phase having sharp transitions, one can write down the expression for the one-loop contribution:
\bea
\bigg[\Delta^{2}_{\zeta,\textbf{One-loop}}(k)\bigg]_{\textbf{USR}_{n}} &=&  \bigg[\Delta^{2}_{\zeta,\textbf{Tree}}(k)\bigg]_{\textbf{SR}_{1}}^{2} \times V_n,\eea
where the factor $V_n$ is defined as:
\bea V_n=\bigg[{\cal T}^{(n)}_{1}+{\cal T}^{(n)}_{2}-  c_{\textbf{USR}_{n}}\bigg],
\eea
where the terms ${\cal T}^{(n)}_{1}$ and ${\cal T}^{(n)}_{2}$ are given by:
\bea
{\cal T}^{(n)}_{1} &=& \frac{1}{4}\bigg[\bigg(\frac{\Delta\eta(\tau_{e_{n}})}{\tilde{c}^{4}_{s}}\bigg)^{2}{\cal J}_{\textbf{USR}_{n}}^{(1)}(\tau_{e_{n}}) - \left(\frac{\Delta\eta(\tau_{s_{n}})}{\tilde{c}^{4}_{s}}\right)^{2}{\cal J}_{\textbf{USR}_{n}}^{(1)}(\tau_{s_{n}})\bigg]. \nonumber \\
{\cal T}^{(n)}_{2} &=& \frac{1}{2} \bigg[\bigg(\frac{\Delta\eta(\tau_{e_{n}})}{\tilde{c}^{4}_{s}}\bigg) {\cal J}_{\textbf{USR}_{n}}^{(2)}(\tau_{e_{n}}) - \bigg(\frac{\Delta\eta(\tau_{e_{n}})}{\tilde{c}^{4}_{s}}\bigg) {\cal J}_{\textbf{USR}_{n}}^{(2)}(\tau_{s_{n}})\bigg].
\eea
  Here, the ratio ${\cal T}^{(n)}_{2} / {\cal T}^{(n)}_{1}\ll 1$, as ${\cal T}^{(n)}_{2}$ represents highly suppressed contributions, and so ${\cal T}^{(n)}_{1}\left(1 + {\cal T}^{(n)}_{2} / {\cal T}^{(n)}_{1}\right) \approx {\cal T}^{(n)}_{1}$. Consequently, the factor $V_n$ is approximated as, $V_n=\left[{\cal T}^{(n)}_{1}-  c_{\textbf{USR}_{n}}\right]$.

  Likewise, the combined one-loop contribution for the SRI and SRII periods is provided by:
  \bea \bigg[\Delta^{2}_{\zeta, {\bf One-loop}}(p)\bigg]_{\bf SR}=\bigg[\Delta^{2}_{\zeta,{\bf Tree}}(p)\bigg]_{\bf SR}\times U_n,\eea
where the definition of quantity $U_n$ is as follows:
\bea U_n=U_{\textbf{SR}_{1}}+U_{\textbf{SR}_{n+1}},\eea
The following terms indicate each of the contributions made by $\textbf{SR}_{1}$ and $\textbf{SR}_{n+1}$:
\bea U_{\textbf{SR}_{1}}&=&\bigg[\Delta^{2}_{\zeta,{\bf Tree}}(p)\bigg]_{\textbf{SR}_{1}}\times\Bigg(1+\frac{2}{15\pi^2}\frac{1}{c^2_{s}p^2_*}\bigg(-\left(1-\frac{1}{c^2_{s}}\right)\epsilon+6\frac{\bar{M}^3_1}{ HM^2_{ pl}}-\frac{4}{3}\frac{M^4_3}{H^2M^2_{ pl}}\bigg)\Bigg)\times\Bigg(c_{\textbf{SR}_{1}}-\frac{4}{3}{\cal J}_{\textbf{SR}_{1}}(\tau_{s_{1}})\Bigg),\\ 
U_{\textbf{SR}_{n+1}}&=&\bigg[\Delta^{2}_{\zeta,{\bf Tree}}(p)\bigg]_{\textbf{SR}_{1}}\times\Bigg(1+\frac{2}{15\pi^2}\frac{1}{c^2_{s}p^2_*}\bigg(-\left(1-\frac{1}{c^2_{s}}\right)\epsilon+6\frac{\bar{M}^3_1}{ HM^2_{ pl}}-\frac{4}{3}\frac{M^4_3}{H^2M^2_{ pl}}\bigg)\Bigg)\times\Bigg(c_{\textbf{SR}_{n+1}}+{\cal J}_{\textbf{SR}_{n+1}}(\tau_{e_{n}})\Bigg),\quad\quad\quad\eea
Now we comment on the explicit structure of the momentum integrals that contribute to the region SRI and SRII, which are given by:
\begin{enumerate}
    \item \underline{\bf $\textbf{SR}_{1}$ integral ${\cal J}_{\textbf{SR}_{1}}$:}\\
    We shall discuss the following integral here, which shows up in the computation of the primordial power spectrum of scalar modes in the $\textbf{SR}_{1}$ region after a one-loop correction. Now let's assess the integral that follows:
\bea\label{intSR1} {\cal J}_{\textbf{SR}_{1}}(\tau_{s_{1}}):&=& \int_{k_{*}}^{k_{s_1}}\frac{dk}{k}(1+k^2c_s ^2 \tau^2)=\bigg[\ln\left(\frac{k_{s_1}}{k_*}\right)+\frac{1}{2}\left(k^2_{s_1}-k^2_*\right)c^2_s\tau^2\bigg]= \bigg[\ln\bigg(\frac{k_{s_1}}{k_{*}}\bigg) + \frac{1}{2}\bigg(\frac{k_{s_1}}{k_{*}}\bigg)^{2} -\frac{1}{2}\bigg].\quad\quad\eea

    \item \underline{\bf $\textbf{SR}_{n+1}$ integral ${\cal J}_{\textbf{SR}_{n+1}}$:}\\
    To begin with, let us express the contribution as a momentum-dependent integral that appears in the SRII region:
\bea  \label{gkk1x} {\cal J}_{\textbf{SR}_{n+
1}}(\tau_{e_{n}}):&=& \left(\frac{k_{e_{n}}}{k_{s_{n}}}\right)^{6}\int_{k_{e_{n}}}^{k_{s_{n+1}}}\frac{dk}{k}\bigg|\alpha^{(2n+1)}_{ k}\left(1+ikc_s\tau\right)e^{-ikc_s\tau}-\beta^{(2n+1)}_{ k}\left(1-ikc_s\tau\right)e^{ikc_s\tau}\bigg|^2 \nonumber \\
&\approx & \bigg\{\ln\left(\frac{k_{s_{n+1}}}{k_{e_{n}}}\right) - \frac{27}{32}\bigg[1-\bigg(\frac{k_{e_n}}{k_{s_{n+1}}}\bigg)^{12}\bigg]\bigg\}\nonumber\\
&\approx& \ln\left(\frac{k_{s_{n+1}}}{k_{e_{n}}}\right).\eea
\end{enumerate}
Then the total one-loop regularized but unrenormalized power spectrum for the scalar modes considering $n$-th sharp transition is described by the following simplified expression:
\bea \bigg[\Delta^{2}_{\zeta, {\bf EFT}}(p)\bigg]_{n}&=&\bigg[\Delta^{2}_{\zeta,{\bf Tree}}(p)\bigg]_{\bf SR_1}+\bigg[\Delta^{2}_{\zeta, {\bf One-loop}}(p)\bigg]_{\bf SR_{n+1}}+\bigg[\Delta^{2}_{\zeta, {\bf One-loop}}(p)\bigg]_{\bf USR_{n+1}}\nonumber\\
&=&\bigg[\Delta^{2}_{\zeta,{\bf Tree}}(p)\bigg]_{\bf SR_1}\times \left(1+U_n+V_n\right).\eea
Below is a representative of the loop diagrams for the total contribution:
\begin{equation} \label{unrenormloop}
\begin{tikzpicture}[baseline={([yshift=-3.5ex]current bounding box.center)},very thick]
  
  \def\radius{1}
  \scalebox{1}{\draw[purple,ultra thick] (0,\radius) circle (\radius);
  \draw[purple,ultra thick] (4.5*\radius,0) circle (\radius);}

  \draw[black, very thick] (-1.5*\radius,0) -- (0,0);
  \draw[blue,fill=blue] (0,0) circle (.5ex);
  \draw[black, very thick] (0,0)  -- (1.5*\radius,0);
  \node at (2*\radius,0) {+};
  \draw[black, very thick] (2.5*\radius,0) -- (3.5*\radius,0); 
  \draw[blue,fill=blue] (3.5*\radius,0) circle (.5ex);
  \draw[blue,fill=blue] (5.5*\radius,0) circle (.5ex);
  \draw[black, very thick] (5.5*\radius,0) -- (6.5*\radius,0);
  

\end{tikzpicture}\quad = \quad \bigg[\Delta^{2}_{\zeta, {\bf EFT}}(p)\bigg]_{n},
\end{equation}
The equation (\ref{unrenormloop}) gives a diagrammatic representation of the total unrenormalized one-loop contributions to the scalar power spectrum. When we fix $n=1$ then the above diagrams represent the contributions from the single sharp transitions. In the case of MSTs we need to take the sum over all $n$ which runs from $1$ to $N$, where $N$ is the maximum transition possible in this setup.

\subsection{Renormalization using standard techniques in Quantum Field Theory}

Our main goal here is to put the computation in a more understandable language so that one can more easily relate it to the typical renormalization methods that are used in the context of Quantum Field Theory. Rather than employing several methods of this kind in this section, we will instead focus only on the method in which the divergences that arise at the level of the unrenormalized/bare action are subtracted by introducing counter-terms. Ultimately, this will result in the renormalized form of the action, where, upon successful completion of this procedure, all potentially damaging divergences, including quadratic UV divergence in particular, may be eliminated totally and logarithmic IR divergences can be smoothed.

Let us start by writing down the expression for the third-order perturbed bare action for comoving curvature perturbation, which is provided by the following expression:
\bea
\label{baction}
         S_{\zeta,{\bf B}}^{(3)} &=& M^{2}_{p}\int d\tau\;d^3x\;\bigg [\left({\bf D}_1\right)_{\bf B}\; \zeta^{\prime} {^2}_{\bf B} \zeta_{\bf B} + \left({\bf D}_2\right)_{\bf B}\;(\partial_i \zeta_{\bf B})^2 \zeta_{\bf B}  -  \left({\bf D}_3\right)_{\bf B}\;\zeta^{\prime}_{\bf B} (\partial_i \zeta_{\bf B}) \bigg (\partial_i \partial ^{-2}\bigg(\frac{\epsilon \zeta^{\prime}_{\bf B}}{c_s ^2}\bigg)\bigg) \nonumber \\ 
        && \quad \quad \quad \quad \quad \quad \quad  - \left({\bf D}_4\right)_{\bf B}\;\left(\zeta^{\prime} {^3}_{\bf B}+\zeta^{\prime}_{\bf B}(\partial_i \zeta_{\bf B})^2 \right) +  \left({\bf D}_5\right)_{\bf B}\;\zeta_{\bf B} \bigg(\partial_i \partial_j \partial^{-2}\bigg (\frac{\epsilon \zeta^{\prime}_{\bf B}}{c_s ^2}\bigg)\bigg)^2 + \underbrace{ \left({\bf D}_6\right)_{\bf B}\zeta^{\prime}_{\bf B} \zeta^2_{\bf B}}_{\textbf{Dominant term}}+.....\bigg],\quad\quad\quad
   \eea 

where we define the bare coupling parameters $\left({\bf D}_i\right)_{\bf B}\forall i=1,2,\cdots,6$ by the following expressions:
\bea  \left({\bf D}_1\right)_{\bf B}&=&\bigg(3(c_s ^2 -1)\epsilon + \epsilon ^2 - \frac{\epsilon ^3}{2}\bigg )a^2,~~~
    \left({\bf D}_2\right)_{\bf B}=\frac{\epsilon}{c_s ^2}\bigg(\epsilon - 2s +1 -c_s ^2 \bigg)a^2, \nonumber\\
    \left({\bf D}_3\right)_{\bf B}&=&\frac{2 \epsilon}{c_s ^2}a^2,~~
    \left({\bf D}_4\right)_{\bf B}=\frac{\epsilon}{aH}\bigg (1-\frac{1}{c_s ^2}\bigg)a^2,~~
    \left({\bf D}_5\right)_{\bf B}=\frac{\epsilon}{2}a^2,~~
    \left({\bf D}_6\right)_{\bf B}=\frac{\epsilon}{2c_s ^2}\partial_{\tau} \bigg(\frac{\eta}{c_s ^2}\bigg)a^2.\eea
    To prevent additional misunderstanding, the bare contributions are expressly indicated by the subscript {\bf B}. 
    
    The rescaling ansatz of the gauge invariant modes, which are very useful in establishing the relationship between the renormalized, unrenormalized/bare, and counter-term contribution, can be used to construct the renormalized version of the third order action for the comoving scalar curvature perturbation. This expression describes the renormalized version of the third-order action:
\bea \zeta_{\bf R}=\zeta_{\bf B}-\zeta_{\bf C}=\sqrt{{\cal Z}^{\rm IR}_{n}}\times \zeta_{\bf B}\quad\quad\quad{\rm where}\quad\quad\quad {\cal Z}^{\rm IR}_{n}:=\left(1+\delta_{{\cal Z}^{\rm IR}_{n}}\right).\eea
Here, the renormalized, bare, and counter-term contributions are denoted by the superscripts {\bf R}, {\bf B}, and {\bf C}. Additionally, it is important to note that, the index $n$ stands for the number of sharp transitions associated with the computation. If we take, $n=1$, then using this one can able to describe the result for a single sharp transition in the present context of the discussion. It is noteworthy to mention that the amount ${\cal Z}^{\rm IR}_{n}$, or more accurately $\delta_{{\cal Z}^{\rm IR}_{n}}$, is usually referred to as the counter-term. In this computation, we must explicitly estimate this quantity by applying the physical renormalization condition. 

We next need to use the recently announced rescaled renormalized form of the gauge invariant scalar curvature perturbation to translate the expression for the second- and third-order unrenormalized/bare action into terms of the renormalized version of the action. By taking the actions listed below, you may accomplish this with ease:
\begin{enumerate}
    \item \underline{\bf Renormalized coupling parameters:}\\ 
By using the previously given ansatz, the renormalized coupling parameters that are present in the third-order perturbed action may be represented as follows in terms of the contributions of the bare and counter-terms:
 \bea \left({\bf D}_i\right)_{\bf R}&=&\left({\bf D}_i\right)_{\bf B}-\left({\bf D}_i\right)_{\bf C}={\cal Z}_{{\bf D}_i}\times \left({\bf D}_i\right)_{\bf B}\quad\quad\quad{\rm where}\quad\quad\quad {\cal Z}_{{\bf D}_i}:=\left(1+\delta_{{\cal Z}_{{\bf D}_i}}\right)\forall~i=1,2,\cdots,6,\eea

  \item \underline{\bf Renormalized operator contributions with couplings:}\\
  Here the individual operators along with the couplings can be written as:
  \bea &&\left({\bf D}_1\right)_{\bf R} \zeta^{\prime} {^2}_{\bf R} \zeta_{\bf R}={\cal Z}_{{\bf D}_1}\left({\cal Z}^{\rm IR}_{n}\right)^{\frac{3}{2}}\times \left({\bf D}_1\right)_{\bf B}\zeta^{\prime} {^2}_{\bf B} \zeta_{\bf B}
     = \left(1+\delta_{{\cal Z}_{{\bf D}_1}}+\frac{3}{2}\delta_{{\cal Z}^{\rm IR}_{n}}+\cdots\right)\times \left({\bf D}_1\right)_{\bf B}\zeta^{\prime} {^2}_{\bf B} \zeta_{\bf B},\nonumber\\
&&\left({\bf D}_2\right)_{\bf R} (\partial_i \zeta_{\bf R})^2 \zeta_{\bf R}={\cal Z}_{{\bf D}_2}\left({\cal Z}^{\rm IR}_{n}\right)^{\frac{3}{2}}\times \left({\bf D}_2\right)_{\bf B}(\partial_i \zeta_{\bf B})^2 \zeta_{\bf B}
     = \left(1+\delta_{{\cal Z}_{{\bf D}_2}}+\frac{3}{2}\delta_{{\cal Z}^{\rm IR}_{n}}+\cdots\right)\times \left({\bf D}_2\right)_{\bf B}(\partial_i \zeta_{\bf B})^2 \zeta_{\bf B},\nonumber\\
    && \left({\bf D}_3\right)_{\bf R} \zeta^{\prime}_{\bf R} (\partial_i \zeta_{\bf R}) \bigg (\partial_i \partial ^{-2}\bigg(\frac{\epsilon \zeta^{\prime}_{\bf R}}{c_s ^2}\bigg)\bigg)={\cal Z}_{{\bf D}_3}\left({\cal Z}^{\rm IR}_{n}\right)^{\frac{3}{2}}\times \left({\bf D}_3\right)_{\bf B}\zeta^{\prime}_{\bf B} (\partial_i \zeta_{\bf B}) \bigg (\partial_i \partial ^{-2}\bigg(\frac{\epsilon \zeta^{\prime}_{\bf B}}{c_s ^2}\bigg)\bigg)\nonumber\\
     &&\quad\quad\quad\quad\quad\quad\quad\quad\quad\quad\quad\quad\quad\quad\quad= \left(1+\delta_{{\cal Z}_{{\bf D}_3}}+\frac{3}{2}\delta_{{\cal Z}^{\rm IR}_{n}}+\cdots\right)\times \left({\bf D}_3\right)_{\bf B}(\partial_i \zeta_{\bf B}) \bigg (\partial_i \partial ^{-2}\bigg(\frac{\epsilon \zeta^{\prime}_{\bf B}}{c_s ^2}\bigg)\bigg),\nonumber\\
     &&\left({\bf D}_4\right)_{\bf R} \left(\zeta^{\prime} {^3}_{\bf R}+\zeta^{\prime}_{\bf R}(\partial_i \zeta_{\bf R})^2 \right)={\cal Z}_{{\bf D}_4}\left({\cal Z}^{\rm IR}_{n}\right)^{\frac{3}{2}}\times \left({\bf D}_4\right)_{\bf B}\left(\zeta^{\prime} {^3}_{\bf B}+\zeta^{\prime}_{\bf B}(\partial_i \zeta_{\bf B})^2 \right)\nonumber\\
     &&\quad\quad\quad\quad\quad\quad\quad\quad\quad\quad\quad\quad= \left(1+\delta_{{\cal Z}_{{\bf D}_4}}+\frac{3}{2}\delta_{{\cal Z}^{\rm IR}_{n}}+\cdots\right)\times \left({\bf D}_4\right)_{\bf B}\left(\zeta^{\prime} {^3}_{\bf B}+\zeta^{\prime}_{\bf B}(\partial_i \zeta_{\bf B})^2 \right),\nonumber\\
&&\left({\bf D}_5\right)_{\bf R} \zeta_{\bf R} \bigg (\partial_i\partial_j \partial ^{-2}\bigg(\frac{\epsilon \zeta^{\prime}_{\bf R}}{c_s ^2}\bigg)\bigg)^2={\cal Z}_{{\bf D}_5}\left({\cal Z}^{\rm IR}_{n}\right)^{\frac{3}{2}}\times \left({\bf D}_5\right)_{\bf B}\zeta_{\bf B}\bigg (\partial_i\partial_j \partial ^{-2}\bigg(\frac{\epsilon \zeta^{\prime}_{\bf B}}{c_s ^2}\bigg)\bigg)^2\nonumber\\
     &&\quad\quad\quad\quad\quad\quad\quad\quad\quad\quad\quad\quad= \left(1+\delta_{{\cal Z}_{{\bf D}_5}}+\frac{3}{2}\delta_{{\cal Z}^{\rm IR}_{n}}+\cdots\right)\times \left({\bf D}_5\right)_{\bf B}\zeta_{\bf B}\bigg (\partial_i\partial_j \partial ^{-2}\bigg(\frac{\epsilon \zeta^{\prime}_{\bf B}}{c_s ^2}\bigg)\bigg)^2\nonumber\\
     &&\left({\bf D}_6\right)_{\bf R} \zeta^{\prime}_{\bf R} \zeta^2_{\bf R}={\cal Z}_{{\bf D}_6}\left({\cal Z}^{\rm IR}_{n}\right)^{\frac{3}{2}}\times \left({\bf D}_6\right)_{\bf B}\zeta^{\prime}_{\bf B} \zeta^2_{\bf B}
     = \left(1+\delta_{{\cal Z}_{{\bf D}_6}}+\frac{3}{2}\delta_{{\cal Z}^{\rm IR}_{n}}+\cdots\right)\times \left({\bf D}_6\right)_{\bf B}\zeta^{\prime}_{\bf B} \zeta^2_{\bf B},
   \eea
   We have discovered that every operator in the third-order perturbed action possesses the universal scaling features based on the previously described study:
    \bea &&{\cal Z}_{{\bf D}_i}\left({\cal Z}^{\rm IR}_{n}\right)^{\frac{3}{2}}\approx \left(1+\delta_{{\cal Z}_{{\bf D}_i}}+\frac{3}{2}\delta_{{\cal Z}^{\rm IR}_{n}}+\cdots\right)\quad\quad\forall \quad i=1,2,\cdots,6.\eea
    In this case, the higher-order components in the associated power-series expansion are represented by the dotted contributions $\cdots$. We have limited our study to first-order terms and disregarded all higher-order minor effects. Consequently, this suggests that we completed the remaining calculations to ascertain the specific contributions of the counter-terms inside the linear domain of the associated expansion. 

\end{enumerate}

Therefore, by following all of the previously stated procedures, one can eventually create the renormalized version of the perturbed actions, which are second- and third-ordered for the gauge invariant comoving curvature perturbation. These equations are as follows:

\bea
\label{raction}
         S_{\zeta,{\bf R}}^{(3)} &=& M^{2}_{p}\int d\tau\;d^3x\;\bigg [\left({\bf D}_1\right)_{\bf R}\; \zeta^{\prime} {^2}_{\bf R} \zeta_{\bf R} + \left({\bf D}_2\right)_{\bf R}\;(\partial_i \zeta_{\bf R})^2 \zeta_{\bf R}  -  \left({\bf D}_3\right)_{\bf R}\;\zeta^{\prime}_{\bf R} (\partial_i \zeta_{\bf R}) \bigg (\partial_i \partial ^{-2}\bigg(\frac{\epsilon \zeta^{\prime}_{\bf R}}{c_s ^2}\bigg)\bigg) \nonumber \\ 
        && \quad \quad \quad \quad \quad \quad \quad  - \left({\bf D}_4\right)_{\bf R}\;\left(\zeta^{\prime} {^3}_{\bf R}+\zeta^{\prime}_{\bf R}(\partial_i \zeta_{\bf R})^2 \right) +  \left({\bf D}_5\right)_{\bf R}\;\zeta_{\bf R} \bigg(\partial_i \partial_j \partial^{-2}\bigg (\frac{\epsilon \zeta^{\prime}_{\bf R}}{c_s ^2}\bigg)\bigg)^2 + \underbrace{ \left({\bf D}_6\right)_{\bf R}\zeta^{\prime}_{\bf R} \zeta^2_{\bf R}}_{\textbf{Dominant term}}+.....\bigg]\nonumber\\
        &=&M^{2}_{p}\int d\tau\;d^3x\;\bigg [\left(1+\delta_{{\cal Z}_{{\bf D}_1}}+\frac{3}{2}\delta_{{\cal Z}^{\rm IR}_{n}}\right)\left({\bf D}_1\right)_{\bf B}\; \zeta^{\prime} {^2}_{\bf B} \zeta_{\bf B} +\left(1+\delta_{{\cal Z}_{{\bf D}_2}}+\frac{3}{2}\delta_{{\cal Z}^{\rm IR}_{n}}\right)\left({\bf D}_2\right)_{\bf B}\;(\partial_i \zeta_{\bf B})^2 \zeta_{\bf B}\nonumber \\ 
        && \quad \quad \quad \quad \quad \quad \quad    -  \left(1+\delta_{{\cal Z}_{{\bf D}_3}}+\frac{3}{2}\delta_{{\cal Z}^{\rm IR}_{n}}\right)\left({\bf D}_3\right)_{\bf B}\;\zeta^{\prime}_{\bf B} (\partial_i \zeta_{\bf B}) \bigg (\partial_i \partial ^{-2}\bigg(\frac{\epsilon \zeta^{\prime}_{\bf B}}{c_s ^2}\bigg)\bigg) \nonumber \\
        && \quad \quad \quad \quad \quad \quad \quad  - \left(1+\delta_{{\cal Z}_{{\bf D}_4}}+\frac{3}{2}\delta_{{\cal Z}^{\rm IR}_{n}}\right)\left({\bf D}_4\right)_{\bf B}\;\left(\zeta^{\prime} {^3}_{\bf B}+\zeta^{\prime}_{\bf B}(\partial_i \zeta_{\bf B})^2 \right) \nonumber \\
        && \quad \quad \quad \quad \quad \quad \quad  +  \left(1+\delta_{{\cal Z}_{{\bf D}_5}}+\frac{3}{2}\delta_{{\cal Z}^{\rm IR}_{n}}\right)\left({\bf D}_5\right)_{\bf B}\;\zeta_{\bf B} \bigg(\partial_i \partial_j \partial^{-2}\bigg (\frac{\epsilon \zeta^{\prime}_{\bf B}}{c_s ^2}\bigg)\bigg)^2\nonumber \\ 
        && \quad \quad \quad \quad \quad \quad \quad  + \underbrace{\left(1+\delta_{{\cal Z}_{{\bf D}_6}}+\frac{3}{2}\delta_{{\cal Z}^{\rm IR}_{n}}\right) \left({\bf D}_6\right)_{\bf B}\zeta^{\prime}_{\bf B} \zeta^2_{\bf B}}_{\textbf{Dominant term}}+.....\bigg],
   \eea 
   Our goal now is to compute the explicit expression for the one-loop corrected scalar power spectrum in the presence of all the counter-terms previously introduced, using the derived form of the renormalized form of the third-order perturbed action for the gauge invariant comoving scalar curvature perturbation as stated in equation (\ref{raction}). This is nothing more than the computation of the renormalized one-loop 1PI effective action corresponding to the two-point amplitude in the description provided by Quantum Field Theory. In order to do this, we employ the previously discussed in-in formalism in the current computing environment. With its help, we were able to derive the following simplified conclusion for each of the distinct phases at the one-loop level:
\bea
    \label{Sr1againA}
        \bigg[\Delta^{2}_{\zeta,\textbf{One-loop}}(k)\bigg]_{\textbf{SR}_{1}} &=& \left(1+\delta_{{\cal Z}^{\rm IR}_{n}}\right)\times\left(1-\frac{2}{15\pi^{2}}\frac{1}{c_{s}^{2}k_{*}^{2}}\left(1-\frac{1}{c_{s}^{2}}\right)\epsilon\right)\nonumber\\
        &&\quad\quad\times \left(\left(\delta_{{\cal Z}_{{\bf D}_1}}+\delta_{{\cal Z}_{{\bf D}_2}}+\delta_{{\cal Z}_{{\bf D}_3}}+\delta_{{\cal Z}_{{\bf D}_4}}+\delta_{{\cal Z}_{{\bf D}_5}}\right)_{\textbf{SR}_{1}}-\frac{4}{3}{\cal J}_{\textbf{SR}_{1}}(\tau_{s_{1}})\bigg[\Delta^{2}_{\zeta,\textbf{Tree}}(k)\bigg]_{\textbf{SR}_{1}}^{2}\right), \\
    \label{usragainB}
        \bigg[\Delta^{2}_{\zeta,\textbf{One-loop}}(k)\bigg]_{\textbf{USR}_{n}} &=&  
        \left(1+\delta_{{\cal Z}^{\rm IR}_{n}}\right)\times\bigg\{\frac{1}{4}\bigg[\Delta^{2}_{\zeta,\textbf{Tree}}(k)\bigg]_{\textbf{SR}_{1}}^{2}\times\bigg[\bigg(\frac{\Delta\eta(\tau_{e_{n}})}{\tilde{c}^{4}_{s}}\bigg)^{2}{\cal J}_{\textbf{USR}_{n}}(\tau_{e_{n}}) - \left(\frac{\Delta\eta(\tau_{s_{n}})}{\tilde{c}^{4}_{s}}\right)^{2}{\cal J}_{\textbf{USR}_{n}}(\tau_{s_{n}})\bigg]\nonumber\\
        &&\quad\quad\quad\quad\quad\quad\quad\quad\quad\quad\quad\quad\quad\quad\quad\quad\quad\quad\quad\quad\quad\quad\quad\quad\quad\quad- \left(\delta_{{\cal Z}_{{\bf D}_6}}\right)_{\textbf{USR}_{n}}\bigg\}, \;\;\quad \\
    \label{sr2againC}
        \bigg[\Delta^{2}_{\zeta,\textbf{One-loop}}(k)\bigg]_{\textbf{SR}_{n+1}} &=& 
\left(1+\delta_{{\cal Z}^{\rm IR}_{n}}\right)\times\left(1-\frac{2}{15\pi^{2}}\frac{1}{c_{s}^{2}k_{*}^{2}}\left(1-\frac{1}{c_{s}^{2}}\right)\epsilon\right)
        \nonumber\\
        &&\quad\quad\times  \bigg(\left(\delta_{{\cal Z}_{{\bf D}_1}}+\delta_{{\cal Z}_{{\bf D}_2}}+\delta_{{\cal Z}_{{\bf D}_3}}+\delta_{{\cal Z}_{{\bf D}_4}}+\delta_{{\cal Z}_{{\bf D}_5}}\right)_{\textbf{SR}_{n+1}}+{\cal J}_{\textbf{SR}_{n+1}}(\tau_{e_{n}})\bigg[\Delta^{2}_{\zeta,\textbf{Tree}}(k)\bigg]_{\textbf{SR}_{1}}^{2}\bigg).\quad\quad
\eea
In addition, we obtain the final compact form of the one-loop corrected renormalized version of the scalar power spectrum for the $n$-th sharp transition, which is characterized by:
\bea
\left[\overline{\Delta_{\zeta,\textbf{EFT}}^{2}(k)}\right]_n &=& \left(1+\delta_{{\cal Z}^{\rm IR}_{n}}\right)\times\bigg\{\bigg[\Delta^{2}_{\zeta,\textbf{Tree}}(k)\bigg]_{\textbf{SR}_{1}}+\left(1-\frac{2}{15\pi^{2}}\frac{1}{c_{s}^{2}k_{*}^{2}}\left(1-\frac{1}{c_{s}^{2}}\right)\epsilon\right)\nonumber\\&&\quad\quad\quad\quad\quad\quad\quad\times \left(\left(\delta_{{\cal Z}_{{\bf D}_1}}+\delta_{{\cal Z}_{{\bf D}_2}}+\delta_{{\cal Z}_{{\bf D}_3}}+\delta_{{\cal Z}_{{\bf D}_4}}+\delta_{{\cal Z}_{{\bf D}_5}}\right)_{\textbf{SR}_{1}}-\frac{4}{3}{\cal J}_{\textbf{SR}_{1}}(\tau_{s_{1}})\bigg[\Delta^{2}_{\zeta,\textbf{Tree}}(k)\bigg]_{\textbf{SR}_{1}}^{2}\right)\nonumber\\
&&\quad\quad\quad\quad\quad+\bigg\{\frac{1}{4}\bigg[\Delta^{2}_{\zeta,\textbf{Tree}}(k)\bigg]_{\textbf{SR}_{1}}^{2}\times\bigg[\bigg(\frac{\Delta\eta(\tau_{e_{n}})}{\tilde{c}^{4}_{s}}\bigg)^{2}{\cal J}_{\textbf{USR}_{n}}(\tau_{e_{n}}) - \left(\frac{\Delta\eta(\tau_{s_{n}})}{\tilde{c}^{4}_{s}}\right)^{2}{\cal J}_{\textbf{USR}_{n}}(\tau_{s_{n}})\bigg]\nonumber\eea\bea
        &&\quad\quad\quad\quad\quad\quad\quad\quad\quad\quad\quad\quad\quad\quad\quad\quad\quad\quad\quad\quad\quad\quad\quad\quad\quad\quad- \left(\delta_{{\cal Z}_{{\bf D}_6}}\right)_{\textbf{USR}_{n}}\bigg\}\nonumber\\
&&\quad\quad\quad\quad\quad+\left(1-\frac{2}{15\pi^{2}}\frac{1}{c_{s}^{2}k_{*}^{2}}\left(1-\frac{1}{c_{s}^{2}}\right)\epsilon\right)\nonumber\\
&&\quad\quad\quad\quad\quad\times\bigg(\left(\delta_{{\cal Z}_{{\bf D}_1}}+\delta_{{\cal Z}_{{\bf D}_2}}+\delta_{{\cal Z}_{{\bf D}_3}}+\delta_{{\cal Z}_{{\bf D}_4}}+\delta_{{\cal Z}_{{\bf D}_5}}\right)_{\textbf{SR}_{n+1}}+{\cal J}_{\textbf{SR}_{n+1}}(\tau_{e_{n}})\bigg[\Delta^{2}_{\zeta,\textbf{Tree}}(k)\bigg]_{\textbf{SR}_{1}}^{2}\bigg)\bigg\},\quad\quad
\eea
which can be written further as:
\bea
\left[\overline{\Delta_{\zeta,\textbf{EFT}}^{2}(k)}\right]_n &=& {\cal Z}^{\rm IR}_{n}{\cal Z}^{\rm UV}_{n}\times \bigg[\Delta^{2}_{\zeta,\textbf{Tree}}(k)\bigg]_{\textbf{SR}_{1}}= {\cal Z}^{\rm IR}_{n}\times \left[\Delta_{\zeta,\textbf{EFT}}^{2}(k)\right]_n,\eea
where IR and UV counter-terms are defined as:
\bea && {\cal Z}^{\rm IR}_{n}=\left(1+\delta_{{\cal Z}^{\rm IR}_{n}}\right),~~~
 {\cal Z}^{\rm UV}_{n}=\left(1+\delta_{{\cal Z}^{\rm UV}_{n}}\right).\eea
 Here we define $\delta_{{\cal Z}^{\rm UV}_{n}}$ by the following expression:
 \bea &&\delta_{{\cal Z}^{\rm UV}_{n}}(\delta_{{\cal Z}_{{\bf D}_1}},\delta_{{\cal Z}_{{\bf D}_2}},\delta_{{\cal Z}_{{\bf D}_3}},\delta_{{\cal Z}_{{\bf D}_4}},\delta_{{\cal Z}_{{\bf D}_5}},\delta_{{\cal Z}_{{\bf D}_6}})\nonumber\\
&=&\bigg[\Delta^{2}_{\zeta,\textbf{Tree}}(k)\bigg]_{\textbf{SR}_{1}}\times\left(1-\frac{2}{15\pi^{2}}\frac{1}{c_{s}^{2}k_{*}^{2}}\left(1-\frac{1}{c_{s}^{2}}\right)\epsilon\right)\nonumber\\
&&\quad\quad\quad\quad\quad\quad\times\left(\frac{1}{\bigg[\Delta^{2}_{\zeta,\textbf{Tree}}(k)\bigg]_{\textbf{SR}_{1}}^{2}}\times\left(\delta_{{\cal Z}_{{\bf D}_1}}+\delta_{{\cal Z}_{{\bf D}_2}}+\delta_{{\cal Z}_{{\bf D}_3}}+\delta_{{\cal Z}_{{\bf D}_4}}+\delta_{{\cal Z}_{{\bf D}_5}}\right)_{\textbf{SR}_{1}}-\frac{4}{3}{\cal J}_{\textbf{SR}_{1}}(\tau_{s_{1}})\right)\nonumber\\
&&\quad\quad\quad\quad\quad+\frac{1}{4}\bigg[\Delta^{2}_{\zeta,\textbf{Tree}}(k)\bigg]_{\textbf{SR}_{1}}\times\Bigg\{\bigg[\bigg(\frac{\Delta\eta(\tau_{e_{n}})}{\tilde{c}^{4}_{s}}\bigg)^{2}{\cal J}_{\textbf{USR}_{n}}(\tau_{e_{n}}) - \left(\frac{\Delta\eta(\tau_{s_{n}})}{\tilde{c}^{4}_{s}}\right)^{2}{\cal J}_{\textbf{USR}_{n}}(\tau_{s_{n}})\bigg]\nonumber\\
        &&\quad\quad\quad\quad\quad\quad\quad\quad\quad\quad\quad\quad\quad\quad\quad\quad\quad\quad\quad\quad\quad\quad\quad\quad\quad\quad- \frac{4}{\bigg[\Delta^{2}_{\zeta,\textbf{Tree}}(k)\bigg]_{\textbf{SR}_{1}}^{2}}\times\left(\delta_{{\cal Z}_{{\bf D}_6}}\right)_{\textbf{USR}_{n}}\Bigg\}\nonumber\\
&&\quad\quad\quad\quad\quad+\bigg[\Delta^{2}_{\zeta,\textbf{Tree}}(k)\bigg]_{\textbf{SR}_{1}}\times\left(1-\frac{2}{15\pi^{2}}\frac{1}{c_{s}^{2}k_{*}^{2}}\left(1-\frac{1}{c_{s}^{2}}\right)\epsilon\right)\nonumber\\
&&\quad\quad\times\left(\frac{1}{\bigg[\Delta^{2}_{\zeta,\textbf{Tree}}(k)\bigg]_{\textbf{SR}_{1}}^{2}}\times\left(\delta_{{\cal Z}_{{\bf D}_1}}+\delta_{{\cal Z}_{{\bf D}_2}}+\delta_{{\cal Z}_{{\bf D}_3}}+\delta_{{\cal Z}_{{\bf D}_4}}+\delta_{{\cal Z}_{{\bf D}_5}}\right)_{\textbf{SR}_{n+1}}+{\cal J}_{\textbf{SR}_{n+1}}(\tau_{e_{n}})\right),\quad\quad
\eea
In order to find the expression for the counter-terms as they appear in the derived result given above, we then apply the renormalization condition. The current topic may be comprehended by explicitly using the Renormalization Group (RG) flow, which will effectively fix the structure of any UV and IR sensitive counter-terms that emerge in the calculation. With the aid of the flow equation and matching beta functions written for the corresponding 1PI one-loop corrected renormalized two-point amplitude in the Fourier space—basically, a representation of the renormalized scalar power spectrum in this particular context—this can be technically accomplished.

Here, the Callan–Symanzik equation in this cosmological configuration may be written as:
\bea \frac{d}{d\ln \mu}\Bigg\{\bigg[\Delta^{2}_{\zeta,\textbf{Tree}}(k)\bigg]_{\textbf{SR}_{1}}\Bigg\}=\frac{d}{d\ln \mu}\Bigg\{\frac{\left[\overline{\Delta_{\zeta,\textbf{EFT}}^{2}(k)}\right]_n}{{\cal Z}^{\rm IR}_{n}{\cal Z}^{\rm UV}_{n}}\Bigg\}=0.\eea
It is now noteworthy to observe that the equivalent total differential operator in this case may be further reduced using the formula that follows:
\bea \frac{d}{d\ln \mu}=\bigg(\frac{\partial}{\partial\ln \mu}+\sum^{6}_{i=1}\beta_{{\bf D}_i}\frac{\partial}{\partial {\bf D}_i}-\gamma_{\rm IR}-\gamma_{\rm UV}\Bigg),\eea
where the beta functions are defined as:
\bea \beta_{{\bf D}_1}&=&\left(\frac{\partial {\bf D}_1}{\partial\ln \mu}\right)= 2\epsilon a^2\Bigg[(\epsilon-\eta)\bigg(3(c^2_s-1)+2\epsilon-\frac{3\epsilon^2}{2}\bigg)+\bigg(3(c^2_s-1)+\epsilon-\frac{\epsilon^2}{2}+6\frac{ s c^2_s}{H}\bigg)\Bigg]\Bigg[1+\left(\epsilon+\frac{s}{H}\right)\Bigg],\\
\beta_{{\bf D}_2}
&=&\left(\frac{\partial {\bf D}_2}{\partial\ln \mu}\right)=\frac{2\epsilon a^2}{c^2_s}\Bigg[(\epsilon-\eta)\bigg(2\epsilon-2s+1-c^2_s\bigg)+\bigg(\epsilon-2s+1-c^2_s\bigg)\left(1-\frac{s}{H}\right)-\left(\frac{\dot{s}}{H}+\frac{sc^2_s}{H}\right)\Bigg]\Bigg[1+\left(\epsilon+\frac{s}{H}\right)\Bigg],\quad\quad\quad\\
\beta_{{\bf D}_3}
&=&\left(\frac{\partial {\bf D}_3}{\partial\ln \mu}\right)=\frac{4\epsilon a^2}{c^2_s}\Bigg(\epsilon-\eta+1-\frac{s}{H}\Bigg)\Bigg[1+\left(\epsilon+\frac{s}{H}\right)\Bigg],\\
\beta_{{\bf D}_4}
&=&\left(\frac{\partial {\bf D}_4}{\partial\ln \mu}\right)=\frac{2a\epsilon}{H}\Bigg[(\epsilon-\eta)\left(1-\frac{1}{c^2_s}\right)+\frac{s}{H c^2_s}+\frac{1}{2}(1+\epsilon H)\left(1-\frac{1}{c^2_s}\right)\Bigg]\Bigg[1+\left(\epsilon+\frac{s}{H}\right)\Bigg],\\
\beta_{{\bf D}_5}
&=&\left(\frac{\partial {\bf D}_5}{\partial\ln \mu}\right)=\epsilon a^2 \left(\epsilon-\eta+1\right)\Bigg[1+\left(\epsilon+\frac{s}{H}\right)\Bigg],\\
\beta_{{\bf D}_6}
&=&\left(\frac{\partial {\bf D}_6}{\partial\ln \mu}\right)=\frac{a^2\epsilon}{c^2_s}\left(\frac{\eta}{c^2_s}\right)^{'}\Bigg[\bigg(\epsilon-\eta+1-\frac{s}{H}\bigg)+\frac{1}{2}\frac{d}{d\ln \mu}\ln \left(\frac{\eta}{c^2_s}\right)^{'}\Bigg]\Bigg[1+\left(\epsilon+\frac{s}{H}\right)\Bigg].
\eea
Additionally, we present two additional parameters that are dependent on the IR and UV counter-terms and are expressed as follows:
\bea && \gamma_{\rm IR}:=\left(\frac{\partial\ln {\cal Z}^{\rm IR}_{n}}{\partial\ln \mu}\right),\quad\quad\quad
 \gamma_{\rm UV}:=\left(\frac{\partial\ln {\cal Z}^{\rm UV}_{n}}{\partial\ln \mu}\right).\eea
 In the end, we obtain the Callan–Symanzik equation for the specified cosmological configuration as follows:
\bea \bigg(\frac{\partial}{\partial\ln \mu}+\sum^{6}_{i=1}\beta_{{\bf D}_i}\frac{\partial}{\partial {\bf D}_i}-\gamma_{\rm IR}-\gamma_{\rm UV}\Bigg)\overline{\Delta_{\zeta,\textbf{EFT}}^{2}(k)}=0.\eea
To further assist in identifying the IR and UV counter-terms in terms of the renormalization scale in the current context, the following flow equations can be computed:
\begin{enumerate}
    \item \underline{\bf Renormalized scalar spectral tilt:}\\ 
    Calculating the renormalized spectral tilt of the scalar power spectrum in terms of the cosmic hierarchical flow may be done as follows:
 \bea 
\left[\overline{n_{\zeta,\textbf{EFT}}(k)-1}\right]_n&=&\frac{d}{d\ln k}\bigg(\ln \left[\overline{\Delta_{\zeta,\textbf{EFT}}^{2}(k)}\right]_n\bigg)\nonumber\\
&=&{\cal Z}^{\rm IR}_{n}\times\Bigg[{\cal Z}^{\rm UV}_{n}\bigg(\bigg[n_{\zeta,\textbf{Tree}}(k)\bigg]_{{\bf SR}_1}-1\bigg)+\bigg(\frac{d{\cal Z}^{\rm UV}_{n}}{d\ln k}\bigg)\bigg(\ln \bigg[\Delta^{2}_{\zeta,\textbf{Tree}}(k)\bigg]_{\textbf{SR}_{1}}\bigg)\Bigg].\eea

       \item \underline{\bf Renormalized running of the scalar spectral tilt:}\\ 
The renormalized running of the spectral tilt of the scalar power spectrum may be calculated as follows in terms of the cosmic hierarchical flow:

    \bea 
\left[\overline{\alpha_{\zeta,\textbf{EFT}}(k)}\right]_n &=&\frac{d}{d\ln k}\bigg(\left[\overline{n_{\zeta,\textbf{EFT}}(k)}\right]_n\bigg)\nonumber\\
&=&{\cal Z}^{\rm IR}_{n}\times\Bigg[{\cal Z}^{\rm UV}_{n}\bigg(\bigg[\alpha_{\zeta,\textbf{Tree}}(k)\bigg]_{{\bf SR}_1}\bigg)+2\bigg(\frac{d{\cal Z}^{\rm UV}_{n}}{d\ln k}\bigg)\bigg(\bigg[n_{\zeta,\textbf{Tree}}(k)\bigg]_{{\bf SR}_1}-1\bigg)\nonumber\\
&&\quad\quad\quad\quad\quad\quad\quad\quad\quad\quad\quad\quad\quad\quad\quad\quad+\bigg(\frac{d^2{\cal Z}^{\rm UV}_{n}}{d\ln k^2}\bigg)\bigg(\ln \bigg[\Delta^{2}_{\zeta,\textbf{Tree}}(k)\bigg]_{\textbf{SR}_{1}}\bigg)\Bigg].\eea

\item \underline{\bf Renormalized running of the running of scalar spectral tilt:}\\ 
The renormalized running of the running of spectral tilt of the scalar power spectrum may be calculated as follows in terms of the cosmic hierarchical flow:

    \bea 
\left[\overline{\beta_{\zeta,\textbf{EFT}}(k)}\right]_n&=&\frac{d}{d\ln k}\bigg(\left[\overline{\alpha_{\zeta,\textbf{EFT}}(k)}\right]_n\bigg)\nonumber\\
&=&{\cal Z}^{\rm IR}_{n}\times\Bigg[{\cal Z}^{\rm UV}_{n}\bigg(\bigg[\beta_{\zeta,\textbf{Tree}}(k)\bigg]_{{\bf SR}_1}\bigg)+2\bigg(\frac{d{\cal Z}^{\rm UV}_{n}}{d\ln k}\bigg)\bigg(\bigg[\alpha_{\zeta,\textbf{Tree}}(k)\bigg]_{{\bf SR}_1}\bigg)\nonumber\\
&&\quad\quad\quad\quad\quad+3\bigg(\frac{d^2{\cal Z}^{\rm UV}_{n}}{d\ln k^2}\bigg)\bigg(\bigg[n_{\zeta,\textbf{Tree}}(k)\bigg]_{{\bf SR}_1}-1\bigg)+\bigg(\frac{d^3{\cal Z}^{\rm UV}_{n}}{d\ln k^3}\bigg)\bigg(\ln \bigg[\Delta^{2}_{\zeta,\textbf{Tree}}(k)\bigg]_{\textbf{SR}_{1}}\bigg)\Bigg].\quad\quad\eea

      \end{enumerate}
      The flow equations outlined above indicate that the two-point amplitude of the primordial power spectrum for the scalar modes, when the IR and UV counter-term effects are present in each individual expression, exhibits scale dependence.

We are now going to apply the renormalization conditions, which will fix the structure of the counter-terms that are susceptible to both UV and IR radiation. The known facts at the CMB pivot scale $k_*$ will now be used for this, which prevents us from taking into account the following essential limitations, which are explained in terms of renormalization conditions:
\begin{enumerate}
    \item \underline{\bf Renormalization condition I:}\\ 
The first renormalization requirement specifies that, at the CMB pivot scale $k_*$, the two-point amplitude of the scalar power spectrum following renormalization must precisely match the tree-level contribution estimated in the first slow-roll phase of our computing setup. Technically speaking, this assertion may be expressed as follows:
  \bea \left[\overline{\Delta_{\zeta,\textbf{EFT}}^{2}(k_*)}\right]_n &=& \bigg[\Delta^{2}_{\zeta,\textbf{Tree}}(k_*)\bigg]_{\textbf{SR}_{1}}.\eea
  This state is entirely justified from a physical standpoint and is in no way ad hoc. Since no discernible quantum effects have been found at the CMB map to yet, such quantum corrections may be safely regarded as missing at the equivalent momentum scale, which is the pivot scale. Regarding the credible cosmic observations, any a-causal aspects occurring outside the horizon (i.e., in the super-Hubble scale) are irrelevant. The CMB displays the result of causal occurrences in terms of the distribution of hot and cool spots in the maps.

 \item  \underline{\bf Renormalization condition II:}\\ 
 The second renormalization requirement is that, following renormalization, the tree-level contribution computed in the first slow-roll phase of our computational setup at the CMB pivot scale $k_*$ must exactly match the logarithmic derivative of the two-point amplitude of the scalar power spectrum with respect to the momentum scale. In the context of primordial cosmology, this logarithmic derivative with respect to the momentum scale essentially computes the scale dependence of the two-point amplitude, which is also referred to as the {\it spectral-tilt} or {\it spectral-index} of the scalar power spectrum. This assertion may be expressed technically as follows:
  \bea 
\left[\overline{n_{\zeta,\textbf{EFT}}(k_*)-1}\right]_n&=&\Bigg[\frac{d}{d\ln k}\bigg(\ln \left[\overline{\Delta_{\zeta,\textbf{EFT}}^{2}(k)}\right]_n\bigg)\Bigg]_{k=k_*}=\bigg(\bigg[n_{\zeta,\textbf{Tree}}(k_*)\bigg]_{{\bf SR}_1}-1\bigg).\eea
The condition indicated above and the condition mentioned above are entirely compatible. In fact, it provides additional information about the scalar power spectrum's structure in the current environment. Despite the tilt being calculated in the CMB pivot scale $k_*$, we are able to further fix the form of the primordial power spectrum computed for the scalar modes because of its non-zero value.

 \item  \underline{\bf Renormalization condition III:}\\
 In our computational setup, the tree-level contribution computed in the first slow-roll phase must exactly match the third renormalization condition, which is the second logarithmic derivative of the two-point amplitude of the scalar power spectrum with respect to the momentum scale. This is essentially the running of the scalar spectral tilt after renormalization. In primordial cosmology, this so-called {\it running of the spectral-tilt} or {\it running of the spectral-index} of the scalar power spectrum is essentially computed as the scale dependence of the two-point amplitude using a logarithmic double derivative with respect to the momentum scale. Said another way, this claim may be expressed technically as:
\bea 
\left[\overline{\alpha_{\zeta,\textbf{EFT}}(k_*)}\right]_n=\Bigg[\frac{d}{d\ln k}\bigg(\left[\overline{n_{\zeta,\textbf{EFT}}(k)}\right]_n\bigg)\Bigg]_{k=k_*}=\bigg(\bigg[\alpha_{\zeta,\textbf{Tree}}(k_*)\bigg]_{{\bf SR}_1}\bigg).\eea
The two conditions that were previously described are entirely compatible with the situation that is given above. Rather of having a tilt, in this situation, it actually informs us more about the structure of the scalar power spectrum. We can fix the shape of the primordial power spectrum computed for the scalar modes in terms of concavity or convexity in the original form of the underlying effective potential or Hubble parameter of the underlying EFT setup, even though the running of the tilt is computed in the CMB pivot scale $k_*$. This is because the value of this parameter is non-zero. A saddle point, inflection point, bump/dip, or other characteristics in the mathematical structure of the effective potential or in the Hubble parameter as seen in the current EFT computation are among the extra features that may exist based on the presence of such running farther.

\item  \underline{\bf Renormalization condition IV:}\\
As for the fourth renormalization condition, it says that the tree-level contribution computed in the first slow-roll phase in our computational setup at the CMB pivot scale $k_*$ must exactly match the third logarithmic derivative of the two-point amplitude of the scalar power spectrum with respect to the momentum scale. This represents the running of the scalar spectral tilt after renormalization. Essentially, in the context of primordial cosmology, this logarithmic triple derivative with respect to the momentum scale computes the further minute scale dependence of the two-point amplitude, which is commonly referred to as the {\it running of the running of spectral-tilt} or {\it running of the running of spectral-index} of the scalar power spectrum. Said another way, this claim may be expressed technically as:
 \bea 
\left[\overline{\beta_{\zeta,\textbf{EFT}}(k_*)}\right]_n=\Bigg[\frac{d}{d\ln k}\bigg(\left[\overline{\alpha_{\zeta,\textbf{EFT}}(k)}\right]_n\bigg)\Bigg]_{k=k_*}=\bigg(\bigg[\beta_{\zeta,\textbf{Tree}}(k_*)\bigg]_{{\bf SR}_1}\bigg).\eea
The three conditions that were previously stated are entirely compatible with the condition that was described above. Rather than providing a running of the spectral tilt, it really provides additional information on the structure of the scalar power spectrum in the current context. While the running of the spectral tilt is calculated in the CMB pivot scale $k_*$, we can further fix the shape of the primordial power spectrum very minutely computed for the scalar modes in terms of concavity or convexity, as previously pointed out in the current version of the EFT computation, because of its non-zero value.

    \end{enumerate}
    The following additional restrictions are placed on the characteristics of the UV and IR sensitive counter-terms as a direct result of the four renormalization requirements previously discussed. These restrictions are listed point-by-point below:
\begin{itemize}
    \item   \underline{\bf Consequence of condition I:}\\
    Following is the constraint condition that expresses the immediate result of the first renormalization condition:
\bea {\cal Z}^{\rm IR}_{n}=  \frac{\small[\overline{\Delta_{\zeta,\textbf{EFT}}^{2}(k_{*})}\small]_{n}}{\small[\Delta_{\zeta,\textbf{EFT}}^{2}(k_{*})\small]_{n}} = \frac{\small[  \Delta_{\zeta,\textbf{Tree}}^{2}(k_{*})\small]_{\textbf{SR}_{1}}}{\small[\Delta_{\zeta,\textbf{EFT}}^{2}(k_{*})\small]_{n}}\quad\quad\quad\Longrightarrow\quad\quad\quad {\cal Z}^{\rm IR}_{n}(k_*){\cal Z}^{\rm UV}_{n}(k_*)=1.\eea
In this case, this connection will be quite useful to explicitly calculate the IR counter-term. It is crucial to remember that the IR counter-term ${\cal Z}^{\rm IR}_{n}$, which we obtain from this relation, is the opposite of the UV counter-term ${\cal Z}^{\rm UV}_{n}$ in the current computation. Therefore, it is also impossible to establish the form of the IR counter-term without fixing the form of the UV counter-term. 

    \item  \underline{\bf Consequence of condition II:}\\
    The following constraint criteria represent the immediate result of the second renormalization condition:
\bea {\cal Z}^{\rm IR}_{n}(k_*){\cal Z}^{\rm UV}_{n}(k_*)=1\quad\quad\quad {\rm and}\quad\quad\quad \bigg(\frac{d{\cal Z}^{\rm UV}_{n}}{d\ln k}\bigg)_{k=k_*}=0.\eea
Upon further inspection, we can see that Condition II is more restricted than Condition I since it already addresses the immediate result of Condition I. Due to an extra restriction arising in terms of the vanishing logarithmic momentum scale dependant derivative computed at the CMB pivot scale $k_*$, this might be useful in figuring out the mathematical structure of the UV counter-term. This result is entirely physically justified in the current computation setting, as the contribution of such words is surprisingly not clearly observable or identifiable on that scale. It is possible to compute the contribution of the IR counter-term and fix the structure of the renormalized scalar spectrum obtained from this EFT configuration immediately after determining the limited structure of the UV counter-term directly.

     \item  \underline{\bf Consequence of condition III:}\\
     The following constraint requirements are the direct result of the third renormalization condition:
 \bea {\cal Z}^{\rm IR}_{n}(k_*){\cal Z}^{\rm UV}_{n}(k_*)=1,\quad\quad\quad \quad\quad\quad \bigg(\frac{d{\cal Z}^{\rm UV}_{n}}{d\ln k}\bigg)_{k=k_*}=0\quad\quad\quad\quad{\rm and}\quad\quad\quad \bigg(\frac{d^2{\cal Z}^{\rm UV}_{n}}{d\ln k^2}\bigg)_{k=k_*}=0.\eea
Compared to the preceding two, it provides more stringent limitations that further limit the behaviour of UV-counter term at the pivot scale, which is reliant on scale.

      \item  \underline{\bf Consequence of condition IV:}\\
      The following constraint criteria represent the direct result of the fourth renormalization condition:
  \bea {\cal Z}^{\rm IR}_{n}(k_*){\cal Z}^{\rm UV}_{n}(k_*)=1,\quad\quad  \bigg(\frac{d{\cal Z}^{\rm UV}_{n}}{d\ln k}\bigg)_{k=k_*}=0\quad\quad\bigg(\frac{d^2{\cal Z}^{\rm UV}_{n}}{d\ln k^2}\bigg)_{k=k_*}=0\quad\quad{\rm and}\quad\quad \bigg(\frac{d^3{\cal Z}^{\rm UV}_{n}}{d\ln k^3}\bigg)_{k=k_*}=0.\quad\eea
It provides even more stringent limitations than the preceding three, severely limiting the UV-counter term at the pivot scale in relation to scale.

    \end{itemize}
    We discovered, upon further analysis of the issue, that the UV counter-term ${\cal Z}^{\rm UV}_{n}$ must be carefully determined in order for it to fulfill the sets of constraint criteria that were previously achieved. This is only achievable if the quadratic divergence contribution is accurately eliminated. The IR counter-term ${\cal Z}^{\rm IR}_{n}$ will have its form automatically fixed after this is accomplished. At this level, however, figuring out the precise form of the UV counter-term ${\cal Z}^{\rm UV}_{n}$ with just the aforementioned limitations at the CMB pivot scale is quite challenging. The primary challenge at the technical level arises from the requirement that counter-terms in ${\bf SR}_1$, ${\bf USR}_n$, and ${\bf SR}_{n+1}$ phases individually eliminate the quadratic divergence contribution. As simple as it may appear, completing the computation we have done so far is rather challenging from a technological standpoint. Herein lies the significance of the following three sections: in each case, the quadratic UV divergences from the ${\bf SR}_1$, ${\bf USR}_n$, and ${\bf SR}_{n+1}$ phases are entirely eliminated thanks to the late time renormalization scheme, or equivalently, the adiabatic renormalization scheme. After completing this, we can use the condition ${\cal Z}^{\rm IR}_{n}(k_*){\cal Z}^{\rm UV}_{n}(k_*)=1$ to quickly ascertain the explicit form of the IR counter-term. This condition, in conjunction with the quadratic divergence free result derived from the late time renormalization scheme, or equivalently the adiabatic renormalization scheme for ${\cal Z}^{\rm UV}_{n}$, has been employed in the context of power spectrum renormalization.

In the next subsections, we will discuss the underlying connection between the power spectrum, adiabatic/wave function, and late time renormalization schemes and the standard approach of adding a counter-term at the level of action, which is utilized in this paper. This will help to clear up any confusion that may arise regarding the renormalization schemes employed or the interconnections among the various tools and techniques used in this paper. We are confident that this kind of explanation will assist readers in comprehending the relevance of the conclusions generated in this work. Point-by-point, let us go over the explanations in depth that are attached below:
\begin{enumerate}
\item[\ding{43}] The counter-term contribution of the third order perturbed action is directly linked to the total eradication of the damaging quadratic UV divergence in the current computation. It is feasible to clearly demonstrate as an immediate result of the computation with this particular section that the sum of the counter-terms $\delta_{{\cal Z}_{{\bf D}_i}}\forall i=1,2,\cdots,6$ for the six operators that were previously described may be represented in terms of a cumulative factor. We characterize this component as the counter-term contribution that, when applied to late-time and wave function/adiabatic renormalization schemes—discussed in the latter part of this review—will entirely eliminate the quadratic UV divergence. Subsections that follow will demonstrate how the cumulative counter-terms in both schemes may be defined in terms of certain combinations of counter-terms that correspond to the six operators described in the third-order perturbed action that are discussed. These links will enable us to eliminate the quadratic UV divergence entirely from the formula for the rectified primordial scalar power spectrum after one loop.

   \item[\ding{43}] However, the counter-term contribution of the third-order perturbed action is directly linked to the coarse grain and smoothing of the logarithmic IR divergence.  One may clearly demonstrate that the single counter term $\delta_{{\cal Z}^{\rm IR}_{n}}$ is the direct result of the calculation using this particular portion. The addition of this factor to the 1PI one-loop corrected two-point amplitude calculation would smooth the logarithmic IR divergence by moving it to a higher order, which is consistent with the higher even loop diagrams that emerge in the perturbative expansion. We shall go into more detail on this topic in the second half of this paper concerning the power spectrum renormalization technique.

\end{enumerate}

\subsection{Renormalized one-loop scalar power spectrum from EFT}

\subsubsection{Adiabatic/Wave function renormalization scheme}

In this section, we primarily want to remove, in a one-loop fashion, the contributions from quadratic UV divergences that manifest in the primordial power spectrum calculation for the comoving curvature perturbation from the SR and USR phases. The sub-horizon region ($-kc_s\tau\gg 1$), where quantum mechanical perturbations play a significant role, is, technically, the source of these UV divergences. The phases previously discussed are described by a non-trivial FLRW backdrop that only has logarithmic divergent contributions left at the late time limit where the scalar modes leave the horizon and reach the super horizon region. Consequently, the quadratic UV divergence may be disregarded in the future. However, it is imperative that we systematically provide a technical foundation that allows us to do so in the current context. This technique does not completely eliminate the impact of quadratic UV divergence at the late time limit. The main goal of this section is to provide the technical computation of a renormalization approach that allows us to automatically eliminate the quadratic divergence contribution by including an appropriate counter-term. Moreover, we have to use a smoothing strategy to eliminate the UV divergences of the underlying cosmological perturbation theory as we have quantized the comoving curvature perturbation and are treating it at the same length of the scalar quantum field. The specific reason for this is because the short-range UV modes appear in the sub-horizon area where quantum fluctuations are accounted for by adiabatic renormalization.

Here, we deal with the Quantum Field Theory of FLRW space-time, where the contributions from the UV divergences seen at different powers are eliminated thanks to the well-known adiabatic regularization and the corresponding renormalization procedure. For a more detailed explanation of the physical implications and potential uses, see references \cite{Durrer:2009ii,Wang:2015zfa,Boyanovsky:2005sh,Marozzi:2011da,Finelli:2007fr,Ford:1977in,Parker:1974qw,Fulling:1974pu,Parker:2012at,Ford:1986sy}. It is straightforward to remove the UV divergence contributions from the underlying Quantum Field Theoretic setup as the sub-horizon quantum fluctuations are recorded by applying the minimal subtraction rule to the relevant UV modes. In this scenario, the minimal subtraction technically corresponds to adding a counter term to the underlying theory in order to remove the UV divergent contribution from the short-range UV modes that are dominant in the sub-horizon area. This method employs a second-order subtraction to sufficiently reduce the effects of quadratic UV divergences, which originate from both the SR and USR periods. Only when considering the contributions of the UV divergent components that appear at higher powers does it become meaningful to eliminate the counter terms beyond the second-order minimum subtraction. Luckily, we don't need to take into account counterterms higher than the second order because the adiabatic renormalization method used in this computation has sufficient power to minimize the quadratic divergence contribution. The purpose of the adiabatic renormalization strategy was to eliminate UV divergences alone in the setting of curved space-time within Quantum Field Theory, especially in FLRW backgrounds. This approach is unable to handle IR divergences. For further information, see references. \cite{Durrer:2009ii,Wang:2015zfa,Boyanovsky:2005sh,Marozzi:2011da,Finelli:2007fr,Ford:1977in,Parker:1974qw,Fulling:1974pu,Parker:2012at,Ford:1986sy}. Notably, the adiabatic renormalization approach is used in this situation to directly renormalize the comoving curvature perturbation modes in the adiabatic limit of the cosmological perturbation.  This is exactly analogous to the renormalization of the wave function that we frequently perform within the framework of quantum field theory. In the present setting, such renormalized modes or the wave function further renormalize the one-loop contribution to the cosmic power spectrum.  Crucially, the WKB approximation strategy is helpful in the present situation in producing a regularized wave function in the adiabatic limit, which is a generalized form of the mode function. Consequently, the regularized version of the mode function may completely remove the contributions of UV divergences from the short-range modes.  The purpose of the generalized structure of the regularized modes is to preserve the universality of the underlying Quantum Field Theoretical setup while enabling the proper handling of UV divergences in any sequence.

A WKB-like adiabatic expansion yields the regularization in adiabatic renormalization by extending the field modes to $m$th order:
\bea 
 \chi_{k}^{(m)} &\equiv& ah_{k}^{(m)} \sim \frac{1}{\sqrt{2 {\cal W}_{k}^{(m)}(\tau)}} \exp{\bigg(-i \int_{\tau_0}^{\tau}{\cal W}_{k}^{(m)}(\tau')d\tau \bigg)}, \\
 \chi_{k}^{(m)^{*}} &\equiv& ah_{k}^{(m)^{*}} \sim \frac{1}{\sqrt{2 {\cal W}_{k}^{(m)}(\tau)}} \exp{\bigg(i \int_{\tau_0}^{\tau}{\cal W}_{k}^{(m)}(\tau')d\tau \bigg)},
\eea
where $a$ is the scale factor and ${\cal W}_{k}^{m}$ is given by:
\bea
{\cal W}_{k}^{(m)} \equiv \omega_{k}^{(0)} + \omega_{k}^{(1)}+\omega_{k}^{(2)}+ \cdots + \omega_{k}^{(m)}.
\eea
The superscripts denote the adiabatic order of the number of time derivatives in $\omega_{k}$. An important thing to note is that the WKB approximation helps with the creation of a regularized wave function over the adiabatic limit in the present scenario. Consequently, the contributions of short-range modes to UV divergence are reduced. We thus consider that the UV divergent contributions appear as $m$-th power in the mode functions in the adiabatic limit. Essentially, the adiabatic renormalization process directly renormalizes the comoving curvature perturbation modes inside the adiabatic limit of cosmic disturbances. First, we should consider that the UV divergences in the adiabatic limiting case are at the $m$-th power, which represents the $m$-th order UV divergences from short-range scalar modes.  In every one of the next SR$_1$, USR$_n$, and SR$_{n+1}$ phases, the comoving scalar curvature perturbation's renormalized mode function might be:
\bea
\label{genWKB}
\zeta_{k, \rm SR_1}^{(m)}(\tau) &=& -\bigg[\frac{c_s H \tau}{2 M_p \sqrt{\epsilon} \sqrt{{\cal W}_{k}^{(m)}(\tau)}}\bigg] \bigg[\alpha_{ k}^{(1)} \exp{\bigg(-i \int_{\tau_{0}}^{\tau} d \tau^{'} {\cal W}_{k}^{(m)}(\tau ^{'})}\bigg) + \beta_{ k}^{(1)} \exp{\bigg(i \int_{\tau_{0}}^{\tau} d \tau^{'} {\cal W}_{k}^{(m)}(\tau ^{'})}\bigg)\bigg], \\
\zeta_{k, \rm USR_{n}}^{(m)}(\tau) &=& -\bigg[\frac{c_s H \tau}{2 M_p \sqrt{\epsilon} \sqrt{{\cal W}_{k}^{(m)}(\tau)}}\bigg] \bigg(\frac{\tau_{0}}{\tau}\bigg)^{3} \bigg[\alpha_{ k}^{(2n)} \exp{\bigg(-i \int_{\tau_{0}}^{\tau}d \tau^{'} {\cal W}_{k}^{(m)}(\tau ^{'})}\bigg) \nonumber \\ 
 && \quad \quad \quad \quad \quad \quad \quad \quad \quad \quad \quad \quad \quad \quad \quad \quad \quad \quad + \beta_{ k}^{(2n)} \exp{\bigg(i \int_{\tau_{0}}^{\tau}d \tau^{'} {\cal W}_{k}^{(m)}(\tau ^{'})}\bigg)\bigg], \\
\zeta_{k, \rm SR_{n+1}}^{(m)}(\tau) &=& -\bigg[\frac{c_s H \tau}{2 M_p \sqrt{\epsilon} \sqrt{{\cal W}_{k}^{(m)}(\tau)}}\bigg] \bigg(\frac{\tau_{0}}{\tau}\bigg)^{3} \bigg[\alpha_{ k}^{(2n+1)} \exp{\bigg(-i \int_{\tau_{0}}^{\tau}d \tau^{'} {\cal W}_{k}^{(m)}(\tau ^{'})}\bigg) \nonumber \\
&& \quad \quad \quad \quad \quad \quad \quad \quad \quad \quad \quad \quad \quad \quad \quad \quad \quad \quad + \beta_{ k}^{(2n+1)} \exp{\bigg(i \int_{\tau_{0}}^{\tau}d \tau^{'} {\cal W}_{k}^{(m)}(\tau ^{'})}\bigg)\bigg]. 
\eea
According to the preceding formulae, m denotes the order of the WKB mode and $n$ the number of sharp transitions. Furthermore, we have defined the generic $m$th order mode function for the SR$_1$ phase without adjusting the Bogoliubov coefficients' related initial Bunch Davies condition. In this case, the features function ${\cal W}_{k}^{(m)}$ stands for the conformal time dependence factor inside adiabatic regularization, which is defined as follows for the $m$th order:
\bea
{\cal W}^{(m)}(\tau) = \sqrt{\bigg(c_s^2 k^2 - \frac{z^{''}}{z}\bigg) - \frac{1}{2}\bigg[\frac{{\cal W}_{k}^{''(m-2)}(\tau)}{{\cal W}_{k}^{m-2}(\tau)} - \frac{3}{2}\bigg(\frac{{\cal W}_{k}^{'(m-2)}(\tau)}{{\cal W}_{k}^{m-2}(\tau)}\bigg)^2\bigg]}, \quad \quad {\rm where} \quad \frac{z''}{z}\approx \frac{2}{\tau ^2}
\eea
The absence of significant alterations to the Bogoliubov coefficients in the USR$_n$ phases can be attributed to the adiabatic limit imposed in the scalar modes. This observation is supported by the validity of adiabatic regularization for the $m$th mode. We shall demonstrate by rigorous computation that the result is not affected by the short-range UV modes and is hence independent of the dynamics of the Bogoliubov coefficients. Suppose that, upon shifting from the original Bunch Davies condition, there are notable modifications to the structure of the underlying vacuum state specified by the Bogoliubov coefficients. It will directly question the usefulness of the well-established adiabatic regularization technique. Here, one may enjoy the elegance of working in the Curved Quasi de-sitter space-time Quantum field theory, which preserves the shape of the underlying shifted quantum vacuum state even in the adiabatic limit of regularization. Keep in mind that these talks apply to SR$_1$, USR$_n$, and SR$_{n+1}$ phases. As previously noted, the UV divergences appear as quadratic terms throughout the SR$_1$, USR$_n$, and SR$_{n+1}$ stages of inflation. Therefore, to eliminate these divergences, a correction of no more than order two would be required.
We thus adjust our explanation to use $m=2$. After considering the aforementioned explanation, the WKB modes under the adiabatic limit are as follows:
\bea
\label{2orderWKB}
\zeta_{ k, \rm SR_1}^{(2)}(\tau) &=& -\bigg[\frac{c_s H \tau}{2 M_p \sqrt{\epsilon} \sqrt{{\cal W}_{k}^{(2)}(\tau)}}\bigg] \bigg[\alpha_{ k}^{(1)} \exp{\bigg(-i \int_{\tau_{0}}^{\tau} d \tau^{'} {\cal W}_{k}^{(2)}(\tau ^{'})}\bigg) + \beta_{ k}^{(1)} \exp{\bigg(i \int_{\tau_{0}}^{\tau} d \tau^{'} {\cal W}_{k}^{(2)}(\tau ^{'})}\bigg)\bigg], \\
\zeta_{ k, \rm USR_{n}}^{(2)}(\tau) &=& -\bigg[\frac{c_s H \tau}{2 M_p \sqrt{\epsilon} \sqrt{{\cal W}_{k}^{(2)}(\tau)}}\bigg] \bigg(\frac{\tau_{0}}{\tau}\bigg)^{3} \bigg[\alpha_{ k}^{(2n)} \exp{\bigg(-i \int_{\tau_{0}}^{\tau}d \tau^{'} {\cal W}_{k}^{(2)}(\tau ^{'})}\bigg) \nonumber \\ 
 && \quad \quad \quad \quad \quad \quad \quad \quad \quad \quad \quad \quad \quad \quad \quad \quad \quad \quad + \beta_{ k}^{(2n)} \exp{\bigg(i \int_{\tau_{0}}^{\tau}d \tau^{'} {\cal W}_{k}^{(2)}(\tau ^{'})}\bigg)\bigg], \\
\zeta_{ k, \rm SR_{n+1}}^{(2)}(\tau) &=& -\bigg[\frac{c_s H \tau}{2 M_p \sqrt{\epsilon} \sqrt{{\cal W}_{k}^{(2)}(\tau)}}\bigg] \bigg(\frac{\tau_{0}}{\tau_e}\bigg)^{3} \bigg[\alpha_{ k}^{(2n+1)} \exp{\bigg(-i \int_{\tau_{0}}^{\tau}d \tau^{'} {\cal W}_{k}^{(2)}(\tau ^{'})}\bigg) \nonumber \\
&& \quad \quad \quad \quad \quad \quad \quad \quad \quad \quad \quad \quad \quad \quad \quad \quad \quad \quad + \beta_{ k}^{(2n+1)} \exp{\bigg(i \int_{\tau_{0}}^{\tau}d \tau^{'} {\cal W}_{k}^{(2)}(\tau ^{'})}\bigg)\bigg]. 
\eea
The second-order characteristic function in this case has the following form:
\bea
{\cal W}_{k}^{(2)} (\tau) = \sqrt{c_s ^2k^2 - \frac{2}{\tau ^2}} \approx c_s k.
\eea
The contribution from the short-range UV modes alone is expressed in the preceding formula using the restriction $-kc_s \tau \rightarrow \infty$. As a consequence of this finding, the second-order comoving curvature perturbation originated WKB mode functions are further reduced to:
\bea
\label{simplewkb2}
\zeta_{ k, \rm SR_1}^{(2)}(\tau) &\approx & -\bigg[\frac{c_s H \tau}{2 M_p \sqrt{\epsilon} \sqrt{c_sk}}\bigg] \bigg[\alpha_{ k}^{(1)} \exp{\bigg(-ic_sk(\tau - \tau_{0})\bigg)} + \beta_{ k}^{(1)} \exp{\bigg(ic_sk(\tau - \tau_{0})\bigg)} \bigg]\\
\zeta_{ k, \rm USR_n}^{(2)}(\tau) &\approx & -\bigg[\frac{c_{s}H \tau}{2 M_p \sqrt{\epsilon} \sqrt{c_sk}}\bigg] \bigg(\frac{\tau_{0}}{\tau}\bigg)^{3} \bigg[\alpha_{ k}^{(2n)}\exp{\bigg(-ic_{s}k(\tau - \tau_{0})\bigg)} + \beta_{ k}^{(2n)} \exp{\bigg(ic_sk(\tau - \tau_{0})\bigg)} \bigg], \\
\zeta_{ k, \rm SR_{n+1}}^{(2)}(\tau) &\approx& -\bigg[\frac{c_s H \tau}{2 M_p \sqrt{\epsilon} \sqrt{c_sk}}\bigg] \bigg(\frac{\tau_{0}}{\tau_e}\bigg)^{3} 
\bigg[\alpha_{ k}^{(2n+1)} \exp{\bigg(-ic_sk(\tau - \tau_{0})\bigg)} + \beta_{ k}^{(2n+1)} \exp{\bigg(ic_sk(\tau - \tau_{0})\bigg)} \bigg], 
\eea
The computation of the formula for the counter terms of these inflationary periods is the primary task of this approach, which we now go to. The counter terms $c_{\rm SR_{1}}(\mu, \mu_0)$, and $c_{\rm USR_{n}}(\mu, \mu_0)$ and $c_{\rm SR_{n+1}}(\mu, \mu_0)$ for the corresponding inflation periods SR$_{1}$, USR$_{n}$, and SR$_{n+1}$ are introduced. These counter-terms are dependent on the adiabatic renormalization scheme-dependent parameters, as previously observed.
\bea
\label{count1}
{\cal Z}_{\bf \zeta, \rm SR_{1}}^{\rm UV} (\mu, \mu_0) &=& \bigg[\Delta^{2}_{\zeta,\textbf{Tree}}(k)\bigg]_{\textbf{SR}_{1}}^{2} \times c_{\rm SR_{1}}(\mu, \mu_0), \\
\label{count2}
{\cal Z}_{\bf \zeta, \rm USR_{n}}^{\rm UV} (\mu, \mu_0) &=& \frac{1}{4}\bigg[\Delta^{2}_{\zeta,\textbf{Tree}}(k)\bigg]_{\textbf{SR}_{1}}^{2} \times c_{\rm USR_{n}}(\mu, \mu_0), \\
\label{count3}
{\cal Z}_{\bf \zeta, \rm SR_{n+1}}^{\rm UV} (\mu, \mu_0) &=& \bigg[\Delta^{2}_{\zeta,\textbf{Tree}}(k)\bigg]_{\textbf{SR}_{1}}^{2} \times c_{\rm SR_{n+1}}(\mu, \mu_0). 
\eea
In the quasi-de sitter background, $\mu$ denotes the renormalization scale of the underlying quantum field theory. In order to carry out the adiabatic regularization technique, $\mu_0$ also denotes the renormalization scale at the conformal time. As long as it makes sense in the given situation, one may use these scales as convenient. Now let us examine the parameters that are reliant on the renormalization technique explicitly:
\bea
c_{\rm SR_{1}}(\mu, \mu_0) &=& \int_{\mu_0}^{\mu}dk\;k^2 c_s^2 \frac{\tau ^2}{k} = \frac{c_s^2}{2}(\mu ^2 - \mu_0 ^2 ) \tau ^2 = \frac{1}{2}\bigg[\bigg(\frac{\mu}{\mu_0}\bigg)^2 - 1 \bigg], \\
c_{\rm USR_{n}}(\mu, \mu_0) &=& \bigg[\left(\frac{\Delta\eta(\tau_{0})}{\tilde{c}^{4}_{s}}\right)^{2}\bigg(\frac{\mu}{\mu_0}\bigg)^6  - \left(\frac{\Delta\eta(\tau_{0})}{\tilde{c}^{4}_{s}}\right)^{2}\bigg] \int_{\mu_0}^{\mu} dk \; c_s ^2 k^2 \frac{\tau ^2}{k} \nonumber \\
&=& \frac{1}{2} \bigg[\left(\frac{\Delta\eta(\tau_{0})}{\tilde{c}^{4}_{s}}\right)^{2}\bigg(\frac{\mu}{\mu_0}\bigg)^6  - \left(\frac{\Delta\eta(\tau_{0})}{\tilde{c}^{4}_{s}}\right)^{2} \bigg]\bigg[\bigg(\frac{\mu}{\mu_0}\bigg)^2 - 1\bigg],\\
c_{\rm SR_{n+1}}(\mu, \mu_0) &=& \int_{\mu_0}^{\mu} dk \; k^2 c_s ^2 \frac{\tau ^2}{k}= \frac{1}{2}\bigg[\bigg(\frac{\mu}{\mu_0}\bigg)^2 - 1 \bigg].
\eea
The relation $-c_s \tau = 1/\mu_0$ was employed in the aforementioned integrals. By replacing the obtained results with the counter-terms of equations (\ref{count1}), (\ref{count2}), and (\ref{count3}), we may derive the following expressions:
\bea
\label{revcount1}
{\cal Z}_{\bf \zeta, \rm SR_{1}}^{\rm UV} (\mu, \mu_0) &=& \bigg[\Delta^{2}_{\zeta,\textbf{Tree}}(k)\bigg]_{\textbf{SR}_{1}}^{2} \times \frac{1}{2}\bigg[\bigg(\frac{\mu}{\mu_0}\bigg)^2 - 1 \bigg], \\
\label{revcount2}
{\cal Z}_{\bf \zeta, \rm USR_{n}}^{\rm UV} (\mu, \mu_0) &=& \frac{1}{8}\bigg[\Delta^{2}_{\zeta,\textbf{Tree}}(k)\bigg]_{\textbf{SR}_{1}}^{2} \times \bigg[\left(\frac{\Delta\eta(\tau_{0})}{\tilde{c}^{4}_{s}}\right)^{2}\bigg(\frac{\mu}{\mu_0}\bigg)^6  - \left(\frac{\Delta\eta(\tau_{0})}{\tilde{c}^{4}_{s}}\right)^{2} \bigg]\bigg[\bigg(\frac{\mu}{\mu_0}\bigg)^2 - 1\bigg], \\
\label{revcount3}
{\cal Z}_{\bf \zeta, \rm SR_{n+1}}^{\rm UV} (\mu, \mu_0) &=& \bigg[\Delta^{2}_{\zeta,\textbf{Tree}}(k)\bigg]_{\textbf{SR}_{1}}^{2} \times \frac{1}{2}\bigg[\bigg\{\bigg(\frac{\mu}{\mu_0}\bigg)^2 - 1 \bigg\}  \bigg].
\eea    
Thus, the adiabatically renormalized total one-loop corrected scalar power spectrum may be expressed as follows:
\bea
\label{adiasr1}
\bigg[\Delta^{2}_{\zeta,\textbf{One-loop}}(k,\mu,\mu_0)\bigg]_{\textbf{SR}_{1}} &=& \bigg[\Delta^{2}_{\zeta,\textbf{Tree}}(k)\bigg]_{\textbf{SR}_{1}}^{2}
\times \bigg[ \left(1-\frac{2}{15\pi^{2}}\frac{1}{c_{s}^{2}k_{*}^{2}}\left(1-\frac{1}{c_{s}^{2}}\right)\epsilon\right) \nonumber \\
&& \quad \quad \quad \quad \quad \bigg(\frac{1}{2}\bigg\{ \bigg(\frac{\mu}{\mu_0}\bigg)^2 - \bigg(\frac{k_{s_1}}{k_{*}}\bigg)^2 \bigg\} - \frac{4}{3} \ln\bigg({\frac{k_{s_1}}{k_{*}}}\bigg) \bigg)\bigg], \\
\label{adiausrn}
\bigg[\Delta^{2}_{\zeta,\textbf{One-loop}}(k,\mu,\mu_0)\bigg]_{\textbf{USR}_{n}} &=& \frac{1}{4}\bigg[\Delta^{2}_{\zeta,\textbf{Tree}}(k)\bigg]_{\textbf{SR}_{1}}^{2} \nonumber \\
&& \quad \times \bigg\{\bigg[\bigg(\frac{\Delta\eta(\tau_{e_{n}})}{\tilde{c}^{4}_{s}}\bigg)^{2}\bigg(\frac{k_{e_n}}{k_{s_n}}\bigg)^6  - \left(\frac{\Delta\eta(\tau_{s_{n}})}{\tilde{c}^{4}_{s}}\right)^{2}\bigg]
\bigg[\ln\bigg({\frac{k_{e_n}}{k_{s_n}}}\bigg) + \frac{1}{2}\bigg\{ \bigg(\frac{k_{e_n}}{k_{s_n}}\bigg)^2 - 1 \bigg\}\bigg] \nonumber \\
&& \quad \quad \quad  - \frac{1}{2}\bigg[\bigg(\frac{\Delta\eta(\tau_{e_{n}})}{\tilde{c}^{4}_{s}}\bigg)^{2} \bigg(\frac{\mu}{\mu_0}\bigg)^6 - \left(\frac{\Delta\eta(\tau_{s_{n}})}{\tilde{c}^{4}_{s}}\right)^{2}\bigg] \bigg[\bigg(\frac{\mu}{\mu_0}\bigg)^2 -1\bigg] \bigg\}, \\
\label{adisrnrest}
\bigg[\Delta^{2}_{\zeta,\textbf{One-loop}}(k,\mu,\mu_0)\bigg]_{\textbf{SR}_{n+1}} &=& \bigg[\Delta^{2}_{\zeta,\textbf{Tree}}(k)\bigg]_{\textbf{SR}_{1}}^{2} \times \bigg[ \left(1-\frac{2}{15\pi^{2}}\frac{1}{c_{s}^{2}k_{*}^{2}}\left(1-\frac{1}{c_{s}^{2}}\right)\epsilon\right) \bigg(\frac{1}{2}\bigg[\bigg(\frac{\mu}{\mu_0}\bigg)^2 - 1 \bigg] \nonumber \\ 
&& \quad \quad \quad \quad \quad  + \bigg(\ln{\bigg(\frac{k_{s_{n+1}}}{k_{e_n}}}\bigg)\bigg)\bigg) \bigg].
\eea
Given a fixed renormalization scale denoted by $\mu$ and a matching reference scale $\mu_0$, the UV divergences will be eliminated, as can be shown from the preceding formulas for the scalar power spectrum of the comoving curvature perturbation. But it is still unclear what will happen to the IR divergences. This leads to a curved space-time quantum field theory that is sensitive to infrared radiation but shielded from ultraviolet light. Perturbative approximations are validated by this kind of IR nature, which is why they should always be followed. If the renormalization scale is fixed in the vicinity of the UV cut-off, then you may always assume some arbitrariness in it, but the perturbativity is preserved throughout time. As a result, we would like to remind you of the cut-off scales that were initially selected in order to evaluate the integrals and show the renormalization scales that we have chosen. 
\begin{enumerate}
    \item \textbf{\underline{Catalogue of Cut-off scales:}} \\
    \begin{align*}
\mathbf{\underline{For\;SR_1}:} & \quad \Lambda_{\rm UV} = k_{s_1}, \quad  \Lambda_{\rm IR} = k_{*}, \\
\mathbf{\underline{For\;USR_n}:} & \quad \Lambda_{\rm UV} = k_{e_n}, \quad  \Lambda_{\rm IR} = k_{s_n}, \\
\mathbf{\underline{For\;SR_{n+1}}:} & \quad \Lambda_{\rm UV} = k_{s_{n+1}}, \quad \Lambda_{\rm IR} = k_{e_n}.
\end{align*}

\item \textbf{\underline{Catalogue of Renormalization scales and necessary constraints:}} \\
\begin{align*}
\mathbf{\underline{For\;SR_1}:} & \quad \mu = k_{s_1} = \Lambda_{\rm UV}^{\rm SR_1}, \quad  \mu_0 = k_{*} = \Lambda_{\rm IR}^{\rm SR_1}, \\
\mathbf{\underline{For\;USR_n}:} & \quad \mu = k_{e_n} =  \Lambda_{\rm UV}^{\rm USR_n}, \quad  \mu_0 = k_{s_n} =  \Lambda_{\rm IR}^{\rm USR_n}, \\
\mathbf{\underline{For\;SR_{n+1}}:} & \quad \mu = \mu_0 = k_{\rm end}.
\end{align*}
\end{enumerate}
Following the application of the aforementioned renormalization scales, the following equations for eqns. (\ref{adiasr1}), (\ref{adiausrn}), and (\ref{adisrnrest}) are obtained: 
\bea
\bigg[\Delta^{2}_{\zeta,\textbf{One-loop}}(k)\bigg]_{\textbf{SR}_{1}} &=& \bigg[\Delta^{2}_{\zeta,\textbf{Tree}}(k)\bigg]_{\textbf{SR}_{1}}^{2}
\times \bigg[ \left(1-\frac{2}{15\pi^{2}}\frac{1}{c_{s}^{2}k_{*}^{2}}\left(1-\frac{1}{c_{s}^{2}}\right)\epsilon\right) \bigg( - \frac{4}{3} \ln\bigg({\frac{k_{s_1}}{k_{*}}}\bigg) \bigg)\bigg], \\
\bigg[\Delta^{2}_{\zeta,\textbf{One-loop}}(k)\bigg]_{\textbf{USR}_{n}}
&=& \frac{1}{4}\bigg[\Delta^{2}_{\zeta,\textbf{Tree}}(k)\bigg]_{\textbf{SR}_{1}}^{2} \bigg[\bigg(\frac{\Delta\eta(\tau_{e_{n}})}{\tilde{c}^{4}_{s}}\bigg)^{2}\bigg(\frac{k_{e_n}}{k_{s_n}}\bigg)^6  - \left(\frac{\Delta\eta(\tau_{s_{n}})}{\tilde{c}^{4}_{s}}\right)^{2}\bigg] \bigg(\ln{\frac{k_{e_n}}{k_{s_n}}}\bigg), \\
\bigg[\Delta^{2}_{\zeta,\textbf{One-loop}}(k)\bigg]_{\textbf{SR}_{n+1}} 
&=& \bigg[\Delta^{2}_{\zeta,\textbf{Tree}}(k)\bigg]_{\textbf{SR}_{1}}^{2} \times \bigg[ \left(1-\frac{2}{15\pi^{2}}\frac{1}{c_{s}^{2}k_{*}^{2}}\left(1-\frac{1}{c_{s}^{2}}\right)\epsilon\right) \bigg( \ln{\bigg(\frac{k_{s_{n+1}}}{k_{e_n}}}\bigg)\bigg) \bigg].
\eea
As a result, under these renormalization scales, the entire one-loop corrected adiabatic renormalized power spectrum goes toward:
\bea
\label{adiapowspec}
\Delta^{2}_{\zeta,\textbf{Total}}(k) &=& \bigg[\Delta^{2}_{\zeta,\textbf{Tree}}(k)\bigg]_{\textbf{SR}_{1}}
\; \bigg\{ 1+ \bigg[\Delta^{2}_{\zeta,\textbf{Tree}}(k)\bigg]_{\textbf{SR}_{1}} \bigg [ \left(1-\frac{2}{15\pi^{2}}\frac{1}{c_{s}^{2}k_{*}^{2}}\left(1-\frac{1}{c_{s}^{2}}\right)\epsilon\right)\times \left(-\frac{4}{3}\ln{\frac{k_{s_1}}{k_{*}}}\right)\bigg] \nonumber \\
&& \quad \quad + \frac{1}{4} \bigg[\Delta^{2}_{\zeta,\textbf{Tree}}(k)\bigg]_{\textbf{SR}_{1}} \times \sum^{N}_{n=1}\bigg(
     \bigg[\bigg(\frac{\Delta\eta(\tau_{e_{n}})}{\tilde{c}^{4}_{s}}\bigg)^{2}\ln{\bigg(\frac{k_{e_n}}{k_{s_n}}\bigg)^6} - \left(\frac{\Delta\eta(\tau_{s_{n}})}{\tilde{c}^{4}_{s}}\right)^{2}\bigg] \ln{\bigg(\frac{k_{e_n}}{k_{s_n}}\bigg)} \bigg)\nonumber \\
     && \quad \quad \quad +  
\bigg[\Delta^{2}_{\zeta,\textbf{Tree}}(k)\bigg]_{\textbf{SR}_{1}}\left(1-\frac{2}{15\pi^{2}}\frac{1}{c_{s}^{2}k_{*}^{2}}\left(1-\frac{1}{c_{s}^{2}}\right)\epsilon\right)
\times \sum^{N}_{n=1} 
\bigg(\ln{\bigg(\frac{k_{s_{n+1}}}{k_{e_n}}}\bigg)\bigg)\bigg\}.
\eea
The aforementioned findings make it clear that the logarithmic IR divergences have maintained the technique, which we will smooth out in our upcoming article on power spectrum renormalization, while the quadratic UV divergences have been eliminated. The order of its amplitude is partially determined by the common factor that emerges from the entire expression for the overall power spectrum. This is SR$_1$'s tree-level contribution, which is stated as follows:
\bea
 \bigg[\Delta^{2}_{\zeta,{\bf Tree}}(k)\bigg]_{\textbf{SR}_{1}} = \displaystyle{ \bigg[\frac{H^2}{8\pi ^2 M_{p}^2 \epsilon c_s}\bigg]_{*}}\left[1+(k/k_{s_{1}})^{2}\right] \xrightarrow{\rm Super-horizon \; scale \; k\ll k_{s_1}} \displaystyle{ \bigg[\frac{H^2}{8\pi ^2 M_{p}^2 \epsilon c_s}\bigg]_{*}}.
\eea
It is important to note that the ultimate outcome of the late-time renormalization scheme, as found in eqn. (\ref{late-tpowspec}), and the complete one loop corrected renormalized power spectrum produced by adiabatic renormalization in eqn. (\ref{adiapowspec}), are identical. This implies that the theory ought to be reliant on the renormalization scheme. In the talks that follow, we will speculatively address this issue as well as a few alternative renormalization approaches.  But first, in order to be consistent and convenient, we need to create certain quantities.
\bea
\label{Utot}
U&=& U_{\textbf{SR}_{1}} + U_{\textbf{SR}_{\textbf{rest}}},\\
\label{Usr1}
U_{\textbf{SR}_{1}} &=&  -\frac{4}{3}\bigg[\Delta^{2}_{\zeta,\textbf{Tree}}(k)\bigg]_{\textbf{SR}_{1}}^{2}\times\left(1-\frac{2}{15\pi^{2}}\frac{1}{c_{s}^{2}k_{*}^{2}}\left(1-\frac{1}{c_{s}^{2}}\right)\epsilon\right)\ln\left(\frac{k_{s_1}}{k_{*}}\right),
\\
\label{Usrrest}
U_{\textbf{SR}_{\textbf{rest}}} &=& \sum^{N}_{n=1}U_{n} = \bigg[\Delta^{2}_{\zeta,\textbf{Tree}}(k)\bigg]_{\textbf{SR}_{1}}^{2}\times\sum^{N}_{n=1}\left(1-\frac{2}{15\pi^{2}}\frac{1}{c_{s}^{2}k_{*}^{2}}\left(1-\frac{1}{c_{s}^{2}}\right)\epsilon\right)\ln\left(\frac{k_{s_{n+1}}}{k_{e_{n}}}\right),
\\
\label{V}
V &=& \sum^{N}_{n=1}V_{n} = \frac{1}{4}\bigg[\Delta^{2}_{\zeta,\textbf{Tree}}(k)\bigg]_{\textbf{SR}_{1}}^{2}\times\sum^{N}_{n=1}\bigg(\frac{\left(\Delta\eta(\tau_{e_{n}})\right)^{2}}{\tilde{c}^{8}_{s}}\left(\frac{k_{e_{n}}}{k_{s_{n}}}\right)^{6} - \frac{\left(\Delta\eta(\tau_{s_{n}})\right)^{2}}{\tilde{c}^{8}_{s}}\bigg )\ln\left(\frac{k_{e_{n}}}{k_{s_{n}}}\right).
\eea
The sum loop contributions from the SR$_1$ and SR$_{n+1}$ inflation periods are represented here by $U$, while the sum loop contributions from the USR$_{n}$ inflation periods are represented by $V$. These above formulae have been developed for any generic $n$ number of sharp transitions, which in our instance we have taken to be $n=6$.  Consequently, the entire one-loop corrected scalar power spectrum may now be rewritten as follows:
\bea
 \label{unrenormps}
\Delta^{2}_{\zeta, {\textbf{EFT}}}(k) = \bigg[\Delta^{2}_{\zeta, {\textbf{Tree}}}(k)\bigg]_{\textbf{SR}_{1}}\bigg\{1+U+V\bigg\}.
\eea
The contributions to the entire one-loop corrected scalar power spectrum from eqn. (\ref{unrenormps}) are shown diagrammatically below:
\begin{equation}
\begin{tikzpicture}[baseline={([yshift=-3.5ex]current bounding box.center)},very thick]
  
  \def\radius{1}
  \scalebox{1}{\draw[cyan,very thick] (0,\radius) circle (\radius);
  \draw[cyan,very thick] (4.5*\radius,0) circle (\radius);}

 \draw[black, very thick] (-4*\radius,0) -- (-2.5*\radius,0);
 \node at (-1.7*\radius,0){+};
  \draw[black, very thick] (-1*\radius,0) -- (0,0);
  \draw[blue,fill=blue] (0,0) circle (.5ex);
  \draw[black, very thick] (0,0)  -- (1*\radius,0);
  \node at (2*\radius,0) {+};
  \draw[black, very thick] (2.5*\radius,0) -- (3.5*\radius,0); 
  \draw[blue,fill=blue] (3.5*\radius,0) circle (.5ex);
  \draw[blue,fill=blue] (5.5*\radius,0) circle (.5ex);
  \draw[black, very thick] (5.5*\radius,0) -- (6.5*\radius,0);
  
\end{tikzpicture}  \quad = \quad {\rm Adiabatically \,\,renormalized} \,\,\Delta^{2}_{\zeta, \textbf{EFT}}(k),
\end{equation}
where the diagrams show the contribution of only one loop to the number $\Delta^{2}_{\zeta, \textbf{EFT}}(k)$ that is derived using the adiabatic renormalization procedure.

Through this computation, we have also discovered some significant information about the relationship between the counter-terms that appear in this context and those that appear in the context of renormalization of the bare action as appearing in the context of Quantum Field Theory in quasi de Sitter space with the gauge invariant coming curvature perturbation. This relationship is represented by the following information:
\bea &&\underline{{\bf SR}_1:}\quad\quad\sum^{6}_{i=1}\delta_{{\cal Z}_{{\bf D}_i}}={\cal Z}_{\bf \zeta, \rm SR_{1}}^{\rm UV}=\left(1+\delta_{{\cal Z}_{\bf \zeta, \rm SR_{1}}^{\rm UV}}\right)\quad{\rm with}\quad\delta_{{\cal Z}_{{\bf D}_6}}=0,\\
&&\underline{{\bf USR}_n:}\quad\quad\delta_{{\cal Z}_{{\bf D}_6}}={\cal Z}_{\bf \zeta, \rm USR_{n}}^{\rm UV}=\left(1+\delta_{{\cal Z}_{\bf \zeta, \rm USR_{n}}^{\rm UV}}\right)\quad {\rm with}\quad \sum^{6}_{i=1}\delta_{{\cal Z}_{{\bf D}_i}}=0,\\
&&\underline{{\bf SR}_{n+1}:}\quad\quad\sum^{6}_{i=1}\delta_{{\cal Z}_{{\bf D}_i}}={\cal Z}_{\bf \zeta, \rm SR_{n+1}}^{\rm UV}=\left(1+\delta_{{\cal Z}_{\bf \zeta, \rm SR_{n+1}}^{\rm UV}}\right)\quad{\rm with}\quad\delta_{{\cal Z}_{{\bf D}_6}}=0.\eea
This yields the following condensed relations:
\bea &&\underline{{\bf SR}_1:}\quad\quad\left(\delta_{{\cal Z}_{{\bf D}_1}}+\delta_{{\cal Z}_{{\bf D}_2}}+\delta_{{\cal Z}_{{\bf D}_3}}+\delta_{{\cal Z}_{{\bf D}_4}}+\delta_{{\cal Z}_{{\bf D}_5}}-1\right)=\delta_{{\cal Z}_{\bf \zeta, \rm SR_{1}}^{\rm UV}},\\
&&\underline{{\bf USR}_n:}\quad\quad\left(\delta_{{\cal Z}_{{\bf D}_6}}-1\right)=\delta_{{\cal Z}_{\bf \zeta, \rm USR_{n}}^{\rm UV}},\\
&&\underline{{\bf SR}_{n+1}:}\quad\quad\left(\delta_{{\cal Z}_{{\bf D}_1}}+\delta_{{\cal Z}_{{\bf D}_2}}+\delta_{{\cal Z}_{{\bf D}_3}}+\delta_{{\cal Z}_{{\bf D}_4}}+\delta_{{\cal Z}_{{\bf D}_5}}-1\right)=\delta_{{\cal Z}_{\bf \zeta, \rm SR_{n+1}}^{\rm UV}}.\eea
Here, we have the following outcomes for the subsequent phases:
\bea \underline{{\bf SR}_1:}\quad\quad {\cal Z}_{\bf \zeta, \rm SR_{1}}^{\rm UV}&=&\left(1+\delta_{{\cal Z}_{\bf \zeta, \rm SR_{1}}^{\rm UV}}\right)\nonumber\\
&=&\left(\delta_{{\cal Z}_{{\bf D}_1}}+\delta_{{\cal Z}_{{\bf D}_2}}+\delta_{{\cal Z}_{{\bf D}_3}}+\delta_{{\cal Z}_{{\bf D}_4}}+\delta_{{\cal Z}_{{\bf D}_5}}\right)\nonumber\\
&=&\left(1-\frac{2}{15\pi^{2}}\frac{1}{c_{s}^{2}k_{*}^{2}}\left(1-\frac{1}{c_{s}^{2}}\right)\epsilon\right)\bigg[\Delta^{2}_{\zeta,\textbf{Tree}}(k)\bigg]^2_{\textbf{SR}_{1}}\times\Bigg\{\frac{1}{2}\bigg[\bigg(\frac{\mu}{\mu_0}\bigg)^2 - 1 \bigg]\Bigg\}_{\mu=k_{s_1},\mu_0=k_*}
\nonumber\\
&=&\left(1-\frac{2}{15\pi^{2}}\frac{1}{c_{s}^{2}k_{*}^{2}}\left(1-\frac{1}{c_{s}^{2}}\right)\epsilon\right)\bigg[\Delta^{2}_{\zeta,\textbf{Tree}}(k)\bigg]^2_{\textbf{SR}_{1}}\times\Bigg\{\frac{1}{2}\bigg[\bigg(\frac{k_{s_1}}{k_*}\bigg)^2 - 1 \bigg]\Bigg\},\\
\underline{{\bf USR}_n:}\quad\quad {\cal Z}_{\bf \zeta, \rm USR_{n}}^{\rm UV}&=&\left(1+\delta_{{\cal Z}_{\bf \zeta, \rm USR_{n}}^{\rm UV}}\right)\nonumber\\
&=&\delta_{{\cal Z}_{{\bf D}_6}}\nonumber\\
&=&\frac{1}{4}\bigg[\bigg(\frac{\Delta\eta(\tau_{e_{n}})}{\tilde{c}^{4}_{s}}\bigg)^{2}\ln{\bigg(\frac{k_{e_n}}{k_{s_n}}\bigg)^6} - \left(\frac{\Delta\eta(\tau_{s_{n}})}{\tilde{c}^{4}_{s}}\right)^{2}\bigg]\bigg[\Delta^{2}_{\zeta,\textbf{Tree}}(k)\bigg]^2_{\textbf{SR}_{1}}\times\bigg\{
\frac{1}{2}\bigg[\bigg(\frac{\mu}{\mu_0}\bigg)^2 - 1 \bigg]\Bigg\}_{\mu=k_{e_n},\mu_0=k_{s_n}}\nonumber\\
&=&\frac{1}{4}\bigg[\bigg(\frac{\Delta\eta(\tau_{e_{n}})}{\tilde{c}^{4}_{s}}\bigg)^{2}\ln{\bigg(\frac{k_{e_n}}{k_{s_n}}\bigg)^6} - \left(\frac{\Delta\eta(\tau_{s_{n}})}{\tilde{c}^{4}_{s}}\right)^{2}\bigg]\bigg[\Delta^{2}_{\zeta,\textbf{Tree}}(k)\bigg]^2_{\textbf{SR}_{1}}\times\bigg\{
\frac{1}{2}\bigg[\bigg(\frac{k_{e_n}}{k_{s_n}}\bigg)^2 - 1 \bigg]\Bigg\},\quad\quad\quad\eea\bea
\underline{{\bf SR}_{n+1}:}\quad\quad {\cal Z}_{\bf \zeta, \rm SR_{n+1}}^{\rm UV}&=&\left(1+\delta_{{\cal Z}_{\bf \zeta, \rm SR_{n+1}}^{\rm UV}}\right)\nonumber\\
&=&\left(\delta_{{\cal Z}_{{\bf D}_1}}+\delta_{{\cal Z}_{{\bf D}_2}}+\delta_{{\cal Z}_{{\bf D}_3}}+\delta_{{\cal Z}_{{\bf D}_4}}+\delta_{{\cal Z}_{{\bf D}_5}}\right)\nonumber\\
&=&\left(1-\frac{2}{15\pi^{2}}\frac{1}{c_{s}^{2}k_{*}^{2}}\left(1-\frac{1}{c_{s}^{2}}\right)\epsilon\right)\bigg[\Delta^{2}_{\zeta,\textbf{Tree}}(k)\bigg]^2_{\textbf{SR}_{1}}\times\Bigg\{\frac{1}{2}\bigg[\bigg(\frac{\mu}{\mu_0}\bigg)^2 - 1 \bigg]\Bigg\}_{\mu=\mu_0=k_{\rm end}}\nonumber\\
&=&0,\eea 
Thus, we obtain the following formula for the UV divergence-free two-point amplitude of the power spectrum: 
\bea
\left[\Delta_{\zeta,\textbf{EFT}}^{2}(k)\right]_n &=& {\cal Z}^{\rm UV}_{n}\times \bigg[\Delta^{2}_{\zeta,\textbf{Tree}}(k)\bigg]_{\textbf{SR}_{1}},\eea
where the following formula provides the value of the factor ${\cal Z}^{\rm UV}_{n}$ for the $n$ th sharp transition:
\bea {\cal Z}^{\rm UV}_{n}&=&\left(1+\delta_{{\cal Z}^{\rm UV}_{n}}\right),\\
&=&\bigg\{ 1+ \bigg[\Delta^{2}_{\zeta,\textbf{Tree}}(k)\bigg]_{\textbf{SR}_{1}} \bigg [ \left(1-\frac{2}{15\pi^{2}}\frac{1}{c_{s}^{2}k_{*}^{2}}\left(1-\frac{1}{c_{s}^{2}}\right)\epsilon\right)\times \left(-\frac{4}{3}\ln{\frac{k_{s_1}}{k_{*}}}\right)\bigg] \nonumber \\
&& \quad \quad + \frac{1}{4} \bigg[\Delta^{2}_{\zeta,\textbf{Tree}}(k)\bigg]_{\textbf{SR}_{1}} \times \bigg(
     \bigg[\bigg(\frac{\Delta\eta(\tau_{e_{n}})}{\tilde{c}^{4}_{s}}\bigg)^{2}\ln{\bigg(\frac{k_{e_n}}{k_{s_n}}\bigg)^6} - \left(\frac{\Delta\eta(\tau_{s_{n}})}{\tilde{c}^{4}_{s}}\right)^{2}\bigg] \ln{\bigg(\frac{k_{e_n}}{k_{s_n}}\bigg)} \bigg)\nonumber \\
     && \quad \quad  +  
\bigg[\Delta^{2}_{\zeta,\textbf{Tree}}(k)\bigg]_{\textbf{SR}_{1}}\left(1-\frac{2}{15\pi^{2}}\frac{1}{c_{s}^{2}k_{*}^{2}}\left(1-\frac{1}{c_{s}^{2}}\right)\epsilon\right)
\times 
\bigg(\ln{\bigg(\frac{k_{s_{n+1}}}{k_{e_n}}}\bigg)\bigg)\bigg\}. \eea
It also establishes that the quadratic UV divergence can be eliminated entirely from the expression for ${\cal Z}^{\rm UV}_{n}$. This identification aids in our comprehension of the relationship that exists between the counter-terms that appear in the bare action (UV sensitive part) and current contexts. Given that the relationship has been established, this identification will be much more useful in bridging the gap between the current adiabatic/wave function scheme and the common renormalization method that may be used in the context of quantum field theory of quasi de Sitter space. Above all, the adiabatic/wave function approach aids in precisely fixing the mathematical structure of the number ${\cal Z}^{\rm UV}_{n}$, within which the post-renormalization total power spectrum for the scalar modes is now found. The only issue is that, given the previously described limitation that appears at the CMB pivot scale, one may compute the IR-counter term ${\cal Z}^{\rm IR}_{n}$ by knowing the structure of ${\cal Z}^{\rm UV}_{n}$. Furthermore, after removing the quadratic divergence contribution, our computations reveal that the unique structure of the term ${\cal Z}^{\rm UV}_{n}$ arises from late-time and adiabatic renormalization techniques. Consequently, the ultimate outcome is found to be independent of schemes, and we are left with the renormalized one-loop spectrum, in which the logarithmic IR divergent contributions characterize the one-loop effect. Due to the unique determination of the structure of the term ${\cal Z}^{\rm UV}_{n}$, the IR-counter term ${\cal Z}^{\rm IR}_{n}$ may now be further determined using the constraint that was previously derived at the CMB pivot scale. The precise structure of the IR counter-term will be fixed with the aid of the calculation carried out in the next section, which entails applying power spectrum renormalization. In reality, the conclusions that follow are only an extension of the unknown outcomes that we were able to calculate using common renormalization procedures as they relate to quantum field theory.

\subsubsection{Late time renormalization scheme}

There are two different kinds of divergences that show up in this computation: UV and IR. As a consequence of the cut-off regularization, the UV divergence manifests as quadratic contributions and the IR one as logarithmic contributions. When contrasted to the IR, the primary problem is the UV regularization. We have specifically employed a fixed UV cutoff scale while executing the loop integrals for this time-dependent computation. As a result, one can assume right away that the momentum integration at a certain time might contain unnecessary momentum modes, and that this kind of overcounting of modes could provide an incorrect conclusion. We will now go into more depth on why this idea is incorrect. In order to illustrate our position, let us begin with the representative momentum integral that is explicitly present in the SRI, USR, and SRII phases when we are dealing with situation with a single sharp transition (note that for the MST we have SR$_1$, USR$_n$, and SR$_{n+1}$ in this case):
\bea {\cal I}_{\bf rep}:=\int^{k_{\rm UV}}_{k_{\rm IR}}\frac{dk}{k}\;\left(A+Bc^2_sk^2\tau^2\right)+C=\bigg[A\ln\left(\frac{k_{\rm UV}}{k_{\rm IR}}\right)+\frac{1}{2}B\tau^2c^2_s\left(k^2_{\rm UV}-k^2_{\rm IR}\right)+C\bigg],\eea 
where the UV and IR momentum cut-offs, which vary for the three phases in question, are represented by the variables $k_{\rm UV}$ and $k_{\rm IR}$. The effective sound speed at the CMB pivot scale, $k=k_*$, is represented by $c_s$ here and is taken into account as a constant for the sake of this calculation. It is also crucial to remember that constant factors $A$ and $B$ vary for the SRI, USR, and SRII phases with single transition (and for MST we have SR$_1$, USR$_n$, and SR$_{n+1}$). In this instance, $C$ plays the same function as counter terms that arise when renormalization is performed. It is significant to notice that the values of the UV and IR momentum cut-offs have not been fixed throughout this demonstration. The momentum loop-integral formula above gives us the following simplified solution when we take into account the super-horizon limit in the previously determined result:
\bea \label{df} {\cal I}_{\bf rep}(\tau\rightarrow 0):=\bigg[\lim_{\tau\rightarrow 0}\bigg\{A\ln\left(\frac{k_{\rm UV}}{k_{\rm IR}}\right)+\frac{1}{2}B\tau^2c^2_s\left(k^2_{\rm UV}-k^2_{\rm IR}\right)\bigg\}+C\bigg]=A\ln\left(\frac{k_{\rm UV}}{k_{\rm IR}}\right).\eea
To fully eliminate the contribution from the quadratic UV divergence from the underlying theoretical setup under examination, the counter term $C$ is fixed at the following value in this case:
\bea \label{plwq} C=-\lim_{\tau\rightarrow 0}\frac{1}{2}B\tau^2c^2_s\left(k^2_{\rm UV}-k^2_{\rm IR}\right)=0.\eea
In establishing the counter term $C$, we have explicitly exploited the fact that $k_{\rm UV}\neq k_{\rm IR}$ and $k_{\rm UV}\gg k_{\rm IR}$.
IR divergence is not deleterious in the current situation, and it is not possible to fully eliminate such contribution from the IR end even when using the recommended counter-term constraint condition in the super-horizon limit. The only constraint condition that helps us soften the behavior of the IR logarithmic divergence in the current computation environment is this one. It looks like the analysis carried out in the references \cite{Kristiano:2022maq,Kristiano:2023scm,Choudhury:2023hvf,Choudhury:2023kdb,Choudhury:2023hfm,Bhattacharya:2023ysp,Motohashi:2023syh,Firouzjahi:2023ahg,Franciolini:2023lgy,Firouzjahi:2023aum,Cheng:2023ikq,Tasinato:2023ukp} is fully compatible with the generic conclusion that was produced, where no over counting of momentum modes is evident. We can clearly see that all of these extraneous, irrelevant momentum modes, which primarily capture the information of the sub-horizon quantum fluctuations after completing momentum integration and accounting for the late time limit (which is crucial from an observational standpoint), completely washed out of the computation at the super-horizon scale and during the re-entry through the horizon, leading to a smoother version of the IR softened result. 

While the values of the UV and IR cut-offs have not been fixed, we have reevaluated the aforementioned integral, taking into account the following trick, in order to provide a full-proof technical justification and to explicitly demonstrate that no overcounting of momentum modes will occur during performing such computations:
\bea \int^{k_{\rm UV}}_{k_{\rm IR}}:=\bigg(\int^{k_{\rm INT}}_{k_{\rm IR}}+\int^{k_{\rm UV}=\Lambda_{\rm UV} a(\tau)/c_s}_{k_{\rm INT}}\bigg),\eea
The UV limit of the integration may be described in terms of the conformal time dependency, which is fairly physically consistent as the related wave number $k_{\rm UV}$ is represented in the comoving scale and all the findings we are calculating in the super-horizon scale at the moment of horizon re-entry. For the goal of better comprehension, we have explicitly included the time-dependent information in the UV limit of the integration $k_{\rm UV}=\Lambda_{\rm UV} a(\tau)/c_s$. In this case, $\Lambda_{\rm UV}$ represents the UV cutoff contribution, which is added after the conformal time dependency in the current context has been extracted. Moreover, it is noteworthy that the bounds of the momentum integral have been split into two components. In essence, the first component handles the finite contribution of the integral, while the second part handles the quadratic UV divergence when time dependence is present at the upper bound of the momentum-dependent loop integration. We may put down the following simplified formulation after using the previously indicated decomposition in the integration:
\bea \label{yh} {\cal I}_{\bf rep}:&=&\bigg\{\bigg(\int^{k_{\rm INT}}_{k_{\rm IR}}+\int^{k_{\rm UV}=\Lambda_{\rm UV} a(\tau)/c_s}_{k_{\rm INT}}\bigg)\frac{dk}{k}\;\left(A+Bc^2_sk^2\tau^2\right)\bigg\}+C\nonumber\\
&=&\bigg[A\bigg\{\ln\left(\frac{k_{\rm INT}}{k_{\rm IR}}\right)+\ln\left(\frac{\Lambda_{\rm UV}}{H}\right)\bigg\}+\frac{1}{2}B\bigg\{\left(\frac{\Lambda_{\rm UV}}{H}\right)^2-1\bigg\}+C\bigg],\eea  
where $a(\tau)/c_sk_{\rm INT}=H^{-1}$ and $Bc^2_s\tau^2(k^2_{\rm INT}-k^2_{\rm IR})/2$ disappear respecting the restriction $k_{\rm INT}\gg k_{\rm IR}$ are used in the computation of the last step of the loop integration. The explicit equation for the counter term $C$ in the current context of discussion may be found by further imposing the restriction from wave function/adiabatic renormalization at the pivot scale $k=k_*$. This simplified formula is as follows:
  \bea C(\mu,\Lambda_{\rm UV})=\bigg[A\bigg\{\ln\left(\frac{\mu}{H}\right)-\ln\left(\frac{\Lambda_{\rm UV}}{H}\right)\bigg\}+\frac{1}{2}B\bigg\{\left(\frac{\mu}{H}\right)^2-\left(\frac{\Lambda_{\rm UV}}{H}\right)^2\bigg\}\bigg].\eea
The renormalization scale linked to the adiabatic renormalization technique is denoted by $\mu$ in this instance.
  By further inserting the previously mentioned result into equation (\ref{yh}), we obtain the one-loop contributions, which is applicable in the SRI, USR, and SRII phases for single sharp transition (and also for MST in the SR$_1$, USR$_n$, and SR$_{n+1}$) independently:
\bea \label{yh1} {\cal I}_{\bf rep}(\mu):
&=&\bigg[A\bigg\{\ln\left(\frac{k_{\rm INT}}{k_{\rm IR}}\right)+\ln\left(\frac{\Lambda_{\rm UV}}{H}\right)\bigg\}+\frac{1}{2}B\bigg\{\left(\frac{\Lambda_{\rm UV}}{H}\right)^2-1\bigg\}\bigg]\nonumber\\
&&\quad\quad\quad\quad\quad\quad+\bigg[A\bigg\{\ln\left(\frac{\mu}{H}\right)-\ln\left(\frac{\Lambda_{\rm UV}}{H}\right)\bigg\}+\frac{1}{2}B\bigg\{\left(\frac{\mu}{H}\right)^2-\left(\frac{\Lambda_{\rm UV}}{H}\right)^2\bigg\}\bigg]\nonumber\\
&=&\bigg[A\bigg\{\ln\left(\frac{k_{\rm INT}}{k_{\rm IR}}\right)+\ln\left(\frac{\mu}{H}\right)\bigg\}+\frac{1}{2}B\bigg\{\left(\frac{\mu}{H}\right)^2-1\bigg\}\bigg].\eea  
Taking into account that the renormalization scale is linked to an adiabatic renormalization scheme at $\mu=H$, we obtain the streamlined solution for the loop-integral as follows:
\bea \label{yh2} {\cal I}_{\bf rep}(\mu=H)
&=&A\ln\left(\frac{k_{\rm INT}}{k_{\rm IR}}\right),\eea 
This is exactly the same result that we have obtained in equation (\ref{df}). It is noteworthy that in this case, the final derived result is independent of the UV cut-off $\Lambda_{\rm UV}$, which we introduced when we included conformal time dependence in the upper limit of the momentum loop integration. The equivalency between the two procedures—late time renormalization and adiabatic/wave function renormalization—is established by this comparative study, which also attests to the accuracy of the calculations made in this work. Most crucially, this technical reason enables us to verify that no superfluous momentum modes are dominating the outcome during computing. Consequently, we can be absolutely certain that the computations made in this study are entirely reliable, as overcounting does not arise in the current context of discussion. One may entirely eliminate the contribution of the risky quadratic UV divergence factor from the one-loop momentum integration by using any of the previously described renormalization techniques. Nevertheless, none of the renormalization strategies listed above are able to eliminate the logarithmic IR divergent contribution. Such IR divergence is not damaging at all, and by using power spectrum renormalization, which accomplishes the aim of course graining as thoroughly covered in the next portion of this study, the associated contribution can be further smoothed.

Finally, using the late-time renormalization scheme, we can express the MST integrals as :
\bea
\lim_{\tau \rightarrow 0}{\cal J}_{\textbf{SR}_{1}}(\tau_{s_{1}}) &=& \ln{\bigg(\frac{k_{s_1}}{k_{*}}}\bigg),  \\
\lim_{\tau \rightarrow 0}{\cal J}_{\textbf{USR}_{n}}(\tau_{s_{n}}) &=& \ln\left(\frac{k_{e_{n}}}{k_{s_{n}}}\right), \\
\lim_{\tau \rightarrow 0}{\cal J}_{\textbf{USR}_{n}}(\tau_{e_{n}})&=&\left(\frac{k_{e_{n}}}{k_{s_{n}}}\right)^{6}{\cal J}_{\textbf{USR}_{n}}(\tau_{s_{n}}), \\
\lim_{\tau \rightarrow 0}{\cal J}_{\textbf{SR}_{n+1}}(\tau_{e_{n}}) &=& \ln\left(\frac{k_{s_{n+1}}}{k_{e_{n}}}\right)
\eea
In the above-mentioned results by substituting $n=1$ one can easily obtain the result for the single transition. After applying these limitations, the overall one-loop corrected late-time renormalized scalar power spectrum is as follows for single transition \footnote{}:
\bea
\label{late-tpowspec}
\Delta^{2}_{\zeta,\textbf{EFT}}(k) &=& \bigg[\Delta^{2}_{\zeta,\textbf{Tree}}(k)\bigg]_{\textbf{SR}}
\; \bigg\{ 1+ \bigg[\Delta^{2}_{\zeta,\textbf{Tree}}(k)\bigg]_{\textbf{SR}} \bigg [ \left(1-\frac{2}{15\pi^{2}}\frac{1}{c_{s}^{2}k_{*}^{2}}\left(1-\frac{1}{c_{s}^{2}}\right)\epsilon\right)\times \left(-\frac{4}{3}\ln{\frac{k_{s}}{k_{*}}}\right)\bigg] \nonumber \\
&& \quad \quad + \frac{1}{4} \bigg[\Delta^{2}_{\zeta,\textbf{Tree}}(k)\bigg]_{\textbf{SR}} 
     \bigg[\bigg(\frac{\Delta\eta(\tau_{e})}{\tilde{c}^{4}_{s}}\bigg)^{2}\ln{\bigg(\frac{k_{e}}{k_{s}}\bigg)^6} - \left(\frac{\Delta\eta(\tau_{s})}{\tilde{c}^{4}_{s}}\right)^{2}\bigg] \ln{\bigg(\frac{k_{e}}{k_{s}}\bigg)} \nonumber \\
     && \quad \quad  +  
\bigg[\Delta^{2}_{\zeta,\textbf{Tree}}(k)\bigg]_{\textbf{SR}}\left(1-\frac{2}{15\pi^{2}}\frac{1}{c_{s}^{2}k_{*}^{2}}\left(1-\frac{1}{c_{s}^{2}}\right)\epsilon\right)
\times 
\ln{\bigg(\frac{k_{s}}{k_{e}}}\bigg)\bigg\}.
\eea
and for MST we have the following result:
\bea
\label{late-tpowspec}
\Delta^{2}_{\zeta,\textbf{EFT}}(k) &=& \bigg[\Delta^{2}_{\zeta,\textbf{Tree}}(k)\bigg]_{\textbf{SR}_{1}}
\; \bigg\{ 1+ \bigg[\Delta^{2}_{\zeta,\textbf{Tree}}(k)\bigg]_{\textbf{SR}_{1}} \bigg [ \left(1-\frac{2}{15\pi^{2}}\frac{1}{c_{s}^{2}k_{*}^{2}}\left(1-\frac{1}{c_{s}^{2}}\right)\epsilon\right)\times \left(-\frac{4}{3}\ln{\frac{k_{s_1}}{k_{*}}}\right)\bigg] \nonumber \\
&& \quad \quad + \frac{1}{4} \bigg[\Delta^{2}_{\zeta,\textbf{Tree}}(k)\bigg]_{\textbf{SR}_{1}}  \sum^{N}_{n=1}\bigg(
     \bigg[\bigg(\frac{\Delta\eta(\tau_{e_{n}})}{\tilde{c}^{4}_{s}}\bigg)^{2}\ln{\bigg(\frac{k_{e_n}}{k_{s_n}}\bigg)^6} - \left(\frac{\Delta\eta(\tau_{s_{n}})}{\tilde{c}^{4}_{s}}\right)^{2}\bigg] \ln{\bigg(\frac{k_{e_n}}{k_{s_n}}\bigg)} \bigg)\nonumber \\
     && \quad \quad  +  
\bigg[\Delta^{2}_{\zeta,\textbf{Tree}}(k)\bigg]_{\textbf{SR}_{1}}\left(1-\frac{2}{15\pi^{2}}\frac{1}{c_{s}^{2}k_{*}^{2}}\left(1-\frac{1}{c_{s}^{2}}\right)\epsilon\right)
\times \sum^{N}_{n=1} 
\bigg(\ln{\bigg(\frac{k_{s_{n+1}}}{k_{e_n}}}\bigg)\bigg)\bigg\}.
\eea
The contributions to the complete one-loop corrected scalar power spectrum from eqn. (\ref{late-tpowspec}) are shown diagrammatically below: 
\begin{equation}
\begin{tikzpicture}[baseline={([yshift=-3.5ex]current bounding box.center)},very thick]
  
  \def\radius{1}
  \scalebox{1}{\draw[red,very thick] (0,\radius) circle (\radius);
  \draw[red,very thick] (4.5*\radius,0) circle (\radius);}

  \draw[black, very thick] (-4*\radius,0) -- 
  (-2.5*\radius,0);
  \node at (-2*\radius,0) {+};
  \draw[black, very thick] (-1.5*\radius,0) -- (0,0);
  \draw[blue,fill=blue] (0,0) circle (.5ex);
  \draw[black, very thick] (0,0)  -- (1.5*\radius,0);
  \node at (2*\radius,0) {+};
  \draw[black, very thick] (2.5*\radius,0) -- (3.5*\radius,0); 
  \draw[blue,fill=blue] (3.5*\radius,0) circle (.5ex);
  \draw[blue,fill=blue] (5.5*\radius,0) circle (.5ex);
  \draw[black, very thick] (5.5*\radius,0) -- (6.5*\radius,0);
  

\end{tikzpicture}\quad = \quad {\rm Late-time \,\,renormalized} \,\,\Delta^{2}_{\zeta,\textbf{EFT}}(k),
\end{equation}
where the late-time renormalization approach yielded $\Delta^{2}_{\zeta,\textbf{EFT}}(k)$, and the diagrams depict the tree and just one-loop contributions to it.

Consequently, you have the identical outcomes as shown below when you compare the loop diagrams for the late-time renormalization with the adiabatic renormalization:
\begin{equation}
\begin{tikzpicture}[baseline={([yshift=-3.5ex]current bounding box.center)},very thick]
  
  \def\radius{1}
  \scalebox{1}{\draw[red,very thick] (-0.6*\radius,0.75*\radius) circle (0.75*\radius);
  \draw[red,very thick] (2.45*\radius,0) circle (0.75*\radius);\draw[cyan,very thick](8.4*\radius,0.75*\radius) circle(0.75*\radius);\draw[cyan,very thick](11.55*\radius,0) circle(0.75*\radius);}

   \draw[black, very thick] (-3*\radius,0) -- (-2.2*\radius,0);
   \node at (-1.8*\radius,0) {+};
  \draw[black, very thick] (-1.4*\radius,0) -- (-0.6*\radius,0);
  \draw[blue,fill=blue] (-0.6*\radius,0) circle (.5ex);
  \draw[black, very thick] (-0.6*\radius,0)  -- (0.2*\radius,0);
  \node at (0.6*\radius,0) {+};
  \draw[black, very thick] (0.9*\radius,0) -- (1.7*\radius,0); 
  \draw[blue,fill=blue] (1.7*\radius,0) circle (.5ex);
  \draw[blue,fill=blue] (3.2*\radius,0) circle (.5ex);
  \draw[black, very thick] (3.2*\radius,0) -- (4*\radius,0);
  \node at (5*\radius,0) {=};
   \draw[black, very thick] (6*\radius,0) -- (6.8*\radius,0);
   \node at (7.2*\radius,0) {+};
   \draw[black, very thick] (7.6*\radius,0) -- (8.4*\radius,0);
  \draw[blue,fill=blue] (8.4*\radius,0) circle (.5ex);
  \draw[black, very thick] (8.4*\radius,0)  -- (9.2*\radius,0);
  \node at (9.6*\radius,0) {+};
  \draw[black, very thick] (10*\radius,0) -- (10.8*\radius,0); 
  \draw[blue,fill=blue] (10.8*\radius,0) circle (.5ex);
  \draw[blue,fill=blue] (12.3*\radius,0) circle (.5ex);
  \draw[black, very thick] (12.3*\radius,0) -- (13.1*\radius,0);
   
\end{tikzpicture}  
\end{equation}

During the current computation, we have also discovered the following significant information about the relationship between the counter-terms that appear in this context and those that appear in the context of renormalization of the bare action as appearing in the context of Quantum Field Theory in quasi de Sitter space with the gauge invariant comoving curvature perturbation. This relationship is represented by:
\bea \sum^{6}_{i=1}\delta_{{\cal Z}_{{\bf D}_i}}=C(\mu=H,\Lambda)=\left(1+\delta_{{\cal Z}_{C}}(\mu=H,\Lambda)\right)=-\Bigg\{A\ln{\bigg(\frac{\Lambda}{H}}\bigg) + \frac{B}{2}\left(\bigg(\frac{\Lambda}{H}\bigg)^2-1\right)\Bigg\},\eea
where $C(\mu=H,\Lambda)$, or more accurately, $\delta_{{\cal Z}_{C}}(\mu=H,\Lambda)$, is the late-time renormalization scheme counter-term. This relation is used to obtain the subsequent relation:
\bea \left(\sum^{6}_{i=1}\delta_{{\cal Z}_{{\bf D}_i}}-1\right)&=&\left(\delta_{{\cal Z}_{{\bf D}_1}}+\delta_{{\cal Z}_{{\bf D}_2}}+\delta_{{\cal Z}_{{\bf D}_3}}+\delta_{{\cal Z}_{{\bf D}_4}}+\delta_{{\cal Z}_{{\bf D}_5}}+\delta_{{\cal Z}_{{\bf D}_6}}-1\right)\nonumber\\
&=&\delta_{{\cal Z}_{C}}(\mu=H,\Lambda)\nonumber\\
&=&-\Bigg\{1+A\ln{\bigg(\frac{\Lambda}{H}}\bigg) + \frac{B}{2}\left(\bigg(\frac{\Lambda}{H}\bigg)^2-1\right)\Bigg\}.\eea
This relation makes it evident how the current scheme is related to the operator counter terms of the third-order action. It is significant to note that in this case, the renormalization scale $\mu=H$ is arbitrary since the Hubble parameter has an intrinsic scale dependency. Because of this, this result provides a generic description of the contributions and relationships between the counter-terms. It is further desired to divide the findings into the sections that are labeled with the symbols ${\rm SR}_1$, ${\rm USR}_n$, and ${\rm SR}_{n+1}$ in order to learn more about the explicit contributions and additional effects. In this case, $n$ represents the number of MST in this current context. In the case of single transition setup the findings will be labeled with the symbols SRI, USR and SRII and considering $n=1$ in the MST result one can further derive the result for the single transition. The following simplified results are obtained by dividing the previously stated results into the respective phases:
\bea &&\underline{{\bf SR}_1:}\quad\quad\sum^{6}_{i=1}\delta_{{\cal Z}_{{\bf D}_i}}=C(\mu=H_{s_1},\Lambda)=\left(1+\delta_{{\cal Z}_{C,{\bf SR}_1}}\right)\quad{\rm with}\quad\delta_{{\cal Z}_{{\bf D}_6}}=0,\\
&&\underline{{\bf USR}_n:}\quad\quad\delta_{{\cal Z}_{{\bf D}_6}}=C(\mu=H_{e_n},\Lambda)=\left(1+\delta_{{\cal Z}_{C,{\bf USR}_n}}\right)\quad {\rm with}\quad \sum^{6}_{i=1}\delta_{{\cal Z}_{{\bf D}_i}}=0,\\
&&\underline{{\bf SR}_{n+1}:}\quad\quad\sum^{6}_{i=1}\delta_{{\cal Z}_{{\bf D}_i}}=C(\mu=H_{\rm end},\Lambda)=\left(1+\delta_{{\cal Z}_{C,{\bf SR}_{n+1}}}\right)\quad{\rm with}\quad\delta_{{\cal Z}_{{\bf D}_6}}=0,\eea
This provides us with the following streamlined relations:
\bea &&\underline{{\bf SR}_1:}\quad\quad\left(\delta_{{\cal Z}_{{\bf D}_1}}+\delta_{{\cal Z}_{{\bf D}_2}}+\delta_{{\cal Z}_{{\bf D}_3}}+\delta_{{\cal Z}_{{\bf D}_4}}+\delta_{{\cal Z}_{{\bf D}_5}}-1\right)=\delta_{{\cal Z}_{C,{\bf SR}_1}},\\
&&\underline{{\bf USR}_n:}\quad\quad\left(\delta_{{\cal Z}_{{\bf D}_6}}-1\right)=\delta_{{\cal Z}_{C,{\bf USR}_n}},\\
&&\underline{{\bf SR}_{n+1}:}\quad\quad\left(\delta_{{\cal Z}_{{\bf D}_1}}+\delta_{{\cal Z}_{{\bf D}_2}}+\delta_{{\cal Z}_{{\bf D}_3}}+\delta_{{\cal Z}_{{\bf D}_4}}+\delta_{{\cal Z}_{{\bf D}_5}}-1\right)=\delta_{{\cal Z}_{C,{\bf SR}_{n+1}}}.\eea
This gives rise to the following results for the counter terms corresponding to the each of the previously mentioned consecutive phases:
\bea \underline{{\bf SR}_1:}\quad\quad C(\mu=H_{s_1},\Lambda)&=&\left(1+\delta_{{\cal Z}_{C,{\bf SR}_1}}\right)\nonumber\\
&=&\left(\delta_{{\cal Z}_{{\bf D}_1}}+\delta_{{\cal Z}_{{\bf D}_2}}+\delta_{{\cal Z}_{{\bf D}_3}}+\delta_{{\cal Z}_{{\bf D}_4}}+\delta_{{\cal Z}_{{\bf D}_5}}\right)\nonumber\\
&=&-\Bigg\{A\ln{\bigg(\frac{\Lambda}{H_{s_1}}}\bigg) + \frac{B}{2}\left(\bigg(\frac{\Lambda}{H_{s_1}}\bigg)^2-1\right)\Bigg\}\nonumber\\
&&{\rm where}\quad\quad A=-\frac{4}{3}B=-\frac{4}{3}\left(1-\frac{2}{15\pi^{2}}\frac{1}{c_{s}^{2}k_{*}^{2}}\left(1-\frac{1}{c_{s}^{2}}\right)\epsilon\right)\bigg[\Delta^{2}_{\zeta,\textbf{Tree}}(k)\bigg]^2_{\textbf{SR}_{1}},\\
\underline{{\bf USR}_n:}\quad\quad C(\mu=H_{e_n},\Lambda)&=&\left(1+\delta_{{\cal Z}_{C,{\bf USR}_n}}\right)\nonumber\\
&=&\delta_{{\cal Z}_{{\bf D}_6}}\nonumber\\
&=&-\Bigg\{A\ln{\bigg(\frac{\Lambda}{H_{e_n}}}\bigg) + \frac{B}{2}\left(\bigg(\frac{\Lambda}{H_{e_n}}\bigg)^2-1\right)\Bigg\}\nonumber\\
&&{\rm where}\quad\quad A=B=\frac{1}{4}\bigg[\bigg(\frac{\Delta\eta(\tau_{e_{n}})}{\tilde{c}^{4}_{s}}\bigg)^{2}\ln{\bigg(\frac{k_{e_n}}{k_{s_n}}\bigg)^6} - \left(\frac{\Delta\eta(\tau_{s_{n}})}{\tilde{c}^{4}_{s}}\right)^{2}\bigg]\bigg[\Delta^{2}_{\zeta,\textbf{Tree}}(k)\bigg]^2_{\textbf{SR}_{1}},\quad\quad\quad\\
\underline{{\bf SR}_{n+1}:}\quad\quad C(\mu=H_{\rm end},\Lambda)&=&\left(1+\delta_{{\cal Z}_{C,{\bf SR}_{n+1}}}\right)\nonumber\\
&=&\left(\delta_{{\cal Z}_{{\bf D}_1}}+\delta_{{\cal Z}_{{\bf D}_2}}+\delta_{{\cal Z}_{{\bf D}_3}}+\delta_{{\cal Z}_{{\bf D}_4}}+\delta_{{\cal Z}_{{\bf D}_5}}\right)\nonumber\\
&=&-\Bigg\{A\ln{\bigg(\frac{\Lambda}{H_{\rm end}}}\bigg) + \frac{B}{2}\left(\bigg(\frac{\Lambda}{H_{\rm end}}\bigg)^2-1\right)\Bigg\}\nonumber\\
&&{\rm where}\quad\quad A=B=\left(1-\frac{2}{15\pi^{2}}\frac{1}{c_{s}^{2}k_{*}^{2}}\left(1-\frac{1}{c_{s}^{2}}\right)\epsilon\right)\bigg[\Delta^{2}_{\zeta,\textbf{Tree}}(k)\bigg]^2_{\textbf{SR}_{1}},\eea
Consequently, we obtain the subsequent formula for the UV divergence-free contribution of the power spectrum:
\bea
\left[\Delta_{\zeta,\textbf{EFT}}^{2}(k)\right]_n &=& {\cal Z}^{\rm UV}_{n}\times \bigg[\Delta^{2}_{\zeta,\textbf{Tree}}(k)\bigg]_{\textbf{SR}_{1}},\eea
where the following formula provides the value of the factor ${\cal Z}^{\rm UV}_{n}$ for the $n$ th sharp transition:
\bea {\cal Z}^{\rm UV}_{n}&=&\left(1+\delta_{{\cal Z}^{\rm UV}_{n}}\right),\\
&=&\bigg\{ 1+ \bigg[\Delta^{2}_{\zeta,\textbf{Tree}}(k)\bigg]_{\textbf{SR}_{1}} \bigg [ \left(1-\frac{2}{15\pi^{2}}\frac{1}{c_{s}^{2}k_{*}^{2}}\left(1-\frac{1}{c_{s}^{2}}\right)\epsilon\right)\times \left(-\frac{4}{3}\ln{\frac{k_{s_1}}{k_{*}}}\right)\bigg] \nonumber \\
&& \quad \quad + \frac{1}{4} \bigg[\Delta^{2}_{\zeta,\textbf{Tree}}(k)\bigg]_{\textbf{SR}_{1}} \times \bigg(
     \bigg[\bigg(\frac{\Delta\eta(\tau_{e_{n}})}{\tilde{c}^{4}_{s}}\bigg)^{2}\ln{\bigg(\frac{k_{e_n}}{k_{s_n}}\bigg)^6} - \left(\frac{\Delta\eta(\tau_{s_{n}})}{\tilde{c}^{4}_{s}}\right)^{2}\bigg] \ln{\bigg(\frac{k_{e_n}}{k_{s_n}}\bigg)} \bigg)\nonumber \\
     && \quad \quad  +  
\bigg[\Delta^{2}_{\zeta,\textbf{Tree}}(k)\bigg]_{\textbf{SR}_{1}}\left(1-\frac{2}{15\pi^{2}}\frac{1}{c_{s}^{2}k_{*}^{2}}\left(1-\frac{1}{c_{s}^{2}}\right)\epsilon\right)
\times 
\bigg(\ln{\bigg(\frac{k_{s_{n+1}}}{k_{e_n}}}\bigg)\bigg)\bigg\}. \eea
This identification shows that the quadratic UV divergence in the expression for ${\cal Z}^{\rm UV}_{n}$ can be eliminated entirely. It also aids in understanding the relationship that exists between the counter-terms that appear in the bare action (UV sensitive part) and current contexts. Now that the relationship has been established, this identification will be much more useful in bridging the gap between the current late-time scheme and the common renormalization method that can be found in the context of quasi-de Sitter space quantum field theory. Most notably, the late-time approach aids in precisely fixing the mathematical structure of the number ${\cal Z}^{\rm UV}_{n}$, which is used to estimate, after renormalization, the total power spectrum for the scalar modes. The only issue is that the IR-counter term ${\cal Z}^{\rm IR}_{n}$ may be computed utilizing the previously described constraint that appears at the CMB pivot scale if one knows the structure of ${\cal Z}^{\rm UV}_{n}$.

\subsubsection{Power spectrum renormalization scheme}

From this point on, we use the power spectrum renormalization approach, which suppresses and improves controllability of logarithmic infrared divergences, so saving us from them. This approach gives us the renormalized form of the scalar power spectrum through the employment of a counter-term, which is calculated by invoking a renormalization condition at the pivot scale $k_{*}$. Every interval with an abrupt transition—that is, $6$ times in our case—has the operation carried out. To generate a finite physical output, the resummation technique must be implemented in the last phase. The power spectrum renormalization carried out under the condition with an arbitrary $N$ number of abrupt transitions will be shown below. Now, let us get started with this same process by rescaling the power spectrum's tree-level contribution as follows:
\bea \label{rescale}
\small[\Delta_{\zeta,\textbf{Tree}}^{2}(k)\small]_{\textbf{SR}_{1}} = N \times \small[\widetilde  \Delta_{\zeta,\textbf{Tree}}^{2}(k)\small]_{\textbf{SR}_{1}}.
\eea
This does not change the main structure of the output; it is done just to make the computation easier. The renormalized form of the scalar power spectrum, where just one abrupt transition is taken into account, may be expressed using the following equation:
\bea \label{neft}
\small[\Delta_{\zeta,\textbf{EFT}}^{2}(k)\small]_{n} = \small[\widetilde  \Delta_{\zeta,\textbf{Tree}}^{2}(k)\small]_{\textbf{SR}_{1}}\{1 + W_{n} + NV_{n}\},
\eea
where the variable $n$ is used to monitor the number of times renormalization occurs, which in the current context runs from $n = 1\;{\rm to}\;6$ summing the $N=6$ MSTs. Also it is important to note that, $W_{n}= U_{\rm SR_1} + NU_{n}$. With this, the renormalized form of the scalar power spectrum for the sum of $n$ abrupt transitions is expressed as follows:
\bea
\Delta_{\zeta,\textbf{EFT}}^{2}(k) = \sum_{n=1}^{N}\; \small[\Delta_{\zeta,\textbf{EFT}}^{2}(k)\small]_{n}.
\eea
Using the rescaling that was previously discussed in eqn. (\ref{rescale}), we can alter the equation above as follows:
\bea
\Delta_{\zeta,\textbf{EFT}}^{2}(k) = \small[\widetilde  \Delta_{\zeta,\textbf{Tree}}^{2}(k)\small]_{\textbf{SR}_{1}}\times\sum_{n=1}^{N}\;\{1 + W_{n} + NV_{n}\}.
\eea
Inside the context of this discussion, the following simplified equation might represent the renormalized form of the 1PI correlation function for any arbitrary $m$-point amplitude determined inside the scope of EFT of Single Field Inflation:
\bea \overline{\Gamma_{{\bf EFT}}[\zeta]}=\sum^{\infty}_{m=2}\frac{i}{m!}\int\prod^{m}_{j=1}d^4x_j\,\overline{\Gamma^{(m)}_{\zeta,{\bf EFT}}(x_i)}\,\zeta(x_j),\eea
It is noteworthy to mention that the $\overline{\Gamma^{(m)}_{\zeta,{\bf EFT}}(x_j)}$ may be further stated as follows in the Fourier space of such a $m$-point renormalized amplitude:
\bea \overline{\Gamma^{(m)}_{\zeta,{\bf EFT}}(x_j)}:=\int \frac{d^4k_j}{(2\pi)^4}\, e^{ik_j. x_j}\,\overline{\Gamma^{(m)}_{\zeta,{\bf EFT}}(k_j,\mu,\mu_0)}\times (2\pi)^4\delta^4\left(\sum^{m}_{j=1}k_j\right)\quad\quad\forall\quad j=1,2,\cdots,m.\eea
where $\mu$ and $\mu_0$ stand for the reference and renormalization scales, respectively. Furthermore, any generic $m$-point renormalized amplitude in the Fourier space may be described as follows in terms of the 1PI effective action:
\bea \overline{\Gamma^{(m)}_{\zeta,{\bf EFT}}(k_1,k_2,k_3,\cdots,k_m,\mu,\mu_0)}=\left({\cal Z}^{\rm IR}_{n}\right)^{\frac{m}{2}}\Gamma^{(m)}_{\zeta,{\bf EFT}}(k_1,k_2,k_3,\cdots,k_m).\eea
Now, we may carry out the renormalization by combining and adding a counter-term for every $n$th contribution that corresponds to a single abrupt transition. The renormalization factor or counter-term, represented as ${\cal Z}^{\rm IR}_{n}$ going forward. The $m$-point renormalized cosmological correlation function may be used to translate this assertion. The following simplified expression can then be used to write the statement in terms of the unrenormalized/bare contribution:
\bea \overline{\langle \zeta_{\bf k_1}\zeta_{\bf k_2}\zeta_{\bf k_3}\cdots\cdots\zeta_{\bf k_m}\rangle_{\zeta,{\bf EFT}}}=\left({\cal Z}^{\rm IR}_{n}\right)^{\frac{m}{2}}\langle \zeta_{\bf k_1}\zeta_{\bf k_2}\zeta_{\bf k_3}\cdots\cdots\zeta_{\bf k_m}\rangle_{\zeta,{\bf EFT}}.\eea
This assertion is made at the $m$-point correlation function level, and it can be readily understood in terms of the relationship that connects the renormalized, unrenormalized/bare, and counter-term contributions in the expression for the gauge invariant comoving scalar curvature perturbation. This expression is as follows:
\bea \zeta^{\bf R}_{\bf k}=\sqrt{{\cal Z}^{\rm IR}_{n}}\times \zeta^{\bf B}_{\bf k}.\eea
We will now limit our study to the renormalization of the scalar power spectrum by fixing $m=2$, which characterizes the two-point amplitude of the cosmic correlation function in the Fourier space, as this is what we are interested in. Following this correction, the two-point amplitude's associated 1PI effective action may be further stated as follows:
\bea \overline{\Gamma^{(2)}_{\zeta,{\bf EFT}}(k_1,k_2,\mu,\mu_0)}={\cal Z}^{\rm IR}_{n}\times\Gamma^{(2)}_{\zeta,{\bf EFT}}(k_1,k_2),\eea
This is further interpretable as follows at the level of the two-point renormalized cosmological correlation function:
\bea \overline{\langle \zeta_{\bf k_1}\zeta_{\bf k_2}\rangle}=\langle \zeta^{\bf R}_{\bf k_1}\zeta^{\bf R}_{\bf k_2}\rangle
&=&{\cal Z}^{\rm IR}_{n}\times \langle \zeta^{\bf B}_{\bf k_1}\zeta^{\bf B}_{\bf k_2}\rangle.\label{corr}\eea
Furthermore, we utilize the following information to put out the formulas for the scalar power spectrum derived from the correlation function between two points:
\bea &&\label{r1}\overline{\langle \zeta_{\bf k_1}\zeta_{\bf k_2}\rangle}=\langle \zeta^{\bf R}_{\bf k_1}\zeta^{\bf R}_{\bf k_2}\rangle=(2\pi)^3\delta^3\left({\bf k_1}+{\bf k_2}\right)\frac{2\pi^2}{k^3_1}\small[\overline{\Delta_{\zeta,\textbf{EFT}}^{2}(k)}\small]_{n},\\
&&\label{r2}\langle \zeta^{\bf B}_{\bf k_1}\zeta^{\bf B}_{\bf k_2}\rangle=(2\pi)^3\delta^3\left({\bf k_1}+{\bf k_2}\right)\frac{2\pi^2}{k^3_1}\small[\Delta_{\zeta,\textbf{EFT}}^{2}(k)\small]_{n}.\eea
This simplified expression governs the renormalized version of the $n$th contribution of the scalar power spectrum, which consists of a single sharp transition. It is derived by using equations (\ref{r1}) and (\ref{r2}) in equation (\ref{corr}) and fixing $k=k_1$ to prevent further confusion caused by nomenclature.
\bea \label{zeft}
\small[\overline{\Delta_{\zeta,\textbf{EFT}}^{2}(k)}\small]_{n} = {\cal Z}^{\rm IR}_{n}\times\;\small[\Delta_{\zeta,\textbf{EFT}}^{2}(k)\small]_{n}.
\eea
This causes the counter-term multiplication to mitigate the logarithmic IR divergences. The following renormalization condition, fixed at the pivot scale $k_{*}$, determines the value of this counter-term:
\bea
\small[\overline{\Delta_{\zeta,\textbf{EFT}}^{2}(k_{*})}\small]_{n} = \small[\widetilde  \Delta_{\zeta,\textbf{Tree}}^{2}(k_{*})\small]_{\textbf{SR}_{1}},
\eea
Due to the aforementioned renormalization requirement, we first follow the definitions in Eqs. (\ref{neft}) and (\ref{zeft}) to ascertain the form of the $n$th counter-term:
\bea
{\cal Z}^{\rm IR}_{n} =  \frac{\small[\overline{\Delta_{\zeta,\textbf{EFT}}^{2}(k_{*})}\small]_{n}}{\small[\Delta_{\zeta,\textbf{EFT}}^{2}(k_{*})\small]_{n}} = \frac{\small[\widetilde  \Delta_{\zeta,\textbf{Tree}}^{2}(k_{*})\small]_{\textbf{SR}_{1}}}{\small[\Delta_{\zeta,\textbf{EFT}}^{2}(k_{*})\small]_{n}} = \frac{1}{1 + W_{n,*} + NV_{n,*}} \approx (1-W_{n,*}-NV_{n,*}+\cdots).
\eea
Any additional higher-order contributions are ignored and the series is terminated after only growing to the first order. Finally, the combined renormalized one-loop corrected scalar power spectrum originating from contributions across all $n$ sharp transitions may be written thanks to this determination of the counter-term:
\bea
\overline{\Delta_{\zeta,\textbf{EFT}}^{2}(k)} = \small[\Delta_{\zeta,\textbf{Tree}}^{2}(k)\small]_{\textbf{SR}_{1}}\bigg\{1 + \frac{1}{N}\sum_{n=1}^{N}{\cal Q}_{n,\textbf{EFT}}\bigg\},
\eea
including the contribution from the $n$th transition and a new EFT dependent term, ${\cal Q}_{n,\textbf{EFT}}$. This new phrase has the following structure:
\bea \label{Qneft}
{\cal Q}_{n,\textbf{EFT}} = \frac{-N\small[\Delta_{\zeta,\textbf{Tree}}^{2}(k)\small]_{\textbf{SR}_{1}}}{\small[\Delta_{\zeta,\textbf{Tree}}^{2}(k_{*})\small]_{\textbf{SR}_{1}}}\{W_{n,*}^{2} + N^{2}V_{n,*}^{2} + \cdots \}.
\eea
As we proceed, in the next sections, to carry out the last required process of resummation over the logarithmic contributions to generate a physically appropriate output, the significance of the expression ${\cal Q}_{n,\textbf{EFT}}$ will become increasingly clear. A universal $N$ number of abrupt transitions can be handled using the approach described in this section. Since $N=6$ is the lowest number that may be used to solve the horizon problem, we have fixed it for the study of this work. The final renormalized one-loop corrected scalar power spectrum exhibits quadratic dependency due to the promotion of the first order logarithmic components included inside the factors $W_{n}$ and $V_{n}$. This is demonstrated by the resultant eqn.(\ref{Qneft}). This is an important finding to keep in mind since it means that the logarithmic divergent terms have gotten even softer by becoming more sub-leading. The following diagrammatic depiction illustrates the loop contributions that are part of eqn.(\ref{Qneft}):
\begin{equation}\label{twoloop}
\begin{tikzpicture}[baseline={([yshift=-3.5ex]current bounding box.center)},very thick]
  
  \def\radius{0.76}
  \scalebox{0.5}{
  \draw[red,very thick] (3*\radius,0) circle (\radius);
  \draw[red,very thick] (5*\radius,0) circle (\radius);
  \draw[red,very thick] (13*\radius,\radius) circle (\radius);
  \draw[red,very thick] (13*\radius,3*\radius) circle (\radius);
  \draw[red,very thick] (21*\radius,0) circle (\radius);}
  \draw[black, very thick] (0,0) -- (\radius,0); 
  \draw[blue,fill=blue] (\radius,0) circle (.3ex);
  \draw[blue,fill=blue] (2*\radius,0) circle (.3ex);
  \draw[blue,fill=blue] (3*\radius,0) circle (.3ex);
  \draw[black, very thick] (3*\radius,0) -- (4*\radius,0);
  \node at (4.5*\radius,0) {+};
  \draw[black, very thick] (5*\radius,0) -- (6.5*\radius,0);
  \draw[blue,fill=blue] (6.5*\radius,0) circle (.3ex);
  \draw[blue,fill=blue] (6.5*\radius,\radius) circle (.3ex);
  \draw[black, very thick] (6.5*\radius,0) -- (8*\radius,0);
  \node at (8.5*\radius,0) {+};
  \draw[black, very thick] (9*\radius,0) -- (10*\radius,0);
  \draw[blue,fill=blue] (10*\radius,0) circle (.3ex);
  \draw[red, very thick] (10*\radius,0) -- (11*\radius,0);
  \draw[blue,fill=blue] (11*\radius,0) circle (.3ex);
  \draw[black, very thick] (11*\radius,0) -- (12*\radius,0);
  \node at (15*\radius,0) {= \;\text{ Two-loop contributions,}};
\end{tikzpicture}
\end{equation}
\begin{equation}\label{fourloop}
    \begin{tikzpicture}[baseline={([yshift=-.5ex]current bounding box.center)},very thick]
    \draw [line width=1pt] (-14.5,0)-- (-14,0);
    \draw[blue,fill=blue] (-14,0) circle (.3ex);
    \draw [red,line width=0.8pt] (-14,0)-- (-13.5,0.5);
    \draw[blue,fill=blue] (-13.5,0.5) circle (.3ex);
    \draw [red,line width=0.8pt] (-14,0)-- (-13.5,-0.5);
    \draw[blue,fill=blue] (-13.5,-0.5) circle (.3ex);
    \draw [red,line width=0.8pt] (-13.5,0.5)-- (-13.5,-0.5);
    \draw [red,line width=0.8pt] (-13.5,0.5)-- (-12.5,0.5);
    \draw [red,line width=0.8pt] (-12.5,0.5)-- (-12.5,-0.5);
    \draw [red,line width=0.8pt] (-13.5,0)-- (-12.5,0);
    \draw[blue,fill=blue] (-13.5,0) circle (.3ex);
    \draw[blue,fill=blue] (-12.5,0) circle (.3ex);
    \draw[blue,fill=blue] (-12.5,0.5) circle (.3ex);
    \draw [red,line width=0.8pt] (-13.5,-0.5)-- (-12.5,-0.5);
    \draw[blue,fill=blue] (-12.5,-0.5) circle (.3ex);
    \draw [red,line width=0.8pt] (-12.5,0.5)-- (-12,0);
    \draw[blue,fill=blue] (-12,0) circle (.3ex);
    \draw[blue,fill=blue] (-12.5,-0.5) circle (.3ex);
    \draw [red,line width=0.8pt] (-12.5,-0.5)-- (-12,0);
    \draw [black,line width=0.8pt] (-12,0)-- (-11.5,0);

    \node at (-11,0) {+};

    \draw [line width=1pt] (-10.5,0)-- (-10,0);
    \draw[blue,fill=blue] (-10,0) circle (.3ex);
    \draw [red,line width=0.8pt] (-10,0)-- (-9.5,0.5);
    \draw [red,line width=0.8pt] (-10,0)-- (-9.5,-0.5);
    \draw [red,line width=0.8pt] (-9.5,0.5)-- (-8.5,0.5);
    \draw [red,line width=0.8pt] (-9.5,-0.5)-- (-8.5,-0.5);
    \draw [red,line width=0.8pt] (-8.5,0.5)-- (-8,0);
    \draw [red,line width=0.8pt] (-8.5,-0.5)-- (-8,0);
    \draw [line width=1pt] (-8,0)-- (-7.5,0);
    \draw [red,line width=0.8pt] (-9.5,0.5)-- (-9,0);
    \draw [red,line width=0.8pt] (-9.5,-0.5)-- (-9,0);
    \draw [red,line width=0.8pt] (-8.5,0.5)-- (-9,0);
    \draw [red,line width=0.8pt] (-8.5,-0.5)-- (-9,0);
    \draw[blue,fill=blue] (-9.5,0.5) circle (.3ex);
    \draw[blue,fill=blue] (-9.5,-0.5) circle (.3ex);
    \draw[blue,fill=blue] (-8.5,0.5) circle (.3ex);
    \draw[blue,fill=blue] (-8.5,-0.5) circle (.3ex);
    \draw[blue,fill=blue] (-9,0) circle (.3ex);
    \draw[blue,fill=blue] (-8,0) circle (.3ex);

    \node at (-7,0) {+};

    \draw [line width=1pt] (-6.5,0)-- (-6,0);
    \draw[blue,fill=blue] (-6,0) circle (.3ex);
    \draw [red,line width=0.8pt] (-6,0)-- (-5.5,0.5);
    \draw[blue,fill=blue] (-5.5,0.5) circle (.3ex);
    \draw [red,line width=0.8pt] (-6,0)-- (-5.5,-0.5);
    \draw[blue,fill=blue] (-5.5,-0.5) circle (.3ex);
    \draw [red,line width=0.8pt] (-5.5,0.5)-- (-5.5,-0.5);
    \draw [red,line width=0.8pt] (-5.5,0.5)-- (-4.5,0.5);
    \draw [red,line width=0.8pt] (-4.5,0.5)-- (-4.5,-0.5);
    \draw [red,line width=0.8pt] (-5,0.5)-- (-5,-0.5);
    \draw [red,line width=0.8pt] (-5.5,-0.5)-- (-4.5,-0.5);
    \draw [red,line width=0.8pt] (-4.5,0.5)-- (-4,0);
    \draw [red,line width=0.8pt] (-4.5,-0.5)-- (-4,0);
    \draw [line width=0.8pt] (-4,-0)-- (-3.5,0);
    \draw[blue,fill=blue] (-5,0.5) circle (.3ex);
    \draw[blue,fill=blue] (-5,-0.5) circle (.3ex);
    \draw[blue,fill=blue] (-4.5,0.5) circle (.3ex);
    \draw[blue,fill=blue] (-4.5,-0.5) circle (.3ex);
    \draw[blue,fill=blue] (-4,0) circle (.3ex);

    \node at (-3,0) {+ $\cdots$};
\end{tikzpicture} = \quad \text{Four-loop contributions.}
\end{equation}
The two-loop and four-loop contributions found inside the term ${\cal Q}_{n,\textbf{EFT}}$ are represented by the aforementioned equations, Eqs.(\ref{twoloop},\ref{fourloop}). It should be noted that many more four-loop and higher-order loop contributions may exist, of which we have only shown a small number of leading order terms; these, however, are more sub-leading than the two-loop contributions provided by the terms $W_{n,*}^{2}$ and $V_{n,*}^{2}$, and as a result, they are indicated inside eqn.(\ref{Qneft}) by using ellipses ($\cdots$).

\subsection{Dynamical Renormalization Group (DRG) resummed scalar power spectrum from EFT}

The purpose of this section is to emphasize how crucial it is to carry out the Dynamical Renormalization Group (DRG) technique \cite{Chen:2016nrs,Baumann:2019ghk,Boyanovsky:1998aa,Boyanovsky:2001ty,Boyanovsky:2003ui,Burgess:2009bs,Dias:2012qy,Chaykov:2022zro,Chaykov:2022pwd}. To successfully capture the quantum effects, the main concept is to fully handle the logarithmically diverging contributions. In order to do this, all of the higher-order loop corrections from a perturbative calculation using scalar modes must be added up in an infinite series. Summing by itself is not necessary, though, unless the series' terms meet strict requirements for convergence on both the super-horizon and horizon crossing scales. Without requiring us to understand the precise behavior of the variables involved in the perturbative expansion, this DRG resummation technique retrieves for us the late time behavior after the convergence is established. The original intention of the Renormalization Group (RG) resummation method was to include the momentum-dependent contributions into the scale-dependent running couplings. This original method is improved upon by the DRG resummation method. The innovation enables to stay inside the tiny coupling domain where the perturbative approximations stay intact within the EFT framework of inflation by expanding the running energy scales to a considerably greater range. at the context of our work, the cosmological $\beta$ functions may be studied because of the DRG resummation at the late-time limit. As a function of the energy scales, the tilt, running, and running of running of the tilt are the well-known characteristics of the scalar power spectrum, and these $\beta$ functions are none other than those features. The primordial scalar power spectrum is not scale invariant, as shown by these physically meaningful entities with a limited existence. As a summary, the DRG resummation approach allows us to know the full series of the logarithmically divergent terms involved without having to worry about the behavior of individual terms involved. This justifies our summing over the secular time and momentum-dependent terms in the perturbation expansion, at the horizon crossing and superhorizon scales.

\subsubsection{For single sharp transition}

We may now safely do the resummation of all higher levels of logarithmic contributions after outlining the basic concept of the DRG technique. We may obtain the power spectrum linked to the scalar modes by resuming it as follows for the case where we have single sharp transition: 
\bea \label{DRG}
\overline{\overline{\Delta_{\zeta,\textbf{EFT}}^{2}(k)}} &=& \bigg[\Delta_{\zeta,\textbf{Tree}}^{2}(k)\bigg]_{\textbf{SR}_{1}}\bigg(1 + {\cal Q}_{\textbf{EFT}} + \frac{1}{2!}{\cal Q}_{\textbf{EFT}}^{2} + \cdots \bigg)\times \bigg\{1+{\cal O}\bigg[\Delta_{\zeta,\textbf{Tree}}^{2}(k_{*})\bigg]^{2}_{\textbf{SR}_{1}}\bigg\} \nonumber\\
&=& \bigg[\Delta_{\zeta,\textbf{Tree}}^{2}(k)\bigg]_{\textbf{SR}_{1}}\exp{\left({\cal Q}_{\textbf{EFT}}\right)}\times \bigg\{1+{\cal O}\bigg[\Delta_{\zeta,\textbf{Tree}}^{2}(k_{*})\bigg]^{2}_{\textbf{SR}_{1}}\bigg\},
\eea
The DRG resummed form of the scalar power spectrum may be summarized as follows in terms of the diagrammatic contributions:
\bea
\overline{\overline{\Delta_{\zeta,\textbf{EFT}}^{2}(k)}} = \begin{tikzpicture}[baseline={([yshift=-.5ex]current bounding box.center)},very thick]
\draw [line width=1pt] (0,0)-- (1,0);
\end{tikzpicture}\times\exp{({\cal F})}\times \bigg\{1+{\cal O}\bigg[\Delta_{\zeta,\textbf{Tree}}^{2}(k_{*})\bigg]^{2}_{\textbf{SR}_{1}}\bigg\},
\eea
which incorporates the tree-level diagram, \begin{tikzpicture}[baseline={([yshift=-.5ex]current bounding box.center)},very thick]
\draw [line width=1pt] (0,0)-- (1,0);
\end{tikzpicture} $\equiv \bigg[\Delta_{\zeta,\textbf{Tree}}^{2}(k)\bigg]_{\textbf{SR}_{1}}$, in addition to the additional Feynman diagrammatic loop representations shown by ${\cal F}$, which may be added together as follows:
\begin{equation}
    \sum {\cal F} =  
\begin{tikzpicture}[baseline={([yshift=-5.5ex]current bounding box.center)},very thick]

  \def\radius{0.5}
  \scalebox{0.5}{
  \draw[red, ultra thick] (3*\radius,0) circle (\radius);
  \draw[red,ultra thick] (5*\radius,0) circle (\radius);
  \draw[red,ultra thick] (13*\radius,\radius) circle (\radius);
  \draw[red,ultra thick] (13*\radius,3*\radius) circle (\radius);
  \draw[red,ultra thick] (21*\radius,0) circle (\radius);}

  \draw[black, very thick] (0,0) -- (\radius,0); 
  \draw[blue,fill=blue] (\radius,0) circle (.3ex);
  \draw[blue,fill=blue] (2*\radius,0) circle (.3ex);
  \draw[blue,fill=blue] (3*\radius,0) circle (.3ex);
  \draw[black, very thick] (3*\radius,0) -- (4*\radius,0);
  \node at (4.5*\radius,0) {+};
  \draw[black, very thick] (5*\radius,0) -- (6.5*\radius,0);
  \draw[blue,fill=blue] (6.5*\radius,0,0) circle (.3ex);
  \draw[blue,fill=blue] (6.5*\radius,\radius) circle (.3ex);
  \draw[black, very thick] (6.5*\radius,0,0) -- (8*\radius,0);
  \node at (8.5*\radius,0) {+};
  \draw[black, very thick] (9*\radius,0) -- (10*\radius,0);
  \draw[blue,fill=blue] (10*\radius,0,0) circle (.3ex);
  \draw[red, ultra thick] (10*\radius,0) -- (11*\radius,0);
  \draw[blue,fill=blue] (11*\radius,0,0) circle (.3ex);
  \draw[black, very thick] (11*\radius,0) -- (12*\radius,0);
  \node at (12.5*\radius,0) {+};
  \draw[black, very thick](13*\radius,0) --(14*\radius,0);
  \draw[blue,fill=blue](14*\radius,0,0) circle (.3ex);
  \draw[red,  thick](14*\radius,0) --(14.5*\radius,\radius);
  \draw[blue,fill=blue](14.5*\radius,\radius) circle (.3ex);
\draw[red, thick](14.5*\radius,\radius) --(15.5*\radius,\radius);
\draw[blue,fill=blue](15.5*\radius,\radius) circle (.3ex);
\draw[red, thick](14*\radius,0) -- (14.5*\radius,-\radius);\draw[blue,fill=blue](14.5*\radius,-\radius) circle (.3ex);
\draw[red,  thick](14.5*\radius,\radius) -- (14.5*\radius,-\radius);
\draw[red, thick](14.5*\radius,-\radius) -- (15.5*\radius,-\radius);
\draw[blue,fill=blue](15.5*\radius,-\radius) circle (.3ex);
\draw[blue,fill=blue](14.5*\radius,0) circle (.3ex);
\draw[red, thick](14.5*\radius,0) -- (15.5*\radius,0);
\draw[red,  thick](15.5*\radius,\radius) -- (15.5*\radius,-\radius);
\draw[blue,fill=blue](15.5*\radius,0) circle (.3ex);
\draw[red, thick](15.5*\radius,\radius) -- (16*\radius,0);
\draw[blue,fill=blue](16*\radius,0) circle (.3ex);
\draw[red, thick](16*\radius,0) -- (15.5*\radius,-\radius);
\draw[black, very thick](16*\radius,0) -- (17*\radius,0);
\node at (17.5*\radius,0) {+};
\draw[black, very thick](18*\radius,0) -- (19*\radius,0);
\draw[blue,fill=blue](19*\radius,0) circle (.3ex);
\draw[red, thick](19*\radius,0) -- (19.5*\radius,\radius);
\draw[red, thick](19*\radius,0) -- (19.5*\radius,-\radius);
\draw[blue,fill=blue](19.5*\radius,\radius) circle (.3ex);
 \draw[blue,fill=blue](19.5*\radius,-\radius) circle (.3ex); 
 \draw[red, thick](19.5*\radius,\radius) -- (20.5*\radius,\radius);
 \draw[red, thick](19.5*\radius,-\radius) -- (20.5*\radius,-\radius);
 \draw[blue,fill=blue](20.5*\radius,\radius) circle (.3ex);
 \draw[blue,fill=blue](20.5*\radius,-\radius) circle (.3ex);
 \draw[red, thick](20.5*\radius,\radius) -- (21*\radius,0);
 \draw[red, thick](20.5*\radius,-\radius) -- (21*\radius,0);
\draw[blue,fill=blue](21*\radius,0) circle (.3ex);
\draw[black,  thick](21*\radius,0) -- (22*\radius,0); 
\draw[red, thick](19.5*\radius,\radius) -- (20.5*\radius,-\radius);
\draw[red, thick](20.5*\radius,\radius) -- (19.5*\radius,-\radius);
\node at (22.5*\radius,0) {+};
\draw[black, thick](23*\radius,0) -- (24*\radius,0);
\draw[red, thick](24*\radius,0) -- (24.5*\radius,\radius);
\draw[red, thick](24*\radius,0) -- (24.5*\radius,-\radius);
\draw[blue,fill=blue](24*\radius,0) circle (.3ex);
\draw[blue,fill=blue](24.5*\radius,\radius) circle (.3ex);
\draw[blue,fill=blue](24.5*\radius,-\radius) circle (.3ex);
\draw[red, thick](24.5*\radius,\radius) -- (25.5*\radius,\radius);
\draw[red,  thick](24.5*\radius,-\radius) -- (25.5*\radius,-\radius);
\draw[red,  thick](25.5*\radius,\radius) -- (26*\radius,0);
\draw[red, thick](26*\radius,0) -- (25.5*\radius,-\radius);
\draw[blue,fill=blue](25.5*\radius,\radius) circle (.3ex);
\draw[blue,fill=blue](26*\radius,0) circle (.3ex);
\draw[blue,fill=blue](25.5*\radius,-\radius) circle (.3ex);
\draw[red,  thick](24.5*\radius,\radius) -- (24.5*\radius,-\radius);
\draw[red, thick](25*\radius,\radius) -- (25*\radius,-\radius);
\draw[red, thick](25.5*\radius,\radius) -- (25.5*\radius,-\radius);
\draw[blue,fill=blue](25*\radius,\radius) circle (.3ex);
\draw[blue,fill=blue](25*\radius,-\radius) circle (.3ex);
\draw[black, thick](26*\radius,0) -- (27*\radius,0); 
\node at (28*\radius,0) {+};
\node at (29*\radius,0) {...};

\end{tikzpicture}
\end{equation}
After applying the stringent convergence conditions in terms of the quantity meeting $|{\cal Q}_{\textbf{EFT}}| \ll 1$, the above equation is the outcome.  It is important to note that the above formula is written assuming that there is only one sharp transition $(n=1)$. The logarithmic IR divergent contributions that are present in the term ${\cal Q}_{\textbf{EFT}}$, which originates from each phase that has a sharp transition in the momentum space, are covered by the DRG resummed version as discussed above. When compared with the renormalized scalar power spectrum version previously stated, such logarithmic contributions are further mitigated by doing the DRG operation. The final resummed form of the one-loop corrected scalar power spectrum may be obtained by equivalently adding the resummed versions corresponding to each sharp transition case, as we are dealing with numerous sharp transitions in this context. The infinite series resummation's contributing terms are all similar to the function of an even-order quantum loop correction found in a perturbative series. Remarkably, we can package their effects to get an exponential function that represents a finite sum without explicitly computing any higher-order terms.

Now that the functions $U$ and $V$ have been defined before in Eqs (\ref{Utot},\ref{V}), let us use their one-loop results to depict this softening. The term ${\cal Q}_{n,\textbf{EFT}}$ from the $n$th sharp transition phase will be used to derive the forthcoming analysis of this softening. The component $(k_{e_{n}}/k_{s_{n}})^{6}\ln{(k_{e_{n}}/k_{s_{n}})}$ in $V$ predominates relative to other factors, thus for the time being, we disregard the function $U$. For the function, these approximations provide the subsequent result:
\bea
{\cal Q}_{n,\textbf{EFT}} \approx -\mathbf{S}(k)\times\small[V_{n,*}^{2}+\cdots\small] = -\mathbf{S}(k)\times\small[\chi_{n,*}({\cal T}^{(n)}_{1,*})^{2}+\cdots\small]  
\eea
when the following new functions are introduced:  
\bea
\mathbf{S}(k) &=& \frac{\bigg[\Delta_{\zeta,\textbf{Tree}}^{2}(k)\bigg]_{\textbf{SR}_{1}}}{\bigg[\Delta_{\zeta,\textbf{Tree}}^{2}(k_{*})\bigg]_{\textbf{SR}_{1}}}, \\
{\cal T}^{(n)}_{1,*} &\approx& \frac{1}{4}\bigg[\Delta_{\zeta,\textbf{Tree}}^{2}(k_{*})\bigg]_{\textbf{SR}_{1}}\bigg(\frac{\Delta\eta(\tau_{e_{n}})}{\tilde{c}^{4}_{s}}\bigg)^{2}\bigg(\frac{k_{e_n}}{k_{s_n}}\bigg)^{6}\times\ln{\left(\frac{k_{e_{n}}}{k_{s_{n}}}\right)}, \\
\chi_{n,*} &=& 1 + \frac{{\cal T}^{(n)}_{2,*}}{{\cal T}^{(n)}_{1,*}},
\eea  
and an additional justification is used to make the phrase ${\cal T}^{(n)}_{1,*}$ simpler:
\bea\displaystyle{\left(\Delta\eta(\tau_{e_{n}})\right)^{2}\bigg(\frac{k_{e_n}}{k_{s_n}}\bigg)^{6}} \gg \left(\Delta\eta(\tau_{s_{n}})\right)^{2}.\eea
The ultimate simplified version of the word is as follows, derived from the foregoing approximations: 
\bea
{\cal Q}_{n,\textbf{EFT}} \approx -\delta_{n,*}\ln^{2}{\left(\frac{k_{e_{n}}}{k_{s_{n}}}\right)},
\eea
where the factor $\delta_{n,*}$ is defined as: 
\bea
\delta_{n,*} = \frac{\chi_{n,*}}{4}\bigg[\Delta_{\zeta,\textbf{Tree}}^{2}(k_{*})\bigg]_{\textbf{SR}_{1}}\bigg(\frac{\Delta\eta(\tau_{e_{n}})}{\tilde{c}^{4}_{s}}\bigg)^{2}\bigg(\frac{k_{e_n}}{k_{s_n}}\bigg)^{6} \ll 1.
\eea
A detailed examination of the aforementioned equation explains why the final exponentiated result, as in eqn.(\ref{DRG}), further softens the divergent logarithmic IR contributions when the DRG resummed result is used. We give a condensed version of the preceding expression in eqn.(\ref{DRG}) for a single abrupt transition $(n=1)$ after using the previously described approximations:
\bea \label{simpleDRG}
\overline{\overline{\Delta_{\zeta,\textbf{EFT}}^{2}(k)}} \approx \bigg[\Delta_{\zeta,\textbf{Tree}}^{2}(k)\bigg]_{\textbf{SR}_{1}}\left(\frac{k_{e_{1}}}{k_{s_{1}}}\right)^{-2.3\delta_{*}}\times\bigg\{1+{\cal O}\bigg[\Delta_{\zeta,\textbf{Tree}}^{2}(k_{*})\bigg]^{2}_{\textbf{SR}_{1}}\bigg\}.
\eea

\subsubsection{For multiple sharp transitions}

To recapitulate the current context of the discussion, we have $n$ sharp transitions in our MST setup. The location of a single USR phase determines the removal of its associated UV divergences and the softening of components in ${\cal Q}_{n,\textbf{EFT}}$ by IR contributions. But since the current scenario also takes into account a background with a constant EoS parameter $w$, it is uncertain whether the characteristics found in the final quantum loop corrected renormalized DRG resummed scalar power spectrum amplitude will be retained in the later portion of the induced gravitational wave spectrum. We thus propose to use phenomenological coarse-graining factors to guarantee that there is no violation of the perturbative approximations across all abrupt transitions. The coarse-graining factors, which must be adjusted specifically for each phase, are equal to the total number of abrupt transitions that are present. However, it's important to realize that the more reliant one becomes on these coarse-graining parameters, the less reliable the underlying setup's forecasts are. With the exception of the initial abrupt transition, our present configuration necessitates the usage of five such coarse-graining factors, whose behavior varies depending on the parameter $w$. Generally speaking, the renormalized scalar power spectrum corresponding to the $n$th transition is modified to yield the following in the final DRG resummed version: 
\bea \label{drgfinal}
\overline{\overline{\Delta_{\zeta,\textbf{EFT}}^{2}(k)}} &=& \bigg[\Delta_{\zeta,\textbf{Tree}}^{2}(k)\bigg]_{\textbf{SR}_{1}} \bigg[\left({\cal Q}_{1,\textbf{EFT}} + \frac{1}{2!}{\cal Q}^{2}_{1,\textbf{EFT}} + \cdots\right) +  \Theta(k-k_{s_{2}})g_{2}(w)\left({\cal Q}_{2,\textbf{EFT}} + \frac{1}{2!}{\cal Q}^{2}_{2,\textbf{EFT}} + \cdots \right) \nonumber\\
&+& \Theta(k-k_{s_{3}})g_{3}(w)\left({\cal Q}_{3,\textbf{EFT}} + \frac{1}{2!}{\cal Q}^{2}_{3,\textbf{EFT}} + \cdots\right) + \cdots + \Theta(k-k_{s_{n}})g_{n}(w)\left({\cal Q}_{n,\textbf{EFT}} + \frac{1}{2!}{\cal Q}^{2}_{n,\textbf{EFT}} + \cdots\right) \bigg]\nonumber\\
&& \quad\quad\quad\quad\quad\quad\quad\quad\quad\quad\quad\quad\quad\quad\quad\quad\quad\quad\quad\quad\quad\quad\quad \times \bigg\{1+{\cal O}\bigg[\Delta_{\zeta,\textbf{Tree}}^{2}(k_{*})\bigg]^{2}_{\textbf{SR}_{1}}\bigg\}, \nonumber\\
&=& \bigg[\Delta_{\zeta,\textbf{Tree}}^{2}(k)\bigg]_{\textbf{SR}_{1}} \left[\exp{\left({\cal Q}_{1,\textbf{EFT}}\right)} + \sum_{n=2}^{N}\Theta(k-k_{s_{n}})g_{n}(w)\exp{\left({\cal Q}_{n,\textbf{EFT}}\right)} \right]\times \bigg\{1+{\cal O}\bigg[\Delta_{\zeta,\textbf{Tree}}^{2}(k_{*})\bigg]^{2}_{\textbf{SR}_{1}}\bigg\},\nonumber\quad\quad \\
&=& \begin{tikzpicture}[baseline={([yshift=-.5ex]current bounding box.center)},very thick]
\draw [line width=1pt] (0,0)-- (1,0);
\end{tikzpicture}\times \left[\exp{({\cal F}_{1})} + \sum_{n=2}^{N}\Theta(k-k_{s_{n}})g_{n}(w)\exp{({\cal F}_{n})} \right]\times \bigg\{1+{\cal O}\bigg[\Delta_{\zeta,\textbf{Tree}}^{2}(k_{*})\bigg]^{2}_{\textbf{SR}_{1}}\bigg\},\quad\quad
\eea
where the diagrammatic form was once more utilized for the tree-level contribution, 
 \begin{tikzpicture}[baseline={([yshift=-.5ex]current bounding box.center)},very thick]
\draw [line width=1pt] (0,0)-- (1,0);
\end{tikzpicture} $\equiv \bigg[\Delta_{\zeta,\textbf{Tree}}^{2}(k)\bigg]_{\textbf{SR}_{1}}$, and for each contribution ${\cal Q}_{n,\textbf{EFT}}$, with the following diagrammatic relations ${\cal F}_n$, we are left with:
\begin{equation}
    \sum {\cal F}_n =  
\begin{tikzpicture}[baseline={([yshift=-5.5ex]current bounding box.center)},very thick]

  \def\radius{0.5}
  \scalebox{0.5}{
  \draw[red, ultra thick] (3*\radius,0) circle (\radius);
  \draw[red,ultra thick] (5*\radius,0) circle (\radius);
  \draw[red,ultra thick] (13*\radius,\radius) circle (\radius);
  \draw[red,ultra thick] (13*\radius,3*\radius) circle (\radius);
  \draw[red,ultra thick] (21*\radius,0) circle (\radius);}

  \draw[black, very thick] (0,0) -- (\radius,0); 
  \draw[blue,fill=blue] (\radius,0) circle (.3ex);
  \draw[blue,fill=blue] (2*\radius,0) circle (.3ex);
  \draw[blue,fill=blue] (3*\radius,0) circle (.3ex);
  \draw[black, very thick] (3*\radius,0) -- (4*\radius,0);
  \node at (4.5*\radius,0) {+};
  \draw[black, very thick] (5*\radius,0) -- (6.5*\radius,0);
  \draw[blue,fill=blue] (6.5*\radius,0,0) circle (.3ex);
  \draw[blue,fill=blue] (6.5*\radius,\radius) circle (.3ex);
  \draw[black, very thick] (6.5*\radius,0,0) -- (8*\radius,0);
  \node at (8.5*\radius,0) {+};
  \draw[black, very thick] (9*\radius,0) -- (10*\radius,0);
  \draw[blue,fill=blue] (10*\radius,0,0) circle (.3ex);
  \draw[red, ultra thick] (10*\radius,0) -- (11*\radius,0);
  \draw[blue,fill=blue] (11*\radius,0,0) circle (.3ex);
  \draw[black, very thick] (11*\radius,0) -- (12*\radius,0);
  \node at (12.5*\radius,0) {+};
  \draw[black, very thick](13*\radius,0) --(14*\radius,0);
  \draw[blue,fill=blue](14*\radius,0,0) circle (.3ex);
  \draw[red,  thick](14*\radius,0) --(14.5*\radius,\radius);
  \draw[blue,fill=blue](14.5*\radius,\radius) circle (.3ex);
\draw[red, thick](14.5*\radius,\radius) --(15.5*\radius,\radius);
\draw[blue,fill=blue](15.5*\radius,\radius) circle (.3ex);
\draw[red, thick](14*\radius,0) -- (14.5*\radius,-\radius);\draw[blue,fill=blue](14.5*\radius,-\radius) circle (.3ex);
\draw[red,  thick](14.5*\radius,\radius) -- (14.5*\radius,-\radius);
\draw[red, thick](14.5*\radius,-\radius) -- (15.5*\radius,-\radius);
\draw[blue,fill=blue](15.5*\radius,-\radius) circle (.3ex);
\draw[blue,fill=blue](14.5*\radius,0) circle (.3ex);
\draw[red, thick](14.5*\radius,0) -- (15.5*\radius,0);
\draw[red,  thick](15.5*\radius,\radius) -- (15.5*\radius,-\radius);
\draw[blue,fill=blue](15.5*\radius,0) circle (.3ex);
\draw[red, thick](15.5*\radius,\radius) -- (16*\radius,0);
\draw[blue,fill=blue](16*\radius,0) circle (.3ex);
\draw[red, thick](16*\radius,0) -- (15.5*\radius,-\radius);
\draw[black, very thick](16*\radius,0) -- (17*\radius,0);
\node at (17.5*\radius,0) {+};
\draw[black, very thick](18*\radius,0) -- (19*\radius,0);
\draw[blue,fill=blue](19*\radius,0) circle (.3ex);
\draw[red, thick](19*\radius,0) -- (19.5*\radius,\radius);
\draw[red, thick](19*\radius,0) -- (19.5*\radius,-\radius);
\draw[blue,fill=blue](19.5*\radius,\radius) circle (.3ex);
 \draw[blue,fill=blue](19.5*\radius,-\radius) circle (.3ex); 
 \draw[red, thick](19.5*\radius,\radius) -- (20.5*\radius,\radius);
 \draw[red, thick](19.5*\radius,-\radius) -- (20.5*\radius,-\radius);
 \draw[blue,fill=blue](20.5*\radius,\radius) circle (.3ex);
 \draw[blue,fill=blue](20.5*\radius,-\radius) circle (.3ex);
 \draw[red, thick](20.5*\radius,\radius) -- (21*\radius,0);
 \draw[red, thick](20.5*\radius,-\radius) -- (21*\radius,0);
\draw[blue,fill=blue](21*\radius,0) circle (.3ex);
\draw[black,  thick](21*\radius,0) -- (22*\radius,0); 
\draw[red, thick](19.5*\radius,\radius) -- (20.5*\radius,-\radius);
\draw[red, thick](20.5*\radius,\radius) -- (19.5*\radius,-\radius);
\node at (22.5*\radius,0) {+}; 
\draw[black, thick](23*\radius,0) -- (24*\radius,0);
\draw[red, thick](24*\radius,0) -- (24.5*\radius,\radius);
\draw[red, thick](24*\radius,0) -- (24.5*\radius,-\radius);
\draw[blue,fill=blue](24*\radius,0) circle (.3ex);
\draw[blue,fill=blue](24.5*\radius,\radius) circle (.3ex);
\draw[blue,fill=blue](24.5*\radius,-\radius) circle (.3ex);
\draw[red, thick](24.5*\radius,\radius) -- (25.5*\radius,\radius);
\draw[red,  thick](24.5*\radius,-\radius) -- (25.5*\radius,-\radius);
\draw[red,  thick](25.5*\radius,\radius) -- (26*\radius,0);
\draw[red, thick](26*\radius,0) -- (25.5*\radius,-\radius);
\draw[blue,fill=blue](25.5*\radius,\radius) circle (.3ex);
\draw[blue,fill=blue](26*\radius,0) circle (.3ex);
\draw[blue,fill=blue](25.5*\radius,-\radius) circle (.3ex);
\draw[red,  thick](24.5*\radius,\radius) -- (24.5*\radius,-\radius);
\draw[red, thick](25*\radius,\radius) -- (25*\radius,-\radius);
\draw[red, thick](25.5*\radius,\radius) -- (25.5*\radius,-\radius);
\draw[blue,fill=blue](25*\radius,\radius) circle (.3ex);
\draw[blue,fill=blue](25*\radius,-\radius) circle (.3ex);
\draw[black, thick](26*\radius,0) -- (27*\radius,0); 
\node at (28*\radius,0) {+ $\cdots$};
\end{tikzpicture}
\end{equation}
where the coarse-graining of the amplitude is carried out by the factors $g_{n}(w)$ in order to preserve the perturbative approximations across the whole wavenumber range. It should be noted that the above equation's notation modifies a comparable number from eqn.(\ref{simpleDRG}) for a single $(n=1)$ transition; as a result, in further discussions, the notation will refer to the current updated version.The first equality shows that, when a Heaviside theta function $\Theta(k-k_{s_{n}})$ acts at the scale of the $n$th transition $k_{s_{n}}$, we may aggregate the DRG resummed outcomes arising from each phase of sharp transition. From the total of $N$ such phases, note the $n$ and $w$ dependency in the components $g_{n}(w)$ for each contribution of the sharp transition phase. 
\bea \label{simpleDRGn}
\overline{\overline{\Delta_{\zeta,\textbf{EFT}}^{2}(k)}} &\approx& \bigg[\Delta_{\zeta,\textbf{Tree}}^{2}(k)\bigg]_{\textbf{SR}_{1}} \bigg[\left(\frac{k_{e_{1}}}{k_{s_{1}}}\right)^{-2.3\delta_{1,*}} + \Theta(k-k_{s_{2}})g_{2}(w)\left(\frac{k_{e_{2}}}{k_{s_{2}}}\right)^{-2.3\delta_{2,*}} + \cdots \nonumber\\
&& \quad\quad\quad\quad\quad\quad\quad\quad\quad \cdots + \Theta(k-k_{s_{n}})g_{n}(w)\left(\frac{k_{e_{n}}}{k_{s_{n}}}\right)^{-2.3\delta_{n,*}} \bigg]
\times \bigg\{1+{\cal O}\bigg[\Delta_{\zeta,\textbf{Tree}}^{2}(k_{*})\bigg]^{2}_{\textbf{SR}_{1}}\bigg\},\nonumber\\
&=& \bigg[\Delta_{\zeta,\textbf{Tree}}^{2}(k)\bigg]_{\textbf{SR}_{1}} \bigg[\left(\frac{k_{e_{1}}}{k_{s_{1}}}\right)^{-2.3\delta_{1,*}} + \sum_{n=1}^{N}\Theta(k-k_{s_{n}})g_{n}(w)\left(\frac{k_{e_{n}}}{k_{s_{n}}}\right)^{-2.3\delta_{n,*}}\bigg] \nonumber \\ 
&& \quad \quad \quad \quad \quad \quad \quad \quad \quad \quad \quad \quad \quad \quad \quad \quad \quad \quad \quad \quad \quad \quad \quad \quad \quad \quad  \times
\bigg\{1+{\cal O}\bigg[\Delta_{\zeta,\textbf{Tree}}^{2}(k_{*})\bigg]^{2}_{\textbf{SR}_{1}}\bigg\}.\quad\quad
\eea

\begin{table}[H]\label{tab1xcc}
\centering 
\begin{tabular}{|l|l|l|}

\hline\hline
\multicolumn{3}{|l|}{\normalsize \textbf{Behaviour of the EoS dependent coarse-graining $g_{n}(w)$}} \\

\hline

\textbf{EoS parameter} & \textbf{Coarse-graining} & \textbf{Condition on $g_{n}(w)$} \\
\hline
$w=1/3$ & Zero & $g_{n}(w=1/3)=1$  \\  \hline

$1/3 \leq w \leq 1$ & Negative & $g_{n+1}(w) < g_{n}(w)$                                           \\ \hline
$0< w < 1/3$ & Positive & $g_{n+1}(w) > g_{n}(w)$                                          \\ \hline
$-0.05 < w \leq 0$ & Positive & $g_{n+1}(w) > g_{n}(w)$                                          \\ \hline
$-0.55 \leq w \leq -0.05$ & Positive & $g_{n+1}(w) > g_{n}(w)$                                       \\ \hline\hline

\end{tabular}
\caption{The type of coarse-graining and how it has been calculated for different EoS metrics.
}
\end{table}
While maintaining our attention on four particular scenarios involving $w=1/3\;,1/3 < w \leq 1\;,0 < w < 1/3\;,{\rm and}\;w \leq 0$ with two subcases, we will here emphasize the nature of the phenomenological coarse-graining parameters dependent on the background EoS $w$. We next go on to illustrate each scenario in the following sequence. The signature of the coarse-graining factor is displayed in Table \ref{tab1xcc}. Applying this factor to the power spectrum for different $w$ scenarios raises it to a level required to maintain perturbativity and obtain the ideal amplitude for obtaining a feasible PBH mass fraction and producing SIGWs consistent with the NANOGrav 15 signal. Point-by-point, the details of the related conversations and their tangible results are attached below:
    \begin{figure*}[ht!]
    	\centering
   {
   \includegraphics[width=18.5cm,height=8.5cm] {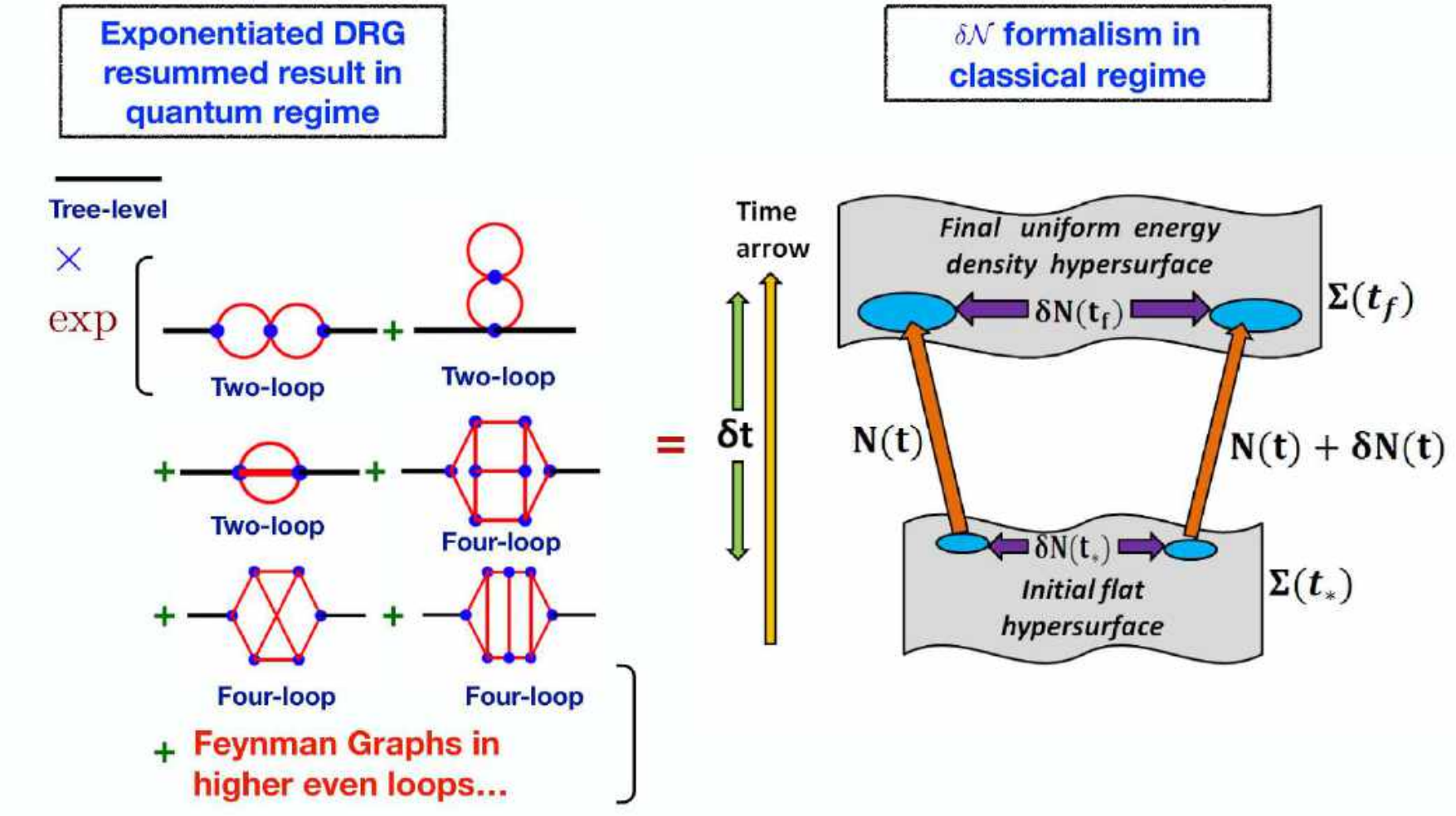}
    }
    	\caption[Optional caption for list of figures]{Illustration demonstrating the equivalentity of the $\delta{\cal N}$ formalism and the DRG resummation technique.
 } 
    	\label{DRGDelN}
    \end{figure*}
\begin{itemize}

\item[$\blacksquare$] \underline{$ w=1/3:$} \\ \\
As will be seen later in subsection \ref{s5}, this situation is argued here to be the most favorable scenario in terms of meeting the perturbative approximations and providing us with results that are closest to the findings from experiments. The key takeaway for this situation is that dealing with $w=1/3$ does not necessitate coarse-graining. This suggests that, for this situation, the final eqn. (\ref{drgfinal}) changes as follows:
\bea \label{finaldrgps1} \overline{\overline{\Delta_{\zeta,\textbf{EFT}}^{2}(k)}} &=& \bigg[\Delta_{\zeta,\textbf{Tree}}^{2}(k)\bigg]_{\textbf{SR}_{1}} \left[\exp{\left({\cal Q}_{1,\textbf{EFT}}\right)} + \sum_{n=2}^{N}\Theta(k-k_{s_{n}})\exp{\left({\cal Q}_{n,\textbf{EFT}}\right)} \right]\nonumber\\
&&\quad\quad\quad\quad\quad\quad\quad\quad\quad\quad\quad\quad\quad\quad\quad\quad\quad\quad\quad\quad\times \bigg\{1+{\cal O}\bigg[\Delta_{\zeta,\textbf{Tree}}^{2}(k_{*})\bigg]^{2}_{\textbf{SR}_{1}}\bigg\},
        \quad\quad \eea 
where $g_{n}(w=1/3)=1$ is set for each coarse-graining factor for the $n$ abrupt transitions in the example above. If $g_{n}(w=1/3)=1$ is deviated from for any $n$, it will indicate limited coarse-graining characteristics. This indicates that there is no coarse-graining. Next, the aforementioned equation is expressed in its simplified form as follows:
\bea
            \overline{\overline{\Delta_{\zeta,\textbf{EFT}}^{2}(k)}} &=& \bigg[\Delta_{\zeta,\textbf{Tree}}^{2}(k)\bigg]_{\textbf{SR}_{1}} \bigg[\left(\frac{k_{e_{1}}}{k_{s_{1}}}\right)^{-2.3\delta_{1,*}} + \sum_{n=1}^{N}\Theta(k-k_{s_{n}})\left(\frac{k_{e_{n}}}{k_{s_{n}}}\right)^{-2.3\delta_{n,*}}\bigg]\nonumber\\
&&\quad\quad\quad\quad\quad\quad\quad\quad\quad\quad\quad\quad\quad\quad\quad\quad\quad\quad\quad\quad\times
            \bigg\{1+{\cal O}\bigg[\Delta_{\zeta,\textbf{Tree}}^{2}(k_{*})\bigg]^{2}_{\textbf{SR}_{1}}\bigg\}.
        \quad\quad \eea
        A hypothesis loses some of its reliability when it incorporates an excessive amount of coarse-graining. No coarse-graining of any type is present in this instance. We shall show via our results that this specific modification for $w=1/3$ fits with results from this instance when the constant scalar power spectrum amplitude property is kept in the produced SIGWs. Based on the results section \ref{s5}, we believe that $w=1/3$ is the optimal scenario to produce a solid theoretical prediction, as it does not need coarse-graining to describe the observationally relevant phenomena of GW experiments and successfully address the overproduction issue. 
        
\item[$\blacksquare$] \underline{$1/3 < w \leq 1:$} \\ \\
As we increase the amount of abrupt transitions for greater wavenumbers, we see in this case a distinct characteristic required for the coarse-graining factors. It can be observed that the peak amplitude is continuously growing, which might result in the perturbative approximations being broken and an excess of PBHs being produced. The total peak amplitude of the $n$th transition must be suppressed by the coarse-graining factors, $g_{n}(w;1/3 < w \leq 1)$, in order to prevent such scenarios. It should specifically behave as follows: for $1/3 < w \leq 1$ and $n=1\;{\rm to}\;N$ with $N$ total transitions, $g_{n+1}(w) < g_{n}(w)$. Most of the negative coarse-graining impact is seen at the final peak.

\item[$\blacksquare$]  \underline{$0 < w < 1/3:$} \\ \\
The opposite aspect from what we saw in the prior situation is needed in this scenario. The perturbative approximations inside each abrupt transition phase remain intact in this case since the peak amplitude continues to decrease, and overproduction does not seem to be a problem. Nonetheless, in order to clarify the fingerprints found in the range examined by the NANOGrav-15 data and further GW experiments detecting at higher frequencies, $f \in {\cal O}(10^{-5}-10^{5}){\rm Hz}$, peak amplitude in that regime is not significant. Thus, for $0 < w < 1/3$ and $n=1\;{\rm to}\;N$ with $N$ total transitions, a ``positive'' coarse-graining feature is needed such that, $g_{n+1}(w) > g_{n}(w)$. The initial peak amplitude needs the least amount to meet the NANOGrav-15 data, while the last peak has the most positive coarse-graining impact.

\item[$\blacksquare$]  \underline{$-0.05 < w \leq 0:$} \\ \\
An intriguing instance is presented by this circumstance. It is unexpected that the peak amplitude is only bigger than the $w=0$ condition for $-0.05 < w < 0$. This can be demonstrated by looking at the findings in section \ref{s5}. Therefore, in contrast, for every $n=1\;{\rm to}\;N$ transition, we must have $g_{n}(-0.05 < w < 0) < g_{n}(w=0)$. Once again, if obtaining observable signatures is the main objective, then this scenario falls under "positive" coarse-graining since the peak amplitudes tend to decline rapidly with an increasing number of transitions, requiring more fine-tuning for the last peak than the first. Although the amplitude in this instance is somewhat smaller than in others, the perturbativity is still present. However, when considering the situation through the eyes of the experimental signatures, it is not very desired.

\item[$\blacksquare$] \underline{$-0.55 \leq w \leq -0.05:$} \\ \\
Overall, the requirements are comparable to the third case for this final remaining period, as permitted by the linearity approximations. When we move lower in $w$, the peak amplitudes need to be fine-tuned since they continue to decrease in magnitude with increasing transitions. This also applies to the "positive" reduction of the values of $w$ and coarse-graining makes the forecasts less reliable.

\end{itemize}

A great way to find the time dependence of the cosmological $n$-point correlation functions is to use the separate universe approach, also known as the $\delta {\cal N}$ formalism to perturbation theory \cite{Dias:2012qy, Burgess:2009bs, Burgess:2014eoa, Burgess:2015ajz, Chaykov:2022zro, Chaykov:2022pwd, Jackson:2023obv}. These correlations are assessed later and end up becoming our observables. In order to compute these observables with any degree of accuracy, one has to know a set of beginning conditions that depend on the subhorizon quantum fluctuations, which turn classical when they cross the horizon. Due to its large-scale nature, the distinct universe technique presents a challenge in obtaining this information. The claims arising from this method for the cosmological correlators assessed at late periods, $k/aH \ll 1$, emphasize that correlators of this kind may be expressed in terms of their value at an earlier horizon-crossing moment, and associated with certain coefficients. As demonstrated subsequently in \cite{Dias:2012qy}, the coefficients in question provide information on the late-time divergent IR adjustments to the required correlators at the lowest order. The most robust method for determining the dynamics of such coefficients is to use the Callan-Symanzik equations, which form the foundation of quantum field theory. We note that the previously described coefficients bearing the IR divergences are plotted as form factors that arise in QCD computations \cite{Dias:2012qy}. In order to provide a finite result and account for the infinite divergent contributions from the late time restriction in the correlators, resummation is thus required. The effects from large scales inside the aforesaid coefficients may be packaged using the renormalization group equations (RGE), and the initial conditions to solve such equations are created at horizon-crossing time by the same correlators. In the end, using just quantum field theoretic techniques, the dynamical renormalization group (DRG) analysis results in an all-order reconstruction of the correlation functions, which were previously necessary from the separate universe approach. We can now turn our attention to the cosmological $\beta$ functions, which are the real physical quantities that have already been released from the one-loop IR divergences at pivot scale following power-spectrum renormalization. These functions are the spectral tilt, running, and running of the running of spectral tilt with the momentum scales. According to our findings, this equivalency leads to an enhancement in the trustworthiness of the $\delta{\cal N}$ method in the superhorizon scales when the results produced from DRG resummation match the observational data. A comparable approach to the logarithmic IR divergences covered in the previous paragraph is demonstrated by Eqn.(\ref{DRG}). Prior to engaging in the final exponentiated form, each term in the series expansion represents the contribution from all even-order loop correction terms. The terms that represent the contributions from each even-order loop correction term are as follows: ${\cal Q}_{\textbf{EFT}}$, which resembles a two-loop contribution; ${\cal Q}_{\textbf{EFT}}^{2}$, which represents a four-loop contribution; and so forth. This behavior demonstrates how the knowledge of the lowest-order terms in the perturbative expansion provided by the DRG resummation enables an all-order reconstruction of the correlations. In the end, we get a far softer version of the logarithmic infrared divergent contributions than what the power spectrum renormalization back in Eqn.(\ref{Qneft}) did. Figure (\ref{DRGDelN}) presents a figure that illustrates this demonstrated equivalency between the two methodologies (DRG and $\delta{\cal N}$).

\subsection{Numerical results: Studying further constraints on PBH mass from causality from EFT framework
}
\subsubsection{For single sharp transition}
    \begin{figure*}[htb!]
    	\centering
    	\subfigure[For $c_s=0.6(<1)$  with $M^4_2/\dot{H}M^2_p\sim -0.89$ (non-canonical and causal).]{
      	\includegraphics[width=8cm,height=6cm] {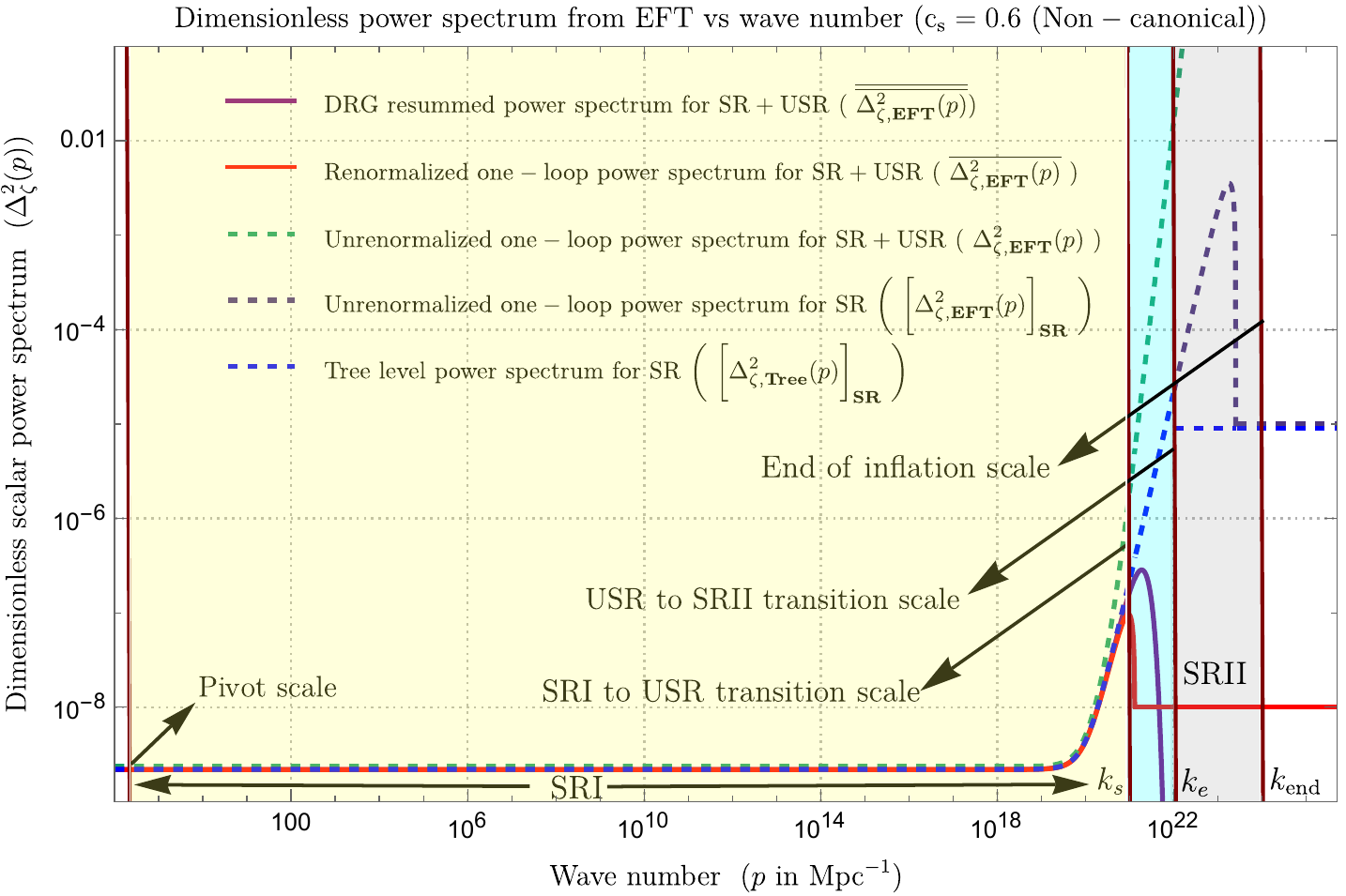}
        \label{G1}
    }
    \subfigure[For $c_s=1$  with $M^4_2/\dot{H}M^2_p\sim 0$ (canonical and causal).]{
       \includegraphics[width=8cm,height=6cm] {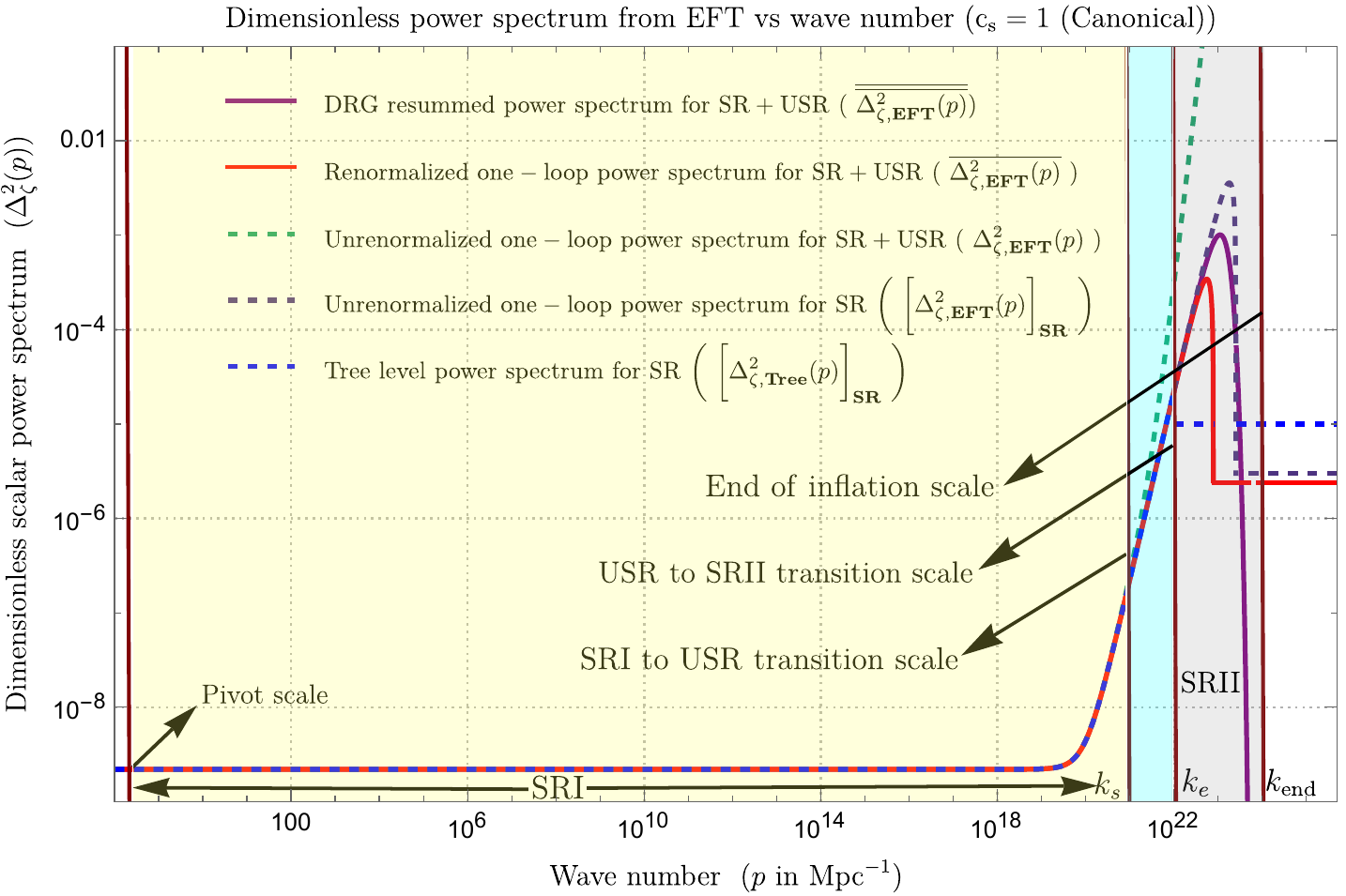}
        \label{G2}
       }
        \subfigure[For $c_s=1.17(>1)$  with $M^4_2/\dot{H}M^2_p\sim 0.13$ (non-canonical and a-causal).]{
       \includegraphics[width=8cm,height=6cm] {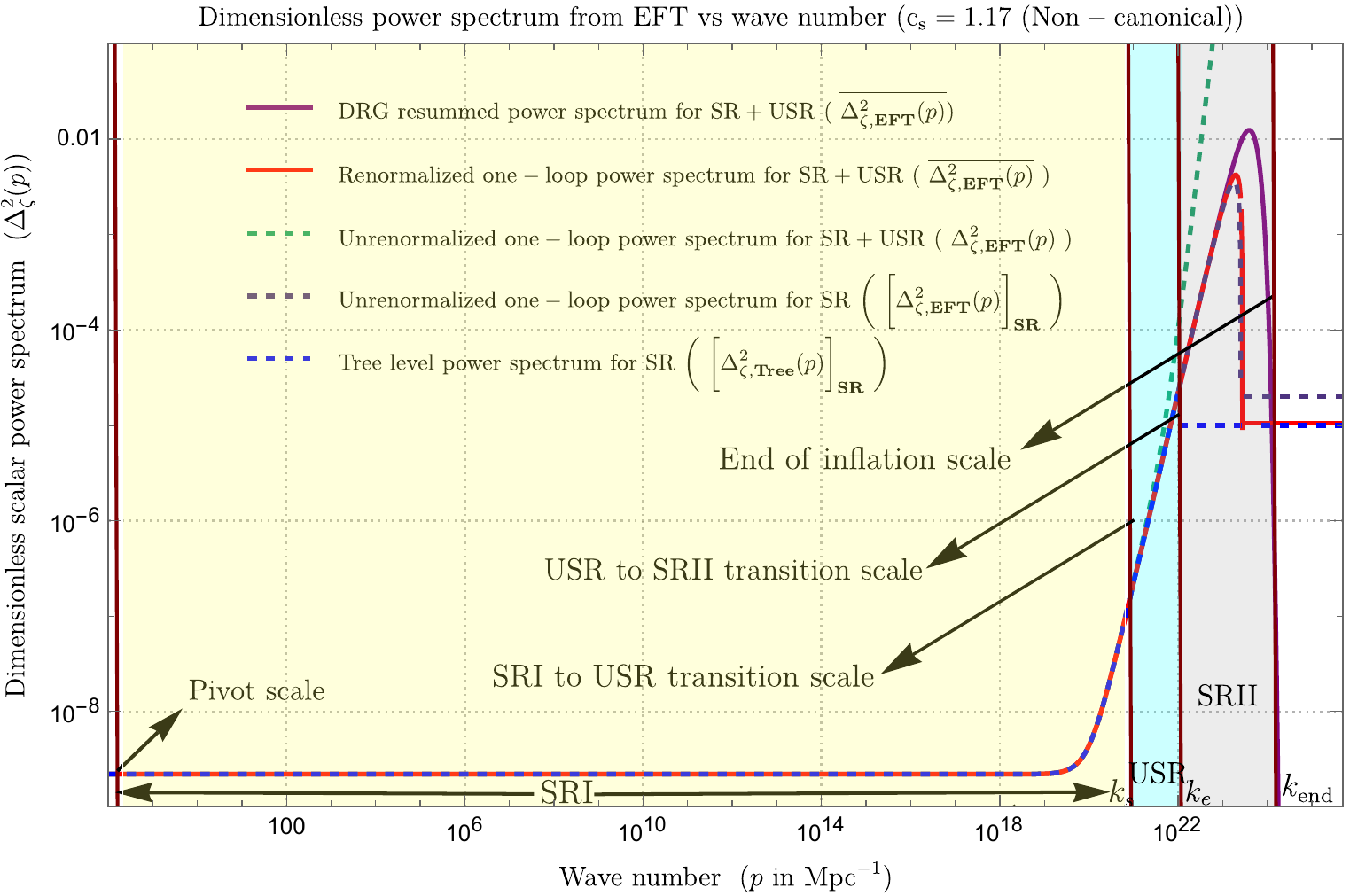}
        \label{G3}
       }
       \subfigure[For $c_s=1.5(>1)$  with $M^4_2/\dot{H}M^2_p\sim 0.28$ (non-canonical and a-causal).]{
       \includegraphics[width=8cm,height=6cm] {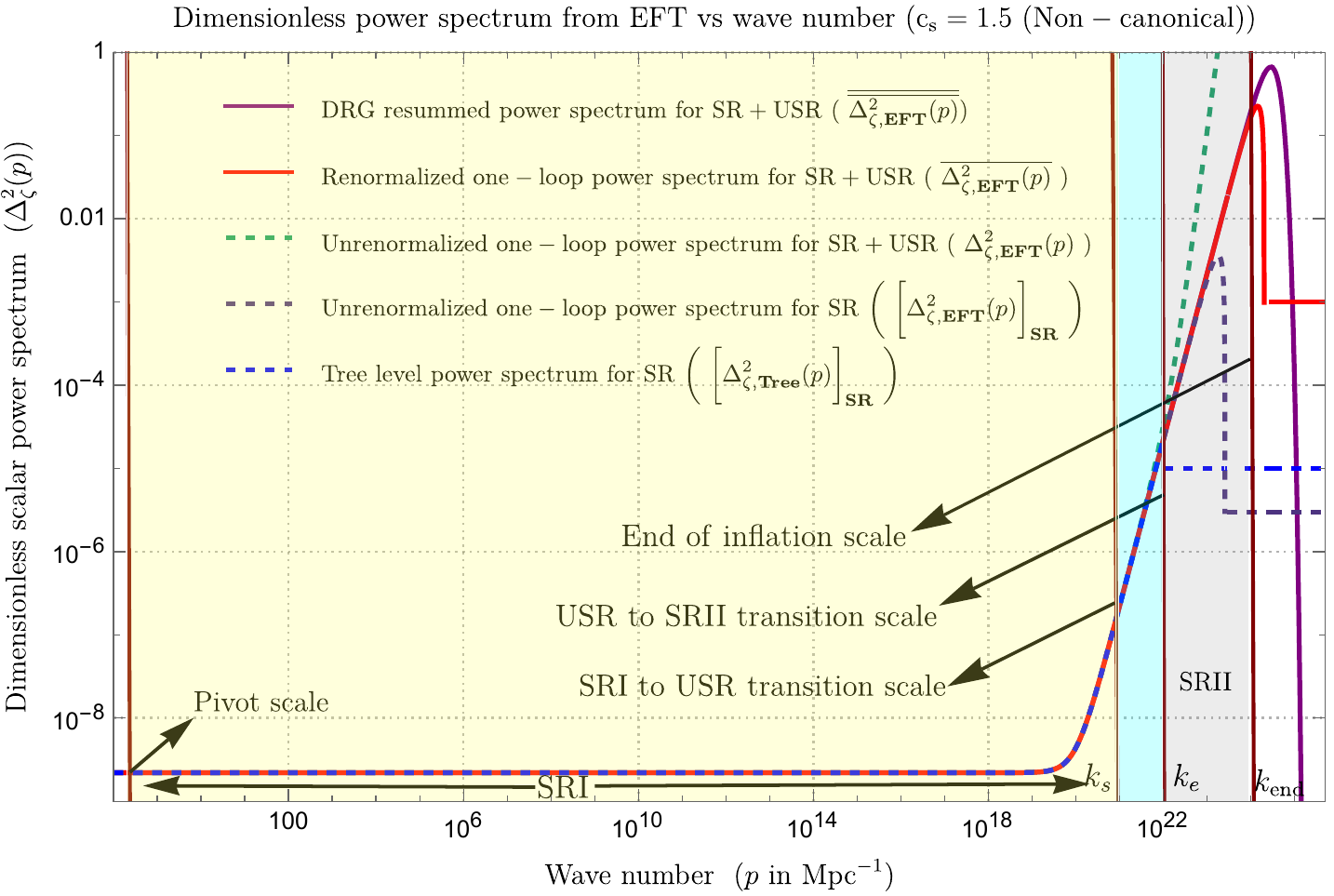}
        \label{G4}
       }
    	\caption[Optional caption for list of figures]{The wave number-dependent behavior of the dimensionless primordial power spectrum for scalar modes. We have examined the behavior of these typical graphs for various settings of the EFT sound speed, $c_s=0.6,1.17,1.5$. Here, we fix the pivot scale at $p_*=0.02\;{\rm Mpc}^{-1}$, the sharp transition scale from SR to USR at $k_s=10^{21}\;{\rm Mpc}^{-1}$, the end of the USR scale at $k_e=10^{22}\;{\rm Mpc}^{-1}$, the end of the inflation scale at $k_{\rm end}=10^{24}\;{\rm Mpc}^{-1}$. For example, $\Delta\eta(\tau_e)=1$, and $\Delta\eta(\tau_s)=-6$. $k_{\rm end}/k_e\approx{\cal O}(100)$ and $k_{\rm UV}/k_{\rm IR}=k_e/k_s\approx{\cal O}(10)$ are the values we discovered in this figure. The permitted range of effective sound speed, as indicated by plots, is $c_s\gtrsim 1$. Of these, $c_s=1.17$ yields the best result for having a ${\cal O}(10^{-2})$ amplitude of the relevant spectrum. 
The amplitude approaches ${\cal O}(1)$ at $c_s=1.5$, when the perturbation theory fails.
 } 
    	\label{Spectrum1}
    \end{figure*}
    \begin{figure*}[htb!]
    	\centering
    	\subfigure[For $c_s=0.6(<1)$  with $M^4_2/\dot{H}M^2_p\sim -0.89$ (non-canonical and causal).]{
      	\includegraphics[width=8cm,height=6cm] {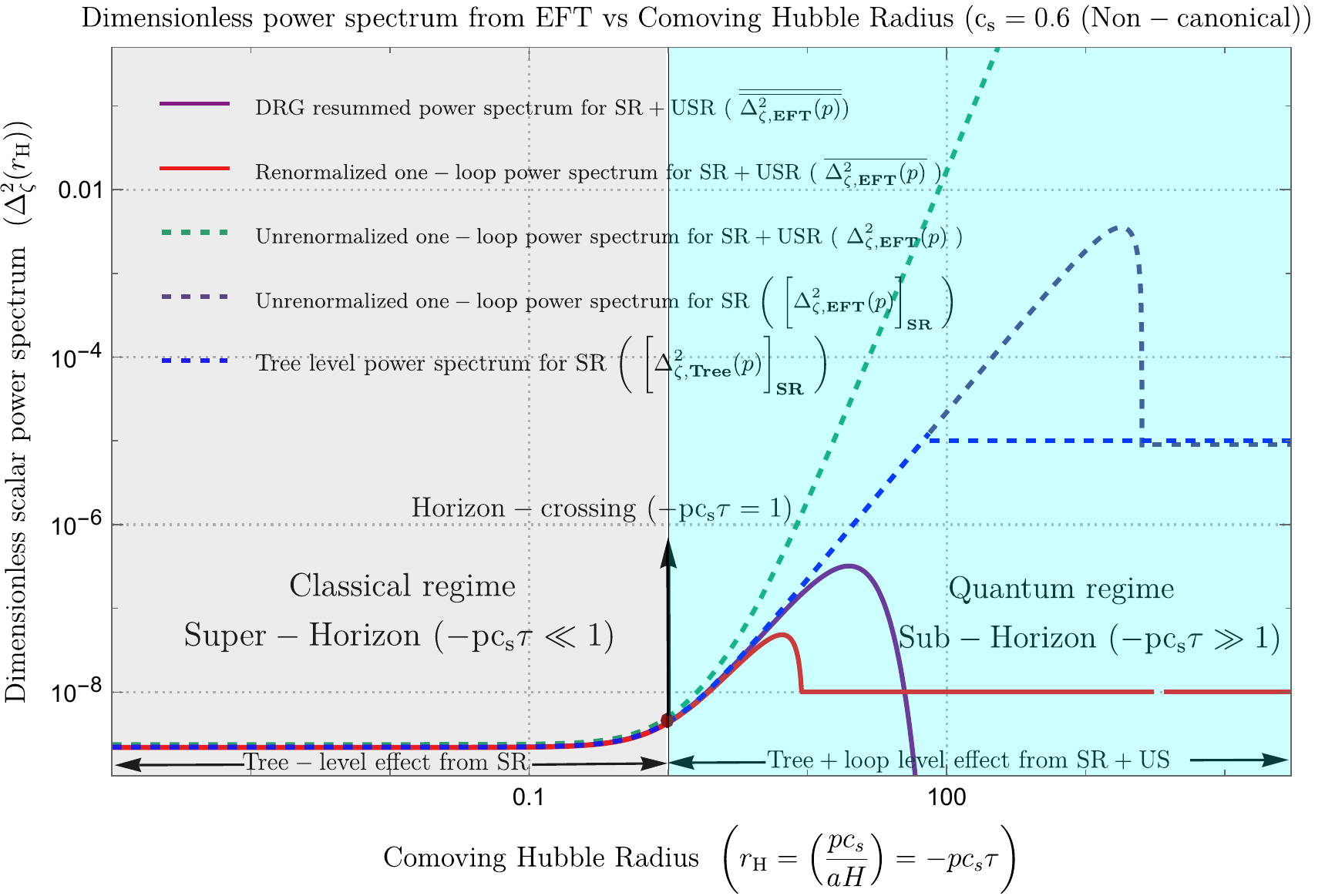}
        \label{H1}
    }
    \subfigure[For $c_s=1$  with $M^4_2/\dot{H}M^2_p\sim 0$ (canonical and causal).]{
       \includegraphics[width=8cm,height=6cm] {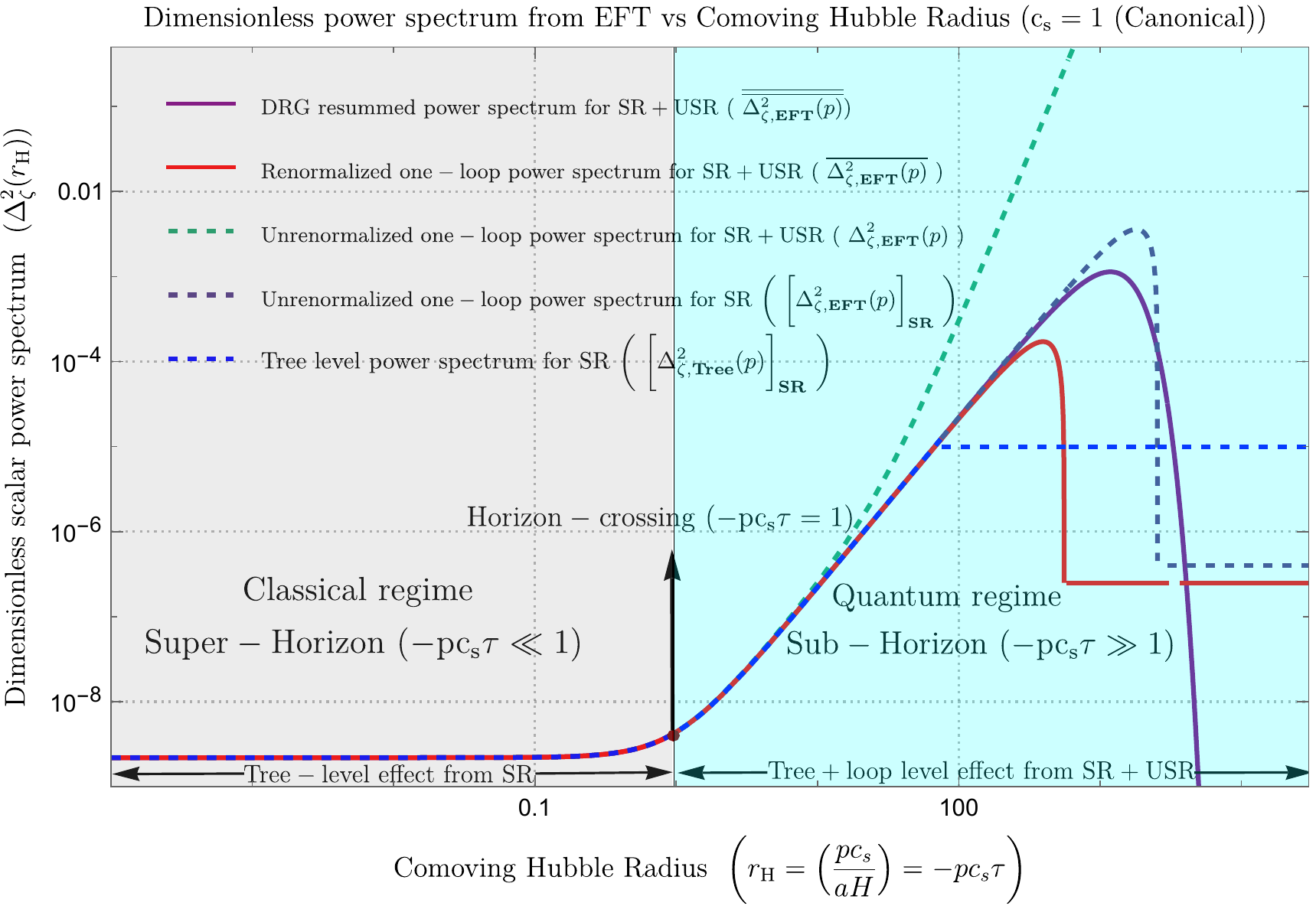}
        \label{H2}
       }
        \subfigure[For $c_s=1.17(>1)$  with $M^4_2/\dot{H}M^2_p\sim 0.13$ (non-canonical and a-causal).]{
       \includegraphics[width=8cm,height=6cm] {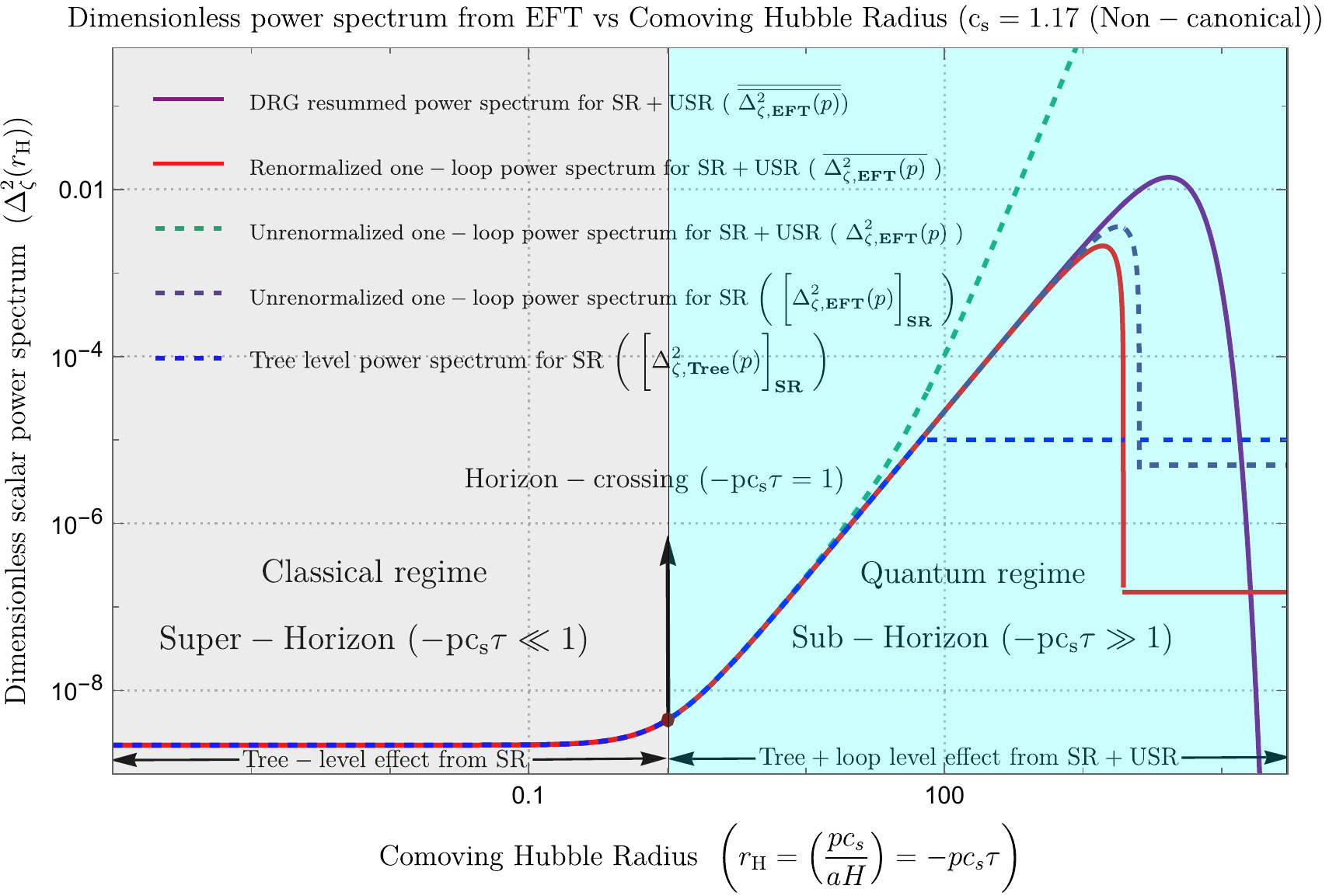}
        \label{H3}
       }
        \subfigure[For $c_s=1.5(>1)$  with $M^4_2/\dot{H}M^2_p\sim 0.28$ (non-canonical and a-causal).]{
       \includegraphics[width=8cm,height=6cm] {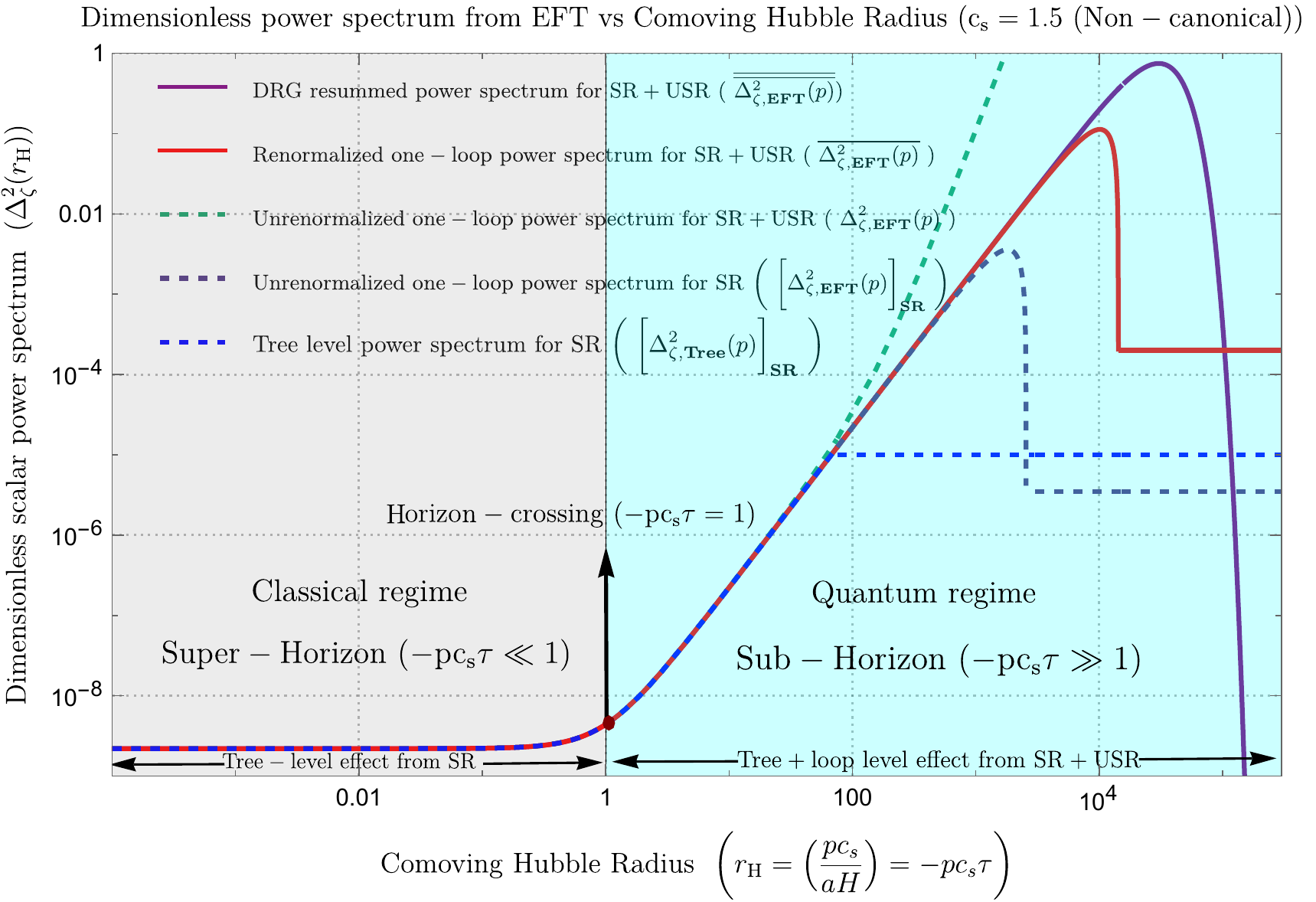}
        \label{H4}
       }
    	\caption[Optional caption for list of figures]{The Comoving Hubble Radius-dependent behavior of the dimensionless primordial power spectrum for scalar modes. We have examined the behavior of these typical graphs for various settings of the EFT sound speed, $c_s=0.6,1.17,1.5$. Here, we fix the pivot scale at $p_*=0.02\;{\rm Mpc}^{-1}$, the sharp transition scale from SR to USR at $k_s=10^{21}\;{\rm Mpc}^{-1}$, the end of the USR scale at $k_e=10^{22}\;{\rm Mpc}^{-1}$, the end of the inflation scale at $k_{\rm end}=10^{24}\;{\rm Mpc}^{-1}$. For example, $\Delta\eta(\tau_e)=1$, and $\Delta\eta(\tau_s)=-6$. $k_{\rm end}/k_e\approx{\cal O}(100)$ and $k_{\rm UV}/k_{\rm IR}=k_e/k_s\approx{\cal O}(10)$ are the values we discovered in this figure. The permitted range of effective sound speed, as indicated by plots, is $c_s\gtrsim 1$. Of these, $c_s=1.17$ yields the best result for having a ${\cal O}(10^{-2})$ amplitude of the relevant spectrum. 
The amplitude approaches ${\cal O}(1)$ at $c_s=1.5$, when the perturbation theory fails.
 } 
    	\label{Spectrum2}
    \end{figure*}

    \begin{figure*}[htb!]
    	\centering
    \subfigure[For $c_s=0.6(<1)$  with $M^4_2/\dot{H}M^2_p\sim -0.89$ (non-canonical and causal).]{
      	\includegraphics[width=8cm,height=6cm] {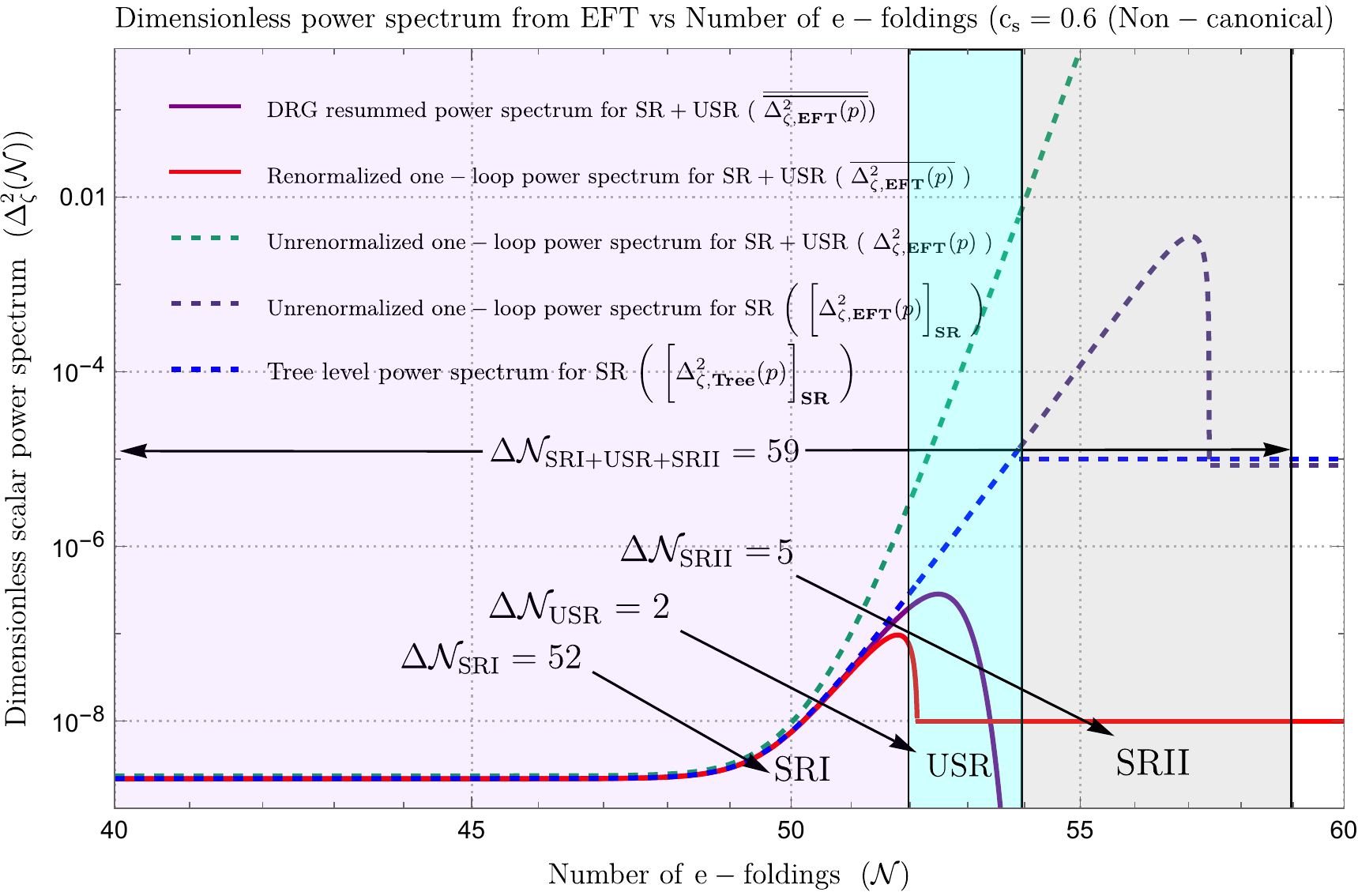}
        \label{I1}
    }
    \subfigure[For $c_s=1$  with $M^4_2/\dot{H}M^2_p\sim 0$ (canonical and causal).]{
       \includegraphics[width=8cm,height=6cm] {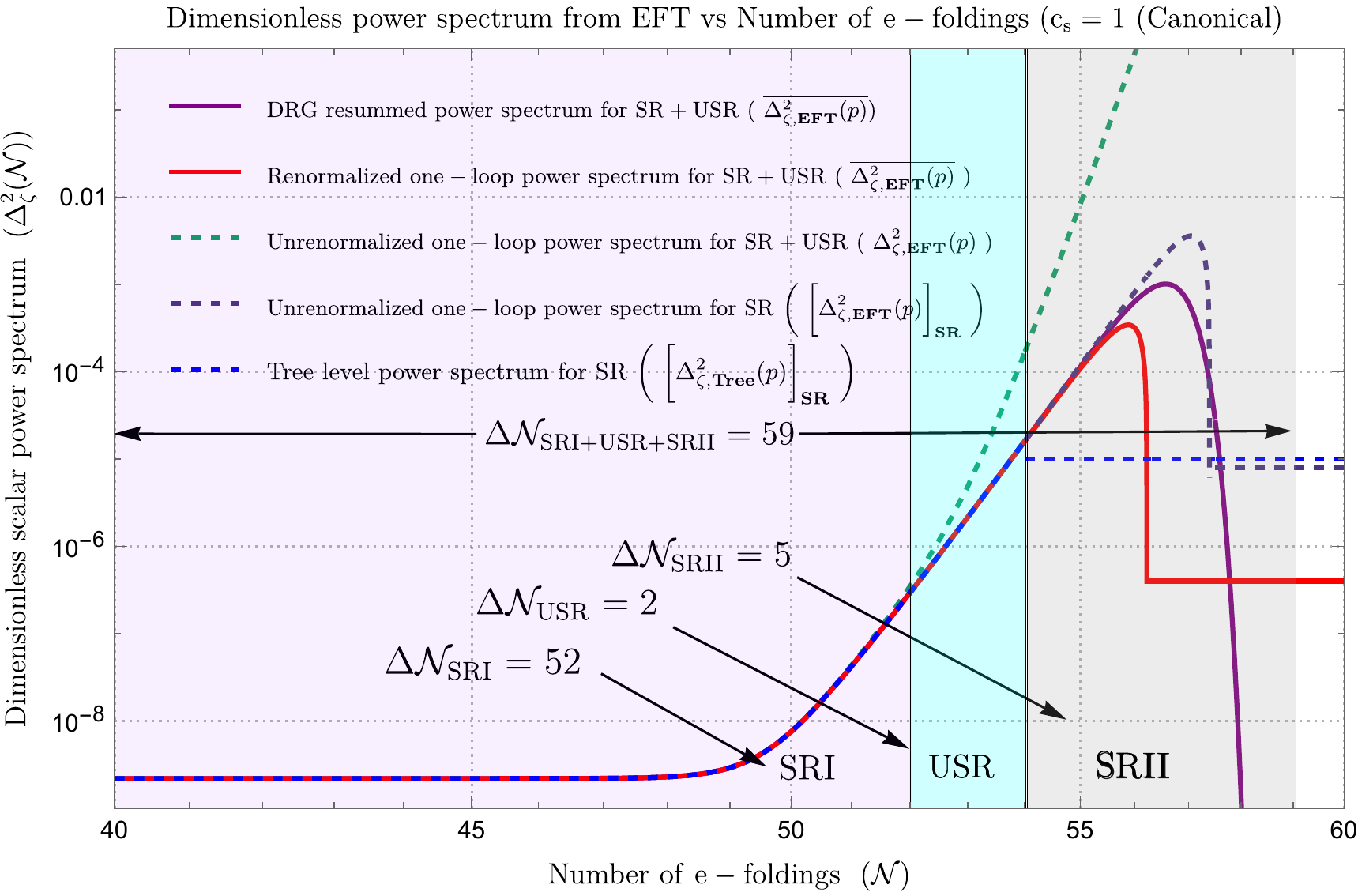}
        \label{I2}
       }
        \subfigure[For $c_s=1.17(>1)$  with $M^4_2/\dot{H}M^2_p\sim 0.13$ (non-canonical and a-causal).]{
       \includegraphics[width=8cm,height=6cm] {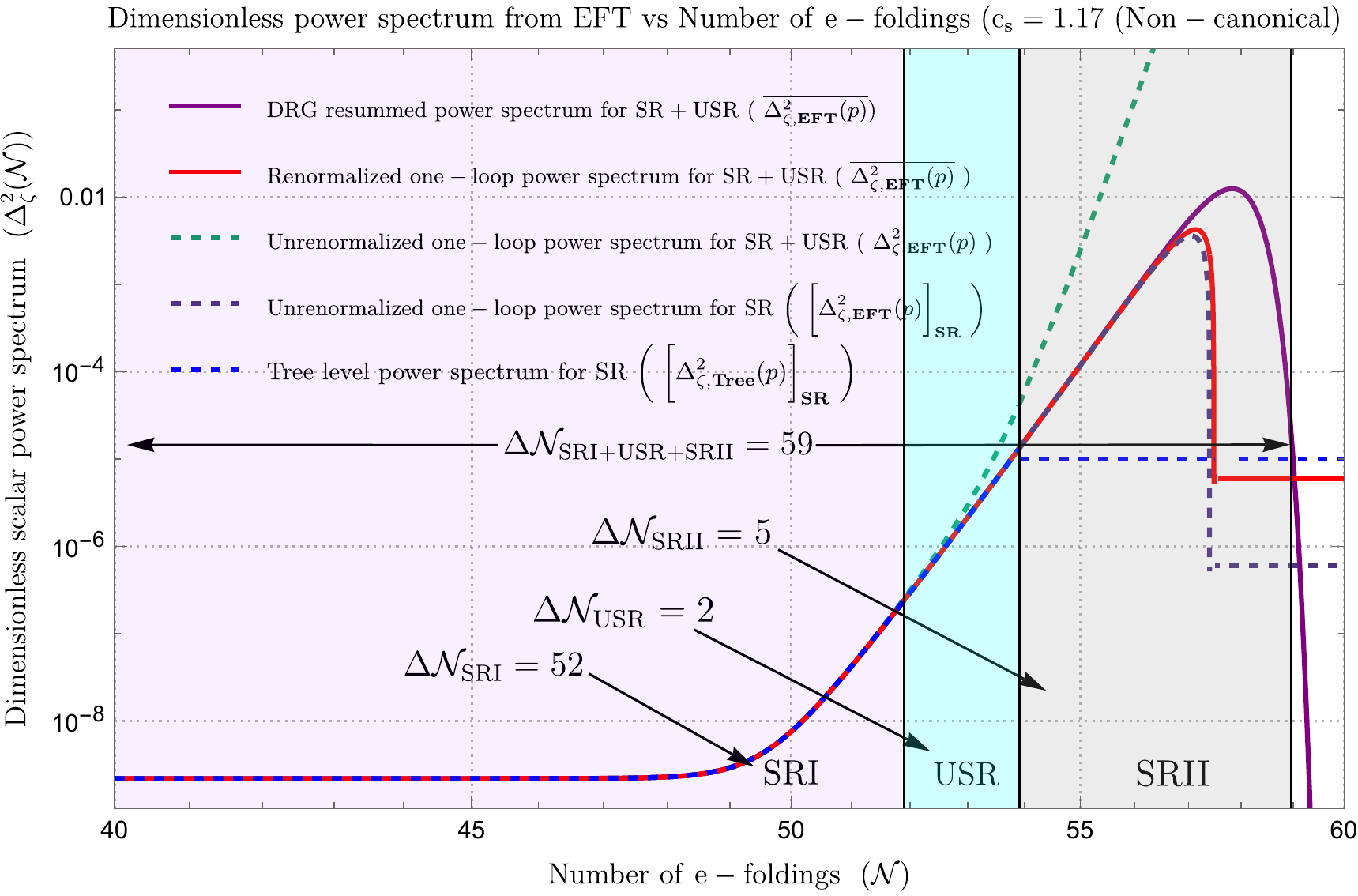}
        \label{I3}
       }
        \subfigure[For $c_s=1.5(>1)$  with $M^4_2/\dot{H}M^2_p\sim 0.28$ (non-canonical and a-causal).]{
       \includegraphics[width=8cm,height=6cm] {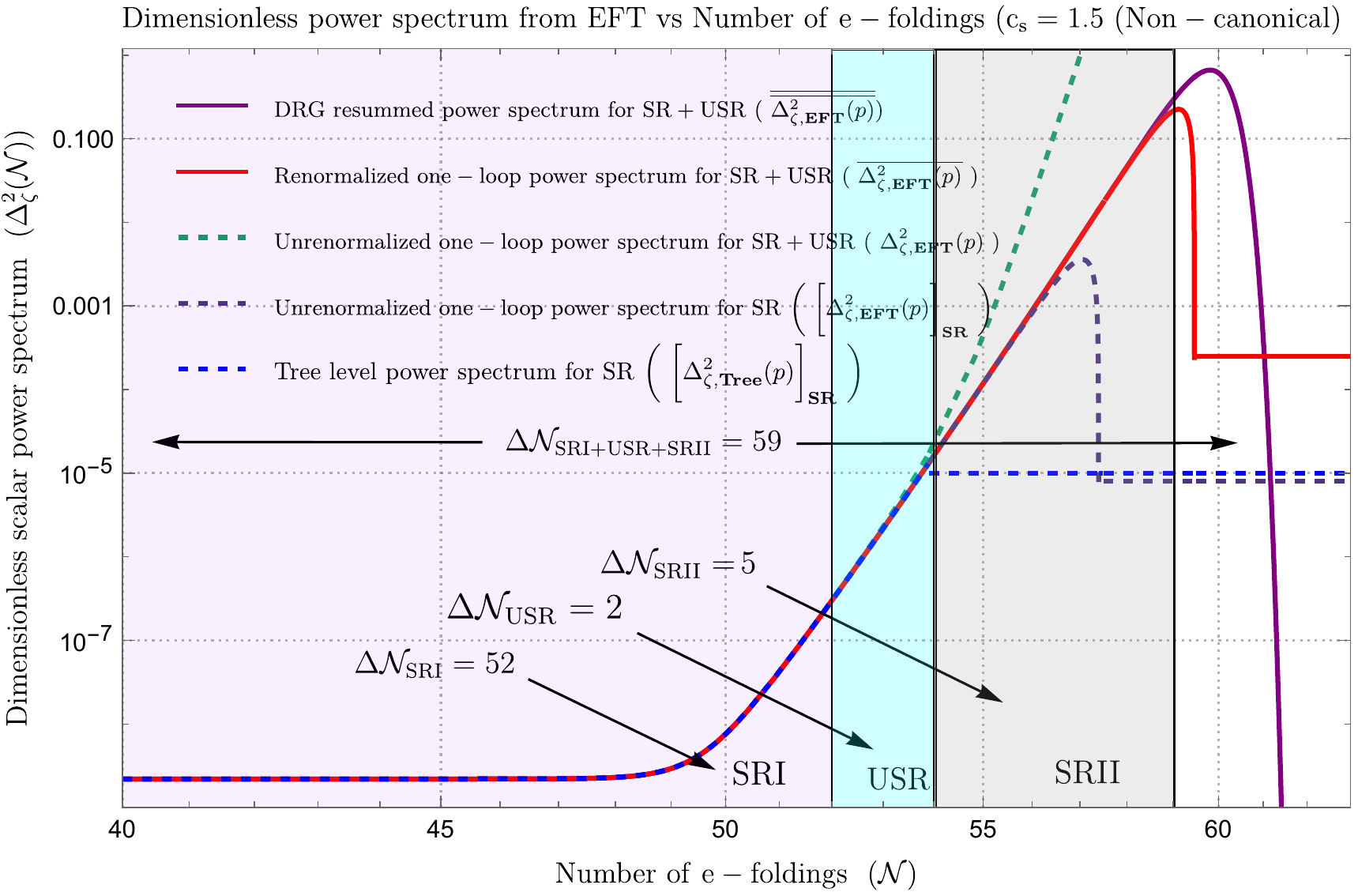}
        \label{I4}
       }
    	\caption[Optional caption for list of figures]{The number of e-foldings-dependent behavior of the dimensionless primordial power spectrum for scalar modes. We have examined the behavior of these typical graphs for various settings of the EFT sound speed, $c_s=0.6,1.17,1.5$. Here, we fix the pivot scale at $p_*=0.02\;{\rm Mpc}^{-1}$, the sharp transition scale from SR to USR at $k_s=10^{21}\;{\rm Mpc}^{-1}$, the end of the USR scale at $k_e=10^{22}\;{\rm Mpc}^{-1}$, the end of the inflation scale at $k_{\rm end}=10^{24}\;{\rm Mpc}^{-1}$. For example, $\Delta\eta(\tau_e)=1$, and $\Delta\eta(\tau_s)=-6$. $k_{\rm end}/k_e\approx{\cal O}(100)$ and $k_{\rm UV}/k_{\rm IR}=k_e/k_s\approx{\cal O}(10)$ are the values we discovered in this figure. The permitted range of effective sound speed, as indicated by plots, is $c_s\gtrsim 1$. Of these, $c_s=1.17$ yields the best result for having a ${\cal O}(10^{-2})$ amplitude of the relevant spectrum. 
The amplitude approaches ${\cal O}(1)$ at $c_s=1.5$, when the perturbation theory fails.} 
    	\label{Spectrum3}
    \end{figure*}
    Figures (\ref{G1}), (\ref{G2}), (\ref{G3}), and (\ref{G4}) illustrate how the dimensionless power spectrum behaves for scalar modes in relation to wave number for various effective sound speed values, such as $c_s=0.6$, $c_s=1$, $c_s=1.17$, and $c_S=1.5$. The tree level, one-loop corrected unrenormalized, renormalized one-loop corrected part, and DRG resummed one-loop corrected part contributions are displayed in each figure. From all of the plots, we were able to determine that the DRG resummed spectrum provides the most accurate interpretation for the current situation, in which the spectrum rapidly decreases at the conclusion of the USR phase. Above all, the resummed spectrum offers the unique characteristics that aid in differentiating between each of the aforementioned contributions. In order to facilitate computation, we set the IR cut-off at $k_s=10^{21}\;{\rm Mpc}^{-1}$, the UV cut-off at $k_e=10^{22}\;{\rm Mpc}^{-1}$, the end of the inflation scale at $k_{\rm end}=10^{24}\;{\rm Mpc}^{-1}$, and the renomalization parameter $c_{\bf SR}=0$ as well as $\Delta\eta(\tau_s)=-6$ and $\Delta\eta(\tau_e)=1$. The constraint $k_{\rm UV}/k_{\rm IR}=k_e/k_s\approx{\cal O}(10)$ and $k_{\rm end}/k_e\approx{\cal O}(100)$ has been upheld. The allowable range of effective sound speed, as indicated by all of these plots, is $c_s\gtrsim 1$. Of these, $c_s=1.17$ yields the best result for having the largest amplitude, ${\cal O}(10^{-2})$, of the corresponding spectrum that is required to produce PBHs from the current setup. The perturbation theory breaks at ${\cal O}(1)$, when the amplitude approaches at $c_s=1.5$. We are unable to continue our research past $c_s=1.5$ as a result. More specifically, the range $1<c_S<1.17$ contains the permitted window of the effective sound speed.

Figures (\ref{H1}), (\ref{H2}), (\ref{H3}), and (\ref{H4}) illustrate the behavior of the dimensionless power spectrum for scalar modes with respect to the Comoving Hubble Radius for various effective sound speed values, namely $c_s=0.6$, $c_s=1$, $c_s=1.17$, and $c_S=1.5$. We have plotted the sizes of the super-horizon, sub-horizon, and horizon crossing points in each plot, indicating the locations where the classical, quantum, and semi-classical effects are most evident. These graphs demonstrate that the effective sound speed range is $1<c_S<1.17$. Of these, $c_s=1.17$ yields the best result in terms of the maximum amplitude, ${\cal O}(10^{-2})$, of the corresponding spectrum required to generate PBHs. 

Lastly, we have shown the behavior of the dimensionless power spectrum for scalar modes with regard to the number of e-foldings in Figures (\ref{I1}), (\ref{I2}), (\ref{I3}), and (\ref{I4}). We conclude from these mentioned plots that:
\bea \Delta {\cal N}_{\rm USR}=\ln(k_e/k_s)\approx\ln(10)\approx 2.\eea 
It suggests that in the current arrangement, only about $2$ e-folds are permitted in the USR phase for the PBH creation. Furthermore, the following gives the permitted e-folds for the SRI and SRII periods:
\bea &&\Delta {\cal N}_{\rm SRI}=\ln(k_s/p_*)\approx 52,\\
&&\Delta {\cal N}_{\rm SRII}=\ln(k_{\rm end}/k_e)\approx 2\ln(10)\approx 5.\eea 
In order to carry out DRG resummation inside the perturbative regime, the SRII phase is restricted. Consequently, the following statement indicates the total number of e-foldings permitted by the current configuration:
\bea \Delta {\cal N}_{\rm Total}=\Delta {\cal N}_{\rm SRI}+\Delta {\cal N}_{\rm USR}+\Delta {\cal N}_{\rm SRII}\sim 52+2+5=59.\eea 
Assuming a sharp transition scale, $k_s=10^{21}\;{\rm Mpc}^{-1}$, $k_e=10^{22}\;{\rm Mpc}^{-1}$, and $k_{\rm end}=10^{24}\;{\rm Mpc}^{-1}$ are fixed for inflation and the end of the USR period. In the next section, we will demonstrate that this option will result in the construction of a tiny mass PBH with enough e-folds for inflation in the current setting.

Considering this now, what would happen if we moved the sharp transition scale, the end of the USR period, and inflation at the scale, $k_e=10^{7}\;{\rm Mpc}^{-1}$, $k_s=10^{6}\;{\rm Mpc}^{-1}$, and $k_{\rm end}=10^{9}\;{\rm Mpc}^{-1}$. The number of e-folds that are permitted for SRI in this scenario may be calculated as follows:
\bea &&\Delta {\cal N}_{\rm SRI}=\ln(k_s/p_*)\approx 18.\eea
The outcomes for SRII and USR will remain the same. This suggests that in the aforementioned example, the formula that follows provides the total number of e-foldings:
\bea \Delta {\cal N}_{\rm Total}=\Delta {\cal N}_{\rm SRI}+\Delta {\cal N}_{\rm USR}+\Delta {\cal N}_{\rm SRII}\sim 18+2+5=25.\eea
The likelihood that this may result in a big mass PBH creation with an inadequate number of e-folds for inflation will be demonstrated in the next section. This makes it possible to rule out this option right away.
\bea &&\Delta {\cal N}_{\rm SRI}=\ln(k_s/p_*)\approx 18.\eea
    \begin{figure*}[htb!]
    	\centering
{
      	\includegraphics[width=14cm,height=10cm] {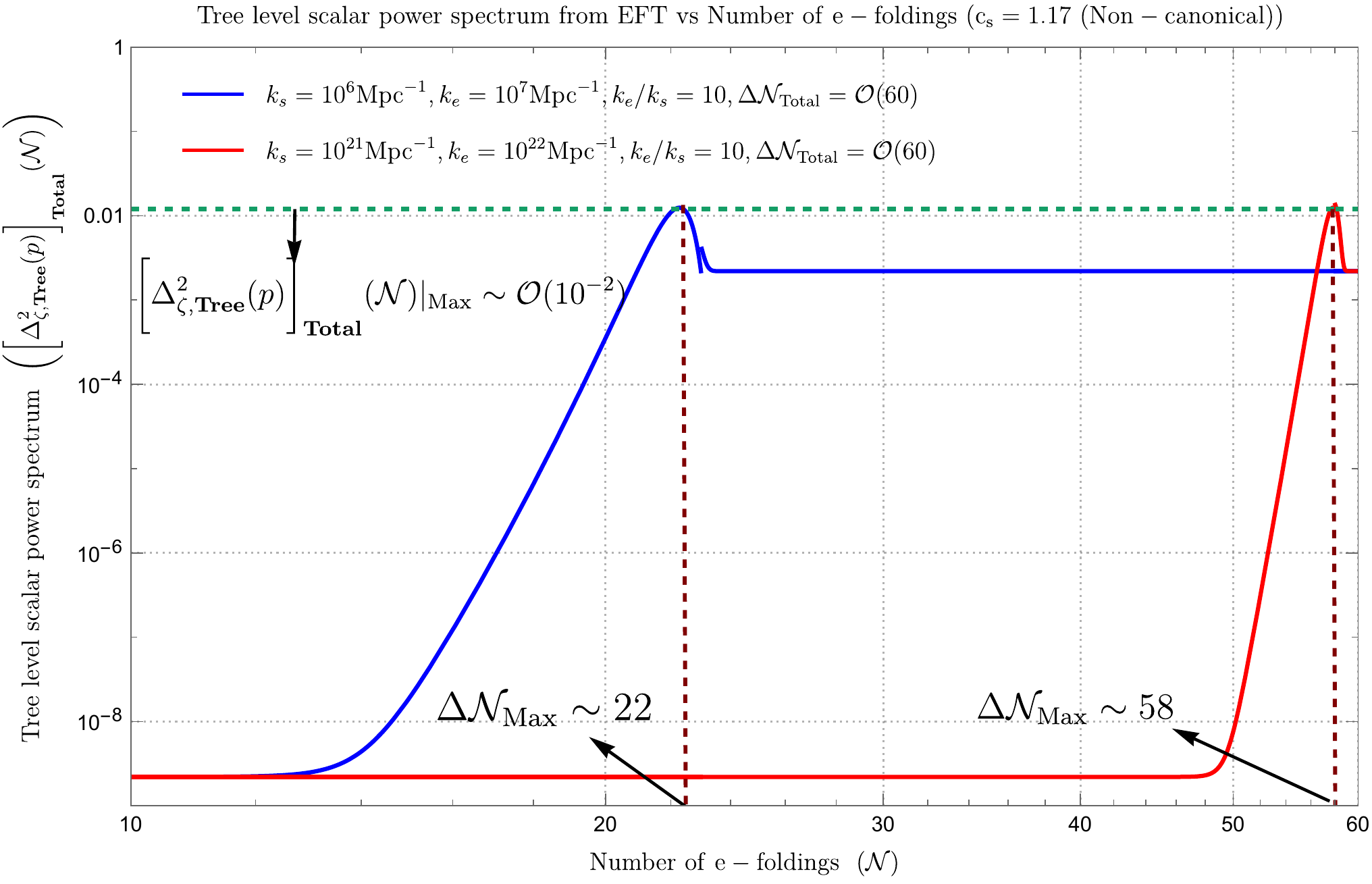}
        \label{D1}
    }
    	\caption[Optional caption for list of figures]{Plotting the behavior of the dimensionless tree level total power spectrum for scalar modes using the current EFT configuration in relation to the number of e-foldings. Fixing the pivot scale at $p_*=0.02\;{\rm Mpc}^{-1}$, we have sharp transition scales at $k_s=10^{6}\;{\rm Mpc}^{-1}$ (for blue) and $k_s=10^{21}\;{\rm Mpc}^{-1}$ (for red) at $c_s=1.17$, where the peak of the spectrum appeared at ${\cal O}(10^{-2})$.
} 
    	\label{SpectrumTree}
    \end{figure*}
    \begin{figure*}[htb!]
    	\centering
{
      	\includegraphics[width=14cm,height=10cm] {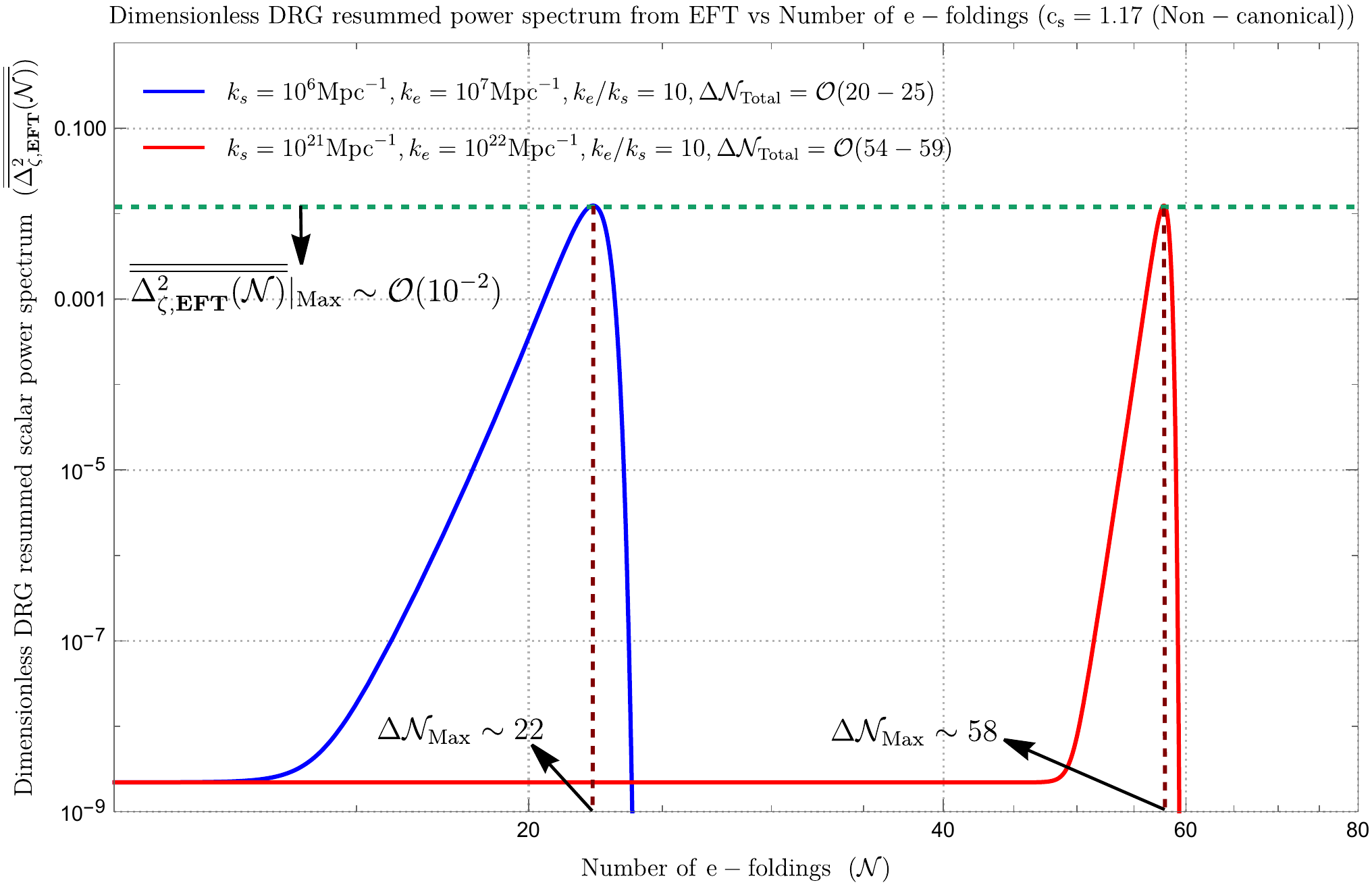}
        \label{D1}
    }
    	\caption[Optional caption for list of figures]{Plotting the behavior of the dimensionless DRG resummed power spectrum for scalar modes in the current EFT configuration against the number of e-foldings. Fixing the pivot scale at $p_*=0.02\;{\rm Mpc}^{-1}$, we have sharp transition scales at $k_s=10^{6}\;{\rm Mpc}^{-1}$ (for blue) and $k_s=10^{21}\;{\rm Mpc}^{-1}$ (for red) at $c_s=1.17$, where the peak of the spectrum appeared at ${\cal O}(10^{-2})$.
} 
    	\label{SpectrumDRG}
    \end{figure*}
    Finally, let us state unequivocally what we have discovered as a result of the point-by-point examination conducted in this paper:
\begin{itemize}
    \item[$\blacksquare$] According to our analysis, large mass PBH with mass $M_{\rm PBH}\sim {\cal O}(10^{29}-10^{30}){\rm kg}$ can be generated within the allowed window of effective sound speed $0.6<c_s<1.17$ if the wave numbers $k_s=10^{6}{\rm Mpc}^{-1}$ and the pivot scale at $p_*=0.02{\rm Mpc}^{-1}$ are fixed. $\Delta{\cal N}_{\rm SR}= \ln(k_s/p_*)\sim 18$ is the maximum number of e-folds permitted in this particular situation for the SR phase. Furthermore, we discovered an unavoidable strong constraint $k_e/k_s=10$, which is required in order to include the perturbative approximation when computing the one-loop contribution from the USR phase in the primordial power spectrum for scalar modes. This further suggests that $\Delta{\cal N}_{\rm USR}= \ln(k_e/k_s)\sim 2$ is the maximum number of e-folds permitted for the USR phase. The total $\Delta{\cal N}_{\rm Total}=\Delta{\cal N}_{SR}+\Delta{\cal N}_{\rm USR}=18+2=20$ e-folds is therefore inadequate to produce inflation from the current configuration. We have discovered that the second SR phase (SRII) and the inflation must strictly adhere to the constraint $k_{\rm end}/k_{\rm e}=10^{2}$ in order to have a correct renormalized and DRG resummed power spectrum. With this information, the number of e-folds during the SRII phase can be calculated as follows: $\Delta{\cal N}_{\rm SRII}= \ln(k_{\rm end}/k_{\rm e})\sim 5$. This suggests, then, that huge mass PBHs can be produced by destroying the number of e-foldings that are required for inflation. This may be quickly rejected because it is plainly not better from a theoretical standpoint. Figures (\ref{SpectrumTree}) and (\ref{SpectrumDRG}) illustrate the behavior of the dimensionless tree-level total and DRG resummed power spectrum for the scalar modes in relation to the number of e-foldings, respectively, to provide a clear justification for this result. For the first (blue) scenario, at $\Delta {\cal N}_{\rm Total}\sim {\cal O}(60)$ with $\Delta {\cal N}_{\rm Max}\sim 22$, we have explicitly demonstrated that the spectrum peak emerges at ${\cal O}(10^{-2})$. As we add the steep transition scale at $k_s=10^{6}{\rm Mpc}^{-1}$, the first (blue) one here reflects the result of that particular situation. The range ${\cal O}(10^{29}-10^{30}){\rm kg}$, where this possibility might produce enormous mass PBHs. For the tree-level situation in this instance, inflation can support a significant number of total e-foldings; at the very least, this option is producing outcomes that are favorable. But when we take into account the DRG restarted normalized power spectrum in the current context of the present discussion, things will significantly alter. The spectrum peak arises at ${\cal O}(10^{-2})$ for the first (blue) scenario at $\Delta {\cal N}_{\rm Total}\sim {\cal O}(20–25)$ with $\Delta {\cal N}_{\rm Max}\sim 22$, as we have explicitly demonstrated in the plot (\ref{SpectrumDRG}) for the DRG resummed case. The first (blue) one in this instance relates to the result of the particular scenario when we introduce the steep transition scale at $k_s=10^{6}{\rm Mpc}^{-1}$. In this case, we have preserved all of the previously described wave number limits, which relates to the end of both the USR phase and inflation. Large mass PBHs within the range ${\cal O}(10^{29}-10^{30}){\rm kg}$ can be produced via this possibility. But as we have already said, this alternative is undesirable in the current context of the debate since inflation is unable to support a significant number of total e-foldings for the DRG ressumed situation.

 \item[$\blacksquare$] However, our analysis reveals that only small mass PBH with mass $M_{\rm PBH}\sim {\cal O}(10^{2}-10^{3}){\rm gm}$ can be generated within the allowed window of effective sound speed $0.6<c_s<1.17$ if we fix the wave numbers $k_s=10^{21}{\rm Mpc}^{-1}$ and the pivot scale at $p_*=0.02{\rm Mpc}^{-1}$. The number of e-folds that are permitted in this particular instance for the SR phase is $\Delta{\cal N}_{\rm SR}= \ln(k_s/p_*)\sim 52$. The same stringent limitation that $k_e/k_s=10$ was discovered in this instance as well. This is required in order to include the perturbative approximation when calculating the one-loop contribution from the USR phase in the primordial power spectrum for scalar modes. This further suggests that $\Delta{\cal N}_{\rm USR}= \ln(k_e/k_s)\sim 2$ is the maximum number of e-folds permitted for the USR phase. To achieve inflation from the current setup, we need to have a total of $\Delta{\cal N}_{\rm Total}=\Delta{\cal N}_{\rm SR}+\Delta{\cal N}_{\rm USR}=52+2=54$ e-folds. In order to obtain an accurate DRG resummed power spectrum, it is necessary to adhere to the constraint that $k_{\rm end}/k_{\rm e}=10^{2}$ for both the inflation and the second SR phase (SRII). This allows for the computation of the number of e-folds during the SRII phase, which can be expressed as $\Delta{\cal N}_{\rm SRII}= \ln(k_{\rm end}/k_{\rm e})\sim 5$. According to this, Framework II will have a total number of permitted e-folds of $\Delta{\cal N}_{\rm Total}=\Delta{\cal N}_{\rm SRI}+\Delta{\cal N}_{\rm USR}+\Delta{\cal N}_{\rm SRII}=52+2+5=59$. This will also be sufficient to achieve inflation from the current setup. This suggests that small mass PBHs can be produced for both frameworks by keeping the inflation requirement in terms of the number of e-foldings constant. From a theoretical standpoint, this is clearly much better and may be taken into consideration for further research. Plotting the behavior of the dimensionless tree-level total and DRG resummed power spectrum for the scalar modes with regard to the number of e-foldings is what we have done in figures (\ref{SpectrumTree}) and (\ref{SpectrumDRG}) to clearly justify this result. The spectrum peak is clearly visible in this figure (\ref{SpectrumTree}), with an appearance at ${\cal O}(10^{-2})$ for the second (red) scenario at $\Delta {\cal N}_{\rm Total}\approx {\cal O}(54-59)$ where $\Delta {\cal N}_{\rm Max}\sim 58$. In this instance, the second (red) one reflects the result of the particular scenario in which the steep transition scale is introduced at $k_s=10^{21}{\rm Mpc}^{-1}$. Small mass PBHs in the range ${\cal O}(10^{2}-10^{3}){\rm gm}$ may be produced via this possibility. Given that this example achieves a significant number of e-foldings for inflation, this option may prove helpful in future research endeavors and meet the requirements at the tree-level case analysis. We have discovered that the tree-level plot's very brief SRII area allows it to support a steady non-vanishing amplitude. Conversely, we have seen a sharply dropping characteristic in the DRG resummed scenario because of the additional strict limits imposed by renormalization and resummation. The plot (\ref{SpectrumDRG}) clearly illustrates that, for the second (red) scenario, the spectrum peak arises at ${\cal O}(10^{-2})$ at $\Delta {\cal N}_{\rm Total}\sim {\cal O}(54-59)$ with $\Delta {\cal N}_{\rm Max}\sim 58$. The result of the particular scenario where we introduce the steep transition scale at $k_s=10^{21}{\rm Mpc}^{-1}$, where we have kept all the previously indicated limits on the wave number, is represented by the second (red) one in this instance. This corresponds to the end of both the USR phase and inflation. Small mass PBHs in the range ${\cal O}(10^{2}-10^{3}){\rm gm}$ may be produced via this possibility. This possibility may be helpful for future research as in this scenario, a sufficient number of e-foldings is accomplished for inflation. 

  \item[$\blacksquare$] Upon closely examining the figures (\ref{SpectrumTree}) and (\ref{SpectrumDRG}), it is evident that a significant shift has occurred in the results obtained from the PBHs generated of the mass range, ${\cal O}(10^{29}-10^{30}){\rm kg}$ at the scale $k_s\sim 10^6{\rm Mpc}^{-1}$ for the tree-level and DRG resummed loop level results. The number of e-foldings $\Delta {\cal N}_{\rm Total}\sim{\cal O}(60)$ that is adequate for the tree-level scenario has been realised. In contrast, we have discovered that the number of e-foldings $\Delta {\cal N}_{\rm Total}\sim{\cal O}(20 - 25)$ becomes insufficient for inflation in the DRG resummed loop level result because of a sharp fall in the spectrum that is directly related to the additional constraints from renormalization and resummation. However, it is also possible to note that, for both the tree-level and DRG resummed loop level results, no such significant shift has been shown in the results from the PBHs created of the mass range, ${\cal O}(10^{2}-10^{3}){\rm gm}$ at the scale $k_s\sim 10^{21}{\rm Mpc}^{-1}$. The most crucial finding in each of these circumstances is that we have obtained $\Delta {\cal N}_{\rm Total}\sim{\cal O}(60)$, the required number of e-foldings. This clearly indicates that, in the context of the discussion at hand, the tree-level result and the DRG resummed result taking into account the loop effects almost give similar consequences for both Frameworks I and II in terms of the PBHs generated of the mass range, ${\cal O}(10^{2}-10^{3}){\rm gm}$. One may confidently accept the acquired result for the mass range of the PBH, ${\cal O}(10^{2}-10^{3}){\rm gm}$, given the comparability between the results produced from the tree-level calculation and the DRG resummed one-loop level computation.

\end{itemize}
\subsubsection{For multiple sharp transitions}
    \begin{figure*}[htb!]
    	\centering
   {
   \includegraphics[width=18cm,height=10cm] {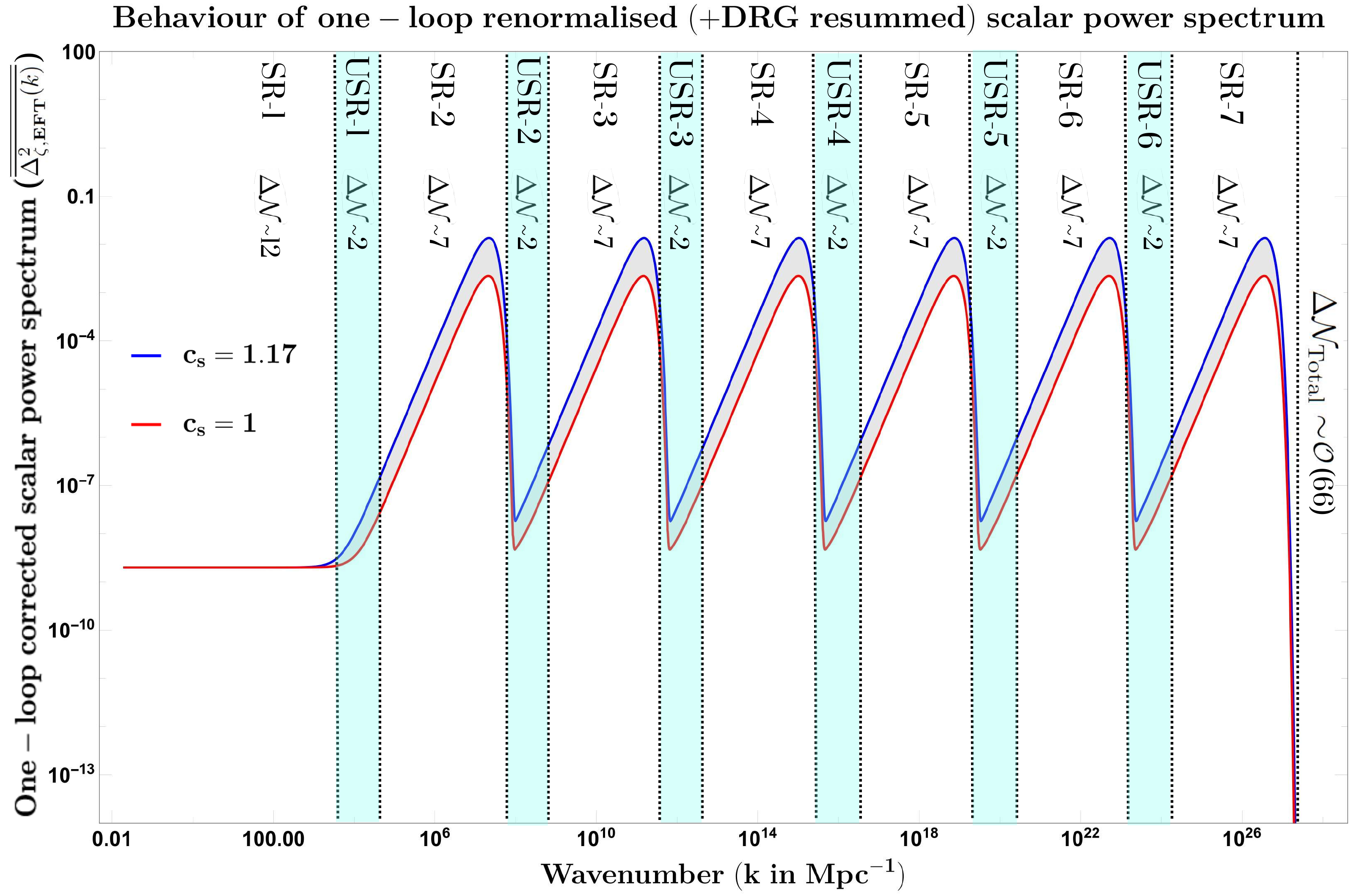}
    }
    	\caption[Optional caption for list of figures]{The aforementioned plot shows the one-loop corrected scalar power spectrum over a broad range of wavenumbers. It is renormalized and resummed using the DRG approach. For every $n=1,2,\cdots,6$, there are distinct markers (cyan bands) on the figure designating the ultra-slow roll (USR$_{n}$) and slow-roll (SR$_{n+1}$) phases. The two instances of effective sound speed, $c_{s}=1$ (red) and $c_{s}=1.17$ (blue), are represented by the behavior of the spectra. The permissible parameter range for the amplitude of enhancements, which should lie between ${\cal O}(10^{-3}-10^{-2})$ for PBH synthesis, is represented by the gray band that separates the two spectra.
} 
    	\label{final1}
    \end{figure*}
    For a broad range of wavenumbers, the behaviour of the one-loop renormalized and DRG resummed scalar power spectrum is shown in figure (\ref{final1}), which is significant for our study in the second part of this letter. This picture shows the example of MSTs evaluated for two distinct effective sound speed values, $c_{s}=1.17$ (blue) and $c_s=1$ (red). The permitted parameter space of our theory is represented by the area between the two spectra. This region is small but very important since results from it not only support the perturbation theory inside it, but they also help to generate scalar induced gravitational waves (SIGWs) and provide an abundance range that makes up for substantial dark matter content. For $n=1,2,\cdots,6$, the pivot scale value is set to $k_{*} = 0.02{\rm Mpc}^{-1}$. The intervals for the following USR and SR phases are then controlled to fulfill $\Delta{\cal N}_{\rm USR_{n}}\approx{\cal O}(2)$ and $\Delta{\cal N}_{\rm SR_{n+1}}\approx{\cal O}(7)$, respectively. Plotting indicates that the scenario with $c_{s}=1.17$, which somewhat violates the causality constraint, is the favoured one since it produces the needed ${\cal O}(10^{-2})$ amplitude for PBH production. The fact that $c_{s}=1.17$ is the number at which the perturbative approximations totally collapse makes it unique as well; the authors go into further detail about this in refs \cite{Choudhury:2023jlt,Choudhury:2023rks}. Furthermore, while the perturbative approximations hold, a close examination of the scenario when $c_{s} < 1$, e.g., $c_{s}=0.6$, indicates that the amplitude of the power spectrum stays much below the threshold needed for PBH production. A detailed discussion of the issue with $c_{s} < 1$ may be found in refs \cite{Choudhury:2023rks,Choudhury:2023jlt}. It is important to note that a minor increase in the parameter value, from $c_{s}=0.6$ to $1$, results in a significant rise in the spectrum amplitude. The graphic shows a series of peaks and troughs with consistent amplitudes across the spectrum. For the effective sound speed, $1 \leq c_{s} \leq 1.17$, the amplitude peak values fall within ${\cal O}(10^{-3}-10^{-2})$, and its dip values fall within ${\cal O}(10^{-8}-10^{-9})$. Due to the powerful DRG resummation mechanism, the last peak amplitude following the last transition into SR$_{7}$ phase declines steeply at ${\cal O}(10^{27}{\rm Mpc}^{-1})$. 
According to \cite{Mishra:2019pzq}, the power spectra created have various bumpy aspects in the potential. Without taking into account any particular structure for the effective potential, we conducted a model-independent investigation using the EFT framework. Regarding MSTs, we've added characteristics to the second SR parameter $\eta$, which might serve the similar function in this particular situation. In contrast, we will employ a certain class of effective potentials in the EFT framework with the canonical and non-canonical single-field inflation models in detail in our upcoming follow-up studies. This includes useful information on the PBH energy density fraction relative to dark matter. Additionally, it will be utilized to produce SIGWs with a broad frequency range. The latter portion of the letter will cover each of these topics. In between the talks, let us not forget the difficulties associated with including several SR and USR stages. It should be noted that we are working with huge quantum fluctuations from $6$ USR phases. Since each of these phases is followed by another SR phase, we must modify the power spectrum amplitude for the scalar modes to reflect each phases predicted behavior. Additionally, one must take into account the validity of the cosmic perturbation across all scales in the theory that covers $k_{*}\sim{\cal O}(0.02\;{\rm Mpc}^{-1})$, the CMB pivot scale to the end of the $k_{\rm end}\sim{\cal O}(10^{27}\;{\rm Mpc}^{-1})$ inflation scale. This correspondence also pertains to limited results from earlier attempts. The writers of the previous works \cite{Choudhury:2023vuj, Choudhury:2023jlt, Choudhury:2023rks} have demonstrated explicitly that in the presence of a single sharp transition, it is not possible to generate large mass PBHs using the present setup along with the condition to achieve the necessary e-foldings of $\Delta{\cal N}_{\rm Total}\approx{\cal O}({60-70})$. In other words, PBHs cannot be generated using this method. The authors of the reference \cite{Choudhury:2023rks} have noted that in the case where the PBH mass is $M_{\rm PBH}\sim{\cal O}(M_{\odot})$, the total e-foldings will be $\Delta{\cal N}_{\rm Total}\approx{\cal O}(25)$, which is quite insufficient to accomplish successful inflation using the corresponding theoretical set-up. However, in order to keep the strong requirement on $\Delta{\cal N}_{\rm Total}$ to end inflation, the only value that can be obtained from this setup, from a cosmological point of view, is $M_{\rm PBH}\sim 10^{2}\;{\rm gm}$, with only one sharp transition at the high momenta scales. As a result, $6$ successive MSTs designed in the SR, USR, and SR sequence have been integrated. We have shown clearly that it is possible to avoid the strong No-go in the obtained PBH mass in this arrangement without sacrificing generality. The analysis done and the results reported in this letter allow us to cover the whole wave number range $k\sim{\cal O}(10^{-2}\;\rm Mpc^{-1} - 10^{27}\;{\rm Mpc^{-1}})$ and obtain the necessary e-foldings to bring inflation to an end. This suggests that the complete spectrum of PBH mass, $M_{\rm PBH}\sim{\cal O}(10^{-31}M_{\odot}-10^{4}M_{\odot})$, can be covered by accounting for $6$ peaks of large amplitude fluctuations that lie in the order of ${\cal O}({10^{-3}-10^{-2}})$ for $1 \leq c_{s} \leq 1.17$. Thus, a designed setup like this avoids the significant limitations on the PBH mass that result from renormalization and DRG resummation. After some thought, we conclude that this finding is significant given the current discussion that has been going on for a few months. However, this is by no means the only achievement that avoids the difficult No-go theorem as suggested in ref \cite{Choudhury:2023vuj,Choudhury:2023jlt,Choudhury:2023rks}. The non-renormalization theorem may be used to circumvent the PBH mass No-go theorem in the context of USR Galileon inflation, which softly breaks the generalized Galilean shift symmetry. This was demonstrated recently by several of the authors. See refs. \cite{Choudhury:2023hvf,Choudhury:2023kdb,Choudhury:2023hfm} for more details. Sharp transitions have also been used in the design of all of the aforementioned frameworks \cite{Choudhury:2023hvf,Choudhury:2023kdb,Choudhury:2023hfm,Kristiano:2022maq,Riotto:2023hoz,Kristiano:2023scm,Franciolini:2023lgy,Cheng:2023ikq,Motohashi:2023syh,Riotto:2023gpm, Firouzjahi:2023ahg,Firouzjahi:2023aum,Franciolini:2023lgy,Cheng:2023ikq,Tasinato:2023ukp,Motohashi:2023syh}. Nevertheless, smooth transitions can also be included \cite{Riotto:2023gpm, Firouzjahi:2023ahg, Firouzjahi:2023aum}. Interestingly, the authors of \cite{Riotto:2023gpm,Firouzjahi:2023ahg,Firouzjahi:2023aum} have noted that using a single smooth transition might provide a way to make enormous PBH masses and avoid the difficulties caused by massive quantum fluctuations, all the while explaining the formation of SIGWs which we will discuss in the later part of this discussion. Because thorough Regularization and Renormalization is required to bolster the hypothesis, the effect of smooth transitions is still up for discussion in the literature. Still, it would be imperative to look for these opportunities for later work. Conversely, our letter presents a thorough analysis of the usefulness of MSTs through the use of methods such DRG resummation and renormalization. In the current study, we have successfully eliminated all of the aforementioned shortcomings by painstaking calculation.

\subsection{Comparison among various outcomes: Sharp vs Smooth Transition}

Allow us to now address the findings of more studies that have been carried out to examine the same topic from an alternative perspective. Let us compare our results with those of other authors in order to be thorough. As observed in the one-loop corrected power spectrum for scalar modes, the authors of ref. \cite{Riotto:2023gpm,Firouzjahi:2023ahg,Firouzjahi:2023aum} recently noted that the effect from large amplitude fluctuation can be neglected by employing a smooth transition from SRI to USR phase and USR to SRII phase. According to these studies, it is also possible to create massive mass PBHs with the size $k_{\rm PBH}=k_s\sim 10^{5}{\rm Mpc}^{-1}$, which can have all the e-foldings required to support inflation. The following references include more study on the same subject \cite{Franciolini:2023lgy,Cheng:2023ikq,Tasinato:2023ukp,Choudhury:2023hvf,Choudhury:2023kdb,Choudhury:2023jlt,Motohashi:2023syh}. In instance, the projected PBH mass and the corresponding PBH abundance are greatly affected by a quick or smooth transition. The one-loop corrected power spectrum is essentially unaffected for sharp transitions if renormalization and resummation are not taken into account \cite{Riotto:2023gpm,Firouzjahi:2023ahg,Firouzjahi:2023aum,Franciolini:2023lgy, Cheng:2023ikq, Tasinato:2023ukp,Choudhury:2023hvf,Choudhury:2023kdb,Motohashi:2023syh}. However, the final outcome becomes very sensitive and does not at all permit big mass PBHs creation from the underlying physical setup \cite{Choudhury:2023vuj,Choudhury:2023jlt}. Also in refs.\cite{Riotto:2023gpm,Firouzjahi:2023ahg,Firouzjahi:2023aum} the authors have shown that the one-loop spectrum is highly suppressed in the present context of the computation in the presence of smooth transitions.
But none of the earlier studies conducted with smooth transitions \cite{Riotto:2023gpm,Firouzjahi:2023ahg,Firouzjahi:2023aum}, nor some of the studies with sharp transitions \cite{Kristiano:2022maq,Riotto:2023hoz,Kristiano:2023scm,Franciolini:2023lgy,Cheng:2023ikq, Motohashi:2023syh} have examined the crucial issues of renormalization and resummation in order to arrive at a final conclusion regarding the insignificant one-loop correction and the production of large mass PBHs. Our research was most successful when we conducted a thorough analysis of renormalization and resummation in an EFT configuration with a rapid transition from SRI to USR phase and USR to SRII phase. This allowed us to reach the ultimate conclusion. Through thorough computation, we proved all of the arguments put out in support of our analysis in the study.

\subsection{Renormalization scheme dependence}

It is crucial to note, in keeping with the discussion from earlier portions of this work, that we have discovered that the renormalization scheme that is being considered affects the explicit mathematical form of the counter term. It is clearly demonstrated by us that the final determined form of the counter terms differs in the late-time and adiabatic/wavefunction renormalization techniques. It was discovered, nevertheless, that the final computed results for the one-loop momentum integrals, at least for the two renormalization schemes mentioned, exhibit exact equivalency. In both cases, quadratic UV divergence can be entirely eliminated, and the result is expressed in terms of logarithmic IR divergent contribution. It is necessary to use the power spectrum renormalization strategy after late-time or adiabatic/wave function renormalization, that is, after the UV divergent quadratic detrimental contribution has been completely eliminated, in order to further smooth out this IR dependency. In this study, we base our final conclusion entirely on the particular schemes of the renormalization carried out in this connected debate. In addition to the aforementioned schemes, the following are also addressed: (1) Late-time (LT) scheme \cite{Choudhury:2023jlt}, (2) Adiabatic-Wave function (AWF) scheme \cite{Choudhury:2023vuj}, and (3) Power Spectrum (PS) scheme \cite{Choudhury:2023vuj,Choudhury:2023jlt}. 
In the literature on quantum field theory of curved space-time, there are several different effective renormalization procedures that include (4) Minimal subtraction (MS) scheme \cite{tHooft:1973mfk,Weinberg:1973xwm,Collins:1984xc} and the related modified minimal subtraction ($\overline{\rm MS}$) scheme \cite{tHooft:1973mfk,Weinberg:1973xwm,Collins:1984xc}, (5) On-shell scheme \cite{Peskin:1995ev}, (6) Bogoliubov-Parasiuk-Hepp-Zimmermann (BPHZ) scheme \cite{Dyson:1949ha,Kraus:1997bi,Piguet:1986ug,Zimmermann:1968mu,Zimmermann:1969jj}, (7) Bogoliubov-Parasiuk-Hepp-Zimmermann-Lowenstein (BPHZL) scheme \cite{Lowenstein:1975rg,Lowenstein:1975ps}, (8) Dimensional Renormalization (DR) scheme \cite{Binetruy:1980xn,Coquereaux:1979eq,Belusca-Maito:2020ala}, (9) Algebraic Renormalization (AR) scheme \cite{Adler:1969er,Batalin:1981jr,Becchi:1973gu,tHooft:1972tcz,Piguet:1995er} and there are many more in the list. In an attempt to prove a strong no-go theorem on the PBH mass, we have not carried out our analysis for the scenarios indicated in (4) to (9) and we have not verified and cross-checked the validity and application of the conclusions made in this study. Conducting this kind of investigation in the near future seems fascinating. Therefore, till this point, we can conclude that we are able to eliminate the quadratic UV divergence, smoothen the logarithmic IR divergence contribution, and ultimately provide a no-go theorem regarding PBH mass by imposing a time constraint on the USR phase's duration in order to preserve the perturbative approximations within the framework examined in this piece. Since the application and accuracy of the proposed no-go theorem on the PBH mass can only be determined after doing the analysis for all classes of the aforementioned renormalization schemes, we are thus refraining from making any more strong claims in this respect. Last but not least, before we wrap up in the following part, there is one more significant issue that we must specifically address. Using the DRG resummation approach, we have supplied the resummed version of the one-loop adjusted power spectrum in our study. This is an enhanced version of the well-known RG resummation method, in which one can construct a finite and controllable amplitude of the scalar power spectrum by performing the sum over all possible secular contributions, i.e., by utilizing the repetitive structure in the higher order loop diagrams. This is completely consistent with cosmological $\beta$-functions and the associated slow-roll hierarchy at the CMB pivot scale of the computation. Consequently, the PBH creation phenomena derived from the DRG resummed spectrum is in full agreement with the Renormalization Group (RG) flow. This is taken into account by the suggested no-go theorem when determining the formation of PBH mass within the current context. We will be delving further into the previously described schemes of renormalization followed by DRG resummation in the near future, so that we may make more pointed remarks on the PBH production associated with the current one-loop correction.

\subsection{PBH formation and effects of quantum loops from Goldstone EFT}

Given that we are operating inside the EFT framework, we will see a few modifications as a result of the effective sound speed $c_{s}$ being included in the mass of PBH calculation. We derive the PBH mass using the same method as in ref.\cite{Mishra:2019pzq}, and the resulting result is comparable, but with the modifications shown below. We begin with the modification $c_{s}(\tau)k_{\rm s,e} = (aH)_{\rm s,e}$ for time $\tau = \tau_{s}\;{\rm and}\;\tau = \tau_{e}$, where we have sharp transitions because of the EFT framework. This indicates that having effective sound speed modifies the horizon crossing condition. The sound speed parameter is parametrized as follows, as was previously discussed: $c_{s}(\tau_{s}) = c_{s}(\tau_{e}) = \tilde{c}_{s} = 1\pm\delta$, where $\delta \ll 1$. Hence, we see that $(aH)_{\rm s,e} = \tilde{c}_{s}k_{\rm s,e}$. Consequently, we have $c_{s}p_{*} = (aH)_{*}$ for the pivot scale $k = p_{*}$ since we also have $c_{s}(\tau_{*}) = c_{s,*} = c_{s}$. Keeping these helpful details in mind, we move on to the PBH mass derivation as follows. When considering the radiation-dominated (RD) epoch, the Hubble scale value adheres to the relation:
\bea \label{Hpbh}
H^{2} &=& \frac{\rho_{r}}{3M_{p}^{2}} = \Omega^{0}_{r}H_{0}^{2}\frac{\rho_{r}}{\rho^{0}_{r}}= \Omega^{0}_{r}h^{2}\times\left(\frac{100 {\rm km}}{\rm s\;Mpc}\right)^{2}\left(\frac{g_{*}}{g^{0}_{*}}\right)\left(\frac{T}{T^{0}}\right)^{4},
\eea
where the energy density-temperature relationship is utilized, and the amount being evaluated at the current time scale is indicated by the notation $``0"$. The following expression for the Hubble scale during the Radiation Dominated (RD) era is obtained by applying the conservation of entropy and making the assumption that the effective number of relative degrees of freedom of the energy and entropy densities, evaluated during the RD era, are almost equal $g_{*} \sim g_{*,s}$:
\bea
H^{2} = \Omega^{0}_{r}h^{2}\left(\frac{100 {\rm km}}{\rm s\;Mpc}\right)^{2}\left(\frac{g_{*}}{g_{0*}}\right)^{-1/3}\left(\frac{g^{0}_{*,s}}{g^{0}_{*}}\right)^{4/3}(1+z)^{4},\quad\eea
where the redshift factor, which is a result of entropy conservation, enters the current computation. Currently, the mass of the PBHs produced by the significant variations that entered during the RD period is dependent on the corresponding mass of the Horizon as follows:
\bea M_{\rm PBH} = \gamma M_{H} = \gamma\left(\frac{4\pi M_{pl}^{2}}{H}\right), \eea
where the critial collapse factor is $\gamma(\sim 0.2)$. The following relation for the mass of PBH may be obtained using the formula for the Hubble scale that was previously determined, the degrees of freedom $g^{0}_{*}=3.38$, $g^{0}_{*,s}=3.94$, and $\Omega^{0}_{r}h^{2}=4.18\times10^{-5}$:
\bea \label{zpbh}
\frac{M_{\rm PBH}}{M_{\odot}}&=&1.55\times10^{24}\times\left(\frac{\gamma}{0.2}\right)\left(\frac{g_{*}}{106.75}\right)^{1/6}(1+z)^{-1/2}, \eea
Leveraging the fact that the Hubble scale is nearly constant during inflation, we apply the established relationship $(aH)_{*} = c_{s}p_{*}$ for the wavenumber $p_{*}$ that exits the Hubble radius at that point. This assumption also leads to the relation $(aH)_{\rm PBH} \sim a_{\rm PBH}H_{*}$ for the wavenumber that exits at the time of PBH formation. We can determine the amount of expansion in terms of the number of e-foldings by using the following expression:
\bea \Delta {\cal N} = \ln\left({\frac{a_{\rm PBH}}{a_{*}}}\right) = \ln\left({\frac{(aH)_{\rm PBH}}{(aH)_{*}}}\right) = \ln\left({\frac{\tilde{c}_{s}k_{\rm PBH}}{c_{s}p_{*}}}
\right).\quad\eea
The scale at which PBH production takes place, $k_{\rm PBH} = k_{s}$, and the effective sound speed at that moment, $\tau=\tau_{s}$, are both utilized here, as is $\tilde{c}_{s}$. The relationship that follows is obtained by using the second equality in the previous equation, $a_{\rm PBH} = 1/(1+z)$, in conjunction with the Hubble scale at PBH formation from eqn. (\ref{Hpbh}) and the mass-redshift relation from eqn. (\ref{zpbh}):
\bea
&&\Delta {\cal N} = 17.33 + \frac{1}{2}\ln\left({\frac{\gamma}{0.2}}\right)-\frac{1}{12}\ln\left({\frac{g_{*}}{106.75}}\right)-\frac{1}{2}\ln\left({\frac{M_{\rm PBH}}{M_{\odot}}}\right). \eea
When we exponentiate the aforementioned expression, we obtain the intended outcome:
\bea
\left(\frac{M_{\rm PBH}}{M_{\odot}}\right)_{c_{s}}&=&1.13\times10^{15}\times\left(\frac{\gamma}{0.2}\right)\left(\frac{g_{*}}{106.75}\right)^{-1/6}\left(\frac{c_{s}p_{*}}{\tilde{c}_{s}k_{s}}\right)^{2},
\eea
where $k_{\rm PBH} = k_{s}$, the formation scale that corresponds with the sharp transition scale in our current computation, is used to quantify the final mass. $\gamma\sim 0.2$ is the efficiency factor, $g_*$ is the relativistic d.o.f., and $M_{\odot}\sim 2\times 10^{30}{\rm kg}$ is the solar mass. The pivot scale is also maintained at $p_*=0.02\;{\rm Mpc}^{-1}$. It is known that $c_s\neq \tilde{c}_s$ and that both $\tilde{c}_s>c_s$ and $\tilde{c}_s<c_s$ are possible given the behavior of the sound seed as described before. Thus, we get a more reduced version of the formula above:
\bea &&\left(\frac{M_{\rm PBH}}{M_{\odot}}\right)_{c_s}=1.13\times 10^{15}\times\bigg(\frac{\gamma}{0.2}\bigg)\bigg(\frac{g_*}{106.75}\bigg)^{-1/6}\times\left(\frac{p_*}{k_s}\right)^{2}\times c^2_s(1\mp 2\delta)\;\approx \left(\frac{M_{\rm PBH}}{M_{\odot}}\right)_{c_s=1}\times c^{2}_s.\quad\eea
Thus, for the purpose of simplicity, the contribution from the tiny fine-tuning component $\delta$ is disregarded. It is noteworthy that \cite{Choudhury:2023vuj} exists for the standard single field $c_s=1$ case.
\bea &&\left(\frac{M_{\rm PBH}}{M_{\odot}}\right)_{c_s=1}=1.13\times 10^{15}\times\bigg(\frac{\gamma}{0.2}\bigg)\bigg(\frac{g_*}{106.75}\bigg)^{-1/6}\left(\frac{p_*}{k_s}\right)^{2}.\quad\quad\eea
Expressing the PBH mass for the EFT configuration in terms of the contribution from the conventional single-field ($c_s=1$) setup serves as the main driving force here. It also makes it easier for us to see that the sole change in this expression's effective sound speed is caused by the EFT configuration, and that the contribution is minimal within a very narrow favored window. Nonetheless, we have made this contribution clear in order to measure the difference when the EFT configuration is present. We fix $\gamma\sim 0.2$, $g_*\sim 106.75$, $p_*=0.02\;{\rm Mpc}^{-1}$, and $k_s=10^{21}{\rm Mpc}^{-1}$ in order to facilitate additional numerical estimation:
\bea \left(\frac{M_{\rm PBH}}{M_{\odot}}\right)_{c_s=1}=4.52\times {10}^{-31},\eea 
using which finally we write:
\bea \left(\frac{M_{\rm PBH}}{M_{\odot}}\right)_{c_s}\approx 4.52\times {10}^{-31}\times c^{2}_s.\quad\eea
    \begin{figure*}[htb!]
    	\centering
{
      	\includegraphics[width=14cm,height=9cm] {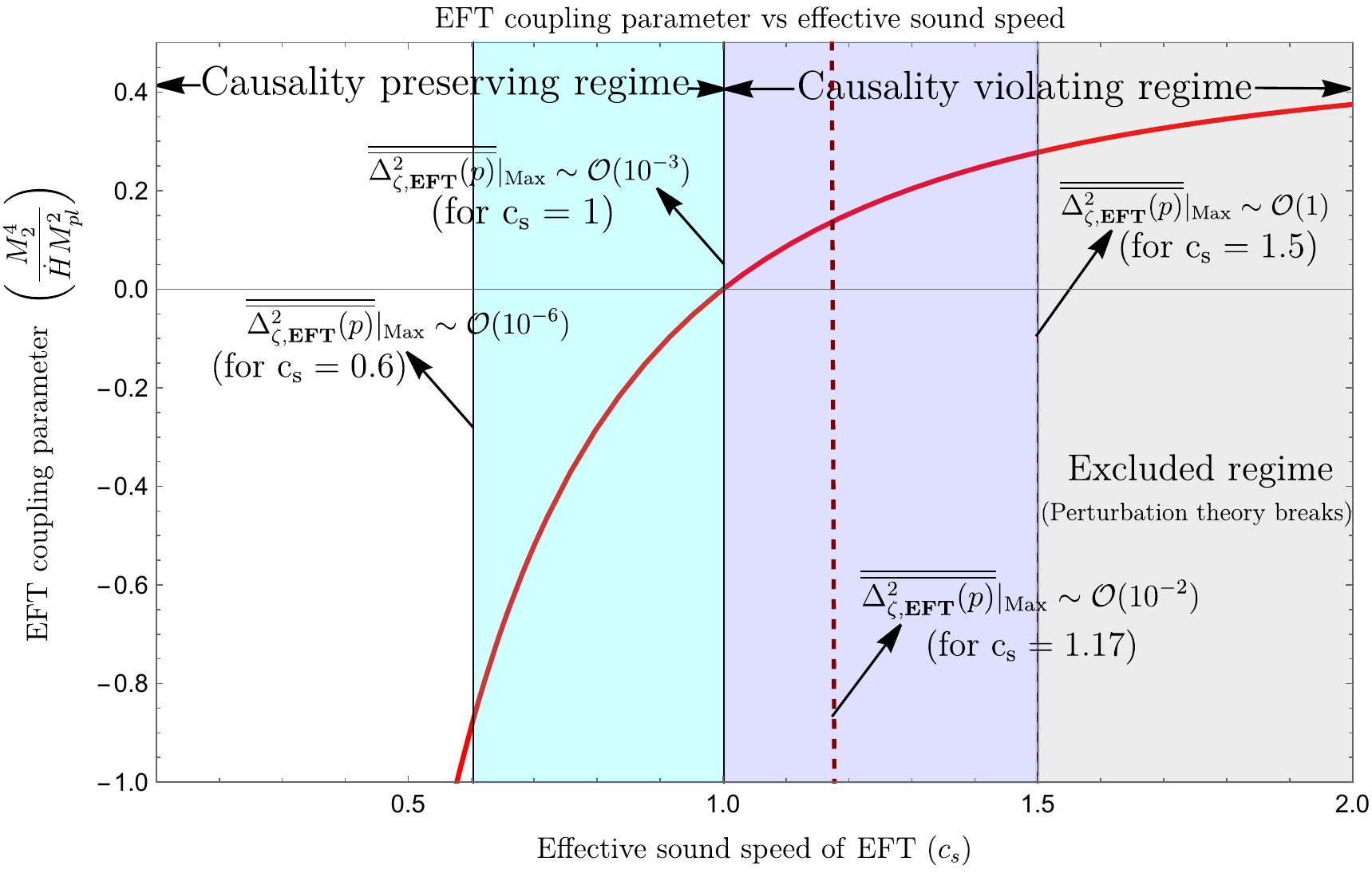}
        \label{KX1c}
    }
    	\caption[Optional caption for list of figures]{Plots showing how the EFT coupling parameter behaves in relation to the effective sound speed $c_s$ are made using the current EFT setup. Plots indicate that the permitted range for effective sound speed is $c_s\gtrsim 1$, and that $c_s=1.17$ yields the best result for an amplitude of ${\cal O}(10^{-2})$ in the associated spectrum. The perturbation theory breaks at ${\cal O}(1)$, when the amplitude approaches at $c_s=1.5$.} 
    	\label{Spectrum5}
    \end{figure*}
    \begin{figure*}[htb!]
    	\centering
    \subfigure[ $M_{\rm PBH}$ vs $c_s$.]{
      	\includegraphics[width=8cm,height=7cm] {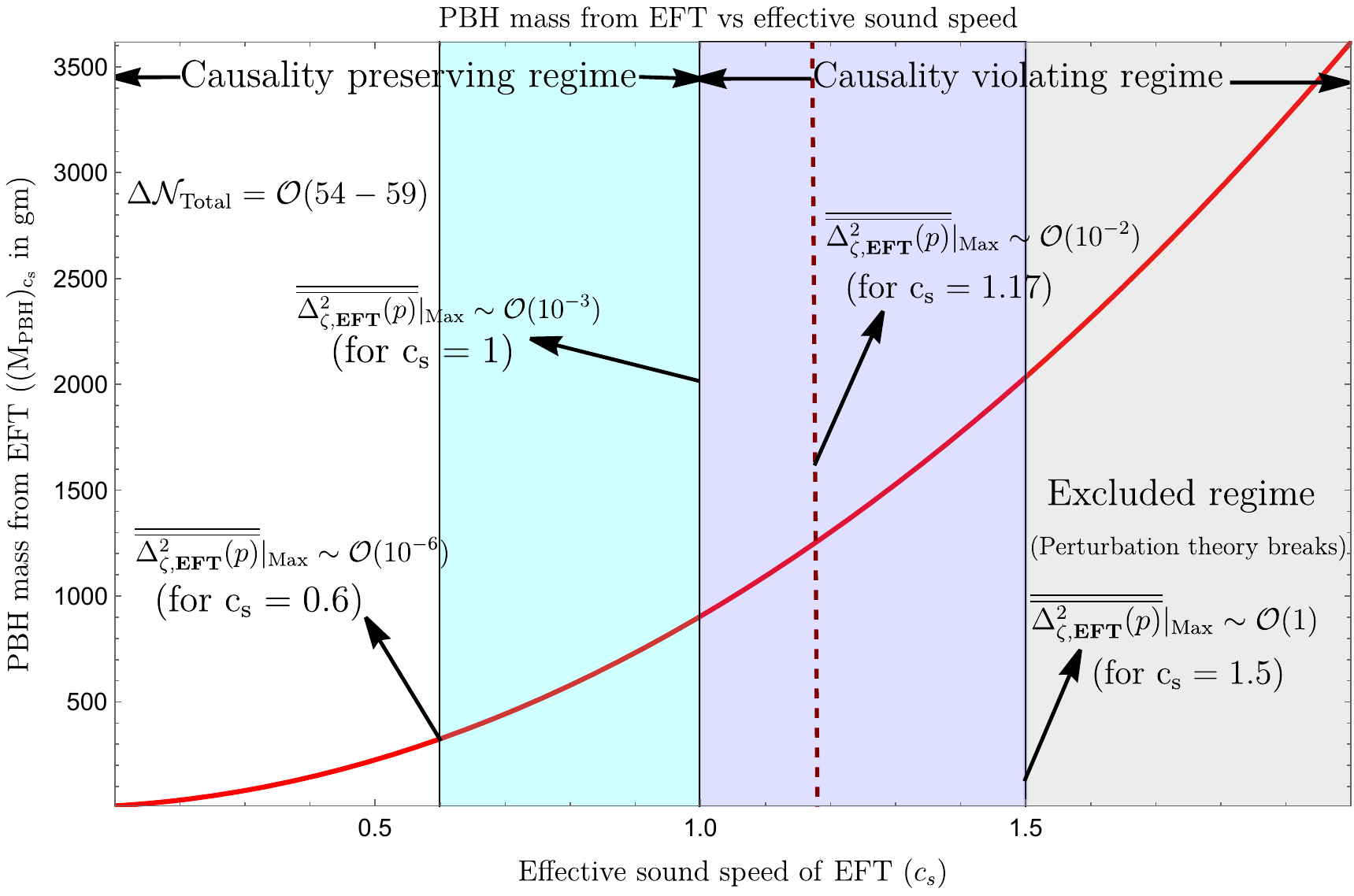}
        \label{Kv1}
    }
    \subfigure[$M_{\rm PBH}$ vs $c_s$.]{
       \includegraphics[width=8cm,height=7cm] {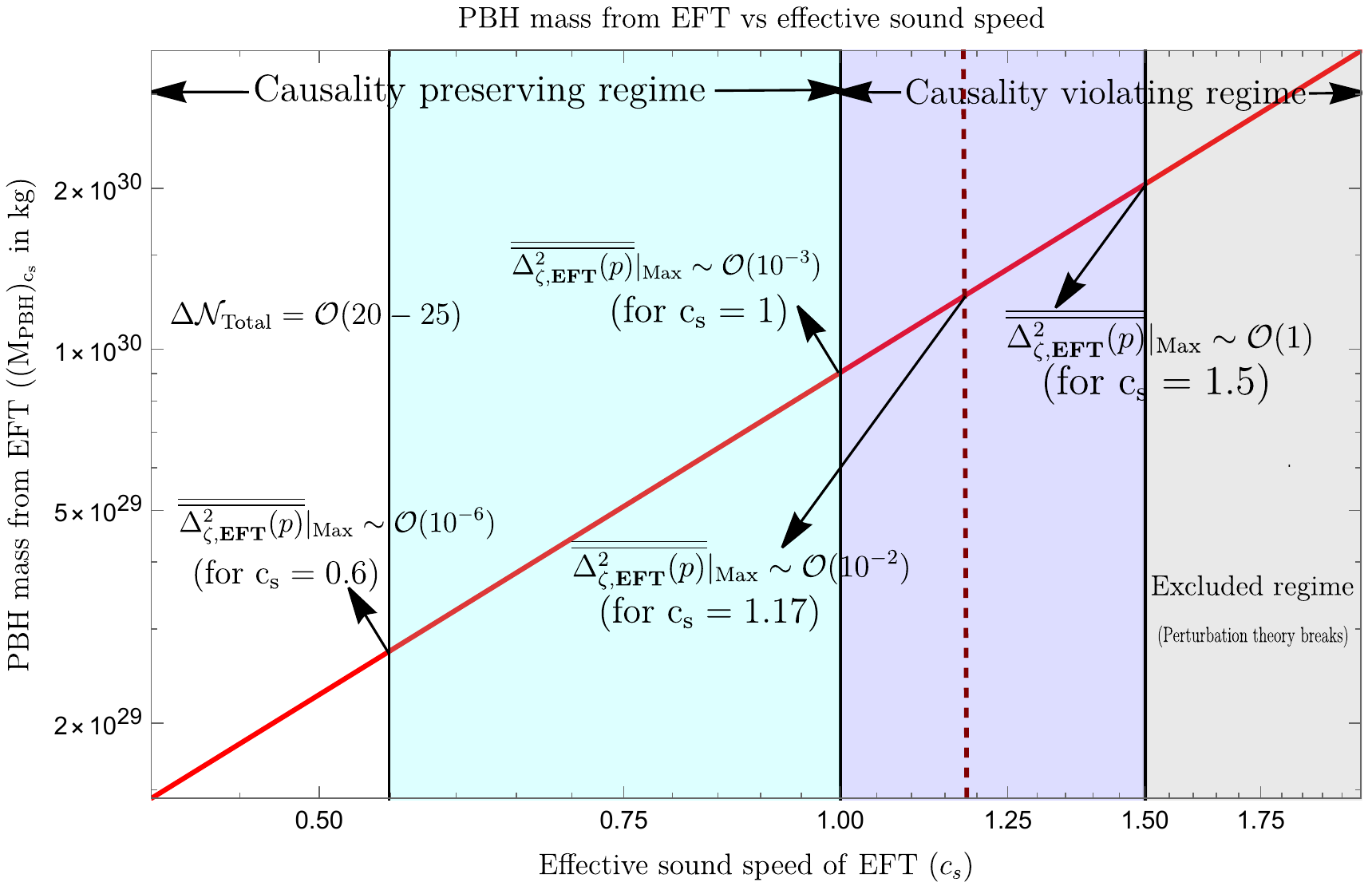}
        \label{Kv2}
       }
    	\caption[Optional caption for list of figures]{Plotting the PBH mass behavior from the current EFT setup with respect to the effective sound speed $c_s$ for two cases, where the total number of e-foldings is fixed at $\Delta{\cal N}_{\rm Total}\sim {\cal O}(54-59)$ and \ref{Kv2} $\Delta{\cal N}_{\rm Total}\sim {\cal O}(20–25)$, respectively. For the first plot, $k_s=10^{21}\;{\rm Mpc}^{-1}$, and the second plot, $k_s=10^{6}\;{\rm Mpc}^{-1}$, we place the pivot scale at $p_*=0.02\;{\rm Mpc}^{-1}$. The SR to USR sharp transition scale is at these locations. Figures indicate that the range of effective sound speed that is permitted is $c_s\gtrsim 1$; of these, $c_s=1.17$ yields the best result for an amplitude of the associated spectrum that is ${\cal O}(10^{-2})$.  Because there are enough e-foldings for inflation, \ref{Kv1} here enables the development of tiny mass PBHs. Conversely, big mass PBH creation with insufficient e-foldings for inflation is supported by \ref{Kv2}.
} 
    	\label{Spectrum6}
    \end{figure*}
    \begin{table}
\footnotesize
\begin{tabular}{|c|c|c|c|c|c|c|c|}
 \hline\hline
 \multicolumn{8}{|c|}{A detailed list of $P(X,\phi)$ models and its connection with Goldstone EFT} \\
 \hline\hline
 Model name & $P(X,\phi)$ function & $\displaystyle \frac{M^4_2}{\dot{H}M^2_{pl}}$ & $\displaystyle \frac{M^3_1}{HM^2_{pl}}$ & $\displaystyle \frac{M^4_3}{H^2M^2_{pl}}$   & $c_s$ & $\displaystyle{\cal M}=\frac{M_{\rm PBH}}{10^2{\rm gm}}$ & Comment\\
 \hline\hline
 & & & & & & &  \\
DBI   & $\displaystyle -\frac{\Lambda^4}{f(\phi)}\sqrt{1-\frac{f(\phi)}{\Lambda^4}X}+\frac{\Lambda^4}{f(\phi)}-V(\phi)$  & $(-0.89)-0$  & $(-0.1)-0.1$ &   $(-0.5)-0.05$ &   $0.6-1$ &  $0.36-9.04$ & Causal (\ding{56})\\
& & & & & & &  \\
\hline\hline
 & & & & & & &  \\
Tachyon   & $\displaystyle -V(\phi)\sqrt{1-2\alpha^{'}X}$  & $(-0.89)-0$  & $(-0.2)-0.1$ &   $(-0.4)-0$ &   $0.6-1$ &  $0.36-9.04$ & Causal (\ding{56})\\
& & & & & & &  \\
\hline\hline
 & & & & & & &  \\
GTachyon   & $\displaystyle -V(\phi)\left(1-2\alpha^{'}X\right)^{q}$\;($q\leq \frac{1}{2}$, $\frac{1}{2}<q<1$)     & $(-0.89)-0$  & $(-0.2)-0$ &   $(-0.4)-0$ &   $0.6-1$ &  $0.36-9.04$ & Causal (\ding{56})
\\
   & \quad\quad\quad\quad\quad\;($q> \frac{1}{2}$)     & $0-0.28$  & $0-0.07$ &   $0-0.09$ &   $1-1.5$ &  $9.04-20.3$ & Acausal (\ding{52})\\
& & & & & & &  \\
\hline\hline
  & & & & & & &  \\
$K$   & $\displaystyle \gamma_n X^n-V(\phi)$\;($1.89>n>1$)  & $(-0.89)-0$  & $(-0.4)-0$ &   $(-0.3)-0$ &   $0.6-1$ &  $0.36-9.04$ & Causal (\ding{56}) \\
  & \quad\quad\quad\quad\quad\;\;($1>n>0.72$)  & $0-0.28$  & $0-0.1$ &   $0-0.2$ &   $1-1.15$ &  $9.04-20.3$ & Acausal (\ding{52})\\
& & & & & & &  \\
\hline\hline
 Canonical& & & & & &  & \\
 Single    & $\displaystyle X-V(\phi)$  & $0$  & $0$ &   $0$ &   $1$ &  $9.04$ & Causal (\ding{52})\\
field & & & & & &  & \\
 \hline\hline
\end{tabular}
\caption{Restrictions on $P(X,\phi)$ theories with respect to PBH creation. Here, we have mentioned the limitations for any general effective potential that the appropriate physical frameworks permit, and we also note that sufficient e-folds of inflation must be reached. The permissible range for the EFT parameters $\displaystyle M^4_2/\dot{H}M^2_{pl}$, $\displaystyle M^3_1/HM^2_{pl}$, and $\displaystyle M^4_3/H^2M^2_{pl}$, as well as the allowed range for the PBH mass from each of the $P(X,\phi)$ models of inflation, have been clearly mentioned. In this case, \ding{56} and \ding{52} correspond to the authorized and not-allowed models of $P(X,\phi)$ models of inflation in the context of small-mass PBH creation. We utilize $k_s=10^{21}{\rm Mpc}^{-1}$ and $k_e=10^{22}{\rm Mpc}^{-1}$ in these estimations, which are the ones that really produce the tiny mass PBHs.  As a result, we have kept the cap on the total number of e-foldings at $\Delta {\cal N}_{\rm Total}\sim {\cal O}(54-59)$, which is unavoidably necessary to get enough inflation in the given situation. Since it cannot produce enough e-foldings for inflation, the potential of big mass PBHs is not investigated in this table. Look at the Appendix \ref{A1a} and \ref{A2a} for more details on this computation.}
\label{tab:1}
\end{table}
Instead of utilizing a pivot scale, the authors of ref.\cite{Sasaki:2018dmp,Kawasaki:2016pql} estimated the PBH mass using the wavenumber at radiation-matter equality $k_{\rm eq}$. The standard single field slow-roll inflation model, where $c_{s}=1$, was also considered in this analysis. It is predicted that the succeeding phases of the derivation would rely on the sound speed since we are performing the analysis in the presence of EFT, as was previously mentioned. We may write down the final expression as follows if we repeat the analysis as done in the stages above, according to the normalization scale of $k_{\rm eq}$ as in ref.\cite{Sasaki:2018dmp,Kawasaki:2016pql}:
\bea
\frac{M_{\rm PBH}}{M_{\odot}}&=& 3.6\left(\frac{\gamma}{0.2}\right)\left(\frac{g_{*}}{106.75}\right)^{-1/6}\left(\frac{\tilde{c}_{s}k_{\rm PBH}}{c_{s}10^{6}{\rm Mpc}^{-1}}\right)^{-2},
\eea
where $\tilde{c}_{s}\;{\rm and}\;c_{s}$ is the scale-dependent effective sound speed. Using the scale $k_{\rm eq}$ to write this expression and $\tilde{\tilde{c}}_{s}$ as the equivalent sound speed, we obtain the following:
\bea
\frac{M_{\rm PBH}}{M_{\odot}}&=& 3.6\left(\frac{\gamma}{0.2}\right)\left(\frac{g_{*}}{106.75}\right)^{-1/6}\times\left(\frac{\tilde{c}_{s}k_{\rm PBH}}{\tilde{\tilde{c}}_{s}k_{\rm eq}}\right)^{-2}\left(\frac{\tilde{\tilde{c}}_{s}k_{\rm eq}}{10^{6}{\rm Mpc}^{-1}}\right)^{-2}. \eea
We obtain the following relation by inserting the values of the parameters $k_{\rm eq} = 0.07\Omega^{0}_{m}h^{2}{\rm Mpc}^{-1}$, where $h\sim 0.674$ and $\Omega^{0}_{m}\sim 0.315$:
\bea
\frac{M_{\rm PBH}}{M_{\odot}}&=& 0.36\times10^{15}\times\left(\frac{\gamma}{0.2}\right)\times\left(\frac{g_{*}}{106.75}\right)^{-1/6}
\left(\frac{\tilde{c}_{s}k_{\rm PBH}}{\tilde{\tilde{c}}_{s}k_{\rm eq}}\right)^{-2}. \eea
We can infer $\tilde{\tilde{c}}_{s} \simeq c_{s}$ in this case because, as we can see from this result, the effective sound speed remains nearly constant within this brief interval due to the wavenumber at the radiation-matter equality scale being very close to the wavenumber at the pivot scale. Hence, we get the following conclusion for the formula for PBH mass when we put in the values for $\gamma \sim 0.2$, $g_{*}=106.75$, and $k_{\rm PBH}=k_{s}\sim 10^{21}{\rm Mpc}^{-1}$ and use the value for $\tilde{c}_{s}=1+\delta$, such that $\delta \ll 1$:
\bea
\left(\frac{M_{\rm PBH}}{M_{\odot}}\right)_{c_{s}} \approx 10^{-31}\times c^{2}_{s}.
\eea
According to this research, the equivalent effective sound speed for the following EFT parameter coupling range fluctuates between $1<c_s<1.17$:
\bea && 0\lesssim M^4_2/\dot{H}M^2_p\lesssim 0.13,\\ 
 && 0\lesssim \bar{M}^3_2/HM^2_p\lesssim 0.08,\\ 
 && 0\lesssim M^4_3/H^2M^2_p\lesssim 0.05.\eea
We have shown the behavior of the PBH mass from the current EFT set up $(M_{\rm PBH})_{c_s}$ and the EFT coupling parameter $M^4_2/\dot{H}M^2_p$ with regard to the effective sound speed $c_s$ in figures (\ref{Spectrum5}), (\ref{Kv1}), and (\ref{Kv2}). The pivot scale is fixed at $p_*=0.02\;{\rm Mpc}^{-1}$, and the sharp transition scale is fixed at $k_s=10^{21}\;{\rm Mpc}^{-1}$ and $k_s=10^{6}\;{\rm Mpc}^{-1}$, respectively, in figure (\ref{Kv1}) and figure (\ref{Kv2}). Each of these figures demonstrates that the permitted range of effective sound speed is $c_s\gtrsim 1$, and that the optimal result to have ${\cal O}(10^{-2})$ amplitude of the associated spectrum is $c_s=1.17$. The amplitude approaches ${\cal O}(1)$ at $c_s=1.5$, at which point the perturbative approximations in this analysis entirely collapse. Through a comparison of figures (\ref{Kv1}) and (\ref{Kv2}), it was discovered that big mass PBHs may be formed with inadequate total e-foldings $\Delta{\cal N}_{\rm Total}\sim {\cal O}(20–25)$ for inflation within the range ${\cal O}(10^{29}-10^{30}){\rm kg}$. However, one can only produce small mass PBHs inside a very narrow window, ${\cal O}(10^{2}-10^{3}){\rm gm}$, in order to accommodate a sufficient total number of e-foldings $\Delta{\cal N}_{\rm Total}\sim {\cal O}(54-59)$ for inflation. The preferred range of effective sound speed in both scenarios will be inside the window $1\leq c_s\leq 1.17$, with $c_s=1.17$ being the only value within which the one-loop adjusted power spectrum may be boosted in amplitude, ${\cal O}(10^{-2})$.

The one-loop contributions yield two additional EFT parameters: $\bar{M}^3_2/HM^2_p$ and $M^4_3/H^2M^2_p$. Since the SR contribution in the one-loop corrected formula for the primordial power spectrum for scalar modes is muted relative to the USR contribution, we discovered that the findings reported in this article are not highly sensitive to both of these parameters. Both of these parameters cannot be fixed from the ground up just by looking at the effect on the one-loop corrected spectrum amplitude. Nonetheless, we are aware that the third-order perturbative action for the scalar curvature perturbation is where the one-loop contribution is calculated, and it is from this action that the three-point function and related impact in the primordial non-Gaussianity can be directly estimated \cite{Maldacena:2002vr,Seery:2005wm,Senatore:2009gt,Chen:2006nt,Chen:2010xka,Chen:2009zp,Chen:2009we,Chen:2008wn,Chen:2006xjb,Choudhury:2012whm,Agarwal:2012mq,Holman:2007na,Creminelli:2005hu,Behbahani:2011it,Smith:2009jr,Cheung:2007sv,Creminelli:2006rz,Creminelli:2006gc,Kalaja:2020mkq, Meerburg:2019qqi,Lee:2016vti,Maldacena:2011nz,Werth:2023pfl}. The mentioned $P(X,\phi)$ theoretical models of inflation contain two EFT parameters, $\bar{M}^3_2/HM^2_p$ and $M^4_3/H^2M^2_p$, which we have placed within a certain range in order to produce viable huge non-Gaussian effects. However, as substantial non-Gaussian effects cannot be produced in the standard single field models of the inflationary paradigm, we have set the EFT parameters for the $P(X,\phi)=X-V$ model to $\bar{M}^3_2/HM^2_p=0$ and $M^4_3/H^2M^2_p=0$. We completed a thorough computation in Appendix \ref{A2a} to support our research in this work and establish a direct link between the Goldstone EFT setup and the $P(X,\phi)$ theoretical models of inflation. It has been demonstrated that the EFT parameter $M^4_2/\dot{H}M^2_p$ may be easily stated in terms of the derivatives of the $P(X,\phi)$ in relation to the background level kinetic contribution $X$. This aids in constraining the EFT parameter $M^4_2/\dot{H}M^2_p$ for a variety of $P(X,\phi)$ single field inflation models, including the standard single field slow roll framework. We have covered the limitations on the different model parameters and the PBH mass derived from the different kinds of $P(X,\phi)$ models of single field inflation in Appendix \ref{A2a}. For the sake of completeness, these results are clearly quoted in Table (\ref{tab:1}). The limitations for each generic effective potential $V(\phi)$ that are permitted by the appropriate underlying physical frameworks are given in this table, and each model must accomplish sufficient e-folds of inflation. The range of permitted EFT parameters, $\displaystyle M^4_2/\dot{H}M^2_{pl}$, $\displaystyle M^3_1/HM^2_{pl}$, and $\displaystyle M^4_3/H^2M^2_{pl}$, as well as the range of permitted PBH masses from each of the $P(X,\phi)$ models of inflation that fall within the effective sound speed range of $0.6<c_s<1.5$, have been explicitly mentioned. Furthermore, based on the highest amount of amplification during PBH generation in the one-loop corrected primordial power spectrum for scalar modes, we observed from this table that the range $1<c_s<1.17$ is more favored where the underlying EFT supports breakdown of causality. In the context of PBH production with tiny mass, we designate the not-allowed and allowed models of $P(X,\phi)$ models of inflation with the symbols \ding{56} and \ding{52} for the sake of greater comprehension. By incorporating the one-loop contribution in the primordial power spectrum, we are unable to strictly generate large mass PBH within the framework of Goldstone EFT, which is easily connected in terms of all known $P(X,\phi)$ models of single field inflation, including both canonical and non-canonical frameworks. Thus, all feasible forms of single-field inflation that can be translated into $P(X,\phi)$ models of the single-field inflationary paradigm are established, and a no-go theorem is thus established. Look at the Appendix \ref{A1a} and \ref{A2a} for more details on this computation.

Using the foregoing results, we calculate the fractional abundance of the PBH created when accounting for the contribution from the one-loop corrected power spectrum, and use this information to make a statement regarding the permitted PBH mass-produced within the framework of EFT of single-field inflation. This necessitates understanding the PBHs formation mass fraction, which is provided as:
\bea f_{\rm PBH}&=&1.68\times 10^{8}\times\left(\frac{\gamma}{0.2}\right)^{1/2}\times\left(\frac{g_{*}}{106.75}\right)^{-1/4}\left(\frac{M_{\rm PBH}}{M_{\odot}}\right)^{-1/2}\beta(M_{\rm PBH}).\quad \eea
As a result, if the density contrast rises to a certain threshold value $\delta_{\rm th} > \delta_c=1/3$, which is necessary for the formation of PBH as the large density perturbations enter the Horizon at a certain scale after inflation, then the probability of that event, calculated using the Press-Schechter formalism, corresponds to:
\bea
\beta(M_{\rm PBH}) &=& \gamma\int_{\delta_{\rm th}}^{1}P(\delta)d\delta= \gamma\int_{\delta_{\rm th}}^{1}\frac{e^{-\delta^{2}/2\sigma_{M_{\rm PBH}}^{2}}}{\sqrt{2\pi}\sigma_{M_{\rm PBH}}}d\delta
\approx \gamma\left(\frac{\sigma_{M_{\rm PBH}}}{\sqrt{2\pi}\delta_{\rm th}}\right)\exp{\left(-\frac{\delta_{\rm th}^{2}}{2\sigma_{M_{\rm PBH}}^{2}}\right)}.\eea
Around the PBH formation scale $R = 1/(\tilde{c}_{s}k_{\rm PBH}) = 1/(aH)_{\rm PBH}$, the quantity $\delta_{\rm th}$ is coarse-grained. Moreover, the variance of the coarse-grained density fluctuations at the mass scale $M_{\rm PBH}$ is represented by $\sigma_{M_{\rm PBH}}$. Effective sound speed parameter appears as a result of current EFT architecture, where $c_{s}=1$ reduces to the classic single-field inflation \cite{Mishra:2019pzq,Sasaki:2018dmp,Kawasaki:2016pql}. These density fluctuations are smoothed across the scale $R$ by means of a Gaussian window function $W(p,R)=\displaystyle{\exp{(-p^{2}R^{2}/2)}}$. The variance that results from this process is calculated as follows:
\bea
\sigma_{M_{\rm PBH}}^{2}&=&\int^{\infty}_{0} d\ln{p}\;\Delta^{2}_{\delta}(p)W^{2}(p,R), 
\eea
Consequently, in the RD period, the power spectrum for density contrast is stated as follows:
\bea \label{power} \Delta^{2}_{\delta}(p)&=&\frac{16}{81}\left(\frac{p}{\tilde{c}_{s}k_{\rm PBH}}\right)^{4}\overline{\overline{\Delta^{2}_{\zeta,{\bf EFT}}(p)}}. \eea
We may now determine the abundance of PBH of a given mass created by the huge amplitude oscillations entering the RD period based on the variance. Furthermore, the calculation above indicates that the variance of the mass fraction and, consequently, the fractional abundance directly carry information about the one-loop correction in the primordial power spectrum.
     \begin{figure}[htb!]
    	\centering
 \subfigure[For $c_s=1$  with $M^4_2/\dot{H}M^2_p\sim -0.89$ (non-canonical and causal).]{
      	\includegraphics[width=8cm,height=7.3cm] {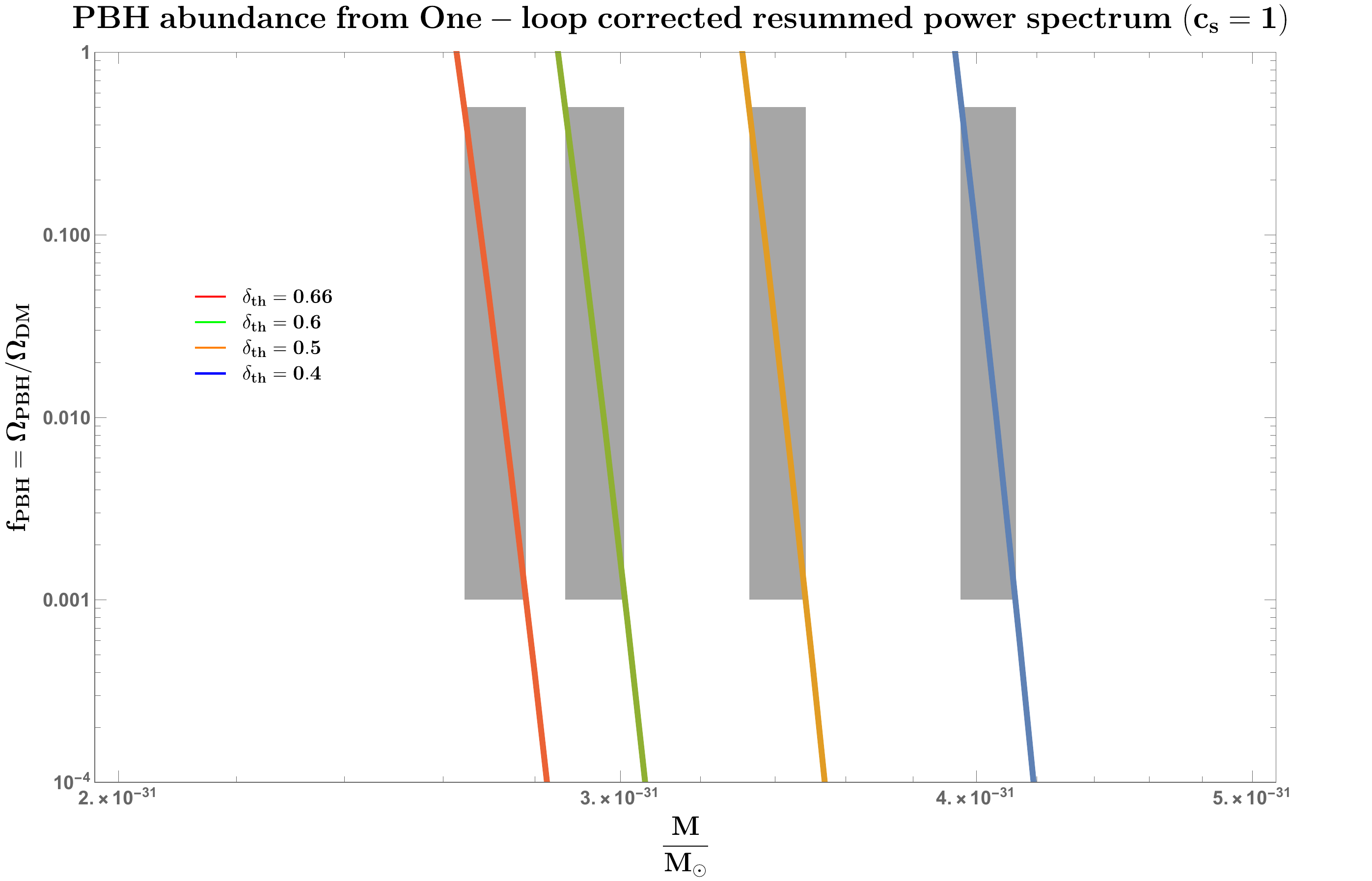}
        \label{ab1}
    }
    \subfigure[For $c_s=1.17(>1)$  with $M^4_2/\dot{H}M^2_p\sim 0.13$ (non-canonical and a-causal).]{
       \includegraphics[width=8cm,height=7.3cm] {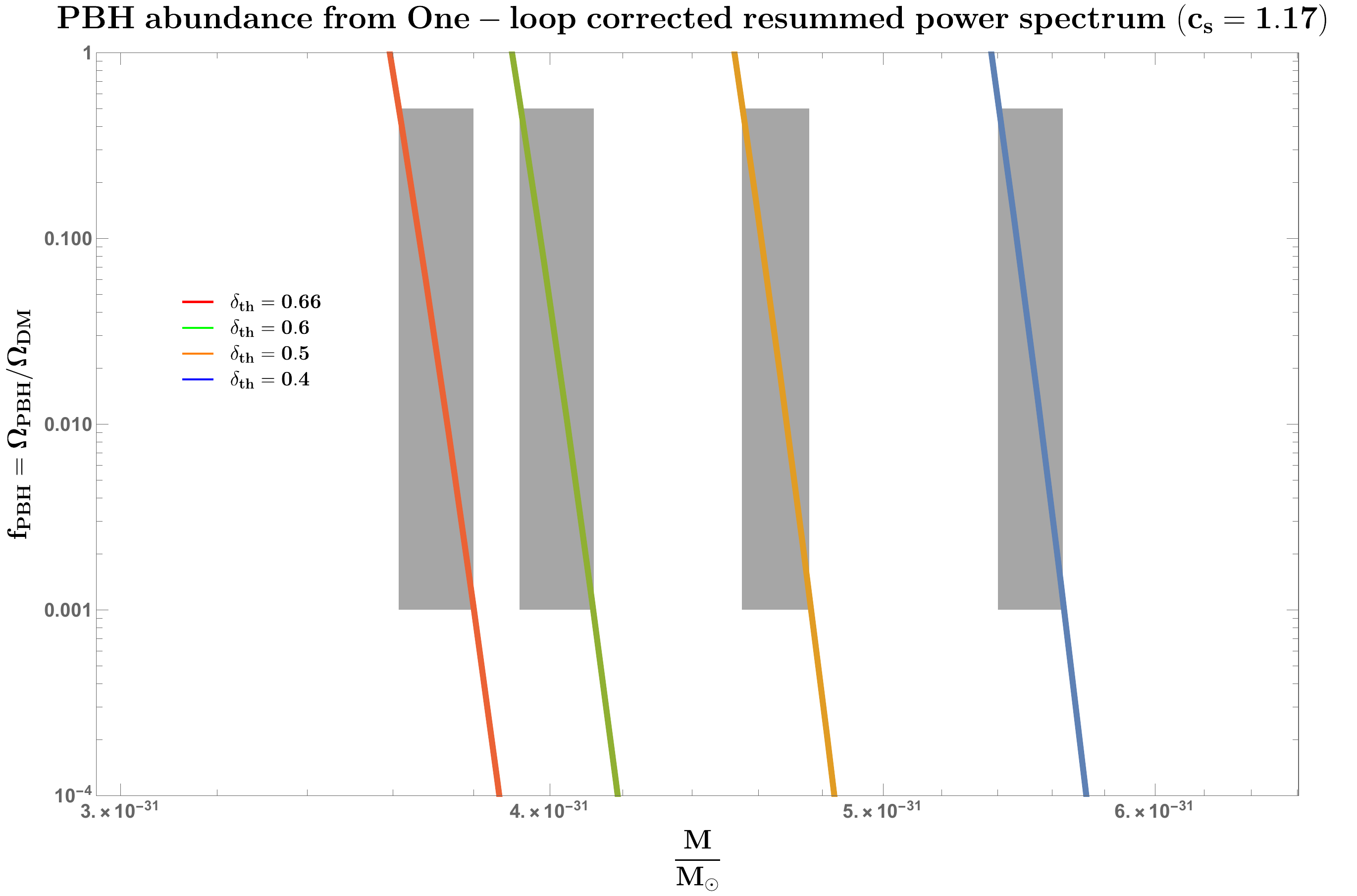}
        \label{ab2}
       }
    	\caption[Optional caption for list of figures]{Behaviour of the abundance of PBHs when plotted against their mass. The effective sound speed is taken to be as $c_{s}=1\;{\rm and}\;1.17$. The gray shaded region suggests the allowed mass region of the PBH produced for the corresponding density contrast threshold where the abundance falls in between $0.001 \leq f \leq 0.5$. Based on the constraint for the threshold of collapse necessary to neglect the non-linearities in the density contrast, the interval $2/5 \leq \delta_{\rm th} \leq 2/3$ is considered and the range of PBH mass, for both the cases of $c_{s}$ values, lies in between $10^{2}{\rm gm} < M_{\rm PBH} \lesssim 10^{3}{\rm gm}$. }
    	\label{abundance}
    \end{figure}
The value for the density contrast threshold must be taken into account when computing the PBH abundance generated by the current EFT framework, since this is the domain where linearities in the density contrast predominate, which is $2/5 \leq \delta_{\rm th} \leq 2/3$, see \cite{Musco:2018rwt}. See refs. \cite{Musco:2020jjb, DeLuca:2023tun} for more information on the recent, significant arguments around the situation of non-linear characteristics in the density contrast and its impact on the ultimate threshold range. In the values located between $1/3 < \delta_{\rm th} < 2/5$ and $2/3 < \delta_{\rm th} < 1$, the density contrasts non-linearities become relevant. The fundamental linking connection in coordinate space between the comoving curvature perturbation and the density contrast is expressed as follows \cite{Harada:2015yda}:
\bea \delta(\mathbf{x}, t) = -\frac{4}{9}\frac{1}{a^{2}H^{2}}\exp{(-2\zeta(\mathbf{x}))}\left(\nabla^{2}\zeta(\mathbf{x})+\frac{1}{2}\partial_{i}\zeta(\mathbf{x})\partial^{i}\zeta(\mathbf{x})\right),\quad
\eea
and the following formula is produced by the leading order term in the quantity $\zeta(\mathbf{x})$:
\bea \delta(\mathbf{x}, t) = -\frac{4}{9a^{2}H^{2}}\nabla^{2}\zeta(\mathbf{x}). \eea
When this formula is converted into the Fourier space, the EFT framework based eqn.(\ref{power}) provides the precise relationship between the power spectrum for the density contrast and the curvature perturbation. The scalar fluctuation component of the metric is likewise first written using the same kind of linear approximation. This also requires examining non-Gaussian characteristics, which is outside the purview of our current examination. In order to do this, it is sufficient to smooth the density contrast, which aids in the calculation of the PBH abundance factor, by having the density contrast profile follow Gaussian statistics and using the same Gaussian window function as previously selected. In the end, there is an underlying non-linear relationship between the curvature perturbation and the density contrast, which, when taken into account, can offer a more thorough understanding of the problems associated with PBH generation. Regarding the presence of this non-linear relation, the following refs.\cite{DeLuca:2019qsy, Young:2019yug} discuss it. The authors find that using perturbation theory alone to generate PBHs is much more difficult and that alternative approaches, such as peak theory and other statistical techniques for the non-Gaussian threshold, are needed to deal with these intrinsic non-Gaussian features and the resulting probability of PBH formation. The study results are now shown in terms of abundance plots for the cases where the effective sound speed is $c_{s} = 1$ and $c_s=1.17$. We derive the permitted range of the resultant PBH mass from the typical plots. $10^{2} {\rm gm} \lesssim M_{\rm PBH}\lesssim 10^{3} {\rm gm}$, or $2/5 \leq \delta_{\rm th} < 2/3$, is the permitted range of PBH masses that we have determined.

We deduce that there is a range of masses of PBH produced in the context of EFT of single-field inflation, which contributes significantly to the dark matter density as observed today, from the figures \ref{ab1}-\ref{ab2}. Under the restriction $2/5 \lesssim \delta_{\rm th} \lesssim 2/3$, the non-linearities in this quantity can be disregarded. This analysis is based on the numerically allowable region for the density contrast, $\delta_{\rm th}$. From the shaded zones in each of the plots created for effective sound speeds $c_s=1$ and $c_s=1.17$, respectively, it can be observed that the mass values are quite sensitive to their specific threshold value. One obvious benefit of using the threshold analysis set of values in our calculation is that it preserves the perturbativity approximation. We discovered that for values closer to the upper bound $\delta_{\rm th}\sim 2/3$, PBHs with $M_{\rm PBH} \sim 10^{2}{\rm gm}$ are generated, whereas for values near the threshold lower bound $\delta_{\rm th}\sim 2/5$, PBHs with $M_{\rm PBH} \sim 10^{3}{\rm gm}$ are produced. 
Since we have to raise the threshold value above the previously mentioned upper bound and take into account the region $\delta_{\rm th} > 1$, which directly corresponds to the breakdown of perturbation theory and is strictly prohibited in the current computation, producing PBH with $M_{\rm PBH} \lesssim 10^{2}{\rm gm}$ is difficult. Furthermore, for the range of PBH masses displayed in the plot, we find that the allowed band of values, for a particular threshold of density contrast, is very small when taking into account a significant abundance region. This suggests that there is a restricted range of PBH masses that can contribute to the dark matter density, where smaller mass PBHs are produced above a higher threshold value. In order to comprehend the PBH production scenario better and ultimately obtain more realistic estimates for the PBH abundance from theory, the non-Gaussianities in the density contrast are fundamental. It is evident from the refs.\cite{Choudhury:2023hvf, Choudhury:2023kdb} that we have made some headway in this regard recently, after an examination of the non-Gaussianities in Galileon theory using the in-in formalism. It is feasible to develop enormous non-Gaussianities in a controlled manner and avoid the \textit{no-go theorem} inside this theory, as we have discovered. Further study, as suggested by \cite{DeLuca:2019qsy, Young:2019yug}, might potentially address the issue of the non-Gaussian characteristics in the density contrast; nevertheless, we will save these studies for our further work.

Further we will summarize the theory in this part and show how the Press-Schechter Formalism is used to calculate the PBH abundance. As previously stated, PBH production happens when the curvature perturbations compress gravitationally during horizon re-entry. In particular, the formation will happen on those Hubble patches for which the inflation-generated overdensities are greater than a certain threshold $\delta_{\rm th}$. The fluctuation of $\delta_{\rm th}$ with respect to the EoS $w$ is now provided by, in accordance with Carr's criteria ($c_s ^2 =1)$ \cite{1975ApJ...201....1C}:
\bea
\label{deltath}
\delta_{\rm th} = \frac{3(1+w)}{5+3w}.
\eea
The dependency of PBH mass on the respective comoving scale of formation for a background with a generic equation of state $w$ is \cite{Alabidi:2013lya}:
\bea
\label{mpbh}
M_{\rm PBH} = 1.13 \times 10^{15} \times \bigg[\frac{\gamma}{0.2}\bigg]\bigg[\frac{g_{*}}{106.75}\bigg]^{-1/6}\bigg[\frac{k_{*}}{k_{\rm s}}\bigg]^{\frac{3(1+w)}{1+3w}} M_{\odot},
\eea
where the wavenumber associated with PBH creation is indicated by $k_{\rm s}$, and the solar mass is represented by $M_{\odot}$. Since the pivot scale $k_{*}$ offers a model-independent method, we will use it as the foundation for our estimates rather than the reheating scale. Here, it is important to note that we perform our analysis within the linear regime, which is defined as follows: we assume a mostly linear behavior for the density contrast in the superhorizon scales, which allows us to calculate the threshold, which falls into the interval $2/5 \leq \delta_{\rm th} \leq 2/3$ \cite{Musco:2020jjb}. Therefore, $-5/9 \leq w \leq 1/3$ is the corresponding range of values of $w$ in this location according to eqn.(\ref{deltath}). In the linear regime, the comoving curvature perturbation and the density contrast approximation on the superhorizon scale are as follows:
\bea
\delta(t,\mathbf{x}) \cong \frac{2(1+w)}{5+3w}\left(\frac{1}{aH}\right)^{2}\nabla^{2}\zeta(\mathbf{x}).
\eea
   
    \begin{figure*}[htb!]
    	\centering
    \subfigure[]{
      	\includegraphics[width=8.5cm,height=7.5cm] {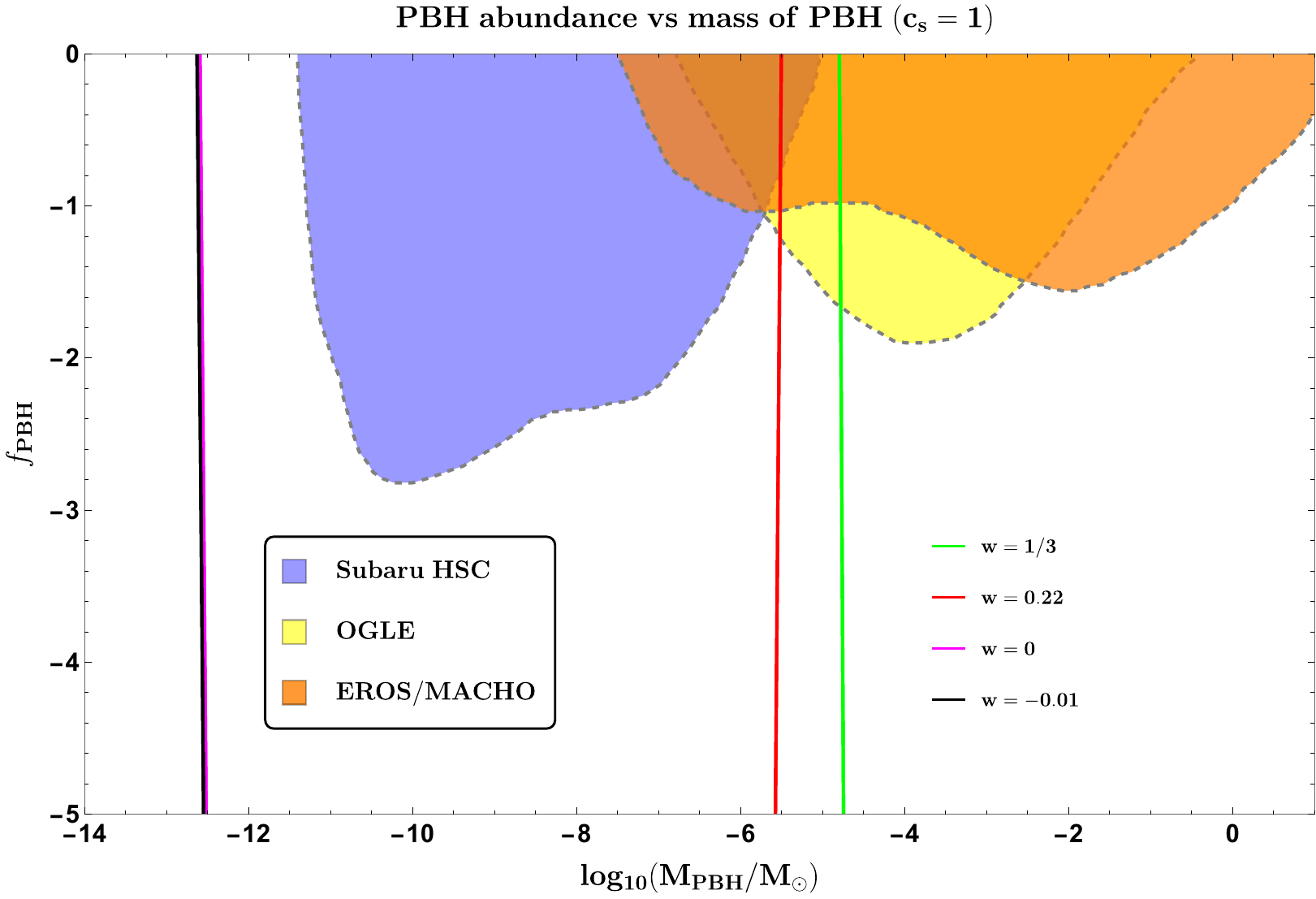}
        \label{microw1}
    }
    \subfigure[]{
        \includegraphics[width=8.5cm,height=7.5cm] {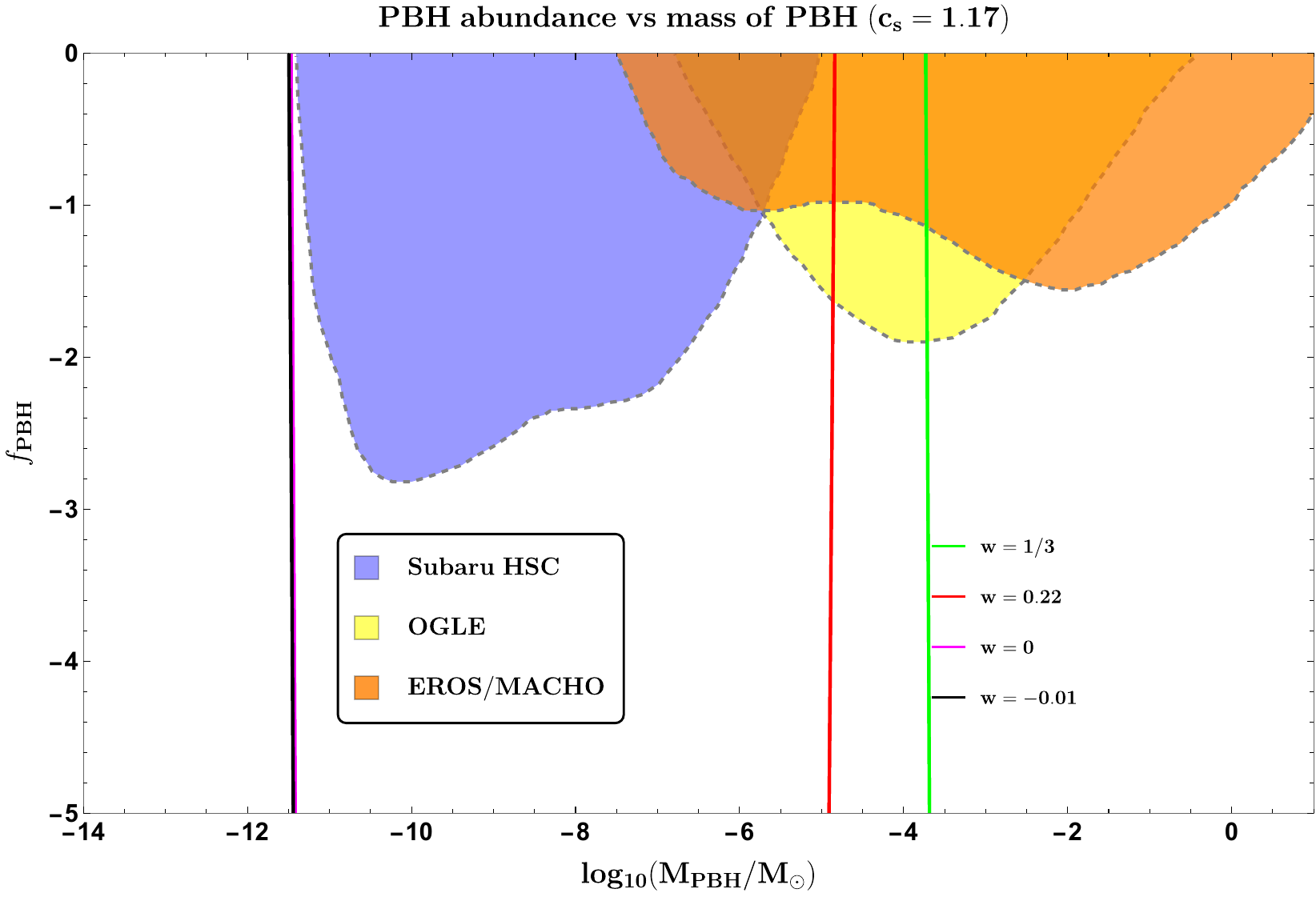}
        \label{microw2}
       }
    	\caption[Optional caption for list of figures]{The way in which PBH abundance changes with mass. With $c_{s}=1$ in the left panel and $c_{s}=1.17$ in the right panel, the effective sound speed $c_{s}$ is fixed. We indicate in blue, yellow, and orange the areas that were not included in the Subaru HSC, OGLE, and EROS/MACHO microlensing tests, respectively. The abundance change at $w=1/3$ and $w=0.22$ for EoS is indicated by green and red lines, respectively. The masses obtained for these values fall within the following range: $M_{\rm PBH} \sim {\cal O}(10^{-6}-10^{-3})M_{\odot}$. The abundance behavior for $w=0$ and $w=-0.01$ is shown by the magenta and black colors, respectively. The graphic illustrates the ranges in which PBH masses corresponding to EoS values $w < 0$ are outside the sensitivities of the microlensing experiments, with $M_{\rm PBH} \lesssim {\cal O}(10^{-13})M_{\odot}$.} 
    	\label{microlens}
    \end{figure*}
Next, we construct the formula for the variance of the smoothed density perturbations on the mass $M$ scale:
\bea
\sigma_{\rm M_{\rm PBH}} ^2 = \bigg[\frac{2(1+w)}{5+3w}\bigg]^2 \int \frac{dk}{k} \; (k
R)^4 \; W^2(kR) \; \overline{\overline{\triangle ^2 _{\zeta, \rm \bf EFT}(k)}}.\eea
As a smoothing function across the PBH formation scales $R=1/(\Tilde{c_s} k_{\rm PBH})$, $W(kR)$ is the Gaussian window function denoted by $\exp{(-k^2 R^2 /4)}$.
At the transition scale between the SR and USR phases, as well as the USR and the subsequent SR phase, the effective sound speed parameter is denoted by the word $\Tilde{c_s}$. Furthermore, the one-loop corrected renormalized DRG resummed scalar power spectrum is denoted by $\overline{\overline{\triangle ^2 _{\zeta, \rm \bf EFT}(k)}}$. We previously computed the PBH mass fraction for the radiation-dominated period with $w=1/3$ using the MST setup in \cite{Bhattacharya:2023ysp}. The $w$ reliance on the present PBH mass fraction, however, is now included and is expressed as follows \cite{Sasaki:2018dmp}:
\bea
\beta_{\rm M_{\rm PBH}} = \gamma \times \frac{\sigma_{\rm M_{\rm PBH}}}{\sqrt{2\pi}\delta_{\rm th}}\exp{\bigg(\frac{-\delta_{\rm th}^2}{2\sigma_{\rm M_{\rm PBH}}^2}\bigg)},
\eea
where the effective factor of collapse, $\gamma \sim 0.2$, aids in defining the portion of mass inside the Hubble Horizon that collapses to generate PBHs. The definition makes it obvious that the variance in the previous equation encodes the $w$ dependency. It is also crucial to make clear that, in generating these formulas, we do not include any non-gaussianities or non-linearities and that the above conclusion for the mass fraction presupposes that we use Gaussian statistics for the density perturbations. They will be taken into account in our further efforts. We now proceed to write the fractional abundance of PBHs as of right now.
\bea
f_{\rm PBH} \equiv \frac{\Omega_{\rm PBH}}{\Omega_{\rm CDM}}= 1.68\times 10^8 \times \bigg(\frac{\gamma}{0.2}\bigg)^{0.5} \times \bigg(\frac{g_{*}}{106.75}\bigg)^{-0.25} \times \left(M_{\rm PBH}\right)^{-\frac{6w}{3(1+w)}}\times \beta_{M_{\rm PBH}},\eea
Our focus now shifts to fig. \ref{microw1} and fig.  \ref{microw2}, which shows how PBHs abundance changes in relation to mass. The abundance is very sensitive to the range of masses at which $f_{\rm PBH} \in (1,10^{-3})$ can be reasonably estimated. We present the variation for the same set of EoS values, $w \in \{1/3,0.22,0,-0.01\}$, in both panels. We find that, for higher EoS values, $w \gtrsim 0.2$, the mass range should fall within the agreeable region $M_{\rm PBH} \sim {\cal O}(10^{-6}-10^{-3})M_{\odot}$, which is also appropriate for the frequencies associated with the NANOGrav-15 signal. The restrictions derived from the set of microlensing tests as seen in the plot nonetheless permit a significant abundance for such values of $w$. Below $w=0$, we find masses for which the abundance is outside the sensitivity range of the presented tests. With black $(w=-0.01)$ to the left of magenta $(w=0)$, the black and magenta lines almost align. We do not go much lower in $w$ since doing so would only result in incredibly low PBH mass values, $M_{\rm PBH} \lesssim {\cal O}(10^{-13})M_{\odot}$. We reaffirm the advantages of a somewhat higher value of $c_{s}$ from above fig. \ref{microw1} and fig. \ref{microw2}, as we achieve a broader mass range with $c_{s}=1.17$ that permits a sizable abundance of PBHs.

\subsection{Scalar Induced Gravity Waves (SIGWs) from Goldstone EFT}

This section aims to review the theory of gravitational waves caused by scalars. In the cosmic perturbation theory, second-order induced GWs are produced by mode couplings of the first-order perturbations to the FLRW metric. In comparison to the CMB fluctuations, this effect produces a considerable amplification in the measured GW spectrum, indicating the presence of a massive scalar disturbance. We are investigating the case when a state with an unknown equation of state and a value of $w$ coexists with the end of inflation. Deep within the horizon in this broad $w$ backdrop, where the modes are mostly sourced, the induced GW formed in the early cosmos can exist. For a generic equation of state and propagation speed, we express the energy density spectrum of the SIGWs as follows:
\bea
\label{Energydensity}
\Omega_{\rm{GW},0}h^2 = 1.62 \times 10^{-5}\;\bigg[\frac{\Omega_{r,0}h^2}{4.18 \times 10^{-5}}\bigg] \bigg[\frac{g_{*}(T_c)}{106.75}\bigg]\bigg[\frac{g_{*,s}(T_c)}{106.75}\bigg]^{-4/3}\Omega_{\rm GW,c}.
\eea
The radiation energy density as measured today is denoted by $\Omega_{r,0}h^2$, and the energy and entropy effective degrees of freedom are denoted by $g_{*},g_{*,s}$. The fractional energy density of the SIGWs assessed at a conformal time in the radiation-dominated universe while the GW density fraction remains constant is denoted by $\Omega_{\rm {GW},c}$. For the scales $k\geq k_{*}$, the SIGW spectrum is represented as \cite{Domenech:2021ztg}:
\bea
\label{omegagw}
\Omega_{\rm {GW},c}= \bigg(\frac{k}{k_{*}}\bigg)^{-2b}\int_{0}^{\infty}dv \int_{|{1-v}|}^{1+v} du \; {\cal T}(u,v,b,c_s) \;\;\overline{\overline{\Del_{\zeta,{\bf EFT}}^{2}(ku)}} \times \overline{\overline{\Del_{\zeta,{\bf EFT}}^{2}(kv)}},
\eea
where the EoS dependent parameter $b$ has the definition $b=(1-3w)/(1+3w)$. Furthermore, as previously stated, the standard pivot scale (CMB) is $k_{*}$. In the expression eqn.(\ref{omegagw}), a general factor of $(k/k_{*})^{-2b}$ occurs. This factor vanishes when you substitute the value of EoS for a radiation-dominated era, i.e., $w=1/3$, in the expression. Furthermore, we have conducted our study using the pivot scale ($k_{*}$) in this paper rather than the reheating scale used in other studies of a similar kind \cite{Domenech:2020ers, Domenech:2021ztg}. Though they may not seem important, its ramifications are rather substantial. The assumption of equilibrium scales lying between the GeV and TeV scales is a significant difficulty with reheating. It is unclear exactly what correct thermalization and underlying microphysics are responsible for this reheating. For exact computations, it cannot be stated to be totally dependable. You may read about earlier thermalization research here \cite{Banerjee:2021lqu,Choudhury:2021tuu,Choudhury:2020yaa} to get some clarification. The majority of the time, reheating is attempted to be included ad hoc, largely from phenomenology, in the literature. While we have nothing against using such a phenomenological approach, it has drawbacks, particularly when it comes to adding model-dependency. Therefore, even when thermalization has been reached and the system is in equilibrium, you still need to exercise additional caution when doing the computation. Given the lack of established theory regarding reheating, we perform our analysis using normalization with respect to the pivot scale ($k_{*}$), as this approach has gained popularity from previous studies and observations (e.g., \cite{Planck:2018jri,Planck:2018vyg,Planck:2015sxf,Choudhury:2015pqa,Choudhury:2014kma,Choudhury:2014sua}). Additionally, taking into account a radiation-dominated era, we achieved correct findings using the pivot scale for our estimation in our prior studies \cite{Choudhury:2023vuj, Choudhury:2023jlt, Choudhury:2023rks, Bhattacharya:2023ysp}.

Currently, \cite{Domenech:2021ztg} provides the transfer function in the Eqn.(\ref{omegagw}) for the general situation of constant values of the EoS $w$ and propagation speed $c_{s}$:
\bea
\label{Transfer func}
{\cal T}(u,v,b,c_s)&=&{\cal N}(b,c_s) \bigg[\frac{4v^2 - (1-u^2+v^2)^2}{4u^2v^2}\bigg]^2 \abs{1-y^2}^b \times \bigg\{\big(P_{b}^{b}(y) + \frac{b+2}{b+1}P_{b+2}^{-b}(y)\big)^2 \Theta(c_s(u+v)-1) \nonumber \\
&& \quad \quad \quad \quad + \frac{4}{\pi^2}\bigg[\big(Q_b ^{-b}(y) + \frac{b+2}{b+1}Q_{b+2}^{-b}(y)\big)^2 \; \Theta(c_s (u+v)-1) \nonumber \\
&& \quad \quad \quad \quad  + \big( {\cal Q}_{b}^{-b}(-y)+ 2\frac{b+2}{b+1}{\cal Q}_{b+2}^{-b}(-y) \; \Theta(1-c_s(u+v))\bigg]\bigg\}.
\eea
Explicit details of the transfer function in the presence of a general equation of state parameter $w$ have been discussed in detail in the previous section of this review. Look at the Appendix \ref{A7a}, \ref{A8a} and \ref{A9a} for more details on this computation. For this reason, we are not going into the depths of the results and their mathematical implications. 
    \begin{figure*}[htb!]
    	\centering
   {
   \includegraphics[width=18.5cm,height=12.5cm] {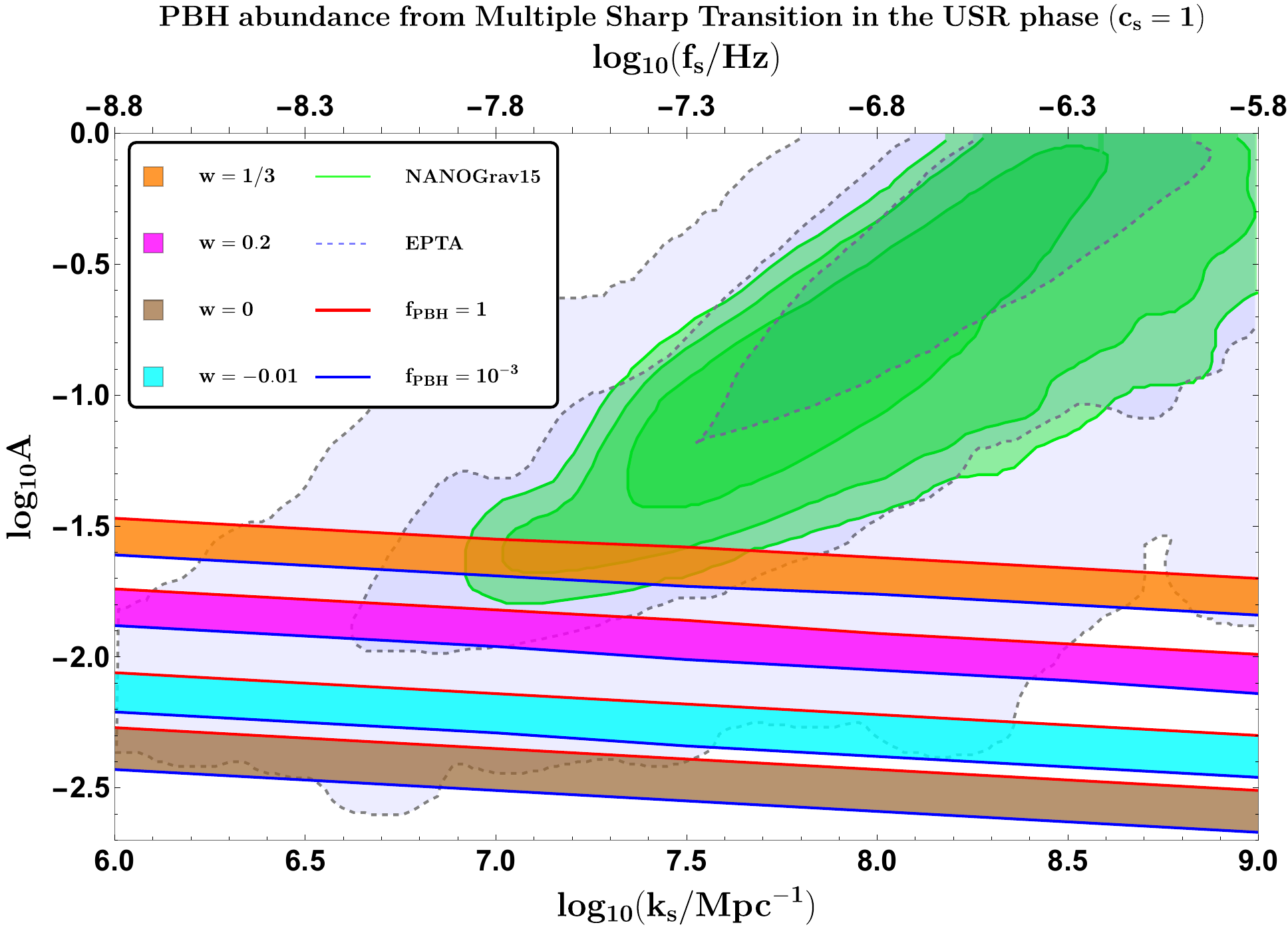}
    }
    	\caption[Optional caption for list of figures]{The transition wavenumber prevents PBH overproduction in the MST configuration, as shown in Figure, which shows the change in amplitude $A$ of the one-loop renormalized and DRG-resummed scalar power spectrum. The fixed effective sound speed parameter takes $c_{s}=1$. The area with considerable abundance $f_{\rm PBH} \in (1,10^{-3})$ is enclosed by red and blue lines. The distinct areas associated with the EoS parameter $w \in \{1/3,0.2,0,-0.01\}$ are shown by the colors orange, magenta, cyan, and brown. The NANOGrav 15 and EPTA data posteriors, which are shown in green and light-blue hues, respectively, are derived from \cite{Franciolini:2023pbf}.
} 
    	\label{overprodcs1}
    \end{figure*}
The results pertaining to the many issues we set out to address will be reviewed in this section. We want to address the overproduction of PBH as one of the most important challenges. When curvature perturbations re-enter the horizon, they cause PBHs to develop by gravitational collapse. The current dark matter abundance may be surpassed by the production of a vast number of PBHs when disturbances are too big. As PBHs are considered potential candidates for relic dark matter, this presents a risk. In contrast to observational evidence, the quantity of PBH surpassing that of dark matter suggests that PBHs make up all of the dark matter that is visible today. Thus, all we need to do is fine-tune our models such that this overproduction problem is addressed. Examining this problem from the perspective of the EoS parameter—which we have examined in this paper—is one approach. According to refs. \cite{Choudhury:2023fwk,Franciolini:2023pbf,Franciolini:2023wun,LISACosmologyWorkingGroup:2023njw, Inui:2023qsd, Chang:2023aba, Gorji:2023ziy,Li:2023xtl, Li:2023qua,Firouzjahi:2023xke} some examples of non-linearity and non-Gaussianity that can be included. Apart from these, we have an additional approach to address the problem, which involves introducing a spectator field \cite{Gorji:2023sil,Ota:2022xni}. The optimal solution, which should also include a thorough analysis, has to consider the Equation of State (EoS) parameter, non-linearities, linearities, and non-Gaussianities. In our next study project, we want to explore the latter component. 
    \begin{figure*}[htb!]
    	\centering
   {
   \includegraphics[width=18.5cm,height=12.5cm] {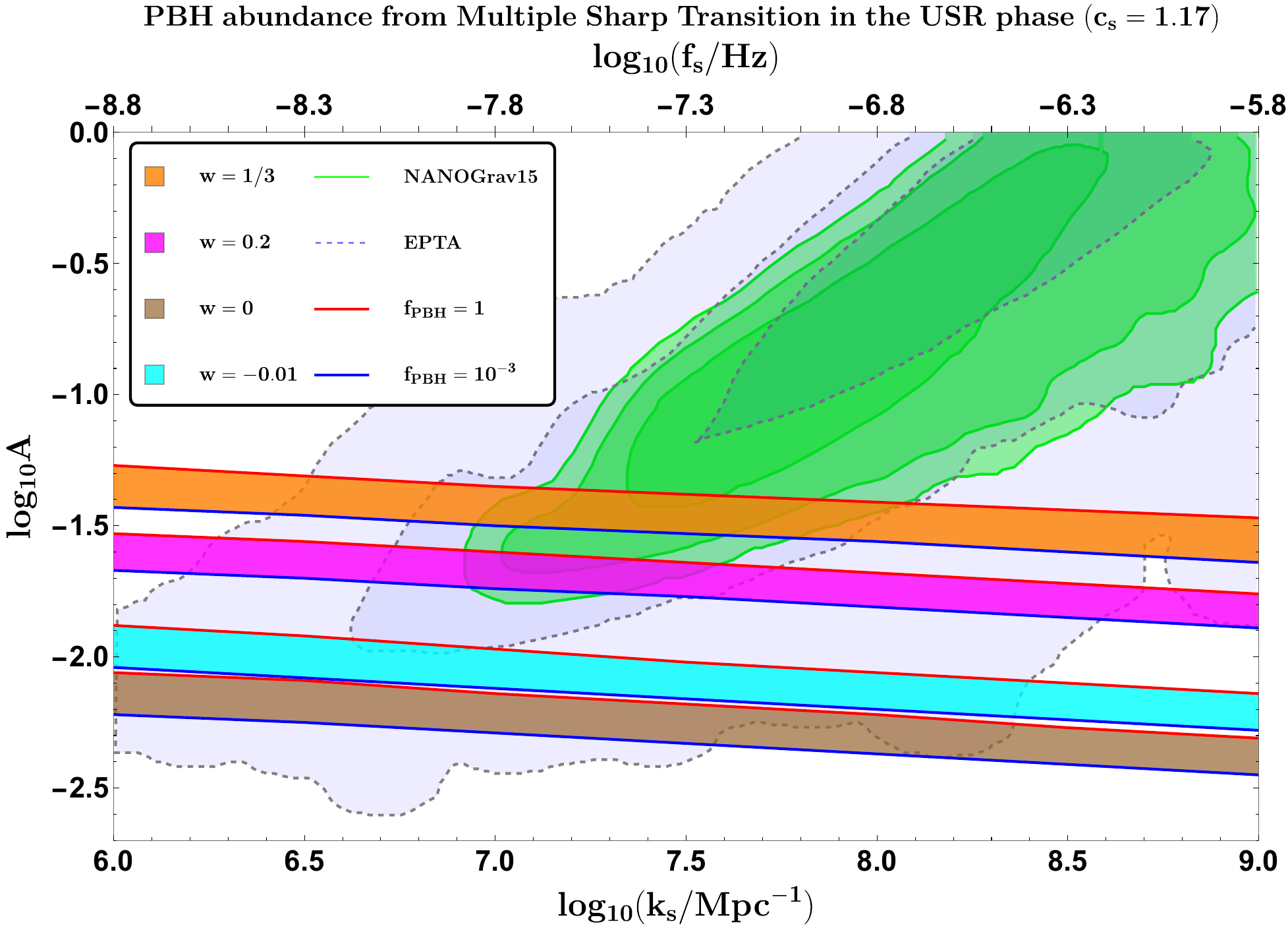}
    }
    	\caption[Optional caption for list of figures]{The transition wavenumber prevents PBH overproduction in the MST configuration, as shown in Figure, which shows the change in amplitude $A$ of the one-loop renormalized and DRG-resummed scalar power spectrum. The fixed effective sound speed parameter takes $c_{s}=1.17$. The area with considerable abundance $f_{\rm PBH} \in (1,10^{-3})$ is enclosed by red and blue lines. The distinct areas associated with the EoS parameter $w \in \{1/3,0.2,0,-0.01\}$ are shown by the colors orange, magenta, cyan, and brown. The NANOGrav 15 and EPTA data posteriors, which are shown in green and light-blue hues, respectively, are derived from \cite{Franciolini:2023pbf}.} 
    	\label{overprodcs117}
    \end{figure*}
    
The dependency of the outcomes on the equation of state parameter, or which value of $w$ will be preferred over the others, is one of the main features of this review. Keeping this in mind, we start with our findings covering the impact of an arbitrary but instantaneous constant EoS parameter $w$ (i.e. constant at a specific slice of time) on the PBH abundance, which form corresponding to the frequencies that also exhibit NANOGrav 15 signal sensitivity. To be more precise, we use Eqn. (\ref{mpbh}) to vary $k_{s}$ within the range of values that fall inside the NANOGrav-15 frequency spectrum, or within, and calculate the range of amplitude that gives us the appropriate PBH abundance for various $w$ values. We then tally this range with the NANOGrav and EPTA data curves. We will analyze the abundance values acquired by microlensing tests as a next step \footnote{When concentrating on the possible galactic origins of dark matter, experiments like Subaru HSC, OGLE, and MACHO/EROS allow for the investigation of the nature of dark matter through the gravitational microlensing phenomena. As prospective dark matter candidates, they currently provide constraints on the mass and abundance of PBHs, including a planet-size mass range of $M_{\rm PBH} \sim {\cal O}(10^{-6}-10^{-2})M_{\odot}$ that is also consistent with the allowed range from the most recent NANOGrav results.} \cite{Niikura:2017zjd,Niikura:2019kqi,EROS-2:2006ryy}. The SIGW spectra results for the particular benchmark values of EoS $w$ parameter will be finally presented. These spectra are compared with the signature of the recent NANOGrav 15 and EPTA signal as well as other ground and space-based experiments \cite{LISA:2017pwj, Kawamura:2011zz, Punturo:2010zz, Reitze:2019iox, Crowder:2005nr, LIGOScientific:2014pky, VIRGO:2014yos, KAGRA:2018plz}.
    \begin{figure*}[htb!]
    	\centering
    \subfigure[]{
      	\includegraphics[width=8.5cm,height=7.5cm] {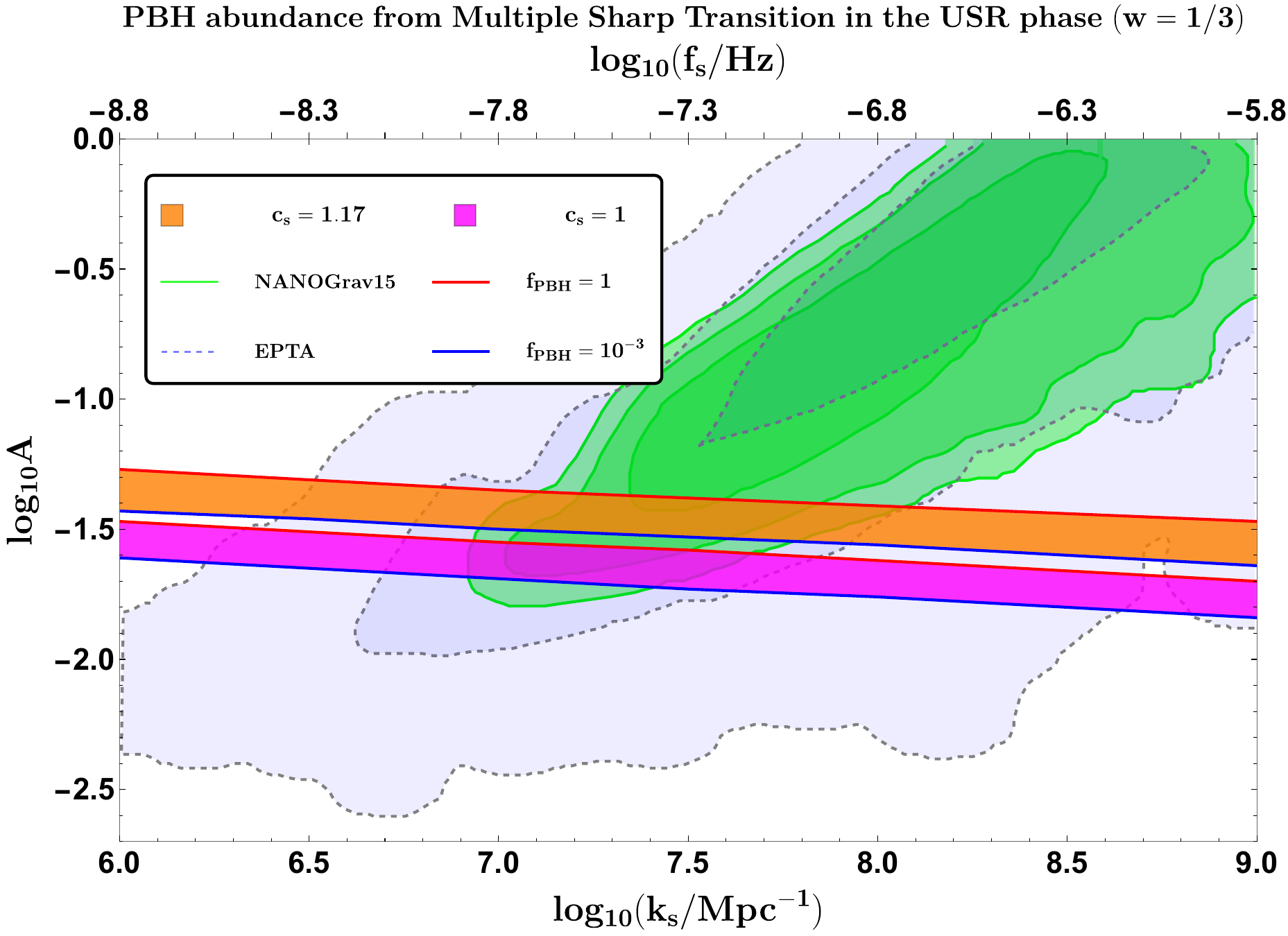}
        \label{w1}
    }
    \subfigure[]{
        \includegraphics[width=8.5cm,height=7.5cm] {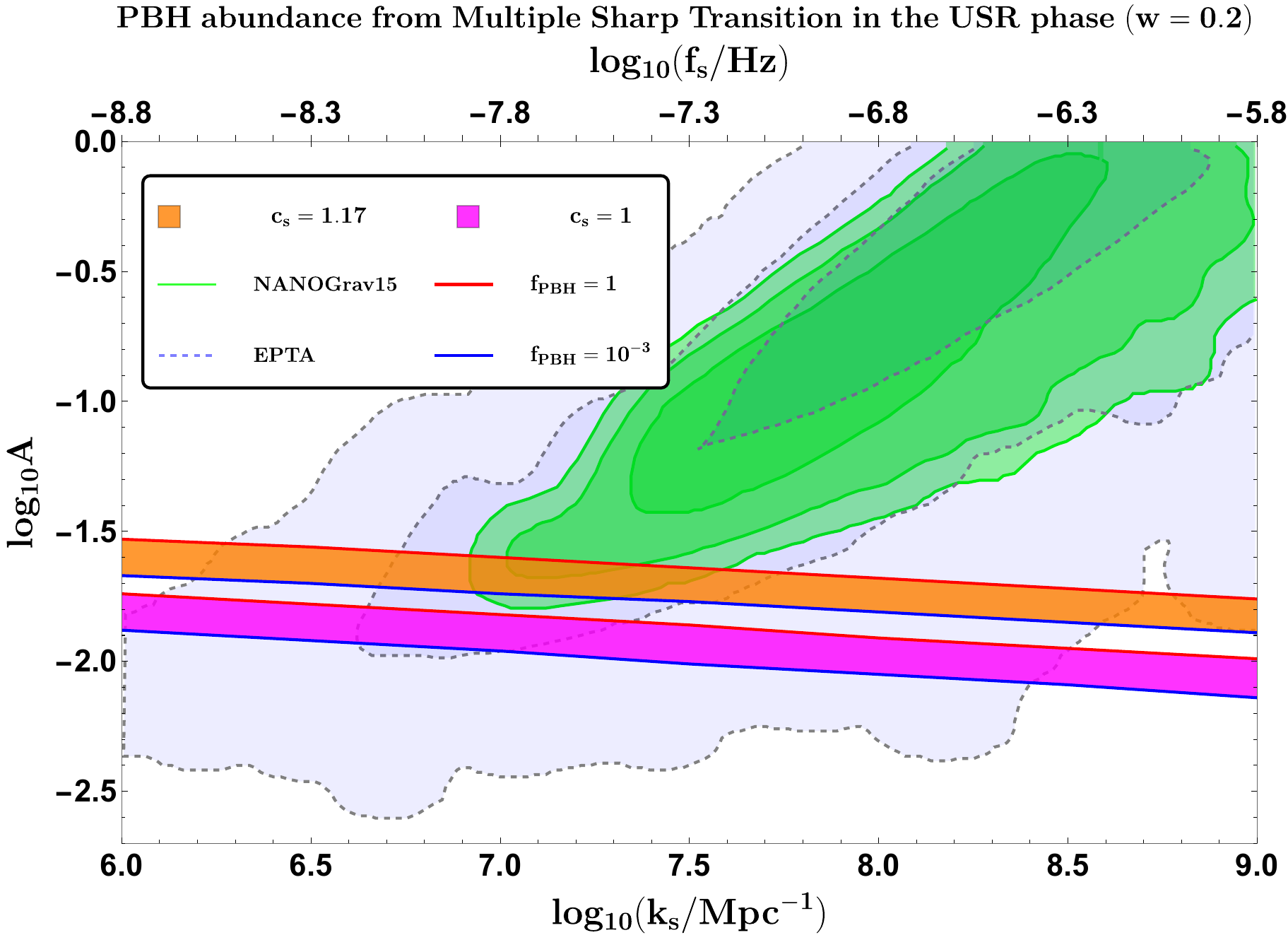}
        \label{w2}
       }
        \subfigure[]{
        \includegraphics[width=8.5cm,height=7.5cm] {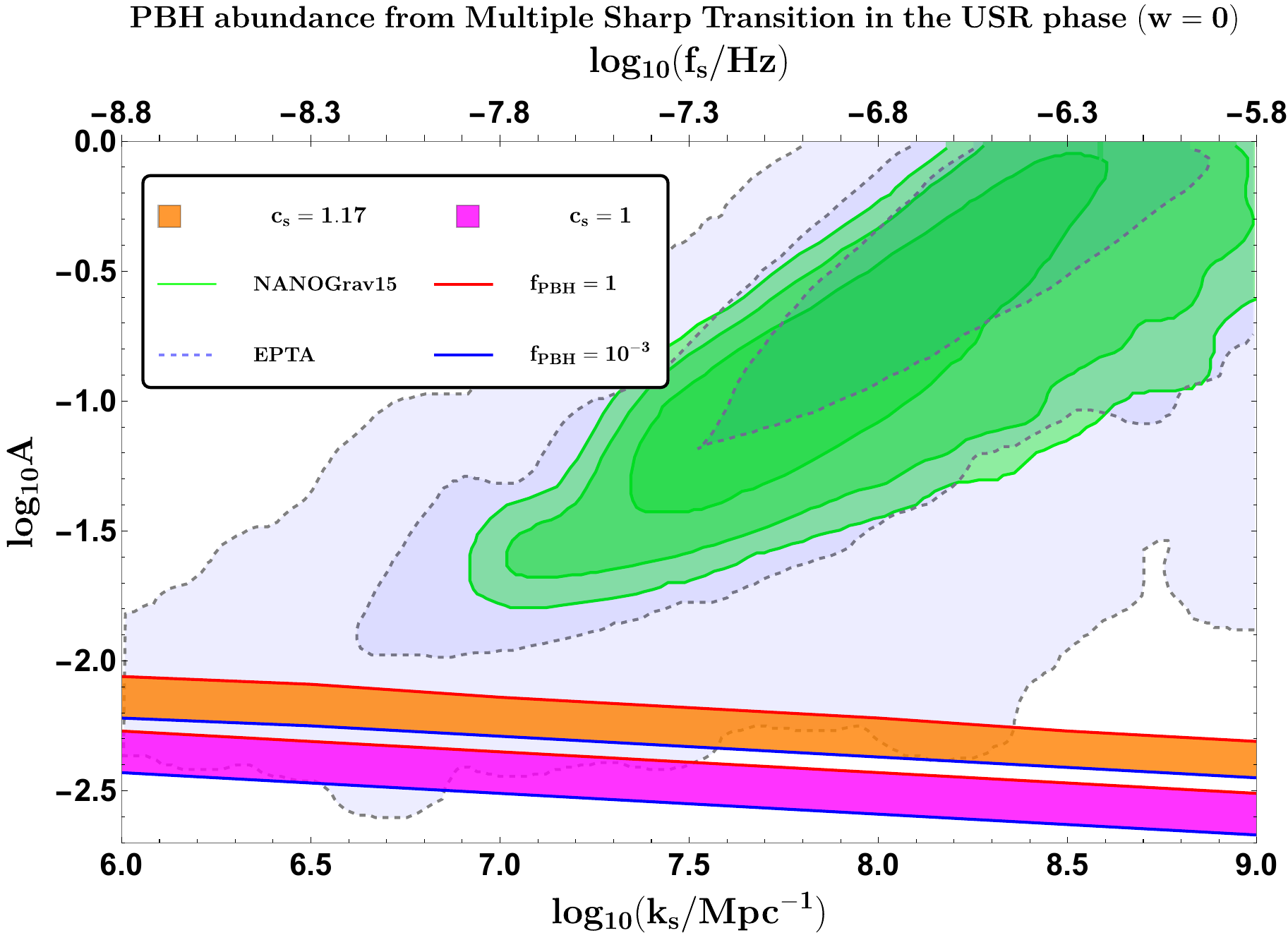}
        \label{w3}
       }
        \subfigure[]{
        \includegraphics[width=8.5cm,height=7.5cm] {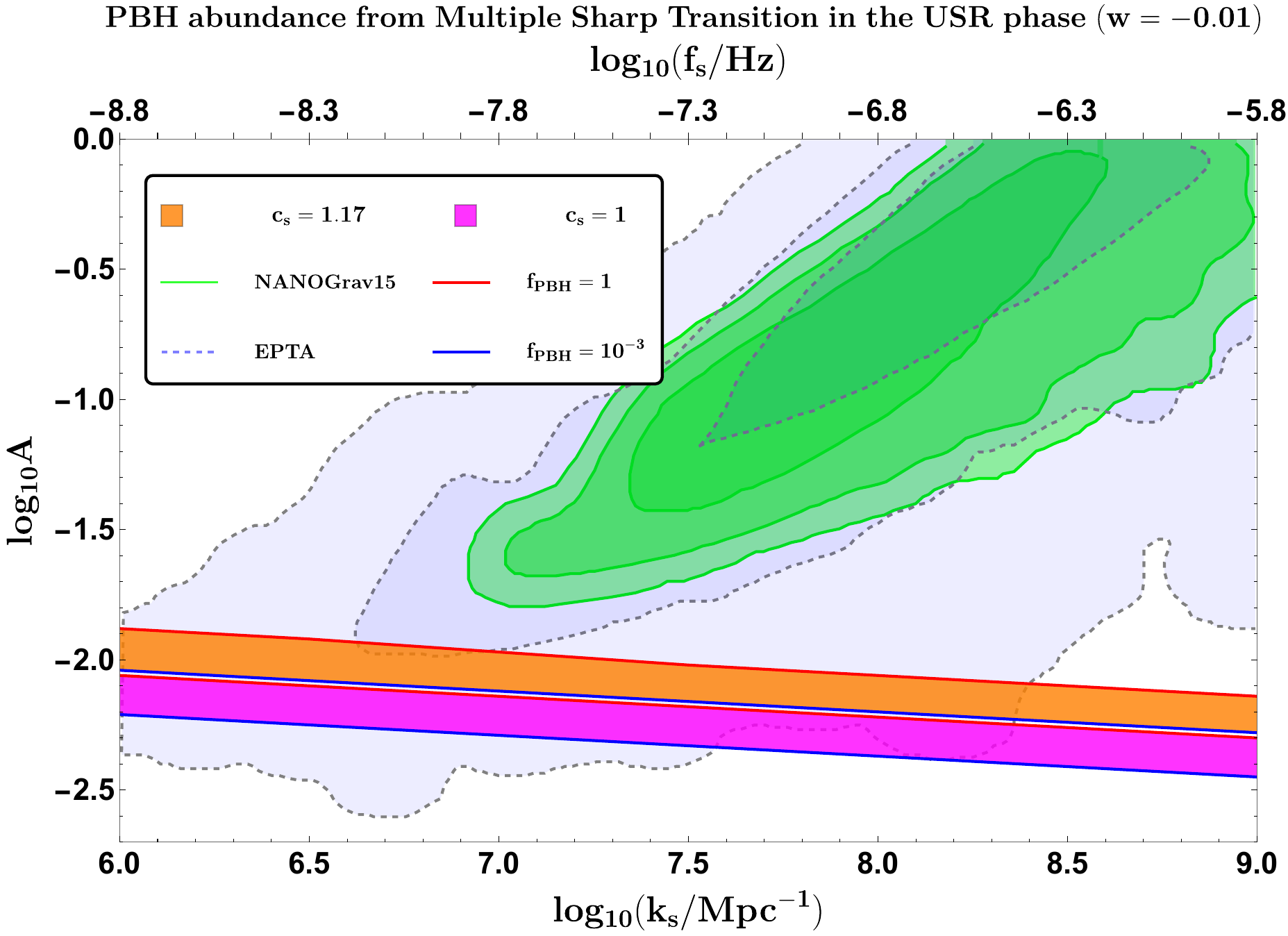}
        \label{w4}
       }
    	\caption[Optional caption for list of figures]{For each value in the benchmark set, $w \in \{1/3,0.2,0,-0.01\}$, the comparative behavior of the amplitude $A$ of the one-loop renormalized and DRG-resummed scalar power spectrum with increasing transition wavenumber is presented for the two examples, $c_{s}=1$ in magenta and $c_{s}=1.17$ in orange.} 
    	\label{overprodcompare}
    \end{figure*}
    
The behavior of the one-loop corrected renormalized and DRG-resummed scalar power spectrum amplitude with increasing frequency in the interval sensitive to the PTA experiments—that is, the NANOGrav 15 and EPTA signals—is depicted in fig. (\ref{overprodcs1}). Here, $c_{s}=1$ is taken as the fixed effective sound speed. We have selected benchmark values for the EoS parameter in this figure, $w \in \{1/3,0.2,0,-0.01\}$, which are permitted under the approximation where the density contrast linearities predominate on super-horizon scales. As you shall see later in the lecture, there is a caveat to the rather arbitrary numbers chosen inside the linear range. Thus, the aforementioned approximation now yields the permitted interval for the threshold density contrast, $2/5 \leq \delta_{\rm th} \leq 2/3$, and subsequently yields, from Eqn.(\ref{deltath}), the interval for the EoS, $-0.55 \leq w \leq 1/3$. It should be noted that, in theory, $\delta_{\rm th}$ can have values in the range $1/3 < \delta_{\rm th} < 1$; nonetheless, the range of $\delta_{\rm th}$ chosen for our study is important since it preserves the linear approximation in the cosmic perturbation theory. To build the perturbation theory outside of this range, non-linear approximation must be incorporated. Now let us examine fig. (\ref{overprodcs1}). In this instance, we observe that the effective sound speed parameter $c_{s}=1$ scenario strongly favors the situation of $w=1/3$. The orange band that corresponds to the NANOGrav 15 signal accords with $2\sigma$. We find ourselves barely outside the $3\sigma$ zone from NANOGrav 15 as we move down in $w$, at $w=0.2$, indicated by the magenta band, but still comfortably inside the $2\sigma$ range of the EPTA signal. It is interesting to note that, in comparison to the $w=-0.01$ case in cyan color, the $w=0$ case in brown color is further eliminated at values near to $w=-0.01$, where the amplitude is $A < {\cal O}(10^{-2})$. This will make sense when we investigate in more detail the amplitude of the total power spectrum and the structure of the SIGW spectrum for these same benchmark values of $w$.

    \begin{figure*}[ht!]
    	\centering
   {
   \includegraphics[width=18.5cm,height=12.5cm] {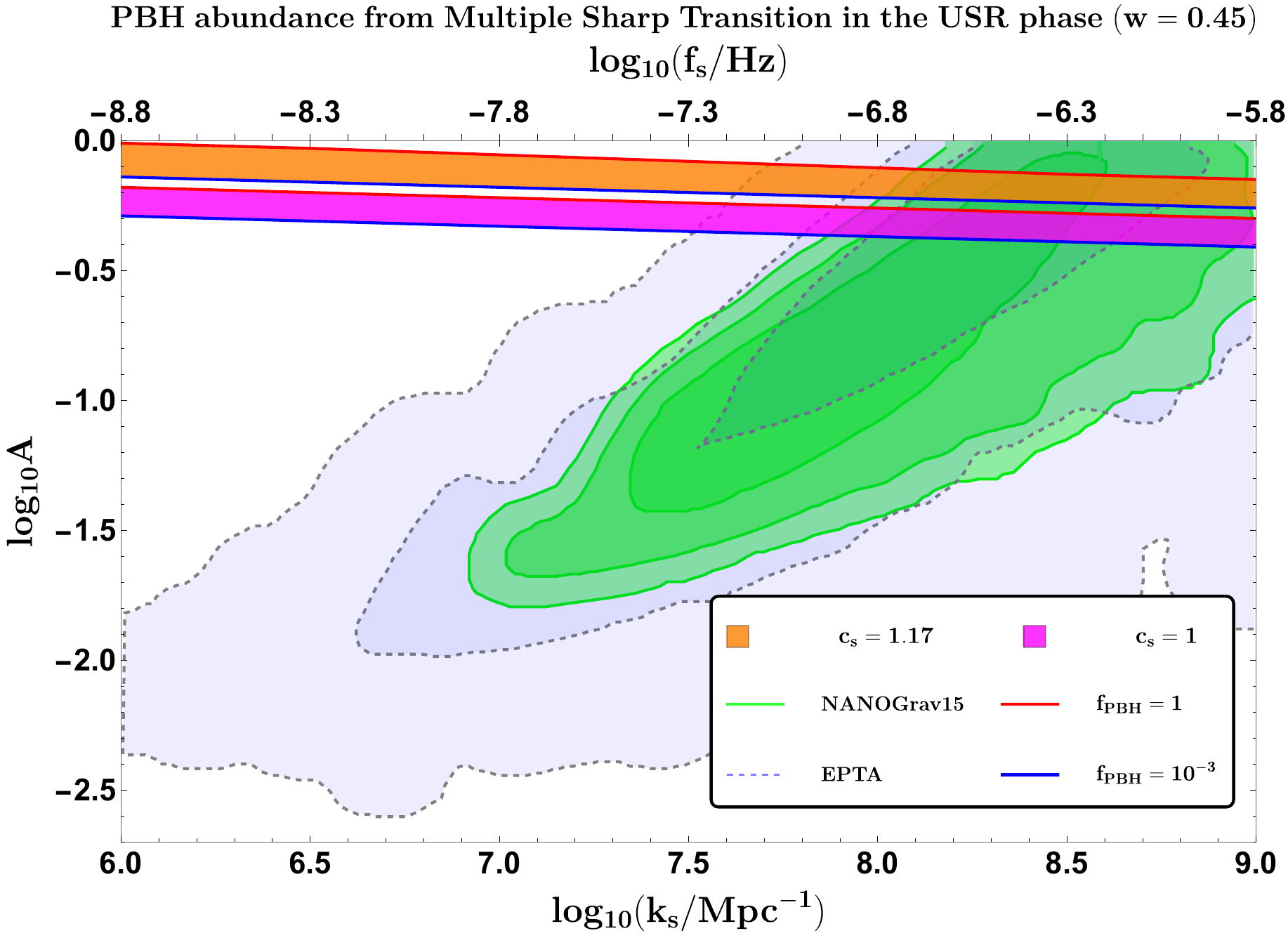}
    }
    	\caption[Optional caption for list of figures]{Figure shows change in the amplitude of the one-loop renormalized and DRG-resummed scalar power spectrum with changing transition wavenumber for the two cases, $c_{s}=1$ in magenta and $c_{s}=1.17$ in orange, for $w=0.45$. The red and blue lines enclose the region corresponding to sizeable abundance $f_{\rm PBH} \in (1,10^{-3})$.} 
    	\label{w5}
    \end{figure*}

Another example of the amplitude behavior of the one-loop corrected renormalized and DRG-resummed scalar power spectrum is shown in fig. (\ref{overprodcs117}), but this time the effective sound speed is set at $c_{s}=1.17$. Compared to the scenario of identical values of EoS $w$ but when the sound speed was set earlier to take $c_{s}=1$ in fig. (\ref{overprodcs1}), note the apparent improvement in the power spectrum amplitude needed to create a large abundance of the PBHs. The PBH abundance found for the $w=1/3$ instance now falls well inside the 1$\sigma$ contour of NANOGrav-15, demonstrating once more that $w=1/3$ is the most preferred condition. Compared to $c_s=1$, the overall bands are moved upward. The concept that a somewhat higher value of $c_{s}=1$ with $c_s=1.17$ gives a more favorable situation to prevent PBH overproduction is reinforced by this graphic. 

After that, fig. (\ref{overprodcompare}) offers a comparative examination of how the scalar power spectrum amplitude behaves at various effective sound speed values while maintaining a constant $w$ value for every panel. Individually speaking, we can see that a higher value of $c_{s}=1.17$ is a more favorable condition for adequately elucidating overproduction. This permits the USR phase to have a suitable amplitude. Let us state clearly now that $w=1/3$ is its maximum value according to our initial approximation of operating in the linear regime, where the interval, $2/5 \leq \delta_{\rm th} \leq 2/3$, holds for the density contrast. But if we attempt to raise $w$ over this range, we are left with the present overproduction problem to consider. If so, the solution is shown in fig.(\ref{w5}), where we increase the value to $w=0.45$ and discover that we may raise the amplitude to much higher values, almost to the point where perturbativity breaks. Even though overproduction may still be prevented in this situation, the results of such calculations are useless since large mass PBH production is not feasible at this NANOGrav-15 scale due to the greatly decreased amplitude. Therefore, we draw the conclusion that raising $w \geq 0.45$ may cause perturbation theory issues, since we would have already gone above the permitted bound established by the linear regime approximation. The aforementioned study supports the hypothesis that $w=1/3$ is the best case scenario to maintain the theoretical limitations and yield superior findings that easily fulfill empirical criteria.
    \begin{figure*}[htb!]
    	\centering
    \subfigure[]{
      	\includegraphics[width=8.5cm,height=7.5cm] {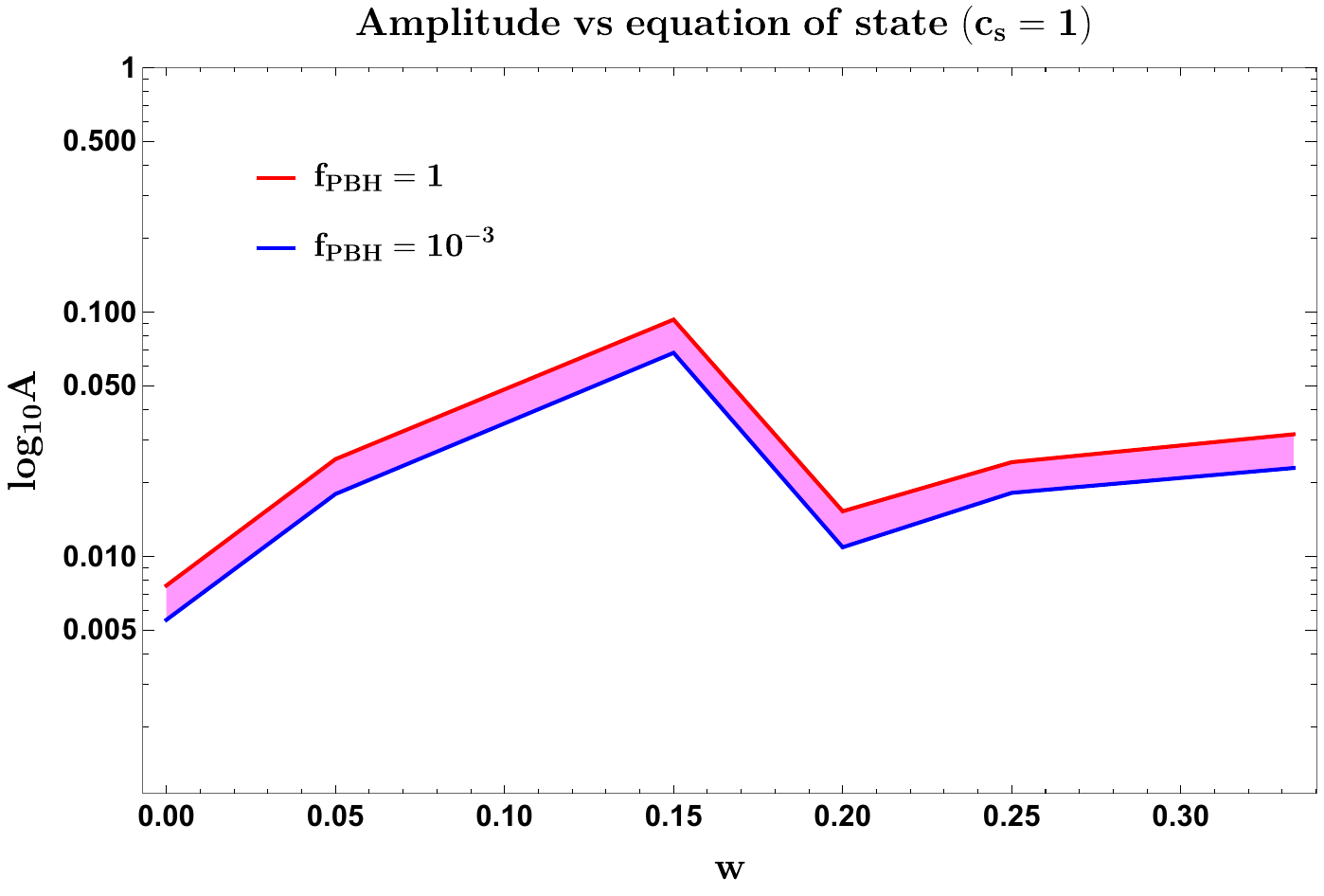}
        \label{ampvswc1}
    }
    \subfigure[]{
        \includegraphics[width=8.5cm,height=7.5cm] {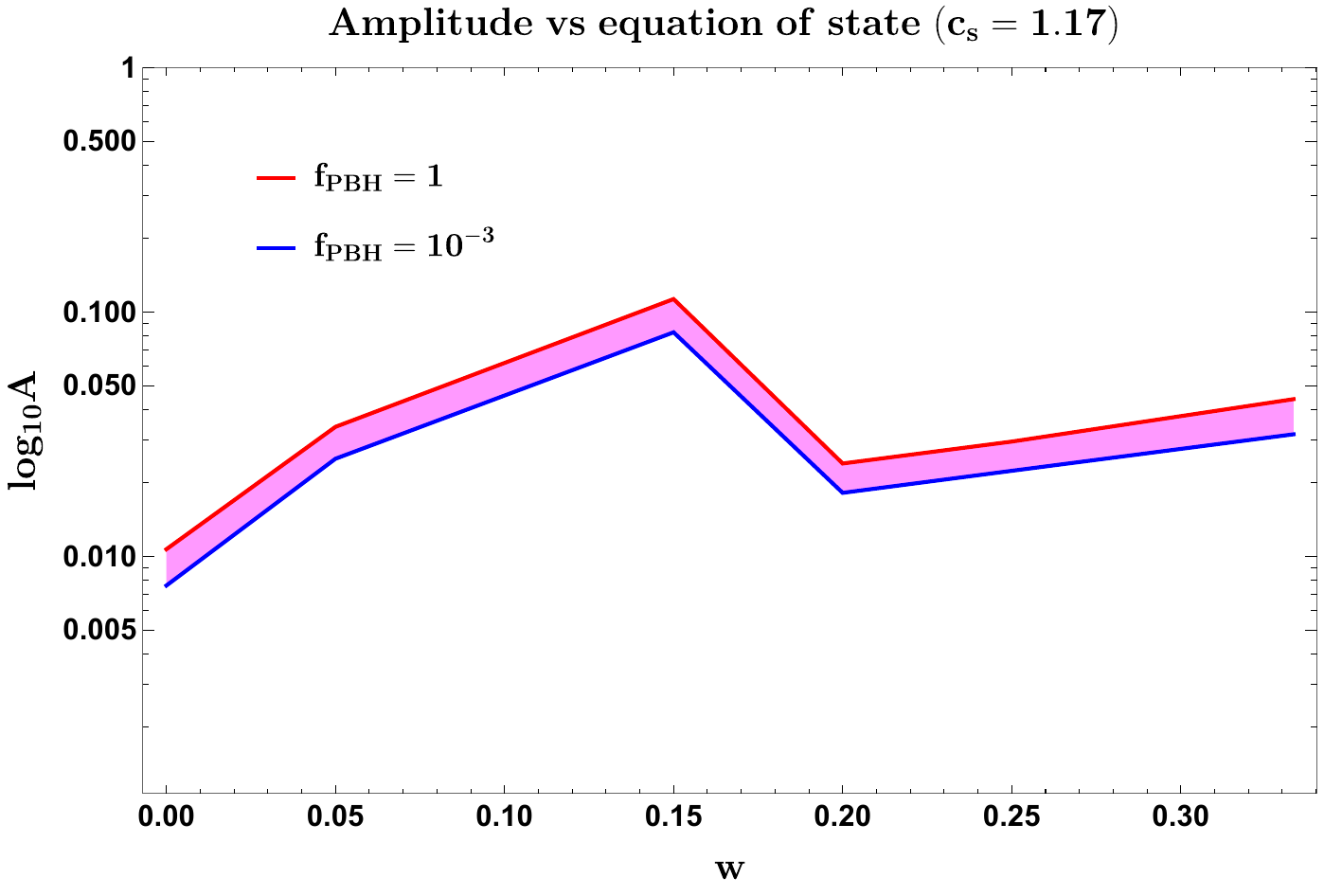}
        \label{ampvswc117}
       }
    	\caption[Optional caption for list of figures]{The behavior of the amplitude $A$ of the one-loop renormalized and DRG-resummed scalar power spectrum with varying EoS $w$, where $k_{\rm PBH} \sim {\cal O}(10^{7}{\rm Mpc^{-1}})$ is satisfied with a fixed transition wavenumber. The permitted zone in magenta where the $f_{\rm PBH} \in (1,10^{-3})$ is satisfied is enclosed by the solid lines in red and blue.
} 
    	\label{ampvsw}
    \end{figure*}

It is evident that kination dominance is not achievable under effective field theory, with the possible exception of certain scenarios such as quintessential inflation, where a catalyst-containing bump may allow for the emergence of this possibility. On the other hand, this is nearly impossible in all other cases.
Thus far, the amplitude of the quantum loop renormalized DRG-resummed scalar power spectrum has shown varying behavior with momentum values that can assist in resolving overproduction. However, the trend in the amplitude is not linear when the EoS parameter $w$ changes. fig.(\ref{ampvsw}) focuses on this issue. We now focus on the region $0 \leq w \leq 1/3$, which shows that the power spectrum amplitude does not necessarily increase monotonically as $w$ increases. The amplitude exhibits a notable relative rise in a particular domain around $w \in (0.1,0.2)$, allowing us to get a sizable abundance of PBHs, $f_{\rm PBH} \in (1,10^{-3})$. As we previously observed from fig.\ref{overprodcompare}, the total amplitude is quite modest to fall within the shown sensitivities of either PTA posterior after attaining values close to $w \sim 0$. Because the amplitude for negative values, $w \sim -0.05$, is bigger than $w=0$, the case of values close to $w = 0$ is important. But when we go much lower, like $w \leq -0.1$, we find severely suppressed amplitudes; however, such findings do not provide any insight into the overproduction issue. As a result, we do not examine the behavior for the remaining range of permitted values, which is $-0.55 \leq w \leq -0.05$. We'll go into more detail about this behavior when we talk about the SIGW spectrum.

For two distinct propagation speeds, $c_s=1$ and $c_s=1.17$, the change in the amplitude of the final scalar power spectrum, with an arbitrary EoS parameter $w$, is represented in Figures \ref{ampvswc1} and \ref{ampvswc117}. We would like to reiterate that there is no significant difference in the two $c_s$ scenarios, even though $c_s=1.17$ shows a higher peak amplitude. This suggests that a little increase in $c_{s}=1$ provides a more favorable environment for inflation. In any case, this conversation emphasizes the comfort of a better end-of-supply scenario. Let us explore the details in more detail now that we have this knowledge. The PBH creation scales within the range that is sensitive to NANOGrav-15 specifically were varied, and amplitudes were estimated so that the PBH abundance falls within a sizable range $f_{\rm PBH} \in (1,10^{-3})$. This resulted in the curves seen in these figures. The characteristics of both figures make it clear that the change in amplitude with the EoS parameter does not follow any particular pattern. The scalar power spectrum peak amplitude rises first, peaks at a certain amount, and then declines. After that, when $w$ rises, the amplitude begins to climb once more. With a further rise in $w$, it is anticipated that the amplitude will continue to grow and defy the perturbativity requirements as $w \rightarrow 1$, or in the event of a kination-dominated era. The scenario of $w \le 0$ is not shown in the graph below, but based on our study, we observed that the peak amplitude increases steadily as we go below $w=0$ to $w\sim-0.05$. However, when we fall even more below $w \sim -0.05$, the peak amplitude once again experiences a safe decline. Such low values of $w$ do not lead to the intended result of creating the large mass PBHs, even when all the requirements of perturbativity are maintained at each transition. Furthermore, note that the analysis is performed for a selected range of $w$. While the underlying theory may predict a certain behavior of this curve, you are free to experiment with even lower values of $w$ to see whether a pattern emerges.

       \begin{figure*}[htb!]
    	\centering
    \subfigure[]{
      	\includegraphics[width=8.5cm,height=7.5cm] {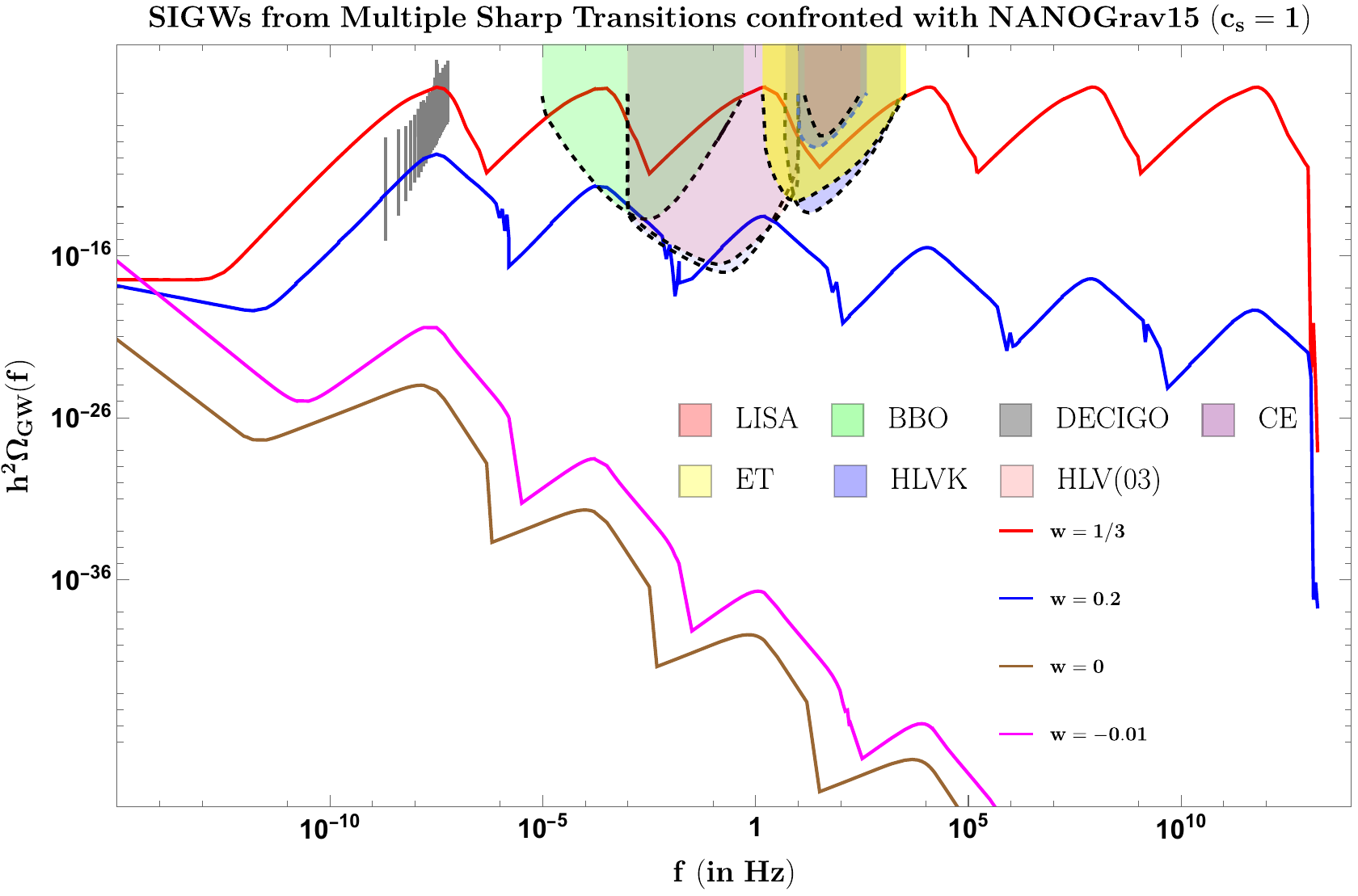}
        \label{SIGWc1NANOa}
    }
    \subfigure[]{
        \includegraphics[width=8.5cm,height=7.5cm] {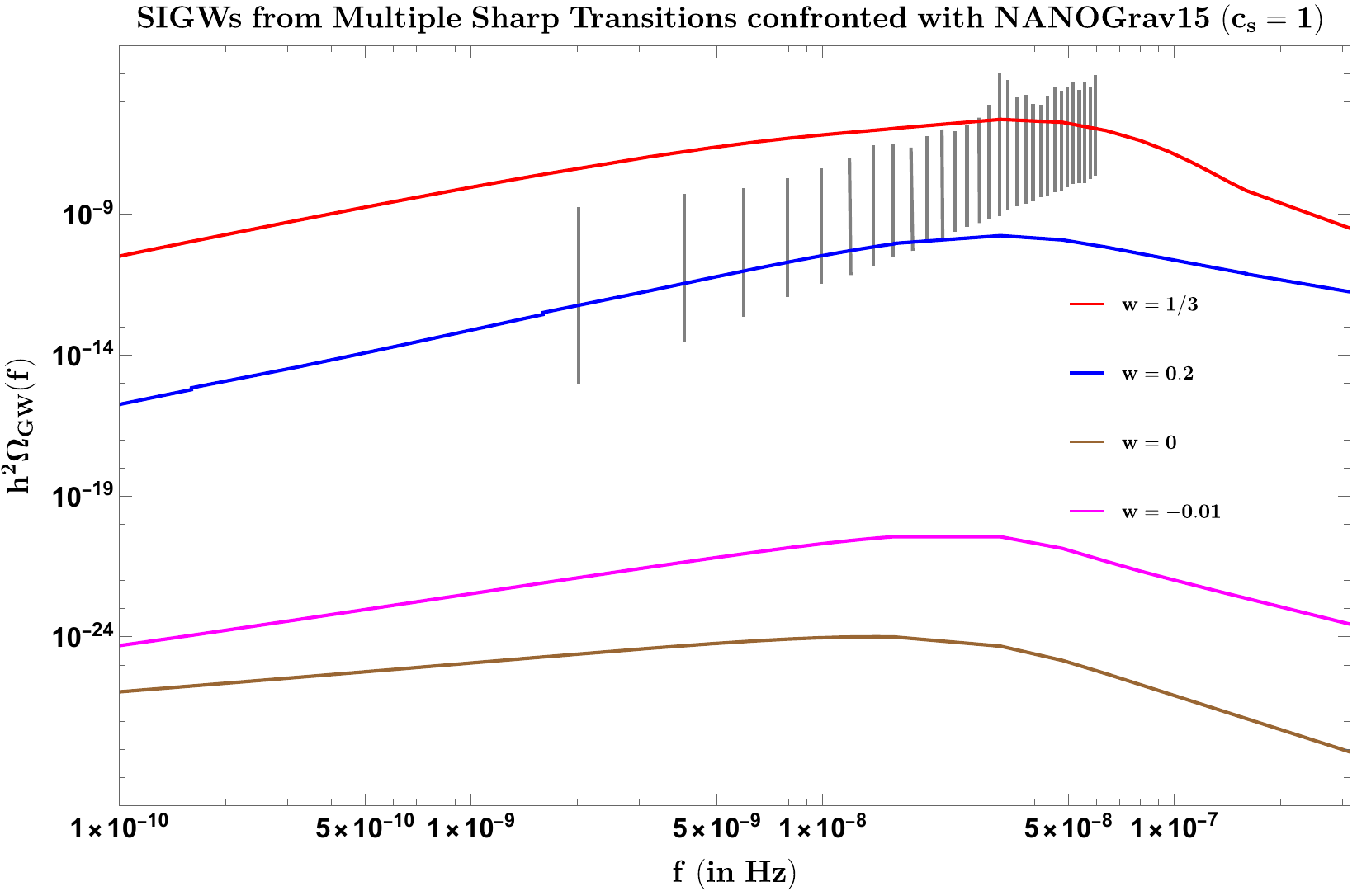}
        \label{SIGWc1NANOb}
       }
    \subfigure[]{
      	\includegraphics[width=8.5cm,height=7.5cm] {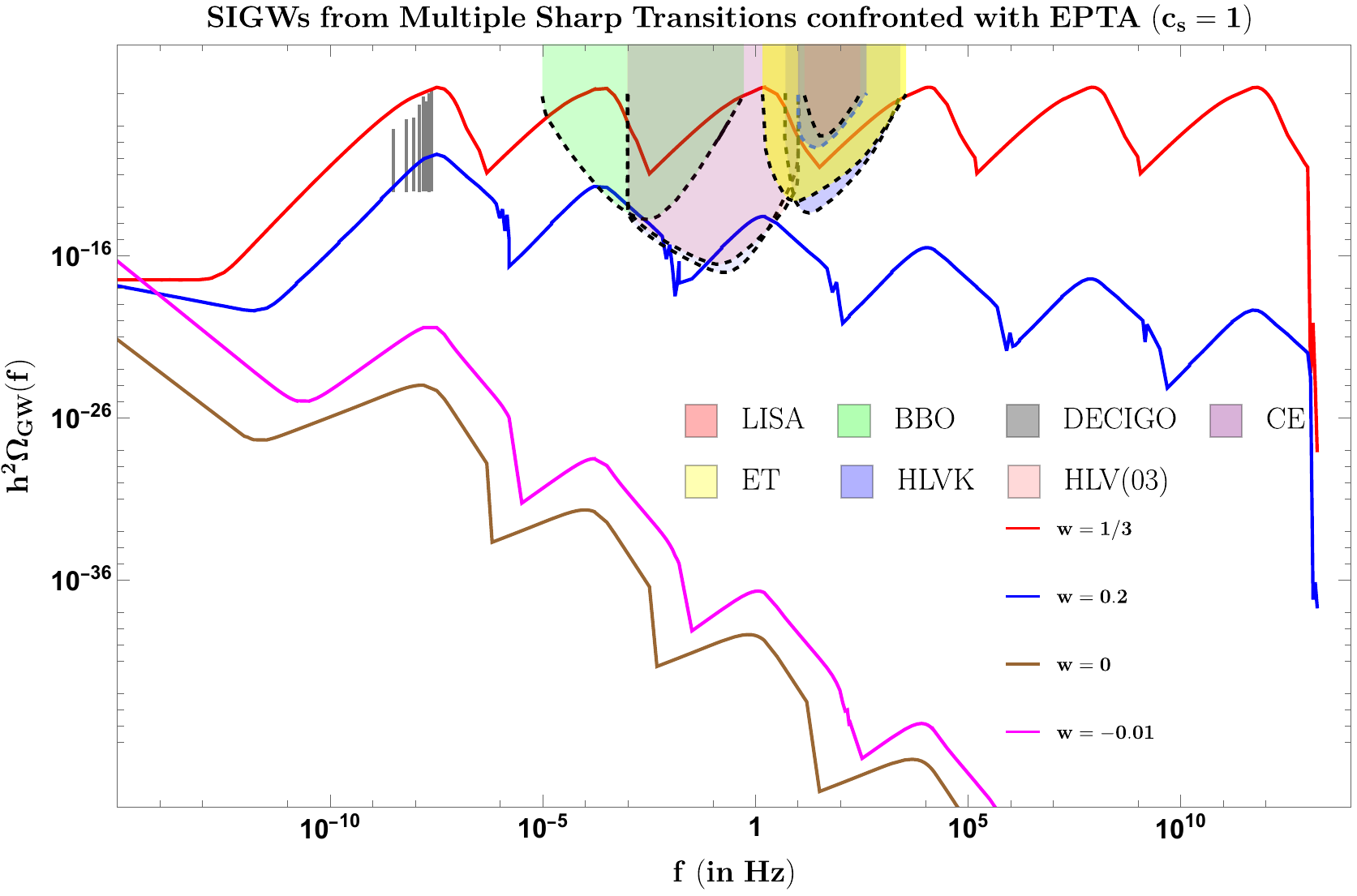}
        \label{SIGWc1EPTAa}
    }
    \subfigure[]{
        \includegraphics[width=8.5cm,height=7.5cm] {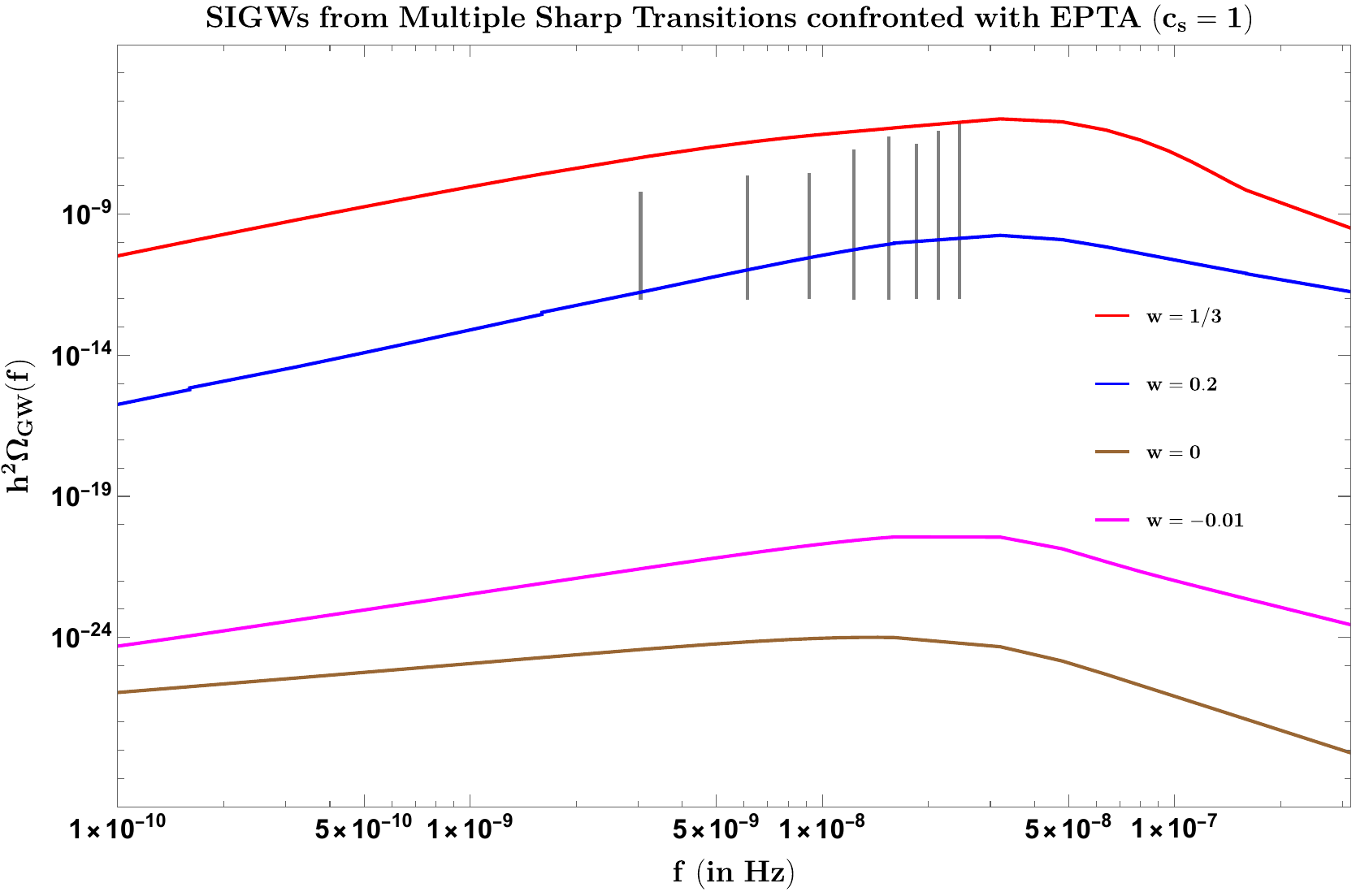}
        \label{SIGWc1EPTAb}
       }
    	\caption[Optional caption for list of figures]{The spectrum of SIGW as a function of its frequency. All the panels feature the value of effective sound speed $c_{s}=1$. The left panel (both top and bottom) shows the complete spectrum, which covers frequencies sensitive to the data from NANOGrav-15, EPTA, and the ground and space-based experiments, which include LISA, DECIGO, BBO, Einstein Telescope (ET), Cosmic Explorer (CE), the HLVK network (aLIGO in Livingstone and Hanford, aVIRGO, and KAGRA), and HLV (O3). The right panel (both top and bottom) focuses primarily on the frequencies involving the NANOGrav 15 and EPTA signals. Red, blue, magenta and brown represent the spectrum corresponding to the EoS values $w \in \{1/3,0.2,0,-0.01\}$, respectively.} 
    	\label{SIGWc1ab}
    \end{figure*}

\begin{figure*}[htb!]
    	\centering
    \subfigure[]{
      	\includegraphics[width=8.5cm,height=7.5cm] {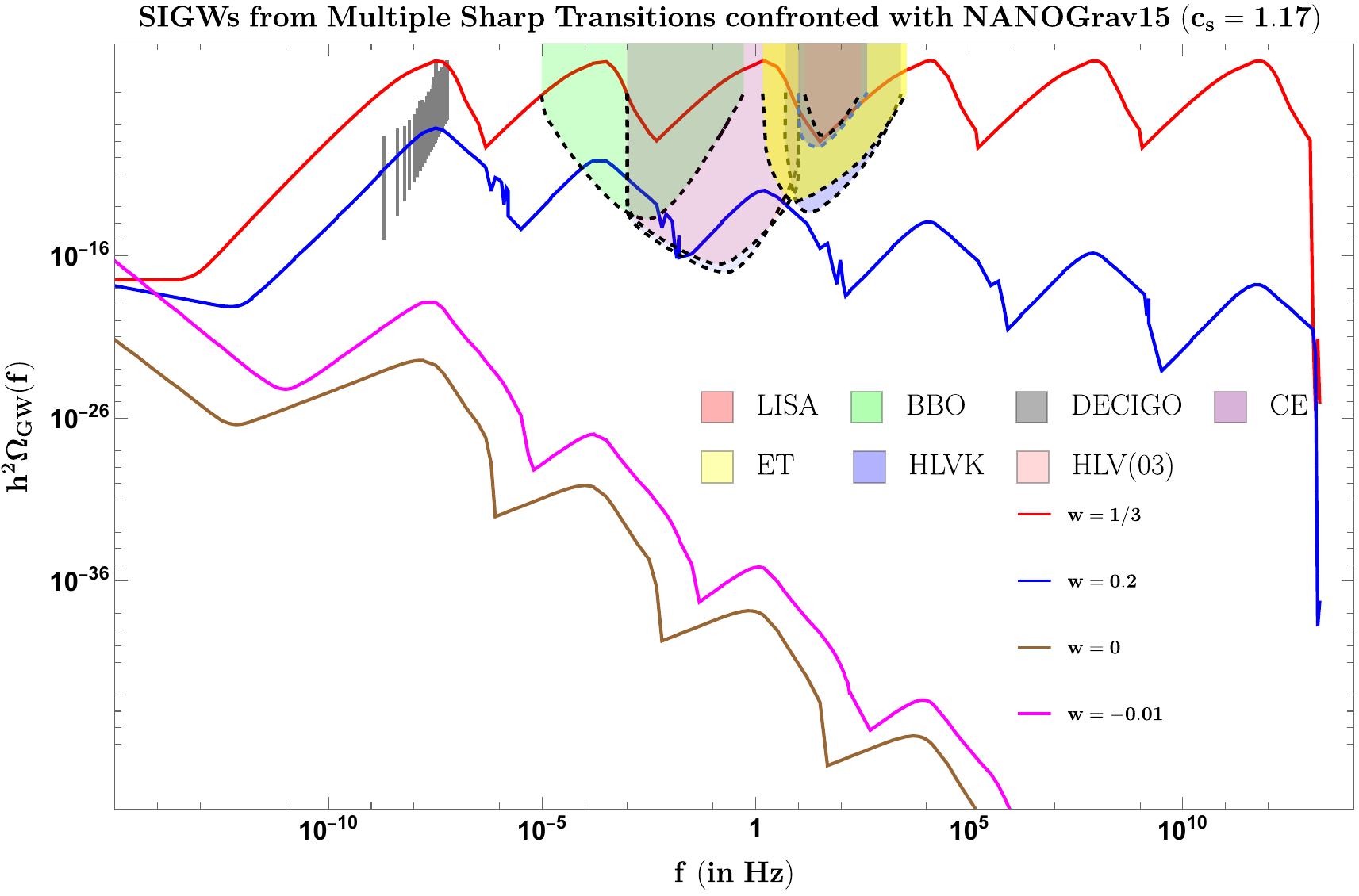}
        \label{SIGWc117NANOa.pdf}
    }
    \subfigure[]{
        \includegraphics[width=8.5cm,height=7.5cm] {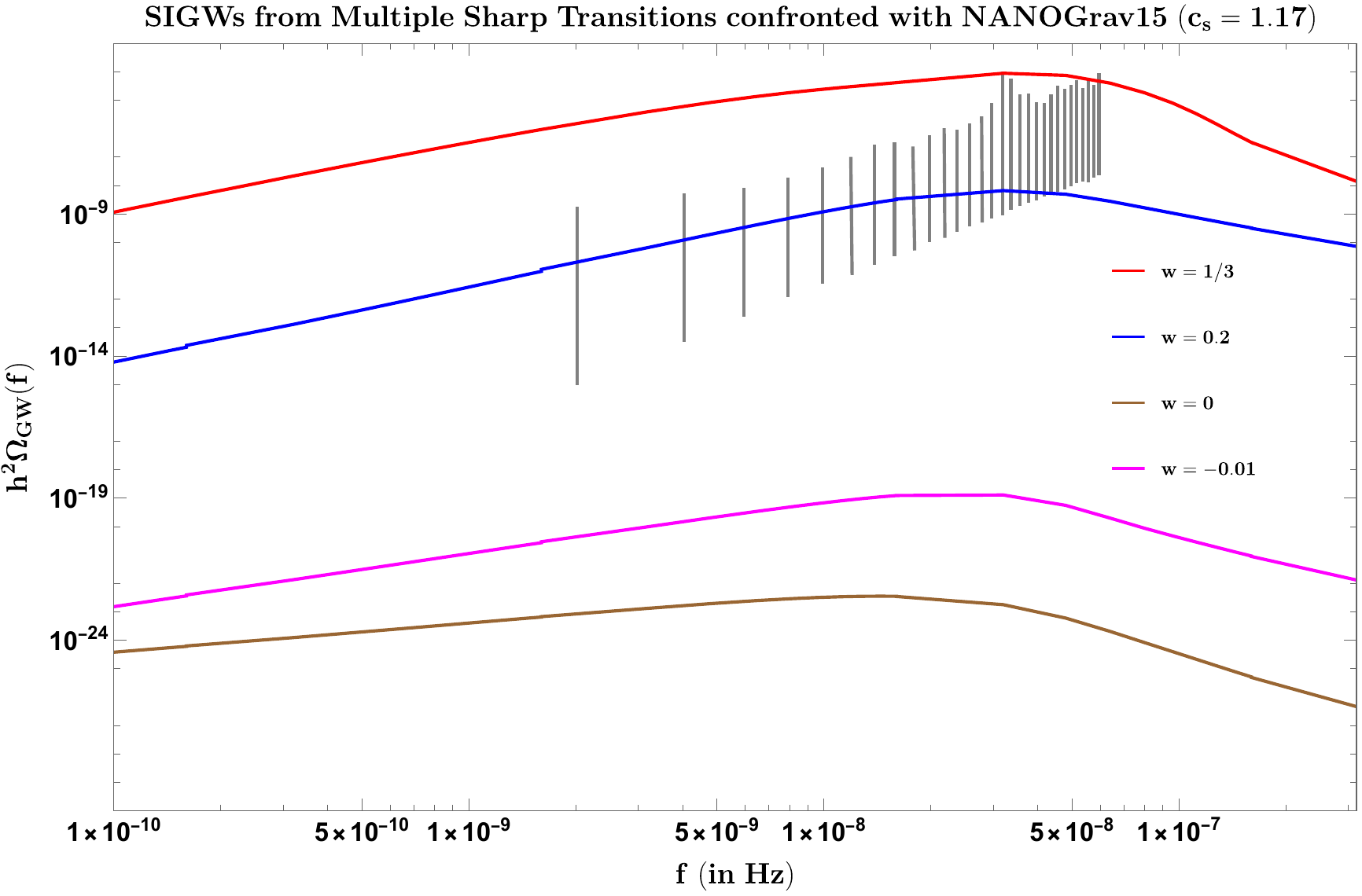}
        \label{SIGWc117NANOb}
       }
    \subfigure[]{
      	\includegraphics[width=8.5cm,height=7.5cm] {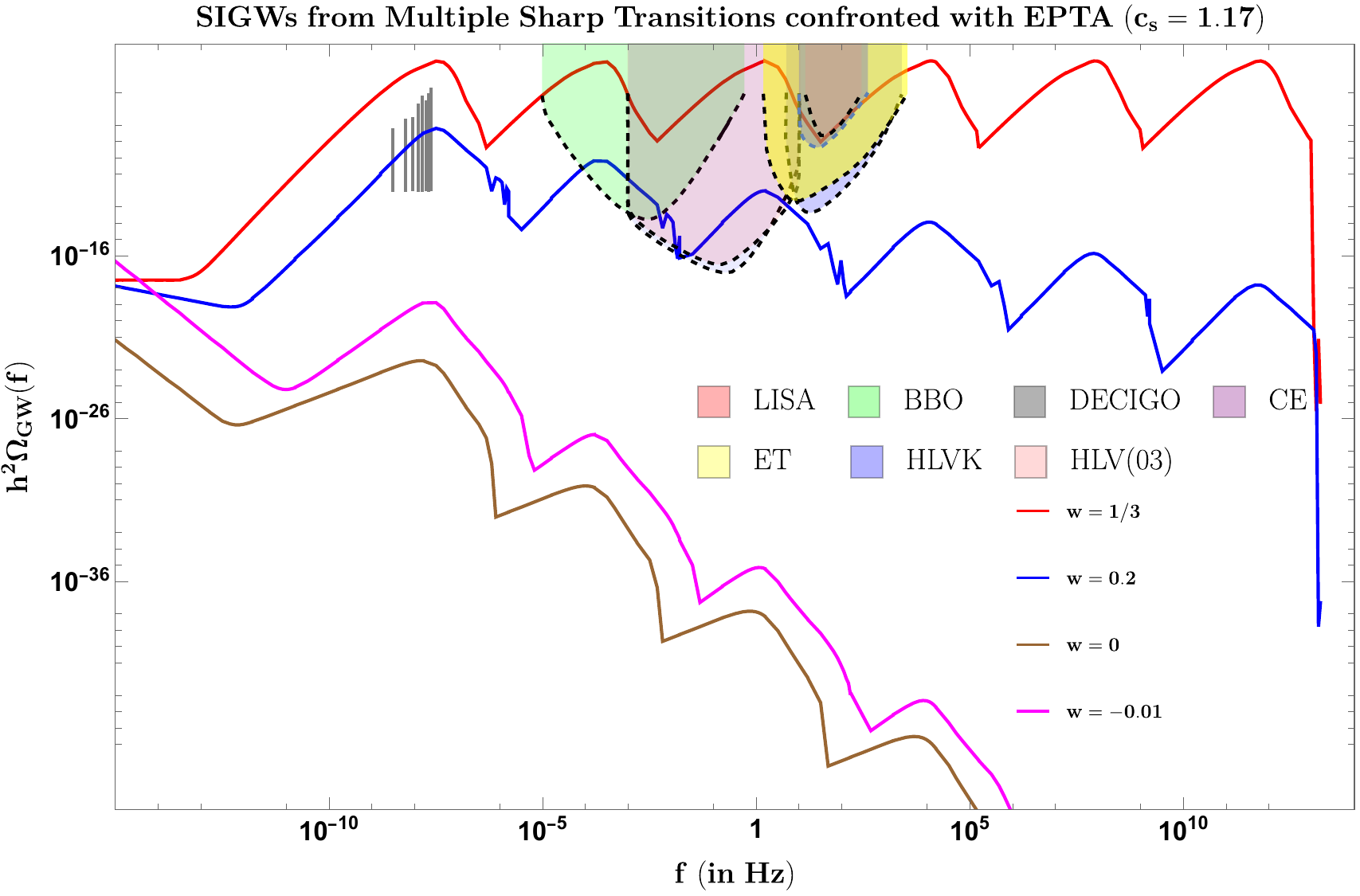}
        \label{SIGWc117EPTAa}
    }
    \subfigure[]{
        \includegraphics[width=8.5cm,height=7.5cm] {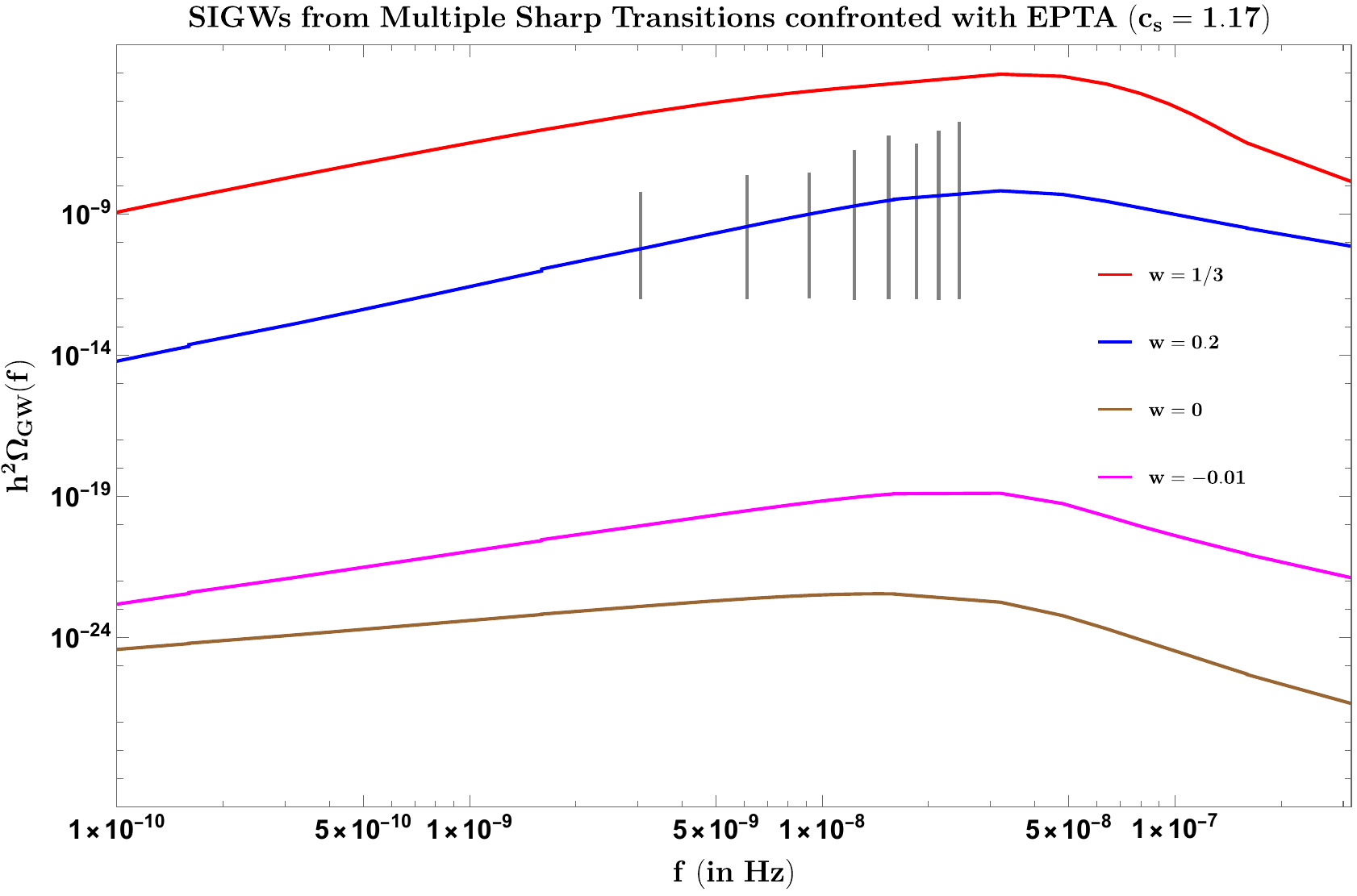}
        \label{SIGWc117EPTAb}
       }
    	\caption[Optional caption for list of figures]{The spectrum of SIGW as a function of its frequency. All the panels feature the value of effective sound speed $c_{s}=1.17$. The left panel (both top and bottom) shows the complete spectrum, which covers frequencies sensitive to the data from NANOGrav-15, EPTA, and the ground and space-based experiments, which include LISA, DECIGO, BBO, Einstein Telescope (ET), Cosmic Explorer (CE), the HLVK network (aLIGO in Livingstone and Hanford, aVIRGO, and KAGRA), and HLV (O3). The right panel (both top and bottom) focuses primarily on the frequencies involving the NANOGrav 15 and EPTA signals. Red, blue, magenta and brown represent the spectrum corresponding to the EoS values $w \in \{1/3,0.2,0,-0.01\}$, respectively.} 
    	\label{SIGWc117ab}
    \end{figure*}

\begin{figure*}
\includegraphics[width=19.5cm,height=12.5cm] {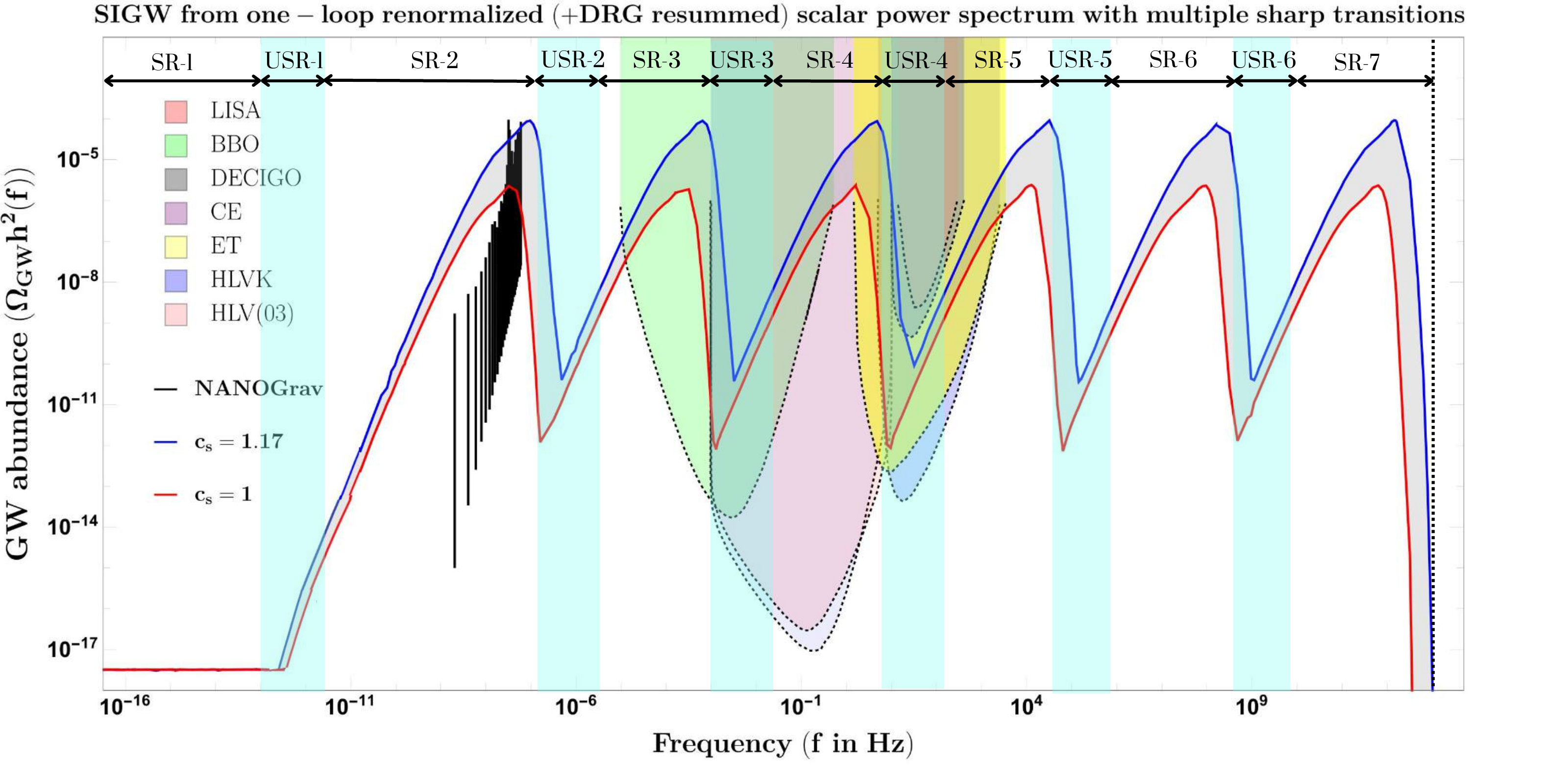}
        \label{w4}
       
    	\caption[Optional caption for list of figures]{SIGW spectrum as a function of the frequency. The spectrum is generated with the fixed EoS condition $w=1/3$. The red and blue lines represent the effective sound speed values $c_{s}=1\;{\rm and }\;1.17$, respectively. The ash-coloured shaded region represents the allowed parameter space for the SIGW signal where $1 \leq c_{s} \leq 1.17$ is satisfied. The first peak is shown to align with the NANOGrav-15 signal and the complete spectrum contains its signature within the sensitivities of the several ground and space-based GW experiments, which include LISA, DECIGO, BBO, Einstein Telescope (ET), Cosmic Explorer (CE), the HLVK network (aLIGO in Livingstone and Hanford, aVIRGO, and KAGRA), and HLV (O3).} 
    	\label{Gw13plot} 
    \end{figure*}

 For certain benchmark values of the EoS parameter, $w \in \{1/3,0.2,0,-0.01\}$, the amplitude variation of many SIGW spectra derived from the one-loop renormalized and DRG-resummed scalar power spectrum is shown in fig. (\ref{SIGWc1ab}). The outcomes are then compared to the NANOGrav-15 and EPTA data with the effective sound speed set at $c_{s}=1.A$. One benefit of having $w=1/3$ is that the generated spectra correlate most closely with the signals from NANOGrav-15 and EPTA, as fig.(\ref{SIGWc1ab}) demonstrates. The $w=1/3$ example similarly exhibits the notable characteristic of consistent peak amplitude. We find that the tail section of the produced spectra aligns with both the EPTA and NANOGrav-15 signals at a value of $w \sim 0.2$. Lowering $w$ results in incredibly low amplitude spectra, which helps rule out the area of $w \lesssim 0$ as a possible explanation for the PTA signals. It is important to note that when addressing overproduction, the amplitude for $w=-0.01$ being larger than $w=0$ translates into the amplitude characteristics shown in fig. (\ref{overprodcs1}). Another intriguing aspect of the graphs for $w \gtrsim 0.2$ is the shift in the spectral tilt upon reaching peak amplitude. When taking into account the benchmark values, $w \in \{1/3,0.2,0,-0.01\}$, the fig. (\ref{SIGWc117ab}) shows a similar change of the SIGW amplitude with the frequency; however, this time, the effective sound speed takes the value $c_{s}=1.17$. For every EoS value in the previously given benchmark set, there is an increase in the SIGW amplitude over the whole frequency range, which corresponds to this change in $c_{s}$. Notably, by staying inside the parameter space permitted by $1 \leq c_{s} \leq 1.17$, the regime near the computed spectrum peak amplitude for $w=1/3$ displays the closest approach to the NANOGrav-15 signal. Going any lower won't be enough to explain the needed characteristics; values of $w \sim 0.2$ are also consistent with both the NANOGrav-15 and EPTA signals. Last but not least, the cumulative features, such as the behavior of the spectrum tilt before and after reaching their peak values for each EoS $w$ scenario, and the overall features across the entire frequency range covering all the listed GW experiments, stay the same, with the exception of the same order of magnitude enhancement in the amplitude of the spectrum for $c_{s}=1.17$ in fig.(\ref{SIGWc117ab}) when compared with fig.(\ref{SIGWc1ab}) for $c_{s}=1$. We illustrate the situation of $w=1/3$ and its benefits from the SIGW spectrum by displaying the figure in fig.(\ref{Gw13plot}). The plot clearly illustrates the behavior of the aforementioned spectrum after each SR and USR interval, the corresponding signature in the sensitivities of the various GW experiments, NANOGrav-15 and other higher-frequency experiments, and a closer examination of the effects caused by the variation in effective sound speed within the interval $1 \leq c_{s} \leq 1.17$. These findings are consistent with the results previously discussed for the $w=1/3$ scenario. The experimentally detected signal of the NANOGrav $15$ data has an amplitude of $\Omega_{\rm GW}h^{2}\sim{\cal O}(10^{-4})$, and we have shown that for both spectra, there exists a resonating behavior at frequency $f\sim{\cal O}(10^{-7})$. The spectra contain locations where peaks lie within the sensitivities of the current and prospective terrestrial and space-based experiments; these regions are emphasized in the figure's background when shown for large frequency values $f\sim{\cal O}(10^{-4}{\rm Hz}-10^{4}{\rm Hz})$. These peaks arise from improvements brought about by MSTs when the system enters several USR (USR$_{n}$) phases. For the value of the parameter, $1 \leq c_{s} \leq 1.17$, the peak amplitude in the spectra falls within the interval, $\Omega_{\rm GW}h^{2}\sim{\cal O}(10^{-4}-10^{-6})$. The spectrum abruptly decreases after the peak values are reached, resulting in a dip-like feature with an amplitude within the interval, $\Omega_{\rm GW}h^{2}\sim{\cal O}(10^{-12}-10^{-10})$, given the parameter's comparable allowable interval, $1 \leq c_{s} \leq 1.17$. The last DRG resummation constraint prior to the end of inflation is responsible for the dramatic reduction in the spectra following the augmentation during the last SR$_{7}$ phase, at $f\sim{\cal O}(10^{12}{\rm Hz})$. The required region is also emphasized in the form of a band between the two spectra. It produces enough enhancements that correspond to signals for the synthesis of PBHs in the mass range of $M_{\rm PBH}\sim{\cal O}(10^{-31}M_{\odot}-10^{4}M_{\odot})$.

\section{Galileon inflation in the light of large fluctuations}
\label{s6}

Massive gravity, sometimes referred to as dRGT, brought Galileon field theory to light in 2010 \cite{deRham:2010ik,deRham:2010kj}. As early as 1939, Pauli-Firz really made the first effort to give the graviton a mass \cite{Fierz:1939ix}. The vDVZ discontinuity, which causes the Fierz-Pauli theory's predictions to not decrease to those of the Einstein theory when the graviton mass is turned zero, was identified as a problem in 1970 \cite{vanDam:1970vg,Zakharov:1970cc}. Promoting the theory to a non-linear backdrop is one way to solve the difficulty, as Vainshtein showed in 1972 \cite{Vainshtein:1972sx}. The Boulware-Deser ghost \cite{Boulware:1972yco} is the inevitable consequence of the latter, and tremendous gravity was all but forgotten until 2010. Nicolis et al. \cite{Nicolis:2008in} created a higher-order derivative scalar field Lagrangian with a distinct structure in Minkowski space-time in 2008. This Lagrangian yields second-order equations of motion that are free from Ostrogradsky instabilities, and it is based on generalized shift symmetry, also known as Galileon symmetry. In ref. \cite{Deffayet:2009wt}, the framework was easily extended to curved space-time. The authors constructed the mass term for graviton in 2010 with the intention of eliminating the Boulware-Deser ghost in dRGT by taking into account the covariant Galileon construct. This meant that in the decoupling limit, which was relevant to local gravity constraints, the longitudinal part of the spin-2 field (scalar field $\phi$) was a Galileon field, which was successfully screened out by the Vainstein mechanism while still adhering to local gravity constraints. It should be noted that while the modern rationale for giving the graviton a tiny mass of the order of $H_0\sim 10^{-33}eV$ was to explain late-time acceleration, the historical motivation {\it a la} Fierz-Pauli was linked to the development of a consistent relativistic equation for massive spin-$2$ field. It's interesting to note that the setup that results is vDVZ discontinuity-free and ghost-free. Regretfully, this paradigm lacks the FLRW cosmology. Next, an attempt was made to solve the problem by substituting the flat Minkowski metric employed in dRGT with the non-trivial fiducial metric, both fixed and dynamical (bi-gravity). But the latter introduces the Higuchi ghost \footnote{If $m_g>\sqrt{2} H$, which is the Higuchi bound, is violated, then it is considered a ghost. In this case, the graviton mass is denoted as $m_g$.} \cite{Higuchi:1986py}. This means that, in addition to a number of other theoretical problems related to large gravity, the scenario is especially inappropriate for late time cosmology. Maybe there is no such thing as a strictly zero graviton mass, or developing a coherent theory of huge gravity is just too difficult.

And yet, from a phenomenological perspective, the Galileon field appears promising on its own, tremendous gravity notwithstanding. Nevertheless, late time acceleration is unrelated to the lowest non-trivial Galileon Lagrangian, which comprises $\mathcal{L}_3\sim (\partial \phi)^2(\Box \phi)$ \cite{Chow:2009fm}. References \cite{Nicolis:2008in} and \cite{Deffayet:2009wt} initially showed that the Galileon Lagrangian \footnote{Taken together with $\mathcal{L}_1$ and $\mathcal{L}_2$ represented by a linear term in $\phi$ and standard kinetic term, makes the complete structure of Galileon Lagrangian in $3+1$-space time dimensions.} leads to a stable de-Sitter solution relevant to late time cosmology. Galileon fields have been used phenomenologically in cosmology, quantum field theory, and gravity as a result of this study. See refs. \cite{Kobayashi:2010wa,Jain:2010ka,Gannouji:2010au,Ali:2010gr,deRham:2011by,Tsujikawa:2010sc,Burrage:2010rs,DeFelice:2010jn,DeFelice:2010gb,Babichev:2010jd,DeFelice:2010pv,DeFelice:2010nf,Hinterbichler:2010xn,Kobayashi:2010cm,Deffayet:2010qz,Burrage:2010cu,Mizuno:2010ag,Nesseris:2010pc,Khoury:2010xi,DeFelice:2010as,Kimura:2010di,Zhou:2010di,Hirano:2010yf,Kamada:2010qe,VanAcoleyen:2011mj,Hirano:2011wj,Li:2011sd,Pujolas:2011he,Kobayashi:2011pc,DeFelice:2011zh,Khoury:2011da,Trodden:2011xh,Burrage:2011bt,Liu:2011ns,Kobayashi:2011nu,PerreaultLevasseur:2011wto,deRham:2011by,Clifton:2011jh,Endlich:2011vg,Brax:2011sv,Gao:2011mz,DeFelice:2011uc,Gao:2011qe,Babichev:2011iz,DeFelice:2011hq,Khoury:2011ay,Qiu:2011cy,Renaux-Petel:2011rmu,DeFelice:2011bh,Kimura:2011td,Wang:2011dt,Kimura:2011dc,DeFelice:2011th,Appleby:2011aa,DeFelice:2011aa,Zhou:2011ix,Goon:2012mu,Shirai:2012iw,Goon:2012dy,deRham:2012az,Ali:2012cv,Liu:2012ww,Choudhury:2012yh,Choudhury:2012whm,Barreira:2012kk,Gubitosi:2012hu,Barreira:2013jma,deFromont:2013iwa,Deffayet:2013lga,Arroja:2013dya,Li:2013tda,Sami:2013ssa,Khoury:2013tda,Burrage:2015lla,Koyama:2015vza,Brax:2015dma,Saltas:2016nkg,Ishak:2018his} for more details. It is noteworthy that Horndeski created a generic scalar field Lagrangian in 1972, which leads to second order equations of motion. This framework is considered to be a generalized Galileon framework \cite{Horndeski:1974wa}. Specifically, recent GW170817 findings \cite{DeFelice:2022xvq, Kubota:2022lbn} place strong constraints on $\mathcal{L}_4$ and $\mathcal{L}_5$ that involve derivative interaction to curvature.

Noteworthy, there has been a resurgence of interest in the study of primordial black holes for a number of reasons \cite{Zeldovich:1967lct,Hawking:1974rv,Carr:1974nx,Carr:1975qj,Chapline:1975ojl,Carr:1993aq,Kawasaki:1997ju,Yokoyama:1998pt,Kawasaki:1998vx,Rubin:2001yw,Khlopov:2002yi,Khlopov:2004sc,Saito:2008em,Khlopov:2008qy,Carr:2009jm,Choudhury:2011jt,Lyth:2011kj,Drees:2011yz,Drees:2011hb,Ezquiaga:2017fvi,Kannike:2017bxn,Hertzberg:2017dkh,Pi:2017gih,Gao:2018pvq,Dalianis:2018frf,Cicoli:2018asa,Ozsoy:2018flq,Byrnes:2018txb,Ballesteros:2018wlw,Belotsky:2018wph,Martin:2019nuw,Ezquiaga:2019ftu,Motohashi:2019rhu,Fu:2019ttf,Ashoorioon:2019xqc,Auclair:2020csm,Vennin:2020kng,Nanopoulos:2020nnh,Gangopadhyay:2021kmf,Inomata:2021uqj,Stamou:2021qdk,Ng:2021hll,Wang:2021kbh,Kawai:2021edk,Solbi:2021rse,Ballesteros:2021fsp,Rigopoulos:2021nhv,Animali:2022otk,Correa:2022ngq,Frolovsky:2022ewg,Escriva:2022duf,Karam:2022nym,Ozsoy:2023ryl,Ivanov:1994pa,Afshordi:2003zb,Frampton:2010sw,Carr:2016drx,Kawasaki:2016pql,Inomata:2017okj,Espinosa:2017sgp,Ballesteros:2017fsr,Sasaki:2018dmp,Ballesteros:2019hus,Dalianis:2019asr,Cheong:2019vzl,Green:2020jor,Carr:2020xqk,Ballesteros:2020qam,Carr:2020gox,Ozsoy:2020kat,Baumann:2007zm,Saito:2008jc,Saito:2009jt,Choudhury:2013woa,Sasaki:2016jop,Raidal:2017mfl,Papanikolaou:2020qtd,Ali-Haimoud:2017rtz,Di:2017ndc,Raidal:2018bbj,Cheng:2018yyr,Vaskonen:2019jpv,Drees:2019xpp,Hall:2020daa,Ballesteros:2020qam,Ragavendra:2020sop,Carr:2020gox,Ozsoy:2020kat,Ashoorioon:2020hln,Ragavendra:2020vud,Papanikolaou:2020qtd,Ragavendra:2021qdu,Wu:2021zta,Kimura:2021sqz,Solbi:2021wbo,Teimoori:2021pte,Cicoli:2022sih,Ashoorioon:2022raz,Papanikolaou:2022chm,Wang:2022nml,Mishra:2019pzq,ZhengRuiFeng:2021zoz,Cohen:2022clv,Arya:2019wck,Bastero-Gil:2021fac,Correa:2022ngq,Gangopadhyay:2021kmf,Cicoli:2022sih,Brown:2017osf,Palma:2020ejf,Geller:2022nkr,Braglia:2022phb,Kawai:2022emp,Frolovsky:2023xid,Aldabergenov:2023yrk,Aoki:2022bvj,Frolovsky:2022qpg,Aldabergenov:2022rfc,Ishikawa:2021xya,Gundhi:2020kzm,Aldabergenov:2020bpt,Cai:2018dig,Fumagalli:2020adf,Cheng:2021lif,Balaji:2022rsy,Qin:2023lgo}. In fact, PBHs may hold the key to solving some of the unsolved mysteries of the early universe, including dark matter and the observed baryon asymmetry in the cosmos, in addition to being candidates for super-massive black holes in galaxies. Apart from its significance to late time cosmology, the Galileon field is notable from a field theoretic perspective and might be of great help in solving the PBH creation conundrum within the context of single field inflation. In fact, only derivative terms should be present in its Lagrangian according to Galileon symmetry; yet, this is generally afflicted by Ostrogradski instabilities. The aforementioned symmetry gives rise to second order equations of motion through a certain arrangement of terms with fewer than two derivatives per field. It's amazing that self-loops do not renormalize the galileons; given that the couplings maintain Galileon symmetry, other field loops also follow the same characteristics. Galileon field theory has this general unique property \footnote{This is valid for derivative coupled theories with unitarity at the background and Ostrogradsky instabilities at the quantum level.}. Regarding PBH creation for single field inflation, the non-renormalizability of the Galileon field may have significant consequences.
Noticeably, PBH creation within the context of single field inflation is now being actively studied \cite{Kristiano:2022maq,Riotto:2023hoz,Choudhury:2023vuj,Choudhury:2023jlt,Kristiano:2023scm,Riotto:2023gpm,Choudhury:2023rks,Firouzjahi:2023aum,Motohashi:2023syh}. One loop adjustments to the power spectrum in $P(X,\phi)$ (where $X\equiv -(\partial_\mu\phi \partial^\mu\phi)/2$) theories have recently been shown to significantly restrict the mass of PBHs created during slow roll to ultra-slow role transitions, which amounts to a no-go result. Galileon theories might potentially circumvent these limitations because of their non-renormalizable nature, making them a great candidate for PBH generation within the context of single field inflation.

\subsection{Galileon Effective Field Theory: A old wine in a new glass}

\subsubsection{Non-Covarinat Galileon Effective Field Theory (NCGEFT)}

In Minkowski space-time, second order equations of motion arise from the Galileon action, a higher derivative scalar field framework that was initially developed in \cite{Nicolis:2008in}.  Later, the authors allowed a non-minimal connection to the gravitational backdrop in Ref.\cite{Deffayet:2009wt}, where they built a ghost free and unitarity maintaining version of the Galileon theory in dynamical space time.  Along with the shift symmetry, which is closely linked to the slow-roll property of inflationary potential, galileon theories are also endowed, at least in planar space, with the following symmetry on the scalar degree of freedom $\phi$ \footnote{The extended form of the standard shift symmetry $ \phi \rightarrow \phi+ c$, known as Galilean symmetry, facilitates the incorporation of any kind of derivative interactions.  It is possible to include mixing of the derivative contributions in the various orders, but this symmetry can't, in theory, commute with the generators formed by the Poincar\'{e} group.    As a result, it needs to be realized non-linearly inside the description of Effective Field Theory (EFT) in terms of the extended Poincar\'{e} group, which is not factorizable.}:
\bea \label{GCS}\phi \rightarrow \phi+ c + b_{\mu}x^{\mu}=\phi+ c + b\cdot x\quad\quad\Longrightarrow\quad\quad \partial_{\mu}\phi\rightarrow\partial_{\mu}\phi+b_{\mu},\eea
where $x^{\mu}$ specifies the equivalent coordinates in $3+1$ space-time dimensions, $b_{\mu}$ represents a vector constant, and $c,$ represents a scalar constant.  It is significant to notice that the space-time translations are represented by the last word, $b_{\mu}x^{\mu}=b\cdot x$.  Because it mimics the coordinate transformation between non-relativistic inertial frames, it gets its name. In flat space at least, it is evident that every term $\partial \partial..\, \phi$ with two or more derivatives is intrinsically Galilean invariant. We shall employ a particularly special set of Galilean invariant terms in this context, which among other things yields a second order equation of motion. These continue to be of great interest and are often explored in literature. 
The Galileon terms under consideration are incredibly rare; only five exist in the four-dimensional space-time environment that is described by the following non-covariant version of the representative Galielon Effective Field Theory (NCGEFT) action and appears alongside the Einstein-Hilbert term in the curved gravitation background:
\bea
S= \int d^4 x \sqrt{-g}\Bigg[\frac{M^2_{pl}}{2}R-V_0+ {\cal L}^{\bf NC}_{\phi}\Bigg],
\eea
where we define:
\be {\cal L}^{\bf NC}_{\phi}=\sum^{5}_{i=1}c_i\mathcal{L}^{\bf NC}_i.\ee
The following equations provide the explicit expression for the $\mathcal{L}^{\bf NC}_i\forall i=1,2,\cdots, 5$:
\bea
{\cal L}^{\bf NC}_1 & = & \phi, \\
{\cal L}^{\bf NC}_2 & = & -\frac{1}{2} \,  \partial \phi \cdot \p \phi, \\
{\cal L}^{\bf NC}_3 & = & - \frac{1}{2} \,  [\Pi_{\phi}] \, \partial \phi \cdot \partial \phi ,\\
{\cal L}^{\bf NC}_4 & = & - \frac{1}{4} \Big\{ [\Pi_{\phi}]^2 \,  \partial \phi \cdot \partial \phi - 2 \,  [\Pi_{\phi}] \, \partial \phi \cdot \Pi_{\phi} \cdot \partial \phi - [\Pi ^2_{\phi} ] \, \partial \phi \cdot \partial \phi + 2 \, \partial \phi \cdot \Pi ^2_{\phi} \cdot \partial\phi \Big\}, \nonumber \\
{\cal L}^{\bf NC}_5 & = & -\frac{1}{5} \Big\{
[\Pi_{\phi}]^3 \, \partial \phi \cdot \partial \phi- 3 [\Pi_{\phi}]^2_{\phi} \,  \partial \phi \cdot \Pi _{\phi}\cdot \partial \phi
-3 [\Pi_{\phi}] [\Pi^2_{\phi}] \,  \partial \phi \cdot \partial \phi
+6 [\Pi_{\phi}] \, \partial \phi \cdot \Pi^2_{\phi} \cdot \partial\phi \nonumber
 \\
&&
\quad\quad\quad\quad\quad\quad\quad\quad\quad\quad\quad\quad\quad+2 [\Pi ^3_{\phi}] \, \partial \phi \cdot \partial \phi
+3 [\Pi ^2_{\phi}] \, \partial \phi \cdot \Pi_{\phi} \cdot \partial \phi
- 6 \,  \partial \phi \cdot \Pi^3_{\phi} \cdot \partial\phi \Big\},
\eea
where, for our purposes, we employ the short-hand notation below:
\be (\partial_\mu \partial_\nu \pi)^n \equiv [ \Pi ^n _{\phi}].\ee
The trace operator is denoted by the brackets $[...]$ in this instance, while the normal Lorentz invariant contraction of space-time indices is denoted by $'\dot'$.  One may write the following as an example:
\be  [\Pi_{\phi}] \, \partial \phi \cdot \partial \phi \equiv \Box\phi \,\partial_{\mu}\phi\partial^{\mu}\phi.\ee
Furthermore, in this context, the generic coefficients $c_i$ serve as a model for the Wilson coefficients in the EFT framework.  The NCGEFT Lagrangian offers additional concepts for consideration. However, in four space-time dimensions, those extra contributions become negligible, and they may all be recast as total derivatives.

The following equation may be written using the NCGEFT Lagrangian set mentioned above:
\be {\cal E}:\equiv \delta_{\phi}{\cal L}^{\bf NC}_{\phi}=\sum^{5}_{i=1}c_i\delta_{\phi}\mathcal{L}^{\bf NC}_i =\sum^{5}_{i=1}c_i {\cal E}_i=-T^{\mu}_{\mu},\ee
Currently, ${\cal E}_i \forall i=1,2,\cdots,5$ are defined as follows:
\bea {\cal E}_1 &=& \delta_{\phi}{\cal L}^{\bf NC}_1=1,\\
{\cal E}_2 &=& \delta_{\phi}{\cal L}^{\bf NC}_2=\Box\phi,\\ 
{\cal E}_3 &=& \delta_{\phi}{\cal L}^{\bf NC}_3=(\Box\phi)^2-(\partial_{\mu}\partial_{\nu}\phi)^2,\\ 
{\cal E}_4 &=& \delta_{\phi}{\cal L}^{\bf NC}_4=(\Box\phi)^3-3\Box\phi (\partial_{\mu}\partial_{\nu}\phi)^2+2(\partial_{\mu}\partial_{\nu}\phi)^3,\\
{\cal E}_4 &=& \delta_{\phi}{\cal L}^{\bf NC}_5=(\Box\phi)^4-6(\Box\phi)^2 (\partial_{\mu}\partial_{\nu}\phi)^2 +8\Box\phi (\partial_{\mu}\partial_{\nu}\phi)^3+3[ \Pi ^2 _{\phi}]^2-6(\partial_{\mu}\partial_{\nu}\phi)^4. \eea
This clearly indicates that the equations of motion are second order, ensuring that the underlying theory is devoid of the Ostrogradski ghost instability.  NCGEFT also results in a configuration where unitarity is preserved at the quantum mechanical level. Numerous extensions of the aforementioned hypothesis have been put forth, utilizing references \cite{Kobayashi:2010wa,Jain:2010ka,Gannouji:2010au,Ali:2010gr,deRham:2011by,Tsujikawa:2010sc,DeFelice:2010gb,DeFelice:2010pv,DeFelice:2010nf,Kobayashi:2010cm,Deffayet:2010qz,Burrage:2010cu,Mizuno:2010ag,Nesseris:2010pc,Khoury:2010xi,DeFelice:2010as,Kimura:2010di,Hirano:2010yf,Kamada:2010qe,Hirano:2011wj,Li:2011sd,Kobayashi:2011pc,DeFelice:2011zh,Burrage:2011bt,Liu:2011ns,Kobayashi:2011nu,PerreaultLevasseur:2011wto,deRham:2011by,Clifton:2011jh,Gao:2011mz,DeFelice:2011uc,Gao:2011qe,DeFelice:2011hq,Qiu:2011cy,Renaux-Petel:2011rmu,DeFelice:2011bh,Wang:2011dt,Kimura:2011dc,DeFelice:2011th,Appleby:2011aa,DeFelice:2011aa,Zhou:2011ix,Shirai:2012iw,deRham:2012az,Ali:2012cv,Liu:2012ww,Choudhury:2012yh,Choudhury:2012whm,Barreira:2012kk,Gubitosi:2012hu,Arroja:2013dya,Sami:2013ssa,Khoury:2013tda,Burrage:2015lla,Koyama:2015vza,Saltas:2016nkg} to study a variety of cosmic events. For this theory to work, the setup must be immersed in a curved de Sitter backdrop. Nevertheless, it was discovered that renormalization in the previously indicated curved background geometry results in significant adjustments for the non-covariant Lagrangians ${\cal L}^{\bf NC}_i \forall i=1,2,\cdots,5$ in order to maintain unitarity.

\subsubsection{Covarinat Galileon Effective Field Theory (CGEFT)}

In order to obtain an inflationary solution from the current NCGEFT configuration, it is necessary to violate the precise Galilean shift symmetry that corresponds to this situation.  Let us now discuss the decoupling limit, $M_{pl}\rightarrow\infty$ and $3H^2M^2_{pl}=V_0$, with $H$ constant. On this boundary, the Galilean shift symmetry remains precise, and the Planck scale in the previously described coefficients suppresses any subsequent soft breaking. It is noteworthy to observe that, even with the Galilean shift symmetry softly broken, the renormalization process does not significantly contribute to the kinetic component or the linear potential term, $V(\phi)=V_0-\lambda^3\phi$, with $c_1=\lambda^3$ for the Galileon.  As a result, the current theoretical framework produces no unexpected results.  For the remaining part of the analysis, we leverage the fact that the coupling with the gravitational sector is strictly bound to be suppressed by the $\Lambda/M_{pl}$ contribution, owing to the soft breaking of the corresponding symmetry. This helps treat the underlying theory in the realistic theoretical regimes.

A more enhanced version of GEFT was developed in the curved space backdrop in ref \cite{Deffayet:2009wt}, taking into account the ghost-free theory from Ostrogradski instability at least in the classical regime. The term Covariantized Galileon Effective Field Theory (CGEFT) is often used to describe this form.  Such a CGEFT may be readily created by starting with a five-dimensional covering theory in a curved backdrop. The relevant action of the theory is provided by:
		\bea \label{CovGal}
S= \int d^4 x \sqrt{-g}\Bigg[\frac{M^2_{pl}}{2}R-V_0+ {\cal L}^{\bf C}_{\phi}\Bigg]\quad\quad\quad{\rm with}\quad\quad\quad {\cal L}^{\bf C}_{\phi}=\sum^{5}_{i=1}c_i\mathcal{L}^{\bf C}_i,
\eea
The following equations provide the explicit expression for the $\mathcal{L}^{\bf C}_i\forall i=1,2,\cdots, 5$:
\bea
{\cal L}^{\bf C}_1 & = & \phi, \\
 {\cal L}^{\bf C}_2&=&-\frac{1}{2} (\grad \phi)^2 ,\\
	{\cal L}^{\bf C}_3&=&\frac{c_3}{\Lambda^3} (\grad \phi)^2 \Box \phi ,\\
	{\cal L}^{\bf C}_4&=& -\frac{c_4}{\Lambda^6} (\grad \phi)^2 \Big\{
					(\Box \phi)^2 - (\grad_\mu \grad_\nu \phi)
					(\grad^\mu \grad^\nu \phi)
					- \frac{1}{4} R (\grad \phi)^2
				\Big\},\\
	{\cal L}^{\bf C}_5&=& \frac{c_5}{\Lambda^9} (\grad \phi)^2 \Big\{
					(\Box \phi)^3 - 3 (\Box \phi)( \grad_\mu \grad_\nu \phi)
					(\grad^\mu	 \grad^\nu \phi)
					\nonumber\\
					&&\quad\quad\quad\quad\quad+ 2 ( \grad_\mu  \grad_\nu \phi)
					(\grad^\nu	 \grad^\alpha \phi)
					(\grad_\alpha \grad^\mu \phi)
					- 6 G_{\mu \nu} \grad^\mu \grad^\alpha \phi
					\grad^\nu \phi \grad_\alpha \phi
				\Big\}.		
	\eea
 The Einstein tensor for the background gravity is represented by $G_{\mu\nu}$ and the Ricci scalar by $R$.  The Galilean symmetry is weakly broken in the current setting by the CGEFT with a curved gravitational backdrop, which is an interesting observation. The coefficients $c_i$ have the same function as previously described prior to covariantization in the Lagrangians discussed above. The construction was done in such a way that these coefficients always appear in a dimensionless manner, which is the only other point to make. Furthermore, $\Lambda$ denotes the fundamental mass scale of CGEFT, which may be seen as the EFT cutoff scale. EFT should not be valid over the aforementioned cutoff scale, according to conventional wisdom. It is crucial to note, nevertheless, that quantum fluctuations may exceed the cutoff scale $\Lambda$ if the Vainshtein effect is active \cite{deRham:2010eu}. We have included the no-minimal coupling with $G_{\mu\nu}$ and $R$ in the covariantized Lagragians ${\cal L}^{\bf C}_4$ and ${\cal L}^{\bf C}_5$, which are suppressed by powers of $H/\Lambda$ in the corresponding contributions. Despite the fact that these components will prove to be negligible in the inflationary regime of interest, where nonlinearities are dominated by Galileon self-interactions, we retain the nonminmal curvature couplings required for covariantization for being comprehensive.
    It may be seen with great attention that the covariantized form of the DGP model can be obtained by fixing the coefficients, $c_4 = 0 = c_5 $. But if we suppose that the Galileon field $\phi$ is relevant only during inflation, then the coefficients $c_i$ are not fixed and are determined independently from cosmological measurements. Several cosmological bounds on $c_2$, $c_3$, and $c_4$ are investigated in ref. \cite{Ali:2010gr}.

\subsection{Covariantized Galileon Effective Field Theory (CGEFT) inflation}

\subsubsection{Covariantized Galileon in de Sitter Background
}
This brings us to the inflationary solution in quasi de Sitter background geometry in the decoupling limit discussed before. Typically, this limit holds true in the current situation when the extra restriction $|\Delta V/V|\ll 1$ is satisfied by the gradual variation in the potential $\Delta V$ in the inflationary effective potential over the inflationary period. Where the Hubble parameter $H$ is not exactly constant and the first slow-roll parameter $\epsilon=-\dot{H}/H^2$ characterizes the small deviation from the exact de Sitter solution, the CGEFT embedded in the quasi de Sitter background in this decoupling limitnig situation is described by the scale factor, $a(t)=\exp(Ht)$.

Now, let us focus just on the Galileon portion of the action. By executing integration by parts and eliminating the boundary terms in the process, we obtain the subsequent action for the background time-dependent homogeneous Galileon field $\bar{\phi}(t)$. This action may be expressed as follows:
\bea S_0=\int d^4x\,a^3 \,\Bigg\{\frac{c_2}{2}\dot{\bar{\phi}}^2_0+\frac{2c_3H}{\Lambda^3}\dot{\bar{\phi}}^3_0+\frac{9c_4H^2}{2\Lambda^6}\dot{\bar{\phi}}^4_0+\frac{6c_5H^3}{\Lambda^9}\dot{\bar{\phi}}^5_0+\lambda^3\bar{\phi}_0\Bigg\},\eea
This follows the definition of the subsequent new coupling constant:
\bea \label{Z} Z\equiv \frac{H\dot{\bar{\phi}}_0}{\Lambda^3},\eea
may be further reformulated into the simplest form shown below:
\bea \label{bcaction} S_0=\int d^4x\,a^3 \,\Bigg\{\dot{\bar{\phi}}^2_0\Bigg(\frac{c_2}{2}+2c_3Z+\frac{9c_4}{2}Z^2+6c_5Z^3\Bigg)+\lambda^3\bar{\phi}_0\Bigg\}.\eea
This leads to the subsequent resolution:
\bea  \label{bcgsoln}
\dot{\bar{\phi}}_0=\frac{\Lambda^3}{12H}\frac{c_2}{c_3}\Bigg[-1+\sqrt{1+\frac{8c_3}{c^2_2}\frac{\lambda^3}{\Lambda^3}}\Bigg]=
\left\{
	\begin{array}{ll}
		\displaystyle \frac{\lambda^3}{3c_2H}\quad\quad\quad & \mbox{when}\quad  Z\ll 1  \;(\rm Weakly-coupled \;solution)  \\ \\
			\displaystyle 
			\displaystyle \sqrt{\frac{\Lambda^3}{18c_3}\frac{\lambda^3}{H^2}}\quad\quad\quad & \mbox{when }  Z\gg 1  \;(\rm Strongly-coupled \;solution)
	\end{array}
\right. \eea
Under weak coupling conditions ($Z\ll 1$), the underlying theory approximates the standard canonical slow-roll inflation. The theory, however, approaches the DGP model in the strong coupling regime ($Z\gg 1$). In order to support inflation, the Galileon interactions become important when $Z\gtrsim 1$, which occurs when the underlying theory interpolates between the weak and strong coupling regime. This is because, in this design in the decoupling limit, the coupling parameter $Z$ has positive powers, which regulate the relative contributions to the lower derivative components with respect to the higher derivative. The most plausible explanation for this observation is that non-minimal coupling with gravity becomes less significant as we approach the decoupling limit because there is no interaction with the gravitational sector. However, we still need to consider non-linear interactions because the Galileon sector contains a variety of derive terms. The mixing contribution from the non-minimal gravitational interactions with Galileon, which is expected to lead to major modifications in the properties of classical single-field inflation, cannot be neglected in the weakly coupled regime ($Z\ll 1$). This circumstance is unimportant to us since, in this study, we only focus on the intermediate regime, when the coupling parameter $Z\gtrsim 1$.

\subsubsection{Underlying connection with the good-old Effective Field Theory of inflation}

\subsubsubsection{Effective action in the Unitary gauge}

It is explicitly demonstrated by the authors in ref. \cite{Cheung:2007st} that the fundamental characteristics of the Effective Field Theory (EFT) setup can correct the structure and perturbation behaviour for an inflationary paradigm in the presence of a quasi-de Sitter background geometrical construction; thus, it can be interpreted in a way that is entirely independent of the model. It is readily transferred to the huge classes of $P(X,\phi)$ theories that are examined in refs \cite{Alishahiha:2004eh,Mazumdar:2001mm,Choudhury:2002xu,Panda:2005sg,Chingangbam:2004ng,Armendariz-Picon:1999hyi,Garriga:1999vw,Choudhury:2017glj,Naskar:2017ekm,Choudhury:2015pqa,Choudhury:2014sua,Choudhury:2014kma,Choudhury:2013iaa,Baumann:2022mni,Baumann:2018muz,Baumann:2015nta,Baumann:2014nda,Baumann:2009ds}. By applying broken time diffeomorphism symmetry and non-linear realization under Lorentz invariance in this setting, a significant theoretical construction is created. Building the classic EFT generalized form of the CGEFT inflationary action that is applicable to minor quantum fluctuations is our main goal in this section. But since this extension involves an extra Galilean symmetry, it is crucial to understand the changes and accompanying limitations that must be carefully considered. 

The EFT setup was created by the authors in ref. \cite{Cheung:2007st} using a single scalar field inflationary paradigm. Subsequently, the effective action is expressed in a particular gravitational gauge in which uniform $\phi$ slices overlap constant time slices. Remarkably, however, no particular form of the effective potential and the kinetic interactions in terms of the scalar field $\phi$ have been employed in ref. \cite{Cheung:2007st}. However, as has recently been demonstrated in refs. \cite{Choudhury:2017glj,Choudhury:2023jlt,Choudhury:2023rks}, the underlying theoretical construction carried out in ref. \cite{Cheung:2007st} precisely mimics a generic $P(X,\phi)$ kind of theories.In the current situation, the generalized version of the EFT action may be built and characterized by the representative action that has the following form:
\bea
		S &=&
		\int \d^4x \; \sqrt{-g} \;
		\Bigg[ \
			\frac{M^2_{pl}}{2} R
			- c(t) g^{00}
			- \Lambda(t)
			+ \frac{1}{2} M^4_2(t) (g^{00}+1)^2
			+ \frac{1}{3} M^4_3(t) (g^{00}+1)^3
		\nonumber\\ && \mbox{}
		\quad\quad\quad\quad\quad\quad- \frac{\bar{M}^3_1(t)}{2} (g^{00}+1) \delta {K^\mu}_{\mu}
		- \frac{\bar{M}^2_2(t)}{2} ({\delta K^\mu}_\mu)^2
		- \frac{\bar{M}^2_3(t)}{2} \delta K^{\mu\nu} \delta K_{\mu\nu}
		\nonumber\\ && \mbox{}\quad\quad\quad\quad\quad\quad
		- \frac{\bar{M}^3_4(t)}{2} (g^{00}+1)^2 \delta {K^\mu}_{\mu}
		- \frac{\bar{M}^2_5(t)}{2} (g^{00}+1) ({\delta K^\mu}_\mu)^2
		- \frac{\bar{M}^2_6(t)}{2} (g^{00}+1)
			\delta K^{\mu\nu} \delta K_{\mu\nu}
		\nonumber\\ && \mbox{}\quad\quad\quad\quad\quad\quad
		- \frac{\bar{M}_7(t)}{2} ({\delta K^\mu}_\mu)^3
		- \frac{\bar{M}_8(t)}{2} ({\delta K^\mu}_\mu)
			(\delta K^{\rho \sigma} \delta K_{\rho\sigma})
		- \frac{\bar{M}_9(t)}{2} \delta K^{\mu \nu} \delta K_{\nu \sigma}
			{\delta K^\sigma}_\mu
		+\cdots \Bigg],
	\eea
It is noteworthy to mention that the time-dependent coefficients, $c(t)$ and $\Lambda(t)$, exhibit precise non-zero values on the quasi de Sitter cosmological background that we are examining for our current analysis. This, in turn, fixes these time-dependent coefficients in relation to the Hubble parameter $H(t)$ during the inflationary episode. Conversely, the background cosmic evolution fails to stabilize the remaining time-dependent coefficients $M_i(t)\forall i=2,3$ and $\bar{M}_i(t)\forall i=1,2,\cdots,9$ and captures the data pertaining to the various theories that are analysed. Accordingly, one should select these parameters with extreme precision for the CGEFT setup. It is predicted that in the current context of debate, CGEFT will exhibit clearly distinguishing qualities. 

It is now necessary to recast the previously discussed action in a more manageable form in order to apply it and express the cosmic perturbation theory order by order for scalar modes formed from perturbation. This may be accomplished by applying the unitary gauge transformation that follows, whereby broken time diffeomorphism is very clearly realized automatically:
\bea t\longrightarrow \tilde{t}=t-\pi(t,{\bf x}).\eea
The equal time hypersurfaces are immediately deformed by the amount $\pi(t,{\bf x})$ as a result of the gauge transformation described above; the deformation parameter is a quantity that depends on space and time. The Goldstone mode is recognized as this particular parameter, and the equivalent method used in this context is called the St$\ddot{u}$ckelberg trick. A common link with the $SU(N)$ gauge theory may be found here. Each operator specified in the previously described extended EFT action changes under the aforementioned gauge transformation, and these transformations comprise a sequence of terms that represent the space-time mixing contributions and spatial and temporal derivatives of the Goldstone modes $\pi(t,{\bf x})$.
See the references for further information. \cite{Choudhury:2017glj,Choudhury:2023rks,Choudhury:2021brg}, where each of these changes is set out in great detail. Using the tools and techniques of the EFT approach, a consistent and beautiful theoretical description of the underlying physical setup is constructed in this way by utilizing the lowest dimensions Wilsonian operators that are compatible with the underlying symmetries in this particular situation.

\subsubsubsection{Decoupling limiting situation}
Within the framework of cosmic perturbation theory, the application of EFT tools and procedures becomes exceedingly challenging without the decoupling limit, which we address in this part. These issues result from the Goldstone modes' mixing contribution with the gravitational sector via the metric. By ignoring the background gravity, the decoupling limit's implementation now treats the perturbation theory more reliably in the presence of significant non-linear contributions from the self-interacting components of the Goldstone modes. This limit can be reached by taking $M_{pl}\rightarrow\infty$ while maintaining a stable Hubble parameter $H$ in the quasi de Sitter cosmic background. In this instance, we provide a brief summary of the main contention.
For the measure variation $\delta g^{00}$, the Ricci scalar will yield the most relevant kinetic term. Consequently, via rescaling the fluctuation:
\bea \delta g^{00} \rightarrow \delta g^{00}_c = M_{pl} \delta g^{00},\eea 
Canonical normalization may now be implemented. The most important kinetic component for the Goldstone modes is produced by the nonlinear operator in the EFT action. It will have the form $M^4 \dot{\pi}^2$ if it has minimum derivatives, where $M$ is a combination of either $\bar{M}_i$ or $M_i$, the coupling parameters. Now $\pi_c=M^2\pi$ further provides the canonically normalized Goldstone modes in this context. For wave numbers $k$ that satisfy the subsequent condition, $k \gtrsim E_{\mathrm{mix}} = \frac{M^2}{M_{pl}}$, unlike the Goldstone kinetic term, a mixing term like $M^4 \dot{\pi} \delta g^{00}$ is negligible at the quadratic level. The situation is similar for cubic terms, where $M^4 \dot{\pi}^2 \delta g^{00}$ is a less important leading blending term than $M^4 \dot{\pi}^3$ in the identical situation. If the most important kinetic component for the Goldstone mode has larger derivatives, or if the prime cubic contributions occur with a physical scale other than $M$, similar arguments can be made. In order for the matching mixing scale $E_{\mathrm{mix}}\ll H$ in the decoupling limit and our projections to be accurate to a relative error of order $E_{\mathrm{mix}}/ H$. In the limit where we work in the ensuing parts, we can presume that the measure remains unchanged. Working in the uniform curvature gauge is the most convenient choice in this situation, as the unperturbed measure is spatially smooth and may be utilized as the background of quasi de Sitter geometry.

Keeping our attention on the decoupling limit, the following equation of motion for the Galileon scalar field in background de Sitter space may be obtained by using the present conservation equation since this situation involves an exact shift symmetry:
\bea \label{currentconserve}
\nabla_{\mu}J^{\mu} = \dot{J}^{t} + 3H\left(J^{t} - \frac{\lambda^{3}}{3H}\right) = 0.
\eea
The temporal component of the Noether current is denoted by $J^{t}$. The identification that follows may be made by comparing the aforementioned equation to the continuity equation discovered for the case of a perfect fluid:
\bea
\rho=J^{t},\quad\quad P=\frac{-\lambda^{3}}{3H}
\eea
where $P$ represents pressure, and $\rho$ represents energy density. The equation of state (EoS) parameter $w$, which is specified as $P=w\rho$, will also be realized with the help of these. Equation (\ref{bcaction}) provides the equation of motion as arising from the background action, which we utilize to include the remaining Galileon EFT coefficients into finding $w$. The following equation \cite{Burrage:2010cu} may be obtained from a modification of the action, $\delta S^{(0)}/\delta\bar{\phi} = 0$:
\bea
\dot{\bar{\phi}}(c_{2}+6c_{3}Z+18c_{4}Z^{2}+30c_{5}Z^{3}) = \frac{\lambda^{3}}{3H},
\eea
Whereas $J^{t}=\lambda^{3}/3H$ is defined as the current $J^{t}$ (or energy density $\rho$) after being determined as a solution of eqn. (\ref{currentconserve}) with additional coefficients. Using the information provided, we can ultimately formulate the equation of state in the Galileon EFT framework as follows:
\bea \label{eosGal}
w = \frac{P}{\rho} = \frac{-\lambda^{3}/3H}{\dot{\bar{\phi}}(c_{2}+6c_{3}Z+18c_{4}Z^{2}+30c_{5}Z^{3}) }.
\eea
This gives us a method to examine the range of values that the Galileon EFT coefficients can have in order to define a certain EoS. In the next conversations, we will endeavor to pinpoint potential ranges that the EFT coefficients $c_{i},\;\forall\;i=1,\cdots,5$ may assume, considering the acute transition framework established for this study. We will go into great depth on how to effectively create an SR/USR/SR-like architecture in Galileon in the next part.

\subsection{Implementation of sharp transition using Galileon EFT}

We examine in this section how the underlying Galileon theory constructed the sharp transition characteristic. We have a configuration in which there is a strong transition between the USR phase and the first slow-roll (SRI) phase, as well as between the USR phase and the second slow-roll phase (SRII). We address the development of this setup explicitly, as it will allow us to precisely examine the formation of PBHs due to the huge improvements in the primordial fluctuations provided by the abrupt transitions and the USR phase. The following slow-roll metrics, which indicate variations from precise de sitter during inflation, are defined as:
\bea \label{srparams}
\epsilon = -\frac{\dot{H}}{H^{2}} = -\frac{d\ln{H}}{d{\cal N}}, \quad\quad \eta \equiv -\frac{\ddot{\phi}}{\dot{\phi}H} = \epsilon - \frac{1}{2}\frac{d\ln{\epsilon}}{d{\cal N}}
\eea
where the number of e-folding is indicated by ${\cal N}$, and the first and second slow-roll parameters are denoted by $\epsilon$ and $\eta$. 
Keep in mind that the Hubble rate $H$ includes details about the corrections resulting from the linear components in the scalar field $\phi$ as well as the non-zero constant term $V_{0}$. The potential with the form $V = V_{0} - \lambda^{3}\phi$ is implied by the slight symmetry breakdown. Using the Friedman equations, we now have:
\bea
H = \frac{1}{M_{pl}}\sqrt{\frac{V}{3}} &=& \frac{1}{M_{pl}}\sqrt{\frac{V_{0}-\lambda^{3}\phi}{3}} \nonumber\\
&=& \frac{1}{M_{pl}}\bigg[\sqrt{\frac{V_{0}}{3}}\bigg(1-\frac{\lambda^{3}}{V_{0}}\phi\bigg)^{1/2}\bigg] \nonumber\\
&=& \frac{1}{M_{pl}}\sqrt{\frac{V_{0}}{3}}\bigg(1-\frac{\lambda^{3}}{2V_{0}}\phi - \frac{\lambda^{6}}{8V_{0}^{2}}\phi^{2} + \cdots\bigg).
\eea
The Hubble parameter $H$ in the quasi-de Sitter spacetime is not a constant; instead, it is expressed using the precise de Sitter Hubble parameter, $H_{\rm dS} = \sqrt{V_{0}/3M_{pl}^{2}}$, as previously mentioned. $V_{0} > 0$ is satisfied by the constant term, and the last line's above expansion is justified by the requirement $\lambda^{3}/V_{0} \ll 1$. The corrections that follow the leading de Sitter contribution $H_{\rm dS}$ are represented by the terms that follow, and this is mirrored in the slow-roll paradigm where the SR parameters are calculated with the aforementioned expansion taken into account.

These slow-roll parameters exhibit varying behaviors in each phase and are related to the background scalar field's smooth solution in eqn.(\ref{bcgsoln}), which involves the coefficients $c_{1},c_{2}$, and $c_{3}$. In this section, we shall propose potential values for these coefficients. Further information on the second-order action for the scalar perturbations is needed in order to restrict the two remaining coefficients, $c_{4},c_{5}$, which we shall discuss later. We conclude by introducing yet another crucial element to take into account: the effective sound speed $c_{s}$. To define this, the following time-dependent coefficients are needed:
\bea \label{coeffA}
{\cal A}&\equiv& \frac{\dot{\bar{\phi}}^2}{2}\Bigg(c_2+12c_3Z+54c_4Z^2+120c_5Z^3\Bigg),\\
\label{coeffB}  {\cal B}&\equiv& 
   \frac{\dot{\bar{\phi}}^2}{2}\Bigg\{c_2+4c_{3}Z\Big(2-\eta\Big)+2c_{4}Z^{2}\Big(13-6\big(\epsilon+2\eta\big)\Big)-24c_5Z^{3}\big(2\epsilon+1\big)\Bigg\}.
\eea
According to the refs.\cite{Choudhury:2023hvf,Choudhury:2023kdb} defines the effective sound speed as follows:
\bea \label{soundspeed}
c_{s}=\sqrt{\frac{\cal B}{\cal A}}.
\eea
where the constant $Z$ is specified in eqn. (\ref{Z}), and $\dot{\bar{\phi}}$ is defined before in eqn. (\ref{bcgsoln}). Using the generic Goldstone EFT framework, we can now discuss the effective sound speed \cite{Cheung:2007st,Choudhury:2017glj,Choudhury:2023hvf,Choudhury:2023jlt,Choudhury:2023rks}:
\bea \label{speedEFT}
c_{s} = \frac{1}{\sqrt{1-\displaystyle{\frac{2M_{2}^{4}}{\dot{H}M_{pl}^{2}}}}},
\eea
The direct relationship between the $c_{i}'$s utilized for ${\cal A}$ and ${\cal B}$ and the EFT coefficient $M_{2}$ is evident upon comparison with eqn. (\ref{soundspeed}):
\bea
\frac{M_{2}^{2}}{M_{pl}} = \sqrt{\frac{\dot{H}}{2}\bigg(1-\frac{{\cal A}}{{\cal B}}\bigg)}
\eea
A later explicit emergence of the sound speed $c_{s}$ has an impact on the tree and one-loop corrected scalar power spectrum calculations. Explicitly present in $c_{s}$ are the Galileon EFT coefficients $c_{i},\;\forall i=1,\cdots,5,$. Therefore, we indirectly incorporate the contributions of the coefficients $c_{i}$ after we set restrictions on $c_{s}$. Nonetheless, we must keep in mind that each of the three stages involved requires the application of limitations. We are able to confine the values of $c_{1},c_{2},c_{3}$ for each phase by requiring the corresponding slow-roll requirements. Therefore, three sets of parameter values are provided by the slow-roll and ultra-slow-roll conditions in the configuration of SRI, USR, and SRII combined. The unitarity and causality criteria on $c_{s}$, along with the knowledge of the necessary tree-level scalar power spectrum amplitude in the three phases, may be used to set the values of the remaining two Galileon EFT coefficients, $c_{4},c_{5}$. The parameter $c_{s}$ appears to have current constraints within $0.024 \leq c_{s} < 1$ \cite{Planck:2015sxf}, given the criteria indicated. The tree-level power spectrum, represented here as $\Delta^{2}_{\zeta,{\bf Tree}}(k)$, has the following values: $\big[\Delta^{2}_{\zeta,{\bf Tree}}\big (k)]_{\rm SRI} \sim {\cal O}(10^{-9})$ in SRI where the CMB sensitive scales appear; $\big[\Delta^{2}_{\zeta,{\bf Tree}}\big (k)]_{\rm USR} \sim {\cal O}(10^{-2})$ in the USR; and $\big[\Delta^{2}_{\zeta,{\bf Tree}}\big (k)]_{\rm SRII} \sim {\cal O}(10^{-5})$ in the SRII.  These kinds of requirements allow us to identify ranges of constraints on the $c_{4},c_{5}$ value parameter space. While $\epsilon$ and $\eta$ are ultimately unaffected by these two values, we may successfully cover the whole set of coefficients $c_{i}$ by utilizing $c_{s}$ and $\Delta^{2}_{\zeta,{\bf Tree}}(k)$ in further discussions of this review.

\subsubsection{Region I: First Slow Roll (SRI) region}

With the aid of the SRI phase, inflation is started when a scalar field gradually reduces the inflationary potential. This slow-roll behaviour is indicated by the two parameters $\epsilon$ and $\eta$, which are previously described in eqn. (\ref{srparams}). The early e-folds investigated close to the pivot scale $k_{*}=0.02{\rm Mpc}^{-1}$, where we achieve the SRI requirements, are matched by the CMB-scale oscillations. We consider $\epsilon$ to be modest and $\eta \simeq 0$ to be a tiny, nearly constant number during SRI. One may write the following for the behaviour of $\epsilon$ based on eqn.(\ref{srparams}):
\bea
\int\frac{d\epsilon}{\epsilon(\epsilon - \eta)} = \int 2d{\cal N}\implies
\ln{\bigg(1 - \frac{\eta}{\epsilon}\bigg)}\Biggr|_{\epsilon_i}^{\epsilon_f} = 2\eta({\cal N}_{f}-{\cal N}_{i}). 
\eea
Whereas the instant of time indicated by e-folds ${\cal N}_{i} = {\cal N}_{*}$ corresponds to the fixed initial condition on $\epsilon_{i}$ for the SRI phase, which is set at the CMB-scale approaching the Horizon with $k_{*}$. Here is the outcome of the expression for $\epsilon$:
\bea \label{epsN1}
\epsilon_{\rm SRI}({\cal N}) = \eta_{\rm SRI}\bigg(1-\bigg(1-\frac{\eta_{\rm SRI}}{\epsilon_{i,\rm SRI}}\bigg)e^{2\eta_{\rm SRI}\Delta{\cal N}}\bigg)^{-1}, \eea
utilising the formula $\Delta{\cal N} = {\cal N}_{f}-{\cal N}_{*}$. For the sake of numerical simplification, we pick $\eta_{\rm SRI} \sim -0.001$ and $\epsilon_{i,\rm SRI} \sim {\cal O}(10^{-3})$. With these, we obtain their respective behaviour in SRI. As anticipated, their behaviour changes very slowly as the number of e-foldings increases. This behaviour will be used with the eqs. (\ref{bcgsoln},\ref{srparams}) to restrict the values of $c_{1},c_{2},$ and $c_{3}$, in SRI. To begin with, we write eqn.(\ref{bcgsoln}) as follows for this purpose:
\bea \label{phiftn}
\dot{\bar{\phi}} = \frac{f(c_{1},c_{2},c_{3},\Lambda)}{H},\quad\quad\quad {\rm where}\quad\quad f(c_{1},c_{2},c_{3},\Lambda) \equiv \frac{\Lambda^3}{12}\frac{c_2}{c_3}\Bigg[-1+\sqrt{1+\frac{8c_3}{c^2_2}\frac{c_{1}}{\Lambda^3}}\Bigg]. 
\eea
\begin{figure*}[htb!]
    	\centering
    \subfigure[]{
      	\includegraphics[width=8.5cm,height=7.5cm]{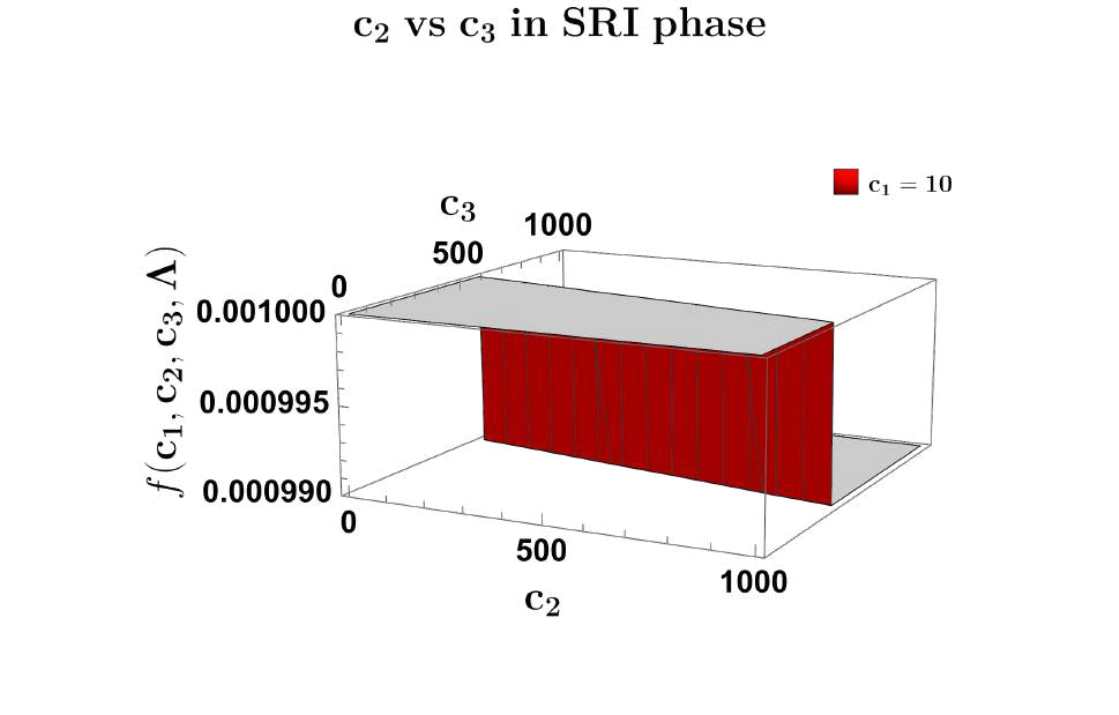}
        \label{sr1c1}
    }
    \subfigure[]{
       \includegraphics[width=8.5cm,height=7.5cm]{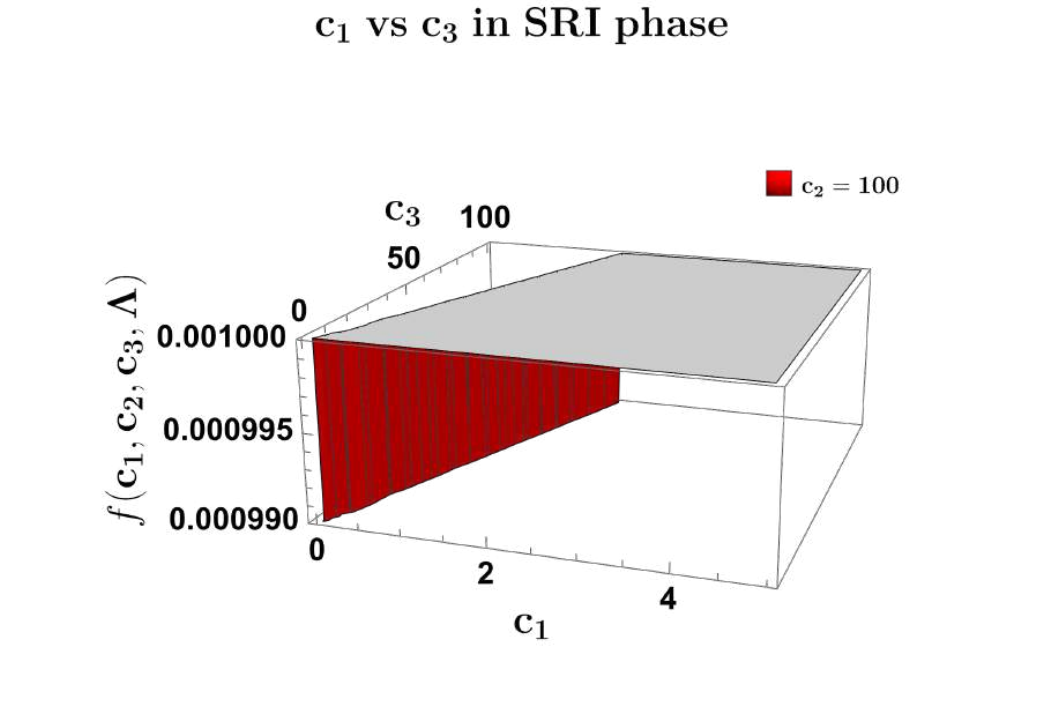}
        \label{sr1c2}
       }\\
   \subfigure[]{
       \includegraphics[width=8.5cm,height=7.5cm]{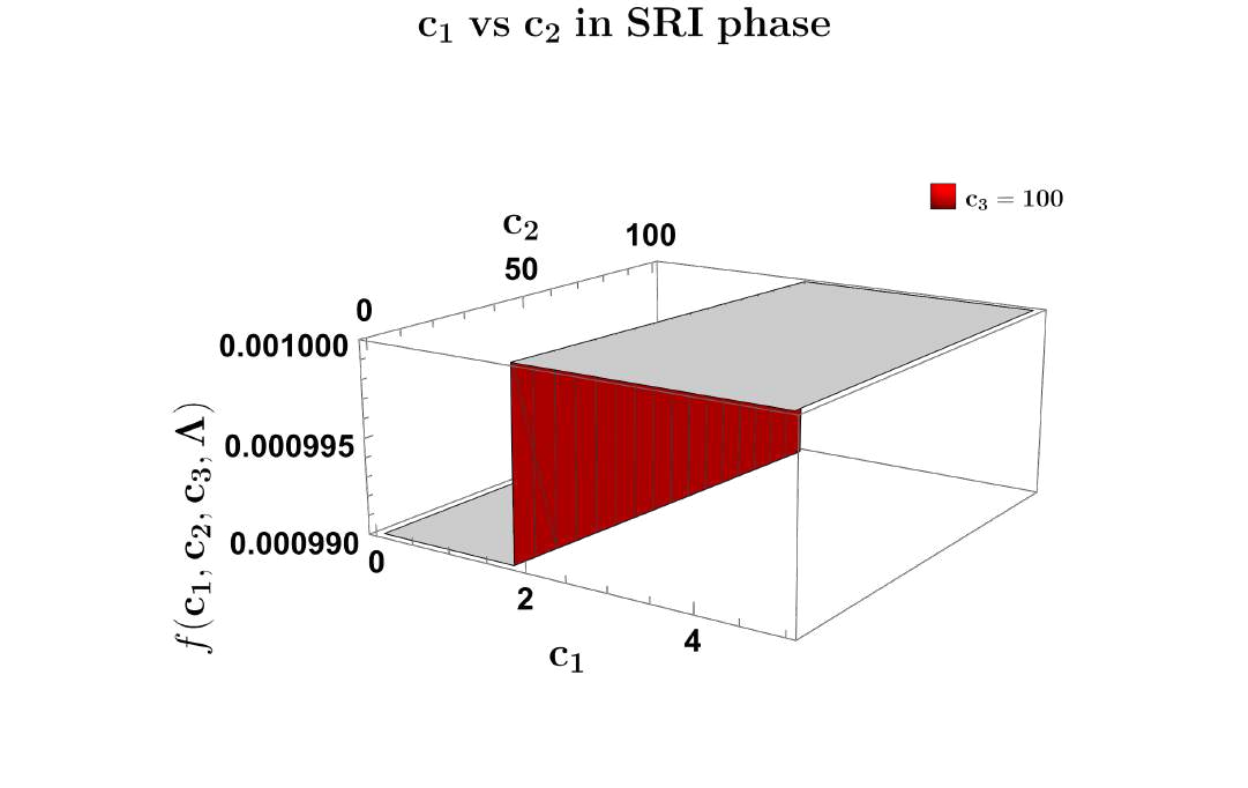}
        \label{sr1c3}
       }\hfill
    
    	\caption[Optional caption for list of figures]{How the three Galileon EFT coefficients, $c_{1},c_{2},c_{3}$, behave in the SRI phase to meet $f\equiv f(c_{1},c_{2},c_{3},\Lambda)$. Since $c_{1}=10$ is fixed in (a), $c_{3}\sim {\cal O}(10^{3})$ is produced, but $c_{2}$ is less susceptible to change in order to achieve $f$. In order to have $c_{1}\sim {\cal O}(10)$, $c_{2}=100$ is fixed in (b), confirming $c_{3}\sim {\cal O}(10^{3})$. In (c), $c_{3}=100$ is constant and less sensitive to $c_{2}$; yet, $f$ is satisfied by $c_{1}\sim {\cal O}(10)$. 
 } 
    	\label{sr1}
    \end{figure*}
For a proper EFT description, this also contains the underlying mass scale $\Lambda$. We settle on $\Lambda \sim {\cal O}(10^{-3}-10^{-1})M_{p}$ for computations in the future. Now, we can get the following relations by combining the above with eqn(\ref{srparams}):
\bea
\eta = -\frac{\ddot{\phi}}{\dot{\phi}H} = -\frac{1}{\dot{\phi}}\frac{d}{d{\cal N}}\bigg(\frac{f}{H}\bigg) = -\frac{1}{f}\bigg(\frac{df}{d{\cal N}} - \frac{f}{H}\frac{dH}{d{\cal N}}\bigg),
\eea
where we have converted from time $t$ to e-folds ${\cal N}$ using the chain rule in the second equality and the fact $d{\cal N}=Hdt$. Due to the nearly constant value of $\eta$, we can solve the differential equation for the function $f$ and obtain the combined behaviour of the coefficients inside a certain interval of e-folds. The following outcome can be obtained by solving the aforementioned equation for SRI with suitable initial conditions:
\bea \label{ftnN1}
f_{\rm SRI}({\cal N}) = \frac{H_{\rm SRI}({\cal N})}{H({\cal N}_{*})}\exp{(-\eta_{\rm SRI}\Delta{\cal N})}. \eea
We note that dependency on the Hubble parameter with e-folds, $H({\cal N})$, is necessary to fully identify $f_{\rm SRI}({\cal N})$ from the preceding formula. The following may be done to do this: Plug the answer from eqn.(\ref{epsN1}) into the second definition of $\epsilon$ from eqn.(\ref{srparams}) to obtain:
\bea
-\frac{dH}{H} = \eta_{\rm SRI}\bigg(1-\bigg(1-\frac{\eta_{\rm SRI}}{\epsilon_{i, \rm SRI}}\bigg)e^{2\eta_{\rm SRI}\Delta{\cal N}}\bigg)^{-1}d{\cal N}, \eea
After we integrate both sides of the aforementioned equation, we get the following outcome:
\bea
\ln{(H)}\Biggr|_{H_{*}}^{H} = -\frac{1}{2}\ln{\bigg(\frac{\exp{(2\eta_{\rm SRI}{\cal N})}}{-\epsilon_{i,\rm SRI}+\exp{(2\eta_{\rm SRI}{\cal N})}(\epsilon_{i,\rm SRI}-\eta_{\rm SRI})}\bigg)}\Biggr|_{{\cal N}_{*}}^{{\cal N}}.
\eea
This simplifies to give us the following formula for the Hubble parameter in terms of the number of e-foldings ${\cal N}$ and other crucial parameters as follows after considering the appropriate integration limits:
\bea \label{HubbleN1}
\frac{H_{\rm SRI}({\cal N})}{H_{*}} = \bigg(\frac{\eta_{\rm SRI}\exp{(2\eta_{\rm SRI}{\cal N})}}{\epsilon_{i, \rm SRI}-\exp{(2\eta_{\rm SRI}{\cal N})}(\epsilon_{i, \rm SRI}-\eta_{\rm SRI})}\bigg)^{-\frac{1}{2}},
\eea
where we opt to maintain ${\cal N_{*}}=0$ for convenience and the phrase $\eta_{\rm SRI}$ appears in the inverse square root numerator after the starting condition is imposed at ${\cal N}={\cal N_{*}}$. The answer from eqn.(\ref{ftnN1}) contains information on the Galileon EFT coefficients, which we can use to determine a potential range of values for the three coefficients, $c_{1},c_{2}$, and $c_{3}$.

The examination of $c_{1},c_{2},c_{3}$ is what we will now start with. Remember that for $Z \gtrsim 1$, there exists a regime where the non-linearities of the Galileon remain significant but the mixing with gravity may be omitted. The function must fulfil $f(c_{1},c_{2},c_{3},\Lambda) \sim \Lambda^{3} \sim {\cal O}(10^{-9}-10^{-3})M_{p}^{3}$, as can be seen from eqs. (\ref{Z},\ref{phiftn}). This comparable pattern is also predicted by the conclusion from eqn.(\ref{ftnN1}) for $f(c_{1},c_{2},c_{3},\Lambda)$ up to the SRI phase functions. In order to canonically normalise the kinetic term in eqn.(\ref{CovGal}), we first fix $c_{2}=1$. For a given $c_{1}$, we obtain a wide range of $c_{3}$ values when $c_{2}$ is constant; however, if we take a specific $c_{3}$ into consideration, the reciprocal of this is not true with $c_{1}$. We may determine the potential ranges for the coefficients by using the previously stated permitted range for the cut-off scale $\Lambda$. Given $\Lambda \sim {\cal O}(10^{-1})M_{p}$ and $c_{1}=\lambda^{3}$, we may get a collection of points for $c_{3}$ inside $c_{3} \sim {\cal O}(10^{-2}-10^{2})$, where the term somewhat breaks Galilean symmetry. The permitted values of $c_{3}$ rise at a faster rate of magnitude for the specified ranges of $c_{1}$ values as we descend below for $\Lambda < {\cal O}(10^{-1})M_{p}$ in comparison to the ranges previously indicated. Whenever $c_{2}$ is varied while $c_{1}$ is fixed, the consequences of changing one of the two values are always found to be negligible. 

Changes in the coefficients $c_{1},c_{2},c_{3}$ enable us to fulfil the behaviour of $f(c_{1},c_{2},c_{3},\Lambda)$ derived from eqn. (\ref{ftnN1}), as seen in fig. (\ref{sr1}). The coefficients' permitted values within the range shown in the plots are indicated by the red surfaces. We conclude that values of $c_{1} \sim (10^{-2}-10^{2})$ enough to obtain the required $f$ based on an overall examination of the three plots included in the figure. Additionally, in the case of $c_{3} \sim {\cal O}(10^{-1}-10^{4})$, a decrease in $c_{3}$ necessitates even lower $c_{1}$ values. Lastly, adjustments to $c_{2}$ have the least impact and do not tighten restrictions on the values of $c_{1},c_{3}$. 

We utilise the effective sound speed $c_{s}$ and the amplitude of the scalar power spectrum $\big[\Delta^{2}_{\zeta,{\bf Tree}} (k)\big]_{\rm SRI} \sim {\cal O}(10^{-9})$ to get the permitted values of the remaining two coefficients, $c_{4}$ and $c_{5}$. Explicit computations in subsequent sections will clarify that the effective sound speed is contained in the analytic formula for the scalar power spectrum.  
We may set limits on the permitted values of the coefficients by using the $c_{4},c_{5}$ information found in both the scalar power spectrum and eqn.(\ref{soundspeed}). To get a range of values, we apply the observational constraint $0.024 \leq c_{s} < 1$.

Following the aforementioned process to produce the values, we see that both coefficients can have both positive and negative signatures and fall within the magnitude range of ${\cal O}(10^{-2})\leq (c_4,c_5)\leq {\cal O}(10^2)$. The numbers $(c_{4},c_{5})\simeq(-2.8,0.27)$ where $c_{2}=1$ and $c_{3}=10$ are determined for the special case of $c_{s}(\tau_{*})=c_{s,*}=0.05$, which we shall select during the computation of the scalar power spectrum and subsequent estimation of the mass of PBH and the GW spectrum.  

Drawing on the previous conversations regarding the range of possibilities for the different Galileon EFT coefficients in the SRI, we will look at some potential values for each of the $c_{i}$'s below. These values can enable the realisation of a certain EoS $w$, for which we provided a formula in eqn.(\ref{eosGal}) utilising these coefficients. 
\begin{table}[H]

\centering
\begin{tabular}{|c|c|c|c|c|c|c|}

\hline\hline
\multicolumn{7}{|c|}{\normalsize \textbf{Galileon EFT coefficients for a given equation of state $w$ in SRI}} \\

\hline

EoS $(w)$ & $f$ &\hspace {0.5cm}  $c_{1}$ \hspace {0.5cm} &\hspace {0.5cm}  $c_{2}$ \hspace {0.5cm} &\hspace {0.5cm}  $c_{3}$ \hspace {0.5cm} &\hspace {0.5cm}  $c_{4}$ \hspace {0.5cm} &\hspace {0.5cm}  $c_{5}$ \hspace {0.5cm}  \\
\hline
$1/3$ & & $0.1$ & $10$ & $50$ & $15$ & $-22.5$ \\ 
$0.25$ & ${\cal O}(10^{-3})$  & $0.5$ & $20$ & $50$ & $23$ & $-46.6$ \\
$0.16$ & & $0.6$ & $25$ & $50$ & $24$ & $-67$ \\ \hline 
\hline

\end{tabular}

\caption{ The table lists potential values that fulfil a certain EoS $w$ for the EFT coefficients, $c_{i}\;\forall\;i=1,\cdots,5$.
 }

\label{tab1eos}

\end{table}

The table \ref{tab1eos} illustrates the many values that the EFT coefficients can assume to achieve a certain EoS $w$ value while also falling within the required ranges to implement the USR phase.

\subsubsection{Region II: Ultra Slow Roll (USR) region}

In this part, we apply the same methodology as for the SRI phase replaced with the USR features to analyse the parameter space of the coefficients $c_{i},\;\forall\;i=1,\cdots,5$. The scalar field experiences an exceptionally flat inflationary potential in the USR regime, resulting in a dramatic increase in the magnitude of the primordial fluctuations produced both during and after the abrupt break from the early SR phase. In our scenario, the USR area is realised by the Galileon EFT coefficients as we are not utilising any model case for the inflationary potential.

During USR, the slow-roll approximation fails, and the SR parameters behave as follows: $\epsilon_{\rm USR} \propto a^{-6}$ is very tiny, whereas $\eta_{\rm USR} \simeq -6$ is very big. Consequently, we determine via eqn.(\ref{srparams}) that the associated $\epsilon_{\rm USR}$ deviates significantly from the preceding value of ${\cal O}(10^{-3})$. The initial conditions for $\epsilon$ are thus going to be $\epsilon_{i,{\rm USR}} = \epsilon_{\rm SRI}({\cal N}_{s})$, where the e-foldings denoting the start and finish of the SRI phase are indicated by ${\cal N}_{s}$. The USR continues to exist between ${\cal N}_{s}$ and ${\cal N}_{e}$ until abruptly shifting into a different SRII phase. Since this wavenumber position corresponding to this moment of time might aid in the generation of solar mass PBHs, we have chosen ${\cal N}_{s}\sim {\cal O}(18)$ for our computations. The relevance of this choice will be evident later on during the study of PBH mass. 

Here, the formula for the parameter $\epsilon$ is changed to:
\bea \label{epsN2}
\epsilon_{\rm USR}({\cal N}) = \eta_{\rm USR}\bigg(1-\bigg(1-\frac{\eta_{\rm USR}}{\epsilon_{i,\rm USR}}\bigg)e^{2\eta_{\rm USR}({\cal N}-18)}\bigg)^{-1}, \eea
When $\eta \simeq -6$, this can produce the sharply declining behaviour for $\epsilon$ in the USR phase. We obtain a new version of eqn.(\ref{HubbleN1}) by inserting the above into eqn.(\ref{srparams}):
\bea \label{HubbleN2}
\frac{H_{\rm USR}({\cal N})}{H_{\rm SRI}({\cal N}_{s})} = \bigg(\frac{\eta_{\rm USR}\exp{(2\eta_{\rm USR}\Delta{\cal N}_{\rm USR})}}{\epsilon_{i,\rm USR}-\exp{(2\eta_{\rm USR}\Delta{\cal N}_{\rm USR})}(\epsilon_{i,\rm USR}-\eta_{\rm USR})}\bigg)^{-\frac{1}{2}},
\eea
\begin{figure*}[htb!]
    	\centering
    \subfigure[]{
      	\includegraphics[width=8.5cm,height=7.5cm]{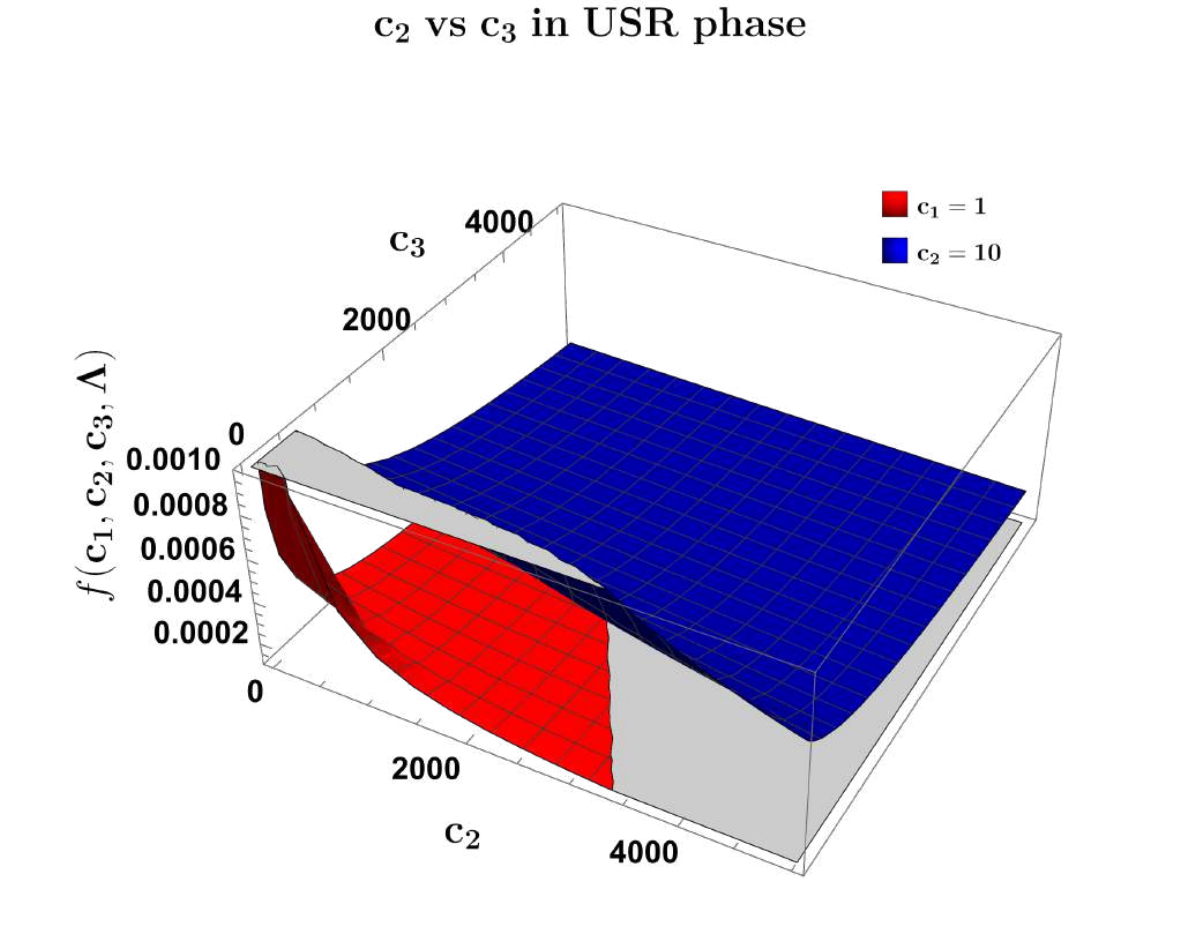}
        \label{usrc1}
    }
    \subfigure[]{
       \includegraphics[width=8.5cm,height=7.5cm]{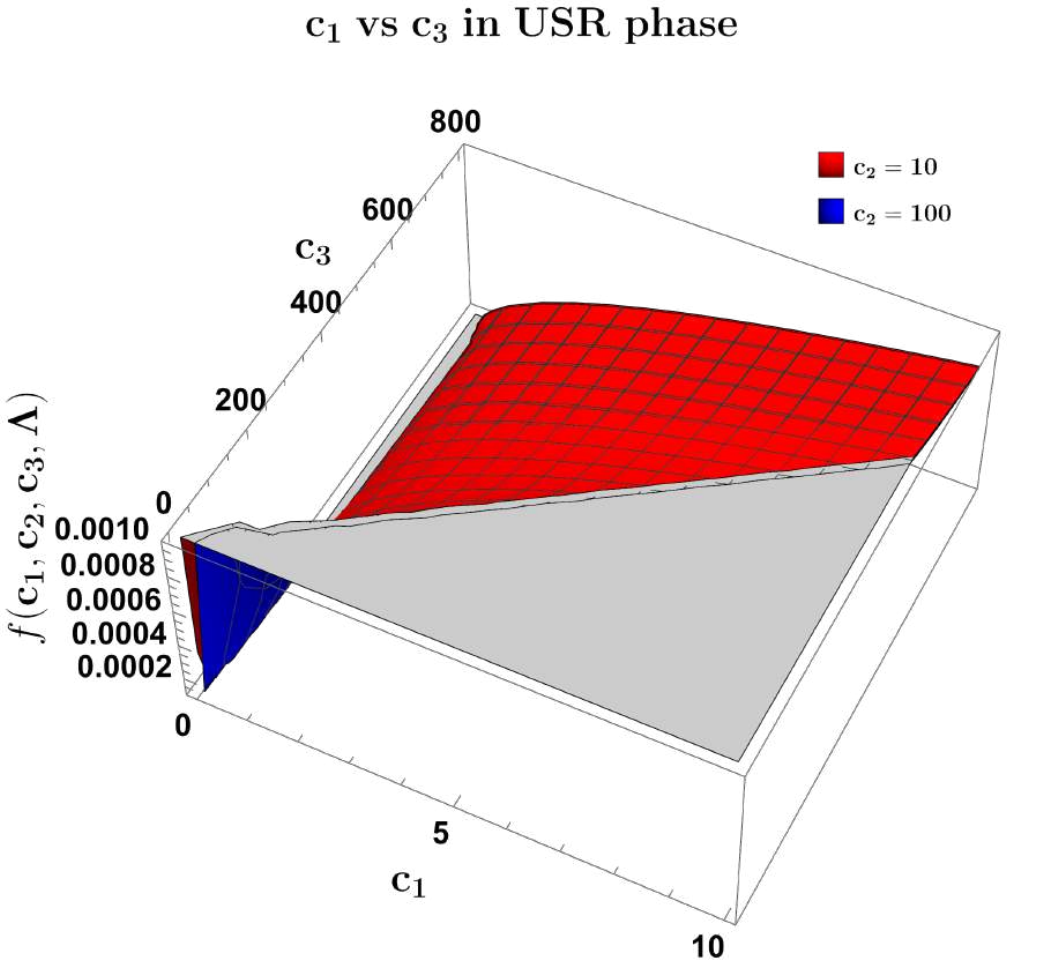}
        \label{usrc2}
       }\\
   \subfigure[]{
       \includegraphics[width=9cm,height=7.5cm]{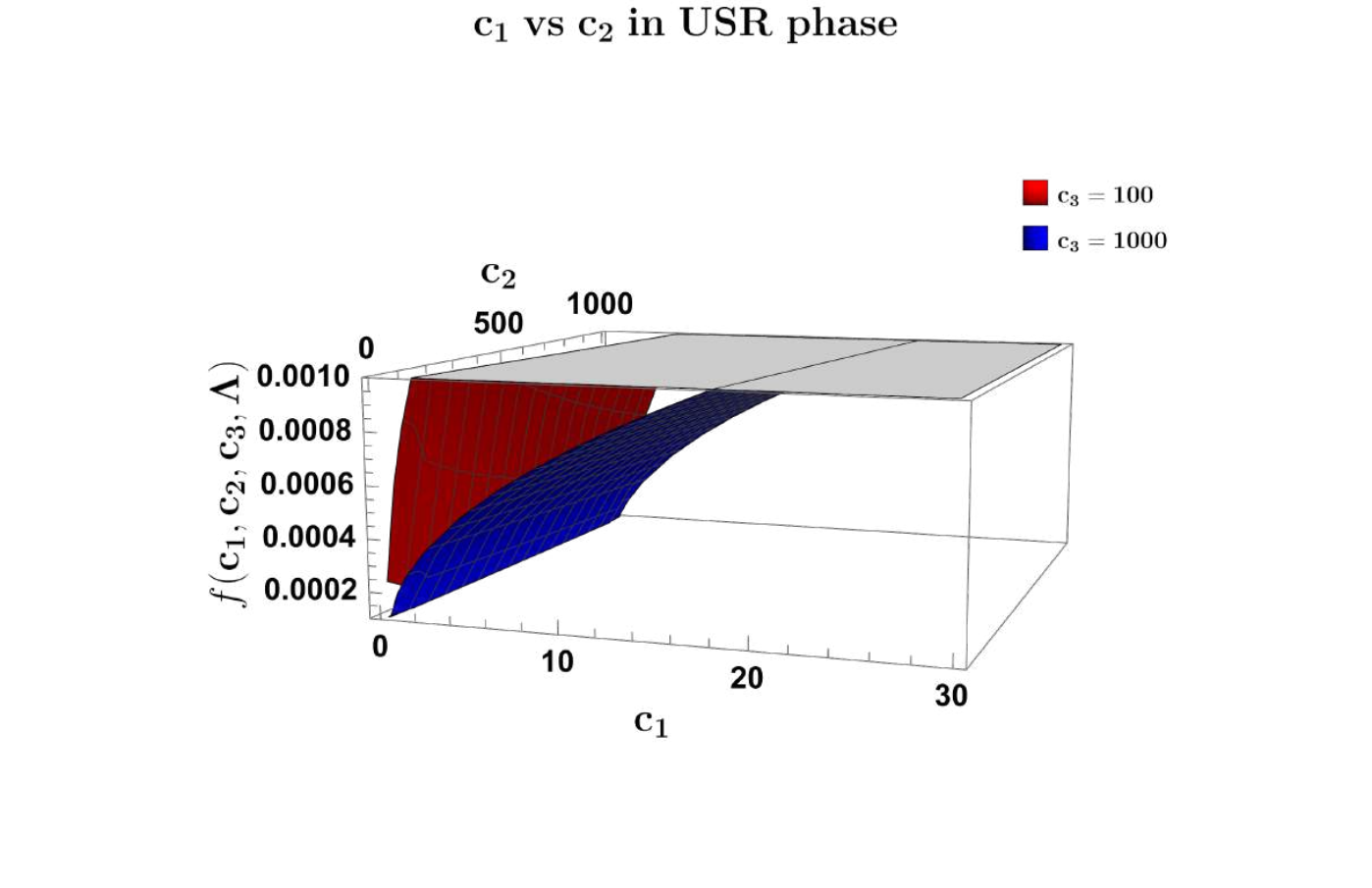}
        \label{usrc3}
       }\hfill
    
    	\caption[Optional caption for list of figures]{Following the USR phase transition, the three Galileon EFT coefficients $c_{1},c_{2},c_{3}$ behaved in a way that satisfied $f\equiv f(c_{1},c_{2},c_{3},\Lambda)$. By keeping $c_{2},c_{3} \sim {\cal O}(10^{-3}),$ $c_{1}\sim {\cal O}(10)$ can provide the desired values of $f$ in (a). In (b), the interval permitted for $c_{1},c_{3}$ remains unchanged when $c_{2}=10$ (red) is changed to $c_{2}=100$ (blue).$c_{3}\sim {\cal O}(10^{2}-10^{3})$ is what we're looking for in (c), where greater $c_{3}$ values translate into larger $c_{1}$ values.
 } 
    	\label{usr1}
    \end{figure*}

\begin{figure*}[htb!]
    	\centering
    \subfigure[]{
      	\includegraphics[width=8.5cm,height=7.5cm]{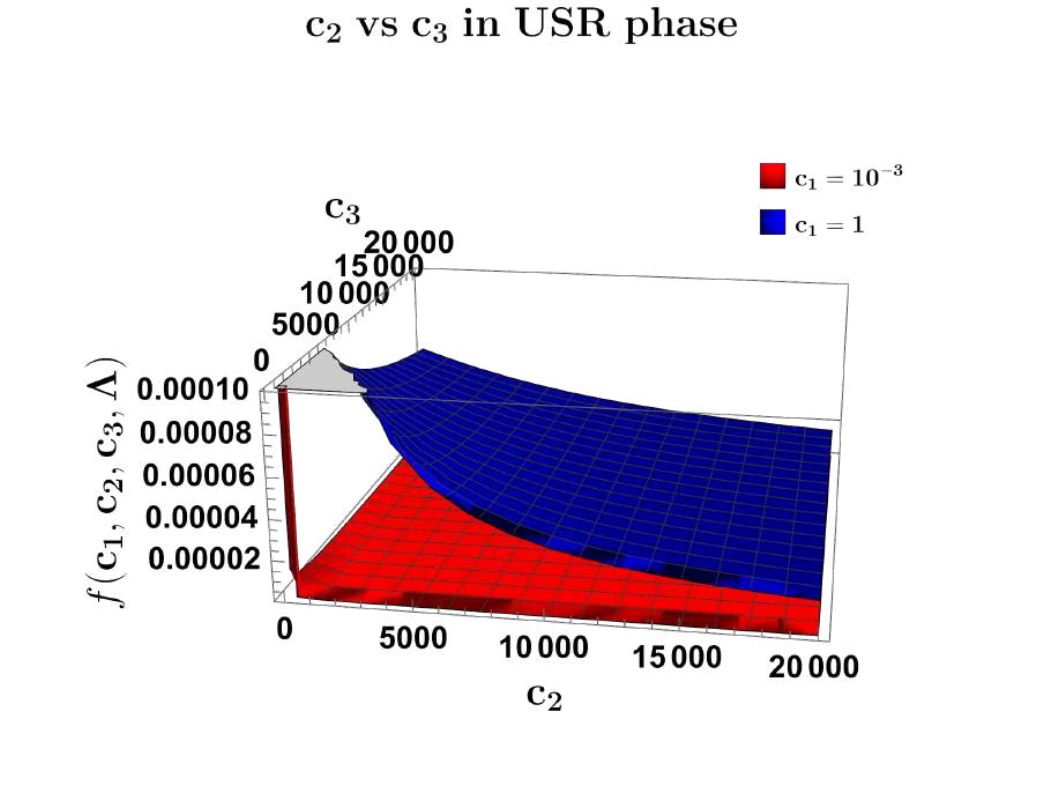}
        \label{usrc11}
    }
    \subfigure[]{
       \includegraphics[width=8.5cm,height=7.5cm]{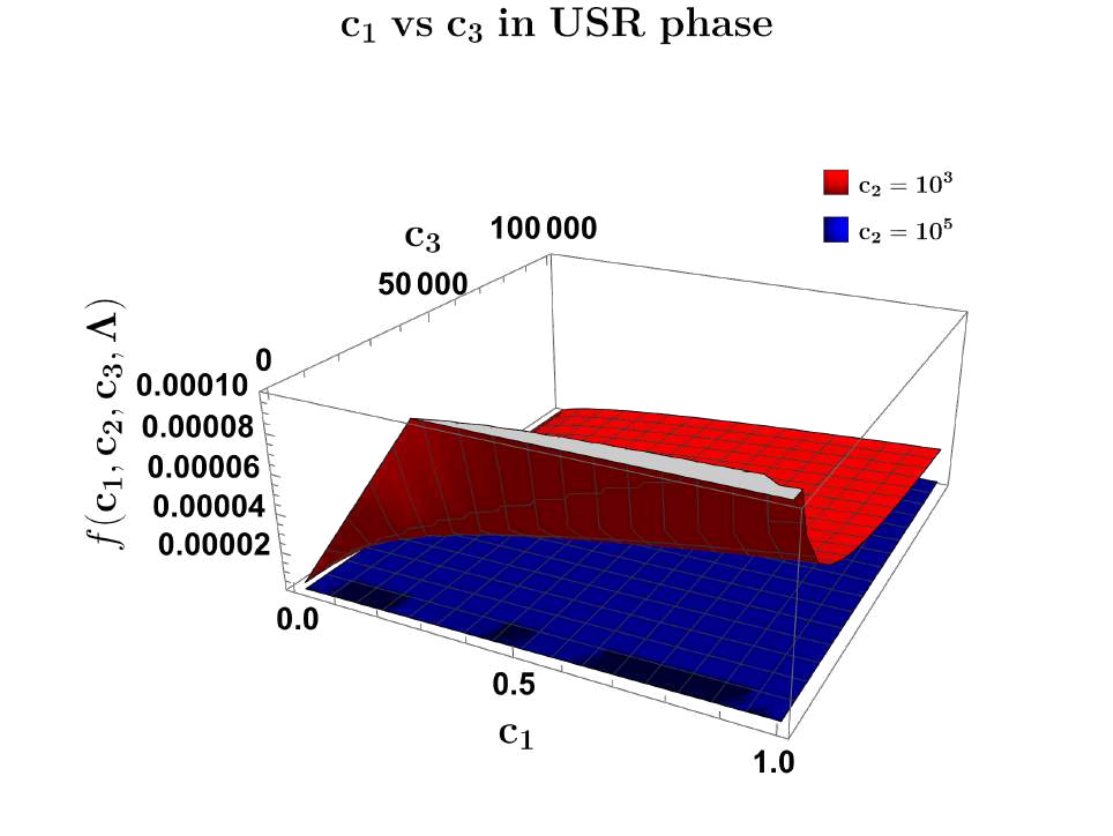}
        \label{usrc21}
       }\\
   \subfigure[]{
       \includegraphics[width=9cm,height=8cm]{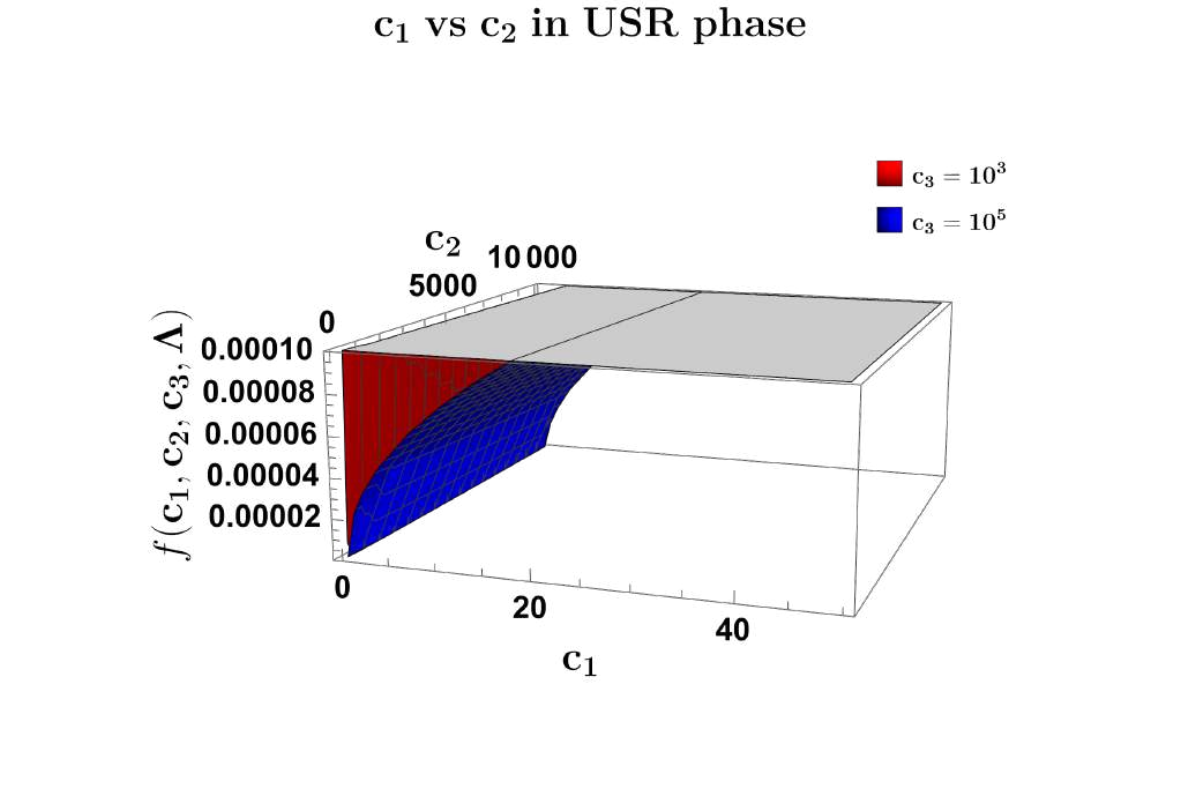}
        \label{usrc31}
       }\hfill
    
    	\caption[Optional caption for list of figures]{Behaviour of the three Galileon EFT coefficients $c_{1},c_{2},c_{3}$ to satisfy $f\equiv f(c_{1},c_{2},c_{3},\Lambda)$ during the USR phase. In (a), $c_{1}=1$ (blue) and $c_{1}=10^{-3}$ (red) are fixed to show the allowed values of $c_{2}$ and $c_{3}$. In (b), $c_{2}=10^{3}$ (red) and $c_{2}=10^{5}$ (blue) are kept fixed. In (c), $c_{3}=10^{3}$ (red) and $c_{1}=10^{5}$ (blue) are kept fixed. } 
    	\label{usr11}
    \end{figure*}

After applying the starting condition at ${\cal N}={\cal N}_{s}$, the word $\eta_{\rm USR}$ appears in the inverse square root numerator. In this case, $\Delta{\cal N}_{\rm USR}={\cal N}-{\cal N}_{s}$, where $\epsilon_{i,\rm USR}$ is the starting condition. This finally yields $f_{\rm USR}({\cal N})$, which is comparable to the expression in eqn.(\ref{ftnN1}):
\bea \label{ftnN2}
f_{\rm USR}({\cal N}) = \frac{H_{\rm USR}({\cal N})}{H_{\rm SRI}({\cal N}_{s})}\exp{(-\eta_{\rm USR}({\cal N}-{\cal N}_{s}))}. \eea
Using $\Lambda \sim {\cal O}(10^{-1})M_{p}$, we will examine how the aforementioned equation behaves for the USR phase in order to restrict $c_{1},c_{2},$, and $c_{3}$. The USR phase continues up to ${\cal N}_{s} = 18$ to ${\cal N}_{e} = 20.4$ according to our numerical analysis. This value is comparable to the perturbativity constraints' needed $\Delta{\cal N}_{\rm USR}=\ln{(k_{e}/k_{s})} \sim {\cal O}(2)$. We concentrate on two scenarios inside the USR in order to more thoroughly examine the permitted values: First, for $f_{\rm USR}$ at the transition instant at ${\cal N}_{s}$; and second, for the number of e-folds remaining to finish USR till ${\cal N}_{e}$ is achieved. In the initial scenario, we see that the value $f_{\rm USR}$ fluctuates at a rate that is far faster than the order of magnitude before it enters the SRI. As a result, we may examine a wide range of values for $c_{1},c_{2},c_{3}$ over a very short interval of e-folds and determine the common, acceptable values after taking each coefficient's change into account. In contrast to the first example, the second one sees a significantly slower change. 

We demonstrate how the EFT coefficients $c_{1},c_{2},c_{3}$ behave when the e-folding moment is close to the sharp transition and the corresponding value of eqn.(\ref{ftnN2}) is taken into account in fig. (\ref{usr1}). The permitted range of values for the coefficients to meet $f$ is displayed by the red and blue surfaces. We see that in order to avoid having to take into account high $c_{2},c_{3}$, smaller values of $c_{1} < {\cal O}(10)$ are strongly preferable. Additionally, we discover that the magnitudes of $c_{2}$ and $c_{3}$ stay constant while being raised in comparison to their earlier values in the SRI case in order to fulfil $f$ close to the transition. In order to satisfy $c_{2},c_{s} \sim {\cal O}(10^{2})$, both coefficients must remain. Considering $c_{1} < {\cal O}(10)$ and $c_{1} \gtrsim {\cal O}(10)$ plots indicate that $c_{2},c_{s} \gtrsim {\cal O}(10^{3})$ are required for both.  Greater coefficient values, such as $c_3$, may indicate that there is an increase in the importance of higher derivative interactions on the scalar field dynamics during that specific interval. These effects originate from the tiny scales stored in the higher derivative operators and tend to grow during the rapid transition at the beginning of USR.

We can now see the coefficients for the remaining USR phase. Figure (\ref{usr11}) illustrates how the Galileon EFT coefficients behave during the USR phase following the abrupt transition. Retaining $c_{1}$ constant, we deduce that greater $c_{1} \gtrsim {\cal O}(1)$ values necessitate even greater $c_{2},c_{3} \sim {\cal O}(10^{5})$ in order to attain the lower end values of $f$. The other panels, in which $c_{2}$ and $c_{3}$ are set separately, provide a closer view of this feature. Lower values of $f$ will never be available until we drastically cut $c_{1}$ in order to decrease $c_{2}$. Therefore, it is inferred that the non-linear interactions inside the Galileon sector dominate even more by raising the size of the coefficients. On the other hand, the suppression of the linear order and quadratic interactions does not change. A lower value of $c_{1}$ indicates a weaker shift symmetry breakdown than in the preceding SRI phase. We first examine the results for different ranges of the coefficients $c_{1},c_{2},c_{3}$. Next, we explain how the other two $c_{4},c_{5}$ coefficients behave in a manner akin to the one described before.

Our approach makes use of the fact that the effective sound speed $c_{s}$ fluctuates sharply at the transition scale, changing from $\tilde{c_{s}}=c_{s}(\tau_{s}) = c_{s}(\tau_{e}) = 1 \pm \delta$, where $\delta \ll 1$, to $c_{s}(\tau)=c_{s,*}$ during the conformal time interval $\tau_{s} \leq \tau \leq \tau_{s}$. A range of values for $c_{4},c_{5}$ will be available to us during the sharp transition instant and in the remaining interval of the USR thanks to this parameterization of the sound speed and the constraint on the amplitude of the scalar power spectrum in the USR as $\big[\Delta^{2}_{\zeta,{\bf Tree}} (k)\big]_{\rm USR} \sim {\cal O}(10^{-2})$. We decide to apply the limitations from $\tilde{c_{s}}=1\pm \delta$ and the power spectrum in the USR, and stay within $c_{2},c_{3} \sim {\cal O}(10-10^{2})$ based on the fig. (\ref{usr1}). Thus, ${\cal O}(10^{-1}) \leq \{c_{4},c_{5}\} \leq {\cal O}(10^{2})$ is the permitted range of values. As we move into the USR, the magnitudes rise and demand larger values of $c_{2},c_{3}$ as well. Although the magnitude is all that is shown by this range, the combined coefficients have both positive and negative fingerprints.

The next $c_{4},c_{5}$ analysis pertains to the situation of the USR's remaining duration. As can be seen from the plots in fig. (\ref{usr11}), there is a reasonable range to generate the values of $f$ as indicated for both $c_{2},c_{3} \lesssim {\cal O}(10^{3})$. We can calculate the interval for $c_{4},c_{5}$ with the use of these values for the effective sound speed and the scalar power spectrum. After entering the USR, $c_{s}(\tau)=c_{s,*}$ has remained constant, with a value limited within the region $0.024 \leq c_{s} < 1$. It is seen that for both $\{ c_{4},c_{5}\}, \sim {\cal O}(10^{2}-10^{3})$. In comparison to the changes in $c_{2},c_{3}$, their values grow rapidly, reaching greater magnitudes for higher $c_{s}$, where both might have positive and negative signatures. The permitted values for $c_{4},c_{5}$ as they vary within ${\cal O}(10^{5}) \leq c_{4},c_{5} \leq {\cal O}(10^{16})$, with $c_{5}$ always a few order of magnitudes higher than $c_{4}$, show how quickly the $\epsilon_{\rm USR}$ decreases for $\eta \sim -6$ as we move into USR. Such behaviour can nonetheless indicate the growth of higher order non-linear interactions in the Galileon sector, even if such big values will similarly stay substantially suppressed by powers of the cut-off scale $\Lambda$.

\begin{table}[H]

\centering
\begin{tabular}{|c|c|c|c|c|c|c|}

\hline\hline
\multicolumn{7}{|c|}{\normalsize \textbf{Galileon EFT coefficients for a given equation of state $w$ in USR }} \\

\hline

EoS $(w)$ & $f$ &\hspace {0.5cm}  $c_{1}$ \hspace {0.5cm} &\hspace {0.5cm}  $c_{2}$ \hspace {0.5cm} &\hspace {0.5cm}  $c_{3}$ \hspace {0.5cm} &\hspace {0.5cm}  $c_{4}$ \hspace {0.5cm} &\hspace {0.5cm}  $c_{5}$ \hspace {0.5cm}  \\
\hline
$1/3$ & & $0.01$ & $100$ & $500$ & $-10^3$ & $-10^5$ \\ 
$0.25$ & $\leq{\cal O}(10^{-4})$  & $0.03$ & $100$ & $500$ & $10^3$ & $-2.9\times 10^5$ \\
$0.16$ & & $0.03$ & $250$ & $700$ & $-10^3$ & $-4.4\times 10^5$ \\ \hline 
\hline

\end{tabular}

\caption{ A potential set of values for the EFT coefficients $c_{i}\;\forall\;i=1,\cdots,5$ fulfilling a particular EoS $w$ is described in the table.
 }

\label{tab2eos}

\end{table}

The values of the Galileon EFT coefficients that can actualize a given EoS $w$ in the USR phase are displayed in the table \ref{tab2eos}. We may recognise that the coefficients continue to fall inside the crucial period for the realisation of the USR phase. In comparison to equivalent changes in either $c_{2}$ or $c_{3}$, the choice of coefficients is significantly more sensitive to lower values of $c_{1}$. Higher values of $c_{4},c_{5}$ may also be reached when $f$ drops inside USR.

\subsubsection{ Region III: Second Slow Roll (SRII) region}

\begin{figure*}[htb!]
    	\centering
    \subfigure[]{
      	\includegraphics[width=8.5cm,height=7.5cm]{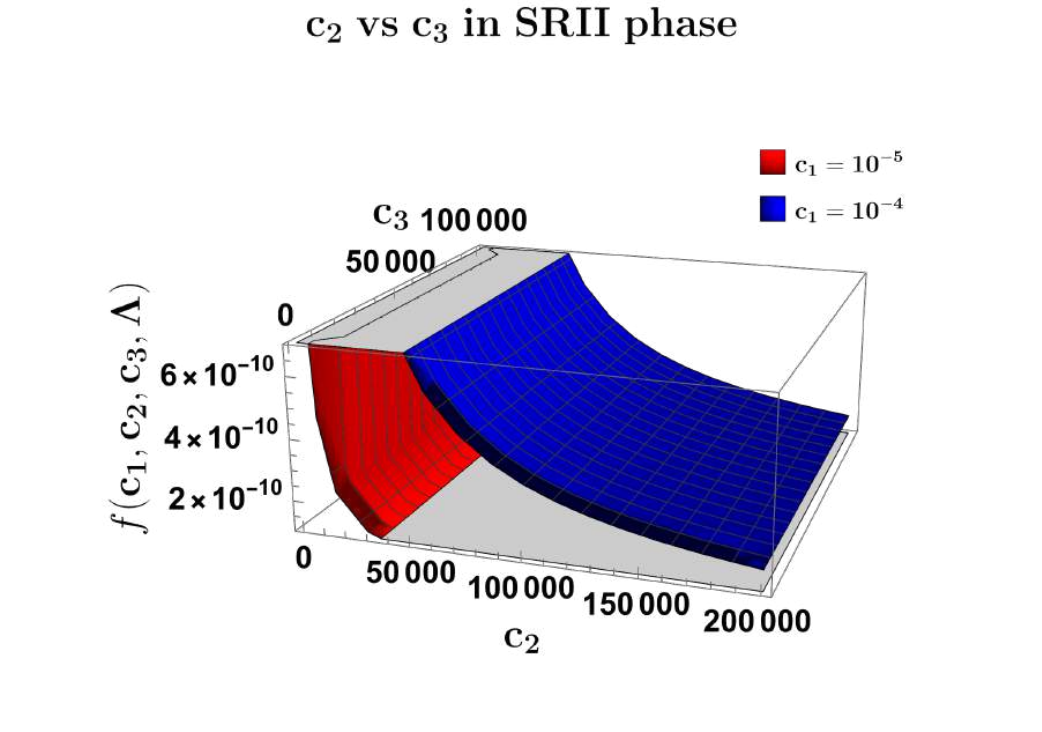}
        \label{sr2c1}
    }
    \subfigure[]{
       \includegraphics[width=8.5cm,height=7.5cm]{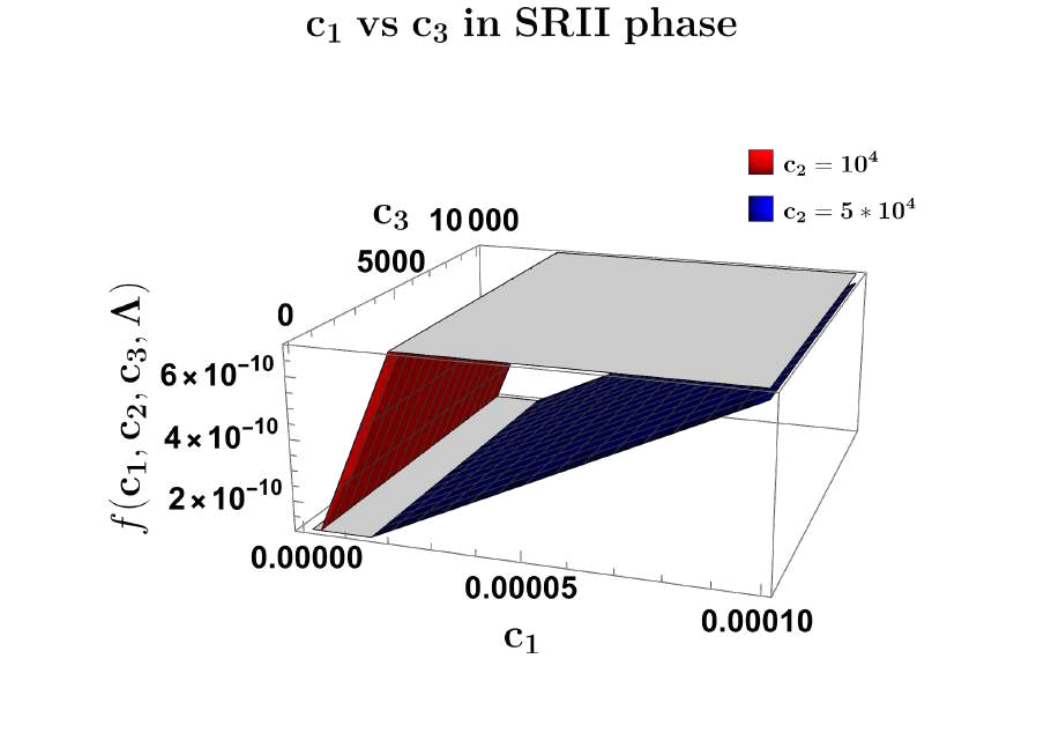}
        \label{sr2c2}
       }\\
   \subfigure[]{
       \includegraphics[width=9cm,height=8cm]{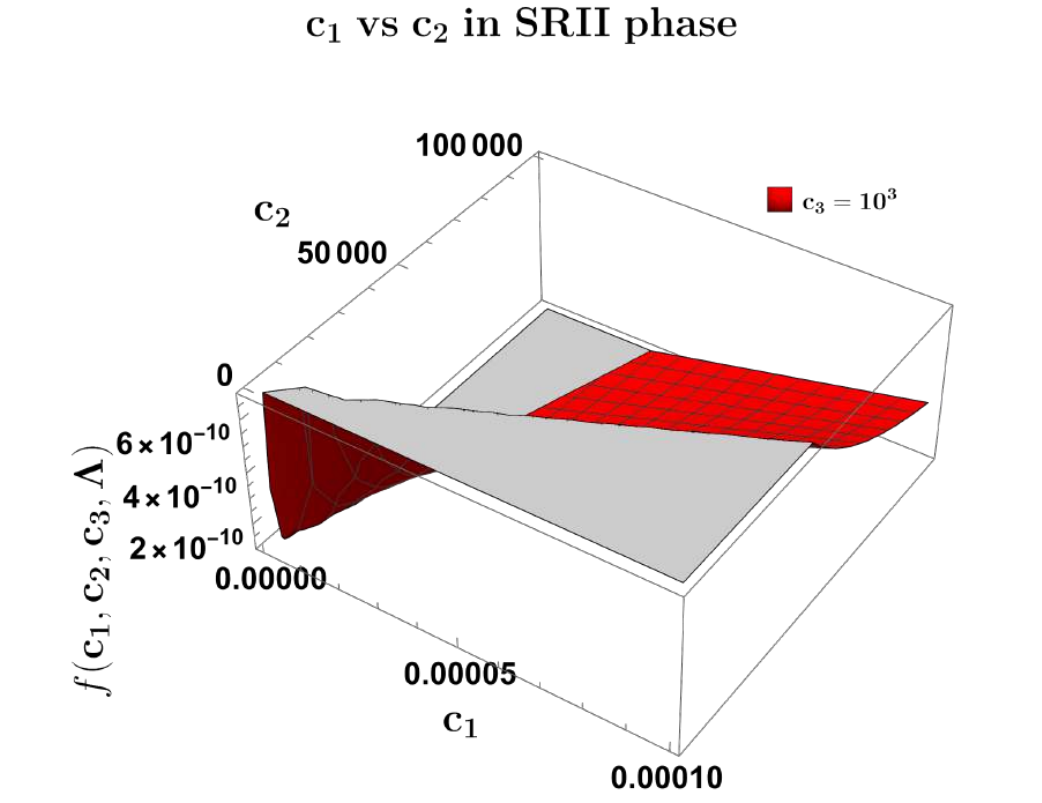}
        \label{sr2c3}
       }\hfill
    
    	\caption[Optional caption for list of figures]{The behaviour of $f\equiv f(c_{1},c_{2},c_{3},\Lambda)$ in the vicinity of the abrupt transition into the SRII phase is determined by the three Galileon EFT coefficients, $c_{1},c_{2},$, and $c_{3}$. To illustrate the permitted values of $c_{2}$ and $c_{3}$, $c_{1}=10^{-5}$ (red) and $c_{1}=10^{-4}$ (blue) are fixed in (a). (b) maintains $c_{2}=5\times 10^{4}$ (blue) and $c_{2}=10^{4}$ (red) constant. We maintain $c_{3}=10^{3}$ (red) constant in (c).   
} 
    	\label{sr2}
    \end{figure*}

In order to understand how these Galileon EFT coefficients might vary, we now perform a similar analysis of the parameter space of the coefficients $c_{i},\;\forall\;i=1,\cdots,5$. We do this by concentrating on the two regimes, which are close to the abrupt transition and during the remaining e-folds of SRII. After the scalar field leaves the USR phase at $\tau=\tau_s$, the SRII phase continues to run until inflation ends, at $\tau=\tau_{rm end}$. After leaving the USR, there is another abrupt transition into the SRII, but this time the potential's characteristics or the previously mentioned coefficients change, causing the slow-roll parameters to start rising in value and eventually approach unity, which is needed to successfully stop inflation. 

Throughout SRII, the initial slow-roll parameter $\epsilon_{\rm SRII}$ acts as a non-constant number and stays exactly proportionate to the $\epsilon_{\rm SRI}$. The second slow-roll parameter, $\eta_{\rm SRII}$, also starts to increase from its initial value in the USR and reaches the end of inflationary value, $\eta_{\rm SRII} \sim -1$. An examination of the notable one-loop adjustments to the tree-level scalar power spectrum also requires an understanding of the abrupt shift in the magnitude of $\eta$ immediately following the transition. In the future, while discussing the scalar power spectrum, the one-loop calculations will be discussed. 

In this stage, we examine $\epsilon_{\rm SRII}({\cal N})$ once again in the following manner:
\bea \label{epsN3}
\epsilon_{\rm SRII}({\cal N}) = \eta_{\rm SRII}\bigg(1-\bigg(1-\frac{\eta_{\rm SRII}}{\epsilon_{i,\rm SRII}}\bigg)e^{2\eta_{\rm SRII}({\cal N}-{\cal N}_{e})}\bigg)^{-1}. \eea
Choosing $\epsilon_{i,\rm SRII} = \epsilon_{\rm USR}({\cal N}_{e})$ as the initial condition, we find that it contains the instant in e-folds ${\cal N}_{e}=20.4$ that indicates the end of USR as well as the moment of sudden transition into the SRII. The value for ${\cal N}_{e}$ was previously mentioned in USR and was determined via the numerical analysis and preservation of perturbativity. The aforementioned equation is then utilised in eqn. (\ref{srparams}), yielding the Hubble parameter in the following form:
\bea \label{HubbleN3}
\frac{H_{\rm SRII}({\cal N})}{H_{\rm USR}({\cal N}_{e})} = \bigg(\frac{\eta_{\rm SRII}\exp{(2\eta_{\rm SRII}\Delta{\cal N}_{\rm SRII})}}{\epsilon_{i,\rm SRII}-\exp{(2\eta_{\rm SRII}\Delta{\cal N}_{\rm SRII})}(\epsilon_{i,\rm SRII}-\eta_{\rm SRII})}\bigg)^{-\frac{1}{2}},
\eea
At ${\cal N}={\cal N}_{e}$, the beginning condition is imposed before the word $\eta_{\rm SRII}$ appears in the inverse square root numerator. The formula for $f_{\rm SRII}({\cal N})$ that is identical to the one in eqn.(\ref{ftnN1}) is eventually given by using $\Delta{\cal N}_{\rm SRII}={\cal N}-{\cal N}_{e}$:
\bea \label{ftnN3}
f_{\rm SRII}({\cal N}) = \frac{H_{\rm SRII}({\cal N})}{H_{\rm USR}({\cal N}_{e})}\exp{(-\eta_{\rm SRII}({\cal N}-{\cal N}_{e}))}. \eea
Now that we have the above formula, we can examine how eqn.(\ref{ftnN3}) behaves in the SRII phase to restrict $c_{1},c_{2},c_{3}$. It is evident from fig. (\ref{sr2}) that ranges reduce to considerably smaller magnitudes for $c_{1}$, where $c_{1}\sim {\cal O}(10^{-5})$, such that $f$ is attained as expected. The sensitivity of $f$ to $c_{1}$ is also revealed by the surface's steepness. It is observed that a drop of one order of magnitude in $c_{1}$ can have a significant impact on the results for $c_{2}$, while scarcely affecting $c_{3}$. Plots in fig. (\ref{usr11}) prior to the transition demonstrate that, in contrast to the previous values, $c_{1}$ declines significantly, while $c_{2}$ grows, with $c_{3}$ not exhibiting a significant variation in magnitude. The subplot with $c_{3}=10^{3}$ fixed validates the behaviour for the other two previously mentioned coefficients. Other values of $c_{3}$, within two orders of magnitude, are not used because their result overlaps with the one present and does not provide additional insight into the matter discussed. 

\begin{figure*}[htb!]
    	\centering
    \subfigure[]{
      	\includegraphics[width=8.5cm,height=7.5cm]{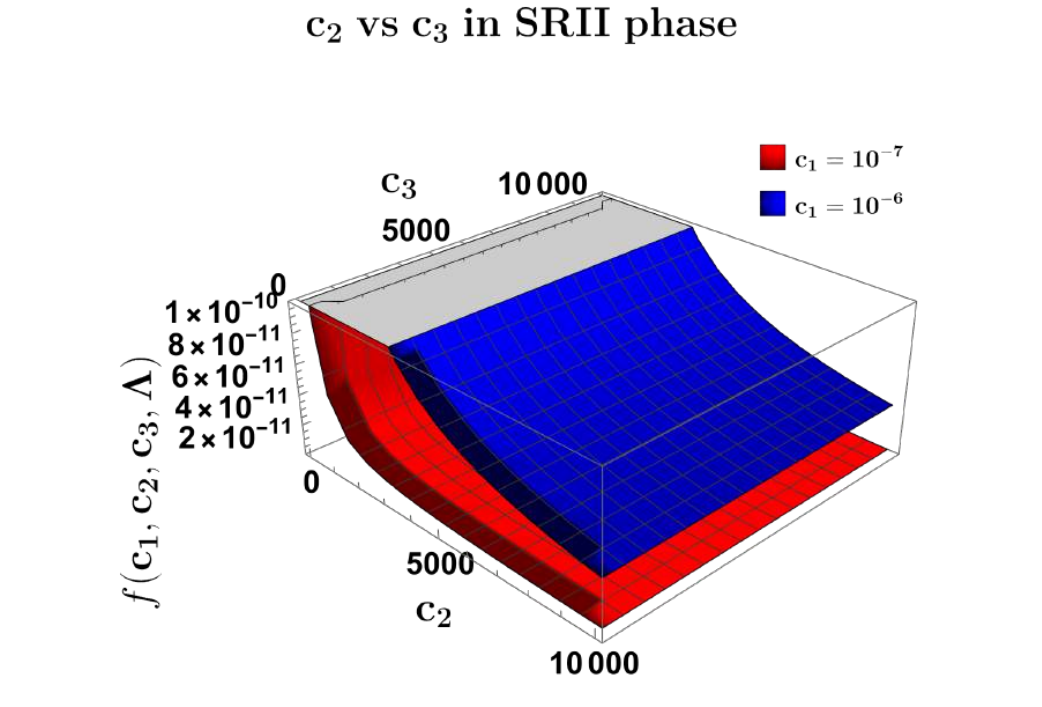}
        \label{sr2c11}
    }
    \subfigure[]{
       \includegraphics[width=8.5cm,height=7.5cm]{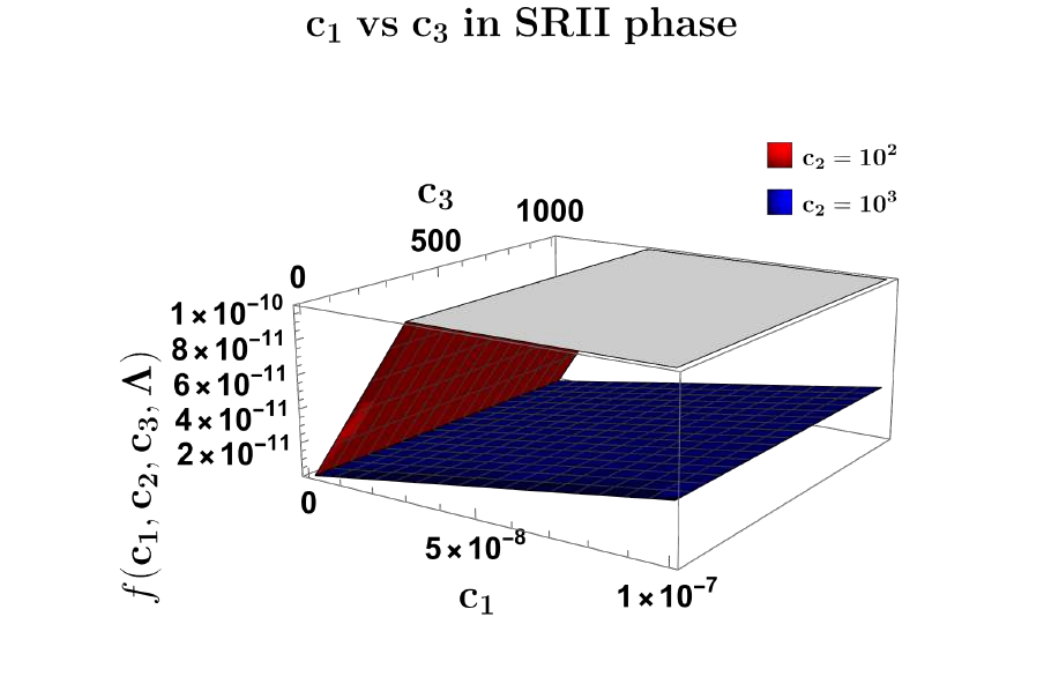}
        \label{sr2c21}
       }\\
   \subfigure[]{
       \includegraphics[width=9cm,height=8cm]{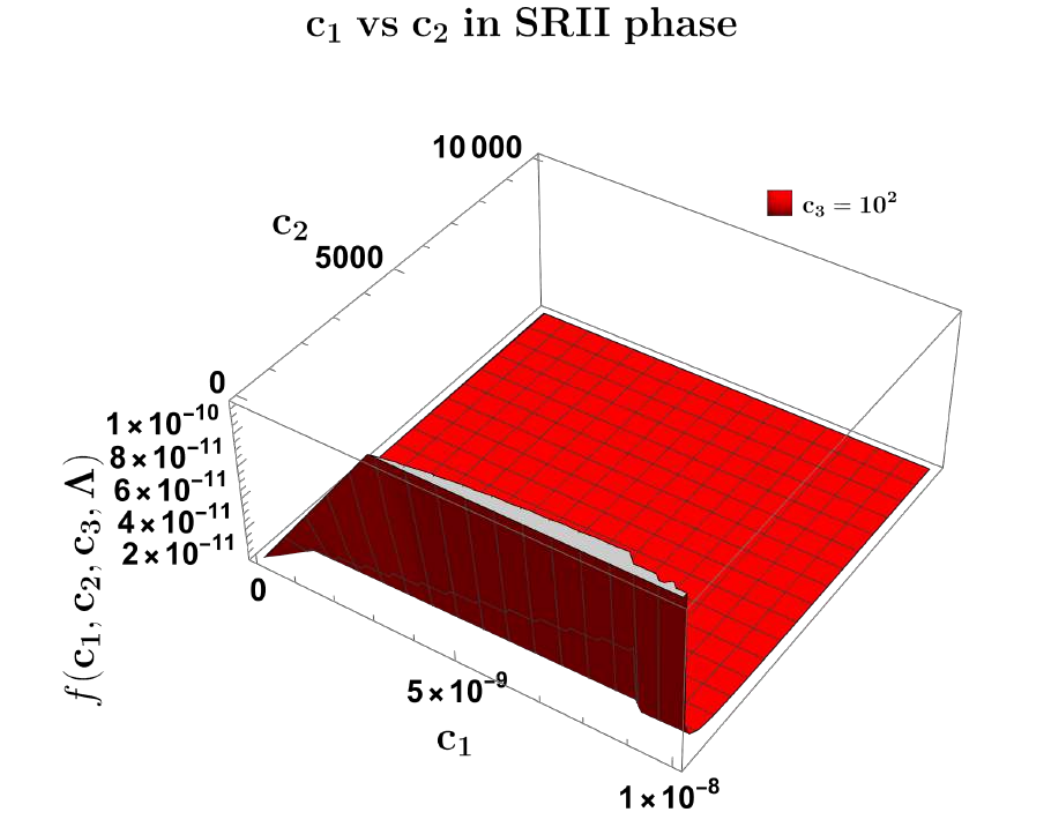}
        \label{sr2c31}
       }\hfill
    
    	\caption[Optional caption for list of figures]{Behaviour of the three Galileon EFT coefficients $c_{1},c_{2},c_{3}$ to satisfy $f\equiv f(c_{1},c_{2},c_{3},\Lambda)$ after the sharp transition into the SRII phase. In (a), $c_{1} = 10^{-6}$ (blue) and $c_{1} = 10^{-7}$ (red) are fixed to achieve the values of $f$. In (b), $c_{2}=10^{2}$ (red) and $c_{2}=10^{3}$ (blue) is kept fixed. In (c), $c_{3}=10^{2}$ is kept fixed. The magnitudes of the parameters provide the respective interval to satisfy the behaviour of $f$ during SRII till the end of inflation. } 
    	\label{sr21}
    \end{figure*}

An examination of the function $f$ across the entire SRII and the effective sound speed $c_{s}$ will yield more constraints on $c_{3}$. In the remaining SRII phase, we first examine the coefficients for $f$. For SRII, the coefficients $c_{1},c_{2},c_{3}$ are easier to grasp with the aid of the figure (\ref{sr21}). A greater reduction in the size of the coefficient $c_{1}$ is required to accommodate the progressively smaller values required for $f$ during SRII, indicating a milder violation of shift symmetry than in the USR. Lower values of $f$ and higher values of $c_{2} \sim {\cal O}(10^{3})$ might result from keeping $c_{2}$ within $c_{2} \gtrsim {\cal O}(10^{2})$ and the range where $c_{1} \lesssim {\cal O}(10^{-7})$ lies.means $c_{1} \gtrsim {\cal O}(10^{-6})$. Throughout the SRII, there are only minor variations in the magnitude of $c_{3}$; $c_{3} \sim {\cal O}(10^{3})$ is still an acceptable number. The examination of the effective sound speed yields further restrictions.

According to our parameterization, the effective sound speed $c_{s}$ during the abrupt transition at the e-fold instant ${\cal N}_{e}$, which indicates the exit from USR and entry into the SRII, takes on the value $c_{s}(\tau_{e}) = \tilde{c_{s}} = 1\pm \delta$, where $\delta \ll 1$. In the remaining SRII phase, we obtain $c_{s}(\tau)=c_{s,*}$ for $\tau_{e} \leq \tau \leq \tau_{\rm end}$, which may impose bounds on $c_{4},c_{5}$ while adhering to the same causality and unitarity requirements as previously stated for $c_{s}$. Finally, we apply an additional restriction on $c_{4},c_{5}$, namely the scalar power spectrum amplitude, where the amplitude satisfies $\big[\Delta^{2}_{\zeta,{\bf Tree}} (k)\big]_{\rm SRII} \sim {\cal O}(10^{-5})$.

We previously observed enormous magnitudes for $c_{4},c_{5}$ when close to the abrupt transition at the conclusion of USR. We see an additional increase in magnitude for $c_{4},c_{5}$ since function $f$ again drops rapidly within a few e-folds after crossing SRII. This type of behaviour only becomes more pronounced when $f$ continues to decrease inside SRII. When compared to their values at the conclusion of the USR phase, both of these begin with values in the interval ${\cal O}(10^{13}) \leq \{ c_{4},c_{5} \} \leq {\cal O}(10^{19})$, which is somewhat higher. There is a sharp increase in values as the SRII goes forward, requiring at most $c_{4} \sim {\cal O}(10^{30})$, while $c_{5} \sim {\cal O}(10^{42})$. These very large coefficient magnitudes are also employed to meet the amplitude of the power spectrum limitation in the SRII phase. This interval for $c_{4},c_{5}$, in relation to eqn.(\ref{CovGal}), demonstrates that, in contrast to the previous case of USR, the large change in $\eta_{\rm USR}$ and corresponding change in $\epsilon_{\rm USR}$ leads to the requirement of higher values of Galileon EFT coefficients. However, the overall coefficient is suppressed by increasingly large powers of the cut-off $\Lambda$.

\begin{table}[H]

\centering
\begin{tabular}{|c|c|c|c|c|c|c|}

\hline\hline
\multicolumn{7}{|c|}{\normalsize \textbf{Galileon EFT coefficients for a given equation of state $w$ in SRII }} \\

\hline

EoS $(w)$ & $f$ &\hspace {0.5cm}  $c_{1}$ \hspace {0.5cm} &\hspace {0.5cm}  $c_{2}$ \hspace {0.5cm} &\hspace {0.5cm}  $c_{3}$ \hspace {0.5cm} &\hspace {0.5cm}  $c_{4}$ \hspace {0.5cm} &\hspace {0.5cm}  $c_{5}$ \hspace {0.5cm}  \\
\hline
$1/3$ &  & $10^{-7}$ & $10^{3}$ & $2\times 10^{3}$ & $-10^{16}$ & $-2.5\times 10^{16}$ \\ 
$0.25$ & $\leq {\cal O}(10^{-10})$  & $10^{-7}$ & $10^{3}$ & $5\times 10^{3}$ & ${\cal O}(10^{29})$ & ${\cal O}(10^{30})$ \\
$0.16$ & & $10^{-7}$ & $10^{3}$ & $7\times 10^{3}$ & ${\cal O}(10^{31})$ & ${\cal O}(10^{33})$ \\ \hline 
\hline

\end{tabular}

\caption{ A collection of potential values for the EFT coefficients $c_{i}\;\forall\;i=1,\cdots,5$ that fulfil a particular EoS $w$ are described in the table.
 }

\label{tab3eos}

\end{table}

The values of the Galileon EFT coefficients that can actualize a given EoS $w$ in the SRII phase are displayed in the table \ref{tab3eos}. Here, while changes in $c_{3}$ are least sensitive in all cases, both $c_{1},c_{2}$ have comparable effects on the other coefficients when they are either raised or lowered. Further reducing $f$, as demonstrated by the table's entries, permits larger values of $c_4,c_5$, also in line with the preceding stages.

Our setup can be constructed more easily once we have a better understanding of how the EFT coefficients behave. Lastly, we can see how the slow-roll parameters depend on the number of e-folds that pass in each phase, as well as how the Hubble parameter behaves and how the effective sound speed changes with conformal time. This information is essential for parameterizing the setup that is needed.

\subsection{Numerical Outcomes: Behaviour of slow-roll parameters in the three consecutive phases}
\begin{figure*}[htb!]
    	\centering
    {
      	\includegraphics[width=17cm,height=8cm]{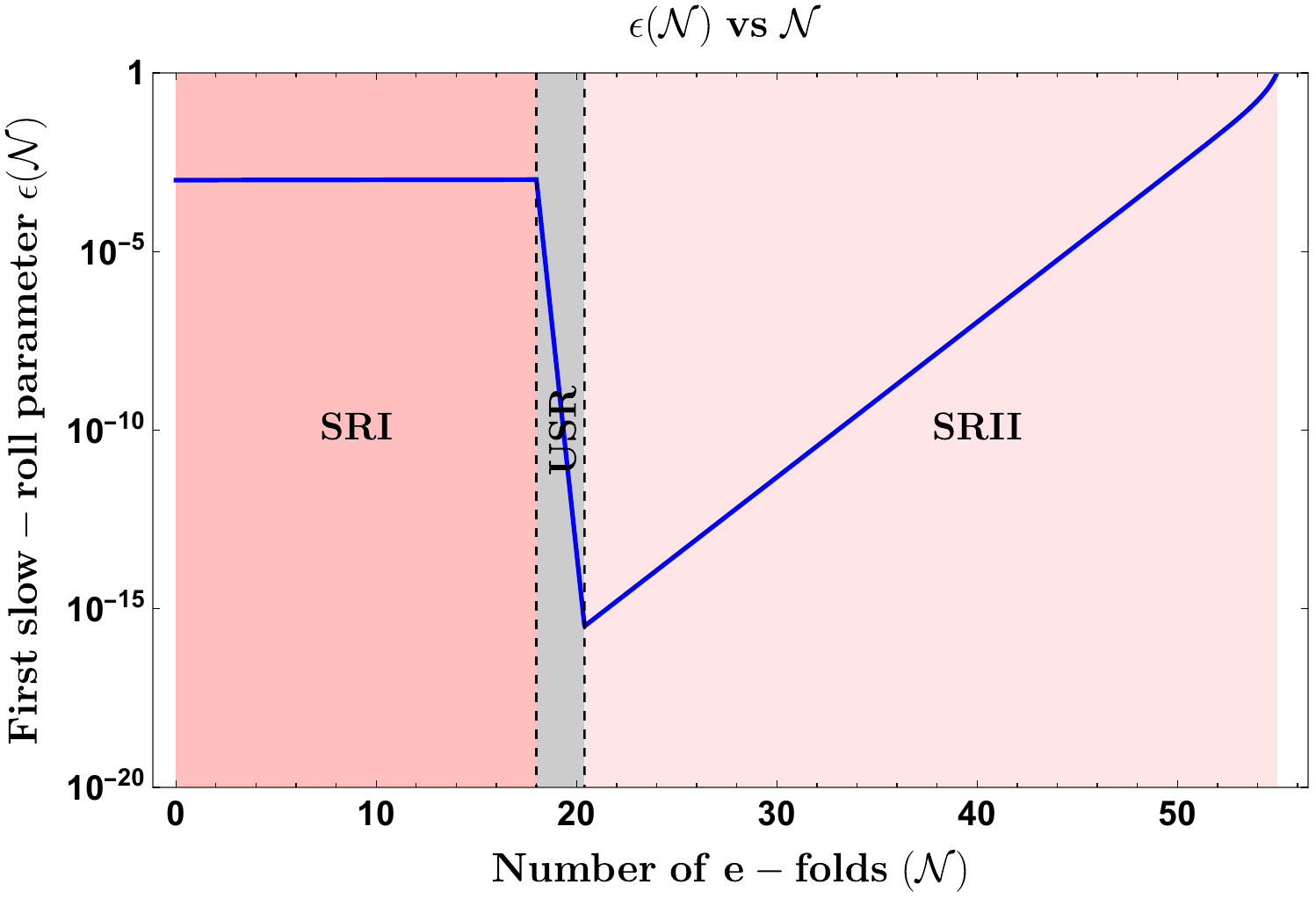}
        \label{epsilon}
    }
    	\caption[Optional caption for list of figures]{Plotting the initial slow-roll parameter $\epsilon$ versus the number of e-folds when abrupt transitions are present. Its behaviour varies across our setup's three phases of interest (SRI, USR, and SRII), with a significant shift occurring as it enters and leaves the USR phase. 
} 
    	\label{s4d1}
    \end{figure*}
Here, we give numerical results for the slow-roll parameters $\epsilon({\cal N})$ and $\eta({\cal N})$ for each of the three inflationary phases, identifying their respective behaviours as functions of the e-folds parameter ${\cal N}$ each time. For the SRI, USR, and SRII phases, we use eqs. (\ref{HubbleN1},\ref{HubbleN2},\ref{HubbleN3}) to determine the corresponding nature of the Hubble parameter $H({\cal N})$. There, we also incorporate our specific parameterization for the effective sound speed $c_{s}$ with the conformal time, which we also mention in this section.

\begin{figure*}[htb!]
    	\centering
    {
       \includegraphics[width=17cm,height=8cm]{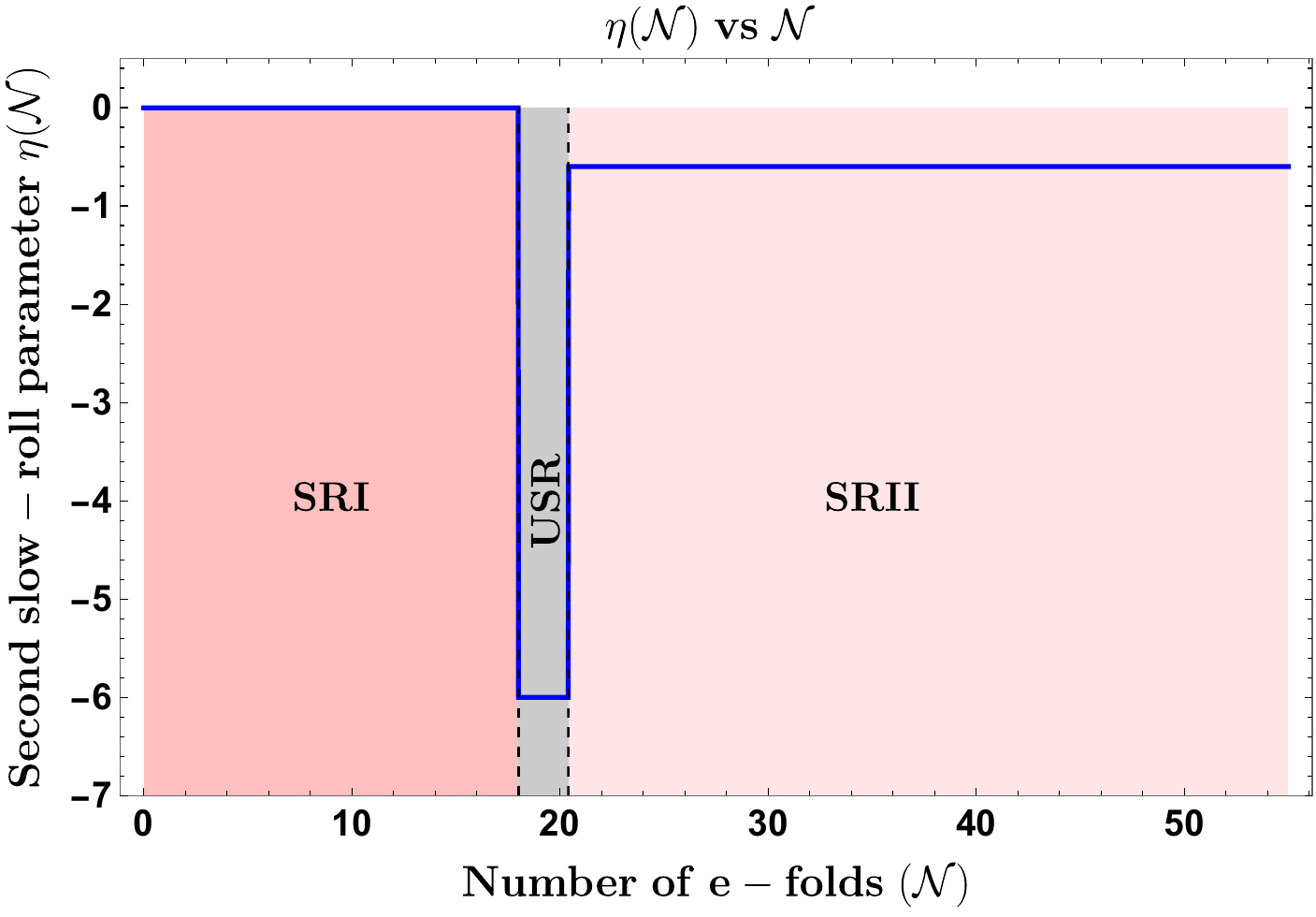}
        \label{eta}
       }
       \caption[Optional caption for list of figures]{In the case of abrupt transitions, the second slow-roll parameter $\eta$ is displayed versus the number of e-folds. Its behaviour in our setup's three phases of interest—SRI, USR, and SRII—varies, and its value abruptly increases during the USR phase. 
}
       \label{s4d2}
    \end{figure*}

In our SRI/USR/SRII system, the development of the first slow-roll parameter during inflation is seen in fig. (\ref{s4d1}).  Starting point: ${\cal N}_{*}=0$ as our reference, where the modes with $k_{*}=0.02 {\rm Mpc^{-1}}$ depart the Horizon, and where we set the starting conditions using $\epsilon_{\rm SRI}({\cal N}_{*}) \sim {\cal O}(10^{-3})$. As we get deeper into the SRI, $\epsilon_{\rm SRI}$ varies in the order of its original value very slowly, practically constant. Next, $\epsilon_{\rm SRI}$ experiences a substantial reduction in magnitude as we approach the first steep transition into the USR. It goes from $\epsilon_{\rm USR} \sim {\cal O}(10^{-3})$ to $\epsilon_{\rm USR} \sim {\cal O}(10^{-15})$.In a few e-folds, $\Delta{\cal N}_{\rm USR} \sim {\cal O}(2)$. The abrupt shift in the $\eta_{\rm USR}$ parameter, which is obvious from fig. (\ref{s4d2}), and the extreme change in $\epsilon_{\rm USR} \propto a^{-6}$ are characteristics of the USR. Right at the conclusion of inflation, the $\epsilon_{\rm SRII}$ value climbs and reaches $\epsilon_{\rm SRII} \sim {\cal O}(1)$ following another abrupt transition during departure from the USR. Using the equations (\ref{epsN1}, \ref{epsN2}, \ref{epsN3}) results in the cumulative nature. 

We show how the $\eta$ parameter behaved during the three periods of interest in the following figure (\ref{s4d2}). We have selected $\eta_{\rm SRI}({\cal N}_{*}) \sim -0.001$ as the initial value. In SRI, where the slow roll approximations are fully valid, $\eta_{\rm SRI}$ appears with a negative signature. Until we come across the abrupt change into the USR, when it simultaneously leaps to acquire $\eta_{\rm USR} \sim -6$, $\eta_{\rm SRI}$ behaves as such. This nearly instantaneous shift in value from $\eta_{\rm SRI}$ to $\eta_{\rm USR}$ indicates that this parameter is sharply transitional, and the Heaviside Theta function is used to express this. Following its departure from the USR, $\eta_{\rm USR}$ returns to a negative value, $|\eta_{\rm SRII}| \sim{\cal O}(1)$, and stays there until inflation stops.  

\begin{figure*}[htb!]
    	\centering
    {
      	\includegraphics[width=17cm,height=8cm]{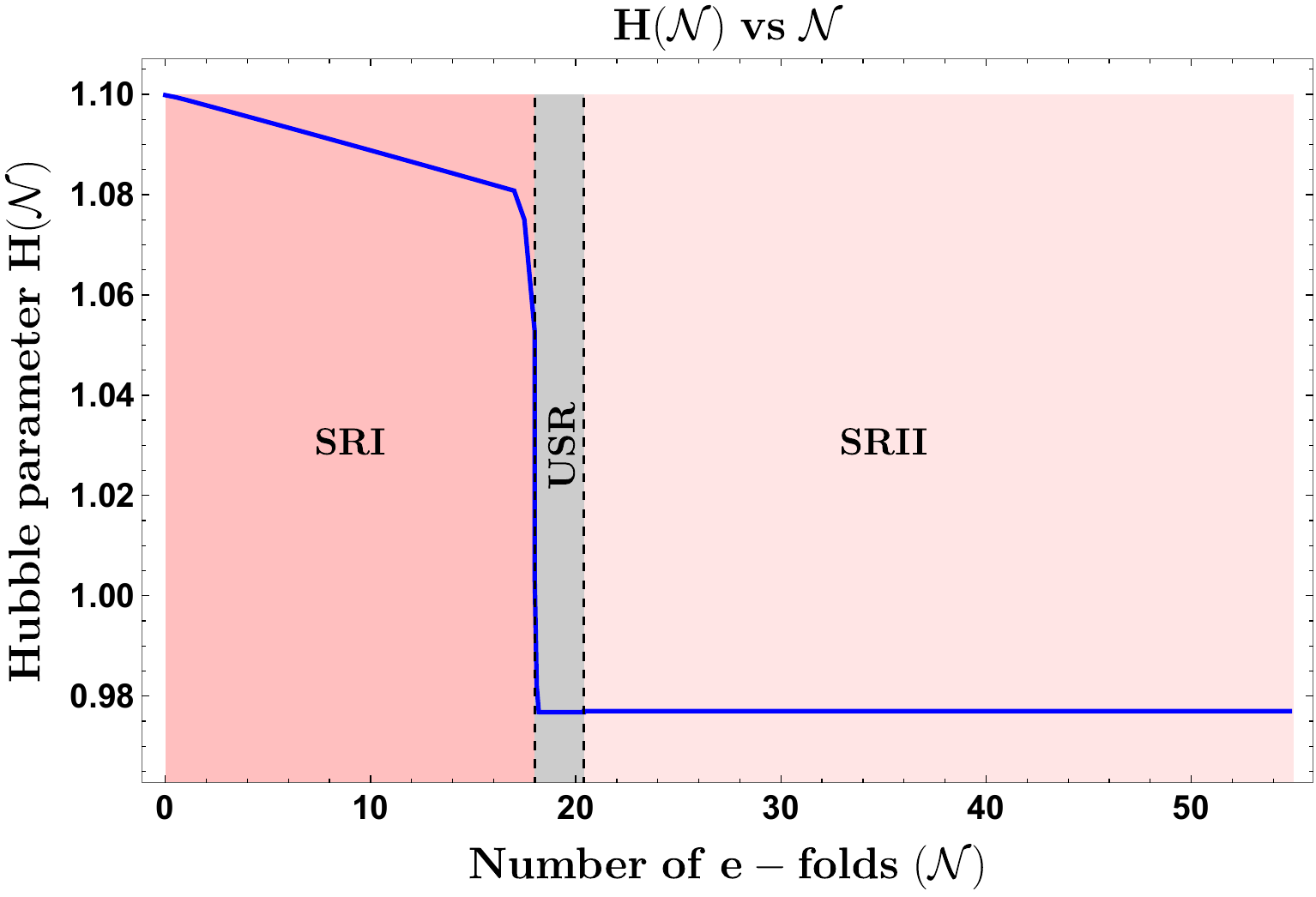}
        \label{hubble}
    }
    \caption[Optional caption for list of figures]{Variation of the Hubble rate $H$ with the number of e-folds. }
    \label{s4d3}
    \end{figure*}
Figure \ref{s4d3} allows us to see how the number of e-foldings changes throughout the course of the three phases of interest, as does the Hubble parameter $H({\cal N})$. We discover that the Hubble rate begins at unity and remains constant until it decreases as we approach the USR, after the imposition of beginning requirements on the slow-roll parameters and the requirement to start with ${\cal N}_{*}=0$. As you can see, the beginning conditions and the incredibly high value of the $\eta_{\rm USR}$ parameter in the USR lead to the order of change in magnitude occurring at the second decimal place, which is quite little. The Hubble rate remains constant until inflation ends farther into the USR, even after it exits and enters the SRII phase.

\begin{figure*}[htb!]
    	\centering
    {
       \includegraphics[width=17cm,height=8cm]{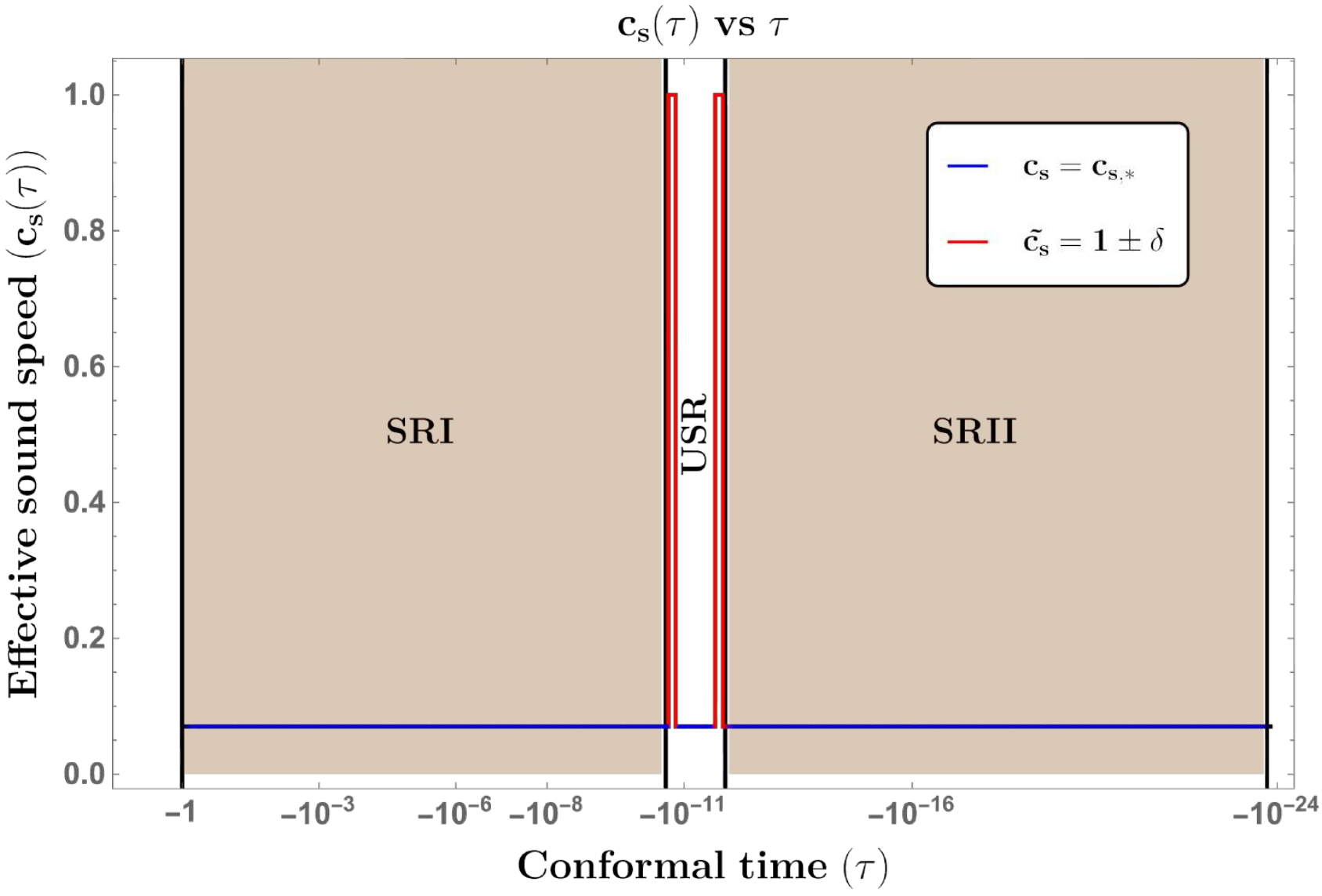}
        \label{sound}
    } 
    \caption[Optional caption for list of figures]{Diagram showing the parameterization used for the conformal time $\tau$ and effective sound speed $c_{s}$. The abrupt jumps in its value that correlate to the sharp transitions are shown by the red colour.
 }
\label{s4d4}
    \end{figure*}
We offer a schematic depiction of the selected parameterization in the final figure (\ref{s4d4}), which allows us to incorporate the abrupt transition feature into our setup. In the current Galileon inflation framework, the abrupt transition in $c_{s}$ to its value $\tilde{c_{s}}=1\pm \delta$ at both $\tau_{s}$ and $\tau_{e}$—highlighted in red—makes it easier to establish USR phase conditions. We have selected this range to confine the parameter space of the EFT coefficients in the Lagrangian eqn. (\ref{CovGal}) based on the observational constraints on the value for $c_{s}$ that have already been previously discussed with $0.024 \leq c_{s} < 1$. Between the length of the interest phases, the value $c_{s,*}$ stays constant, and for our upcoming computations, we have selected $c_{s,*}=0.05$. 

We illustrate the benefits and ramifications of integrating a sudden change in direction in our arrangement with the USR. Heaviside Theta functions $\Theta(\tau-\tau_{s})$ and $\Theta(\tau-\tau_{e})$, which are used at the exact point of transition between $\tau_{s}$ and $\tau_{e}$ (or ${\cal N}_{s}$ and ${\cal N}_{e}$), allow to execute the sharp transition throughout. The purpose of this function is to show the characteristics of the $\eta$ parameter, which is subsequently utilised to produce the features that are evident for the $\epsilon$ parameter. Currently, we select a certain parameterization for the $\eta$, which like this:
 \bea
\eta(\tau) &=&  \eta_{\rm SRI}\Theta(\tau-(\tau_{*}-\tau_{s})) +  \eta_{\rm USR}\Theta(\tau-(\tau_{s}-\tau_{e})) + - \Delta\eta(\tau)\{\Theta(\tau-\tau_{s}) - \Theta(\tau-\tau_{e})\} + \eta_{\rm SRII}\Theta(\tau-(\tau_{e}-\tau_{\rm end})),   \nonumber\\ 
&=& \eta_{\rm SRI}(\tau_{*} < \tau < \tau_{s}) + \eta_{\rm USR}(\tau_{s} < \tau < \tau_{e}) - \Delta\eta(\tau)\{\Theta(\tau-\tau_{s}) - \Theta(\tau-\tau_{e})\} + \eta_{\rm SRII}(\tau_{e} < \tau < \tau_{\rm end}).
\eea
Upon taking the conformal time derivative of this $\eta$ parameterization at the transition moments, we may derive the following:
\bea
\bigg(\frac{\eta(\tau)}{c^{2}_{s}}\bigg)' &=& \underbrace{\bigg(\frac{\eta_{\rm SRI}(\tau_{*} < \tau < \tau_{s})}{c^{2}_{s,*}}\bigg)'}_{=0} + \underbrace{\bigg(\frac{\eta_{\rm USR}(\tau_{s} < \tau < \tau_{e})}{c^{2}_{s,*}}\bigg)'}_{=0} \nonumber\\
&& \quad\quad\quad\quad\quad\quad\quad\quad\quad\quad - \frac{\Delta\eta(\tau)}{\tilde{c_{s}}^{2}}\{\delta(\tau-\tau_{s}) - \delta(\tau-\tau_{e})\} + \underbrace{\bigg(\frac{\eta_{\rm SRII}(\tau_{e} < \tau < \tau_{\rm end})}{c^{2}_{s,*}}\bigg)'}_{=0},
\eea
where a conformal time derivative is indicated by the prime notation. When discussing the non-renormalization theorem, one will observe that the dominating interaction term with the coefficient $(\eta/c_{s}^{2})'$ is of the type $\zeta'\zeta^{2}$ when computing one-loop corrections. We can see from the preceding equation that this introduces improvements at the transitions that resemble Dirac deltas. Luckily, such terms in the final computations of the one-loop contributions are avoided by the process of delicately breaking the Galilean shift symmetry. Nevertheless, the way the $\eta$ parameter appears in our analysis is evident from the above construction, with sharp transitions observable in the fig.(\ref{s4d2}) at the two transition moments $\tau=\tau_{s}\;( {\cal N}={\cal N}_{s})$ and $\tau=\tau_{e}\;( {\cal N}={\cal N}_{e})$. In an effort to better understand the effects of quantum corrections on the scalar power spectrum and PBH production, attempts have been made to integrate the SR/USR and USR/SR scenarios using a smooth transition \cite{Riotto:2023hoz,Riotto:2023gpm,Firouzjahi:2023ahg,Firouzjahi:2023aum}. Additionally, the impact of a bump/dip-like feature in the inflationary potential \cite{Mishra:2019pzq} has also been studied.

For each of the three phases, the scalar power spectrum amplitude was a crucial factor that was taken into consideration to restrict the coefficients of the higher-derivative interaction components. After applying the one-loop adjustments for each phase specified in our setup, we present the procedures for building the total power spectrum for the scalar modes in the next section.

\subsection{Semi-Classical modes from Cosmological Perturbation}

It is necessary to execute perturbation theory up to second order in the comoving curvature perturbation in order to get the power spectrum associated with the scalar modes using eqn. (\ref{CovGal}). The evolution equation for the scalar modes in the Fourier space is given by this process, and the solutions to it subsequently assist us in constructing the scalar power spectrum for each of the three phases—SRI, USR, and SRII. The Israel junction conditions, boundary conditions at the intersections of the abrupt transition, are necessary to accurately calculate the solutions for each of the three phases. In this part, we investigate this approach and give the equations for the overall scalar power spectrum in terms of the distinct contributions from each phase.  

In our quasi de Sitter background, where we omit any effects arising from mixing with the gravity sector, the second order action for the comoving curvature modes from eqn.(\ref{CovGal}) is as follows:
\bea
S^{(2)}_{\zeta} = \int d\tau\;\frac{d^{3}\mbf{k}}{(2\pi)^{3}}a(\tau)^2\frac{\cal A}{H^{2}}\left(|\zeta_{k}^{'}(\tau)|^{2} - c_{s}^{2}k^{2}|\zeta_{k}(\tau)|^{2}\right) = \int d\tau\;\frac{d^{3}\mbf{k}}{(2\pi)^{3}}a(\tau)^2\frac{\cal B}{c_{s}^{2}H^{2}}\left(|\zeta_{k}^{'}(\tau)|^{2} - c_{s}^{2}k^{2}|\zeta_{k}(\tau)|^{2}\right),
\eea
where eqs. (\ref{coeffA},\ref{coeffB}) previously determine the time-dependent coefficients ${\cal A},\;{\cal B}$, $H$ is the Hubble parameter (for our background spacetime, which is not quite a constant), and the effective sound speed $c_{s}$ back in eqn.(\ref{soundspeed}). The Mukhanov-Sasaki equation, also known as the evolution equation for the comoving curvature perturbation modes $\zeta_{\mbf{k}}$, may be obtained with ease by using the action mentioned above. Its form is as follows:
\bea \label{MS}
\zeta^{''}_{ k}(\tau)+2\frac{z^{'}(\tau)}{z(\tau)}\zeta^{'}_{ k}(\tau) +c^2_sk^2\zeta_{k}(\tau)=0.
\eea
whereby $z(\tau) = a\sqrt{2{\cal A}}/H^{2}$ is used. The three phases' curvature perturbation modes can result by solving the preceding equation; we will find their answers in the next section. It is important to note that the variable $c_{s}$ includes the specific parameterization for carrying out each of the three stages. We will also discuss this variable and its associated coefficients in the section that follows. This variable will show up in every mode solution during the phases. 

The upcoming part uses the previously solved eqn. (\ref{MS}) to identify the most general mode solutions in each of the three phases of interest in our setup. These solutions are then reduced by selecting appropriate starting quantum vacuum state conditions to provide the desired form for our calculations.

\subsubsection{Region I: First Slow Roll (SRI) region}

The formula that follows provides the general mode solution for eqn. (\ref{MS}) and the associated canonically conjugate momentum during the SRI phase that works inside the conformal time window, $\tau_{*} \leq \tau < \tau_{s}$, or in e-foldings, ${\cal N}_{*} \leq {\cal N} < {\cal N}_{s}$.
\bea \label{s5a}
    \zeta_{k}(\tau)&=& \left(\frac{iH^{2}}{2\sqrt{\cal A}}\right)\frac{1}{(c_{s}k)^{3/2}}\times\left\{\alpha^{(1)}_{k}\left(1+ikc_{s}\tau\right)\exp{\left(-ikc_{s}\tau\right)} - \beta^{(1)}_{k}\left(1-ikc_{s}\tau\right)\exp{\left(ikc_{s}\tau\right)}\right\},\\
\label{s5b}
\Pi_{k}(\tau) &=&\zeta^{'}_{k}(\tau)= \left(\frac{iH^{2}}{2\sqrt{\cal A}}\right)\frac{1}{(c_{s}k)^{3/2}}\times\frac{k^{2}c_{s}^{2}\tau^{2}}{\tau}\left\{\alpha^{(1)}_{k}\exp{\left(-ikc_{s}\tau\right)} - \beta^{(1)}_{k}\exp{\left(ikc_{s}\tau\right)} \right\}.
\eea
\begin{figure*}[htb!]
    	\centering
    \subfigure[]{
      	\includegraphics[width=8.5cm,height=7.5cm]{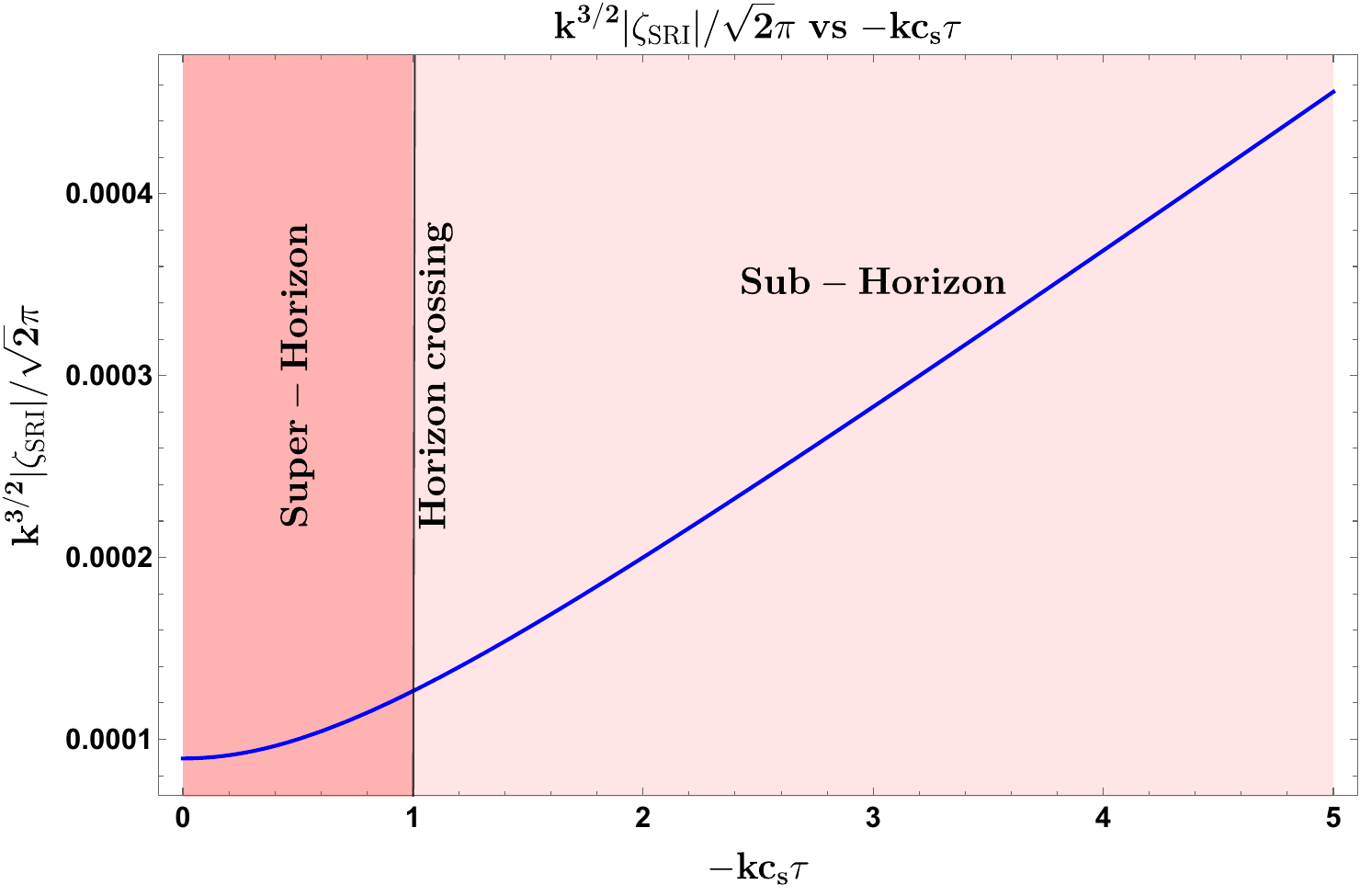}
        \label{zetasr1}
    }
    \subfigure[]{
       \includegraphics[width=8.5cm,height=7.5cm]{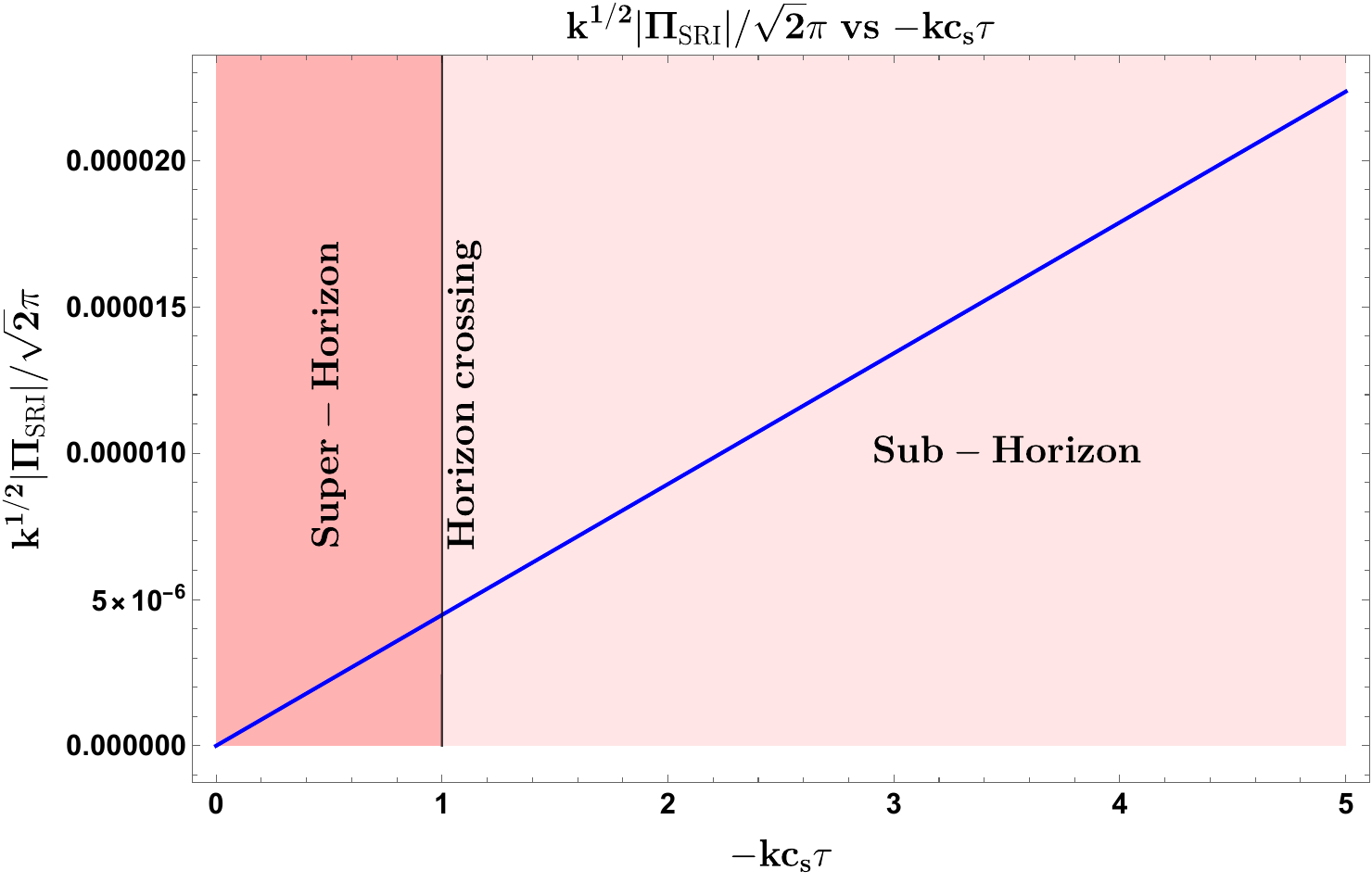}
        \label{pisr1}
    }
    	\caption[Optional caption for list of figures]{$k^{3/2}\abs{\zeta_{\mbf{k}}}/\sqrt{2}\pi$ (\textit{left panel}) and the conjugate momenta $k^{1/2}\abs{\Pi_{\mbf{k}}}/\sqrt{2}\pi$ (\textit{right panel}) plotted against $-kc_{s}\tau$ in the SRI phase. The expressions show monotonic behaviour since their start from the super-Horizon limit, ($-kc_{s}\tau \ll 1$), crossing the horizon ($-kc_{s}\tau = 1$), till their evolution in the sub-Horizon regime ($-kc_{s}\tau \gg 1$). } 
    	\label{modessr1}
    \end{figure*}
The Bogoliubov coefficients, $\alpha^{(1)}_{k}$ and $\beta^{(1)}_{k}$, are utilised to ascertain the initial vacuum state conditions for our mode solution. $c_{s}(\tau)=c_{s,*}$ is fulfilled for the aforementioned solutions, which comprise the curvature perturbation mode and its conjugate momenta, during the SRI phase. For these coefficients, we use the most often used Bunch-Davies initial vacuum condition, which is expressed as follows:
\bea \label{s5a1}
    \alpha^{(1)}_{ k}=1,\\
    \label{s5a2}    \beta^{(1)}_{ k}=0. 
    \eea
The mode solution already acquired is reduced to a form appropriate for our next computations by selecting the vacuum mentioned above. The aforementioned solution's final form is shown as follows:
\bea \label{s51} \zeta_{k}(\tau) &=& \left(\frac{iH^{2}}{2\sqrt{\cal A}}\right)\frac{1}{(c_{s}k)^{3/2}}\left(1+ikc_{s}\tau\right)\exp{(-ikc_{s}\tau)}, \\
\label{s52} \Pi_{k}(\tau) &=&\zeta^{'}_{k}(\tau)= \left(\frac{iH^{2}}{2\sqrt{\cal A}}\right)\frac{1}{(c_{s}k)^{3/2}}\times\frac{k^{2}c_{s}^{2}\tau^{2}}{\tau}\exp{\left(-ikc_{s}\tau\right)}.
\eea
During this stage, $\eta_{\rm SRI}$ likewise has a stable constant value with a negative signature, while the parameter $\epsilon_{\rm SRI}$ continues to change slowly. 

Equation (\ref{s51},\ref{s52}) is used to illustrate the behaviour of the mode solution and associated conjugate momenta with varying $-kc_{s}\tau$ in fig. (\ref{modessr1}). Both solutions develop gradually but more rapidly as they approach the horizon crossing at $-kc_{s}\tau=1$, starting with the solutions located far in the super-Horizon. Deep into the sub-Horizon phase, the solutions continue to grow slowly after crossing. From the curvature perturbation solution in the super-Horizon, the conjugate momenta solution is quite tiny, and it stays that way throughout its evolution in the sub-Horizon.  

\subsubsection{Region II: Ultra Slow Roll (USR) region}
\begin{figure*}[htb!]
    	\centering
    \subfigure[]{
      	\includegraphics[width=8.5cm,height=7.5cm]{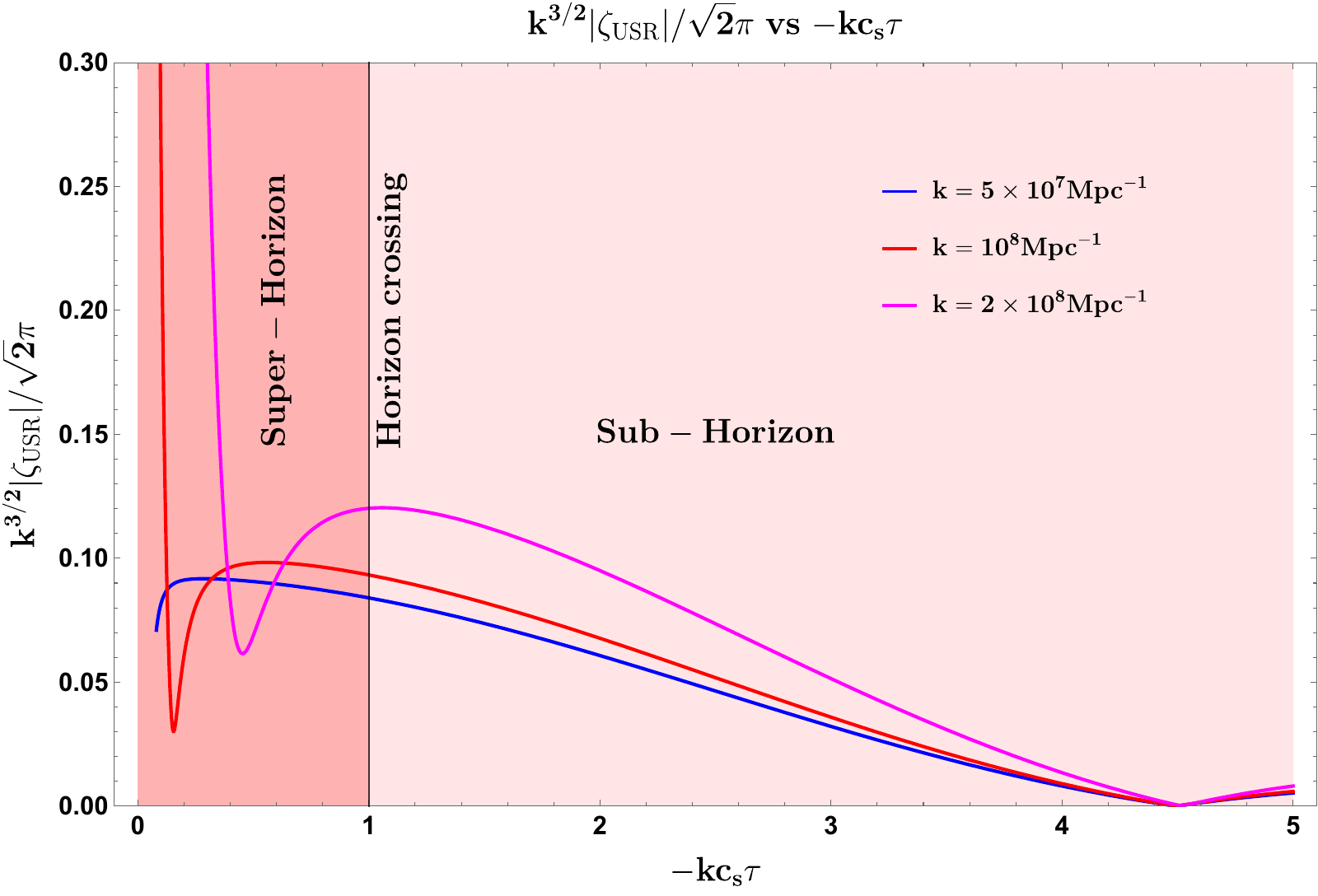}
        \label{zetausr}
    }
    \subfigure[]{
       \includegraphics[width=8.5cm,height=7.5cm]{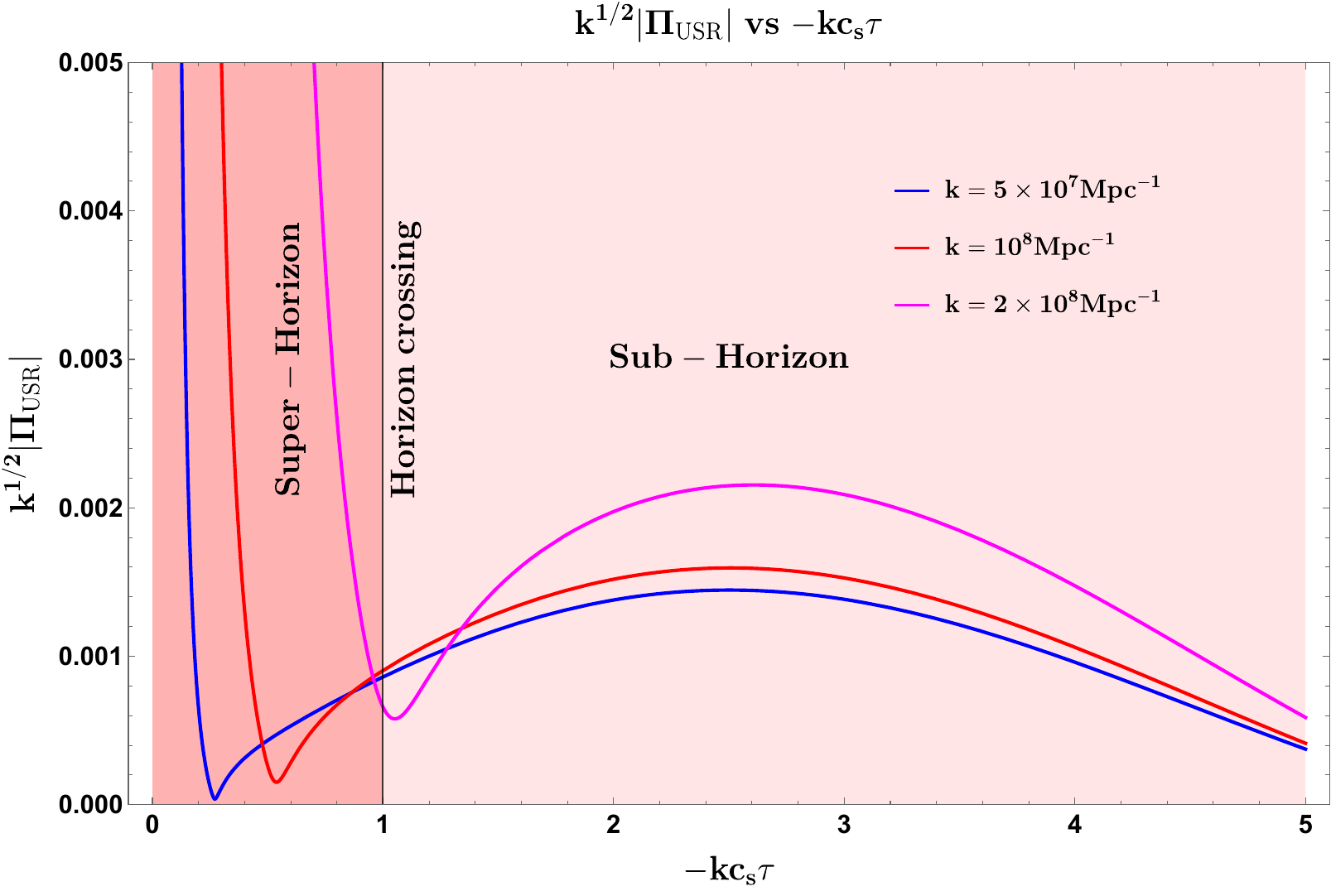}
        \label{piusr}
    }
    	\caption[Optional caption for list of figures]{Plotting against $-kc_{s}\tau$ in the USR phase are $k^{3/2}\abs{\zeta_{\mbf{k}}}/\sqrt{2}\pi$ (\textit{left panel}) and $k^{1/2}\abs{\Pi_{\mbf{k}}}/\sqrt{2}\pi$ (\textit{right panel}). The mode solution behaviour is depicted on the left, moving between the sub-Horizon ($-kc_{s}\tau \gg 1$) and super-Horizon ($-kc_{s}\tau \ll 1$) regimes, as well as crossing the horizon ($-kc_{s}\tau = 1$). Likewise, the conjugate momenta of the mode solution's progression are depicted in the right across each of the three previously described regimes. When mode $k=5\times 10^{7}{\rm Mpc}^{-1}$ (blue), $k=10^{8}{\rm Mpc}^{-1}$ (red), and $k=2\times 10^{8}{\rm Mpc}^{-1}$ (magenta) are fixed, both panels display evolution. } 
    	\label{modesusr}
    \end{figure*}
In this USR phase, the following expression provides the general solution for the curvature perturbation modes and the corresponding canonically conjugate momentum allowed within the conformal time interval: $\tau_{s} \leq \tau \leq \tau_{e}$; in e-foldings, this is ${\cal N}_{s} \leq {\cal N}{e}$:
\bea \label{s53}
    \zeta_{k}(\tau)&=&\left(\frac{iH^{2}}{2\sqrt{\cal A}}\right)\left(\frac{\tau_{s}}{\tau}\right)^{3}\frac{1}{(c_{s}k)^{3/2}}\times \left\{\alpha^{(2)}_{ k}\left(1+ikc_{s}\tau\right)\exp{\left(-ikc_{s}\tau\right)} - \beta^{(2)}_{ k}\left(1-ikc_{s}\tau\right)\exp{\left(ikc_{s}\tau\right)} \right\}, \\
\label{s54}
\Pi_{k}(\tau) &=&\zeta^{'}_{k}(\tau)= \left(\frac{iH^{2}}{2\sqrt{\cal A}}\right)\frac{1}{(c_{s}k)^{3/2}}\frac{\tau_{s}^{3}}{\tau^{4}}\left\{\alpha^{(2)}_{k}(k^{2}c_{s}^{2}\tau^{2}-3(1+ikc_{s}\tau))\exp{\left(-ikc_{s}\tau\right)}\right.\nonumber\\
&&\left.\quad\quad\quad\quad\quad\quad\quad\quad\quad\quad\quad\quad\quad\quad- \beta^{(2)}_{k}(k^{2}c_{s}^{2}\tau^{2}-3(1-ikc_{s}\tau))\exp{\left(ikc_{s}\tau\right)} \right\}.
\quad\quad \eea
The solutions provided above for the modes and their conjugate momenta result in the introduction of two additional Bogoliubov coefficients. The parameters $c_{s}=c_{s,*}$ and $\beta^{(2)}_{k}$, as well as $\alpha^{(2)}_{k}$, stay satisfied throughout the phase, but at the time of the two sharp transitions, it takes on the value of $c_{s}=\tilde{c_{s}}=1\pm \delta$.
Additionally, we must use the $\epsilon_{\rm USR}$ definition in this case as:
\bea
\epsilon_{\rm USR}(\tau) = \epsilon_{\rm SRI}\left(\frac{\tau}{\tau_{s}}\right)^{6}.
\eea
Once the modes at the conformal time of sharp transition $\tau=\tau_{s}$ are subjected to the continuity and differentiability boundary conditions—collectively referred to as the Israel junction conditions—the new Bogoliubov coefficients are found. The following is the expression for these coefficients in their final form:
\bea
    \label{s5b1}\alpha^{(2)}_{k}&=&1+\frac{3k_{s}^{3}}{2 i k^{3}}\left(1+\left(\frac{k}{k_{s}}\right)^{2}\right),\\
    \label{s5b2}\beta^{(2)}_{ k}&=&\frac{3k_{s}^{3}}{2 i k^{3}}\left(1-i\left(\frac{k}{k_{s}}\right)^{2}\right)^{2}\; \exp{\left(2i\frac{k}{k_{s}}\right)}.
\eea
In this case, the wavenumber at the abrupt transition scale at conformal time $\tau_{s}$ is denoted by $k_{s}$. For the current USR phase, the shifted quantum vacuum state described by the new $\alpha^{(2)}_k$ and $\beta^{(2)}_k$ differs from the previously selected Bunch-Davies vacuum state and will continue to be important for the scalar power spectrum analysis. Due to our choice of transition, the $\eta_{\rm USR}$ parameter experiences a dramatic spike to the value $\eta_{\rm USR} \sim -6$, while the $\epsilon_{\rm USR}$ parameter takes on extremely tiny values. 

Using eqn. (\ref{s53},\ref{s54}) to change $-kc_{s}\tau$, the behaviour for the mode solution and associated conjugate momenta are shown in fig. (\ref{modesusr}). Since our variable of interest is $-kc_{s}\tau$, which further introduces a wavenumber dependency in the plots, this analysis necessitates setting $-k_{s}c_{s}\tau_{s} \sim {\cal O}(-0.01)$ for the nature depicted. When concentrating on the super-Horizon in the left panel, the curvature perturbation solution approaches $-kc_{s}\tau \ll 1$ asymptotically. The evolution of many wavenumbers from the super-Horizon to the sub-Horizon regime has been depicted. The solution peaks as it approaches the horizon crossing and then trails off as it approaches the sub-horizon. In the super-Horizon, lower wavenumbers peak distant from the horizon crossing, whereas larger wavenumbers peak near to it. The modes exhibit strongly suppressed amplitude oscillations as we move farther into the horizon, which is caused by the Bogoliubov coefficients. In the right panel, we observe a slightly similar behaviour for the conjugate momenta: they reach their maximum value after entering the sub-Horizon, asymptote abruptly in the super-Horizon, and decline rapidly at the horizon crossing. Here, we see that as we move within the sub-Horizon, the amplitudes get bigger for larger wavenumbers and vice versa for lesser wavenumbers. Until the point of crossing, the conjugate momenta in the super-Horizon exhibit a significant amplitude.

\subsubsection{ Region III: Second Slow Roll (SRII) region}

\begin{figure*}[htb!]
    	\centering
    \subfigure[]{
      	\includegraphics[width=8.5cm,height=7.5cm]{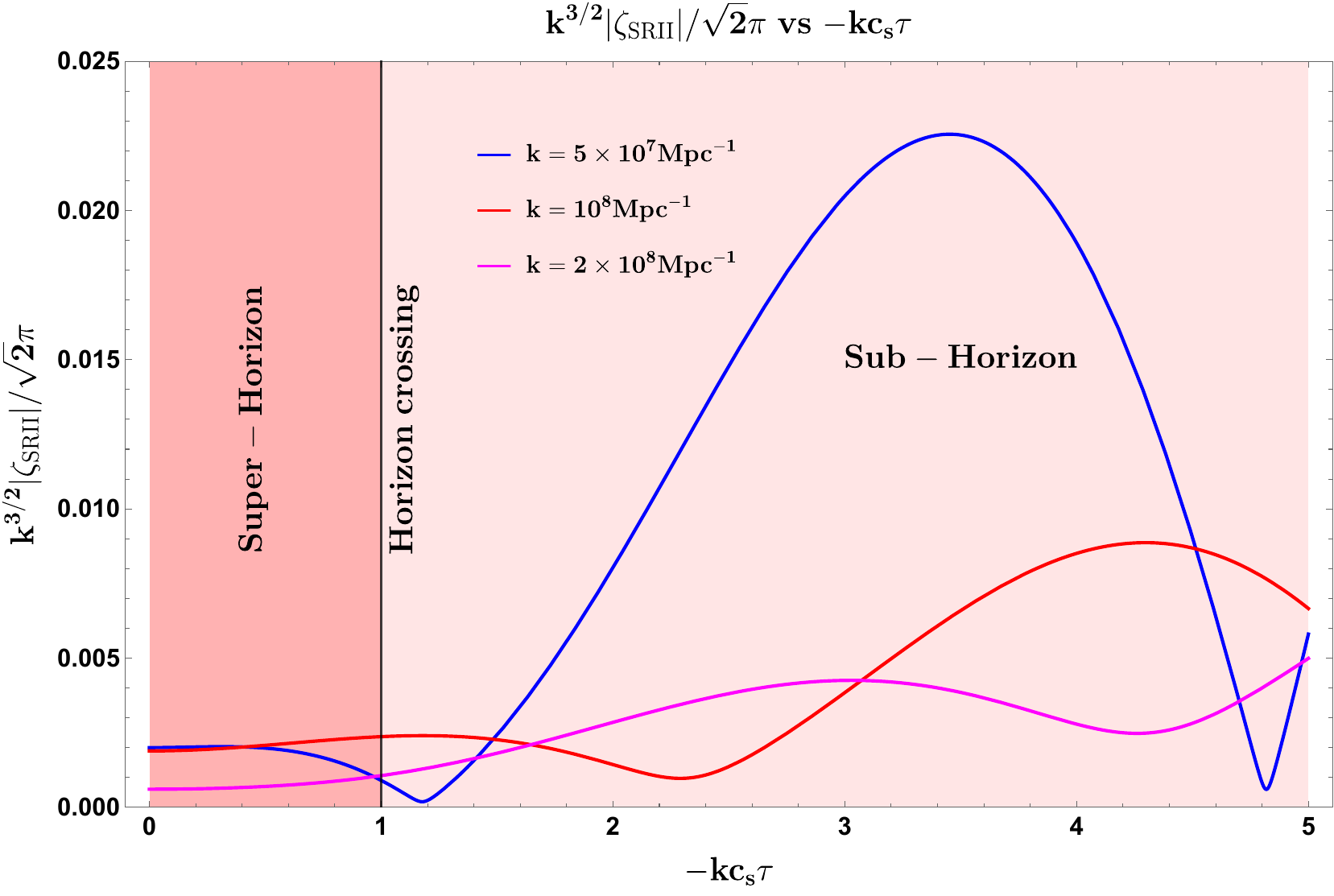}
        \label{zetasr2}
    }
    \subfigure[]{
       \includegraphics[width=8.5cm,height=7.5cm]{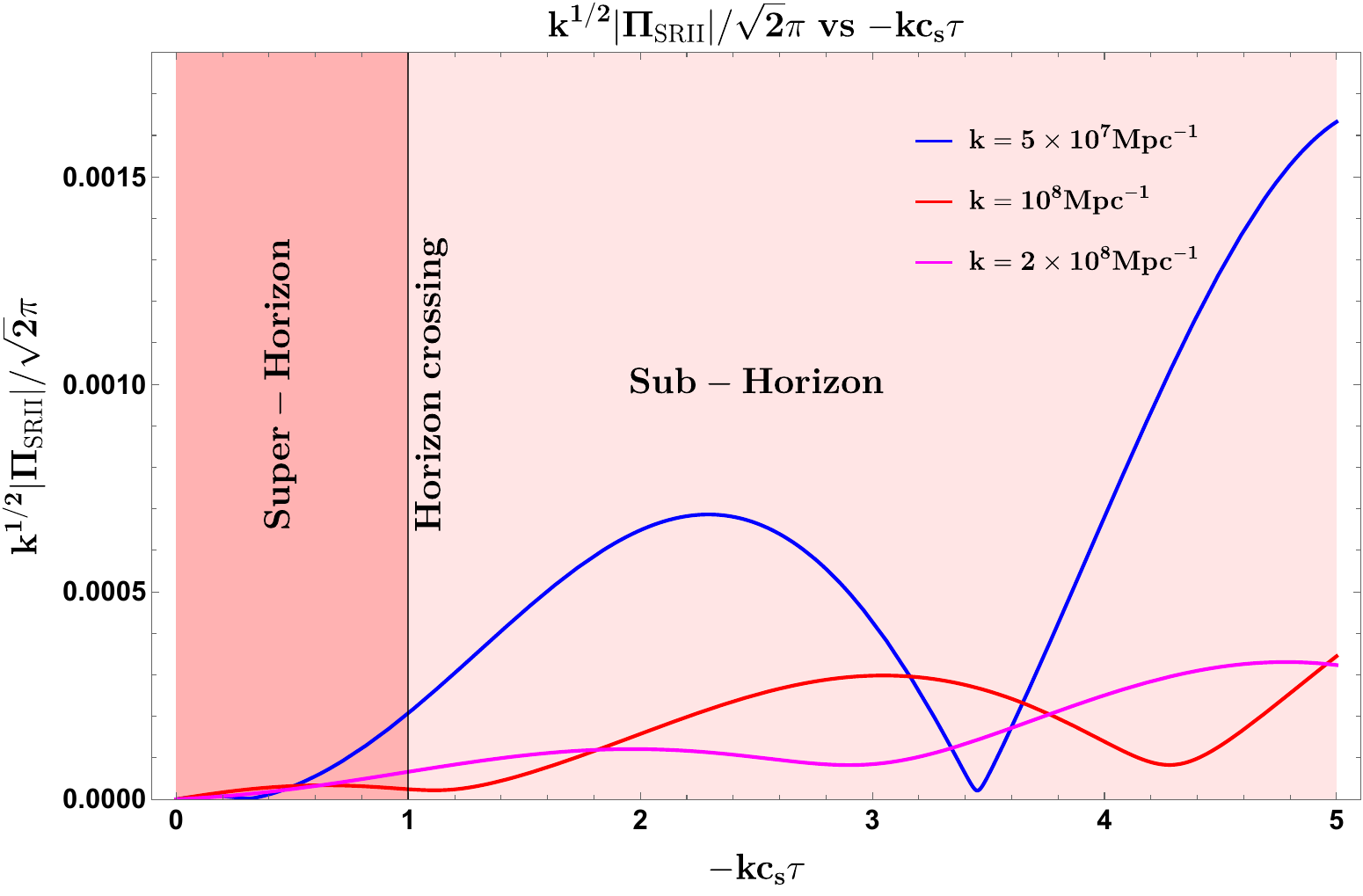}
        \label{pisr2}
    }
    	\caption[Optional caption for list of figures]{Plotting against $-kc_{s}\tau$ in the SRII phase are $k^{3/2}\abs{\zeta_{\mbf{k}}}/\sqrt{2}\pi$ (\textit{left panel}) and $k^{1/2}\abs{\Pi_{\mbf{k}}}/\sqrt{2}\pi$ (\textit{right panel}). The mode solution behaviour is depicted on the left, moving between the sub-Horizon ($-kc_{s}\tau \gg 1$) and super-Horizon ($-kc_{s}\tau \ll 1$) regimes, as well as crossing the horizon ($-kc_{s}\tau = 1$). Likewise, the conjugate momenta of the mode solution's progression are depicted in the right across each of the three previously described regimes. When the modes $k=5\times 10^{7}{\rm Mpc}^{-1}$ (blue), $k=10^{8}{\rm Mpc}^{-1}$ (red), and $k=2\times 10^{8}{\rm Mpc}^{-1}$ (magenta) are fixed, the evolution is displayed in both panels. } 
    	\label{modessr2}
    \end{figure*}
This phase is still the final in our setup, operating inside the conformal time frame $\tau_{e} \leq \tau \leq \tau_{\rm end}$, or in e-foldings ${\cal N}_{e} \leq {\cal N} \leq {\cal N}_{\rm end}$, where the inflation ends at time $\tau_{\rm end}$. Herein lies the universal solution for the development of curvature perturbation modes and the related canonically conjugate momentum created during this phase:
\bea \label{s55}
    \zeta_{k}(\tau)&=&\left(\frac{iH^{2}}{2\sqrt{\cal A}}\right)\left(\frac{\tau_{s}}{\tau_{e}}\right)^{3}\frac{1}{(c_{s}k)^{3/2}}\times \left\{\alpha^{(3)}_{ k}\left(1+ikc_{s}\tau\right)\exp{\left(-ikc_{s}\tau\right)} - \beta^{(3)}_{ k}\left(1-ikc_{s}\tau\right)\exp{\left(ikc_{s}\tau\right)} \right\}. \\
\label{s56}
\Pi_{k}(\tau) &=&\zeta^{'}_{k}(\tau)= \left(\frac{iH^{2}}{2\sqrt{\cal A}}\right)\frac{1}{(c_{s}k)^{3/2}}\left(\frac{\tau_{s}}{\tau_{e}}\right)^{3}\times \frac{k^{2}c_{s}^{2}\tau^{2}}{\tau}\left\{\alpha^{(3)}_{k}\exp{\left(-ikc_{s}\tau\right)} - \beta^{(3)}_{k}\exp{\left(ikc_{s}\tau\right)} \right\},
\eea
which reveals the new set of Bogoliubov coefficients, $\alpha^{(3)}_{k}$ and $\beta^{(3)}_{k}$. It also shows that the parameter $c_{s}=\tilde{c_{s}}$ is only satisfied at $\tau=\tau_{e}$, and that $c_{s}=c_{s,*}$ is the value that follows after the value returns. The following definition of $\epsilon$ must also be used in order to complete the previous solution:
\bea
\epsilon_{\rm SRII}(\tau) = \epsilon_{\rm SRI}\left(\frac{\tau_{e}}{\tau_{s}}\right)^{6}.
\eea
The Bogoliubov coefficients must be solved by utilising the Israel junction conditions at the boundary, where the conformal time $\tau=\tau_{e}$ is used to determine which modes escape the USR and which enter SRII. Another abrupt change occurs at this point, as the quantum vacuum state moves away from the circumstances of USR. The expressions that are produced are as follows: 
\bea
    \label{s5c1} \alpha^{(3)}_{ k}&=&-\frac{k_{s}^{3}k_{e}^{3}}{4k^6}\Bigg[9 \left(-\frac{k}{k_e}+i\right)^2 \left(\frac{k}{k_s}+i\right)^2 \exp{\left(2 i k
   \left(\frac{1}{k_s}-\frac{1}{k_e}\right)\right)}\nonumber\\
    &&\quad\quad\quad\quad\quad\quad\quad\quad\quad\quad\quad\quad\quad\quad-
    \left\{\left(\frac{k}{k_{e}}\right)^2\left(-2\frac{k}{k_{e}}-3i\right)-3i\right\}\left\{\left(\frac{k}{k_{s}}\right)^2\left(-2\frac{k}{k_{s}}+3i\right)+3i\right\}\Bigg],\\
    \label{s5c2}
    \beta^{(3)}_{ k}&=&\frac{3k_{s}^{3}k_{e}^{3}}{4k^6}\Bigg[\left(\frac{k}{k_{s}}+i\right)^2\left\{\left(\frac{k}{k_{e}}\right)^{2}\left(2i\frac{k}{k_{e}}\right)+3\right\}\exp{\left(2i\frac{k}{k_{s}}\right)}\nonumber\\
    &&\quad\quad\quad\quad\quad\quad\quad\quad\quad\quad\quad\quad\quad\quad\quad\quad+i\left(\frac{k}{k_{e}}+i\right)^2\left\{3i+\left(\frac{k}{k_{s}}\right)^{2}\left(-2\frac{k}{k_{s}}+3i\right)\right\}\exp{\left(2i\frac{k}{k_{e}}\right)}\Bigg].
\eea
where the wavenumbers corresponding to the steep transition scales are $k_{s}$ and $k_{e}$. At the conclusion of inflation, the $\epsilon_{\rm SRII}$ parameter deviates from constant behaviour and increases from its prior values in the USR to ${\cal O}(1)$. At $\tau=\tau_{e}$ in the USR, the $\eta_{\rm SRII}$ parameter abruptly jumps from $\eta_{\rm USR} \sim -6$ to $\eta_{\rm SRII} \sim -1$, where it stays until inflation stops.

\subsection{ Tree level power spectrum from comoving curvature perturbation}

We covered the semi-classical solutions for the comoving curvature perturbation modes for each of the three phases in the previous section. The tree-level scalar power spectrum may now be calculated using the knowledge of the scalar modes and the Bogoliubov coefficients that correspond with them. Quantizing the curvature perturbation modes and calculating the tree-level contribution to the two-point correlation function are necessary steps in building this power spectrum. In this process, a set of creation and annihilation operators $\hat{a}_{\mbf{k}}$ and $\hat{a}^{\dagger}_{\mbf{k}}$ are introduced. These operators, when applied to the quantum vacuum state $\ket{0}$ of the Hilbert space, either produce an excited state or destroy it. Now that they are elevated to operators, the curvature perturbation modes may be expressed as follows:
\bea
\hat{\zeta}_{\mbf{k}}(\tau) = \zeta_{k}\hat{a}_{\mbf{k}} + \zeta^{*}_{k}a^{\dagger}_{-\mbf{k}}\quad\quad\quad\quad \hat{\Pi}_{\mbf{k}}(\tau) = \Pi_{k}\hat{a}_{\mbf{k}} + \Pi^{*}_{k}a^{\dagger}_{-\mbf{k}}
\eea
using $\Pi_{k} = \partial_{\tau}\zeta_{k}$ as the conjugate momentum. 
now, the two-point correlation function may be expressed as follows:
\bea \langle \hat{\zeta}_{\bf k}\hat{\zeta}_{{\bf k}^{'}}\rangle_{{\bf Tree}} =(2\pi)^{3}\;\delta^{3}\left({\bf k}+{\bf k}^{'}\right)\frac{2\pi^2}{k^3}\Delta^{2}_{\zeta,{\bf Tree}}(k),\quad\quad \text{where} \quad \Delta^{2}_{\zeta,{\bf Tree}}(k) =  \frac{k^{3}}{2\pi^{2}}|{\zeta}_{\bf k}(\tau)|^{2}_{\tau\rightarrow 0}.
\eea
It gives us the pertinent tree-level contribution determined by using $\tau \rightarrow 0$ as the late-time limit. For a given tree-level scalar power spectrum, the dimensionless term is $\Delta^{2}_{\zeta,{\bf Tree}}(k)$. Using the mode solutions previously indicated in eqs. (\ref{s51},\ref{s53},\ref{s55}) for the three phases of interest in our arrangement, we can now discuss the tree-level scalar power spectrum. The dimensionless scalar power spectrum can be expressed in the following final form:
\bea  \label{treespec}
\Delta^{2}_{\zeta,{\bf Tree}}(k)
&=& \displaystyle
\displaystyle\left\{
	\begin{array}{ll}
		\displaystyle\Delta_{\bf {Tree}}^{\text{SRI}}(k)=\left(\frac{H^{4}}{8\pi^{2}{\cal A} c^3_s}\right)_* \Bigg\{1+\Bigg(\frac{k}{k_s}\Bigg)^2\Bigg\}& \mbox{when}\quad  k\leq k_s  \;(\rm SRI)  \\  
			\displaystyle 
			\displaystyle\Delta_{\bf {Tree}}^{\text{USR}}(k)=\left(\frac{H^{4}}{8\pi^{2}{\cal A} c^3_s}\right)_* \left(\frac{k_e}{k_s }\right)^{6}\left|\alpha^{(2)}_{\bf k}-\beta^{(2)}_{\bf k}\right|^2 & \mbox{when }  k_s\leq k\leq k_e  \;(\rm USR)\\ 
   \displaystyle 
			\displaystyle\Delta_{\bf {Tree}}^{\text{SRII}}(k)=\left(\frac{H^{4}}{8\pi^{2}{\cal A} c^3_s}\right)_* \left(\frac{k_e}{k_s }\right)^{6}\left|\alpha^{(3)}_{\bf k}-\beta^{(3)}_{\bf k}\right|^2 & \mbox{when }  k_e\leq k\leq k_{\rm end}  \;(\rm SRII) 
	\end{array}
\right. \eea
where we apply the super-Horizon limit conditions for the modes that obey $-kc_{s}\tau \ll 1$ and cross the Hubble horizon. Based on the aforementioned, we describe below the overall tree-level scalar power spectrum, which is the consequence of the various phases' contributions being connected by a Heaviside Theta function to indicate the existence of abrupt transitions in our configuration: 
\bea \label{totpower} \left[\Delta^{2}_{\zeta,{\bf Tree}}(k)\right]_{\bf Total} &=&\Delta_{\bf {Tree}}^{\text{SRI}}(k)+\Delta_{\bf {Tree}}^{\text{USR}}(k)\Theta(k-k_s)+\Delta_{\bf {Tree}}^{\text{SRII}}(k)\Theta(k-k_e).
\eea
In the next sections, additional analysis will need the whole power spectrum. 
The amplitude of the aforementioned scalar power spectrum in each of the three phases is another limitation on the Galileon EFT coefficients $c_{4},c_{5}$, as we have already noted in the analysis of the preceding section. This is something we must reiterate here.

\subsection{Primordial non-Gaussianity from ultra slow-roll phase}

\subsubsection{General introduction to Primordial Non-Gaussianity from comoving scalar perturbation modes}

We give a broad overview of primordial non-Gaussianities in this section. 
They are locally expressed as deviations from the Gaussian distribution of the primordial density perturbations, and they are defined using the Gaussian random fields $\zeta_{g}(x)$, which meet the Gaussian statistics. As the first vacuum fluctuations from inflation leave the Horizon, they are elevated to classical disturbances. Each point in the super-horizon regime can be addressed independently locally, and its development will be dictated by initial conditions expressed as Gaussian distributions. It is now possible to express the observed curvature perturbations, up to first order, as a term linear in the perturbations that the Gaussian random fields describe. As a result, the non-linearities will indicate whether non-Gaussianities are present in the observations. Higher-order correlation functions would have to avoid being zero in order to achieve this. In this study, we have used the bispectrum or the three-point correlation function for the gauge invariant scalar modes to assess non-Gaussianities. The following statement serves as a mathematical representation of their non-linear nature:
\bea 
\zeta(\bold{x}) = \zeta_{g}(\bold{x}) + \frac{3}{5}f_{\text{NL}}\big(\zeta^{2}_{g}(\bold{x}) - \langle \zeta^{2}_{g}(\bold{x}) \rangle\big)+\cdots\cdots
\eea
where the primordial comoving curvature perturbation is denoted by $\zeta(\bold{x})$. Here, $\fnl$ stands for the model-dependent local amplitude of the non-Gaussianity. The value of $\fnl = -0.9 \pm 5.1$ at $68\%$ level of confidence is derived from CMB measurements, according to the most recent observational estimates \cite{Planck:2019kim}. A physically definite assertion cannot now be made due to the high error bars in the measurements. Nevertheless, the new observations are predicted to offer improvements over the existing estimates by an order of magnitude using data from the upcoming surveys \cite{Achucarro:2022qrl,CMB-S4:2023zem}.

Separating those density perturbations into a large-scale component—which shows noticeable changes at scales similar to the long modes' wavelengths—and a small-scale component—which already exhibits significant deviations on scales smaller than the long modes—provides another intriguing aspect of understanding primordial non-Gaussianities. Additionally, this may be expressed in the following manner using the aforementioned decomposition, $\zeta_{g} = \zeta_{g,L} + \zeta_{g,S}$ and $\zeta = \zeta_{L} + \zeta_{S}$: 
\bea \zeta_{S}(\bold{x}) &=& \zeta_{g,S}(\bold{x})\left(1 + \frac{6}{5}\fnl\zeta_{g,L}(\bold{x})\right) + \frac{3}{5}f_{\text{NL}}\zeta_{g,S}(\bold{x})^{2},\\
\zeta_{L}(\bold{x}) &=& \zeta_{g,L}(\bold{x}) + \frac{3}{5}\fnl\zeta_{g,L}(\bold{x})^{2}.
\eea
where the link between the small- and large-scale cosmic fluctuations is non-trivial as given by the first equation. The topic raised by these discussions is crucial: Is there a theoretical model that can demonstrate the production of non-Gaussianities of the order $|f_{\text{NL}}| \sim {\cal O}(1)$? This question takes on additional significance as it can provide valuable insights into the origin of structure in the very early universe. These features are discovered while examining the cosmological perturbations created in the very early universe and, astonishingly, their observational imprints are also discernible in the form of the CMB data. 

In accordance with the standard single field inflation model, Maldacena arrived at a consistency condition \cite{Maldacena:2002vr}:
\bea f_{\rm NL}=\frac{5}{12}\left(1-n_s\right),\eea
This is sometimes called a {\it no-go theorem} and states that, in the squeezed limit, the amplitude of non-Gaussianity, $\fnl$, is connected to the spectral index of the primordial power spectrum of scalar modes and is estimated to be ${\cal O}(10^{-2})$. This is a very small value when one considers the detection of primordial non-Gaussianities in cosmological observations.

Producing a significant number of non-Gaussianities from inflation single field models is undoubtedly a difficult undertaking. In order to accomplish this intriguing goal, we focus entirely on studying the generation of large non-Gaussianities in this paper by explicitly breaking the {\it no-go theorem} of Maldacena in the squeezed limit of the three-point cosmological correlation function for the scalar modes, utilising the underlying theory of Galileon inflation. A technique that enables us to obtain a significant amount of non-Gaussian amplitude from this calculation will be presented in the next section. In order to achieve the previously described goal, we will present the three phases: SRI, USR, and SRII, as well as the abrupt changes in phase that occur at the limits of the USR phase near its start and finish. By using our computation, we will demonstrate clearly that the cosmological three-point function can produce a large non-Gaussianity amplitude from sharp transitions and transition scale positions described in terms of small wave numbers. This is achieved by breaking the previously mentioned {\it no-go theorem} in the squeezed limit, specifically in the USR and SRII phases.

\subsubsection{Computing the three-point function and the associated bispectrum from scalar modes}

In this part, we compute the tree-level three-point correlation function and, correspondingly, the related bispectrum for each of the three SRI, USR, and SRII areas using the mode information that was acquired in the previous section. The validity of the consistency condition for single-field inflation in the current Galileon inflation theory is likewise our main concern, covering all three areas with particular emphasis on the USR region, which will provide insight into the extent of this condition when taking PBH creation into account. Therefore, we investigate the squeezed limit behaviour of the bispectrum precisely. The popular technique referred to as the {\it Schwinger-Keldysh formalism}, or {\it in-in formalism}, is utilised to compute the bispectrum and related non-Gaussianity amplitude. It is also used to introduce the expectation value for the necessary three-point correlation function in conjunction with the third-order action of Covariantized Galileon Theory (CGT).

\subsubsubsection{The Schwinger-Keldysh formalism}

We are more interested in the expectation of operators at a given instant in time, and the corresponding boundary conditions for the scalar modes are imposed only at very early times, so the approach from {\it in-in formalism} is more appropriate to evaluate the expectation value of the necessary three-point correlation function. The formalism known as {\it in-in} is called for this reason primarily. 

Next, using this approach, the time-dependent expectation value for a certain operator in the interaction picture is provided as follows:
\bea \bra{\Omega(\tau)}Q^{I}(\tau)\ket{\Omega(\tau)}=\bra{0}\bigg[\overline{T}\exp\Bigg(\displaystyle{i\int_{-\infty(1-i\epsilon)}^{\tau}H_{\text{Int}}^{I}(\tau')d\tau'}\Bigg)\bigg]Q^{I}(\tau)\bigg[T\exp\Bigg(\displaystyle{-i\int_{-\infty(1+i\epsilon)}^{\tau}H_{\text{Int}}^{I}(\tau')d\tau'}\Bigg)\bigg]\ket{0}. \eea
where the Heisenberg picture operator is denoted by $Q^{I}(\tau)$, and the far-in the past interacting and free theory vacuums are represented by $\ket{\Omega}$ and $\ket{0}$, respectively and it is represented by:
\bea \ket{\Omega(\tau)}=\bigg[T\exp\Bigg(\displaystyle{-i\int_{-\infty(1+i\epsilon)}^{\tau}H_{\text{Int}}^{I}(\tau')d\tau'}\Bigg)\bigg]\ket{0}.\eea
Additionally, $T$ and $\overline{T}$ indicate to us whether the operators are time- or anti-time-ordered. The $\tau_{0} \rightarrow -\infty$ takes the early time limit, and the $i\epsilon$ prescription regularises the integral after that. In conclusion, the interaction image includes the operators and the fields constituting the interaction Hamiltonian, as shown by the superscript ${I}$. To handle the non-linearities that come from the interactions in the overall Hamiltonian and enter the equations of motion, this image is presented. For any operator, the expected value may be calculated by first developing the fields from the early past to the relevant point in time, and then from that point back to the beginning of time.

In order to obtain the tree-level contribution, which is what we are interested in for our additional calculations, we now extend this equation and pick the leading order term when expanding in $H_{\text{Int}}^{I}$. Next, the leading order word is provided by:
\bea \label{a3} \bra{\Omega(\tau)}Q^{I}(\tau)\ket{\Omega(\tau)} = -i\int_{-\infty}^{\tau}d\tau'\bra{0}[Q^{I}(\tau),H_{\text{Int}}^{I}(\tau')]\ket{0} = 2\times\text{Im}\left[\int_{-\infty}^{\tau}d\tau'\bra{0}Q^{I}(\tau)H^{I}_{\text{Int}}(\tau')\ket{0}\right],
\eea
where the Hermiticity condition is applied in the last equality to further reduce the commutator form. Regarding the subject matter of this work, $Q^{I}(\tau)$ represents the curvature perturbation's three-point correlation, expressed as follows: $\hat{\zeta}(\bf{k}_1,\tau)\hat{\zeta}(\bf{k}_2,\tau)\hat{\zeta}(\bf{k}_3,\tau)$.

The evaluation of the correlation function starts with the use of the interaction Hamiltonian, which is also formed by the interaction picture fields. We do this by performing contractions into a product of green's functions using the commutation relations for the fields for the quantized curvature perturbation. The absence of an analogue for the Feynman propagator in an inflationary background is the key distinction between the conventional {\it in-out} and the {\it in-in formalism} in this case. Therefore, while executing contractions, we must have it in mind.

In order to assess correlation functions using the Wick contraction approach, the fields are first ordered normally, with the negative frequency operators being on the left and the positive frequency modes being retained to the right of the product in order to annihilate the vacuum. This arrangement will be similar under normal ordering for both the in-out and in-in cases. When we next begin to perform the contraction, we encounter the previously mentioned expression for the two-point cosmological correlation of the quantized curvature perturbation $\langle\hat{\zeta}(\bf{k},\tau)\hat{\zeta}(\bf{k'},\tau)\rangle$ which turns out to be a real quantity, more equivalently an absolute value squared of the mode functions, contrary to the imaginary quantity of the Feynman propagator in case of the {\it in-out formalism}. Combining both normal ordering and what we learned from the behavior of contractions from above, we can say that the expectation value of a string of field operators evaluated using {\it Wick's theorem} is written as:
\bea \bra{0}\hat{\zeta_1}\hat{\zeta_2}\cdots\hat{\zeta_n}\ket{0} = \bra{0}:\hat{\zeta_1}\hat{\zeta_2}\cdots\hat{\zeta_n}:\ket{0} + \textit{all possible contractions}.
\eea
where \textit{all possible contractions} and $\zeta_i = \zeta(\bold{k}_i,\tau_i)$ denote the inclusion of one contraction term for each possible pairing of the $n$ operators. We will explicitly calculate the three-point correlation function for each of the various SRI, USR, and SRII areas using the techniques we've learned here. To be able to perform the calculations in the future, we must first comprehend the Covariantized Galileon Theory's third-order action. As a result, we go into further depth on this move for the CGT in the next section.

\subsubsubsection{Third order action of comoving curvature perturbation and Non-renormalization theorem}

Note that in a generic field theoretic situation, the non-renormalization \footnote{One should distinguish between the non-renormalization theorem and a theory's non-renormalizability. As an example, four Fermi-theory illustrates how the number of counter terms needed in a theory that is not renormalizable is not finite, while the former suggests that couplings in the underlying theory are stable under radiative corrections and do not run. The structure of divergences that arises in lower orders of perturbation theory, on the other hand, keeps recurring in higher orders in renormalizable theories, so that a finite number of counter terms is sufficient.
} theorem suggests that there are no radiative corrections applied to the underlying theory couplings. Remarkably, this Galileon symmetry-related quality holds true in this scenario more strongly than it does in the other theories; for instance, a little symmetry rupture maintains the aforementioned property. Transitioning from the de Sitter to the quasi-de Sitter regime achieves a modest breach of Galileon symmetry. Second, as long as the Galilean symmetry is maintained, the Galileon operators remain unrenormalized even when linked to strong external fields (See refs. \cite{Burrage:2010cu,Goon:2016ihr} for details).

Before going into the technical intricacies of the supporting information, let us first discuss the fundamental cause of the lack of radiative adjustments to the power spectrum. For standard single field inflation \footnote{For typical single field inflation, the third order action is given by \cite{Choudhury:2023vuj,Choudhury:2023jlt,Choudhury:2023rks}:
\bea &&S^{(3)}_{\zeta}=\int d\tau\;  d^3x\;  M^2_{ pl}a^2\; \bigg(\left(3\left(c^2_s-1\right)\epsilon+\epsilon^2\right)\zeta^{'2}\zeta+\frac{\epsilon}{c^2_s}\bigg(\epsilon+1-c^2_s\bigg)\left(\partial_i\zeta\right)^2\zeta-\frac{2\epsilon}{c^2_s}\zeta^{'}\left(\partial_i\zeta\right)\left(\partial_i\partial^{-2}\left(\frac{\epsilon\zeta^{'}}{c^2_s}\right)\right)\nonumber\\
&&\quad\quad\quad\quad\quad\quad\quad\quad\quad-\frac{1}{aH}\left(1-\frac{1}{c^2_{s}}\right)\epsilon \bigg(\zeta^{'3}+\zeta^{'}(\partial_{i}\zeta)^2\bigg)
 +\frac{1}{2}\epsilon\zeta\left(\partial_i\partial_j\partial^{-2}\left(\frac{\epsilon\zeta^{'}}{c^2_s}\right)\right)^2+\underbrace{\frac{1}{2c^2_s}\epsilon\partial_{\tau}\left(\frac{\eta}{c^2_s}\right)\zeta^{'}\zeta^{2}}+\cdots\bigg),\quad\quad\eea
 where the considerable one loop contribution to the power spectrum for the scalar modes is exclusively attributable to the final highlighted component in the preceding formula. This kind of contribution is strictly absent in the case of Galileon because of its slightly violated Galilean symmetry, protecting the power spectrum from the radiative adjustments. We have covered this topic in great length throughout this article and the remainder of the paper.
}, the significant one-loop contribution to the power spectrum is caused by the $\zeta^{'}\zeta^2$ operator in the third-order action for comoving curvature perturbation (with other terms in the action giving insignificant contribution \cite{Choudhury:2023vuj,Choudhury:2023jlt,
Choudhury:2023rks}. This operator has a coefficient proportional to $\eta'$, which is large during a sharp transition. The non-renormalization theorem is respected in single-field Galileon inflation with soft Galilean symmetry breakdown when the aforementioned operator is missing. Loop adjustments are therefore negligible in this instance.  As stated in terms of the field in equation (\ref{GCS}), in fact, under Galilean symmetry, the comoving curvature perturbation is converted by the following expression:

\bea \zeta\rightarrow\zeta-\frac{H}{\dot{\bar{\phi}}_0}\left(b\cdot \delta x\right),\eea
This may be used to change the curvature perturbation's temporal and spatial derivatives as:
\bea \zeta^{'}\rightarrow\zeta^{'}-\frac{H}{\dot{\bar{\phi}}_0}b_0, \quad\quad\quad\quad\quad \partial_i\zeta\rightarrow \partial_i\zeta-\frac{H}{\dot{\bar{\phi}}_0}b_i,\quad\quad\quad\quad\quad \partial^2\zeta\rightarrow \partial^2\zeta.\eea
This may be found in every third order action contribution. It is evident from this that the Galilean symmetry is slightly violated in $\zeta$, $\zeta^{'}$, and $\partial_{i}\zeta$. This is a necessary requirement to use the inflationary paradigm in the current debate setting. However, we can also see that the phrase $\partial^2\zeta$ is entirely protected by Galilean symmetry. This means that in order to gently disrupt the Galilean symmetry, such a contribution has to couple with some additional contributions. Owing to the transformation's properties at the curvature perturbation level, as well as its spatial and temporal derivatives, it is evident that some terms can be eliminated through field redefinition, while others can be expressed as total derivative terms that vanish at the boundary, allowing for very small breaking of the Galilean symmetry to perform inflation. The following contributions have been made: $\zeta^{'2}\zeta$, $\left(\partial_i\zeta\right)^2\zeta$, and $\zeta^{'}\left(\partial_i\zeta\right)\left(\partial_i\partial^{-2}\left(\epsilon\zeta^{'}\right)\right)$. The third order action lacks $\zeta\partial_{\tau}\left(\partial_{i}\zeta\right)^2$, $\zeta\left(\partial_i\partial_j\partial^{-2}\left(\epsilon\zeta^{'}\right)\right)^2$, and $\zeta^{'}\zeta^{2}$ because of the extremely mild breakdown of the Galilean symmetry. The last term, $\zeta^{'}\zeta^{2}$, is the most important of all of these contributions during the USR phase because of its coefficient $\partial_{\tau}\left(\eta/c^2_s\right)$, which makes a large contribution at both the SRI to USR and USR to SRII transition points. Even yet, the contribution from SRI to the USR transition is far greater than the contribution from USR to the SRII transition. These contributions have been demonstrated in refs. \cite{} to have enormous one-loop effects, which make them particularly detrimental to PBH production. Since this contribution is not present when the Galilean shift symmetry is softly broken, it is crucial to clarify how exactly one may use this word, which is considered redundant in the current context. The transformation of $\zeta^{'}\zeta^{2}$ under the previously given transformation is as follows:
\bea \zeta^{'}\zeta^{2}\rightarrow \bigg(\zeta^{'}-\frac{H}{\dot{\bar{\phi}}_0}b_0\bigg)\bigg(\zeta-\frac{H}{\dot{\bar{\phi}}_0}\left(b\cdot \delta x\right)\bigg)^2\sim \frac{1}{3}\left(b\cdot \delta x\right)\partial_{\tau}\left(\zeta^2\right)\sim 0 \quad ({\rm at \; boundary}).\eea
In this case, the contributions $\left(\frac{H}{\dot{\bar{\phi}}_0}\right)^2\left(b\cdot \delta x\right)\zeta^{'}$, $\left(\frac{H}{\dot{\bar{\phi}}_0}\right)^3 b_0\left(b\cdot \delta x\right)^2$, $\left(\frac{H}{\dot{\bar{\phi}}_0}\right)^2 b_0\left(b\cdot \delta x\right)$, $\left(\frac{H}{\dot{\bar{\phi}}_0}\right)\left(b\cdot \delta x\right)\zeta^{'}$ are not going to be involved in the third order action. By employing field redefinition and discarding some contributions at the boundary, one may promptly include them into the coefficients of the second-order perturbed action or eliminate them entirely since the boundary has a whole derivative structure. Thus, $\zeta^{'3}$, $\zeta^{'2}\left(\partial^2\zeta\right)$, $\zeta^{'}\left(\partial_i\zeta\right)^2$, and $\left(\partial_i\zeta\right)^2\left(\partial^2\zeta\right)$ are the contributions that will survive because of the small amount of soft Galilean symmetry break. These can be physically understood as the bulk self interactions of the Galileon in terms of the scalar curvature perturbation. After executing the previously mentioned transformation, several additional terms for each of the surviving contributions mentioned above arise. Of these, some can be thrown away because of their total derivative structure at the boundary, while others can be absorbed by using field redefinition. The remaining contributions, which are proportional to the quadratic or cubic in the amount of slight breaking of Galilean symmetry, can be conveniently ignored in this construction. By carrying out the cosmic perturbation theory in full, it is necessary to create the third order action using such contributions. To learn more about these facets, refer to ref. \cite{Burrage:2010cu}, where the writers have meticulously crafted the third order action by permitting a minor degree of gently disrupting Galilean symmetry.
The following computation will be performed in the current scenario under the assumption that the curvature perturbation grows the CGEFT action in third order. For the same, the following expression may be used to represent the third order action:
\bea &&S^{(3)}_{\zeta}=\int d\tau\;  d^3x\;  \frac{a^2}{H^3}\; \bigg[\frac{{\cal G}_1}{a}\zeta^{'3}+\frac{{\cal G}_2}{a^2}\zeta^{'2}\left(\partial^2\zeta\right)+\frac{{\cal G}_3}{a}\zeta^{'}\left(\partial_i\zeta\right)^2+\frac{{\cal G}_4}{a^2}\left(\partial_i\zeta\right)^2\left(\partial^2\zeta\right)\bigg].\eea
In the third order perturbed action, the coupling parameters ${\cal G}_i\forall i=1,2,3,4$ are represented by the following expressions:
\bea {\cal G}_1:&\equiv& \frac{2H\dot{\bar{\phi}}^3_0}{\Lambda^3}\Bigg(c_3+9c_4Z+30c_5Z^2\Bigg),\\
      {\cal G}_2:&\equiv& -\frac{2\dot{\bar{\phi}}^3_0}{\Lambda^3}\Bigg(c_3+6c_4Z+18c_5Z^2\Bigg),\\ 
       {\cal G}_3:&\equiv& -\frac{2H\dot{\bar{\phi}}^3_0}{\Lambda^3}\Bigg(c_3+7c_4Z+18c_5Z^2\Bigg)+\frac{2\dot{\bar{\phi}}^2_0\ddot{\bar{\phi}}_0}{\Lambda^3}\Bigg(c_3+6c_4Z+18c_5Z^2\Bigg)\nonumber\\
       &=&-\frac{2H\dot{\bar{\phi}}^3_0}{\Lambda^3}\Bigg(c_3+7c_4Z+18c_5Z^2\Bigg)-\frac{2\dot{\bar{\phi}}^3_0H\eta}{\Lambda^3}\Bigg(c_3+6c_4Z+18c_5Z^2\Bigg),\\
        {\cal G}_4:&\equiv& \frac{\dot{\bar{\phi}}^3_0}{\Lambda^3}\bigg\{c_3+3c_4Z+6c_5\bigg[Z^2+\frac{\dot{H}\dot{\bar{\phi}}^2_0}{\Lambda^6}\bigg]\bigg\}+\frac{3\dot{\bar{\phi}}^3_0\ddot{\bar{\phi}}_0}{\Lambda^6}\bigg\{c_4+4c_5Z\bigg\}\nonumber\\
        &=&\frac{\dot{\bar{\phi}}^3_0}{\Lambda^3}\bigg\{c_3+3c_4Z+6c_5\bigg[Z^2+\frac{\dot{H}\dot{\bar{\phi}}^2_0}{\Lambda^6}\bigg]\bigg\}-\frac{3\dot{\bar{\phi}}^4_0H\eta}{\Lambda^6}\bigg\{c_4+4c_5Z\bigg\},\eea
        In this case, equation (\ref{Z}) already defines the component $Z$ before. In order to account for the dominance of Galileon self couplings over all other contributions, and to ignore the mixing contributions from the gravitational background in the decoupling limit, one should examine the area where $Z\gtrsim 1$ occurs. This study limits the scope of our analysis by taking into account $Z\sim 1$, where nonlinear Galileon interactions exist but do not pose a threat to the remaining calculations in this research. Accordingly, one may regulate such contributions at the perturbative level of computation in the region $Z\sim 1$. In the remainder of the review, we are interested in calculating the effect one-loop contribution from the third order action for the comoving curvature perturbation; however, with the same action, it could be quite interesting to compute the Bispectrum and Trispectrum from SRI, USR, and SRII phases and discuss the primordial non-Gaussian \cite{Maldacena:2002vr,Seery:2005wm,Senatore:2009gt,Chen:2006nt,Chen:2010xka,Chen:2009zp,Chen:2009we,Chen:2008wn,Chen:2006xjb,Choudhury:2012whm,Agarwal:2012mq,Holman:2007na,Creminelli:2005hu,Behbahani:2011it,Smith:2009jr,Cheung:2007sv,Creminelli:2006rz,Creminelli:2006gc,Kalaja:2020mkq,Meerburg:2019qqi,Lee:2016vti,Maldacena:2011nz,Werth:2023pfl} features in the current context. 

The coefficients that occur in front of the operators may now be expressed in terms of CGEFT coefficients using the classic EFT construction:
\bea &&\frac{M^4_3}{H^2M^2_{pl}}=\frac{3}{4}\frac{1}{a^2H^4}\left({\cal G}_3-{\cal G}_1\right),\\
&&\frac{\bar{M}^3_1}{HM^2_{pl}}=\frac{2}{3}\bigg[\left(1-\frac{1}{c^2_s}\right)\epsilon+\frac{1}{a^2H^4}{\cal G}_3\bigg].
\eea
Similarly, by comparing term by term in the third order action, one may fix the other two coefficients, ${\cal G}_2$ and ${\cal G}_4$, in terms of the EFT coefficients. 

All of the above-mentioned inputs will now be used especially to extract the correction from the one-loop quantum effect. Three alternative scenarios for the first and second slow-roll parameters, $\epsilon$ and $\eta$, which must be carefully considered during calculation, are shown in the contributions described as appearing in the third order action:
\begin{enumerate}
    \item \underline{\bf Region I (SRI):} The following describes how the two slow roll parameters behave in the first slow-roll regime (SRI):
    \bea \epsilon\sim {\rm Constant},\quad\quad\quad \eta\sim 0\quad\quad\quad {\rm where}\quad\quad\tau<\tau_s.\eea
    The SRI area is represented here by the conformal time scale $\tau<\tau_s$, where the SR to USR transition takes place at $\tau=\tau_s$ scale.

    \item \underline{\bf Region II (USR):} The two slow roll parameters listed above behave as follows under the Ultra Slow-Roll Regime (USR):
    \bea \epsilon(\tau)=\epsilon  \;\left(\frac{\tau}{\tau_s}\right)^{6},\quad\quad\quad \eta\sim -6\quad\quad\quad {\rm where}\quad\quad\tau_s\leq\tau\leq \tau_e.\eea
    The slow-roll parameter in this case is $\epsilon$, as it appears in the phase SRI. The behaviour is handled here at the SRI to USR transition point $\tau=\tau_s$.

    \item \underline{\bf Region III (SRII):} The two slow roll parameters listed above behave as follows in the second slow-roll regime (SRII):
    \bea \epsilon(\tau)=\epsilon  \;\left(\frac{\tau_e}{\tau_s}\right)^{6},\quad\quad\quad \eta\sim 0\quad\quad\quad {\rm where}\quad\quad\tau_e\leq\tau\leq \tau_{\rm end}.\eea
    Here, $\epsilon$ represents the slow-roll parameter as it appears in the SRI phase, as was previously described. The behaviour is handled here at the USR to SRII transition point $\tau=\tau_e$.

\end{enumerate}

It is evident from the formulation of these three phases that the first slow-roll parameter $\epsilon$ behaves smoothly at the transition points from SRI to USR and USR to SRII, at $\tau=\tau_s$ and $\tau=\tau_e$, respectively. Conversely, the functional form that follows may be used to parametrize the second slow-roll parameter $\eta$ around the transition locations that have been mentioned:
\bea \eta(\tau)=-6-\Delta\eta\left[\Theta(\tau-\tau_s)-\Theta(\tau-\tau_e)\right].\eea
It suggests that $\eta\sim -6-\Delta\eta=0$ in the SRII area since $\Delta\eta\sim -6$ must be taken. This parameter yields $\eta\sim 0$ in the SRI area, where we have $\tau<\tau_s$, similar to the SRII region. The design of this parametrization makes the following contribution emerge if we calculate the time derivative of the second slow-roll parameter $\eta$ at the transition points:
\bea \eta^{'}(\tau)=-\Delta\eta\left[\delta(\tau-\tau_s)-\delta(\tau-\tau_e)\right].\eea
But such terms are not allowed in the third order action discussed above due to the soft disruption of Galilean symmetry. This word specifically appears in references. when there was no such underlying symmetry \cite{Kristiano:2022maq,Riotto:2023hoz,Choudhury:2023vuj,Choudhury:2023jlt,Kristiano:2023scm,Riotto:2023gpm,Choudhury:2023rks,Firouzjahi:2023aum,Motohashi:2023syh}. Thus, we do not need to explicitly handle the derivative of the second slow-roll parameter $\eta$ at the transition locations in our article. However, the design makes it evident that at the specified transition points, the behaviour of the second slow-roll parameter, $\eta$, is not smooth. Hence, in order to maintain consistency in our subsequent analysis, we have employed the previously specified parametrization of the second slow-roll parameter $\eta$ at the transition locations. 

In order to maintain track of the final result and its overall contribution added to the tree-level result, it is important to discuss the strengths of the four types of cubic self interactions in the SRI, USR, and SRII region before moving on to the more technical details of the one-loop contributions in the next subsection. Due to vanishing contributions from the second slow-roll parameter $\eta$ in both SRI and SRII phases, all four of these interactions are suppressed. For the sake of thoroughness, however, we shall compute these contributions. However, the last two terms of the representative third order action in the USR phase become dominant over the other two terms with the coupling coefficients, ${\cal G}_1$ and ${\cal G}_2$, because of the presence of the second slow-roll parameter $\eta$ in the two coupling parameters, ${\cal G}_3$ and ${\cal G}_4$. We shall clearly demonstrate in the next subsection that the final two terms of the third order action that arise during the USR period will dominate any enhancement that results from the one-loop contribution.
Given that we do not have any terms that involve the time derivative of the second slow-roll parameter $\eta$ at the transition points from SRI to USR and USR to SRII, the corresponding enhancement of the power spectrum in the USR period in the one-loop contribution will be sufficiently suppressed in comparison to the findings found in the refs. \cite{Kristiano:2022maq,Riotto:2023hoz,Choudhury:2023vuj,Choudhury:2023jlt,Kristiano:2023scm,Riotto:2023gpm,Choudhury:2023rks,Firouzjahi:2023aum,Motohashi:2023syh} where this kind of contribution is present. Even yet, the one-loop contributions from the SRI and SRII areas are significantly lower than the suppressed contribution during the USR era. All of these alternatives will be thoroughly examined in the upcoming subsection.

\subsubsubsection{Local non-Gaussianity from three point function and the associated Bispectrum computation}

From this point on, we start our in-depth examination of the tree-level three-point cosmological correlation function computation, commonly referred to as the Bispectrum. As the lowest order measure of the departure from conventional Gaussian statistics, this function is significant. We start this part by introducing the generic three-point function as follows, which comes from the third-order action in the curvature perturbations presented using the Covariantized Galileon Theory in the previous section and works with the {\it in-in formalism} which is also explained before the cubic action:
\bea \label{tpt} \langle\hat{\zeta}_{\bf k_{1}}\hat{\zeta}_{\bf k_{2}}\hat{\zeta}_{\bf k_{3}}\rangle:&&=\left\langle\bigg[\overline{T}\exp\bigg(i\int^{\tau}_{-\infty(1-i\epsilon)}d\tau^{'}\;H_{\rm int}(\tau^{'})\bigg)\bigg]\;\;\hat{\zeta}_{\bf k_{1}}(\tau)\hat{\zeta}_{\bf k_{2}}(\tau)\hat{\zeta}_{\bf k_{3}}(\tau)\right.\nonumber\\
&& \left.
\;\;\quad\quad\quad\quad\quad\quad\quad\quad\quad\quad\quad\quad\quad\quad\times\bigg[{T}\exp\bigg(-i\int^{\tau}_{-\infty(1+i\epsilon)}d\tau^{''}\;H_{\rm int}(\tau^{''})\bigg)\bigg]\right\rangle_{\tau\rightarrow 0},\eea
the unitary operators, which are composed of the time integral of the interacting Hamiltonian, which in this instance is defined as follows, have anti-time and time ordering represented by $\bar{T}$ and $T$, respectively.
\bea && H_{\rm int}(\tau)=-\int d^3x\; \frac{a(\tau)^2}{H^3}\; \bigg[\frac{{\cal G}_1}{a}\zeta^{'3}+\frac{{\cal G}_2}{a^2}\zeta^{'2}\left(\partial^2\zeta\right)+\frac{{\cal G}_3}{a}\zeta^{'}\left(\partial_i\zeta\right)^2+\frac{{\cal G}_4}{a^2}\left(\partial_i\zeta\right)^2\left(\partial^2\zeta\right)\bigg].\eea
The perturbed cubic action retains the same coefficients ${\cal G}_i, \forall i=1,2,3,4$.
Afterwards, the Hamiltonian's interactions are represented by the following contribution to the three-point function, which comes from every diagram:
\bea &&\label{g}\langle\zeta_{\bold{k}_{1}}\zeta_{\bold{k}_{2}}\zeta_{\bold{k}_{3}}\rangle = \langle\zeta_{\bold{k}_{1}}\zeta_{\bold{k}_{2}}\zeta_{\bold{k}_{3}}\rangle_{\zeta'^{3}} + \langle\zeta_{\bold{k}_{1}}\zeta_{\bold{k}_{2}}\zeta_{\bold{k}_{3}}\rangle_{\zeta'^{2}(\partial^{2}\zeta)} + \langle\zeta_{\bold{k}_{1}}\zeta_{\bold{k}_{2}}\zeta_{\bold{k}_{3}}\rangle_{\zeta'(\partial_{i}\zeta)^{2}} + \langle\zeta_{\bold{k}_{1}}\zeta_{\bold{k}_{2}}\zeta_{\bold{k}_{3}}\rangle_{(\partial_{i}\zeta)^{2}(\partial^{2}\zeta)}.
\eea
See equation (\ref{inin}) and related section \ref{s1} for more details on this aspect.
Specifically estimating these contributions in each of the three phases—SRI, USR, and SRII—would be our primary responsibility. We compute them using the equation derived from the formula for the expectation of any operator up to the leading order in $H_{\text{Int}}(\tau)$, which was expanded perturbatively. We then indicate the general equation from which the correlation functions are formed for each interaction operator in order to maintain track of formulae that we will use in the future:
\bea \langle\hat{\zeta}_{\bf k_{1}}\hat{\zeta}_{\bf k_{2}}\hat{\zeta}_{\bold{k}_{3}}\rangle&=&\text{2 $\times$ Im}\left[\zeta_{\bold{k}_{1}}(\tau)\zeta_{\bold{k}_{2}}(\tau)\zeta_{\bold{k}_{3}}(\tau)\int^{\tau}_{-\infty}d\tau_{1}d^{3}x\frac{a(\tau_1)^2}{H^3}Q(x,\tau_{1})\right]\nonumber\\
&=&\text{2 $\times$ Im}\left[\zeta_{\bold{k}_{1}}(\tau)\zeta_{\bold{k}_{2}}(\tau)\zeta_{\bold{k}_{3}}(\tau)\int\frac{d^{3}\bold{q}_{1}}{(2\pi)^3}\frac{d^{3}\bold{q}_{2}}{(2\pi)^3}\frac{d^{3}\bold{q}_{3}}{(2\pi)^3}\int^{\tau}_{-\infty}d\tau_{1}\frac{a(\tau_{1})^2}{H^3}Q_{\bold{q}}(\tau_{1})\exp(-i(\bold{q}_{1}+\bold{q}_{2}+\bold{q}_{3}).\bold{x})\right].\quad\quad\quad
\eea
Here we use the following Fourier transform ansatz:
\be Q(x, \tau_{1}) = \int^{\tau}_{-\infty}d\tau_{1}Q_{\bold{q}}(\tau_{1})\exp(-i(\bold{q}_{1}+\bold{q}_{2}+\bold{q}_{3}).\bold{x}),\ee
is employed, and the operators $\displaystyle{\frac{{\cal G}_1}{a}\zeta^{'3}}$,$\displaystyle{\frac{{\cal G}_2}{a^2}\zeta^{'2}\left(\bold{q}_{i}^{2}\zeta\right)}$,$\displaystyle{\frac{{\cal G}_3}{a}\zeta^{'}\left(\bold{q}_{i}.\bold{q}_{j}\zeta^2\right)}$, and $\displaystyle{\frac{{\cal G}_4}{a^2}((\bold{q}_{i}.\bold{q}_{j})\bold{q}_{k}^{2})\left(\zeta\right)^2\left(\partial^2\zeta\right)}$ are represented by $Q_{q}(\tau_{1})$, where the mode functions for the relevant areas are $\zeta \equiv \zeta_{\bold{q}}(\tau_{1})$ and $a \equiv a(\tau_{1})$. In all three zones, this expression will be applied to every operator. We can use the wick contraction method to contract operators that are outside the time integral with the interactions ones inside the integral, but first we need to perform additional calculations to gather information about the behaviour of mode functions in the three regions, which has already been assessed in the previous sections. The evaluation of the interaction operators' contribution for every SRI, USR, and SRII area will be demonstrated fully in the next subsections. Look at the Appendix \ref{A4a} for more details on this computation.

Additionally, the integral across conformal time will be analysed by splitting it with regard to the region of concern while computing in each of the aforementioned regions:
\bea {\bf Conformal\;time\;integral:}\quad\quad\quad\quad\lim_{\tau\rightarrow 0}\int^{\tau}_{-\infty}:\equiv \underbrace{\Bigg(\int^{\tau_s}_{-\infty}\Bigg)}_{\bf SRI}+\underbrace{\Bigg(\int^{\tau_e}_{\tau_s}\Bigg)}_{\bf USR}+\underbrace{\Bigg(\int^{\tau_{\rm end}\rightarrow 0}_{\tau_e}\Bigg)}_{\bf SRII}.\eea

\subsubsubsection{Total bispectrum and associated non-Gaussian amplitude}

\begin{itemize}
    \item[$\blacksquare$] \underline{\textbf{A. Bispectrum and associated non-Gaussian amplitude for region I: SRI}}\\ \\
The bispectrum computation for each operator in the SRI area has now started. The specified region is for the conformal time range $-\infty < \tau < \tau_{s}$, where an abrupt transition from the SRI to the USR region happens at $\tau_{s}$. In this area, the second slow-roll parameter, $\eta$, is essentially zero, whereas only the first one, $\epsilon$, is finite (a constant, to be exact). We shall demonstrate from the analysis in this section that the non-Gaussian amplitude achieved here will have a very minimal contribution to the SRI phase, and in the squeezed limiting situation, it will be compatible with the {\it no-go theorem} of Maldacena. Before we continue, it should be noted that the authors of \cite{Burrage:2010cu} discovered non-zero findings for the non-Gaussianity amplitude, $\fnl$, under the equilateral limit, but they came to the conclusion that the value of $\fnl$ decays to zero for the squeezed limit. Our findings for SRI in this section demonstrate that, in the absence of any USR and SRII phases, the value of $\fnl$ likewise goes to zero under the limit $\tau_{s} \rightarrow 0$. As a consequence, our findings concur with those of the writers in \cite{Burrage:2010cu}. 

 Appendix \ref{A4a} contains a full study of the contributions made by each operator as well as their total contribution to the tree-level scalar three-point correlation function. The sum of the individual contributions is expressed as the combined contribution, which is as follows:
   \bea \langle\hat{\zeta}_{\bold{k}_{1}}\hat{\zeta}_{\bold{k}_{2}}\hat{\zeta}_{\bold{k}_{3}}\rangle_{\text{SRI}} = (2\pi)^{3}\delta^{3}(\bold{k}_{1}+\bold{k}_{2}+\bold{k}_{3}){B}^{\text{SRI}}_{\zeta\zeta\zeta}(k_{1},k_{2},k_{3}).
\eea
where the total of each operator's separate contributions is included in the RHS as follows:
\bea \label{c1BI} {B}^{\text{SRI}}_{\zeta\zeta\zeta}(k_{1},k_{2},k_{3}) = \sum_{i=1}^{4}{B}_{Q}^{\text{SRI}}(k_1,k_2,k_3).
\eea
where $Q$ stands for the four interaction operators, allowing us to obtain the following explicit contributions for each one:
\bea &&{B}^{\text{SRI}}_{\zeta^{'3}} = \frac{H^{12}}{(4{\cal A})^{3}}\frac{{\cal G}_{1}}{H^4}\frac{6}{(k_{1}^{3}k_{2}^{3}k_{3}^{3})}\frac{k_{1}^2 k_{2}^2 k_{3}^2}{c_{s}^{6} K^3}\left\{\left(\frac{K^{2}}{k_{s}^{2}}-2\right)\cos\left(\frac{K}{k_{s}}\right)-2\frac{K}{k_{s}}\sin\left(\frac{K}{k_{s}}\right)\right\}. \\&&{B}^{\text{SRI}}_{\zeta^{'2}(\partial^{2}\zeta)} = \frac{H^{12}}{(4{\cal A})^{3}}\frac{{\cal G}_{2}}{H^3}\frac{4}{(k_{1}^{3}k_{2}^{3}k_{3}^{3})}\displaystyle{\frac{k_{1}^2 k_{2}^2 k_{3}^2}{c_{s}^{8} K^3}\left\{\left(12-6\frac{K^{2}}{k_{s}^2}\right)\cos\left(\frac{K}{k_{s}}\right) - \left(\frac{K^{3}}{k_{s}^3}-12\frac{K}{k_{s}}\right)\sin\left(\frac{K}{k_{s}}\right)\right\}}. \\ &&{B}^{\text{SRI}}_{\zeta^{'}(\partial_{i}\zeta)^{2}} = \frac{H^{12}}{(4{\cal A})^{3}}\frac{{\cal G}_{3}}{H^4}\frac{4}{k_{1}^{3}k_{2}^{3}k_{3}^{3}}\frac{k_{1}k_{2}k_{3}}{c_{s}^{8}K^{3}k_{s}^{2}}\Bigg\{\bigg(18 k_1 k_2 k_3 K+ 2 k_1^3\left(k_2+k_3\right) +2k_2^3\left(k_3+k_1\right) + 2k_3^3\left(k_1+k_2\right)\nonumber\\
   &&\quad\quad\quad\quad\quad\quad + 2\left(k_1^2 k_2^2+k_2^2 k_3^2+k_3^2 k_1^2\right)\bigg) k_s \sin
   \left(\frac{K}{k_s}\right)-3 k_1 k_2 k_3 K^{2} \cos
   \left(\frac{K}{k_s}\right)+\bigg(k_1^3 + k_2^3+k_3^3 \nonumber\\
   &&\quad\quad\quad\quad\quad\quad  + 5k_1^2\left(k_2+k_3\right) +5 k_2^2\left(k_3+k_1\right)+5k_3^2 \left(k_1+k_2\right) + 18 k_1 k_2 k_3\bigg) k_s^2 \cos
   \left(\frac{K}{k_s}\right)\Bigg\}.\\
   && {B}^{\text{SRI}}_{(\partial_{i}\zeta)^{2}(\partial^{2}\zeta)} = \frac{H^{12}}{(4{\cal A})^{3}}\frac{{\cal G}_{4}}{H^3}\frac{6}{k_{1}^{3}k_{2}^{3}k_{3}^{3}}
\frac{k_{1}k_{2}k_{3}}{c_{s}^{10}K^{3}}\left\{k_{1}k_{2}k_{3}\frac{K^{3}}{k_{s}^{3}}\left(\sin{\frac{K}{k_s}}\right)+\frac{K^{2}}{k_{s}^{2}}\bigg(k_{1}^{2}(k_{2}+k_{3})+k_{2}^{2}(k_{3}+k_{1}) +k_{3}^{2}(k_{2}+k_{1}) \right.\nonumber\\
&&\left. \quad\quad\quad\quad\quad + 6k_{1}k_{2}k_{3}\bigg)\left(\cos{\frac{K}{k_s}}\right) -\frac{1}{k_{s}}\bigg(k_{1}^{4}+ k_{2}^{4}+k_{3}^{4}+6k_{1}^{3}(k_{2}+k_{3})+6k_{2}^{3}(k_{1}+k_{3}) + 6k_{3}^{3}(k_{1}+k_{2}) + 28k_{1}k_{2}k_{3}K \right. \nonumber\\
&& \left. \quad\quad\quad\quad\quad  + 10k_{1}^{2}k_{2}^{2}+8k_{2}^{2}k_{3}^{2}+10k_{1}^{2}k_{3}^{2} \bigg)\left(\sin{\frac{K}{k_{s}}}\right)-2\bigg(k_{1}^{3}+k_{2}^{3}+k_{3}^{3}+4k_{1}^{2}(k_{2}+k_{3})+ 4k_{2}^{2}(k_{3}+k_{1}) +4k_{3}^{2}(k_{1}+k_{2}) \right.\nonumber\\
&& \left. \quad\quad\quad\quad\quad + 12k_{1}k_{2}k_{3}\bigg)\left(\cos{\frac{K}{k_{s}}}\right) \right\}. \eea
Once the specific contributions from each interaction operator to the tree-level three-point function have been discussed, we use the following equation to further assess the associated values of the non-Gaussian amplitude $\fnl$:
\bea \label{c1Bf} {B}^{\text{SRI}}(k_1,k_2,k_3) = \frac{6}{5}f^{\text{SRI}}_{\text{NL}}{\times}(2\pi^{2})^{2}\bigg[\frac{\Delta_{\bf {Tree}}^{\text{SRI}}(k_1)\Delta_{\bf {Tree}}^{\text{SRI}}(k_2)}{k_1^{3}k_2^{3}}+\frac{\Delta_{\bf {Tree}}^{\text{SRI}}(k_2)\Delta_{\bf {Tree}}^{\text{SRI}}(k_3)}{k_2^{3}k_3^{3}}+\frac{\Delta_{\bf {Tree}}^{\text{SRI}}(k_3)\Delta_{\bf {Tree}}^{\text{SRI}}(k_1)}{k_3^{3}k_1^{3}}\bigg].\quad\quad\quad
\eea
To extract information on the non-Gaussianity amplitude $\fnl$, the same factorization is also used in the USR and SRII areas.
It is also vital to remember that the dimensionless power spectrum in the SRI area is used to write the aforementioned formula. Look at the Appendix \ref{A4a} for more details on this computation.

 \item[$\blacksquare$] \underline{\textbf{B. Bispectrum and associated non-Gaussian amplitude computation for region II: USR}}\\ \\
 In this part, we determine the explicit formulation of the bispectrum for the scalar modes in the USR area, thereby continuing our investigation of it. In the conformal time period $\tau_{s} < \tau < \tau_{e}$, this area is defined. At $\tau_{s}$, there are strong transitions between phases SRI and USR, and at $\tau_{e}$, there are transitions from SRI to SRII. During the USR phase, $\epsilon$ is not a constant parameter. The parameter $\eta$ is likewise not a constant; instead, it has the value $\eta \sim -6$. This means that it relies on the conformal time through the relation as described previously when addressing the modes for USR. 
These facts will affect the behavior of the bispectrum strength and the conformal time dependence of the Bogoliubov coefficients.

In Appendix \ref{A4a}, we discuss the comprehensive examination of all the various contributions made by each operator separately as well as the sum of their outcomes. The findings for the tree-level scalar three-point correlation function in the USR area are discussed here as the total contribution from each operator utilizing the particular functions; further information about these functions may be found in the appendix. 

The following may be used to express the tree-level contribution to the three-point function that each operator in the USR area made:
\bea \langle\hat{\zeta}_{\bold{k}_{1}}\hat{\zeta}_{\bold{k}_{2}}\hat{\zeta}_{\bold{k}_{3}}\rangle_{\text{USR}} = (2\pi)^{3}\delta^{3}(\bold{k}_{1}+\bold{k}_{2}+\bold{k}_{3}){B}^{\text{USR}}_{\zeta\zeta\zeta}(k_{1},k_{2},k_{3}).
\eea
where the individual contributions to the tree-level bispectrum value add up to the RHS:
\bea \label{c2BI} {B}^{\text{USR}}_{\zeta\zeta\zeta}(k_{1},k_{2},k_{3}) = \sum_{i=1}^{4}{B}_{Q}^{\text{USR}}(k_1,k_2,k_3).
\eea
The explicit contributions from each operator separately are expressed as follows, where $Q$ stands for the $4$ interaction operators.
The following is the entire tree-level contribution to the three-point correlation function as determined by the first operator:
\bea \label{c2r1}{B}^{\text{USR}}_{\zeta^{'3}} &=& \frac{H^{12}}{(4{\cal A})^{3}}\frac{{\cal G}_{1}}{H^4}\frac{-2}{(k_{1}^{3}k_{2}^{3}k_{3}^{3})}\left[\zeta_{\bold{k_{1}}}(\tau_{s})\zeta_{\bold{k_{2}}}(\tau_{s})\zeta_{\bold{k_{3}}}(\tau_{s})\right]\Bigg\{\bigg((\alpha_{k_{1}}^{(2)*}\alpha_{k_{2}}^{(2)*}\alpha_{k_{3}}^{(2)*})(\bold{I}_{1})_{1}({\cal K}_{1},k_1,k_2,k_3) - (\alpha_{k_{1}}^{(2)*}\beta_{k_{2}}^{(2)*}\alpha_{k_{3}}^{(2)*}) \nonumber\\ 
&&\times (\bold{I}_{1})_{2}({\cal K}_{2},k_1,-k_2,k_3) -(\alpha_{k_{1}}^{(2)*}\alpha_{k_{2}}^{(2)*}\beta_{k_{3}}^{(2)*})(\bold{I}_{1})_{3}({\cal K}_{1},k_1,k_2,-k_3) - (\beta_{k_{1}}^{(2)*}\alpha_{k_{2}}^{(2)*}\alpha_{k_{3}}^{(2)*})(\bold{I}_{1})_{4}({\cal K}_{4},-k_1,k_2,k_3) - \text{c.c} \bigg)\nonumber\\
&&+\bigg( (\alpha_{k_{1}}^{(2)*}\alpha_{k_{2}}^{(2)*}\alpha_{k_{3}}^{(2)*})\bigg(k_3^{2}(\bold{I}_{2})_{1}({\cal K}_{1},-k_1,-k_2)  + k_1^{2}(\bold{I}_{2})_{2}({\cal K}_{1},-k_2,-k_3) + k_2^{2}(\bold{I}_{2})_{3}({\cal K}_{1},-k_1,-k_3) \bigg)\nonumber\\
&& - (\alpha_{k_{1}}^{(2)*}\beta_{k_{2}}^{(2)*}\alpha_{k_{3}}^{(2)*})\bigg( k_3^{2}(\bold{I}_{2})_{4}({\cal K}_{3},k_1,-k_2) + k_1^{2}(\bold{I}_{2})_{5}({\cal K}_{3},k_2,-k_3)
+ k_2^{2}(\bold{I}_{2})_{6}({\cal K}_{3},k_3,-k_1) \bigg)  \nonumber\\
&& -(\alpha_{k_{1}}^{(2)*}\alpha_{k_{2}}^{(2)*}\beta_{k_{3}}^{(2)*})\bigg(k_3^{2}(\bold{I}_{2})_{7}({\cal K}_{2},-k_1,k_2) + k_1^{2}(\bold{I}_{2})_{8}({\cal K}_{2},-k_2,k_3) + k_2^{2}(\bold{I}_{2})_{9}({\cal K}_{2},-k_3,k_1)\bigg)\nonumber\\ 
&& - (\beta_{k_{1}}^{(2)*}\alpha_{k_{2}}^{(2)*}\alpha_{k_{3}}^{(2)*})\bigg(k_3^{2}(\bold{I}_{2})_{10}({\cal K}_{4},-k_1,-k_2) +  k_1^{2}(\bold{I}_{2})_{11}({\cal K}_{4},-k_2,-k_3) + k_2^{2}(\bold{I}_{2})_{12}({\cal K}_{4},-k_3,-k_1)\bigg) - \text{c.c} \bigg)\nonumber\\
&&+\bigg( (\alpha_{k_{1}}^{(2)*}\alpha_{k_{2}}^{(2)*}\alpha_{k_{3}}^{(2)*})\bigg(k_2^{2}k_3^{2}(\bold{I}_{3})_{1}({\cal K}_{1},-k_1) + k_3^{2}k_1^{2}(\bold{I}_{3})_{2}({\cal K}_{1},-k_2) +
k_1^{2}k_2^{2}(\bold{I}_{3})_{3}({\cal K}_{1},-k_3)\bigg) -
(\alpha_{k_{1}}^{(2)*}\beta_{k_{2}}^{(2)*}\alpha_{k_{3}}^{(2)*}) \nonumber\\
&&\bigg(k_2^{2}k_3^{2}(\bold{I}_{3})_{4}({\cal K}_{2},-k_1)+k_3^{2}k_1^{2}(\bold{I}_{3})_{5}({\cal K}_{2},k_2) + k_1^{2}k_2^{2}(\bold{I}_{3})_{6}({\cal K}_{2},-k_3)\bigg) -(\alpha_{k_{1}}^{(2)*}\alpha_{k_{2}}^{(2)*}\beta_{k_{3}}^{(2)*})\bigg(k_2^{2}k_3^{2}(\bold{I}_{3})_{7}({\cal K}_{3},-k_1)\nonumber\\
&&+k_1^{2}k_3^{2}(\bold{I}_{3})_{8}({\cal K}_{3},-k_2) + k_1^{2}k_2^{2}(\bold{I}_{3})_{9}({\cal K}_{3},k_3)\bigg) - (\beta_{k_{1}}^{(2)*}\alpha_{k_{2}}^{(2)*}\alpha_{k_{3}}^{(2)*})\bigg(k_3^{2}k_2^{2}(\bold{I}_{3})_{10}({\cal K}_{4},k_1) +k_1^{2}k_3^{2}(\bold{I}_{3})_{11}({\cal K}_{4},-k_2) \nonumber\\
&& +k_1^{2}k_2^{2}(\bold{I}_{3})_{12}({\cal K}_{4},-k_3)\bigg) - \text{c.c} \bigg) +k^2_1k^2_2k^2_3\bigg((\alpha_{k_{1}}^{(2)*}\alpha_{k_{2}}^{(2)*}\alpha_{k_{3}}^{(2)*})(\bold{I}_{4})_{1}({\cal K}_{1}) - (\alpha_{k_{1}}^{(2)*}\beta_{k_{2}}^{(2)*}\alpha_{k_{3}}^{(2)*})(\bold{I}_{4})_{2}({\cal K}_{2}) - \nonumber\\
&&(\alpha_{k_{1}}^{(2)*}\alpha_{k_{2}}^{(2)*}\beta_{k_{3}}^{(2)*})(\bold{I}_{4})_{3}({\cal K}_{3}) - (\beta_{k_{1}}^{(2)*}\alpha_{k_{2}}^{(2)*}\alpha_{k_{3}}^{(2)*})(\bold{I}_{4})_{4}({\cal K}_{4}) - \text{c.c}\bigg) + \text{2 Perms.}\Bigg\}.
\eea
The following is the overall contribution to the three-point, tree-level correlation function derived from the second operator:
\bea \label{c2r2} {B}^{\text{USR}}_{\zeta^{'2}\partial^{2}\zeta} &=& \frac{H^{12}}{(4{\cal A})^{3}}\frac{{\cal G}_{2}}{H^3}\frac{4}{(k_{1}^{3}k_{2}^{3}k_{3}^{3})}\left[\zeta_{\bold{k_{1}}}(\tau_{s})\zeta_{\bold{k_{2}}}(\tau_{s})\zeta_{\bold{k_{3}}}(\tau_{s})\right]\Bigg\{\bigg( (\alpha_{k_{1}}^{(2)*}\alpha_{k_{2}}^{(2)*}\alpha_{k_{3}}^{(2)*})(\bold{I}_{1})_{1}({\cal K}_{1},k_1,k_2,k_3) - (\alpha_{k_{1}}^{(2)*}\beta_{k_{2}}^{(2)*}\alpha_{k_{3}}^{(2)*}) \nonumber\\ 
&&\times (\bold{I}_{1})_{2}({\cal K}_{2},k_1,-k_2,k_3)-(\alpha_{k_{1}}^{(2)*}\alpha_{k_{2}}^{(2)*}\beta_{k_{3}}^{(2)*})(\bold{I}_{1})_{3}({\cal K}_{3},k_1,k_2,-k_3) - (\beta_{k_{1}}^{(2)*}\alpha_{k_{2}}^{(2)*}\alpha_{k_{3}}^{(2)*})(\bold{I}_{1})_{4}({\cal K}_{4},-k_1,k_2,k_3)- \text{c.c} \bigg)\nonumber\\
&&+\bigg((\alpha_{k_{1}}^{(2)*}\alpha_{k_{2}}^{(2)*}\alpha_{k_{3}}^{(2)*}) \bigg(k^{2}_2(\bold{I}_{2})_{1}({\cal K}_{1},-k_1,-k_3) + k^{2}_1(\bold{I}_{2})_{2}({\cal K}_{1},-k_2,-k_3)\bigg) -
(\alpha_{k_{1}}^{(2)*}\beta_{k_{2}}^{(2)*}\alpha_{k_{3}}^{(2)*}) \bigg(k^{2}_2(\bold{I}_{2})_{3}({\cal K}_{2},-k_1,-k_3)\nonumber\\
&&+ k^{2}_1(\bold{I}_{2})_{4}({\cal K}_{2},k_2,-k_3) \bigg) -(\alpha_{k_{1}}^{(2)*}\alpha_{k_{2}}^{(2)*}\beta_{k_{3}}^{(2)*})\bigg(k^{2}_2(\bold{I}_{2})_{5}({\cal K}_{3},-k_1,k_3) + k^{2}_1(\bold{I}_{2})_{6}({\cal K}_{3},-k_2,k_3)\bigg) -(\beta_{k_{1}}^{(2)*}\alpha_{k_{2}}^{(2)*}\alpha_{k_{3}}^{(2)*})\nonumber\\
&&\bigg(k_2^{2}(\bold{I}_{2})_{7}({\cal K}_{4},k_1,-k_3) +k_{1}^{2}(\bold{I}_{2})_{8}({\cal K}_{4},-k_2,-k_3)\bigg) - \text{c.c} \bigg) + (k_1^{2}k_2^{2})\bigg((\alpha_{k_{1}}^{(2)*}\alpha_{k_{2}}^{(2)*}\alpha_{k_{3}}^{(2)*})(\bold{I}_{3})_{1}({\cal K}_{1},-k_3)\nonumber\\
&& - (\alpha_{k_{1}}^{(2)*}\beta_{k_{2}}^{(2)*}\alpha_{k_{3}}^{(2)*})(\bold{I}_{3})_{2}({\cal K}_{2},-k_3) - (\alpha_{k_{1}}^{(2)*}\alpha_{k_{2}}^{(2)*}\beta_{k_{3}}^{(2)*})(\bold{I}_{3})_{3}({\cal K}_{3},k_3) - (\beta_{k_{1}}^{(2)*}\alpha_{k_{2}}^{(2)*}\alpha_{k_{3}}^{(2)*})(\bold{I}_{3})_{4}({\cal K}_{4},-k_3)\nonumber\\
&& - \text{c.c}\bigg) + \text{2 Perms.} \Bigg\}
\eea
The overall contribution to the three-point, tree-level correlation function may be obtained from the third operator in the following way:
\bea \label{c2r3} {B}^{\text{USR}}_{\zeta^{'}(\partial_{i}\zeta)^{2}} &=& \frac{H^{12}}{(4{\cal A})^{3}}\frac{{\cal G}_{3}}{H^4}\frac{-4}{(k_{1}^{3}k_{2}^{3}k_{3}^{3})}\left[\zeta_{\bold{k_{1}}}(\tau_{s})\zeta_{\bold{k_{2}}}(\tau_{s})\zeta_{\bold{k_{3}}}(\tau_{s})\right]\Bigg\{\frac{-k_1 k_2}{3}\bigg( (\alpha_{k_{1}}^{(2)*}\alpha_{k_{2}}^{(2)*}\alpha_{k_{3}}^{(2)*})(\bold{I}_{1})_{1}({\cal K}_{1},k_1,k_2,k_3) - (\alpha_{k_{1}}^{(2)*}\beta_{k_{2}}^{(2)*}\alpha_{k_{3}}^{(2)*}) \nonumber\\
&&\times (\bold{I}_{1})_{2}({\cal K}_{2},k_1,-k_2,k_3)-(\alpha_{k_{1}}^{(2)*}\alpha_{k_{2}}^{(2)*}\beta_{k_{3}}^{(2)*})(\bold{I}_{1})_{3}({\cal K}_{3},k_3,k_2,-k_3) - (\beta_{k_{1}}^{(2)*}\alpha_{k_{2}}^{(2)*}\alpha_{k_{3}}^{(2)*})(\bold{I}_{1})_{4}({\cal K}_{4},-k_1,k_2,k_3) \nonumber\\
&&- \text{c.c}\bigg)-\frac{k_1 k_2 k_{3}^{2}}{3}\bigg((\alpha_{k_{1}}^{(2)*}\alpha_{k_{2}}^{(2)*}\alpha_{k_{3}}^{(2)*})  (\bold{I}_{2})_{1}({\cal K}_{1},-k_1,-k_2) -
(\alpha_{k_{1}}^{(2)*}\beta_{k_{2}}^{(2)*}\alpha_{k_{3}}^{(2)*})(\bold{I}_{2})_{2}({\cal K}_{2},-k_1,k_2)\nonumber\\
&& - (\alpha_{k_{1}}^{(2)*}\alpha_{k_{2}}^{(2)*}\beta_{k_{3}}^{(2)*})(\bold{I}_{2})_{3}({\cal K}_{3},-k_1,-k_2) - (\beta_{k_{1}}^{(2)*}\alpha_{k_{2}}^{(2)*}\alpha_{k_{3}}^{(2)*})(\bold{I}_{2})_{4}({\cal K}_{4},k_1,-k_2) - \text{c.c} \bigg) + \text{2 Perms.}\Bigg\}
\eea
The overall contribution to the three-point, tree-level correlation function may be obtained from the fourth operator as follows:
\bea \label{c2r4} {B}^{\text{USR}}_{\partial^{2}\zeta(\partial_{i}\zeta)^{2}} &=& \frac{H^{12}}{(2{\cal A})^{3}}\frac{{\cal G}_{4}}{H^3}\frac{2}{(k_{1}^{3}k_{2}^{3}k_{3}^{3})}\left[\zeta_{\bold{k_{1}}}(\tau_{s})\zeta_{\bold{k_{2}}}(\tau_{s})\zeta_{\bold{k_{3}}}(\tau_{s})\right]\Bigg\{ k_{1}k_{2}k_{3}^{2}\bigg( (\alpha_{k_{1}}^{(2)*}\alpha_{k_{2}}^{(2)*}\alpha_{k_{3}}^{(2)*})(\bold{I}_{1})_{1}({\cal K}_{1},k_1,k_2,k_3) - (\alpha_{k_{1}}^{(2)*}\beta_{k_{2}}^{(2)*}\alpha_{k_{3}}^{(2)*})\nonumber\\ 
&&\times (\bold{I}_{1})_{2}({\cal K}_{2},k_1,-k_2,k_3) -(\alpha_{k_{1}}^{(2)*}\alpha_{k_{2}}^{(2)*}\beta_{k_{3}}^{(2)*})(\bold{I}_{1})_{3}({\cal K}_{3},k_1,k_2,-k_3) - (\beta_{k_{1}}^{(2)*}\alpha_{k_{2}}^{(2)*}\alpha_{k_{3}}^{(2)*})(\bold{I}_{1})_{4}({\cal K}_{4},-k_1,k_2,k_3)\nonumber\\
&& - \text{c.c} \bigg) + \text{2 Perms} \Bigg\}
\eea
The complex conjugate of each of the preceding components, $(\text{c.c})$, in the aforementioned equations accounts for the contributions made by the negative exponential integrals. We address additional $2$ permutations in momentum variables while presenting the numerical findings. We may utilize the connection to further assess $\fnl$ by utilizing the formula for the tree-level bispectrum in the USR region:
\bea \label{c2Bf} {B}^{\text{USR}}(k_1,k_2,k_3) = \frac{6}{5}f^{\text{USR}}_{\text{NL}}{\times}(2\pi^{2})^{2}\bigg[\frac{\Delta_{\bf {Tree}}^{\text{USR}}(k_1)\Delta_{\bf {Tree}}^{\text{USR}}(k_2)}{k_1^{3}k_2^{3}}+\frac{\Delta_{\bf {Tree}}^{\text{USR}}(k_2)\Delta_{\bf {Tree}}^{\text{USR}}(k_3)}{k_2^{3}k_3^{3}}+\frac{\Delta_{\bf {Tree}}^{\text{USR}}(k_3)\Delta_{\bf {Tree}}^{\text{USR}}(k_1)}{k_3^{3}k_1^{3}}\bigg].\quad\quad\quad
\eea
where we have applied the dimensionless power spectrum in the USR area. It would be laborious and not very instructive to put here the exact calculation of the squeezed limit for the bispectrum in the USR area and the associated non-Gaussianity amplitude $\fnl$ in the preceding equations. Therefore, we analyze various effective sound speed values and do a numerical analysis for the amplitude $\fnl$ with regard to the wave numbers in the USR area to provide the findings for the case of the squeezed limit. Look at the Appendix \ref{A4a} for more details on this computation.

\item[$\blacksquare$] \underline{\textbf{C. Bispectrum and associated non-Gaussian amplitude computation for region III: SRII}}\\ \\
Here, we determine the explicit manifestation of the bispectrum in the last SRII area, completing our study for it. The conformal time interval $\tau_{e}< \tau <\tau_{end}$ defines this area. $\tau_{e}$ denotes the USR to SRII region transition, and $\tau_{end}$ denotes the end of the SRII phase, which will ultimately be brought to zero in the late time limit.

We discuss in appendix \ref{A4a} the comprehensive examination of all the various contributions from each operator and their summative outcomes. We discuss the findings for the SRII region's tree-level scalar three-point correlation function in this section, which represents the total contribution from each operator's particular results in the appendix.

The following may be used to express the tree-level contribution to the three-point function that each operator in the SRII region makes:
\bea \langle\hat{\zeta}_{\bold{k}_{1}}\hat{\zeta}_{\bold{k}_{2}}\hat{\zeta}_{\bold{k}_{3}}\rangle_{\text{SRII}} = (2\pi)^{3}\delta^{3}(\bold{k}_{1}+\bold{k}_{2}+\bold{k}_{3}){B}^{\text{SRII}}_{\zeta\zeta\zeta}(k_{1},k_{2},k_{3}).
\eea
where the individual contributions to the tree-level bispectrum value add up to the RHS:
\bea \label{c3BI} {B}^{\text{SRII}}_{\zeta\zeta\zeta}(k_{1},k_{2},k_{3}) = \sum_{i=1}^{4}{B}_{Q}^{\text{SRII}}(k_1,k_2,k_3).
\eea
In this case, $Q$ stands for the $4$ interaction operators, and the explicit contributions made by each operator separately are expressed as follows. The following represents the contribution of the first operator to the tree-level three-point correlation function:
\bea \label{c3r1} B_{\zeta^{'3}}^{\text{SRII}} &=& \frac{H^{12}}{(4{\cal A})^{3}}\frac{{\cal G}_{1}}{H^4}\frac{-2}{(k_{1}^{3}k_{2}^{3}k_{3}^{3})}\left[\zeta_{\bold{k_{1}}}(\tau_{e})\zeta_{\bold{k_{2}}}(\tau_{e})\zeta_{\bold{k_{3}}}(\tau_{e})\right]\frac{k_{e}^{18}}{k_{s}^{18}}(k_1^{2}k_2^{2}k_3^{2})\Bigg\{(\alpha_{k_{1}}^{(2)*}\alpha_{k_{2}}^{(2)*}\alpha_{k_{3}}^{(2)*})(\bold{J}_{1})_{1}({\cal K}_{1})-(\alpha_{k_{1}}^{(2)*}\beta_{k_{2}}^{(2)*}\alpha_{k_{3}}^{(2)*})\nonumber\\
&& \times(\bold{J}_{1})_{2}({\cal K}_{2})-(\alpha_{k_{1}}^{(2)*}\alpha_{k_{2}}^{(2)*}\beta_{k_{3}}^{(2)*})(\bold{J}_{1})_{3}({\cal K}_{3})-(\beta_{k_{1}}^{(2)*}\alpha_{k_{2}}^{(2)*}\alpha_{k_{3}}^{(2)*})(\bold{J}_{1})_{4}({\cal K}_{4}) - \text{c.c} + \text{2 Perms.}\Bigg\}
\eea
The contribution for the second operator is provided as follows:
\bea  \label{c3r2} B_{\zeta^{'2}\partial^{2}\zeta}^{\text{SRII}} &=& \frac{H^{12}}{(4{\cal A})^{3}}\frac{{\cal G}_{2}}{H^3}\frac{4}{(k_{1}^{3}k_{2}^{3}k_{3}^{3})}\left[\zeta_{\bold{k_{1}}}(\tau_{e})\zeta_{\bold{k_{2}}}(\tau_{e})\zeta_{\bold{k_{3}}}(\tau_{e})\right]\frac{k_{e}^{18}}{k_{s}^{18}}\Bigg\{(\alpha_{k_{1}}^{(2)*}\alpha_{k_{2}}^{(2)*}\alpha_{k_{3}}^{(2)*})k_{1}^{2}k_{2}^{2}(\bold{J}_{2})_{1}({\cal K}_{1},-k_{3})-(\alpha_{k_{1}}^{(2)*}\beta_{k_{2}}^{(2)*}\alpha_{k_{3}}^{(2)*})\nonumber\\
&& \times k_{1}^{2}k_{2}^{2}(\bold{J}_{2})_{2}({\cal K}_{2},-k_3)-(\alpha_{k_{1}}^{(2)*}\alpha_{k_{2}}^{(2)*}\beta_{k_{3}}^{(2)*})k_{1}^{2}k_{2}^{2}(\bold{J}_{2})_{3}({\cal K}_{3},k_3)-(\beta_{k_{1}}^{(2)*}\alpha_{k_{2}}^{(2)*}\alpha_{k_{3}}^{(2)*})k_{1}^{2}k_{2}^{2}(\bold{J}_{2})_{4}({\cal K}_{4},-k_3)\nonumber\\
&& - \text{c.c} +  \text{2 Perms.} \Bigg\}
\eea
The third operator receives their contribution in the following way:
\bea  \label{c3r3} B_{\zeta^{'}(\partial_{i}\zeta)^{2}}^{\text{SRII}} &=& \frac{H^{12}}{(4{\cal A})^{3}}\frac{{\cal G}_{3}}{H^4}\frac{-4}{(k_{1}^{3}k_{2}^{3}k_{3}^{3})}\left[\zeta_{\bold{k_{1}}}(\tau_{e})\zeta_{\bold{k_{2}}}(\tau_{e})\zeta_{\bold{k_{3}}}(\tau_{e})\right]\frac{k_{e}^{18}}{k_{s}^{18}}\Bigg\{(\alpha_{k_{1}}^{(2)*}\alpha_{k_{2}}^{(2)*}\alpha_{k_{3}}^{(2)*})k_{3}^{2}(\bold{J}_{3})_{1}({\cal K}_{1},-k_{1},-k_{2})-(\alpha_{k_{1}}^{(2)*}\beta_{k_{2}}^{(2)*}\alpha_{k_{3}}^{(2)*})\nonumber\\
&& \times k_{3}^{2}(\bold{J}_{3})_{2}({\cal K}_{2},-k_1,k_2)-(\alpha_{k_{1}}^{(2)*}\alpha_{k_{2}}^{(2)*}\beta_{k_{3}}^{(2)*})k_{3}^{2}(\bold{J}_{3})_{3}({\cal K}_{3},-k_1,-k_2)-(\beta_{k_{1}}^{(2)*}\alpha_{k_{2}}^{(2)*}\alpha_{k_{3}}^{(2)*})k_{3}^{2}(\bold{J}_{3})_{4}({\cal K}_{4},k_1,-k_2)\nonumber\\
&&  - \text{c.c} +  \text{2 Perms.} \Bigg\}
\eea
The fourth operator receives its contribution in the following way:
\bea \label{c3r4} B_{(\partial^{2}\zeta)(\partial_{i}\zeta)^{2}}^{\text{SRII}} &=& \frac{H^{12}}{(4{\cal A})^{3}}\frac{{\cal G}_{4}}{H^3}\frac{2}{(k_{1}^{3}k_{2}^{3}k_{3}^{3})}\left[\zeta_{\bold{k_{1}}}(\tau_{e})\zeta_{\bold{k_{2}}}(\tau_{e})\zeta_{\bold{k_{3}}}(\tau_{e})\right]\frac{k_{e}^{18}}{k_{s}^{18}}\Bigg\{(\alpha_{k_{1}}^{(2)*}\alpha_{k_{2}}^{(2)*}\alpha_{k_{3}}^{(2)*})(\bold{J}_{4})_{1}({\cal K}_{1},k_{1},k_{2},k_{3})-(\alpha_{k_{1}}^{(2)*}\beta_{k_{2}}^{(2)*}\alpha_{k_{3}}^{(2)*})\nonumber\\
&& \times (\bold{J}_{4})_{2}({\cal K}_{2},k_1,-k_2,k_3) -(\alpha_{k_{1}}^{(2)*}\alpha_{k_{2}}^{(2)*}\beta_{k_{3}}^{(2)*})(\bold{J}_{4})_{3}({\cal K}_{3},k_1,k_2,-k_3)-(\beta_{k_{1}}^{(2)*}\alpha_{k_{2}}^{(2)*}\alpha_{k_{3}}^{(2)*})(\bold{J}_{4})_{4}({\cal K}_{4},-k_1,k_2,k_3)\nonumber\\
&& - \text{c.c} +  \text{2 Perms.} \Bigg\}
\eea
The contributions from each of the negative exponential integrals are shown in the preceding formulae (c.c), and these findings also account for the components derived from the other $2$ permutations of the momentum variables. We may utilise the connection to further assess $\fnl$ by using the formula for the tree-level bispectrum in the SRII region:
\bea \label{c3Bf} {B}^{\text{SRII}}(k_1,k_2,k_3) = \frac{6}{5}f^{\text{SRII}}_{\text{NL}}{\times}(2\pi^{2})^{2}\bigg[\frac{\Delta_{\bf {Tree}}^{\text{SRII}}(k_1)\Delta_{\bf {Tree}}^{\text{SRII}}(k_2)}{k_1^{3}k_2^{3}}+\frac{\Delta_{\bf {Tree}}^{\text{SRII}}(k_2)\Delta_{\bf {Tree}}^{\text{SRII}}(k_3)}{k_2^{3}k_3^{3}}+\frac{\Delta_{\bf {Tree}}^{\text{SRII}}(k_3)\Delta_{\bf {Tree}}^{\text{SRII}}(k_1)}{k_3^{3}k_1^{3}}\bigg].\quad\quad\quad
\eea
where we have applied the dimensionless power spectrum in the SRII area. Similar to how the USR region was handled, the explicit calculations for the squeezed limit in the aforementioned equations for the bispectrum and the associated non-Gaussianity amplitude $\fnl$ in the SRII region are handled by carrying out a numerical analysis for the amplitude $\fnl$ with respect to the wave numbers in the SRII region, for various values of the effective sound speed. Look at the Appendix \ref{A4a} for more details on this computation.

\end{itemize}

The combined version from the various bispectrum contributions and the corresponding non-Gaussian amplitudes from the three-point function of the scalar modes—which were calculated for each of the three regions—are shown in this section. The behaviour of the amplitude at the transition sites, $\tau_s$ for SRI to USR and $\tau_e$ for USR to SRII transitions, would require significant caution if we were to do this. The following is how the distinct transition aspects between the three zones are handled at the corresponding wavenumbers: 
\bea \label{c4B} 
{B}^{\text{Total}}_{\zeta\zeta\zeta}(k_{1},k_{2},k_{3}) = {B}^{\text{SRI}}_{\zeta\zeta\zeta}(k_{1},k_{2},k_{3}) + \Theta(k-k_{s}){B}^{\text{USR}}_{\zeta\zeta\zeta}(k_{1},k_{2},k_{3}) + \Theta(k-k_{e}){B}^{\text{SRII}}_{\zeta\zeta\zeta}(k_{1},k_{2},k_{3}) \eea
When aggregating the individual bispectrum findings, we resort to the Eqs. (\ref{c1BI},\ref{c2BI},\ref{c3BI}) to determine the overall bispectrum, which is represented by this amount. In the end, this value will be helpful in providing us with the overall behaviour of the non-Gaussian amplitude over all wavenumbers. In this formula, the various contributions are likewise carefully joined at their respective wavenumbers using Heaviside Theta functions, much as in the case of the total tree-level scalar power spectrum in Eqn.(\ref{totpower}).

In the squeezed limit scenario, when one of the momenta is significantly shorter (long wavelength) than the others, namely $k_{1} \rightarrow 0$ and $k_{2} \approx k_{3} = k$, the overall contribution to the tree-level bispectrum may be expressed as follows:
\bea \label{c4Bsq}
B^{\text{Total}}_{\zeta\zeta\zeta}(k_{1} \rightarrow 0, k, k) &=& B^{\text{SRI}}_{\zeta\zeta\zeta}(k_{1} \rightarrow 0, k, k | k \leq k_{s}) + \Theta(k-k_{s})B^{\text{USR}}_{\zeta\zeta\zeta}(k_{1} \rightarrow 0, k, k | k_{s} \leq k \leq k_{e})\nonumber\\
&& + \Theta(k-k_{e})B^{\text{SRII}}_{\zeta\zeta\zeta}(k_{1} \rightarrow 0, k, k | k_{e} \leq k \leq k_{\text{end}}).
\eea 
We now utilise this to mention the formula for the total amplitude of non-Gaussianity $\fnl$ in the squeezed limit, which is as follows:
\bea \label{c4fsq}
f^{\text{Total}}_{\text{NL}}(k) = f^{\text{SRI}}_{\text{NL}}(k \leq k_{s}) + \Theta(k-k_{s})f^{\text{USR}}_{\text{NL}}(k_{s} \leq k \leq k_{e}) + \Theta(k-k_{e})f^{\text{SRII}}_{\text{NL}}(k_{e} \leq k \leq k_{\text{end}}).
\eea
The subsequent part will analyse the data pertaining to this amount, beginning with the individual contributions and concluding with the ultimate behaviour of the entire non-Gaussian amplitude across all wavenumbers meticulously connected using Heaviside Theta functions.

\subsubsection{Numerical Results}
Here, we report our findings from the investigation of the squeezed limit's non-Gaussian amplitude, $\fnl$. The plots that illustrate the behaviour of the parameter $\fnl$ with respect to the wave number ($k$) for various effective sound speed values ($c_{s}$) originating from each of the four interaction operators separately and, ultimately, their combined contributions separately for each of the three regions (SRI, USR, and SRII) will be discussed first. Here, we present our results on the study of the non-Gaussian amplitude, $\fnl$, of the squeezed limit. First, we will talk about the plots that show how the parameter $\fnl$ behaves in relation to the wave number ($k$) for different effective sound speed values ($c_{s}$) that come from the four interaction operators separately and, in the end, their combined contributions separately for each of the three regions (SRI, USR, and SRII).

\subsubsubsection{Results obtained from region I: SRI}

    \begin{figure*}[htb!]
    	\centering
    \subfigure[]{
      	\includegraphics[width=8.5cm,height=7cm] {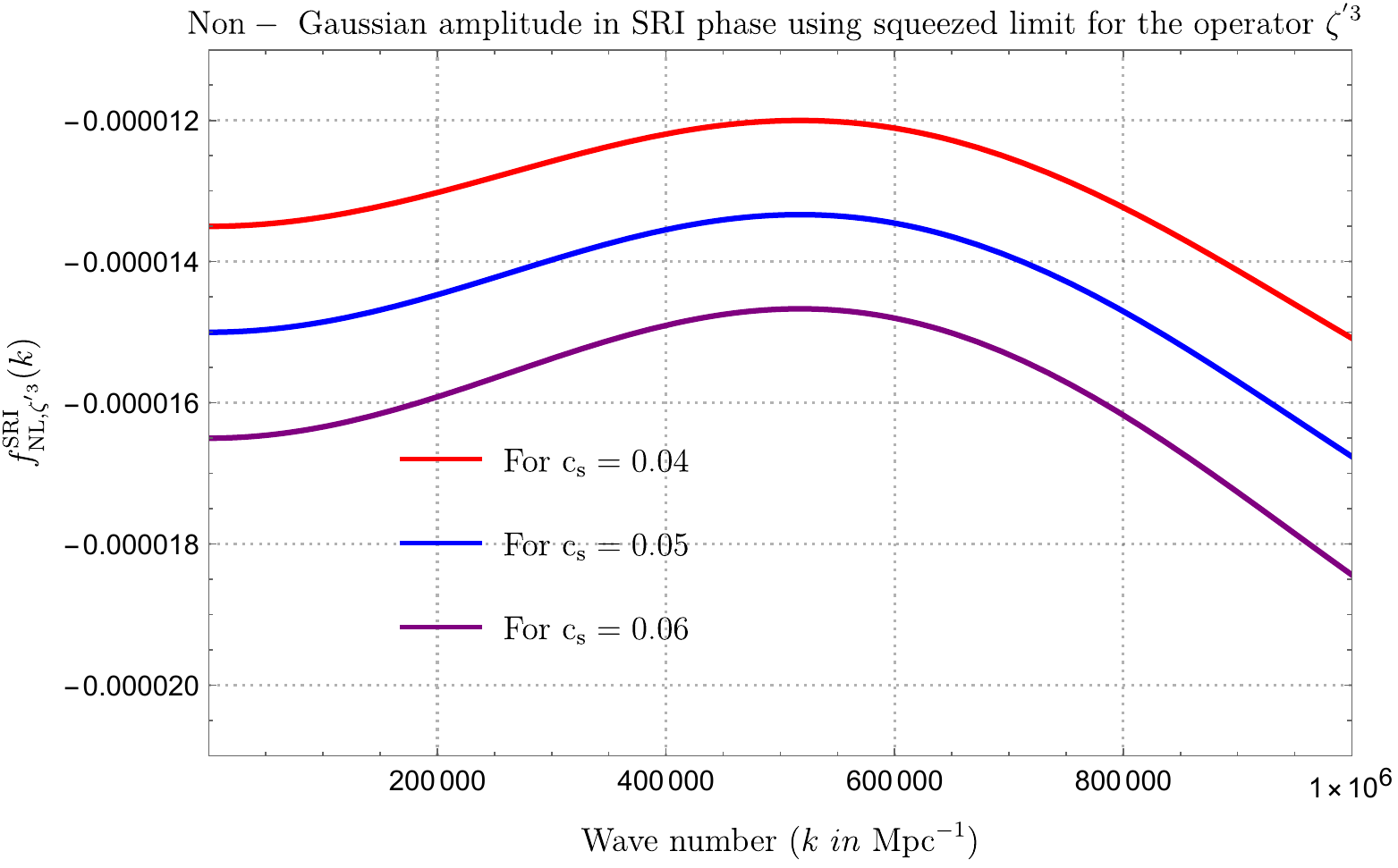}
        \label{I1}
    }
    \subfigure[]{
       \includegraphics[width=8.5cm,height=7cm] {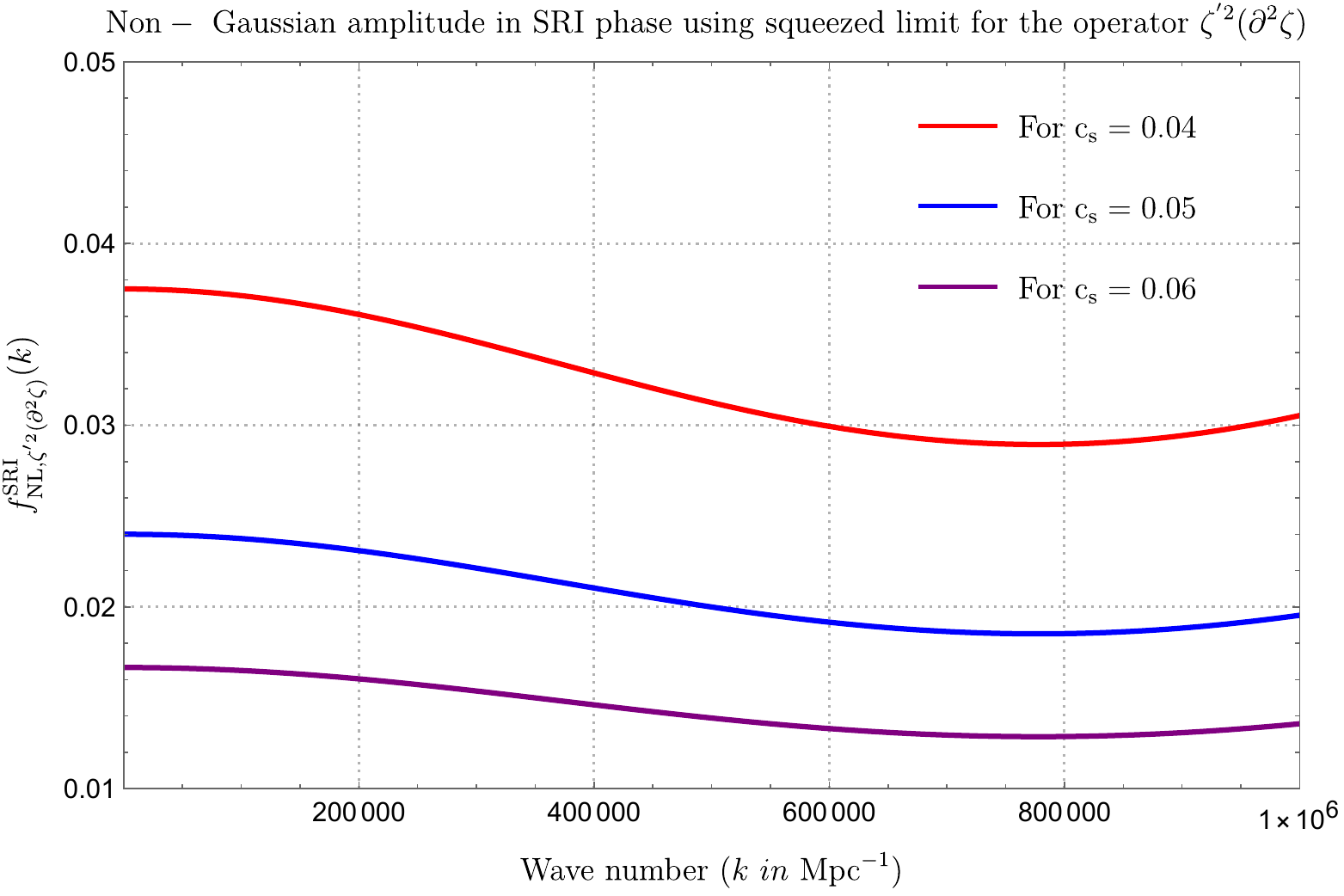}
        \label{I2}
       }
        \subfigure[]{
       \includegraphics[width=8.5cm,height=7cm] {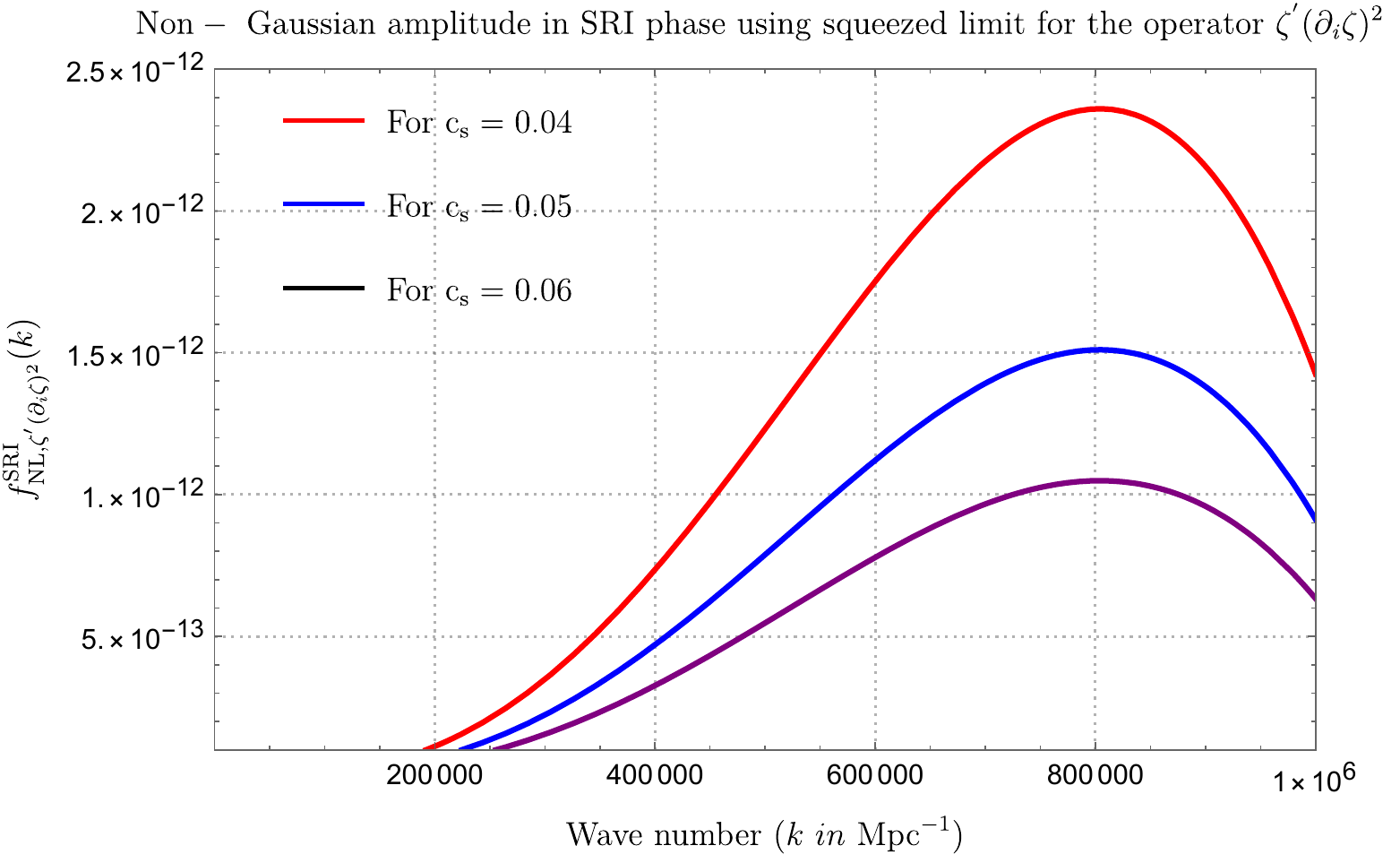}
        \label{I3}
       }
        \subfigure[]{
       \includegraphics[width=8.5cm,height=7cm] {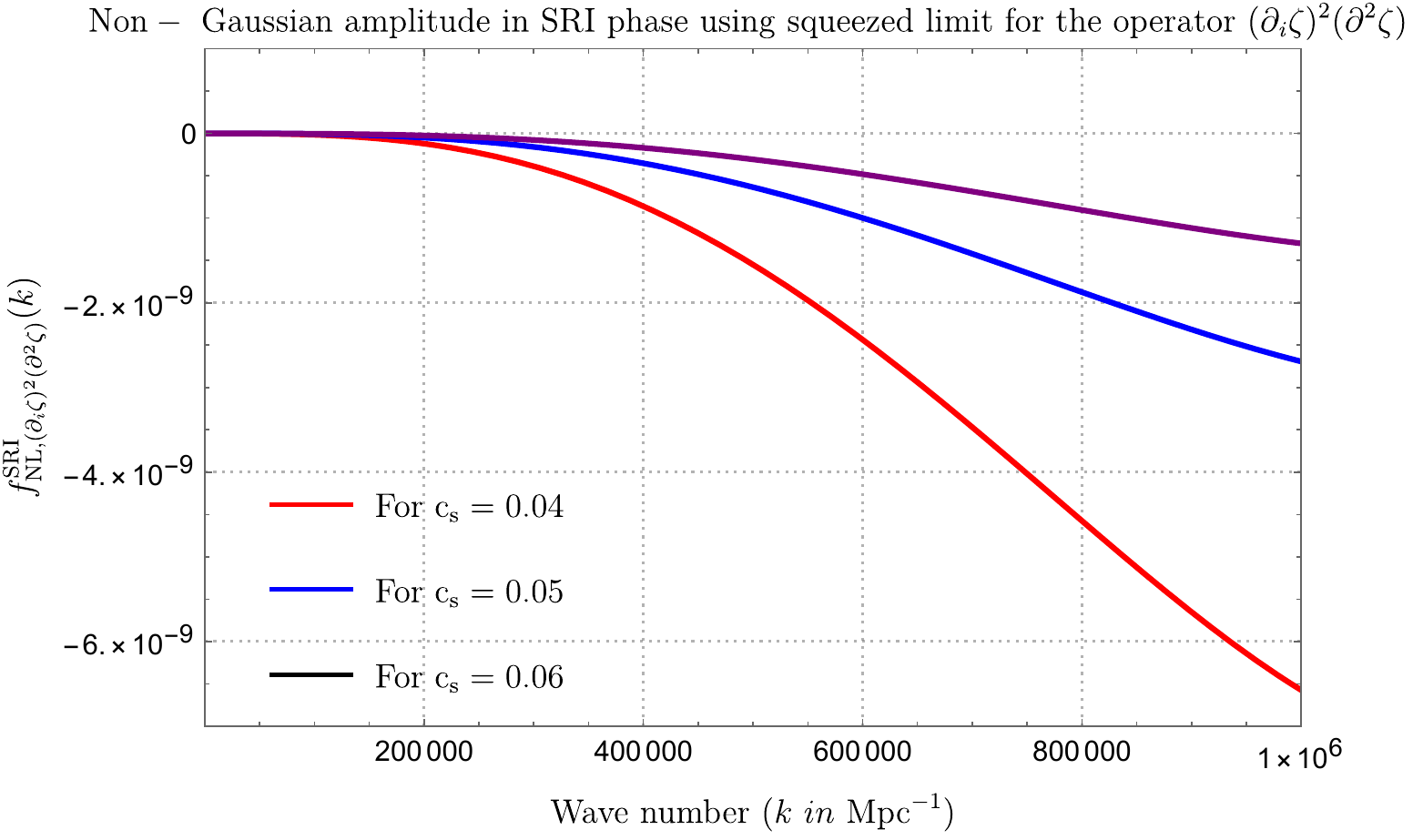}
        \label{I4}
       }
    	\caption[Optional caption for list of figures]{Plotting the contributions of different operators to the non-Gaussian amplitude $f_{\text{NL}}$ in the SRI area as a function of wave number is shown in the figure. The following numbers are selected to represent the effective sound speed parameter $c_{s}$. $c_{s} = 0.05,0.06, 0.04,$. The contributions from the first and second operators are shown in the graphs in the upper row. Contributions from the third and fourth operators are shown in the plots in the bottom row.
} 
    	\label{fNLSRI:a}
    \end{figure*}
    \begin{figure*}[htb!]
    	\centering
{
      	\includegraphics[width=18cm,height=12.5cm] {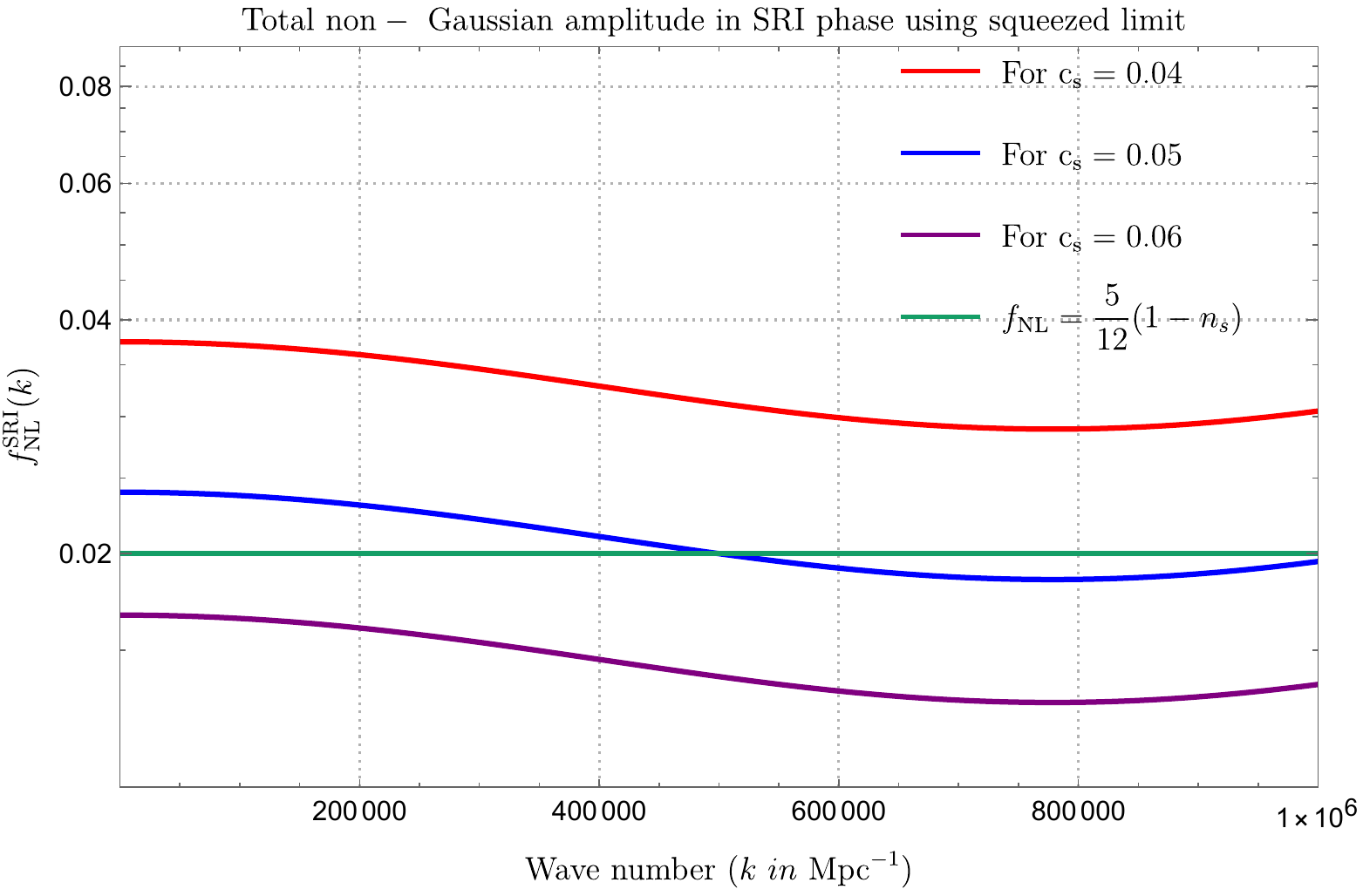}
        \label{K1}
    }
    	\caption[Optional caption for list of figures]{Plot shows the entire operator contribution to the non-Gaussian amplitude $f_{\text{NL}}$ under the squeezed limit, as a function of wave number in the SRI phase. The effective sound speed parameter $c_{s}$ is determined to have the following values: $c_{s} = 0.04,0.05,0.06$. In this instance, we discover that the value indicated by the consistency requirement and $c_{s} = 0.05$ have the best agreement.
} 
    	\label{fNLSRI:b}
    \end{figure*}

    We provide our results in this paragraph using example graphs that show how the non-Gaussian amplitude $\fnl$ varies in the SRI area as a function of wave number ($k$). We first demonstrate each operator's unique contribution to this behaviour, and then we present the collective behaviour of all the operators.

    The contribution of the first operator $\zeta^{'3}$ towards $\fnl$ in plot (\ref{I1}) is less than that of the second operator, which makes the largest overall contribution in the SRI, plot (\ref{I2}), as can be observed from the figure in Fig. (\ref{fNLSRI:a}). The least significant contribution in plots (\ref{I3},\ref{I4}) comes from the operators third $\zeta^{'}(\partial_{i}\zeta)^{2}$ and fourth $(\partial_{i}\zeta)^{2}\partial^{2}\zeta$. In particular, for $c_{s}=0.05$, the result is in precise agreement with the consistency relation. The results from the plot (\ref{I2}) are extremely near to what the value of $\fnl$ is determined from the Maldacena's consistency relation in the squeezed limit, i.e., $\fnl \sim {\cal O}(10^{-2})$. When $c_{s}$ is raised, all individual contributions—aside from those from the first operator—tend to go towards zero. Because of the analytic structure of the individual terms and their reliance on $c_{s}$, where the causality and unitarity requirements are precisely preserved throughout the analysis, this particular behaviour results.

    The sum of the contributions from each operator is shown in fig. (\ref{fNLSRI:b}), where the $\fnl$ amplitude is plotted against the wave number, taking into account a range of various sound speed values to look for variations in the findings. Based on the plot analysis shown in fig. (\ref{fNLSRI:a}), it can be observed that the overall contribution in Fig. (\ref{fNLSRI:b}) resembles that of the second operator, but the inclusion of contributions from the first operator is visible in the outcome. The coefficients ${\cal G}_{i}$, $\forall i=1,2,3,4$, capture the fact that the second operator contributes so much, in addition to the first operator's negligible but noticeable contribution and the other two operators' least significant impacts. Ultimately, based on the overall findings in fig. (\ref{fNLSRI:b}), it is evident that, for $c_{s}=0.05$, the consistency relation derived from Maldacena's \textit{no-go theorem} is strictly valid. Even minor variations around this value do not result in a violation of the consistency relation, as would be expected in the case of the SRI region.

\subsubsubsection{Results obtained from region II: USR}

    \begin{figure*}[htb!]
    	\centering
    \subfigure[]{
      	\includegraphics[width=8.5cm,height=7.5cm] {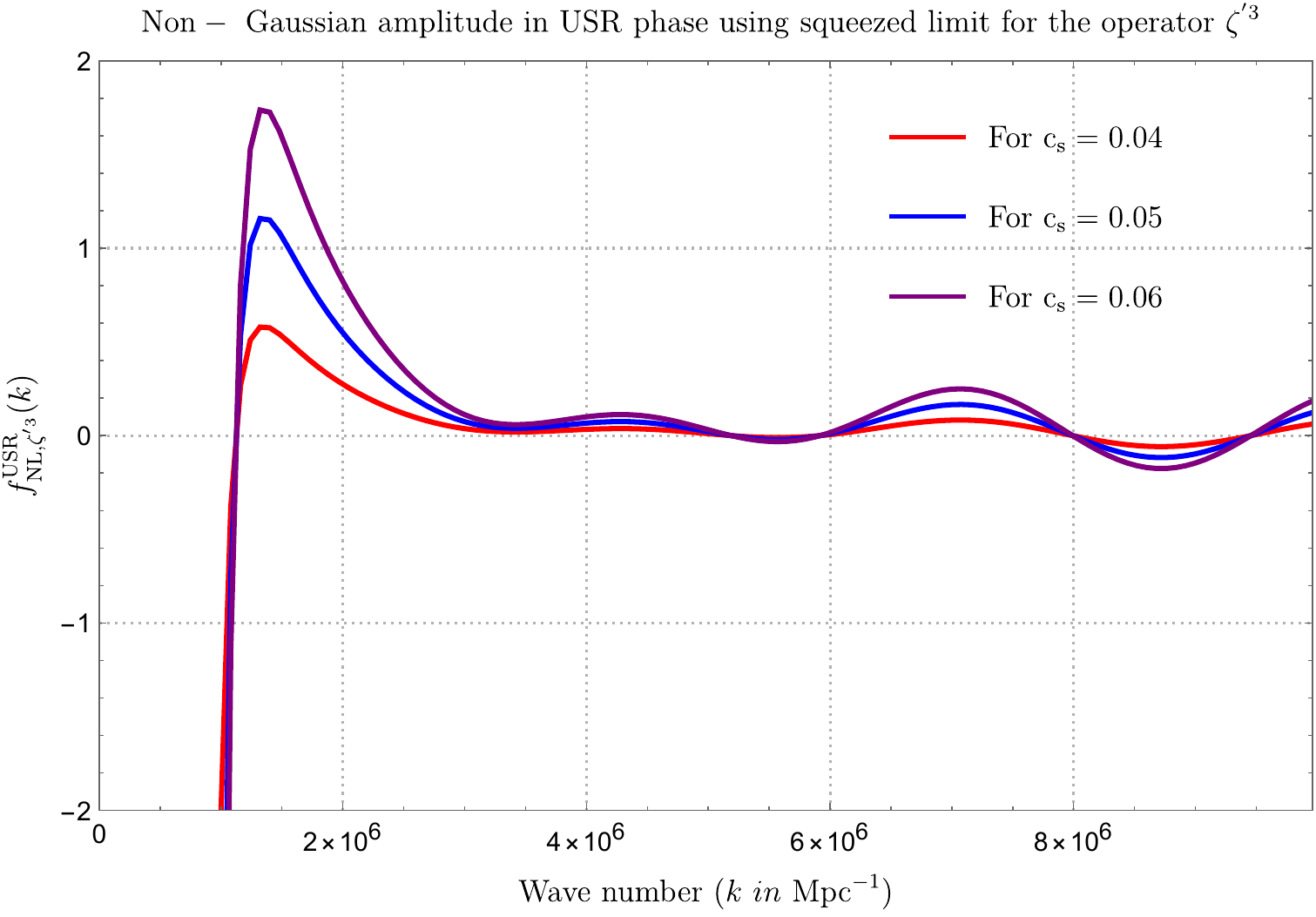}
        \label{IU1}
    }
    \subfigure[]{
       \includegraphics[width=8.5cm,height=7.5cm] {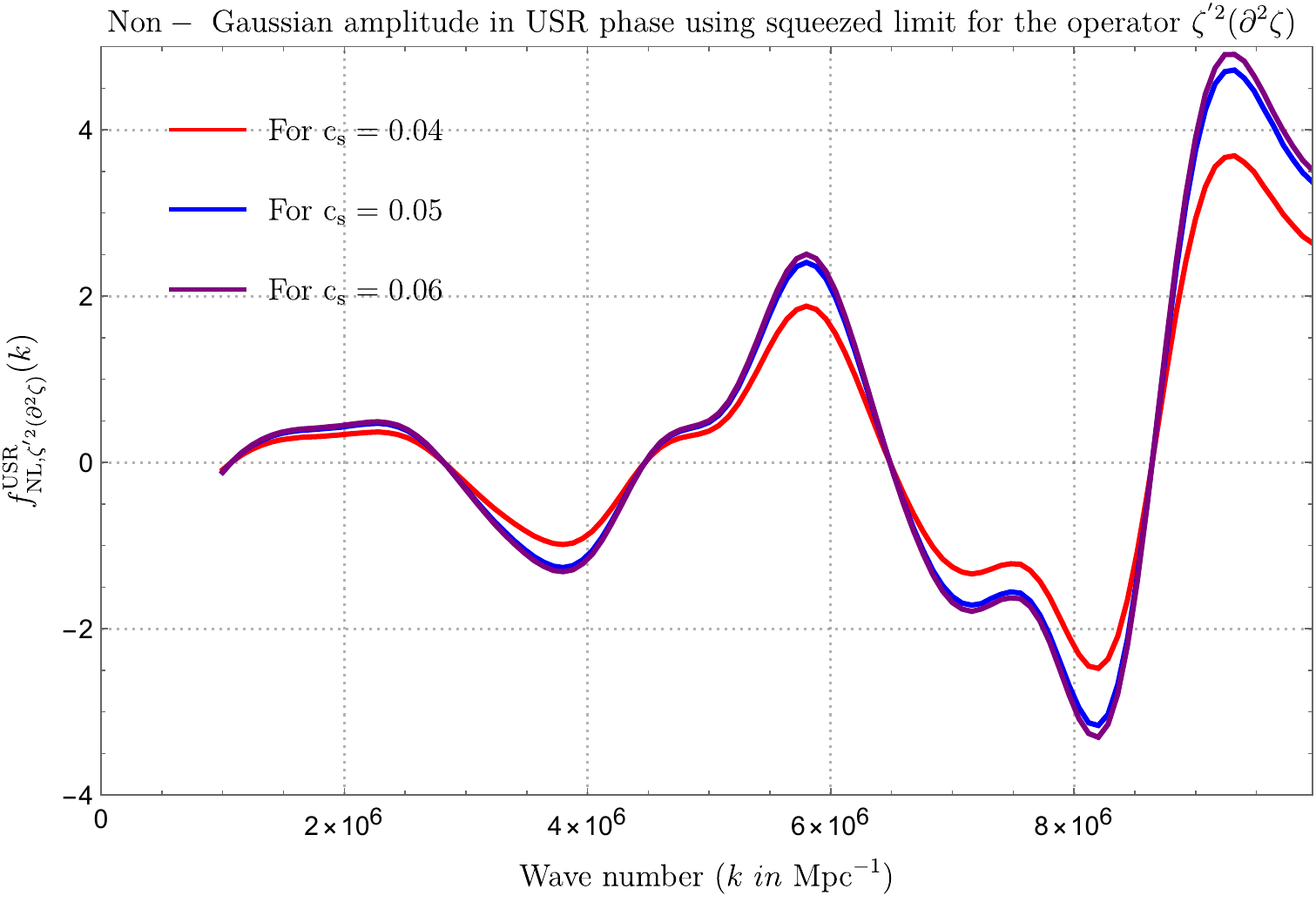}
        \label{IU2}
       }
        \subfigure[]{
       \includegraphics[width=8.5cm,height=7.5cm] {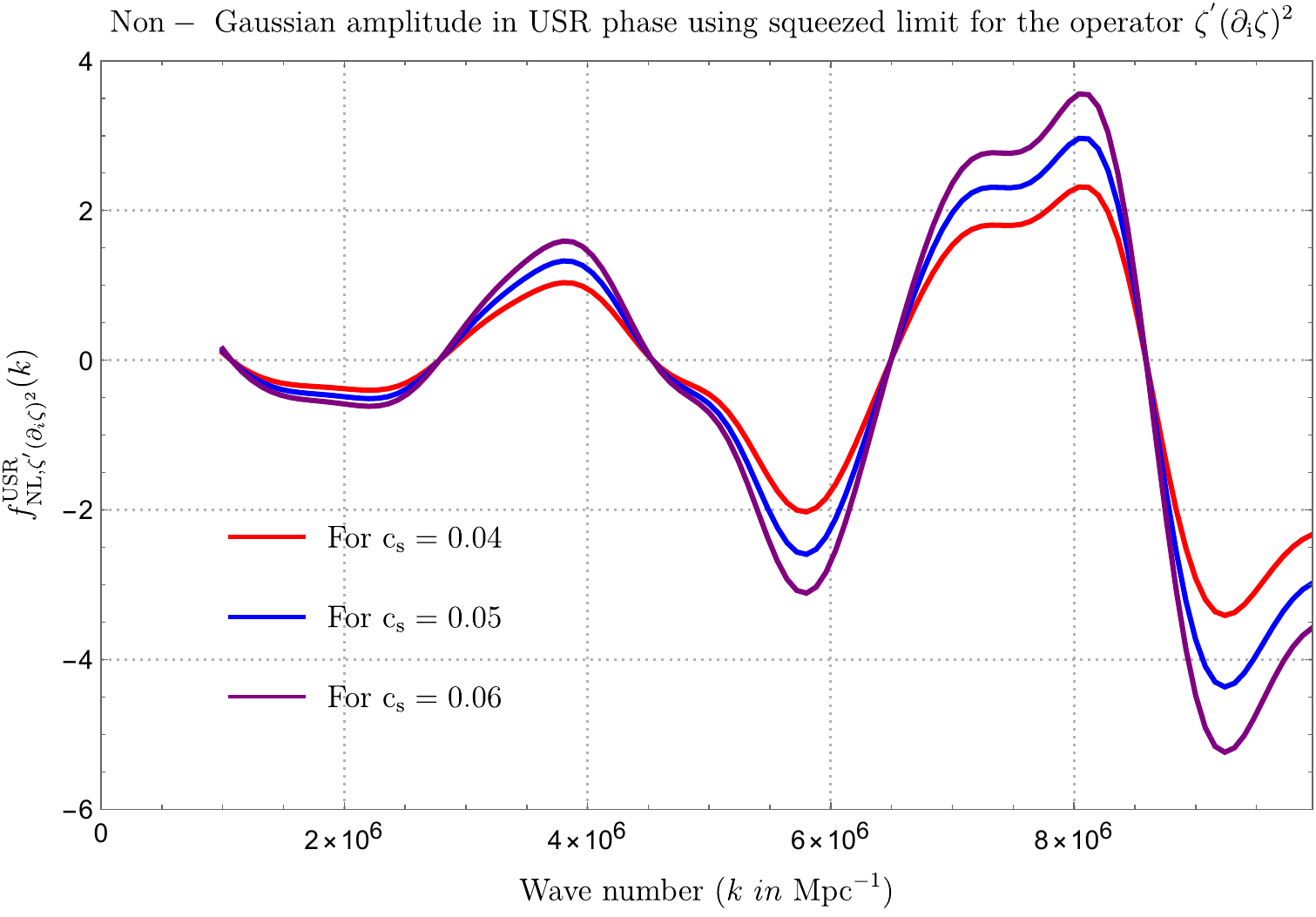}
        \label{IU3}
       }
        \subfigure[]{
       \includegraphics[width=8.5cm,height=7.5cm] {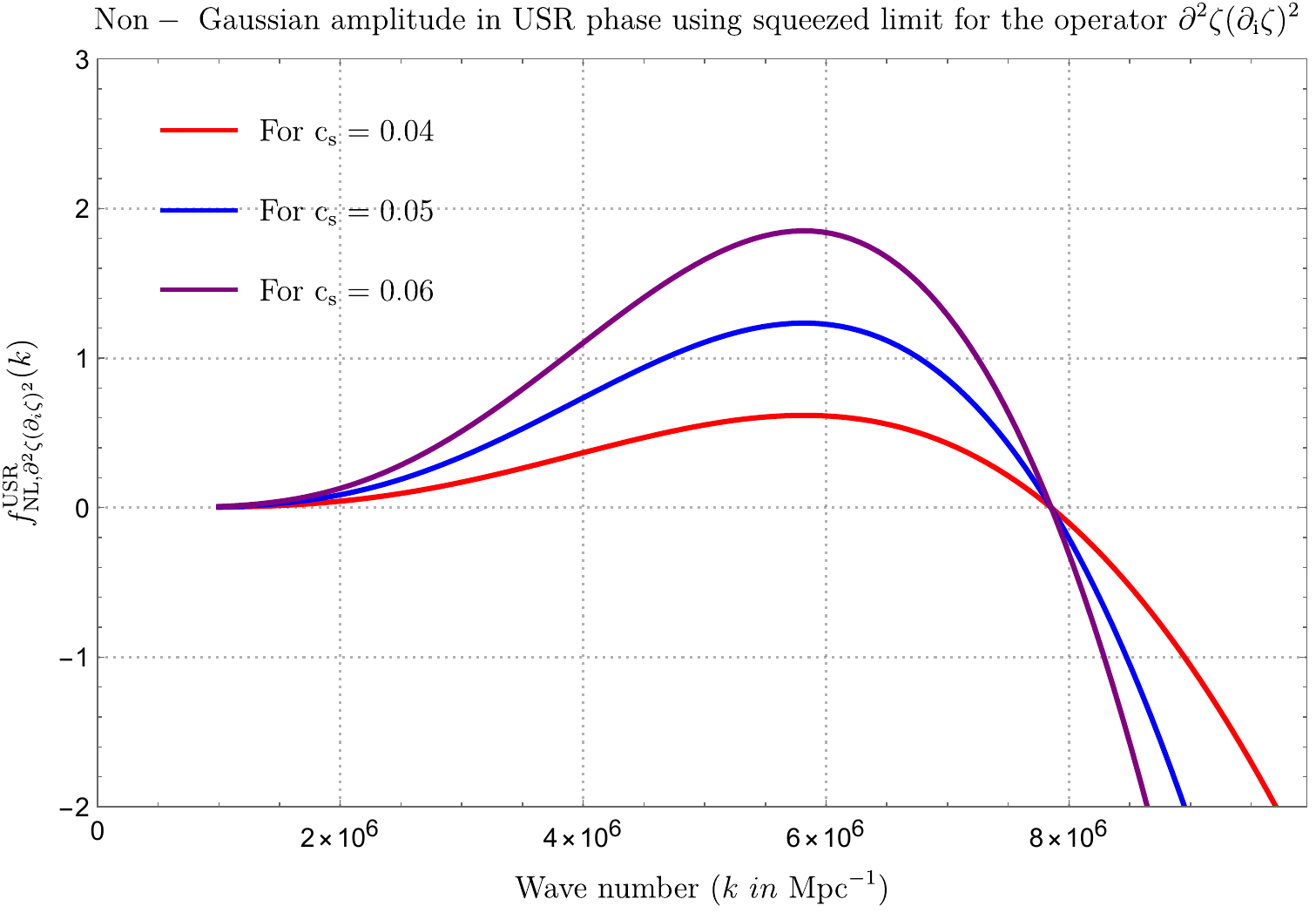}
        \label{IU4}
       }
    	\caption[Optional caption for list of figures]{Plotting the individual operator contributions to the non-Gaussian amplitude $f_{\text{NL}}$ in the USR area as a function of wave number is shown. The following numbers are selected to represent the effective sound speed parameter $c_{s}$. $c_{s} = 0.05,0.06, 0.04,$. The contributions from the first and second operators are shown in the graphs in the upper row. Contributions from the third and fourth operators are shown in the plots in the bottom row.
} 
    	\label{fNLUSR:a}
    \end{figure*}

    \begin{figure*}[htb!]
    	\centering
   {
      	\includegraphics[width=18cm,height=12.5cm] {
      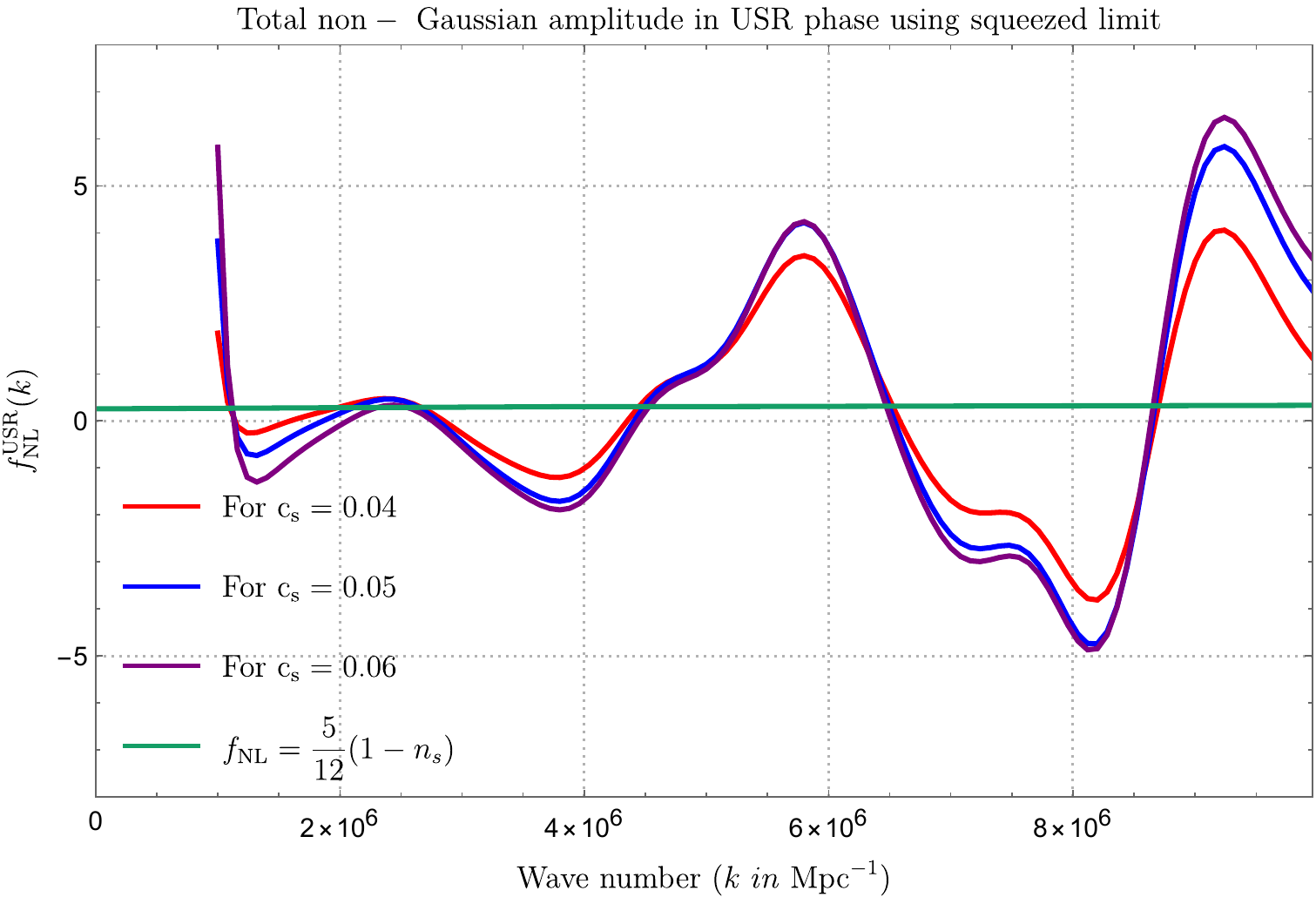}
        \label{KU1}
    }
    	\caption[Optional caption for list of figures]{Plot displays, under the squeezed limit, the total contribution of all operators to the non-Gaussian amplitude $f_{\text{NL}}$ as a function of wave number in the USR phase. The following numbers are selected to represent the effective sound speed parameter $c_{s}$. We have $c_{s} = 0.04,0.05,0.06$. We discover that this is a violation of the consistency condition's anticipated value.
} 
    	\label{fNLUSR:b}
    \end{figure*}
    This subsection contains the typical graphs that show how the non-Gaussian amplitude $f_{\text{NL}}$ varies in the USR area as a function of wave number. Similar to before, we first display each operator's unique contribution to this behaviour before providing the overall behaviour of all the operators. 

    The value provided by the consistency relation in the squeezed limit, i.e., $\fnl \sim {\cal O}(10^{-2})$, is substantially lower than the contributions from individual operators in the USR area to the non-Gaussian amplitude, $\fnl$, as can be seen in figure Fig. (\ref{fNLUSR:a}). This conduct demonstrates a blatant disregard for the {\it Maldacena's consistency relation} in the USR area. Unlike the SRI region, where a small number of operators had a noticeable influence, contributions from all of the operators are necessary to understand the general behaviour in this region. The numbers in these plots offer us the necessary boost for the creation of PBH and in the related primordial power spectrum amplitude, and they are well within the limitations for sustaining the perturbativity approximation. As compared to the contributions from the second and third operators in plots (\ref{IU2}, \ref{IU3}), the first operator in plot (\ref{IU1}) contributes oscillations that are less significant, and the fourth operator in plot (\ref{IU4}) does not clearly display an oscillatory nature. Additionally, all of the operators exhibit a sudden spike or decrease in the amplitude's behaviour, either at the start of the USR phase (for $k_s \sim {\cal O}(10^6){\rm Mpc^{-1}}$) or towards the phase's conclusion (for $k_e \sim {\cal O}(10^7){\rm Mpc^{-1}}$). This leads us to the conclusion that there will be a noticeable change when moving from the SRI to the USR region or the USR to the SRII zone. The charts do not noticeably differ from one another for varying sound speed values, and the unitarity and causality criteria are completely upheld. All of these typical statistics clearly show that the behaviour is more enhanced when we increase the sound speed values than it was for the prior values.

We show the overall contribution of all operators to the non-Gaussian amplitude $f_{\rm NL}$ in the plot in Fig. (\ref{fNLUSR:b}). Following their addition, the values precisely at the location of the transition wave numbers $k_{s}$ and $k_{e}$ exhibit a distinct sharp behaviour. The values for $f_{\rm NL}$ in the USR region are substantially bigger than that, suggesting a violation of the consistency relation. The green line represents the value derived from the consistency relation. To regulate the amplitude's behaviour, the coefficients ${\cal G}_{i}$ $\forall i=1,2,3,4$ are changed appropriately.

\subsubsubsection{Results obtained from region III: SRII}

    \begin{figure*}[htb!]
    	\centering
    \subfigure[]{
      	\includegraphics[width=8.5cm,height=7.5cm] {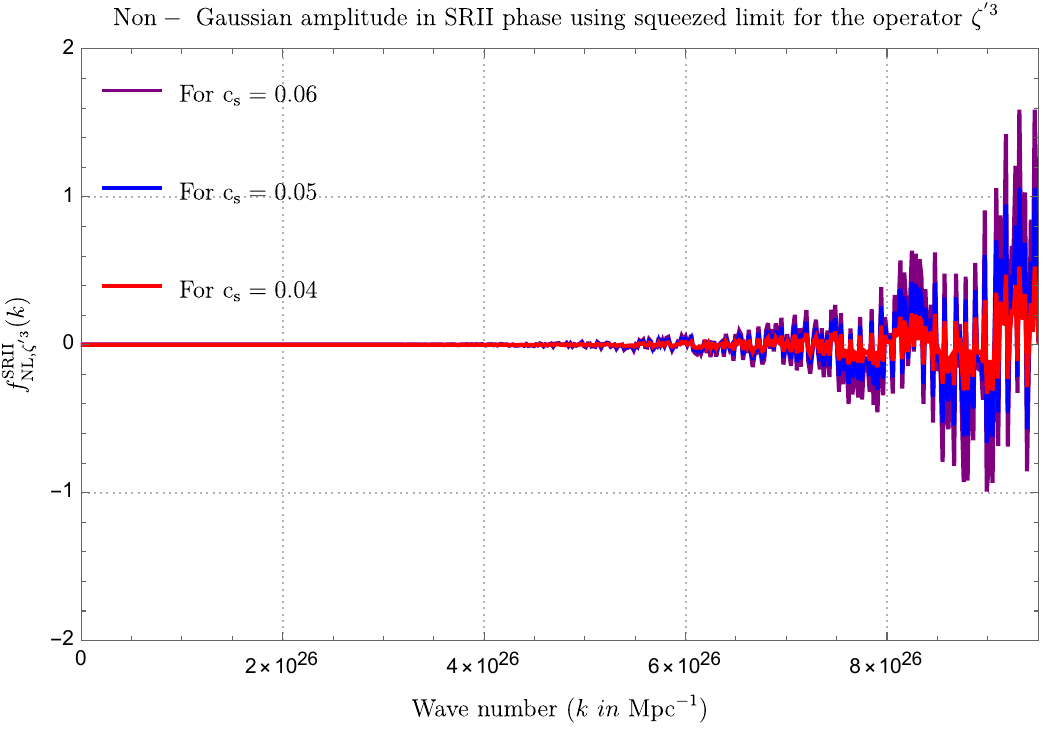}
        \label{Is21}
    }
    \subfigure[]{
       \includegraphics[width=8.5cm,height=7.5cm] {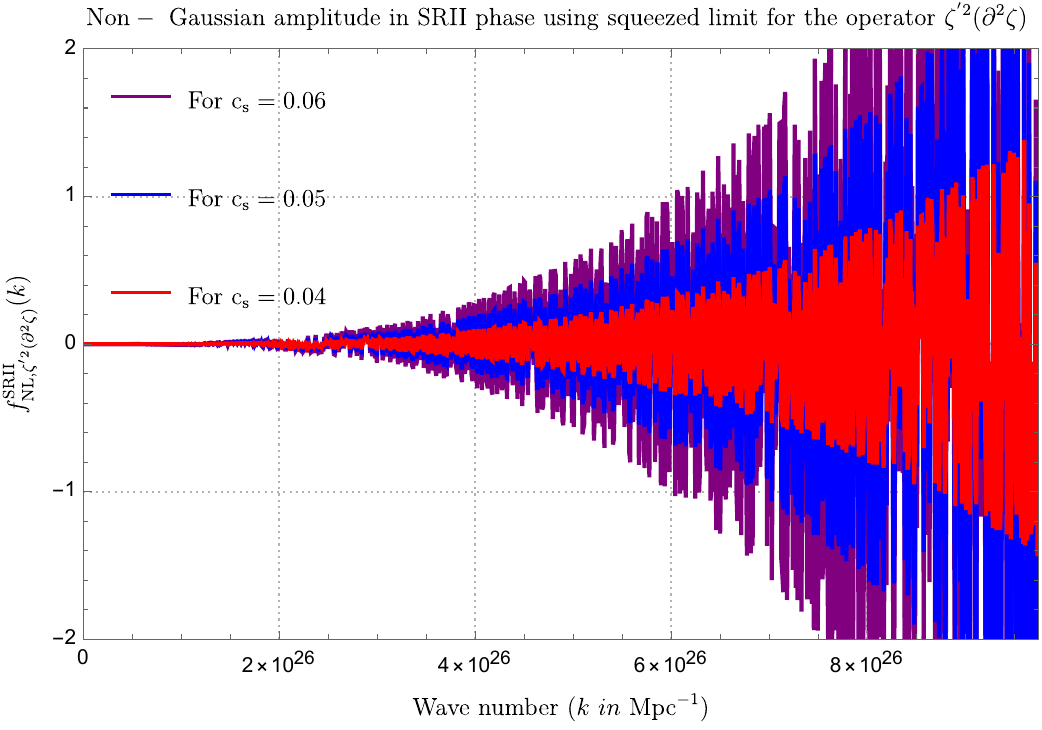}
        \label{Is22}
       }
        \subfigure[]{
       \includegraphics[width=8.5cm,height=7.5cm] {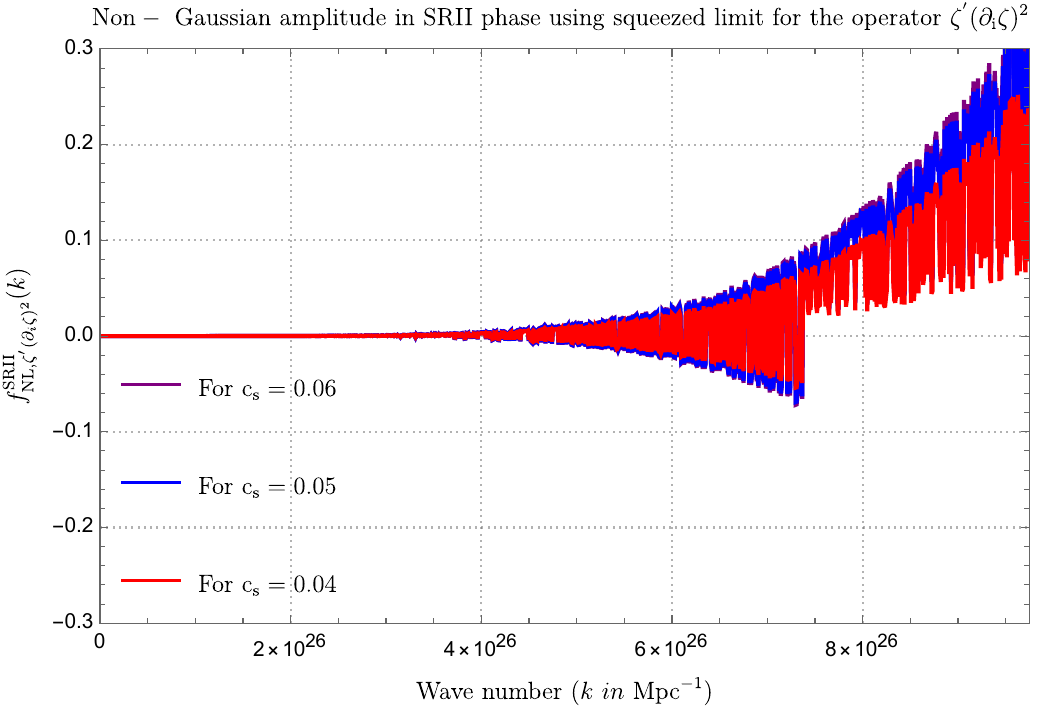}
        \label{Is23}
       }
        \subfigure[]{
       \includegraphics[width=8.5cm,height=7.5cm] {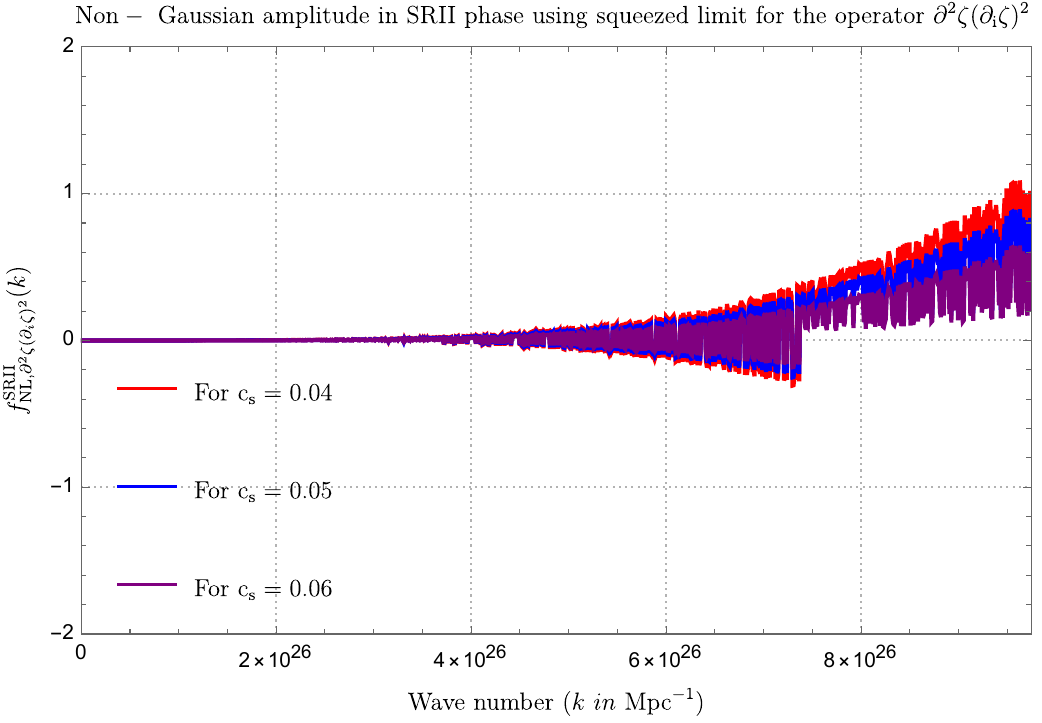}
        \label{Is24}
       }
    	\caption[Optional caption for list of figures]{ The non-Gaussian amplitude $f_{\text{NL}}$ in the SRII area is plotted as a function of wave number, showing the contributions of various operators. The following values are used for the effective sound speed parameter $c_{s}$. We have $c_{s} = 0.04,0.05,0.06$. The first and second operators' contributions are shown in the graphs in the top row. The third and fourth operators' contributions are shown in the graphs in the bottom row.The non-Gaussian amplitude $f_{\text{NL}}$ in the SRII area is plotted as a function of wave number, showing the contributions of various operators. The following values are used for the effective sound speed parameter $c_{s}$. We have $c_{s} = 0.04,0.05,0.06$. The contributions from the third and fourth operators are shown in the graphs in the bottom row.} 
    	\label{fNLSRII:a}
    \end{figure*}
    \begin{figure*}[htb!]
    	\centering
   {
      	\includegraphics[width=18cm,height=8.5cm] {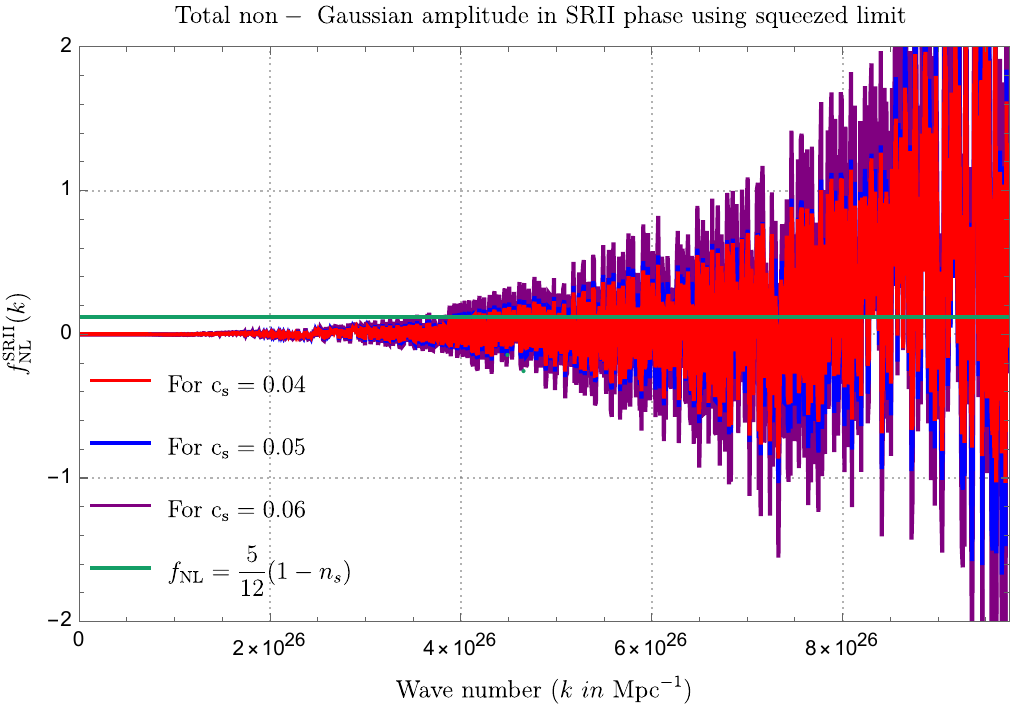}
        \label{Ks21}
    }
      	\caption[Optional caption for list of figures]{The non-Gaussian amplitude $f_{\text{NL}}$ is plotted as a function of wave number in the SRII phase under the squeezed limit, showing the combined contribution of all operators. The effective sound speed parameter $c_{s}$ is determined to have the following values: $c_{s} = 0.04,0.05,0.06$. It is evident from this that the consistency requirement is once more broken.
} 
    	\label{fNLSRII:b}
    \end{figure*}
The non-Gaussian amplitude $f_{\text{NL}}$ and its change with wave number in the second slow-roll (SRII) area are plotted in this subsection. We first demonstrate the unique contributions made by each operator to this behaviour, and then we present the collective behaviour of all the operators.

As can be observed from the figure Fig. (\ref{fNLSRII:a}), these contributions show a larger deviation from the value obtained from the consistency condition, i.e., $\fnl \sim {\cal O}(10^{-2})$, than their corresponding operator plots in the SRI region. This is to be expected as the SRII region's fluctuations are already more pronounced after passing through the USR phase, and the region's vacuum state is also of a non-Bunch-Davies kind, which permits substantial non-Gaussianities. The figures make it evident that the amplitude must be greater than the values in the SRI area while remaining less than those in the USR region. Furthermore, it appears that the fluctuations only get much larger when $k_{\text{end}} \geq 6 \times 10^{26}$ Mpc$^{-1}$, which is the end of inflation, is achieved. Regarding individual contributions, the fluctuations from the plot (\ref{Is21}) tend to be less rapid and grow quickly when $c_{s}$ is raised, however the fluctuations from the plot (\ref{Is22}) are more violent and increase violently as $c_{s}$ is increased. Similar circumstances apply to the graphs (\ref{Is23}, \ref{Is24}), where we only see positive oscillations beyond a particular wave number. On the other hand, increasing $c_{s}$ only causes the oscillations to diminish for the fourth operator. The analytic connection for this operator and its reliance on the parameter $c_{s}$ make this clear.

The combined behaviour of all the operators is given in Fig. (\ref{fNLSRII:b}), where the parameter $\fnl$ is plotted against the wave number while taking into account the same set of varied sound speed values to check for variations in the findings. As the parameter $c_{s}$ is raised, the result finally exhibits the same violent oscillations that are observed in the individual contributions, with values increasing. The plot makes the departure from the consistency criterion evident as well.

    \begin{figure*}[htb!]
    	\centering
   {
      	\includegraphics[width=18cm,height=8.5cm] {
      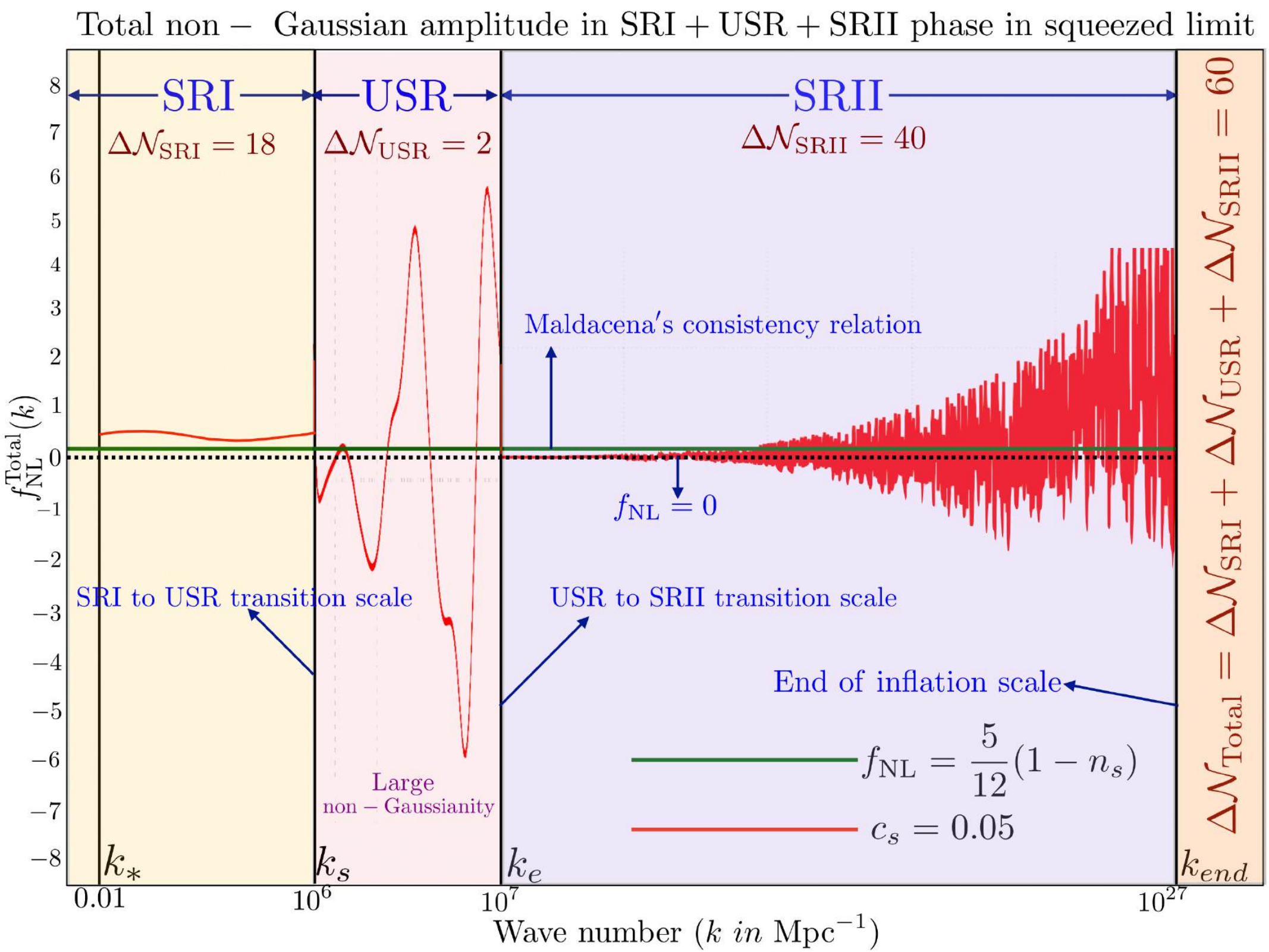}
        \label{KUPBH}
    }
    	\caption[Optional caption for list of figures]{A representative graphic showing the three phases' cumulative behavior—SRI, USR, and SRII—as a function of wave number in the squeezed limit, leading to the non-Gaussian amplitude $f_{\text{NL}}$. As $c_{s} = 0.05$, the effective sound speed $c_{s}$ is fixed. $k_{*} = 0.01$ Mpc$^{-1}$ is the fixed pivot scale; $k_{s} = 10^{6}$ Mpc$^{-1}$ is the fixed SRI to USR transition; $k_{e} = 10^{7}$ Mpc$^{-1}$ is the fixed USR to SRII transition; and $k_{\text{end}} = 10^{27}$ Mpc$^{-1}$ is the fixed inflation end. The USR phase exhibits significant non-Gaussianity, with the maximal non-Gaussianity value being found at a scale of $\sim$ 8 $\times$ 10$^{6}$ Mpc$^{-1}$. Lastly, our research yields a sufficient number of 60 e-foldings.
}
        \label{totPBH}
    \end{figure*}
    Diagram illustrating the cumulative behaviour of the three phases—SRI, USR, and SRII—in the squeezed limit as a function of wave number, leading to the non-Gaussian amplitude $f_{\text{NL}}$ is depicted in figure (\ref{totPBH}). Here $c_{s} = 0.05$ is the fixed effective sound speed $c_{s}$. The fixed values for the pivot scale, SRI to USR transition, USR to SRII transition, and end of inflation are as follows: $k_{\text{end}} = 10^{7}$ Mpc$^{-1}$, $k_{s} = 10^{6}$ Mpc$^{-1}$, and $k_{e} = 10^{7}$ Mpc$^{-1}$. Within the USR phase, there is a significant amount of non-Gaussianity. The maximal non-Gaussianity value is determined at a scale of $\sim$ 8 $\times$ 10$^{6}$ Mpc$^{-1}$. Finally, the results of our study yield a sufficient number of 60 e-foldings.

\subsubsubsection{Bispectrum related shape function and the corresponding numerical plots
}
    \begin{figure*}[htb!]
    	\centering
   {
      	\includegraphics[width=16cm,height=8.5cm] {
    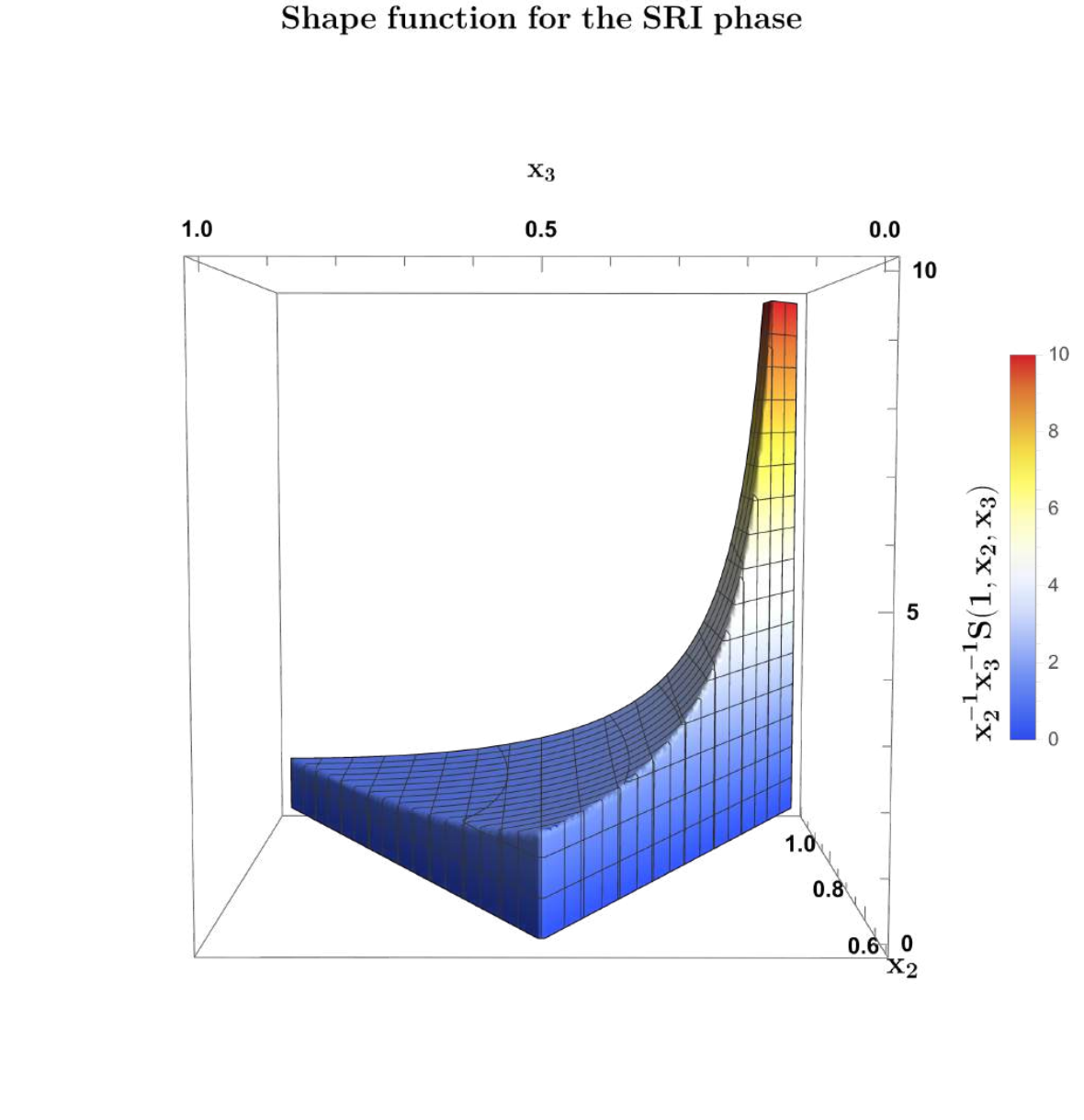}
        \label{shape1}
    }
    	\caption[Optional caption for list of figures]{The function $x_{2}^{-1}x_{3}^{-1}S(1,x_{2},x_{3})$ is plotted for the SRI area. In the equilateral limit, $x_{2}=x_{3}=1$, the shape is normalised to $1$, and it vanishes beyond the area $1-x_{2} \leq x_{3} \leq x_{2}$.
}  
        \label{SRIshape}
    \end{figure*}

    \begin{figure*}[htb!]
    	\centering
   {
      	\includegraphics[width=16cm,height=8.5cm] {
    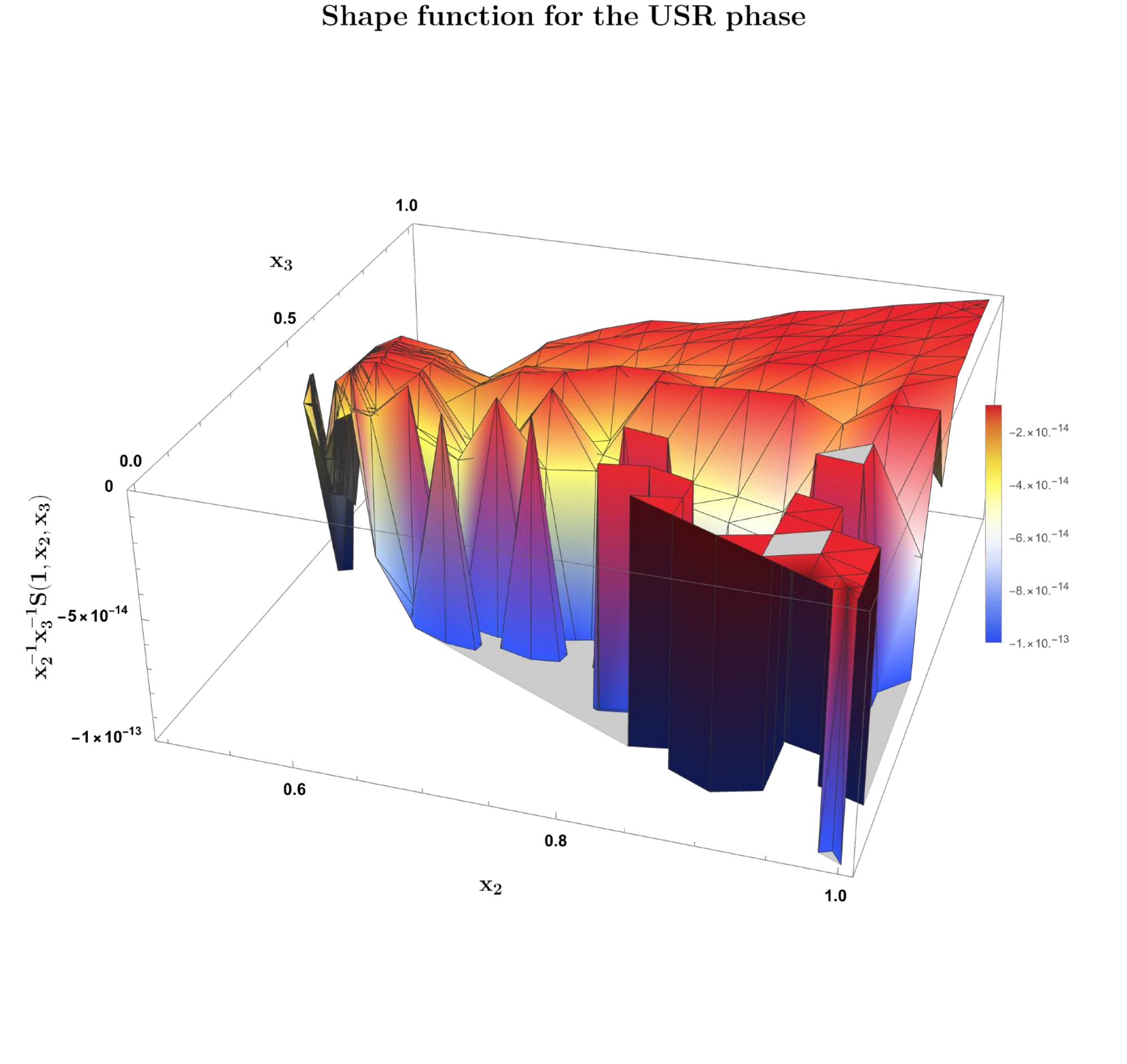}
        \label{shape2}
    }
    	\caption[Optional caption for list of figures]{The function $x_{2}^{-1}x_{3}^{-1}S(1,x_{2},x_{3})$ is plotted for the USR area. Outside of the region $1-x_{2} \leq x_{3} \leq x_{2}$, the form is made to disappear.
}  
        \label{USRshape}
    \end{figure*}

    \begin{figure*}[htb!]
    	\centering
   {
      	\includegraphics[width=16cm,height=8.5cm] {
    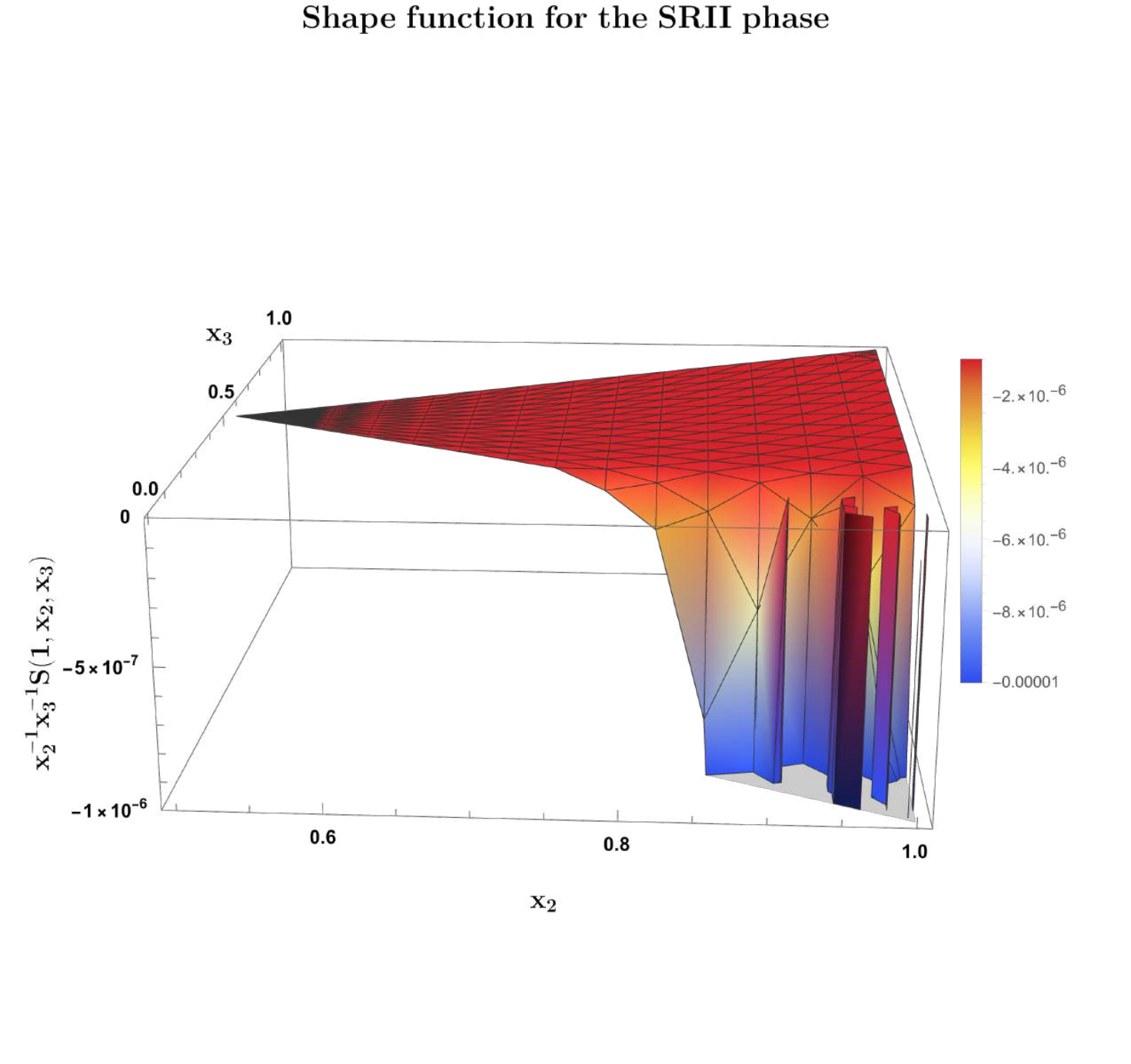}
        \label{shape3}
    }
    	\caption[Optional caption for list of figures]{The function $x_{2}^{-1}x_{3}^{-1}S(1,x_{2},x_{3})$ is plotted for the SRII area. Outside of the region $1-x_{2} \leq x_{3} \leq x_{2}$, the form is made to disappear.
}  
        \label{SRIIshape}
    \end{figure*}
Using the previously disclosed explicitly determined form for the three-point correlation functions, we presented our findings for the behaviour of the non-Gaussiantiy parameter, $f_{\rm NL}$, over the three areas in the preceding subsections. In this part, we discuss in more detail the form function—another important bispectrum property—and look at its findings for each of the three phases.

The definition of the bispectrum informs us that it is a function of the three momenta, ${\bf k_{1}}, {\bf k_{2}}$, and ${\bf k_{3}}$, such that those three momenta, depending on their magnitudes relative to each other, form different triangular configurations under the constraints imposed by translation invariance (homogeneity) and rotation invariance (isotropy). Information on the non-Gaussianity source may be found in the triangle's final shape. Accordingly, its qualitative analysis also contributes to the validation of the $f_{\rm NL}$ parameter determination \cite{Babich:2004gb,Chen:2006nt}. 

The alternative version that was utilised to define the bispectrum is now presented as follows:
\bea \label{shapeftn}
{B}_{\zeta\zeta\zeta}(k_{1},k_{2},k_{3}) = {S}(k_{1},k_{2},k_{3})\times\frac{6}{5}f_{\rm NL}(2\pi^{2})^{2}\prod\limits_{i=1}^{3}\frac{1}{k_{i}^{3}}.
\eea 
The aforementioned formula stems from the current bispectrum relation, which is involved in the tree-level scalar power spectrum in Eqs. (\ref{c1Bf},\ref{c2Bf},\ref{c3Bf}), and which in the end allows the quantity $f_{\rm NL}$ to be calculated for each region.

This specific version of $S(k_{1},k_{2},k_{3})$ is referred to as the bispectrum. It treats the momentum ratios, i.e., $x_{2} \equiv k_{2}/k_{1}$ and $x_{3} \equiv k_{3}/k_{1}$, as its variables, while keeping the overall momentum, $K=k_{1}+k_{2}+k_{3}$, constant. Depending on how the function and momenta are related, the form goes by several names. One of these is the \textit{squeezed} configuration, in which $k_{1} \ll k_{2} \sim k_{3}$, the momenta of which is relatively tiny in relation to the other two. This configuration becomes equal owing to momentum conservation. A further configuration is the \textit{equilateral}, which $k_{1}=k_{2}=k_{3}$ maintains the equality of all three momenta. The $3$D plot of the shape function is necessary in order to visualise these specific forms. There, we begin with the supposition that $x_{2} \leq x_{3}$, which helps to prevent thinking about repetition in configuration, and the arguments must abide by the triangle inequality $x_{2}+x_{3}\geq 1$. Since the definition employed in Eqn.(\ref{shapeftn}) requires the bispectrum to be determined as a homogeneous function of degree $-6$, the real quantity to $3$D plot is the dimensionless quantity, $x_{2}^{-1}x_{3}^{-1}S(1,x_{2},x_{3})$. Additionally, in order to prevent the same configuration from being shown again, we set the region outside the interval, $1-x_{2} \leq x_{3} \leq x_{2}$, to zero.

We now examine the different forms that were found for the three areas. The bispectrum, which begins with the SRI phase, shows interactions between several Fourier modes that come from the expression found in Eqn. (\ref{shapeftn}). On the other hand, the most relevant characteristic that can be seen in the image Fig.(\ref{SRIshape}) is where the shape peaks near the squeezed limit, at $x_{3} \rightarrow 0$ and $x_{2} \rightarrow 1$. This limit is associated with the situation where one of the modes grows far longer than the others, $k_{3} \rightarrow 0$ in this example, and freezes out well before it forms a backdrop for the other two modes to evolve against. The above form is the result of the non-linearities developing when they are outside the horizon. In this case, the shape function behaviour supports our previously established result for the non-Gaussianity $f_{\rm NL}$, in Fig.(\ref{fNLSRI:b}), for the SRI area, since the consistency criterion from Maldacena's theorem remains true.

Moving forward, we create the shape of the USR region next. Interesting form behaviour is seen in picture Fig. (\ref{USRshape}). The violet hue shows how steeply the form collapses at the squeezed limits, $x_{3} \rightarrow 0$ and $x_{2} \rightarrow 1$. The correlations in the squeezed limit are greatly muted since this property is exactly the opposite of what is required to establish the consistency criterion. The shape also exhibits jagged and discontinuous areas, particularly as one approaches the squeezed limit. These regions, as previously shown in fig.(\ref{fNLUSR:b}), indicate severe and fast changes in the non-Gaussianity $f_{\rm NL}$. The values of the form steadily grow when $x_{2} \rightarrow 1$ and $x_{3} \rightarrow 1$, which is also the equilateral limit. There is a divergence from the consistency criteria when there is a USR phase since it is associated with a non-attractor characteristic in the theory. Further, because the modes in our theory are higher derivative operators, there will eventually be more interactions between the modes that cross the horizon simultaneously; beyond that, however, their interactions will become insignificant. The equilateral limit's apparent bispectrum form is mostly due to this property.

A shape representation of the SRII phase is shown in figure Fig. (\ref{SRIIshape}). This form is nearly flat until it drastically shifts at the squeezed limit, also maximising near the equilateral limit, in a manner reminiscent to the USR instance. It is no longer guaranteed that the consistency requirement will hold in both the USR and SRII phases since they have an underlying quantum vacuum structure that is gradually pushed away from the Bunch-Davies vacuum state. Large non-Gaussianities in the USR and SRII, as also shown in fig. (\ref{fNLUSR:b},\ref{fNLSRII:b}), are therefore a result of the previously described vacuum characteristics and higher interactions, as indicated by the bispectrum shape in the equilateral limit.

\subsubsubsection{Cumulative results obtained from all regions: SRI + USR + SRII}

The typical figure in Fig. (\ref{totPBH}) depicts the combination of the particular behaviour of $f_{\text{NL}}$, as observed in the Figs, and explains the behaviour of the non-Gaussianity amplitude $f_{\text{NL}}$ through all three areas, SRI, USR, and SRII.combined into a single plot (\ref{fNLSRI:b},\ref{fNLUSR:b},\ref{fNLSRII:b}).

The constants $k_s = 10^{6}$ Mpc$^{-1}$ and $k_e = 10^{7}$ Mpc$^{-1}$ are fixed in the single field slow-roll model; this constraint results from the one-loop computations. This is necessary to get huge mass PBH. Currently, the relation $M_{\text{PBH}} \propto 1/k_{s}^{2}$ is satisfied by the mass for PBH. The issue occurs when $k_{s}$ is minimal because, in the case of single-field slow-roll models, this leads to the generation of significant solar mass PBH but also reduces the number of e-foldings. When the quantum loop effects are taken into account, we can only obtain 20–25 e-folds. Consequently, by increasing the values to $k_s = 10^{21}$ Mpc$^{-1}$ and $k_e = 10^{22}$ Mpc$^{-1}$, where such big values are necessary to have inflation, only PBH with tiny $M_{\text{PBH}}$ can be created. Because of the proportionality connection previously mentioned, we therefore obtain $M_{\text{PBH}} \sim {\cal O}(10^{2})$ gm in this framework by increasing the wave number towards greater values. Due to the necessity of using renormalization and resummation procedures, the single field slow-roll model has this issue. This means that the total number of e-foldings, $\Delta {\cal N}_{\rm Total}\sim 25$, and the duration of the SRII area must both be extremely minimal. Thus, in order to generate big $M_{\text{PBH}}$, total e-foldings would need to be $25$; if, as is currently the case, the aim is to have adequate inflation in $60$ e-foldings, then the single field models of inflation can only produce little $M_{\text{PBH}}$. Refer to the references \cite{Choudhury:2023vuj,Choudhury:2023jlt,Choudhury:2023rks,Choudhury:2023hvf} for further information, wherein the specifics of single field inflation have been examined and appropriately justified.

Both of the possibilities that might produce both big and small mass PBHs are accommodated in the current Galileon inflationary paradigm. Furthermore, the Galileon theory allows for the modification of the SRII duration, whereas the USR duration, $\Delta {\cal N}_{\rm USR}\sim 2$, remains constant across almost all framework types. A steep transition scale from SRI to USR may be used to create huge $M_{\text{PBH}}$ if the SRII phase's duration is extended to get the $60$ e-foldings. The USR to SRII sharp transition scale is $k_e = 10^{7}$ Mpc$^{-1}$, whereas $k_s = 10^{6}$ Mpc$^{-1}$. The SRII lifetime will be quite little if we shift $k_s = 10^{21}$ Mpc$^{-1}$ and $k_e = 10^{22}$ Mpc$^{-1}$. In the context of Galileon inflation, we can create only small $M_{\text{PBH}}$. 

The third-order perturbed action, which is also utilised to compute the loop effects, is utilised to compute the non-Gaussianities \cite{Choudhury:2023vuj,Choudhury:2023jlt,Choudhury:2023rks,Choudhury:2023hvf}. Resummation is also not essential since the features of the non-renormalization theorem in the current Galileon theory do not call for a renormalized version of the power spectrum. As a result, the Galileon theory does not contain the prior approaches that provided severe limitations and did not permit a protracted SRII phase, allowing us to produce enormous $M_{\text{PBH}}$. A significant new feature has to be added to the theory in order for it to generate big non-Gaussianities, which may also be used to identify and aid in the development of massive $M_{\text{PBH}}$. When there is a sharp transition from one region to another, there is a significant deviation from Gaussianity. The final results would indicate that the amplitude is proportional to the inverse power of those transition wave numbers, so knowing the position of both transitions—from SRI to USR and from USR to SRII—is crucial. Therefore, the overall contribution will be suppressed if $k_{s}$ and $k_{e}$ in the denominator rise in value. The enhancement is considerable at wave numbers $k_s = 10^{6}$ Mpc$^{-1}$ and $k_e = 10^{7}$ Mpc$^{-1}$. In summary, only huge non-Gaussianities are generated for big $M_{\text{PBH}}$, and for minuscule $M_{\text{PBH}}$, the non-Gaussianities would be insignificant.

As of right now, the $M_{\text{PBH}}$ proportionality relation is $M_{\text{PBH}} \propto 1/k_{s}^{2}$, and the $\fnl$ proportionality relation is currently adjusted to become proportional to inverse powers of both $k_{s}$ and $k_{e}$. The result for PBH synthesis is not compelling if we take the assumption that massive quantum fluctuations can generate an amplitude of order ${\cal O}(10^{-4})$ instead of the real ${\cal O}(10^{-2})$ amplitude needed for PBH generation. Substantial non-Gaussianities are insufficient, even in cases when the amplitude of the power spectrum is minimal. Calculating $M_{\text{PBH}}$ and $f_{\rm NL}$ requires knowing the exact position of the transition wave numbers.

\begin{table}

\scriptsize

\begin{tabular}{|c|c|c|c|c|}

 \hline\hline

 \multicolumn{5}{|c|}{\footnotesize A comparison between the One-loop power spectrum and Tree-level non-Gaussian effects in Galileon theory for all three phases} \\

 \hline\hline

 \bf{Phase} & \bf{One-loop corrected} & \bf{Non-Gaussian amplitude} & \bf{Allowed Wave number} & \bf{Highlights and findings}\\
 & \bf{Power spectrum amplitude} ($\Delta^{2}_{\zeta}(k)$) & $(f_{\text{NL}})$ & ($k$ Mpc$^{-1}$) &\\

 \hline\hline

& & & & \\

& & & $k_{*} \leq k \leq k_{s}$ & No large non-Gaussianities observed. \\
\bf{SRI} & ${\cal O}(10^{-9})$  & $(0.01,0.05)$  & where $k_{*} = 0.01$Mpc$^{-1}$,  & Non-Gaussianity of ${\cal O}(10^{-2})$ observed.\\
& & &   and $k_{s} = 10^{6}$Mpc$^{-1}.$  & Consistency condition (\ding{52})\\
& & &   & Total e-folds achieved: $\Delta{\cal N}_{\rm SRI} = 18$. \\

& & & & \\

\hline\hline

& & & & \\

&  &  &  & Sharp transition observed at \\
& & & $k_{s} \leq k \leq k_{e}$  & the beginning and end of the phase. \\
\bf{USR}  & ${\cal O}(10^{-2})$ & $(-5,5)$ &   where $k_{s} = 10^{6}$Mpc$^{-1}$,  & Large non-Gaussianities $\sim$ ${\cal O}(1)$.\\
& & &   and $k_{e} = 10^{7}$Mpc$^{-1}.$  & Favourable for large $M_{\text{PBH}}$. \\
& & &   & Consistency condition (\ding{56}) \\
& & &   & Total e-folds required to maintain \\
& & &   & perturbativity approximation: $\Delta{\cal N}_{\rm USR} = 2$. \\

& & & & \\

\hline\hline

& & & & \\

&  &  &  & Sharp transition observed while \\
& & & $k_{e} \leq k \leq k_{end}$ & exiting the USR phase in beginning. \\
\bf{SRII} & ${\cal O}(10^{-5})$ & $(-1,2)$ & where $k_{e} = 10^{7}$Mpc$^{-1}$,  & Large non-Gaussianities $\sim$ ${\cal O}(1)$.\\
& & &   and $k_{\text{end}} = 10^{27}$Mpc$^{-1}.$  & Consistency condition (\ding{56}) \\
& & &   & Rapid oscillatory behaviour near $k_{\text{end}}$. \\
& & &   & Total e-folds required to complete \\
& & &   & inflation: $\Delta{\cal N}_{\rm SRII} = 40$.
\\

& & & & \\

\hline\hline




\end{tabular}

\caption{An analysis comparing tree-level bispectrum amplitudes with one-loop corrected power spectrum. Along with the corresponding wave numbers involved, we include each of their values for each of the three phases. The key findings from our examination of the bispectrum's data are also included, covering the quantity of non-Gaussianities generated, the non-Gaussian amplitude's behaviour, the total number of e-folds needed, and the reliability of the consistency condition for each phase as represented by \ding{52} and \ding{56}.
}

\label{tab:1}

\end{table}

Now, we explicitly state the salient findings and their physiological implications from the point-by-point study carried out in this review:
\begin{enumerate}
    \item \underline{\textbf{Implication of Non-renormalization theorem}:}\\ \\  This theorem is the salient point of the Galileon theory. We will not be able to simultaneously attain big $M_{\text{PBH}}$ and sustain $60$ e-folds of expansion in the case of single field slow-roll models because of the existence of quantum loop effects and the requirement to complete inflation with the necessary amount of e-foldings, i.e., $60$ e-folds. Additional restrictions come from the requirement for renormalization and resummation processes. However, in Galileon theory, these limitations do not exist, and we are able to have an SRII time that is long enough to finish inflation. Later on, while moving from SRI to USR and USR to SRII areas, a new sharp transition feature helps to build huge non-Gaussianities, which in turn helps to form massive $M_{\text{PBH}}$.

    \item \underline{\textbf{Mass of PBH}:}\\ \\ Our focus in this review is mainly on the scenario of big mass PBHs, which are more intriguing from a cosmological standpoint as they can contribute significantly to the dark matter that exists today. The relation $M_{\rm PBH} \propto 1/k_s^2$ \cite{Choudhury:2023vuj, Choudhury:2023jlt, Choudhury:2023rks} indicates that only small $M_{\rm PBH}$ can be generated when working with canonical and non-canonical single field models of inflation in the EFT framework due to the strong constraints placed on the USR's position as a result of the need to perform renormalization and resummation procedures.
But according to the non-renormalization theorem, quantum loop corrections and associated limitations do not apply in the Galileon theories, allowing us to move the transition scale at the point of interest \cite{Choudhury:2023hvf}. As we move it in the direction of the higher values, we see smaller $M_{\rm PBH}$, and in the direction of the smaller value, we see larger $M_{\rm PBH}$. In response, we will see substantial non-Gaussianity regardless of the USR location. Luckily, we also witness the related situation of big non-Gaussianity in the USR phase, as we demand the case of large $M_{\rm PBH}$ in the current study.

    \item \underline{\textbf{Controlling Non-Gaussianities}:}\\ \\
    The quantity of non-Gaussianity increases significantly in the USR area in the current framework in order to construct PBH with the aid of a sharp transition window fulfilling $k_{e}/k_{s} \sim {\cal O}(10)$. This must be regulated using the coefficients ${\cal G}_{i}$, $\forall i=1,2,3,4.$ Large $M_{\text{PBH}}$ can be obtained if the transition wave numbers are of lower order, which would allow for controlled enhancement management. 
We can only anticipate tiny $M_{\text{PBH}}$ with modest non-Gaussianities if the same wave numbers are shifted to much larger orders, which would need extremely fine-tuning of the coefficients and perhaps violate the power spectrum limitations. Therefore, we have lesser values of the transition wave numbers in the current theory in order to have regulated non-Gaussianities. The quantity of non-Gaussianity from the SRI region is of the order ${\cal O}(10^{-2})$, and from the USR and SRII region, it is of the order ${\cal O}(1)$, according to our results in Fig. (\ref{totPBH}). As a result, the order of $|\fnl| \sim {\cal O}(1)$ represents the cumulative average of the amount of non-Gaussianity generated throughout all three areas. The amplitude $\fnl$ bound from Maldacena's \textit{no-go theorem}, $|\fnl| \sim {\cal O}(10^{-2})$, is clearly met in the SRI area by these results, but it is significantly violated in the USR and SRII sectors.

\end{enumerate}

The amplitude of the one-loop adjusted power spectrum as determined in \cite{Choudhury:2023hvf} and the degree of non-Gaussianity that we achieve for each of the three phases—SRI, USR, and SRII—in our current study are compared in the table \ref{tab:1}. It also draws attention to the salient characteristics of the behaviour of the amplitude $\fnl$ during phase transitions and the validity of the consistency condition, both of which are derived from our findings and illustrated by our plots in Figs.(\ref{totPBH},\ref{fNLSRI:b},\ref{fNLUSR:b},\ref{fNLSRII:b}). 

\subsubsubsection{Comparison with other works}

We demonstrate in this review that the consistency constraint between the spectral tilt and the degree of non-Gaussianity in single-field inflation models of inflation under the squeezed limit is violated. This violation, which only attractor models of inflation respect, is properly displayed inside our underlying framework of Galileon theory, including a USR phase that allows a non-attractor development of the background. A fast transition from a steady roll to a short-lived USR phase causes a dramatic violation as the amplitude rises quickly. We have conducted an extensive investigation including the comoving curvature perturbation dynamics in the superhorizon domain in order to ascertain precisely how the overall non-Gaussianity strength is generated in different phases. The final conclusion is the non-Gaussianity behaviour with wavenumber in the squeezed limit using the current relations, Eqs. (\ref{c1Bf},\ref{c2Bf},\ref{c3Bf}).

A thorough computation for the bispectrum is the foundation of the current investigation. A broad explanation of the underlying causes of non-Gaussianity for single-field inflation models with a non-attractor development of the background is given in the paper by Palma \textit{et al.} \cite{Mooij:2015yka}. They have provided a modified consistency condition that will hold throughout the USR phase, based only on their consideration of the symmetries involved in their justifications for this dramatic breach of the consistency requirement. The new conclusion that is suggested follows a similar methodology used in order to generate the original consistency result, which was re-derived mostly based on symmetry considerations. There is no updated version of the consistency criterion introduced by our investigation. Their main finding in \cite{Mooij:2015yka} reduces to $f_{\rm NL} = 5/2$ in the limit of $\epsilon\rightarrow 0$ and $\eta\rightarrow -6$, and when a violation of the consistency condition occurs, it agrees with the previously derived relation between the bispectrum and the power spectrum in the squeezed limit \cite{Namjoo:2012aa, Martin:2012pe}. Large non-Gaussianity in the USR is shown by the value of $f_{\rm NL} = 5/2$, and the magnitude also fits with the predictions made by our computations. However, our study also demonstrates that a maximum of $f_{\rm NL} \sim -6$ may be reached. The results obtained by Palma \textit{et al.} \cite{Mooij:2015yka} and from the current study both suggest that the superhorizon development of the curvature perturbation modes is the cause of this violation. 

The authors of a different study by Cai \textit{et al.} \cite{Cai:2018dkf} talk about how the non-gaussianity produced in the USR phase is altered by the two types of transitions—smooth and sharp—from one SR area to another. Compared to the other smooth instance, we have specifically focused on the abrupt form of transition, which has a dramatically different influence on the evolution of perturbations. The non-attractor phase of background evolution is also relatively minimal, $\Delta{\cal N}_{\rm USR}\sim {\cal O}(2)$, made clear by the construct in our work, as compared to the total e-foldings needed for successful inflation that is obtained following the inclusion of another SRII phase. The current setup lacks a relaxation phase, as evidenced by the instantaneous transitions from the SRI to the non-attractor USR and the USR to the SRII phase. This is also demonstrated by the shift in the $\eta$ parameter, which goes from $\eta \rightarrow 0$ in the SRI to $\eta \rightarrow -6$ in the USR and from $\eta \rightarrow -6$ to $\eta \rightarrow 0$ from the USR to SRII. If we were to consider the scenario where an initial non-attractor phase starts and transitions into another SR phase for such a step function transition feature, as per the analysis in \cite{Cai:2018dkf}, this would change the non-gaussianity in the non-attractor phase and prevent it from returning to its value $f_{\rm NL}=5/2$ because the mode functions do not freeze. In one of our setups, the two slow-roll phases (SRI and SRII) are connected by an abrupt transition. We have discovered that the consistency criterion is broken and stays that way because of these abrupt transitions once we see the change from the original SRI. This finding deviates from the \cite{Cai:2018dkf} authors' conclusion. Their work focuses on the influence of transition characteristics on non-gaussianity in canonical non-attractor and particular models of the $P(X,\phi)$ class. The Galileon theory, a beyond $P(X,\phi)$ type of model, is our underlying model. Large non-gaussianity, $f_{\rm NL}\sim -6$, is observed in this Covariantized Galileon Theory (CGT) driven USR phase as a result of this abrupt transition. This feature also suppresses the amount of non-gaussianity to $f_{\rm NL} \sim 2$ in the SRII phase, which still falls into the large non-gaussianity category and violates the consistency condition. We infer that, in the driven USR phase model, the produced non-gaussianity will always be larger than in the slow-roll case and will stay larger even after we transition into a new SRII phase. The effective sound speed $c_{s}$, which is parameterized to peak abruptly around the transition scales in our model, is also closely tied to this scenario. This characteristic leads to a significant result: the theory is compatible and supportive with the data from the most recent NANOGrav 15 signal \cite{NANOGrav:2023gor}, and it also tags the substantial non-gaussianity with PBH generation in the USR. Similar studies have been done in \cite{Choudhury:2023kam,Choudhury:2023hfm,Choudhury:2023fwk,Ferrante:2022mui,Gorji:2023sil}. These studies emphasise the benefits of this large non-gaussianity feature, $f_{\rm NL}>1$, when observing with the NANOGrav 15 signal and when addressing the severe PBH overproduction issue. These results demonstrate that our model is a good fit for corroborating the findings from the observational side.

\subsection{Cut-off regularized one-loop power spectrum for comoving curvature perturbation from CGEFT}

In this segment, we examine the pivotal function of the extremely slight disruption of the Galilean shift symmetry in eliminating the deleterious effects of the third order action. As a result, quantum loop corrections arise that are substantial and akin to their tree-level counterparts, as observed in the instances of the classic EFT of inflation and $P(X,\phi)$ inflation models. Due to the lack of these terms, quantum loop effects are suppressed in the current Galileon inflation and may be viewed as a subdominant correction to the relevant spectrum's tree-level equivalent. Next, we go over the explicit function of the in-in formalism in computing the one-loop contributions in the SRI, USR, and SRII phases, respectively. Furthermore, the cut-off regularised one-loop adjustments from the SRI, USR, and SRII phases are explicitly computed. Subsequently, we have included all tree level corrections and one loop corrections derived from SRI, USR, and SRII phases, respectively, in the entire formula for the one-loop corrected power spectrum for comoving curvature perturbation. 

In order to maintain track of the final result and its overall contribution added to the tree-level result, it is important to discuss the strengths of the four types of cubic self interactions in the SRI, USR, and SRII region before moving on to the more technical details of the one-loop contributions in the next subsection. Due to vanishing contributions from the second slow-roll parameter $\eta$ in both SRI and SRII phases, all four of these interactions are suppressed. For the sake of thoroughness, however, we shall compute these contributions. However, the last two terms of the representative third order action in the USR phase become dominant over the other two terms with the coupling coefficients, ${\cal G}_1$ and ${\cal G}_2$, because of the presence of the second slow-roll parameter $\eta$ in the two coupling parameters, ${\cal G}_3$ and ${\cal G}_4$. We shall clearly demonstrate in the next subsection that the final two terms of the third order action that arise during the USR period will dominate any enhancement that results from the one-loop contribution. 
Given that we do not have any terms that involve the time derivative of the second slow-roll parameter $\eta$ at the transition points from SRI to USR and USR to SRII, the corresponding enhancement of the power spectrum in the USR period in the one-loop contribution will be sufficiently suppressed in comparison to the findings found in the refs. \cite{Choudhury:2023vuj,Choudhury:2023jlt,Choudhury:2023rks,Choudhury:2024ybk} where this kind of contribution is present. Even yet, the one-loop contributions from the SRI and SRII areas are significantly lower than the suppressed contribution during the USR era. All of these alternatives will be thoroughly examined in the upcoming subsection.

\subsubsection{The direct In-In formalism for the one-loop corrected two-point function and Non-renormalization theorem}

This section of the review will be devoted to a thorough explanation of each word that emerges from our study using the CGEFT framework. To this end, we employ the well-known in-in formula.  This results in the two-point function that follows at $\tau\rightarrow 0$, which is provided by:
\bea \label{tpt} \langle\hat{\zeta}_{\bf p}\hat{\zeta}_{-{\bf p}}\rangle:&=&\left\langle\bigg[\overline{T}\exp\bigg(i\int^{\tau}_{-\infty(1-i\epsilon)}d\tau^{'}\;H_{\rm int}(\tau^{'})\bigg)\bigg]\;\;\hat{\zeta}_{\bf p}(\tau)\hat{\zeta}_{-{\bf p}}(\tau)
\;\;\bigg[{T}\exp\bigg(-i\int^{\tau}_{-\infty(1+i\epsilon)}d\tau^{''}\;H_{\rm int}(\tau^{''})\bigg)\bigg]\right\rangle_{\tau\rightarrow 0}.\quad\quad \eea
The statement above, which is composed of the time integral of the interaction Hamiltonian and is represented by the following expression, uses $T$ and $\bar{T}$ to denote the time ordering and anti-time ordering of the unitary operators:
\bea && H_{\rm int}(\tau)=-\int d^3x\; \frac{a^2}{H^3}\; \bigg[\frac{{\cal G}_1}{a}\zeta^{'3}+\frac{{\cal G}_2}{a^2}\zeta^{'2}\left(\partial^2\zeta\right)+\frac{{\cal G}_3}{a}\zeta^{'}\left(\partial_i\zeta\right)^2+\frac{{\cal G}_4}{a^2}\left(\partial_i\zeta\right)^2\left(\partial^2\zeta\right)\bigg],\eea
where the preceding subsection defines each of these coefficients ${\cal G}_i\forall i=1,2,3,4$.

The simplified formula that results from extending equation (\ref{tpt}) order by order and gathering the non-trival terms that will aid in the one-loop correction of the two-point correlation function of the comoving curvature perturbation is as follows:
\bea  &&\label{g}\langle\hat{\zeta}_{\bf p}\hat{\zeta}_{-{\bf p}}\rangle=\langle\hat{\zeta}_{\bf p}\hat{\zeta}_{-{\bf p}}\rangle_{(0,0)}+\langle\hat{\zeta}_{\bf p}\hat{\zeta}_{-{\bf p}}\rangle_{(0,2)}+\langle\hat{\zeta}_{\bf p}\hat{\zeta}_{-{\bf p}}\rangle^{\dagger}_{(0,2)}+\langle\hat{\zeta}_{\bf p}\hat{\zeta}_{-{\bf p}}\rangle_{(1,1)},
\eea
where the subsequent terms physically reflect the one-loop correction to the tree-level contribution of the primordial two-point correlation function, while the first terms represent the tree-level contribution. We have previously computed the first term in the earlier part. Our major goal at this point is to calculate the final three elements in the above equation. It is helpful to note that these contributions have expressions in terms of the interaction Hamiltonian, which may be found by using the following formulas, for future computing needs:
\bea &&\label{A1}\langle\hat{\zeta}_{\bf p}\hat{\zeta}_{-{\bf p}}\rangle_{(0,2)}=\lim_{\tau\rightarrow 0}\left[\int^{\tau}_{-\infty}d\tau_1\;\int^{\tau}_{-\infty}d\tau_2\;\langle \hat{\zeta}_{\bf p}(\tau)\hat{\zeta}_{-{\bf p}}(\tau)H_{\rm int}(\tau_1)H_{\rm int}(\tau_2)\rangle\right],\\
 &&\label{A2}\langle\hat{\zeta}_{\bf p}\hat{\zeta}_{-{\bf p}}\rangle^{\dagger}_{(0,2)}=\lim_{\tau\rightarrow 0}\left[\int^{\tau}_{-\infty}d\tau_1\;\int^{\tau}_{-\infty}d\tau_2\;\langle \hat{\zeta}_{\bf p}(\tau)\hat{\zeta}_{-{\bf p}}(\tau)H_{\rm int}(\tau_1)H_{\rm int}(\tau_2)\rangle^{\dagger}\right],\\
  &&\label{A3}\langle\hat{\zeta}_{\bf p}\hat{\zeta}_{-{\bf p}}\rangle_{(1,1)}=\lim_{\tau\rightarrow 0}\left[\int^{\tau}_{-\infty}d\tau_1\;\int^{\tau}_{-\infty}d\tau_2\;\langle H_{\rm int}(\tau_1)\hat{\zeta}_{\bf p}(\tau)\hat{\zeta}_{-{\bf p}}(\tau)H_{\rm int}(\tau_2)\rangle\right].\eea
  See equation (\ref{inin}) and related section \ref{s1} for more details on this aspect.
  Furthermore, the one-loop contributions to the two-point primordial cosmic correlation function may be quantified using the following formulas in terms of the individual cubic self interactions:
\bea \label{A11}\langle\hat{\zeta}_{\bf p}\hat{\zeta}_{-{\bf p}}\rangle_{(0,2)}&=&\sum^{4}_{i=1}{\bf Z}^{(1)}_i,\\ 
\label{A22}\langle\hat{\zeta}_{\bf p}\hat{\zeta}_{-{\bf p}}\rangle^{\dagger}_{(0,2)}&=&\sum^{4}_{i=1}{\bf Z}^{(2)}_i,\\
\label{A33}\langle\hat{\zeta}_{\bf p}\hat{\zeta}_{-{\bf p}}\rangle^{\dagger}_{(1,1)}&=&\sum^{4}_{i=1}{\bf Z}^{(3)}_i,\eea
where the following formulae explain the factors ${\bf Z}^{(1)}_i\forall i=1,2,3,4$, ${\bf Z}^{(2)}_i\forall i=1,2,3,4$, and ${\bf Z}^{(3)}_i\forall i=1,2,3,4$ are defined in the Appendix \ref{A5a}. During this computation, we use the crucial fact that:
\bea |{\bf K}|=|{\bf k}_1+{\bf k}_2+{\bf k}_3|=|{\bf k}_4+{\bf k}_5+{\bf k}_6|=\sqrt{k^2_1+k^2_2+k^3_3}=\sqrt{k^2_4+k^2_5+k^2_6}.\eea
The integral across conformal time and the momentum scales must also be addressed by splitting them into the three zones that were previously mentioned:
\bea {\bf Conformal\;time\;integral:}\quad\quad\quad\quad\lim_{\tau\rightarrow 0}\int^{\tau}_{-\infty}:\equiv \underbrace{\Bigg(\int^{\tau_s}_{-\infty}\Bigg)}_{\bf SRI}+\underbrace{\Bigg(\int^{\tau_e}_{\tau_s}\Bigg)}_{\bf USR}+\underbrace{\Bigg(\int^{\tau_{\rm end}\rightarrow 0}_{\tau_e}\Bigg)}_{\bf SRII},\eea
and 
\bea {\bf Momentum\;integral:}\quad\quad\quad\quad\int^{\infty}_{0}:\equiv \underbrace{\Bigg(\int^{k_s}_{k_*}\Bigg)}_{\bf SRI}+\underbrace{\Bigg(\int^{k_e}_{k_s}\Bigg)}_{\bf USR}+\underbrace{\Bigg(\int^{k_{\rm end}\rightarrow 0}_{k_e}\Bigg)}_{\bf SRII},\eea
where the IR and UV cut-offs, which are essential for extracting the finite contribution from the current calculation, fulfill the function of the finite limits of the integration in the description above. The presence of these cut-off regulators does, however, allow for some divergent effects in the final output, which must be eliminated by renormalization and resummation, procedures that we shall go into further detail about in the second part of this study. Currently, our main goal is to explicitly calculate the curvature perturbation cut-off controlled expressions for the two-point primordial cosmological correlators in the three successive areas (SRI, USR, and SRII, respectively). We will examine these facts in more depth in the next subsection.

\subsubsection{Computation of the cut-off regularized one-loop correction to the tree-level power spectrum}

By applying the cut-off regularization approach in the aforementioned locations, we compute the explicit contributions from the one-loop correction in the ensuing subsection. Prior to delving into the technical aspects of this section, let us first note that the one-loop result may be reduced using the following equation by employing all conceivable kinds of Wick contractions:
\bea \label{OLP} \langle\langle\hat{\zeta}_{\bf p}\hat{\zeta}_{-{\bf p}}\rangle\rangle_{\bf One-loop}&=&\langle\langle\hat{\zeta}_{\bf p}\hat{\zeta}_{-{\bf p}}\rangle\rangle_{(1,1)}+2{\rm Re}\bigg[\langle\langle\hat{\zeta}_{\bf p}\hat{\zeta}_{-{\bf p}}\rangle\rangle_{(0,2)}\bigg]\nonumber\\
 &=&\sum^{4}_{i=1}{\bf Z}^{(3)}_i+2{\rm Re}\Bigg[\sum^{4}_{i=1}{\bf Z}^{(1)}_i\Bigg], \eea 
 which in the three areas under discussion is simpler to assess.

\subsubsubsection{Result for the Region I (SRI)}

The following is the computation of the one-loop contribution to the primordial power spectrum derived from the comoving curvature perturbation in the SRI region:
\bea \label{OLP1} \Bigg[\Delta^{2}_{\zeta,{\bf One-loop}}(p)\Bigg]_{\bf SRI}&=&\Bigg[\Delta^{2}_{\zeta,{\bf Tree}}(p)\Bigg]^2_{\bf SRI}\Bigg\{c_{\bf SRI}-\frac{1}{8{\cal A}^2_*\pi^4}\sum^{4}_{i=1}\widetilde{\cal G}_{i,{\bf SRI}} {\bf F}_{i,{\bf SRI}}(k_s,k_*)\Bigg\}\nonumber\\
 &=&\Bigg[\Delta^{2}_{\zeta,{\bf Tree}}(p)\Bigg]^2_{\bf SRI}\Bigg\{c_{\bf SRI}-\frac{c^4_s}{8{\cal B}^2_*\pi^6}\sum^{4}_{i=1}\widetilde{\cal G}_{i,{\bf SRI}} {\bf F}_{i,{\bf SRI}}(k_s,k_*)\Bigg\}.\eea 
where the following gives the power spectrum in the SRI region:
\bea \Bigg[\Delta^{2}_{\zeta,{\bf Tree}}(p)\Bigg]_{\bf SRI}&=&\left(\frac{H^{4}}{8\pi^{2}{\cal A} c^3_s}\right)_*\Bigg\{1+\Bigg(\frac{p}{k_s}\Bigg)^2\Bigg\}=\left(\frac{H^{4}}{8\pi^{2}{\cal B} c_s}\right)_*\Bigg\{1+\Bigg(\frac{p}{k_s}\Bigg)^2\Bigg\}\quad\quad{\rm where}\quad p<k_S.\eea
Here, for the current computational purpose in the SRI phase, $c_{\bf SRI}$ denotes the regularization scheme dependent parameter.
In terms of CGEFT couplings, the coefficients and the formula for momentum dependent functions are explicitly mentioned in the Appendix.

         The following formulae determine the parameters ${\cal A}_*$ and ${\cal B}_*$ at the pivot scale $\tau=\tau_*$:
\bea {\cal A}_*&\equiv& {\cal A}(\tau_*)=\frac{\dot{\bar{\phi}}^2_0(\tau_*)}{2}\Bigg(c_2+12c_3Z_*+54c_4Z^2_*+120c_5Z^3_*\Bigg),\\
    {\cal B}_*&\equiv& {\cal B}(\tau_*)\nonumber\\&=& \frac{\dot{\bar{\phi}}^2_0(\tau_*)}{2}\Bigg\{c_2+4c_3\Bigg(2Z_*-\frac{H(\tau_*)\dot{\bar{\phi}}_0(\tau_*)}{\Lambda^3}\eta_*\Bigg)+2c_4\Bigg[13Z^2_*-\frac{6}{\Lambda^6}\dot{\bar{\phi}}^2_0(\tau_*)H^2(\tau_*)\big(\epsilon(\tau_*)+2\eta(\tau_*)\big)\Bigg]\nonumber\\
    &&\quad\quad\quad\quad\quad\quad\quad\quad\quad\quad\quad\quad-\frac{24c_5}{\Lambda^9}H^3(\tau_*)\dot{\bar{\phi}}^3_0(\tau_*)\big(2\epsilon(\tau_*)+1\big)\Bigg\}.\quad\quad\eea
which, by utilizing the following relationship, are quite helpful in fixing the effective sound speed at that scale:
         \bea c_s=c_s(\tau_*)=\sqrt{\frac{{\cal B}_*}{{\cal A}_*}}.\eea
Furthermore, the explicit formulation for the coupling parameter $Z_*$ at the pivot scale $\tau=\tau_*$ that appears in the SRI phase should be mentioned. It is as follows:
\bea Z_*\equiv Z(\tau_*)= \frac{H(\tau_*)\dot{\bar{\phi}}_0(\tau_*)}{\Lambda^3},\eea
when the expression $\dot{\bar{\phi}}_0(\tau_*)$ is given by:
  \bea \dot{\bar{\phi}}_0(\tau_*)=\frac{\Lambda^3}{12H(\tau_*)}\frac{c_2}{c_3}\Bigg[-1+\sqrt{1+\frac{8c_3}{c^2_2}\frac{\lambda^3}{\Lambda^3}}\Bigg].\eea

\subsubsubsection{Result for the Region II (USR)}

The one-loop contribution to the primordial power spectrum computed from the comoving curvature perturbation in the USR region is computed as:
 \bea \label{OLP2} \Bigg[\Delta^{2}_{\zeta,{\bf One-loop}}(p)\Bigg]_{\bf USR}&=&\Bigg[\Delta^{2}_{\zeta,{\bf Tree}}(p)\Bigg]^2_{\bf SRI}\Bigg\{c_{\bf USR}+\frac{1}{8{\cal A}^2_*\pi^4}\sum^{4}_{i=1}\widetilde{\cal G}_{i,{\bf USR}} {\bf F}_{i,{\bf USR}}(k_e,k_s)\Bigg\}\Theta(p-k_s)\nonumber\\
 &=&\Bigg[\Delta^{2}_{\zeta,{\bf Tree}}(p)\Bigg]^2_{\bf SRI}\Bigg\{c_{\bf USR}+\frac{c^4_s}{8{\cal B}^2_*\pi^6}\sum^{4}_{i=1}\widetilde{\cal G}_{i,{\bf USR}} {\bf F}_{i,{\bf USR}}(k_e,k_s)\Bigg\}\Theta(p-k_s).\eea 
 when there has previously been prior definition of the power spectrum in the SRI region. The regularisation scheme dependant parameter for the current computational purpose in the USR phase is represented by the symbol $c_{\bf USR}$ in this case. In the range $\tau_s\leq \tau \leq \tau_e$, the CGEFT couplings, the coefficients and the formula for momentum-dependent functions are explicitly mentioned in the Appendix.

         Not to be overlooked is the clear formulation provided by the following for the coupling parameters $Z_e$ and $Z_s$ at the end of the USR and SRI to USR transition scales, $\tau=\tau_e$ and $\tau=\tau_s$:
 \bea && Z_e\equiv Z(\tau_e)= \frac{H(\tau_e)\dot{\bar{\phi}}_0(\tau_e)}{\Lambda^3},\\
         && Z_s\equiv Z(\tau_s)= \frac{H(\tau_s)\dot{\bar{\phi}}_0(\tau_s)}{\Lambda^3},\eea
          where $\dot{\bar{\phi}}_0(\tau_e)$ and $\dot{\bar{\phi}}_0(\tau_s)$ is given by:
          \bea &&\dot{\bar{\phi}}_0(\tau_e)=\frac{\Lambda^3}{12H(\tau_e)}\frac{c_2}{c_3}\Bigg[-1+\sqrt{1+\frac{8c_3}{c^2_2}\frac{\lambda^3}{\Lambda^3}}\Bigg], \\ &&\dot{\bar{\phi}}_0(\tau_s)=\frac{\Lambda^3}{12H(\tau_s)}\frac{c_2}{c_3}\Bigg[-1+\sqrt{1+\frac{8c_3}{c^2_2}\frac{\lambda^3}{\Lambda^3}}\Bigg].\eea
\subsubsubsection{Result for the Region III (SRII)}

The following formula is used to calculate the one-loop contribution to the primordial power spectrum derived from the comoving curvature perturbation in the SRII region:
\bea \label{OLP3} \Bigg[\Delta^{2}_{\zeta,{\bf One-loop}}(p)\Bigg]_{\bf SRII}&=&\Bigg[\Delta^{2}_{\zeta,{\bf Tree}}(p)\Bigg]^2_{\bf SRI}\Bigg\{c_{\bf SRII}+\frac{1}{8{\cal A}^2_*\pi^4}\sum^{4}_{i=1}\widetilde{\cal G}_{i,{\bf SRII}} {\bf F}_{i,{\bf SRII}}(k_{\rm end},k_e)\Bigg\}\Theta(p-k_e)\nonumber\\
 &=&\Bigg[\Delta^{2}_{\zeta,{\bf Tree}}(p)\Bigg]^2_{\bf SRI}\Bigg\{c_{\bf SRII}+\frac{c^4_s}{8{\cal B}^2_*\pi^6}\sum^{4}_{i=1}\widetilde{\cal G}_{i,{\bf SRII}} {\bf F}_{i,{\bf SRII}}(k_{\rm end},k_e)\Bigg\}\Theta(p-k_e),\eea 
 when there has previously been prior definition of the power spectrum in the SRI region.The regularisation scheme dependant parameter for the current computational purpose in the SRII phase is represented by the symbol $c_{\bf SRII}$. In terms of CGEFT couplings, the coefficients and the formula for momentum dependent functions are explicitly mentioned in the Appendix inside the range $\tau_{e}\leq \tau \leq \tau_{\rm end}$.
        
    It is also crucial to note that the coupling parameters $Z_{\rm end}$ and $Z_e$ at the end of SRII and USR scales, $\tau=\tau_{\rm end}$ and $\tau=\tau_e$, are explicitly expressed as follows:
 \bea && Z_{\rm end}\equiv Z(\tau_{\rm end})= \frac{H(\tau_{\rm end})\dot{\bar{\phi}}_0(\tau_{\rm end})}{\Lambda^3},\quad\quad 
         Z_e\equiv Z(\tau_e)= \frac{H(\tau_e)\dot{\bar{\phi}}_0(\tau_e)}{\Lambda^3},\eea
         where $\dot{\bar{\phi}}_0(\tau_{\rm end})$ and $\dot{\bar{\phi}}_0(\tau_e)$ is given by:
\bea && \dot{\bar{\phi}}_0(\tau_{\rm end})=\frac{\Lambda^3}{12H(\tau_{\rm end})}\frac{c_2}{c_3}\Bigg[-1+\sqrt{1+\frac{8c_3}{c^2_2}\frac{\lambda^3}{\Lambda^3}}\Bigg], \\
         &&\dot{\bar{\phi}}_0(\tau_e)=\frac{\Lambda^3}{12H(\tau_e)}\frac{c_2}{c_3}\Bigg[-1+\sqrt{1+\frac{8c_3}{c^2_2}\frac{\lambda^3}{\Lambda^3}}\Bigg].\eea

\subsection{Final result for the cut-off regularized one-loop corrected total power spectrum}

We will now provide the overall cut-off regularised one-loop adjusted primordial power spectrum in this subsection, which was calculated using a comoving curvature perturbation. We have taken great care to account for the specific contributions from SRI, USR, and SRII. The following outcome is obtained by adding together each one-loop contribution to the tree level contribution:
\bea \Bigg[\Delta^{2}_{\zeta}(p)\Bigg]_{\bf Total}
&=&\Bigg[\Delta^{2}_{\zeta,{\bf Tree}}(p)\Bigg]_{\bf SRI}+\Bigg[\Delta^{2}_{\zeta,{\bf Tree}}(p)\Bigg]_{\bf USR}\Theta(p-k_s)+\Bigg[\Delta^{2}_{\zeta,{\bf Tree}}(k)\Bigg]_{\bf SRII}\Theta(p-k_e)\nonumber\\
&&\quad\quad\quad+ \Bigg[\Delta^{2}_{\zeta,{\bf One-loop}}(p)\Bigg]_{\bf SRI}+\Bigg[\Delta^{2}_{\zeta,{\bf One-loop}}(p)\Bigg]_{\bf USR}\Theta(p-k_s)+\Bigg[\Delta^{2}_{\zeta,{\bf One-loop}}(p)\Bigg]_{\bf SRII}\Theta(p-k_e)\nonumber\\
&=& \Bigg(\Bigg[\Delta^{2}_{\zeta,{\bf Tree}}(p)\Bigg]_{\bf SRI}+ \Bigg[\Delta^{2}_{\zeta,{\bf One-loop}}(p)\Bigg]_{\bf SRI}\Bigg)\nonumber\\
&&\quad\quad\quad\quad\quad+\Bigg(\Bigg[\Delta^{2}_{\zeta,{\bf Tree}}(p)\Bigg]_{\bf USR}+ \Bigg[\Delta^{2}_{\zeta,{\bf One-loop}}(p)\Bigg]_{\bf USR}\Bigg)\Theta(p-k_s)\nonumber\\
&&\nonumber\\
&&\quad\quad\quad\quad\quad\quad\quad\quad\quad+\Bigg(\Bigg[\Delta^{2}_{\zeta,{\bf Tree}}(p)\Bigg]_{\bf SRII}+ \Bigg[\Delta^{2}_{\zeta,{\bf One-loop}}(p)\Bigg]_{\bf SRII}\Bigg)\Theta(p-k_e)\nonumber\eea\bea
&\approx&\Bigg[\Delta^{2}_{\zeta,{\bf Tree}}(p)\Bigg]_{\bf SRI}\Bigg\{1+\left(\frac{k_e}{k_s }\right)^{6}\Bigg[\left|\alpha^{(2)}_{\bf k}-\beta^{(2)}_{\bf k}\right|^2\Theta(p-k_s)+\left|\alpha^{(3)}_{\bf k}-\beta^{(3)}_{\bf k}\right|^2\Theta(k-k_e)\Bigg]\nonumber\\
&&\quad\quad\quad\quad\quad\quad\quad\quad\quad\quad\quad+\Bigg[\Delta^{2}_{\zeta,{\bf Tree}}(p)\Bigg]_{\bf SRI}\Bigg\{c_{\bf SRI}-\frac{1}{8{\cal A}^2_*\pi^4}\sum^{4}_{i=1}\widetilde{\cal G}_{i,{\bf SRI}} {\bf F}_{i,{\bf SRI}}(k_s,k_*)\Bigg\}\nonumber\\
&&\quad\quad\quad\quad\quad\quad\quad\quad\quad\quad\quad+\Bigg[\Delta^{2}_{\zeta,{\bf Tree}}(p)\Bigg]_{\bf SRI}\Bigg\{c_{\bf USR}+\frac{1}{8{\cal A}^2_*\pi^4}\sum^{4}_{i=1}\widetilde{\cal G}_{i,{\bf USR}} {\bf F}_{i,{\bf USR}}(k_e,k_s)\Bigg\}\Theta(p-k_s)\nonumber\\
&&\quad\quad\quad\quad\quad\quad\quad\quad\quad\quad\quad+\Bigg[\Delta^{2}_{\zeta,{\bf Tree}}(p)\Bigg]_{\bf SRI}\Bigg\{c_{\bf SRII}+\frac{1}{8{\cal A}^2_*\pi^4}\sum^{4}_{i=1}\widetilde{\cal G}_{i,{\bf SRII}} {\bf F}_{i,{\bf SRII}}(k_{\rm end},k_e)\Bigg\}\Theta(p-k_e)\Bigg\}\nonumber\\
&\approx&\Bigg[\Delta^{2}_{\zeta,{\bf Tree}}(p)\Bigg]_{\bf SRI}\Bigg\{1+\left(\frac{k_e}{k_s }\right)^{6}\Bigg[\left|\alpha^{(2)}_{\bf k}-\beta^{(2)}_{\bf k}\right|^2\Theta(p-k_s)+\left|\alpha^{(3)}_{\bf k}-\beta^{(3)}_{\bf k}\right|^2\Theta(k-k_e)\Bigg]\nonumber\\
&&\quad\quad\quad\quad\quad\quad+\Bigg[\Delta^{2}_{\zeta,{\bf Tree}}(p)\Bigg]_{\bf SRI}\Bigg\{\bigg(c_{\bf SRI}+c_{\bf USR}\Theta(p-k_s)+c_{\bf SRII}\Theta(p-k_e)\bigg)\nonumber\\
&&\quad\quad\quad\quad\quad\quad\quad-\frac{1}{8{\cal A}^2_*\pi^4}\sum^{4}_{i=1}\bigg(\widetilde{\cal G}_{i,{\bf SRI}} {\bf F}_{i,{\bf SRI}}(k_s,k_*)-\widetilde{\cal G}_{i,{\bf USR}} {\bf F}_{i,{\bf USR}}(k_e,k_s)\Theta(p-k_s)\nonumber\\
&&\quad\quad\quad\quad\quad\quad\quad\quad\quad\quad\quad\quad\quad\quad\quad-\widetilde{\cal G}_{i,{\bf SRII}} {\bf F}_{i,{\bf SRII}}(k_{\rm end},k_e)\Theta(p-k_e)\Bigg)\Bigg\}
\nonumber\\
&=&\Bigg[\Delta^{2}_{\zeta,{\bf Tree}}(p)\Bigg]_{\bf SRI}\Bigg\{1+\left(\frac{k_e}{k_s }\right)^{6}\Bigg[\left|\alpha^{(2)}_{\bf k}-\beta^{(2)}_{\bf k}\right|^2\Theta(p-k_s)+\left|\alpha^{(3)}_{\bf k}-\beta^{(3)}_{\bf k}\right|^2\Theta(k-k_e)\Bigg]\nonumber\\
&&\quad\quad\quad\quad\quad\quad+\Bigg[\Delta^{2}_{\zeta,{\bf Tree}}(p)\Bigg]_{\bf SRI}\Bigg\{\bigg(c_{\bf SRI}+c_{\bf USR}\Theta(p-k_s)+c_{\bf SRII}\Theta(p-k_e)\bigg)\nonumber\\
&&\quad\quad\quad\quad\quad\quad\quad-\frac{c^4_s}{8{\cal B}^2_*\pi^6}\sum^{4}_{i=1}\bigg(\widetilde{\cal G}_{i,{\bf SRI}} {\bf F}_{i,{\bf SRI}}(k_s,k_*)-\widetilde{\cal G}_{i,{\bf USR}} {\bf F}_{i,{\bf USR}}(k_e,k_s)\Theta(p-k_s)\nonumber\\
&&\quad\quad\quad\quad\quad\quad\quad\quad\quad\quad\quad\quad\quad\quad\quad-\widetilde{\cal G}_{i,{\bf SRII}} {\bf F}_{i,{\bf SRII}}(k_{\rm end},k_e)\Theta(p-k_e)\Bigg)\Bigg\}.\eea
This may be recast in the following simple form to represent the power spectrum in the SRI region:
\bea \Bigg[\Delta^{2}_{\zeta,{\bf Tree}}(p)\Bigg]_{\bf SRI}&=&\left(\frac{H^{4}}{8\pi^{2}{\cal A} c^3_s}\right)_*\Bigg\{1+\Bigg(\frac{p}{k_s}\Bigg)^2\Bigg\}=\left(\frac{H^{4}}{8\pi^{2}{\cal B} c_s}\right)_*\Bigg\{1+\Bigg(\frac{p}{k_s}\Bigg)^2\Bigg\}.\eea
All momentum-dependent, time-dependent fixed functions, coupling parameters, and Bogoliubov coefficients in the SRII and USR phases have already been discussed clearly in the current context.

One can further rewrite the spectrum in the following simplified format:
\bea \label{s4dtot}  
\Bigg[\Delta^{2}_{\zeta}(k)\Bigg]_{\bf Total}
&=&  \displaystyle \Bigg[\Delta^{2}_{\zeta,\bf {Tree}}(k)\Bigg]_{\rm \textbf{SRI}} +  \Bigg[\Delta^{2}_{\zeta,\bf {Tree}}(k)\Bigg]_{\rm \textbf{USR}}\Theta(k-k_{s})   + \Bigg[\Delta^{2}_{\zeta,\bf {Tree}}(k)\Bigg]_{\rm \textbf{SRII}}\Theta(k-k_{e}) + {\cal Q}_{c},\nonumber\\
&\approx& A\Bigg[\Delta^{2}_{\zeta,\bf {Tree}}(k)\Bigg]_{\rm \textbf{SRI}}\left\{\left(\frac{k_{s}}{k_{e}}\right)^{6}(1+{\cal Q}_{c}) + \left(\big|\alpha_{\bf k}^{(2)}-\beta_{\bf k}^{(2)}\big|^2 \;\Theta(k-k_s)+\big|\alpha_{\bf k}^{(3)}-\beta_{\bf k}^{(3)}\big|^2 \;\Theta(k-k_e)\right)\right\},\quad\quad\quad
\eea
where the one-loop corrections, which have the following form, are labelled by the word ${\cal Q}_{c}$:
\bea \label{quantcorr}
{\cal Q}_{c} &=& \Bigg[\Delta^{2}_{\zeta,\bf {One-Loop}}(k)\Bigg]_{\rm \textbf{SRI}} + \Bigg[\Delta^{2}_{\zeta,\bf {One-Loop}}(k)\Bigg]_{\rm \textbf{USR}}\Theta(k-k_{s}) + \Bigg[\Delta^{2}_{\zeta,\bf {One-Loop}}(k)\Bigg]_{\rm \textbf{SRII}}\Theta(k-k_{e})\nonumber\\
&=& \Bigg[\Delta^{2}_{\zeta,\bf {Tree}}(k)\Bigg]_{\rm \textbf{SRI}}\times \frac{1}{8{\cal A}^{2}_{*}\pi^{4}}\bigg\{-\sum^{4}_{i=1}{\cal G}_{i,\mbf{SRI}}\mbf{F}_{i,\mbf{SRI}}(k_{s},k_{*}) + \sum^{4}_{i=1}{\cal G}_{i,\mbf{USR}}\mbf{F}_{i,\mbf{USR}}(k_{e},k_{s})\;\Theta(k-k_{s}) \nonumber\\
&+& \sum^{4}_{i=1}{\cal G}_{i,\mbf{SRII}}\mbf{F}_{i,\mbf{SRII}}(k_{\rm end},k_{e})\;\Theta(k-k_{e})\bigg\}.
\eea
Here we define the amplitude as: 
\bea A=\displaystyle{\left(\frac{H^{4}}{8\pi^{2}{\cal A}c_{s}^{3}}\right)_{*}\left(\frac{k_{e}}{k_{s}}\right)^{6}}.\eea
It is noteworthy to acknowledge that there exist concerns regarding the necessity of a nontrivial potential. Since potential terms are not shift symmetry invariant, they must be seen as irrelevant physical operators that result in small modifications. However, if they do exist, operators with mass dimensions smaller than four would often be subject to massive renormalizations, which would diminish the model's significance. In actuality, there is no way around this issue because a smooth finish to the inflationary phase requires the presence of an effective potential composed at least of quadratic order contribution. Remarkably, both the linear and quadratic contributions, $\phi$ and $\phi^2$, are protected by a non-renormalization theorem and may thus be safely seen as the entirely unimportant deformations of the Galilean shift symmetry that apply to the underlying theoretical framework. This may be demonstrated with ease using the one particle irreducible effective action technique in the context of the route integral, as demonstrated in the ref. \cite{Burrage:2010cu}. No additional operators in the underlying theory are permitted by such a strong non-renormalization theorem to break the symmetry at the quantum mechanical level. As previously noted for the comoving curvature perturbation variable, this strong non-renormalization theorem, proven at the level of scalar field, is likewise propagated to the quantum fluctuations caused by the terms appearing in the second and third order perturbed action. Therefore, it follows that all self interactions are safeguarded by various types of radiative quantum corrections in the presence of Galilean shift symmetry or in the case of its extremely slight breaking at the level of cosmic disruption. Furthermore, any correlation functions derived from the cosmological perturbations created via covariant Galileon must be non-renormalizable yet stable under radiative quantum corrections, owing to the strong non-renormalization theorem. As such, the fundamental idea of renormalization and resummation has no place in the current theoretical paradigm. Because of this, the cut off regularised total one-loop adjusted primordial power spectrum that is calculated by comving scalar curvature perturbation will be more than adequate, and we will draw our final conclusion using this outcome. It is abundantly clear from the deduced structure of the one-loop corrections as derived from the SRI, USR, and SRII phases that the precise structure of the momentum-dependent components involved in the calculation is free of both logarithmic and quadratic divergences. The absence of the term $\eta^{'}\zeta^{'}\zeta^{2}$ in the third order action, which would very slightly break the Galilean shift symmetry, and the underlying structure of the CGEFT theory explain why the aforementioned divergences are completely absent in the primordial power spectrum. This is a remarkable finding from the current computation. The remaining components that show up in the one-loop correction are severely power law suppressed, and some of them have limited oscillation amplitudes and are highly oscillatory. The extremely slight correction these contributions will provide to the tree level amplitude of the primordial power spectrum is entirely beneficial for the current computational goal. Due to this, the cosmic perturbation will remain valid during the SRI, USR, and SRII phases, and all perturbative approximations will hold true flawlessly when the one-loop correction is being computed. As one of the prerequisites for the formation of PBH, the augmentation of the spectrum from ${\cal O}(10^{-9})$ (SRI) to ${\cal O}(10^{-2})$ (USR) is expected to occur as a result of the greatly suppressed one-loop correction. The ratio $k_e/k_s$ must be rigorously kept at ${\cal O}(10)$ in order to preserve the perturbative approximations and properly enhance the tree level power spectrum. This essentially limits the number of e-foldings to $2$ in the USR regime of the calculation. This indicates that the current calculation does not allow for a protracted USR phase. However, in order to manufacture large mass PBHs, the location of the SRI to USR transition scale must be kept at $k_s\leq 10^{6}$. This indicates that $k_e\leq 10^{7}$ must be restricted in order to keep $k_e/k_s\sim {\cal O}(10)$ consistently. This aids in the production of solar mass PBHs and beyond. We will go into the specifics of the aforementioned discoveries and how they affect the creation of PBHs in the following sections. Above all, we will thoroughly examine the limitations in order to build a large mass PBH in the context of covarinatized Galileon-induced single field inflation. Look at the Appendix \ref{A6a} for more details on this computation.

\subsubsection{Numerical results: Constraints on the PBH mass and evaporation time scale}

    \begin{figure*}[htb!]
    	\centering
{
      	\includegraphics[width=18cm,height=14cm] {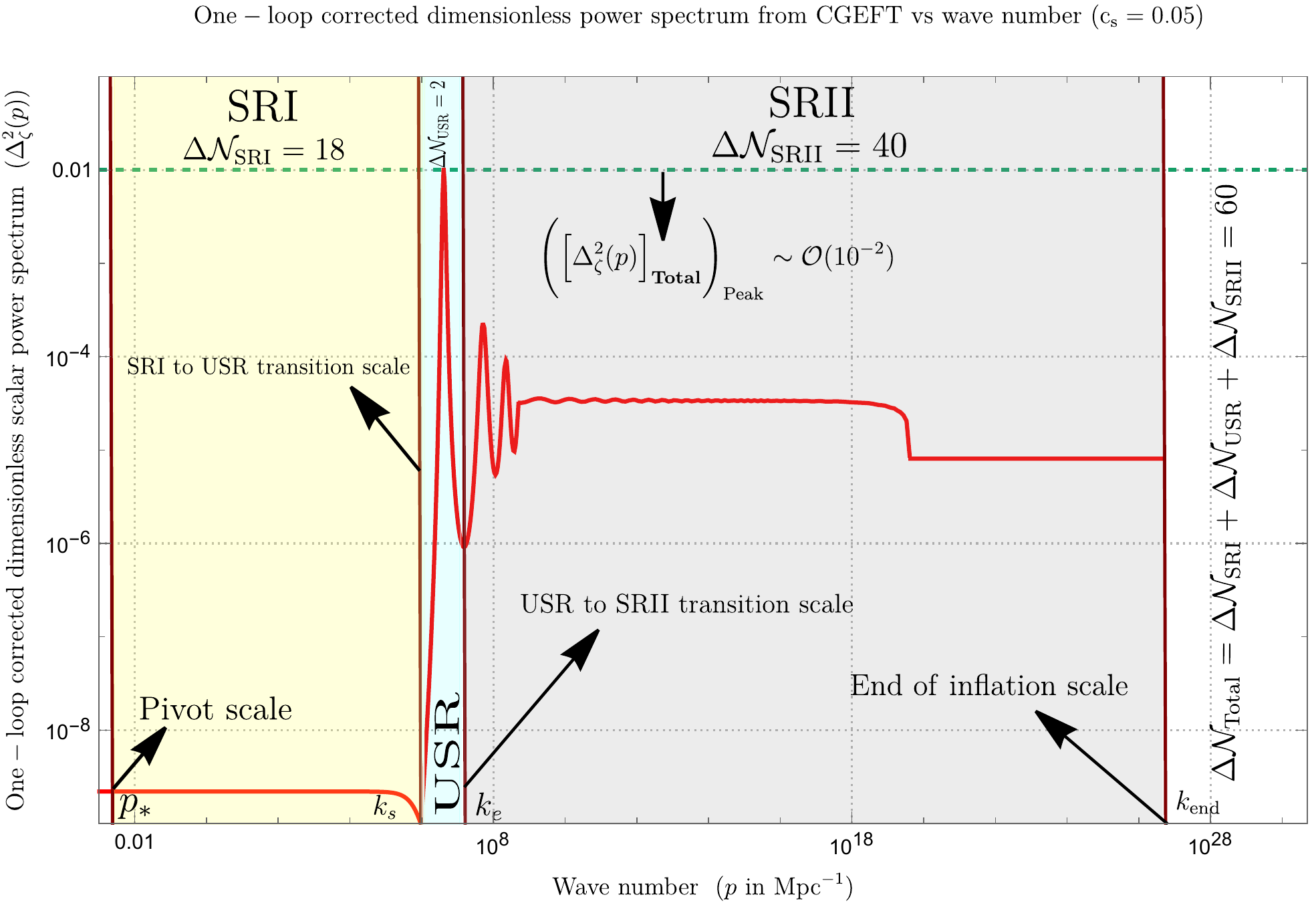}
    }
    	\caption[Optional caption for list of figures]{The wave number-dependent behaviour of the dimensionless primordial power spectrum for scalar modes. In this instance, $c_s=0.05$ is the fixed effective sound speed. The pivot scale is fixed at $p_*=0.02\;{\rm Mpc}^{-1}$, the SRI to USR transition scale at $k_s=10^{6}\;{\rm Mpc}^{-1}$, the end of the USR scale at $k_e=10^{7}\;{\rm Mpc}^{-1}$, the end of the inflation scale at $k_{\rm end}=10^{27}\;{\rm Mpc}^{-1}$, the regularisation parameters, $c_{\bf SRI}=0$; $c_{\bf USR}=0$ and $c_{\bf SRII}=0A$. The values $\Delta\eta(\tau_e)=1$ and $\Delta\eta(\tau_s)=-6$ are also utilised. We have determined from this plot that $k_{\rm end}/k_e\approx{\cal O}(10^{20})$ and $k_{\rm UV}/k_{\rm IR}=k_e/k_s\approx{\cal O}(10)$. The matching spectrum's peak amplitude ${\cal O}(10^{-2})$ is reached at $5k_s\sim 5\times 10^{6}{\rm Mpc}^{-1}$. With this study, a total of $60$ e-foldings is obtained, which is adequate for inflation.
  } 
    	\label{GL}
    \end{figure*}
With regard to the wave number and a fixed value of the effective sound speed $c_s=0.05$ for the Galileon, the behaviour of the dimensionless power spectrum for comoving curvature perturbation is explicitly displayed in figure (\ref{GL}). We have simply demonstrated the behaviour of the one-loop corrected cut off regularised power spectrum, where the cumulative effects from the phases SRI, USR, and SRII are encoded, as renormalization and resummation are not required for Galileon because of its strong non-renormalization theorem. We discovered that the power spectrum reaches a minimum value shortly before the SRI to USR transition, and then accelerates sharply to a peak value, ${\cal O}(10^{-2})$ at $5\times 10^{6}{\rm Mpc}^{-1}$ momentum scale, which appears in the mid point in the USR region. The $\Delta{\cal N}_{\rm Peak}\sim {\cal O}(20)$ number of e-foldings corresponds to the peak value of the amplitude of the power spectrum. The amplitude begins to drastically decrease after reaching the greatest enhancement in the corresponding power spectrum, and in the current setup, it ends with a value of ${\cal O}(10^{-6})$ at the conclusion of the USR phase. Furthermore, the power spectrum begins to oscillate at a very high frequency with a decreasing overall amplitude due to sinusoidal contributions. It will saturate in a very short period of time to a value with a very small oscillating envelop that will persist for a longer period of time. After that, a sharp fall is seen, and it saturates once more to the amplitude ${\cal O}(10^{-5})$, which will persist until the end of the SRII phase, when the inflation also ends.

For numerical purposes, we fix the termination of USR at $k_e=10^{7}\;{\rm Mpc}^{-1}$ and the IR cut-off at $k_s=10^{6}\;{\rm Mpc}^{-1}$. Also, the end of inflation scale is fixed at $k_{\rm end}=10^{27}\;{\rm Mpc}^{-1}$ (where we fix the UV cut-off). Additionally, we fix the parameters of the regularisation,  
   $c_{\bf SRI} = 0$, $c_{\bf USR} = 0$; $c_{\bf SRII}=0$, and $\Delta\eta(\tau_e)=1$ as well as $\Delta\eta(\tau_s)=-6$. In order to accomplish inflation, we fix $k_{\rm end}/k_e\approx{\cal O}(10^{20})$ and $k_{\rm UV}/k_{\rm IR}=k_e/k_s\approx{\cal O}(10)$. This ensures that the perturbative approximations hold true perfectly. Our study is valid within the domain $0.024\leq c_s<1$, meaning that Galileon causality is rigorously preserved inside the scope of EFT, even if we have given the figure for the sound speed $c_s=0.05$. 

   We deduced from this plot that the maximum number of e-foldings permitted during the USR phase is:
\bea \Delta {\cal N}_{\rm USR}=\ln(k_e/k_s)\approx\ln(10^{7}/10^{6})=\ln(10)\approx 2.\eea
This, in the current Galileon inflationary paradigm, provides crucial information for the production of PBHs. Additionally, the SRI and SRII periods' permitted number of e-foldings is provided by:
\bea &&\Delta {\cal N}_{\rm SRI}=\ln(k_s/p_*)\approx \ln(10^{6}/0.02)\sim 18,\\
&&\Delta {\cal N}_{\rm SRII}=\ln(k_{\rm end}/k_e)\approx \ln(10^{27}/10^{7})\approx 40.\eea 
Therefore, the total number of e-foldings that Galileon inflation permits is as follows:
\bea \Delta {\cal N}_{\rm Total}=\Delta {\cal N}_{\rm SRI}+\Delta {\cal N}_{\rm USR}+\Delta {\cal N}_{\rm SRII}\sim 18+2+40=60.\eea 
This scenario might result in the construction of a massive PBH with an adequate number of e-folds for inflation inside the context of a single field with Galileon.

We discovered that a longer USR duration is completely prohibited for PBH generation in the context of Galileon inflation. For the underlying CGEFT, the PBH mass in this area may be approximated in terms of the effective sound speed $c_s$:
\bea \label{PBH1}M_{\rm PBH}&=&1.13\times 10^{15}\times\bigg(\frac{\gamma}{0.2}\bigg)\bigg(\frac{g_*}{106.75}\bigg)^{-1/6}\bigg(\frac{k_s}{p_*}\bigg)^{-2}M_{\odot}\times c^{2}_s\nonumber\\
&=&0.46\times\bigg(\frac{\gamma}{0.2}\bigg)\bigg(\frac{g_*}{106.75}\bigg)^{-1/6}M_{\odot}\times c^{2}_s
\nonumber\\
&\approx& {\cal O}(10^{29}-10^{30}){\rm kg}\approx {\cal O}(M_{\odot}),\eea
where the solar mass is denoted by $M_{\odot}\sim 2\times 10^{30}{\rm kg}$; the SRI to USR transition scale is represented by $k_s=10^{6}\;{\rm Mpc}^{-1}$; the pivot scale is represented by $p_*=0.02\;{\rm Mpc}^{-1}$; the effective sound speed is $0.024\leq c_s<1$, $\gamma\sim 0.2$, and the relativistic d.o.f. For SUSY d.o.f., $g_*\sim 226$ and for SM, $g_*\sim 106.75$. 

Lastly, using the Galileon inflationary paradigm, the evaporation time scale of the produced huge mass PBHs may be calculated as follows:
\bea t^{\rm evap}_{\rm PBH}&=&{10}^{64}\bigg(\frac{M_{\rm PBH}}{M_{\odot}}\bigg)^{3}{\rm years}\approx {10}^{64}{\rm years}.\eea 
Even though there are few e-foldings in the span of PBH formation ($\Delta {\cal N}_{\rm USR}\sim 2$), the estimated mass of PBHs in this context is quite big ($\sim M_{\odot})$, and the associated evaporation time scale is likewise very large ($\sim {10}^{64}{\rm years}$).

\subsection{PBH formation}

\subsubsection{In linear regime}

The generation of PBHs via the process of enormous density fluctuations collapsing into a cosmic backdrop with equation of state (EoS) $w$ is the focus of this section. Recently, the importance of the EoS parameter $w$ has drawn some attention \cite{Liu:2023pau,Liu:2023hpw,Balaji:2023ehk,Domenech:2021ztg,Domenech:2020ers,Domenech:2019quo,Altavista:2023zhw}. This is because it can be used to derive information about the physics of the early Universe in the pre-BBN era, using the most recent conclusive signature of a stochastic gravitational wave background (SGWB) reported by the PTA collaborations. Given the current state of uncertainty regarding the primordial content of the Universe, assuming an arbitrary EoS background where even the effective sound speed $c_{s}$ for the fluctuations is unknown offers an exciting chance to investigate the implications of such parameters in the current theory and validate their signatures with observational data. Among the several scenarios that may be used to represent the theoretical results for PBH production and GW generation, we will concentrate on the effect of the parameter $w$ on PBH production and further connect it to the induced GWs from the underlying framework of Galileon inflation. 

For the purpose of comprehending PBH generation in an age of continuous EoS $w$, we would rather utilise the conventional threshold statistical method. According to this technique, when a particular threshold condition on the perturbation overdensity is fulfilled, the primordial density fluctuations must gravitationally collapse and produce PBHs. This section will go into more detail on the amplitude of the scalar power spectrum, which is relevant in this situation. The Press-Schechter formalism, which has been altered to include the constant EoS $w$, will be our primary tool of choice.

The mass that was present in the Horizon at its development is still proportionate to the mass of the produced PBH. Nevertheless, the perturbation overdensities must meet the threshold requirement $\delta\rho/\rho \equiv \delta>\delta_{\rm th}$ in order to start the formation. As a result, the relation for the threshold is as follows: Carr's criteria of $c_{s}^{2}=1$ \cite{1975ApJ...201....1C}, which we employ:
\bea
\label{deltath}
\delta_{\rm th} = \frac{3(1+w)}{5+3w}.
\eea
Additionally, we make the assumption that, in the super-Horizon regime, the density contrast and the comoving curvature perturbation are approximately linear:
\bea \label{deltalinear}
\delta(t,\mathbf{x}) \cong \frac{2(1+w)}{5+3w}\left(\frac{1}{aH}\right)^{2}\nabla^{2}\zeta(\mathbf{x}).
\eea
The PBH production analysis in the aforementioned relation with non-linearities is available in refs. \cite{Ferrante:2022mui,Franciolini:2023pbf,Franciolini:2023wun}. $w$ modifies the generated PBH's final mass in the way that is demonstrated in \cite{Alabidi:2013lya}:
\bea \label{mpbh}
M_{\rm PBH} = 1.13 \times 10^{15} \times \bigg(\frac{\gamma}{0.2}\bigg)\bigg(\frac{g_{*}}{106.75}\bigg)^{-1/6}\bigg(\frac{k_{*}}{k_{\rm s}}\bigg)^{\frac{3(1+w)}{1+3w}} M_{\odot}, \eea
where the solar mass is denoted by $M_{\odot}$, the pivot scale value is labelled by $k_{*}=0.02{\rm Mpc^{-1}}$, and the efficiency factor of collapse is $\gamma \sim 0.2$. The estimation of the variance in the primordial overdensity distribution is necessary to determine the PBH abundance. Here's how to compute this variance: 
\bea
\sigma_{\rm M_{\rm PBH}} ^2 = \bigg(\frac{2(1+w)}{5+3w}\bigg)^2 \int \frac{dk}{k} \; (k
R)^4 \; W^2(kR) \;\left[\Delta^{2} _{\zeta }(k)\right]_{\bf Total}.\eea
There we observe the importance of the previously specified amplitude $A$ of the whole scalar power spectrum in eqn.(\ref{s4dtot}). As may be seen from our numerical results for the abundance mentioned in later sections, the change in the amplitude $A$ is sensitive to the variance estimations. Here, $W(kR)$ represents the Gaussian smoothing function across the PBH formation scales, $R=1/(\Tilde{c_s} k_{s})$, which is provided by $\exp{(-k^2 R^2 /4)}$. The acceptable threshold regime where the collapse of perturbations, creating a significant abundance of PBH, is attained based on the initial form of the power spectrum is limited by the assumption of working with the linear relation in eqn.(\ref{deltalinear}). Numerous numerical studies have been conducted on this regime, and the results show that $2/5 \leq \delta_{\rm th} \leq 2/3$ \cite{Musco:2020jjb}. The range $-5/9 \leq w \leq 1/3$ is examined in terms of $w$. We want to utilise this estimate to determine the appropriate regime that can contribute to producing the required PBH abundance and the induced GW signature consistent with the recent NANOGrav15 finding. The PBHs' mass fraction \cite{Sasaki:2018dmp} now looks like this:
\bea
\beta(M_{\rm PBH})= \gamma \frac{\sigma_{\rm M_{\rm PBH}}}{\sqrt{2\pi}\delta_{\rm th}}\exp{\bigg(-\frac{\delta_{\rm th}^2}{2\sigma_{\rm M_{\rm PBH}}^2}\bigg)}.
\eea
The mass fraction now includes the $w$ dependency and mass resulting from the variance. The mass fraction, which is a function of Gaussian statistics for $\delta$, reflects our decision to ignore any non-linear contributions in the density contrast. Then, the equation for the PBHs current abundance is expressed as follows:
\bea
f_{\rm PBH} \equiv \frac{\Omega_{\rm PBH}}{\Omega_{\rm CDM}}= 1.68\times 10^{8} \bigg(\frac{\gamma}{0.2}\bigg)^{1/2} \bigg(\frac{g_{*}}{106.75}\bigg)^{-1/4} \left(M_{\rm PBH}\right)^{-\frac{6w}{3(1+w)}}\times \beta(M_{\rm PBH}).\eea
In this case, the relativistic degrees of freedom are represented by $g_{*}=106.75$. It is observed that the wavenumber and frequency are related by $f \simeq 1.6\times 10^{-15}(k/{\rm Mpc}^{-1})$. We estimate the abundance and calculate the acceptable range for the EoS $w$ that can still give a sizable abundance after staying within the numerically permissible range of the threshold using the Galileon scalar power spectrum.
\begin{figure*}[htb!]
    	\centering
    {
        \includegraphics[width=17cm,height=12cm]{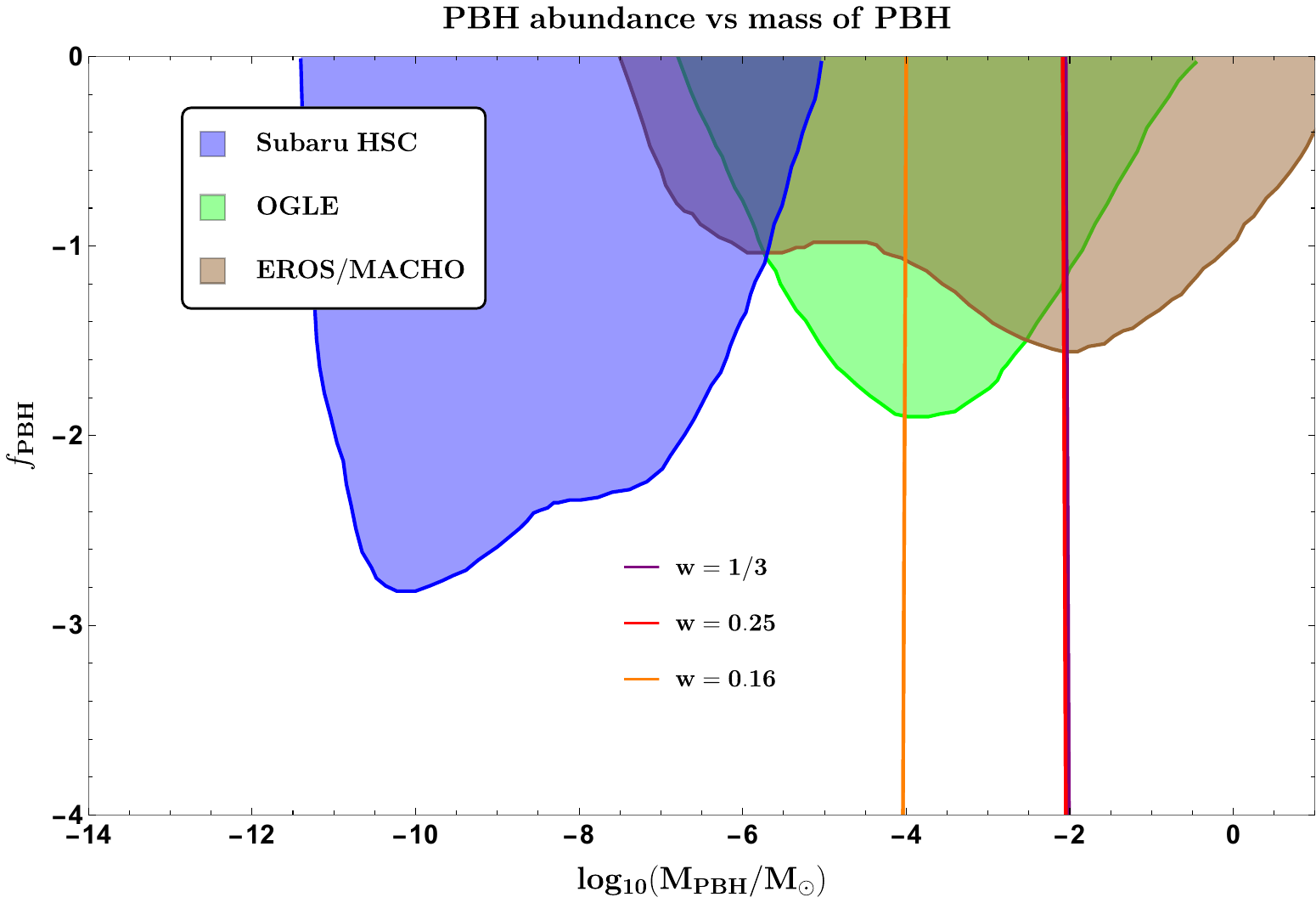}
        \label{fpbhvsmpbh}
    } 
    \caption[Optional caption for list of figures]{PBH abundance expressed in solar mass units as a function of mass. The current microlensing limitations from Subaru HSC (blue), OGLE (green), and EROS/MACHO (brown) are shown by the blue, green, and brown shaded background contours, respectively \cite{Niikura:2017zjd,Niikura:2019kqi,EROS-2:2006ryy}. Purple ($w=1/3$), red ($w=0.25$), and orange ($w=0.16$) are used to illustrate the different EoS $w$ scenarios. These cases can produce a sizable abundance of $f_{\rm PBH} \in (10^{-3},1)$. For PBH formation, $k_{s}\sim {\cal O}(10^{7}){\rm Mpc^{-1}}$ is the transition scale.
 }
\label{s6d1}
    \end{figure*}

Figure (\ref{s6d1}) shows the abundance behaviour for different masses of PBHs, each of which corresponds to a different constant value of $w$. We find that PBHs close to solar mass are created for $w=1/3$, which, when constrained by the microlensing experiment results, is within the region of a sizable abundance. For a backdrop with $w=0.25$, where we find the generation of identical near solar mass PBH within our framework, one arrives at the same result. By further decreasing the $w$ value, we discovered that $M_{\rm PBH} \sim {\cal O}(10^{-4})$ may also be generated for $w=0.16$, providing sufficient abundance to be classified as a viable dark matter candidate. Since the resultant SIGW spectrum does not closely match the SGWB signal received by the PTA, we do not analyse situations for $w$ less than $w=0.16$.

Here we further present a comparative study between the approaches that were employed prior to accounting for any variation in the EoS parameter, mainly focusing on the radiation-dominated (RD) era $w=1/3$, and the modifications that take place following the inclusion of an arbitrary but constant $w$ background in our PBH formation calculations. Using the Press-Schechter formalism is not the only way to evaluate the mass fraction and get good estimates of the PBH abundance related to the frequency of the NANOGrav-15 data when researching PBHs under the conventional scenario of assuming an RD-era for the Universe. The main causes are the linearity approximations in the super-Horizon, as previously mentioned in eqn.(\ref{deltalinear}), and the Gaussian distribution assumptions for the density contrast. To improve the estimation of PBH abundance without overproducing them, it is necessary to incorporate non-linearities and non-gaussianities. This calls for a modification of the customary threshold statistics, with the compaction function formalism offering a potential substitute. References \cite{Franciolini:2023pbf,Ferrante:2022mui,Choudhury:2023fwk} discuss the relevance of non-Gaussianities in relation to PBH overproduction using the compaction function. As for the current focus of our study, we consider a scenario in which a constant $w$ background predominates during PBH creation in the very early Universe, and we explore the potential of preventing overproduction. Our results demonstrate that the $w$-Press-Schechter formalism is a good fit for meeting our needs of producing a significant number of PBHs while preventing overproduction. It is still unclear how to create a more solid picture that incorporates the $w$ parameter into the compaction function formalism and non-linearities, which might support the findings from both the theoretical and empirical perspectives. To enable a more precise estimate of the mass fraction, one may also think of computing the higher point non-Gaussian correlations, by which the corresponding higher-order non-Gaussianity parameters can further confine the range of values. To get the intended PBH abundance, however, we experiment with the EoS $w$ in this study.

\subsubsection{In non-linear regime}

We now discuss the main problem relating to PBH abundance and important overproduction difficulties. It is necessary to appropriately take into account the non-linearities between the comoving curvature and the disturbances in the density contrast field, as provided by \cite{Musco:2018rwt}, in order to estimate the abundance with accuracy:
\bea \label{NLrelation}  \delta(r, t) &=& -\frac{2}{3}\frac{u}{(aH)^2}e^{-2\zeta(r)} \bigg[\zeta^{''}(r) + \frac{2}{r}\zeta^{'}(r) + \frac{1}{2}\zeta^{'}(r)^{2}\bigg],\quad\quad \eea
where the equation of state parameter $w$ is connected to $u = 3(1+w)/(5+3w)$ \cite{Polnarev:2006aa}. For the remainder of the computation, we will additionally utilise $w=1/3$ for the RD era. On super-horizon scales, the above equation makes the assumption of long wavelength approximation and spherical symmetry for the locally affected area. The density contrast field will subsequently be obtained by translating the additional primordial NGs into the conserved curvature perturbation random field $\zeta(r)$ via the previously stated non-linear connection. The well-known ansatz, which represents the quadratic NG model, is our emphasis in this review:
\bea \zeta = \zeta_{G} + \frac{3}{5}f_{\rm NL}\zeta_{G}^{2}, \eea
where the quantity of NG in the theory is measured by $f_{\rm NL}$, and Gaussian statistics are followed by $\zeta_{G} \equiv \zeta_{G}(r)$. In order to precisely estimate PBH abundance in the presence of the aforementioned traits, we now apply the threshold statistics technique to the compaction function. The SM contains information on the compaction function.  The specification of the compaction function ${\cal C}(r,t)$ then obtains a time-independent behaviour valid in the super-horizon scales from the usage of eqn.(\ref{NLrelation}), leading to the following expression:
\bea
\label{Cg}
{\cal C}(r) = {\cal C}_{G}(r)\frac{d\zeta}{d\zeta_{G}} - \frac{1}{4 u}\left({\cal C}_{G}(r)\frac{d\zeta}{d\zeta_{G}}\right)^{2}.
\eea 
This is further expressed as ${\cal C}_{G}(r) = -2ur\zeta^{'}_{G}(r)$. Both $\zeta_{G}(r)$ and ${\cal C}_{G}(r)$, which are Gaussian variable derivatives, are Gaussian variables included in the connection above. According to \cite{Ferrante:2022mui, Franciolini:2023pbf}, the overall proportion of dark matter in PBHs originates from the relation after integrating across a range of horizon masses ($M_{H}$):
\bea \label{fpbh} f_{\rm PBH} &=& \frac{1}{\Omega_{\rm DM}}\int d\;\ln{M_{H}}\left(\frac{M_{H}}{M_{\odot}}\right)^{-\frac{1}{2}} \times\left(\frac{g_{*}}{106.75}\right)^{\frac{3}{4}}\left(\frac{g_{*s}}{106.75}\right)^{-1}\left(\frac{\beta_{\rm NG}(M_{H})}{7.9 \times 10^{-10}}\right),
\eea
where the dark matter density of the cosmos is represented by $\Omega_{\rm DM} \simeq 0.264$, and the degrees of freedom for effective energy and entropy are denoted by $g_{*},\;g_{*s}$. The creation of PBH is well known to exhibit exponential sensitivity to the tail of the Probability Distribution Function (PDF) of density fluctuations, where non-Gaussian effects are prevalent \cite{DeLuca:2022rfz, Taoso:2021uvl,Atal:2018neu, Young:2013oia,Byrnes:2012yx,Bullock:1996at}. The two Gaussian random variables in the current situation, $\cal{C}_G$ and $\zeta_G$, have non-zero auto and cross-correlations, which result in the two-dimensional joint PDF:
\bea
   \label{PDF}
 P_G({\cal C}_G,\zeta_G) &=& \frac{1}{2\pi\sigma_c \sigma_{r}\sqrt{1-\gamma_{\rm cr}^2}}\exp{\left(-\frac{\zeta_G ^2}{2\sigma_r ^2}\right)}\times \exp{\bigg[-\frac{1}{2(1-\gamma_{\rm cr}^2)}\bigg(\frac{{\cal C}_G}{\sigma_c}-\frac{\gamma_{\rm cr}\zeta_G}{\sigma_r}\bigg)^2\bigg]},
   \eea
The correlation coefficient is represented by the expression $\gamma_{cr}=\sigma^{2}_{cr}/(\sigma_{c}\sigma_{r})$. When taking into account the domain established by the threshold statistics on the compaction function \cite{Ferrante:2022mui}, this PDF assists in determining the necessary mass fraction of PBHs:
\bea   
\label{Beta}
\beta_{\rm NG}(M_{H}) = \int_{\cal D}{\cal K}({\cal C}-{\cal C}_{\rm th})^{\gamma}P_{G}({\cal C}_G, \zeta_G) d{\cal C}_G\;d\zeta_{G}, \eea
for the PBH mass generated during horizon re-entry \cite{Choptuik:1992jv, Evans:1994pj}, where the \textit{critical scaling relation} is included through ${\cal K}({\cal C}-{\cal C}_{\rm th})^{\gamma}$. The simulations yielded $\gamma \sim 0.36$ for the RD era and values for the constant ${\cal K}\sim {\cal O}(1-10)$ and the threshold ${\cal C}_{\rm th}$, which are based on \cite{Musco:2020jjb}. The integration domain is ${\cal D} = \{{\cal C}(r) \geq {\cal C}_{\rm th} \wedge {\cal C}_{G}(d\zeta/d\zeta_{G}) \leq 2u\}$. This entails maximising the compaction function, the specifics of which we will explain in the SM. We now present the definitions for the different correlations found in the 2D joint PDF \cite{Franciolini:2023pbf,Young:2022phe}:
    \begin{figure*}[htb!]
    	\centering
   { 
   \includegraphics[width=18cm,height=6cm] {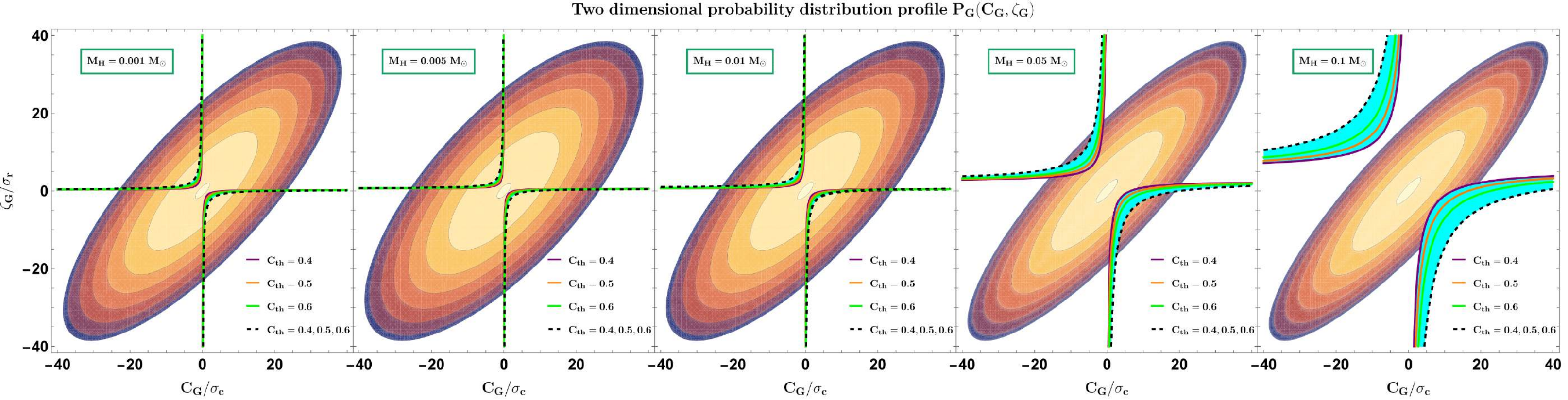}
    }
 	\caption[Optional caption for list of figures]{ 2D PDF logarithmic plot for different horizon masses $M_H$, with many threshold density contrast values, ${\cal C}_{\rm th}={0.4,0.5,0.6}$. The contour lines shown against the Gaussian variables, ${\cal C}_G,{\cal \zeta}_G$, represent $P_G ({\cal C}_G,{\cal \zeta}_G)$. The parameter space, or integration domain to get a sizable PBH mass fraction, is shown by the shaded area between the two dotted lines. For the scalar power spectrum, all figures are obtained with an amplitude of $A=10^{-2}$.
} 
    	\label{2dpdffig}
    \end{figure*}
\bea \label{sigcr}
\sigma_{cr}^{2} &=& \frac{2u}{3} \int_{0}^{\infty}\frac{dk}{k}(k r_{m})^{2}W_{g}(k,r_{m})W_{s}(k,r_{m})\tilde{\Delta}^{2}_{\zeta}(k),\quad
\\
\label{sigc} \sigma_{c}^{2} &=& \left(\frac{2u}{3}\right)^{2}\int_{0}^{\infty}\frac{dk}{k}(k r_{m})^{4}W_{g}^{2}(k,r_{m})\tilde{\Delta}^{2}_{\zeta}(k),
\\
\label{sigr} \sigma_{r}^{2} &=& \int_{0}^{\infty}\frac{dk}{k}W^{2}_{s}(k,r_{m})\tilde{\Delta}^{2}_{\zeta}(k),
\eea
where it is decided to use the Gaussian smoothing functions $W_{g}(k,r)$ and $W_{s}(k,r)$, i.e., $\exp{(-k^{2}r^{2}/2)}$. This is often not done, particularly when using the so-called spherical-shell window function $W_{s}(k,r)$. Choosing this option is advantageous for our theory since including the top-hat or other sinusoidal functions causes significant oscillations that are ineffective in mitigating the small-scale disruptions. Utilising the smoothing characteristic of the radiation transfer function $T(k,\tau) = 3(\sin{l}-l\cos{l})/l^{3}$, with $l=k\tau/\sqrt{3}$ and $\tau=1/aH$, we established the new power spectrum form as $\tilde{\Delta}^{2}_{\zeta}(k) = T^{2}(k,r_{m})\Delta^{2}_{\zeta}(k)$. It is significant to remember that the horizon re-entry scale, $r_{m} = (\tilde{c_{s}}k_{H})^{-1}$, corresponds to the wavenumber during PBH creation and the scale where the compaction function maximises. This is where eqs. (\ref{sigcr}-\ref{sigr}) are assessed. The mass fraction exhibits $M_{H}$ dependency due to the relationship between $\tilde{c_{s}}k_{H} \propto 1/\sqrt{M_{H}}$ \cite{Sasaki:2018dmp,Kawasaki:2016pql}. See SM for further information.

The contour plots of the 2D-PDF for various values of the horizon mass are shown in Fig. (\ref{2dpdffig}). The permitted domains of integration for the mass fraction in eqn.(\ref{Beta}) are shown by the purple, orange, and green coloured lines. Every domain has a single common border. We note that the domain becomes increasingly compressed and shifts away from the contour centre as the threshold on the compaction function, ${\cal C}_{\rm th}$, rises. The likelihood of having a sizable abundance of that specific PBH mass—which is somewhat less than the horizon mass upon re-entry—increases with the strength of the PDF's domain support. Significant behaviour is also evident in the contours, as $\gamma_{cr}$ grows both above and below the mass, $M_{H} \sim {\cal O}(10^{-3}M_{\odot})$, indicating strong connections between the two Gaussian variables. It is unlikely that our approach will prevent the overproduction of near solar mass PBHs for masses where $M_{H} > {\cal O}(0.1 M_{\odot})$ since the domain does not overlap with the PDF. 
    \begin{figure}[htb!]
    	\centering
   {
   \includegraphics[width=12cm,height=7cm] {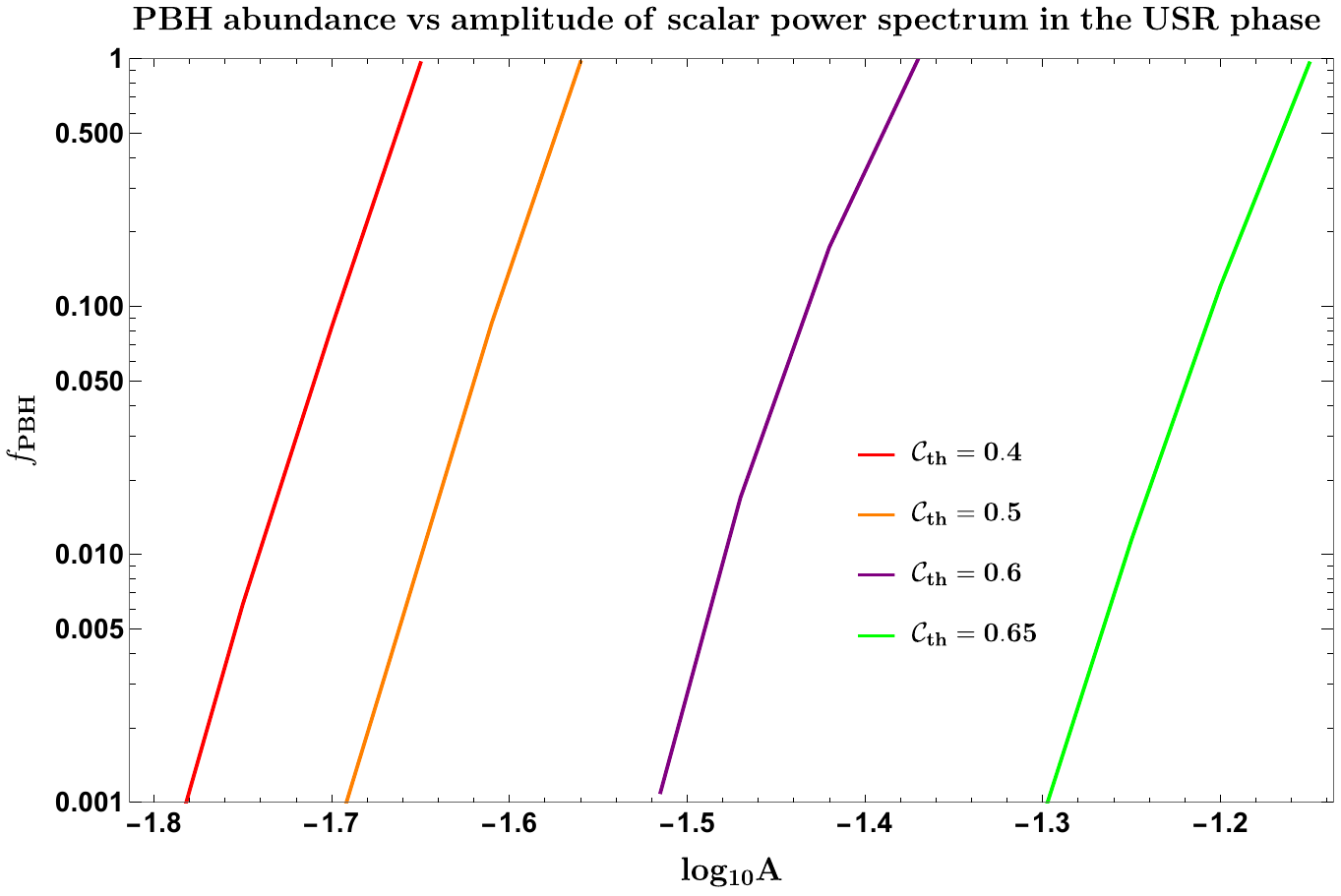}
    }
    	\caption[Optional caption for list of figures]{PBH abundance dependence in the USR phase on the magnitude of the scalar power spectrum of the curvature perturbation. The compaction function's threshold values, ${\cal C}_{\rm th}=\{0.4,0.5,0.6,0.65\}$, are represented by the red, orange, blue, and green lines, in that order. For this plot, the transition scale is set at $k_{s}=10^{7}{\rm Mpc}^{-1}$.
} 
    	\label{fpbhvamp}
    \end{figure}


    \begin{figure}[htb!]
    	\centering
   {
   \includegraphics[width=12cm,height=8cm] {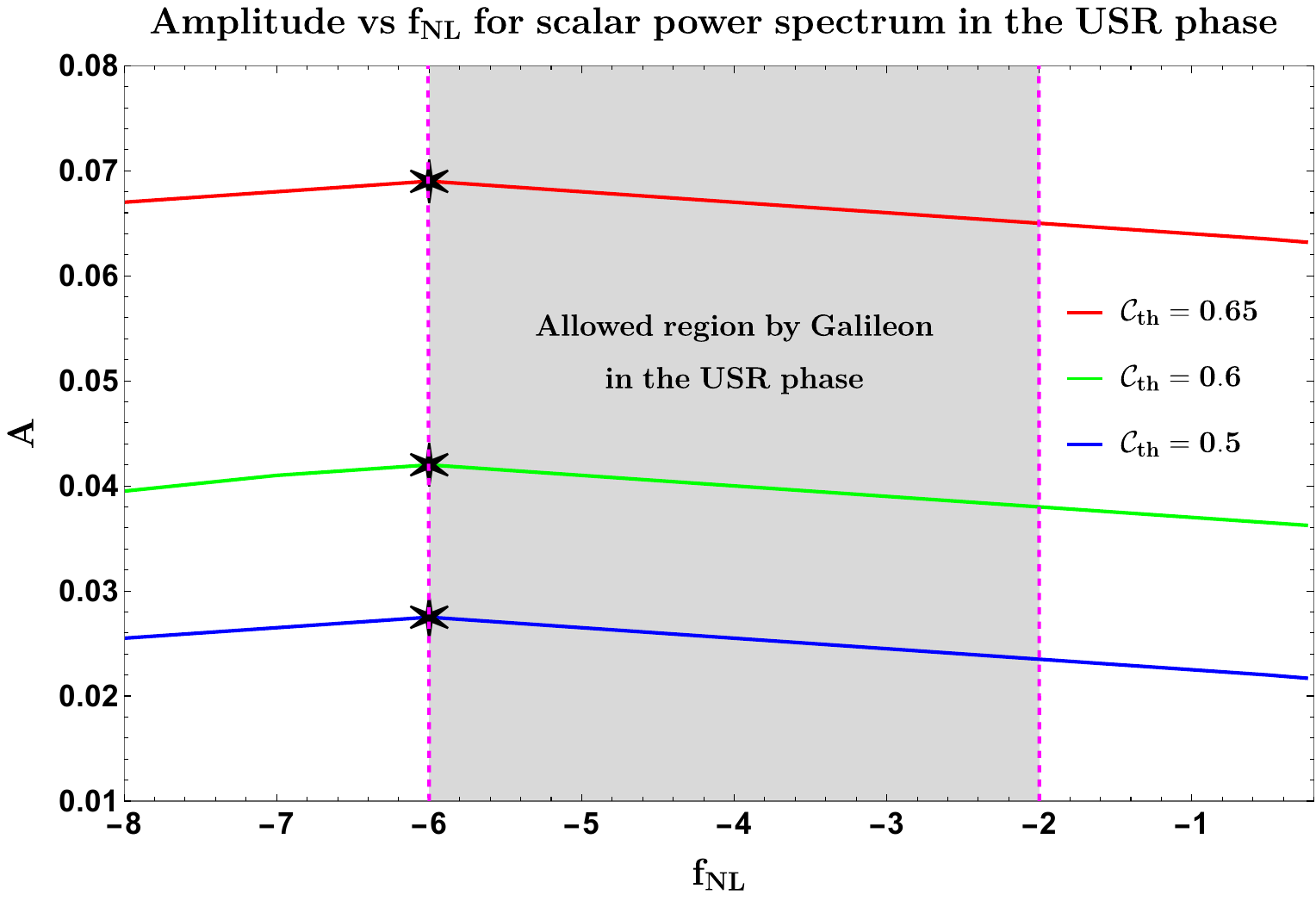}
    }
    	\caption[Optional caption for list of figures]{Behaviour of the USR phase peak amplitude with varying values of $f_{\rm NL}$, the negative NG. While the USR transition wavenumber $k_{s}=10^{7}{\rm Mpc}^{-1}$ and the ensuing fractional abundance $f_{\rm PBH}=1$ are maintained fixed, various values of ${\cal C}_{\rm th}=\{0.65,0.6,0.5\}$ are taken into consideration and shown by red, green, and blue lines, respectively. The amplitude value that corresponds to the highest permitted abundance under the current circumstances for $f_{\rm NL}=-6$ is represented by the black star. The theoretically permitted NG values in Galileon theory during the USR phase are displayed in the gray-shaded area with the equation $-6 \leq f_{\rm NL} \leq -2$.
} 
    	\label{ampvsfnl}
    \end{figure}
The amplitude changes for specific values of the compaction threshold to produce the fractional abundance between the interval $f_{\rm PBH} \in (10^{-3},1)$ when the transition scale $k_{s}=10^{7}{\rm Mpc}^{-1}$ is fixed. This is shown in the fig. (\ref{fpbhvamp}) above. Larger amplitudes of the scalar power spectrum are needed to seed sizable abundances of black holes, as one might expect, when one increases the threshold. Additionally, the plot shows how sensitive the abundance is to the amplitude, fluctuating rapidly and reaching unity in a remarkably limited range of amplitude values. In this case, the study has been limited to the previously described threshold value regime, ${\cal C}_{\rm th}$, where the theory of cosmic perturbation holds.

We now want to enhance our explanation of findings in which NG is crucial in establishing the final fractional abundance of PBH and the power spectrum amplitude. To reach a maximum feasible abundance of unity, the plot in fig. (\ref{ampvsfnl}) shows how the peak amplitude of the power spectrum varies for various values of negative local NG in the USR phase. massive oscillations that produce higher-mass black holes can be sparked by massive negative NG. The masses of PBH generated with the transition scale set at $k_{s}=10^{7}{\rm Mpc}^{-1}$ have been the focus of our attention here. It is possible to generate $M_{\rm PBH} \sim {\cal O}(10^{-2}M_{\odot})$ on this scale. We find that the amplitude in the USR phase decreases to saturate the abundance of the aforementioned PBH masses as we tend to lower the quantity of NG ($f_{\rm NL}$) below $-6$. Conversely, increasing $f_{\rm NL}$ over $-6$ will lead to a greater abundance of masses in the permitted PBH mass spectrum's bottom end, requiring less amplitude to form. Galileon theory currently prohibits NG values outside of the interval $-6 \leq f_{\rm NL} \leq -2$; the theoretically permitted zone is indicated by the gray-shaded area. The most satisfactory results for the agreement of the generated SIGWs with the PTA data were obtained by us when we took into consideration $f_{\rm NL}=-6$ and its related power spectrum amplitudes shown using a black star marker for different thresholds, ${\cal C}_{\rm th} = \{0.5,0.6,0.65\}$.  

    \begin{figure}[htb!]
    	\centering
   {
   \includegraphics[width=18cm,height=11cm] {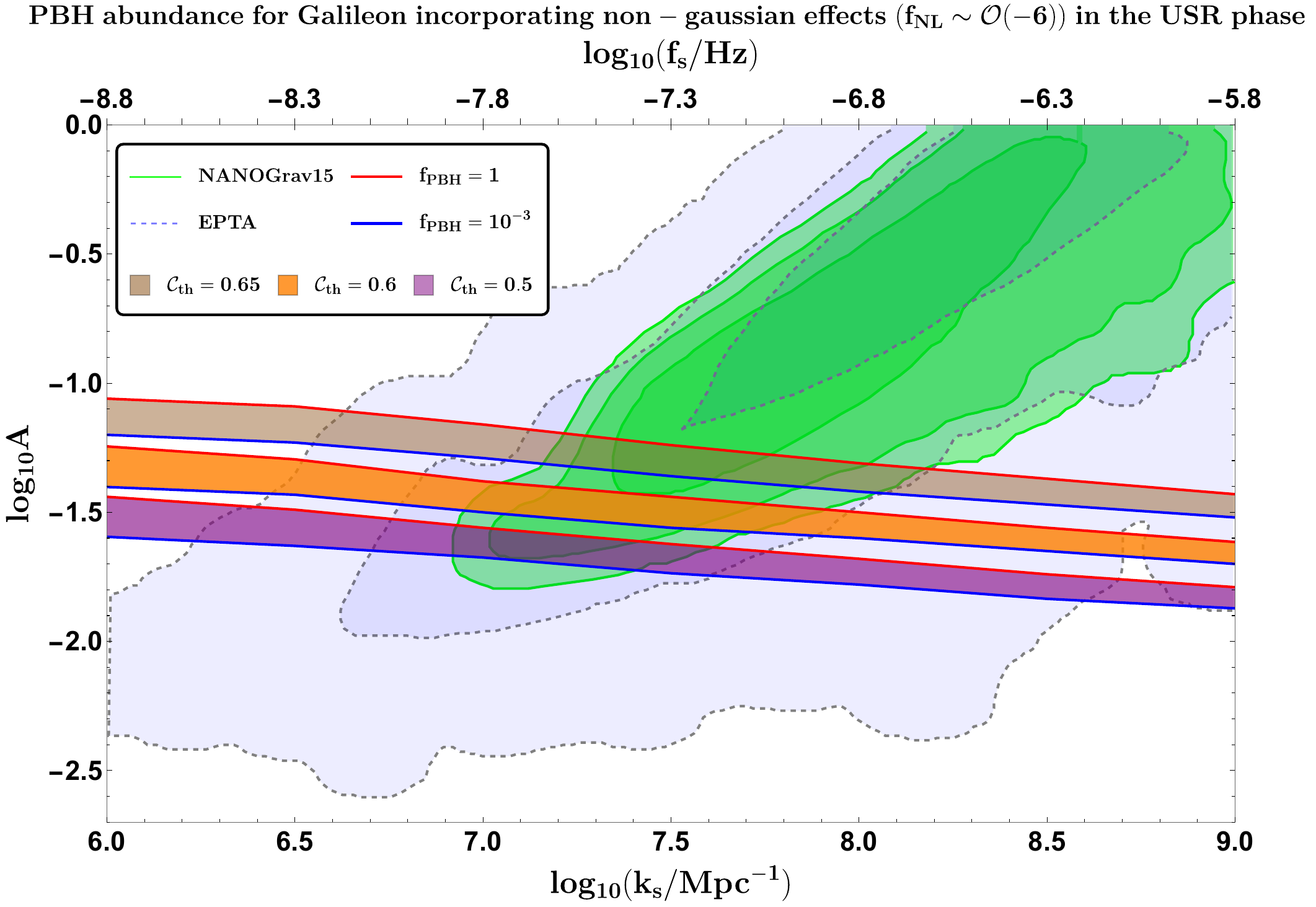}
    }
    	\caption[Optional caption for list of figures]{PBH abundance calculated using Galileon theory for various volume-averaged density contrast threshold values. $\mathcal{C}_{\rm th}=\{0.65,0.6,0.5\}$ represents the threshold values that are corresponding to the brown, orange, and purple coloured bands, respectively. The zone of abundance $f_{\rm PBH} \in (1, 10^{-3})$ is delimited by the red and blue borders of both bands. The green and light-blue posteriors are from \cite{Franciolini:2023pbf}, and they stand for the NANOGrav 15 and EPTA, respectively.
} 
    	\label{pbhfig}
    \end{figure}
Understanding the mass fraction and its permitted domain is necessary to estimate the PBH abundance. Using eqn.(\ref{fpbh}), as shown in fig.(\ref{pbhfig}), we now analyse our results.
The scalar power spectrum amplitude fluctuation in the USR phase, $A$, is depicted in the figure together with the transition wavenumber $k_{s}$, allowing one to calculate the overall abundance in the range $f_{\rm PBH} \in (1, 10^{-3})$. In the case of ${\cal C}_{\rm th}=0.65$, the brown band can be pushed to larger amplitude values; this is possible as long as $\Delta{\cal N}_{\rm USR} \sim {\cal O}(2)$ is satisfied, and there is no overproduction that favours the SIGW interpretation of the NANOGrav 15 data within $1\sigma$. The NANOGrav 15 signal's $1\sigma$ contour is touched by the orange no overproduction zone in the case of ${\cal C}_{\rm th}=0.6$. According to our idea, in the case of a big negative NG, the amplitude is such that the band remains close to the best-fit zone. When we approach ${\cal C}_{\rm th}=0.5$, the amplitude of the purple band decreases while remaining within $2\sigma$. This behaviour makes sense since raising the threshold would need using greater amplitude in order to get the same PBH mass. It is important to remember that with increasing wavenumbers, the band's amplitude diminishes. Due to the inverse relationship between PBH mass and transition wavenumber squared, smaller mass PBHs can produce significant abundances with a smaller amplitude of the scalar power spectrum. 

Based on our analysis of PBH abundance, we find that the single-field Galileon theory uses an additional USR feature during inflation whose duration respects the established perturbativity conditions, in contrast to using a two-field model, such as the curvaton, to generate the PBH forming curvature perturbations \cite{Choudhury:2023hvf,Choudhury:2023kdb,Choudhury:2023hfm}. It adds to the overall quantum one-loop corrections and adequately improves the scalar power spectrum for PBH production. Without running the danger of overproduction, the obtained amplitude corresponds to a sizable PBH abundance.
The curvaton model study in \cite{Franciolini:2023pbf, Ferrante:2022mui} takes into account the NGs in detail to produce an appropriate abundance, which results in the amplitude of the power spectrum that is sufficient to reduce the tension between SIGWs and NANOGrav 15. However, in the presence of the quantum loop effects, Galileon adds additional bigger negative NGs, such that the resultant amplitude produces the SIGWs that highly correlate with the NANOGrav 15 data.

We now present some general discussions regarding our analysis based on the non-linearities, NGs, and the necessary perturbativity conditions (we refer to supplementary material as SM). 
\begin{itemize}[label=$\bullet$]

\item \underline{\textbf{Inclusion of non-linearities}}:\\ \\ Although the analysis presented in this review is wide, it does not discount the apparent non-linearities in the density contrast field that may result from curvature disturbances that behave non-linearly on super-horizon scales. Given that it offers a more accurate estimate of the density contrast formation threshold in the presence of non-linear changes, we opted for the compaction function \cite{Musco:2020jjb, Musco:2018rwt}. Using the crucial scaling relation and the integration domain that is reliant on the threshold value and the effects caused by non-linearities, we have used this estimate in this study to assess the PBH mass fraction.  Additionally, these non-linearities do not usually become unmanageable due to their impacts, as demonstrated by our earlier results and which will also be discussed in a moment when we analyse additional results (see fig.\ref{fpbhvamp}-\ref{ampvsfnl}).

\item  \underline{\textbf{Effects of non-Gaussianities}}: \\ \\ These come from two crucial sources in our current understanding. First, even if one assumes Gaussian statistics for the curvature perturbation, the non-Gaussian effects propagate from the non-linearities in the density contrast field. The second has to do with PBH creation, when amplitude of the scalar power spectrum is adequately enhanced by a short-lived slow-roll violation. For a more realistic understanding of the production of PBHs and their fractional abundance for the overall dark matter composition of the present, these sources must be taken into account. Our usage of the quadratic model for the curvature perturbation, which is then translated into the compaction function, contains the local primordial NGs. Thus, the quantity of NG ($f_{\rm NL}$) present eventually determines the PBH abundance and the scalar power spectrum amplitude, which will be further discussed in the results of this section.

\item \underline{\textbf{Preserving perturbativity}}:\\ \\
In our configuration, a USR phase that produces the significant curvature perturbations is subject to periodical constraints. The key criterion $\Delta{\cal N}_{\rm USR}\sim {\cal O}(2)$ on the e-foldings of the USR, which ensures that the regulated creation of NGs and the cosmological perturbative approximations are both met, has been upheld throughout our study. We limit our analysis to $2/5 \leq \delta_{\rm th} \leq 2/3$ for the volume-averaged density contrast threshold and the compaction threshold ${\cal C}_{\rm th}$ at the moment of horizon crossing, which also requires another perturbative approximation. During the long-wavelength approximation on super-horizon scales, non-linear effects are still manageable inside this window. As can be shown from the next explanations of our findings, suitable solutions are obtained by maintaining perturbativity.

\item  \underline{\textbf{Respecting Causality and Unitarity}}:\\ \\ The current configuration exhibits a conformal time-dependent behaviour for the effective sound speed parameter, $c_{s}$. Its value at the pivot scale is $c_{s}(\tau_{*})=c_{s}$, where the conformal time for the pivot scale $k_{*}=0.02 {\rm Mpc}^{-1}$ is denoted by $\tau_{*}$. By meeting the observational constraint of $0.024 \leq c_{s} \leq 1$, we were able to maintain the unitarity and causality of the SIGWs produced by Galileon theory during our study \cite{Planck:2015sxf}.

\item  \underline{\textbf{Mass of PBH and NGs}}:\\ \\ In the current scenario, a USR phase is used to induce the transient slow-roll violation necessary for PBH formation. Specifically, the critical parameter dictating the final mass of the PBH is $k_{s}$, the wavenumber linked to the slow-roll to USR phase transition. According to the dependence, $\tilde{c_{s}}k_{s} \propto 1/\sqrt{M_{\rm PBH}}$. Not matter where the USR is located precisely, our arrangement maintains significant improvements in the NGs. The transition scale in this study is set at $k_{s}=10^{7}{\rm Mpc}^{-1}$. This results in the creation of $M_{\rm PBH} \sim {\cal O}(0.01)M_{\odot}$ and also produces significant negative NGs \cite{Choudhury:2023kdb}.

\end{itemize}

\subsection{Scalar Induced Gravity Waves (SIGWs) from CGEFT}

In this part, we address the theory of gravitational waves (SIGW) caused by scalars and produced in the presence of a generic cosmic background with a constant EoS $w$. We first investigate the underlying theoretical setting while building the required mathematical framework, and then we apply the scalar power spectrum of Galileon theory and analyse the resultant GW spectrum. The generation of SIGWs is studied in this part by looking at their resultant spectrum, using the generic equations for the kernel (or transfer) function discussed in the preceding sections. When the Universe eventually recovers the hot Big Bang scenario, the $w$-SIGW phenomenon corresponds to an arbitrary background EoS $w$, which is postulated during the last phases of inflation. In order to visualise the induced GW spectrum by sourcing from the scalar modes sub-Horizon during such a broad cosmic backdrop, we adopt this scenario for our Galileon inflation setup. Thus, with a generic $w$ backdrop, the induced GW spectrum is expressed as follows:
\bea
\label{GWdensity}
\Omega_{\rm{GW},0}h^2 = 1.62 \times 10^{-5}\;\bigg(\frac{\Omega_{r,0}h^2}{4.18 \times 10^{-5}}\bigg) \bigg(\frac{g_{*}(T_c)}{106.75}\bigg)\bigg(\frac{g_{*,s}(T_c)}{106.75}\bigg)^{-4/3}\Omega_{\rm GW,c},
\eea
where the radiation energy density as measured today is denoted by $\Omega_{r,0}h^2$, and the energy and entropy effective degrees of freedom are represented by $g_{*},g_{*,s}$. When such generated GWs behave as freely propagating GWs throughout the radiation-dominated period, indicated by the instant ``c'', the number $\Omega_{\rm GW,c}$ reflects the GW energy density fraction. 

Using the kernel functions for the modes and scales that fulfil $k \geq k_{*}$, we further describe the energy density $\Omega_{\rm GW,c}$ as follows:
\bea
\label{omegac}
\Omega_{\rm {GW},c}&=& \frac{k^{2}}{12a^{2}H^{2}}\times\overline{\Delta^{2}_{h}(k,\tau)}  \nonumber\\
&=& \bigg(\frac{k}{k_{*}}\bigg)^{-2b}\int_{0}^{\infty}dv \int_{|{1-v}|}^{1+v} du \; {\cal T}(u,v,w,c_s) \;\;\Big[\Delta^{2}_{\zeta}(ku)\Big]_{\bf Total} \times \Big[\Delta^{2}_{\zeta}(kv)\Big]_{\bf Total},
\eea
with the propagation speed $c_{s}$ being constant and the transfer function for a constant EoS background. The pivot scale in our setup, which we will use for our next GW spectrum analysis, is denoted by the value $k_{*}$. To ensure clarity, we refer to the transfer function that was ultimately utilised, ${\cal T}(u,v,w,c_{s})$, which is the result of applying eqn.(\ref{kernelavg}) and correcting for additional multiplicative factors found in eqn.(\ref{tensorpspec}):
\bea \label{transfer}
{\cal T}(u,v,w,c_{s}) &=& (b+1)^{-2(b+1)}\frac{4^{2b}}{3c^{4}_{s}}\bigg[\frac{3(1+w)}{1+3w}\bigg]^{2}\Gamma^{4}(b+3/2)\bigg[\frac{4v^2 - (1-u^2 +v^2 )^2}{4u^2v^2}\bigg]^{2}\bigg(\frac{Z}{2uv}\bigg)^{2b} \nonumber\\
&& \quad\quad\quad\quad \times\bigg\{\bigg(P_{-b}^{b}(y) + \frac{3(1+w)}{2}P_{b+2}^{-b}(y)\bigg)^{2}\Theta(c_{s}(u+v)-1) \nonumber\\
&& \quad\quad\quad\quad + \frac{4}{\pi^{2}}\bigg(Q_{-b}^{b}(y) + \frac{3(1+w)}{2}Q_{-b}^{b+2}(y)\bigg)^{2}\Theta(c_{s}(u+v)-1) \nonumber\\
&& \quad\quad\quad\quad + \frac{4}{\pi^{2}}\bigg({\cal Q}^{-b}_{b}(-y) + 3(1+w){\cal Q}^{-b}_{b+2}(-y)\bigg)^{2}\Theta(1-c_{s}(u+v)) \bigg\}.
\eea
Utilising the transfer function mentioned above, we can assess the GW density and examine the spectrum's behaviour for different EoS situations for different cases of $b$ or $w$ values. We note a key aspect about the suitability of this theory for producing induced GWs from a generic $w$ background in the current Galileon EFT setup before moving on to more results. As we noted at the beginning of the section, examining the self-interactions in Galileon theory requires dealing with the decoupling limit. This constraint allows us to successfully ignore the gravity sector mixing effects, which would require a whole new formulation of the development of tensor modes in generated gravitational waves and complicate the study overall. The foregoing analysis for the metric perturbations and its application with the Galileon EFT is therefore validated by the decoupling limit, which saves us by concentrating on the Galileon self-interactions separately rather than considering couplings with the gravity sector. Look at the Appendix \ref{A7a}, \ref{A8a} and \ref{A9a} for more details on this computation.

\subsection{Issues related to overproduction}

We have attempted to identify its energy density spectrum and explore the formation of induced GWs in a generic EoS $w$ context in the preceding sections. A PBH counterpart can be connected to the large boosts resulting from scalar mode couplings in the very early Universe. The observation of SIGW may also indicate the possibility of producing near solar-mass PBH in the very early Universe due to the elevated power spectrum that is sensitive to NANOGrav15 scales.

\subsubsection{Overproduction problem}

Regarding the issue of PBH overproduction, there has been a lot of progress recently. Many people investigated the SGWB signal that the PTA partnerships published to determine whether it had astronomical or cosmic origins. The SIGW model is one of the many potential outcomes that most closely matches the PTA data, yet it was shortly proposed that this scenario could have an overproduction issue. Many people have actively sought an explanation and remedies for this problem; take references into consideration.For the relevant literature, see \cite{Ferrante:2023bgz,Franciolini:2023pbf,Gow:2023zzp,Gorji:2023sil,Firouzjahi:2023xke}. The major issue here is that attaining a sizable near-solar-mass PBH abundance that matches the frequencies investigated by the NANOGrav15 signal ultimately results in a conflict with the SIGW interpretation of the same SGWB signal, which is supported by the PTA collaborations. Since the augmented amplitude becomes $\sim {\cal O}(1)$ during the domain associated with PBH generation, forcing higher statistical agreement with PTA would inevitably result in a collapse of perturbation theory. From the standpoint of closely matching the data, the aforementioned makes the problem more problematic. In order to avoid the overproduction, it either asks for a substitute that can more closely resemble the SIGW interpretation of the data or offers a substitute interpretation of the signal itself. The authors in \cite{Franciolini:2023pbf,Ferrante:2022mui} incorporate non-linearities in the density contrast existing while in the super-Horizon regime and the influence of non-Gaussianities to strengthen the assertions concerning overproduction. The problem becomes more stiff and requires more attention because of these theoretically required components and a detailed study across numerous models. Further basic problems about the precise theoretical calculations for the PBH abundance remain, such as the appropriate density contrast threshold range and higher-order estimations of the super-Horizon abundance. These still need to be shown in practice, though, and it's unlikely that they will materially change the findings of the current research.  

As was already indicated, a number of writers have recently attempted to offer appealing solutions to the overproduction problem for PBHs. Among these resolutions is a potential curvaton scenario which is responsible for the curvature perturbations at PBH-relevant scales, and which might exist as a spectator field in the very early Universe. Furthermore, the reported signal might also be caused by an additional tensor spectator field. Due to the participation of non-Gaussianities, which can become extremely sensitive to the scalar power spectrum amplitude, the presence of non-attractor characteristics in the theory of concern is also essential for having abundant PBH generation. In this part, we focus on the non-Gaussian aspects of the primordial density perturbations, which are crucial in understanding PBH generation. Furthermore, given the overproduction problem, the rationale behind selecting this scenario will be covered in this part, as we are working with an arbitrary EoS value.

\subsubsection{Resolution with primordial non-Gaussianity}

A common scenario entails augmenting inflation with an ultra-slow roll (USR) phase. This introduces substantial quantum fluctuations that produce the first curvature perturbations, which eventually collapse into PBHs or, in some situations, give birth to induced giant waves. Although we assume the power spectrum profile for the curvature perturbations to be Gaussian for simplicity's sake, including non-Gaussianity enables a thorough and in-depth examination. In order to compute the PBHs abundance, for example, non-Gaussianity significantly affects the tail of the probability distribution function (PDF) \cite{Kawaguchi:2023mgk,DeLuca:2022rfz,Taoso:2021uvl,Atal:2018neu,Young:2013oia,Byrnes:2012yx,Bullock:1996at}. A comprehensive examination of the PDFs associated with single-field inflation, as reported in Ref.\cite{Pi:2022ysn}, indicates the existence of a logarithmic relation for e-fold fluctuations, $\delta N$, when a non-attractor regime is present during inflation. The predominance of these logarithms in $\delta N$ can cause the PDF to rapidly develop exponential tails, which can have a substantial impact on the PBH mass fraction. In the context of Galileon theory, we addressed the non-linearity and non-Gaussianity related to the curvature perturbations in Ref.\cite{Choudhury:2023fwk} by using the threshold statistics of the compaction function in linear cosmological perturbation theory. Large negative local non-Gaussianity related to the primordial disturbances was taken into consideration. While taking into consideration $f_{\rm NL} \sim -6$, our results imply that PBH masses up to $M_{\rm PBH} \sim {\cal O}(10^{-3}-10^{-2} M_{\odot})$ might avoid PBH overproduction.

\subsubsection{Resolution with equation of state}
\begin{figure*}[htb!]
    	\centering
    \subfigure[]{
      	\includegraphics[width=8.5cm,height=7.5cm]{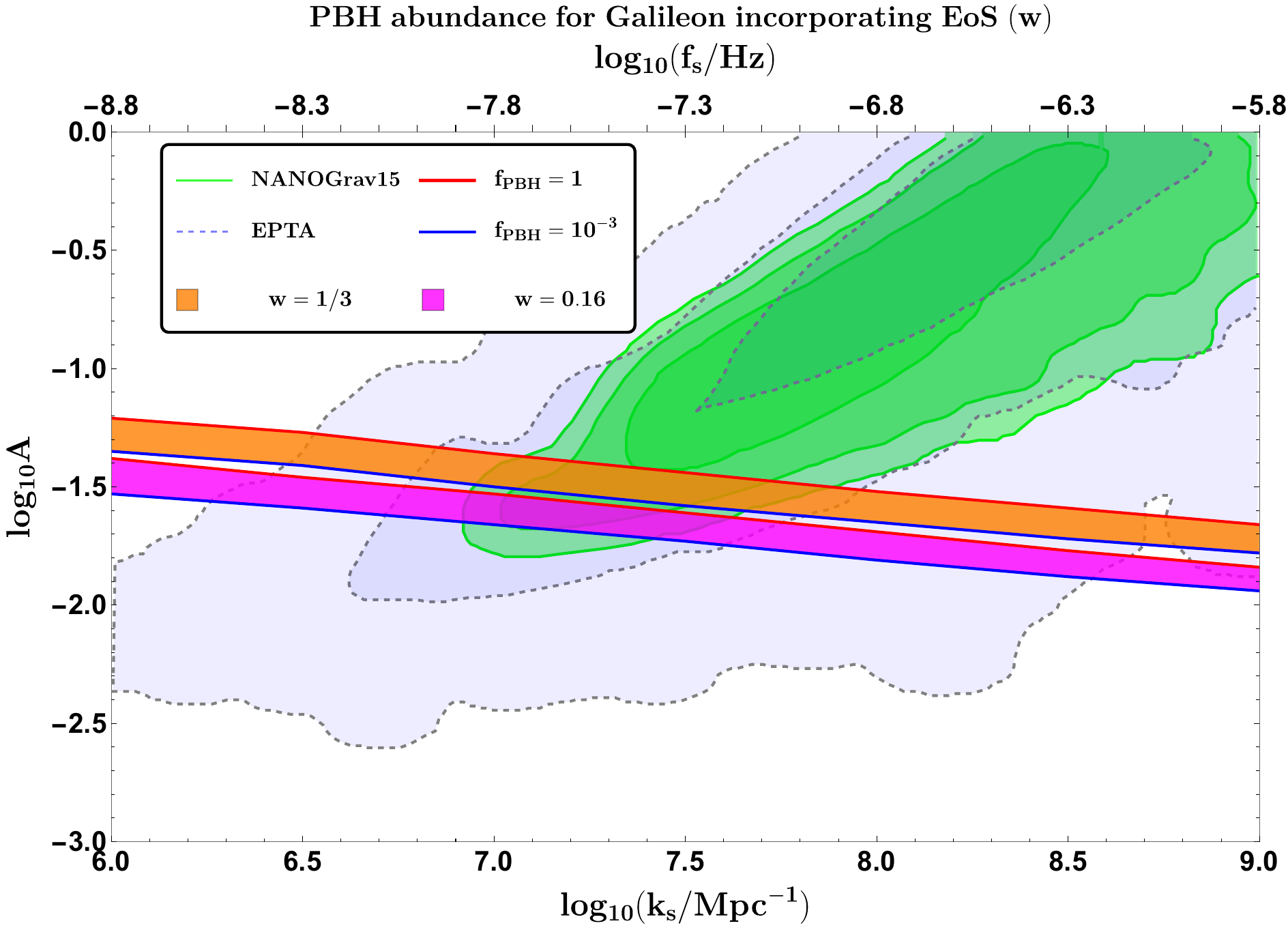}
        \label{wgalfpbh}
    }
    \subfigure[]{
       \includegraphics[width=8.5cm,height=7.5cm]{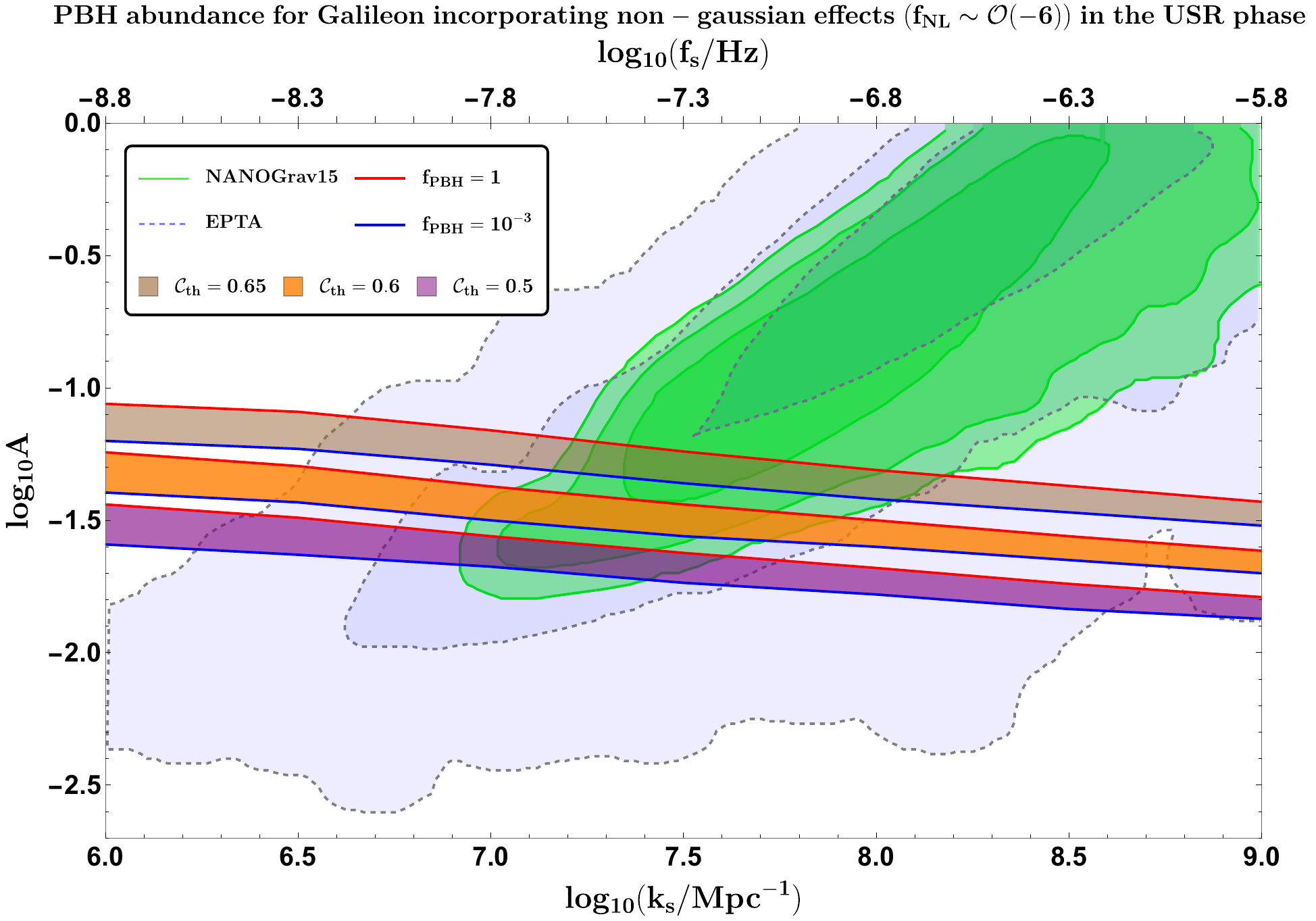}
        \label{overprod}
    }
    	\caption[Optional caption for list of figures]{Comparing the use of the EoS parameter with non-Gaussianity in order to address PBH overproduction. \textit{Left panel}: For EoS parameters $w=1/3$ and $w=0.16$, respectively, the orange and magenta coloured bands reflect locations of large PBH abundance, $f_{\rm PBH} \in (10^{-3},1)$. In the right panel: To obtain $f_{\rm PBH} \in (10^{-3},1)$, non-gaussian effects created in the USR are combined with different values of the compaction function threshold, ${\cal C}_{\rm th} \in \{0.65,0.6,0.5\}$. For $f_{\rm PBH}=1$ and $f_{\rm PBH}=10^{-3}$, respectively, the red and blue lines are used. The NANOGrav15 and EPTA signal sensitivity curves are shown by the filled contours with light blue and green backgrounds, respectively, as reported in \cite{Franciolini:2023pbf}. 
} 
    	\label{compare1}
    \end{figure*}
This offers an attractive platform for investigation, allowing us to travel through potential PBH formation scenarios by including different constant EoS factors. In this study, we have studied scenarios when the EoS parameter $w$ has discrete values, that is, $w\in (-5/9, 1/3)$. This is because the perturbation theory operates inside the linear regime. Additionally, we concentrate our considerations on the cases in which we maintain $c_{s}^{2}=1$ for the propagation speed in eqn.(\ref{transfer}). The selection for $c_{s}$ was explicated before in detail. Since the critical collapse hypothesis provides a well-known mechanism for PBH creation, the case of an RD period has been selected till now. Nevertheless, we take use of the freedom that comes with not knowing about the EoS in the very early Universe to study the potential impacts of the huge fluctuations that enter such a period on the induced GWs and PBH production that follows. We may obtain a more comprehensive understanding of the cosmic backdrop, which is a potential site for several cosmological processes, by comparing the outcomes of our research with the latest data.  We can strengthen the argument for $w=1/3$ in comparison to the impacts of other values of $w$ by our work with Galileon. This approach has obvious implications for the mass fraction of the near solar-mass PBH and the SIGW spectrum.

We offer the figure (\ref{compare1}) to conduct a comparison study between the two resolution approaches that were previously described. The study is presented in the left panel, where the amplitude of the total scalar power spectrum associated with a significant abundance is analysed by taking into account various EoS values. The EoS parameter $w$ is found to reside within $w \sim (0.16,1/3)$, allowing us to attain $f_{\rm PBH} \in (10^{-3},1)$. When faced with NANOGrav15, the overproduction avoiding situations fall within $1\sigma-2\sigma$. We do not find adequate abundance for PBH masses corresponding to the scales of the NANOGrav15 signal for $w$ values below the stated range. In the right panel, we employ the findings from \cite{Choudhury:2023fwk} as a baseline against the current analysis. In order to estimate the PBH abundance for the findings on the right, we take into account the impacts of non-Gaussianity and apply the compaction function technique. There, the evaluation of $f_{\rm PBH}$ depends on two important factors: the compaction threshold and the non-Gaussianity amplitude. The scalar power spectrum amplitude exhibits a high degree of sensitivity to threshold variations, with the optimal agreement with the NANOGrav15 data found at the extreme value of ${\cal C}_{\rm th}=0.65$. By employing both techniques for the resolution, we discover that, although substantial negative non-gaussian effects might offer a more insightful interpretation of the observed signal while also preventing overproduction, areas that avoid overproduction are not eliminated beyond $2\sigma$.

We wish to draw attention to the significant correlation that exists between the compaction threshold ${\cal C}_{\rm th}$ and the EoS $w$. Remember the relationship between the parameter $w$ and the density contrast threshold $\delta_{\rm th}$ from the relation, $\delta_{\rm th} = 3(1+w)/(5+w)$. The analysis we do with various values of $w$ takes the density contrast values into account as well. For instance, $w=1/3$ and $\delta_{\rm th}=2/3$ are equivalent, as are $w=0.25$ and $\delta_{\rm th}=0.652$, $w=0.16$ and $\delta_{\rm th}=0.635$, and so forth. However, we need alternative values of the compaction threshold, ${\cal C}_{\rm th}$, when working with non-linearities and the compaction function formalism. For example, ${\cal C}_{\rm th} = \{0.65,\;0.6,\;0.5\}$, as selected in the right panel of fig.(\ref{compare1}). Therefore, we are essentially dealing with changes in two distinct threshold variables: (a) the density contrast threshold when considering the super-Hubble scales linear regime approximation, see eqn.(\ref{deltalinear}), where non-Gaussianities are absent; and (b) the compaction threshold when taking into account the non-Gaussianities that are present in the super-Hubble and that come from this in the comoving curvature perturbation along with the non-Gaussianities, $f_{\rm NL}$. Moreover, it is important to note that we can use the same interval found for the density contrast under linear approximations with the compaction function because the equality between the compaction threshold and the density contrast threshold can only be established once the PBH forming modes re-enter the Horizon. This leads to the conclusion that the compaction threshold ${\cal C}_{\rm th}$ and the EoS parameter $w$ have a one-to-one connection.

We underline once more that our analysis is only valid when considering the density contrast within the linear domain; for more information, see eqn.(\ref{deltalinear}). According to figs.(\ref{compare1})(left panel) $\&$ (\ref{wGalNANO},\ref{wGalEPTA}), the Press-Schechter formalism modified with the EoS $w$ predicts $w\simeq 1/3$ as the best choice to handle the overproduction issue under this assumption. The current analysis using the EoS is unsatisfactory because it does not include non-linear statistics of the density contrast and does not incorporate non-Gaussianities, $f_{\rm NL}$, into the examination of the PBH mass fraction. Due to slow-roll violation, curvature perturbations during PBH generation typically do not obey Gaussian statistics. As a result, a local perturbative expansion of the following form is typically selected to account for this:
\bea
\zeta({\bf x}) = \zeta_{G}({\bf x}) + \frac{3}{5}f_{\rm NL}(\zeta_{G}^{2}({\bf x}) - \langle\zeta_{G}^{2}({\bf x})\rangle) + \cdots.
\eea 
Such a non-Gaussianity factor can have a significant effect on PBH creation.
To wrap off our conversation, we want to draw attention to the specific non-linear relationship that was previously discussed.
\bea \label{NLdensity}
\delta({\bf x},t) &=& -\frac{4}{9}\left(\frac{1}{aH}\right)^{2}e^{-2\zeta({\bf x})}\bigg[\nabla^{2}\zeta({\bf x}) + \frac{1}{2}\partial_{i}\zeta({\bf x})\partial_{i}\zeta({\bf x})\bigg].
\eea
It compels us to take into account the non-Gaussian statistics for $\delta({\bf x},t)$. Combined with the local non-Gaussianity in the curvature perturbation, $f_{\rm NL}$, this might modify the PBH mass fraction analysis considerably \cite{Choudhury:2023fwk, Young:2019yug, DeLuca:2019qsy}.

\subsection{Numerical outcomes and related discussions}           
\begin{figure*}[htb!]
    	\centering
    {
        \includegraphics[width=15cm,height=9.5cm]{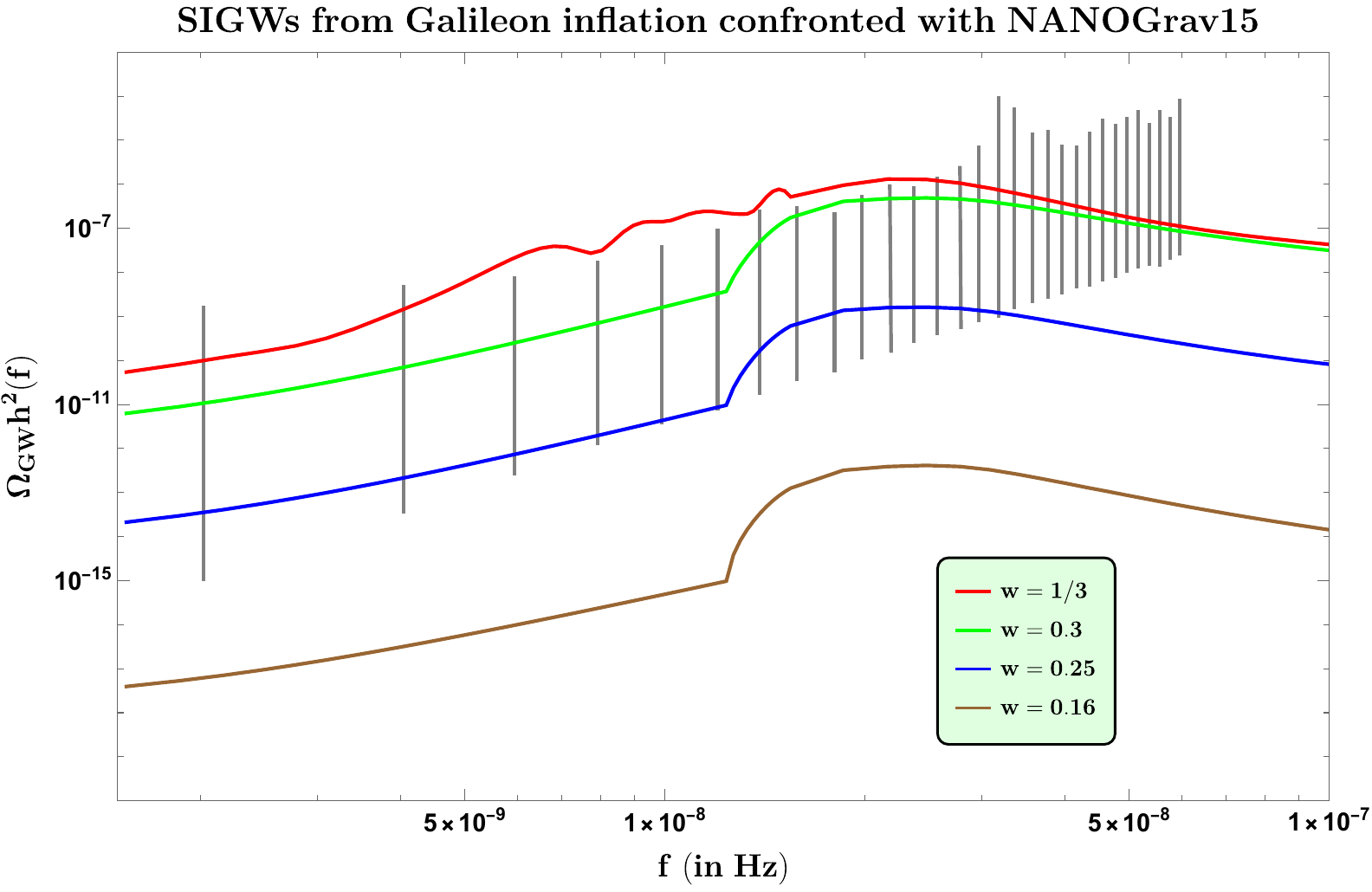}
        \label{wGalnano}
    }
    \caption[Optional caption for list of figures]{The NANOGrav15 signal is encountered in the spectrum of GW energy density as a function of frequency for different EoS $w$ values. Plotting of the spectra is done for the following values: $w=1/3$ (red), $w=0.3$ (green), $w=0.25$ (blue), and $w=0.16$ (brown). It is used to get the transfer function from eqn.(\ref{rdtransfer}) for $w=1/3$, and eqn.(\ref{transfer}) for general $w$.
    }
\label{wGalNANO}
    \end{figure*}

\begin{figure*}[htb!]
    	\centering
    {
        \includegraphics[width=15cm,height=9.5cm]{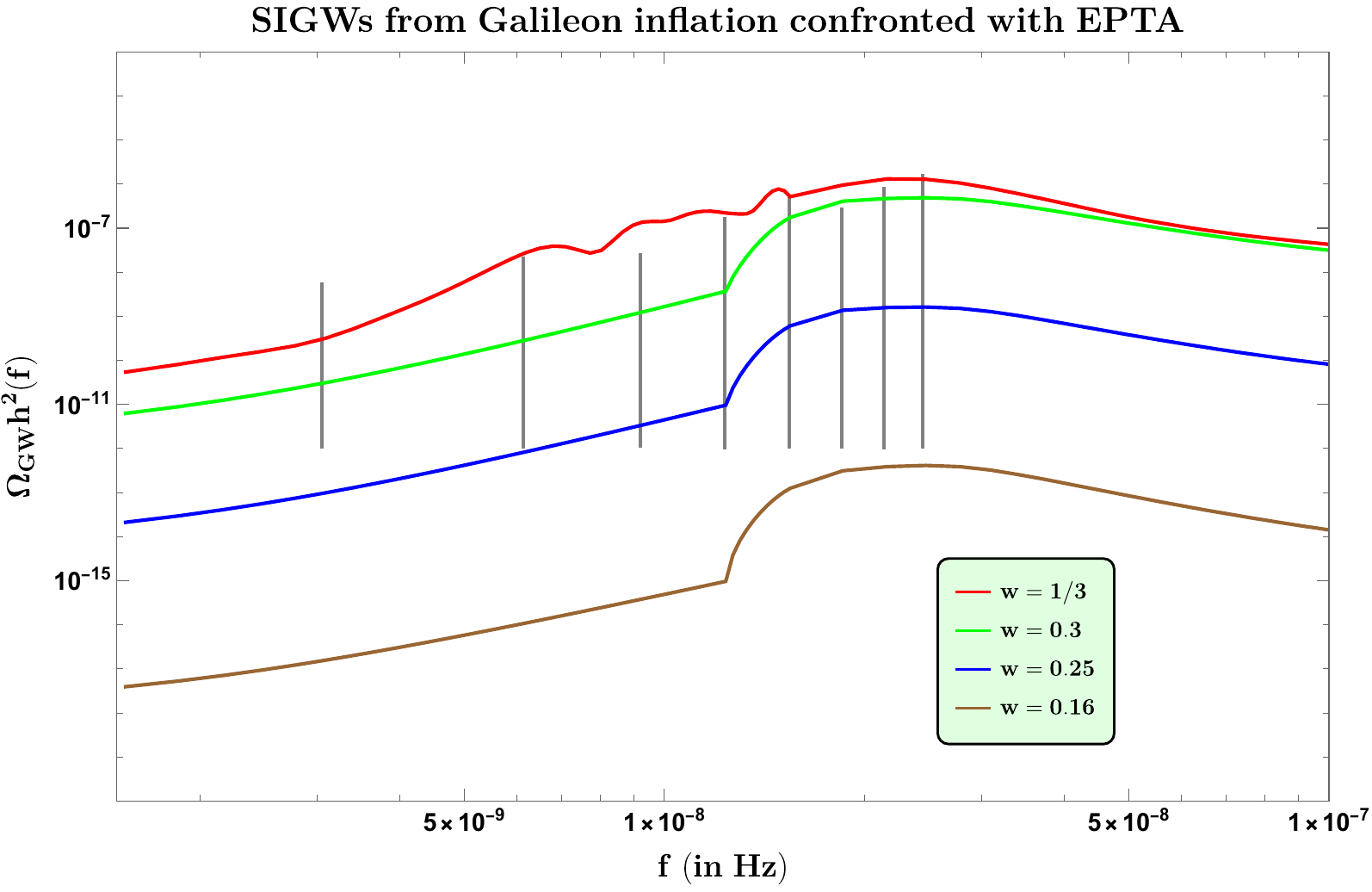}
        \label{wGalepta}
    }
    \caption[Optional caption for list of figures]{The GW energy density spectrum for different EoS $w$ values when faced with the EPTA signal is represented as a function of frequency. Values of $w=1/3$ (red), $w=0.3$ (green), $w=0.25$ (blue), and $w=0.16$ (brown) are displayed on the spectra, accordingly. Although the transfer function from eqn.(\ref{transfer}) is utilised for $w=1/3$, eqn.(\ref{transfer}) is used for $w$ in general. 
    }
\label{wGalEPTA}
    \end{figure*}

\begin{figure*}[htb!]
    	\centering
    {
        \includegraphics[width=15cm,height=9.5cm]{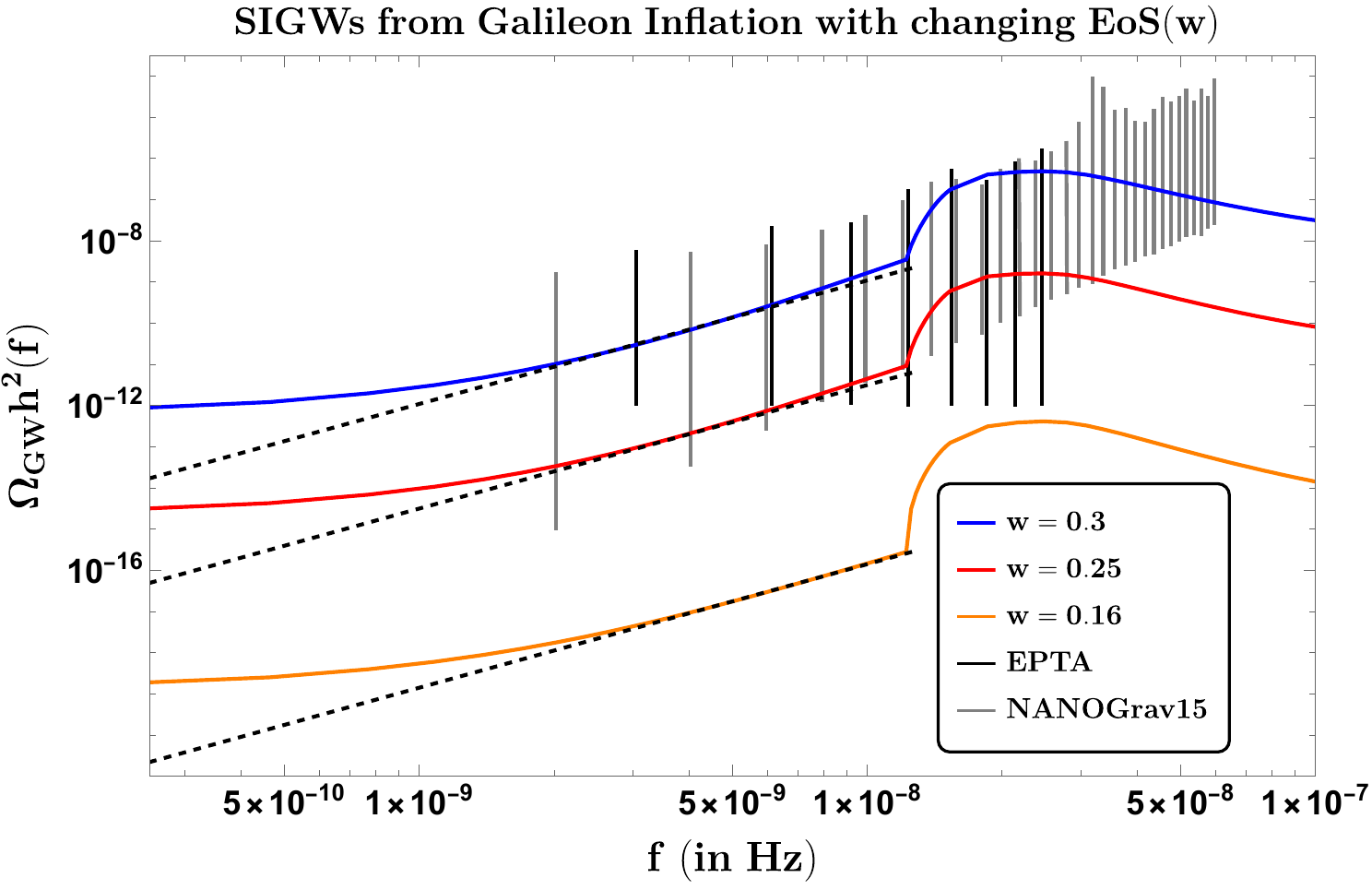}
        \label{IRsigw}
    }
    \caption[Optional caption for list of figures]{ The GW energy density spectrum for different EoS $w$ values as a function of frequency and in comparison to the NANOGrav15 and EPTA signal. The IR tail of the relation $\Omega_{\rm GW}h{^2}\propto k^{3}$ is described by the dashed-black line. Plotting of the spectra shows values $w=0.3$ (blue), $w=0.25$ (red), and $w=0.16$ (orange), with grey solid lines representing the PTA (NANOGrav5 and EPTA) signals present in the background, respectively. 
    }
\label{GalIR}
    \end{figure*}

\begin{figure*}[htb!]
    	\centering
    \subfigure[]{
      	\includegraphics[width=8.6cm,height=7.5cm]{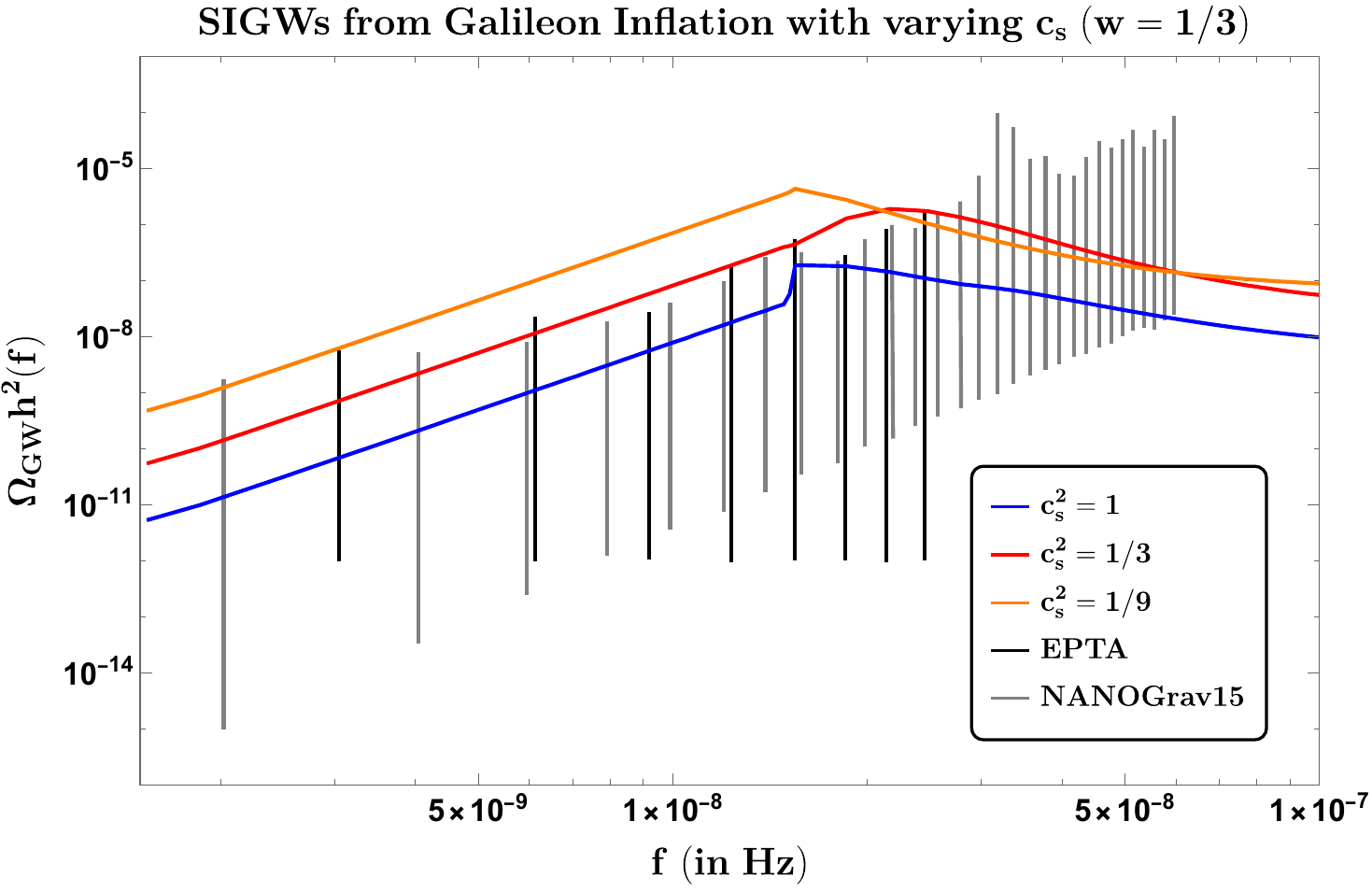}
        \label{SIGWcs1}
    }
    \subfigure[]{
       \includegraphics[width=8.6cm,height=7.5cm]{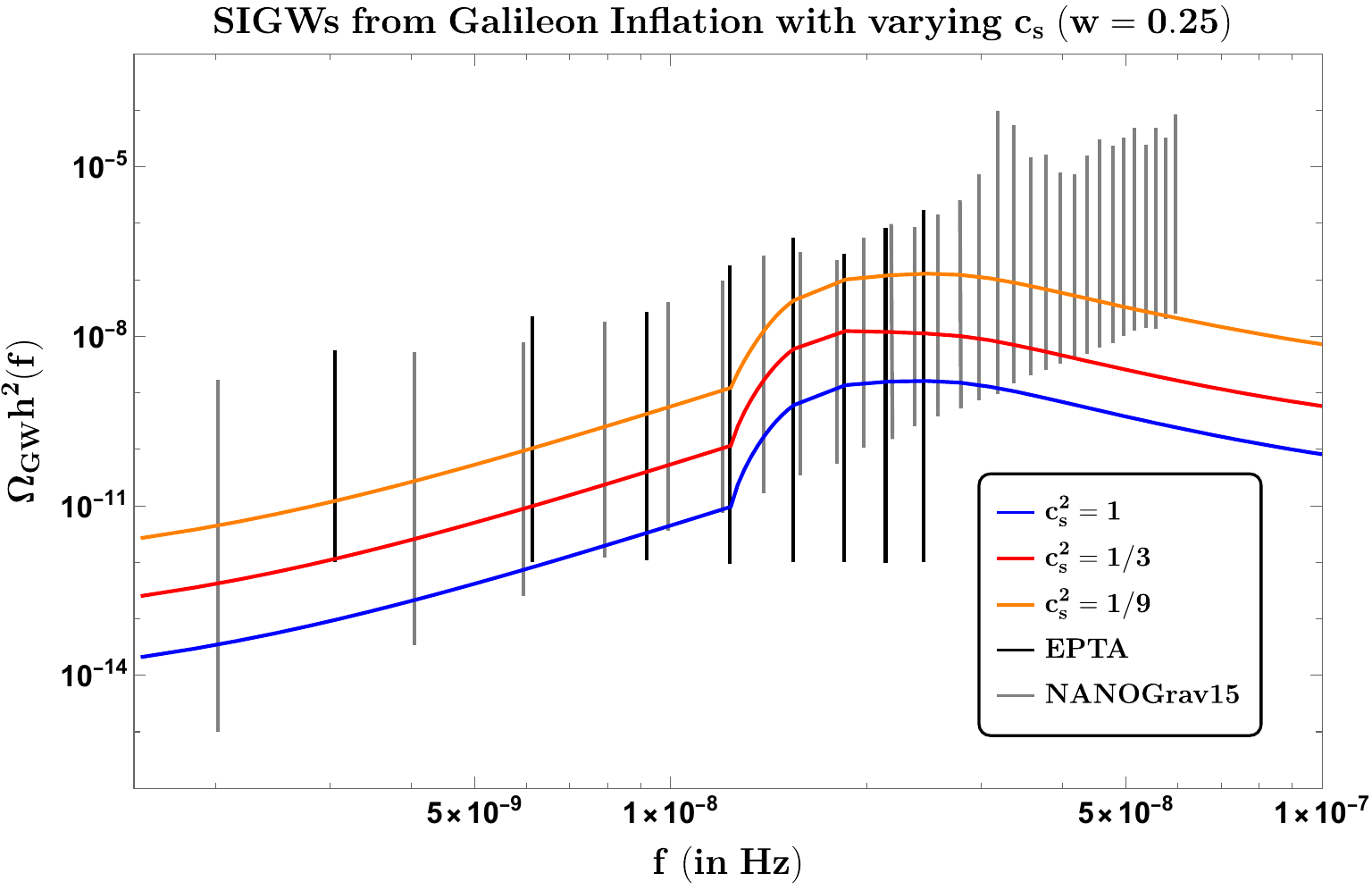}
        \label{SIGWcs2}
    }
    	\caption[Optional caption for list of figures]{SIGW spectra plotted once for EoS $w=1/3$ on the left and $w=0.25$ on the right, with different values of the propagation speed $c_{s}$. For both EoS scenarios and with the PTA (NANOGrav15 and EPTA) signals present in the background as grey and black solid lines, respectively, the values of $c_{s}$ are selected with $c_{s}^2=1$ (blue), $c_{s}^2=1/3$ (red), and $c_{s}^2=1/9$ (orange). We see that the high-frequency tail for $w=1/3$ varies with various values of $c_{s}$, but $w=0.25$ does not much vary, with the exception of a rise in the amplitude of the SIGW spectrum for both as $c_{s}$ continues to decrease. } 
    	\label{SIGWcs}
    \end{figure*}

\begin{figure*}[htb!]
    	\centering
    {
        \includegraphics[width=18cm,height=12cm]{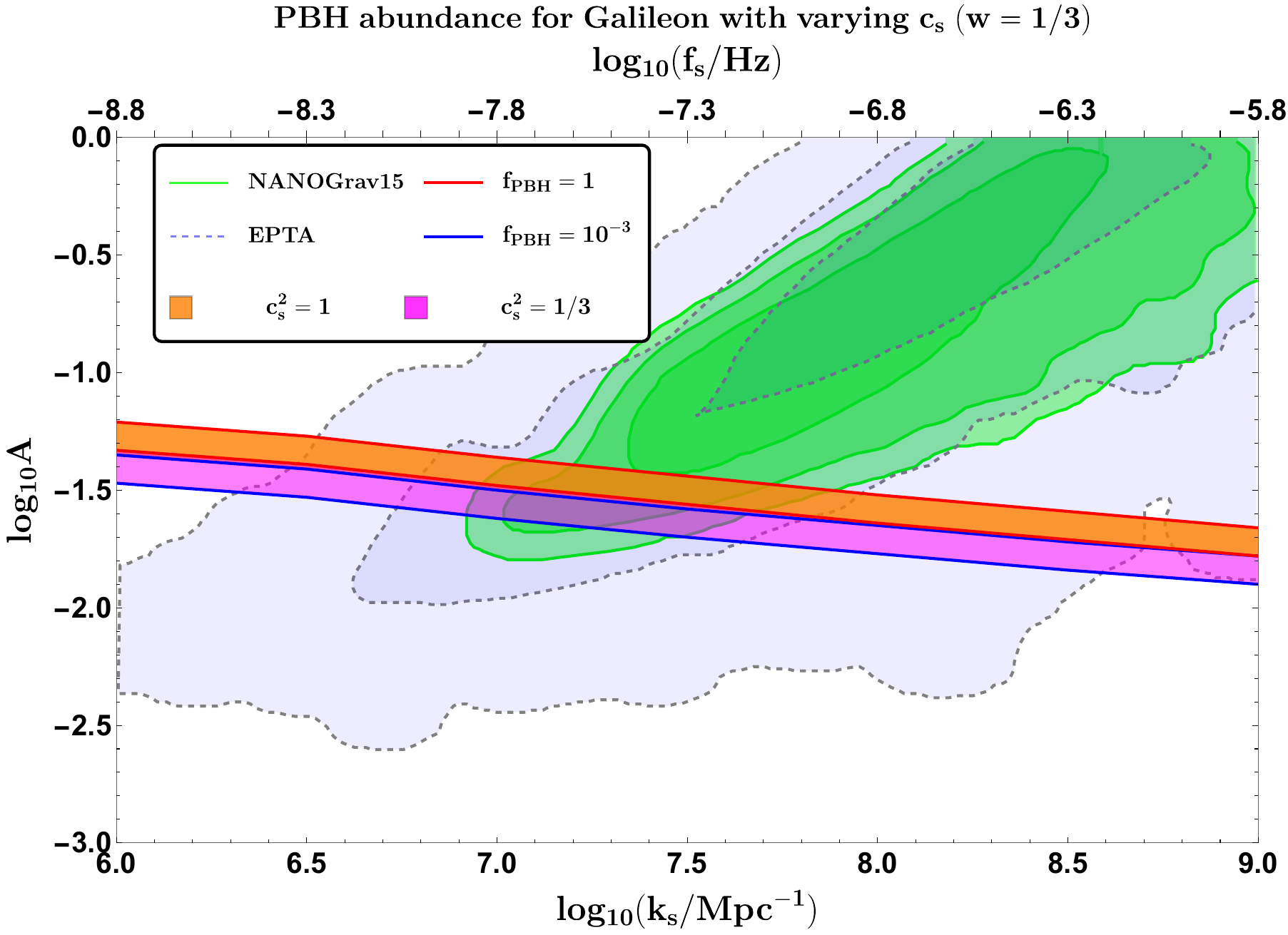}
        \label{pbhabundcs}
    } 
    \caption[Optional caption for list of figures]{Total scalar power spectrum $A$ amplitude as a function of the transition wavenumber $k_{s}$ when the EoS is fixed at $w=1/3$. If $c_{s}^2=1$ is selected, the orange band corresponds to the region of considerable abundance, $f_{\rm PBH}\in (10^{-3},1)$; if $c_{s}^2=1/3$ is selected, the magenta band corresponds to the region of sizeable abundance. While $c_{s}^2$ is reduced, the amplitude $A$ rises, maintaining the indicated zones of abundance inside the PTA signal's 2$\sigma$ contour. The sensitivity curves for the NANOGrav15 and EPTA signals are shown by the filled contours with light blue and green backgrounds, respectively, as reported in \cite{Franciolini:2023pbf}. }
\label{abundancecs}
    \end{figure*}

\begin{figure*}[htb!]
    	\centering
    {
        \includegraphics[width=18cm,height=12cm]{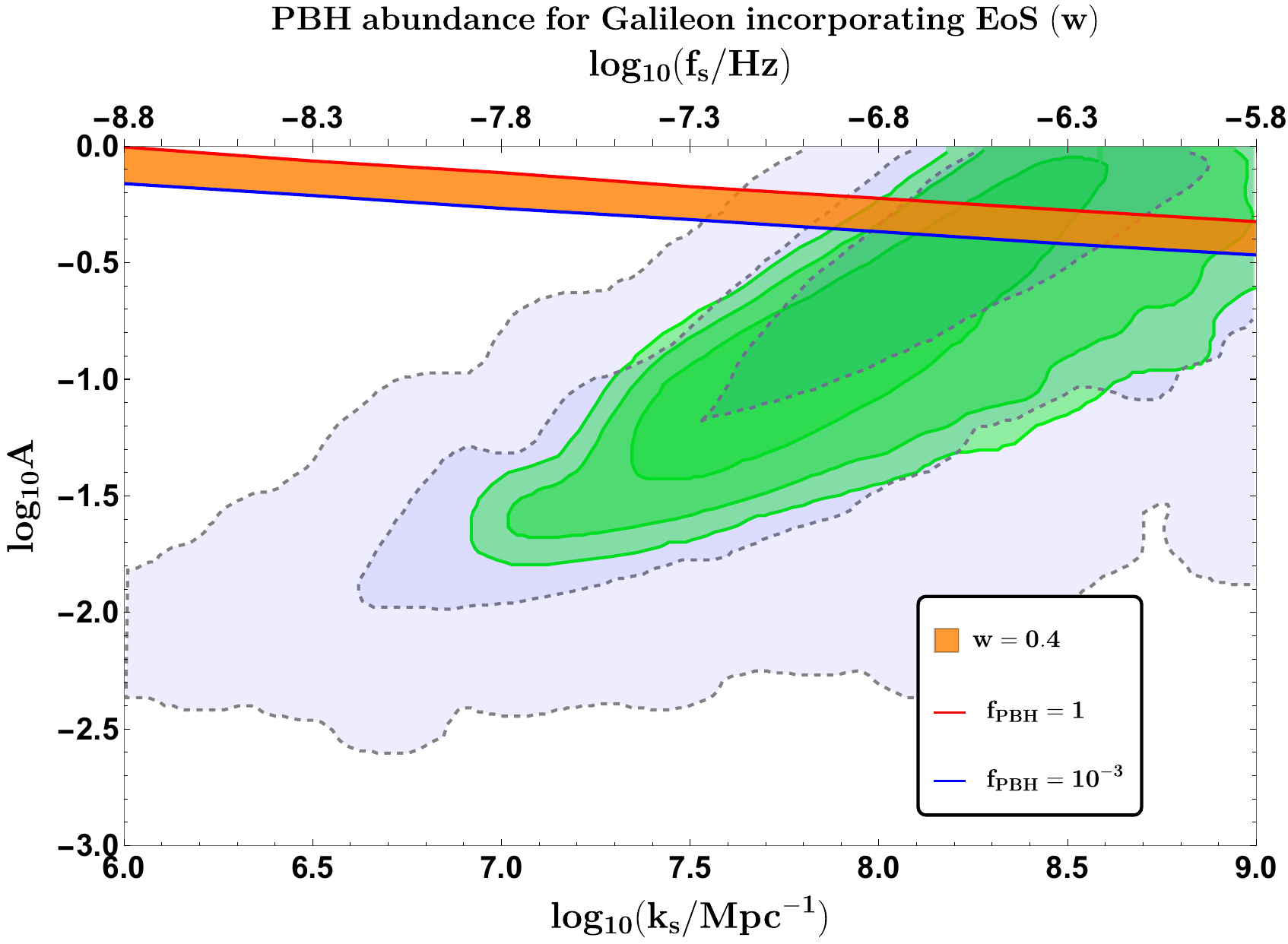}
        \label{wGal2}
    } 
    \caption[Optional caption for list of figures]{ Maximum value of the total scalar power spectrum $A$ in relation to the USR's transition wavenumber $k_{s}$. This uses a fixed EoS setting of $w=0.4$. The sensitivity curves for the NANOGrav15 and EPTA signals are shown by the filled contours with light blue and green backgrounds, respectively, as reported in \cite{Franciolini:2023pbf}.  }
\label{wGal2}
    \end{figure*}

In this part, we investigate a variable baseline EoS $w$ and give the findings for the energy density spectrum of GWs produced within the current framework of Galileon inflation. For this procedure, we apply eqn.(\ref{transfer}), and we analyse our findings. 

Plots for the SIGW spectra in the case of several constant EoS values are shown in figs. (\ref{wGalNANO},\ref{wGalEPTA}). NANOGrav15 and EPTA signals, which were recently noticed, are shown by the grey vertical bars in the backdrop. When $w=1/3$ is set in the kernel, the resulting spectrum for the RD era is shown in red. For the observational signal with maximum spectrum amplitude, the RD era is still consistent. The GW spectrum amplitude rapidly decreases as we move down in $w$ values. The area above the blue spectrum is still reachable from the NANOGrav15 signal up to $w \gtrsim 0.24$. When $w=0.16$ is used, it is evident that moving below this number results in a spectrum that is even lower than the signal amplitudes, with its spectrum shown in brown. We have already noted in fig.(\ref{wgalfpbh}) that, at $w=0.16$, the magenta band that prevents overproduction is still separated from the NANOGrav15 posteriors by around $2\sigma$. Based on the figure above, we can thus deduce that a lower value of $w \sim 0.24$ will not accord with the PTA signal; hence, the only region that might be useful to comprehend the cosmic background in the very early Universe and the formation of SIGWs is the region between $w \in (1/3,0.24)$.

An other important feature of the SIGWs produced in a universe that is filled with an arbitrary EoS $w$ fluid is how their spectrum behaves in the infrared (IR) band. We illustrate this particular behaviour in fig. (\ref{GalIR}) over a range of EoS values and with $c_{s}^{2}=1$ fixed. The GW spectrum's approximate $k^3$ scaling is seen by the dashed-black line there. A global infrared scaling ratio $\Omega_{\rm GW}\propto k^{3}$ was obtained for the tensor modes re-entering the universe in a radiation-dominated epoch in \cite{Cai:2019cdl}. Our goal is to illustrate how the IR scaling behaviour of the SIGW spectrum is affected by changes in EoS values and Galileon inflation, and to compare this behaviour to the scaling relation that was determined during the radiation period. In the IR scenario, an analytic analysis of the GW spectrum with wavenumbers $k$ considerably smaller than the peak scale is not conceivable due to the extraordinarily complicated structure of the scalar power spectrum in the current Galileon theory. The scaling along the wavenumbers smaller than the peak location of the GW spectrum is around $k^{3}$ till $f\sim 2\times 10^{-9}{\rm Hz}$, as we can see in fig. (\ref{GalIR}). The GW spectrum does not decrease and instead becomes flatter in nature as we go down at frequency $f\lesssim 10^{-9}{\rm Hz}$ (nano-Hz). It is worth noting that the scaling behaviour remains nearly constant for all the $w$ values examined in this investigation, except when we go below the nano-Hz frequency range.The spectra exhibit near-parity even at lower frequencies, with the amplitude being the sole variation. This suggests that the GWs from the various EoS diminish in intensity at the same rate in the infrared. 

For a constant value of the EoS $w$, we examine the effects that an arbitrary propagation speed $c_{s}$ may have on the produced SIGWs. Changes in the produced SIGWs for $w=1/3$ on the left and $w=0.25$ on the right are highlighted in the figure fig.(\ref{SIGWcs}), where $c_{s}^{2}\in \{1,1/3,1/9\}$ is taken into account for both cases. A comparable alteration with varying $c_{s}$ is the SIGW spectrum changing to bigger amplitudes when $c_{s}$ lowers in comparison to the situation of $c_{s}^{2}=1$. When $c_{s}^2=1/3$ is used, the transfer function in eqn.(\ref{rdtransfer}) is taken into account, and the result matches the radiation-dominated spectrum for $w=1/3$. The curve in red represents the SIGWs created when this occurs. With only amplitude differences, the tails in the low-frequency area of the signal for the $w=1/3$ instance appear to be comparable to one another, which is an intriguing observation. Significant alterations only transpire beyond the frequency that corresponds to the transition wavenumber $k_{s}$. Subsequently, we see the variations in the high-frequency tail behaviour caused by altering $c_{s}$; the tail for $c_{s}^2=1/9$ seeks to cross over the tail of $c_{s}^2=1/3$. For this reason, arbitrary $c_{s}$ and a background EoS of $w=1/3 $ might have an impact on the SIGW spectrum that can be significant for observations at the higher frequencies tails. Differentiating $c_{s}$ will mostly affect the spectrum amplitude rather than the tail areas around the PTA signal in the event of a distinct $w=0.25$ situation. All over the frequency range, the overall effect on the amplitude appears to be identical.

When the EoS is held constant at $w=1/3$ in fig.(\ref{abundancecs}), the effect of having variable $c_{s}$ is also investigated in the context of overproduction of PBHs. Orange and magenta bands are used to emphasise the $f_{\rm PBH}\in (10^{-3},1)$ region of substantial abundance. Red and blue lines indicate the top and lower bounds of this $f_{\rm PBH}$ interval. The previously examined scenario, in which the propagation speed $c_{s}^2=1$ is fixed, is shown by the orange bands. When we concentrate on having $c_{s}^2=w=1/3$, the case gets more intriguing. In this scenario, the criterion for applying the threshold for analysing PBH abundance from eqn.(\ref{deltath}) does not apply. To supply enough abundance for the situations of interest, we need to select the density contrast threshold value sensibly. The linear regime relation in eqn.(\ref{deltalinear}) shows that the threshold analysis is sensitive to the form of the primordial spectrum, which directly affects the density contrast. According to fig. (\ref{GL}), in the current context, the USR phase is responsible for the peak in the overall primordial power spectrum, with an amplitude of $A\sim {\cal O}(10^{-2})$. We identify the most important contributions to PBH abundance from the behaviour of the tiny scales in the USR phase, as the PBH mass fraction is exponentially sensitive to the amplitude of the primordial power spectrum. Since the shape has a tiny, non-vanishing width during the interval but maintains a sharp peak in the USR, we approximate the spectrum behaviour in the neighbourhood of a log-normal power spectrum peaked in the limit of a finite but small width. Because of the aforementioned justifications, we decide to employ the value of $\delta_{\rm th}=0.59$ that was discovered for a power spectrum with these characteristics under the circumstance that $c_{s}^2=w=1/3$ \cite{Musco:2020jjb}, providing us with a reliable approximation to carry out the abundance analysis using these parameters. The scalar power spectrum amplitude is seen to be preferred to have a greater value when $c_{s}^{2}=1$ is bigger than when $c_{s}^2=1/3$ with the new matching threshold value.

We clarify a crucial decision in our analysis of employing the NANOGrav and EPTA posteriors in light of the arguments up to this point; background contours are shown in fig. (\ref{compare1}). A few presumptions based on our power spectrum and the consequent GW spectrum drove the decision. Our power spectrum in the USR strongly resembles a log-normal power spectrum, providing a sharp peak with a finite breadth, as discussed above based on fig. (\ref{GL}). We estimate aspects of our power spectrum for such circumstances as being comparable to a log-normal spectrum, since the USR area is the region of spectrum contributing most substantially to the PBH abundance and the peak amplitude in the GW spectrum. By using the same, we have also demonstrated in fig. (\ref{GalIR}) that, up until we stay inside the NANOGrav signal frequency range, the GW spectrum for different EoS $(w)$ values exhibits a causality tail, $\Omega_{\rm GW}\propto k^{3}$, in the IR. A distinctive outcome of using a log-normal spectrum in the IR limit is also a tail feature like this one. We may benefit from the posteriors generated for a log-normal power spectrum thanks to the aforementioned approximations. Thus, we think that developing the posteriors in our study does not need an explicit probability analysis. We want to expand on this analysis in the future by delving further into the PBH overproduction problem and doing comprehensive numerical modeling. 

An emphasis on EoS bigger than the RD era $w=1/3$ is given in Fig. (\ref{wGal2}). To illustrate the impact of rising EoS beyond the RD period on the amplitude of the entire scalar power spectrum, we choose $w=0.4$. We observe that the amplitude may be significantly altered by slightly exceeding the bound on $w$. For the NANOGrav15, the total amplitude approaches violating the assumptions of perturbation theory inside the wavenumber interval. For $f_{\rm PBH} \in (10^{-3},1)$, the criteria are shown by the orange colour area. We infer from this plot that scenarios with damage perturbativity approximations of $w > 1/3$ are not appropriate to study other related occurrences, such as SIGW and PBH creation.  
Prior to wrapping up this part, we discuss the key components of fig.(\ref{compare1}). The inclusion of the non-gaussian character of density fluctuations in the analysis offers a comprehensive method for determining the mass fraction of PBHs. After accounting for $f_{\rm NL} \sim -6$, the threshold statistics on compaction function are used in the right panel of fig. (\ref{overprod}) to accurately estimate PBH abundance. This method yields a relatively close agreement with the NANOGrav15 signal compared to what is observed when using the $w-$Press-Schechter used in fig. (\ref{wgalfpbh}). Given the statistical aspects, the $w-$Press-Schechter formalism remains a useful tool for assessing the PBH mass fraction and, consequently, offers empirically supported insights into the nature of $w$ and its impact on processes.

   \section{Stochastic Single Field Inflation in the light of large fluctuations}
\label{s7}

   One of the most prominent models for the early universe is cosmological inflation, which offers a mechanism for the seeding of primordial quantum fluctuations into the large-scale structures of today. These fluctuations, usually related to a scalar field involved in inflation, go through a transition from being originally in the quantum realm to entering the large-scale, classical regime. Previously proposed \cite{Starobinsky:1986fx}, the stochastic inflationary paradigm is used to investigate the dynamics of large-scale fluctuations impacted by noise terms originating from the quantum-to-classical transition of the tiny wavelength modes of primordial fluctuations. Refer to the following references for further information on soft de Sitter Effective Theory (SdSET), which reinforces the fundamentals of the stochastic inflationary paradigm: \cite{Gorbenko:2019rza,Cohen:2021fzf,Cohen:2022clv,Green:2022ovz,Cohen:2021jbo,Cohen:2020php}. The creation of objects referred to as primordial black holes (PBHs) is an intriguing byproduct of the primordial perturbations in the early universe, namely those formed at tiny scales towards the end of inflation. PBH production \cite{Zeldovich:1967lct,Hawking:1974rv,Carr:1974nx,Carr:1975qj,Chapline:1975ojl,Carr:1993aq,Choudhury:2011jt,Yokoyama:1998pt,Kawasaki:1998vx,Rubin:2001yw,Khlopov:2002yi,Khlopov:2004sc,Saito:2008em,Khlopov:2008qy,Carr:2009jm,Choudhury:2011jt,Lyth:2011kj,Drees:2011yz,Drees:2011hb,Ezquiaga:2017fvi,Kannike:2017bxn,Hertzberg:2017dkh,Pi:2017gih,Gao:2018pvq,Dalianis:2018frf,Cicoli:2018asa,Ozsoy:2018flq,Byrnes:2018txb,Ballesteros:2018wlw,Belotsky:2018wph,Martin:2019nuw,Ezquiaga:2019ftu,Motohashi:2019rhu,Fu:2019ttf,Ashoorioon:2019xqc,Auclair:2020csm,Vennin:2020kng,Nanopoulos:2020nnh,Inomata:2021uqj,Stamou:2021qdk,Ng:2021hll,Wang:2021kbh,Kawai:2021edk,Solbi:2021rse,Ballesteros:2021fsp,Rigopoulos:2021nhv,Animali:2022otk,Frolovsky:2022ewg,Escriva:2022duf,Ozsoy:2023ryl,Ivanov:1994pa,Afshordi:2003zb,Frampton:2010sw,Carr:2016drx,Kawasaki:2016pql,Inomata:2017okj,Espinosa:2017sgp,Ballesteros:2017fsr,Sasaki:2018dmp,Ballesteros:2019hus,Dalianis:2019asr,Cheong:2019vzl,Green:2020jor,Carr:2020xqk,Ballesteros:2020qam,Carr:2020gox,Ozsoy:2020kat,Baumann:2007zm,Saito:2008jc,Saito:2009jt,Choudhury:2013woa,Sasaki:2016jop,Raidal:2017mfl,Papanikolaou:2020qtd,Ali-Haimoud:2017rtz,Di:2017ndc,Raidal:2018bbj,Cheng:2018yyr,Vaskonen:2019jpv,Drees:2019xpp,Hall:2020daa,Ballesteros:2020qam,Carr:2020gox,Ozsoy:2020kat,Ashoorioon:2020hln,Papanikolaou:2020qtd,Wu:2021zta,Kimura:2021sqz,Solbi:2021wbo,Teimoori:2021pte,Cicoli:2022sih,Ashoorioon:2022raz,Papanikolaou:2022chm,Papanikolaou:2023crz,Wang:2022nml,ZhengRuiFeng:2021zoz,Cohen:2022clv,Cicoli:2022sih,Brown:2017osf,Palma:2020ejf,Geller:2022nkr,Braglia:2022phb,Frolovsky:2023xid,Aldabergenov:2023yrk,Aoki:2022bvj,Frolovsky:2022qpg,Aldabergenov:2022rfc,Ishikawa:2021xya,Gundhi:2020kzm,Aldabergenov:2020bpt,Cai:2018dig,Cheng:2021lif,Balaji:2022rsy,Qin:2023lgo,Riotto:2023hoz,Riotto:2023gpm,Papanikolaou:2022did,Choudhury:2011jt,Choudhury:2023vuj, Choudhury:2023jlt, Choudhury:2023rks,Choudhury:2023hvf,Choudhury:2023kdb,Choudhury:2023hfm,Bhattacharya:2023ysp,Choudhury:2023fwk,Choudhury:2023fjs,Choudhury:2024one,Harada:2013epa,Harada:2017fjm,Kokubu:2018fxy,Gu:2023mmd,Saburov:2023buy,Stamou:2023vxu,Libanore:2023ovr,Friedlander:2023qmc,Chen:2023lou,Cai:2023uhc,Karam:2023haj,Iacconi:2023slv,Gehrman:2023esa,Padilla:2023lbv,Xie:2023cwi,Meng:2022low,Qiu:2022klm,Mu:2022dku,Fu:2022ypp,Davies:2023hhn,Firouzjahi:2023ahg,Firouzjahi:2023aum, Iacconi:2023ggt,Davies:2023hhn,Jackson:2023obv,Riotto:2024ayo} is associated with enormous fluctuations at lower scales that, upon re-entry into the horizon, give rise to regions of underdensities and overdensities in the universe's substance. These regions then gravitationally collapse after passing a threshold, giving rise to PBHs. The stochastic inflation formalism has been applied to a wide range of settings recently, and it has shown great potential in the study of PBH production \cite{Vennin:2024yzl,Animali:2024jiz,LISACosmologyWorkingGroup:2023njw,Animali:2022otk,Ezquiaga:2022qpw, Jackson:2022unc, Tada:2021zzj,Pattison:2021oen,Ando:2020fjm,Vennin:2020kng,Ezquiaga:2019ftu,Pattison:2019hef,Noorbala:2018zlv,Pattison:2017mbe,Grain:2017dqa,Hardwick:2017fjo}. PBHs are a compelling candidate for dark matter, and their generating fluctuations may also contribute to the generation of primordial gravitational waves, the fingerprints of which may be found \cite{NANOGrav:2023gor, NANOGrav:2023hde, NANOGrav:2023ctt, NANOGrav:2023hvm, NANOGrav:2023hfp, NANOGrav:2023tcn, NANOGrav:2023pdq, NANOGrav:2023icp,EPTA:2023fyk, EPTA:2023sfo, EPTA:2023akd, EPTA:2023gyr, EPTA:2023xxk, EPTA:2023xiy,Reardon:2023gzh, Reardon:2023zen, Zic:2023gta,Xu:2023wog,LISACosmologyWorkingGroup:2023njw, Inomata:2023zup,Choudhury:2023hfm,Bhattacharya:2023ysp,Choudhury:2023fwk,Choudhury:2023fjs,Franciolini:2023pbf,Inomata:2023zup,Wang:2023ost,Balaji:2023ehk,Gorji:2023sil,Choudhury:2023kam,Yi:2023mbm,Cai:2023dls,Cai:2023uhc,Huang:2023chx,Huang:2023mwy,Frosina:2023nxu,Zhu:2023faa,Cheung:2023ihl,Gouttenoire:2023bqy,Salvio:2023ynn,Yi:2023npi,Di:2017ndc,Ballesteros:2018wlw,Sasaki:2018dmp,Fu:2019ttf,Ballesteros:2020qam,Papanikolaou:2020qtd,Inomata:2021uqj,Wang:2022nml,Ashoorioon:2022raz,Frolovsky:2022ewg,Domenech:2021ztg,Yuan:2021qgz,Chen:2019xse,Cang:2022jyc,Heydari:2023rmq,Bhaumik:2023wmw,Chen:2024gqn}. As a result, interest in PBHs has grown quickly. One of the proposed processes for PBH development is that the scalar-field fluctuations involved in PBH production are greatly enhanced by an almost flat zone that emerges at the tiny scales of the inflationary potential. This type of inflationary regime is also known as an ultra-slow roll (USR) phase, when, following horizon escape, quantum diffusion effects become equally significant and contribute to the overall dynamics of the large-scale classical perturbations. Stochastic inflation has been used to examine a wide range of models with this region and its effects on PBH production \cite{Vennin:2024yzl,Animali:2024jiz,LISACosmologyWorkingGroup:2023njw,Animali:2022otk,Ezquiaga:2022qpw, Jackson:2022unc, Tada:2021zzj,Pattison:2021oen,Ando:2020fjm,Vennin:2020kng,Ezquiaga:2019ftu,Pattison:2019hef,Noorbala:2018zlv,Pattison:2017mbe,Grain:2017dqa,Hardwick:2017fjo,Mishra:2023lhe}. In this work, we construct a soft de Sitter Effective Field Theory (EFT) formulation of stochastic single-field inflation, without explicitly introducing any scalar field into the framework, with the goal of model-independently generalising this image. When the gauge-invariant variable, or the comoving curvature perturbation $\zeta$ in the EFT of inflation, is separated into its long and short wavelength halves, the low-energy component remains, and this is what is meant to be understood by the term "soft." Following horizon crossing, the short-wavelength components would eventually join the long-wavelength dynamics after experiencing coarse-graining from stochastic influences. A comprehensive technical description of the UV full theory may be built in a model-independent way thanks to the EFT of inflation framework \cite{Weinberg:2008hq,Cheung:2007st,Choudhury:2017glj,Delacretaz:2016nhw,Naskar:2017ekm}. This allows us to explore the dynamics of the metric perturbations around a quasi de Sitter spacetime without being concerned with the scalar field being driven by a particular model potential. The relevant EFT action is constructed around symmetries that appear in the higher-dimensional operators that make up the action's structure. We follow the well-known St$\ddot{u}$ckelberg method when working with the unitary gauge. A new scalar degree of freedom, called the Goldstone mode, arises when the gauge invariance in the EFT action is restored. This mode transforms non-linearly under the broken-time diffeomorphism symmetry. 
In contrast, stochastic inflation addresses the dynamics of large-scale metric perturbations that are influenced by classical noises that arise from stochastic processes in the vicinity of horizon crossing. We demonstrate how to include the fundamental ideas of EFT into the process of obtaining the evolution equation for the same large-scale comoving curvature perturbations, denoted as $\zeta$. These perturbations arise from the conversion of the initial quantum fluctuations following the end of inflation. When there are random effects close to the horizon-crossing moment, the Ultra-Violet (UV) component of the fluctuations experiences what is known as coarse-graining. Furthermore not precisely defined, this instant varies according to the coarse-graining window function selected. A suitable instant in the time coordinate indicates the horizon crossing in the scenario when a Heaviside Theta function is used as the window. This window function results in noise terms that are called \textit{white noise}, which leads to a Markovian description of the system. Conversely, a window function defined by the use of a certain profile causes the sounds to be categorised as \textit{coloured noise}, which in turn defines a non-Markovian system. The horizon crossing continues into the super-Hubble scales, and the stochastic Langevin equation controls the evolution of the coarse-grained curvature perturbations, which are also known as the Infra-Red (IR) component. When looking into PBH formation, these equations become significant.

Many techniques, such as the Press-Schechter formalism, peak theory, or the recently developed compaction function approach, \cite{Choudhury:2023jlt,Bhattacharya:2023ysp,Choudhury:2023fwk,Choudhury:2023fjs,Choudhury:2024one,Franciolini:2023wun,DeLuca:2023tun,DeLuca:2022rfz,Musco:2021sva,Musco:2020jjb,Kalaja:2019uju,Kehagias:2019eil,Young:2019yug,Musco:2018rwt,Bardeen:1985tr,Green:2004wb,Ianniccari:2024bkh,Franciolini:2023pbf,Ferrante:2022mui,Ferrante:2023bgz}, have their favoured regimes of application under particular situations. The process for PBH production has been explored relatively thoroughly. According to the Press-Schechter method, the initial $\zeta$ profile is assumed to be Gaussian in nature. PBH formation happens when large curvature perturbations surpass a threshold, $\zeta_{\rm th}\sim {\cal O}(1)$; in this case, it is not possible to disregard the effects of quantum diffusion in the USR region during the process. In fact, during PBH generation, the assumption of Gaussian statistics for the curvature perturbation fails, and this suggests that one cannot expect a perturbative treatment to provide an adequate understanding. This further implies that in order for the distribution to be most influenced by large perturbations, it must acquire deviations from Gaussianity and substantial tail characteristics. The non-perturbative, stochastic-$\delta N$ formalism has been proposed in many works to address this development \cite{Enqvist:2008kt,Fujita:2013cna,Fujita:2014tja,Vennin:2015hra}. This formalism successfully enables us to relate the statistics of the curvature perturbation distribution to that of the amount of integrated expansion, $N$, which we solve in this paper taking perturbative corrections from the different regimes (classical and quantum). We adopt this strategy and demonstrate how the EFT framework, based on the previously calculated version of the Langevin equations, modifies the Fokker-Planck equation for the probability distribution function (PDF) of the expansion variable in the USR. We execute this investigation from the standpoint of the stochastic-$\delta N$ formalism. This study of stochastic effects in a USR regime has also been undertaken in some previous studies \cite{Firouzjahi:2018vet,Ballesteros:2020sre}.

The characteristics that characterise the slow-roll conditions required for inflation bear the brunt of the weight of realising any given theory, as we are not addressing any particular model of a scalar field driven by a potential. The Hubble rate is a crucial quantity in the EFT of inflation, allowing us to express the different perturbative correction components originating from higher-derivative operators as part of the overall theory. In order to clarify this realisation, we demonstrate how our three-phase scenario—which consists of a USR phase sandwiched between two slow-roll phases—realizes the requirements on the slow-roll parameters and the Hubble rate. Working with the second-order perturbed action inside the EFT framework allows for the extraction of information about the curvature perturbation's mode solutions. For the three-phase situation, this solution may then be worked out separately. It involves a particular parameterization of the phase transition, which is chosen to be sharp in this case. We emphasise, however, that the final conclusion must remain unchanged regardless of how smoothly or sharply the shift is made. The final result that we quote in this study will not alter, even though it could show more coarse-graining or smoother behaviour. The author in \cite{Choudhury:2024ybk} has recently highlighted the similarity between selecting a sharp and seamless transition for a three-phase scenario, which is the same as this study, and the reason why the end findings stay the same regardless of the kind of transition. Many writers have engaged in heated dispute about the precise nature of the shift in references. \cite{Choudhury:2023vuj,Choudhury:2023jlt,Choudhury:2023rks,Choudhury:2023hvf,Choudhury:2024ybk,Bhattacharya:2023ysp,Kristiano:2022maq,Kristiano:2023scm,Riotto:2023gpm,Firouzjahi:2023ahg,Firouzjahi:2023aum,Franciolini:2023lgy,Taoso:2021uvl,Choudhury:2024jlz,Choudhury:2024dei}. While taking quantum loop corrections to the overall power spectrum in the three phases, the argument centres on the purported function of transition in managing the perturbation enhancement, getting reflected in the scalar power spectrum, and their repercussions. Additionally, the EFT description contains an effective sound speed parameter $c_{s}$, where the non-canonical single-field models of inflation are denoted by $c_{s}\ne 1$ and the canonical single-field models by $c_{s}=1$. In order to implement the three-phase configuration employed in this study, this parameter is also essential. In the current setting, we further discuss the effects of the effective sound speed by first defining the precise parameterization that was applied and then going over the modifications that its potential values may have made to the different auto-correlation and cross-correlation components of the power spectrum while still adhering to the unitarity and causality constraints from experiments. We discover that the aforementioned physical criteria do not need to be broken, and perturbativity arguments hold true, leading to the sufficient amplitude required for PBH synthesis. We also discuss additional aspects arising from the stochastic effects in the system, particularly in the vicinity of the phase transitions.

\subsection{Underlying physical motivation and approach}

The rationale for attempting to find an EFT generalisation of the stochastic single-field inflation paradigm is discussed in this section. Our goal is to outline the process by which, in the current situation, such a generalisation can occur. A strong motivation for the current formulation is to comprehend the super-Hubble dynamics of the scalar perturbations and address the emergence of logarithmic infrared divergences during one-loop calculations, which deteriorate exponentially with time. As a potential solution, we investigate the theory of stochastic inflation, which has demonstrated significant utility in solving IR-related problems in dS space \cite{Enqvist:2008kt,Podolsky:2008qq,Finelli:2008zg,Seery:2010kh,Garbrecht:2014dca, Gorbenko:2019rza, Baumgart:2019clc,Mirbabayi:2019qtx,Cohen:2020php}. Transforming the issue and concentrating on the temporal evolution of the PDF of the long-wavelength component of the scalar perturbations is a crucial step in this direction. The Fokker-Planck equation controls the PDF's temporal development. In \cite{Cohen:2021fzf}, the authors derive their conclusions using the Soft de Sitter Effective Theory (SdSET), providing a detailed explanation of how to formally realise the higher-derivative adjustments to the PDF's evolution equation. It also draws attention to an important detail about the corrections to stochastic inflation that have been discussed: these corrections can be obtained effectively by a proper Dynamical Renormalization Group (DRG) analysis \cite{Boyanovsky:1998aa,Boyanovsky:2001ty,Boyanovsky:2003ui,Burgess:2009bs,Dias:2012qy,Chen:2016nrs,Baumann:2019ghk, Burgess:2009bs,Chaykov:2022zro,Chaykov:2022pwd}, involving resummation of the logarithmic IR divergences at all orders in the loop calculations. Methods like the ones listed above are a major source of inspiration for the current work's spirit and provide a solid basis upon which we might expand our investigation of more issues in the future. We will not be dealing with an explicit scalar field, $\phi(t,{\bf x})$, which is based on a particular model of potential $V(\phi)$ during inflation for the time being. Instead, we will be entirely focused on the dynamics of the scalar metric perturbations, $\zeta(t,{\bf x})$. An incredibly general framework for analysing such processes is offered by the EFT of inflation. The effective sound speed parameter $c_{s}$, whose value $c_{s}=1$ denotes canonical single-field inflationary models and $c_{s}\ne 1$ denotes non-canonical single-field models, must also be considered within this EFT framework. With our models satisfying the causality, unitarity, and lack of superluminal propagation constraints, we can use the current image to test for any violation of the following experimental constraint: $0.024 \leq c_{s}< 1$, \cite{Planck:2018jri}. The peak of the power spectrum amplitude may be further restricted to ${\cal O}(10^{-3}-10^{-2})$ by keeping inside this interval of $c_{s}$. This allows us to confine to smaller values of the stochastic parameter $\sigma$, where $\sigma\ll 1$. PBH generation is feasible at these amplitudes, and we can prevent overproduction of PBH by keeping the proportion of PBH in the present below unity. With our discussions on the numerical results for the power spectrum in this work, we want to give a clear image of the following qualities. 

Currently, when dealing with a scalar field, their perturbations, $\delta\phi(t,{\bf x})$, are significant in a cosmological context, particularly when calculating correlation functions of any kind. However, if we additionally include the presence of a gravitational background for the system, the scalar perturbation technique is not the most appropriate one, as it does not result in a fully gauge-invariant formulation of the following concept. The first step is to think of the scalar field as split into two wavelength components: the long (soft/IR) and the short (hard/UV). 
In order to construct a gauge-invariant description of the entire process, we must first take into account a division of $\zeta$ into its long (soft/IR) and short (hard/UV) components as encountered in the stochastic inflation framework. This issue appears when we focus more on the perturbations of the scalar field as the main variable of interest. Using the St$\ddot{u}$ckelberg technique, the EFT picture offers a safer way to do the aforementioned analysis in a model-independent manner. The resulting Goldstone modes function as the comoving curvature perturbation, which is a gauge-invariant quantity. Since the underlying $\lambda\phi^4$ potential is used to perform Quantum Field Theory in de Sitter space, without first focusing on the scalar field perturbations, this problem does not arise in the presentations of \cite{Gorbenko:2019rza,Cohen:2021fzf}. Additionally, we think that the results of earlier studies that addressed the IR divergences issue using $\lambda\phi^4$ theory and, in a similar vein, using the SdSET technique after employing the UV and IR split of the scalar field, may also be retrieved in the language of the curvature perturbations.  

When we start thinking about $\zeta$, a perturbative extension of the EFT action that incorporates higher-derivative operators for further calculations, will be made up of mixtures of $\zeta$'s spatial and temporal derivatives, each accompanied by the random effects. The conventional Schwinger-Keldysh (\textit{in-in}) approach for evaluating higher-point correlation functions will now be characterised in terms of the stochastic parameter $\sigma$ for each higher-order interaction following its quantization, which will have significant ramifications. The regularisation and renormalization procedures must then be incorporated in order to move forward with this modified process of calculating cosmological correlations with the stochastic effects. This is followed by a DRG analysis to meet the ultimate goal of eliminating any undesired UV divergences and softening the logarithmic IR divergences originating from the quantum loops. The most difficult part of the aforementioned process is the last stage in doing the DRG, which is accounting for stochasticity. 

A finite outcome for the correlation functions is ultimately provided by the DRG approach, as previously introduced \cite{Chen:2016nrs,Baumann:2019ghk,Boyanovsky:1998aa,Boyanovsky:2001ty,Boyanovsky:2003ui, Burgess:2009bs,Dias:2012qy,Chaykov:2022zro,Chaykov:2022pwd}. This involves an all-order resummation of the higher-point loop diagrams, which have a similar diagrammatic and analytic structure. This approach softens the logarithmic IR divergences at the super-Hubble scales in cosmic correlations and enables one to capture the quantum effects of the diagrams up to all orders in the perturbative expansion. Currently, each interaction component in the computations has these stochastic effects due to the inclusion of the parameter $\sigma$, and it becomes difficult to keep track of this at all orders of the conceivable loop diagrams. In this case, $\sigma$ functions as a regulator, serving to coarse-grain the contributions and causing the logarithmic IR divergences at all orders of quantum loops in the SdSET to soften. Every diagram has the same parameter $\sigma$; nevertheless, this means limiting its effect so that it doesn't spread to every other diagram and provide an insignificant result. One can choose to work with the stochastic-$\delta N$ formalism and avoid such complications. By selecting this option, we may avoid doing the previously stated DRG analysis with stochastic effects. Instead, the stochastic-$\delta N$ analysis essentially imitates the analysis for the scalar perturbation modes in a similar way. 
The equivalency between the DRG method and the classical $\delta N$ formalism was thoroughly worked out in \cite{Dias:2012qy}. Additionally, with the development of \cite{Cohen:2021fzf,Burgess:2015ajz,Burgess:2014eoa,Burgess:2009bs}, it is anticipated that the DRG analysis in the presence of stochastic effects and the independent execution of the stochastic-$\delta N$ analysis will converge to yield results as previously discussed.

\subsection{The Stochastic EFT of single field inflation}

The process of setting up Effective Field Theory (EFT) entails creating an effective action that is valid below a UV or extremely high energy cut-off scale. The energy scale over which the underlying theory's effective description fails to hold is established by this cut-off. Little is now understood about the potential physics that may define a full theory in the ultraviolet. Nevertheless, any physics that may exist above such a UV cut-off $\Lambda$ should show up in the effective action that is employed to carry out any analysis at lower energies. Terms that are constructed to respect the fundamental symmetries of the theory that endure at these low energies are used in the creation of this action.

Our goal is to investigate the theory of fluctuations over a time-dependent backdrop using this EFT framework. To this end, we start with inflation driven by a scalar field $\phi(t,{\bf x})$. Though its perturbations $\delta\phi$ show broken temporal diffeomorphism symmetry, they remain scalars under the spatial diffeomorphisms, despite the scalar field itself obeying complete diffeomorphism symmetry. With respect to time-diffeomorphisms, the non-linear transformation for the scalar perturbations is expressed as follows:
\bea
t \rightarrow t+\xi^0(t, {\bf x}), \quad x^i \rightarrow x^i \quad \forall\;i=1,2,3 \implies
\del\phi \rightarrow \del\phi+\dot{\phi}_0(t) \xi^0(t, {\bf x}),
\eea
where, in a homogeneous isotropic FLRW space-time, $\phi_0(t)$ is the time-dependent background scalar field and $\xi_0(t,{\bf x})$ is the time-diffeomorphism parameter. In order to move on, we decide to use the unitary gauge, which results in the constraint $\phi(t,{\bf x}) = \phi_0 (t)$, which zeros out the inflationary perturbations. In doing so, the metric consumes the scalar perturbations variable, increasing the number of $3$ physical degrees of freedom: This precisely reproduces the occurrence of spontaneous symmetry breaking in the $SU(N)$ non-abelian gauge theory, with $1$ for the scalar mode and $2$ helicities.
In our theory, the spatially flat FLRW space-time with the following metric has a quasi de Sitter solution:
\bea 
ds^2 = a^2(\tau) ( -d\tau^2 + d{\bf x}^2)\quad\quad {\rm where}\quad\quad a(\tau)=-\frac{1}{H\tau}\quad\quad -\infty<\tau<0.
\eea 
The metric $g_{\mu \nu}$ and its derivatives, the Riemann tensor ($ R_{\mu\nu\alpha\beta}$), the Ricci tensor ($R_{\mu\nu}$), and the Ricci scalar $(R)$, are needed to build the effective action in the current context. The temporally perturbed component of the metric, $\del g^{00} = (g^{00} + 1)$, is another crucial parameter that is required. This word is essential to the construction since it respects the spatial diffeomorphisms. Converting the conformal time to the physical time is still a relevant identity for the quasi de Sitter background solution:
\bea
t=\frac{1}{H}\ln{\Big(-\frac{1}{H\tau }\Big)}, \quad 0 < t < \infty \quad {\rm and}\quad -\infty<\tau<0
\eea
where the scale factor for de Sitter is $a(\tau) = - 1/H\tau$. Another variable invariant under the spatial diffeomorphisms is the extrinsic curvature of the constant time-slice surfaces. The action is structured at a fixed time slice by incorporating an expansion in powers of its temporally perturbed component, i.e.
$\del K_{\mu\nu}= (K_{\mu\nu}-a^2 H h_{\mu\nu}),$
is demonstrated to offer the most versatile and efficient Lagrangian. The extrinsic curvature in this case is denoted by $K_{\mu\nu}$, the unit normal is denoted by $n_\mu$, and the induced metric on the three-dimensional hyper-surface is denoted by $h_{\mu\nu}$. These definitions are relevant here: 
\bea 
&& h_{\mu\nu}=g_{\mu\nu}+n_\mu n_\nu ,\quad 
n_\mu = \frac{\partial_\mu t}{\sqrt{-g_{\mu\nu}\partial_\mu t  \partial_\nu t}}, \quad K_{\mu\nu} = h_\mu ^ \sigma \grad _\sigma n_\nu.
\eea
Now that the overall most effective action that will assist us in computing the complete power spectrum with the one-loop quantum corrections included has been presented, let us do so:
\bea 
\label{eftaction}
S &=&\int d^4x \sqrt{-g}\Bigg[ \frac{M_p^2}{2} R  + M_{p}^{2}\dot{H}g^{00} - M_{p}^{2}(3H^{2} + \dot{H}) + \frac{M_2^4(t)}{2!} (g^{00}+1)^2
+\frac{M_3^4(t)}{3!} (g^{00}+1)^3 \nonumber \\
&& \quad\quad\quad\quad\quad\quad\quad\quad\quad\quad\quad\quad\quad\quad -
\frac{\Bar{M}_1^{3}(t)}{2}(g^{00}+1)\del K_\mu ^\mu 
-\frac{\Bar{M}_2^{2}(t)}{2}(\del K_\mu ^\mu )^{2} 
-\frac{\Bar{M}_3^{2}(t)}{2}\del K_\mu ^\nu \del K_\nu ^\mu  \Bigg]. \eea 
The Wilson coefficients in this case are represented by $M_2(t) , M_3(t) ,\Bar{M_1}(t), \Bar{M_2}(t) , \Bar{M_3}(t) $, which need to be adjusted by additional analysis.

\subsection{Second order perturbed action from Goldstone EFT in decoupling limit}

It is considerably simpler to study the Goldstone modes or disturbances in the current EFT inflation setting when one works at the low energy scale limit where gravity tends to decouple. Here, we apply this particular limiting approach to the investigation of the scalar perturbations during inflation in the three areas involved in our overall setup, which we will refer to as the comoving curvature perturbation. The decoupling limit is specified where $E_{mix}=\sqrt{\dot{H}}$, meaning that the mixing contributions between the metric fluctuations and the Goldstone mode are to be ignored. Writing the second-order Goldstone action as, after applying this restriction, is possible. 
\bea \label{EFTpi}
S_{\pi}^{(2)} \approx \int d^4x \,  
  a^3\Bigg\{ -M_p^2\dot{H} \Bigg( \dot{\pi}^2  - \frac{1}{a^2}(\partial_i \pi )^2 \Bigg) + 2M_2 ^4 \dot{\pi}^2 \Bigg\} = \int d^4 x \,  a^3 \Bigg( \frac{-M_p ^2 \dot{H}}{c_s ^2}\Bigg)\Bigg[ \dot{\pi}^2 - c_s ^2 \frac{(\partial_i \pi)^2 }{a^2}\Bigg]
\eea
where the EFT Wilson coefficient may be used to construct the effective sound speed parameter, $c_s$,
\bea
c_s = \frac{1}{\sqrt{1-\displaystyle{\frac{2M_2 ^4 }{\dot{H}M_p^2 }}}}.
\eea 
The spatial component $g_{ij}$ of the perturbed metric is now understood to be defined as, 
\bea
g_{ij}\sim a^2(t)\{(1+2\zeta (t,{\bf x }))\del_{ij}\} \quad  \forall \quad  i=1,2,3,
\eea
It takes $a(t)=e^{H t}$ as the scale factor for a quasi de Sitter space-time. When broken time diffeomorphisms are present, $a(t)$ now changes as:
\bea
a(t)\rightarrow a(t-\pi(t,{\bf x })) = a(t) - H \pi(t,{\bf x})a(t)+.....\approx a(t)(1-H\pi(t,{\bf x})),
\eea 
which, when changed, provides us with: 
\bea
a^2(t)(1-H\pi(t,{\bf x}))^2 \approx  a^2(t)(1-2H\pi(t,{\bf x})) = a^2(t)(1+2\zeta(t,{\bf x})),
\eea 
and this implies that the Goldstone mode $\pi(t,{\bf x})$ may be used to express the comoving scalar curvature perturbation $\zeta(t,{\bf x})$ as $\zeta(t,{\bf x}) \approx -H \pi(t,{\bf x})$.
With the aid of eqn.(\ref{EFTpi}), we can now express the second-order perturbed action in terms of the variable $\zeta(t,{\bf x})$. While working with conformal time $(\tau)$ as opposed to the physical time $(t)$ is still convenient,
\bea  \label{s2zeta}
S_\zeta ^ {(2)} = \int {\cal L}_{\zeta}^{(2)}d\tau = M_p ^2 \int d\tau  \, d^3x  \,  a^2 \Big( \frac{\epsilon}{c_s ^2}\Big)\Bigg[ \zeta'\,^2 - c_{s}^2(\partial_{i}\zeta)^{2} \Bigg].
\eea

\subsection{Implementing stochasticity within EFT setup}

We outlined the overall EFT setting of inflation in the preceding section. This will serve as the foundation for our future study, which will use stochastic inflation formalism to primarily focus on the dynamics of perturbations. In order to incorporate stochasticity into the EFT formalism, one must first analyse the stochastic nature of fluctuations using Hamilton's equations, or more specifically, the Langevin equations. From there, one must use the Fokker-Planck equation to evolve the distribution function of the curvature perturbations. We shall use the EFT formalism to infer the Fokker-Planck equation in the following sections.

\subsubsection{Hamilton's equation: The way towards formulating Langevin equation}
\begin{figure*}[htb!]
    	\centering
    {
       \includegraphics[width=19cm,height=12.5cm]{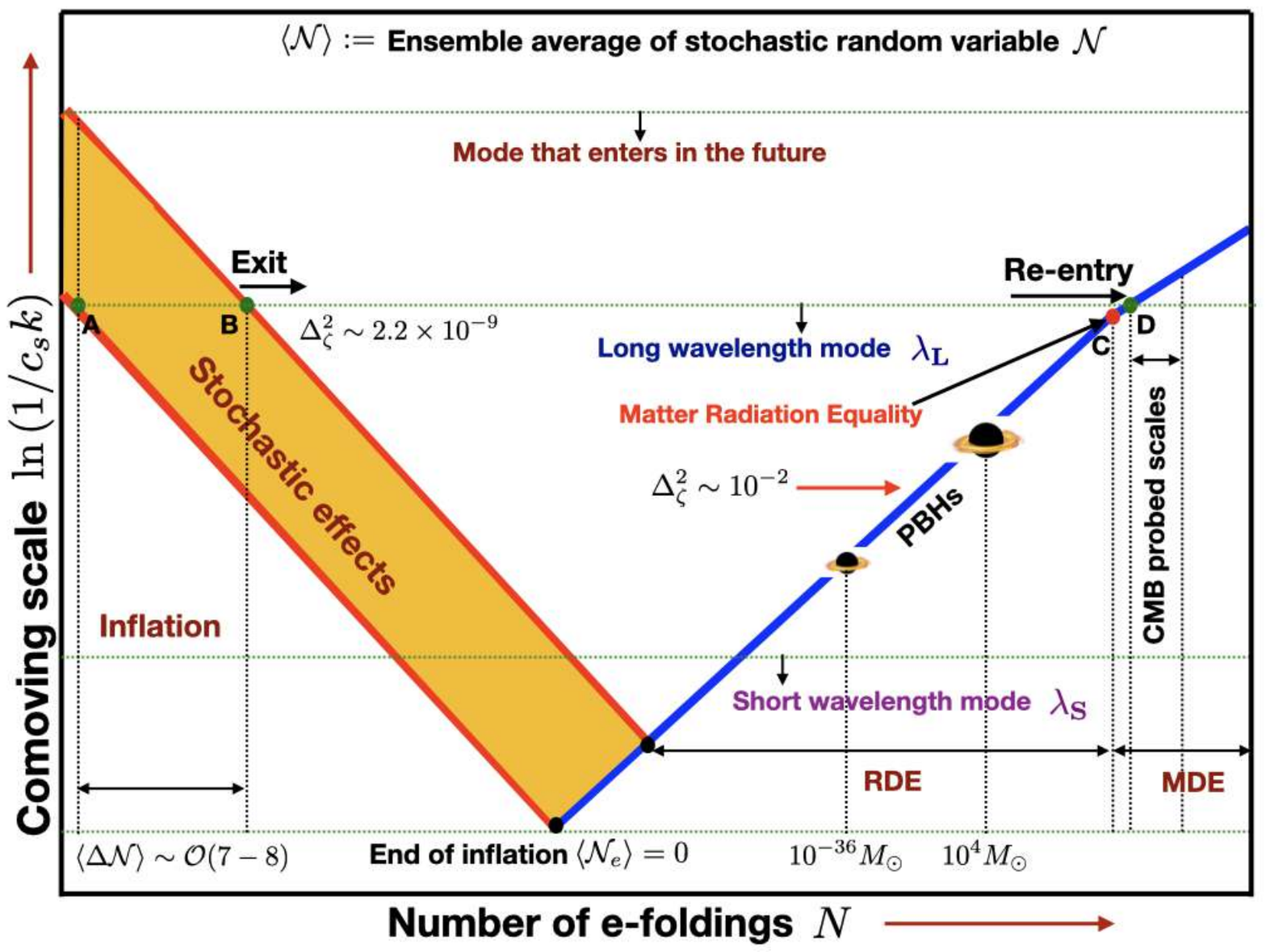}
        \label{CS}
    } 
    \caption[Optional caption for list of figures]{Diagrammatic representation of modes that go from the Sub-Horizon during inflation, experience random effects at the Horizon crossing, and then return to the Horizon. The CMB scales are related with long wavelength $\lambda_{L}$ modes, which escape at positions $A$ and $B$ and re-enter at $D$. Given that $C$ is quite near to the CMB re-entry scale, it is designated as the radiation-matter equality. In the radiation-dominated era (RDE) with variable masses, the short wavelength $\lambda_{s}$ is related with the tiny length scales towards the end of inflation that cause subsequent collapse into PBHs. A stochastic variable ${\cal N}$ is the number of e-folds that pass between an initial and final set of circumstances. }
\label{stochasticdiag2}
    \end{figure*}
An efficient theoretical method for examining the long-wavelength components of inflationary quantum fluctuations is stochastic inflation. To do this, these variations are coarse-grain-sized across a set scale, which is somewhat greater than the Hubble radius during the inflationary era. In the Fourier space, the fixed scale is as follows:
\bea k_{\sigma}c_{s}=\sigma aH,
\eea
where the coarse-grained sector of the quantum fluctuations is contributed by the modes $k$, $- kc_{s}\tau \ll \sigma$ (with $\sigma \ll 1$), and this scale acts as a cut-off for these modes. The long-wavelength or infrared (IR) sector grows in size as inflation progresses when variations from the small-wavelength or UV (ultraviolet) sector leave the Hubble radius and join it. The Langevin equation provides a classical stochastic theory description of the dynamics of the resultant coarse-grained sector of the quantum field. The classical drift term and the effects of the quantum noise, which merge into the classical noise term after the horizon exit, are included in this stochastic differential equation.

Regarding the Hamiltonian formulation in stochastic inflation, we consult \cite{Grain:2017dqa,Vennin:2020kng}). The UV and IR sectors of the quantum field causing inflation would need to be identified before we could produce the Langevin equations from the Hamiltonian equations. We commence using a similar method as described in \cite{Grain:2017dqa,Vennin:2020kng} in order to get the Langevin equations in the EFT setting where our variable of interest remains the comoving curvature perturbation $\zeta(t,{\bf x})$. We would begin with the second-order perturbed action for $\zeta(t,{\bf x})$, as stated in eqn. (\ref{s2zeta}). From this, we may compute the conjugate momentum variable as follows:
\bea
\tilde{\Pi}_{\zeta} = \frac{\partial {\cal L}_{\zeta}^{(2)} }{\partial \zeta'} = \frac{2M_{p}^{2}a^{2}\epsilon}{c_{s}^{2}}\zeta'.
\eea
This may be used to assess the associated Hamiltonian density following a Legendre transformation in the manner described below:
\bea
{\cal H}_{\zeta}^{(2)} &=& \tilde{\Pi}_{\zeta}\zeta' - {\cal L}^{(2)}_{\zeta} = M_{p}^{2}\int d^{3}x\;a^2 \Big( \frac{\epsilon}{c_s ^2}\Big) \Bigg[\frac{c_{s}^{4}}{4M_{p}^{4}a^{4}\epsilon^{2}}\tilde{\Pi}^{2}_{\zeta} + c_{s}^2(\partial_{i}\zeta)^{2} \Bigg].
\eea
With this Hamiltonian density, we can now analyse the following Hamilton's equations of motion: 
\bea
\label{Ham1}
-\tilde{\Pi}_{\zeta}' &=& \frac{\partial{\cal H}_{\zeta}^{(2)}}{\partial\zeta} = 0,\\
\label{Ham2}
\zeta' &=& \frac{\partial{\cal H}_{\zeta}^{(2)}}{\partial\tilde{\Pi}_{\zeta}} = \frac{c_{s}^2}{2M_{p}^{2}a^{2}\epsilon}\tilde{\Pi}_{\zeta}.
\eea
The variable $\zeta$ is shown to be a cyclic co-ordinate from the Lagrangian ${\cal L}^{(2)}_{\zeta}$, where the prime indicates derivative with respect to the conformal time. We modify our time variable selection to the e-folds $N$ instead of the conformal time, which results in the eqn. (\ref{Ham2}) possess the subsequent structure:
\bea \label{Langevin1}
\frac{d\zeta}{dN} &=& \frac{c_{s}^2}{2M_{p}^{2}Ha^{3}\epsilon}\tilde{\Pi}_{\zeta} = \Pi_{\zeta}, \\
\label{Langevin2}
\frac{d\Pi_{\zeta}}{dN} &=& \frac{1}{2M_{p}^{2}}\frac{d}{dN}\Bigg[\frac{c_{s}^2}{Ha^{3}\epsilon}\tilde{\Pi}_{\zeta}\Bigg] 
= \frac{c_{s}^2}{2M_{p}^{2}Ha^{3}\epsilon}\tilde{\Pi}_{\zeta}\Bigg[2s + \epsilon -3 - \eta \Bigg]= -(3-\epsilon)\Bigg[1 - \frac{2(s-\frac{\eta }{2})}{(3-\epsilon)} \Bigg]\Pi_{\zeta}.
\eea
In the last equality of the equation, $dN/d\tau = Ha$, is utilised as a conversion between the two time variables. In (\ref{Langevin1}), the tilde is eliminated and the conjugate momentum is renamed. We execute the derivative of the conjugate momentum with the e-folds $N$, where the third and last lines employ the eqn., to produce the other eqn. (\ref{Langevin2}). (\ref{Ham1}) and $\Pi_{\zeta}$, respectively, as well as the slow-roll parameters' descriptions shown below:
\bea \label{slowrollparams}
s = \frac{d\ln c_{s}}{dN},\quad\quad \epsilon = -\frac{d\ln{H}}{dN},\quad\quad \eta = \epsilon-\frac{1}{2}\frac{d\ln{\epsilon}}{dN}.
\eea
The aforementioned equations (\ref{Langevin1} and (\ref{Langevin2}) will also allow us to ascertain the Langevin equation that regulates the dynamics of the coarse-grained supper-Hubble fields when white noise is present. The evolution history of the long- and short-wavelength modes is shown in fig. \ref{stochasticdiag2}. These modes experience stochastic effects from the very beginning of their voyage within the Horizon until they get \textit{classicalized} and continue their adventure in the super-Hubble scales. In e-foldings, various wavelength modes re-enter at different times. While the long-wavelength modes are responsible for the CMB observations, the short-wavelength modes may have had a role in the development of the primordial black hole. The orange-colored zone that describes the stochastic effects is characterised by the value of the stochastic parameter $\sigma$. While the band shortens and the stochastic or coarse-graining effects tend to reduce fast with rising $\sigma\sim 1$, the band gets bigger and similarly indicates an increase in the duration of stochastic effects when $\sigma\ll 1$ is maintained. 

We first introduce the quantum operator image and then give the classical version of the equations on the super-Hubble scales. In order to derive the Langevin equations for the coarse-grained components of the initial quantum fields, it is helpful to break down the perturbations into their UV and IR components. This may be achieved in the following way:
\bea
\Hat{\Gamma} = \Hat{\Gamma}_{\bf IR} + \Hat{\Gamma}_{\bf UV}\quad\quad {\rm where}\quad \Hat{\Gamma}_{\bf IR}= \{\Hat{\zeta},\Hat{\Pi}_{\zeta}\}\quad {\rm and} \quad \Hat{\Gamma}_{\bf UV}= \{\Hat{\zeta}_{s},\Hat{\Pi}_{\zeta_{s}}\},
\eea
where small-wavelength components are indicated by the subscript $s$. These ultraviolet components are expandable in Fourier modes and are chosen if $k > k_{\sigma}$ or if they are less than the cut-off scale:
\bea 
\Hat{\zeta_{s}} &=& \int_\mathbb{R^3}\frac{d^3 k }{(2\pi)^3 }W\Bigg(\frac{k}{k_\sigma}\Bigg)\Bigg[ \Hat{a}_{{\bf k}}\zeta_k(\tau) e^{-i {\bf k}.{\bf x}} + \Hat{a}_{{\bf k}}^{\dagger}\zeta_k^*(\tau) e^{i{\bf k}.{\bf x}}\Bigg],\\
\Hat{\Pi}_{\zeta_{s}} &=& \int_\mathbb{R^3}\frac{d^3 k}{(2\pi)^3}W\Bigg(\frac{k}{k_\sigma}\Bigg)\Bigg[ \Hat{a}_{{\bf k}}\Pi_{\zeta_k}(\tau) e^{-i {\bf k}.{\bf x}} + \Hat{a}_{{\bf k}}^{\dagger}\Pi_{\zeta_k}^*(\tau) e^{i{\bf k}.{\bf x}}\Bigg].
\eea 
These include $a_{\bf k}$ and $a_{\bf k}^{\dagger}$, the annihilation and creation operators, respectively, satisfying the standard canonical commutation relations:
\bea [\Hat{a_{\bf k}}, \Hat{a_{\bf k'}}^{\dagger}] =\del^3({\bf k}-{\bf k'}), \quad\quad [\Hat{a_{\bf k}}, \Hat{a_{\bf k'}}] = [\Hat{a_{\bf k}^{\dagger}}, \Hat{a_{\bf k'}^{\dagger}}] = 0.\eea 
In this case, the window function $W$ functions as follows:
\begin{equation}
 W = \begin{cases}
  1  & k > k_{\sigma} \\
0                                                        & k < k_{\sigma},
  \end{cases} \\
\end{equation}
Thus, modes that contribute to the UV or small-wavelength region are chosen. In accordance with \cite{Grain:2017dqa,Vennin:2020kng}, the Langevin equations derived from Hamilton's equation may be achieved using the aforementioned field decomposition and the UV mode equations that subsequently contribute to the noise components. 
\bea \label{eftlangevin}
\frac{d\hat{\zeta}}{dN} &=&  \hat{\Pi}_{\zeta} + 
 \hat{\xi}_\zeta(N),\\ 
\frac{d\hat{\Pi}_\zeta }{dN} &=& -(3-\epsilon)\hat{\Pi}_\zeta \Bigg[1 -\frac{2(s-\frac{\eta }{2})}{(3-\epsilon)} \Bigg] + \hat{\xi}_{\pi_\zeta}(N),
\eea 
The quantum white noise terms are shown by the values $\hat{\xi}_\zeta(N)$ and $\hat{\xi}_{\pi_\zeta}(N)$. These are derived from the continuous outflow of UV modes into the IR modes.
\bea 
\Hat{\xi}_\zeta &=& -\int_\mathbb{R^3}\frac{d^3 k }{(2\pi)^3 }\frac{d}{dN}W\Bigg(\frac{k}{k_\sigma}\Bigg)\Bigg[ \Hat{a}_{{\bf k}}\zeta_k(\tau) e^{-i {\bf k}.{\bf x}} + \Hat{a}_{{\bf k}}^{\dagger}\zeta_k^*(\tau) e^{i{\bf k}.{\bf x}}\Bigg],\\
\Hat{\xi}_{\pi_{\zeta}} &=& -\int_\mathbb{R^3}\frac{d^3 k }{(2\pi)^3 }\frac{d}{dN}W\Bigg(\frac{k}{k_\sigma}\Bigg)\Bigg[ \Hat{a}_{{\bf k}}\Pi_{\zeta_k}(\tau) e^{-i {\bf k}.{\bf x}} + \Hat{a}_{{\bf k}}^{\dagger}\Pi_{\zeta_k}^*(\tau) e^{i{\bf k}.{\bf x}}\Bigg],
\eea
and a Heaviside function has been selected as the window function $W$'s form for convenience:
\bea 
W\Bigg(\frac{k}{k_\sigma}\Bigg) = \Theta\Bigg(\frac{k}{k_\sigma} - 1\Bigg)=\Theta\Bigg(\frac{k}{\sigma aH}-1\Bigg),
\eea
such that $k_{\sigma}=\sigma aH$. For the stochastic course graining parameter, $\sigma$ serves as a physical representation. After that, we calculate the formula for the window function's derivative with respect to the number of e-foldings, which will be very helpful in figuring out the quantized version of the white noise, as was previously stated clearly. Here we have it:
\bea \frac{d}{dN}W\Bigg(\frac{k}{k_\sigma}\Bigg)=\frac{k}{k_{\sigma}}\left(\epsilon-1\right) \frac{d}{d\tau}W\Bigg(\frac{k}{k_\sigma}\Bigg)=\frac{k}{k_{\sigma}}\left(\epsilon-1\right)k_{\sigma}\delta(k-k_{\sigma})=k\left(\epsilon-1\right)\delta(k-k_{\sigma}).\eea
The Dirac Delta distribution is produced by taking the derivative of this function with e-folds, which enforces a severe cut off between the IR and UV modes. $\delta(k-k_{\sigma}(N))$ providing us contributions to the noises following integration across the several $k_{\sigma}$ cut-off scales. Due to the stochastic character of the noise factors, the system may be described as probabilistic and examined using the Langevin Equations. By transitioning into the appropriate Fokker-Planck equation, a second-order partial differential equation whose formulation we cover in the following section, the equations may be solved analytically.

\subsubsection{Fokker-Planck equation from the Langevin equation}

The main topic of this section is a quick explanation of how the Langevin equation is used to get the Fokker-Planck equation. The Fokker-Planck is a useful tool for characterising the evolution of the field variables' probability distribution in the phase space as they change from an initial condition at any point in time to a final field configuration at a later point in time, typically chosen to mark the end of inflation. To begin, we need to utilise a probability rate that will allow the system to begin at a specific initial field configuration, land at a tiny distance into a new field configuration, and then increase the time parameter by a tiny amount. In order to do this, we define the transition probability rate $W_{\Delta {\bf \Gamma}}({\bf \Gamma},N)$ using the following formula for the field variables, ${\bf \Gamma} = \{\zeta , \Pi_\zeta\}$: 
\bea
W_{\Delta {\bf \Gamma}}({\bf \Gamma},N)\del N = P({\bf \Gamma} +\Delta {\bf \Gamma},N +\delta N|{\bf \Gamma}, N),
\eea
This is equivalent to the likelihood that, in a $\del N $ infinitesimal increment of time, the system with ${\bf \Gamma}$ at time $N$ develops to ${\bf \Gamma}+\Delta {\bf \Gamma}$ at $ N+\del N $. Equation for the probability $P({\bf \Gamma},N)$ to reach a field configuration ${\bf \Gamma}$ from an initial configuration ${\bf \Gamma}_{\rm in}$ may be obtained using this: 
\bea \label{Prob1}
\frac{\partial}{\partial N}P({\bf \Gamma},N)=\int d\Delta{\bf \Gamma}\Bigg[W_{\Delta{\bf \Gamma}}({\bf \Gamma}-\Delta {\bf \Gamma},N)P({\bf \Gamma}-\Delta {\bf \Gamma},N)-W_{-\Delta {\bf \Gamma}}( {\bf \Gamma},N)P( {\bf \Gamma},N)\Bigg].
\eea
As we move from ${\bf \Gamma}$ to ${\bf \Gamma}-\Delta {\bf \Gamma}$ to ${\bf \Gamma}$, the first term increases and the second term decreases. These changes are then integrated across the increments $\Delta {\bf \Gamma}$. When we apply Taylor expansion to the initial term in the integrand of equation (\ref{Prob1}), we observe:
\bea
W_{\Delta{\bf \Gamma}}({\bf \Gamma}-\Delta {\bf \Gamma},N)P({\bf \Gamma}-\Delta {\bf \Gamma},N)=W_{\Delta{\bf \Gamma}}({\bf \Gamma},N)P({\bf \Gamma},N)+\Bigg( -\Delta \Gamma_i\frac{\partial}{\partial \Gamma_i}+ \frac{1}{2}\Delta \Gamma_i \Delta \Gamma_j \frac{\partial^2}{\partial \Gamma _i \partial \Gamma_j}\nonumber \\
+........+\frac{(-1)^l}{l!}\Delta \Gamma_i \Delta \Gamma_j......\Delta \Gamma_l\frac{\partial^l}{\partial \Gamma_i \partial \Gamma_j...\partial \Gamma _l}+.....\Bigg)[W_{\Delta{\bf \Gamma}}({\bf \Gamma},N)P({\bf \Gamma},N)].
\eea
and the sum of the dummy indices $\Gamma_{i}=\{\zeta_{i},\Pi_{\zeta,i}\}$ between the field variables is obtained. $\Delta {\bf \Gamma}\rightarrow - \Delta {\bf \Gamma}$ can be used to change the integration variable of the second term in eqn. (\ref{Prob1}), allowing for the following expression: 
\bea \label{Prob2}
\frac{\partial}{\partial N}P({\bf \Gamma},N)=\sum_{l=1}^\infty\frac{(-1)^l}{l!}\frac{\partial^l}{\partial \Gamma_i \partial \Gamma_j...\partial \Gamma _l}[a_{ij....l}({\bf \Gamma},N)P({\bf \Gamma},N)],
\eea
where the labels for the different field moments of ${\bf \Gamma}$ are: 
\bea \label{Probmoment}
a_{i j....l}({\bf \Gamma},N)=\int d\Delta {\bf \Gamma }\;\Delta \Gamma_i \Delta \Gamma_j....\Delta \Gamma_l W_{\Delta {\bf \Gamma}}({\bf \Gamma },N).
\eea
The Fokker-Planck differential equation will eventually need to be determined, and these moments will be crucial. We use the Langevin equation, found in eqn. (\ref{eftlangevin}), to assess these moments that correspond to the field increments. ${\bf\Gamma} = \{\zeta, \Pi_\zeta\}$, where ${\bf\Gamma}$ value evolves between $N$ and $N+\delta N$:
\bea \label{evolveGamma}
{\bf\Gamma}(N +\del N) = {\bf\Gamma}(N)+ F({\bf\Gamma})\del N +G({\bf\Gamma}).\int_N ^{N+\del N}\xi(\tilde{N})d\tilde{N}.
\eea
where $G({\bf \Gamma})$ denotes the portion that, after being squared, creates the noise matrix element resulting from the various potential noise correlators, and $F({\bf \Gamma})$ represents the classical portion of the motion:
\bea \Sigma_{f_1,g_1}\equiv (G^{2})_{f,g}\del(\tau_1 - \tau_2), \quad\quad\quad {\rm where},\quad\quad\quad (G^{2})_{f,g}\equiv \Sigma_{f,g}(\tau_1),\eea
where the condition of white noise is represented by the function $\delta(\tau_{1}-\tau_{2})$. In the subsequent section, further important details on these matrix members are expanded. We now propose a new notation in terms of a parameter $\alpha$ for the purposes of the present derivation:
\bea \label{meanGamma}
{\bf \Gamma}_\alpha (N)=(1-\alpha){\bf \Gamma}(N)+\alpha {\bf \Gamma}(N+\del N).
\eea
It enables the functions $F({\bf \Gamma})$ and $G({\bf \Gamma})$ to be evaluated at the point of interest with respect to specific weights, in this case $(1-\alpha)$ and $\alpha$. Between $\alpha\in [0,1]$ is the possible range for the parameter $\alpha$. Equation (\ref{evolveGamma}) is transformed as follows according to this definition:
\bea \label{evolveGamma2}
{\bf \Gamma}(N +\del N) = {\bf \Gamma}(N)+ F({\bf \Gamma}_\alpha (N))\del N +G({\bf \Gamma}_\alpha (N)).\int_N ^{N+\del N}\xi(\tilde{N})d\tilde{N},
\eea
It is evident from the use of new notation that, for any arbitrary $\alpha$, we may solve eqn.(\ref{evolveGamma2}) in terms of ${\bf \Gamma}(N)$. In order to solve, we employ a perturbative method, as follows:
\bea \del {\bf \Gamma}= {\bf \Gamma}(N+\del N)-{\bf \Gamma}(N),\eea
as well as Taylor expanding functions $F$ and $G$ with the variable in the phase space expressed as:
\bea {\bf \Gamma}_\alpha = {\bf \Gamma}+\alpha \del {\bf \Gamma},\eea
we can express eqn.(\ref{meanGamma}) as follows:
\bea
F_{m}[{\bf \Gamma}_\alpha (N)]= F_{m}({\bf \Gamma})+\alpha \del \Gamma_i\frac{\partial}{\partial \Gamma_i}F_{m}({\bf \Gamma})+\frac{\alpha^2}{2}\del \Gamma_i\del \Gamma_j \frac{\partial^2}{\partial \Gamma_i \partial \Gamma_j}F_{m}({\bf \Gamma})+.....
\eea
where the next equation will soon make the subscript $m$ evident. It is also possible to do an analogous analysis for $G[{\Gamma_\alpha}(N)]$. The $m$th component of $\delta{\bf \Gamma}$ is given by these expansions for $F$ and $G$ in eqn. (\ref{evolveGamma2}), which must form a series in powers of $\delta N$:
\bea 
\delta\Gamma_{m} = F_{m}({\bf \Gamma})\del N + G_{mi}({\bf \Gamma})\int _N ^{N
+\del N}\xi_i(\tilde{N})d\tilde{N}+\alpha G_{pq}({\bf \Gamma})\frac{\partial G_{mj}({\bf \Gamma})}{\partial \Gamma_p}\int_{N} ^{N
+\del N}\xi_q(\tilde{N})d\tilde{N}\int_{N} ^{N
+\del N}\xi_j(\tilde{N})d\tilde{N}+\cdots,
\eea 
In the expansion, the words higher-order are denoted by the ellipses `$\cdots$'. Now that we have this formula along with eqn.(\ref{Probmoment}), we can at last start calculating the moments of field displacement in the manner shown below:
\bea
&& a_i({\bf \Gamma})=\lim_{\del N\rightarrow 0}\frac{\langle \del \Gamma_i\rangle}{\del N}=\Bigg( F_i({\bf \Gamma})+\alpha G_{mj}({\bf \Gamma})\frac{\partial G_{ij}({\bf \Gamma})}{\partial \Gamma_m}\Bigg) \\
&& a_{ij}({\bf \Gamma})=\lim_{\del N \rightarrow 0}\frac{\langle \del \Gamma_i\del \Gamma_j\rangle}{\del N} = \Bigg( G_{im}({\bf \Gamma})G_{jm}({\bf \Gamma}) \Bigg),\\
&&a_{ij.....l}({\bf \Gamma})=0.
\eea 
where the situation of white noise is
implemented by making use of the following condition:
\bea \langle\xi_{i}(N)\xi_{j}(\tilde{N})\rangle = \delta_{ij}\delta(N-\tilde{N}).\eea
Judging from the foregoing, we may conclude that only the first and second moments are non-vanishing. Adding them to the PDF's evolution, eqn. (\ref{Prob2}), yields the required Fokker-Planck equation.

Ultimately, the Fokker-Planck Equation that corresponds to the Langevin Equation may be expressed as follows in terms of the variables ${\bf\Gamma} = \{\zeta, \Pi_\zeta\}$:
\bea
\frac{\partial P({\bf \Gamma},N)}{\partial N}=\mathcal{L}_{\rm FP}({\bf \Gamma})  P({\bf \Gamma},N).
\eea 
For the probability density function $P({\bf\Gamma},N)$, the Fokker-Planck operator is denoted by $\mathcal{L}_{\rm FP}({\bf\Gamma})$. This Fokker-Planck operator has the following form because of the non-zero moments:
\bea
\mathcal{L}_{\rm FP}({\bf\Gamma})=\Bigg\{-\Bigg(F_i({\bf \Gamma})+\alpha G_{lj}({\bf \Gamma})\frac{\partial G_{ij}({\bf \Gamma})}{\partial \Gamma_l}\Bigg)\frac{\partial}{\partial \Gamma_i} 
 + \frac{1}{2}G_{il}({\bf \Gamma} )G_{jl}({\bf \Gamma})\frac{\partial^2}{\partial\Gamma_i\partial\Gamma_j}\Bigg\}.
\eea
The differential equation that results from the adjoint Fokker-Planck equation is similarly helpful and is as follows:
\bea
\frac{\partial}{\partial N}P_{{\bf \Gamma}}(N)= -\mathcal{L}^\dagger_{\rm FP}({\bf \Gamma})P_{{\bf \Gamma}}(N).
\eea
When the following operation is integrated by parts using the Fokker-Planck operator, the equivalent adjoint Fokker-Planck operator is defined:
\bea 
\int d{\bf \Gamma} f_1({\bf \Gamma}) \Bigg[\mathcal{L}_{\rm {FP}}({\bf \Gamma}).f_2({\bf \Gamma})\Bigg] =\int d{\bf \Gamma} \Bigg[\mathcal{L}_{\rm FP}^\dagger.f_1({\bf \Gamma})\Bigg ] f_2({\bf \Gamma}).
\eea 

For the purpose of managing and analysing stochastic differential equations, there are two key descriptions:
\begin{itemize}
    \item \underline{\textbf{It\^{o} prescription $(\alpha=0)$}:}
    \bea
\mathcal{L}^{\dagger , \text{It\^{o}}}_{\rm FP}({\bf\Gamma}) = \Bigg\{F_i(\Gamma)\frac{\partial}{\partial \Gamma_i}  +\frac{1}{2}G_{il}(\Gamma)G_{jl}(\Gamma)\frac{\partial^2}{\partial\Gamma_i \partial\Gamma_j}\Bigg\},
\eea
    \item  \underline{\textbf{Stratonovich prescription $(\alpha=\frac{1}{2})$ }:}
    \bea
\mathcal{L}^{\dagger ,\rm  {Stratonovich}}_{\rm {FP}}({\bf\Gamma})  = \Bigg\{ F_i(\Gamma)\frac{\partial}{\partial \Phi_i}+ \frac{1}{2}G_{ij}(\Gamma)\frac{\partial G_{lj}(\Gamma)}{\partial \Gamma _l}\frac{\partial }{\partial \Gamma_i} +\frac{1}{2}G_{il}(\Gamma)G_{jl}(\Gamma)\frac{\partial^2}{\partial\Gamma_i \partial\Gamma_j} \Bigg\}.
\eea
\end{itemize}
Given $ G=0 $, that is, no stochastic noise, the $\alpha$ term is meaningless since deterministic differential equations don't depend on a prescription. However, in situations when there are several inflating fields on curved field spaces, the prescription parameter $\alpha$ may become important. We are going to work in It\^{o}'s prescription, $(\alpha=0)$, for the sake of this study. Then, the adjoint Fokker-Planck operator can be expressed as follows:
\bea
\mathcal{L}^{\dagger , \text{It\^{o}}}_{\rm FP}({\bf\Gamma}) = \Bigg(F_i \frac{\partial }{\partial \Gamma_i}+\frac{1}{2}\Sigma_{ij}\frac{\partial ^2}{\partial \Gamma_i \partial \Gamma_j}\Bigg),
\eea
where $F_i$ represents the classical drift terms and ${\bf \Gamma} = \{\zeta,\Pi_\zeta \}$ is the same as previously:
\bea
F_i=\Bigg\{\Pi_\zeta, -(3-\epsilon)\Pi_\zeta \Bigg[1 -\frac{2(s- \frac{\eta }{2})}{(3-\epsilon)} \Bigg]
\Bigg \}.
\eea 
In the latter part of this study, we will explicitly compute the noise correlation matrix elements $\Sigma_{i j}$ for our underlying theoretical framework.

\subsection{Realising ultra slow-roll phase within the framework of EFT of Stochastic Single Field Inflation}

In this part, we study how, for a stochastic single field inflation framework, the ultra slow-roll phase—sandwiched between the slow-roll, SRI, and SRII phases—can be realised. The method we take is to start with a specific parameterization of the second slow-roll parameter $\eta$ throughout the three phases of interest, and then use that to determine how the other two parameters, $\epsilon$ and the Hubble expansion $H$, behave. This will help us understand how a USR phase might be represented visually by examining its impacts on the parameters that define and contribute to the inflationary observables. In eqn. (\ref{slowrollparams}), the formulae for the slow-roll parameters with regard to the conformal time are already defined. The definition for $\epsilon$, assuming constant $\eta$, provides us with:
\bea
\eta = \epsilon-\frac{1}{2}\frac{d\ln{\epsilon}}{dN} \implies \int_{\epsilon_{i}}^{\epsilon}\frac{d\epsilon}{2\epsilon(\epsilon-\eta)} = \int_{N_{i}}^N dN, 
\eea
This yields the following formula after integration using a few starting condition values, $(\epsilon_{i},N_{i})$:
\bea \label{epsilonphases}
\epsilon(N) = \eta\bigg(1-\bigg(1-\frac{\eta}{\epsilon_{i}}\bigg)e^{2\eta\Delta N}\bigg)^{-1}\sim \left\{
	\begin{array}{ll}
		\displaystyle  \epsilon(N) & \mbox{when}\quad  N_{*} \leq N \leq N_{s}  \;(\rm SRI),  \\  
			\displaystyle 
			\displaystyle \epsilon(N_{s})e^{-2\eta(N)(N-N_{s})} & \mbox{when}\quad  N_{s} \leq N \leq N_{e}  \;(\rm USR), \\ 
   \displaystyle 
			\displaystyle \epsilon(N_{s})e^{-2\eta(N_{e})(\Delta N_{es})}e^{-2\eta(N)(N-N_{e})} & \mbox{when}\quad  N_{e} \leq N \leq N_{\rm end}  \;(\rm SRII).
	\end{array} \right.
\eea
By choosing suitable limiting behaviours for the values of the slow-roll parameters at each phase, the aforementioned equations are produced. The whole behaviour is displayed in Fig. (\ref{slowrollparams}). The initial condition values for the USR and SRII, respectively, when joining them together are $\epsilon(N_{s})$ and $\epsilon(N_{s})e^{-2\eta(N_{e})(\Delta N_{es})}$, where $\Delta N_{es}=N_{e}-N_{s}$. The sum of the values of the last $\epsilon$ in each region remains the only thing that it is:
\bea
\epsilon(N)= \epsilon(N\leq N_{s}) + \epsilon(N_{s}\leq N\leq N_{e}) + \epsilon(N_{e}\leq N\leq N_{\rm end}).
\eea
The $\eta$ parameter in the current work is constant in value during each phase, therefore we may write:
\bea
\eta(N) = \left\{
	\begin{array}{ll}
		\displaystyle  \eta_{\rm I}(N) & \mbox{when}\quad  N_{*} \leq N \leq N_{s}  \;(\rm SRI),  \\  
			\displaystyle 
			\displaystyle \eta_{\rm II}(N) & \mbox{when}\quad  N_{s} \leq N \leq N_{e}  \;(\rm USR), \\ 
   \displaystyle 
			\displaystyle \eta_{\rm III}(N) & \mbox{when}\quad  N_{e} \leq N \leq N_{\rm end}  \;(\rm SRII), 
	\end{array} \right.
\eea
where various constants $\eta_{\rm I},\;\eta_{\rm II},\;\eta_{\rm III}$ are used. The following may be stated by joining for each of the three phases using a Heaviside Theta function:
\bea
\eta(N) = \eta_{\rm I}(N\leq N_{s}) + \Theta(N-N_{s})\eta_{\rm II}(N_{s}\leq N\leq N_{e}) + \Theta(N-N_{e})\eta_{\rm III}(N_{e}\leq N\leq N_{\rm end}).
\eea
A similar method uses the definition of $\epsilon$ to determine the Hubble parameter, as shown below:
\bea
\epsilon(N) = -\frac{d\ln{H}}{dN} \implies H(N) = H_{i}\exp{\bigg(-\int_{N_{i}}^{N}\epsilon(N)dN\bigg)}.
\eea
Hence, for each phase, the Hubble expansion's initial condition value is $H_{i}\equiv H(N_{i})$. This allows the Hubble to be stated for each phase as follows:
\bea
H(N) = \left\{
	\begin{array}{ll}
		\displaystyle  H(N_{*})\exp{\bigg(-\int_{N_{*}}^{N}\epsilon(N)dN\bigg)} & \mbox{when}\quad  N_{*} \leq N \leq N_{s}  \;(\rm SRI),  \\  
			\displaystyle 
			\displaystyle H(N_{s})\exp{\bigg(-\int_{N_{s}}^{N}\epsilon(N)dN\bigg)} & \mbox{when}\quad  N_{s} \leq N \leq N_{e}  \;(\rm USR), \\ 
   \displaystyle 
			\displaystyle H(N_{e})\exp{\bigg(-\int_{N_{e}}^{N}\epsilon(N)dN\bigg)} & \mbox{when}\quad  N_{e} \leq N \leq N_{\rm end}  \;(\rm SRII).
	\end{array} \right.
\eea
Thus after combining the contributions for each phase independently, the final formula for the Hubble expansion as a function of the e-folds is as follows:
\bea
H(N) = H(N_{*})\exp{\bigg(-\int_{N_{*}}^{N}\epsilon(N)dN\bigg)} + H(N_{s})\exp{\bigg(-\int_{N_{s}}^{N}\epsilon(N)dN\bigg)} + H(N_{e})\exp{\bigg(-\int_{N_{e}}^{N}\epsilon(N)dN\bigg)},
\eea
where the starting conditions $H(N_{*}),\;H(N_{s}),\;H(N_{e})$ are derived from the continuity of values between the various phases.
\begin{figure*}[ht!]
    	\centering
    \subfigure[]{
      	\includegraphics[width=8.5cm,height=7.5cm]{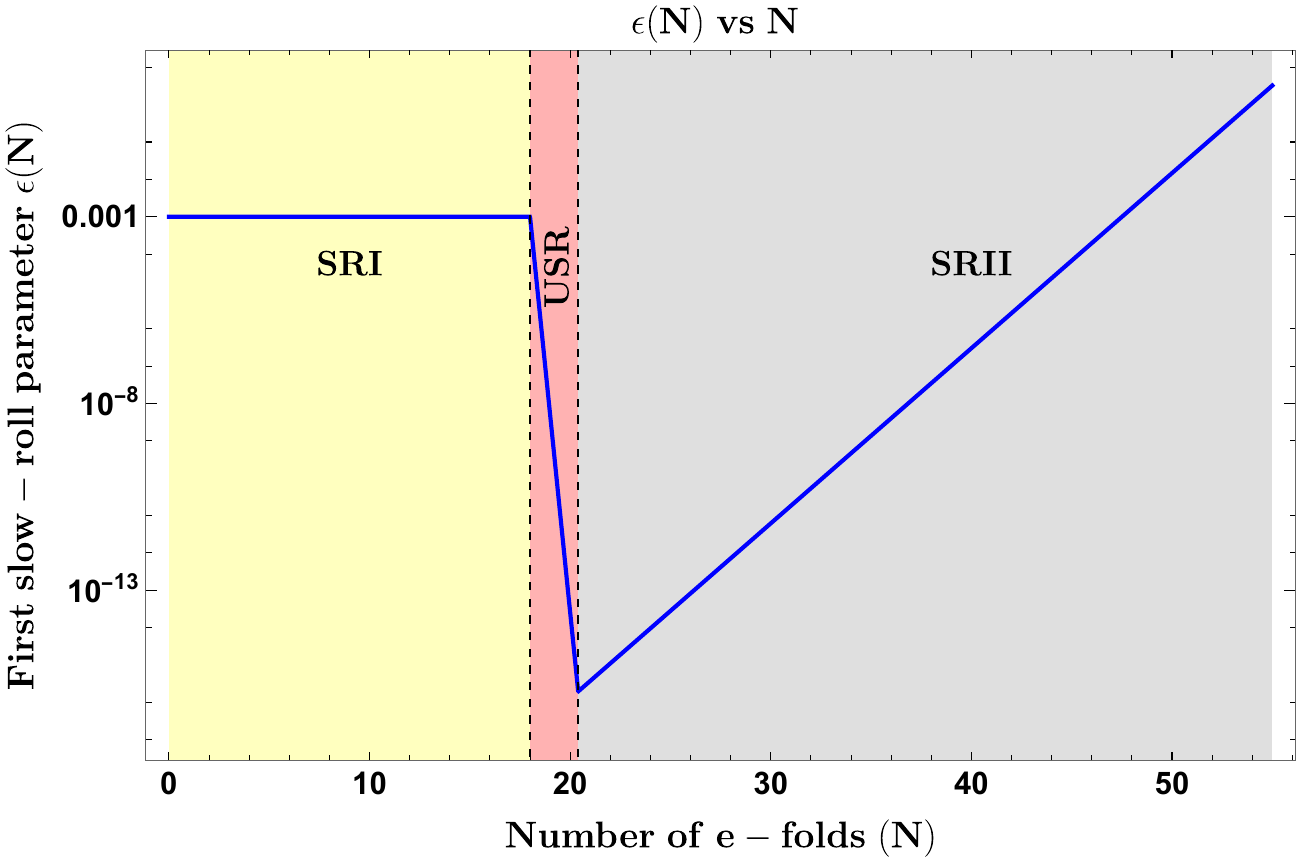}
        \label{epsilons}
    }
    \subfigure[]{
        \includegraphics[width=8.5cm,height=7.5cm]{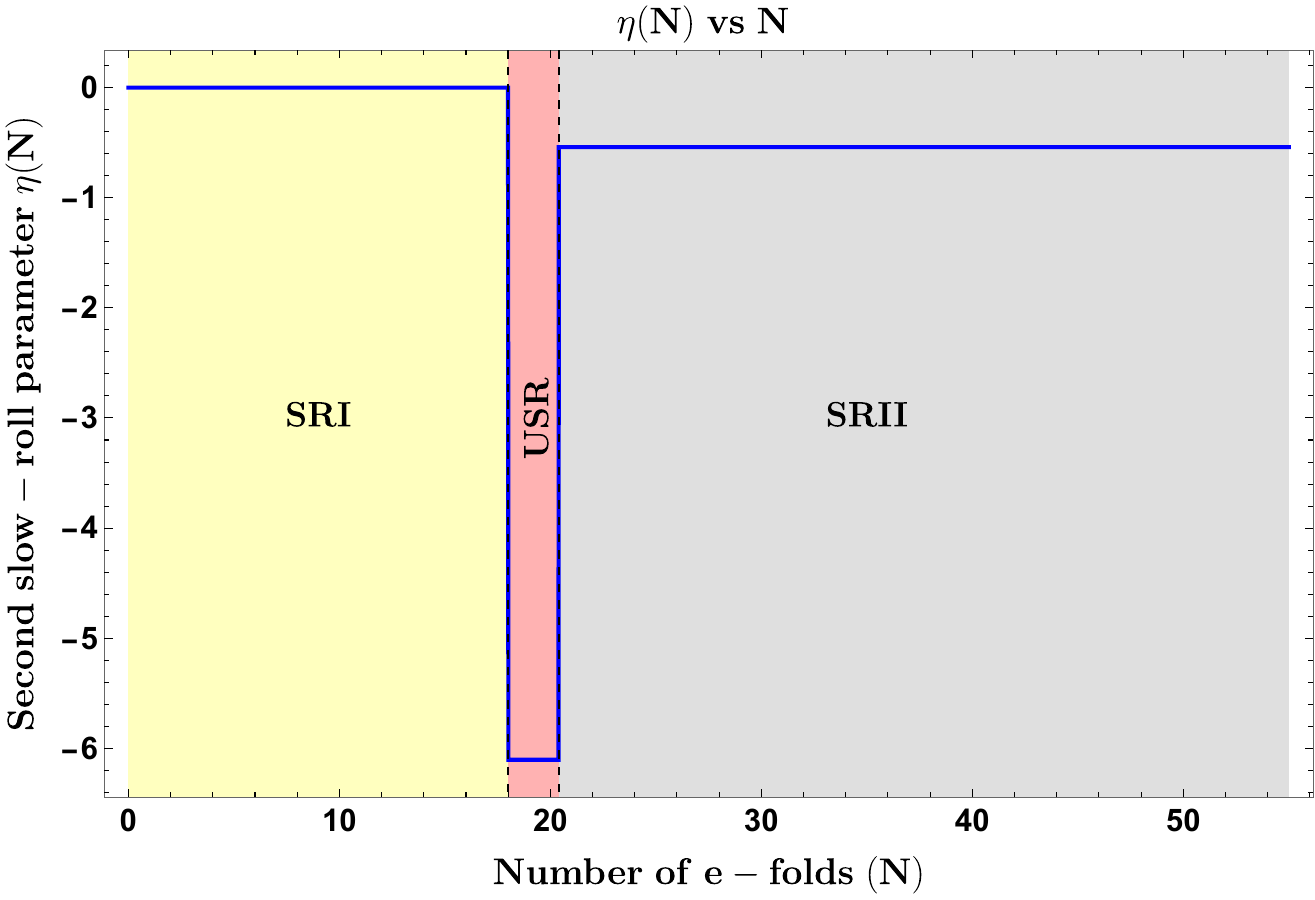}
        \label{etas}
    }
    \subfigure[]{
        \includegraphics[width=8.5cm,height=7.5cm]{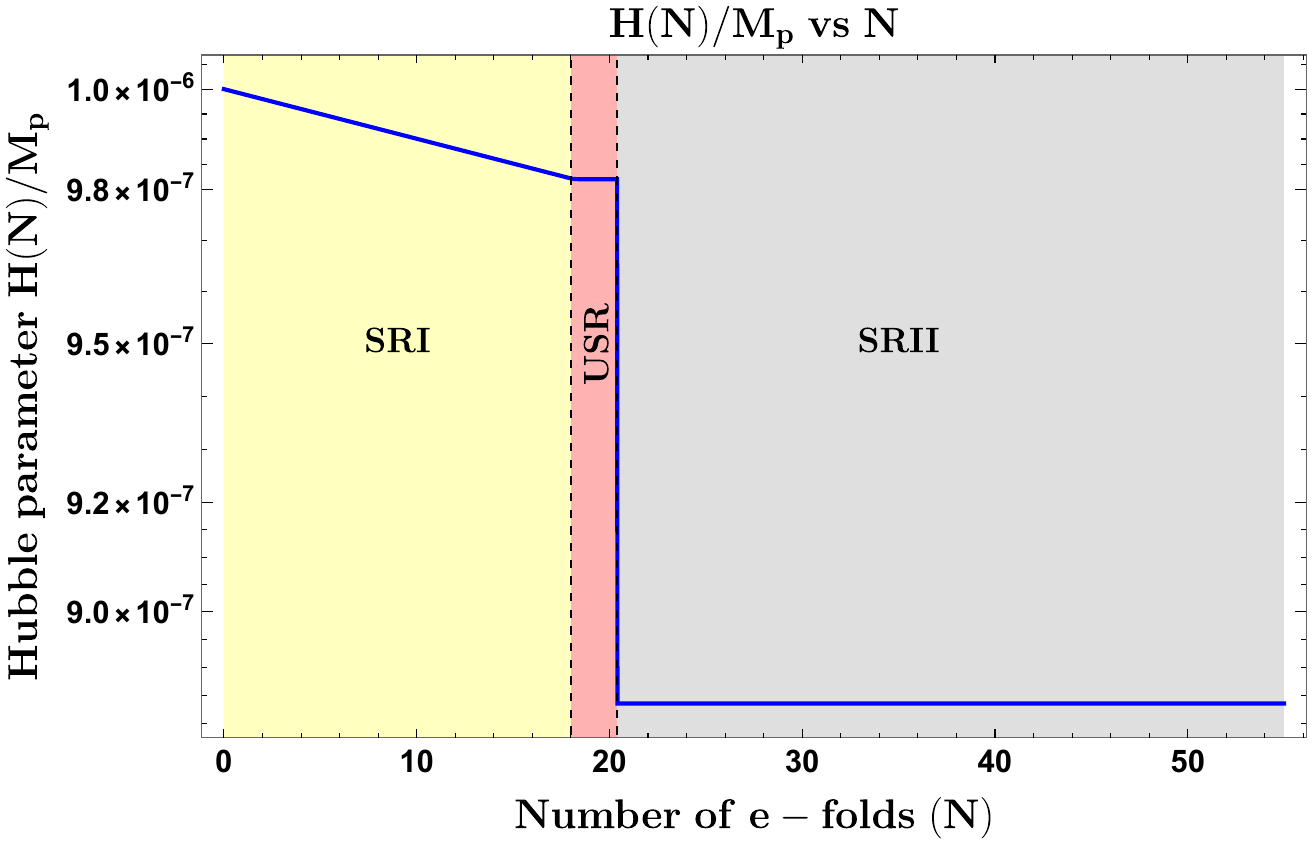}
        \label{hubbles}
    }
    \subfigure[]{
        \includegraphics[width=8.5cm,height=7.5cm]{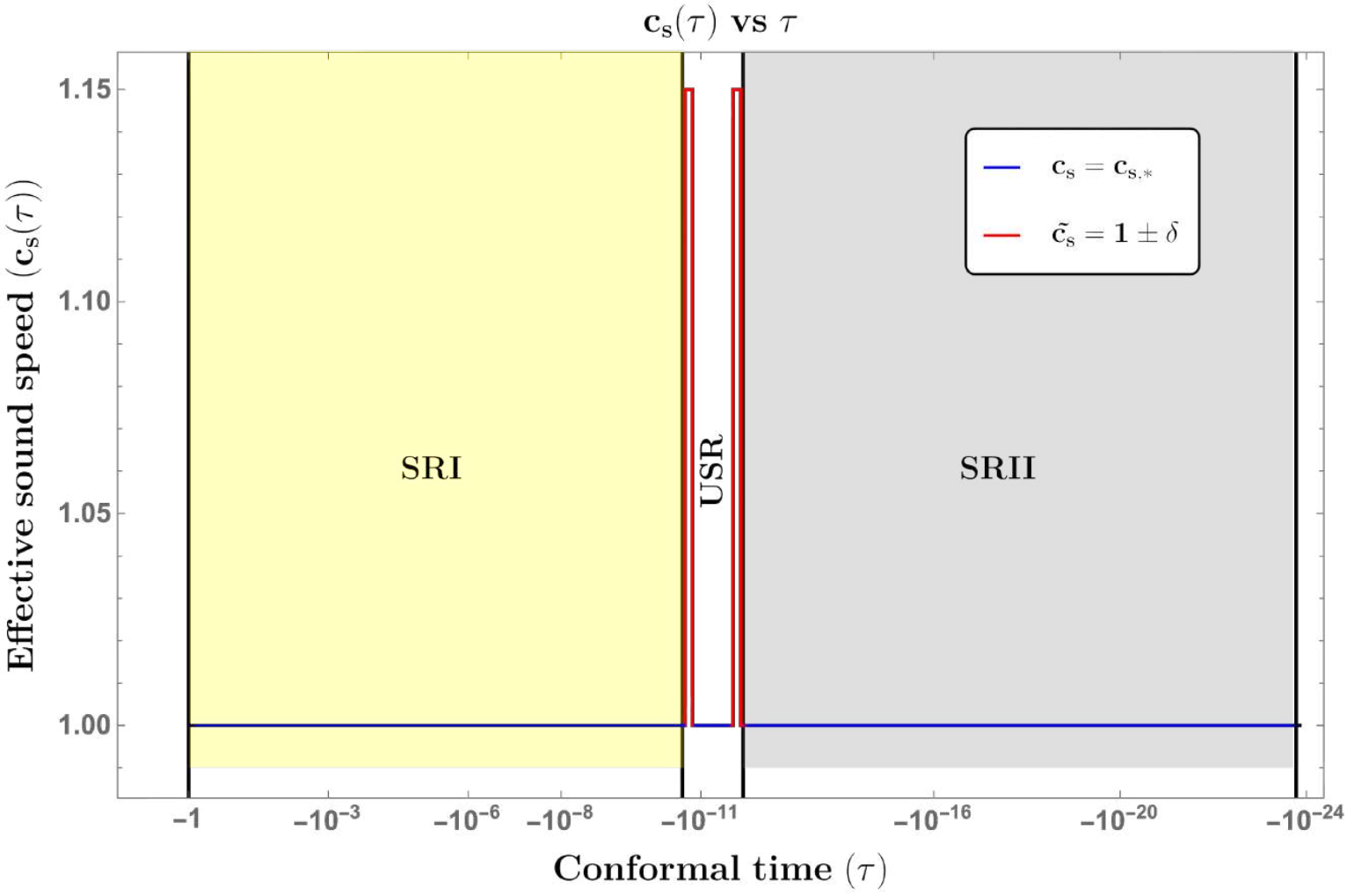}
        \label{sounds}
        }
    	\caption[Optional caption for list of figures]{As a function of the e-foldings $N$, the slow-roll and Hubble parameters are realised during inflation in the presence of a USR phase. The $\epsilon(N)$ parameter is displayed in the \textit{top-left} panel, the $\eta(N)$ parameter is displayed in the \textit{top-right} panel throughout the setup stages, and the Hubble parameter $H(N)$ is displayed in the \textit{bottom-left} panel throughout each phase. A schematic representing the effective sound speed $c_{s}$ during the three stages is shown in the \textit{bottom-right} panel.
 } 
    	\label{realisation}
    \end{figure*}

The graphs in figs. \ref{etas} and \ref{epsilons} are now followed. $\eta$ acts virtually like a negatively tiny and constant number at the beginning of the SRI, which begins at the moment $N_{*}$ and when slow-roll requirements are required to be satisfied. Similarly, $\epsilon$ likewise remains a small constant quantity. This continues until, after a certain number of e-folds $N_{s}$, a variation from conventional slow-roll arguments is encountered. The following phase begins at this instant, during which a severe breach of the slow-roll criterion may occur. For the sake of this work, the transition at $N_{s}$ is selected to be sharp in nature. It is represented by a Heaviside Theta function $\Theta(k-k_{s})$, where the associated transition wavenumber is $k_{s}$. Similar to how $\epsilon$ experiences a sharp decline in magnitude, $\eta$ likewise experiences a significant rise in magnitude in the USR. The aforementioned behaviour continues until the moment $N_{e}$, with the matching wavenumber $k_{e}$, marks the end of the USR duration. The slow-roll circumstances begin to return shortly after another abrupt transition of the kind $\Theta(k-k_{e})$ is discovered to occur. The SRII's parameters $\epsilon$ and $\eta$ begin to rise for the whole phase until they reach magnitudes of ${\cal O}(1)$. The inflationary phase ends at the corresponding instant in e-folds $N_{\rm end}$. The overall change in e-foldings from the starting moment $N_*$ must fall within the range $\Delta N_{\rm Total}=N_{\rm end}-N_{*}\sim {\cal O}(55-60)$  in order for inflation to be effective in the setup. The overall behaviour of the Hubble parameter in the dimensionless unit is plottedd in fig. \ref{hubbles}.

For the current setup, the effective sound speed $c_{s}$ parameterization is shown in the plot in fig. \ref{sounds}. In the first slow-roll phase, we decide to have $c_{s}(\tau=\tau_{*})=c_{s,*}=1$, where $c_{s,*}$ is the appropriate value at the pivot scale. A standard stochastic single-field inflation model is denoted by the condition $c_{s}=1$. When the effective sound speed approaches the point of sharp transition, at conformal time $\tau=\tau_{s}$ with $N=N_{s}$, it shifts from its previous value to a new one, denoted above as $\tilde{c_{s}}=1\pm \delta$, where a fine-tuning variable $\delta\ll 1$. The abrupt change in amplitude between the SRI and USR phases is what drives their abrupt changeover. The sound speed returns to $c_{s}$ throughout the remainder of the USR ($\tau_{s}\leq \tau \leq\tau_{e}$). Once again, we witness the value experiencing a sudden change to $c_{s}=\tilde{c_{s}}=1\pm \delta$ at the instant of the second sharp transition, $\tau=\tau_{e}$. Subsequently, the sound speed reverts to $c_{s}$ until the end of inflation is reached at the conformal time instant $\tau=\tau_{\rm end}$ or $N=N_{\rm end}$.

It is also possible to write the previously given parameterization of the effective sound speed for the three areas, including at the abrupt transition points, in the following way:
\bea
c_{s}(\tau) = \left\{
	\begin{array}{ll}
		\displaystyle  c_{s,*} & \mbox{when}\quad  \tau_{*} \leq \tau < \tau_{s}  \;(\rm SRI),  \\
        \displaystyle 
			\displaystyle \tilde{c_{s}} = 1\pm \delta & \mbox{when}\quad  \tau = \tau_{s}  \;(\rm USR),\\
			\displaystyle 
			\displaystyle c_{s,*} & \mbox{when}\quad  \tau_{s} < \tau < \tau_{e}  \;(\rm USR), \\
        \displaystyle 
			\displaystyle \tilde{c_{s}} = 1\pm \delta & \mbox{when}\quad  \tau = \tau_{e}  \;(\rm SRII),\\
        \displaystyle 
			\displaystyle c_{s,*} & \mbox{when}\quad  \tau_{e} < N < \tau_{\rm end}  \;(\rm SRII), 
	\end{array} \right.
\eea
where the value at the pivot scale instance, as shown in the parameterization explanation above, is $c_{s,*}=c_{s}(\tau=\tau_{*})$.

\subsection{Modes representing comoving  scalar curvature perturbation}

The main objective of this part is to leverage the underlying EFT in inflation formalism to create the solutions for the comoving curvature perturbation in a quasi de Sitter backdrop. To carefully investigate the behaviour of the mode solutions for the three phases that make up our setup of SR, USR, and SRII, we utilise the decoupling limit. Once the general formulae are known, they will be useful in building the different components of the scalar power spectrum by allowing correlations between the conjugate momentum variable and the comoving curvature perturbation to be found. 

This section examines the MS equation answers for each of the three separate inflationary stages in our scenario.
Subsequently, the provided solutions will be crucial for our Power Spectrum analysis and for determining the relevant noise matrix elements. The Mukhanov-Sasaki (MS) equation may be obtained by varying the action equation (\ref{s2zeta}), and solving it in Fourier space yields the curvature perturbation modes for various phases. The MS equation has the following form in the Fourier space:
\bea \label{MSfourier}
\bigg(\partial^2_{\tau}+2 \frac{z'(\tau)}{z(\tau)}\partial_{\tau}+c_s ^2 k^2\bigg)\zeta_{\bf k}(\tau) = 0.
\eea 
The {\it Mukhanov-Sasaki} variable, denoted by $z(\tau)$ in this case, may be found in the following expression:
\bea z(\tau):=\frac{a(\tau)\sqrt{2\epsilon}}{c_s}\quad\quad\quad {\rm where}\quad\quad \quad  a(\tau)=-\frac{1}{H\tau}\quad\quad\quad {\rm with} -\infty<\tau<0.\eea
The second order differential equation given above may now be solved by using the following findings, which are very helpful:
\bea  &&\frac{z'(\tau)}{z(\tau)}={\cal H}\left(1-\eta+\epsilon-s\right)=-\frac{1}{\tau}\left(1-\eta+\epsilon-s\right)\quad\quad\quad{\rm where}\quad\quad\quad s:=\frac{c^{'}_s}{{\cal H}c_s},\\
&&\frac{z''(\tau)}{z(\tau)}={\cal H}^2\left(1-\eta+\epsilon-s\right)^2+{\cal H}^{'}\left(1-\eta+\epsilon-s\right)=2{\cal H}^2\left(1-\frac{\eta}{2}+\frac{3}{2}\epsilon-\frac{3}{2}s\right)=\frac{1}{\tau^2}\left(\nu^2--\frac{1}{4}\right),\quad\quad\quad\eea
where the following statement defines the parameter $\nu$:
\bea \nu:=\frac{3}{2}+3\epsilon-\eta-3s.\eea
In this case, $\nu=3/2$ is the de Sitter limiting solution in the current context of discussion, and $s$ is an additional slow-roll parameter in each of the three phases.

\subsubsection{First Slow Roll phase (SRI)}

The following is the generic solution for a quasi de Sitter background and an arbitrary starting quantum vacuum state to the MS equation (refMSfourier) in the first slow-roll or SRI region:
\bea \label{modezetaSRI}
{\bf \zeta}_{\bf SRI}=\frac{2^{\nu-\frac{3}{2}} c_s H (-k c_s \tau )^{\frac{3}{2}-\nu}}{i \sqrt{2 \epsilon}(k c_s)^{\frac{3}{2}}\sqrt{2} M_p}\Bigg|\frac{\Gamma(\nu)}{\Gamma(\frac{3}{2})}\Bigg |\Bigg\{\alpha_1 (1+i k c_s\tau) e^{-i(kc_s\tau+\frac{\pi}{2}(\nu+\frac{1}{2}))}-\beta_1(1-i k c_s \tau)e^{i(k c_s\tau+\frac{\pi}{2}(\nu+\frac{1}{2}))}\Bigg\}.
\eea
One may derive the mode function of the canonically conjugate momentum by differentiating $\zeta_{\rm SRI}$:
\bea \label{modepiSRI}
{\bf \Pi_\zeta}_{\bf  SRI} = {\bf \zeta'}_{\bf SRI} = \frac{2^{\nu-\frac{3}{2}} c_s H (-k c_s \tau )^{\frac{3}{2}-\nu}}{i \tau  \sqrt{2 \epsilon}(k c_s)^{\frac{3}{2}}\sqrt{2} M_p }\Bigg|\frac{\Gamma(\nu)}{\Gamma(\frac{3}{2})}\Bigg|\Bigg [\alpha_1 \Bigg\{\Bigg(\frac{3}{2}-\nu \Bigg)(1+i k c_s \tau ) + k^2 c_s^2 \tau^2 \Bigg\}e^{-i(k c_s\tau+\frac{\pi}{2}(\nu+\frac{1}{2}))}\nonumber\\
-\beta_1 \Bigg\{\Bigg(\frac{3}{2}-\nu \Bigg)(1 - i k c_s \tau ) + k^2 c_s^2 \tau^2 \Bigg\} e^{i(k c_s\tau+\frac{\pi}{2}(\nu+\frac{1}{2}))}\Bigg],
\eea
where the Bogoliubov coefficients in the SRI area in the aforementioned solutions are $\alpha_1$ and $\beta_1$. These coefficients can be set by selecting an appropriate quantum vacuum state as beginning conditions. As long as $\tau < \tau_s$, the SRI phase continues. The Super Slow Roll (USR) is reached by the SRI at $\tau = \tau_s$, and the nature of this transition will be crucial for the remaining study. When it comes to SRI, $\epsilon$ is nearly constant and changes very slowly over time, whereas $\eta$, the second slow roll parameter, is extremely tiny and nearly constant:
\bea
&&\epsilon = -\frac{\dot{H}}{H^2} = \Bigg(1-\frac{\mathcal{H}'}{\mathcal{H}^2}\Bigg),\\
&& \eta = \epsilon-\frac{1}{2}\frac{\epsilon'}{\epsilon\mathcal{H}}.
\eea
In the SRI period, the relevant Bogoliubov coefficients are provided by the following formula if we select the well-known Euclidean quantum vacuum state, also known as the Bunch-Davies vacuum state:
 \bea
    \alpha_{1} = 1 ,\quad  \beta_{1} = 0.
    \eea 
    Upon replacing SRI with the specified Bogoliubov coefficient values, the expression for the comoving curvature perturbation in the initial vacuum state of Bunch Davies may be reformulated as follows:
\bea
{\bf \zeta}_{{\bf SRI},{\bf BD}}=\frac{2^{\nu-\frac{3}{2}} \, c_s \, H (-k c_s \tau )^{\frac{3}{2}-\nu}}{i \sqrt{2 \epsilon}(k c_s)^{3/2}\sqrt{2} M_p}\Bigg|\frac{\Gamma(\nu)}{\Gamma(\frac{3}{2})}\Bigg |(1+i k c_s\tau) e^{-i(kc_s\tau+\frac{\pi}{2}(\nu+\frac{1}{2}))}.
    \eea 
    The momentum that is canonically conjugate under the same Bunch-Davies beginning conditions is like this:
\bea
{\bf \Pi_\zeta}_{{\bf SRI},{\bf BD}} = 
{\bf \zeta'}_{{\bf SRI},{\bf BD}} = \frac{2^{\nu-\frac{3}{2}} c_s H (-k c_s \tau )^{\frac{3}{2}-\nu}}{i \tau  \sqrt{2 \epsilon}(k c_s)^{\frac{3}{2}}\sqrt{2} M_p }\Bigg|\frac{\Gamma(\nu)}{\Gamma(\frac{3}{2})}\Bigg|\Bigg\{\Bigg(\frac{3}{2}-\nu \Bigg)(1+i k c_s \tau ) + k^2 c_s^2 \tau^2 \Bigg\}e^{-i(k c_s\tau+\frac{\pi}{2}(\nu+\frac{1}{2}))}.
\eea
One may obtain the simplified formulations for the precise de Sitter space-time situation by extending the preceding obtained conclusion and applying the limiting case of $\nu = 3/2$. This simpler form is attained by the comoving curvature perturbation in this limit:
\bea
{\bf \zeta}_{{\bf SRI},{\bf dS}}=\frac{i c_s H }{\sqrt{2 \epsilon}(k c_s)^{\frac{3}{2}}\sqrt{2} M_p}(1+i k c_s\tau) e^{-ik c_s\tau}.
\eea
Likewise, the following is how the canonically conjugate momentum's precise de Sitter solution appears:
\bea 
{\bf \Pi_\zeta}_{{\bf SRI},{\bf dS}} = {\bf \zeta'}_{{\bf SRI},{\bf dS}} = \frac{i c_s H }{ \tau  \sqrt{2 \epsilon}(k c_s)^{\frac{3}{2}}\sqrt{2} M_p }\; k^2 c_s^2 \tau^2 \;e^{-ik c_s\tau}.
\eea 
\begin{figure*}[ht!]
    	\centering
    \subfigure[]{
      	\includegraphics[width=8.5cm,height=7.5cm]{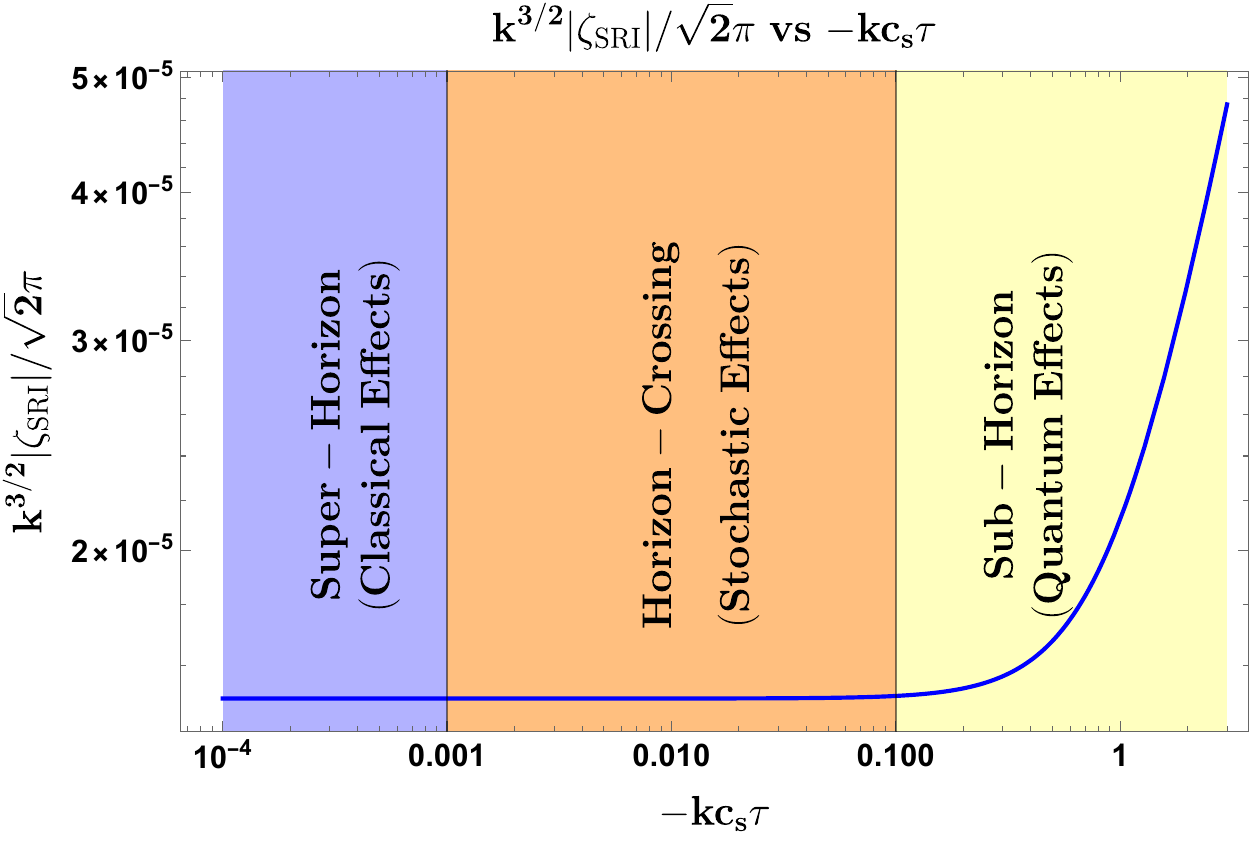}
        \label{zetasr1mode}
    }
    \subfigure[]{
        \includegraphics[width=8.5cm,height=7.5cm]{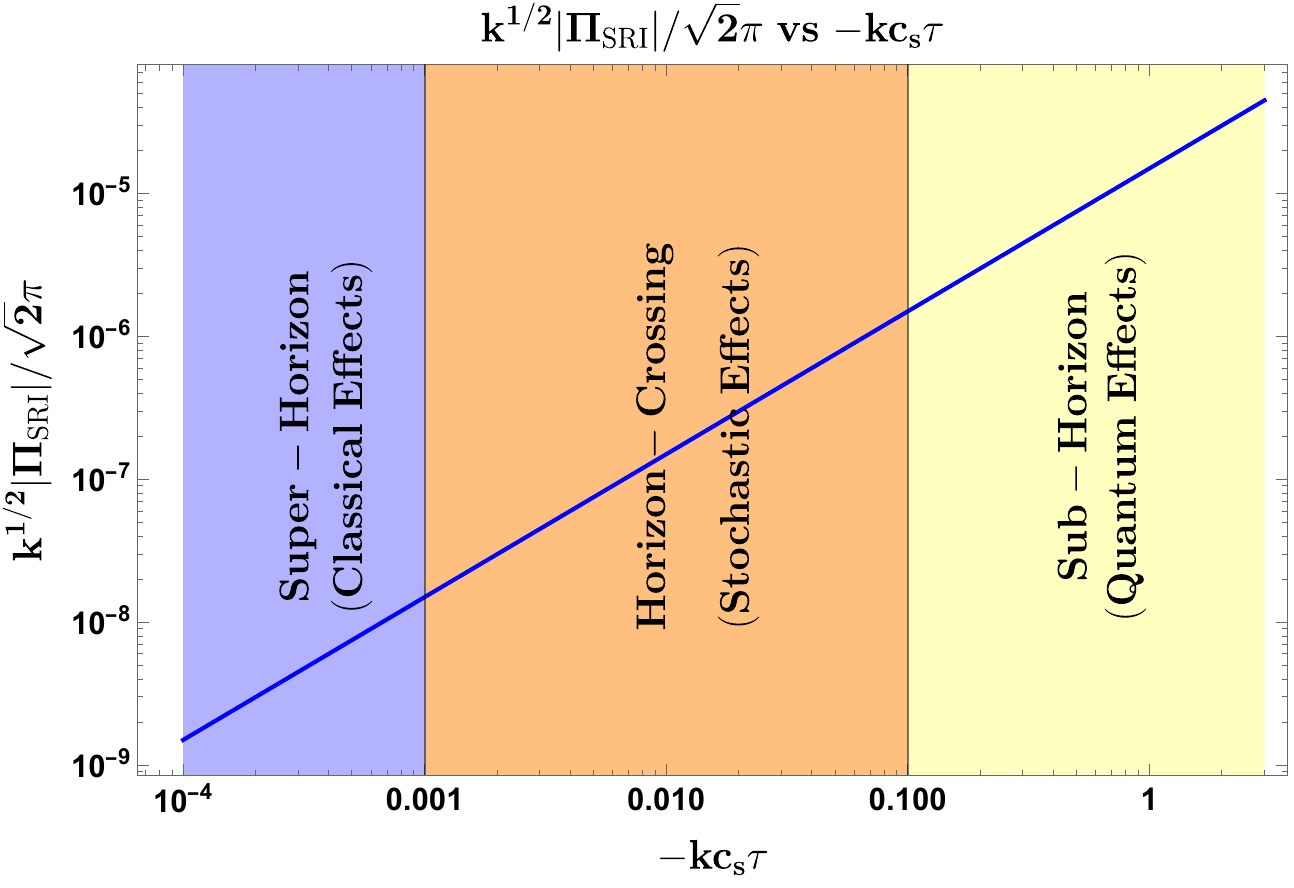}
        \label{pisr1mode}
    }
    	\caption[Optional caption for list of figures]{Behaviour of the conjugate momentum mode and curvature perturbation in the SRI as a function of $-kc_{s}\tau$. The development of $k^{3/2}|\zeta_{\rm SRI}|/\sqrt{2}\pi$ is displayed in the \textit{left-panel} following the selection of the Bunch-Davies initial conditions $(\alpha_{1}=1,\beta_{1}=0)$ and $\nu\sim 3/2$. In the same way, the evolution of $|k^{1/2}\Pi_{\rm SRI}|/\sqrt{2}\pi$ is shown in the \textit{right-panel}. The stochastic effects where $-kc_{s}\tau = \sigma \in (0.001,0.1)$ are shown by the orange coloured area. The coarse-graining factor and explanation for the modes' quantum-to-classical transition is provided by the stochastic parameter $\sigma$. The classical and quantum effects are highlighted by the yellow and blue shaded sections, respectively. 
} 
    	\label{sr1dSmodes}
    \end{figure*}
The development of the conjugate momenta of the scalar curvature perturbation mode and the dimensionless variable $-kc_{s}\tau$ are described in fig. \ref{sr1dSmodes}. Selecting $(\alpha_1=1,\beta_1=0)$ as the Bunch-Davies starting vacuum conditions and using $\nu\sim 3/2$ as the limiting case led to the solution that was employed. In \ref{zetasr1mode}, we see that the modulus $|\zeta_{\rm SRI}|$ for a certain mode behaves as follows while it is within the Horizon, its amplitude decreases as it approaches the border when stochastic influences begin to act. A coarse-graining factor, $\sigma$ ($\sigma\ll 1$), is added to the stochastic inflation formalism to create a region in the super-Horizon where short-wavelength modes transit and witness a quantum-to-classical transition before being finally treated as long-wavelength modes in the super-Horizon. In the super-Horizon, the number $|\zeta_{\rm SRI}|$ turns into a constant when classical influences take charge. $|\Pi_{\rm SRI}|$, a conjugate momenta-related quantity, is displayed in \ref{pisr1mode}. It shows a sharp decline in value over the course of the three quantum-to-classical transition regimes as the conjugate momenta mode ultimately becomes super-Horizon, where it is most suppressed.

\subsubsection{Ultra Slow Roll phase (USR)}

During the conformal time period $\tau_s \leq \tau \leq \tau_e$, the USR phase is in operation. Here, $\tau_s$ represents the time when SRI stops and USR begins, and $\tau_e$ represents the time when USR finishes and the subsequent SRII begins.  In terms of $\epsilon$, the slow-roll parameter $\epsilon $ in the USR is now highly suppressed and time-dependent, meaning that:
\bea \label{epsilonusr}
\epsilon(\tau) = \epsilon \Bigg(\frac{a(\tau_s)}{a(\tau )} \Bigg )^6 =  \epsilon \Bigg(\frac{\tau}{\tau_s} \Bigg)^6.
\eea 
The way that the second slow-roll parameter, $\eta(\tau)$, behaves in the USR over time is significant. It does so in the following ways:
\bea \label{etausr}
\eta(\tau)\sim \eta -6 \approx {\cal O}(-6).
\eea
It should be mentioned that we have thought about a sudden change from the SRI to the USR. Now, the following simplified form may be used to define the general solution for the MS equation (\ref{MSfourier}) in the USR area using an arbitrary quantum vacuum state:
\bea \label{modezetaUSR}
{\bf \zeta}_{\bf USR}=\frac{2^{\nu-\frac{3}{2}} c_s   H (-k c_s \tau )^{\frac{3}{2}-\nu}}{i \sqrt{2 \epsilon}(k c_s)^{\frac{3}{2}}\sqrt{2} M_p}\Bigg[\frac{\tau_s}{\tau}\Bigg]^3\Bigg|\frac{\Gamma(\nu)}{\Gamma(\frac{3}{2})}\Bigg |\Bigg\{\alpha_2 (1+i k c_s\tau) e^{-i(kc_s\tau+\frac{\pi}{2}(\nu+\frac{1}{2}))}-\beta_2(1-i k c_s \tau)e^{i(k c_s\tau+\frac{\pi}{2}(\nu+\frac{1}{2}))}\Bigg\}.\quad\quad
\eea
Additionally, by differentiating $\zeta_{\bf USR}$, the appropriate canonically conjugate momentum mode function may be constructed:
\bea 
\label{modepiUSR}
{\bf \Pi_\zeta}_{\bf  USR} = {\bf \zeta'}_{\bf USR} = \frac{2^{\nu-\frac{3}{2}} c_s  H (-k c_s \tau )^{\frac{3}{2}-\nu} 
 \tau_s^3 }{i  \sqrt{2 \epsilon}(k c_s)^{\frac{3}{2}}\sqrt{2} M_p  \, \tau^4} 
  \Bigg|\frac{\Gamma(\nu)}{\Gamma(\frac{3}{2})}\Bigg|\Bigg[ 
 \Bigg\{k^2 c_s^2 \tau^2 + (1+i k c_s \tau )\Bigg(\Bigg(\frac{3}{2}-\nu\Bigg)-3\Bigg)\Bigg\} \alpha_2 e^{-i(k c_s\tau+\frac{\pi}{2}(\nu+\frac{1}{2}))}\nonumber \\ 
 - \Bigg\{k^2 c_s^2 \tau^2 + (1- i k c_s \tau )\Bigg(\Bigg(\frac{3}{2}-\nu\Bigg)-3\Bigg)\Bigg\}\beta_2 e^{i(k c_s\tau+\frac{\pi}{2}(\nu + \frac{1}{2}))}\Bigg],\quad\quad
\eea
where the Bogoliubov coefficients in the USR area are $\alpha_2$ and $\beta_2$, which may be obtained in terms of $\alpha_1$ and $\beta_1$ by applying the two boundary conditions, which are read as Israel junction conditions at time scale $\tau = \tau_s$.

According to the first requirement, the scalar modes found are continuous at the abrupt transition point between SRI and USR, $\tau=\tau_s$:
\bea
[\zeta(\tau)]_{\bf SRI,\tau=\tau_s} = [\zeta(\tau)]_{\bf USR,\tau=\tau_s}.
\eea 
The second requirement suggests that momentum modes are continuous at the steep $\tau=\tau_s$ transition point between USR and SRI:
\bea
[\zeta'(\tau)]_{{\bf SRI},\, {\bf \tau=\tau_s}} = [\zeta'(\tau)]_{\bf USR, \, \tau=\tau_s}.
\eea 
In the USR area, we get two constraint equations for Bogoliubov coefficients by applying the two junction conditions mentioned above:
\bea \label{alpha2qds}
&& \alpha_2 = \frac{1}{2 k^3 \tau_s^3 c_s^3} \Bigg\{ \Bigg(3 i + 3 i  k^2 c_s^2 \tau_s ^2  + 2  k^3 c_s^3  \tau_s ^3  \Bigg  )\alpha_1 - \Bigg(3 i +6  k c_s \tau_s  -3 i k^2  c_s^2 \tau_s ^2    \Bigg) \beta_1 e^{i \left(2  k \tau_s  c_s +  \pi  \left(\nu +\frac{1}{2}\right)\right)}\Bigg\},
 \\
&&  \label{beta2qds} \beta_2 = \frac{1}{2 k ^3 c_s ^3 \tau_s ^3} \Bigg\{ \Bigg( 3i -6 k c_s \tau_s -3i k^2 c_s ^2 \tau_s^2 \Bigg) \alpha_1 e^{-i(\pi(\nu+\frac{1}{2})+ 2k c_s \tau_s)}- \Bigg(3i +3 i k^2 c_s ^2 \tau_s ^ 2 - 2 k^3c_s^3 \tau_s ^ 3  \Bigg )\beta_1
 \Bigg\}.
\eea 
When the USR Bogoliubov coefficients are fitted with the Bunch Davies starting vacuum state $(\alpha_1=1,, \beta_1=0)$, the $\alpha_2$ and $\beta_2$ are expressed in their simplified form:
\bea \label{alpha2bd}
&& \alpha_{{2},{\bf BD}} =  \frac{1}{2 k^3 \tau_s^3 c_s^3}  \Bigg(3 i + 3 i  k^2 c_s^2 \tau_s ^2  + 2  k^3 c_s^3  \tau_s ^3  \Bigg  ), \\
&& \label{beta2bd}  \beta_{{2},{\bf BD}} = \frac{1}{2 k ^3 c_s ^3 \tau_s ^3}  \Bigg( 3i -6 k c_s \tau_s -3i k^2 c_s ^2 \tau_s^2 \Bigg) e^{-i(\pi(\nu+\frac{1}{2})+ 2k c_s \tau_s)}.
\eea 
By extending the application of the limiting case $\nu=\frac{3}{2}$ in the Bunch Davies vacuum derived result, the simplified equation for USR Bogoliubov coefficients in the de Sitter space case can be obtained:
\bea  \label{alpha2ds}
&& \alpha_{{2},{\bf dS}} =  \frac{1}{2 k^3 \tau_s^3 c_s^3}  \Bigg(3 i + 3 i  k^2 c_s^2 \tau_s ^2  + 2  k^3 c_s^3  \tau_s ^3  \Bigg  ), \\
&& \label{beta2ds} \beta_{{2},{\bf dS}} = \frac{1}{2 k ^3 c_s ^3 \tau_s ^3}  \Bigg( 3i -6 k c_s \tau_s -3i k^2 c_s ^2 \tau_s^2 \Bigg) e^{- 2 i k c_s \tau_s}.
\eea 
The de Sitter result, which may also be expressed as follows, is derived from the mode function in the limiting situation $\nu=\frac{3}{2}$ in the Bunch Davies vacuum:
\bea
{\bf \zeta}_{{\bf USR},{\bf dS} }=\frac{i c_s   H }{ \sqrt{2 \epsilon}(k c_s)^{\frac{3}{2}}\sqrt{2} M_p}\Bigg[\frac{\tau_s}{\tau}\Bigg]^3\Bigg\{\alpha_{{2},{\bf dS}} (1+i k c_s\tau) e^{-ikc_s\tau}-\beta_{{2},{\bf dS}}(1-i k c_s \tau)e^{ik c_s\tau}\Bigg\}.
\eea
For the example given above, the momentum mode function is:
\bea
{\bf \Pi_\zeta}_{{\bf USR},{\bf dS} } = {\bf \zeta'}_{{\bf USR},{\bf dS} } = \frac{ i c_s  H  
 \tau_s^3 }{ \sqrt{2 \epsilon}(k c_s)^{\frac{3}{2}}\sqrt{2} M_p  \, \tau^4} 
  \Bigg[ 
 \Bigg\{k^2 c_s^2 \tau^2 -3(1+i k c_s \tau )\Bigg\} \alpha_{{2},{\bf dS}} e^{-ik c_s\tau}\nonumber \\
 - \Bigg\{k^2 c_s^2 \tau^2 -3 (1- i k c_s \tau )\Bigg\}\beta_{{2},{\bf dS}} e^{i k c_s\tau}\Bigg].
\eea 
\begin{figure*}[ht!]
    	\centering
    \subfigure[]{
      	\includegraphics[width=8.5cm,height=7.5cm]{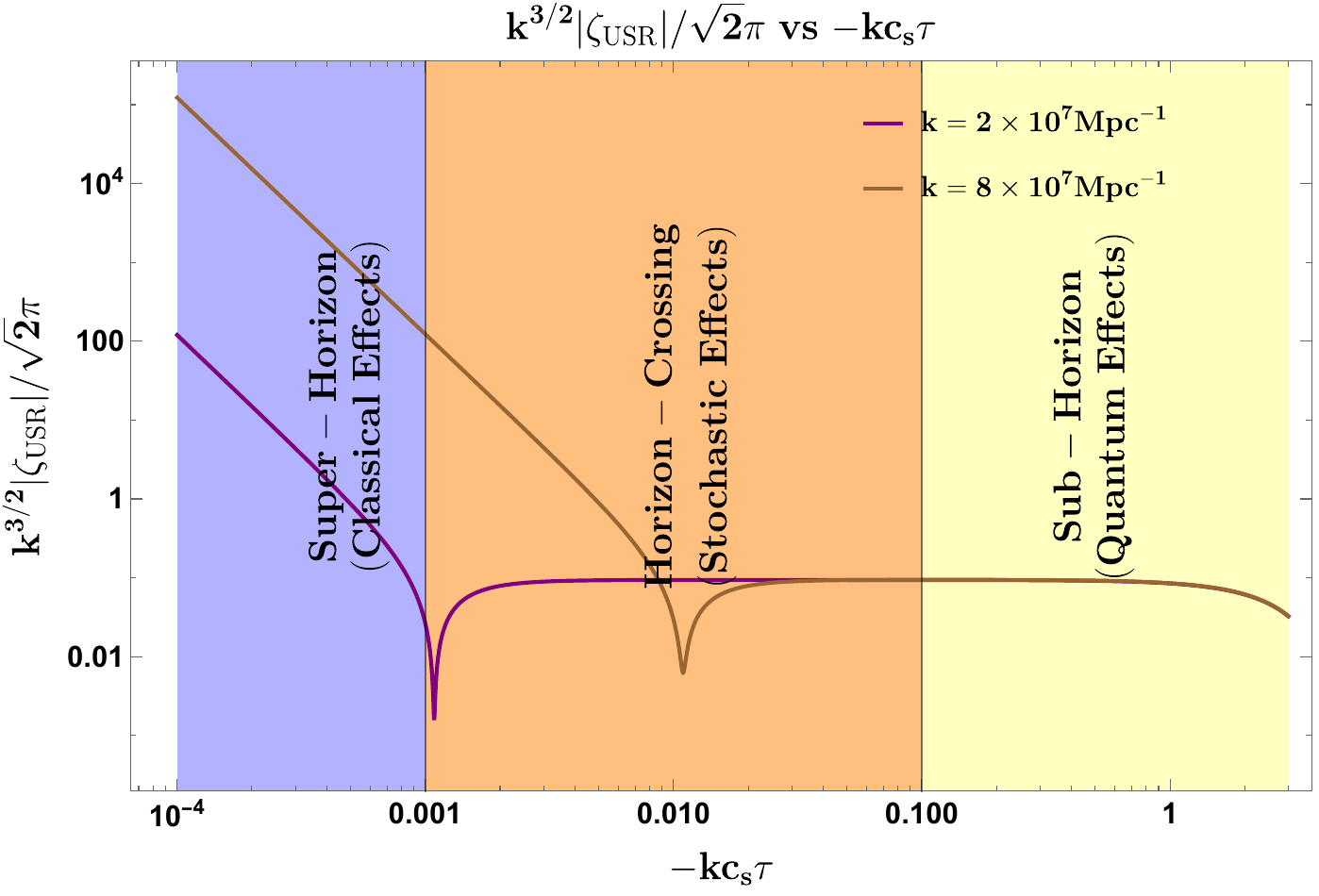}
        \label{zetausrmode}
    }
    \subfigure[]{
        \includegraphics[width=8.5cm,height=7.5cm]{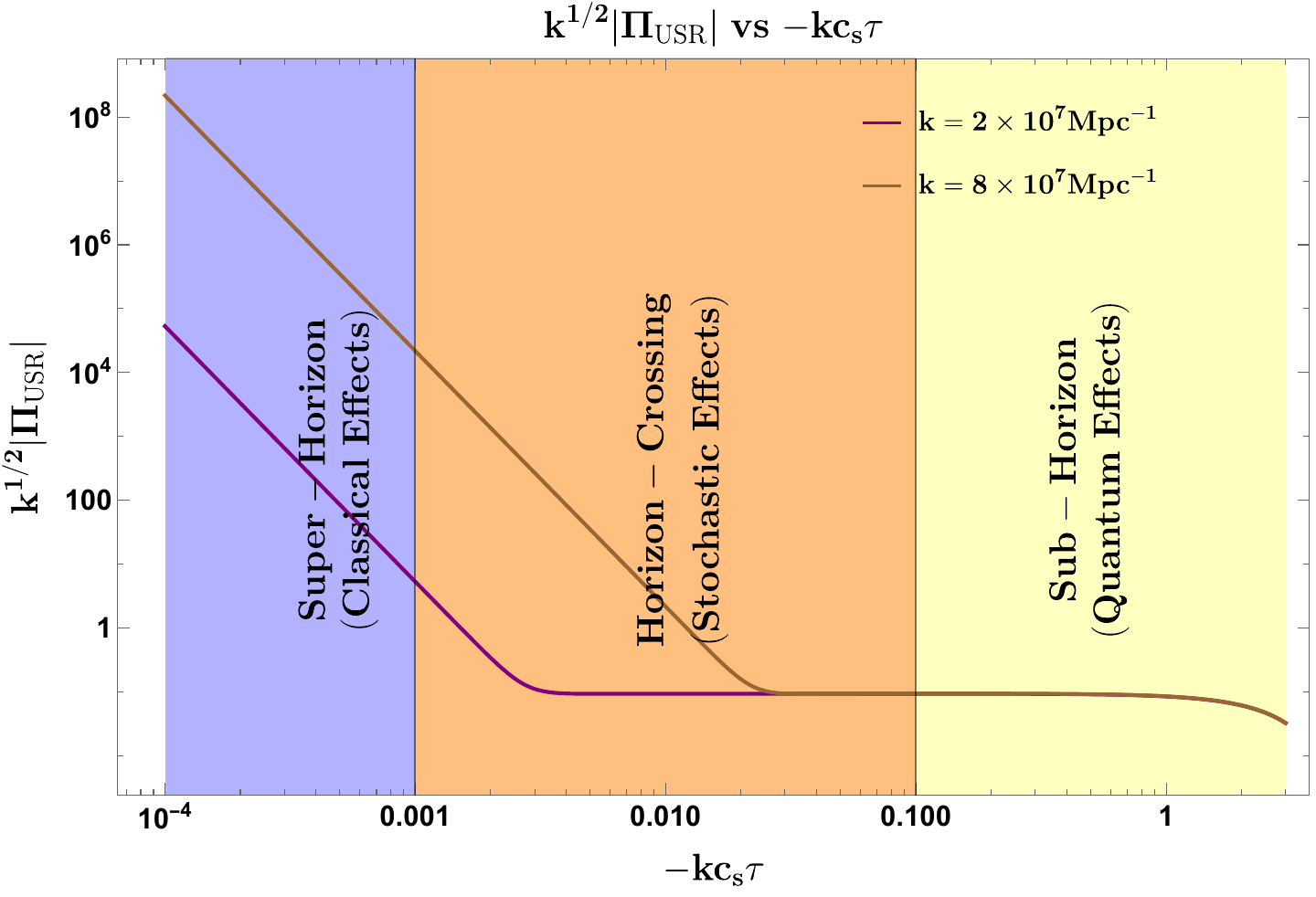}
        \label{piusrmode}
    }
    	\caption[Optional caption for list of figures]{Behaviour of the conjugate momentum mode and curvature perturbation in the USR as a function of $-kc_{s}\tau$. The development of $k^{3/2}|\zeta_{\rm USR}|/\sqrt{2}\pi$ is displayed in the \textit{left-panel} following the selection of the Bunch-Davies initial conditions $(\alpha_{2,{\bf dS}},\beta_{2,{\bf dS}})$ and $\nu\sim 3/2$. In the same way, the evolution of $|k^{1/2}\Pi_{\rm USR}|/\sqrt{2}\pi$ is displayed in the \textit{right-panel}. The stochastic effects where $-kc_{s}\tau = \sigma \in (0.001,0.1)$ are shown by the orange coloured area. When the modes change from quantum to classical near Horizon corssing, the stochastic parameter $\sigma$ serves as a coarse-graining parameter. The classical and quantum effects are highlighted by the yellow and blue shaded sections, respectively. 
} 
    	\label{usrdSmodes}
    \end{figure*}
The evolution of the conjugate momenta and the scalar curvature perturbation mode in the USR as a function of the dimensionless variable $-kc_{s}\tau$ are shown in figure \ref{usrdSmodes}. As in SRI, we proceed by selecting the Bunch-Davies starting condition for the Bogoliubov coefficients in this phase and choosing the limit $\nu\sim 3/2$. This gives rise to the coefficients $(\alpha_{2,{\bf dS}},\beta_{2,{\bf dS}})$; eqn. (\ref{alpha2ds},\ref{beta2ds}). Here, we graph the behaviour for various wavenumbers, where the mode solutions' use of the quantity $k_{s}c_{s}\tau_{s}$ introduces an extra momentum dependency. Following the selection of $k_{s}c_{s}\tau_{s}\sim {\cal O}(-0.01)$, the figure is displayed. We observe from \ref{zetausrmode} that while in the sub-Horizon, the magnitude $|\zeta_{\rm USR}|$ does not change. A variety of short-scale modes experience stochastic effects close to the Horizon crossover, and they go through a quantum-to-classical transition before ultimately arriving in the super-Horizon domain. While the quantity $|\zeta_{\rm USR}|$ tends to increase significantly in this regime, the strength of the modulus is sufficient to give $k^{3}|\zeta_{\rm USR}|^{2}\sim {\cal O}(10^{-3}-10^{-2})$ when considering the behaviour of the modes near the Horizon crossing, including stochastic effects. The evolution of the conjugate momenta mode in the USR is seen in fig. \ref{piusrmode}. This time, $|\Pi_{\rm USR}|$ is large, indicating that the conjugate momenta are important. In the presence of stochastic influences during Horizon crossing, the mode begins to intensify and continues monotonically as they get longer-wavelength in the super-Horizon scales.

\subsubsection{Second Slow Roll phase (SRII)}

Throughout the conformal time scale, the second slow roll phase continues. Whereas $\tau_e$ denotes the conclusion of the USR phase and the start of the SRII phase, $\tau_{\rm end}$ denotes the end of the inflationary paradigm. $\tau_e\leq\tau \leq \tau_{\rm end} $. The following describes the relationship between the first and second slow roll parameters:
\bea 
&& \epsilon(\tau)= \epsilon \Bigg (\frac{a(\tau_s)}{a(\tau_e)}\Bigg)^6 = \epsilon \Bigg(\frac{\tau_e}{\tau_s}\Bigg)^6\longrightarrow \epsilon(\tau=\tau_{end})=1,  \\
&& \eta(\tau) =\eta \longrightarrow \eta(\tau=\tau_{end})\sim\mathcal{O}(-1).
\eea 
in which the slow roll parameter in the SRI area is $\epsilon$.  We examined the abrupt change at the border between the SRI and USR in the part before; in this section, we also looked at the abrupt change from the USR to the SRII area. It is possible to derive the constraint relation for Bogoliubov coefficients in the SRII area by imposing the two boundary requirements known as the Israel junction condition. For SRII area, the mode solution is:
\bea \label{modezetaSR2}
{\bf \zeta}_{\bf SRII}=\frac{2^{\nu-\frac{3}{2}} c_s   H (-k c_s \tau )^{\frac{3}{2}-\nu}}{i \sqrt{2 \epsilon}(k c_s)^{\frac{3}{2}}\sqrt{2} M_p}\Bigg[\frac{\tau_s}{\tau_e}\Bigg]^3\Bigg|\frac{\Gamma(\nu)}{\Gamma(\frac{3}{2})}\Bigg |\Bigg\{\alpha_3 (1+i k c_s\tau) e^{-i\left (kc_s\tau+\frac{\pi}{2}(\nu+\frac{1}{2})\right )}-\beta_3(1-i k c_s \tau)e^{i\left(k c_s\tau+\frac{\pi}{2}(\nu+\frac{1}{2})\right)}\Bigg\}.\quad\quad
\eea
By differentiating $\zeta_{\bf SRII}$, one may derive the canonically conjugate momentum mode function, which can be represented as follows:
\bea \label{modepiSR2}
{\bf \Pi_\zeta}_{\bf  SRII} = {\bf \zeta'}_{\bf SRII} =  \frac{2^{\nu-\frac{3}{2}} c_s  H (-k c_s \tau )^{\frac{3}{2}-\nu}}{i \sqrt{2 \epsilon}(k c_s)^{\frac{3}{2}}\sqrt{2} M_p \, \tau } 
 \Bigg[\frac{\tau_s}{\tau_e}\Bigg]^3 \Bigg|\frac{\Gamma(\nu)}{\Gamma(\frac{3}{2})}\Bigg|\Bigg \{\alpha_3 \Bigg\{k^2 c_s^2 \tau^2 + (1+i k c_s \tau )\Bigg(\frac{3}{2}-\nu\Bigg)\Bigg\} e^{-i\left(k c_s\tau+\frac{\pi}{2}(\nu+\frac{1}{2})\right)}\nonumber \\
 -\beta_3 \Bigg\{k^2 c_s^2 \tau^2 + (1- i k c_s \tau )\Bigg(\frac{3}{2}-\nu\Bigg)\Bigg\} e^{i \left (k c_s\tau+\frac{\pi}{2}(\nu+\frac{1}{2})\right )}\Bigg\},\quad\quad
\eea
where the Bogoliubov coefficients are denoted by $\alpha_3$ and $\beta_3$. The Israel junction conditions, also known as boundary conditions, allow one to determine the constraint relation for these coefficients in terms of $\alpha_2$ and $\beta_2$. In contrast to USR, the fundamental vacuum state in SRII is different. There is a transition at $\tau=\tau_e$.

The scalar modes derived from USR and SRII are implied to be continuous at the abrupt transition point, $\tau = \tau_e$, by the first boundary condition:
\bea
[\zeta(\tau)]_{{\bf USR},{\tau=\tau_e}}= [\zeta(\tau)]_{{\bf SRII},{\tau=\tau_e}}.\eea
According to the second boundary condition, at the transition point, the momentum modes for the scalar perturbation derived from USR and SRII become continuous:
\bea
[\zeta'(\tau)]_{{\bf USR},{\bf \tau = \tau_e}}= [\zeta'(\tau)]_{{\bf SRII},{\bf \tau = \tau_e}}.
\eea
Following the imposition of these constraints, two constraint equations for Bogoliubov coefficients in the SRII region are obtained:
\bea \label{alpha3qds}
 && \alpha _3 =  \frac{1}{2 k^3 \tau_e^3 c_s^3}\Bigg\{\left(-3 i -3 i  k^2 \tau_e^2 c_s^2 +2  k^3 \tau_e^3 c_s^3 \right)\alpha _2 - \left(-3 i -6  k \tau_e c_s   +3 i  k^2 \tau_e^2 c_s^2 \right)\beta _2 e^{\left(2 i k \tau_e c_s+i \pi  \left(\nu +\frac{1}{2}\right)\right)}\Bigg\},\\
&&  \label{beta3qds}  \beta _3 =  \frac{1}{2 k^3 \tau_e^3 c_s^3}\Bigg\{ \left(-3 i  +6  k \tau_e c_s +3 i k^2 \tau_e^2 c_s^2 \right)\alpha _2  e^{-\left(2 i k \tau_e c_s + i \pi  \left(\nu +\frac{1}{2}\right)\right)}+ \left(2  k^3 \tau_e^3 c_s^3 +3 i  k^2 \tau_e^2 c_s^2 +3 i  \right)\beta _2\Bigg\}, 
\eea 
where the Bogoliubov coefficients in the USR area are denoted by $\alpha_2$ and $\beta_2$, respectively, in eqs. (\ref{alpha2qds}) and (\ref{beta2qds}). Changing these values in equations (\ref{beta3qds}) and (\ref{alpha3qds}):
\bea
  \alpha _3 &=&  \frac{1}{(2 k^3 \tau_e^3 c_s^3)(2 k^3 \tau_s^3 c_s^3)}\Bigg\{\left(-3 i -3 i  k^2 \tau_e^2 c_s^2 +2  k^3 \tau_e^3 c_s^3 \right) \Bigg\{ \left(3 i + 3 i  k^2 c_s^2 \tau_s ^2  + 2  k^3 c_s^3  \tau_s ^3   \right )\alpha_1   - \left(3 i +6  k c_s \tau_s  -3 i k^2  c_s^2 \tau_s ^2   \right)
 \nonumber\\ 
&& \quad\quad\quad\quad\quad\quad\quad\quad\quad\quad\quad \times \beta_1  e^{i \left(2  k \tau_s  c_s +  \pi  \left(\nu +\frac{1}{2}\right)\right)}\Bigg\}
 \quad  - \left(-3 i -6  k \tau_e c_s   +3 i  k^2 \tau_e^2 c_s^2 \right) \quad \Bigg\{ \left( 3i -6 k c_s \tau_s -3i k^2 c_s ^2 \tau_s^2 \right)  
 \nonumber \\
  &&\quad\quad\quad \quad\quad\quad \quad\quad\quad\quad\quad \times \alpha_1 e^{-i(\pi(\nu+\frac{1}{2})+ 2k c_s \tau_s)} - \left (3i +3 i k^2 c_s ^2 \tau_s ^ 2 - 2 k^3c_s^3 \tau_s ^ 3  \right )\beta_1 \quad 
 \Bigg\} e^{\left(2 i k \tau_e c_s+i \pi  \left(\nu +\frac{1}{2}\right)\right)}\Bigg\},
 \\
 \beta_3 &=& \frac{1}{(2 k^3 \tau_e^3 c_s^3)(2 k^3 \tau_s^3 c_s^3)}\Bigg\{ \left(-3 i  +6  k \tau_e c_s +3 i k^2 \tau_e^2 c_s^2\right) \Bigg\{ \left(3 i + 3 i  k^2 c_s^2 \tau_s ^2  + 2  k^3 c_s^3  \tau_s ^3  \right)\alpha_1 - \left(3 i +6  k c_s \tau_s  -3 i k^2  c_s^2 \tau_s ^2    \right) \nonumber \\
 && \quad \quad\quad\quad\quad\quad\quad \quad\quad\times \beta_1 e^{i \left(2  k \tau_s  c_s +  \pi  \left(\nu +\frac{1}{2}\right)\right)}\Bigg\}  e^{-\left(2 i k \tau_e c_s + i \pi  \left(\nu +\frac{1}{2}\right)\right)}+ \left(2  k^3 \tau_e^3 c_s^3 +3 i  k^2 \tau_e^2 c_s^2 +3 i  \right )  \nonumber \\
 &&  \quad \quad\quad\quad\quad\quad\quad\quad\quad\times\Bigg\{ \left( 3i -6 k c_s \tau_s -3i k^2 c_s ^2 \tau_s^2 \right) \alpha_1 e^{-i\left(\pi(\nu+\frac{1}{2})+ 2k c_s \tau_s\right)}
 - \left(3i +3 i k^2 c_s ^2 \tau_s ^ 2 - 2 k^3c_s^3 \tau_s ^ 3  \right )\beta_1
 \Bigg\}\Bigg\}.
\quad\quad\quad \eea 
By incorporating the Bunch Davies initial vacuum state $(\alpha_1 = 1, \beta_1 = 0)$ into the SRII Bogoliubov coefficients, we can obtain the simplified form of $\alpha_3$ and $\beta_3$. However, to do so, we must substitute $\alpha_2$ and $\beta_2$ from eqs. (\ref{alpha2bd}) and (\ref{beta2bd}), respectively, which have already been solved in the Bunch Davies initial vacuum state:
\bea
 \alpha _{3, {\bf BD }} &=& \frac{1}{(2 k^3 \tau_e^3 c_s^3)(2 k^3 \tau_s^3 c_s^3)}\Bigg\{\left(-3 i -3 i  k^2 \tau_e^2 c_s^2 +2  k^3 \tau_e^3 c_s^3 \right) \left(3 i + 3 i  k^2 c_s^2 \tau_s ^2  + 2  k^3 c_s^3  \tau_s ^3  \right  )  \nonumber\\ &&
\quad\quad\quad\quad\quad\quad\quad\quad\quad\quad - \left(-3 i -6  k \tau_e c_s   +3 i  k^2 \tau_e^2 c_s^2 \right)
 \left( 3i -6 k c_s \tau_s -3i k^2 c_s ^2 \tau_s^2 \right) e^{2 i k c_s( \tau_e -\tau_s) }\Bigg\},
 \\
 \beta _{3,{\bf BD}} &=&  \frac{1}{(2 k^3 \tau_e^3 c_s^3)(2 k ^3 c_s ^3 \tau_s ^3)}\Bigg\{ \left(-3 i  +6  k \tau_e c_s +3 i k^2 \tau_e^2 c_s^2\right) \left(3 i + 3 i  k^2 c_s^2 \tau_s ^2  + 2  k^3 c_s^3  \tau_s ^3  \right) e^{-\left(2 i k \tau_e c_s + i \pi  \left(\nu +\frac{1}{2}\right)\right)} \nonumber \\ 
&& +\left(2  k^3 \tau_e^3 c_s^3 + 3 i  k^2 \tau_e^2 c_s^2 +3 i \right )  \left( 3i -6 k c_s \tau_s -3i k^2 c_s ^2 \tau_s^2 \right) e^{-i\left(\pi(\nu+\frac{1}{2})+ 2k c_s \tau_s \right)}\Bigg\}.
\eea 
By extending the use of the limiting case $\nu=\frac{3}{2}$ in the Bunch Davies vacuum derived result, the simplified equation for SRII Bogoliubov coefficients in the de Sitter space case may be obtained:
\bea \label{alpha3ds}
 \alpha _{{3},{\bf dS}} &=&  \frac{1}{(2 k^3 \tau_e^3 c_s^3)(2 k^3 \tau_s^3 c_s^3)}\Bigg\{\left(-3 i -3 i  k^2 \tau_e^2 c_s^2 +2  k^3 \tau_e^3 c_s^3 \right)   \left(3 i + 3 i  k^2 c_s^2 \tau_s ^2  + 2  k^3 c_s^3  \tau_s ^3  \right ) \nonumber \\ &&
\quad\quad\quad\quad\quad\quad\quad\quad\quad\quad - \left(-3 i -6  k \tau_e c_s   +3 i  k^2 \tau_e^2 c_s^2 \right) 
  \left( 3i -6 k c_s \tau_s -3i k^2 c_s ^2 \tau_s^2 \right ) e^{ i 2  k c_s( \tau_e-\tau_s)} \Bigg\},
  \\ \label{beta3ds}
\beta _{{3},{\bf dS}} &=&  \frac{1}{(2 k^3 \tau_e^3 c_s^3)(2 k^3 \tau_s^3 c_s^3 )}\Bigg\{ \left(-3 i  +6  k \tau_e c_s +3 i k^2 \tau_e^2 c_s^2\right) \left(3 i + 3 i  k^2 c_s^2 \tau_s ^2  + 2  k^3 c_s^3  \tau_s ^3  \right  )  e^{-2 i k \tau_e c_s }\nonumber\\ &&   
 \quad \quad\quad\quad\quad\quad\quad\quad\quad\quad 
 + \left(2  k^3 \tau_e^3 c_s^3 +3 i  k^2 \tau_e^2 c_s^2 +3 i  \right) \left( 3 i - 6 k c_s \tau_s -3i k^2 c_s ^2 \tau_s^2 \right) e^{- 2 i k c_s \tau_s}\Bigg\}.
\eea 
In the limiting scenario $\nu=\frac{3}{2}$ in the Bunch Davies vacuum, the mode function yields the de Sitter result, which is further expressed as follows: 
\bea
{\bf \zeta}_{{\bf SRII},{\bf dS}}=\frac{ i c_s   H }{ \sqrt{2 \epsilon}(k c_s)^{\frac{3}{2}}\sqrt{2} M_p}\Bigg[\frac{\tau_s}{\tau_e}\Bigg]^3\Bigg\{\alpha_{{3},{\bf dS}} (1+i k c_s\tau ) e^{-i kc_s\tau}-\beta_{{3},{\bf dS}}(1-i k c_s \tau )e^{i k c_s\tau}\Bigg\}.
\eea
In the limiting scenario $\nu=\frac{3}{2}$ in the Bunch Davies vacuum, the canonically conjugate momentum mode function yields the de Sitter conclusion, which is further expressed as follows:
\bea
{\bf \Pi_\zeta}_{{\bf  SRII},{\bf dS}} = {\bf \zeta'}_{{\bf  SRII},{\bf dS}} =  \frac{i c_s  H }{ \sqrt{2 \epsilon}(k c_s)^{\frac{3}{2}}\sqrt{2} M_p \, \tau } 
 \Bigg[\frac{\tau_s}{\tau_e}\Bigg]^3 (k^2 c_s^2 \tau^2 )\Bigg \{\alpha_{{3}, {\bf dS}}e^{-ik c_s \tau}
 -\beta_{{3},{\bf dS}}  e^{i k c_s \tau }\Bigg\}.
\eea

\begin{figure*}[ht!]
    	\centering
    \subfigure[]{
      	\includegraphics[width=8.5cm,height=7.5cm]{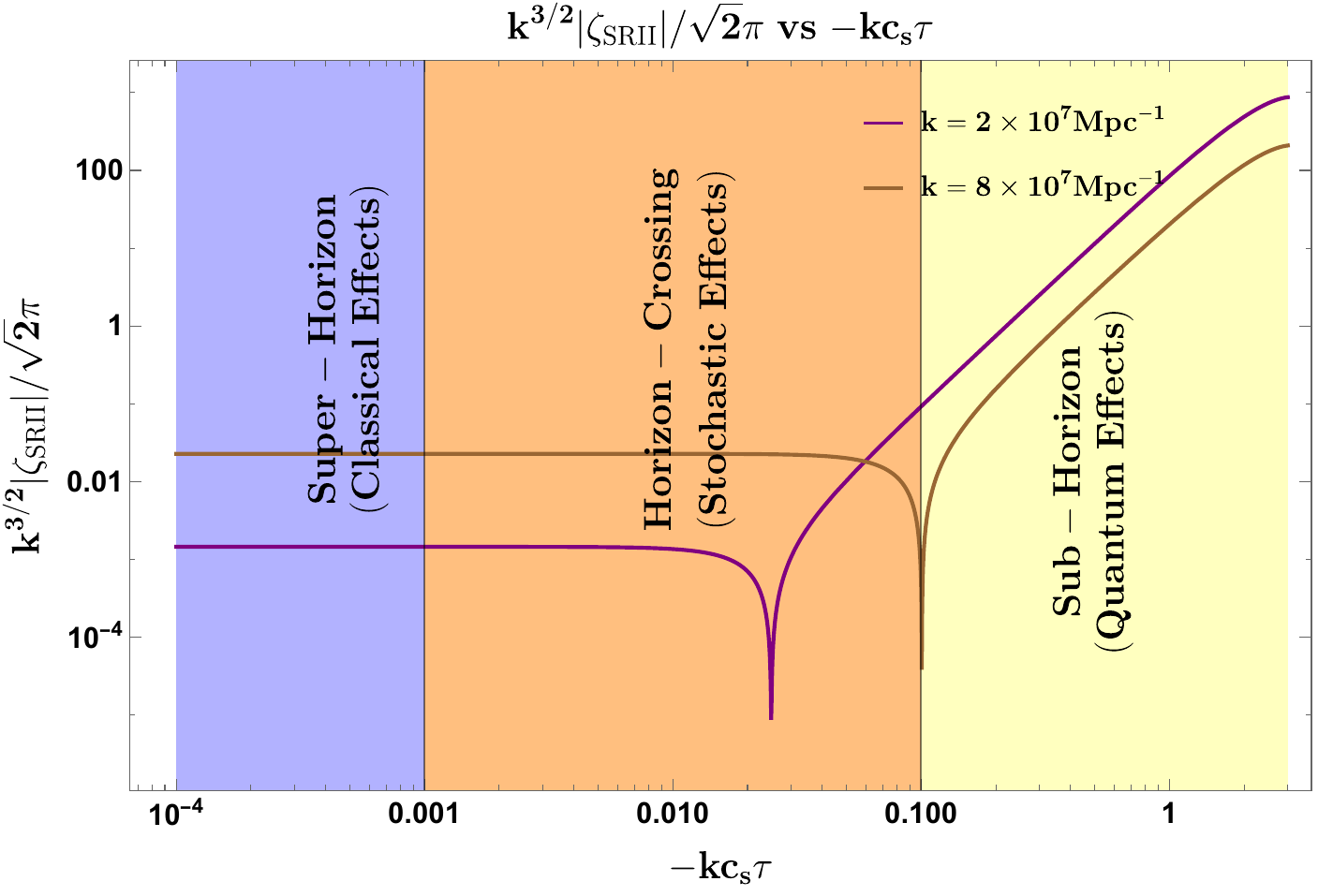}
        \label{zetasr2mode}
    }
    \subfigure[]{
        \includegraphics[width=8.5cm,height=7.5cm]{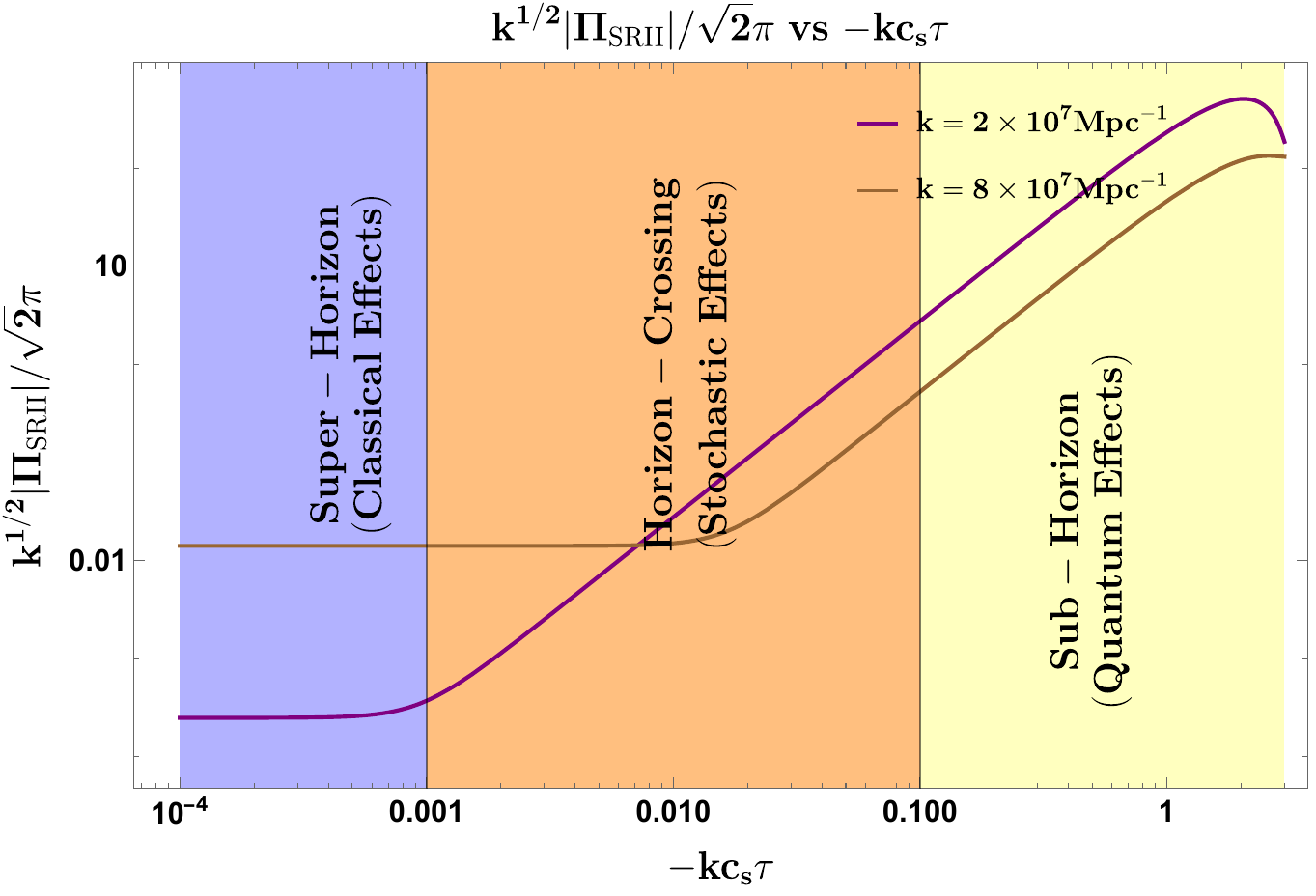}
        \label{pisr2mode}
    }
    	\caption[Optional caption for list of figures]{Behaviour of the conjugate momentum mode and curvature perturbation in the SRII as a function of $-kc_{s}\tau$. The development of $k^{3/2}|\zeta_{\rm SRII}|/\sqrt{2}\pi$ is displayed in the \textit{left-panel} following the selection of the Bunch-Davies initial conditions $(\alpha_{3,{\bf dS}},\beta_{3,{\bf dS}})$ and $\nu\sim 3/2$. In the same way, the evolution of $|k^{1/2}\Pi_{\rm SRII}|/\sqrt{2}\pi$ is shown in the \textit{right-panel}. The stochastic effects where $-kc_{s}\tau = \sigma \in (0.001,0.1)$ are shown by the orange coloured area. When the modes change from quantum to classical near Horizon corssing, the stochastic parameter $\sigma$ serves as a coarse-graining parameter. The classical and quantum effects are highlighted by the yellow and blue shaded sections, respectively.
 } 
    	\label{sr2dSmodes}
    \end{figure*}

The evolution of the conjugate momenta and the scalar curvature perturbation mode in the SRII as a function of the dimensionless variable $-kc_{s}\tau$ is seen in figure \ref{sr2dSmodes}. We also use the limiting case of de Sitter $\nu\sim 3/2$ and the same Bunch-Davies starting conditions here, which yields the Bogoliubov Bogoliubov coefficients $(\alpha_{3,{\bf dS}},\beta_{3,{\bf dS}})$; see eqn. (\ref{alpha3ds},\ref{beta3ds}). Quantities of the type $k_{s}c_{s}\tau_{s}$ and $k_{e}c_{s}\tau_{e}$, which we have selected to be of order ${\cal O}(-0.1)$, are what cause the extra wavenumber impact when they are included in the mode solutions. The number $|\zeta_{\rm SRII}|$ has a bigger magnitude while in the sub-Horizon regime, as can be observed from \ref{zetasr2mode}. As the intensity gets closer to the Horizon-crossing, the significance of the stochastic effects increases and the magnitude eventually stabilises. The mode undergoes a quantum-to-classical transition and then enters the super-Horizon phase, where it remains constant throughout, like SRI only with a larger magnitude of $|\zeta_{\rm SRII}|$. The development of the associated conjugate momenta is seen in fig. \ref{pisr2mode}. The magnitude of $|\Pi_{\rm SRII}|$ is sufficiently big while inside the Horizon, but it is still modest in comparison to $|\zeta_{\rm SRII}|$, and it keeps becoming smaller even when stochastic effects are present. As a mode moves from the quantum to the classical regime, its strength decays at varying rates. The strength of $|\Pi_{\rm SRII}|$ assumes constant values and remains constant until the modes enter the super-Horizon phase.

\subsection{ Tree level Power Spectrum and Noise Matrix Elements }
The appropriate tree-level contribution to the two-point cosmic correlation function for comoving scalar curvature disturbance may be represented as follows: $-kc_{s}\tau \rightarrow \sigma$. This comoving curvature perturbation occurs at a late time scale:
\bea
\langle\hat{\zeta_{\bf k}} \hat{\zeta_{\bf k'}}\rangle_{\bf Tree} &=& (2 \pi)^3 \, \del^3({\bf k}+{\bf k'})P_{\zeta\zeta} ^{\bf Tree}(k), \\
\langle\hat{\zeta_{\bf k}}\hat{\Pi_{{\zeta}{\bf k'}}}\rangle_{\bf Tree}&=& (2 \pi)^3 \, \del^3({\bf k}+{\bf k'})P_{\zeta \Pi_{\zeta}} ^ {\bf Tree}(k), \\
\langle\hat{\Pi_{{\zeta}{\bf k'}}}\hat{\zeta_{\bf k}}\rangle_{\bf Tree}&=& (2 \pi)^3 \, \del^3({\bf k}+{\bf k'})P_{\Pi_{\zeta} \zeta } ^ {\bf Tree}(k), \\
\langle\hat{\Pi_{{\zeta}{\bf k}}}\hat{\Pi_{{\zeta}{\bf k'}}}\rangle_{\bf Tree}&=& (2 \pi)^3 \, \del^3({\bf k}+{\bf k'})P_{\Pi_{\zeta}} ^ {\bf Tree}(k).
\eea 
In this instance, $P_{\zeta \Pi_\zeta}^{\bf Tree}(k),\, P_{\Pi_\zeta \zeta}^{\bf Tree}(k),\, \, P_{\Pi_\zeta \zeta}^{\bf Tree}(k),\, \, P_{\Pi_\zeta \Pi_\zeta}^{\bf Tree}(k)$ denotes a dimensionful power spectrum in Fourier space, which is assessable using: 
\bea
 P_{\zeta \zeta}^{\bf Tree}(k) &=& 
\langle\hat{\zeta_{\bf k}} \hat{\zeta_{\bf -k}}\rangle_{(0,0)} = [\zeta_{\bf k}(\tau) \zeta_{\bf -k}(\tau) ]_{-kc_{s}\tau \rightarrow \sigma}= |\zeta_{\bf k}(\tau)|_{-kc_{s}\tau \rightarrow \sigma }^2 \\
 P_{\zeta \Pi_\zeta}^{\bf Tree}(k)&=& \langle\hat{\zeta_{\bf k}}\hat{\Pi_{\zeta_{\bf -k}}}\rangle_{(0,0)}=[\zeta_{\bf k}(\tau ) \Pi_{\zeta_{\bf -k}}(\tau)]_{-kc_{s}\tau \rightarrow \sigma}=(\zeta_{\bf k}(\tau) \Pi_{\zeta_{\bf k}}^*(\tau) )_{-kc_{s}\tau \rightarrow \sigma}\\
  P_{\Pi_\zeta \zeta }^{\bf Tree}(k)&=& \langle\hat{\Pi_{\zeta_{\bf k}}}\hat{\zeta_{\bf -k}}\rangle_{(0,0)}=[\Pi_{\zeta_{\bf k}}(\tau) \zeta_{\bf -k}(\tau ) ]_{-kc_{s}\tau \rightarrow \sigma}=(\Pi_{\zeta_{\bf k}}(\tau) \zeta_{\bf k}^*(\tau)  )_{-kc_{s}\tau \rightarrow \sigma }\\
 P_{{\Pi_\zeta}{ \Pi_\zeta }}^{\bf Tree}(k)&=& \langle\hat{\Pi_{\zeta_{\bf k}}}\hat{\Pi_{\zeta_{\bf -k}}}\rangle_{(0,0)} = [\Pi_{\zeta_{\bf k}}(\tau) \Pi_{\zeta_{\bf -k}}(\tau ) ]_{-kc_{s}\tau \rightarrow \sigma}=|\Pi_{\zeta_{\bf k}}(\tau)  |^2_{-kc_{s}\tau \rightarrow \sigma }
\eea 
For practical reasons, it is convenient to work with the dimensionless power spectrum in Fourier space and to link cosmic measurements. The power spectrum's dimensionless form may be written as follows:
\bea
\Delta^2_{\zeta\zeta,{\bf Tree} }(k)&=&\frac{k^3}{2 \pi^2}P_{\zeta\zeta}^{\bf Tree}(k) = \frac{k^3}{2 \pi^2}|\zeta|^2_{-kc_{s}\tau \rightarrow \sigma }\\
\Delta^2_{\zeta \Pi_\zeta,{\bf Tree} }(k)&=&\frac{k^3}{2 \pi^2} \tau P_{\zeta \Pi_\zeta}^{\bf Tree}(k) = \frac{k^3}{2 \pi^2} \tau (\zeta \Pi_{\zeta}^*)_{-kc_{s}\tau \rightarrow \sigma}\\
\Delta^2_{\Pi_\zeta \zeta ,{\bf Tree} }(k)&=&\frac{k^3}{2 \pi^2} \tau P_{\Pi_\zeta \zeta }^{\bf Tree}(k) = \frac{k^3}{2 \pi^2} \tau (\Pi_{\zeta} \zeta^*)_{-kc_{s}\tau \rightarrow \sigma}\\
\Delta^2_{\Pi_\zeta \Pi_\zeta ,{\bf Tree} }(k)&=&\frac{k^3}{2 \pi^2} \tau^2  P_{\Pi_\zeta \Pi_\zeta }^{\bf Tree}(k) = \frac{k^3}{2 \pi^2} \tau^2 |\Pi_{\zeta}|^2 _{-kc_{s}\tau \rightarrow \sigma }
\eea 
Utilising the Fourier modes of the comoving curvature perturbation for each of the three distinct areas yields the power spectrum components. The instant at which the wavenumbers fulfil $-k c_s \tau = \sigma$ is known as the moment of horizon crossing in stochastic inflation. The stochastic tuning parameter $\sigma$ statistically inserts stochasticity into the noise matrix components and the power spectrum.

By using the statistical characteristics of the quantum noise, a relationship may be obtained between the dimensionless scalar power spectrum and the noise matrix components. For the fields $\zeta_{k}$ and ${\Pi_\zeta}_{k}$, when they begin from the vacuum state, we adhere to the assumption for their quantum starting conditions. Thus, all the statistical characteristics of the noise may be discovered by examining their two-point correlation matrix, which is provided as follows, as working at the leading order in perturbation theory requires Gaussian statistics for the modes:
\bea  \label{noisecorrlmatrix}
\Sigma({\bf x_1},\tau_1;{\bf x_2},\tau_2)=\begin{pmatrix}
\langle 0 | \Hat{\xi}_\zeta({\bf x_1},\tau_1)\Hat{\xi}_\zeta({\bf x_2},\tau_2)|0\rangle & \langle 0 | \Hat{\xi}_\zeta({\bf x_1},\tau_1)\Hat{\xi}_{\Pi_{\zeta}}({\bf x_2},\tau_2)|0\rangle \\
\langle 0 |\Hat{\xi}_{\Pi_{\zeta}}({\bf x_1},\tau_1)\Hat{\xi}_\zeta({\bf x_2},\tau_2)|0\rangle & \langle 0 |\Hat{\xi}_{\Pi_{\zeta}}({\bf x_1},\tau_1)\Hat{\xi}_{\Pi_{\zeta}}({\bf x_2},\tau_2)|0\rangle
\end{pmatrix},
\eea 
where vector quantities are indicated by the boldface ${\bf x}$.
Our technique is valid in both the classical and quantum regimes, as demonstrated by the relationship we find between the noise matrix components and the power spectrum using stochastic canonical quantization at the correlation level. 
Once the annihilation and creation operators are applied to the vacuum state $\ket{0}$, each member in the correlation matrix above may be represented as follows:
\bea 
\Sigma_{{f_1},{g_1}}= \int_{\mathbb{R}^3}\frac{d^3 k }{(2\pi)^3}\frac{d}{d\tau}W\Bigg[\frac{k}{k_\sigma(\tau_1)}\Bigg]\frac{d}{d\tau}W\Bigg[ \frac{k}{k_\sigma(\tau_2)}\Bigg]f_k(\tau_1)g_k^*(\tau_2) e^{i{\bf k}.({\bf x_2}-{\bf x_1})},
\eea
where we use the fact that:
\bea \Sigma_{{f_1},{g_1}}=\langle 0 | \xi_f({\bf x_1},\tau_1)\xi_g({\bf x_2},\tau_2)|0\rangle.\eea
Moreover, $\{f, g\}$ may be $\{\zeta, \Pi_\zeta\}$. The order of subscripts $f$ and $g$ matters because $[\xi_\zeta, \xi_{\Pi_\zeta}]\ne 0$. The angular integral over ${\bf k}/k $ is easier to estimate in the above because of the reliance of mode functions $\zeta_{k}$ and ${\Pi_\zeta}_{k}$ on the norm of ${\bf k}$. Thus, we derive the following expression:
\bea
\Sigma_{f_1, g_1} = \int_{\mathbb{R}^+} \frac{k^{2} dk}{2\pi^2}\frac{d}{d\tau}W\Bigg[\frac{k}{k_\sigma(\tau_1)}\Bigg]\frac{d}{d\tau}W\Bigg[\frac{k}{k_\sigma(\tau_2)}\Bigg]f_k(\tau_1)g_k ^*(\tau_2)\frac{\sin(k|{\bf x_2}-{\bf x_1}|)}{k|{\bf x_2}-{\bf x_1}|},
\eea
where the window function $W(k/k_{\sigma})$ functions as follows: $W(k/k_\sigma)= \Theta(k/k_\sigma - 1)$, which is conveniently selected to be a Heaviside Theta function. The integrand of the above equation, when its time derivative is taken as the Dirac delta distribution, integrates $\del(k-k_\sigma(\tau_1) )\del(k-k_\sigma(\tau_2))$, yielding $\del(\tau_1 - \tau_2)$, which denotes the existence of white noise. This brings us to the following expression:
\bea \label{noiseelement}
\Sigma_{{f_1},{ g_1}}= \frac{1}{6 \pi^2}\frac{dk_{\sigma}^3(\tau)}{d\tau }\Bigg|_{\tau_1}f_{k=k_{\sigma}(\tau)}g_{k=k_{\sigma}(\tau)}^*\frac{\sin[k_{\sigma}(\tau_1)|{\bf x_2}-{\bf x_1}|]}{k_{\sigma}(\tau_1)|{\bf x_2}-{\bf x_2}|}\delta(\tau_1 - \tau_2).
\eea 
Our attention is on the situation when there is maximum spatial correlation between the sounds, which occurs at ${\bf x_1}\rightarrow {\bf x_2}$. This leads to:
\bea \lim_{{\bf x_1}\rightarrow {\bf x_2}} \frac{\sin[k_{\sigma}(\tau_1)|{\bf x_2}-{\bf x_1}|]}{k_{\sigma}(\tau_1)|{\bf x_2}-{\bf x_1}|}=1.\eea
Only when there is an equivalent amount of white noise do the correlations stay non-zero. So, the noise correlation matrix may be expressed as follows:
\bea
\Sigma_{f_1,g_1}\equiv \Sigma_{f,g}(\tau_1)\del(\tau_1 - \tau_2).
\eea 
Furthermore, one may use the following description of the power spectrum element between the quantum fluctuations by employing this important assertion about the noises:
\bea 
P_{f,g}(k;\tau)=\frac{k^3}{2\pi^2} f_{k}(\tau)g_{k}^*(\tau),
\eea 
and this immediately results in the formula that follows, which includes the time-dependent noise matrix element:
\bea \label{noisepower}
\Sigma_{f,g}(\tau) = \frac{d\ln{(k_\sigma(\tau))}}{d\tau}P_{f,g}(k_\sigma(\tau);\tau).
\eea 
The following connection from the dimensionless power spectrum may be used to compute the noise matrix elements, which are the correlators of the noise correlation matrix elements. The expression employed here is found in eq.(\ref{noisepower}), where $f$ and $g$ correspond to $\zeta$ and $\Pi_\zeta$, respectively. These relations are obtained from the derivation presented in the earlier subsection:
\bea
\Sigma_{\zeta\zeta}&=& (1-\epsilon)\Delta^2_{\zeta\zeta,{\bf Tree} }(k)= (1-\epsilon)\frac{k^3}{2 \pi^2}P_{\zeta\zeta}^{\bf Tree}(k) = (1-\epsilon) \frac{k^3}{2 \pi^2}|\zeta|^2_{-kc_{s}\tau \rightarrow \sigma},\\
\Sigma_{\zeta \Pi_\zeta}&=&(1-\epsilon)\Delta^2_{\zeta \Pi_\zeta,{\bf Tree} }(k)=(1-\epsilon) \frac{k^3}{2 \pi^2} \tau P_{\zeta \Pi_\zeta}^{\bf Tree}(k) =(1-\epsilon) \frac{k^3}{2 \pi^2} \tau (\zeta \Pi_{\zeta}^*)_{-kc_{s}\tau \rightarrow \sigma },\\
\Sigma_{ \Pi_\zeta \zeta }&=&(1-\epsilon) \Delta^2_{\Pi_\zeta \zeta ,{\bf Tree} }(k)=(1-\epsilon) \frac{k^3}{2 \pi^2} \tau P_{\Pi_\zeta \zeta }^{\bf Tree}(k) = (1-\epsilon) \frac{k^3}{2 \pi^2} \tau (\Pi_{\zeta} \zeta^*)_{-kc_{s}\tau \rightarrow \sigma },\\
\Sigma_{\Pi_\zeta\Pi_\zeta }&=&(1-\epsilon) \Delta^2_{\Pi_\zeta \Pi_\zeta ,{\bf Tree} }(k) = (1-\epsilon) \frac{k^3}{2 \pi^2} \tau^2  P_{\Pi_\zeta \Pi_\zeta }^{\bf Tree}(k) =  (1-\epsilon) \frac{k^3}{2 \pi^2} \tau^2 |\Pi_{\zeta}|^2 _{-kc_{s}\tau \rightarrow \sigma }.
\eea

\subsubsection{Results in SRI region}

The analytic equations of the scalar power spectrum with stochastic effects are shown in this section for a generic quasi de Sitter background spacetime and any starting quantum vacuum condition.

\subsubsubsection{Power spectrum in SRI}
Equations (\ref{modezetaSRI}) and (\ref{modepiSRI}) that give the general mode function solutions for the first slow-roll (SRI) phase are used to derive the following expressions for the various elements of the scalar power spectrum. For every wavenumber $k$, the power spectrum elements are assessed at the Horizon crossing condition as $-kc_{s}\tau=\sigma$, where $\sigma$ represents the stochastic, coarse-graining parameter:
\bea
 \Delta_{\zeta\zeta}^2&=&2^{2\nu-3}  (\sigma)^{3-2\nu} \frac{H^2}{8\pi^2\epsilon  c_s M_p^2}\Bigg|\frac{\Gamma{(\nu)}}{\Gamma{(\frac{3}{2})}}\Bigg|^2(1+\sigma^2)\Bigg|\alpha_1 e^{-i(\frac{\pi}{2}(\nu+\frac{1}{2})-\sigma)}-\frac{(1+i \sigma)}{(1-i \sigma)}\beta_1 e^{i(\frac{\pi}{2}(\nu+\frac{1}{2})-\sigma)}\Bigg|^2,
\\ 
\Delta_{\zeta\Pi_\zeta}^2&=&2^{2\nu-3}  (\sigma)^{3-2\nu} \frac{H^2}{8\pi^2\epsilon  c_s M_p^2}\Bigg|\frac{\Gamma{(\nu)}}{\Gamma{(\frac{3}{2})}}\Bigg|^2\Bigg[\Bigg(\frac{3}{2}-\nu \Bigg) (1+\sigma^2)+\sigma^2\Bigg]\Bigg\{\Bigg|\alpha_1 e^{-i(\frac{\pi}{2}(\nu+\frac{1}{2})-\sigma)}-\beta_1 e^{i(\frac{\pi}{2}(\nu+\frac{1}{2})-\sigma)}\Bigg|^2\nonumber
 \\&& \quad\quad\quad\quad\quad\quad\quad\quad
+\frac{2\Bigg(\frac{3}{2}-\nu\Bigg)\sigma^2}{\Bigg\{\Bigg(\frac{3}{2}-\nu \Bigg) (1+\sigma^2)+\sigma^2\Bigg\}}\Bigg\{\alpha_1^*\beta_1 e^{2i(\frac{\pi}{2}(\nu+\frac{1}{2})-\sigma)}+\beta_1^*\alpha_1 e^{-2i(\frac{\pi}{2}(\nu+\frac{1}{2})-\sigma)} \Bigg\} \Bigg\},
\\
\Delta_{\Pi_\zeta\zeta}^2&=&2^{2\nu-3}  (\sigma)^{3-2\nu} \frac{H^2}{8\pi^2\epsilon  c_s M_p^2}\Bigg|\frac{\Gamma{(\nu)}}{\Gamma{(\frac{3}{2})}}\Bigg|^2\Bigg[\Bigg(\frac{3}{2}-\nu \Bigg) (1+\sigma^2)+\sigma^2\Bigg]\Bigg\{\Bigg|\alpha_1 e^{-i(\frac{\pi}{2}(\nu+\frac{1}{2})-\sigma)}-\beta_1 e^{i(\frac{\pi}{2}(\nu+\frac{1}{2})-\sigma)}\Bigg|^2\nonumber
 \\&& \quad\quad\quad\quad\quad\quad\quad\quad
+ \frac{2\Bigg(\frac{3}{2}-\nu\Bigg)\sigma^2}{\Bigg\{\Bigg(\frac{3}{2}-\nu \Bigg) (1+\sigma^2)+\sigma^2\Bigg\}}\Bigg\{\alpha_1^*\beta_1 e^{2i(\frac{\pi}{2}(\nu+\frac{1}{2})-\sigma)}+\beta_1^*\alpha_1 e^{-2i(\frac{\pi}{2}(\nu+\frac{1}{2})-\sigma)} \Bigg\}\Bigg\},
\\
\Delta_{\Pi_\zeta\Pi_\zeta}^2&=&2^{2\nu-3}(\sigma)^{3-2\nu}\frac{H^2}{8\pi^2 \epsilon  
 c_s M_p^2}\Bigg|\frac{\Gamma{(\nu)}}{\Gamma{(\frac{3}{2})}}\Bigg|^2 \Bigg\{ \Bigg(\Bigg(\frac{3}{2}-\nu\Bigg)+\sigma^2 \Bigg)^2+\Bigg(\frac{3}{2}-\nu\Bigg)^2 \sigma^2 \Bigg \} \nonumber
 \\ && \quad\quad\quad\quad\quad\quad\quad\quad\quad
 \times\Bigg|\alpha_1 e^{-i(\frac{\pi}{2}(\nu+\frac{1}{2})-\sigma)}-\beta_1 \frac{\Bigg\{\Bigg(\frac{3}{2}-\nu\Bigg)(1+i \sigma )+\sigma^2 \Bigg \}}{\Bigg\{\Bigg(\frac{3}{2}-\nu\Bigg)(1-i \sigma)+\sigma^2 \Bigg \}} e^{i(k c_s\tau+\frac{\pi}{2}(\nu+\frac{1}{2}))}\Bigg|^2.
\eea
The aforementioned power spectrum elements may be represented in terms of the Bunch Davies initial vacuum state, where $\alpha_1=1$ and $\beta_1=0$, which is a more popular choice of beginning condition when discussing inflationary observables. The general de Sitter condition phrases are given below:
\bea \label{pspecsr1BD}
\Delta_{{\zeta\zeta},{\bf BD}}^2&=&2^{2\nu-3}  (\sigma)^{3-2\nu} \frac{H^2}{8\pi^2\epsilon  c_s M_p^2}\Bigg|\frac{\Gamma{(\nu)}}{\Gamma{(\frac{3}{2})}}\Bigg|^2(1+\sigma^2),
\\ 
\Delta_{{\zeta\Pi_\zeta},{\bf BD}}^2&=&2^{2\nu-3}  (\sigma)^{3-2\nu} \frac{H^2}{8\pi^2\epsilon  c_s M_p^2}\Bigg|\frac{\Gamma{(\nu)}}{\Gamma{(\frac{3}{2})}}\Bigg|^2\Bigg[\Bigg(\frac{3}{2}-\nu \Bigg) (1+\sigma^2)+\sigma^2\Bigg],
\\
\Delta_{{\Pi_\zeta\zeta},{\bf BD}}^2&=&2^{2\nu-3}  (\sigma)^{3-2\nu} \frac{H^2}{8\pi^2\epsilon  c_s M_p^2}\Bigg|\frac{\Gamma{(\nu)}}{\Gamma{(\frac{3}{2})}}\Bigg|^2\Bigg[\Bigg(\frac{3}{2}-\nu \Bigg) (1+\sigma^2)+\sigma^2\Bigg],
\\
\Delta_{{\Pi_\zeta\Pi_\zeta},{\bf BD}}^2&=&2^{2\nu-3}(\sigma)^{3-2\nu}\frac{H^2}{8\pi^2 \epsilon  c_s M_p^2}\Bigg|\frac{\Gamma{(\nu)}}{\Gamma{(\frac{3}{2})}}\Bigg|^2 \Bigg[\Bigg(\Bigg(\frac{3}{2}-\nu\Bigg)+\sigma^2 \Bigg)^2+\Bigg(\frac{3}{2}-\nu\Bigg)^2 \sigma^2 \Bigg]. 
\eea
One may obtain the simpler equation for Power spectrum elements in the case of de Sitter space by further incorporating the limiting number $\nu=\frac{3}{2}$ in the Bunch Davies vacuum derived result:
\bea \label{pspecsr1dS}
\Delta_{{\zeta\zeta},{\bf dS}}^2&=&   \frac{H^2}{8\pi^2\epsilon  c_s M_p^2}(1+\sigma^2),
\\
\Delta_{{\zeta\Pi_\zeta},{\bf dS}}^2&=&  \frac{H^2}{8\pi^2\epsilon  c_s M_p^2}\sigma^2,
\\
\Delta_{{\Pi_\zeta\zeta},{\bf dS}}^2&=&  \frac{H^2}{8\pi^2\epsilon  c_s M_p^2}\sigma^2,
\\
\Delta_{{\Pi_\zeta\Pi_\zeta},{\bf dS}}^2&=&\frac{H^2}{8\pi^2 \epsilon  c_s M_p^2} \sigma^4.
\eea

\subsubsubsection{Noise Matrix elements in SRI}
The following formulae may be used to write the noise correlation matrix elements in the SRI for an arbitrary de Sitter background spacetime, which is characterised by $\nu$' and a generic starting quantum vacuum state defined with coefficients $(\alpha_1,\beta_1)$:
\bea 
\Sigma_{\zeta\zeta}&=&(1-\epsilon) 2^{2\nu-3}  (\sigma)^{3-2\nu} \frac{H^2}{8\pi^2\epsilon  c_s M_p^2}\Bigg|\frac{\Gamma{(\nu)}}{\Gamma{(\frac{3}{2})}}\Bigg|^2(1+\sigma^2)\Bigg|\alpha_1 e^{-i(\frac{\pi}{2}(\nu+\frac{1}{2})-\sigma)}-\frac{(1+i \sigma)}{(1-i \sigma)}\beta_1 e^{i(\frac{\pi}{2}(\nu+\frac{1}{2})-\sigma)}\Bigg|^2,
\\
\Sigma_{\zeta\Pi_\zeta}&=&(1-\epsilon) 2^{2\nu-3}  (\sigma)^{3-2\nu} \frac{H^2}{8\pi^2\epsilon  c_s M_p^2}\Bigg|\frac{\Gamma{(\nu)}}{\Gamma{(\frac{3}{2})}}\Bigg|^2\Bigg[\Bigg(\frac{3}{2}-\nu \Bigg) (1+\sigma^2)+\sigma^2\Bigg]\Bigg\{\Bigg|\alpha_1 e^{-i(\frac{\pi}{2}(\nu+\frac{1}{2})-\sigma)}-\beta_1 e^{i(\frac{\pi}{2}(\nu+\frac{1}{2})-\sigma)}\Bigg|^2\nonumber
 \\  && \quad \quad \quad \quad \quad \quad \quad \quad \quad 
+\frac{2\Bigg(\frac{3}{2}-\nu\Bigg)\sigma^2}{\Bigg\{\Bigg(\frac{3}{2}-\nu \Bigg) (1+\sigma^2)+\sigma^2\Bigg\}}\Bigg\{\alpha_1^*\beta_1 e^{2i(\frac{\pi}{2}(\nu+\frac{1}{2})-\sigma)}+\beta_1^*\alpha_1 e^{-2i(\frac{\pi}{2}(\nu+\frac{1}{2})-\sigma)} \Bigg\}\Bigg\},
\\
\Sigma_{\Pi_\zeta\zeta}&=&(1-\epsilon) 2^{2\nu-3}  (\sigma)^{3-2\nu} \frac{H^2}{8\pi^2\epsilon  c_s M_p^2}\Bigg|\frac{\Gamma{(\nu)}}{\Gamma{(\frac{3}{2})}}\Bigg|^2\Bigg[\Bigg(\frac{3}{2}-\nu \Bigg) (1+\sigma^2)+\sigma^2\Bigg]\Bigg\{\Bigg|\alpha_1 e^{-i(\frac{\pi}{2}(\nu+\frac{1}{2})-\sigma)}-\beta_1 e^{i(\frac{\pi}{2}(\nu+\frac{1}{2})-\sigma)}\Bigg|^2\nonumber
 \\ && \quad \quad \quad \quad \quad \quad \quad \quad \quad 
+ \frac{2\Bigg(\frac{3}{2}-\nu\Bigg)\sigma^2}{\Bigg\{\Bigg(\frac{3}{2}-\nu \Bigg) (1+\sigma^2)+\sigma^2\Bigg\}}\Bigg\{\alpha_1^*\beta_1 e^{2i(\frac{\pi}{2}(\nu+\frac{1}{2})-\sigma)}+\beta_1^*\alpha_1 e^{-2i(\frac{\pi}{2}(\nu+\frac{1}{2})-\sigma)} \Bigg\}\Bigg\},
\eea\bea
\Sigma_{\Pi_\zeta\Pi_\zeta}&=&(1-\epsilon) 2^{2\nu-3}(\sigma)^{3-2\nu}\frac{H^2}{8\pi^2 \epsilon  
 c_s M_p^2}\Bigg|\frac{\Gamma{(\nu)}}{\Gamma{(\frac{3}{2})}}\Bigg|^2 \Bigg[\Bigg\{\Bigg(\frac{3}{2}-\nu\Bigg)+\sigma^2 \Bigg\}^2+\Bigg(\frac{3}{2}-\nu\Bigg)^2 \sigma^2 \Bigg ] \nonumber
 \\ && \quad \quad \quad \quad \quad \quad  \quad \quad \quad \times
 \Bigg|\alpha_1 e^{-i(\frac{\pi}{2}(\nu+\frac{1}{2})-\sigma)}-\beta_1 \frac{\Bigg\{\Bigg(\frac{3}{2}-\nu\Bigg)(1+i \sigma )+\sigma^2 \Bigg \}}{\Bigg\{\Bigg(\frac{3}{2}-\nu\Bigg)(1-i \sigma)+\sigma^2 \Bigg \}} e^{i(\frac{\pi}{2}(\nu+\frac{1}{2})-\sigma)}\Bigg|^2.
\eea
By changing the values of $\alpha_1 = 1$ and $\beta_1 = 0$ in the set of noise matrix elements above, one may describe the noise correlation matrix elements in the initial vacuum state of Bunch Davies:
\bea 
\Sigma_{{\zeta\zeta},{\bf BD}}&=&(1-\epsilon) 2^{2\nu-3}  (\sigma)^{3-2\nu} \frac{H^2}{8\pi^2\epsilon  c_s M_p^2}\Bigg|\frac{\Gamma{(\nu)}}{\Gamma{(\frac{3}{2})}}\Bigg|^2(1+\sigma^2),
\\
\Sigma_{{\zeta\Pi_\zeta},{\bf BD}}&=&(1-\epsilon) 2^{2\nu-3}  (\sigma)^{3-2\nu} \frac{H^2}{8\pi^2\epsilon  c_s M_p^2}\Bigg|\frac{\Gamma{(\nu)}}{\Gamma{(\frac{3}{2})}}\Bigg|^2\Bigg[\Bigg(\frac{3}{2}-\nu \Bigg) (1+\sigma^2)+\sigma^2\Bigg],
\\
\Sigma_{{\Pi_\zeta\zeta},{\bf BD}}&=&(1-\epsilon) 2^{2\nu-3}  (\sigma)^{3-2\nu} \frac{H^2}{8\pi^2\epsilon  c_s M_p^2}\Bigg|\frac{\Gamma{(\nu)}}{\Gamma{(\frac{3}{2})}}\Bigg|^2\Bigg[\Bigg(\frac{3}{2}-\nu \Bigg) (1+\sigma^2)+\sigma^2\Bigg],
\\
  \Sigma_{{\Pi_\zeta\Pi_\zeta},{\bf BD}}&=&(1-\epsilon) 2^{2\nu-3}(\sigma)^{3-2\nu}\frac{H^2}{8\pi^2 \epsilon  
 c_s M_p^2}\Bigg|\frac{\Gamma{(\nu)}}{\Gamma{(\frac{3}{2})}}\Bigg|^2 \Bigg[ \Bigg\{\Bigg(\frac{3}{2}-\nu\Bigg)+\sigma^2 \Bigg\}^2+\Bigg(\frac{3}{2}-\nu\Bigg)^2 \sigma^2 \Bigg].
\eea
The simplified equation for noise correlation matrix elements in the de Sitter space situation may be obtained by further incorporating the limiting value $\nu = \frac{3}{2}$ in the result of the Bunch Davies vacuum generated process:
\bea 
\Sigma_{{\zeta\zeta},{\bf dS}}&=&(1-\epsilon)   \frac{H^2}{8\pi^2\epsilon  c_s M_p^2}(1+\sigma^2),
\\
\Sigma_{{\zeta\Pi_\zeta},{\bf dS}}&=&(1-\epsilon)  \frac{H^2}{8\pi^2\epsilon  c_s M_p^2}\sigma^2,
\\
\Sigma_{{\Pi_\zeta\zeta},{\bf dS}}&=&(1-\epsilon)\frac{H^2}{8\pi^2\epsilon  c_s M_p^2} \sigma^2,
\\
  \Sigma_{{\Pi_\zeta\Pi_\zeta},{\bf dS}}&=&(1-\epsilon) \frac{H^2}{8\pi^2 \epsilon  
 c_s M_p^2}\sigma^4.
\eea

\subsubsection{Results in USR region}

\subsubsubsection{Power spectrum in USR}

Equations (\ref{modezetaUSR}) and (\ref{modepiUSR}) that give the general mode function solutions for the ultra slow-roll (USR) phase are used to derive the future expressions for the various elements of the scalar power spectrum. At the Horizon crossing condition, the power spectrum elements for any wavenumber $k$ are assessed as $-kc_{s}\tau=\sigma$, where $\sigma$ represents the stochastic, coarse-graining parameter:
\bea
\Delta_{\zeta\zeta}^2&=&2^{2\nu-3}  (\sigma)^{3-2\nu} \frac{H^2}{8\pi^2\epsilon  c_s M_p^2} 
 \Bigg(\frac{k_e}{k_s}\Bigg)^6 \Bigg|\frac{\Gamma{(\nu)}}{\Gamma{(\frac{3}{2})}}\Bigg|^2(1+\sigma^2)\Bigg|\alpha_2 e^{-i(\frac{\pi}{2}(\nu+\frac{1}{2})-\sigma)}-\frac{(1+i \sigma)}{(1-i \sigma)}\beta_2 e^{i(\frac{\pi}{2}(\nu+\frac{1}{2})-\sigma)}\Bigg|^2,
 \\ \Delta_{\zeta\Pi_\zeta}^2&=&2^{2\nu-3}  (\sigma)^{3-2\nu} \frac{H^2}{8\pi^2\epsilon  c_s M_p^2}\Bigg(\frac{k_e}{k_s}\Bigg)^6 
 \Bigg|\frac{\Gamma{(\nu)}}{\Gamma{(\frac{3}{2})}}\Bigg|^2\Bigg[\Bigg(\Bigg(\frac{3}{2}-\nu \Bigg)-3\Bigg) (1+\sigma^2)+\sigma^2\Bigg]\Bigg\{\Bigg|\alpha_2 e^{-i(\frac{\pi}{2}(\nu+\frac{1}{2})-\sigma)}-\beta_2 e^{i(\frac{\pi}{2}(\nu+\frac{1}{2})-\sigma)}\Bigg|^2\nonumber
 \\ && \quad \quad \quad \quad \quad \quad \quad\quad  \quad \quad \quad 
+\frac{2\Bigg(\frac{3}{2}+\nu\Bigg)\sigma^2}{\Bigg\{\Bigg(\frac{3}{2}+\nu \Bigg) (1+\sigma^2)-\sigma^2\Bigg\}}\Bigg\{\alpha_2^*\beta_2 e^{2i(\frac{\pi}{2}(\nu+\frac{1}{2})-\sigma)}+\beta_2^*\alpha_2 e^{-2i(\frac{\pi}{2}(\nu+\frac{1}{2})-\sigma)} \Bigg\}\Bigg\},
\eea\bea
\Delta_{\Pi_\zeta\zeta}^2&=&2^{2\nu-3}  (\sigma)^{3-2\nu} \frac{H^2}{8\pi^2\epsilon  c_s M_p^2}\Bigg(\frac{k_e}{k_s}\Bigg)^6 
\Bigg|\frac{\Gamma{(\nu)}}{\Gamma{(\frac{3}{2})}}\Bigg|^2\Bigg[\Bigg(\Bigg(\frac{3}{2}-\nu \Bigg) -3\Bigg)(1+\sigma^2)+\sigma^2\Bigg]\Bigg\{\Bigg|\alpha_2 e^{-i(\frac{\pi}{2}(\nu+\frac{1}{2})-\sigma)}-\beta_2 e^{i(\frac{\pi}{2}(\nu+\frac{1}{2})-\sigma)}\Bigg|^2\nonumber
 \\&&\quad \quad \quad \quad \quad \quad \quad \quad \quad \quad \quad 
+ \frac{2\Bigg(\frac{3}{2}+\nu\Bigg)\sigma^2}{\Bigg\{\Bigg(\frac{3}{2}+\nu \Bigg) (1+\sigma^2)-\sigma^2\Bigg\}}\Bigg\{\alpha_2^*\beta_2 e^{2i(\frac{\pi}{2}(\nu+\frac{1}{2})-\sigma)}+\beta_2^*\alpha_2 e^{-2i(\frac{\pi}{2}(\nu+\frac{1}{2})-\sigma)} \Bigg\}\Bigg\},
\\
\Delta_{\Pi_\zeta\Pi_\zeta}^2&=&2^{2\nu-3}(\sigma)^{3-2\nu}\frac{H^2}{8\pi^2 \epsilon  
 c_s M_p^2}\Bigg(\frac{k_e}{k_s}\Bigg)^6 
 \Bigg|\frac{\Gamma{(\nu)}}{\Gamma{(\frac{3}{2})}}\Bigg|^2 \Bigg[\Bigg\{\Bigg(\frac{3}{2}+ \nu\Bigg)- \sigma^2 \Bigg\}^2+\Bigg(\frac{3}{2}+\nu\Bigg)^2 \sigma^2 \Bigg ] \nonumber
 \\&& \quad \quad\quad\quad\quad\quad\quad\quad\quad\quad\quad\quad \times
 \Bigg|\alpha_2 e^{-i(\frac{\pi}{2}(\nu+\frac{1}{2})-\sigma)}-\beta_2 \frac{\Bigg\{\Bigg(\frac{3}{2}+ \nu\Bigg)(1+i \sigma )- \sigma^2 \Bigg \}}{\Bigg\{\Bigg(\frac{3}{2}+ \nu\Bigg)(1-i \sigma)- \sigma^2 \Bigg \}} e^{i(\frac{\pi}{2}(\nu+\frac{1}{2})-\sigma)}\Bigg|^2.
\eea
Under the selection of an initial Bunch Davies vacuum condition, the results with an arbitrary $\nu$' have their Bogoliubov coefficients changed from $(\alpha_2,\beta_2)$ to $(\alpha_{2,\bf BD},\beta_{2,\bf BD})$. In the case of exact de Sitter space, one can obtain the simplified expression for power spectrum elements with the new Bogoliubov coefficients $(\alpha_{2, {\bf dS}},\beta_{2, {\bf dS}})$ by further implementing the limiting value $\nu=3/2 $ in the derived result of $(\alpha_{2,{\bf BD}},\beta_{2,{\bf BD}})$ for the USR:
\bea \label{pspecusrdS}
\Delta_{{\zeta\zeta},{\bf dS}}^2&=&  \frac{H^2}{8\pi^2\epsilon  c_s M_p^2} 
 \Bigg(\frac{k_e}{k_s}\Bigg)^6 (1+\sigma^2)\Bigg|\alpha_{2,{\bf dS}} e^{i\sigma}-\frac{(1+i \sigma)}{(1-i \sigma)}\beta_{2,{\bf dS}} e^{-i\sigma}\Bigg|^2,
 \\
 \Delta_{{\zeta\Pi_\zeta},{\bf dS}}^2&=&\frac{H^2}{8\pi^2\epsilon  c_s M_p^2}\Bigg(\frac{k_e}{k_s}\Bigg)^6 
 \Bigg[-3 (1+\sigma^2)+\sigma^2\Bigg]\Bigg\{\Bigg|\alpha_{2,{\bf dS}} e^{i\sigma}-\beta_{2,{\bf dS}} e^{-i\sigma}\Bigg|^2\nonumber
 \\&& \quad \quad \quad \quad \quad\quad  \quad \quad \quad \quad \quad 
+\frac{6\sigma^2}{\Bigg\{3 (1+\sigma^2)-\sigma^2\Bigg\}}\Bigg\{\alpha_{2,{\bf dS}}^*\beta_{2,{\bf dS}} e^{-2i\sigma}+\beta_{2,{\bf dS}}^*\alpha_{2,{\bf dS}} e^{2i\sigma} \Bigg\}\Bigg\},
\\
\Delta_{{\Pi_\zeta\zeta},{\bf dS}}^2&=& \frac{H^2}{8\pi^2\epsilon  c_s M_p^2}\Bigg(\frac{k_e}{k_s}\Bigg)^6 
\Bigg[-3(1+\sigma^2)+\sigma^2\Bigg]\Bigg\{\Bigg|\alpha_{2,{\bf dS}} e^{i\sigma}-\beta_{2,{\bf dS}} e^{-i\sigma}\Bigg|^2\nonumber
\\&&\quad \quad \quad \quad \quad \quad \quad \quad \quad \quad \quad 
+ \frac{6\sigma^2}{\Bigg\{3(1+\sigma^2)-\sigma^2\Bigg\}}\Bigg\{\alpha_{2,{\bf dS}}^*\beta_{2,{\bf dS}} e^{-2i\sigma}+\beta_{2,{\bf dS}}^* \alpha_{2,{\bf dS}} e^{2i\sigma} \Bigg\}\Bigg\},
\\
\Delta_{{\Pi_\zeta\Pi_\zeta},{\bf dS}}^2&=&\frac{H^2}{8\pi^2 \epsilon  
 c_s M_p^2}\Bigg(\frac{k_e}{k_s}\Bigg)^6 
  \Bigg[\Bigg\{3- \sigma^2 \Bigg\}^2+9\sigma^2 \Bigg ]
 \Bigg|\alpha_{2,{\bf dS}} e^{i\sigma}-\beta_{2,{\bf dS}} \frac{\Bigg\{3(1+i \sigma )- \sigma^2 \Bigg \}}{\Bigg\{3(1-i \sigma)- \sigma^2 \Bigg \}} e^{-i\sigma}\Bigg|^2.
\eea
\subsubsubsection{Noise Matrix elements in USR}

With a generic starting quantum vacuum state described with coefficients $(\alpha_2,\beta_2)$ and an arbitrary de Sitter background spacetime characterised by $\nu$', the noise correlation matrix elements in the USR are expressed as follows:
\bea
\Sigma_{\zeta\zeta}&=&(1-\epsilon)2^{2\nu-3}  (\sigma)^{3-2\nu} \frac{H^2}{8\pi^2\epsilon  c_s M_p^2} 
 \Bigg(\frac{k_e}{k_s}\Bigg)^6 \Bigg|\frac{\Gamma{(\nu)}}{\Gamma{(\frac{3}{2})}}\Bigg|^2(1+\sigma^2)\Bigg|\alpha_2 e^{-i(\frac{\pi}{2}(\nu+\frac{1}{2})-\sigma)}-\frac{(1+i \sigma)}{(1-i \sigma)}\beta_2 e^{i(\frac{\pi}{2}(\nu+\frac{1}{2})-\sigma)}\Bigg|^2,
\eea\bea
\Sigma_{\zeta\Pi_\zeta}&=&(1-\epsilon) 2^{2\nu-3}  (\sigma)^{3-2\nu} \frac{H^2}{8\pi^2\epsilon  c_s M_p^2}\Bigg(\frac{k_e}{k_s}\Bigg)^6 
 \Bigg|\frac{\Gamma{(\nu)}}{\Gamma{(\frac{3}{2})}}\Bigg|^2\Bigg\{\Bigg(\Bigg(\frac{3}{2}-\nu \Bigg)-3\Bigg) (1+\sigma^2)+\sigma^2\Bigg\}\Bigg\{\Bigg|\alpha_2 e^{-i(\frac{\pi}{2}(\nu+\frac{1}{2})-\sigma)}\nonumber
 \\&&
 -\beta_2 e^{i(\frac{\pi}{2}(\nu+\frac{1}{2})-\sigma)}\Bigg|^2
+\frac{2\Bigg(\frac{3}{2}+\nu\Bigg)\sigma^2}{\Bigg\{\Bigg(\frac{3}{2}+\nu \Bigg) (1+\sigma^2)-\sigma^2\Bigg\}}\Bigg\{\alpha_2^*\beta_2 e^{2i(\frac{\pi}{2}(\nu+\frac{1}{2})-\sigma)}
 +\beta_2^*\alpha_2 e^{-2i(\frac{\pi}{2}(\nu+\frac{1}{2})-\sigma)} \Bigg\}\Bigg\},
\\
\Sigma_{\Pi_\zeta\zeta}&=&(1-\epsilon) 2^{2\nu-3}  (\sigma)^{3-2\nu} \frac{H^2}{8\pi^2\epsilon  c_s M_p^2}\Bigg(\frac{k_e}{k_s}\Bigg)^6 
 \Bigg|\frac{\Gamma{(\nu)}}{\Gamma{(\frac{3}{2})}}\Bigg|^2\Bigg\{\Bigg(\Bigg(\frac{3}{2}-\nu \Bigg) -3\Bigg)(1+\sigma^2)+\sigma^2\Bigg\}\Bigg\{\Bigg|\alpha_2 e^{-i(\frac{\pi}{2}(\nu+\frac{1}{2})-\sigma)}\nonumber
 \\&&
 -\beta_2 e^{i(\frac{\pi}{2}(\nu+\frac{1}{2})-\sigma)}\Bigg|^2
+ \frac{2\Bigg(\frac{3}{2}+\nu\Bigg)\sigma^2}{\Bigg\{\Bigg(\frac{3}{2}+\nu \Bigg) (1+\sigma^2)-\sigma^2\Bigg\}}\Bigg\{\alpha_2^*\beta_2 e^{2i(\frac{\pi}{2}(\nu+\frac{1}{2})-\sigma)}+\beta_2^*\alpha_2 e^{-2i(\frac{\pi}{2}(\nu+\frac{1}{2})-\sigma)} \Bigg\}\Bigg\},
\\
\Sigma_{\Pi_\zeta\Pi_\zeta}&=&(1-\epsilon) 2^{2\nu-3}(\sigma)^{3-2\nu}\frac{H^2}{8\pi^2 \epsilon  
 c_s M_p^2}\Bigg(\frac{k_e}{k_s}\Bigg)^6 
 \Bigg|\frac{\Gamma{(\nu)}}{\Gamma{(\frac{3}{2})}}\Bigg|^2 \Bigg\{\Bigg(\Bigg(\frac{3}{2}+ \nu\Bigg)- \sigma^2 \Bigg)^2+\Bigg(\frac{3}{2}+\nu\Bigg)^2 \sigma^2 \Bigg \} \nonumber
\\ && \quad \quad\quad\quad\quad
 \Bigg|\alpha_2 e^{-i(\frac{\pi}{2}(\nu+\frac{1}{2})-\sigma)}-\beta_2 \frac{\Bigg\{\Bigg(\frac{3}{2}+ \nu\Bigg)(1+i \sigma )- \sigma^2 \Bigg \}}{\Bigg\{\Bigg(\frac{3}{2}+ \nu\Bigg)(1-i \sigma)- \sigma^2 \Bigg \}} e^{i(\frac{\pi}{2}(\nu+\frac{1}{2})-\sigma)}\Bigg|^2.
\eea
As in the previous SRI case, we further implement the Bunch Davies vacuum conditions and then apply the limiting value of $\nu=3/2$ in the result of the Bunch Davies vacuum-derived process to obtain the new set $(\alpha_{2, \bf dS},\beta_{2, \bf dS})$. The noise matrix elements in the exact de Sitter space can then be expressed more simply as follows:
\bea
\Sigma_{{\zeta\zeta},{\bf dS}}&=&(1-\epsilon) \frac{H^2}{8\pi^2\epsilon  c_s M_p^2} 
 \Bigg(\frac{k_e}{k_s}\Bigg)^6 (1+\sigma^2)\Bigg|\alpha_{2,{\bf dS}} e^{i\sigma}-\frac{(1+i \sigma)}{(1-i \sigma)}\beta_{2,{\bf dS}} e^{-i\sigma}\Bigg|^2,
\\
\Sigma_{{\zeta\Pi_\zeta},{\bf dS}}&=&(1-\epsilon)  \frac{H^2}{8\pi^2\epsilon  c_s M_p^2}\Bigg(\frac{k_e}{k_s}\Bigg)^6 
\Bigg[-3 (1+\sigma^2)+\sigma^2\Bigg]\Bigg\{\Bigg|\alpha_{2,{\bf dS}} e^{i\sigma}
 -\beta_{2,{\bf dS}} e^{-i\sigma}\Bigg|^2   \nonumber
 \\&& \quad\quad\quad\quad\quad\quad\quad\quad\quad \quad \quad \quad 
+\frac{6\sigma^2}{\Bigg\{3 (1+\sigma^2)-\sigma^2\Bigg\}}\Bigg\{\alpha_{2,{\bf dS}}^*\beta_{2,{\bf dS}}e^{-2i\sigma}
 +\beta_{2,{\bf dS}}^*\alpha_{2,{\bf dS}} e^{2i\sigma} \Bigg\}\Bigg\},
\\
\Sigma_{{\Pi_\zeta\zeta},{\bf dS}}&=&(1-\epsilon) \frac{H^2}{8\pi^2\epsilon  c_s M_p^2}\Bigg(\frac{k_e}{k_s}\Bigg)^6 
 \Bigg[-3(1+\sigma^2)+\sigma^2\Bigg]\Bigg\{\Bigg|\alpha_{2,{\bf dS}} e^{i\sigma}
 -\beta_{2,{\bf dS}} e^{-i\sigma}\Bigg|^2 \nonumber
 \\&& \quad  \quad  \quad  \quad \quad\quad\quad\quad\quad\quad\quad\quad
+ \frac{6\sigma^2}{\Bigg\{3(1+\sigma^2)-\sigma^2\Bigg\}}\Bigg\{\alpha_{2,{\bf dS}}^*\beta_{2,{\bf dS}} e^{-2i\sigma}+\beta_{2,{\bf dS}}^*\alpha_{2,{\bf dS}} e^{2i\sigma} \Bigg\}\Bigg\},
\\
\Sigma_{{\Pi_\zeta\Pi_\zeta},{\bf dS}}&=&(1-\epsilon) \frac{H^2}{8\pi^2 \epsilon  
 c_s M_p^2}\Bigg(\frac{k_e}{k_s}\Bigg)^6 
  \Bigg[\Bigg(3- \sigma^2 \Bigg)^2+ 9 \sigma^2 \Bigg ]
 \Bigg|\alpha_{2,{\bf dS}} e^{i\sigma}-\beta_{2,{\bf dS}} \frac{\Bigg\{3(1+i \sigma )- \sigma^2 \Bigg \}}{\Bigg\{3(1-i \sigma)- \sigma^2 \Bigg \}} e^{-i\sigma}\Bigg|^2.
\eea
 
\subsubsection{Results in SRII region}

\subsubsubsection{Power spectrum in SRII}

Equations (\ref{modezetaSR2}) and (\ref{modepiSR2}) that provide generic mode function solutions for the second slow-roll (SRII) phase are used to compute the future expressions for the various elements of the scalar power spectrum. For every given wavenumber $k$, the power spectrum elements are assessed at the Horizon crossing condition as $-kc_{s}\tau=\sigma$, where $\sigma$ represents the stochastic, coarse-graining parameter:
\bea
\Delta_{\zeta\zeta}^2&=&2^{2\nu-3}  (\sigma)^{3-2\nu} \frac{H^2}{8\pi^2\epsilon  c_s M_p^2} 
 \Bigg(\frac{k_e}{k_s}\Bigg)^6 \Bigg|\frac{\Gamma{(\nu)}}{\Gamma{(\frac{3}{2})}}\Bigg|^2(1+\sigma^2)\Bigg|\alpha_3 e^{-i(\frac{\pi}{2}(\nu+\frac{1}{2})-\sigma)}-\frac{(1+i \sigma)}{(1-i \sigma)}\beta_3 e^{i(\frac{\pi}{2}(\nu+\frac{1}{2})-\sigma)}\Bigg|^2,
\\
\Delta_{\zeta\Pi_\zeta}^2&=&2^{2\nu-3}  (\sigma)^{3-2\nu} \frac{H^2}{8\pi^2\epsilon  c_s M_p^2}\Bigg(\frac{k_e}{k_s}\Bigg)^6 
 \Bigg|\frac{\Gamma{(\nu)}}{\Gamma{(\frac{3}{2})}}\Bigg|^2\Bigg\{\Bigg(\frac{3}{2}-\nu \Bigg)(1+\sigma^2)+\sigma^2\Bigg\}\Bigg\{\Bigg|\alpha_3 e^{-i(\frac{\pi}{2}(\nu+\frac{1}{2})-\sigma)}-\beta_3 e^{i(\frac{\pi}{2}(\nu+\frac{1}{2})-\sigma)}\Bigg|^2\nonumber
 \\&& \quad \quad \quad \quad \quad \quad \quad \quad \quad 
+\frac{2\Bigg(\frac{3}{2}-\nu\Bigg)\sigma^2}{\Bigg\{\Bigg(\frac{3}{2}-\nu \Bigg) (1+\sigma^2)+\sigma^2\Bigg\}}\Bigg\{\alpha_3^*\beta_3 e^{2i(\frac{\pi}{2}(\nu+\frac{1}{2})-\sigma)}+\beta_3^*\alpha_3 e^{-2i(\frac{\pi}{2}(\nu+\frac{1}{2})-\sigma)} \Bigg\}\Bigg\},
\eea\bea
\Delta_{\Pi_\zeta\zeta}^2&=&2^{2\nu-3}  (\sigma)^{3-2\nu} \frac{H^2}{8\pi^2\epsilon  c_s M_p^2}\Bigg(\frac{k_e}{k_s}\Bigg)^6 
 \Bigg|\frac{\Gamma{(\nu)}}{\Gamma{(\frac{3}{2})}}\Bigg|^2\Bigg\{\Bigg(\frac{3}{2}-\nu \Bigg)(1+\sigma^2)+\sigma^2\Bigg\}\Bigg\{\Bigg|\alpha_3 e^{-i(\frac{\pi}{2}(\nu+\frac{1}{2})-\sigma)}-\beta_3 e^{i(\frac{\pi}{2}(\nu+\frac{1}{2})-\sigma)}\Bigg|^2\nonumber
 \\ && \quad \quad \quad \quad \quad \quad \quad \quad \quad 
+ \frac{2\Bigg(\frac{3}{2}-\nu\Bigg)\sigma^2}{\Bigg\{\Bigg(\frac{3}{2}-\nu \Bigg) (1+\sigma^2)+\sigma^2\Bigg\}}\Bigg\{\alpha_3^*\beta_3 e^{2i(\frac{\pi}{2}(\nu+\frac{1}{2})-\sigma)}+\beta_3^*\alpha_3 e^{-2i(\frac{\pi}{2}(\nu+\frac{1}{2})-\sigma)} \Bigg\}\Bigg\},
\\
\Delta_{\Pi_\zeta\Pi_\zeta}^2&=&2^{2\nu-3}(\sigma)^{3-2\nu}\frac{H^2}{8\pi^2 \epsilon  
 c_s M_p^2}\Bigg(\frac{k_e} {k_s}\Bigg)^6 
 \Bigg|\frac{\Gamma{(\nu)}}{\Gamma{(\frac{3}{2})}}\Bigg|^2 \Bigg[\Bigg(\Bigg(\frac{3}{2}- \nu\Bigg)+ \sigma^2 \Bigg)^2+\Bigg(\frac{3}{2}-\nu\Bigg)^2 \sigma^2 \Bigg ] \nonumber
 \\ && \quad \quad \quad \quad \quad \quad \quad \quad \quad \quad \quad 
 \Bigg|\alpha_3 e^{-i(\frac{\pi}{2}(\nu+\frac{1}{2})-\sigma)}-\beta_3 \frac{\Bigg\{\Bigg(\frac{3}{2}- \nu\Bigg)(1+i \sigma )+ \sigma^2 \Bigg \}}{\Bigg\{\Bigg(\frac{3}{2}- \nu\Bigg)(1-i \sigma)+ \sigma^2 \Bigg \}} e^{i(\frac{\pi}{2}(\nu+\frac{1}{2})-\sigma)}\Bigg|^2.
\eea
Under the selection of an initial Bunch Davies vacuum condition, the results with an arbitrary $nu$' have their Bogoliubov coefficients changed from $(\alpha_3,\beta_3)$ to $(\alpha_{3,\bf BD},\beta_{3,\bf BD})$. In the case of exact de Sitter space, one can obtain the simplified expression for power spectrum elements with the new Bogoliubov coefficients $(\alpha_{3,{\bf dS}},\beta_{3,{\bf dS}})$ by further implementing the limiting value $\nu=3/2 $ in the derived result of $(\alpha_{3,{\bf BD}},\beta_{3,{\bf BD}})$ for the SRII:
\bea \label{pspecsr2dS}
\Delta_{{\zeta\zeta},{\bf dS}}^2&=&\frac{H^2}{8\pi^2\epsilon  c_s M_p^2} 
\Bigg(\frac{k_e}{k_s}\Bigg)^6 (1+\sigma^2)\Bigg|\alpha_{3,{\bf dS}} e^{i\sigma}-\frac{(1+i \sigma)}{(1-i \sigma)}\beta_{3,{\bf dS}} e^{-i\sigma}\Bigg|^2,
\\
\Delta_{{\zeta\Pi_\zeta},{\bf dS}}^2&=& \frac{H^2}{8\pi^2\epsilon  c_s M_p^2}\Bigg(\frac{k_e}{k_s}\Bigg)^6 
\sigma^2 \Bigg|\alpha_{3,{\bf dS}} e^{i\sigma}-\beta_{3,{\bf dS}} e^{-i\sigma}\Bigg|^2,
\\
\Delta_{{\Pi_\zeta\zeta},{\bf dS}}^2&=& \frac{H^2}{8\pi^2\epsilon  c_s M_p^2}\Bigg(\frac{k_e}{k_s}\Bigg)^6 
\sigma^2\Bigg|\alpha_{3,{\bf dS}} e^{i\sigma}-\beta_{3,{\bf dS}} e^{-i\sigma}\Bigg|^2,
\\
\Delta_{{\Pi_\zeta\Pi_\zeta},{\bf dS}}^2&=&\frac{H^2}{8\pi^2 \epsilon  
 c_s M_p^2}\Bigg(\frac{k_e} {k_s}\Bigg)^6 
 \sigma^4 
 \Bigg|\alpha_{3,{\bf dS}} e^{i\sigma}-\beta_{3,{\bf dS}}  e^{-i\sigma}\Bigg|^2.
\eea

\subsubsubsection{Noise Matrix elements in SRII}

With a generic starting quantum vacuum state defined by coefficients $(\alpha_3,\beta_3)$ and an arbitrary de Sitter background spacetime denoted by $\nu$', the noise correlation matrix elements in the SRII may be expressed as follows:
\bea
\Sigma_{\zeta\zeta}&=&(1-\epsilon) 2^{2\nu-3}  (\sigma)^{3-2\nu} \frac{H^2}{8\pi^2\epsilon  c_s M_p^2} 
 \Bigg(\frac{k_e}{k_s}\Bigg)^6 \Bigg|\frac{\Gamma{(\nu)}}{\Gamma{(\frac{3}{2})}}\Bigg|^2(1+\sigma^2)\Bigg|\alpha_3 e^{-i(\frac{\pi}{2}(\nu+\frac{1}{2})-\sigma)}-\frac{(1+i \sigma)}{(1-i \sigma)}\beta_3 e^{i(\frac{\pi}{2}(\nu+\frac{1}{2})-\sigma)}\Bigg|^2,
\\
\Sigma_{\zeta\Pi_\zeta}&=&(1-\epsilon) 2^{2\nu-3}  (\sigma)^{3-2\nu} \frac{H^2}{8\pi^2\epsilon  c_s M_p^2}\Bigg(\frac{k_e}{k_s}\Bigg)^6 
 \Bigg|\frac{\Gamma{(\nu)}}{\Gamma{(\frac{3}{2})}}\Bigg|^2\Bigg\{\Bigg(\frac{3}{2}-\nu \Bigg)(1+\sigma^2)+\sigma^2\Bigg\}\Bigg\{\Bigg|\alpha_3 e^{-i(\frac{\pi}{2}(\nu+\frac{1}{2})-\sigma)}-\beta_3 e^{i(\frac{\pi}{2}(\nu+\frac{1}{2})-\sigma)}\Bigg|^2, \nonumber
 \\  && \quad \quad \quad \quad \quad \quad \quad \quad \quad \quad \quad 
+\frac{2\Bigg(\frac{3}{2}-\nu\Bigg)\sigma^2}{\Bigg\{\Bigg(\frac{3}{2}-\nu \Bigg) (1+\sigma^2)+\sigma^2\Bigg\}}\Bigg\{\alpha_3^*\beta_3 e^{2i(\frac{\pi}{2}(\nu+\frac{1}{2})-\sigma)}+\beta_3^*\alpha_3 e^{-2i(\frac{\pi}{2}(\nu+\frac{1}{2})-\sigma)} \Bigg\}\Bigg\},
\eea\bea
\Sigma_{\Pi_\zeta\zeta}&=&(1-\epsilon) 2^{2\nu-3}  (\sigma)^{3-2\nu} \frac{H^2}{8\pi^2\epsilon  c_s M_p^2}\Bigg(\frac{k_e}{k_s}\Bigg)^6 
 \Bigg|\frac{\Gamma{(\nu)}}{\Gamma{(\frac{3}{2})}}\Bigg|^2\Bigg\{\Bigg(\frac{3}{2}-\nu \Bigg)(1+\sigma^2)+\sigma^2\Bigg\}\Bigg\{\Bigg|\alpha_3 e^{-i(\frac{\pi}{2}(\nu+\frac{1}{2})-\sigma)}-\beta_3 e^{i(\frac{\pi}{2}(\nu+\frac{1}{2})-\sigma)}\Bigg|^{2}\nonumber
 \\ && \quad \quad \quad \quad \quad \quad \quad \quad \quad \quad \quad 
+ \frac{2\Bigg(\frac{3}{2}-\nu\Bigg)\sigma^2}{\Bigg\{\Bigg(\frac{3}{2}-\nu \Bigg) (1+\sigma^2)+\sigma^2\Bigg\}}\Bigg\{\alpha_3^*\beta_3 e^{2i(\frac{\pi}{2}(\nu+\frac{1}{2})-\sigma)}+\beta_3^*\alpha_3 e^{-2i(\frac{\pi}{2}(\nu+\frac{1}{2})-\sigma)} \Bigg\}\Bigg\},\quad\quad
\\
\Sigma_{\Pi_\zeta\Pi_\zeta}&=&(1-\epsilon) 2^{2\nu-3}(\sigma)^{3-2\nu}\frac{H^2}{8\pi^2 \epsilon  
 c_s M_p^2}\Bigg(\frac{k_e}{k_s}\Bigg)^6 
 \Bigg|\frac{\Gamma{(\nu)}}{\Gamma{(\frac{3}{2})}}\Bigg|^2 \Bigg[\Bigg(\Bigg(\frac{3}{2}- \nu\Bigg)+ \sigma^2 \Bigg)^2+\Bigg(\frac{3}{2}-\nu\Bigg)^2 \sigma^2 \Bigg ] \nonumber
 \\ &&  \quad \quad \quad \quad \quad \quad \quad \quad \quad \quad \quad \quad \quad \quad 
 \Bigg|\alpha_3 e^{-i(\frac{\pi}{2}(\nu+\frac{1}{2})-\sigma)}-\beta_3 \frac{\Bigg\{\Bigg(\frac{3}{2}- \nu\Bigg)(1+i \sigma )+ \sigma^2 \Bigg \}}{\Bigg\{\Bigg(\frac{3}{2}- \nu\Bigg)(1-i \sigma)+ \sigma^2 \Bigg \}} e^{i(\frac{\pi}{2}(\nu+\frac{1}{2})-\sigma)}\Bigg|^{2}.
\eea
The new set $(\alpha_{3, \bf dS},\beta_{3, \bf dS})$ is obtained by further implementing the Bunch Davies vacuum conditions and applying the limiting value of $\nu=3/2$ in the outcome of the Bunch Davies vacuum derived procedure. The simplified equation for noise matrix elements in precise de Sitter space is as follows:
\bea
\Sigma_{{\zeta\zeta},{\bf dS}}&=&(1-\epsilon) \frac{H^2}{8\pi^2\epsilon  c_s M_p^2} 
 \Bigg(\frac{k_e}{k_s}\Bigg)^6 (1+\sigma^2)\Bigg|\alpha_{3,{\bf dS}} e^{i\sigma}-\frac{(1+i \sigma)}{(1-i \sigma)}\beta_{3,{\bf dS}} e^{-i\sigma}\Bigg|^2,
\\
\Sigma_{{\zeta\Pi_\zeta},{\bf dS}}&=&(1-\epsilon) \frac{H^2}{8\pi^2\epsilon  c_s M_p^2}\Bigg(\frac{k_e}{k_s}\Bigg)^6 
 \sigma^2 \Bigg|\alpha_{3,{\bf dS}} e^{i\sigma}-\beta_{3,{\bf dS}} e^{-i\sigma}\Bigg|^2, 
\\
\Sigma_{{\Pi_\zeta\zeta},{\bf dS}}&=&(1-\epsilon)  \frac{H^2}{8\pi^2\epsilon  c_s M_p^2}\Bigg(\frac{k_e}{k_s}\Bigg)^6 
\sigma^2 \Bigg|\alpha_{3,{\bf dS}} e^{i\sigma}-\beta_{3,{\bf dS}} e^{-i\sigma}\Bigg|^2,
\\
\Sigma_{{\Pi_\zeta\Pi_\zeta},{\bf dS}}&=&(1-\epsilon) \frac{H^2}{8\pi^2 \epsilon  
 c_s M_p^2}\Bigg(\frac{k_e}{k_s}\Bigg)^6 
  \sigma^4  
 \Bigg|\alpha_{3,{\bf dS}} e^{i\sigma}-\beta_{3,{\bf dS}} e^{-i\sigma}\Bigg|^2.
\eea

\subsection{Stochastic-$\delta {\cal N}$ formalism and its applications in EFT of Stochastic Single Field Inflation}

Because it can directly relate the probability distribution of the stochastic e-folding variable ${\cal N}$ to the statistics of the curvature perturbation variable $\zeta$, the $\delta N$-formalism \cite{Sugiyama:2012tj,Dias:2012qy,Naruko:2012fe,Takamizu:2013gy,Abolhasani:2013zya,Clesse:2013jra,Chen:2013eea,Choudhury:2014uxa,vandeBruck:2014ata,Dias:2014msa,Garriga:2015tea,Choudhury:2015hvr,Choudhury:2016wlj,Choudhury:2017cos,Choudhury:2018glz, Starobinsky:1985ibc,Sasaki:1995aw,Sasaki:1998ug,Lyth:2005fi,Lyth:2004gb,Abolhasani:2018gyz,Passaglia:2018ixg} has become a prominent technique to compute cosmological correlations in the super-Hubble scales and has found interesting applications in the framework of stochastic inflation \cite{Cruces:2021iwq,Prokopec:2019srf,PerreaultLevasseur:2014ziv,Cruces:2022imf,Fujita:2014tja,Tada:2023fvd,Honda:2023unh,Asadi:2023flu,Mishra:2023lhe,Rigopoulos:2022gso,Escriva:2022duf,Tomberg:2022mkt,Ahmadi:2022lsm,Talebian:2022jkb,Figueroa:2021zah,Rigopoulos:2021nhv,De:2020hdo,Firouzjahi:2020jrj,Vennin:2020kng,Bounakis:2020jdx,Firouzjahi:2018vet,Tada:2016pmk,Assadullahi:2016gkk,PerreaultLevasseur:2014ziv}. At leading order, the variations in the e-folds across various patches of homogeneous FLRW Universes are associated with the quantity of curvature perturbations produced at a final time, often selected towards the end of the super-Hubble scales. 

Typically, we begin with the presumption that the spacetime is isotropic, locally homogeneous, and unperturbed, and that it is governed by the metric:
\bea
ds^{2} = -dt^{2} + a^{2}(t)\delta_{ij}dx^{i}dx^{j},
\eea
where $a(t)\sim \exp(Ht)$ (where $\dot{H}\ne 0$ such that $\epsilon=-\dot{H}/H^2$ exists) and $t$ represent the cosmic time. The Hubble parameter is still the sole relevant scale to work in the unperturbed Universes scenario. Any quantity of interest is taken to be sufficiently smooth on the scales of order $k^{-1}$ in the presence of some smoothing. The ensuing anisotropies can occur perturbatively in powers of ${\cal O}(k/aH)$. The perturbations arise as the related coordinate scale $k^{-1}$ approaches the Hubble radius, $k \sim aH$.

Next, we think about selecting a certain gauge to eliminate the gauge redundancy. Selecting the uniform density gauge with uniform energy density for the spatial slices of a defined time $t$ is one way to do this. With this selection, the perturbed metric may then be written as follows, taking into account only the scalar perturbations at this level:
\bea
ds^{2} = -dt^{2} + a^{2}(t)e^{2\zeta(t,{\bf x})}dx^{i}dx^{j},
\eea
\begin{figure*}[htb!]
    	\centering
    {
       \includegraphics[width=17cm,height=12cm]{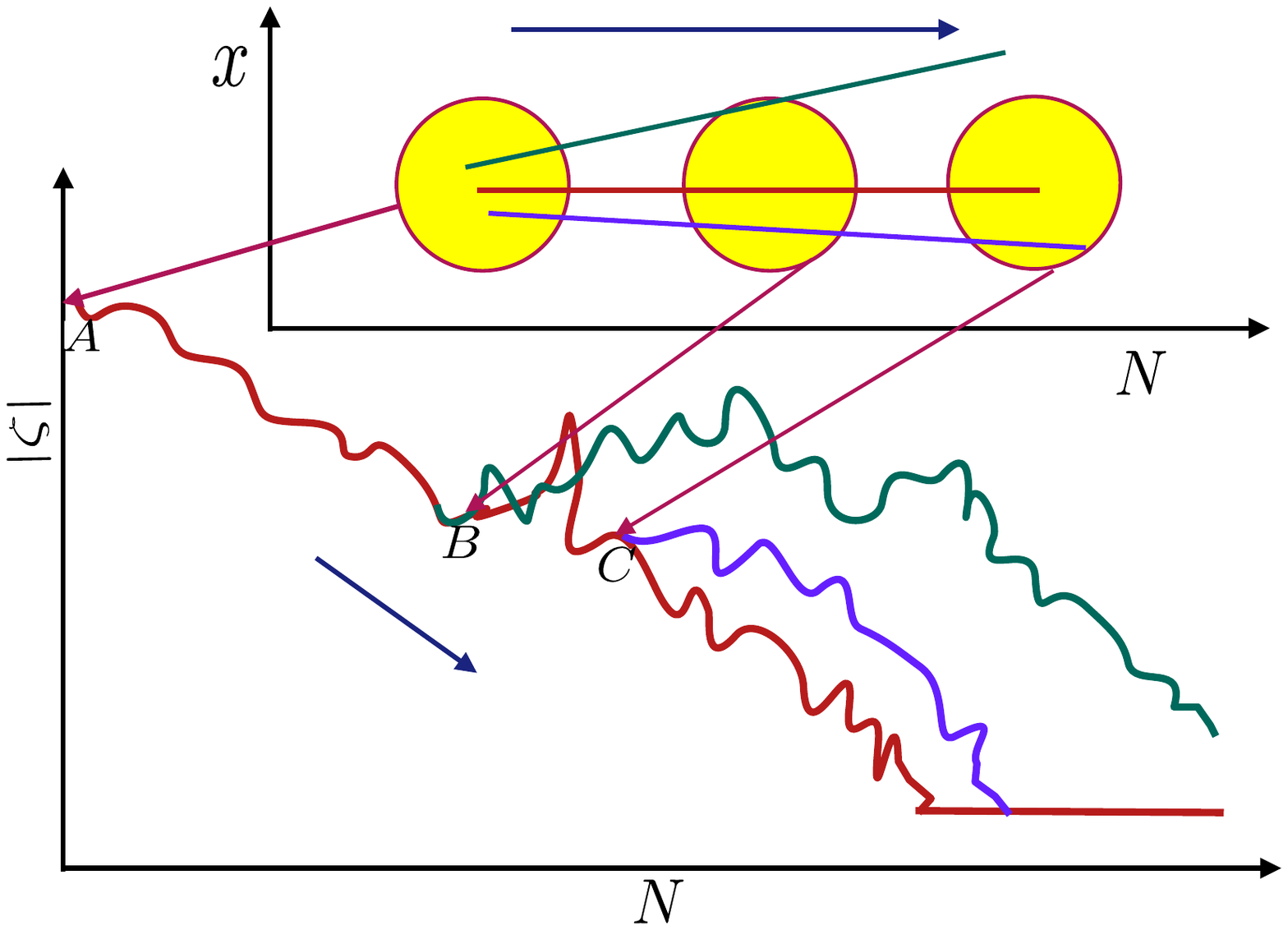}
        \label{CGCP}
    } 
    \caption[Optional caption for list of figures]{A diagrammatic representation of the perturbation of coarse-grained curvature change using e-folds $N$. The coarse-grained $\zeta$ value at various spatial positions is displayed in the bottom graphs. Initially, at $A$, a collection of spatial locations reside inside the same Hubble radius of the centre point, within which the white noises are correlated and whose progression with time is red. The other points' temporal branching development is depicted by the green and blue lines. After a given amount of time, distinct points at $B$ and $C$ travel outside of the original Hubble radius and cause the white noises to behave differently going forward. }
\label{deltaNschematic1}
    \end{figure*}
in which the curvature disturbance that permeates the whole visible Universe is represented by the new variable $\zeta(t,{\bf x})$. The perturbation may be expressed as $\tilde{a}(t,{\bf x}) = a(t)e^{2\zeta(t,{\bf x})}$, and this form of the above-perturbed metric enables the introduction of a local scale factor made up of a global time-dependent component. This new scale factor allows us to include the amount of expansion realised between a final, constant $t$ hypersurface ($\Omega_{f}$) assigned with some uniform-density and an initially flat, constant $t=t_{i}$ hypersurface ($\Omega_{i}$):
\bea
N(t,{\bf x}) = \ln{\bigg(\frac{\tilde{a}(t,{\bf x})}{a(t_{i})}\bigg)}.
\eea
The amount of curvature disturbance experienced at a spatial location $x_{i}$, at instant $t$, in respect to the e-foldings elapsed up to that instant, may then be expressed using the preceding formula:
\bea
\zeta(t,{\bf x}_{i}) = N(t,{\bf x}_{i}) - \bar{N}(t),\quad\quad \text{where}\;\bar{N}(t) = \ln{\bigg(\frac{a(t)}{a(t_{i})}\bigg)},
\eea
where the amount of expansion for unperturbed, FLRW Universes is indicated by $N(t,{\bf x}_i)$, and the unperturbed quantity is shown by the bar notation.

In this section, we will quickly review the advantages of merging the stochastic formalism in inflation with the previously discussed $\delta N$-formalism. Our focus is on a group of FLRW Universes that have developed from a certain starting point on the phase space variables, rather than just one. When these variables are merged, they may be represented as a phase space vector ${\Gamma}_{i}=\{\zeta_{i},\Pi_{\zeta,i}\}$, where the different components are labelled with an index. To express the stochastic formalism in terms of curvature perturbation, one must build an effective theory with low energy for the long-wavelength or IR component of the initial primordial fluctuations. These fluctuations are described as $k_{\sigma}=\sigma aH$, where $\sigma\ll 1$, and are coarse-grained at a certain fixed scale, around the Hubble radius. The resultant scalar curvature perturbation is as follows: 
\bea
\zeta_{\rm cg}({\bf x}) = \int_\mathbb{R^3}\frac{d^3 k }{(2\pi)^3 }\zeta_{k}e^{-i{\bf k}.{\bf x}},
\eea
where the IR modes in the coarse-grained curvature perturbation, or modes with wavenumber $k<k_{\sigma}$, are taken into account during integration.
An increasing number of short-scale modes engage in the zone of stochastic effects as the Horizon size decreases, being 'classicalized' and ultimately entering the IR sector. This leads to the emergence of classical sounds, which influence the Langevin equation-described super-Hubble mode dynamics. As a result, there are fluctuations in the number of e-folds realised for each point. The noises only become correlated when they are within the Hubble radius, emphasising their Markovian nature. Over this radius, the spatial points begin to evolve independently of one another with distinct noises. 

For now, we concentrate on a single point in space. From an initial condition to a final hypersurface, the quantity realised along the worldline trajectory of the single point becomes a stochastic variable, denoted by ${\cal N}$. The curvature perturbation generated at a coarse-grained spatial point between the scale crossing at an initial conformal time $\tau_{i}$ and the scale crossing out at a final conformal time $\tau_{f}$ can be related to the perturbation in the e-folding realised along the same worldline in between $\tau_{i}$ and $\tau_{f}$ using the $\delta N$-formalism discussed earlier in this section:
\bea \label{coarsezeta}
\zeta_{\rm cg}({\bf x}) = {\cal N}({\bf x})-\langle{\cal N}\rangle = \delta {\cal N},
\eea
where the statistical average obtained from solving the Langevin equation for several realisations at a certain spatial position is shown by the angle brackets. We stress that the realised e-folds naturally experience variations owing to random disturbances; hence, ${\cal N}$ is a stochastic variable whose statistical features may be evaluated afterwards. These fluctuations are nothing more than the comoving curvature perturbation that the $\delta N$-formalism produces at the final hypersurface. The spatial point evolution during inflation is depicted in fig. \ref{deltaNschematic1}, along with the points at $B$ and $C$ where Gaussian random noises begin to impact the points independently. All spatial locations inside a Hubble patch (yellow circles) have the same coarse-grained value of the curvature perturbation (yellow circles), except from points $A$ to $B$ and $C$, after which they develop in a statistically separate way (shown by green and blue lines).
\begin{table}[H]
\centering
\begin{tabular}{|c|c|c|}

\hline\hline
\multicolumn{3}{|c|}{\normalsize \textbf{Comparison between the $\delta N$ (without stochasticity) and Stochastic-$\delta N$ formalism }} \\

\hline

 & \bf{$\delta N$-formalism} & \bf{Stochastic-$\delta N$ formalism} \\
\hline
$1).$ & e-folds of evolution vary ($\delta N$) for each point $(\bf{x})$  &  e-folds for a specific point $(\bf{x})$ on the final slice \\
& on the final hypersurface slice. & receive quantum fluctuations throughout evolution.\\ \hline
$2).$ & Curvature perturbations on the final slice  &  Curvature perturbations requires coarse-graining\\
& directly related to $\delta N$ in the super-Horizon. & before relating to e-folds statistics in the super-Horizon. \\ \hline
$3).$ & $\zeta(t,{\bf x}_{i}) = N(t,{\bf x}_{i}) - \bar{N}(t) = \delta N$,  &   $\zeta_{\rm cg}({\bf x}) = {\cal N}({\bf x})-\langle{\cal N}\rangle = \delta {\cal N}$,   \\
& $N$: e-fold amount for expanding FLRW Universes  &  ${\cal N}$: stochastic variable under Langevin equations for $\zeta_{\rm cg}(\bf{x})$ \\ \hline
$4).$ & No influence of quantum noises on evolution history. & Quantum noises \textit{classicalize} after cut-off and affect $\zeta_{\rm cg}$ evolution.                              \\ \hline
\hline

\end{tabular}
\caption{Study of the differences between the two frameworks: stochastic-$\delta N$ and $\delta N$ (without stochasticity). }
\label{tabcomparison}

\end{table}
The salient features of the stochastic-$\delta N$ formalism and the standard $\delta N$ (sans stochasticity) are delineated in Table \ref{tabcomparison}. In lieu of solving the Langevin equation over several geographical sites, we demonstrate above how the potential of the $\delta N$-formalism may be harnessed using stochastic formalism. A variable that may be readily calculated from this stochastic-$\delta N$ formalism is the power spectrum $\Delta^{2}_{f,g}$, which we will now discuss. For the e-folding fluctuation $\delta N$, the dimensionless power spectrum may be expressed as follows using Fourier mode decomposition, where $\{f,g\}$ denotes the phase space variables $\{\zeta,\Pi_{\zeta}\}$:
\bea
\Delta^{2}_{\delta N}(k) = \frac{k^{3}}{2\pi^{2}}\int\;d^{3}{\bf x}\int\;d^{3}{\bf x'}\langle\delta N({\bf x'})\delta N({\bf x})\rangle e^{-i{\bf k}.({\bf x}-{\bf x'})}.
\eea
This, when flipped, yields the variance in e-folds as follows:
\bea \label{varPspec}
\langle\delta{\cal N}^{2}\rangle = \langle({\cal N}-\langle{\cal N}\rangle)^{2}\rangle = \langle{\cal N}^{2}\rangle - \langle{\cal N}\rangle^{2}
&=& \int\frac{d^{3}{\bf k}}{(2\pi)^{3}}\int\frac{d^{3}{\bf k'}}{(2\pi)^{3}}\langle\delta N_{\bf k}\delta N_{\bf k'}\rangle \nonumber\\
&=& \int\frac{d^{3}{\bf k}}{(2\pi)^{3}}\int\frac{dk'}{k'}\frac{k'^{3}}{2\pi^{2}}\langle\delta N_{\bf k}\delta N_{\bf k'}\rangle \nonumber\\
&=& \int\frac{d^{3}{\bf k}}{(2\pi)^{3}}\int\frac{dk'}{k'}\Delta^{2}_{\delta N}(k')\delta^{(3)}({\bf k}+{\bf k'}) \nonumber\\
&=& \int_{k}^{k_{f}}\frac{dk}{k}\Delta^{2}_{\delta N}(k) \nonumber\\
&=& \int_{\ln{k_{f}}-\langle{\cal N}\rangle}^{\ln{k_{f}}}\Delta^{2}_{\delta N}(N)dN,
\eea

wherein the power spectrum definition is implemented in the second line, and the final equality conversion is carried out for a given wavenumber $k=|{\bf k}|$ using $\ln{(k)}=\ln{(aH)}$, and the average e-foldings is equal to $\langle{\cal N}\rangle=\ln{(a_{f}H/aH)}=\ln{(k_{f}/k)}$, assuming constant Hubble parameter $H$. The stochastic variable $\langle{\cal N}\rangle$ on the right is averaged to produce the statistical quantity known as variance on the left. Finally, using the information provided, the dimensionless power spectrum for the curvature perturbation may be expressed as follows:
\bea \label{pspecdeltaN}
\Delta^{2}_{\zeta\zeta}(k) = \Delta^{2}_{\delta N}(k) = \frac{d}{d\langle{\cal N}\rangle }\langle\delta{\cal N}^{2}\rangle \big|_{\langle{\cal N}\rangle=\ln{(k_{f}/k)}},
\eea
where the above employs all the modes $k$ crossing out the Hubble radius between some starting instant and the end of inflation, and the first equality arises from eqn. (\ref{coarsezeta}). Similarly, we can also express several additional higher-order correlation functions in terms of the higher-order statistical moments of $\langle\delta{\cal N}\rangle$. In position space and around a Gaussian component, the local non-gaussianity is typically expressed as an expansion ansatz for the curvature perturbation $\zeta({\bf x})$:
\bea
\zeta({\bf x}) = \zeta_{g}({\bf x}) + \frac{3}{5}f_{\rm NL}\zeta^{2}_{g}({\bf x}) + \frac{9}{25}g_{\rm NL}\zeta^{3}_{g}({\bf x}) + {\cal O}(\zeta^{4}_{g}({\bf x})) + \cdots
\eea
The perturbation component following Gaussian statistics is denoted by $\zeta_{g}({\bf x})$. Our focus is on estimating the different non-Gaussian parameters presented above, namely $f_{\rm NL},g_{\rm NL},\tau_{\rm NL}$, utilising the stochastic-$\delta N$ formalism and the statistics of the e-folds ${\cal N}$. 

Now let us consider the situation of the bispectrum and $f_{\rm NL}$, the non-Gaussianity parameter. The three-point correlation function $\langle\zeta_{\bf k_1}\zeta_{\bf k_2}\zeta_{\bf k_3}\rangle$ is defined in respect to the bispectrum. A similar method for the bispectrum, this time doubly integrated, yields the third-moment of the e-folds $\langle\delta{\cal N}^{3}\rangle$, just as the variance is connected to the inverse Fourier mode of the power spectrum, as in eqn. (\ref{varPspec}). We can write to illustrate this:
\bea \label{thirdmoment}
\langle\delta{\cal N}^{3}\rangle = \langle({\cal N}-\langle{\cal N}\rangle)^{3}\rangle &=& \langle{\cal N}^{3}\rangle - 3\langle{\cal N}\rangle\langle{\cal N}^2\rangle + 2\langle{\cal N}\rangle^{3}\nonumber\\
&=& \int\frac{d^{3}{\bf k_1}}{(2\pi)^{3}}\int\frac{d^{3}{\bf k_2}}{(2\pi)^{3}}
\int\frac{d^{3}{\bf k_3}}{(2\pi)^{3}}\langle\delta N_{\bf k_1}\delta N_{\bf k_2}\delta N_{\bf k_3}\rangle, \nonumber\\
&=& \int\frac{d^{3}{\bf k_1}}{(2\pi)^{3}}\int\frac{d^{3}{\bf k_2}}{(2\pi)^{3}}\int\frac{d^{3}{\bf k_3}}{(2\pi)^{3}}(2\pi)^{3}{\cal B}_{\delta N}(k_1,k_2,k_3)\delta^{(3)}({\bf k_1}+{\bf k_2}+{\bf k_3}),\nonumber\\
&=& \int\frac{d^{3}{\bf k_1}}{(2\pi)^{3}}\int\frac{d^{3}{\bf k_2}}{(2\pi)^{3}}{\cal B}_{\delta N}(k_1,k_2,k_3)\big|_{{\bf k_1}+{\bf k_2}+{\bf k_3}=0},
\eea
This, because of a triangle restriction from the delta function, demands additional volume integral across a tetrahedral area. For the time being, nevertheless, we note that the bispectrum provides double the integrated contribution to the third moment of e-folds $\langle\delta{\cal N}^{3}\rangle$. As a result, we can again construct the formula for the Bispectrum using eqn. (\ref{coarsezeta}):
\bea
{\cal B}_{\zeta\zeta\zeta}(k_1,k_2,k_3) \propto \frac{d\langle\delta{\cal N}^{3}\rangle^2}{d\langle{\cal N}\rangle^2},
\eea
The relationship between the bispectrum and the parameter $f_{\rm NL}$ is now as follows:
\bea \label{bispecPower}
{\cal B}_{\zeta\zeta\zeta}(k_1,k_2,k_3) = \frac{6}{5}f_{\rm NL}\left[\Delta^{2}_{\zeta\zeta}(k_1)\Delta^{2}_{\zeta\zeta}(k_2) + \Delta^{2}_{\zeta\zeta}(k_1)\Delta^{2}_{\zeta\zeta}(k_3) + \Delta^{2}_{\zeta\zeta}(k_2)\Delta^{2}_{\zeta\zeta}(k_3)\right],
\eea
where the relation for $f_{\rm NL}$ may be considered using eqn. (\ref{pspecdeltaN}):
\bea
f_{\rm NL} = \frac{5}{36}\frac{d\langle\delta{\cal N}^{3}\rangle^2}{d\langle{\cal N}\rangle^2}\big(\Delta^{2}_{\zeta\zeta}(k)\big)^{-2} = \frac{5}{36}\frac{d\langle\delta{\cal N}^{3}\rangle^2}{d\langle{\cal N}\rangle^2}\bigg(\frac{d\langle\delta{\cal N}^{2}\rangle}{d\langle{\cal N}\rangle}\bigg)^{-2},
\eea
It is the bispectrum divided by the power spectrum, with convention providing the outside factor of $5/36$. Similar to this, the trispectrum's formulation entails calculating the third derivative of the e-folds $\langle\delta{\cal N}^{4}\rangle$ with respect to $\langle{\cal N}\rangle$, as it now gets three integrated contributions. It is associated with the four-point correlation function, which we begin with the fourth moment of e-folds, $\langle\zeta_{\bf k_1}\zeta_{\bf k_2}\zeta_{\bf k_3}\zeta_{\bf k_4}\rangle$:
\bea \label{fourthmoment}
\langle\delta{\cal N}^{4}\rangle = \langle({\cal N}-\langle{\cal N}\rangle)^{4}\rangle &=& \langle{\cal N}^{4}\rangle - 4\langle{\cal N}\rangle\langle{\cal N}^3\rangle + 6\langle{\cal N}\rangle^{2}\langle{\cal N}^{2}\rangle - 3\langle{\cal N}\rangle^{4},\nonumber\\
&=& \int\frac{d^{3}{\bf k_1}}{(2\pi)^{3}}\int\frac{d^{3}{\bf k_2}}{(2\pi)^{3}}
\int\frac{d^{3}{\bf k_3}}{(2\pi)^{3}}\int\frac{d^{3}{\bf k_4}}{(2\pi)^{3}}\langle\delta N_{\bf k_1}\delta N_{\bf k_2}\delta N_{\bf k_3}\delta N_{\bf k_4}\rangle,\nonumber\\
&=& \int\frac{d^{3}{\bf k_1}}{(2\pi)^{3}}\int\frac{d^{3}{\bf k_2}}{(2\pi)^{3}}\int\frac{d^{3}{\bf k_3}}{(2\pi)^{3}}\int\frac{d^{3}{\bf k_4}}{(2\pi)^{3}}(2\pi)^{3}T_{\delta N}(k_1,k_2,k_3,k_4)\delta^{(3)}({\bf k_1}+{\bf k_2}+{\bf k_3}+{\bf k_4}),\quad\quad\quad\nonumber\\
&=& \int\frac{d^{3}{\bf k_1}}{(2\pi)^{3}}\int\frac{d^{3}{\bf k_2}}{(2\pi)^{3}}\int\frac{d^{3}{\bf k_3}}{(2\pi)^{3}}T_{\delta N}(k_1,k_2,k_3,k_4)\big|_{{\bf k_1}+{\bf k_2}+{\bf k_3}+{\bf k_4}=0},
\eea
for the linked portion of the four-point function, where the relation has been utilised. In conjunction with the power spectrum, the linked component of the trispectrum for the curvature perturbations is now defined as follows:
\bea \label{trispecPower}
T_{\zeta\zeta\zeta\zeta}(k_1,k_2,k_3,k_4) &=& \tau_{\rm NL}\big[\Delta^{2}_{\zeta\zeta}(k_{13})\Delta^{2}_{\zeta\zeta}(k_{3})\Delta^{2}_{\zeta\zeta}(k_{4}) + 11\;\text{perms.} \big]\nonumber\\
&& + \frac{54}{25}g_{\rm NL}\big[\Delta^{2}_{\zeta\zeta}(k_{2})\Delta^{2}_{\zeta\zeta}(k_{3})\Delta^{2}_{\zeta\zeta}(k_{4}) + 3\;\text{perms.} \big].
\eea
The trispectrum got its name because of its cubic reliance on the power spectrum and the notation for magnitude $k_{ij} = |{\bf k}_i + {\bf k}_j|$. The linked component of the trispectrum yields two separate non-linearity parameters, $\tau_{\rm NL}$ and $g_{\rm NL}$, which we see to develop depending on their $k$ dependency. Given that the magnitude of $k_{ij}=k_{ji}$ stays constant, $3$ different options for the indices in $k_{ij}$ account for the $12$ permutations. Using the momentum conserving delta function, we obtain $k_{12}=k_{34},\;k_{23}=k_{14},\;k_{13}=k_{24}$.
Once more, the conservation of momentum leads to the $4$ permutations. In terms of the derivatives of the fourth moment, we can finally write the following relations for the non-Gaussianity parameters from eqns. (\ref{coarsezeta},\ref{pspecdeltaN},\ref{trispecPower}):
\bea
\tau_{\rm NL} &=& \frac{1}{36}\bigg(\frac{d^2\langle\delta{\cal N}^3\rangle}{d\langle{\cal N}\rangle^2}\bigg)^{2}\big(\Delta^{2}_{\zeta\zeta}(k)\big)^{-4} = \frac{1}{36}\bigg(\frac{d^2\langle\delta{\cal N}^3\rangle}{d\langle{\cal N}\rangle^2}\bigg)^{2}\bigg(\frac{d\langle\delta{\cal N}^{2}\rangle}{d\langle{\cal N}\rangle}\bigg)^{-4},\\
g_{\rm NL} &=& \frac{d\langle\delta{\cal N}^{4}\rangle^3}{d\langle{\cal N}\rangle^3}\big(\Delta^{2}_{\zeta\zeta}(k)\big)^{-3} = \frac{d\langle\delta{\cal N}^{4}\rangle^3}{d\langle{\cal N}\rangle^3}\bigg(\frac{d\langle\delta{\cal N}^{2}\rangle}{d\langle{\cal N}\rangle}\bigg)^{-3},
\eea
where convention-driven factors once more account for the factor $1/36$.

We use the preceding advancements to explain the probability distribution driving the inflation duration. We have now outlined the stochastic-$\delta N$ formalism in the context of EFT of single field inflation and its applications into computing higher-order correlation functions. When focused on the drift-dominated phase during inflation, computation of this distribution offers significant advantages in terms of determining the non-Gaussianity parameters and determining the mass fraction of the PBH. Our goal is to carry out these computations using the stochastic-$\delta N$ formalism in conjunction with the EFT image.

\subsection{Probability Distribution Function from Fokker-Planck equation }

In order to assist us in the analysis of the probability distribution functions, we derive the Fokker-Planck equation in its entirety in this section. This definition of the probability density function is for the stochastic e-folds variable $cal N$ elapsed between the start and conclusion of the evolution. The Fokker-Planck equation allows us to investigate the evolution of the probability density function from one point to another in the field space. This entails the probability distribution function $P_{\bf \Gamma}({\cal N})$ being affected by the adjoint Fokker-Planck operator in the following way:
\bea 
\frac{\partial }{\partial {\cal N}}P_{\bf \Gamma}({\cal N})=\Bigg(F_i \frac{\partial }{\partial \Phi_i}+\frac{1}{2}\Sigma_{ij}\frac{\partial ^2}{\partial \Gamma_i \partial \Gamma_j}\Bigg) P_{\bf \Gamma}({\cal N}), 
\eea
where the following can be used to extend the Adjoint Fokker-Planck operator:
\bea
F_i \frac{\partial }{\partial \Gamma_i}+\frac{1}{2}\Sigma_{ij}\frac{\partial ^2}{\partial \Gamma_i \partial \Gamma_j} &=& \Pi_{\zeta}\frac{\partial}{\partial \zeta} - (3-\epsilon)\Pi_{\zeta}\bigg[1-\frac{2(s-\frac{\eta}{2})}{3-\epsilon}\bigg]\frac{\partial}{\partial\Pi_{\zeta}}  \nonumber\\
&& \quad\quad\quad\quad\quad\quad\quad + \frac{1}{2}\bigg(\Sigma_{\zeta\zeta}\frac{\partial^{2}}{\partial\zeta^{2}}  + \Sigma_{\zeta\Pi_{\zeta}}\frac{\partial^{2}}{\partial\zeta\partial\Pi_{\zeta}} + \Sigma_{\Pi_{\zeta}\zeta}\frac{\partial^{2}}{\partial\Pi_{\zeta}\partial\zeta} + \Sigma_{\Pi_{\zeta}\Pi_{\zeta}}\frac{\partial^{2}}{\partial\Pi_{\zeta}\partial\Pi_{\zeta}}\bigg).
\eea
For the time being, we examine this equation using the super-Hubble conditions, $\sigma \ll 1$, which makes the noise matrix components $\Sigma_{\zeta\Pi_{\zeta}},\;\Sigma_{\Pi_{\zeta}\zeta},\Sigma_{\Pi_{\zeta}\Pi_{\zeta}}$ essentially insignificant. In order to streamline the Fokker-Planck operator's representation, we rescale the coarse-grained variables $\{\zeta,\Pi_{\zeta}\}$ into the new variables using the subsequent method:
\bea \label{newphasevars}
\zeta = f,\quad\quad \Pi_{\zeta} = -3y,
\eea
Using this, we see how the different differential operators that are present are transformed as follows:
\bea
&&\frac{1}{2}\Sigma_{\zeta\zeta}\frac{\partial^{2}}{\partial\zeta^{2}} = \frac{1}{\mu^{2}}\frac{\partial^{2}}{\partial f^{2}},\\
&&\Pi_{\zeta}\frac{\partial}{\partial\Pi_{\zeta}} = y\frac{\partial}{\partial y},\\
&&\Pi_{\zeta}\frac{\partial}{\partial\zeta} = -3y\frac{\partial}{\partial f}.
\eea
Now that the variables have been rescaled, we may rewrite the Fokker-Planck operator as follows:
\bea
\frac{\partial }{\partial \mathcal{N} }P_{\bf \Gamma}(\mathcal{N})&=& \bigg[\frac{1}{\mu^{2}}\frac{\partial^{2}}{\partial f^{2}} - 3\bigg\{y\frac{\partial}{\partial f} + \bigg(1-\frac{\epsilon}{3}\bigg)\bigg(1-\frac{2(s-\frac{\eta}{2})}{3-\epsilon}\bigg)y\frac{\partial}{\partial y} \bigg\}\bigg]P_{\bf \Gamma}({\cal N}),
\eea
whereby we also carry out the subsequent series of re-definitions for clarity:
\bea \label{EFTparam}
C = \bigg(1-\frac{\epsilon}{3}\bigg)\bigg(1-\frac{2(s-\frac{\eta}{2})}{3-\epsilon}\bigg),\quad\quad Cf = x,\quad\quad \left(\frac{C}{\mu}\right)^{2} = \frac{1}{\tilde{\mu}^{2}}
\eea
and the result of these modifications is the version of our interest based on the Fokker-Planck equation:
\bea \label{FPE-EFT}
\frac{\partial }{\partial \mathcal{N} }P_{\bf \Gamma}(\mathcal{N})&=& \bigg[\frac{1}{\tilde{\mu}^{2}}\frac{\partial^{2}}{\partial x^{2}} - 3Cy\bigg\{\frac{\partial}{\partial x} + \frac{\partial}{\partial y}\bigg\} \bigg]P_{\bf \Gamma}({\cal N}). 
\eea
Observe that the coefficient $C$, which we will hereafter refer to as the characteristic parameter, contains the changes resulting from an underlying EFT configuration. This coefficient will exhibit observable changes when we comprehend the development of the probability distribution in the following section. Additionally, the coefficient $\mu$, which contains the auto-correlated power spectrum element for $\zeta$, feels the shift from $C$. While we are dealing with non-canonical single-field inflation models for $C\ne 1$, in the situation when $C=1$ simplifies to the canonical single-field inflation scenario. We examine the technique in the next part that will allow us to compute this PDF for a generic EFT configuration based on the conditions that are of interest.

\subsection{Characteristic function and Initial Conditions}

We see the evolution of the associated PDF guided by the Fokker Planck equation for the e-folds ${\cal N}$, which are our stochastic variable of interest in the stochastic inflation framework. Information about the moments that are linked with this PDF is necessary to completely understand its nature, as these moments affect its statistical characteristics. Here, we go over the characteristic function approach that allows us to reconstruct the PDF in its entirety and assess the moments of any degree for a particular PDF, $P_{\Gamma_i}({\cal N})$. 

With respect to the phase space ${\Gamma_i} = \{\zeta,\Pi_{\zeta}\}$ and the variable ${\cal N}$, the probability distribution is transformed by the characteristic function from a function of $3$ variables to a function of only the phase space variables ${\Gamma_i}$. The definition of it is the following: it is the ${\cal N}$ co-ordinate Fourier transform of the original PDF:
\bea \label{characterDefn}
\chi(t;x,y) = \int_{-\infty}^{+\infty}e^{it{\cal N}}P_{{\Gamma_i}}({\cal N})d{\cal N},
\eea
where $t$, a complex quantity in general, serves as a dummy variable. The different statistical moments connected to the complete PDF $P_{\Gamma_i}({\cal N})$ may be obtained with the help of this function, $\chi(t;{\Gamma_i})$. Conversely, the PDF may be obtained as follows from the characteristic function:
\bea \label{characterFourier}
P_{{\Gamma_i}}({\cal N}) = \frac{1}{2\pi}\int_{-\infty}^{+\infty}e^{-it{\cal N}}\chi(t;{\Gamma_i})dt.
\eea
By extending $\chi(t;{\Gamma_i})$ around $t=0$, one may extract the appropriate information for the moments. This observation can be made as follows:
\bea \label{chimoments}
\chi(t;{\Gamma_i}) = \sum_{n=0}^{\infty}\frac{(it)^{n}}{n!}\langle{\cal N}^{n}({\Gamma_i})\rangle \implies \langle{\cal N}^{n}({\Gamma_i})\rangle = i^{-n}\frac{\partial^{n}}{\partial t^{n}}\chi(t;{\Gamma_i})|_{t=0},
\eea
where the $n$th moment of the complete PDF, $\langle{\cal N}^{n}({\Gamma_i})\rangle$, is indicated by the index $n$. The characteristic function also complies with the following when the definition of PDF from eqn. (\ref{characterFourier}) is entered into the Fokker-Planck eqn. (\ref{FPE-EFT}):
\bea \label{FPEcharacter}
\bigg[\frac{1}{\tilde{\mu}^{2}}\frac{\partial^{2}}{\partial x^{2}} - 3Cy\bigg\{\frac{\partial}{\partial x} + \frac{\partial}{\partial y}\bigg\} + it\bigg]\chi(t;{\Gamma_i}) = 0.
\eea
The function $\chi(t;{\Gamma_i})$ can be identified by solving this equation. The characteristic function solutions under certain $y$ parameter circumstances and in the presence of specified boundary conditions required to adequately handle the aforementioned two-dimensional partial differential equation will be analysed in the following sections.

\subsection{Diffusion Dominated Regime: Methods to solve for the Characteristic function}

We address the domain where quantum diffusion effects dominate the dynamics in this section. In addition, the conjugate momentum variable, as shown in eqn.(\ref{newphasevars}), assumes extremely tiny values in this regime, $0< y\ll 1$. Since we will see characteristic characteristics in the upper tail of the PDF—exactly the region of interest for PBH formation—this regime is important from the standpoint of more precisely analysing PBH. Additionally, as it diminishes in this domain, it may be seen from the standpoint of the conjugate field momenta. As a result, diffusion effects predominantly control the behaviour of the scalar perturbations and PDF characteristics. It is necessary to solve the partial differential equation in eqn.(\ref{FPEcharacter}) in order to produce the PDF. In order to construct the PDF, we must first solve for the characteristic function $(\chi)$. This section covers the associated methods.

In light of this, we begin with the following variable re-definition:
\bea
u= x-y,\quad\quad\quad v=y,
\eea
that reduces the characteristic function equation to the following simpler form:
\bea \label{FPEcharacterDiff}
\bigg[\frac{1}{\tilde{\mu}^{2}}\frac{\partial^{2}}{\partial u^{2}} - 3Cv\frac{\partial}{\partial v} + it\bigg]\chi(t;u,v)=0.
\eea
Now that the variables in the aforementioned equation can be separated, it can take an ansatz of the following form:
\bea \label{chiansatz}
\chi(t;u,v) = \sum_{n=0}^{\infty}v^{n}U_{n}(t;u),
\eea
and in this case, we make additional use of the fact that the function $R(t;r)$ permits oscillatory solutions, as shown by the form of the new partial differential equation, eqn.(\ref{FPEcharacterDiff}):
\bea
\frac{\partial^{2}}{\partial r^{2}}U_{n}(t;u) + \omega_{n}^{2}(t)U_{n}(t;u) = 0.
\eea
When the impacts of the EFT parameter are included in the frequency as follows:
\bea \label{omega1}
\omega_{n}^{2}(t) = (it-3Cn)\mu^{2}.
\eea
We now construct the required boundary conditions that may characterise the behaviour of the perturbations as they enter and depart the diffusion-dominated zone, allowing us to fully utilise the ansatz and get $\chi(t;u,v)$. In the first case, the perturbations are almost completely unaffected by quantum diffusion processes, and the perturbations are still absent under the influence of the USR. In the second case, on the other hand, the diffusion effects control the overall dynamics and determine how the PDF behaves at large perturbation values. The characteristic function is used to write the previously specified criteria for the converted phase space variables ${\Gamma_i} = \{u,v\}$ as follows:
\bea 
&& \label{bdrycondn1}\underline{\bf Boundary \;Condition\; I:}\quad\quad\quad\quad\chi(t;\Gamma_i)|_{u+v=0} = 1,\\
&& \label{bdrycondn2}\underline{\bf Boundary \;Condition\; II:}\quad\quad\quad \left.\frac{\partial\chi(t;\Gamma_i)}{\partial u}\right\vert_{u+v=1} = 0.
\eea
\begin{figure*}[ht!]
    	\centering
    {
       \includegraphics[width=18cm,height=12cm]{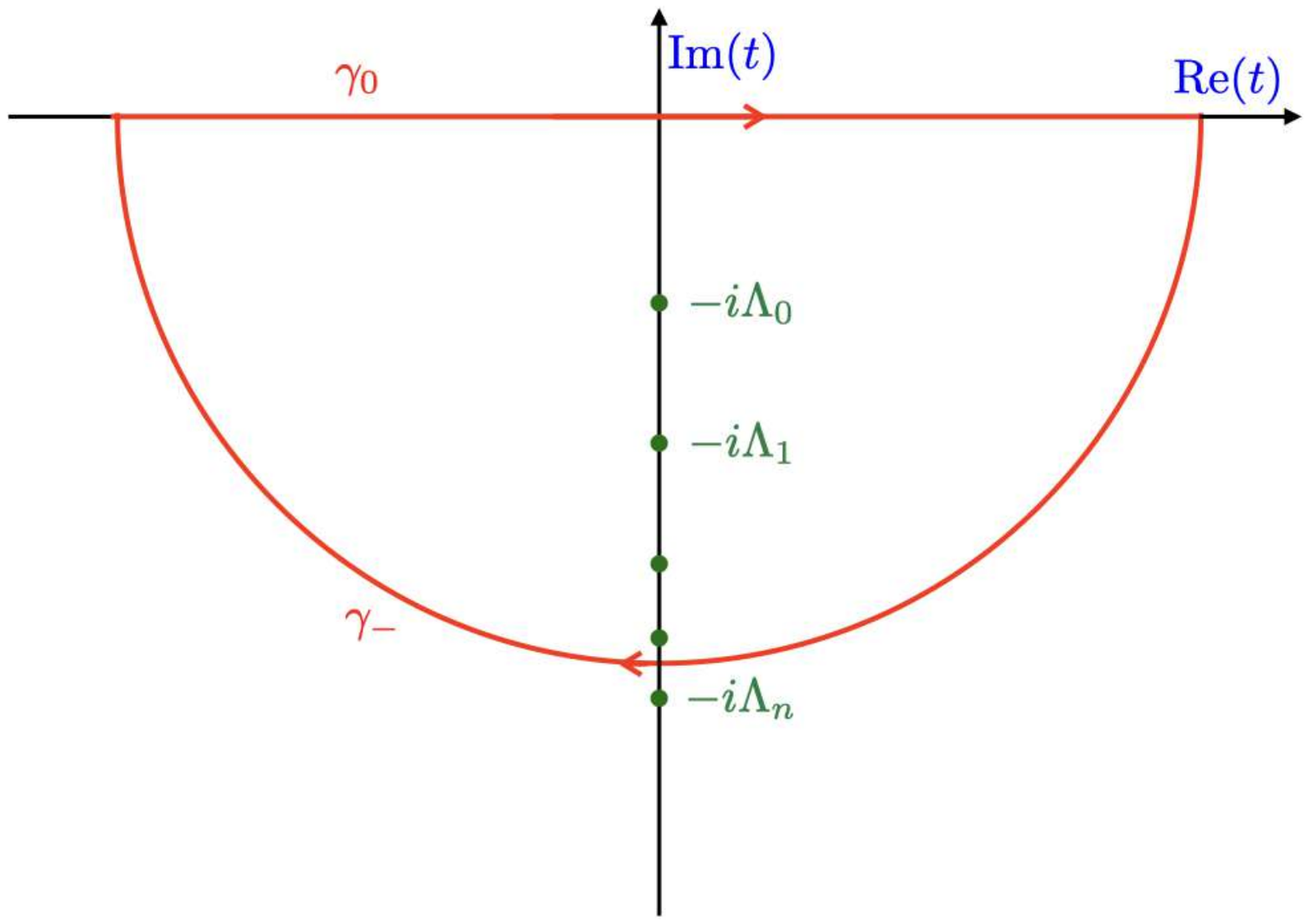}
        \label{CS}
    } 
    \caption[Optional caption for list of figures]{On the imaginary $t$ axis, the closed contour $\gamma_{0}\cup\gamma_{-}$ in the complex plane contains the poles of the characteristic function. The PDF $P_{\Gamma}({\cal N})$ is produced at these poles by using the residue theorem.  }
\label{complecontour}
    \end{figure*}
These conditions deal with the value of $\zeta$ when they are expressed for the additional variables, $(u,v)$. Smaller values of $\zeta$ are indicated by the first condition, whereas bigger values of $\zeta$ are indicated by the second. However, because of the nature of the conditions involving both additional variables, the solution now incorporates mode mixing, and we analyse their impact on the solution obtained from the series ansatz. Given that it exhibits oscillatory behaviour, the function $U_{n}(t;u)$ may be solved as follows:
\bea
U_{n}(t;u) = R_{n}(t)\cos{((1-u)\omega_{n}(t))} + Q_{n}(t)\sin{((1-u)\omega_{n}(t))}.
\eea
When applying the requirements from eqns. (\ref{bdrycondn1}) and (\ref{bdrycondn2}) to this solution, we obtain the following limitations on the series:
\bea
&&\sum_{n=0}^{\infty}v^{n}\big[R_{n}\cos{((1+v)\omega_{n}(t))} + Q_{n}(t)\sin{((1+v)\omega_{n}(t))}\big] = 1,\nonumber\\
&&\sum_{n=0}^{\infty}\omega_{n}v^{n}\big[R_{n}\sin{(\omega_{n}v)} - Q_{n}\cos{(\omega_{n}v)}\big] = 0.
\eea
This allows one to create recurrence relations for the different coefficients $R_{n},\;Q_{n}$ for each order in index $n$ by series extending the oscillatory functions. Using the aforementioned recurrence relations, we perform our analysis for each order in the series until we reach $n=3$, which enables us to create the next-to-next-to-next leading order (NNNNLO) version of the PDF. To do this, we need the explicit form of these coefficients. In order to build the second order (NNLO) contribution, we refer to the coefficients up to $m=2$ as follows:
\bea \label{recurrNNLO}
R_{0} &=& \frac{1}{\cos{(\omega_{0})}},
\nonumber\\
R_{1} &=& \frac{\omega_{0}}{\omega_{1}}\frac{\omega_{1}\sin{(\omega_{0})}-\omega_{0}\sin{(\omega_{1})}}{\cos{(\omega_{0})}\cos{(\omega_{1})}}, \nonumber\\
R_{2} &=& \frac{\omega_{1}\sin{(\omega_{1})}P_{1}-\omega_{1}\cos{(\omega_{1})}Q_{1}-\sin{(\omega_{2})}Q_{2}+\frac{\omega_{0}^{2}}{2}\cos{(\omega_{0})}P_{0}}{\cos{(\omega_{2}})}, 
\\  Q_{0} &=& 0,\nonumber\\
    Q_{1} &=& \frac{\omega_{0}^{2}}{\omega_{1}}P_{0},\nonumber\\
    Q_{2} &=& \frac{\omega_{1}^{2}}{\omega_{2}}P_{1},\eea
    and those pertaining to the third order are discussed in the NNNLO section that follows.

We employ the strategy described in \cite{Ezquiaga:2019ftu}, which involves expressing the characteristic function as a series expansion \footnote{The eigenvalue problem technique is an alternative way for calculating the tail expansion of the PDF $P_{\Gamma_i}({\cal N})$; comprehensive descriptions of this approach are available in \cite{Ezquiaga:2019ftu}. The PDE in equation (\ref{FPE-EFT}) is treated as a heat equation, enabling the use of techniques meant to solve for differential equations resembling diffusion. In order to account for the selected model potential's shape, a newly specified scalar-product formulation requires the introduction of a set of eigenfunctions for the adjoint Fokker-Planck operator, which maintains its Hermitian nature. With a different set of boundary conditions, the resultant eigenfunction obeys an equation like that for the characteristic function, eqn. (\ref{FPEcharacter}).
} in the format shown below:
\bea \label{chiexpand}
\chi(t;{\Gamma_i}) = \sum_{m}\sum_{n}\frac{r_{n}^{(m)}({\Gamma_i})}{\Lambda_{n}^{(m)}-it} + h(t;{\Gamma_i}),
\eea
in which the poles, $\Lambda_{n}^{(m)}$, always stay independent of the set of values in ${\Gamma_i}$, as we shall demonstrate in the following step, and the usage of ${\Gamma_i}$ shows the residues' dependency on the phase space variables, or the initial conditions. The regular component of the expansion is the function $h(t;{\Gamma_i})$, which, as we shall see, will not be discussed further. The residue theorem is applied at the poles on the imaginary $t$-axis that are surrounded by a contour $\gamma_{0}\cup\gamma_{-}$, as shown in fig. \ref{complecontour}. The terms $R_{n},\;Q_{n}$ accept simple poles in their structure, as can be shown from their formulas above. Consequently, an expansion of the type in eqn.(\ref{chiexpand}) works for our purpose for the characteristic function $\chi(t;{\Gamma_i})$. The poles line up with the circumstance:
\bea
\cos{(\omega_{m}(t))}=0,\quad\quad \text{hence,}\quad \omega_{m}(t) = \left(n+\frac{1}{2}\right)\pi.
\eea
When the number of unique places is denoted by the integer $n$, and we have a pole of order $m$. By using eqn. (\ref{omega1}), the equation above may be transformed to provide the pole expression, which is as follows:
\bea \label{poles}
\Lambda_{n}^{(m)} = 3Cm + \bigg[\frac{\pi}{\tilde{\mu}}\left(n+\frac{1}{2}\right)\bigg]^{2},
\eea
and here, as shown by eqn. (\ref{EFTparam}), we can observe the influence of the typical parameter. The residues, denoted by $r_{n}^{(m)}({\Gamma_i})$ in this case, are another quantity of significance. To evaluate them, invert the expression in eqn.(\ref{chiexpand}) and take the time-derivative as follows:
\bea \label{residue}
r_{n}^{(m)}({\Gamma_i}) = -i\bigg[\left.\frac{\partial}{\partial t}\chi^{-1}(t;{\Gamma_i})\right\vert_{t=-i\Lambda_{n}^{(m)}}\bigg]^{-1}.
\eea
If we combine the Fourier transform equation in eqn.(\ref{characterFourier}) with the previously mentioned information about the poles and residues, we can create the PDF as follows:
\bea \label{PDFfinal}
P_{\Gamma_i}({\cal N}) = \sum_{m=0,1,\cdots}\sum_{n=0}^{\infty}r_{n}^{(m)}({\Gamma_i})\exp{(-\Lambda_{n}^{(m)}{\cal N})}.
\eea
with the outer total increasing to the target order $m$th order. The value of the poles $\Lambda_{n}^{(m)}$ determines the exponential decay of the PDF at each order in the expansion, and for high values of the e-folds ${\cal N}$, the lowest pole $\Lambda_{0}$ and its corresponding residue $r_{0}^{(m)}$ provide the only significant contribution. As previously stated, the final PDF expansion is independent of the regular function in the expansion, eqn. (\ref{chiexpand}). Thus, utilising the foregoing, we can construct the entire PDF at the $m$th order by applying the characteristic function and determining its residues corresponding to the simple poles at each consecutive level in the expansion.

\subsubsection{Calculation of PDF at LO}

The computation of the probability distribution for the diffusion-dominated regime at leading order (LO) is covered in this section. We employ the method from the previous section, which is based on the characteristic function, and utilise its series expansion to determine the residues and the associated decay factors that include the poles. We only need to employ the first set of coefficients in eqn. (\ref{recurrNNLO}), namely $(R_{0},Q_{0})$, for the leading order. These suggest that because we are operating in the limit where $(v\rightarrow 0,\;u\rightarrow x)$, the drift contributions are entirely disregarded, and we may proceed to determine its PDF by using the associated LO characteristic function. Utilising the following LO function is the first task:
\bea \label{chiLO1}
\chi^{\rm LO}(t;x,0) = \frac{\cos{((1-x)\omega_{0})}}{\cos{(\omega_{0})}},
\eea
then use $\Lambda_{n}^{(0)}$, the $m=0$ poles, in eqn. (\ref{poles}) to obtain the residues using eqn. (\ref{residue}). Take note of the fact that at LO, the characteristic parameter $C$ for the EFT is not directly involved. After then, the residues collected are given by,
\bea \label{resLO}
r_{n,0}^{(m=0)} = (2n+1)\frac{\pi}{\tilde{\mu}^{2}}\sin{\bigg((2n+1)\frac{\pi x}{2}\bigg)}.
\eea
This, when combined with the poles, yields the PDF in eqn. (\ref{PDFfinal}):
\bea \label{pdfLO}
P_{\Gamma_{i}}^{\rm LO}({\cal N}) &=& \sum_{n=0}^{\infty}r_{n,0}^{(m=0)}\exp{(-\Lambda_{n}^{(0)}{\cal N})}= -\frac{\pi}{2\tilde{\mu}^2}{\cal V}'_{2}\bigg(\frac{\pi}{2}x,\exp{\bigg(-\frac{\pi^2}{\tilde{\mu}^2}{\cal N}\bigg)}\bigg),
\eea
where the elliptic theta function is denoted by ${\cal V}_2$:
\bea
{\cal V}_{2}(\alpha,x)=2\sum_{n=0}^{\infty}x^{(n+\frac{1}{2})^2}\cos{((2n+1)\alpha)}.
\eea
Furthermore, the function's derivative with respect to its first input is indicated by the notation ${\cal V}'_{2}$. In this case, when ${\cal N}$ is taken extremely big, the contribution from $n=0$ dominates most. The PDF for the next-to-leading order (NLO) is created in a similar manner. With the connection in eqn. (\ref{chimoments}), one may also compute the first and second moments from the aforementioned PDF at the leading order. When we address the primordial black hole abundance in the upcoming sections, the understanding of $\langle {\cal N}\rangle$ will come in handy. The following is the equation for these moments of the number of e-folds ${\cal N}$ at LO:
\bea \label{meanLO}
\langle{\cal N}(x,y=0)\rangle_{\rm LO} &=& \tilde{\mu}^{2}x\bigg(1-\frac{x}{2}\bigg),\\
\langle{\cal N}^{2}(x,y=0)\rangle_{\rm LO} &=& \frac{1}{3}\tilde{\mu}^{4}x(2-x^{2}+x^{3}).
\eea
It is evident from the start that the mean value is more susceptible to variations in $\tilde{\mu}$. In order to gain further insight into this sensitivity, let's assess the variance in the leading order, which provides us with:
\bea
\sigma_{\rm LO}^{2} = \langle{\cal N}^{2}(x,y=0)\rangle_{\rm LO} - \langle{\cal N}(x,y=0)\rangle_{\rm LO}^{2} = \frac{1}{6}\tilde{\mu}^{4}x(2+x^{2}-2x)(2-x),
\eea
As $\tilde{\mu}$ is slightly raised, the variance becomes more sensitive.

\subsubsection{Calculation of PDF at NLO}

Here, we calculate the pure diffusion limit result from the preceding section's initial sub-dominant adjustment. To accurately get the characteristic function and subsequent residues at the next-to-leading order in our $y$-expansion, we must employ the second set of coefficients, $(R_{1},Q_{1})$, in eqn. (\ref{recurrNNLO}). This is the function that we have at NLO:
\bea \label{chiNLO1}
\chi^{\rm NLO}(t;\Gamma_{i}) = U_{0}(t;u) + vU_{1}(t;u),
\eea
and making additional use of the equations' formulae. (\ref{poles},\ref{residue}) with information up to the first order ${\cal O}(y)$, as follows, for each residue at each order using the characteristic function mentioned above:
\bea \label{resNLO0}
r_{n,1}^{(m=0)} &=& (2n+1)\frac{\pi}{\tilde{\mu}^{2}}\bigg[\sin{\left((2n+1)\frac{\pi x}{2}\right)} - y(2n+1)\frac{\pi}{2}\cos{\left((2n+1)\frac{\pi x}{2}\right)}\bigg] \nonumber\\
&+& y(2n+1)\frac{\pi}{2}\frac{1}{\omega^{(0)}_{n,1}\cos{(\omega^{(0)}_{n,1})}}\bigg[\omega^{(0)}_{n,1}\cos{\omega^{(0)}_{n,1}(1-x)}-(-1)^{n}(2n+1)\frac{\pi}{2}\sin{(x\; \omega^{(0)}_{n,1})}\bigg],
\eea
the following is what the zero-order frequency at NLO notation means:
\bea
\omega^{(0)}_{n,1} = \sqrt{\left(n+\frac{1}{2}\right)^{2}\pi^{2}-3C\tilde{\mu}^{2}}.
\eea
The aforementioned residues are connected to $\Lambda_{n}^{(0)}$, the zero-order pole at NLO. Likewise, the residues for the first order poles, $\Lambda_{n}^{(1)}$, have the structure shown below:
\bea \label{resNLO1}
r_{n,1}^{(m=1)} &=& \frac{(-1)^{n+1}2y}{\tilde{\mu}^{2}}\frac{\sin{\left((n+\frac{1}{2})\pi x\right)}}{\cos{\left(\sqrt{\left(n+\frac{1}{2}\right)^{2}\pi^{2}+3C\tilde{\mu}^{2}}\right)}}\bigg[\left(n+\frac{1}{2}\right)^{2}\pi^{2}+3C\tilde{\mu}^{2}\nonumber\\
&& \quad\quad\quad\quad\quad -(-1)^{n}\frac{\pi}{2}(2n+1)\sqrt{\left(n+\frac{1}{2}\right)^{2}\pi^{2}+3C\tilde{\mu}^{2}}\sin{\bigg(\sqrt{\left(n+\frac{1}{2}\right)^{2}\pi^{2}+3C\tilde{\mu}^{2}}\bigg)}\bigg].
\eea
Equations (\ref{resNLO0}, \ref{resNLO1}) are combined to create the final PDF from eqn. (\ref{PDFfinal}), which yields:
\bea \label{PDFN1}
P_{\Gamma_{i}}^{\rm NLO}({\cal N}) = \sum_{n=0}^{\infty}\big[r_{n,1}^{(m=0)}+r_{n,1}^{(m=1)}e^{-3C{\cal N}}\big]e^{-\Lambda_{n}^{(0)}{\cal N}}.
\eea
Where we have another version of the poles, 
\bea \Lambda_{n}^{(1)} = 3C + \Lambda_{n}^{(0)},\eea contributing to the decay rate, in addition to the set of poles for $m=0$. Additionally, depending on its value, which stays near to $1$, the characteristic parameter $C$ changes the decay feature. The evaluation of the initial moment or mean number $\langle{\cal N}\rangle$ might provide the following results, just as in the case of LO:
\bea \label{meanNLO}
\langle{\cal N}(\Gamma)\rangle_{\rm NLO} = \tilde{\mu}^{2}x\bigg(1-\frac{x}{2}\bigg) + \tilde{\mu}^{2}y\bigg(x-1 + \frac{\cosh{[\sqrt{3C}\tilde{\mu}(1-x)]}}{\cosh{[\sqrt{3C}\tilde{\mu}]}} - \frac{\sinh{[\sqrt{3C}\tilde{\mu}x]}}{\sqrt{3C}\tilde{\mu}\cosh{[\sqrt{3C}\tilde{\mu}]}} \bigg).
\eea
This, up to ${\cal O}(y)$, comprises the linear correction terms. This phrase will be used for PBH analysis in the next parts.

\subsubsection{Calculation of PDF at NNLO}

The computation of the probability distribution for the diffusion-dominated regime at the next-to-next-leading order (NNLO) is the subject of this discussion. Essentially, we utilise the series ansatz in equation (\ref{chiansatz}) that has been trimmed to the quadratic terms in order to determine the corresponding adjustments. To start, we would need to be aware of the residues and their matching poles. Here, $(R_{2},Q_{2})$, the last set of coefficients in eqn. (\ref{recurrNNLO}), will be useful. The form of the characteristic function at NNLO is as follows:
\bea \label{chiNNLO1}
\chi^{\rm NLO}(t;\Gamma_{i}) = U_{0}(t;u) + vU_{1}(t;u) + v^{2}U_{2}(t;u).
\eea
Subsequently, to build the PDF, the residues at orders $m=0,1,2$, each enlarged to the second order ${\cal O}(y^{2})$, are needed, as found in equation (\ref{residue}). Because of the total intricacy of each residue, we do not directly describe its analytic form. The numerical analysis section will provide a detailed discussion of the acquired numbers for $P_{\Gamma_{i}}^{\rm NNLO}(\cal N)$. The following expression may be made using the PDF's final form:
\bea \label{PDFN2}
P_{\Gamma_i}^{\rm NNLO}({\cal N}) = \sum_{n=0}^{\infty}\big[r_{n,2}^{(0)}+ r_{n,2}^{(1)}e^{-3C{\cal N}} + r_{n,2}^{(2)}e^{-6C{\cal N}} \big]e^{-\Lambda_{n}^{(0)}{\cal N}}.
\eea
Along with the characteristic function in the aforementioned eqn. (\ref{chiNNLO1}), we can also compute the corrections to the mean number of e-folds at NNLO using the eqn. (\ref{chimoments}). As a result, we may express the mean as follows:
\bea \label{meanNNLO}
\langle{\cal N}(\Gamma)\rangle_{\rm NNLO} &=& \tilde{\mu}^{2}x\bigg(1-\frac{x}{2}\bigg) + \tilde{\mu}^{2}y(-1+x-y) + \frac{\tilde{\mu}^{2}y^2}{2}\bigg(\frac{\cosh{[\sqrt{6C}\tilde{\mu}(1-x)]}}{\cosh{[\sqrt{3C}\tilde{\mu}]}\cosh{[\sqrt{6C}\tilde{\mu}]}}\bigg)\nonumber\\
&&\times\bigg(-2+\cosh{[\sqrt{3C}\tilde{\mu}]}+\sqrt{6C}\tilde{\mu}\sinh{[\sqrt{6C}\tilde{\mu}]} - (2\sqrt{3C}\tilde{\mu}^{2}+\sqrt{2}\sinh{[\sqrt{6C}\tilde{\mu}]})\sinh{[\sqrt{3C}\tilde{\mu}]}\bigg)\nonumber\\
&&+\frac{\tilde{\mu}^{2}y^2}{\sqrt{2}}\frac{\sinh{[\sqrt{6C}\tilde{\mu}(1-x)]}}{\sqrt{3C}\tilde{\mu}^2} - \tilde{\mu}y\sinh{[\sqrt{3C}\tilde{\mu}(1-x+y)]} -\frac{\tilde{\mu}^{2}y^2}{\sqrt{2}}\sinh{[\sqrt{6C}\tilde{\mu}(1-x)]}\tanh{[\sqrt{3C}\tilde{\mu}]} \nonumber\\
&& + \tilde{\mu}^{2}y\cosh{[\sqrt{3C}\tilde{\mu}(1-x+y)]}\bigg(\frac{1}{\cosh{[\sqrt{3C}\tilde{\mu}]}}-\frac{\tanh{[\sqrt{3C}\tilde{\mu}]}}{\sqrt{3C}\tilde{\mu}}\bigg),
\eea 
where terms are only taken into account up to ${\cal O}(y^{2})$.

\subsubsection{Calculation of PDF at NNNLO}

We now describe the recurrence connection between the coefficients, which is necessary to fully determine the PDF:
\bea
R_{3} &=& \frac{1}{\cos{(\omega_{3})}}\bigg[\frac{1}{2}\omega_{1}^{2}(R_{1}\cos{(\omega_{1})} + Q_{1}\sin{(\omega_{1})}) + \omega_{2}R_{2}\sin{(\omega_{2})}-\frac{1}{6}\omega_{0}^{3}R_{0}\sin{(\omega_{0})}-\omega_{2}Q_{2}\cos{(\omega_{2})}-Q_{3}\sin{(\omega_{3})}\bigg],\nonumber\\
Q_{3} &=& \frac{1}{\cos{(\omega_{3})}}\bigg[\omega_{2}^{2}R_{2}+\frac{1}{2}\omega_{1}^{3}Q_{1} - \frac{1}{6}\omega_{0}^{4}R_{0}\bigg].
\eea
At $t=-i\Lambda_{n}^{(3)}$, the simple pole resulting from $\cos{(\omega_{3})}$ is clearly visible. We can already see that as we move above NNLO and these combined with the prior values of the coefficients in eqn, the formulations grow increasingly complicated.The characteristic function required to construct the whole NNNLO PDF using eqs is (\ref{recurrNNLO}).(\ref{PDFfinal}), (\ref{chiansatz}), and (\ref{residue}).

In relation to the poles, the index $m$ now ranges from $m=0,1,2,3$, with the final $m=3$, or \bea \Lambda_{n}^{(3)} = \Lambda_{n}^{(0)} + 3Cm,\eea providing the smallest order of contribution while the dominant contribution, which appears to be more significant as we approach large values of ${\cal N}$, is obtained at $m=0$. Since the residues, $r_{n}^{(m)}\;\forall\;m=0,1,2,3$, are quite complex and informative in and of themselves, we do not discuss their whole form here. Instead, we just reveal their contribution in the form of the PDF realised after estimating them. Using eqn.(\ref{PDFfinal}), the final NNNLO PDF looks like this:
\bea \label{PDFN3}
P_{\Gamma_i}^{\rm NNNLO}({\cal N}) = \sum_{n=0}^{\infty}\big[r_{n,3}^{(0)}+ r_{n,3}^{(1)}e^{-3C{\cal N}} + r_{n,3}^{(2)}e^{-6C{\cal N}} + r_{n,3}^{(3)}e^{-9C{\cal N}} \big]e^{-\Lambda_{n}^{(0)}{\cal N}}.
\eea
The initial phase space variables, ${\Gamma_i}$, do affect the residues, particularly the term $r_{n}^{(0)}$ that controls the amplitude; however, the decay from the dominating exponential, $\exp{(-\Lambda_{n}^{0}{\cal N})}$, is not affected by the initial phase space variables. 
The equation in eqn. (\ref{poles}) also indicates that the poles maintain their good separation along the imaginary $t$-axis when $\tilde{\mu} \gtrsim 1$. The mean number of e-folds expression now becomes considerably more complex and long, and it is represented by the following relation:
\bea \label{meanNNNLO}
\langle{\cal N}(\Gamma)\rangle_{\rm NNNLO} &=& \frac{\tilde{\mu}^{2}}{2}\bigg\{1-(1-x+y)^{2} + \frac{12\tilde{\mu} ^2 y\left(\sqrt{3C\tilde{\mu} ^2}-\sinh \big[\sqrt{3C\tilde{\mu} ^2}\big]\right) 
   \cosh \big[\sqrt{3C\tilde{\mu} ^2} (x-y-1)\big]}{\sqrt{3C\tilde{\mu} ^2}\cosh\big[\sqrt{3C\tilde{\mu} ^2}\big]}\nonumber\\
&-& 3\tilde{\mu}^2 y^2 \cosh \big[\sqrt{6C \tilde{\mu} ^2} (x-y-1)\big]\nonumber\\
&\times& \bigg(\frac{4-2\sqrt{6C\tilde{\mu} ^2}
   \sinh \big[\sqrt{6C \tilde{\mu} ^2}\big]+2 \sinh \big[\sqrt{c \tilde{\mu} ^2}\big] \left(2 \sqrt{3C\tilde{\mu} ^2}+\sqrt{2} \sinh
   \big[\sqrt{6C \tilde{\mu} ^2}\big]\right)-2 \cosh \big[\sqrt{3C\tilde{\mu} ^2}\big]}{\cosh\big[\sqrt{3C \tilde{\mu} ^2}\big] \cosh\big[\sqrt{6C\tilde{\mu} ^2}\big]} \bigg)\nonumber\\
&-& \frac{12 y\sqrt{3C \tilde{\mu} ^2} \sinh \big[\sqrt{3C\tilde{\mu}^2}(x-y-1)\big]}{3C}+6 \sqrt{2}\tilde{\mu} ^2 y^2 \left(\sqrt{3C \tilde{\mu} ^2}-\sinh \big[\sqrt{3C \tilde{\mu} ^2}\big]\right)
   \frac{\sinh\big[\sqrt{6C\tilde{\mu} ^2} (x-y-1)\big]}{\cosh\big[\sqrt{3C\tilde{\mu} ^2}\big]}\nonumber\\
&+& \frac{2 \tilde{\mu} ^2 y^3\cosh
   \big[\sqrt{3} (x-1) \sqrt{3C \tilde{\mu} ^2}\big]}{\sqrt{3C \tilde{\mu} ^2}\cosh\big[\sqrt{6C \tilde{\mu} ^2}\big]\cosh\big[\sqrt{9C \tilde{\mu}^2}\big]}\bigg[6 c\tilde{\mu} ^2 \left(\sqrt{3C \tilde{\mu} ^2}-\sinh \big[\sqrt{3C \tilde{\mu} ^2}\big]\right) \frac{\cosh ^2\big[\sqrt{6C \tilde{\mu} ^2}\big]}{\cosh\big[\sqrt{3C\tilde{\mu} ^2}\big]}\nonumber\\
&+& 6 \left(3\sqrt{c^2 \tilde{\mu} ^4}\sinh \big[\sqrt{3C \tilde{\mu} ^2}\big]-\left(3C \tilde{\mu} ^2\right)^{3/2}\right) \frac{\sinh ^2\big[
   \sqrt{6C \tilde{\mu} ^2}\big]}{\cosh\big[\sqrt{3C \tilde{\mu} ^2}\big]}
   -3C\tilde{\mu} ^2 \left(3 \sqrt{3C \tilde{\mu} ^2}-\sqrt{3} \sinh \big[3\sqrt{C\tilde{\mu} ^2}\big]\right)\nonumber\\
&\times&\cosh \big[\sqrt{6C
   \tilde{\mu} ^2}\big] -6\sqrt{3}C \tilde{\mu} ^2 \sinh \big[3 \sqrt{C\tilde{\mu} ^2}\big]\left(2 \sqrt{3C \tilde{\mu} ^2} \tanh \big[\sqrt{3C \tilde{\mu} ^2}\big] +\frac{2}{\cosh\big[\sqrt{3C\tilde{\mu} ^2}\big]}
-1\right)\nonumber\\
&+& \sqrt{2} \sinh \big[\sqrt{6C\tilde{\mu} ^2}\big] \bigg(6C\tilde{\mu} ^2 \left(3 \sqrt{3C\tilde{\mu} ^2}-\sqrt{3} \sinh \big[3
\sqrt{C\tilde{\mu} ^2}\big]\right)\tanh \big[\sqrt{3C \tilde{\mu} ^2}\big]\nonumber\\
&+& \frac{\left(18\sqrt{C^2 \tilde{\mu} ^4}+2 \sqrt{3} \left(3C\tilde{\mu} ^2\right)^{3/2} \sinh \big[3
\sqrt{C\tilde{\mu} ^2}\big]\right)}{\cosh\big[\sqrt{c \tilde{\mu} ^2}\big]}-9\sqrt{C^2\tilde{\mu}
   ^4}\bigg)\bigg]+ 2 \sqrt{3}\tilde{\mu}^2 y^3 \sqrt{3C\tilde{\mu} ^2} \sinh \big[3(x-1) \sqrt{C \tilde{\mu} ^2}\big] \nonumber\eea\bea
&\times& \bigg(1-2 \sqrt{2} \tanh \big[\sqrt{3C\tilde{\mu} ^2}\big] \tanh \big[\sqrt{6C \tilde{\mu} ^2}\big] +\frac{1}{\cosh\big[\sqrt{6C
   \tilde{\mu} ^2}\big]}\bigg(2-4\sqrt{3C\tilde{\mu} ^2} \tanh \big[\sqrt{3C\tilde{\mu} ^2}\big]\nonumber\\
&-& \frac{4}{\cosh\big[\sqrt{3C\tilde{\mu}
   ^2}\big]} \bigg)+ \frac{2\sqrt{6C \tilde{\mu} ^2} \tanh \big[\sqrt{6C \tilde{\mu} ^2}\big]}{\cosh\big[\sqrt{3C \tilde{\mu}
   ^2}\big]}\bigg)
   \bigg\}.
\eea
With the help of eqn. (\ref{PDFN3}), we can now build the PDF by incorporating a greater number of higher-order terms into the entire expansion. This gives us additional insight into the upper tail region of the PDF, which indicates a rise in its non-Gaussian nature and has significant implications for the prospects of PBH formation.

\subsection{Drift Dominated Regime: Methods to Solve for the characteristic and calculating moments}

This part deals with the other side condition in which the quantum diffusion effects continue to be subdominant. As a result, the behaviour of the scalar perturbations $\zeta$ across the USR is primarily controlled in a classical limit when the Langevin equation's drift effects become crucial. Diffusion processes, which contribute to most of the e-folding realisations for given $\Gamma_{i}$ between certain beginning and final conditions, do not influence the scalar perturbations here primarily.

In order to solve eqn. (\ref{FPEcharacter}), a partial differential equation, the constraints on the variables require that one operate under the limit where $y\gg 1$. This limit also implies high conjugate canonical momenta in terms of the phase space variables, as shown by eqn. (\ref{newphasevars}). We will learn about the properties surrounding the distribution's maximum in terms of the PDF because this study is mostly classical in nature. To precisely investigate the structure of the whole PDF away from the maximum, one may build the higher-order moments, which include the skewness and kurtosis, starting with the mean value of the stochastic e-folds variable ${\cal N}$, or the first moment of the PDF from eqn. (\ref{chimoments}). 

For the collection of coarse-grained variables ${\Gamma_{i}} = \{\zeta_i,\Pi_i\}$, we address here the general approach to solve for the different moments of the PDF $P_{\Gamma_{i}}({\cal N})$. The PDF is developed up to the next-to-next-to-next-to-leading order in the variable $y$. We begin by working at the leading order in $y$. This approach is referred to as the characteristics technique. The characteristic curves in the phase space are obtained by solving linear ordinary differential equations for each of the characteristic lines that are introduced by parameterizing our phase space variables $\Gamma= \{x,y\}$ with an arbitrary parameter. We will now demonstrate how these solutions, starting with the solution leading order in $y$, are a practical way to solve for the original PDE in eqn. (\ref{FPEcharacter}).

\subsubsection{Calculation of PDF at LO}

The phase space variables are first expressed as $\Gamma(p)=\{x(p),y(p)\}$, in which $p$ denotes a parameter defined for the corresponding set of characteristic lines in phase space. We must use the PDE in eqn. (\ref{FPEcharacter}) to solve these lines. To do this, write the difference as follows:
\bea
\frac{d\chi(t;\Gamma(p))}{dp}= \bigg(\frac{dx}{dp}\frac{\partial}{\partial x}+ \frac{dy}{dp}\frac{\partial}{\partial y}\bigg)\chi(t;\Gamma(p)),
\eea
If the following equations can be solved, the RHS resembles the differential operator in the PDE equation (\ref{FPEcharacter}):
\bea \label{solutionODE}
\frac{dx(p)}{dp}=\frac{dy(p)}{dp}=-3C,
\eea
and using the ensuing family of curves, we are able to express the leading order PDE in the form of an ordinary differential equation, which fully disregards any influence from the initial diffusion component in equation (\ref{FPEcharacter}):
\bea \label{chiLO}
\bigg(\frac{d}{dp}+\frac{it}{y_{0}-3Cp}\bigg)\chi(t;\Gamma(p))= 0,
\eea
where the solution curve for the equation (\ref{solutionODE}) parameterized by $p$ is formed by using the solution $y= y_{0}-3Cp$, given some starting condition value $y_{0}$. It is simple to solve the aforementioned first-order ODE to obtain:
\bea
\chi(t;\Gamma(p)) = \chi_{0}(t;\Gamma_{0})\big[y_{0}-3Cp\big]^{\frac{it}{3C}},
\eea
where $C$, the characteristic parameter, is stated clearly. After renaming the parameterization using the difference $x_{0}-y_{0}$, we obtain $y=-3Cp$ and $x_{0}=x-y$ for the original variables after setting one constant, $y_{0}=0$. As a result, the solution for the initial condition now depends solely on $y$. By converting back into $\{x,y\}$ and using the same boundary conditions as in Equations (\ref{bdrycondn1}) and (\ref{bdrycondn2}), we obtain the following from the first condition:
\bea
\chi(t;\Gamma)|_{x=0} = 1,\quad\implies\quad\;\chi_{0}(-y)=y^{-\frac{it}{3C}},
\eea
This makes it simple to reduce the final solution to the following expression:
\bea \label{chiLOsol}
\chi^{\rm LO}(t;\Gamma(p)) = \bigg(1-\frac{x}{y}\bigg)^{-\frac{it}{3C}},
\eea
We just need to apply the Fourier transform eqn.(\ref{characterFourier}) in order to continue constructing the PDF at LO:
\bea \label{pdfLO1}
P_{\Gamma}^{\rm LO}({\cal N})= \delta({\cal N}-\langle{\cal N}_{\rm LO}\rangle),
\eea
where the stochastic variable ${\cal N}$ yields the following first and second moments:
\bea
\langle{\cal N}(\Gamma)\rangle_{\rm LO} &=& -\frac{1}{3C}\ln{\bigg(1-\frac{x}{y}\bigg)}.\\
\langle{\cal N}^2(\Gamma)\rangle_{\rm LO} &=& \langle{\cal N}(\Gamma)\rangle^2_{\rm LO},
\eea
and based on the foregoing, we deduce that,
\bea \sigma_{\rm LO}^{2} = \langle{\cal N}^2(\Gamma)\rangle_{\rm LO} - \langle{\cal N}(\Gamma)\rangle^2_{\rm LO} = 0.\eea
The skewness and kurtosis, among other higher-order parameters that characterise departures from Gaussianity, are also absent, and as a result, the parameters $f_{\rm NL},\;g_{\rm NL},\;\tau_{\rm NL}$ also disappear. One key observation on the fluctuations in the e-folds is provided by the PDF eqn. (\ref{pdfLO}), which is that they essentially lead to zero curvature perturbations since, according to the stochastic-$\delta N$ formalism, ${\cal N}-\langle{\cal N}\rangle=\zeta_{\rm cg}$.

\subsubsection{Calculation of PDF at NLO}

This section builds on the methodology used in the previous section to derive the characteristic function at the next-to-leading order (NLO), which will provide additional corrections to properties like the mean number of e-folds and introduce us to additional new statistical properties like the variance of the stochastic number of e-folds, among many other significant measures of interest. As we mentioned at the conclusion of the last study, we must move beyond the LO findings in order to integrate the curvature perturbations. For this reason, the present talks of NLO corrections are where we start.  

We start by looking at eqn. (\ref{FPEcharacter}), but this time we will let the diffusion effects brought about by the first term to just work on the leading order solution that was found before. The other terms will only have a negligible influence on $\chi^{\rm LO}$. We compose:
\bea \chi = \chi^{\rm LO} + \chi^{\rm NLO},\eea 
And the sub-dominant correction to the LO result is represented by the term $\chi^{\rm NLO}$. Equation (\ref{FPEcharacter}) is satisfied by the NLO contribution when the above is substituted.
\bea \label{chiNLO}
\bigg(-3Cy\bigg[\frac{\partial}{\partial x}+ \frac{\partial}{\partial y}\bigg] + it\bigg)\chi^{\rm NLO}(t;\Gamma(p)) = -\frac{1}{y^{2}}\frac{it}{3C\tilde{\mu}^{2}}\bigg(1+\frac{it}{3C}\bigg)\bigg(1-\frac{x}{y}\bigg)^{-\frac{it}{3C}-2}.
\eea
It is noteworthy that the quantum diffusion effects operate at the NLO in a sub-dominant manner. Consequently, the diffusion operator acting on the LO solution serves as the source term for the NLO drift solution of the characteristic function. For $\{x(p),y(p)\}$, the characteristic lines are still the same. Together with this, we have a linear ODE of the following form for $\chi^{\rm NLO}$:
\bea
\bigg(\frac{d}{dp} + \frac{it}{y}\bigg)\chi^{\rm NLO}(t;\Gamma(p)) = -\frac{1}{y^{3}}\frac{it}{3C\tilde{\mu}^{2}}\bigg(1+\frac{it}{3C}\bigg)\bigg(1-\frac{x}{y}\bigg)^{-\frac{it}{3C}-2},
\eea
yielding a generic solution for us to solve:
\bea \label{gensolNLO}
\chi^{\rm NLO}(t;\Gamma(p)) = \bigg(\chi_{1}(y-x) + \frac{it}{9\tilde{\mu}^{2}C^{2}}\bigg(1+\frac{it}{3C}\bigg)\ln{(y)}(y-x)^{-i\frac{it}{3C}-2}\bigg)y^{\frac{it}{3C}},
\eea
Using the boundary condition $\chi(t;\Gamma)|_{x=0}=1$ once more, we obtain the following solution for the starting condition:
\bea
\chi_{1}(y-x) = (y-x)^{-\frac{it}{3C}}\bigg(1-\frac{it}{9C^{2}\tilde{\mu}^{2}(y-x)^{2}}\bigg(1+\frac{it}{3C}\bigg)\ln{(y-x)}\bigg),
\eea
Hence, we obtain the following when we apply it to the generic NLO solution equation (\ref{gensolNLO}):
\bea \label{chiNLOsol}
\chi^{\rm NLO}(t;\Gamma) = \bigg(1-\frac{x}{y}\bigg)^{-\frac{it}{3C}}\bigg[1-\frac{it}{9C^{2}\tilde{\mu}^{2}}\bigg(1+\frac{it}{3C}\bigg)\frac{\ln{\big(1-\frac{x}{y}\big)}}{(y-x)^{2}}\bigg].
\eea
The LO result is corrected by ${\cal O}(\tilde{\mu}^{-2}y^{-2})$ in the above. Given that ${\cal N}$ eventually becomes a stochastic variable, this solution can also provide us the PDF at NLO order and related statistical moments. Firstly, we examine the PDF resulting from the inverse Fourier transform of $\chi^{\rm NLO}(t;\Gamma)$, whose structure makes use of derivatives of the Dirac delta distribution in the following way:
\bea \label{pdfNLO1}
P^{\rm NLO}_{\Gamma}({\cal N}) &=& P^{\rm LO}_{\Gamma}({\cal N}) + \frac{1}{9C^2\tilde{\mu}^2(y-x)^2}\bigg(\delta^{(1)}({\cal N}-\langle{\cal N}_{\rm LO}\rangle) - \frac{1}{3C}\delta^{(2)}({\cal N}-\langle{\cal N}_{\rm LO}\rangle)\bigg)\ln{\bigg(1-\frac{x}{y}\bigg)}\nonumber\\
&=& P^{\rm LO}_{\Gamma}({\cal N}) - \frac{1}{9C^2\tilde{\mu}^2(y-x)^2}\bigg(\frac{1}{\cal N} + \frac{2}{3C{\cal N}^2}\bigg)\delta({\cal N}-\langle{\cal N}_{\rm LO}\rangle)\ln{\bigg(1-\frac{x}{y}\bigg)},
\eea
where the Dirac delta's derivative characteristics are employed in the second line, and the superscripts $(i)$ indicate the $i$th derivative of the Dirac delta with regard to the e-folds ${\cal N}$. We can also rapidly ascertain the number of e-folds' mean value and variation from this result. The following criteria are used to evaluate the mean value at NLO:
\bea
\langle{\cal N}(\Gamma)\rangle_{\rm NLO} = \langle{\cal N}(\Gamma)\rangle_{\rm LO}\bigg(1+\frac{1}{3C^2\tilde{\mu}^2(y-x)^2}\bigg), 
\eea
and similarly, in order to evaluate the variance at NLO, one must first determine the second instant, which is determined to be:
\bea
\langle{\cal N}^{2}(\Gamma)\rangle_{\rm NLO} = \frac{2\langle{\cal N}\rangle_{\rm LO}}{9C^2\tilde{\mu}^2(y-x)^2}\bigg[1-\bigg(1+\frac{3}{2}C\tilde{\mu}^2(y-x)^2\bigg)\ln{\bigg(1-\frac{x}{y}\bigg)}\bigg],
\eea
which is used to analyze the variance to the following:
\bea \label{delta2nlo}
\sigma_{\rm NLO}^{2} = \langle\delta{\cal N}^2(\Gamma)\rangle_{\rm NLO}&=&  \langle{\cal N}^{2}(\Gamma)\rangle_{\rm NLO} - \langle{\cal N}(\Gamma)\rangle_{\rm NLO}^{2}\nonumber\\
&=& \frac{2\langle{\cal N}(\Gamma)\rangle_{\rm LO}}{9C^2\tilde{\mu}^2(y-x)^2}\bigg(1-\frac{\langle{\cal N}(\Gamma)\rangle_{\rm LO}}{2C^2\tilde{\mu}^2(y-x)^2}\bigg).
\eea
where the primary contribution is made by the first term. The power spectrum and other non-Gaussianity parameters are examples of significant quantities of observational relevance that may be calculated with the use of several statistical moments. We now examine the third instant generated at NLO order for any adjustments. Here, we receive the following for the third instant at NLO:
\bea
\langle{\cal N}^3\rangle_{\rm NLO} = \frac{2-\ln{\big(1-\frac{x}{y}\big)}(1+C\tilde{\mu}^2(x-y)^2)}{27C^4\tilde{\mu}^2(x-y)^2}\ln^2{\bigg(1-\frac{x}{y}\bigg)},
\eea
when it is shown to be the primary contributor to the third instant away from the mean:
\bea \label{delta3nlo}
\langle\delta{\cal N}^3\rangle_{\rm NLO} = \langle({\cal N}-\langle{\cal N}\rangle)^3\rangle_{\rm NLO} &=& \langle{\cal N}^3\rangle_{\rm NLO} + 2\langle{\cal N}\rangle_{\rm NLO}^3 - 3\langle{\cal N}\rangle_{\rm NLO}\langle{\cal N}^2\rangle_{\rm NLO}, \nonumber\\
&=& -\frac{2\langle{\cal N}(\Gamma)\rangle_{\rm LO}^2}{9\;C^3\tilde{\mu}^4(x-y)^4},
\eea
and the above appears to be more repressed in ${\cal O}((\tilde{\mu}y)^{-2})$, the order of expansion taken into consideration for NLO. Suppressed contributions also result from looking for adjustments to the fourth instant from the mean. We first need the fourth instant listed below:
\bea
\langle{\cal N}^4\rangle_{\rm NLO} = \frac{-12+\ln{\big(1-\frac{x}{y}\big)}(4+3C\tilde{\mu}^2(x-y)^2)}{243C^5\tilde{\mu}^2(x-y)^2}\ln^3{\bigg(1-\frac{x}{y}\bigg)},
\eea
with the help of which the primary contribution to the interest quantity is determined as:
\bea \label{delta4nlo}
\langle\delta{\cal N}^4\rangle_{\rm NLO} = \langle({\cal N}-\langle{\cal N}\rangle)^4\rangle_{\rm NLO} &=& \langle{\cal N}^4\rangle_{\rm NLO} - 4\langle{\cal N}\rangle_{\rm NLO}\langle{\cal N}^{3}\rangle_{\rm NLO} + 6\langle{\cal N}\rangle_{\rm NLO}^{2}\langle{\cal N}^2\rangle_{\rm NLO}- 3\langle{\cal N}\rangle_{\rm NLO}^{4}, \nonumber\\
&=& \frac{4\langle{\cal N}(\Gamma)\rangle_{\rm LO}^3}{27\;C^4\tilde{\mu}^6(x-y)^6}.
\eea
The skewness and kurtosis of the distribution may be used to calculate deviations from a Gaussian distribution from these higher-order statistical moments. The following is a definition of these:
\bea
\gamma = \frac{\langle({\cal N}-\langle{\cal N}\rangle)^3\rangle}{\sigma^3}, \quad\quad\quad \kappa = \frac{\langle({\cal N}-\langle{\cal N}\rangle)^4\rangle}{\sigma^4},
\eea
then using the following eqns (\ref{delta2nlo}, \ref{delta3nlo}, \ref{delta4nlo}) at NLO, we calculate the skewness and kurtosis:
\bea
\gamma_{\rm NLO} &=& \frac{\langle({\cal N}-\langle{\cal N}\rangle)^3\rangle_{\rm NLO}}{\sigma_{\rm NLO}^3} = -3C\sqrt{\frac{\langle{\cal N}\rangle_{\rm LO}}{2C^2(x-y)^2\tilde{\mu}^2}},\\
\kappa_{\rm NLO} &=& \frac{\langle({\cal N}-\langle{\cal N}\rangle)^4\rangle_{\rm NLO}}{\sigma_{\rm NLO}^4} = \frac{3\langle{\cal N}\rangle_{\rm LO}}{(x-y)^2\tilde{\mu}^2},
\eea
The power spectrum may be evaluated using eqn. (\ref{delta2nlo}) as follows, starting with the formulas for different values of statistical and observational importance in light of the foregoing results:
\bea
\Delta^{2}_{\zeta\zeta,\rm NLO} = \frac{d\langle\delta{\cal N}^2\rangle}{d\langle{\cal N}\rangle} = \frac{2}{9C^2\tilde{\mu}^2(y-x)^2}.
\eea
This stands for the contribution of the leading order. Next, we use eqns. (\ref{delta3nlo}) and (\ref{delta2nlo}) to find the non-Gaussianity parameter $f_{\rm NL}$ as follows:
\bea \label{fnlNLO}
f_{\rm NL} = \frac{5}{36}\frac{1}{[\Delta^{2}_{\zeta\zeta,\rm NLO}]^2}\frac{d^2\langle\delta{\cal N}^3\rangle}{d\langle{\cal N}\rangle^2} = -\frac{5C}{4}.
\eea
Likewise, $g_{\rm NL}$ at NLO and $\tau_{\rm NL}$, the two non-Gaussian parameters, are ascertained as follows:
\bea \label{tauNLO}
\tau_{\rm NL} &=& \frac{1}{36}\frac{1}{[\Delta^{2}_{\zeta\zeta,\rm NLO}]^4}\bigg(\frac{d^2\langle\delta{\cal N}^3\rangle}{d\langle{\cal N}\rangle^2}\bigg)^2 = \frac{9}{4}C^2,\\
g_{\rm NL} &=&  \frac{1}{[\Delta^{2}_{\zeta\zeta,\rm NLO}]^3}\frac{d^3\langle\delta{\cal N}^4\rangle}{d\langle{\cal N}\rangle^3} = 18C^2.
\eea
The information above provides some fascinating insights on the type of non-Gaussianity that may be anticipated at the NLO. For further information, see references. \cite{Maldacena:2002vr,Alishahiha:2004eh,Mazumdar:2001mm,Choudhury:2002xu,Panda:2005sg,Chingangbam:2004ng,Armendariz-Picon:1999hyi,Garriga:1999vw,Burrage:2010cu,Choudhury:2012yh,Choudhury:2012whm,Chen:2010xka,Chen:2006nt,Chen:2009zp,Chen:2009we,Chen:2008wn,Chen:2006xjb,Chen:2013aj,Chen:2012ge,Chen:2009bc,Creminelli:2010ba,Kobayashi:2010cm,Mizuno:2010ag,Burrage:2011hd,Kobayashi:2011pc,DeFelice:2011zh,Renaux-Petel:2011lur,DeFelice:2011uc,Gao:2011qe,deRham:2012az,Ohashi:2012wf,DeFelice:2013ar,Arroja:2013dya,Choudhury:2013qza,Pirtskhalava:2015zwa,Baumann:2009ds,Senatore:2016aui,Baumann:2018muz,Das:2023cum,Choudhury:2011sq,Choudhury:2012yh,Choudhury:2012ib,Choudhury:2012whm,Choudhury:2013jya,Choudhury:2013zna,Esposito:2019jkb,Goldstein:2022hgr,Arkani-Hamed:2015bza,Arkani-Hamed:2023kig,Green:2023ids,Baumann:2021fxj,Baumann:2020dch,Baumann:2019oyu,Meerburg:2019qqi,Arkani-Hamed:2018kmz}. Because of the expression in eqn. (\ref{delta3nlo}), the $f_{\rm NL}$ has a negative signature. Consequently, when operating at the NLO expansion in the diffusion-dominated regime, one can anticipate obtaining ${\cal O}(1)$ negative non-Gaussianity, which is dependent on the underlying EFT structure defined by the parameter $C$. For the canonical stochastic single-field inflation, the situation $C=1$ yields $f_{\rm NL}=-5/4$, a number that defies the bound derived from Maldcena's consistency requirement, $f_{\rm NL}=(5/12)(1-n_{s})$, where the spectral tilt is represented by $n_{s}\ll 1$ in \cite{Maldacena:2002vr}. With the release of the most current NANOGrav15 data, large negative non-Gaussianity of $f_{\rm NL}\lesssim {\cal O}(-1)$ has shown to be of great value in solving the PBH overproduction problem; for recent efforts to address this issue, see \cite{Franciolini:2023pbf,Franciolini:2023wun,Inomata:2023zup, Inui:2023qsd,Chang:2023aba,Gorji:2023ziy,Li:2023xtl,Firouzjahi:2023xke,Gorji:2023sil,Raatikainen:2023bzk,Choudhury:2023fwk,Choudhury:2023fjs,Ferrante:2023bgz,Gorji:2023sil}. The relationship $\tau_{\rm NL}=(6f_{\rm NL}/5)^{2}$ \cite{Suyama:2007bg}, which only holds true for single-field inflation models, is maintained by the relation in eqn. (\ref{tauNLO}). When dealing with many fields, the equality sign in the relation is changed to an inequality, resulting in the expression $\tau_{\rm NL}< (6f_{\rm NL}/5)^{2}$. Regarding $f_{\rm NL}$, the case $C=1$ yields $g_{\rm NL}=18$ and predicts a big $\tau_{\rm NL}=9/4$.

\subsubsection{Calculation of PDF at NNLO}

Here, we go on to the next-to-next-to-leading order, which is one order higher in the characteristic function. We see some notable non-Gaussian traits here, and the method is identical to the prior NLO instance. Accordingly, we express the characteristic function that satisfies eqn. (\ref{FPEcharacter}) as follows:
\bea \chi=  \chi^{\rm LO} + \chi^{\rm NLO} + \chi^{\rm NNLO},\eea
that, after removing the diffusion factors, now has an additional sub-dominant contribution, which we need to assess. With this, we can get the following equation for $\chi_{\rm NNLO}$:
\bea
\bigg(\frac{d}{dp} + \frac{it}{y}\bigg)\chi^{\rm NNLO}(t;\Gamma(p)) = -\frac{1}{y\tilde{\mu}^{2}}\frac{\partial^2}{\partial x^2}\chi^{\rm NLO},
\eea
where the previously mentioned characteristic lines—which remain unchanged—have been utilized, along with the equations (\ref{chiLO},\ref{chiNLO}). Depending on the function in eqn. (\ref{chiNLOsol}), the expression on the RHS gets increasingly complicated. We continue to provide the RHS in the following condensed form:
\bea
-\frac{1}{y\tilde{\mu}^{2}}\frac{\partial^2}{\partial x^2}\chi^{\rm NLO} &=& \frac{1}{3\tilde{\mu}^{2}}\bigg(1-\frac{x}{y}\bigg)^{-\frac{it}{3C}-2}\bigg[3b\bigg(-\frac{1}{y^{3}}-\frac{4\big(1-\frac{x}{y}\big)}{Cy^{2}}+\frac{6\big(1-\frac{x}{y}\big)^{2}\ln{\big(1-\frac{x}{y}\big)}}{yC^2}\bigg)\nonumber\\
&-& 2itb\bigg(\frac{1}{Cy^3}-\frac{2\big(1-\frac{x}{y}\big)\ln{\big(1-\frac{x}{y}\big)}}{C^{2}y^2}\bigg)-it\bigg(1+\frac{it}{3C}\bigg)\frac{1-b\ln{\big(1-\frac{x}{y}\big)}}{Cy^{3}}\bigg],
\eea
where the definition of the parameter $b$ is:
\bea b=\frac{it}{9C^{4}\tilde{\mu}^{2}}\left(1+\frac{it}{3C}\right).\eea
From this, we derive an analogous linear ODE for $\chi^{\rm NNLO}$. After continuing the same process of solving the ODE and figuring out the initial condition function at NNLO, we finally get the following equation, which leads us to mention the PDF directly:
\bea \label{chiNNLOsol}
\chi^{\rm NNLO}(t;\Gamma) &=& \bigg(1-\frac{x}{y}\bigg)^{-\frac{it}{3c}}\bigg[1-\frac{it(1+\frac{it}{3C})\ln{\big(1-\frac{x}{y}\big)}}{9(x-y)^{2}C^{2}\tilde{\mu}^2}\nonumber\\
&&\quad\quad\quad\quad\quad\quad\quad\quad\quad- \frac{it(1+\frac{it}{3C})\ln{\big(1-\frac{x}{y}\big)^2}\big(90C^2+12iCt+(t-9Ci)(t-6Ci)\big)}{486(x-y)^{4}C^{4}\tilde{\mu}^4}\bigg]. \quad
\eea
Hence, according to eqn.(\ref{chiNLOsol}), the final element within the second bracket is a fresh addition to the other terms that already compose $\chi^{\rm NLO}$. A series expansion in powers of ${(C\tilde{\mu}y)^{-2}}$ is produced by the characteristic function expression, as can be shown. For the above $\chi^{\rm NNLO}$, the expansion ends at ${\cal O}((C\tilde{\mu}y)^{-4})$. Similar to how $\chi^{\rm NLO}$ was represented in the preceding example, the PDF corresponding to the inverse Fourier transform of the function $\chi^{\rm NNLO}$ may also be expressed using the Dirac delta derivatives. The PDF appears in the following final form:
\bea \label{pdfNNLO1}
P_{\Gamma}^{\rm NNLO}({\cal N}) &=& P_{\Gamma}^{\rm NLO}({\cal N}) \nonumber\\
&&- \frac{1}{486\;{C}^4\tilde{\mu}^4(y-x)^4}\ln{\bigg(1-\frac{x}{y}\bigg)}\bigg[-90\delta^{(1)}({\cal N}-\langle{\cal N}_{\rm LO}\rangle)+ 40C\delta^{(2)}({\cal N}-\langle{\cal N}_{\rm LO}\rangle)
\nonumber\\
&&- 4\delta^{(3)}({\cal N}-\langle{\cal N}_{\rm LO}\rangle)+\ln{\bigg(1-\frac{x}{y}\bigg)}\bigg(54C^2\delta^{(1)}({\cal N}-\langle{\cal N}_{\rm LO}\rangle)- 33C\delta^{(2)}({\cal N}-\langle{\cal N}_{\rm LO}\rangle)\nonumber\\
&&+ 6\delta^{(3)}({\cal N}-\langle{\cal N}_{\rm LO}\rangle)- \frac{1}{3C}\delta^{(4)}({\cal N}-\langle{\cal N}_{\rm LO}\rangle) \bigg)\bigg],\nonumber\\
&=& P_{\Gamma}^{\rm NLO}({\cal N}) \nonumber\\
&&- \frac{1}{486\;{C}^4\tilde{\mu}^4(y-x)^4}\delta({\cal N}-\langle{\cal N}_{\rm LO}\rangle)\bigg[\frac{90}{\cal N}+ \frac{80C}{{\cal N}^2}+ \frac{12C}{{\cal N}^3}+  \ln{\bigg(1-\frac{x}{y}\bigg)}\nonumber\\
&&\times \bigg(-\frac{54C^2}{\cal N}-\frac{66C}{{\cal N}^2}- \frac{18}{{\cal N}^3}-\frac{4}{3C{\cal N}^4}\bigg)\bigg]\ln{\bigg(1-\frac{x}{y}\bigg)},
\eea
Further higher-order corrections calculated for the mean number of ${\cal N}$ may be gathered and expressed as follows using Eqn.(\ref{chiNNLOsol}) and Eqn.(\ref{chimoments}):
\bea
\langle{\cal N}(\Gamma)\rangle_{\rm NNLO} = \langle{\cal N}(\Gamma)\rangle_{\rm LO}\bigg(1+\frac{1}{3C^2\tilde{\mu}^2(y-x)^2} + \frac{9C^2\langle{\cal N}(\Gamma)\rangle_{\rm LO}+5}{9C^3\tilde{\mu}^4(y-x)^4} \bigg). 
\eea
Likewise, in order to estimate the variance for the present NNLO at order ${\cal O}((\tilde{\mu}y)^{-4})$, the second moment must be provided as follows:
\bea
\langle{\cal N}^{2}(\Gamma)\rangle_{\rm NNLO} = \frac{14C+3\langle{\cal N}\rangle_{\rm LO}(10+11C^2)}{27C^3(x-y)^4\tilde{\mu}^2}\langle{\cal N}\rangle_{\rm LO}.
\eea
This supplies the subsequent order correction to the variance:
\bea \label{delta2nnlo}
\sigma_{\rm NNLO}^{2} = \langle\delta{\cal N}^2(\Gamma)\rangle_{\rm NNLO}&=& -\frac{2}{27C^3\tilde{\mu}^2(y-x)^2}\ln{\bigg(1-\frac{x}{y}\bigg)} + \frac{(11C-1)}{81C^4\tilde{\mu}^4(x-y)^4}\ln^2{\bigg(1-\frac{x}{y}\bigg)},\nonumber\\
&=& \frac{2\langle{\cal N}(\Gamma)\rangle_{\rm LO}}{9C^2\tilde{\mu}^2(y-x)^2}+  \frac{(11C-1)\langle{\cal N}(\Gamma)\rangle^{2}_{\rm LO}}{9C^2\tilde{\mu}^4(x-y)^4},
\eea
In this instance, the third moment from the probability distribution's mean is determined using equations (\ref{pdfNNLO1}) and (\ref{chimoments}). Our contribution comes from first analyzing the third instant itself, which is as follows:
\bea
\langle{\cal N}^3(\Gamma)\rangle_{\rm NNLO} = \frac{4\langle{\cal N}(\Gamma)\rangle_{\rm LO}}{27C^4\tilde{\mu^4}(x-y)^4}\bigg[1+15C^2\langle{\cal N}(\Gamma)\rangle_{\rm LO}\bigg],
\eea
Moreover, at leading order in ${\cal O}((\tilde{\mu}y)^{-4})$, the third moment corrections emerge as follows after removing additional higher powers of logarithms:
\bea \label{delta3nnlo}
\langle\delta{\cal N}^3(\Gamma)\rangle_{\rm NNLO} = \langle({\cal N}-\langle{\cal N}\rangle)^3\rangle_{\rm NNLO} &=& \frac{1}{81\;C^5\tilde{\mu}^4(x-y)^4}2\ln{\bigg(1-\frac{x}{y}\bigg)}\bigg[(-1+3C)\ln{\bigg(1-\frac{x}{y}\bigg)}-2\bigg],\nonumber\\
&=& \frac{4\langle{\cal N}(\Gamma)\rangle_{\rm LO}}{27\;C^4\tilde{\mu}^4(x-y)^4}\bigg[1+\frac{3C}{2}(3C-1)\langle{\cal N}(\Gamma)\rangle_{\rm LO}\bigg], 
\eea
This takes into account the impact of the characteristic parameter $C$. Analogously, corrections to the fourth moment away from the mean can be assessed. Once higher logarithm multiples are eliminated, we obtain the following for the fourth moment:
\bea
\langle{\cal N}^{4}(\Gamma)\rangle_{\rm NNLO} = \frac{4(4+C)\langle{\cal N}(\Gamma)\rangle_{\rm LO}^2}{27C^4\tilde{\mu^4}(x-y)^4}
\eea
Additionally, we have the following to the leading order in ${\cal O}((\tilde{\mu}y)^{-4})$:
\bea \label{delta4nnlo}
\langle\delta{\cal N}^4(\Gamma)\rangle_{\rm NNLO} = \langle({\cal N}-\langle{\cal N}\rangle)^4\rangle_{\rm NNLO} &=& \frac{4}{243\;C^6\tilde{\mu}^4(x-y)^4}\ln^2{\bigg(1-\frac{x}{y}\bigg)}\bigg[8+C+\ln{\bigg(1-\frac{x}{y}\bigg)}(4+40C)\bigg],\nonumber\\
&=& \frac{4\langle{\cal N}(\Gamma)\rangle_{\rm LO}^2}{27\;C^4\tilde{\mu}^4(x-y)^4}\bigg[8+C+3C\langle{\cal N}(\Gamma)\rangle_{\rm LO}(4+40C)\bigg].
\eea
From this, we may once more proceed to ascertain the distribution's skewness and kurtosis characteristics at NNLO. For the current distribution, they are defined as follows:
\bea
\gamma_{\rm NNLO} &=& \frac{\langle({\cal N}-\langle{\cal N}\rangle)^3\rangle_{\rm NNLO}}{\sigma_{\rm NNLO}^3} = \frac{1}{\langle{\cal N}\rangle_{\rm LO}^2}\frac{2(2+3\langle{\cal N}\rangle_{\rm LO}C(C-1))}{C(11C-1)^{3/2}}(x-y)^{2}\tilde{\mu}^2,\\
\kappa_{\rm NNLO} &=& \frac{\langle({\cal N}-\langle{\cal N}\rangle)^4\rangle_{\rm NNLO}}{\sigma_{\rm NNLO}^4} = \frac{1}{\langle{\cal N}\rangle^{2}_{\rm LO}}\frac{12C(x-y)^4\tilde{\mu}^4}{(11C-1)^2}.
\eea
Using the variance adjustments that were previously determined at the NNLO and found in equation (\ref{delta2nnlo}), we now estimate the corrected version of the power spectrum:
\bea
\Delta^{2}_{\zeta\zeta,\rm NNLO} = \frac{d\langle\delta{\cal N}^2\rangle}{d\langle{\cal N}\rangle} =  \frac{2}{9C^2\tilde{\mu}^2(y-x)^2} + \frac{2(11C-1)\langle{\cal N}(\Gamma)\rangle_{\rm LO}}{9C^2\tilde{\mu}^4(x-y)^4}.
\eea
We use eqns (\ref{delta3nnlo},\ref{delta4nnlo}) to assess the remaining set of non-Gaussianity parameters in the following ways:
\bea \label{fnlNNLO}
f_{\rm NL} = \frac{5}{36}\frac{1}{[\Delta^{2}_{\zeta\zeta,\rm NNLO}]^2}\frac{d^2\langle\delta{\cal N}^3\rangle}{d\langle{\cal N}\rangle^2} &=& \frac{5}{36}\frac{3C-1}{2C^3}\frac{8}{9\tilde{\mu}^4(x-y)^4}\frac{81\tilde{\mu}^4(x-y)^4}{4}\bigg(1+\frac{(11C-1)\langle{\cal N}\rangle_{\rm LO}}{\tilde{\mu}^{2}(x-y)^{2}}\bigg)^{-2},\nonumber\\
&=& \frac{5}{2}\frac{3C-1}{2C^3} - \frac{2(11C-1)\langle{\cal N}\rangle_{\rm LO}}{\tilde{\mu}^{2}(x-y)^{2}},
\eea
where the term ${\cal O}(\tilde{\mu}^{-2}y^{-2})$ is a suppressed correction from the power spectrum, and we estimate the result at leading order in the series expansion since $\tilde{\mu}y \gg 1$ is a required requirement to obtain a drift-dominated situation. Following us are:
\bea \label{tauNNLO}
\tau_{\rm NL} &=&  \frac{1}{36}\frac{1}{[\Delta^{2}_{\zeta\zeta,\rm NNLO}]^4}\bigg(\frac{d^2\langle\delta{\cal N}^3\rangle}{d\langle{\cal N}\rangle^2}\bigg)^{2} = \bigg(\frac{3C-1}{2C^3}\frac{8}{9\tilde{\mu}^4(x-y)^4}\bigg)^{2}\bigg(\frac{81\tilde{\mu}^4(x-y)^4}{4}\bigg)^{2}\bigg(1+\frac{(11C-1)\langle{\cal N}\rangle_{\rm LO}}{\tilde{\mu}^{2}(x-y)^{2}}\bigg)^{-4},\nonumber\\
&=& \frac{9}{4}\bigg(\frac{3C-1}{C^{3}}\bigg)^{2}\bigg(1-\frac{4(11C-1)\langle{\cal N}\rangle_{\rm LO}}{\tilde{\mu}^{2}(x-y)^{2}}\bigg),\\
g_{\rm NL} &=& \frac{1}{[\Delta^{2}_{\zeta\zeta,\rm NNLO}]^3}\frac{d^3\langle\delta{\cal N}^4\rangle}{d\langle{\cal N}\rangle^3} = \frac{24C(4+40C)}{9}\frac{81C^2\tilde{\mu}^2(x-y)^2}{4}\bigg(1+\frac{(11C-1)\langle{\cal N}\rangle_{\rm LO}}{\tilde{\mu}^{2}(x-y)^{2}}\bigg)^{-3},\nonumber\\
&=& 54C^3(4+40C)\bigg(1- \frac{3(11C-1)\langle{\cal N}\rangle_{\rm LO}}{\tilde{\mu}^{2}(x-y)^{2}}\bigg).
\eea
where a comparable phrase leading order in the growth is taken into consideration, as previously said. In comparison to the $C=1$ estimate at NLO, eqn. (\ref{fnlNLO}), the case $C=1$ in eqn. (\ref{fnlNNLO}) yields at leading order $f_{\rm NL}\sim 5/2$, which is positive and has a higher amplitude. Furthermore, Maldacena's constraint for canonical single-field inflation models is violated by the existing non-Gaussianity level. It is important to note that when $C<1/3$, we begin to observe negative $f_{\rm NL}$ at NNLO, suggesting that some non-canonical stochastic single-field models may be able to attain negative non-Gaussianity, and therefore provide a potential candidate to address the PBH overproduction issue as well. We obtain $\tau_{\rm NL} \sim 9 $ at leading order for $C=1$ in eqn. (\ref{tauNNLO}), with only minor adjustments made to its value from $C=1$ in eqn. (\ref{tauNNLO}). At NNLO, the Suyama-Yamaguchi constraint $\tau_{\rm NL}=(6f_{\rm NL}/5)^{2}$ is still valid.

\subsubsection{Calculation of PDF at NNNLO}

In the last section, we go over the characteristic function results in the next-to-next-to-next-leading order. With the diffusion terms as follows, the current characteristic function appears as a further higher-order term in the series expansion that satisfies eqn. (\ref{FPEcharacter}). 
\bea \chi=  \chi^{\rm LO} + \chi^{\rm NLO} + \chi^{\rm NNLO} + \chi^{\rm NNNLO}.\eea
Since it is not feasible to obtain the NNNLO contribution from the method of characteristics in this section, we refer to \cite{Pattison:2021oen}, which describes a general series expansion method designed to obtain the characteristic function terms to any degree in the drift-dominated or classical regime.

Rather of going into detail about the PDF (which is not very informative by itself), we just discuss the adjustments that have to be made from its structure to the statistical moments that now interest us. First, we adjust the mean value.
\bea
\langle{\cal N}(\Gamma)\rangle_{\rm NNNLO} = \langle{\cal N}(\Gamma)\rangle_{\rm NNLO}+ \langle{\cal N}(\Gamma)\rangle_{\rm LO}\frac{60\langle{\cal N}(\Gamma)\rangle^2_{\rm LO}+77\langle{\cal N}(\Gamma)\rangle_{\rm LO}+17}{9C^4\tilde{\mu}^6(y-x)^6}, 
\eea
where the adjustment occurs at ${\cal O}((\tilde{\mu}y)^{-6})$ this time. A adjustment at the same order in the expansion for the following statistical measure, the distribution's variance, looks like this:
\bea \label{delta2nnnlo}
\sigma_{\rm NNNLO}^{2} = \langle\delta{\cal N}^2(\Gamma)\rangle_{\rm NNNLO}= \frac{2\langle{\cal N}(\Gamma)\rangle_{\rm LO}}{81\;C^4\tilde{\mu}^6(y-x)^6}\bigg(28+143\;C\langle{\cal N}(\Gamma)\rangle_{\rm LO}+384\;C^2\langle{\cal N}(\Gamma)\rangle^{2}_{\rm LO}\bigg).
\eea
The equation for the third moment, which follows, is necessary in order to make corrections to it away from the mean, at NNNLO, and in the expansion order ${\cal O}((C\tilde{\mu}y)^{-6})$:
\bea
\langle{\cal N}^3\rangle_{\rm NNNLO} = \frac{\langle{\cal N}(\Gamma)\rangle_{\rm LO}}{81\;C^5\tilde{\mu}^6(y-x)^6}\bigg(76+1056\;C\langle{\cal N}(\Gamma)\rangle_{\rm LO}+3798\;C^2\langle{\cal N}(\Gamma)\rangle^{2}_{\rm LO}\bigg).
\eea
This allows us to derive the expression:
\bea \label{delta3nnnlo}
\langle\delta{\cal N}^3(\Gamma)\rangle_{\rm NNNLO}= \frac{4\langle{\cal N}(\Gamma)\rangle_{\rm LO}}{27\;C^4\tilde{\mu}^4(y-x)^4}\bigg[1+3C\langle{\cal N}(\Gamma)\rangle_{\rm LO} + \frac{19+ 120\;C\langle{\cal N}(\Gamma)\rangle_{\rm LO}+ 132\;C^2\langle{\cal N}(\Gamma)\rangle^{2}_{\rm LO}}{3\;C\tilde{\mu}^2(y-x)^2}\bigg].
\eea
Lastly, it is also necessary to first determine the fourth instant in order to make modifications to the fourth moment away from the mean at NNNLO:
\bea
\langle{\cal N}^4\rangle_{\rm NNNLO} = \frac{4\langle{\cal N}(\Gamma)\rangle_{\rm LO}}{243\;C^6\tilde{\mu}^6(y-x)^6}\bigg(10+ 366\;C\langle{\cal N}(\Gamma)\rangle_{\rm LO}+ 2667\;C^2\langle{\cal N}(\Gamma)\rangle^{2}_{\rm LO}\bigg),
\eea
possessing this allows us to obtain the following form:
\bea \label{delta4nnnlo}
\langle\delta{\cal N}^4(\Gamma)\rangle_{\rm NNNLO}= \frac{4\langle{\cal N}(\Gamma)\rangle_{\rm LO}}{27\;C^4\tilde{\mu}^4(y-x)^4}\bigg[\langle{\cal N}(\Gamma)\rangle_{\rm LO} + \frac{2(5+ 63\;C\langle{\cal N}(\Gamma)\rangle_{\rm LO}+ 105\;C^2\langle{\cal N}(\Gamma)\rangle^{2}_{\rm LO})}{9\;C^2\tilde{\mu}^2(y-x)^2}\bigg].
\eea
Finally, we discuss the skewness and kurtosis of the distribution at NNNLO using the data above. They are listed below:
\bea
\gamma_{\rm NNNLO} &=& \frac{\langle({\cal N}-\langle{\cal N}\rangle)^3\rangle_{\rm NNNLO}}{\sigma_{\rm NNNLO}^3} = 9\sqrt{2}C^3\tilde{\mu}^3(x-y)^3 \frac{19 +120C\langle{\cal N}\rangle_{\rm LO} +132C^2\langle{\cal N}\rangle_{\rm LO}^2}{\sqrt{\langle{\cal N}\rangle_{\rm LO}}(28+ 143C\langle{\cal N}\rangle_{\rm LO} + 384C^2\langle{\cal N}\rangle_{\rm LO}^2)^{3/2}},\\
\kappa_{\rm NNNLO} &=& \frac{\langle({\cal N}-\langle{\cal N}\rangle)^4\rangle_{\rm NNNLO}}{\sigma_{\rm NNNLO}^4} = \frac{54C^2(x-y)^6\tilde{\mu}^6}{\langle{\cal N}\rangle_{\rm LO}}\frac{5+ 63C\langle{\cal N}\rangle_{\rm LO}+ 105C^2\langle{\cal N}\rangle^{2}_{\rm LO}}{28+ 143C\langle{\cal N}\rangle_{\rm LO} + 384C^2\langle{\cal N}\rangle_{\rm LO}^2}.
\eea
Once the addition from eqn. (\ref{delta2nnnlo}) is applied, the rectified power spectrum may be assessed as follows:
\bea
\Delta^{2}_{\zeta\zeta, \rm NNNLO} = \frac{d\langle\delta{\cal N}^2\rangle}{d\langle{\cal N}\rangle} &=& \frac{2}{9C^2\tilde{\mu}^2(y-x)^2} + \frac{2(11C-1)\langle{\cal N}(\Gamma)\rangle_{\rm LO}}{9C^2\tilde{\mu}^4(y-x)^4} \nonumber\\
&+& \frac{4}{81C^4\tilde{\mu}^6(y-x)^6}\bigg(14+ 143C\langle{\cal N}(\Gamma)\rangle_{\rm LO}+ 576C^2\langle{\cal N}(\Gamma)\rangle^{2}_{\rm LO}\bigg),
\eea
finally calculated to provide the following set of non-Gaussianity characteristics at the NNNLO:
\bea
\label{fnlNNNLO}
f_{\rm NL} &=& \frac{5}{36}\frac{1}{[\Delta^{2}_{\zeta\zeta,\rm NNNLO}]^2}\frac{d^2\langle\delta{\cal N}^3\rangle}{d\langle{\cal N}\rangle^2} = \bigg[\frac{5C}{2} + \frac{10(10+33C\langle{\cal N}\rangle_{\rm LO})}{3\tilde{\mu}^{2}(y-x)^{2}}\bigg]\bigg(1+ \frac{g_{1}(C,\langle{\cal N}\rangle_{\rm LO})}{\tilde{\mu}^{2}(y-x)^{2}} + \frac{2g_{2}(C,\langle{\cal N}\rangle_{\rm LO})}{9C^{2}\tilde{\mu}^{4}(y-x)^{4}}\bigg)^{-2},\\
\label{tauNNNLO}
\tau_{\rm NL}&=& \frac{1}{36}\frac{1}{[\Delta^{2}_{\zeta\zeta,\rm NNNLO}]^4}\bigg(\frac{d^2\langle\delta{\cal N}^3\rangle}{d\langle{\cal N}\rangle^2}\bigg)^{2} = \bigg[\frac{C}{2}+\frac{2(10+33C\langle{\cal N}\rangle_{\rm LO})}{3\tilde{\mu}^{2}(y-x)^{2}}\bigg]^{2}\bigg(1+ \frac{g_{1}(C,\langle{\cal N}\rangle_{\rm LO})}{\tilde{\mu}^{2}(y-x)^{2}} + \frac{2g_{2}(C,\langle{\cal N}\rangle_{\rm LO})}{9C^{2}\tilde{\mu}^{4}(y-x)^{4}}\bigg)^{-4},\quad\quad\\
g_{\rm NL}&=& \frac{1}{[\Delta^{2}_{\zeta\zeta,\rm NNNLO}]^3}\frac{d^3\langle\delta{\cal N}^4\rangle}{d\langle{\cal N}\rangle^3} = 1890C^2\bigg(1+ \frac{g_{1}(C,\langle{\cal N}\rangle_{\rm LO})}{\tilde{\mu}^{2}(y-x)^{2}} + \frac{2g_{2}(C,\langle{\cal N}\rangle_{\rm LO})}{9C^{2}\tilde{\mu}^{4}(y-x)^{4}}\bigg)^{-3}.
\eea
where the following formulas describe the coefficients, $g_{1}(C,\langle{\cal N}\rangle_{\rm LO})$, and $g_{2}(C,\langle{\cal N}\rangle_{\rm LO})$:
\bea && g_{1}(C,\langle{\cal N}\rangle_{\rm LO})=(11C-1)\langle{\cal N}\rangle_{\rm LO},\\
&& g_{2}(C,\langle{\cal N}\rangle_{\rm LO})= 14+143C\langle{\cal N}\rangle_{\rm LO}+576C^{2}\langle{\cal N}\rangle_{\rm LO}^{2}.\eea
We can see the impact of modifications to the leading order result in the formulas for the different non-Gaussianity indicators above. When considering $f_{\rm NL}$, at $C=1$, one observes a marginally larger magnitude than the leading order value of $f_{\rm NL}=5/2$, which originates from the ${\cal O}(\tilde{\mu}^{-2}y^{-2})$ suppressed term with positive signature. It does not appear that negative non-Gaussianity exists in this instance. At this order of analysis, the aforementioned equations (\ref{fnlNNNLO},\ref{tauNNNLO}) can also preserve the Suyama-Yamaguchi relation, $\tau_{\rm NL}=(6f_{\rm NL}/5)^{2}$. With $g_{\rm NL}\gg \tau_{\rm NL}$ at their leading order, the other two parameters, $\tau_{\rm NL}$ and $g_{\rm NL}$, assume high positive magnitudes.

\subsection{Spectral distortion and its implications}

In this part, we address spectral distortions, a distinct yet crucial physical consequence shown as inhomogeneities in the cosmic microwave background distributions. Depending on the mass of the generated PBH, these effects, which are associated with the collapse phase of their production mechanism, might be observed as large aberrations in the CMB spectrum. Based on the redshift intervals and the range of processes occurring in the corresponding history, these distortions are categorized in the literature along with their capacity to cause temperature variations in the standard black-body distribution of the CMB photons. These distortion effects seem to provide important indications from the pre-recombination era. For a thorough discussion of these effects and how they relate to the creation of PBH, see \cite{zeldovich1969interaction,Sunyaev:1970eu,Illarionov:1975ei,Hu:1992dc,Chluba:2012we,Khatri:2012tw,Chluba:2003exb,Stebbins:2007ve,Chluba:2011hw,Pitrou:2014ota,Hooper:2023nnl,Deng:2021edw}.

The two primary categories of significant spectral distortions under investigation are the $y$ and $\mu$ type distortions. The history of the CMB spectrum's creation includes an era called the $\mu$-era, during which the $\mu$-type distortions are caused by redshifts that are in between, $2\times 10^{5}\leq z\leq 2\times 10^{6}$. Past $z \geq 2\times 10^{6}$, we enter the high-temperature thermalization period, when mechanisms like as Compton scattering, double Compton scattering, and bremsstrahlung are able to sustain a state of thermal equilibrium, suggesting much reduced distortion effects. The $\mu$-era is named because at this time, deviations from a black-body spectrum begin to manifest. Compton scattering mechanisms continue to operate, but they produce a Bose-Einstein spectrum instead, which is defined by a non-zero chemical potential $\mu\ne 0$. Even the Compton scattering effects start to lose their effectiveness in preserving a Bose-Einstein distribution below the redshift, $z\leq 2\times 10^{5}$. As a result, equilibrium is lost, and in the late time scales, close to the CMB period, temperature disparities become more sensitive. The $y$-type distortions are the name given to them.

According to \cite{Hooper:2023nnl}, they may be calculated in terms of the wavenumber from the scalar power spectrum of the curvature perturbations:
\bea \label{mudistortion}
\mu &\simeq& 2.2\int_{k_{i}}^{\infty}\frac{dk}{k}\Delta^{2}_{\zeta\zeta}(k)\bigg\{\exp{\bigg[-\frac{k}{5400\;{\rm Mpc^{-1}}}\bigg]}-\exp{\bigg[-\bigg(\frac{k}{31.6\;{\rm Mpc^{-1}}}\bigg)^{2}\bigg]}\bigg\},\\
\label{ydistortion}
y &\simeq& 0.4\int_{k_{i}}^{\infty}\frac{dk}{k}\Delta^{2}_{\zeta\zeta}(k)\exp{\bigg[-\bigg(\frac{k}{31.6\;{\rm Mpc^{-1}}}\bigg)^{2}\bigg]},
\eea
when one uses $k_{i}=1\;{\rm Mpc^{-1}}$. The power spectrum defined above, $\Delta^{2}_{\zeta\zeta}(k)$, is the total power spectrum obtained for each of the three phases in the equations. As in (\ref{pspecsr1dS}, \ref{pspecusrdS}, \ref{pspecsr2dS}). When the wavenumber for PBH formation fulfills $k_{\rm PBH}\sim {\cal O}(10^{7}{\rm Mpc^{-1}})$, the two distortion estimations from the aforementioned total power spectrum give us values, $\mu\sim 2.35\times 10^{-8}$ and $y\sim 2.78\times 10^{-9}$. The PBH mass and transition wavenumber are correlated as follows:
\bea \label{pbhmass}
\frac{M_{\rm PBH}}{M_{\odot}} = 1.13 \times 10^{15} \times \bigg(\frac{\gamma}{0.2}\bigg)\bigg(\frac{g_{*}}{106.75}\bigg)^{-1/6}\bigg(\frac{k_{*}}{k_{\rm PBH}}\bigg)^{2},
\eea
where $k_{*}\sim 0.02{\rm Mpc^{-1}}$ is located. We apply the aforementioned formula in conjunction with the definitions in eqn. (\ref{mudistortion},\ref{ydistortion}) to get $\mu$ and $y$ as functions of the PBH mass in order to explore the permitted PBH mass range. In a separate numerical outcome section, we then compare the results to the observational limitations on their values that are currently in place. Here, it's helpful to keep in mind $A\sim {\cal O}(10^{-2})$, the amplitude of the whole scalar power spectrum that achieves the necessary levels for PBH generation. Using the previously described power spectrum equation, the following expression for the amplitude may be extracted:
\bea
\Delta^{2}_{\zeta\zeta}(k) &=& \big[\Delta^{2}_{\zeta\zeta,{\bf dS}}(k<k_{s})\big]_{\bf SRI} + \Theta(k-k_{s})\big[\Delta^{2}_{\zeta\zeta,{\bf dS}}(k_{s}\leq k < k_{e})\big]_{\bf USR} + \Theta(k-k_{e})\big[\Delta^{2}_{\zeta\zeta,{\bf dS}}(k_{e}\leq k< k_{\rm end})\big]_{\bf SRII},\nonumber\\
&=& A(k_s)\bigg[1+\bigg(\frac{k}{k_{s}}\bigg)^{2}\bigg](1+\sigma^{2})\times \bigg\{\bigg(\frac{k_{s}}{k_{e}}\bigg)^{6}\nonumber\\
&& \quad\quad+ \bigg(\Theta(k-k_{s})\Big|\alpha_{2,{\bf dS}} e^{i\sigma}-\frac{(1+i \sigma)}{(1-i \sigma)}\beta_{2,{\bf dS}} e^{-i\sigma}\Big|^2 + \Theta(k-k_{e})\Big|\alpha_{3,{\bf dS}} e^{i\sigma}-\frac{(1+i \sigma)}{(1-i \sigma)}\beta_{3,{\bf dS}} e^{-i\sigma}\Big|^2\bigg)\bigg\},
\eea
the amplitude is expressed as follows:
\bea
A(k_{s}) = \bigg(\frac{H^2}{8\pi^2\epsilon  c_s M_p^2}\bigg)_{*}\bigg(\frac{k_{e}}{k_{s}}\bigg)^{6}.
\eea
Consequently, as in eqn. (\ref{pbhmass}), the wavenumber associated with the abrupt transition into the USR that leads to PBH creation is denoted by $k_{*}\simeq 0.02{\rm Mpc^{-1}}$ and $k_{s}$. The $*$ notation suggests evaluation at the pivot scale. The $k-$dependent element outside the power spectrum is present at all scales, and before we analyze a specific regime, the wavenumber dependency in the total power spectrum above originates from the Bogoliubov coefficients for the USR and SRII.

\subsection{PBH Formation from EFT of Stochastic Single Field Inflation using stochastic-$\delta N$ formalism}

In the context of stochastic inflation, this section focuses on the study of primordial black hole generation. The overall mechanism of PBH production in the early Universe is the first topic we address. Following our formation description, we turn to the analysis of the PBH mass fraction in the context of stochastic inflation. Here, we show how the probability distribution analysis from the earlier sections can be used to derive numerical values for the PBH density fraction today and its ratio to the current dark matter density in the form of PBH abundance.

\subsubsection{Formation Mechanism}

When stochastic influences play a significant role and likely to enable the rise of curvature perturbations to sufficiently high values, we investigate the PBH creation mechanism. This is mostly caused by disturbances at tiny scales that re-enter the Hubble radius during inflation. If these perturbations are significant enough, they can cause gravitational collapse of the Hubble patches, which will produce PBHs. The following is the usual method for calculating the resultant PBH mass fraction. The curvature perturbation $\zeta$ is first distributed according to a Gaussian distribution $P(\zeta)$. Given such a distribution, the likelihood of PBH creation or the PBH mass fraction may be obtained by integrating its value over a threshold magnitude of disturbances, such as $\zeta_{\rm th}$, inside a Hubble patch.
\bea \label{betazeta}
\beta(M_{\rm PBH}) \sim \int_{\zeta_{\rm th}}P(\zeta)d\zeta,
\eea
Currently, PBH production is an uncommon occurrence. For dominating quantum diffusion processes, curvature perturbations are greatly enhanced until the threshold condition $\zeta_{\rm th}\sim {\cal O}(1)$ is satisfied. Therefore, the non-Gaussian structure of their distribution, in the extreme tail area, has the greatest impact on such big variations. Large improvements during inflation are made possible by the ultra-slow roll phase for tiny scales, which is where quantum diffusion processes take over as the primary driving force behind PBH production. Because of non-Gaussian tail features, the expected nature of the probability distribution function for the curvature perturbations is no longer Gaussian. The implications of the obtained PDFs for PBH mass fraction calculations will be ascertained by referring to the outcomes of the preceding sections in the diffusion-dominated regime. 

Our methodology also benefits from the use of the stochastic variable ${\cal N}$. As previously noted, given certain starting values of the phase space variables $\Gamma$, the different distributions are computed to realise the stochastic period ${\cal N}$. In order to transfer the statistics of ${\cal N}$ to those of the curvature perturbation $\zeta$ on the super-Hubble scales, one can work with the stochastic-$\delta N$ formalism. The following is the identification:
\bea \label{deltaN}
\zeta_{\rm cg}({\bf x}) = \delta {\cal N} =  {\cal N}({\bf x}) -\langle{\cal N}\rangle.
\eea
Consequently, for a given set of beginning and end conditions on phase space variables, the subscript "cg" specifies the coarse-grained value of the curvature perturbations in terms of the modes that enter and depart the Hubble radius. We will use the eqns. (\ref{betazeta}, \ref{deltaN}) to estimate the resultant PBH mass fraction. The PBH mass generated is connected to the size of the mode involved in the coarse-graining.

\subsubsection{Mass Fraction of PBH $(\beta)$ }

We utilize the PDF findings, as indicated in eqns, to compute the PBH mass fraction.Apply the stochastic-$\delta N$ formalism to (\ref{pdfLO},\ref{PDFN1},\ref{PDFN2},\ref{PDFN3}). First, we note that the functions describing the aforementioned PDFs are associated with the stochastic e-folds ${\cal N}$. We write the same definition after eqn.(\ref{deltaN}), since the mass fraction is defined from the curvature perturbation distribution as in eqn.(\ref{betazeta}) as:
\bea
\beta \sim \int_{\zeta_{\rm th}+\langle{\cal N}\rangle}^{\infty}P_{\Gamma}({\cal N})d{\cal N},
\eea
which needs to know the average value of the e-folds from the different distribution functions that were previously determined using equations (\ref{meanLO},\ref{meanNLO},\ref{meanNNLO},\ref{meanNNNLO}). For $\zeta_{\rm th}\sim {\cal O}(1)$, the threshold value is maintained. Due to the rapid decay of the higher-order poles' contributions in the PDF, the above integral is better approximated with a summation version of the same. Using the PDF's series formula, we can find the following relation for the mass fraction:
\bea
\beta_{m} \sim \sum_{i=0}^{m}\sum_{n=0}^{\infty}\frac{r_{n,m}^{(i)}}{\Lambda_{n}^{(i)}}\exp{(-\Lambda_{n}^{(i)}[\langle{\cal N}\rangle_{m} + \zeta_{\rm th}])},
\eea
where the collection of poles at $i$-th order, $\Lambda_{n}^{(i)}$, is defined in eqn, and the subscript $(m)$ denotes the order of perturbative expansion under consideration.(\ref{poles}). The analytical formulae for the residues $r_{n,m}^{(i)}$, which were previously evaluated at each order in the equations, are used in the above. with the exception of the NNLO and NNNLO residues, which have not previously been discussed since they are long and do not provide illumination on their own (\ref{resLO}, \ref{resNLO0}, \ref{resNLO1}). The PBH fraction will be calculated in the small $y$ limit using the aforementioned method, and we will discuss its usefulness based on the outcomes at each order in the $y$-expansion. 

Here, the residue equation from the previous eqns may be used to mention the analytical result of the mass fraction $\beta$ at the leading order. (\ref{resLO}), as shown by:
\bea
\beta_{\rm LO} &=& \sum_{n=0}^{\infty}\frac{r_{n,0}^{(0)}}{\Lambda_{n}^{(0)}}\exp{(-\Lambda_{n}^{(0)}[\langle{\cal N}\rangle_{\rm LO} + \zeta_{\rm th}])},\nonumber\\
&=& \frac{4}{\pi}\sum_{n=0}^{\infty}\frac{1}{2n+1}\sin{\left(\frac{(2n+1)\pi x}{2}\right)}\exp{\bigg\{-\pi^{2}\left(n+\frac{1}{2}\right)^{2}\left[x\left(1-\frac{x}{2}\right)+\frac{\zeta_{\rm th}}{\tilde{\mu}^{2}}\right]\bigg\}},
\eea
where the equation for the mean e-folds from eqn. (\ref{meanLO}) is put into practice. Likewise, one may write: for the NLO:
\bea
\beta_{\rm NLO} &=& \sum_{i=0}^{m=1}\sum_{n=0}^{\infty}\frac{r_{n,1}^{(i)}}{\Lambda_{n}^{(i)}}\exp{(-\Lambda_{n}^{(i)}[\langle{\cal N}\rangle_{\rm NLO} + \zeta_{\rm th}])},\nonumber\\
&=& \sum_{n=0}^{\infty}\bigg(\frac{r_{n,1}^{(0)}}{\Lambda_{n}^{(0)}} + \frac{r_{n,1}^{(1)}\exp{(-3C[\langle{\cal N}\rangle_{\rm NLO} + \zeta_{\rm th}])}}{\Lambda_{n}^{(0)}+3C} \bigg)\exp{(-\Lambda_{n}^{(0)}[\langle{\cal N}\rangle_{\rm NLO} +\zeta_{\rm th}])}.
\eea
This contains the mean obtained from eqn. (\ref{meanNLO}) at NLO. Similarly, we may use the ensuing higher-order contributions in the perturbative expansion to assess the mass fraction. After that, the outcome for NNLO is expressed as:
\bea
\beta_{\rm NNLO} &=& \sum_{n=0}^{\infty}\bigg(\frac{r_{n,2}^{(0)}}{\Lambda_{n}^{(0)}} + \frac{r_{n,2}^{(1)}\exp{(-3C[\langle{\cal N}\rangle_{\rm NNLO} + \zeta_{\rm th}])}}{\Lambda_{n}^{(0)}+3C} \nonumber\\
&&\quad\quad\quad\quad\quad\quad\quad\quad\quad\quad\quad\quad + \frac{r_{n,2}^{(2)}\exp{(-6C[\langle{\cal N}\rangle_{\rm NNLO} + \zeta_{\rm th}])}}{\Lambda_{n}^{(0)}+6C} \bigg)\exp{(-\Lambda_{n}^{(0)}[\langle{\cal N}\rangle_{\rm NNLO} +\zeta_{\rm th}])}.
\eea
Finally, we may express the PBH mass fraction for the NNNLO as follows:
\bea
\beta_{\rm NNNLO} &=& \sum_{n=0}^{\infty}\bigg(\frac{r_{n,3}^{(0)}}{\Lambda_{n}^{(0)}} + \frac{r_{n,3}^{(1)}\exp{(-3C[\langle{\cal N}\rangle_{\rm NNNLO} + \zeta_{\rm th}])}}{\Lambda_{n}^{(0)}+3C} + \frac{r_{n,3}^{(2)}\exp{(-6C[\langle{\cal N}\rangle_{\rm NNNLO} + \zeta_{\rm th}])}}{\Lambda_{n}^{(0)}+6C}\nonumber\\
&&\quad\quad\quad\quad\quad\quad\quad\quad\quad\quad\quad + \frac{r_{n,3}^{(3)}\exp{(-9C[\langle{\cal N}\rangle_{\rm NNNLO} + \zeta_{\rm th}])}}{\Lambda_{n}^{(0)}+9C} \bigg)\exp{(-\Lambda_{n}^{(0)}[\langle{\cal N}\rangle_{\rm NNNLO} +\zeta_{\rm th}])},
\eea

\subsubsection{Abundance of PBH $(f_{PBH})$}

The fraction $f_{\rm PBH}$ is used to indicate the current day fraction of PBH density in relation to the current day dark matter density. When these amplified fluctuations re-enter the Horizon during inflation, the presence of diffusion effects that predominate the small scale domain creates the ideal conditions for PBH formation. The percentage of current total dark matter that is present in this form may be determined by studying the abundance of these PBH, which is an essential quantity to examine. The idea of the PBH mass fraction was explained in depth in the preceding section, along with the nuances pertaining to the distribution function of curvature perturbations in the context of stochastic inflation. The abundance $f_{\rm PBH}$ may also be calculated with the same mass fraction $\beta(M_{\rm PBH})$ by using the following relation:
\bea
\label{fPbh}
f_{\rm PBH}&=&1.68\times 10^8 \bigg(\frac{\gamma}{0.2}\bigg)^{\frac{1}{2}}\bigg(\frac{g_{*}}{106.75}\bigg)^{-\frac{1}{4}}\bigg(\frac{M_{\rm PBH}}{M_{\odot}}\bigg)^{-\frac{1}{2}}\times \beta(M_{\rm PBH}).
\eea
where the wavenumber of PBH production, $k_{\rm PBH}$, is used to describe the mass of PBH in solar mass units, $M_{\odot}$, using the relation:
\bea
\frac{M_{\rm PBH}}{M_{\odot}} = 1.13 \times 10^{15} \times \bigg(\frac{\gamma}{0.2}\bigg)\bigg(\frac{g_{*}}{106.75}\bigg)^{-1/6}\bigg(\frac{k_{*}}{k_{\rm PBH}}\bigg)^{2}.
\eea
Furthermore, the pivot scale is fixed at $k_{*}\simeq 0.02\;{\rm Mpc^{-1}}$.
The PBH abundance in a sizable window $f_{\rm PBH}\leq 1$ is mostly determined by the threshold $\zeta_{\rm th}\sim {\cal O}(1)$. We shall display the behavior of $f_{\rm PBH}$ as a function of the PBH masses in the numerical result section. The PBH masses are computed from the mass fraction at each order in the perturbative expansion for diffusion.

\subsection{Numerical Outcomes}

This section covers a variety of findings from the numerical examination of significant observational characteristics associated with PBH generation and the scalar power spectrum previously examined for each phase in our setup. We first go over the power spectrum at the tree-level for each of the three phases of interest and highlight how it relates to the noise matrix element. By utilizing this power spectrum, we assess the consequences of spectral distortion for various PBH mass ranges and investigate the significance of observational limitations on the power spectrum's amplitude.  Next, we examine how the probability distribution function behaves in relation to the number of e-folds ${\cal N}$ at every rank in the perturbative expansion. Subsequently, we examine the PBH mass fraction and current PBH abundance in the same perturbative investigation that was previously conducted for the diffusion-dominated regime.

\subsubsection{Outcomes of tree-level power spectrum and noise matrix elements}

\begin{figure*}[ht!]
    	\centering
    \subfigure[]{
      	\includegraphics[width=8.5cm,height=7.5cm]{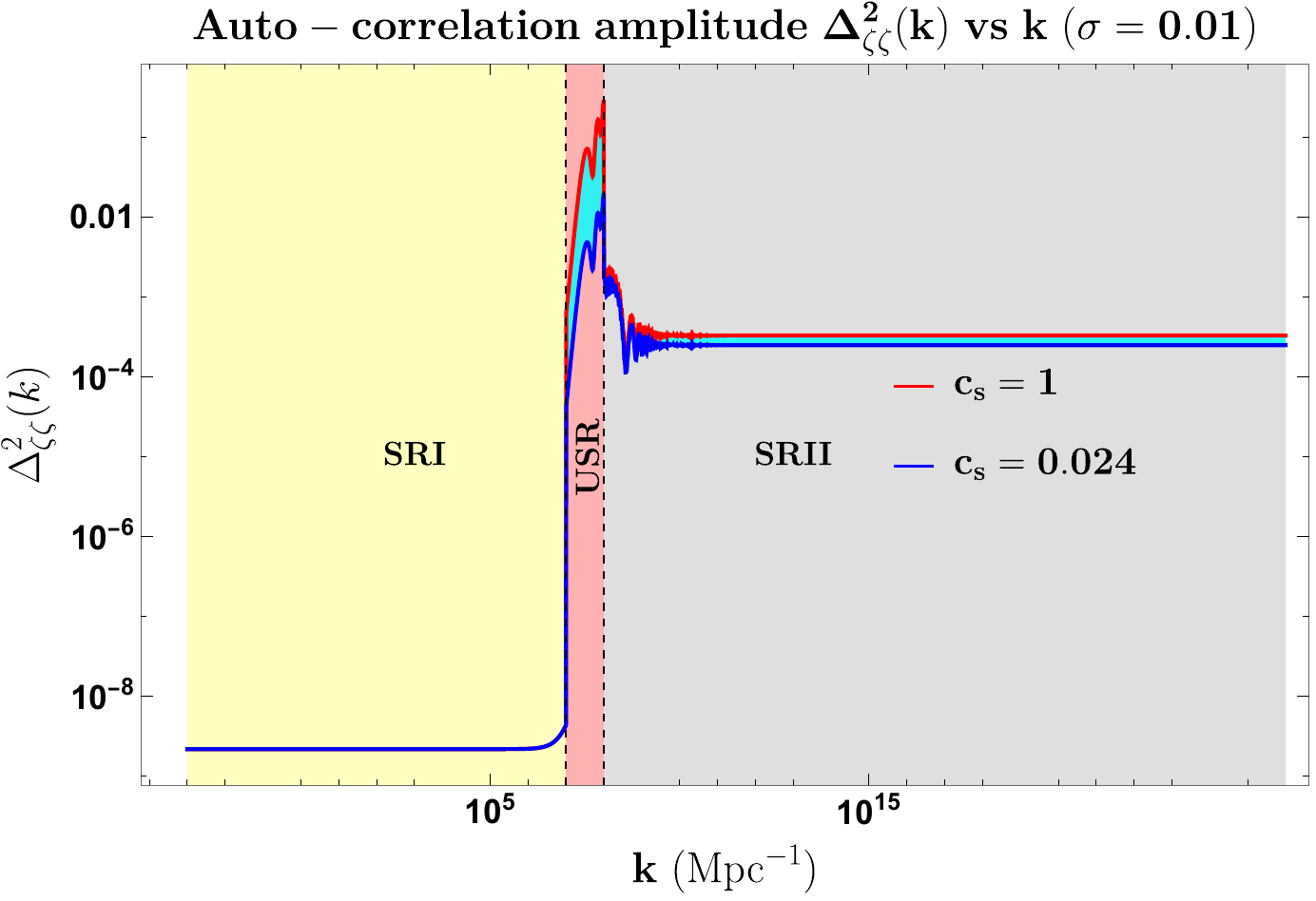}
        \label{pspeczz1}
    }
    \subfigure[]{
        \includegraphics[width=8.5cm,height=7.5cm]{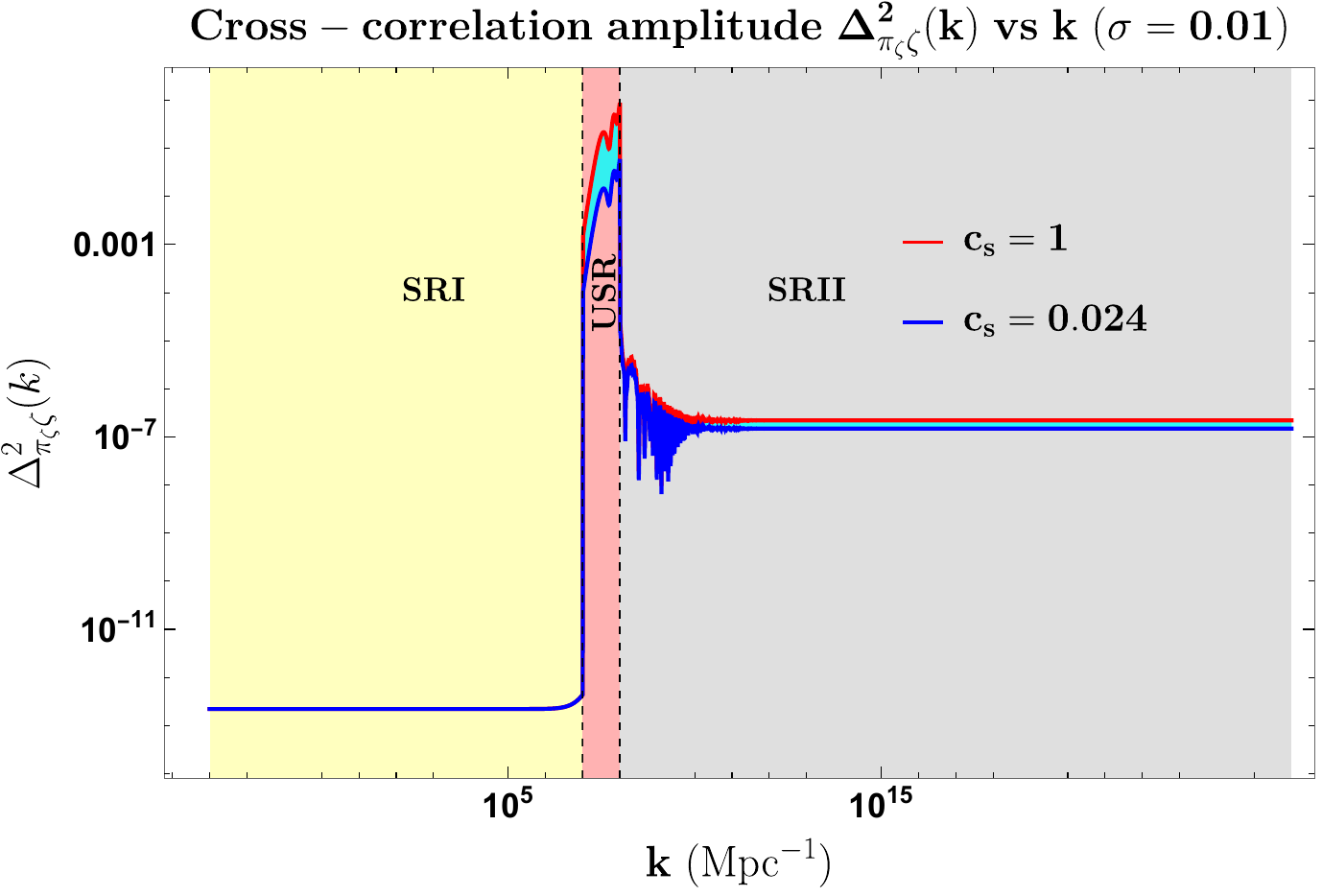}
        \label{pspecpz1}
    }
       \subfigure[]{
        \includegraphics[width=8.5cm,height=7.5cm]{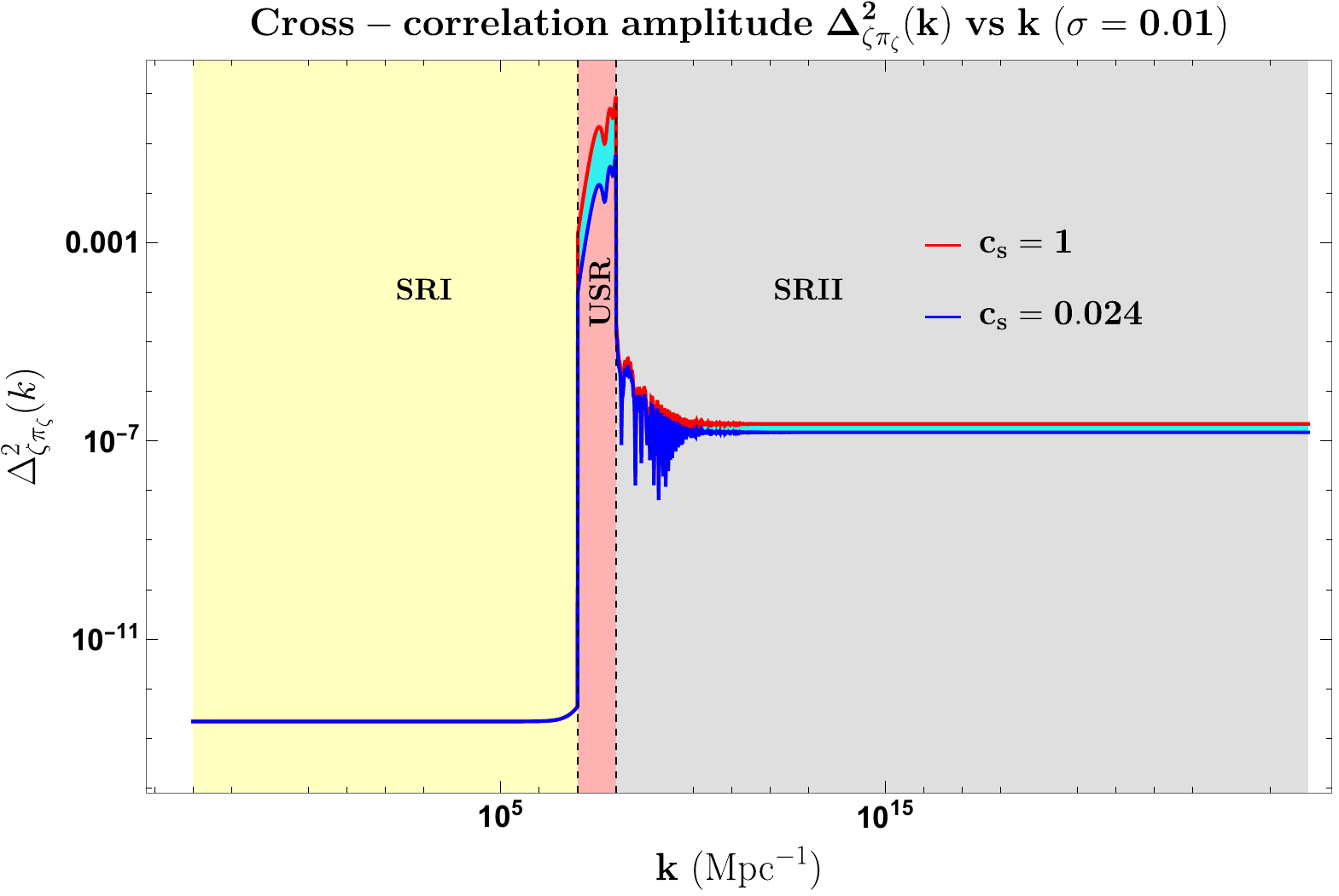}
        \label{pspeczp1}
    }
    \subfigure[]{
        \includegraphics[width=8.5cm,height=7.5cm]{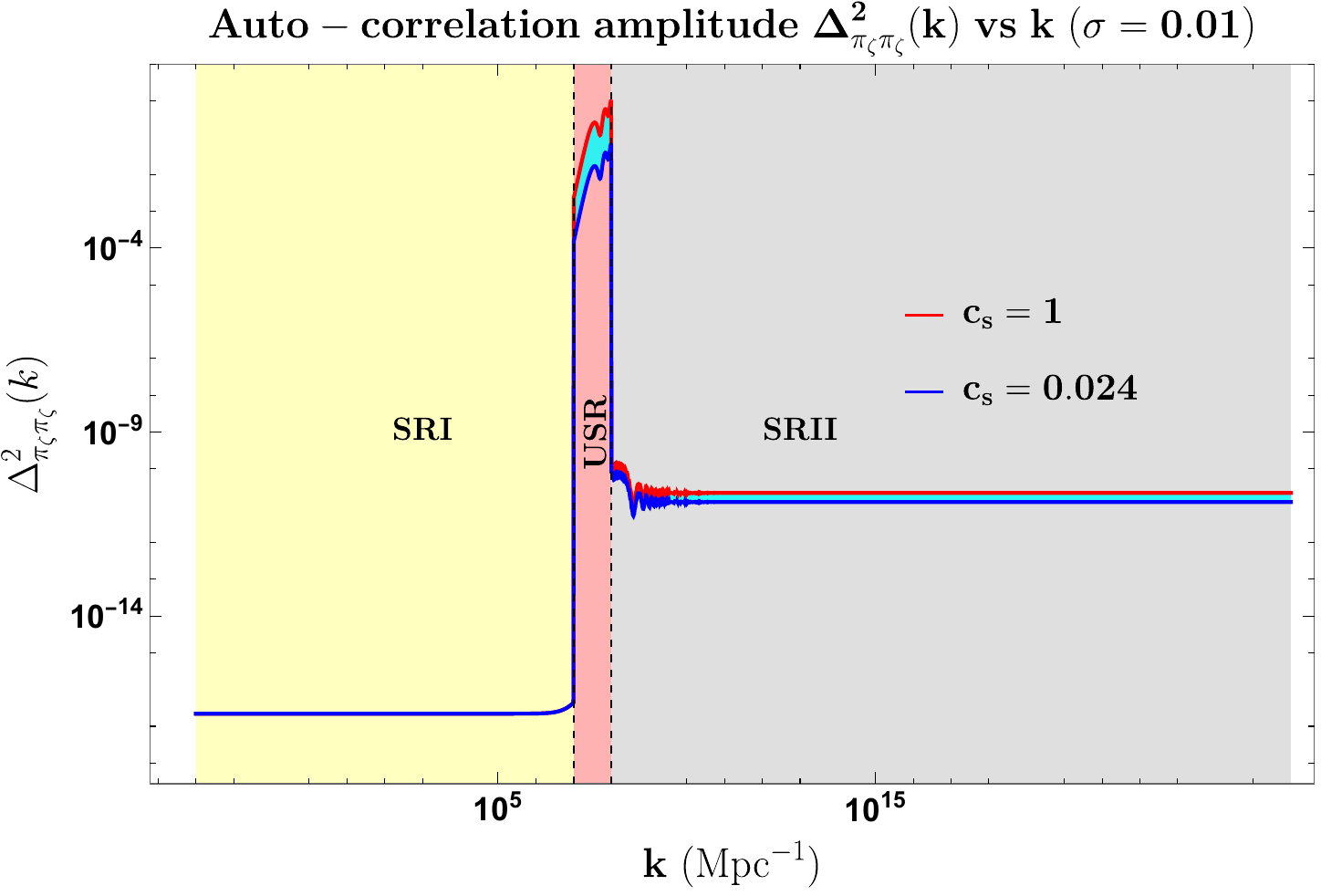}
        \label{pspecpp1}
       }
    	\caption[Optional caption for list of figures]{Plot showing different parts of the power spectrum $\Delta^{2}_{fg}$ as a function of wavenumber $k$, where $f,g\;\in\{\zeta,\Pi_{\zeta}\}$. The elements $\Delta^{2}_{\zeta\zeta}$, $\Delta^{2}_{\zeta\Pi_{\zeta}}$, and $\Delta^{2}_{\Pi_{\zeta}\zeta}$ are displayed in the \textit{top-left}, \textit{top-right}, and \textit{bottom-left}, respectively, and $\Delta^{2}_{\zeta\Pi_{\zeta}}$, and $\Delta^{2}_{\Pi_{\zeta}\zeta}$ in the \textit{bottom-right} panels. $\sigma=0.01$ is the fixed value of the stochastic parameter. The effective sound speed parameter, $c_{s}=c_{s,*}=\{0.024,1\}$, is set at the pivot scale. We display each correlation for these two values in red ($c_{s}=1$) and blue ($c_{s}=0.024$). Distinct shading indicates the behavior of each element in the three phases: SRI, USR, and SRII. 
} 
    	\label{spectrumplots}
    \end{figure*}
The behavior of the different power spectrum elements as a function of wavenumber for the three phases of interest in our setup—SRI, USR, and SRII—is shown in this section.

Based on fig. (\ref{spectrumplots}), we see that, up until slow-roll criteria are fulfilled, the auto-correlation $\Delta^{2}_{\Pi_{\zeta}\Pi_{\zeta}}$, as shown in \ref{pspecpp1}, is the most suppressed in magnitude. This takes place in both the SRI and SRII phases respectively. Similar to this, with an amplitude of ${\cal O}(10^{-9})$ in the SRI and ${\cal O}(10^{-3})$ in the SRII phases, the auto-correlation $\Delta^{2}_{\zeta\zeta}$, see \ref{pspeczz1}, is the most prominent in both the SRI and SRII phases. It is more crucial to pay attention to how all of the auto and cross-correlations behave during the USR period. The USR exhibits significant increases in correlation amplitudes up to order ${\cal O}(10^{-1}-1)$. Notably, the correlation amplitude $\Delta^{2}_{\zeta\zeta}$ is comparatively less than the other two components. If we want to investigate the impact of such substantial increases being carried into the quantum correction computations, this USR finding has significant ramifications.   
Rapid oscillations occur in the amplitude shortly after a sharp exit from the USR at $k=k_{e}$, and they continue to be larger in the cross-correlations, $\Delta^{2}_{\zeta\Pi_{\zeta}}$ and $\Delta^{2}_{\Pi_{\zeta}\zeta}$, than in the auto-correlations, $\Delta^{2}_{\zeta\zeta}$ and $\Delta^{2}_{\Pi_{\zeta}\Pi_{\zeta}}$. These oscillations eventually die out to give a constant magnitude.
We highlight that the new set of Bogoliubov coefficients, $(\alpha_{2},\beta_{2})$ and $(\alpha_{3},\beta_{3})$ in eqn. (\ref{alpha2qds},\ref{beta2qds},\ref{alpha3qds},\ref{beta3qds}), which contains complex phase factors after solving for the boundary conditions at each sharp transition $k=k_{s}$ and $k=k_{e}$, is the reason for the oscillatory nature of the power spectrum in the USR and later into the SRII. 

In fig. (\ref{spectrumplots}), the function of the effective sound speed, $c_{s}$, is also presented. For two choices of $c_{s}=0.024,1$, we display each correlation together with the upper and lower bounds derived from observational requirements for causality and unitarity \cite{Planck:2018jri}. As discussed in the comments on fig. \ref{sounds}, the red curve represents the standard stochastic single-field inflation situation, where $c_{s}=c_{s,*}=1$. In this instance, $c_{s,*}$ denotes its value set at the pivot scale. The upper limit, denoted by $c_{s,*}=1$, is the value at which the causality and unitarity restrictions are broken. The amplitude of the power spectrum reaches $\Delta^{2}_{\zeta\zeta}\gtrsim {\cal O}(10^{-2})$, which is more than enough to generate PBH. As $c_{s,*}$ is reduced to its minimum, the blue curve appears at $c_{s,*}=0.024$. The cyan-shaded area between the two curves symbolizes the potential range of values originating from distinct $c_{s}$ that may exist for other non-canonical stochastic single-field models. An significant inference from this is that the perturbativity assumptions are maintained by lowering the amplitude of both the auto-correlations and cross-correlations in the USR phase, which maintains the causality and unitarity restrictions. The value of the stochastic parameter, $\sigma=0.01$, remains constant. From the standpoint of PBH generation, the lowest amplitude in the USR may still be found within $\Delta^{2}_{\zeta\zeta}\sim {\cal O}(10^{-3})$. Furthermore, we point out that the largest influence from changes in $c_{s}$ is only observed in the USR; in contrast, the changes in $c_{s}$ are negligible in the two SRI and SRII phases, and the least significant in SRI. 

We now highlight the occurrence of first-order phase transitions and out-of-equilibrium characteristics in the current cosmological framework of stochastic inflation. In the current setting, primordial curvature perturbations and their corresponding momenta act as free energies and must meet specific boundary requirements. In the first-order phase transitions, the free energy displays a discontinuity in its first derivative. The continuity and differentiability constraints imposed at the critical points of every sharp transition are known as the Israel junction conditions, and they play a function akin to that of phase transitions in statistical physics. Consequently, distinct sets of Bogoliubov coefficients from each successive phase transition that begins from the SRI phase are included into the mode solutions and their conjugate momenta. A stochastic coarse-graining parameter $\sigma$, which promotes the combining of the several phases near the abrupt transitions, is one of the stochastic effects that the framework of stochastic inflation encourages to be present at the Horizon crossing moment.
Stochasticity causes the potential auto- and cross-correlations between the primordial fluctuations to become significant at the encounter of each sharp transition, which individually introduces out-of-equilibrium effects for the various phases near their Horizon crossing. The ultimate result is the disappearance of a specific example of the phase transition from quantum to classical. To maintain perturbativity in the underlying theoretical framework, the stochastic parameter $\sigma$ must always meet a magnitude of order $\sigma \ll 1$. When studying quantum loop corrections, $\sigma$ plays an increasingly important role as a coarse-graining factor or regulator. They can also be seen in the structure of logarithmic IR divergences, where they can emerge.
In the past, IR-related problems in dS space have drawn attention to the stochastic inflation theory \cite{Podolsky:2008qq,Finelli:2008zg,Seery:2010kh,Garbrecht:2014dca,Burgess:2015ajz,Gorbenko:2019rza,Baumgart:2019clc,Mirbabayi:2019qtx,Cohen:2020php}. When no stochastic features are present, the auto-correlation $\Delta^{2}_{\zeta\zeta}$, which is observationally significant, dominates the contribution. The other auto-correlation, $\Delta^{2}_{\Pi_{\zeta}\Pi_{\zeta}}$, and the cross-correlations, $\Delta^{2}_{\Pi_{\zeta}\zeta}$ and $\Delta^{2}_{\zeta\Pi_{\zeta}}$, appear to be highly suppressed.
\begin{figure*}[ht!]
    	\centering
    \subfigure[]{
      	\includegraphics[width=8.5cm,height=7.5cm]{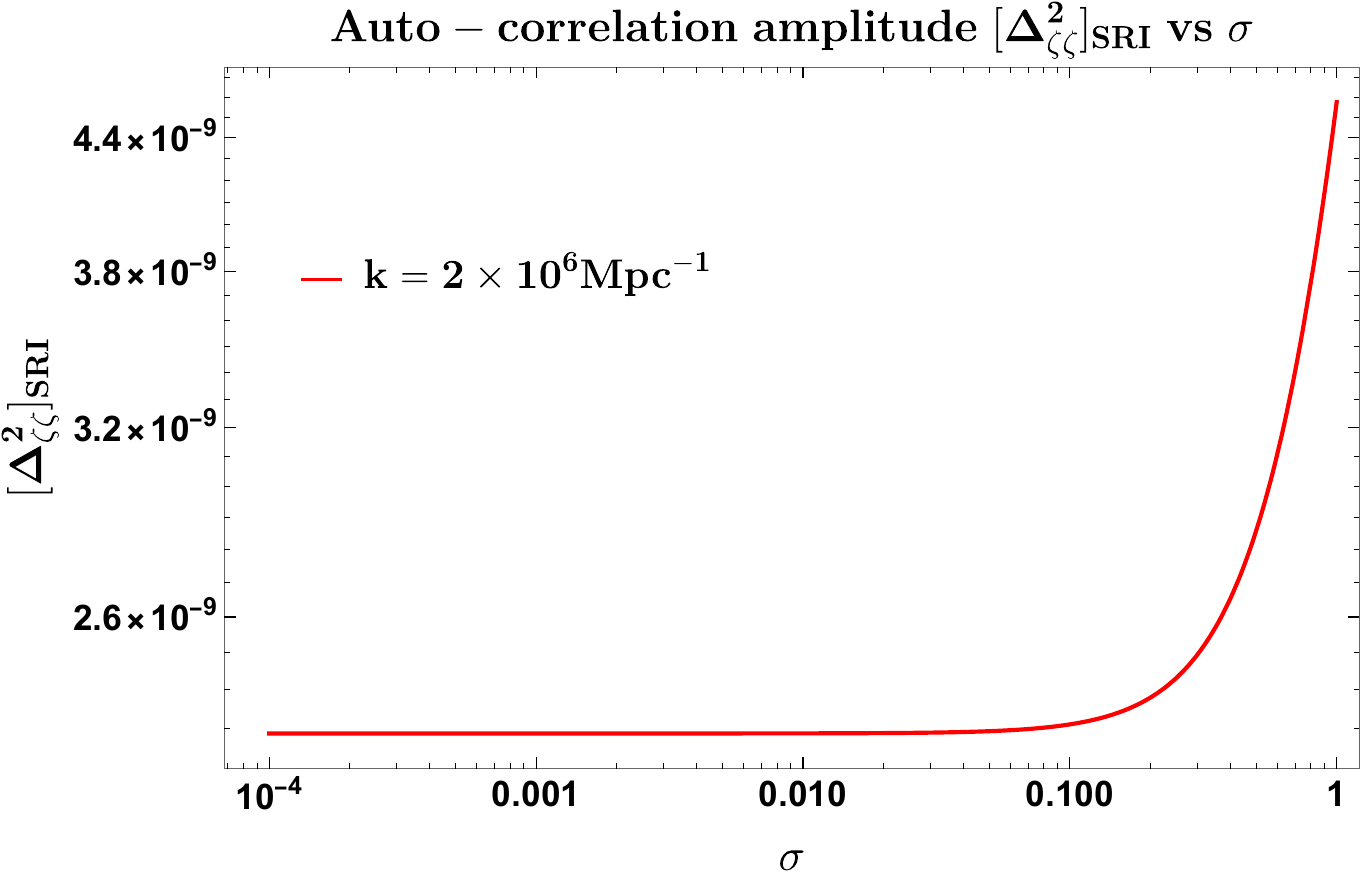}
        \label{sr1pspecsigma}
    }
    \subfigure[]{
        \includegraphics[width=8.5cm,height=7.5cm]{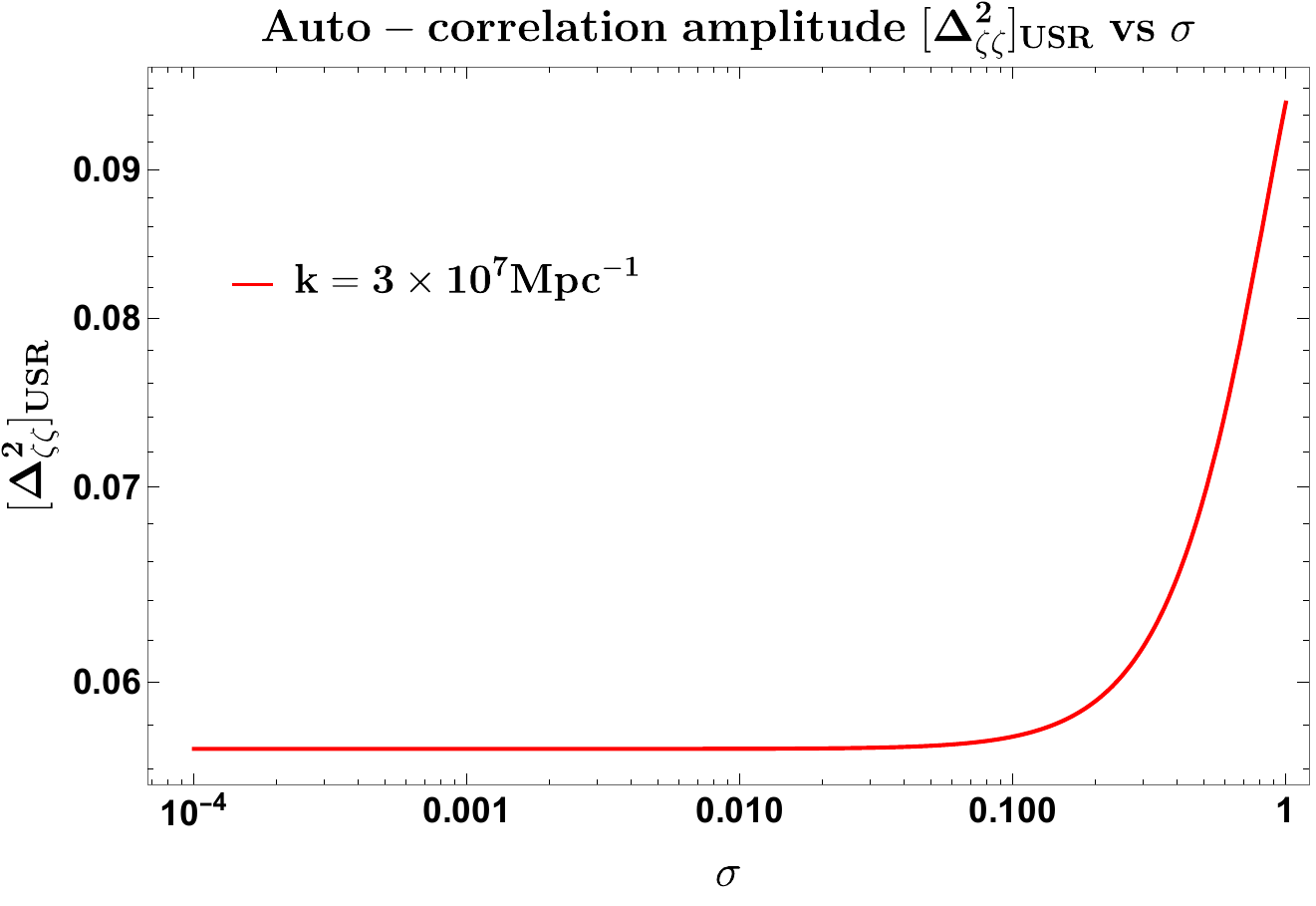}
        \label{usrpspecsigma}
    }
       \subfigure[]{
        \includegraphics[width=8.5cm,height=7.5cm]{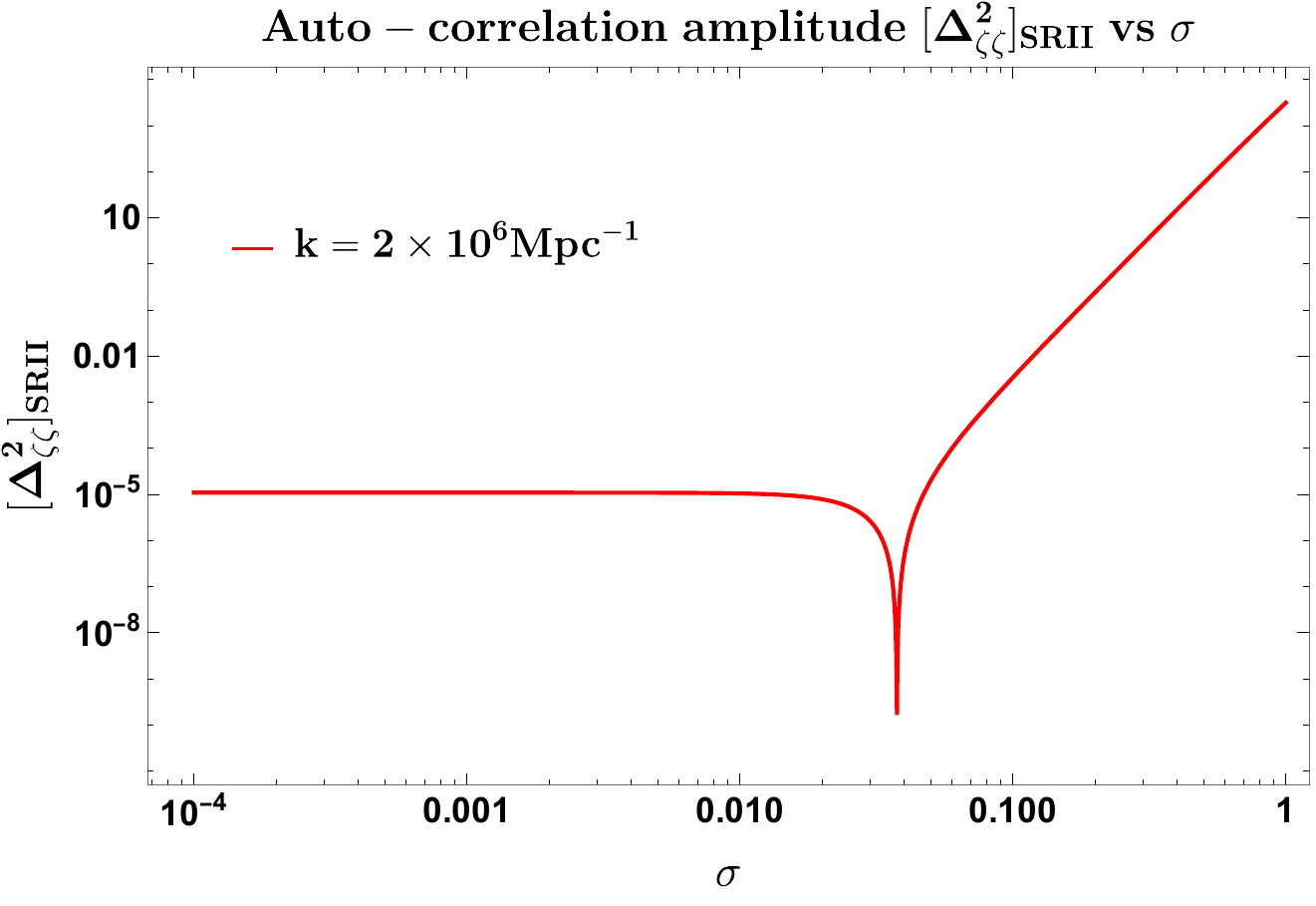}
        \label{sr2pspecsigma}
    }
    	\caption[Optional caption for list of figures]{The behavior of the auto-correlation amplitude $\Delta^{2}_{\zeta\zeta}(k)$ in relation to the change in the stochastic parameter $\sigma$ for each phase. For both the SRI and USR phases, the \textit{top row} is displayed, whereas the \textit{bottom} is for the SRII phase. Amplitudes of the correlation in each phase stay fulfilled for $\sigma\in (10^{-4},0.1)$, and vary fast when $\sigma>0.1$ is taken into account.
  } 
    	\label{spectrumplotsigma}
    \end{figure*}

Figure \ref{spectrumplotsigma} illustrates the behavior of the $\Delta^{2}_{\zeta\zeta}$ auto-correlation for each phase in our setup, with respect to the stochastic parameter $\sigma$. Point-by-point explanation of the meaning of $\sigma$ in each step is appended below:
\begin{itemize}
    \item \underline{\textbf{In SRI:}} The amplitude of the power spectrum in the SRI for a given wavenumber mode, $k<{\cal O}(10^{7}{\rm Mpc^{-1}})$, is found to remain within its magnitude at the pivot scale, $[\Delta^{2}_{\zeta\zeta}(k)]_{\rm SRI}\sim 2.2\times 10^{-9}$, when we take into account the parameter interval, $\sigma\in (10^{-4},0.1)$. As the contribution from the existence of $\sigma$ becomes more significant, the power spectrum amplitude increases beyond the values $\sigma>0.1$. We must proceed with caution, though, since it is evident from eqn. (\ref{pspecsr1dS}) that the pivot scale value given here has the incorrect amplitude and that there is no smooth crossover between $\sigma\ll 1$ and $\sigma\sim 1$.

    \item \underline{\textbf{In USR:}} Since it allows us to produce huge mass PBH, $M_{\rm PBH}\sim {\cal O}(10^{-2}-1\;M_{\odot})$, we maintain a fixed wavenumber $k$ near the transition scale of $k_{s}\sim {\cal O}(10^{7}){\rm Mpc^{-1}}$. Once this is met, the USR's power spectrum amplitude $[\Delta^{2}_{\zeta\zeta}(k)]_{\rm USR}\sim{\cal O}(10^{-2})$ is large enough to produce PBHs. We see that when $\sigma\in (10^{-4},0.1)$, the amplitude nearly always stays at the same value until $\sigma>0.1$, at which point it increases to a magnitude of $\sim{\cal O}(10^{-1})$. This range of the stochastic parameter $\sigma$ keeps the amplitude from exceeding $\sim {\cal O}(1)$, the perturbativity requirement, and also doesn't significantly increase the amplitude, which might easily result in PBH overproduction.

    \item \underline{\textbf{In SRII:}} Like the first two phases, the corresponding power spectrum amplitude in the SRII for a particular mode remains at $[\Delta^{2}_{\zeta\zeta}(k)]_{\rm SRII}\sim{\cal O}(10^{-5})$ until the interval of interest stays at $\sigma\in (10^{-4},1)$. We reiterate that, under the identical conditions, a limiting situation of $\sigma\sim 1$ cannot be physically understood with the behavior for lower values of $\sigma$. Nevertheless, the total amplitude grows at a higher rate subsequently for $\sigma>0.1$. When $\sigma>0.1$ is taken into account, the amplitude rapidly violates perturbativity. We conclude that the power spectrum can still effectively monitor the effects of coarse-graining at $\sigma<0.1$.  

    \end{itemize}
Drawing on the previous conversations, we discover that $\sigma> 0.1$ does not yield significant results when compared to estimates of theoretical and observable significance, such as power spectrum amplitude at pivot scale, required enhancement for PBH production without producing too many of them, and not violating perturbativity constraints. However, any modifications resulting from $\sigma\ll 1$ remain minuscule. Therefore, we may conclude that any value of $\sigma<0.1$ will do for our purposes, and we are unable to explain this situation in conjunction with the case of $\sigma>0.1$.

As previously seen in eqn. (\ref{noisepower}), a critical distinction may be drawn with direct reference to the relationship between the noise matrix components and the power spectrum elements. We obtain the noise matrix auto-correlations and cross-correlations corresponding to the relevant power spectrum correlations in previous sections as a consequence of the same method, after multiplying by a factor of $(1-\epsilon)$. Since the slow-roll parameter $\epsilon$ is still much less than unity in the SRI and SRII but vanishingly tiny in the USR, it can be concluded that the plots in fig. (\ref{spectrumplots}) also accurately depict the behavior of related noise auto and cross-correlations. The \textit{quantum kicks} originating from the UV modes after they had classicalized in the distant super-Horizon (or IR) regions are represented by these noise correlations. The conjugate momentum fluctuations correlations, $\Sigma_{\Pi_{\zeta}\zeta}$, $\Sigma_{\zeta\Pi_{\zeta}}$, and $\Sigma_{\Pi_{\zeta}\Pi_{\zeta}}$, produce sub-dominant noise amplitudes in the SRI and SRII regime, whereas the $\Sigma_{\zeta\zeta}$ correlation accounts for the majority of the noise contributions. However, all the noise correlations stated are as prominent in the USR, suggesting large quantum shocks influencing the IR dynamics.

Quantum loop effects were previously examined by writers in \cite{Choudhury:2023vuj,Choudhury:2023jlt,Choudhury:2023rks,Choudhury:2023hvf}, without incorporating any stochastic elements. A stochastic parameter $\sigma$ is present in the theory and acts as a regulator or a coarse-graining component. Correct one-loop calculations could lead to a final result where the underlying logarithmic IR divergences are smoothed out more effectively than in calculations done without such a stochastic regulator, provided regularization and renormalization techniques are used. In the future, we want to work closely toward achieving this objective. In the context of single-field inflation, when a single-sharp transition takes place, PBH creation is constrained. Using the Dynamical Renormalization Group (DRG) analysis \cite{Chen:2016nrs,Baumann:2019ghk,Boyanovsky:1998aa,Boyanovsky:2001ty,Boyanovsky:2003ui,Burgess:2015ajz,Burgess:2014eoa,Burgess:2009bs,Dias:2012qy,Chaykov:2022zro,Chaykov:2022pwd}, $\Delta{\cal N}\sim {\cal O}(20–25)$ (unsuccessful inflation) and strong constraints on the total number of e-folds arising from the necessary renormalization procedures and a resummation of logarithmic IR divergences at all orders in the loop calculations prohibit the formation of any $M_{\rm PBH}\sim {\cal O}(M_{\odot})$. It is demonstrated to avoid this strong restriction and eventually make possible the generation of large solar mass PBHs, unlike some previous attempts, such as the presence of multiple sharp transitions (MSTs) \cite{Bhattacharya:2023ysp,Choudhury:2023fjs} and within the framework of single-field Galileon theory with a single sharp transition \cite{Choudhury:2023hvf,Choudhury:2023kdb,Choudhury:2023hfm,Choudhury:2024one}, where the non-renormalization theorem plays the most critical role in tackling the PBH mass constraints. The current stochastic inflation formalism is far superior because it produces $M_{\rm PBH}\sim {\cal O}(M_{\odot})$ with only one transition and no additional symmetry effects. This means that the stochastic regulator $\sigma$ has the capacity to circumvent the restrictions on PBH mass that have been discussed.

\subsubsection{Outcomes of spectral distortions}

\begin{figure*}[htb!]
    	\centering
    \subfigure[]{
      	\includegraphics[width=8.5cm,height=7.5cm]{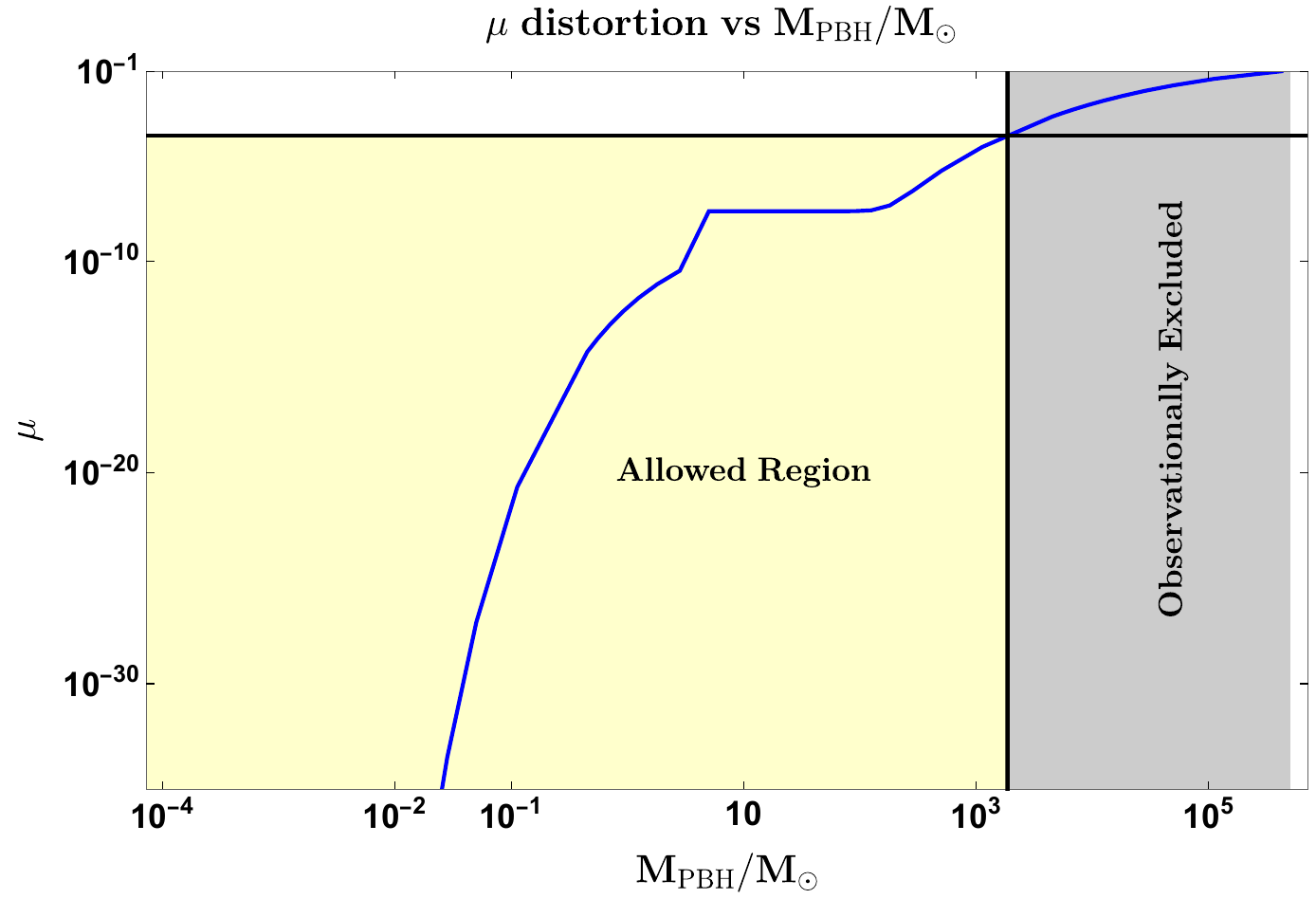}
        \label{mudistort}
    }
    \subfigure[]{
        \includegraphics[width=8.5cm,height=7.5cm]{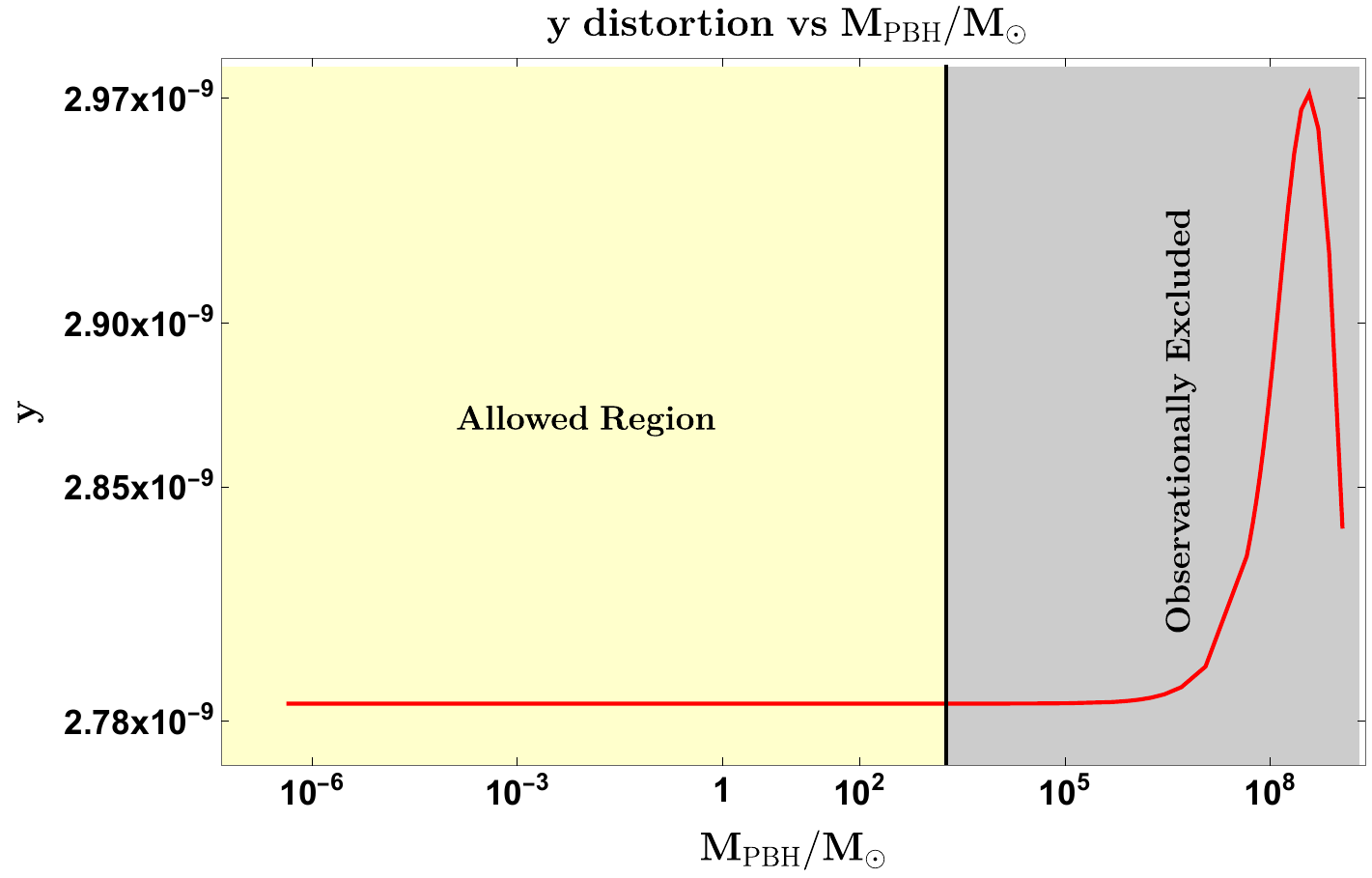}
        \label{ydistort}
       }
    	\caption[Optional caption for list of figures]{The $\mu$- and y-distortions (\textit{left-panel} and \textit{right-panel}) in Figure are shown as functions of the PBH mass $M_{\rm PBH}$ (in $M_{\odot}$). The permitted values of the distortion effects following limitations from the COBE/FIRAS data are shown by the yellow shaded zone. The values that the same observations prohibit are shown by the gray-shaded areas.
 } 
    	\label{distortion}
    \end{figure*}

\begin{figure*}[htb!]
    	\centering
    {
       \includegraphics[width=17cm,height=10cm]{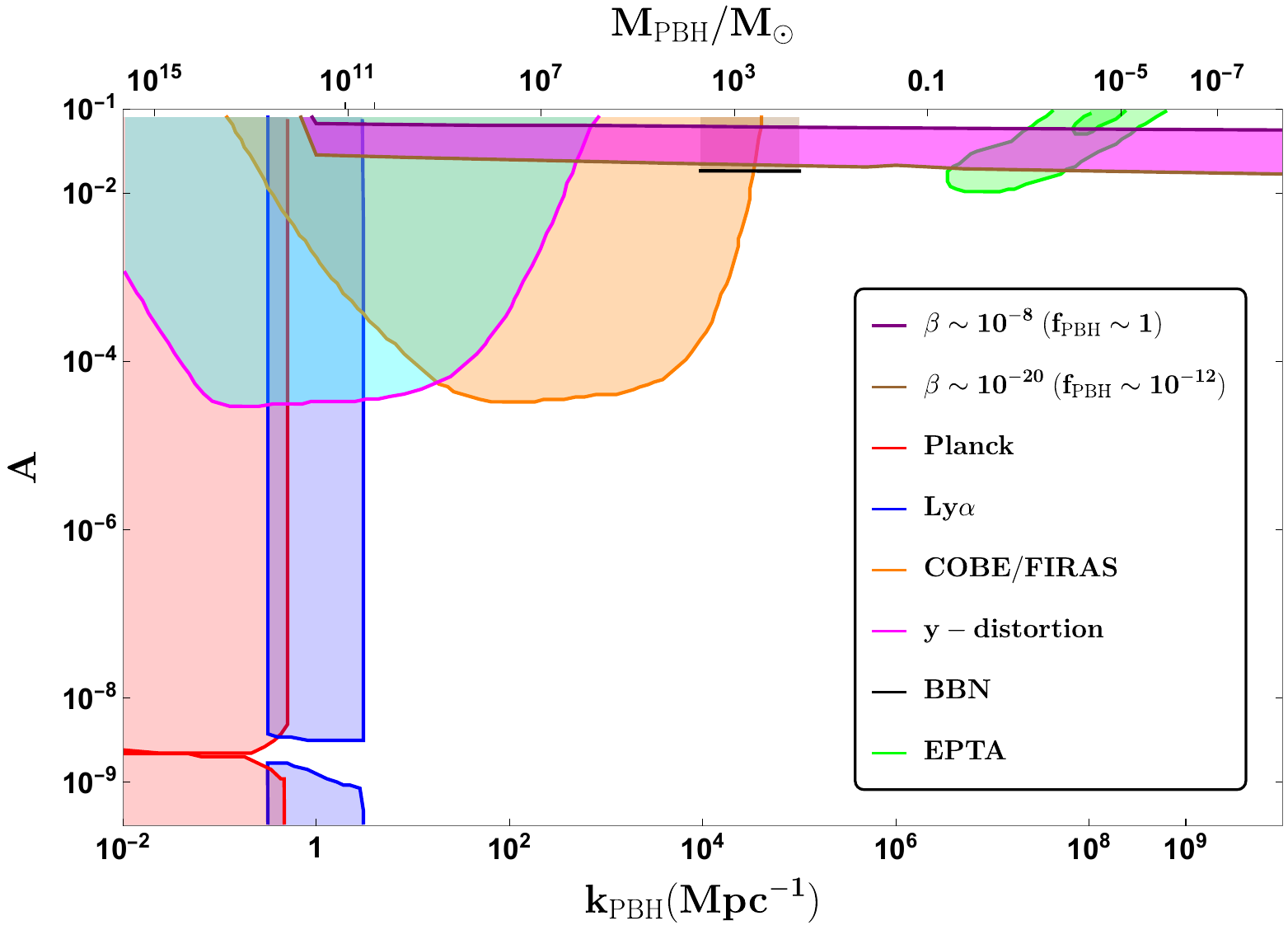}
        \label{ampdistortion}
    } 
    \caption[Optional caption for list of figures]{Functioning from the PBH forming wavenumber $k_{s}$, the amplitude of the scalar power spectrum needed to attain a non-negligible PBH mass fraction is shown. Constrained amplitudes from the large-scale CMB temperature anisotropies (red) \cite{Planck:2018jri}, Lyman-$\alpha$ forest (blue) \cite{bird2011minimally}, COBE/FIRAS (orange) and the y-distortion (cyan) effects \cite{Cyr:2023pgw}, BBN (black) \cite{Jeong:2014gna}, and the $1$ and $2\sigma$ contours reported by the pulsar timing array collaboration (green) \cite{EPTA:2023xxk} are all part of the background. The area of amplitude where the initial PBH mass fraction from the Press-Schechter formalism lies in the window $\beta\sim {\cal O}(10^{-20}-10^{-8})$ is highlighted by the magenta-colored band, which is delimited by the brown and purple lines.
 }
\label{amplitudedistortion}
    \end{figure*}
In this part, the numerical results for the estimations of the two $y$ and $\mu$ distortion effects are discussed. We adhere to the formulas found in eqns. ($k_{\rm PBH}$ corresponding to the wavenumber, $M_{\rm PBH}$), and use (\ref{mudistortion}, \ref{ydistortion}) to get different estimates as a function of the PBH mass. 

The $\mu$ distortion estimations as a function of $M_{\rm PBH}$ are shown in fig. \ref{mudistort}. The strongest restrictions that are currently available come from the COBE/FIRAS data, which means that the $\mu$ distortion needs to fulfill $|\mu|\simeq 9\times 10^{-5}$. Due to the banned $\mu$ values, the excluded mass range is highlighted in the graphic by the gray colored area. For masses smaller than $M_{\rm PBH}\lesssim {\cal O}(M_{}\odot)$, we observe that the $\mu$ values decline considerably more quickly, causing such masses to fall inside the permitted zone. The black vertical line corresponds to the corresponding mass value, while the horizontal line denotes the previously indicated upper bound on $\mu$. This indicates that PBHs with $M_{\rm PBH}\gtrsim 1.8\times 10^{3}\;M_{\odot}$ are correspondingly removed from the $\mu$ constraints.
The behavior of the $y$ distortion values as a function of the PBH masses is shown in fig. \ref{ydistort}. After $M_{\rm PBH}\lesssim {\cal O}(10^{7}M_{\odot})$, the $y$-distortion does not seem as a substantial impact, and its behavior might alter in an unanticipated way. From the COBE/FIRAS data, the distortion is constrained to $|y|\simeq 1.5\times 10^{-5}$.

The scalar power spectrum amplitude must be in the range of $A\sim {\cal O}(10^{-2})$ in order to produce PBH production. We employ the identical equations, found in \ref{mudistortion} and \ref{ydistortion}, to ascertain the amplitude required to maintain the current limits on distortion effects. Simultaneously, the initial PBH mass fraction that results should also not be insignificant. For the same reason, we select $\beta$ to reside within the window $\sim {\cal O}(10^{-20}-10^{-8})$. 

The findings for the power spectrum amplitude $A$ as a function of the PBH producing wavenumber $k_{s}$ are shown in figure (\ref{amplitudedistortion}). The initial mass fraction area, $\beta\sim {\cal O}(10^{-20}-10^{-8})$, is highlighted by the magenta colored band. The interval is bordered by the purple and brown lines, respectively. Several contours of observational restriction on the amplitude $A$ are present in the background. At bigger scales, the primordial power spectrum amplitude is already severely constrained by CMB anisotropy. It can be observed that the COBE/FIRAS (in orange) and the y-distortion constraint (in magenta) exhibit constraints related to distortion effects that lead to the disfavoring of PBHs with masses $M_{\rm PBH}\gtrsim 1.8\times 10^{3}M_{\odot}$ when utilizing the amplitude of $A\sim {\cal O}(10^{-2})$. The notion of restricting the amplitude of the primordial power spectrum by observing CMB spectral aberrations on bigger scales was firstly established in \cite{Chluba:2012gq,Chluba:2012we}. In light of the recently observed stochastic gravitational wave background signal, \cite{Cyr:2023pgw} presents a detailed analysis on improving these distortion effects constraints and their connections. The updated constraints from $\mu-$ and $y-$ type distortions on the power spectrum amplitude are displayed in the orange and cyan curves. The amplitude is additionally constrained by the Big Bang Nucleosynthesis (BBN), which is shown by orange limitations at the right end of the distortion contour. The energy dissipation of the accompanying waves ultimately results in restrictions on the primordial power spectrum from overproduction of primordial deuterium and helium estimations at scales after BBN but before spectral-distortion era, $k\simeq 10^{4}-10^{5}\;{\rm Mpc^{-1}}$. To attain significant initial abundance, PBH with bigger mass or smaller wavenumbers need larger amplitudes, $A\sim {\cal O}(10^{-2}-10^{-1})$. The amplitude only tends to rise even faster for $M_{\rm PBH}\gtrsim {\cal O}(10^{11}M_{\odot})$. Exceeding the upper purple line results in an excess of PBHs. It is important to note that when analyzing the distortion estimates using the total power spectrum from equations (\ref{pspecsr1dS}, \ref{pspecusrdS}, \ref{pspecsr2dS}), the computations assume the use of an initial Gaussian profile for the curvature perturbation $\zeta$.

\subsubsection{Outcomes of PDF}
\begin{figure*}[htb!]
    	\centering
    \subfigure[]{
      	\includegraphics[width=8.5cm,height=7.5cm]{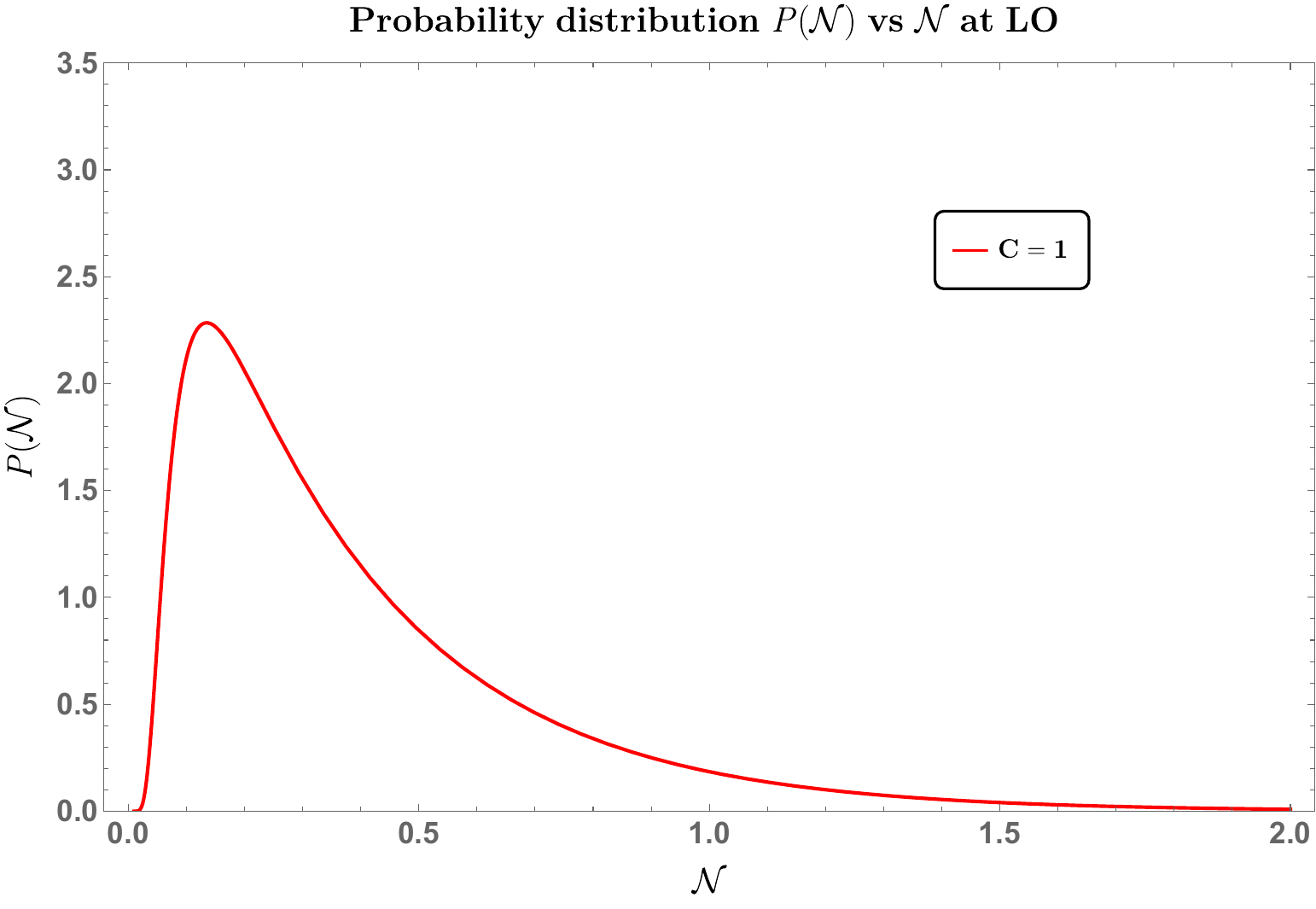}
        \label{N0PDF}
    }
    \subfigure[]{
        \includegraphics[width=8.5cm,height=7.5cm]{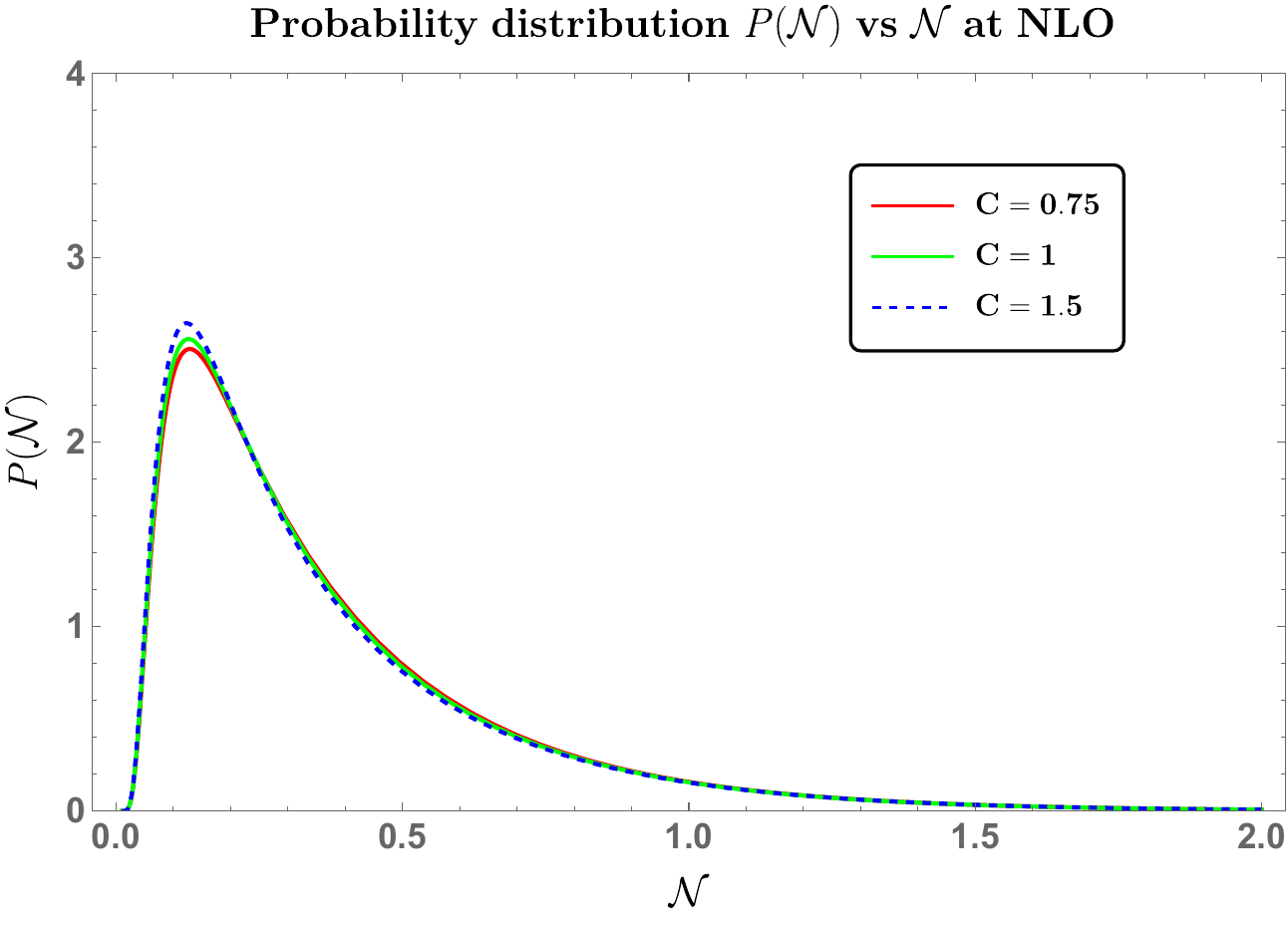}
        \label{N1PDF}
       }
    \subfigure[]{
      	\includegraphics[width=8.5cm,height=7.5cm]{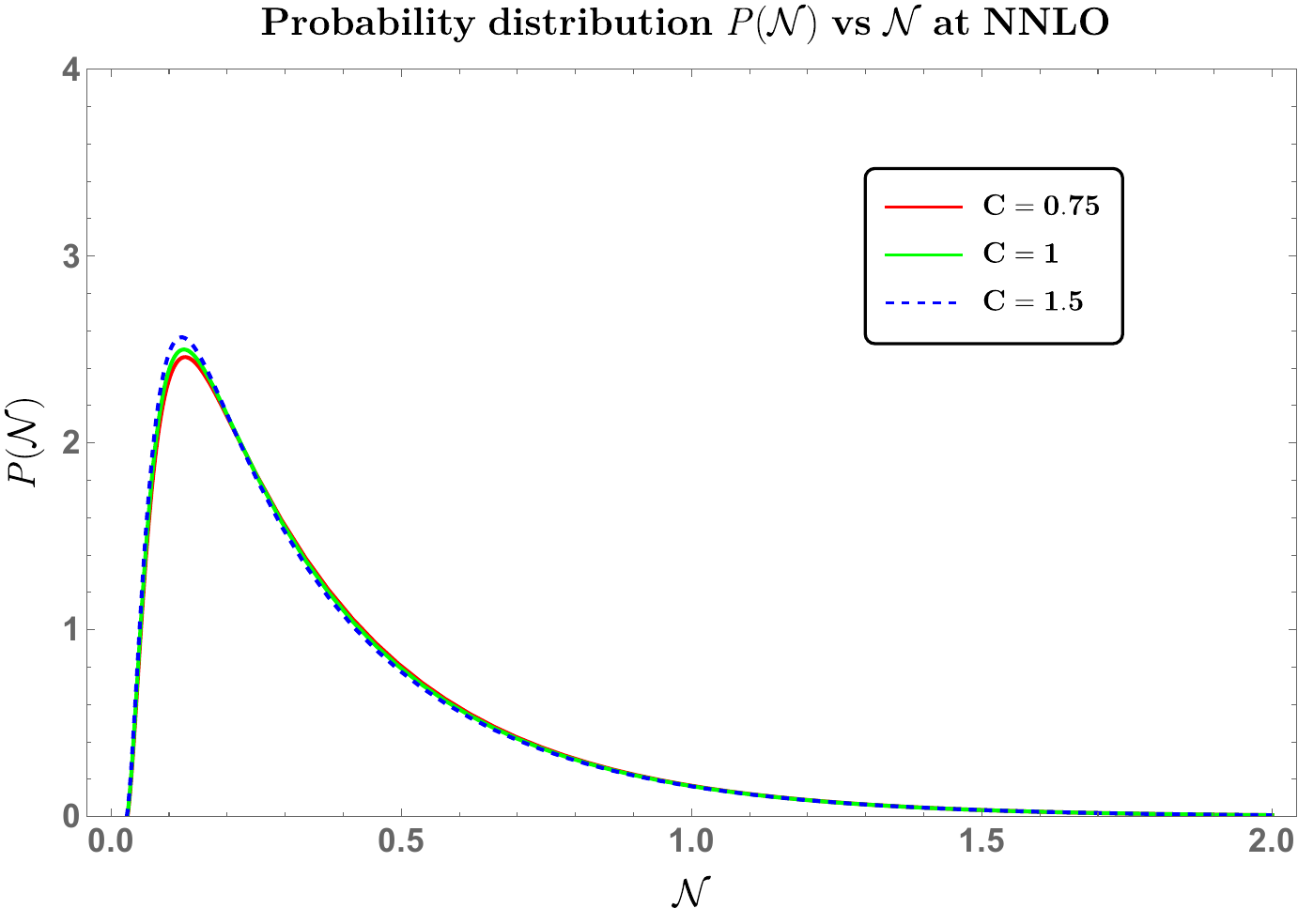}
        \label{N2PDF}
    }
    \subfigure[]{
        \includegraphics[width=8.5cm,height=7.5cm]{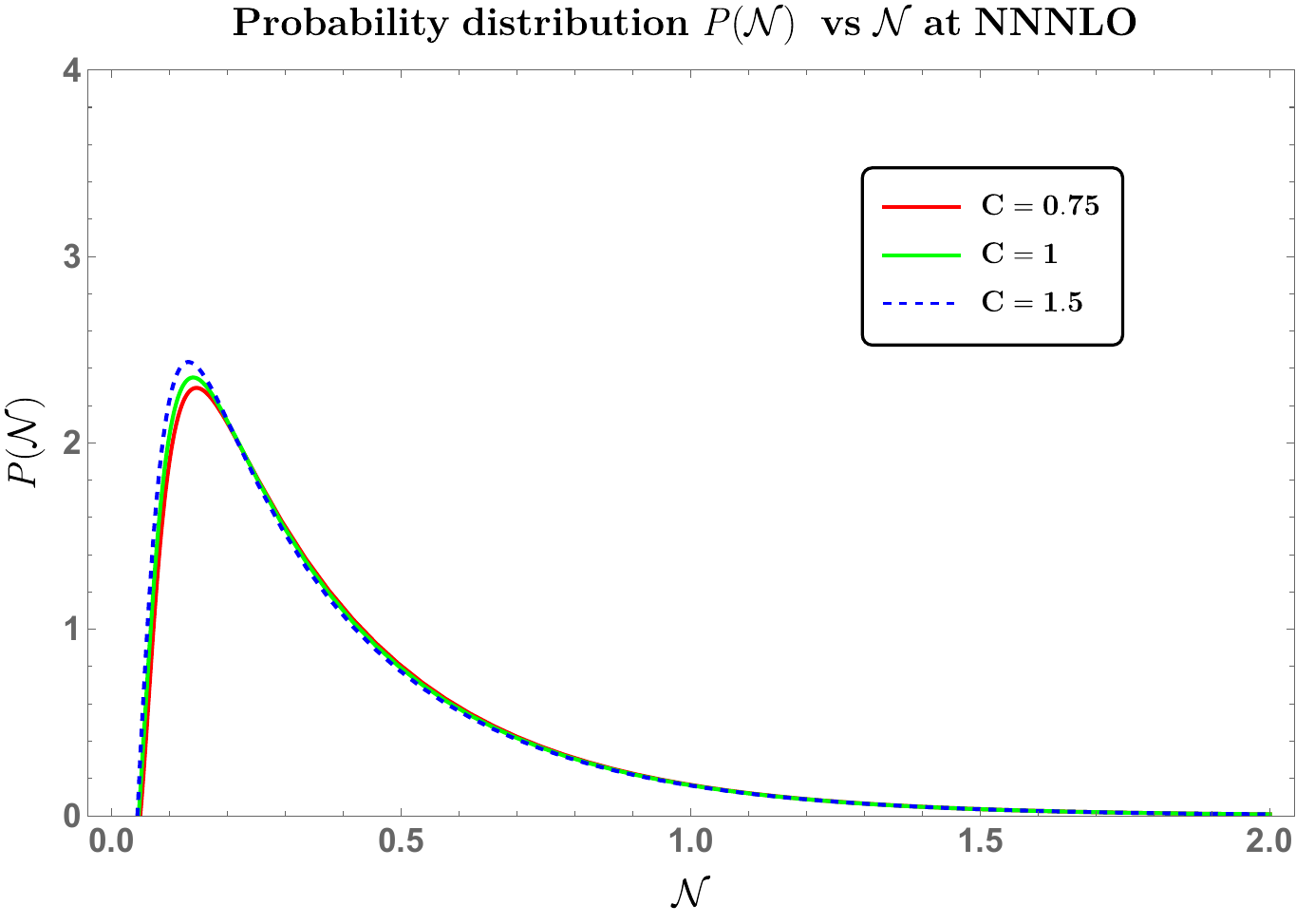}
        \label{N3PDF}
       }
    	\caption[Optional caption for list of figures]{The Probability Distribution Function's behavior as the number of e-folds ${\cal N}$ changes. Different $y$ values are displayed with PDF at the leading order (LO) in the \textit{top-left} panel. The PDF is displayed for various values of the characteristic parameter $C=\{0.75,1,1.5\}$ in red, green, and blue, respectively, at the next-to-leading order (NLO) in the \textit{top-right} panel, the next-to-next-to-leading order (NLO) in the \textit{bottom-left} panel, and the next-to-next-to-next-to-leading order (NLO) in the \textit{bottom-right} panel. 
 } 
    	\label{diffPDF}
    \end{figure*}

\begin{figure*}[htb!]
    	\centering
    \subfigure[]{
      	\includegraphics[width=8.5cm,height=7.5cm]{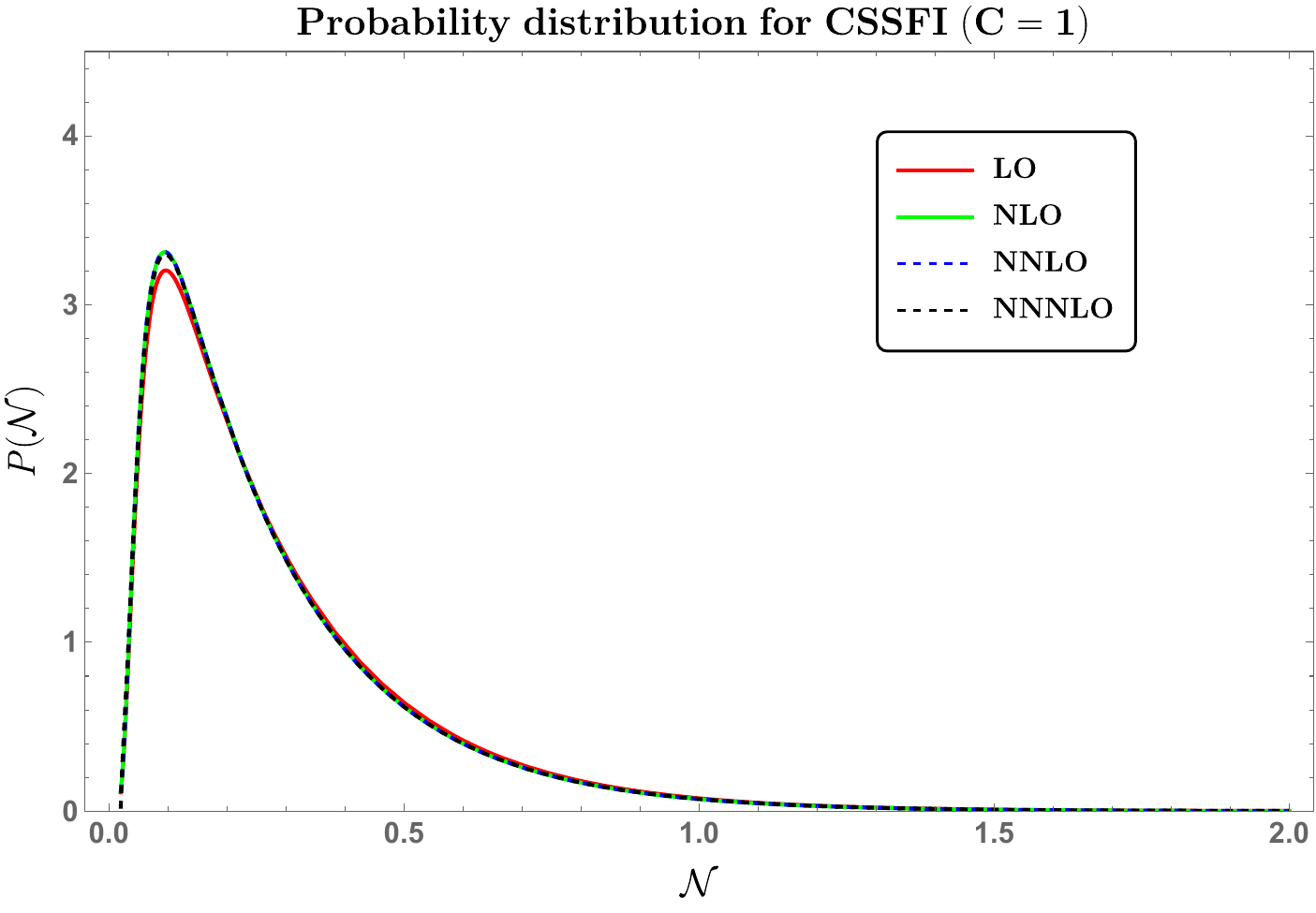}
        \label{cssfi}
    }
    \subfigure[]{
        \includegraphics[width=8.5cm,height=7.5cm]{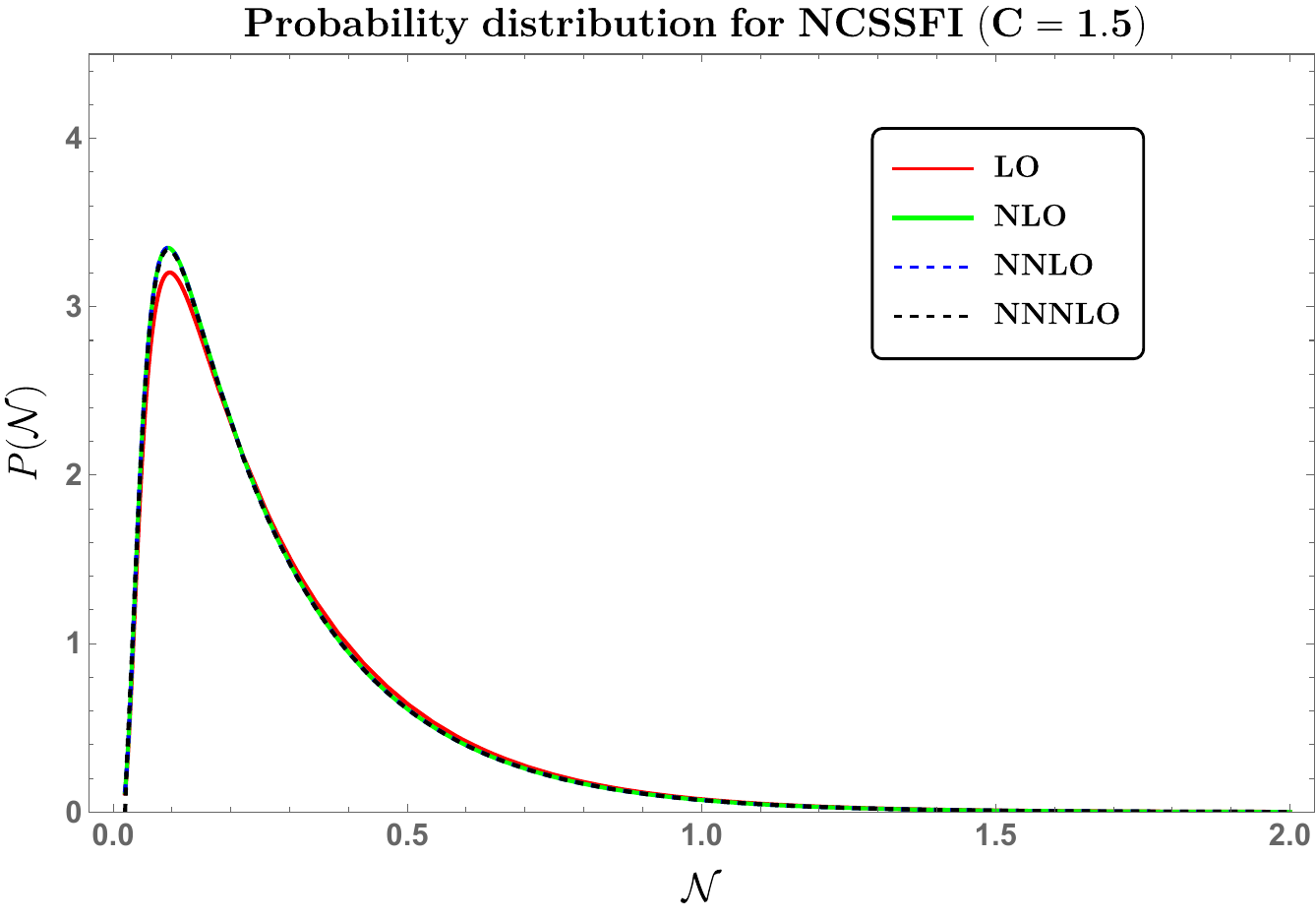}
        \label{ncssfi}
       }
    	\caption[Optional caption for list of figures]{The PDF is shown against the stochastic variable ${\cal N}$, encompassing all perturbative expansion orders up to NNNLO. Similar to the \textit{right-panel}, which examines the case of non-canonical single-field inflation $(C=1.5)$, the \textit{left-panel} examines the case of canonical stochastic single-field inflation $(C=1)$ superposed with all the different PDF from the perturbative expansion.
 } 
    	\label{allSSFI}
    \end{figure*}
In the diffusion-dominated regime, we examine the results of our PDF computations up to NNNLO order in this section. Although the PDF expressions at the NNLO and NNNLO have not been examined in detail, we offer their graphical representations and go over their properties here. As previously described in eqn. (\ref{newphasevars}), the initial phase space variables $(x,y)$ are associated with the coarse-grained curvature perturbation and its corresponding momentum variable.

The analysis at leading order (LO) is where we start. The characteristic parameter $C$ does not appear to be involved in the residues or the poles in the previous equation (\ref{pdfLO}). It can be seen from eqn. (\ref{poles}) that higher-order poles must exist, at least when $m=1$, in order to explicitly witness $C$. The distribution from eqn. (\ref{pdfLO}) is plotted against the stochastic e-folding number, ${\cal N}$, with a given $C=1$, as shown in fig. \ref{N0PDF}. There is no noticeable change in the features when $y$ is changed because this is at the leading order in the $y$ expansion. Here, the remaining parameters are set to $\tilde{\mu}= 0.9$ and $x=1$. As we move to the next order in the expansion, we see that, as shown in fig. (\ref{N1PDF}), we now have two sets of poles that add up to the entire PDF at NLO. You may also obtain the analytical expression from eqn. (\ref{pdfNLO1}). Observe how the curves corresponding to various non-canonical single-field inflation situations differ from one another, with $C$ being a finite and variable quantity. The EFT description is not unwanted because of the relative difference, which is not so great. The red and dotted blue curves indicate that we have decided to maintain $C$ near $1$ in this instance. This time, there is a discrepancy to the first decimal place between the amplitude and the LO case, which has increased. In this case, $y=0.15$ is fixed, but $x=1$ and $\tilde{\mu}=0.9$ are not. The leading terms in the series expansion of eqn. (\ref{pdfLO1}) roughly represent the majority of the PDF in this regime of $y$ values, whereas the remainder decay extremely fast. There is the PDF at the NNLO farther on in the bottom-left figure \ref{N2PDF}. Although the analytic formulation for this PDF is not previously stated, the distribution shows that, when different values of $C$ are taken into account, its behavior stays mostly unchanged from the prior PDFs, with only minor relative changes at the maximum. The difference between the peak amplitude and the PDF at NLO is now pushed to the second decimal point. The set of parameters, $x=1$, $y=0.15$, and $\tilde{\mu}=1$, remains constant. Finally, we investigate the distribution at NNNLO in the bottom-right panel of fig. (\ref{N2PDF}). For different $C$ values within the permitted range for a valid EFT formulation, the peak amplitude clearly varies, and the amplitude change with respect to NNLO now happens at the first decimal place, lowering the estimate. This indicates that, after the remaining parameters, $x=1$, $y=0.15$, and $\tilde{\mu}=1$, have been determined, the perturbativity argument should be preserved throughout the analysis.

We show that the distribution functions at successive orders sufficiently maintain the perturbativity criteria in the diffusion-dominated domain. The $C$ value of interest only appears close to the peak amplitude areas in the observed variations for different EFT descriptions, and the extent of the variances does not significantly alter the PDF's overall characteristics. The left and right tails exhibit comparable patterns at all orders displayed in fig.(\ref{diffPDF}), with only minor variations apparent in the vicinity of the peak amplitude. 

Keeping $y=0.073$, $x=1$, and $\tilde{\mu}=0.76$ unchanged, the figure \ref{allSSFI} shows the behavior of all the superimposed distribution functions, which captures various higher-order corrections in the diffusion-dominated regime in the small $y$ limit. We will subsequently provide a justification for the parameter values chosen, which are linked to the expected PBH abundance characteristics in the future sections. We see practically minor changes in the PDF values after superimposing the PDFs. This time, relative differences in amplitude are pushed farther to the second and third decimal places, indicating that the expansion variable $y$ is likewise kept considerably lower than the analysis in fig. (\ref{diffPDF}). Figure \ref{cssfi} depicts the canonical stochastic single-field inflation scenario with $C=1$, whereas fig. \ref{ncssfi} shows the non-canonical stochastic single-field inflation situation with $C=1.5$. In addition to the tail characteristics being comparable across orders, the peak amplitudes stay near to each other, maintaining perturbativity. With the exception of minor variations in the peak amplitude at the second decimal places, the distributions for the non-canonical scenario similarly stay mostly unchanged. Furthermore, there is a little behavioral difference in the fall around ${\cal N}\sim 0.5$ between the LO and all subsequent orders taken together. The poles $\Lambda_{n}^{(0)}$ rapidly provide the comparable dominating characteristics in the PDF at all orders in the fig. \ref{allSSFI}, and the entire combination of dominant contribution from the leading reside $r^{(0)}_{n,i}$, where $i$ indicates the order of expansion. 
\begin{figure*}[htb!]
    	\centering
    \subfigure[]{
      	\includegraphics[width=8.5cm,height=7.5cm]{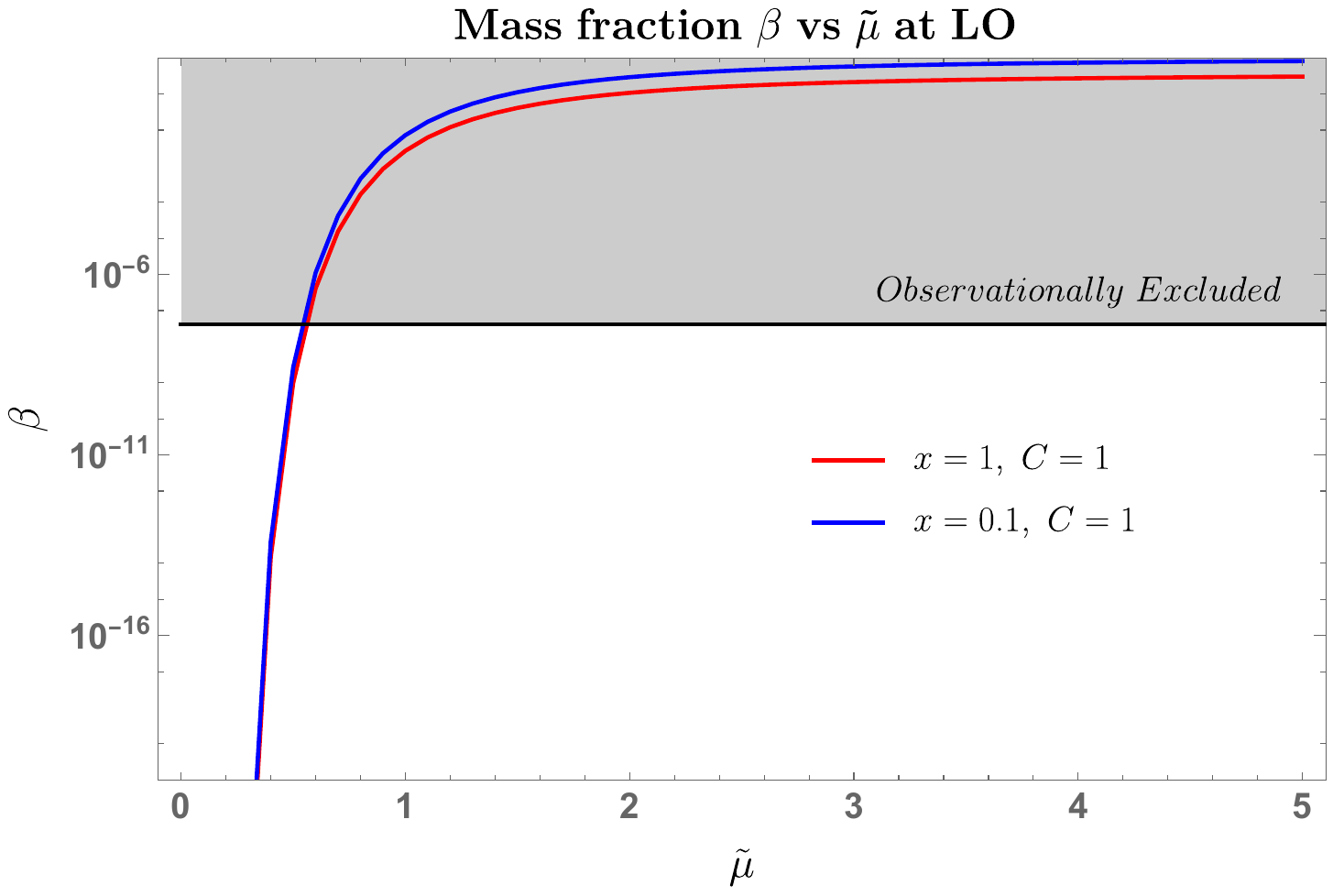}
        \label{massfracLO}
    }
    \subfigure[]{
        \includegraphics[width=8.5cm,height=7.5cm]{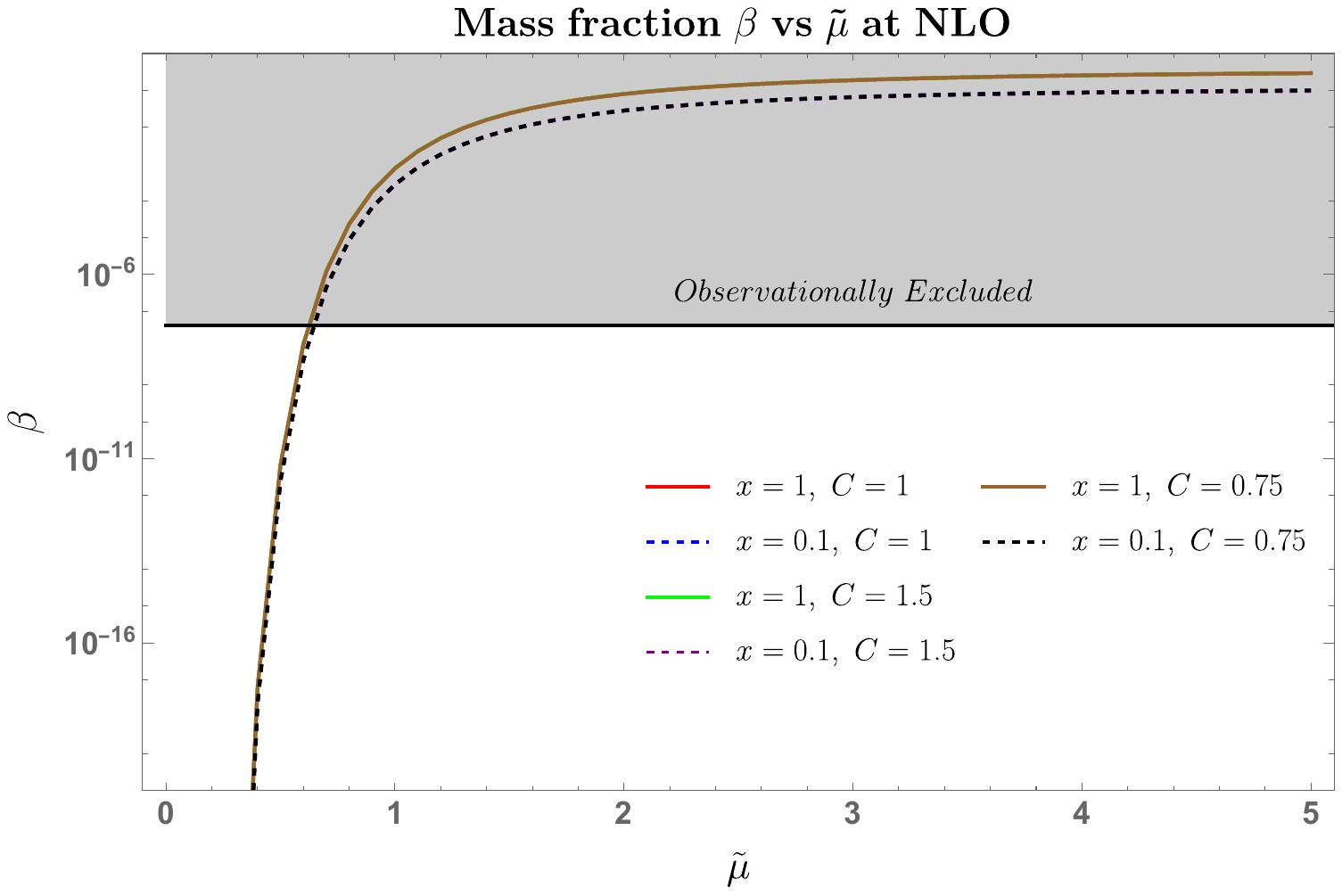}
        \label{mfracNLO}
    }
       \subfigure[]{
        \includegraphics[width=8.5cm,height=7.5cm]{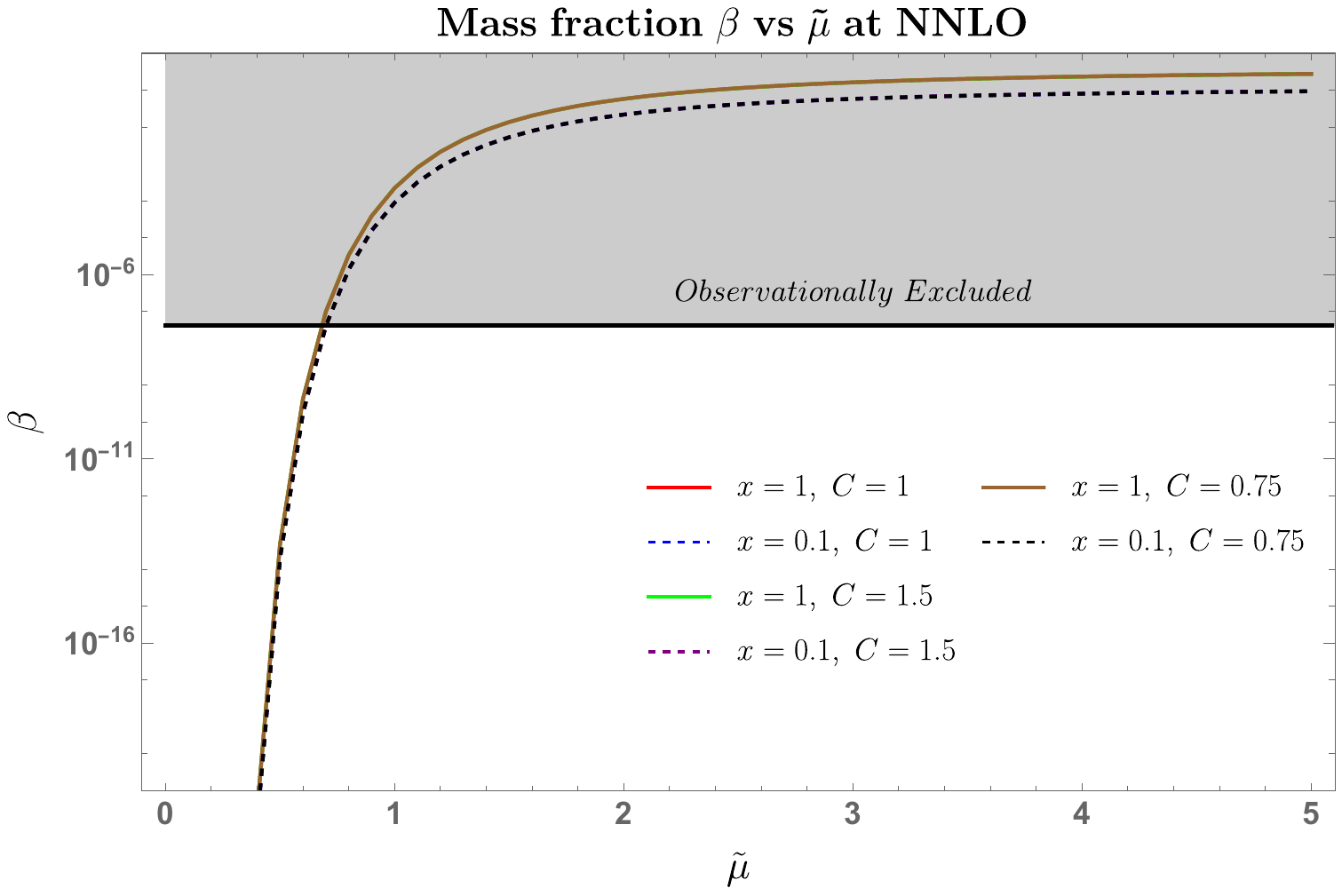}
        \label{massfracNNLO}
    }
    \subfigure[]{
        \includegraphics[width=8.5cm,height=7.5cm]{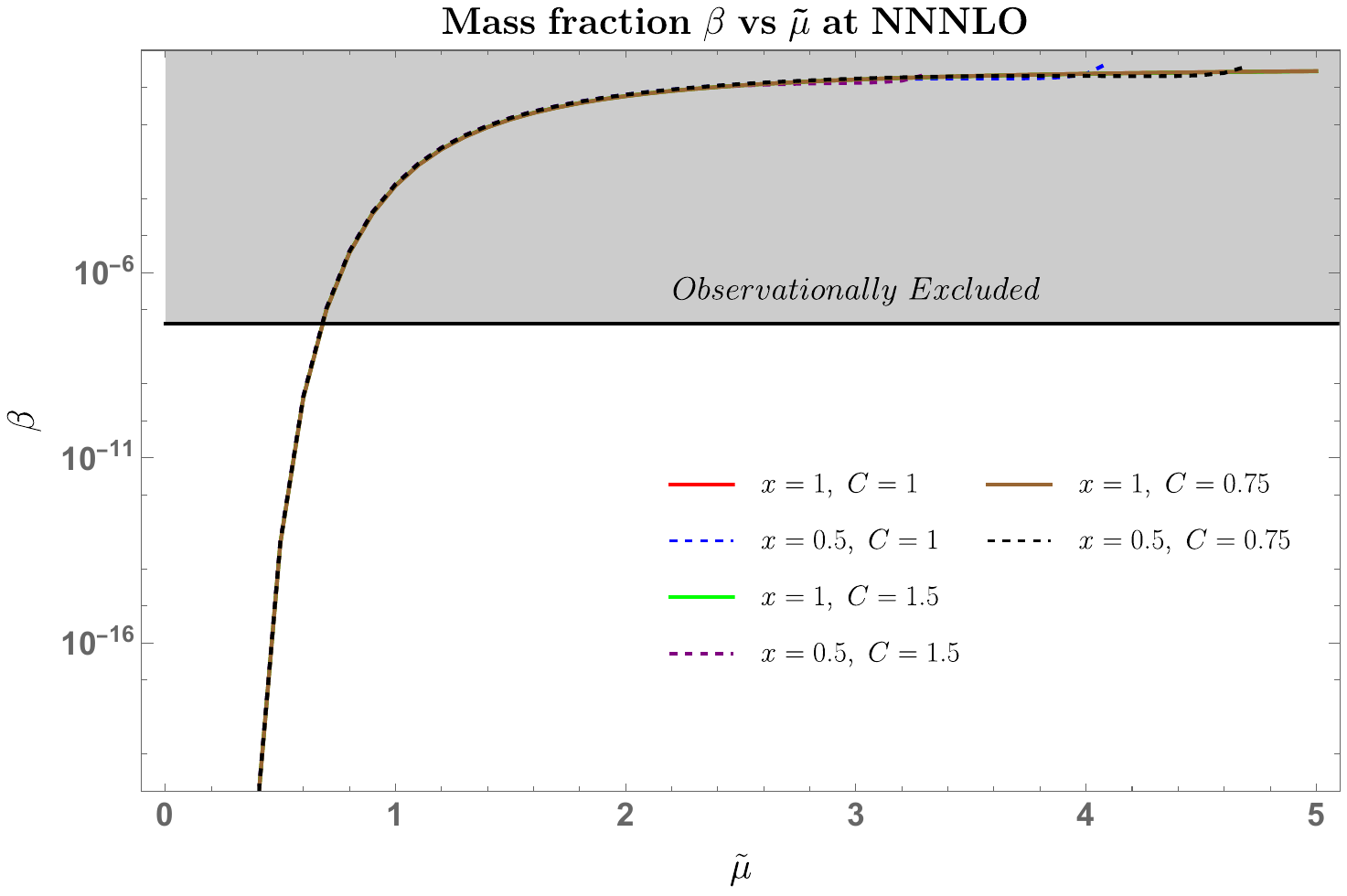}
        \label{massfracNNNLO}
       }
    	\caption[Optional caption for list of figures]{Mass fraction $\beta$ in relation to $\tilde{\mu}$ fluctuation. The mass fraction is less susceptible to variations in $x$ and more sensitive to $\mu$ values near or below $1$. Without a variable $C$, the \textit{top-left} panel shows the behavior of $\beta$ for PDF at LO, $P_{\Gamma}^{\rm LO}({\cal N})$. The $\beta$ behavior for PDF at NLO is shown in the \textit{top-right} panel. For PDF at NNLO, $P_{\Gamma}^{\rm NLO}({\cal N})$, the \textit{bottom-left} panel shows the behavior of $\beta$; for PDF at NNNLO, $P_{\Gamma}^{\rm NNLO}({\cal N})$, and for PDF at NNNLO, $P_{\Gamma}^{\rm NNNLO}({\cal N})$, the \textit{bottom-right} panel shows the behavior of $\beta$. Several examples are displayed, where $C \in\{0.75,1,1.5\}$. The values of $\beta\gtrsim {\cal O}(10^{-8})$, which are in the range of $10^{16}$g to $10^{50}$g (or $M_{\rm PBH}\sim {\cal O}(10^{-17}-10^{16})M_{\odot}$), are highlighted in the gray shaded region. These values are excluded by current observations for heavy mass PBH.
 } 
    	\label{betaplot0123}
    \end{figure*}

\begin{figure*}[htb!]
    	\centering
    \subfigure[]{
        \includegraphics[width=8.5cm,height=7.5cm]{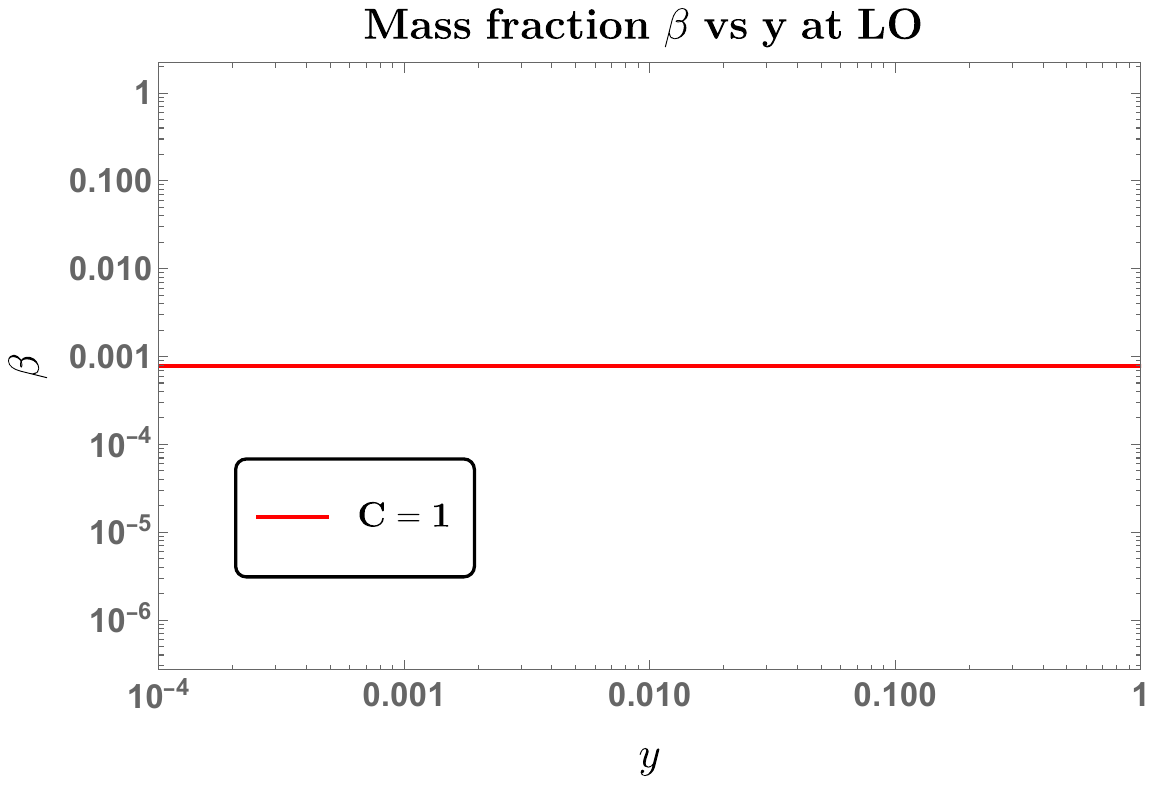}
        \label{massfracLOy}
    }
    \subfigure[]{
        \includegraphics[width=8.5cm,height=7.5cm]{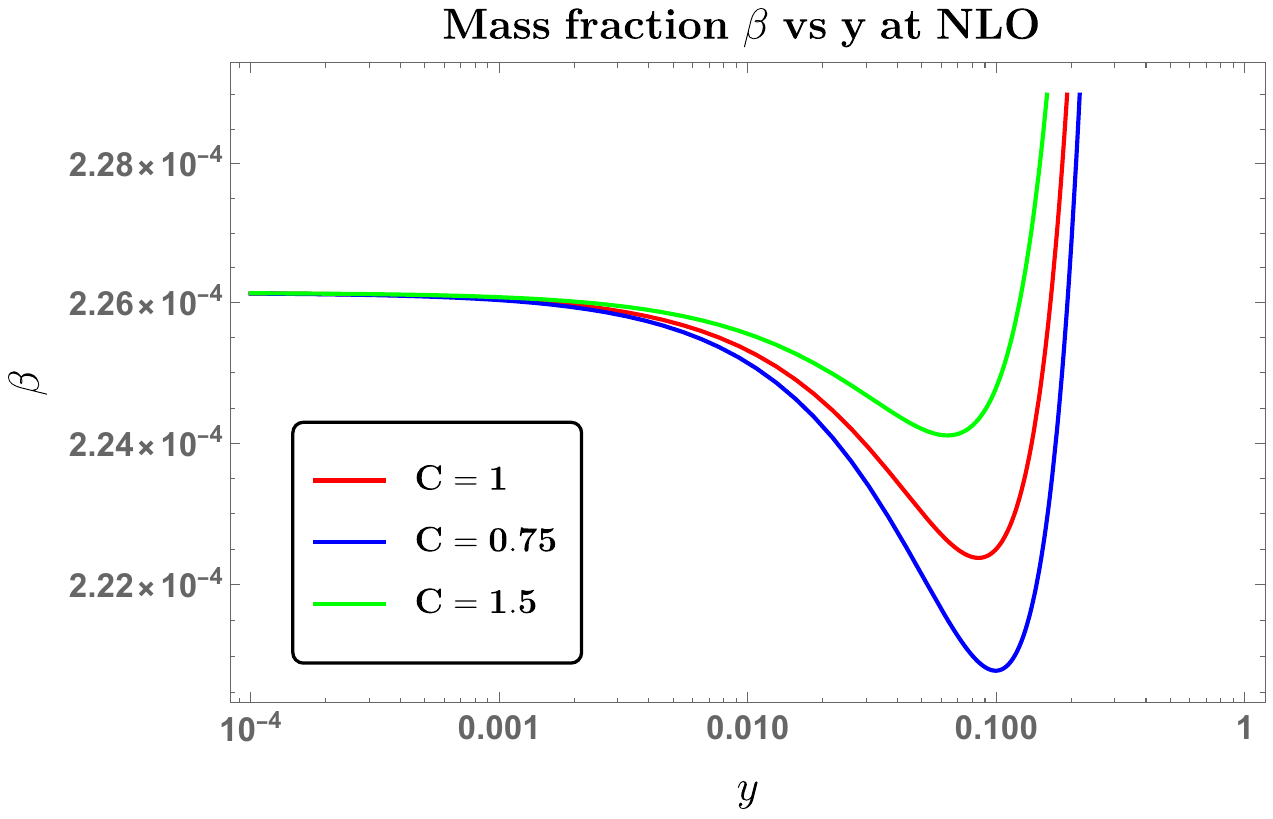}
        \label{mfracNLOy}
    }
       \subfigure[]{
        \includegraphics[width=8.5cm,height=7.5cm]{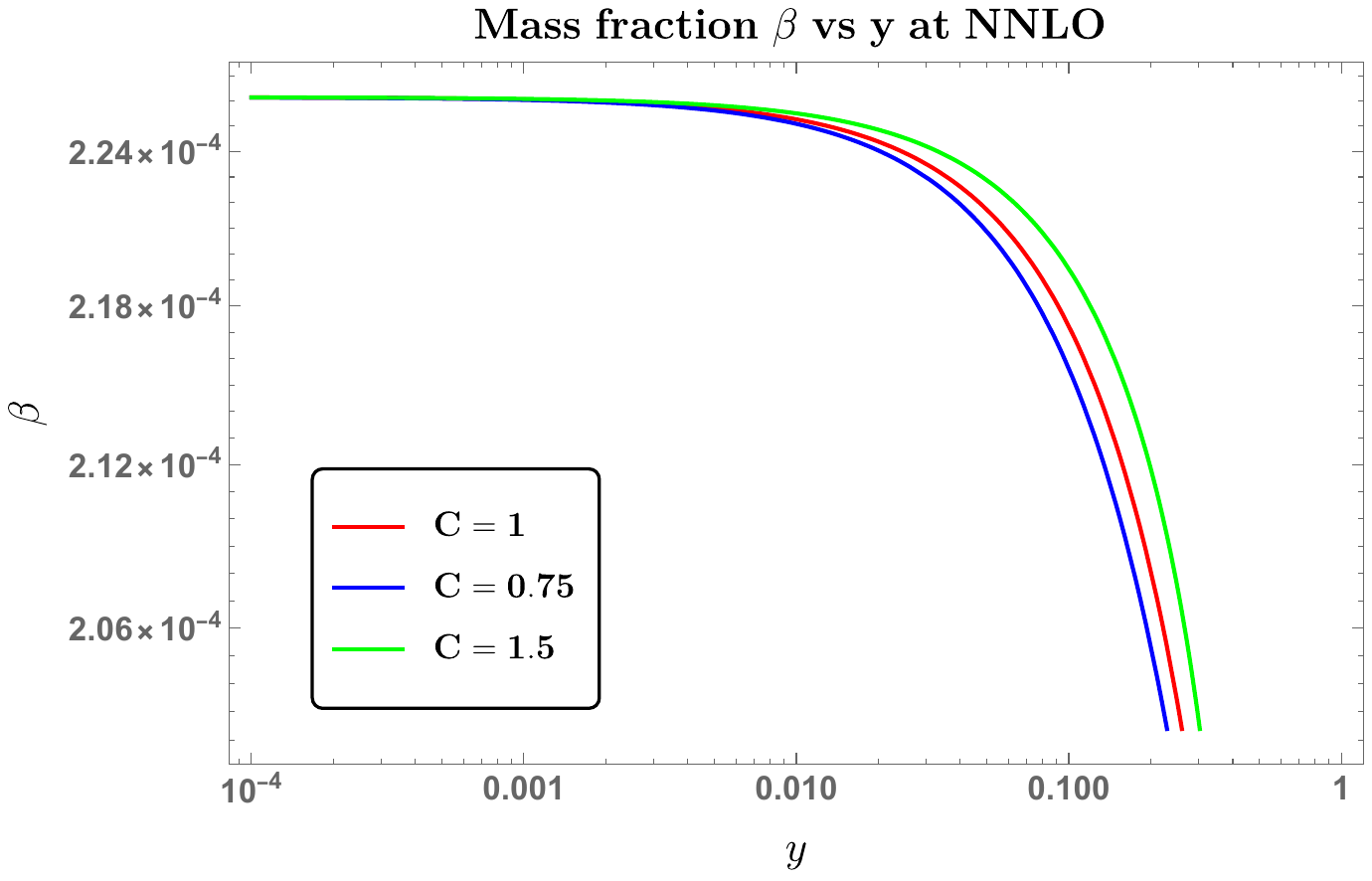}
        \label{massfracNNLOy}
    }
    \subfigure[]{
        \includegraphics[width=8.5cm,height=7.5cm]{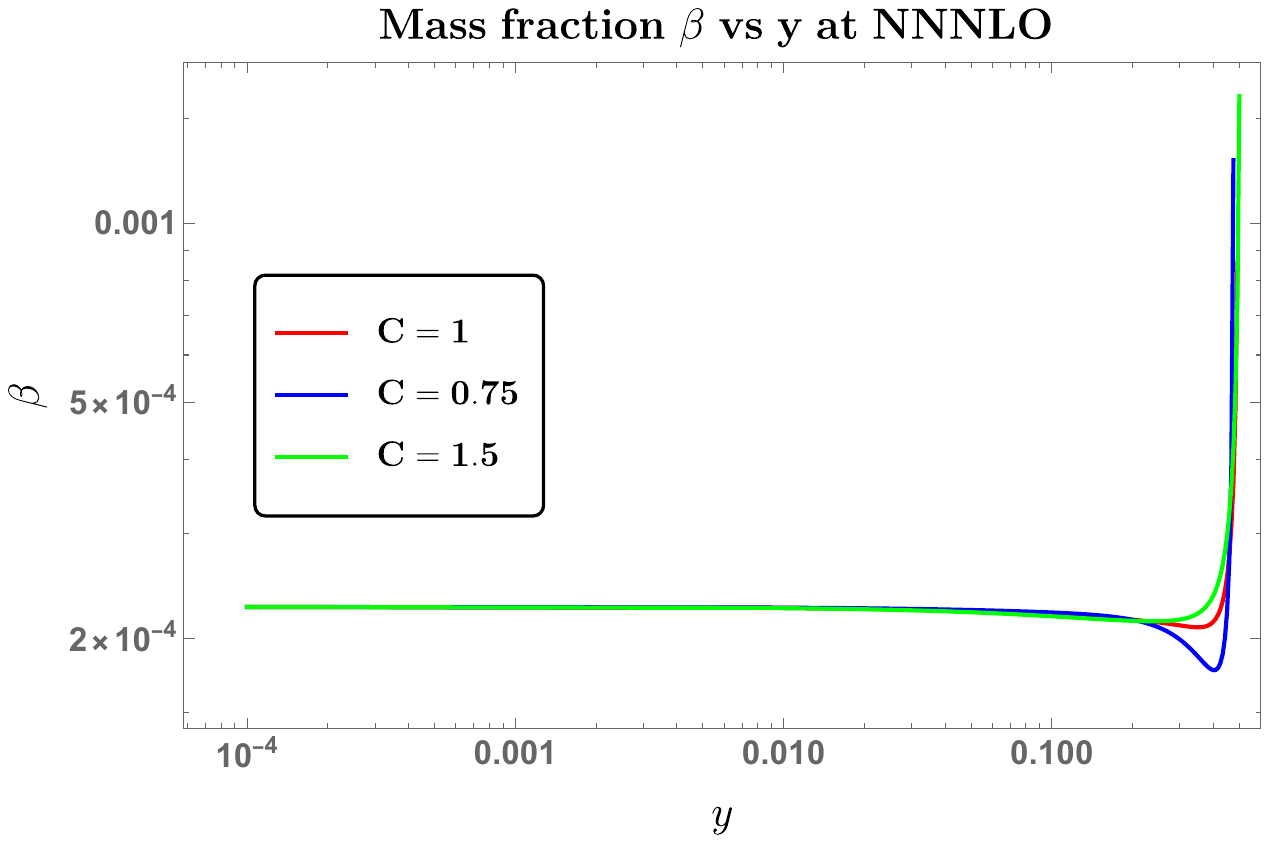}
        \label{massfracNNNLOy}
       }
    	\caption[Optional caption for list of figures]{Mass fraction $\beta$ in opposition to $y$ fluctuation.  The behavior of $\beta$ at LO with varying $y$ is shown in the \textit{top-left} panel; variable $C\ne 1$ is not included. The \textit{bottom-left} panel shows how $\beta$ behaves when $y$ changes at NNLO, the \textit{bottom-right} panel shows how $\beta$ behaves when $y$ changes at NNNLO, and the \textit{top-right} panel shows how $\beta$ behaves at NLO with $y$. $\tilde{\mu}=1$ and $\zeta_{\rm th}\sim {\cal O}(1)$ are held unchanged, and many examples with values of $C=1$ (red), $C=0.75$ (blue), and $C=1.5$ (green) are displayed.
 } 
    	\label{betaplot0123y}
    \end{figure*}

Based on the analytical analysis of the distribution functions for the diffusion-dominated regime, we examine the results for the PBH mass fraction and abundance in this section. The behavior of the PBH mass fraction for varying values of $\tilde{\mu}$ is shown in fig. (\ref{betaplot0123}). In this case, $\beta$ changes sharply in value for $\tilde{\mu}\leq 1$, and the mass fraction rapidly saturates to values near but below $1$ as $\tilde{\mu}$ rises over $1$. $\beta$ is not significantly affected by changes in $x$, and the effects are only discernible when $\tilde{\mu}\geq 1$. When PDF at leading order is taken into account, the left-panel displays $\beta$. It shows the standard stochastic single-field inflation scenario. $\beta$ is highlighted in the right panel, with the PDF's next-to-leading order. While the impacts of different $C$ are not large enough to be discernible, we see that for $\tilde{\mu}\geq 1$, $\beta$ at NLO changes less as $x$ decreases than $\beta$ at LO. At the same $\tilde{\mu}$ value, the PDF at NLO predicts a lower mass fraction than the PDF at LO. As a result, for $\tilde{\mu}\leq 1$, mass fraction is shown to fall by at least one order of magnitude. 

The behavior of the mass fraction at NNLO may be obtained by concentrating on the next order in the perturbative expansion, as shown in the bottom-left figure \ref{massfracNNLO}. For $\tilde{\mu}\leq 1$, there is an additional order of magnitude fall in the mass fraction. For lower $\tilde{\mu}$ values, there has also been an improvement in the rate of decline of $\beta$. When we take into account the PBH abundance that results from the mass fraction, this will have important ramifications. As in the earlier instance at NLO, the contributions from the characteristic parameter $C$ to the mass fraction are not large enough to make a meaningful distinction. Finally, the PDF at NNNLO experiences some intriguing modifications, as does its derived $\beta$ behavior. 
A positive magnitude of $\beta$ persists for $x=0.5$ up to a specific value of $\tilde{\mu}>1$, beyond which the numerical values provide negative results, as can be shown in fig. \ref{massfracNNNLO} for $\beta$ at NNNLO. Therefore, it is not possible to reduce $x$ to very small values because doing so would just cause $\beta$ to become negative more quickly and for values of $\mu\sim 1$. Once $x$ is decreased from $x=1$, the relative change has likewise nearly disappeared. It is more interesting to see the impact of the characteristic parameter $C$ in this case. In contrast to reducing $C$ below $C=1$, which causes the $\beta$ to stretch to greater $\tilde{\mu}$ until it once more goes to negative values, increasing $C$ beyond $C=1$ together with $x<1$ causes the related $\beta$ to become negative for lower values of $\tilde{\mu}$. Taking into account $x\gtrsim 1$, no such problem occurs. 

We investigate how the mass fraction $\beta$ behaves at each order in the perturbative expansion with varying values of $y$ or the coarse-grained conjugate momentum variable, as previously described in eqn. (\ref{newphasevars}), as shown in fig. \ref{betaplot0123y}. Since the properties of the diffusion-dominated regime are used to study this behavior (see section \ref{s11}), the limit of $y\ll 1$ would therefore offer a much more acceptable interval from which the PBH mass fraction, including higher-order corrections from NNNLO analysis, may be reliably determined. Any modification with $y$ in the top-left panel \ref{massfracLOy} would not provide any meaningful insights into the behavior of the mass fraction since the leading order (LO) scenario is an expansion independent of $y$. Observing the NLO situation, we obtain extra components linear in $y$ for our PDF. Consequently, we notice minor modifications to $\beta$ from the top-right panel of fig. \ref{mfracNLOy}. The relative separation increases when we put $y> 0.01$ in terms of different $C$ values. The mass fraction then rapidly increases until it reaches $y\sim 1$, changing by at least one order of magnitude. Consideration of the NNLO example, fig. \ref{massfracNNLOy}, alters the impact of sub-dominant adjustments.
Here, when $y>0.01$ is taken into account, the mass fraction $\beta$ continues to decrease in magnitude. In contrast to the higher values, $C\geq 1$, the mass fraction for the lower values, $C<1$, falls earlier. Finally, we analyze the impact of the adjustment up to NNNLO to $\beta$ using fig. \ref{massfracNNNLOy}. This time, throughout a wide range of $y<0.1$, the mass fraction remains constant. The relative difference only becomes substantial if we opt to have $y>0.5$, even when many $C$ are taken into consideration. When we approach $y\sim 1$, the case of the conventional stochastic single-field $(C=1)$ changes significantly, but otherwise does not reveal many distinctions. For various $C$ values, the mass fraction converges to exhibit comparable behavior in the $(y\ll 1)$ and $y\sim 1$ limits. 

We sum up by stating that, upon comparison between each higher-order adjustment to the mass fraction $\beta$, the interval $y\sim {\cal O}(0.01)$ coincides well. The behavior might vary significantly for $y\geq 0.1$, contingent on the expansion order taken into consideration.

\subsubsection{Outcomes from PBH abundance $f_{PBH}$}

This section covers the observed results for the PBH abundance as a function of PBH mass after $\zeta_{\rm th}\sim {\cal O}(1)$ is reached as a threshold for the amount of curvature perturbations. 

\begin{figure*}[htb!]
    	\centering
    \subfigure[]{
      	\includegraphics[width=8.5cm,height=7.5cm]{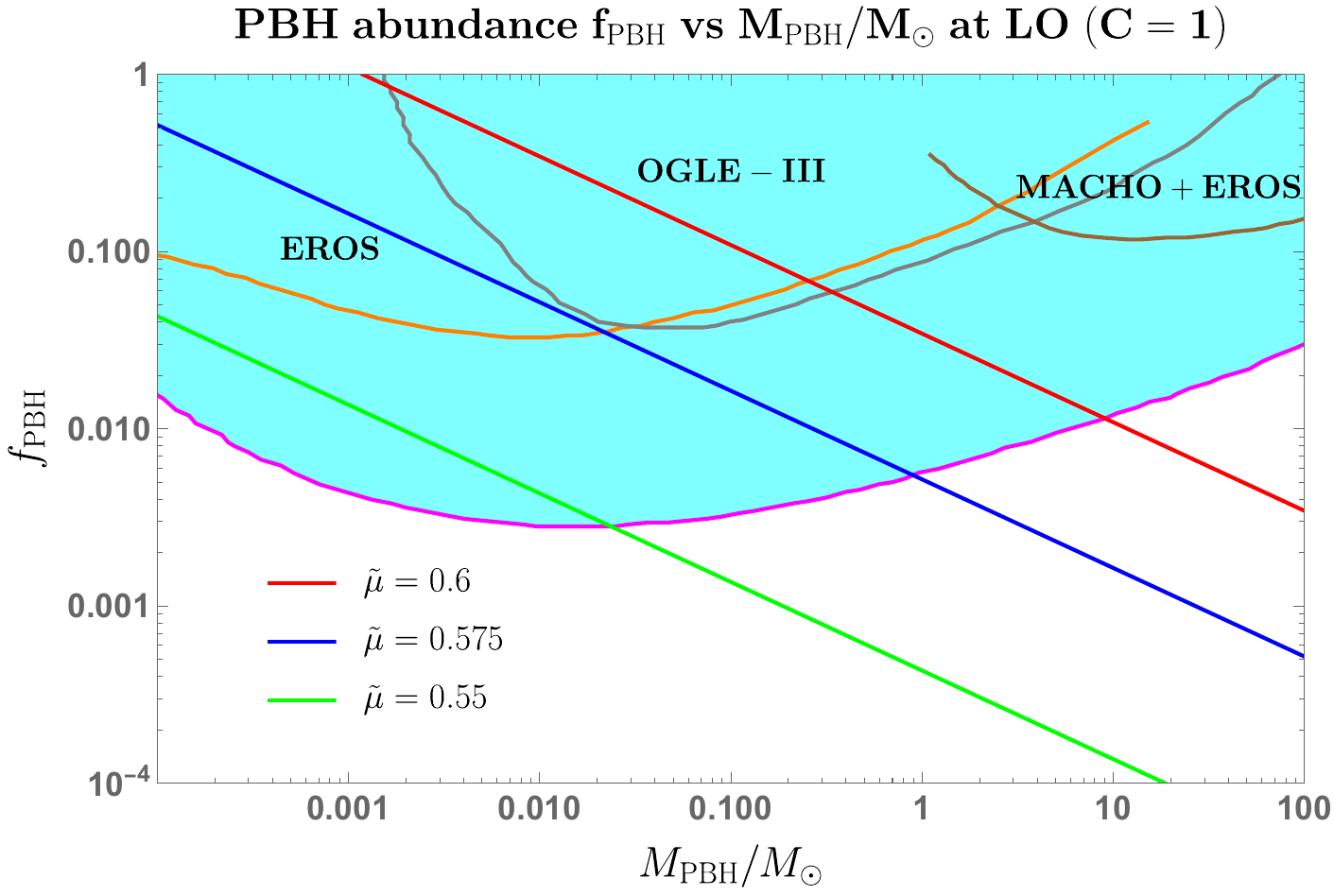}
        \label{fpbhLO}
    }
    \subfigure[]{
        \includegraphics[width=8.5cm,height=7.5cm]{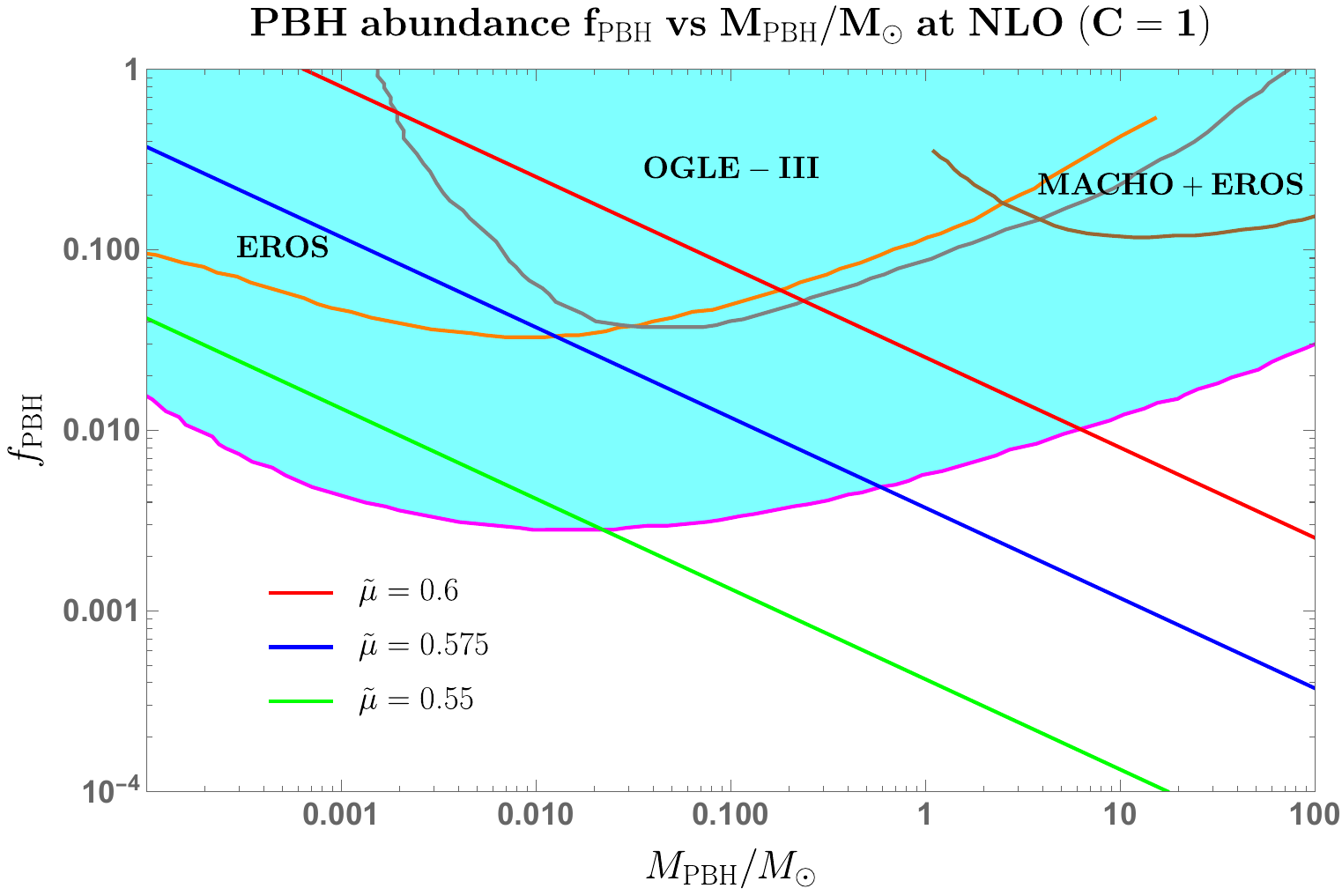}
        \label{fpbhNLO}
    }
       \subfigure[]{
        \includegraphics[width=8.5cm,height=7.5cm]{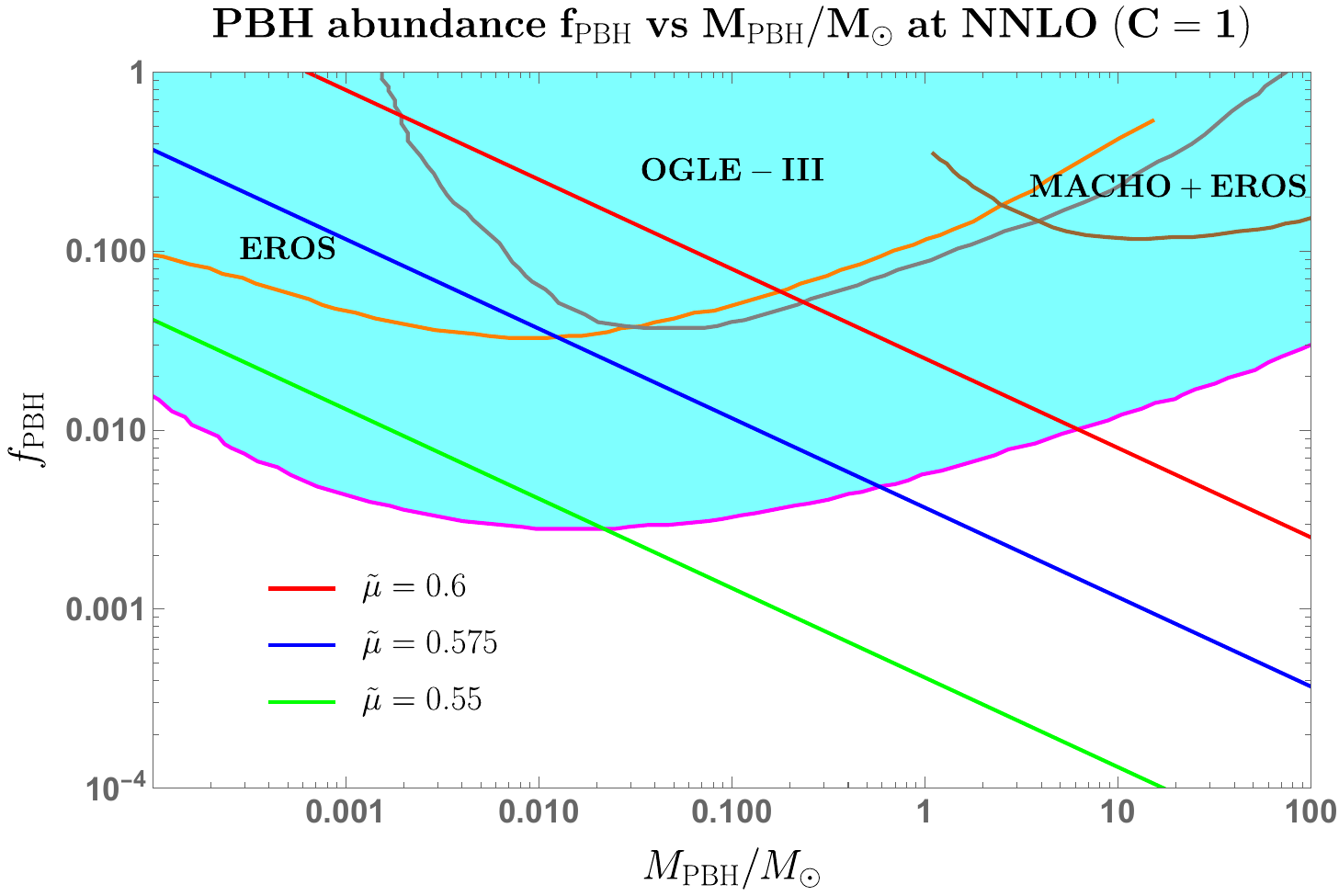}
        \label{fpbhNNLO}
    }
    \subfigure[]{
        \includegraphics[width=8.5cm,height=7.5cm]{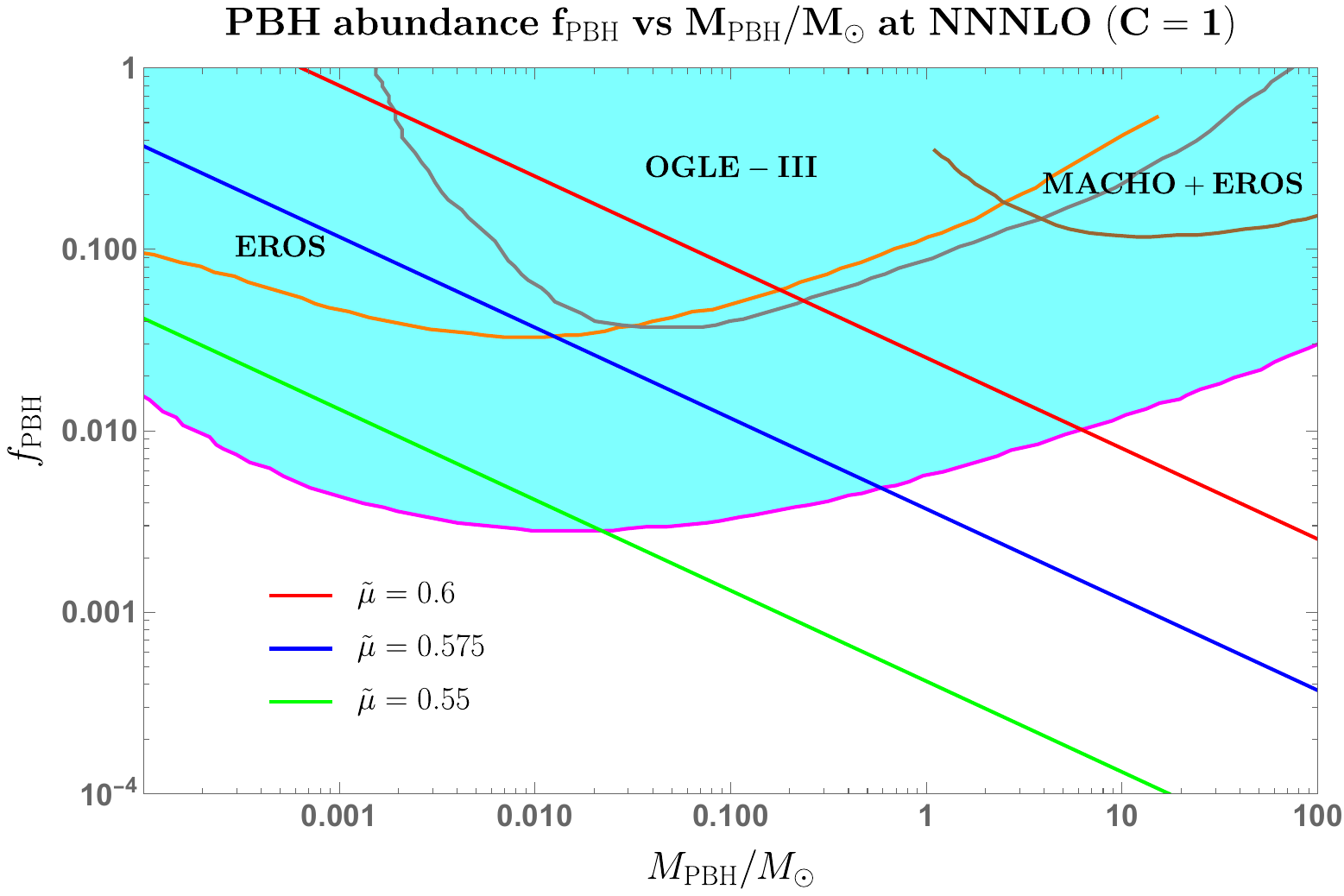}
        \label{fpbhNNNLO}
       }
    	\caption[Optional caption for list of figures]{PBH abundance (in $M_{\odot}$) as a function of mass $f_{\rm PBH}$. The parameters $x=1$, $y=0.06$, and $C=1$ have fixed values. The values of $\mu\in \{0.55,0.575,0.6\}$ vary for each plot. The recently determined $95\%$ upper limits on PBH abundance from microlensing events are highlighted in the cyan-colored zone. The stringent limitations on $f_{\rm PBH}$ \cite{Mroz:2024mse} are marked by the magenta boundary, which also incorporates limits from other surveys of dark matter, such as OGLE-III (gray) \cite{wyrzykowski2011ogle}, EROS (orange) \cite{EROS-2:2006ryy}, and MACHO+EROS (brown) \cite{Blaineau:2022nhy}.
 } 
    	\label{fpbhplot0123}
    \end{figure*}
\begin{figure*}[htb!]
    	\centering
    \subfigure[]{
      	\includegraphics[width=8.5cm,height=7.5cm]{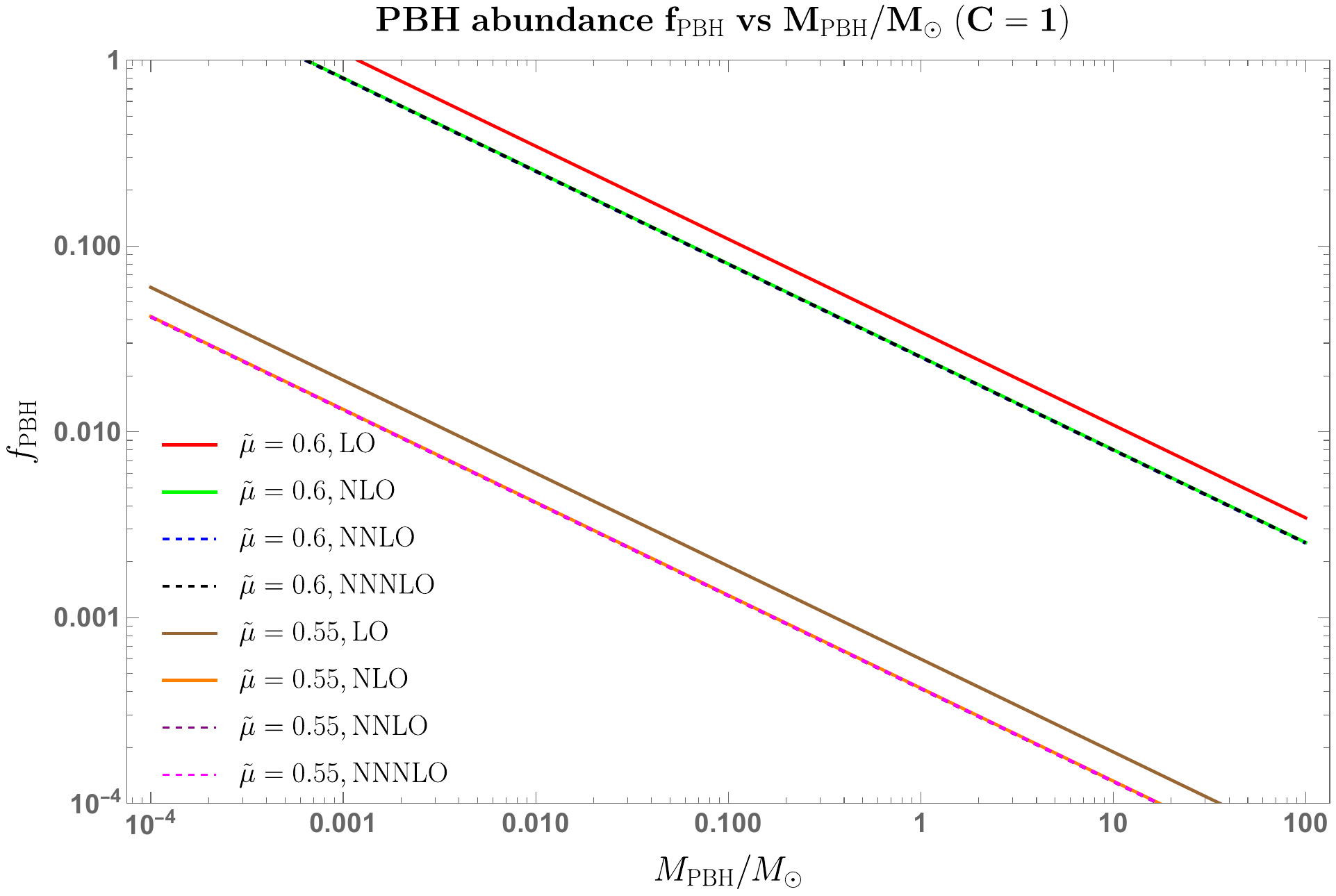}
        \label{fpbhALL}
    }
    \subfigure[]{
        \includegraphics[width=8.5cm,height=7.5cm]{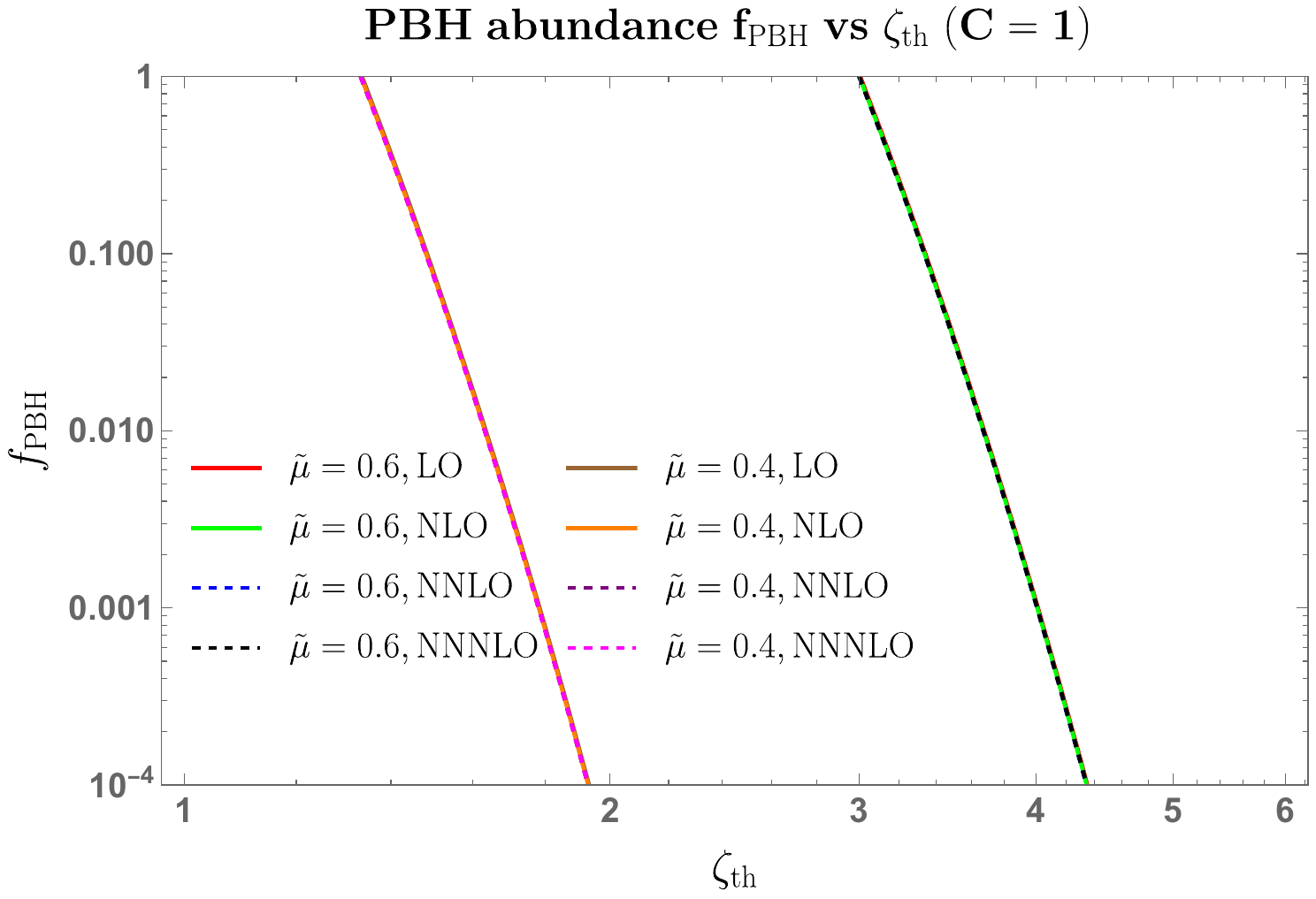}
        \label{fpbhALL2}
    }
    	\caption[Optional caption for list of figures]{\textit{left-panel} displays PBH abundance $f_{\rm PBH}$ as function of the mass (in $M_{\odot}$) including all orders of the perturbative expansion. \textit{right-panel} shows $f_{\rm PBH}$ as function of the threshold $\zeta_{\rm th}$ including all order of the perturbative expansion. The figures are for a fixed set of $\tilde{\mu} \in \{0.55,0.6\}$ and $C=1$. } 
    	\label{fpbhplotab}
    \end{figure*}

Figure (\ref{fpbhplot0123}) displays the abundance of a spectrum of PBH masses, which range from $M_{\rm PBH}\sim {\cal O}(10^{-4}M_{\odot})$ to $M_{\rm PBH}\sim {\cal O}(10^{2}M_{\odot})$. When $x=1$, $y=0.06$, and $C=1$ are constant and $\zeta_{\rm th}\sim {\cal O}(1)$ is met throughout, the charts demonstrate how sensitive the abundance is to the values of $\tilde{\mu}$. The distribution function at different orders results in the situation of a broad enough mass range that can attain considerable PBH abundance, as we can see from the different sub-figures, \ref{fpbhLO}, \ref{fpbhNLO}, \ref{fpbhNNLO}, and \ref{fpbhNNNLO}. Their impact on the abundance decreases with increasing perturbation expansion, with very tiny deviations seen over the mass range under investigation. We incorporate the findings from several micorlensing studies for the proportion of total dark matter that is dispersed as PBHs. The region confined by the Optical Gravitational Lensing Experiment (OGLE) data in their $20$ years run, which consists of the OGLE-III $(2001-2009)$ and OGLE-IV $(2010-2020)$ runs, is marked by the cyan colored region with a solid magenta line. Refer to the research in \cite{Mroz:2024wag,Mroz:2024mse} for further information on the observation setup and appropriate data production analysis. We conclude from the stringent limitations that at most $10\%$ of the total dark matter may be composed of huge $M_{\rm PBH}\sim {\cal O}(1-100)M_{\odot}$.    
A clearer view of the sensitivity to the PDF adjustments at each order is provided by the plot in fig. \ref{fpbhALL}. Compared to the next-to-leading order (NLO) corrections from the accompanying PDF, the leading order (LO) contribution forecasts a greater magnitude of abundance, and the following adjustments do not significantly alter the picture. 

Another crucial parameter to adjust is the threshold curvature perturbation $\zeta_{\rm th}$. Variations in its value demonstrate a strong sensitivity to the range of potential $\tilde{\mu}$ values; bigger values of $\tilde{\mu}\geq 0.5$ indicate a greater sensitivity to $\zeta_{\rm th}$, leading to a notably larger abundance than smaller values of $\tilde{\mu}\leq 0.5$. The impact of varying $\zeta_{\rm th}$ with a given $\tilde{\mu}$ and fixed PBH mass, $M_{\rm PBH}\sim {\cal O}(10^{-3}M_{\odot})$, is presented in the right-panel of figure \ref{fpbhALL2}. This includes corrections from every scenario in the perturbative analysis of the PDF. The effects of threshold changes are shown to be independent of the order of correction taken into account in the PDF expansion. This is also true for the PBH abundance, which exhibits a similar declining pattern when the threshold is raised. We may also infer from the slope of \ref{fpbhALL} that, for a given $C$ value, altering $\tilde{\mu}$ impacts the abundance for various masses in the same way. In summary, we find that further perturbative adjustments to the PDF result in essentially insignificant changes to the abundance estimates across a wide range of PBH masses, and that the quantity $f_{\rm PBH}$ is extremely sensitive to variations in $\tilde{\mu}$, which impacts all PBH masses equally up to a given threshold $\zeta_{\rm th}$.


Using the language of conventional, classical $\delta N$-formalism, a large body of recent and past studies has concentrated on using the stochastic inflation formalism and its more recent variant. This section presents a comparative study of the conclusions drawn from earlier research using the stochastic-$\delta N$ formalism. We emphasize the differences in the perturbative analysis of the probability distribution driving the stochastic e-folding ${\cal N}$, and we focus on how these conclusions relate to primordial black holes.  The early development of the stochastic-$\delta N$ formalism was documented in refs.\cite{Fujita:2013cna,Fujita:2014tja,Vennin:2015hra}. These studies cleared the path by presenting a novel approach for computing the curvature perturbation's correlation functions following the use of the $\delta N$-formalism technique and numerous Langevin equation realizations. Stochastic-$\delta N$ approaches have also been applied to the case when quantum diffusion effects are important for a certain time during inflation, as in some of the works \cite{Animali:2022otk,Pattison:2021oen,Ezquiaga:2019ftu,Ballesteros:2020sre,Firouzjahi:2018vet}.
Large primordial curvature fluctuations are also produced by these processes, and they start the development of PBHs after inflation. The e-folds fluctuate as a result of these stochastic effects, and the stochastic-$\delta N$ formalism aids in obtaining the statistics of the curvature perturbation in terms of the e-folding statistics. PBH creation events are uncommon, and their abundance is significantly influenced by differences in the curvature perturbation distribution's tail. The method for obtaining explicit moments of the variable ${\cal N}$, which is then analyzed till the next-to-leading order (NLO), is presented in the paper \cite{Pattison:2021oen}. They have also conducted a thorough analysis on the implications of higher-order corrections on PBH mass fraction. Our study expands on previously established understandings to provide a field-independent picture of the whole analysis and visualizes the implications of taking even higher-order correction into account up to the next-to-next-to-next-to-leading order (NNNLO). The end result is a description of stochastic single field inflation in effective field theory (EFT), where the EFT effects are associated with a new parameter called the characteristic parameter $C$, as seen in eqn. (\ref{EFTparam}). Given the behavior of slow-roll parameters in many regimes of interest, including the ultra-slow roll during inflation, the range of $C$ is assumed to be near to unity. Our results for the distribution function and PBH mass fraction analysis coincide with prior discussions presented in \cite{Pattison:2021oen} for the standard stochastic single-field case, or when we set $C=1$. We next proceed to a more thorough higher-order PDF analysis, taking into consideration the changes resulting from the $C$ parameter throughout, for both the drift and diffusion-dominated regimes. Building on the work of \cite{Pattison:2021oen}, the generated PDF provides adjustments to different non-Gaussianity parameters for the drift regime at each analytical order. Unlike prior efforts with canonical single-field models, these corrections are computed considering $C$ dependency. We evaluate the behavior of the PDF shape around its mean by metrics like skewness and kurtosis for each correction order. Our work, mainly influenced by \cite{Ezquiaga:2019ftu,Pattison:2021oen}, casts earlier approaches into the EFT language and again investigates higher-order corrections in the diffusion-dominated regime, emphasizing the tail aspects. The diffusion regime is the result of a thorough examination that takes into account the permitted correction in relation to the expansion order and its variable under the current circumstances. Our thorough examination of the PDF up to NNNLO reveals that the perturbative approximations continue to be entirely consistent. In line with the previous endeavors, we also examine the consequences of the diffusion regime PDF analysis on the PBH mass fraction and abundance, taking into account a $C$ dependent feature. Detailed analysis of the stochastic effects in the potential auto- and cross-correlated power spectrum amplitudes and their intrinsic relationship with the corresponding noise matrix elements is another important component of our study, albeit it looks differently. A universal quasi de Sitter spacetime with arbitrary starting quantum vacuum state conditions is the subject of the analytic investigation. We emphasize in detail the significance of each correlation in the current configuration and the influence of stochasticity, in particular the regulatory aspects of the coarse-graining parameter $\sigma$.

\section{Conclusion}
\label{s8}
In this review, we thoroughly examined the 
the implications of large primordial fluctuation for production of PBHs  and secondary gravity waves taking into account the loop corrections to quantum correlations
We utilized the EFT setup, which enables us to draw general model-independent conclusions and allows to keep the debate broad and applicable to all classes of $P(X,\phi)$ theories. 
Sections \ref{s1}, \ref{s2}, \ref{s3} and \ref{s4} are devoted to a concise and comprehensive review of the background material, namely, conceptual and technical issues related to inflation and computational framework  of quantum correlations, mechanisms for large fluctuation production and secondary gravity waves.

The key conclusions drawn in the preceding section \ref{s5} that determine the PBH mass are outlined below. 
 (1) The size of   amplitude peak of power spectrum; (2) the location of the sharp transition  from SR to USR, denoted by $k_s$; and (3) the ratio, $k_e/k_s\sim \mathcal{O}(10)$, which amounts to approximately equal to the $2$ e-folds required by the validity of perturbation theory. If we limit our consideration to the tree-level computation, we can have PBH production in a large range of masses that might meet the conditions of both cosmology (dark matter and baryon asymmetry) and astrophysics (supermassive black holes). 
Nevertheless, the possibilities of the production of PBHs with generic masses are severely restricted by the presence of one-loop corrections to power spectrum. We considered the range of $k_e$ and $k_s$ dictated by the applicability of perturbative treatment for the permissible range of sound speed, namely, $1<c_s<1.17$ (equivalent to gravitational coupling, $0\lesssim M^4_2/\dot{H}M^2_p\lesssim 0.13$). In this case, the need for high PBH masses reduces the likelihood that inflation would occur, resulting in a far less number of e-folds than those required to address the causality problem. The corresponding PBH masses are minuscule, falling within a very narrow window of $10^{2}{\rm gm}\lesssim M_{\rm PBH}\lesssim 10^{3} \rm gm$, of barely any astrophysical/cosmological relevance, if we stick to the general number of e-folds ($\Delta \mathcal{N}\sim 60$).
It should be highlighted that our results are based on more accurate computation rather than just one loop contribution; in fact, we used the DRG re-summed power spectrum. We pointed out that the canonical and a-causal frameworks ($c_s\geq 1$) are both permitted, while the a-causal framework is favored since it maximizes the PBH power spectrum's enhancement to the peak value of $\mathcal{O}(10^{-2})$. These results firmly exclude the possibility of producing large mass PBHs using the generic model-independent EFT framework applicable to any single field inflation models, whether canonical and non-canonical. 
Finally, a resummed spectrum is computed through the application of the DRG resummation technique. Computation of the quantum loop corrections in the presence of alternative renormalization schemes, followed by the DRG resummation approach, is necessary for a deeper comprehension and a more robust assessment of the accuracy and application in a wider view.

Our goals in this section \ref{s5} of the review were twofold: first,  successful 
generation of a range of PBH masses evading the current No-go theorem for heavy mass PBHs. Secondly, a thorough discussion of the PBH production process taking into account renormalization,  followed by the  DRG resummation and the creation of the SIGW spectrum.
This review also addresses the uncertainty surrounding the precise nature of the equation of state parameter (EoS) $w$ immediately prior to the beginning of the BBN and its implications for the  PBH overproduction problem  keeping in mind the consistency with  the observed GW energy density spectrum from the PTA collaboration (which includes both NANOGrav-15 and EPTA).  A single-field inflation model with multiple sharp transitions (MSTs) in the EFT framework makes up the theoretical foundation for the underlying discussion. After integrating the renormalization and resummation procedures, the model is utilized to examine the formation of PBHs. To this effect, Dynamical Renormalization Group (DRG) approach and wave function/adiabatic renormalization—are detailed in great depth. In view of recent findings by the pulsar timing array (PTA) cooperation, we discuss a PBH formation scenario that includes a wide range of EoS parameter. Based upon SIGW interpretation of the PTA signal, the analysis reveals an interval of $0.2 \leq w \leq 1/3$, where a large PBH abundance, $f_{\rm PBH} \in (10^{-3},1)$, is detected. 
The regularized-renormalized-resummed scalar power spectrum compatible with perturbativity condition, in the range $1 \leq c_{s} \leq 1.17$ can successfully address the  PBH overproduction provided that EoS dependent scalar generated GW agrees  with PTA evidence.  We pointed out that the multiple sharp transitions within the EFT framework gives rise  to  SIGWs in agreement with NANOGrav 15 data as well as  findings of other established observatories, thereby avoiding a strong No-go theorem for the generation of large mass primordial black holes.

We further examined the PBH production in the framework of single field Galileon inflation in section \ref{s6}. We have shown that in all classes of single field inflationary frameworks (e.g., $P(X,\phi)$ inflation, EFT of inflation, which covers various canonical and non-canonical models) to explain the formation of large mass PBHs, the no-go theorem  can be evaded within the framework of Galileon inflation. In order to support this assertion, let us emphasize the following: (1) In the considered case, perturbation theory holds perfectly and the one-loop quantum effects are sub-dominant over the tree level amplitude of the primordial power spectrum. This is mostly due to the comoving curvature perturbation, $\zeta\rightarrow\zeta-{H}/{\dot{\bar{\phi}}_0}\left(b\cdot \delta x\right)$, which is produced by the underlying EFT framework and exhibits a very weak breach of Galilean shift symmetry. The third order action for the comoving curvature perturbation does not include the cubic interaction term $\eta^{'}\zeta^{'}\zeta^{2}$ because of this slight breaking of Galilean symmetry. This is because the contribution can be recast as a total derivative term at the boundary, which results in a negligible contribution that can be absorbed in the field redefinition. In the technical section of the conversation, this point was emphasized. The abrupt change from SRI to USR has no negative effects for the current computational goal since the other terms in the third-order action do not contain any terms that contain the time derivatives of the second slow-roll parameter $\eta$. We indicated that neither logarithmic nor quadratic divergences are present in the one-loop contributions calculated from the SRI, USR, and SRII areas, respectively, because of the lack of the particular cubic self interaction factor. The remaining contributions that show up in the one-loop contributions are either composed of strongly oscillating sinusoidal components with limited tiny amplitudes or severely power law suppressed. Our investigation revealed that the one-loop contribution is subdominant over the outcome derived from the tree-level equivalent, as a consequence. (2) Since Galileon EFT has a strong non-renormalization theorem, these effects have no bearing on the one-loop adjusted power spectrum in this case, so there are no severe limitations from renormalization and ressumation. (3) Remarkably, establishing the SRI to USR transition scale position at $k_s\sim 10^{6}{\rm Mpc}^{-1}$, which is essential for the development of significant fluctuations—which are ${\cal O}(10^{-2})$—does not present any problems. (4) As a result, massive  PBHs have evaporation time scale of ${10}^{64}{\rm years}$, or ${\cal O}(M_{\odot})$. (5) When there are no rigorous limitations on the SRI to USR transition scale position shifting, it is possible to produce PBHs with incredibly large masses $M_{\rm PBH}\gg M_{\odot}$ with a wide evaporation time scale. However, by moving the SRI to USR transition scale to a very large value, it is also possible to construct extremely tiny mass PBHs $M_{\rm PBH}\ll M_{\odot}$. This suggests that PBHs of any mass may be created using Galileon inflation evading the negative outcome caused by quantum loop corrections on the primordial power spectrum.
As for the realization,  we have confirmed that an ultra slow roll phase may be explicitly constructed in the effective field theory of single-field Galileon inflation. We have examined PBH creation and SIGW production in this configuration. We also thoroughly investigated, without explicitly utilizing the notion of primordial non-Gaussianity, the PBH overproduction problem that was recently found.  In addition to creating a comfortable abundance of near-solar-mass black holes, $10^{-3} \lesssim f_{\rm PBH} \lesssim 1$, we have demonstrated that the setup under discussion generates SIGWs well compatible with the PTA signal. Additionally, we have presented a thorough analysis of the production of significant primordial non-Gaussianities in the context of Galileon inflation during the slow-roll (SR) to ultra-slow-roll (USR) transitions. We discovered that we are able to generate the non-Gaussianity amplitude of the following order: $|\fnl| \sim {\cal O}(10^{-2})$ in the SRI, $-5 < \fnl < 5$ in the USR, and $-2 < \fnl < 2$ in the SRII phases. This is because the USR phase has sharp transitions that last for $\Delta {\cal N}_{\rm USR} \sim 2$ e-folds. We may therefore obtain a cumulative average value of $|\fnl| \sim {\cal O}(1)$. This suggests that, in the squeezed limit, our results strictly meet Maldacena's no-go theorem only for SRI, while they strictly fail the same criterion in the USR and SRII phases.Our findings on the formation of huge mass primordial black holes and big non-Gaussianities are supported by the non-renormalization theorem in the Galileon theory. We demonstrate that both outcomes rely on the precise locations of the transition wave numbers set at low scales. Next, we have presented the role of the explicit effect of $f_{\rm NL}$, the negative local non-Gaussianity, in suppressing the primordial black hole (PBH) abundance in the single-field model of Galileon inflation. The scalar power spectrum must be substantially enhanced for PBH creation, which has a substantial impact on their abundance. The pulsar timing array (PTA) data may be explained by the scalar-induced gravitational wave (SIGW) creation, which is also sensitive to the related frequencies in the nHz regime. Using the curvature perturbation improvements that provide $f_{\rm NL} \sim {\cal O}(-6)$, we show in our study based on threshold statistics on the compaction function that Galileon theory not only avoids PBH overproduction but also produces SIGWs that closely match the PTA data.

We conclude this review considering  the application of the Effective Field Theory (EFT) model-independent single-field framework to the production process of primordial black holes (PBHs) through the extension to stochastic inflation in section \ref{s7}. We demonstrate how the current EFT picture is significantly altered by the Langevin equations for the "soft" modes in the quasi de Sitter background, which are described by the Infra-Red (IR) contributions of scalar perturbations and the ensuing Fokker-Planck equation that drives the probability distribution for the stochastic duration ${\cal N}$. Using the stochastic-$\delta N$ formalism, an explicit perturbative analysis of the distribution function is carried out in higher orders for both the quantum-diffusion dominated regime and the classical-drift regime. The technical details
included in the  Appendices should suffice for the smooth reading of the review.

\section*{Acknowledgement}
We express our sincere thanks to our colleagues, Gautam Mandal, Rong-Gen Cai, Anzhong Wang, Jun'ichi Yokoyama, Hassan Firouzjahi, A. Dolgov, S. Tsujikawa, Roy Maartens, A Toporensky, V. Sahni, R. Kaul, Debajyoti Choudhury, L. Sriramkumar, Rajeev Kumar Jain, R. Gannouji, N. Dadhich and Mayukh Raj Gangopadhyay for useful comments and discussions. 
We thank Mohit K. Sharma
for his help in drawing some of the figures included in the draft.
SC would like to thank The North American Nanohertz Observatory for Gravitational Waves (NANOGrav) collaboration and the National Academy of Sciences (NASI), Prayagraj, India, for being elected as an associate member and the member of the academy respectively. 
SC would like to acknowledge the inputs from The North American Nanohertz Observatory for Gravitational Waves (NANOGrav) collaboration members for useful comments and discussions, which helped improve the presentation of the review. SC would like to especially thank Soumitra SenGupta and Supratik Pal for inviting to IACS, Kolkata, and ISI, Kolkata, during the work. Furthermore, SC thanks Supratik Pal and his students for inviting to give an inaugural plenary talk at the discussion meeting titled, {\it Cosmo Mingle}, where part of the work was presented. SC would also like to thank all the members of Quantum Aspects of the Space-Time \& Matter
(QASTM) for elaborative discussions. MS is supported by Science
and Engineering Research Board (SERB), DST, Government of India under the Grant Agreement number CRG/2022/004120 (Core Research Grant). MS is
also partially supported by the Ministry of Education
and Science of the Republic of Kazakhstan, Grant No.
0118RK00935, and CAS President’s International Fellowship Initiative (PIFI). 

\newpage
\appendix

\section*{Appendix}

\section{In-out vacuums, Bogoliubov transformation and particle creation}
\label{Ap}

Let us briefly discuss the distinction between "in" and "out" vacuum states, $|0_{in}\rangle~;~ |0_{out}\rangle$ and Bogoliubov transformation. To this effect,
let us the consider scalar field operator expressed through mode functions and annihilation and creation operators,
\begin{equation}
\label{FE1}
    \hat{\phi}({\bf x},t)=\frac{1}{({2\pi})^{3}}\int{\hat{\phi}_{\bf k}(t) e^{{i\bf k}.{\bf x}}{d^3{\bf k}}}= \frac{1}{({2\pi})^{3}}\int{(v_k(t)\hat{a}_{\bf k} +v^*_k(t)\hat{a}^\dag_{-\bf k}   ) e^{{i\bf k}.{\bf x}}{d^3{\bf k}}}
\end{equation}
In case interaction is absent, $\phi$ obeys Klein-Gordon equation and the mode function  $v_k(t)$ correspondingly satisfies the equation of harmonic oscillator (\ref{FE}). If  interaction is present(for instance, self interaction in $\lambda \phi^4$ theory), field equation
\bea
(\partial_\mu \partial^\mu+m^2){\phi}=\lambda {\phi}^3
\eea
is non-linear  and problem is tackled perturbatively order by order. Simplification does occur in  asymptotic regions, $t\to -\infty$ (in region) and $t\to +\infty$ (out region) where interaction switches off adiabatically allowing us to determine the mode functions in the corresponding regions. The underlying problem is non-linear and there exists no linear relation between "in" and "out" modes in this case.
A similar phenomenon, where the "in" and "out" modes are connected by linear relation, is caused by the coupling of the quantum field to a classical source $\hat{\rm a}$ {\it la} {\it Schwinger process}, 
\bea
 \mathcal{L}  =-\frac{1}{2}\partial^\mu\phi\partial_\mu \phi-\frac{1} 
 {2}m^2 \phi^2+\frac{1}{2}gJ(t) \phi^2
\eea
giving rise to equation of a parametric oscillator,
\bea
\ddot{v}_k(t)+\omega^2_k(t)v_k(t)=0;~~~\omega^2_k(t)=k^2+m^2+g J(t)
\eea
where  $J(t)$ is a classical source;  $g$ is a coupling constant and the Lagrangian is quadratic in $\phi$.
In case, the source is such that the interaction switches off adiabatically in the "in"  and "out" regions, one can unambiguously define the vacuum state in these regions as discussed above. 
Such a situation is mimicked by the non-trivial nature of the background in cosmology, with a distinction that there is no "out" vacuum state in that case, to be discussed in the subsection to follow, see Refs.\cite{Mandal:2015kxi,Birrell:1982ix}.
In the asymptotic regions, we have,

\bea
&& \ddot{v}^{in}_k(t)+\omega'^2_{ k} v_k^{in}(t)=0;~~~~~         v_k^{in}(t)=\frac{1}{\sqrt{2\omega'_{ k}}}e^{\pm i\omega'_{ k }t},~~~~t\to -\infty\\
&& \ddot{v}^{out}_k(t)+\omega^2_k v_k^{out}(t)=0;~~~~ v_k^{out}(t)=\frac{1}{\sqrt{2\omega_k}}e^{\pm i\omega_k t},~~t\to +\infty
\eea
where, in general, $\omega_k'$ and $\omega_k$ corresponding to "in" and "out" regions are different. 
In the "in" region, for two independent field solutions $v_k^{in}~\& ~v_k^{*in}$, we can always choose Woronskian normalization such that the annihilation and creation operators
obey standard commutation relation,
\bea
[\hat{a}^{in}_{\bf k},\hat{a}^{\dag in}_{\bf k'}    ] =\delta({\bf k}-{\bf k'});~~~
   \hat{a}^{in}_{\bf k}|0_{in}\rangle=0
\eea
and $|in>$ states can be built in the Fock space using the annihilation and creation operators. We can do the same procedure for the "out" region,
\bea
[\hat{a}^{out}_{\bf k},\hat{a}^{\dag out}_{\bf k'}    ] =\delta({\bf k}-{\bf k'});~~~
   \hat{a}^{out}_k|0_{out}\rangle=0
\eea
and build the corresponding $|out>$
states. The field operator for a given mode can be represented through "in"("out") 
 modes and corresponding annihilation and creation operators,
\bea
\hat{\phi}_{\bf k}(t)=v^{in}_k(t)\hat{a}^{in}_{\bf k} +v^{*in}_k(t)\hat{a}^{\dag in}_{-\bf k}   =v^{out}_k(t)\hat{a}^{out}_{\bf k} +v^{*out}_k(t)\hat{a}^{\dag out}_{-\bf k}.   
\eea
Finally, we can express the "in" mode functions through "out" basis,
\bea
\label{IOM}
v_k^{in}(t)=\alpha_k v_k^{out}(t)+\beta_kv_k^{* out}(t),
\eea
which is naturally reflected on the 
annihilation and creation operators and allows us  to relate them in the "in" and "out" regions,
\bea
&& a_{\bf k}^{out}=\alpha_k \hat{a}_{\bf k}^{in}+\beta_k \hat{a}_{-\bf k}^{\dag in},\\
&& a^{\dag out}_{\bf k}=\alpha^*_k \hat{a}_{\bf k}^{\dag in}+\beta^*_k \hat{a}_{-\bf k}^{in},
\eea
where $\alpha_k$ and $\beta_k$ are known as Bogoliubov coefficients and the relation is as Bogoliubov transformation. Consistency of commutation relations in the "in" and "out" regions impose a condition on Bogoliubov coefficients, 
\bea
\label{IOC}
|\alpha_k|^2-|\beta_k|^2=1.
\eea
Let us reiterate that the "out" state is fundamentally different from the "in" state; the former carries the legacy of the interaction present between the asymptotic regions. The Fock states in the asymptotic regions are built from $|0_{in}\rangle$ and $|0_{out}\rangle$ using the annihilation and creation operators defined in these regions. Surprisingly, the vacuum states, in general, are also different when interaction is present, which results in particle creation. Indeed,
\bea
\langle 0_{in}|\hat{a}^{\dag in}_{\bf k}\hat{a}^{in}_{\bf k}|0_{in}\rangle=\langle 0_{out}|\hat{a}^{\dag out}_{\bf k}\hat{a}^{out}_{\bf k}|0_{out}\rangle=0,
\eea
where $\hat{N}_{\bf k}=\hat{a}^\dag_{\bf k} \hat{a}_{\bf k}$
designates particle number operator. Let us emphasize that $|0_{in}\rangle$ designates the Heisenberg vacuum state which does not evolve; the average particle number density in this state computed in the "out" region is non-vanishing,
\bea
 n_k\equiv \langle 0_{in}|\hat{a}^{\dag out}_{\bf k}\hat{a}^{ out}_{\bf k}|0_{in}\rangle=
\langle 0_{in}| (  \alpha^*_k \hat{a}_{\bf k}^{\dag in}+\beta^*_k \hat{a}_{-\bf k}^{in})(\alpha_k \hat{a}_{\bf k}^{in}+\beta_k \hat{a}_{-\bf k}^{\dag in}   )|0_{in}\rangle=|\beta_k|^2.
\eea
In the absence of interaction, $\alpha_k=1~\&~\beta_k=0$; otherwise, we have particle creation.
We can also estimate the vacuum fluctuations of the field with respect to the $|0_{out}\rangle$,
\bea
\langle\phi({\bf x},t)\phi({\bf y},t)\rangle&\equiv&
\langle 0_{in}|\phi({\bf x},t)\phi({\bf y},t)|0_{in}\rangle-
\langle 0_{out}|\phi({\bf x},t)\phi({\bf y},t)|0_{out}\rangle\nonumber\\
&=&\frac{1}{( 2\pi)^3}\int \left(\frac{1}{2\omega'_k}-\frac{1}{2\omega_k}\right)\; e^{i{\bf k}.({\bf x}-{\bf y})}\;d^3{\bf k}\nonumber\\
&=&\frac{1}{( 2\pi)^3}\int\left(|v_k^{in}|^2-\frac{1}{2\omega_k}\right)\; e^{i{\bf k}.({\bf x}-{\bf y})}\;d^3{\bf k}
\eea
Expressing the $v_k^{in}$  through "out" mode functions using (\ref{IOM}),  making use of the consistency condition, (\ref{IOC}) and ignoring the fast oscillating terms($t\to \infty$ in "out" region), one arrives at an important relation,
\bea
&& \langle\phi({\bf x},t)\phi({\bf y},t)\rangle=\frac{1}{( 2\pi)^3}
\int{{|\beta_k|^2}\; e^{i{\bf k}.({\bf x}-{\bf y})}\;\frac{d^3{\bf k}}{\omega_k}}=\frac{1}{(2\pi)^2}  \int dk~{|\beta_k|^2}~\frac{k^2}{\omega_{ k}}~\frac{\sin(k|{\bf x}-{\bf y}|)}{k|{\bf x}-{\bf y}|},\\
&&\langle\phi^2({\bf x},t)\rangle=\frac{1}{( 2\pi)^3}
\int{{|\beta_k|^2}\; \frac{d^3{\bf k}}{\omega_k}},
\eea
which shows that vacuum fluctuation depends upon the number density of created particles. In case, interaction is absent, the Bogoliubov transformation is trivial. Let us remark that particle creation in quantum field in Minkowski space time caused by the non-linear interaction term in the Hamiltonian, can not be captured the Bogoliubov transformation which is linear.

\section{Connecting generalized $P(X, \phi)$ theory with Goldstone modes of EFT}
\label{A1a}

Our goal in this part is to establish a direct link between the Goldstone modes of the EFT set-up and the generalized $P(X,\phi)$ theory. These modes arise when the breaking of temporal diffeomorphism occurs under unitary gauge of cosmic perturbation. For an unperturbed space-time, one may write the following for the inflaton perturbation $\delta\phi(t,{\bf x})$, as the contributions after mixing of the metric fluctuations can be readily omitted in the decoupling limit:
\bea X=\bar{X}+\delta X,\eea
where we use the following to express the background and perturbation on $X$:
\bea \bar{X}=-\frac{1}{2}\dot{\bar{\phi}}^2_0(t)\quad\quad{\rm and}\quad\quad \delta X=\dot{\bar{\phi}}_0(t)\partial_{t}\left(\delta \phi(t,{\bf x})\right)-\frac{1}{2}(\partial_{\mu}\left(\delta \phi(t,{\bf x})\right))^2.\eea
Additionally, utilizing the connection that connects the Goldstone mode $\pi(t,{\bf x})$ to the inflaton perturbation $\delta\phi(t,{\bf x})$, as follows:
\bea \delta\phi(t,{\bf x})=\dot{\bar{\phi}}_0(t)\pi(t,{\bf x}).\eea
The simplified outcome of the perturbation on $X$ is as follows, and it is obtained by disregarding the contributions from the derivative $\dot{\bar{\phi}}_0(t)$ (slow-roll suppressed):
\bea \delta X\approx 2\bar{X}\bigg[\dot{\pi}-\frac{1}{2}\left(\partial_{\mu}\pi\right)^2\bigg]=2\bar{X}\bigg[\dot{\pi}+\frac{1}{2}\dot{\pi}^2-\frac{1}{2a^2}\left(\partial_{i}\pi\right)^2\bigg].\eea
Next, we examine the function $P(X,\phi)$ through its Taylor series expansion, yielding the following:
\bea P(X,\phi)=P(\bar{X},\bar{\phi}_0)+P_{,\bar{X}}\delta X+\frac{1}{2!}P_{,\bar{X}\bar{X}}\left(\delta X\right)^2+\frac{1}{3!}P_{,\bar{X}\bar{X}\bar{X}}\left(\delta X\right)^3+\cdots.\eea
It has the following expression when stated in terms of the Goldstone fluctuation:
\bea P(X,\phi)=-\bar{X}P_{,\bar{X}}\left(\partial_{\mu}\pi\right)^2+2\bar{X}^2P_{,\bar{X}\bar{X}}\bigg(\dot{\pi}^2-\dot{\pi}\left(\partial_{\mu}\pi\right)^2+\cdots\bigg)+\frac{4}{3}\bar{X}^3 P_{,\bar{X}\bar{X}\bar{X}}\bigg(\dot{\pi}^3+\cdots\bigg)+\cdots,\eea
Using the equation of motion for the background now, we obtain:
\bea \label{back} \bar{X}P_{,\bar{X}}=-M^2_{pl}\dot{H}.\eea
This will adjust the Taylor series' initial expansion coefficient as previously described. Nevertheless, it is not possible to determine the remaining coefficient in the series expansion indicated above. These coefficients can be constrained by applying a certain cosmological paradigm.

The aforementioned $P(X,\phi)$ function can also be recast in terms of the streamlined Goldstone action as follows:
\bea \label{pxpi}
P(X,\phi)=-M^2_{pl}\dot{H}\left(\partial_{\mu}\pi\right)^2+2M^4_2\bigg(\dot{\pi}^2-\dot{\pi}\left(\partial_{\mu}\pi\right)^2+\cdots\bigg)+\frac{4}{3}M^3_4\bigg(\dot{\pi}^3+\cdots\bigg)+\cdots,\eea
whereby the derivatives coming from $P(X,\phi)$ function may be used to represent all of the coefficients except the first term, which is already determined due to the background equation of motion, as follows:
\bea M^4_n=\bar{X}^n \left(\frac{d^nP}{d\bar{X}^n}\right)\quad\forall n=2,3,4,\cdots.\eea
For this reason, we may start writing right away:
\bea \label{coeff} M^4_2=\bar{X}^2 \left(\frac{d^2P}{d\bar{X}^2}\right)=\bar{X}^2P_{,\bar{X}\bar{X}},\quad M^4_3=\bar{X}^3 \left(\frac{d^3P}{d\bar{X}^3}\right)=\bar{X}^3P_{,\bar{X}\bar{X}\bar{X}}.\eea
Now, in order to facilitate our connection with a few well-known physical frameworks, let us discuss the numerical values of these EFT coefficients:
\begin{itemize}
    \item Firstly, the basic single scalar field slow roll inflationary paradigm with the kinetic term $X$ and the effective potential $V(\phi)$ can be explained if one fixes the coefficients as $M_2=0$ and $M_3=0$. In addition, we must employ the decoupling limit approximation, that leads to $\dot{H}\rightarrow 0$ and $M_{pl}\rightarrow \infty$ to satisfy the following constraint:
    \bea M^2_{pl}\dot{H}={\rm Constant}.\eea
In this instance, all perturbations are considered fully Gaussian, meaning that neither minor non-Gaussian nor any contributions in the higher order show up in the one-loop adjusted power spectrum or when calculating higher point cosmological correlation functions.

\item Given that both are connected to the Goldstone operator $\dot{\pi}^2$ in the current discussion, it is evident from the structure of the resulting equation (\ref{pxpi}) that the $M_2$ coefficient gets treated as a correction over the leading order coefficient $-M^2_{pl}\dot{H}$ in this computation. Nevertheless, just the fixed coefficient dependent on the background, $M^2_{pl}\dot{H}$, is linked to the spatial component of the gradient operator of the Goldstone mode, $(\partial_i\pi)^2/a^2$. Such contributions do not come with the coefficient $M_2$ attached. The time diffeomorphism symmetry breaking directly results in this, and consequently, a non-trivial effective sound speed $c_s$ is formed. The contribution at second order to the Goldstone action can then be expressed in terms of this simplified expression:
\bea {\cal L}^{(2)}_{\pi}=-\frac{M^2_{pl}\dot{H}}{c^2_s}\bigg(\dot{\pi}^2-c^2_s\frac{(\partial_i\pi)^2}{a^2}\bigg),\eea
whereby $c_s$, the effective sound speed parameter, is introduced as follows:
\bea \label{cs1} c_s\equiv \frac{1}{\displaystyle \sqrt{1-\frac{2M^4_2}{\dot{H}M^2_{pl}}}}.\eea

\end{itemize}
The sound speed thus can now be expressed using the $P(X,\phi)$ function and its derivatives as follows, after substituting the value of $M^2_{pl}\dot{H}$ in the background equation of motion, given by equation (\ref{back}), and using the higher order Taylor expansion coefficients as mentioned in equation (\ref{coeff}):
\bea \label{cs2} c_s=\frac{1}{\displaystyle \sqrt{1+\frac{2\bar{X}^2P_{,\bar{X}\bar{X}}}{\bar{X}P_{,\bar{X}}}}}=\sqrt{\frac{P_{,\bar{X}}}{P_{,\bar{X}}+2\bar{X}P_{,\bar{X}\bar{X}}}}. \eea
The EFT coefficient $M_2$ may now be expressed using the derivatives of the $P(X,\phi)$ function by comparing equations (\ref{cs1}) and (\ref{cs2}). This expression is as follows:
\be \frac{M^4_2}{\dot{H}M^2_{pl}}=\frac{1}{2}\Bigg(1-\frac{P_{,\bar{X}}}{P_{,\bar{X}}+2\bar{X}P_{,\bar{X}\bar{X}}}\Bigg)=\Bigg(\frac{\bar{X}P_{,\bar{X}\bar{X}}}{P_{,\bar{X}}+2\bar{X}P_{,\bar{X}\bar{X}}}\Bigg).\ee
The aforementioned formula is quite useful in establishing a connection, in terms of coefficients, between a given $P(X,\phi)$ theory and the Goldstone EFT action. It also explains how can limitations on the sound speed limit the model parameters.

\section{PBH mass constraints on generalized $P(X, \phi)$ theory }
\label{A2a}

In order to understand the limits on PBH mass and to visually represent the obvious relationship between $P(X,\phi)$ models and the EFT model for Goldstone modes as present in the context of single field inflation, we will now go deeper into this matter.

\subsection{Dirac-Born-Infeld (DBI) model}
First, let's look at the Dirac-Born-Infeld (DBI) model, for which the following represents the $P(X,\phi)$ function:
\bea P(X,\phi)=-\frac{\Lambda^4}{f(\phi)}\sqrt{1-\frac{f(\phi)}{\Lambda^4}X}+\frac{\Lambda^4}{f(\phi)}-V(\phi)\quad\quad {\rm where}\quad\quad X=-\frac{1}{2}\left(\partial_{\mu}\phi\right)^2.\eea
Here, the values of $f(\phi)$, $\Lambda^4$, and $V(\phi)$ rely on the specifics of the $D3$ brane's underlying String Theory framework. The EFT coefficient $M_2$ and the effective sound speed $c_s$ may be calculated from this model as follows:
 \bea c_s=\frac{1}{\displaystyle \sqrt{1-2\frac{f(\bar{\phi}_0)}{\Lambda^4}\bar{X}}}\quad\quad {\rm and}\quad \quad \frac{M^4_2}{\dot{H}M^2_{pl}}=\frac{f(\bar{\phi}_0)}{\Lambda^4}\bar{X}.\eea 
The effective sound speed for this model is predicted to be less than unity, and based on our study presented in this work, we observed that $c_s>0.6$. It suggests that the DBI inflation model is subject to the following constraint:
 \bea -0.89\frac{\Lambda^4}{f(\bar{\phi}_0)}<\bar{X}<0.\eea
The following statement now provides the estimated PBH mass in context of the DBI model:
\bea \left(M_{\rm PBH}\right)_{\rm DBI}=9.04\times 10^2\times \frac{1}{\displaystyle \bigg(1-2\frac{f(\bar{\phi}_0)}{\Lambda^4}\bar{X}\bigg)}\times \bigg(\frac{\gamma}{0.2}\bigg)\bigg(\frac{g_*}{106.75}\bigg)^{-1/6}{\rm gm},\eea 
where $p_*=0.02\;{\rm Mpc}^{-1}$ is the pivot scale, and $k_s=10^{21}\;{\rm Mpc}^{-1}$ is fixed. Now, using the DBI inflationary model and adjusting the efficiency factor $\gamma=0.2$ and degrees of freedom $g_*=106.75$, we obtain the PBH mass constraint as follows:
\bea 0.36\times 10^2{\rm gm}<\left(M_{\rm PBH}\right)_{\rm DBI}<9.04\times 10^2{\rm gm}.\eea

\subsection{Tachyon model} 
The Tachyon model is the subject of our next discussion, and for it, the $P(X,\phi)$ function is expressed as follows:
\bea P(X,\phi)=-V(\phi)\sqrt{1-2\alpha^{'}X}\quad\quad {\rm where}\quad\quad X=-\frac{1}{2}\left(\partial_{\mu}\phi\right)^2.\eea
The string tension $T$ is inversely represented by the Regge slope parameter $\alpha^{'}$ in this case, and the features of the String Theory model under discussion determine $V(\phi)$. The EFT coefficient $M_2$ and the effective sound speed $c_s$ may be calculated from this model as follows:
 \bea c_s=\frac{1}{\displaystyle \sqrt{1-2\alpha^{'}\bar{X}}}\quad\quad {\rm and}\quad \quad \frac{M^4_2}{\dot{H}M^2_{pl}}=\alpha^{'}\bar{X}.\eea 
As anticipated, the effective sound speed for this model must be smaller than unity; yet, based on the study we conducted for this work, $c_s>0.6$. That suggests the Tachyon inflation model is subject to the following constraint:
 \bea -\frac{0.89}{\alpha^{'}}<\bar{X}<0.\eea
The following equation provides the estimate of PBH mass in the context of the Tachyon model:
\bea \left(M_{\rm PBH}\right)_{\rm Tachyon}=9.04\times 10^2\times \frac{1}{\displaystyle \bigg(1-2\alpha^{'}\bar{X}\bigg)}\times \bigg(\frac{\gamma}{0.2}\bigg)\bigg(\frac{g_*}{106.75}\bigg)^{-1/6}{\rm gm},\eea 
where $p_*=0.02\;{\rm Mpc}^{-1}$ is the pivot scale, and $k_s=10^{21}\;{\rm Mpc}^{-1}$ is fixed. Now, using the Tachyon inflationary model to set the efficiency factor $\gamma=0.2$ and degrees of freedom $g_*=106.75$, we obtain the PBH mass constraint as follows:
\bea 0.36\times 10^2{\rm gm}<\left(M_{\rm PBH}\right)_{\rm Tachyon}<9.04\times 10^2{\rm gm}.\eea

\subsection{GTachyon model} 
Subsequently, we will talk about the GTachyon model, for which the $P(X,\phi)$ function has the following expression:
 \bea P(X,\phi)=-V(\phi)\left(1-2\alpha^{'}X\right)^{q}\quad\quad {\rm where}\quad\quad X=-\frac{1}{2}\left(\partial_{\mu}\phi\right)^2.\eea
We take this as the generalized form of the Tachyon model that may be obtained from the setup of String Theory. Under this concept, inflation can occur in the following branches of the index $q$ values: (1) $q<1/2$, (2) $1/2<q<1$, and (3) $q>1$ under the current setting. The outcome for $q=1/2$ will precisely line up with the Tachyon model's predictions, which were previously described. This model allows for the evaluation of the EFT coefficient $M_2$ and the effective sound speed $c_s$:
\bea c_s=\sqrt{\frac{ \displaystyle 1+2\alpha^{'}(1-2q)\bar{X}}{\displaystyle 1-2\alpha^{'}\bar{X} }}\quad\quad {\rm and}\quad \quad \frac{M^4_2}{\dot{H}M^2_{pl}}=\bigg(\frac{2\alpha^{'}(1-q)\bar{X}}{1+2\alpha^{'}(1-2q)\bar{X}}\bigg).\eea 
It is anticipated that for (1) $q<1/2$ and (2) $1/2<q<1$, the effective sound speed for this model must be smaller than unity. Based on our study in this work, we discovered that $c_s>0.6$. It suggests that the following restriction is true for the $q$ parameter in the GTachyon model of inflation, which is defined in the following regions: $q<1/2$ and $1/2<q<1$. This constraint is given by:
\bea -\frac{0.89}{2\alpha^{'}\left[\left(1-q\right)+0.89\left(1-2q\right)\right]}<\bar{X}<0 \quad\quad\quad\quad {\rm for}\quad q<\frac{1}{2}\quad {\rm and}\quad \frac{1}{2}<q<1.\eea
It can, however, defy causality and travel faster than the speed of sound when $q>1$. It suggests that the GTachyon inflation model is subject to the following restriction for the $q$ parameter, which is defined in the area $q>1$, as follows:
 \bea 0<\bar{X}<\frac{0.28}{2\alpha^{'}\left[\left(1-q\right)-0.28\left(1-2q\right)\right]} \quad\quad\quad\quad {\rm for}\quad q>1.\eea
The estimation of PBH mass inside the GTachyon model may be expressed as follows, with respect to any permitted values of the $q$ parameter for GTachyon:
\bea \left(M_{\rm PBH}\right)_{\rm GTachyon}=9.04\times 10^2\times \bigg(\frac{ \displaystyle 1+2\alpha^{'}(1-2q)\bar{X}}{\displaystyle 1-2\alpha^{'}\bar{X} }\bigg)\times \bigg(\frac{\gamma}{0.2}\bigg)\bigg(\frac{g_*}{106.75}\bigg)^{-1/6}{\rm gm},\eea 
where the pivot scale is set at $p_*=0.02\;{\rm Mpc}^{-1}$ and $k_s=10^{21}\;{\rm Mpc}^{-1}$. Upon adjusting the efficiency factor $\gamma=0.2$ and degrees of freedom $g_*=106.75$, the GTachyon inflationary model yields the subsequent constraint on the PBH mass:
\bea && 0.36\times 10^2{\rm gm}<\left(M_{\rm PBH}\right)_{\rm GTachyon}<9.04\times 10^2{\rm gm}\quad\quad\quad\quad {\rm for}\quad q<\frac{1}{2}\quad {\rm and}\quad \frac{1}{2}<q<1,\\
&& 9.04\times 10^2{\rm gm}<\left(M_{\rm PBH}\right)_{\rm GTachyon}<2.03\times 10^3{\rm gm}\quad\quad\quad\quad {\rm for}\quad q>1.\eea

\subsection{$K$ inflation model} 
Subsequently, we will examine the $K$ inflation model, in which the $P(X,\phi)$ function may be expressed as follows:
 \bea P(X,\phi)=K(X)-V(\phi)\quad\quad {\rm where}\quad\quad X=-\frac{1}{2}\left(\partial_{\mu}\phi\right)^2.\eea
In this case, the effective potential $V(\phi)$ is supplied and will correspond to inflation in a single field. In contrast, the generalized kinetic term $K(X)$ is a generic function of $X$, and it is compatible with observational limitations to allow this function with any arbitrary form. In this instance, we examine a particular illustration of this function, denoted by the subsequent relation:
    \bea K(X)=\gamma_n X^n,\eea
where the $n$-th contribution to the kinetic interaction is represented by the integer $n$ and the overall dimensionfull constant $\gamma_n$. The simplest model, which appears as a monomial kind of kinetic interaction in the present discussion, does not consider the summation over index $n$.

This model allows for the evaluation of the EFT coefficient $M_2$ and the effective sound speed $c_s$:
\bea c_s=\frac{1}{\sqrt{2n-1}}\quad\quad {\rm and}\quad \quad \frac{M^4_2}{\dot{H}M^2_{pl}}=\bigg(\frac{n-1}{2n-1}\bigg).\eea 
Here, there is an extra limitation on this index $n$ that must be satisfied in order to have a genuine, non-trivial effective sound speed:
    \bea n>\frac{1}{2}.\eea
It is anticipated that for a given value of $n$, the effective sound speed for this model must be smaller than unity; nevertheless, based on our study presented in this work, $c_s>0.6$. It suggests that the $K$-inflation model is subject to the following restriction, which is provided by:
 \bea 1.89>n>1.\eea
On the other hand, causality is broken and sound speed is exceeded beyond unity when:
 \bea 1>n>0.72.\eea
With respect to any permitted value of $n$ parameter, the estimation of PBH mass within the framework of the $K$-inflation model is provided by the subsequent expression:
\bea \left(M_{\rm PBH}\right)_{\rm K}=9.04\times 10^2\times \bigg(\frac{ 1}{\displaystyle 2n-1}\bigg)\times \bigg(\frac{\gamma}{0.2}\bigg)\bigg(\frac{g_*}{106.75}\bigg)^{-1/6}{\rm gm},\eea 
where the pivot scale is set at $p_*=0.02\;{\rm Mpc}^{-1}$ and $k_s=10^{21}\;{\rm Mpc}^{-1}$. Upon adjusting the efficiency factor $\gamma=0.2$ and degrees of freedom $g_*=106.75$, the GTachyon inflationary model yields the subsequent constraint on the PBH mass:
\bea && 9.04\times 10^2{\rm gm}<\left(M_{\rm PBH}\right)_{\rm K}<2.51\times 10^3{\rm gm}\quad\quad\quad\quad {\rm for}\quad 1>n>0.68,\\
&& 3.96\times 10^2{\rm gm}<\left(M_{\rm PBH}\right)_{\rm K}<9.04\times 10^2{\rm gm}\quad\quad\quad\quad {\rm for}\quad 1.64>n>1.\eea

    \subsection{Canonical single field model}
Lastly, we talk about the standard single field inflation model, in which the expression presented below represents the $P(X,\phi)$ function:
 \bea P(X,\phi)=X-V(\phi)\quad\quad {\rm where}\quad\quad X=-\frac{1}{2}\left(\partial_{\mu}\phi\right)^2.\eea
In this case, the effective potential $V(\phi)$ is supplied and will correspond to inflation in a single field. 
This model allows for the evaluation of the EFT coefficient $M_2$ and the effective sound speed $c_s$:
    \bea c_s=1\quad\quad {\rm and}\quad \quad \frac{M^4_2}{\dot{H}M^2_{pl}}=0.\eea 
The estimate of PBH mass in the framework of the standard single-field inflation model is now provided by the following expression:
\bea \left(M_{\rm PBH}\right)_{\rm K}=9.04\times 10^2\times \bigg(\frac{\gamma}{0.2}\bigg)\bigg(\frac{g_*}{106.75}\bigg)^{-1/6}{\rm gm},\eea 
where the pivot scale is set at $p_*=0.02\;{\rm Mpc}^{-1}$ and $k_s=10^{21}\;{\rm Mpc}^{-1}$. After adjusting for the efficiency factor $\gamma=0.2$ and degrees of freedom $g_*=106.75$, the PBH mass from the standard single field inflationary model is as follows:
\bea \left(M_{\rm PBH}\right)_{\rm K}=9.04\times 10^2{\rm gm}.\eea

\section{One-loop momentum integrals for EFT of single field inflation}
\label{A3a}

We want to explicitly analyze the momentum integral contributions in this appendix. Specifically, we use them to measure the one-loop contribution to the primordial scalar power spectrum from the SRI, SRII, and USR area at the late time scale, $\tau\rightarrow 0$. 

\subsection{Region I: First Slow Roll (SRI) region}

We will examine the following integral that arises when the one-loop correction to the primordial power spectrum of scalar modes in the SRI area is computed. Let's assess the integral that follows:
\bea\label{intSR1} {\bf K}(\tau):=\int^{k_e}_{p_*}\frac{dk}{k}\;\left|{\cal M}_{k}(\tau)\right|^{2},\eea
where the following is the definition of a new function ${\cal M}_{ k}(\tau)$:
		\bea {\cal M}_{ k}(\tau)&=&\left(1+ikc_s\tau\right)\; e^{-ikc_s\tau}.\eea
The following outcome is obtained by replacing the precise version of the previously mentioned function ${\cal M}_{ k}(\tau)$ in equation (\ref{intSR1}):
\bea{\bf K}(\tau)=\int^{k_e}_{p_*}\frac{dk}{k}\;\left(1+k^2c^2_s\tau^2\right)=\bigg[\ln\left(\frac{k_e}{p_*}\right)+\frac{1}{2}\left(k^2_e-p^2_*\right)c^2_s\tau^2\bigg].\eea
The following simplified result was then discovered in the super-horizon late time limiting scale:
\bea {\bf K}(\tau_e)=\ln\left(\frac{k_e}{p_*}\right)=\ln\left(\frac{k_{\rm UV}}{p_*}\right).\eea
In this instance, $p_*$ denotes the pivot scale in the SRI phase, which in the current framework under discussion is anticipated to be $p_*\ll k_s$. The final one-loop outcome in the SRI region will solely be influenced by the logarithmically divergent contribution indicated previously.

\subsection{Region II: Ultra Slow Roll (USR) region}

\subsubsection{First contribution}
Initially, we will express the contribution using a momentum-dependent integral that is present for the USR area:
\bea  \label{gk1z} &&{\bf I}(\tau):=\int^{k_e}_{k_s}\frac{dk}{k}\;\left|{\cal G}_{ k}(\tau)\right|^{2},\eea
where the following is the definition of a new function, ${\cal G}_{ k}(\tau)$:
\bea \label{hhgxz} {\cal G}_{\bf k}(\tau)&=&\bigg[\alpha^{(2)}_{ k}\left(1+ikc_s\tau\right)\; e^{-ikc_s\tau}-\beta^{(2)}_{ k}\left(1-ikc_s\tau\right)\; e^{ikc_s\tau}\bigg].\eea
In this instance, the following equations represent the Bogoliubov coefficients in the USR area, $\beta^{(2)}_{ k}$ and $\alpha^{(2)}_{ k}$:
 \bea \alpha^{(2)}_{ k}&=&1-\frac{3}{2ik^{3}c^{3}_s\tau^{3}_s}\left(1+k^{2}c^{2}_s\tau^{2}_s\right),
  \quad\quad\quad
\beta^{(2)}_{ k}=-\frac{3}{2ik^{3}c^{3}_s\tau^{3}_s}\left(1+ikc_s\tau_s\right)^{2}\; e^{-2ikc_s\tau_s}.\eea
Following the substitution of the function ${\cal G}_{ k}(\tau)$ in its particular form in equation (\ref{gk1z}), we obtain the simplified relation as follows:
\bea \label{gk2z} &&{\bf I}(\tau)= {\bf I}_1(\tau)+{\bf I}_2(\tau)+{\bf I}_3(\tau)+{\bf I}_4(\tau),\eea
where ${\bf I}_i(\tau)\forall i=1,2,3,4$, the four individual contributions, may be explicitly calculated and defined as follows:
\bea {\bf I}_1(\tau)&=&\int^{k_e}_{k_s}\frac{dk}{k}\;\bigg(1+\frac{9}{4}\frac{\left(1+k^2c^2_s\tau^2_s\right)^2}{k^6c^6_s\tau^6_s}\bigg) \left(1+k^2c^2_s\tau^2\right),\\
{\bf I}_2(\tau)&=&\int^{k_e}_{k_s}\frac{dk}{k}\;\bigg(\frac{9}{4}\frac{\left(1+k^2c^2_s\tau^2_s\right)^2}{k^6c^6_s\tau^6_s}\bigg) \left(1+k^2c^2_s\tau^2\right),\\
{\bf I}_3(\tau)&=&-\int^{k_e}_{k_s}\frac{dk}{k}\;\bigg(1+\frac{3}{2i}\frac{\left(1+k^2c^2_s\tau^2_s\right)}{k^3c^3_s\tau^3_s}\bigg)\bigg(-\frac{3}{2i} \frac{\left(1+ikc_s\tau_s\right)^2}{k^3c^3_s\tau^3_s}\bigg)\left(1-ikc_s\tau\right)^2\; e^{2ikc_s(\tau-\tau_s)},\\
{\bf I}_4(\tau)&=&-\int^{k_e}_{k_s}\frac{dk}{k}\;\bigg(1-\frac{3}{2i}\frac{\left(1+k^2c^2_s\tau^2_s\right)}{k^3c^3_s\tau^3_s}\bigg)\bigg(\frac{3}{2i} \frac{\left(1-ikc_s\tau_s\right)^2}{k^3c^3_s\tau^3_s}\bigg)\left(1+ikc_s\tau\right)^2\; e^{-2ikc_s(\tau-\tau_s)}.\eea
Upon executing many algebraic operations, we obtain the subsequent outcomes for the aforementioned integrals:
\bea \label{c1x}{\bf I}_1(\tau)&=&\bigg[\frac{1}{2}\left(k^2_e-k^2_s\right)c^2_s\tau^2+\left(1+\frac{9}{4}\left(\frac{\tau}{\tau_s}\right)^2\right) \ln\left(\frac{k_e}{k_s}\right)-\frac{9}{8}\frac{1}{c^4_s\tau^4_s}\bigg(1+\left(\frac{\tau}{\tau_s}\right)^2\bigg)\bigg(\frac{1}{k^4_e}-\frac{1}{k^4_s}\bigg)\nonumber\\
&&\quad\quad\quad\quad\quad\quad\quad\quad\quad\quad\quad\quad -\frac{9}{8}\frac{1}{c^2_s\tau^2_s}\bigg(1+2\left(\frac{\tau}{\tau_s}\right)^2\bigg)\bigg(\frac{1}{k^2_e}-\frac{1}{k^2_s}\bigg)-\frac{3}{8}\frac{1}{c^6_s\tau^6_s}\bigg(\frac{1}{k^6_e}-\frac{1}{k^6_s}\bigg)\bigg],\\
\label{c2x}{\bf I}_2(\tau)&=&\bigg[\frac{9}{4}\left(\frac{\tau}{\tau_s}\right)^2 \ln\left(\frac{k_e}{k_s}\right)-\frac{9}{8}\frac{1}{c^4_s\tau^4_s}\bigg(1+\frac{1}{2}\left(\frac{\tau}{\tau_s}\right)^2\bigg)\bigg(\frac{1}{k^4_e}-\frac{1}{k^4_s}\bigg)\nonumber\\
&&\quad\quad\quad\quad\quad\quad\quad\quad\quad\quad\quad\quad -\frac{9}{8}\frac{1}{c^2_s\tau^2_s}\bigg(1+2\left(\frac{\tau}{\tau_s}\right)^2\bigg)\bigg(\frac{1}{k^2_e}-\frac{1}{k^2_s}\bigg)-\frac{3}{8}\frac{1}{c^6_s\tau^6_s}\bigg(\frac{1}{k^6_e}-\frac{1}{k^6_s}\bigg)\bigg],\\
\label{c3x}{\bf I}_3(\tau)&=&-\frac{1}{16 c^6_s \tau _s^6}\bigg[\bigg(36 c^6_s \tau^2 \tau _s^4+8 c^6_s \tau _s^6-8 c^6_s \tau^6 \bigg)\bigg(\text{Ei}\left(2 i c_s k_e \left(\tau-\tau _s\right)\right)-\text{Ei}\left(2 i c_s k_s \left(\tau-\tau _s\right)\right)\bigg)\nonumber\\
&&+ \bigg(\frac{12 c^6_s \tau^2 \tau _s^5}{\tau-\tau _s}\bigg(e^{2 i c_s k_e \left(\tau-\tau _s\right)}-e^{2 i c_s k_s \left(\tau-\tau _s\right)}\bigg)\nonumber\\
&&+4 i c^5_s \left(\tau^4 \tau _s+\tau^3 \tau _s^2+\tau^2 \tau _s^3+7 \tau \tau _s^4+\tau^5+\tau _s^5\right)\bigg(\frac{e^{2 i c_s k_s \left(\tau-\tau _s\right)}}{k_s}-\frac{e^{2 i c_s k_e \left(\tau-\tau _s\right)}}{k_e}\bigg)\nonumber\\
&&+2 c^4_s \left(2 \tau^3 \tau _s+3 \tau^2 \tau _s^2+28 \tau \tau _s^3+\tau^4-7 \tau _s^4\right)\bigg(\frac{e^{2 i c_s k_s \left(\tau-\tau _s\right)}}{k^2_s}-\frac{e^{2 i c_s k_e \left(\tau-\tau _s\right)}}{k^2_e}\bigg)\nonumber\\
&&-2 i c^3_s \left(3 \tau^2 \tau _s+6 \tau \tau _s^2+\tau^3-14 \tau _s^3\right)\bigg(\frac{e^{2 i c_s k_s \left(\tau-\tau _s\right)}}{k^3_s}-\frac{e^{2 i c_s k_e \left(\tau-\tau _s\right)}}{k^3_e}\bigg)\nonumber\\
&&-3 c^2_s \left(-8 \tau \tau _s+\tau^2-2 \tau _s^2\right)\bigg(\frac{e^{2 i c_s k_s \left(\tau-\tau _s\right)}}{k^4_s}-\frac{e^{2 i c_s k_e \left(\tau-\tau _s\right)}}{k^4_e}\bigg)\nonumber\\
&&-12 i c_s \left(\tau-\tau _s\right)\bigg(\frac{e^{2 i c_s k_s \left(\tau-\tau _s\right)}}{k^5_s}-\frac{e^{2 i c_s k_e \left(\tau-\tau _s\right)}}{k^5_e}\bigg)+6\bigg(\frac{e^{2 i c_s k_s \left(\tau-\tau _s\right)}}{k^6_s}-\frac{e^{2 i c_s k_e \left(\tau-\tau _s\right)}}{k^6_e}\bigg)\bigg)\bigg],\\
\label{c4x}{\bf I}_4(\tau)&=&-\frac{1}{16 c^6_s \tau _s^6}\bigg[\bigg(36 c^6_s \tau^2 \tau _s^4+8 c^6_s \tau _s^6-8 c^6_s \tau^6 \bigg)\bigg(\text{Ei}\left(-2 i c_s k_e \left(\tau-\tau _s\right)\right)-\text{Ei}\left(-2 i c_s k_s \left(\tau-\tau _s\right)\right)\bigg)\nonumber\\
&&+ \bigg(\frac{12 c^6_s \tau^2 \tau _s^5}{\tau-\tau _s}\bigg(e^{-2 i c_s k_e \left(\tau-\tau _s\right)}-e^{-2 i c_s k_s \left(\tau-\tau _s\right)}\bigg)\nonumber\\
&&-4 i c_s^5 \left(\tau ^4 \tau _s+\tau ^3 \tau _s^2+\tau ^2 \tau _s^3+7 \tau  \tau _s^4+\tau _s^5+\tau ^5\right)\bigg(\frac{e^{-2 i c_s k_s \left(\tau-\tau _s\right)}}{k_s}-\frac{e^{-2 i c_s k_e \left(\tau-\tau _s\right)}}{k_e}\bigg)\nonumber\\
&&+2 c_s^4 \left(2 \tau ^3 \tau _s+3 \tau ^2 \tau _s^2+28 \tau  \tau _s^3-7 \tau _s^4+\tau ^4\right)\bigg(\frac{e^{-2 i c_s k_s \left(\tau-\tau _s\right)}}{k^2_s}-\frac{e^{-2 i c_s k_e \left(\tau-\tau _s\right)}}{k^2_e}\bigg)\nonumber\\
&&+2 i c_s^3 \left(3 \tau ^2 \tau _s+6 \tau  \tau _s^2-14 \tau _s^3+\tau ^3\right)\bigg(\frac{e^{-2 i c_s k_s \left(\tau-\tau _s\right)}}{k^3_s}-\frac{e^{-2 i c_s k_e \left(\tau-\tau _s\right)}}{k^3_e}\bigg)\nonumber\\
&&-3 c_s^2 \left(-8 \tau  \tau _s-2 \tau _s^2+\tau ^2\right)\bigg(\frac{e^{-2 i c_s k_s \left(\tau-\tau _s\right)}}{k^4_s}-\frac{e^{-2 i c_s k_e \left(\tau-\tau _s\right)}}{k^4_e}\bigg)\nonumber\\
&&+12 i c_s \left(\tau -\tau _s\right)\bigg(\frac{e^{-2 i c_s k_s \left(\tau-\tau _s\right)}}{k^5_s}-\frac{e^{-2 i c_s k_e \left(\tau-\tau _s\right)}}{k^5_e}\bigg)+6\bigg(\frac{e^{-2 i c_s k_s \left(\tau-\tau _s\right)}}{k^6_s}-\frac{e^{-2 i c_s k_e \left(\tau-\tau _s\right)}}{k^6_e}\bigg)\bigg)\bigg].\eea
In order to comprehend the behaviour of the findings obtained in the USR regime, we will now sum up all of the contributions obtained in equations (\ref{c1x}), (\ref{c2x}), (\ref{c3x}), and (\ref{c4x}) together. This will lead to the results as follows after simplification:
\bea \label{c11}{\bf I}(\tau)&=&{\bf I}_1(\tau)+{\bf I}_2(\tau)+{\bf I}_3(\tau)+{\bf I}_4(\tau)\nonumber\\
&=&\Bigg\{\bigg[\frac{1}{2}\left(k^2_e-k^2_s\right)c^2_s\tau^2+\left(1+\frac{9}{2}\left(\frac{\tau}{\tau_s}\right)^2\right) \ln\left(\frac{k_e}{k_s}\right)-\frac{9}{4}\frac{1}{c^4_s\tau^4_s}\bigg(1+\frac{3}{4}\left(\frac{\tau}{\tau_s}\right)^2\bigg)\bigg(\frac{1}{k^4_e}-\frac{1}{k^4_s}\bigg)\nonumber\\
&&\quad\quad\quad\quad\quad\quad\quad\quad\quad\quad\quad\quad -\frac{9}{4}\frac{1}{c^2_s\tau^2_s}\bigg(1+2\left(\frac{\tau}{\tau_s}\right)^2\bigg)\bigg(\frac{1}{k^2_e}-\frac{1}{k^2_s}\bigg)-\frac{3}{4}\frac{1}{c^6_s\tau^6_s}\bigg(\frac{1}{k^6_e}-\frac{1}{k^6_s}\bigg)\bigg]\nonumber\eea
\bea
&&-\frac{1}{16 c^6_s \tau _s^6}\bigg[\bigg(8 c^6_s \tau^6-36 c^6_s \tau^2 \tau _s^4-8 c^6_s \tau _s^6 \bigg)\nonumber\\
&&\quad\quad\quad\quad\times\bigg(\text{Ei}\left(2 i c_s k_s \left(\tau-\tau _s\right)\right)+\text{Ei}\left(-2 i c_s k_s \left(\tau-\tau _s\right)\right)-\text{Ei}\left(2 i c_s k_e \left(\tau-\tau _s\right)\right)-\text{Ei}\left(-2 i c_s k_e \left(\tau-\tau _s\right)\right)\bigg)\nonumber\\
&&+ \bigg(\frac{24 c^6_s \tau^2 \tau _s^5}{\tau-\tau _s}\bigg(\cos \left(c_s k_e \left(\tau-\tau _s\right)\right)-\cos \left(c_s k_s \left(\tau-\tau _s\right)\right)\bigg)\nonumber\\
&&-8 c_s^5 \left(\tau ^4 \tau _s+\tau ^3 \tau _s^2+\tau ^2 \tau _s^3+7 \tau  \tau _s^4+\tau _s^5+\tau ^5\right)\bigg(\frac{\sin \left(c_s k_s \left(\tau-\tau _s\right)\right)}{k_s}-\frac{\sin \left(c_s k_e \left(\tau-\tau _s\right)\right)}{k_e}\bigg)\nonumber\\
&&+4 c_s^4 \left(2 \tau ^3 \tau _s+3 \tau ^2 \tau _s^2+28 \tau  \tau _s^3-7 \tau _s^4+\tau ^4\right)\bigg(\frac{\cos \left(c_s k_s \left(\tau-\tau _s\right)\right)}{k^2_s}-\frac{\cos \left(c_s k_e \left(\tau-\tau _s\right)\right)}{k^2_e}\bigg)\nonumber\\
&&+4 c_s^3 \left(3 \tau ^2 \tau _s+6 \tau  \tau _s^2-14 \tau _s^3+\tau ^3\right)\bigg(\frac{\sin \left(c_s k_s \left(\tau-\tau _s\right)\right)}{k^3_s}-\frac{\sin \left(c_s k_e \left(\tau-\tau _s\right)\right)}{k^3_e}\bigg)\nonumber\\
&&-6 c_s^2 \left(-8 \tau  \tau _s-2 \tau _s^2+\tau ^2\right)\bigg(\frac{\cos \left(c_s k_s \left(\tau-\tau _s\right)\right)}{k^4_s}-\frac{\cos \left(c_s k_e \left(\tau-\tau _s\right)\right)}{k^4_e}\bigg)\nonumber\\
&&+24 c_s \left(\tau -\tau _s\right)\bigg(\frac{\sin \left(c_s k_s \left(\tau-\tau _s\right)\right)}{k^5_s}-\frac{\sin \left(c_s k_e \left(\tau-\tau _s\right)\right)}{k^5_e}\bigg)\nonumber\\
&&+6\bigg(\frac{\cos \left(c_s k_s \left(\tau-\tau _s\right)\right)}{k^6_s}-\frac{\cos \left(c_s k_e \left(\tau-\tau _s\right)\right)}{k^6_e}\bigg)\bigg)\bigg]\Bigg\}.\eea
We have discovered the following result in the limiting case of scales at late time in the super-horizon, which will help with the one-loop integral:
\bea {\bf I}(\tau_e)={\bf I}(\tau_s)\approx {\cal O}(1)+\ln\left(\frac{k_e}{k_s}\right)\approx \ln\left(\frac{k_e}{k_s}\right)=\ln\left(\frac{k_{\rm UV}}{k_{\rm IR}}\right).\eea
In order to get this streamlined outcome, we have disregarded the contributions from the oscillatory and inverse power law components of $(k_s/k_e)(\ll 1)$, as none of these can meaningfully alter the general behavior of the scalar power spectrum throughout the USR regime. Furthermore, for the same reason, a contribution from ${\cal O}(1)$ is ignored. One other significant factor that is vital for ignoring the ${\cal O}(1)$ contribution from the previously discussed finding is that it comes from a deeper physical base.
The ${\cal O}(1)$ impact is shown to be inconsistent with the contributions that emerge in the sub-horizon, horizon crossing, and super-horizon. Since the sub-horizon scale, where quantum effects predominate, absolutely lacks such an impact. Such an effect appears from the super-horizon scale to the horizon crossover. However, such input from our computation may be promptly disregarded in order to have a consistent, trustworthy contribution across all sizes. The only significant influence is from the logarithmically divergent contribution, which is ultimately shown in the one-loop corrected primordial power spectrum results for scalar perturbation modes during the USR era. Importantly, we have limited the momentum integration within a range, $k_s<k<k_e$, by introducing two well-known physical cut-offs: the IR cut-off $k_{\rm IR}=k_s$ and the UV cut-off $k_{\rm UV}=k_e$. This enables us to calculate the finite contributions from each of the mentioned integrals.

\subsubsection{Second contribution}
The sub-leading contribution may be expressed as follows using a momentum-dependent integral that appears in the USR region:
\bea  \label{gslk1xz} &&{\bf F}(\tau):=\int^{k_e}_{k_s}\frac{dk}{k}\;|{\cal G}_{ k}(\tau)|^2\bigg(\frac{d\ln|{\cal G}_{ k}(\tau)|^2 }{d\ln k}\bigg)=\int^{k_e}_{k_s}d\ln k\;\bigg(\frac{d|{\cal G}_{ k}(\tau)|^2 }{d\ln k}\bigg)=\Bigg[|{\cal G}_{ k}(\tau)|^2\Bigg]^{k_e}_{k_s},\eea
where the function ${\cal G}_{ k}$ has previously been defined in equation (\ref{hhgxz}). Equation (\ref{gslk1xz}) was solved using the particular expression of the function ${\cal G}_{ k}(\tau)$, and the resulting simplified relation was found:
\bea \label{gksim2xz} &&{\bf F}(\tau)= {\bf F}_1(\tau)+{\bf F}_2(\tau)+{\bf F}_3(\tau)+{\bf F}_4(\tau),\eea
where ${\bf F}_i(\tau)\forall i=1,2,3,4$, the four individual contributions, may be explicitly calculated and defined as follows:
\bea {\bf F}_1(\tau)&=&\Bigg[\bigg(1+\frac{9}{4}\frac{\left(1+k^2_ec^2_s\tau^2_s\right)^2}{k^6_ec^6_s\tau^6_s}\bigg) \left(1+k^2_ec^2_s\tau^2\right)-\bigg(1+\frac{9}{4}\frac{\left(1+k^2_sc^2_s\tau^2_s\right)^2}{k^6_sc^6_s\tau^6_s}\bigg) \left(1+k^2_sc^2_s\tau^2\right)\Bigg],\\
{\bf F}_2(\tau)&=&\Bigg[\bigg(\frac{9}{4}\frac{\left(1+k^2_ec^2_s\tau^2_s\right)^2}{k^6_ec^6_s\tau^6_s}\bigg) \left(1+k^2_ec^2_s\tau^2\right)-\bigg(\frac{9}{4}\frac{\left(1+k^2_sc^2_s\tau^2_s\right)^2}{k^6_sc^6_s\tau^6_s}\bigg) \left(1+k^2_sc^2_s\tau^2\right)\Bigg],\\
{\bf F}_3(\tau)&=&\Bigg[\bigg(1+\frac{3}{2i}\frac{\left(1+k^2_sc^2_s\tau^2_s\right)}{k^3_sc^3_s\tau^3_s}\bigg)\bigg(-\frac{3}{2i} \frac{\left(1+ik_sc_s\tau_s\right)^2}{k^3_sc^3_s\tau^3_s}\bigg)\left(1-ik_sc_s\tau\right)^2\; e^{2ik_sc_s(\tau-\tau_s)}\nonumber\\
&&\quad\quad\quad\quad\quad\quad-\bigg(1+\frac{3}{2i}\frac{\left(1+k^2_ec^2_s\tau^2_s\right)}{k^3_ec^3_s\tau^3_s}\bigg)\bigg(-\frac{3}{2i} \frac{\left(1+ik_ec_s\tau_s\right)^2}{k^3_ec^3_s\tau^3_s}\bigg)\left(1-ik_ec_s\tau\right)^2\; e^{2ik_ec_s(\tau-\tau_s)}\Bigg],\\
{\bf F}_4(\tau)&=&\Bigg[\bigg(1-\frac{3}{2i}\frac{\left(1+k^2_sc^2_s\tau^2_s\right)}{k^3_sc^3_s\tau^3_s}\bigg)\bigg(\frac{3}{2i} \frac{\left(1-ik_sc_s\tau_s\right)^2}{k^3_sc^3_s\tau^3_s}\bigg)\left(1+ik_sc_s\tau\right)^2\; e^{-2ik_sc_s(\tau-\tau_s)}\nonumber\\
&&\quad\quad\quad\quad\quad\quad-\bigg(1-\frac{3}{2i}\frac{\left(1+k^2_ec^2_s\tau^2_s\right)}{k^3_ec^3_s\tau^3_s}\bigg)\bigg(\frac{3}{2i} \frac{\left(1-ik_ec_s\tau_s\right)^2}{k^3_ec^3_s\tau^3_s}\bigg)\left(1+ik_ec_s\tau\right)^2\; e^{-2ik_ec_s(\tau-\tau_s)}\Bigg].\eea
For the aforementioned functions, we have now discovered the following findings on the super-horizon late time limiting scale:
\bea {\bf F}_1&=&\frac{9}{4}\Bigg[\frac{\left(1+k^2_ec^2_s\tau^2_s\right)^2}{k^6_ec^6_s\tau^6_s}-\frac{\left(1+k^2_sc^2_s\tau^2_s\right)^2}{k^6_sc^6_s\tau^6_s}\Bigg],\\
{\bf F}_2&=&\frac{9}{4}\Bigg[\frac{\left(1+k^2_ec^2_s\tau^2_s\right)^2}{k^6_ec^6_s\tau^6_s}-\frac{\left(1+k^2_sc^2_s\tau^2_s\right)^2}{k^6_sc^6_s\tau^6_s}\Bigg],\\
{\bf F}_3&=&-\Bigg[\bigg(1+\frac{3}{2i}\frac{\left(1+k^2_ec^2_s\tau^2_s\right)}{k^3_ec^3_s\tau^3_s}\bigg)\bigg(-\frac{3}{2i} \frac{\left(1+ik_ec_s\tau_s\right)^2}{k^3_ec^3_s\tau^3_s}\bigg)\; e^{-2ik_ec_s\tau_s}\nonumber\\
&&\quad\quad\quad\quad\quad\quad-\bigg(1+\frac{3}{2i}\frac{\left(1+k^2_sc^2_s\tau^2_s\right)}{k^3_sc^3_s\tau^3_s}\bigg)\bigg(-\frac{3}{2i} \frac{\left(1+ik_sc_s\tau_s\right)^2}{k^3_sc^3_s\tau^3_s}\bigg)\; e^{-2ik_sc_s\tau_s}\Bigg],\\
{\bf F}_4&=&-\Bigg[\bigg(1-\frac{3}{2i}\frac{\left(1+k^2_ec^2_s\tau^2_s\right)}{k^3_ec^3_s\tau^3_s}\bigg)\bigg(\frac{3}{2i} \frac{\left(1-ik_ec_s\tau_s\right)^2}{k^3_ec^3_s\tau^3_s}\bigg)\; e^{2ik_ec_s\tau_s}\nonumber\\
&&\quad\quad\quad\quad\quad\quad-\bigg(1-\frac{3}{2i}\frac{\left(1+k^2_sc^2_s\tau^2_s\right)}{k^3_sc^3_s\tau^3_s}\bigg)\bigg(\frac{3}{2i} \frac{\left(1-ik_sc_s\tau_s\right)^2}{k^3_sc^3_s\tau^3_s}\bigg)\; e^{2ik_sc_s\tau_s}\Bigg].\eea
Summing up these contributions for the super-Horizon scale and doing some algebra, we get the following result after simplification:
\bea {\bf F}(\tau_e)={\bf F}(\tau_s)=-{\cal O}(1).\eea
Because none of the contributions from the oscillatory components or the inverse power law terms of $(k_s/k_e)(\ll 1)$ will be able to significantly modify the overall behaviour of the scalar power spectrum at the sub-leading order in the USR regime, we have ignored them in order to arrive at this simplified result. Above all, this specific contribution is entirely devoid of any kind of divergencies, which is why it is suppressed when compared to the leading logarithmically divergent contribution, and the effective sound speed is $c_s<1$. Conversely, when $c_s>1$, as calculated in the preceding subsection, it can be demonstrated that the second contribution outweighs the first.

\subsubsection{Relative strength of these contributions}

In order to comprehend the difference in amplitude between these two contributions, we will now calculate the ratio of the momentum integrals, which can be obtained using the super-horizon scale formula shown below: 
\bea \frac{{\bf F}(\tau_e)}{{\bf I}(\tau_e)}=\frac{{\bf F}(\tau_s)}{{\bf I}(\tau_s)}=\frac{{\bf F}}{{\bf I}}\approx\frac{{\cal O}(1)}{\displaystyle\ln\left(\frac{k_s}{k_e}\right)}.\eea
We select $k_e=10^{22}{\rm Mpc}^{-1}$ and $k_s=10^{21}{\rm Mpc}^{-1}$ for numerical purposes, and $k_s/k_e=10^{-1}\ll 1$ for each. As a result, we obtain the subsequent numerical input from the matching ratio for the specified wave number selections, $k_e$ and $k_s$:
\bea \frac{{\bf F}(\tau_e)}{{\bf I}(\tau_e)}=\frac{{\bf F}(\tau_s)}{{\bf I}(\tau_s)}=\frac{{\bf F}}{{\bf I}}\approx-0.43<1.\eea
The integral ${\bf F}$ is suppressed relative to the contribution from the integral ${\bf I}$, which is accurately justified by this calculation.
Furthermore, this estimate indicates that ${\bf I}$ will get an additional adjustment from the contribution from ${\bf F}$. Nonetheless, it appears that the adjustment is somewhat misleading, even though it is not by much. In order to provide an accurate estimate, one must consider the pre-factors that are included in the equation for the one-loop contribution originating from the USR regime. Let's put down the contribution that we have cited in the text section specifically to demonstrate this. We represent it from the expression as follows:
\bea {\bf Z}_1:&=&\frac{1}{4}\bigg[\Delta^{2}_{\zeta,{\bf Tree}}(p)\bigg]_{\bf SR}\times \Bigg(\frac{\left(\Delta\eta(\tau_e)\right)^2}{c^8_s} \left(\frac{k_e}{k_s}\right)^{6}-\frac{\left(\Delta\eta(\tau_s)\right)^2}{c^8_s}\Bigg)\times{\bf I},\\
{\bf Z}_2:&=&\frac{1}{2}\bigg[\Delta^{2}_{\zeta,{\bf Tree}}(p)\bigg]_{\bf SR}\times\bigg(\frac{\left(\Delta\eta(\tau_e)\right)}{c^4_s}\left(\frac{k_e}{k_s}\right)^{6}-\frac{\left(\Delta\eta(\tau_s)\right)}{c^4_s}\bigg)\times{\bf F}.\eea
In order to appropriately handle the pre-factors, we should now use the ratio of ${\bf Z}_2$ and ${\bf Z}_1$, which can be obtained using the following equation, rather than calculating the ratio of ${\bf F}$ and ${\bf I}$:
\bea \frac{{\bf Z}_2}{{\bf Z}_1}=2\times \frac{\displaystyle \bigg(\frac{\left(\Delta\eta(\tau_e)\right)}{c^4_s}\left(\frac{k_e}{k_s}\right)^{6}-\frac{\left(\Delta\eta(\tau_s)\right)}{c^4_s}\bigg)}{\displaystyle \Bigg(\frac{\left(\Delta\eta(\tau_e)\right)^2}{c^8_s} \left(\frac{k_e}{k_s}\right)^{6}-\frac{\left(\Delta\eta(\tau_s)\right)^2}{c^8_s}\Bigg)}\times \frac{{\bf F}}{{\bf I}}.\eea
Here, we choose $\Delta\eta(\tau_e)=1$ and $\Delta\eta(\tau_s)=-6$ for numerical purposes. For $k_e=10^{22}{\rm Mpc}^{-1}$ and $k_s=10^{21}{\rm Mpc}^{-1}$, we have $k_s/k_e=10^{-1}\ll 1$. These are the values we pick. Furthermore, we take into account that the sound speed will fall inside the widow, $0.6<c_s<1.5$. In order to examine the impact of sound speed, we will explore three distinct intervals: $0.6< c_s< 1$, $c_s=1$, and $1< c_s< 1.5$. Next, in the two successive areas, we discovered the following limitations on this ratio:
\bea &&\underline{{\bf For}\;\; 0.6< c_s< 1 :}\quad\quad 0.12<\frac{{\bf Z}_2}{{\bf Z}_1}<0.86,\\
&&\underline{{\bf For}\;\; c_s= 1 :}\quad\quad\quad\quad\quad\quad\quad\;\;\;\frac{{\bf Z}_2}{{\bf Z}_1}\sim 0.86,\\
&&\underline{{\bf For}\;\; 1< c_s< 1.5 :}\quad\quad 0.86<\frac{{\bf Z}_2}{{\bf Z}_1}<4.35.\eea
\subsection{ Region III: Second Slow Roll (SRII) region}
Initially, we express the contribution using a momentum-dependent integral that is present in the SRII area:
\bea  \label{gkk1xz} &&{\bf O}(\tau):=\left(\frac{\tau_s}{\tau_e}\right)^6\int^{k_{\rm end}}_{k_e}\frac{dk}{k}\;\left|{\cal X}_{k}(\tau)\right|^{2},\eea
where the following is the definition of a new function, ${\cal X}_{ k}(\tau)$:
\bea {\cal X}_{ k}(\tau)&=&\bigg[\alpha^{(3)}_{ k}\left(1+ikc_s\tau\right)\; e^{-ikc_s\tau}-\beta^{(3)}_{k}\left(1-ikc_s\tau\right)\; e^{ikc_s\tau}\bigg].\eea
In this case, the Bogoliubov coefficients in the SRII area, $\alpha^{(3)}_{ k}$ and $\beta^{(3)}_{ k}$, are as follows:
\bea \alpha^{(3)}_{ k}&=&-\frac{1}{4k^6c^6_s\tau^3_s\tau^3_e}\Bigg[9\left(kc_s\tau_s-i\right)^2\left(kc_s\tau_e+i\right)^2 e^{2ikc_s(\tau_e-\tau_s)}\nonumber\\
&&\quad\quad\quad\quad\quad\quad\quad\quad\quad\quad\quad\quad\quad\quad\quad\quad+
\left\{3i-k^2c^2_s\tau^2_e\left(2kc_s\tau_e-3i\right)\right\}\left\{3i+k^2c^2_s\tau^2_s\left(2kc_s\tau_s+3i\right)\right\}\Bigg],\\
\beta^{(3)}_{ k}&=&\frac{3}{4k^6c^6_s\tau^3_s\tau^3_e}\Bigg[\left(kc_s\tau_s-i\right)^2\left\{k^2c^2_s\tau^2_e\left(3-2ikc_s\tau_e\right)+3\right\}e^{-2ikc_s\tau_s}\nonumber\\
&&\quad\quad\quad\quad\quad\quad\quad\quad\quad\quad\quad\quad\quad\quad\quad\quad+i\left(kc_s\tau_e-i\right)^2\left\{k^2c^2_s\tau^2_s\left(2kc_s\tau_s+3i\right)+3i\right\}e^{-2ikc_s\tau_e}\Bigg].\eea
We obtained the following simplified relation by replacing the particular form of the function ${\cal X}_{ k}(\tau)$ in equation (\ref{gkk1xz}):
\bea \label{gk21xz} &&{\bf O}(\tau)= \left(\frac{\tau_s}{\tau_e}\right)^6\Bigg[{\bf O}_1(\tau)+{\bf O}_2(\tau)+{\bf O}_3(\tau)+{\bf O}_4(\tau)\Bigg].\eea
The four distinct contributions ${\bf O}_i(\tau)\forall i=1,2,3,4$ may be clearly calculated and defined as follows:
\bea {\bf O}_1(\tau)&=&\int^{k_{\rm end}}_{k_e}\frac{dk}{k}\;\frac{1}{16k^{12}c^{12}_s\tau^6_s\tau^6_e}\Bigg|\Bigg[9\left(kc_s\tau_s+i\right)^2\left(kc_s\tau_e-i\right)^2 e^{2ikc_s(\tau_e-\tau_s)}\nonumber\\
&&\quad\quad\quad\quad\quad\quad\quad\quad+
\left\{3i-k^2c^2_s\tau^2_e\left(2kc_s\tau_e-3i\right)\right\}\left\{3i+k^2c^2_s\tau^2_s\left(2kc_s\tau_s+3i\right)\right\}\Bigg]\Bigg|^{2} \left(1+k^2c^2_s\tau^2\right),\\
{\bf O}_2(\tau)&=&\int^{k_{\rm end}}_{k_e}\frac{dk}{k}\;\frac{9}{16k^{12}c^{12}_s\tau^6_s\tau^6_e}\Bigg|\Bigg[\left(kc_s\tau_s-i\right)^2\left\{3+k^2c^2_s\tau^2_e\left(3-2ikc_s\tau_e\right)\right\}e^{-2ikc_s\tau_s}\nonumber\\
&&\quad\quad\quad\quad\quad\quad\quad\quad+i\left(kc_s\tau_e-i\right)^2\left\{k^2c^2_s\tau^2_s\left(2kc_s\tau_s+3i\right)+3i\right\}e^{-2ikc_s\tau_e}\Bigg]\Bigg|^{2} \left(1+k^2c^2_s\tau^2\right),\\
{\bf O}_3(\tau)&=&\int^{k_{\rm end}}_{k_e}\frac{dk}{k}\;\frac{3}{16k^{12}c^{12}_s\tau^6_s\tau^6_e}\Bigg[9\left(kc_s\tau_s+i\right)^2\left(kc_s\tau_e-i\right)^2 e^{2ikc_s(\tau_e-\tau_s)}\nonumber\\
&&\quad\quad\quad\quad\quad\quad\quad\quad+
\left\{3i-k^2c^2_s\tau^2_s\left(2kc_s\tau_e-3i\right)\right\}\left\{3i+k^2c^2_s\tau^2_s\left(2kc_s\tau_s+3i\right)\right\}\Bigg]\nonumber\eea
\bea
&&\quad\quad\quad\quad\quad\quad\quad\quad\times\Bigg[\left(kc_s\tau_s+i\right)^2\left\{3+k^2c^2_s\tau^2_e\left(3+2ikc_s\tau_e\right)\right\}e^{2ikc_s\tau_s}\nonumber\\
&&\quad\quad\quad\quad\quad\quad\quad\quad+i\left(kc_s\tau_e+i\right)^2\left\{k^2c^2_s\tau^2_s\left(2kc_s\tau_s-3i\right)-3i\right\}e^{2ikc_s\tau_e}\Bigg]\left(1-ikc_s\tau\right)^2\; e^{2ikc_s(\tau-\tau_s)},\\
{\bf O}_4(\tau)&=&\int^{k_{\rm end}}_{k_e}\frac{dk}{k}\;\frac{3}{16k^{12}c^{12}_s\tau^6_s\tau^6_e}\Bigg[9\left(kc_s\tau_s+-i\right)^2\left(kc_s\tau_e+i\right)^2 e^{-2ikc_s(\tau_e-\tau_s)}\nonumber\\
&&\quad\quad\quad\quad\quad\quad\quad\quad-
\left\{3i+k^2c^2_s\tau^2_e\left(2kc_s\tau_e+3i\right)\right\}\left\{k^2c^2_s\tau^2_s\left(2kc_s\tau_s-3i\right)-3i\right\}\Bigg]\nonumber\\
&&\quad\quad\quad\quad\quad\quad\quad\quad\times\Bigg[\left(kc_s\tau_s-i\right)^2\left\{3+k^2c^2_s\tau^2_e\left(3-2ikc_s\tau_e\right)\right\}e^{-2ikc_s\tau_s}\nonumber\\
&&\quad\quad\quad\quad\quad\quad\quad\quad+i\left(kc_s\tau_e-i\right)^2\left\{k^2c^2_s\tau^2_s\left(2kc_s\tau_s+3i\right)+3i\right\}e^{-2ikc_s\tau_e}\Bigg]\left(1+ikc_s\tau\right)^2\; e^{-2ikc_s(\tau-\tau_s)}.\quad\quad\quad\quad\eea
It is easy to compute explicit conclusions from the integral described above. Nevertheless, we refrain from quoting the specific results in full here because they need laborious phrasing. We have now discovered the following result in the limiting case of scales at late time in the super-horizon, which will help with the one-loop integral:
\bea {\bf O}(\tau_e)={\bf O}(\tau_{\rm end})\approx -\ln \left(\frac{k_e}{k_{\rm end}}\right)-\frac{27}{32}\Bigg\{1-\left(\frac{k_e}{k_{\rm end}}\right)^{12}\Bigg\}.\eea
For this reason, $k_e/k_{\rm end}\ll 1$ and similar contributions are severely suppressed in the SRII phase $k_e\ll k_{\rm end}$. We select $k_e=10^{22}{\rm Mpc}^{-1}$ and $k_{\rm end}=10^{24}{\rm Mpc}^{-1}$ for numerical purposes; for these, we have $k_e/k_{\rm end}=10^{-2}\ll 1$. The ultimate one-loop outcome in the SRII area will be governed by power law and logarithmic contributions. However, the logarithmic component predominates over the power law term because of the restriction $k_e/k_{\rm end}\ll 1$. As a result, we obtain the subsequent numerical input from the integral corresponding to the specified wave numbers selections, $k_e$ and $k_{\rm end}$.
\bea {\bf O}(\tau_e)={\bf O}(\tau_{\rm end})\approx 3.76.\eea

\section{Detailed computation of the bispectrum for Galileon inflation }
\label{A4a}

\subsection{Detailed computation in the region I: SRI}

The subsection provides the necessary comprehensive analysis to determine the integrals for obtaining the tree-level contribution to the three-point correlation function. We first provide a generic description for the three-point function that will be utilized thereafter, as well as for the assessment of correlation functions for the SRII and USR areas in the future, and calculating contributions from specific operators.

We start with the first operator, $\displaystyle{a(\tau_{1}){\cal G}_1}\zeta^{'3}$, and apply it to Eq.(\ref{a3}), where we write the expression as follows following the usual contractions:
\bea \label{c1a} \langle\hat{\zeta}_{\bf k_{1}}\hat{\zeta}_{\bf k_{2}}\hat{\zeta}_{\bf k_{3}}\rangle_{\zeta^{'3}} &=& \text{2 $\times$ Im}\Bigg[\zeta_{\bold{k}_{1}}(\tau)\zeta_{\bold{k}_{2}}(\tau)\zeta_{\bold{k}_{3}}(\tau)\int\frac{d^{3}\bold{q}_{1}}{(2\pi)^3}\frac{d^{3}\bold{q}_{2}}{(2\pi)^3}\frac{d^{3}\bold{q}_{3}}{(2\pi)^3}\int^{\tau_{s}}_{-\infty}d\tau_{1}d^{3}x\displaystyle{\frac{a(\tau_{1}){\cal G}_1(\tau_1)}{H^3}\partial_{\tau}\zeta^{*}_{\bold{q}_{1}}(\tau_{1})\partial_{\tau}\zeta^{*}_{\bold{q}_{2}}(\tau_{1})\partial_{\tau}\zeta^{*}_{\bold{q}_{3}}(\tau_{1})}\nonumber\\
&& \quad\quad\quad\times \displaystyle{e^{-i(\bold{q}_{1}+\bold{q}_{2}+\bold{q}_{3}).\bold{x}}}(2\pi)^{9}\delta^{3}(\bold{k}_{1}+\bold{q}_{1})\delta^{3}(\bold{k}_{2}+\bold{q}_{2})\delta^{3}(\bold{k}_{3}+\bold{q}_{3}) + \text{1} \longleftrightarrow \text{2} + \text{1} \longleftrightarrow \text{3} \Bigg],\eea
where the final two elements represent the $2$ potential contractions with modes outside of the time integral. Wick contraction between two modes with distinct momenta gives birth to the delta function. Now, as the other interaction operators we define here are also crucial for several computations, analogous expressions may be created for them as well:
\bea \label{c1b} \langle\hat{\zeta}_{\bold{k}_{1}}\hat{\zeta}_{\bold{k}_{2}}\hat{\zeta}_{\bold{k}_{3}}\rangle_{\zeta^{'2}\partial^{2}\zeta} &=& \text{2$\times$2$\times$Im}\Bigg[\zeta_{\bold{k}_{1}}(\tau)\zeta_{\bold{k}_{2}}(\tau)\zeta_{\bold{k}_{3}}(\tau)\int\frac{d^{3}\bold{q}_{1}}{(2\pi)^3}\frac{d^{3}\bold{q}_{2}}{(2\pi)^3}\frac{d^{3}\bold{q}_{3}}{(2\pi)^3}\int^{\tau_{s}}_{-\infty}d\tau_{1}d^{3}x\displaystyle{\frac{{\cal G}_{2}(\tau_1)}{H^3}}\partial_{\tau}\zeta^{*}_{\bold{q}_{1}}(\tau_{1})\partial_{\tau}\zeta^{*}_{\bold{q}_{2}}(\tau_{1})\zeta^{*}_{\bold{q}_{3}}(\tau_{1})\nonumber\\
&& \quad\quad\quad\times (\bold{q}^{2}_{3})\displaystyle{e^{-i(\bold{q}_{1}+\bold{q}_{2}+\bold{q}_{3}).\bold{x}}}(2\pi)^{9}\delta^{3}(\bold{k}_{1}+\bold{q}_{1})\delta^{3}(\bold{k}_{2}+\bold{q}_{2})\delta^{3}(\bold{k}_{3}+\bold{q}_{3}) + \text{1} \longleftrightarrow \text{2} + \text{1} \longleftrightarrow \text{3} \bigg].\quad\quad
\eea
\bea \label{c1c} \langle\hat{\zeta}_{\bold{k}_{1}}\hat{\zeta}_{\bold{k}_{2}}\hat{\zeta}_{\bold{k}_{3}}\rangle_{\zeta^{'}(\partial_{i}\zeta)^{2}} &=& \text{2$\times$2$\times$Im}\Bigg[\zeta_{\bold{k}_{1}}(\tau)\zeta_{\bold{k}_{2}}(\tau)\zeta_{\bold{k}_{3}}(\tau)\int\frac{d^{3}\bold{q}_{1}}{(2\pi)^3}\frac{d^{3}\bold{q}_{2}}{(2\pi)^3}\frac{d^{3}\bold{q}_{3}}{(2\pi)^3}\int^{\tau_{s}}_{-\infty}d\tau_{1}d^{3}x\displaystyle{\frac{a(\tau_1){\cal G}_{3}(\tau_1)}{H^3}\zeta^{*}_{\bold{q}_{1}}(\tau_{1})\zeta^{*}_{\bold{q}_{2}}(\tau_{1})\partial_{\tau}\zeta^{*}_{\bold{q}_{3}}(\tau_{1})}\nonumber\\
&& \quad\quad\times (\bold{q}_{1}.\bold{q}_{2})\displaystyle{e^{-i(\bold{q}_{1}+\bold{q}_{2}+\bold{q}_{3}).\bold{x}}}(2\pi)^{9}\delta^{3}(\bold{k}_{1}+\bold{q}_{1})\delta^{3}(\bold{k}_{2}+\bold{q}_{2})\delta^{3}(\bold{k}_{3}+\bold{q}_{3}) + \text{1} \longleftrightarrow \text{2} + \text{1} \longleftrightarrow \text{3} \bigg].\quad\quad
\eea
\bea \label{c1d} \langle\hat{\zeta}_{\bold{k}_{1}}\hat{\zeta}_{\bold{k}_{2}}\hat{\zeta}_{\bold{k}_{3}}\rangle_{(\partial_{i}\zeta)^{2}(\partial^{2}\zeta)} &=& \text{2$\times$Im}\Bigg[\zeta_{\bold{k}_{1}}(\tau)\zeta_{\bold{k}_{2}}(\tau)\zeta_{\bold{k}_{3}}(\tau)\int\frac{d^{3}\bold{q}_{1}}{(2\pi)^3}\frac{d^{3}\bold{q}_{2}}{(2\pi)^3}\frac{d^{3}\bold{q}_{3}}{(2\pi)^3}\int^{\tau_{s}}_{-\infty}d\tau_{1}d^{3}x\displaystyle{\frac{{\cal G}_{4}(\tau_1)}{H^3}\zeta^{*}_{\bold{q}_{1}}(\tau_{1})\zeta^{*}_{\bold{q}_{2}}(\tau_{1})\zeta^{*}_{\bold{q}_{3}}(\tau_{1})}\nonumber\\
&& \times((\bold{q}_{1}.\bold{q}_{2})\bold{q}^{2}_{3})\displaystyle{e^{-i(\bold{q}_{1}+\bold{q}_{2}+\bold{q}_{3}).\bold{x}}}(2\pi)^{9}\delta^{3}(\bold{k}_{1}+\bold{q}_{1})\delta^{3}(\bold{k}_{2}+\bold{q}_{2})\delta^{3}(\bold{k}_{3}+\bold{q}_{3}) + \text{1} \longleftrightarrow \text{2} + \text{1} \longleftrightarrow \text{3} \bigg].\quad\quad
\eea
where the additional $2$ multiplied before the expressions denotes the same terms arising from permutations, respectively, contracting with two $\partial_{\tau}\zeta_{k}^{*}$ or $\zeta^{*}_{k}$ terms. The time derivative of the mode in the SRI area should also be discussed:
\bea \Pi_{k} = \partial_{\tau}\zeta_{k} = \left(\frac{iH^{2}}{2\sqrt{\cal{A}}}\right)\frac{1}{(c_{s}k)^{3/2}}(k^{2}c_{s}^{2}\tau)e^{-ikc_{s}\tau}.
\eea
With this knowledge, we can now clearly state the contributions that each interaction operator makes to the three-point correlation function. The formula for \underline{\textbf{Operator 1}: $a(\tau_{1})\zeta^{'3}$} is provided by (\ref{c1a}):
\bea \langle\hat{\zeta}_{\bold{k}_{1}}\hat{\zeta}_{\bold{k}_{2}}\hat{\zeta}_{\bold{k}_{3}}\rangle_{\zeta^{'3}} &=& \text{2 $\times$ Im}\Bigg[\frac{H^{12}}{(4{\cal A})^{3}}\zeta_{\bold{k}_{1}}(\tau)\zeta_{\bold{k}_{2}}(\tau)\zeta_{\bold{k}_{3}}(\tau)\int^{\tau_{s}}_{-\infty}d\tau_{1}\displaystyle{\frac{a(\tau_{1}){\cal G}_1(\tau_1)}{H^3 c_{s}^{3}}k_{1}^{2}k_{2}^{2}k_{3}^2\tau_{1}^{3}\exp{(ic_{s}(k_{1}+k_{2}+k_{3})\tau_{1})}}\nonumber\\
&& \quad\quad\quad \times (2\pi)^{3}\delta^{3}(\bold{k}_{1}+\bold{k}_{2}+\bold{k}_{3}) + \text{1} \longleftrightarrow \text{2} + \text{1} \longleftrightarrow \text{3} \Bigg]_{\tau \rightarrow 0}.
\eea
Here, the following condition is applied while performing the integral over the momentum-delta functions:
\be \int\d^{3}x e^{-i(\bold{k}_{1}+\bold{k}_{2}+\bold{k}_{3}).\bold{x}} = (2\pi)^{3}\delta^{3}(\bold{k}_{1}+\bold{k}_{2}+\bold{k}_{3}).\ee 
The correlation is evaluated in the limit $\tau \rightarrow 0$ using the shortcut $K=k_{1}+k_{2}+k_{3}$, where $k_{i} = |\bold{k}_{i}|$ $\forall i=1,2,3$, and the relation $\tau_{s} = -1/(k_{s}c_{s})$, $\tau=-1/(a(\tau)H)$. By doing this, we are also able to utilize the shortcut of absorbing the delta function for momentum conservation as follows:
\begin{align*}
    \langle\hat{\zeta}_{\bold{k}_{1}}\hat{\zeta}_{\bold{k}_{2}}\hat{\zeta}_{\bold{k}_{3}}\rangle = (2\pi)^{3}\delta{(\bold{K})}\langle\hat{\zeta}_{\bold{k}_{1}}\hat{\zeta}_{\bold{k}_{2}}\hat{\zeta}_{\bold{k}_{3}}\rangle^{'}
\end{align*}
where we use the fact that, $\bold{K} = \bold{k}_{1}+\bold{k}_{2}+\bold{k}_{3}$ and $\Tilde{K} = |\bold{K}| = \sqrt{k_{1}^{2} + k_{2}^{2} + k_{3}^{2}}$.

After accounting for the aforementioned replacements, we are left with the following integral, which evaluates to:
\bea \label{c1Ia} \displaystyle{\int_{-\infty}^{\tau_{s}}d\tau_{1}k_{1}^{2}k_{2}^{2}k_{3}^2\tau_{1}^{2}c_{s}^{-3}\exp{(ic_{s}K\tau_{1})} = \frac{-ik_{1}^2 k_{2}^2 k_{3}^2}{c_{s}^{6} K^3} \left(\frac{K^2}{k_{s}^2}-\frac{2i K}{k_{s}}-2\right) \exp{\left(-i\frac{K}{k_{s}}\right)}}.\eea
The final result follows after substituting the above and just using the imaginary component derived from the same integral:
\bea \label{c1r1} \langle\hat{\zeta}_{\bold{k}_{1}}\hat{\zeta}_{\bold{k}_{2}}\hat{\zeta}_{\bold{k}_{3}}\rangle_{\zeta^{'3}}^{'} =  \frac{H^{12}}{(4{\cal A})^{3}}\frac{{\cal G}_{1}}{H^4}\frac{6}{(k_{1}^{3}k_{2}^{3}k_{3}^{3})}\frac{k_{1}^2 k_{2}^2 k_{3}^2}{c_{s}^{6} K^3}\left\{\left(\frac{K^{2}}{k_{s}^{2}}-2\right)\cos\left(\frac{K}{k_{s}}\right)-2\frac{K}{k_{s}}\sin\left(\frac{K}{k_{s}}\right)\right\}.
\eea
This provides us with the first interaction's contribution to the SRI region's three-point correlation function. It is possible to analyze the contributions from the other operators by using a similar analysis that takes the replacements into account.

After accounting for all the permutations in the momentum variables, the formula from (\ref{c1b}) yields the following conclusion for \underline{\textbf{Operator 2}: $\zeta^{'2}(\partial^{2}\zeta)$}:
\bea \label{c1r2} \langle\hat{\zeta}_{\bold{k}_{1}}\hat{\zeta}_{\bold{k}_{2}}\hat{\zeta}_{\bold{k}_{3}}\rangle_{\zeta^{'2}(\partial^{2}\zeta)}^{'} = \frac{H^{12}}{(4{\cal A})^{3}}\frac{{\cal G}_{2}}{H^3}\frac{4}{(k_{1}^{3}k_{2}^{3}k_{3}^{3})}\displaystyle{\frac{k_{1}^2 k_{2}^2 k_{3}^2}{c_{s}^{8} K^3}\left\{\left(12-6\frac{K^{2}}{k_{s}^2}\right)\cos\left(\frac{K}{k_{s}}\right) - \left(\frac{K^{3}}{k_{s}^3}-12\frac{K}{k_{s}}\right)\sin\left(\frac{K}{k_{s}}\right)\right\}}.\eea

After accounting for all the permutations in the momentum variables, the formula from (\ref{c1c}) yields the following conclusion for \underline{\textbf{Operator 3}: $a(\tau_{1})\zeta^{'}(\partial_{i}\zeta)^{2}$}:
\bea \label{c1r3} \langle\hat{\zeta}_{\bold{k}_{1}}\hat{\zeta}_{\bold{k}_{2}}\hat{\zeta}_{\bold{k}_{3}}\rangle_{\zeta^{'}(\partial_{i}\zeta)^{2}}^{'} &=& \frac{H^{12}}{(4{\cal A})^{3}}\frac{{\cal G}_{3}}{H^4}\frac{4}{k_{1}^{3}k_{2}^{3}k_{3}^{3}}
\frac{k_{1}k_{2}k_{3}}{c_{s}^{8}K^{3}k_{s}^{2}}\Bigg\{\bigg(16 k_2 k_3 k_1 K+ 2 k_1^3\left(k_2+k_3\right) +2k_3^3\left(k_1+k_2\right)+ 2k_2^3\left(k_1+k_3\right)\nonumber\\
   && + 2\left(k_1^2 k_2^2+k_3^2 k_2^2+k_1^2 k_3^2\right)\bigg) k_s \sin
   \left(\frac{K}{k_s}\right)-3 k_1 k_2 k_3 K^{2} \cos
   \left(\frac{K}{k_s}\right)+\bigg(k_1^3+ 5\left(k_2+k_3\right) k_1^2 \nonumber\\
   && +18 k_2 k_3 k_1+k_2^3+k_3^3+5
   \left(k_1+k_2\right) k_3^2+5 k_2^2 \left(k_1+k_3\right)\bigg) k_s^2 \cos
   \left(\frac{K}{k_s}\right)\Bigg\}. \eea

   After accounting for all the permutations in the momentum variables, the formula from (\ref{c1d}) yields the following conclusion for \underline{\textbf{Operator 4}: $\partial^{2}\zeta(\partial_{i}\zeta)^{2}$}:
\bea && \label{c1r4}\langle\hat{\zeta}_{\bold{k}_{1}}\hat{\zeta}_{\bold{k}_{2}}\hat{\zeta}_{\bold{k}_{3}}\rangle_{\partial^{2}\zeta(\partial_{i}\zeta)^{2}}^{'} = \frac{H^{12}}{(4{\cal A})^{3}}\frac{{\cal G}_{4}}{H^3}\frac{6}{k_{1}^{3}k_{2}^{3}k_{3}^{3}}
\frac{k_{1}k_{2}k_{3}}{c_{s}^{10}K^{3}}\left\{k_{1}k_{2}k_{3}\frac{K^{3}}{k_{s}^{3}}\left(\sin{\frac{K}{k_s}}\right)+\frac{K^{2}}{k_{s}^{2}}\bigg(6k_{1}k_{2}k_{3} + k_{1}^{2}(k_{2}+k_{3})+k_{2}^{2}(k_{3}+k_{1})\right.\nonumber\\
&&\left. \quad\quad\quad\quad\quad\quad\quad\quad\quad +k_{3}^{2}(k_{2}+k_{1})\bigg)\left(\cos{\frac{K}{k_s}}\right) -\frac{1}{k_{s}}\bigg(k_{1}^{4}+ k_{2}^{4}+k_{3}^{4}+6k_{1}^{3}(k_{2}+k_{3})+6k_{2}^{3}(k_{1}+k_{3}) + 6k_{3}^{3}(k_{1}+k_{2})\right. \nonumber\\
&& \left. \quad\quad\quad\quad\quad\quad\quad\quad\quad +28k_{1}k_{2}k_{3}K+ 10k_{1}^{2}k_{2}^{2}+8k_{2}^{2}k_{3}^{2}+10k_{1}^{2}k_{3}^{2} \bigg)\left(\sin{\frac{K}{k_{s}}}\right)-2\bigg(k_{1}^{3}+k_{2}^{3}+k_{3}^{3}+ 12k_{1}k_{2}k_{3} + 4k_{1}^{2}(k_{2}+k_{3}) \right.\nonumber\\
&& \left. \quad\quad\quad\quad\quad\quad\quad\quad\quad + 4k_{2}^{2}(k_{3}+k_{1}) +4k_{3}^{2}(k_{1}+k_{2})\bigg)\left(\cos{\frac{K}{k_{s}}}\right) \right\}.
\eea
The bispectrum in the SRI area, or the tree-level, three-point correlation function, results from our collection of all the contributions made by the interaction operators. (\ref{c1r1},\ref{c1r2},\ref{c1r3},\ref{c1r4}) contributions in order to obtain the overall bispectrum contribution for the SRI region. 

\subsection{Detailed computation in the region II: USR}

The results of our calculation of the bispectrum at the tree level in the USR region will now be analyzed in relation to the detailed expressions of the set of momentum-dependent functions we provide in this section. It will be beneficial to specifically note the time-derivative of the mode function in the USR area before going over the calculations:
\bea \Pi^{*}_{k} = \partial_{\tau}\zeta^{*}_{k} = \left(\frac{\tau_{s}}{\tau}\right)^{9}\left\{\alpha_{k}^{(2)*}\left(k^{2}c_{s}^{2}\tau-\frac{3}{\tau}(1-ikc_{s}\tau)\right)\exp{(ikc_{s}\tau)}+\beta_{k}^{(2)*}\left(\frac{3}{\tau}(1+ikc_{s}\tau)-k^{2}c_{s}^{2}\tau\right)\exp{(-ikc_{s}\tau)}\right\},\quad\quad\quad
\eea
where the curvature perturbation time-derivative's complex conjugate is written, and $k = |\bold{k}|$. While it is also included at the front, it is not included here for clarity: $(-iH^{2})/(2\sqrt{{\cal A}})$.

Depending on the variables involved, the time-derivatives in various operators significantly increase the amount of integrals to assess and produce distinct kinds of integrals. Each of these integrals will result in specific types, and the others will vary in some way, as we will also indicate. All outcomes are expressed in terms of the functions covered in this section. The various sets of functions have been grouped according to the type of integral that corresponds to each category. Additionally, we have thought about scaling all of the equations using the wave number $k_{e}$ and have subsequently made adjustments to the expressions to ensure that they have the required right dimensions. Because of this, it is now much easier to understand the functions' physical effects based on the wave numbers' powers.

We have identified a few factors that will be shown in the following results: The following linear combinations of the magnitude of the three momenta are represented by $\displaystyle{x_{e,i} = {\cal K}_{i}/{k_{e}}}$, $\displaystyle{x_{s.i} = {\cal K}_{i}/{k_{s}}}$, and ${\cal K}_i$, where $i = 1,2,3,4$. $k_1, k_2, k_3$,
\begin{align} \label{c2k}
    {\cal K}_1 = k_1+k_2+k_3, \quad{\cal K}_2 = k_1-k_2+k_3, \quad{\cal K}_3 = k_1+k_2-k_3, \quad{\cal K}_4 = -k_1+k_2+k_3.
\end{align} 
We describe a specific number of contributions and explain how each one is interconnected. The findings of the previous section explicitly take the total contribution from all the integrals in the USR region combined.

We utilize the formula in Eq.(\ref{c1a}) to calculate the bispectrum contribution from \underline{\textbf{Operator 1}: $a(\tau_{1})\zeta^{'3}$}. For this operator, the following kinds of integrals are available:
\begin{itemize}
    \item \underline{\textbf{Integral of the first kind}}:
The appropriate set of functions for an integral of the first kind are as follows:
 \begin{subequations}
    \begin{align}
    {\cal P}_{1,1}({\cal K}_{i}, k_{1},k_{2},k_{3}) &= \frac{-11{\cal K}_{i}^3}{k_e^3}+\frac{132 {\cal K}_{i} \left(k_2 k_3+k_1 k_2+k_1 k_3\right)}{k_e^3}-\frac{1320 k_1 k_2 k_3}{k_e^3}.\\
    {\cal Q}_{1,1}({\cal K}_{i}) &= \frac{i{\cal K}_{i}^8}{k_e^8}-\frac{{\cal K}_{i}^7}{k_e^7} -\frac{2 i{\cal K}_{i}^6}{k_e^6} + \frac{6 {\cal K}_{i}^5}{k_e^5} + \frac{24 i {\cal K}_{i}^4}{k_e^4} - \frac{120 {\cal K}_{i}^3}{k_e^3} - \frac{720 i {\cal K}_{i}^2}{k_e^{2}} + \frac{5040 {\cal K}_{i}}{k_e} + 40320 i.\\
    {\cal R}_{1,1}({\cal K}_{i}) &= k_e^3 \bigg\{\frac{{\cal K}_{i}^9 \exp{(ix_{s,i})} \left(\text{Ei}\left(-ix_{s,i}\right)-\text{Ei}\left(-ix_{e,i}\right)\right)}{k_s^{12}}-\frac{i
    {\cal K}_{i}^8}{k_s^{11}}+\frac{{\cal K}_{i}^7}{k_s^{10}}+\frac{2 i{\cal K}_{i}^6}{k_s^9}-\frac{6 {\cal K}_{i}^5}{k_s^8}-\frac{24 i
    {\cal K}_{i}^4}{k_s^7}\notag\\
    & \quad\quad\quad\quad +\frac{120 {\cal K}_{i}^3}{k_s^6} 
    +\frac{720 i {\cal K}_{i}^2}{k_s^5}-\frac{5040
    {\cal K}_{i}}{k_s^4}-\frac{40320 i}{k_s^3}\bigg\}.
    \end{align} 
    \label{c211f}
    \end{subequations}
    The following form is used to express the outcome of this integral using the functions mentioned above:
\bea \label{c211aI} (\bold{I_{1}})_{j}({\cal K}_i,k_1,k_2,k_3) &=& -\frac{k_e^{12} \exp{(-i x_{e,i})}}{17740800 c_{s}^6 k_s^9}\bigg\{{\cal P}_{1,1}({\cal K}_i,k_1,k_2,k_3){\cal Q}_{1,1}({\cal K}_i) + 39916800 \nonumber\\
   && -\frac{362880 \left(132 \left(k_2 k_3+k_1 k_2+k_1 k_3\right)-11 {\cal K}_i^{2}\right)}{k_e^2} + i\frac{39916800
   {\cal K}_i}{k_e}  \nonumber\\
   && + \frac{k_s^{12}}{k_e^{12}}\exp{(i(x_{e,i}-x_{s,i}))}\bigg({\cal P}_{1,1}({\cal K}_i){\cal R}_{1,1}({\cal K}_i)-39916800 -\frac{39916800 i {\cal K}_{i}}{k_s} \nonumber\\
   && + \frac{362880 \left(132 \left(k_2 k_3+k_1 k_2+k_1 k_3\right)-11
   {\cal K}_{i}^{2}\right)}{k_s^2} \bigg) \bigg\}
    \eea
     \begin{enumerate}
        \item Consider \textbf{Integral 1a}:
         \bea \label{c211a} \displaystyle{\int_{\tau_{s}}^{\tau_{e}}d\tau_{1}\frac{-27\tau_{s}^{9}}{c_s^{9}\tau_{1}^{13}}\exp{(ic_{s}{\cal K}_{1}\tau_{1})}}(1-ik_{1}c_{s}\tau_{1})(1-ik_{2}c_{s}\tau_{1})(1-ik_{3}c_{s}\tau_{1}) = (\bold{I_{1}})_{1}({\cal K}_1,k_1,k_2,k_3).
    \eea 
    Using the above-described generic formula, the result for this integral is written. Results in the form of the preceding equation are also obtained for other integrals of the same sort, for various ${\cal K}_i$. The momentum variables in the integral have the same signature as those in the component ${\cal K}_{i}$. For this, let's look at another instance.
\item For a  
    similar \textbf{Integral 1b}: \bea\displaystyle{\int_{\tau_{s}}^{\tau_{e}}d\tau_{1}\frac{27\tau_{s}^{9}}{c_{s}^{9}\tau_{1}^{13}}\exp{(ic_{s}{\cal K}_{2}\tau_{1})}}(1-ik_{1}c_{s}\tau_{1})(1+ik_{2}c_{s}\tau_{1})(1-ik_{3}c_{s}\tau_{1}) = -(\bold{I_{1}})_{2}({\cal K}_2,k_1,-k_2,k_3). 
    \eea  
    \item For \textbf{Integral 1c}:
    \bea \label{c211b} \displaystyle{\int_{\tau_{s}}^{\tau_{e}}d\tau_{1}\frac{-27\tau_{s}^{9}}{c_{s}^{9}\tau_{1}^{13}}\exp{(-ic_{s}{\cal K}_{3}\tau_{1})}}(1+ik_{1}c_{s}\tau_{1})(1+ik_{2}c_{s}\tau_{1})(1-ik_{3}c_{s}\tau_{1}) = (\bold{I_{1}})_{3}^{*}({\cal K}_{3},k_1,k_2,-k_3).
    \eea
    For the factor ${\cal K}_{i}$, the notation of maintaining the greatest number of positive momentum variables is utilized. Integrals with negative exponential factors will result from this. Equation (\ref{c211aI}) yields the complex conjugate of the general result, which is used to express the result for the above integral. Like those in factor ${\cal K}_{i}$, the momentum signature in the RHS argument.

        \end{enumerate}
A total of $4$ first-kind integrals with positive and negative exponentials will be present.
\item \underline{\textbf{Integral of the second kind}}:
The appropriate set of functions for an integral of the second kind is as follows:
 \begin{subequations}
    \begin{align} 
    {\cal P}_{1,2}({\cal K}_{i}, k_{a},k_{b}) &= \frac{{\cal K}_{i}^2}{k_e^2}+\frac{10 {\cal K}_{i} \left(k_a + k_b\right)}{k_e^2}+\frac{90 k_a k_b}{k_e^2}.\\
    {\cal Q}_{1,2}({\cal K}_{i}) &= -\frac{i{\cal K}_{i}^7}{k_e^7} +\frac{{\cal K}_{i}^6}{k_e^6} + \frac{2i {\cal K}_{i}^5}{k_e^5} - \frac{6{\cal K}_{i}^4}{k_e^4} - \frac{24i{\cal K}_{i}^3}{k_e^3} + \frac{120 {\cal K}_{i}^2}{k_e^{2}} +  \frac{720 i{\cal K}_{i}}{k_e} -5040.\\
    {\cal R}_{1,2}({\cal K}_{i}) &= k_e^2 \bigg\{ \frac{k^8 (\exp{ix_{s,i}}) \left(\text{Ei}\left(-ix_{e,i}\right)-\text{Ei}\left(-ix_{s,i}\right)\right)}{k_s^{10}}+\frac{i
   {\cal K}_{i}^7}{k_s^9}-\frac{k^6}{k_s^8}-\frac{2 i {\cal K}_{i}^5}{k_s^7}+\frac{6 {\cal K}_{i}^4}{k_s^6}+\frac{24 i {\cal K}_{i}^3}{k_s^5}-\frac{120
   {\cal K}_{i}^2}{k_s^4}\notag\\
   &-\frac{720 i {\cal K}_{i}}{k_s^3}+\frac{5040}{k_s^2}\bigg\}.
    \end{align}
    \label{c22f}
    \end{subequations}
     where $a,b=1,2,3$. The following form is used to express the outcome of this integral using the functions mentioned above:
\bea \label{c212aI} (\bold{I_{2}})_{j}({\cal K}_i,k_a,k_b) &=& \frac{k_e^{10} \exp{(-i x_{e,i})}}{403200 c_{s}^6 k_s^9}\bigg\{{\cal P}_{1,2}({\cal K}_i,k_a,k_b){\cal Q}_{1,2}({\cal K}_i) +\frac{40320 i \left(10\left(k_a+k_b\right) - {\cal K}_i\right)}{k_e} \nonumber\\
    && + 362880 + \frac{k_s^{10}}{k_e^{10}}\exp{(i(x_{e,i}-x_{s,i}))}\bigg({\cal P}_{1,2}({\cal K}_i){\cal R}_{1,2}({\cal K}_i)-362880 \nonumber\\
    && +\frac{40320 i\left(-10(k_a+k_b) + {\cal K}_{i}\right)}{k_s}\bigg) \bigg\}.
    \eea
    \begin{enumerate} 
    \item Consider \textbf{Integral 2a}: \bea \label{c212a} \displaystyle{\int_{\tau_{s}}^{\tau_{e}}d\tau_{1}\frac{9\tau_{s}^{9}}{\tau_{1}^{12}c_{s}^{9}}\exp{(ic_{s}{\cal K}_{1}\tau_{1})}}(1-ik_{1}c_{s}\tau_{1})(k_{3}^{2}c_{s}^{2}\tau_{1})(1-ik_{2}c_{s}\tau_{1}) = k_{3}^{2}(\bold{I_{2}})_{1}({\cal K}_1,-k_1,-k_2).
        \eea
where $k_a = -k_1$, $k_b = -k_2$, and ${\cal K}_i = {\cal K}_1$ result from the fact that both $k_1,k_3$ have negative signatures inside the integral. The generic formula in the above equation yields the result. We also illustrate a similar another integral.

 \item For a similar \textbf{Integral 2b}:
    \bea \displaystyle{\int_{\tau_{s}}^{\tau_{e}}d\tau_{1}\frac{-9\tau_{s}^{9}}{\tau_{1}^{12}c_{s}^{9}}\exp{(ic_{s}{\cal K}_{2}\tau_{1})}}(k_{1}^{2}c_{s}^{2}\tau_{1})(1+ik_{2}c_{s}\tau_{1})(1-ik_{3}c_{s}\tau_{1}) = -k_{1}^{2}(\bold{I_{2}})_{5}({\cal K}_2,k_2,-k_3).
    \eea
In the integral, $k_a = k_2$ and $k_b=-k_3$ rely on their respective signatures.
\item For \textbf{Integral 2c}:
    \bea  \displaystyle{\int_{\tau_{s}}^{\tau_{e}}d\tau_{1}\frac{9\tau_{s}^{9}}{\tau_{1}^{12}c_{s}^{9}}\exp{(-ic_{s}{\cal K}_{3}\tau_{1})}}(k_{1}^{2}c_{s}^{2}\tau_{1})(1+ik_{2}c_{s}\tau_{1})(1-ik_{3}c_{s}\tau_{1}) = k_{1}^{2}(\bold{I_{2}})^{*}_{9}({\cal K}_3,k_2,-k_3). \eea
    As seen for the integral above, the integrals with negative exponential are connected to the formula in Eq. (\ref{c212aI}) via a complex conjugate.

        \end{enumerate}
        Twelve integrals of the second kind, with a positive and negative exponential in each, will be present. 
    \item \underline{\textbf{Integral of the third kind}}:
The appropriate set of functions for an integral of the third kind are as follows:
 \begin{subequations}
    \begin{align} 
    {\cal P}_{1,3}({\cal K}_{i}, k_{a}) &= \frac{{\cal K}_{i}}{k_e}+\frac{8 k_a}{k_e}.\\
    {\cal Q}_{1,3}({\cal K}_{i}) &= \frac{i{\cal K}^6}{k_e^6} - \frac{{\cal K}_{i}^5}{k_e^5} - \frac{2i{\cal K}_{i}^4}{k_e^4} + \frac{6{\cal K}_{i}^3}{k_e^3} + \frac{24i{\cal K}_{i}^2}{k_e^2}  - \frac{120{\cal K}_{i}}{k_e} - 720 i.\\
    {\cal R}_{1,3}({\cal K}_{i}) &= k_e \bigg\{ \frac{{\cal K}_{i}^7 \exp{(ix_{s,i})} \left(\text{Ei}\left(-ix_{e,i}\right)-\text{Ei}\left(-ix_{s,i}\right)\right)}{k_s^{8}}+\frac{i
   {\cal K}_{i}^6 }{k_s^7}-\frac{{\cal K}_{i}^5}{k_s^6}-\frac{2 i {\cal K}_{i}^4}{k_s^5}+\frac{6{\cal K}_{i}^3}{k_s^4}+\frac{24 i{\cal K}_{i}^2}{k_s^3}\notag\\
   &-\frac{120{\cal K}_{i}}{k_{s}^{2}} -\frac{720 i}{k_s} \bigg\}.
    \end{align}
    \label{c23f}
    \end{subequations}
     where $a=1,2,3$. The following form is used to express the outcome of this integral using the functions mentioned above:
 \bea \label{c213aI} (\bold{I_{3}})_{j}({\cal K}_i,k_a) &=& -\frac{k_e^{8} \exp{(-i x_{e,i})}}{13440 c_{s}^6 k_s^9}\bigg\{{\cal P}_{1,3}({\cal K}_i,k_a){\cal Q}_{1,3}({\cal K}_i) + 5040 \nonumber\\
   && - \frac{k_s^{8}}{k_e^{8}}\exp{(i(x_{e,i}-x_{s,i}))}\bigg({\cal P}_{1,3}({\cal K}_1){\cal R}_{1,3}({\cal K}_i)+5040\bigg)\bigg\}.
    \eea 
\begin{enumerate} 
    \item Consider \textbf{Integral 3a}: \bea \label{c213a} \displaystyle{\int_{\tau_{s}}^{\tau_{e}}d\tau_{1}\frac{-3\tau_{s}^{9}}{\tau_{1}^{11}c_{s}^{9}}\exp{(ic_{s}{\cal K}_{1}\tau_{1})}}(k_{2}^{2}c_{s}^{2}\tau_{1})(k_{3}^{2}c_{s}^{2}\tau_{1})(1-ik_{1}c_{s}\tau_{1}) = k_{2}^{2}k_{3}^{2}(\bold{I_{3}})_{1}({\cal K}_1,-k_1).
    \eea
The momentum variable's signature inside the integral provides the basis for the aforementioned finding, $k_a = -k_1$. The generic formula in Eq.(\ref{c213aI}) yields the result. Think of another illustration of this.
 \item For a similar \textbf{Integral 3b}:
     \bea \displaystyle{\int_{\tau_{s}}^{\tau_{e}}d\tau_{1}\frac{3\tau_{s}^{9}}{\tau_{1}^{11}c_{s}^{9}}\exp{(ic_{s}{\cal K}_{2}\tau_{1})}}(k_{1}^{2}c_{s}^{2}\tau_{1})(k_{3}^{2}c_{s}^{2}\tau_{1})(1+ik_{2}c_{s}\tau_{1}) =  -k_1^{2}k_{3}^{2}(\bold{I_{3}})_{6}({\cal K}_2,k_2).
    \eea
    \item For \textbf{Integral 3c}:
     \bea \label{c213c} \displaystyle{\int_{\tau_{s}}^{\tau_{e}}d\tau_{1}\frac{-3\tau_{s}^{9}}{\tau_{1}^{11}c_{s}^{9}}\exp{(-ic_{s}{\cal K}_{3}\tau_{1})}}(k_{1}^{2}c_{s}^{2}\tau_{1})(k_{2}^{2}c_{s}^{2}\tau_{1})(1+ik_{3}c_{s}\tau_{1}) = k_1^{2}k_{2}^{2}(\bold{I_{3}})^{*}_{9}({\cal K}_3,k_3).
    \eea
    As seen for the integral above, the negative exponential integrals are connected to the formula in Eq. (\ref{c213aI}) via a complex conjugate.

    \end{enumerate}
Such third-kind integrals will total $12$, and each will have an exponential component that is both positive and negative.
    \item \underline{\textbf{Integral of the fourth kind}}:
The appropriate set of functions for the integral of the fourth kind are as follows:
 \begin{subequations}
    \begin{align} 
    {\cal Q}_{1,4}({\cal K}_{i}) &= \frac{i{\cal K}^5}{k_e^5} - \frac{{\cal K}_{i}^4}{k_e^4} - \frac{2i{\cal K}_{i}^3}{k_e^3} + \frac{6{\cal K}_{i}^2}{k_e^2} + i\frac{24{\cal K}_{i}}{k_e} - 120.\\
    {\cal R}_{1,4}({\cal K}_{i}) &= \bigg\{ \frac{{\cal K}_{i}^6 \exp{(ix_{s,i})} \left(\text{Ei}\left(-ix_{e,i}\right)-\text{Ei}\left(-ix_{s,i}\right)\right)}{k_s^{6}}-\frac{i
   {\cal K}_{i}^5}{k_s^5}+\frac{{\cal K}_{i}^4}{k_s^4}+\frac{2 i {\cal K}_{i}^3}{k_s^3}-\frac{6{\cal K}_{i}^2}{k_s^2}-\frac{24 i{\cal K}_{i}}{k_s} +120\bigg\}.
    \end{align}
    \label{c24f}
    \end{subequations}
    The following form is used to express the outcome of this integral using the functions mentioned above: 
\bea \label{c214aI} (\bold{I_{4}})_{j}({\cal K}_i) &=& -\frac{k_e^{6} \exp{(-i x_{e,i})}}{720c_{s}^6 k_s^9}\bigg\{{\cal Q}_{1,4}({\cal K}_i) + \frac{k_s^{6}}{k_e^{6}}\exp{(i(x_{e,i}-x_{s,i}))}{\cal R}_{1,4}({\cal K}_i)\bigg\}.
    \eea   
    \begin{enumerate}
    \item Consider \textbf{Integral 4a}:
    \bea \label{c214a} \displaystyle{\int_{\tau_{s}}^{\tau_{e}}d\tau_{1}\frac{\tau_{s}^{9}}{c_{s}^{3}\tau_{1}^{7}}k_{1}^{2}k_{2}^{2}k_{3}^2\exp{(ic_{s}{\cal K}_{1}\tau_{1})}} = k_{1}^{2}k_{2}^{2}k_{3}^{2}(\bold{I_{4}})_{1}({\cal K}_1).
    \eea
For any such integral, the argument depends simply on the component ${\cal K}_{i}$.  The generic formula in Eq. (\ref{c214aI}) yields the result. 
 \item For \textbf{Integral 4b}:
    \bea \label{c214b} \displaystyle{\int_{\tau_{s}}^{\tau_{e}}d\tau_{1}\frac{\tau_{s}^{9}}{c_{s}^{3}\tau_{1}^{7}}k_{1}^{2}k_{2}^{2}k_{3}^{2}\exp{(-ic_{s}{\cal K}_{3}\tau_{1})}} = k_{1}^{2}k_{2}^{2}k_{3}^{2}(\bold{I_{4}})^{*}_{3}({\cal K}_3).
    \eea
    As seen for the integral above, the integrals with exponential are connected to the formula in Eq. (\ref{c214aI}) via a complex conjugate.

    \end{enumerate}
An exponential component of both positive and negative will be present in each of the $4$ fourth-kind integrals.
    \end{itemize}

    We apply the formula in Eq.(\ref{c1b}) to calculate the bispectrum contribution for \underline{\textbf{Operator 2}: $\displaystyle{\zeta^{'2}\partial^{2}\zeta}$}. For this operator, the following integral types are available:
\begin{itemize}
    \item \underline{\textbf{Integral of the first kind}}:
The appropriate set of functions for an integral of the first sort are as follows:
\begin{subequations}
    \begin{align}
    {\cal P}_{2,1}({\cal K}_{i}, k_{1},k_{2},k_{3}) &= \frac{-9{\cal K}_{i}^3}{k_e^3}+\frac{90 {\cal K}_{i} \left(k_2 k_3+k_1 k_2+k_1 k_3\right)}{k_e^3}-\frac{720 k_1 k_2 k_3}{k_e^3}.\\
    {\cal Q}_{2,1}({\cal K}_{i}) &= \frac{ -i{\cal K}_{i}^6}{k_e^6}+\frac{{\cal K}_{i}^5}{k_e^5} +\frac{2 i{\cal K}_{i}^4}{k_e^4} - \frac{6{\cal K}_{i}^3}{k_e^3} - \frac{24 i{\cal K}_{i}^2}{k_e^2} + \frac{120{\cal K}_{i}}{k_e} + 720i.\\
    {\cal R}_{2,1}({\cal K}_{i}) &= k_e^3 \bigg\{\frac{{\cal K}_{i}^7 \exp{(ix_{s,i})} \left(\text{Ei}\left(-ix_{e,i}\right)-\text{Ei}\left(-ix_{s,i}\right)\right)}{k_s^{10}}+\frac{i
    {\cal K}_{i}^6}{k_s^{9}}-\frac{{\cal K}_{i}^5}{k_s^{8}}-\frac{2i{\cal K}_{i}^4}{k_s^7}+\frac{6{\cal K}_{i}^3}{k_s^6}+\frac{24i
    {\cal K}_{i}^2}{k_s^5}-\frac{120{\cal K}_{i}}{k_s^4}\notag\\
    &-\frac{720 i}{k_s^3}\bigg\}.
    \end{align} 
    \label{c221f}
    \end{subequations}
    The following form is used to express the outcome of this integral using the functions mentioned above:
\bea \label{c221aI} (\bold{I_{1}})_{j}({\cal K}_i,k_1,k_2,k_3) &=& \frac{k_e^{10} \exp{(-i x_{e,i})}}{403200 c_{s}^8 k_s^9}\bigg\{{\cal P}_{2,1}({\cal K}_i,k_1,k_2,k_3){\cal Q}_{2,1}({\cal K}_i) + i 362880\frac{{\cal K}_{i}}{k_e} \nonumber\\
    && -\frac{5040 \left(90\left(k_2 k_3+k_1 k_2+k_1 k_3\right)-9 {\cal K}_i^{2}\right)}{k_e^2} + 362880 \nonumber\\
    && + \frac{k_s^{10}}{k_e^{10}}\exp{(i(x_{e,i}-x_{s,i}))}\bigg({\cal P}_{2,1}({\cal K}_i){\cal R}_{2,1}({\cal K}_i)-362880 -i362880\frac{{\cal K}_i}{k_s} \nonumber\\
    && + \frac{5040\left(90\left(k_2 k_3+k_1 k_2+k_1 k_3\right)-9 {\cal K}_i^{2}\right)}{k_e^2} \bigg) \bigg\}.
    \eea  
 \begin{enumerate}
        \item Consider \textbf{Integral 1a}:
        \bea \label{c221a} \displaystyle{\int_{\tau_{s}}^{\tau_{e}}d\tau_{1}\frac{9\tau_{s}^{9}}{\tau_{1}^{11}c_{s}^{9}}\exp{(ic_{s}{\cal K}_{1}\tau_{1})}}(1-ik_{1}c_{s}\tau_{1})(1-ik_{2}c_{s}\tau_{1})(1-ik_{3}c_{s}\tau_{1}) = (\bold{I_{1}})_{1}({\cal K}_1,k_1,k_2,k_3).
    \eea 
For each such integral, the signatures of the momentum arguments rely on their corresponding signatures in ${\cal K}_{i}$. Eq.(\ref{c221aI}) provides the formula for expressing the outcome.
 \item For \textbf{Integral 1b}:
    \bea \label{c221b} \displaystyle{\int_{\tau_{s}}^{\tau_{e}}d\tau_{1}\frac{9\tau_{s}^{9}}{\tau_{1}^{11}c_{s}^{9}}\exp{(-ic_{s}{\cal K}_{3}\tau_{1})}}(1+ik_{1}c_{s}\tau_{1})(1+ik_{2}c_{s}\tau_{1})(1-ik_{3}c_{s}\tau_{1}) = (\bold{I_{1}})^{*}_{3}({\cal K}_3,k_1,k_2,-k_3).
    \eea 
    Within the component ${\cal K}_{i}$, the signatures of the momenta in the result coincide accordingly. The expression in Eq. (\ref{c221aI}) is connected to the result of this integral by a complex conjugate, just like it is for the previous integral.

    \end{enumerate}
This type of integral will total $4$, and each one will have an exponential component that is both positive and negative.
\item  \underline{\textbf{Integral of the second kind}}:
The appropriate set of functions for an integral of the second sort is as follows:
 \begin{subequations}
    \begin{align} 
    {\cal P}_{2,2}({\cal K}_{i}, k_{a},k_{b}) &= \frac{{\cal K}_{i}^2}{k_e^2}+\frac{8 {\cal K}_{i} \left(k_a + k_b\right)}{k_e^2}+\frac{56 k_a k_b}{k_e^2}.\\
    {\cal Q}_{2,2}({\cal K}_{i}) &= \frac{i{\cal K}_{i}^5}{k_e^5} -\frac{{\cal K}_{i}^4}{k_e^4} - \frac{2i{\cal K}_{i}^3}{k_e^3} + \frac{6{\cal K}_{i}^2}{k_e^2} + \frac{24i{\cal K}_{i}}{k_e} - 120.\\
    {\cal R}_{2,2}({\cal K}_{i}) &= k_e^2 \bigg\{ -\frac{k^6 (\exp{ix_{s,i}}) \left(\text{Ei}\left(-ix_{e,i}\right)-\text{Ei}\left(-ix_{s,i}\right)\right)}{k_s^{8}}-\frac{i
   {\cal K}_{i}^5}{k_s^7}+\frac{{\cal K}_{i}^4}{k_s^6}+\frac{2 i{\cal K}_{i}^3}{k_s^5}-\frac{6 {\cal K}_{i}^2}{k_s^4}-\frac{24 i{\cal K}_{i}}{k_s^3}+\frac{120}{k_s^2}\bigg\}.
    \end{align}
    \label{c222f}
    \end{subequations}
     where $a, b=1, 2, 3$. The following form is used to express the outcome of this integral using the functions mentioned above:
\bea \label{c222aI} (\bold{I_{2}})_{j}({\cal K}_i,k_a,k_b) &=& -\frac{k_e^{8} \exp{(-i x_{e,i})}}{13440 c_{s}^8 k_s^9}\bigg\{{\cal P}_{2,2}({\cal K}_i,k_a,k_b){\cal Q}_{2,2}({\cal K}_i) -\frac{720i \left({\cal K}_i + 8\left(k_a+k_b\right)\right)}{k_e} \nonumber\\
    && + 5040 + \frac{k_s^{8}}{k_e^{8}}\exp{(i(x_{e,i}-x_{s,i}))}\bigg({\cal P}_{2,2}({\cal K}_i){\cal R}_{2,2}({\cal K}_i)-5040 \nonumber\\
    && +\frac{720 i\left(8(k_a+k_b) + {\cal K}_{i}\right)}{k_s}\bigg) \bigg\}.
    \eea 
    \begin{enumerate}
    \item Consider \textbf{Integral 2a}: \bea \label{c222a} \displaystyle{\int_{\tau_{s}}^{\tau_{e}}d\tau_{1}\frac{-3\tau_{s}^{9}}{\tau_{1}^{10}c_{s}^{9}}\exp{(ic_{s}{\cal K}_{1}\tau_{1})}}(k_{2}^2c_{s}^2\tau_{1})(1-ik_{1}c_{s}\tau_{1})(1-ik_{3}c_{s}\tau_{1}) = k_{2}^{2}(\bold{I_{2}})_{1}({\cal K}_1,-k_1,-k_3).
    \eea 
For each such integral, the momentum arguments signature inside the result is dependent upon their signature when inside the integral.
\item For \textbf{Integral 2b}:
    \bea \label{c222b} \displaystyle{\int_{\tau_{s}}^{\tau_{e}}d\tau_{1}\frac{-3\tau_{s}^{9}}{\tau_{1}^{10}}\exp{(-ic_{s}{\cal K}_{3}\tau_{1})}}(k_{2}^2c_{s}^2\tau_{1})(1+ik_{1}c_{s}\tau_{1})(1-ik_{3}c_{s}\tau_{1}) = k_{2}^{2}(\bold{I_{2}})^{*}_{5}({\cal K}_3,k_1, -k_3).
    \eea 
    As seen with the foregoing integral, the outcome of such integrals is connected to the formula in Eq. (\ref{c222aI}) via a complex conjugate.

    \end{enumerate}
This type of integral will total $8$, with a positive and negative exponential component for each.

\item  \underline{\textbf{Integral of the third kind}}:
The appropriate set of functions for an integral of the third kind are as follows:
\begin{subequations}
    \begin{align} 
    {\cal P}_{2,3}({\cal K}_{i}, k_{a}) &= \frac{{\cal K}_{i}}{k_e}+\frac{6 k_a}{k_e}\\
    {\cal Q}_{2,3}({\cal K}_{i}) &= -\frac{i{\cal K}_{i}^4}{k_e^4} + \frac{{\cal K}_{i}^3}{k_e^3} + \frac{2i {\cal K}_{i}^2}{k_e^2} - \frac{6{\cal K}_{i}}{k_e} - 24i\\
    {\cal R}_{2,3}({\cal K}_{i}) &= k_e \bigg\{ -\frac{{\cal K}_{i}^5 \exp{(ix_{s,i})} \left(\text{Ei}\left(-ix_{e,i}\right)-\text{Ei}\left(-ix_{s,i}\right)\right)}{k_s^{6}}-\frac{i {\cal K}_{i}^4}{k_s^5}+\frac{{\cal K}_{i}^3}{k_s^4}+\frac{2 i{\cal K}_{i}^2}{k_s^3}-\frac{6 {\cal K}_{i}}{k_s^2}-\frac{24 i}{k_s}\bigg\}.
    \end{align}
    \label{c223f}
    \end{subequations}
    where $a=1, 2, 3$. The following form is used to express the outcome of this integral using the functions mentioned above:
\bea \label{c223aI} (\bold{I_{3}})_{j}({\cal K}_i,k_a) &=& \frac{k_e^{6} \exp{(-i x_{e,i})}}{720 c_{s}^8 k_s^9}\bigg\{{\cal P}_{2,3}({\cal K}_i,k_a){\cal Q}_{2,3}({\cal K}_i) + 120 \nonumber\\
    &&-\frac{k_s^{6}}{k_e^{6}}\exp{(i(x_{e,i}-x_{s,i}))}\bigg({\cal Q}_{2,3}({\cal K}_1){\cal R}_{2,3}({\cal K}_i)+120 \bigg)\bigg\}. \eea
    \begin{enumerate}
    \item Consider \textbf{Integral 3a}: \bea \label{c223a}\displaystyle{\int_{\tau_{s}}^{\tau_{e}}d\tau_{1}\frac{\tau_{s}^{9}}{\tau_{1}^{9}c_{s}^{9}}\exp{(ic_{s}{\cal K}_{1}\tau_{1})}}(k_{1}^2c_{s}^2\tau_{1})(k_{2}^2c_{s}^2\tau_{1})(1-ik_{3}c_{s}\tau_{1}) = k_{1}^{2}k_{2}^{2}(\bold{I_{3}})_{1}({\cal K}_1,-k_3).
    \eea 
The accompanying values within for each such integral determine the momentum variables signature within the result.
 \item For \textbf{Integral 3b}:
    \bea \label{c223b} \displaystyle{\int_{\tau_{s}}^{\tau_{e}}d\tau_{1}\frac{\tau_{s}^{9}}{\tau_{1}^{9}c_{s}^{9}}\exp{(-ic_{s}{\cal K}_{3}\tau_{1})}}(k_{1}^2c_{s}^2\tau_{1})(k_{2}^2c_{s}^2\tau_{1})(1-ik_{3}c_{s}\tau_{1}) =  k_{1}^{2}k_{2}^{2}(\bold{I_{3}})^{*}_{3}({\cal K}_1, -k_3).
    \eea
    As seen for the foregoing integral, the outcome of such integrals is connected to the expression in Eq. (\ref{c223aI}) via a complex conjugate.

    \end{enumerate}
A total of $4$ second-kind integrals with positive and negative exponential factors will be present.

    \end{itemize}
We apply the formula in Eq.(\ref{c1c}) to calculate the bispectrum contribution for \underline{\textbf{Operator 3}: $\displaystyle{a(\tau_{1})\zeta^{'}(\partial_{i}\zeta)^{2}}$}. For this operator, the following integral types are available:
\begin{itemize}
    \item   \underline{\textbf{Integral of the first kind}}:
    \begin{enumerate}
        \item Consider \textbf{Integral 1a}:
        \bea \label{c231a} \displaystyle{\int_{\tau_{s}}^{\tau_{e}}d\tau_{1}\frac{-3\tau_{s}^{9}}{\tau_{1}^{11}c_{s}^{9}}\exp{(ic_{s}{\cal K}_{1}\tau_{1})}}(1-ik_{1}c_{s}\tau_{1})(1-ik_{2}c_{s}\tau_{1})(1-ik_{3}c_{s}\tau_{1}) = (\bold{I_{2}})_{1}({\cal K}_1,k_1,k_2,k_3).
    \eea 
This integral's form is precisely the same as the one specified in Eq. (\ref{c221a}). Therefore, we can write the result for the form of integrals in Eq.(\ref{c231a}) using the same functions that were previously defined in Eq.(\ref{c221f}) with the same general solution that was described in Eq.(\ref{c221aI}). The momentum arguments' signatures rely on each other's signatures in ${\cal K}_{i}$.
  \item For \textbf{Integral 1b}:
        \bea \displaystyle{\int_{\tau_{s}}^{\tau_{e}}d\tau_{1}\frac{3\tau_{s}^{9}}{\tau_{1}^{11}c_{s}^{9}}\exp{(-ic_{s}{\cal K}_{4}\tau_{1})}}(1-ik_{1}c_{s}\tau_{1})(1+ik_{2}c_{s}\tau_{1})(1+ik_{3}c_{s}\tau_{1}) = -(\bold{I_{1}})^{*}_{4}({\cal K}_4,-k_1,k_2,k_3).
    \eea
    Additionally, this integral and the one stated in Eq. (\ref{c221b}) are absolutely equal. Therefore, we can express the outcomes of such integrals by utilizing the identical functions that were previously described in Eq. (\ref{c221f}), but now using a complex conjugate of the general expression in Eq. (\ref{c221aI}).

    \end{enumerate}
This first form of integral will have a total of $4$ integrals, each with a positive and negative exponential component. 

 \item  \underline{\textbf{Integral of the second kind}}:
    \begin{enumerate}
        \item Consider \textbf{Integral 2a}: \bea \label{c232a}\displaystyle{\int_{\tau_{s}}^{\tau_{e}}d\tau_{1}\frac{\tau_{s}^{9}}{\tau_{1}^{10}c_{s}^{9}}\exp{(ic_{s}{\cal K}_{1}\tau_{1})}}(1-ik_{1}c_{s}\tau_{1})(1-ik_{2}c_{s}\tau_{1})(k_{3}^{2}c_{s}^{2}\tau_{1}) =  k_{3}^{2}(\bold{I_{1}})_{1}({\cal K}_1,-k_1,-k_2).
        \eea
This integral's form is precisely the same as the one specified in Equation (\ref{c222a}). Therefore, we can write the result of the integrals of the form in Eq. (\ref{c232a}) using the same functions as stated before Eq. (\ref{c222f}) along with the general solution defined in Eq. (\ref{c222aI}). The momentum arguments' signatures rely on each other's signatures inside the integral.
\item For \textbf{Integral 2b}:
       \bea \displaystyle{\int_{\tau_{s}}^{\tau_{e}}d\tau_{1}\frac{-\tau_{s}^{9}}{\tau_{1}^{10}c_{s}^{9}}\exp{(ic_{s}{\cal K}_{3}\tau_{1})}}(1+ik_{1}c_{s}\tau_{1})(1+ik_{2}c_{s}\tau_{1})(k_{3}^{2}c_{s}^{2}\tau_{1})= -k_{3}^{2}(\bold{I_{1}})^{*}_{1}({\cal K}_1,k_1,k_2).
       \eea
        \end{enumerate}
        This first form of integral will have a total of $4$ integrals, each with a positive and negative exponential component.

    \end{itemize}
We utilize the formula in Eq.(\ref{c1d}) to calculate the bispectrum contributions for \underline{\textbf{Operator 4}: $\displaystyle{(\partial^{2}\zeta)(\partial_{i}\zeta)^{2}}$}. We only have one type of integral here. 
\begin{itemize}
     \item Consider \underline{\textbf{Integral of the first kind}}:
The equivalent set of functions for an integral of this type are as follows:
\begin{subequations}
    \begin{align}
    {\cal P}_{4,1}({\cal K}_{i}, k_{1},k_{2},k_{3}) &= \frac{-7{\cal K}_{i}^3}{k_e^3}+\frac{56 {\cal K}_{i} \left(k_2 k_3+k_1 k_2+k_1 k_3\right)}{k_e^3}-\frac{336 k_1 k_2 k_3}{k_e^3}.\\
    {\cal Q}_{2,1}({\cal K}_{i}) &= \frac{ i{\cal K}_{i}^4}{k_e^4}-\frac{{\cal K}_{i}^3}{k_e^3} -\frac{2 i{\cal K}_{i}^2}{k_e^2} + \frac{6{\cal K}_{i}}{k_e} + 24i. \\
    {\cal R}_{2,1}({\cal K}_{i}) &= k_e^3 \bigg\{\frac{{\cal K}_{i}^5 \exp{(ix_{s,i})} \left(\text{Ei}\left(-ix_{e,i}\right)-\text{Ei}\left(-ix_{s,i}\right)\right)}{k_s^{8}}+\frac{i
    {\cal K}_{i}^4}{k_s^{7}}-\frac{{\cal K}_{i}^3}{k_s^{6}}-\frac{2i{\cal K}_{i}^2}{k_s^5}+\frac{6{\cal K}_{i}}{k_s^4}+\frac{24i}{k_s^3}\bigg\}.
\end{align} 
\label{c241f}
\end{subequations}
The following form is used to express the outcome of this integral using the functions mentioned above:
 \bea \label{c224aI} (\bold{I_{1}})_{j}({\cal K}_i,k_1,k_2,k_3) &=& \frac{k_e^{8} \exp{(-i x_{e,i})}}{40320 c_{s}^{10} k_s^9}\bigg\{{\cal P}_{4,1}({\cal K}_i,k_1,k_2,k_3){\cal Q}_{4,1}({\cal K}_i) + i 5040\frac{{\cal K}_{i}}{k_e} \nonumber\\
    && -\frac{ 120\left(56\left(k_2 k_3+k_1 k_2+k_1 k_3\right)-7 {\cal K}_i^{2}\right)}{k_e^2} + 5040  \nonumber\\
    && -\frac{k_s^{8}}{k_e^{8}}\exp{(i(x_{e,i}-x_{s,i}))}\bigg({\cal Q}_{4,1}({\cal K}_i){\cal R}_{4,1}({\cal K}_i)+5040 +i\frac{5040
    {\cal K}_i}{k_s} \nonumber\\
    && - \frac{120\left(56\left(k_2 k_3+k_1 k_2+k_1 k_3\right)-7 {\cal K}_i^{2}\right)}{k_s^2} \bigg) \bigg\}. \eea
    \begin{enumerate}
    \item \textbf{Integral 1a}: \bea \label{c224a} \displaystyle{\int_{\tau_{s}}^{\tau_{e}}d\tau_{1}\frac{\tau_{s}^{9}}{\tau_{1}^{9}c_{s}^{9}}\exp{(ic_{s}{\cal K}_{1}\tau_{1})}}(1-ik_{1}c_{s}\tau_{1})(1-ik_{2}c_{s}\tau_{1})(1-ik_{3}c_{s}\tau_{1}) = (\bold{I_{1}})_{1}({\cal K}_1,k_1,k_2,k_3). \eea
The signatures of the momentum arguments in the corresponding outcomes of such integrals rely on ${\cal K}_{i}$.
 \item For \textbf{Integral 1b}:
    \bea \label{c224b}\displaystyle{\int_{\tau_{s}}^{\tau_{e}}d\tau_{1}\frac{\tau_{s}^{9}}{\tau_{1}^{9}c_{s}^{9}}\exp{(-ic_{s}{\cal K}_{3}\tau_{1})}}(1+ik_{1}c_{s}\tau_{1})(1+ik_{2}c_{s}\tau_{1})(1-ik_{3}c_{s}\tau_{1}) = (\bold{I_{1}})^{*}_{3}({\cal K}_3,k_1,k_2,-k_3).
    \eea
    As seen with the previous integral, the result of this integral is connected to the expression in Eq. (\ref{c224aI}) via a complex conjugate.

    \end{enumerate}
Four integrals of the fourth kind, each having a positive and negative exponential component, will amount $4$.

     \end{itemize}

\subsection{Detailed computation in the region III: SRII}

The detailed calculations for the contributions made by each operator to the three-point correlation function at tree-level in the SRII area are shown in this subsection. Since the time-derivative of the mode function would be highly useful for additional computations, we mention it in the SRII phase. This is provided by:
\bea \Pi_{k}^{*} = \partial_{\tau}\zeta_{k}^{*} = \left(\frac{-iH^{2}}{2\sqrt{\cal A}}\right)\left(\frac{\tau_{s}}{\tau_{e}}\right)^{3}\frac{1}{(c_{s}k)^{3/2}}\times\left(k^{2}c_{s}^{2}\tau\right)\left(\alpha_{k}^{(3)*}\exp{(ikc_{s}\tau)}-\beta_{k}^{(3)*}\exp{(-ikc_{s}\tau)}\right).
\eea
where $\bold{k}|$ = k. The factor ${\cal K}_{i}$ in the ensuing integrals is the same as the factor from Eq. (\ref{c2k}).

To calculate the contribution to the bispectrum, we begin with \underline{\textbf{Operator 1}: $\displaystyle{a(\tau_{1})\zeta^{'3}}$} and use the method found in Eq.(\ref{c1a}).
\begin{enumerate}
\item Consider \textbf{Integral 1a}:
\bea \label{c31a} 
\displaystyle{\int_{\tau_{e}}^{\tau_{end}\rightarrow 0}d\tau_{1}k_{1}^{2}k_{2}^{2}k_{3}^{2}c_{s}^{-3}\tau_{1}^{2}\exp{(ic_{s}{\cal K}_{1}\tau_{1})}} = k_1^2 k_2^2 k_3^2(\bold{J}_{1})_{1}({\cal K}_{1}).
\eea 
And the following generic formula is used to express the outcome of the aforementioned integral: 
\bea \label{c31aI} 
(\bold{J}_{1})_{i}({\cal K}_{i}) = \frac{1}{c_{s}^6 {\cal K}_{i}^3}\left\{\left(2 i-\exp{\left(-\frac{i {\cal K}_{i}}{k_e}\right)} \left(2 i-\frac{{\cal K}_{i}}{k_e} \left(2+\frac{i{\cal K}_{i}}{k_e}\right)\right)\right)\right\}
\eea
By altering the element ${\cal K}_{i}$ in the above general formula, other integrals that resemble the above will be connected.
\item For \textbf{Integral 1b}:
\bea \label{c31b} 
\displaystyle{\int_{\tau_{e}}^{\tau_{end}\rightarrow 0}d\tau_{1}k_{1}^{2}k_{2}^{2}k_{3}^{2}c_{s}^{-3}\tau_{1}^{2}\exp{(-ic_{s}{\cal K}_{4}\tau_{1})}} = k_{1}^{2}k_{2}^{2}k_{3}^{2}(\bold{J}_{1})^{*}_{4}({\cal K}_{4}).
\eea
Its outcome may be expressed in terms of the general formula's complex conjugate in Eq. (\ref{c31aI}).

\end{enumerate}
Different ${\cal K}_{i}$'s and positive and negative exponential factors will be included in the $4$ first-kind integrals.

To calculate the contribution to the bispectrum for \underline{\textbf{Operator 2}: $\displaystyle{\zeta^{'2}\partial^{2}\zeta}$}, we employ the method found in Eq.(\ref{c1b}).
\begin{enumerate}
    \item Consider \textbf{Integral 2a}:
    \bea \label{c32a} 
    \displaystyle{\int_{\tau_{e}}^{\tau_{end}\rightarrow 0}d\tau_{1}k_{1}^{2}k_{2}^{2}(1-ik_{3}c_{s}\tau_{1})c_{s}^{-5}\tau_{1}^{2}\exp{(ic_{s}{\cal K}_{1}\tau_{1})}} = k_1^{2}k_2^{2}(\bold{J}_{2})_{1}({\cal K}_{1},-k_3).\eea
The following generic formula is used to express the outcome of the aforementioned integral: 
\bea \label{c32aI}
(\bold{J}_{2})_{i}({\cal K}_{i},k_a) &=& \frac{1}{c^8 {\cal K}_{i}^4}\Bigg\{\left({\cal K}_{i}\left(2 i-\exp{\left(-\frac{i {\cal K}_{i}}{k_e}\right)} \left(2 i-\frac{{\cal K}_{i}}{k_e}\left(2+\frac{i{\cal K}_{i}}{k_e}\right)\right)\right)\right)\nonumber\\
&&\quad\quad\quad\quad\quad\quad\quad\quad\quad\quad +i k_a \left(-6+\exp{\left(-\frac{i {\cal K}_{i}}{k_e}\right)}\left(6+\frac{{\cal K}_{i}}{k_e}\left(6 i-\frac{{\cal K}_{i}}{k_e}\left(3+\frac{i {\cal K}_{i}}{k_e}\right)\right)\right)\right)\Bigg\}.\quad\quad\quad
\eea 
Similar integrals like the one above can be connected by altering the component ${\cal K}_{i}$ and the momentum variable signatures in the general formula above, which are dependent on the values of the corresponding variables inside the integral. $k_{a} = -k_{3}$ in the example above.
 \item For \textbf{Integral 2b}:
    \bea \label{c32b}
    \displaystyle{\int_{\tau_{e}}^{\tau_{end}\rightarrow 0}d\tau_{1}k_{1}^{2}k_{2}^{2}(1+ik_{3}c_{s}\tau_{1})c_{s}^{-5}\tau_{1}^{2}\exp{(-ic_{s}{\cal K}_{2}\tau_{1})}} = k_1^{2}k_2^{2}(\bold{J}_{2})^{*}_{2}({\cal K}_{2}).\eea
    Equation (\ref{c32aI}) expresses the outcome of this integral by taking the complex conjugate of the generic formula.

    \end{enumerate}
    There will be $4$ second-kind integrals with varying ${\cal K}_{i}$ and positive and negative exponential components.

We apply the formula in Eq.(\ref{c1c}) to calculate the bispectrum contribution for \underline{\textbf{Operator 3}: $\displaystyle{a(\tau_{1})\zeta^{'}(\partial_{i}\zeta)^{2}}$}. 
\begin{enumerate}
    \item Consider \textbf{Integral 3a}:
    \bea \label{c33a} 
    \displaystyle{\int_{\tau_{e}}^{\tau_{end}\rightarrow 0}d\tau_{1}(1-ik_{1}c_{s}\tau_{1})(1-ik_{2}c_{s}\tau_{1})k_{3}^{2}c_{s}^{-7}\exp{(ic_{s}{\cal K}_{1}\tau_{1})}} = k_{3}^{2}(\bold{J}_{3})_{1}({\cal K}_{1},k_1,k_2,k_3).
    \eea
The following generic formula is used to express the outcome of the aforementioned integral: 
\bea \label{c33aI} 
     (\bold{J}_{3})_{i}({\cal K}_{i},k_1,k_2,k_3) &=& -\frac{1}{c^{8}{\cal K}_{i}{}^3}\Bigg\{i \left(2 k_1^2+3 \left(2 k_2+k_3\right) k_1+\left(k_2+k_3\right) \left(2
    k_2+k_3\right)\right)-\exp{\left(-\frac{i{\cal K}_{i}}{k_e}\right)} \nonumber\\
    &&\bigg(-i k_1 k_2\frac{ {\cal K}_{i}{}^2}{k_e^2} +i \left(2 k_1^2+3 \left(2 k_2+k_3\right)
    k_1+\left(k_2+k_3\right) \left(2 k_2+k_3\right)\right)\nonumber\\
    && -{\cal K}_{i}\frac{\left(k_1^2+\left(4 k_2+k_3\right) k_1+k_2
    \left(k_2+k_3\right)\right)}{k_e}\bigg)\Bigg\}.\eea
    Similar integrals to the one above can be connected by altering the component ${\cal K}_{i}$ and the momentum variables' signature in the general formula above, which also depends on the values of the corresponding variables inside ${\cal K}_{i}$.
   \item For \textbf{Integral 3b}:
     \bea \label{c33b} 
    \displaystyle{\int_{\tau_{e}}^{\tau_{end}\rightarrow 0}d\tau_{1}(1+ik_{1}c_{s}\tau_{1})(1+ik_{2}c_{s}\tau_{1})k_{3}^{2}c_{s}^{-7}\exp{(-ic_{s}{\cal K}_{3}\tau_{1})}} = k_{3}^{2}(\bold{J}_{3})^{*}_{3}({\cal K}_{3},k_1,k_2,-k_3).
     \eea
One way to express the outcome of this integral is using the complex conjugate of the generic formula found in Eq. (\ref{c33aI}).

    \end{enumerate}
    There will be $4$ third-kind integrals, with varying ${\cal K}_{i}$ and positive and negative exponential components.

We apply the formula in Eq.(\ref{c1d}) to calculate the bispectrum contribution for \underline{\textbf{Operator 4}: $\partial^{2}\zeta(\partial_{i}\zeta)^{2}$}. 
\begin{enumerate}
    \item Consider \textbf{Integral 4a}:
    \bea \label{c34a} 
    \displaystyle{\int_{\tau_{e}}^{\tau_{end}\rightarrow 0}d\tau_{1}(1-ik_{1}c_{s}\tau_{1})(1-ik_{2}c_{s}\tau_{1})(1-ik_{3}c_{s}\tau_{1})c_{s}^{-9}\exp{(ic_{s}{\cal K}_{2}\tau_{1})}} = (\bold{J}_{4})_{1}({\cal K}_{2},k_1,k_2,k_3).
    \eea
The outcome gets stated using the generic formula that follows:
\bea \label{c34aI} (\bold{J}_{4})_{j}({\cal K}_{i},k_1,k_2,k_3) &=& \frac{i}{c^{10}{\cal K}_{i}^4}\Bigg\{\bigg(-2 \left(k_1^3+4 \left(k_2+k_3\right) k_1^2+4 \left(k_2^2+3 k_3 k_2+k_3^2\right) k_1+\left(k_2+k_3\right) \left(k_2^2+3 k_3 k_2+k_3^2\right)\right)\nonumber\\
    && +\exp{\left(-\frac{i {\cal K}_{i}}{k_e}\right)} \bigg(-\frac{{\cal K}_{i}^2\left(\left(k_2+k_3\right) k_1^2+\left(k_2^2+6 k_3 k_2+k_3^2\right) k_1+k_2 k_3 \left(k_2+k_3\right)\right)}{k_e^2}-i k_1 k_2 k_3 \frac{{\cal K}_{i}{}^3}{k_e^3}\nonumber\\
    && +\frac{i {\cal K}_{i} \left(k_1^3+5 \left(k_2+k_3\right) k_1^2+\left(5 k_2^2+18 k_3 k_2+5 k_3^2\right) k_1+\left(k_2+k_3\right)\left(k_2^2+4 k_3 k_2+k_3^2\right)\right)}{k_e}\nonumber\\
    && +2 \left(k_1^3+4 \left(k_2+k_3\right) k_1^2+4 \left(k_2^2+3 k_3 k_2+k_3^2\right) k_1+\left(k_2+k_3\right) \left(k_2^2+3 k_3 k_2+k_3^2\right)\right)\bigg)\bigg)\Bigg\}.
    \eea
In the generic formula above, different integrals that resemble the above will be connected by altering the factor ${\cal K}_{i}$ and the momentum variable signatures.
   \item For \textbf{Integral 4b}:
     \bea \label{c34b} 
    \displaystyle{\int_{\tau_{e}}^{\tau_{end}\rightarrow 0}d\tau_{1}(1+ik_{1}c_{s}\tau_{1})(1+ik_{2}c_{s}\tau_{1})(1-ik_{3}c_{s}\tau_{1})c_{s}^{-9}\exp{(-ic_{s}{\cal K}_{3}\tau_{1})}} = k_{3}^{2}(\bold{J}_{4})^{*}_{3}({\cal K}_{3},k_1,k_2,-k_3).
     \eea
     One way to express the outcome of this integral is using the complex conjugate of the generic formula found in Eq. (\ref{c34aI}).

    \end{enumerate}
    There will be $4$ fourth-kind integrals with varying ${\cal K}_{i}$ and positive and negative exponential components.
    
\section{One-loop correction to the two-point function in presence of cubic self interaction for Galileon inflation}
\label{A5a}

 The one-loop contributions to the primordial two-point cosmic correlation function can be calculated using the following formulas, which are based on the individual cubic self-interactions:
\bea \label{A11}\langle\hat{\zeta}_{\bf p}\hat{\zeta}_{-{\bf p}}\rangle_{(0,2)}&=&\sum^{4}_{i=1}{\bf Z}^{(1)}_i,\\ 
\label{A22}\langle\hat{\zeta}_{\bf p}\hat{\zeta}_{-{\bf p}}\rangle^{\dagger}_{(0,2)}&=&\sum^{4}_{i=1}{\bf Z}^{(2)}_i,\\
\label{A33}\langle\hat{\zeta}_{\bf p}\hat{\zeta}_{-{\bf p}}\rangle^{\dagger}_{(1,1)}&=&\sum^{4}_{i=1}{\bf Z}^{(3)}_i,\eea
where the following formulae explain the factors ${\bf Z}^{(1)}_i\forall i=1,2,3,4$, ${\bf Z}^{(2)}_i\forall i=1,2,3,4$, and ${\bf Z}^{(3)}_i\forall i=1,2,3,4$:
\begin{eqnarray} \label{A11a}{\bf Z}^{(1)}_1:&=&\lim_{\tau\rightarrow 0}\Bigg[\int^{0}_{-\infty}d\tau_1\frac{a^2(\tau_1)}{H^3(\tau_1)}\frac{{\cal G}_1(\tau_1)}{a(\tau_1)}\;\int^{0}_{-\infty}d\tau_2\;\frac{a^2(\tau_2)}{H^3(\tau_2)}\frac{{\cal G}_1(\tau_2)}{a(\tau_2)}\nonumber\\
  &&\quad\quad\quad\quad\times\int \frac{d^{3}{\bf k}_1}{(2\pi)^3} \int \frac{d^{3}{\bf k}_2}{(2\pi)^3} \int \frac{d^{3}{\bf k}_3}{(2\pi)^3} \int \frac{d^{3}{\bf k}_4}{(2\pi)^3} \int \frac{d^{3}{\bf k}_5}{(2\pi)^3} \int \frac{d^{3}{\bf k}_6}{(2\pi)^3}\nonumber\\
  &&\quad\quad\quad\quad\times \delta^3\bigg({\bf k}_1+{\bf k}_2+{\bf k}_3\bigg) \delta^3\bigg({\bf k}_4+{\bf k}_5+{\bf k}_6\bigg)\nonumber\\
  &&\quad\quad\quad\quad\times \langle \hat{\zeta}_{\bf p}(\tau)\hat{\zeta}_{-{\bf p}}(\tau)\hat{\zeta}^{'}_{{\bf k}_1}(\tau_1)\hat{\zeta}^{'}_{{\bf k}_2}(\tau_1)\hat{\zeta}^{'}_{{\bf k}_3}(\tau_1)\hat{\zeta}^{'}_{{\bf k}_4}(\tau_2)\hat{\zeta}^{'}_{{\bf k}_5}(\tau_2)\hat{\zeta}^{'}_{{\bf k}_6}(\tau_2)\rangle\Bigg],\\ 
  \label{A11b}{\bf Z}^{(1)}_2:&=&\lim_{\tau\rightarrow 0}\Bigg[\int^{0}_{-\infty}d\tau_1\frac{a^2(\tau_1)}{H^3(\tau_1)}\frac{{\cal G}_2(\tau_1)}{a^2(\tau_1)}\;\int^{0}_{-\infty}d\tau_2\;\frac{a^2(\tau_2)}{H^3(\tau_2)}\frac{{\cal G}_2(\tau_2)}{a^2(\tau_2)}\nonumber\\
  &&\quad\quad\quad\quad\times\int \frac{d^{3}{\bf k}_1}{(2\pi)^3} \int \frac{d^{3}{\bf k}_2}{(2\pi)^3} \int \frac{d^{3}{\bf k}_3}{(2\pi)^3} \int \frac{d^{3}{\bf k}_4}{(2\pi)^3} \int \frac{d^{3}{\bf k}_5}{(2\pi)^3} \int \frac{d^{3}{\bf k}_6}{(2\pi)^3}\nonumber\\
  &&\quad\quad\quad\quad\times \delta^3\bigg({\bf k}_1+{\bf k}_2+{\bf k}_3\bigg) \delta^3\bigg({\bf k}_4+{\bf k}_5+{\bf k}_6\bigg)|{\bf K}|^4\nonumber\\
  &&\quad\quad\quad\quad\times \langle \hat{\zeta}_{\bf p}(\tau)\hat{\zeta}_{-{\bf p}}(\tau)\hat{\zeta}^{'}_{{\bf k}_1}(\tau_1)\hat{\zeta}^{'}_{{\bf k}_2}(\tau_1)\hat{\zeta}_{{\bf k}_3}(\tau_1)\hat{\zeta}^{'}_{{\bf k}_4}(\tau_2)\hat{\zeta}^{'}_{{\bf k}_5}(\tau_2)\hat{\zeta}_{{\bf k}_6}(\tau_2)\rangle\Bigg],\\
 \label{A11c}{\bf Z}^{(1)}_3:&=&\lim_{\tau\rightarrow 0}\Bigg[\int^{0}_{-\infty}d\tau_1\frac{a^2(\tau_1)}{H^3(\tau_1)}\frac{{\cal G}_3(\tau_1)}{a(\tau_1)}\;\int^{0}_{-\infty}d\tau_2\;\frac{a^2(\tau_2)}{H^3(\tau_2)}\frac{{\cal G}_3(\tau_2)}{a(\tau_2)}\nonumber\\
  &&\quad\quad\quad\quad\times\int \frac{d^{3}{\bf k}_1}{(2\pi)^3} \int \frac{d^{3}{\bf k}_2}{(2\pi)^3} \int \frac{d^{3}{\bf k}_3}{(2\pi)^3} \int \frac{d^{3}{\bf k}_4}{(2\pi)^3} \int \frac{d^{3}{\bf k}_5}{(2\pi)^3} \int \frac{d^{3}{\bf k}_6}{(2\pi)^3}\nonumber\\
  &&\quad\quad\quad\quad\times \delta^3\bigg({\bf k}_1+{\bf k}_2+{\bf k}_3\bigg) \delta^3\bigg({\bf k}_4+{\bf k}_5+{\bf k}_6\bigg)\left({\bf k}_2\cdot{\bf k}_3\right)\left({\bf k}_5\cdot{\bf k}_6\right)\nonumber\\
  &&\quad\quad\quad\quad\times \langle \hat{\zeta}_{\bf p}(\tau)\hat{\zeta}_{-{\bf p}}(\tau)\hat{\zeta}^{'}_{{\bf k}_1}(\tau_1)\hat{\zeta}_{{\bf k}_2}(\tau_1)\hat{\zeta}_{{\bf k}_3}(\tau_1)\hat{\zeta}^{'}_{{\bf k}_4}(\tau_2)\hat{\zeta}_{{\bf k}_5}(\tau_2)\hat{\zeta}_{{\bf k}_6}(\tau_2)\rangle\Bigg],\\
   \label{A11d}{\bf Z}^{(1)}_4:&=&\lim_{\tau\rightarrow 0}\Bigg[\int^{0}_{-\infty}d\tau_1\frac{a^2(\tau_1)}{H^3(\tau_1)}\frac{{\cal G}_4(\tau_1)}{a^2(\tau_1)}\;\int^{0}_{-\infty}d\tau_2\;\frac{a^2(\tau_2)}{H^3(\tau_2)}\frac{{\cal G}_4(\tau_2)}{a^2(\tau_2)}\nonumber\\
  &&\quad\quad\quad\quad\times\int \frac{d^{3}{\bf k}_1}{(2\pi)^3} \int \frac{d^{3}{\bf k}_2}{(2\pi)^3} \int \frac{d^{3}{\bf k}_3}{(2\pi)^3} \int \frac{d^{3}{\bf k}_4}{(2\pi)^3} \int \frac{d^{3}{\bf k}_5}{(2\pi)^3} \int \frac{d^{3}{\bf k}_6}{(2\pi)^3}\nonumber\\
  &&\quad\quad\quad\quad\times \delta^3\bigg({\bf k}_1+{\bf k}_2+{\bf k}_3\bigg) \delta^3\bigg({\bf k}_4+{\bf k}_5+{\bf k}_6\bigg)\left({\bf k}_1\cdot{\bf k}_2\right)\left({\bf k}_4\cdot{\bf k}_5\right)|{\bf K}|^4\nonumber\\
  &&\quad\quad\quad\quad\times \langle \hat{\zeta}_{\bf p}(\tau)\hat{\zeta}_{-{\bf p}}(\tau)\hat{\zeta}_{{\bf k}_1}(\tau_1)\hat{\zeta}_{{\bf k}_2}(\tau_1)\hat{\zeta}_{{\bf k}_3}(\tau_1)\hat{\zeta}_{{\bf k}_4}(\tau_2)\hat{\zeta}_{{\bf k}_5}(\tau_2)\hat{\zeta}_{{\bf k}_6}(\tau_2)\rangle\Bigg],\end{eqnarray}
  and
\begin{eqnarray} \label{A22a}{\bf Z}^{(2)}_1:&=&\lim_{\tau\rightarrow 0}\Bigg[\int^{0}_{-\infty}d\tau_1\frac{a^2(\tau_1)}{H^3(\tau_1)}\frac{{\cal G}_1(\tau_1)}{a(\tau_1)}\;\int^{0}_{-\infty}d\tau_2\;\frac{a^2(\tau_2)}{H^3(\tau_2)}\frac{{\cal G}_1(\tau_2)}{a(\tau_2)}\nonumber\\
  &&\quad\quad\quad\quad\times\int \frac{d^{3}{\bf k}_1}{(2\pi)^3} \int \frac{d^{3}{\bf k}_2}{(2\pi)^3} \int \frac{d^{3}{\bf k}_3}{(2\pi)^3} \int \frac{d^{3}{\bf k}_4}{(2\pi)^3} \int \frac{d^{3}{\bf k}_5}{(2\pi)^3} \int \frac{d^{3}{\bf k}_6}{(2\pi)^3}\nonumber\\
  &&\quad\quad\quad\quad\times \delta^3\bigg({\bf k}_1+{\bf k}_2+{\bf k}_3\bigg) \delta^3\bigg({\bf k}_4+{\bf k}_5+{\bf k}_6\bigg)\nonumber\\
  &&\quad\quad\quad\quad\times \langle \hat{\zeta}_{\bf p}(\tau)\hat{\zeta}_{-{\bf p}}(\tau)\hat{\zeta}^{'}_{{\bf k}_1}(\tau_1)\hat{\zeta}^{'}_{{\bf k}_2}(\tau_1)\hat{\zeta}^{'}_{{\bf k}_3}(\tau_1)\hat{\zeta}^{'}_{{\bf k}_4}(\tau_2)\hat{\zeta}^{'}_{{\bf k}_5}(\tau_2)\hat{\zeta}^{'}_{{\bf k}_6}(\tau_2)\rangle^{\dagger}\Bigg]=\big[{\bf Z}^{(1)}_1\big]^{\dagger},\\
  \label{A22b}{\bf Z}^{(2)}_2:&=&\lim_{\tau\rightarrow 0}\Bigg[\int^{0}_{-\infty}d\tau_1\frac{a^2(\tau_1)}{H^3(\tau_1)}\frac{{\cal G}_2(\tau_1)}{a^2(\tau_1)}\;\int^{0}_{-\infty}d\tau_2\;\frac{a^2(\tau_2)}{H^3(\tau_2)}\frac{{\cal G}_2(\tau_2)}{a^2(\tau_2)}\nonumber\\
  &&\quad\quad\quad\quad\times\int \frac{d^{3}{\bf k}_1}{(2\pi)^3} \int \frac{d^{3}{\bf k}_2}{(2\pi)^3} \int \frac{d^{3}{\bf k}_3}{(2\pi)^3} \int \frac{d^{3}{\bf k}_4}{(2\pi)^3} \int \frac{d^{3}{\bf k}_5}{(2\pi)^3} \int \frac{d^{3}{\bf k}_6}{(2\pi)^3}\nonumber\\
  &&\quad\quad\quad\quad\times \delta^3\bigg({\bf k}_1+{\bf k}_2+{\bf k}_3\bigg) \delta^3\bigg({\bf k}_4+{\bf k}_5+{\bf k}_6\bigg)|{\bf K}|^4\nonumber\\
  &&\quad\quad\quad\quad\times \langle \hat{\zeta}_{\bf p}(\tau)\hat{\zeta}_{-{\bf p}}(\tau)\hat{\zeta}^{'}_{{\bf k}_1}(\tau_1)\hat{\zeta}^{'}_{{\bf k}_2}(\tau_1)\hat{\zeta}_{{\bf k}_3}(\tau_1)\hat{\zeta}^{'}_{{\bf k}_4}(\tau_2)\hat{\zeta}^{'}_{{\bf k}_5}(\tau_2)\hat{\zeta}_{{\bf k}_6}(\tau_2)\rangle^{\dagger}\Bigg]=\big[{\bf Z}^{(1)}_2\big]^{\dagger},\\
 \label{A22c}{\bf Z}^{(2)}_3:&=&\lim_{\tau\rightarrow 0}\Bigg[\int^{0}_{-\infty}d\tau_1\frac{a^2(\tau_1)}{H^3(\tau_1)}\frac{{\cal G}_3(\tau_1)}{a(\tau_1)}\;\int^{0}_{-\infty}d\tau_2\;\frac{a^2(\tau_2)}{H^3(\tau_2)}\frac{{\cal G}_3(\tau_2)}{a(\tau_2)}\nonumber\\
  &&\quad\quad\quad\quad\times\int \frac{d^{3}{\bf k}_1}{(2\pi)^3} \int \frac{d^{3}{\bf k}_2}{(2\pi)^3} \int \frac{d^{3}{\bf k}_3}{(2\pi)^3} \int \frac{d^{3}{\bf k}_4}{(2\pi)^3} \int \frac{d^{3}{\bf k}_5}{(2\pi)^3} \int \frac{d^{3}{\bf k}_6}{(2\pi)^3}\nonumber\\
  &&\quad\quad\quad\quad\times \delta^3\bigg({\bf k}_1+{\bf k}_2+{\bf k}_3\bigg) \delta^3\bigg({\bf k}_4+{\bf k}_5+{\bf k}_6\bigg)\left({\bf k}_2\cdot{\bf k}_3\right)\left({\bf k}_5\cdot{\bf k}_6\right)\nonumber\\
  &&\quad\quad\quad\quad\times \langle \hat{\zeta}_{\bf p}(\tau)\hat{\zeta}_{-{\bf p}}(\tau)\hat{\zeta}^{'}_{{\bf k}_1}(\tau_1)\hat{\zeta}_{{\bf k}_2}(\tau_1)\hat{\zeta}_{{\bf k}_3}(\tau_1)\hat{\zeta}^{'}_{{\bf k}_4}(\tau_2)\hat{\zeta}_{{\bf k}_5}(\tau_2)\hat{\zeta}_{{\bf k}_6}(\tau_2)\rangle^{\dagger}\Bigg]=\big[{\bf Z}^{(1)}_3\big]^{\dagger},\\
   \label{A22d}{\bf Z}^{(2)}_4:&=&\lim_{\tau\rightarrow 0}\Bigg[\int^{0}_{-\infty}d\tau_1\frac{a^2(\tau_1)}{H^3(\tau_1)}\frac{{\cal G}_4(\tau_1)}{a^2(\tau_1)}\;\int^{0}_{-\infty}d\tau_2\;\frac{a^2(\tau_2)}{H^3(\tau_2)}\frac{{\cal G}_4(\tau_2)}{a^2(\tau_2)}\nonumber\\
  &&\quad\quad\quad\quad\times\int \frac{d^{3}{\bf k}_1}{(2\pi)^3} \int \frac{d^{3}{\bf k}_2}{(2\pi)^3} \int \frac{d^{3}{\bf k}_3}{(2\pi)^3} \int \frac{d^{3}{\bf k}_4}{(2\pi)^3} \int \frac{d^{3}{\bf k}_5}{(2\pi)^3} \int \frac{d^{3}{\bf k}_6}{(2\pi)^3}\nonumber\\
  &&\quad\quad\quad\quad\times \delta^3\bigg({\bf k}_1+{\bf k}_2+{\bf k}_3\bigg) \delta^3\bigg({\bf k}_4+{\bf k}_5+{\bf k}_6\bigg)\left({\bf k}_1\cdot{\bf k}_2\right)\left({\bf k}_4\cdot{\bf k}_5\right)|{\bf K}|^4\nonumber\\
  &&\quad\quad\quad\quad\times \langle \hat{\zeta}_{\bf p}(\tau)\hat{\zeta}_{-{\bf p}}(\tau)\hat{\zeta}_{{\bf k}_1}(\tau_1)\hat{\zeta}_{{\bf k}_2}(\tau_1)\hat{\zeta}_{{\bf k}_3}(\tau_1)\hat{\zeta}_{{\bf k}_4}(\tau_2)\hat{\zeta}_{{\bf k}_5}(\tau_2)\hat{\zeta}_{{\bf k}_6}(\tau_2)\rangle^{\dagger}\Bigg]=\big[{\bf Z}^{(1)}_4\big]^{\dagger},\end{eqnarray}
  and

\begin{eqnarray} \label{A33a}{\bf Z}^{(3)}_1:&=&\lim_{\tau\rightarrow 0}\Bigg[\int^{0}_{-\infty}d\tau_1\frac{a^2(\tau_1)}{H^3(\tau_1)}\frac{{\cal G}_1(\tau_1)}{a(\tau_1)}\;\int^{0}_{-\infty}d\tau_2\;\frac{a^2(\tau_2)}{H^3(\tau_2)}\frac{{\cal G}_1(\tau_2)}{a(\tau_2)}\nonumber\\
  &&\quad\quad\quad\quad\times\int \frac{d^{3}{\bf k}_1}{(2\pi)^3} \int \frac{d^{3}{\bf k}_2}{(2\pi)^3} \int \frac{d^{3}{\bf k}_3}{(2\pi)^3} \int \frac{d^{3}{\bf k}_4}{(2\pi)^3} \int \frac{d^{3}{\bf k}_5}{(2\pi)^3} \int \frac{d^{3}{\bf k}_6}{(2\pi)^3}\nonumber\\
  &&\quad\quad\quad\quad\times \delta^3\bigg({\bf k}_1+{\bf k}_2+{\bf k}_3\bigg) \delta^3\bigg({\bf k}_4+{\bf k}_5+{\bf k}_6\bigg)\nonumber\\
  &&\quad\quad\quad\quad\times \langle  \hat{\zeta}^{'}_{{\bf k}_1}(\tau_1)\hat{\zeta}^{'}_{{\bf k}_2}(\tau_1)\hat{\zeta}^{'}_{{\bf k}_3}(\tau_1)
 \hat{\zeta}_{\bf p}(\tau)\hat{\zeta}_{-{\bf p}}(\tau)\hat{\zeta}^{'}_{{\bf k}_4}(\tau_2)\hat{\zeta}^{'}_{{\bf k}_5}(\tau_2)\hat{\zeta}^{'}_{{\bf k}_6}(\tau_2)\rangle\Bigg],\eea\bea
  \label{A33b}{\bf Z}^{(3)}_2:&=&\lim_{\tau\rightarrow 0}\Bigg[\int^{0}_{-\infty}d\tau_1\frac{a^2(\tau_1)}{H^3(\tau_1)}\frac{{\cal G}_2(\tau_1)}{a^2(\tau_1)}\;\int^{0}_{-\infty}d\tau_2\;\frac{a^2(\tau_2)}{H^3(\tau_2)}\frac{{\cal G}_2(\tau_2)}{a^2(\tau_2)}\nonumber\\
  &&\quad\quad\quad\quad\times\int \frac{d^{3}{\bf k}_1}{(2\pi)^3} \int \frac{d^{3}{\bf k}_2}{(2\pi)^3} \int \frac{d^{3}{\bf k}_3}{(2\pi)^3} \int \frac{d^{3}{\bf k}_4}{(2\pi)^3} \int \frac{d^{3}{\bf k}_5}{(2\pi)^3} \int \frac{d^{3}{\bf k}_6}{(2\pi)^3}\nonumber\\
  &&\quad\quad\quad\quad\times \delta^3\bigg({\bf k}_1+{\bf k}_2+{\bf k}_3\bigg) \delta^3\bigg({\bf k}_4+{\bf k}_5+{\bf k}_6\bigg)|{\bf K}|^4\nonumber\\
  &&\quad\quad\quad\quad\times \langle \hat{\zeta}^{'}_{{\bf k}_1}(\tau_1)\hat{\zeta}^{'}_{{\bf k}_2}(\tau_1)\hat{\zeta}_{{\bf k}_3}(\tau_1)\hat{\zeta}_{\bf p}(\tau)\hat{\zeta}_{-{\bf p}}(\tau)\hat{\zeta}^{'}_{{\bf k}_4}(\tau_2)\hat{\zeta}^{'}_{{\bf k}_5}(\tau_2)\hat{\zeta}_{{\bf k}_6}(\tau_2)\rangle\Bigg],\eea\bea
 \label{A33c}{\bf Z}^{(3)}_3:&=&\lim_{\tau\rightarrow 0}\Bigg[\int^{0}_{-\infty}d\tau_1\frac{a^2(\tau_1)}{H^3(\tau_1)}\frac{{\cal G}_3(\tau_1)}{a(\tau_1)}\;\int^{0}_{-\infty}d\tau_2\;\frac{a^2(\tau_2)}{H^3(\tau_2)}\frac{{\cal G}_3(\tau_2)}{a(\tau_2)}\nonumber\\
  &&\quad\quad\quad\quad\times\int \frac{d^{3}{\bf k}_1}{(2\pi)^3} \int \frac{d^{3}{\bf k}_2}{(2\pi)^3} \int \frac{d^{3}{\bf k}_3}{(2\pi)^3} \int \frac{d^{3}{\bf k}_4}{(2\pi)^3} \int \frac{d^{3}{\bf k}_5}{(2\pi)^3} \int \frac{d^{3}{\bf k}_6}{(2\pi)^3}\nonumber\\
  &&\quad\quad\quad\quad\times \delta^3\bigg({\bf k}_1+{\bf k}_2+{\bf k}_3\bigg) \delta^3\bigg({\bf k}_4+{\bf k}_5+{\bf k}_6\bigg)\left({\bf k}_2\cdot{\bf k}_3\right)\left({\bf k}_5\cdot{\bf k}_6\right)\nonumber\\
  &&\quad\quad\quad\quad\times \langle \hat{\zeta}^{'}_{{\bf k}_1}(\tau_1)\hat{\zeta}_{{\bf k}_2}(\tau_1)\hat{\zeta}_{{\bf k}_3}(\tau_1)\hat{\zeta}_{\bf p}(\tau)\hat{\zeta}_{-{\bf p}}(\tau)\hat{\zeta}^{'}_{{\bf k}_4}(\tau_2)\hat{\zeta}_{{\bf k}_5}(\tau_2)\hat{\zeta}_{{\bf k}_6}(\tau_2)\rangle\Bigg],\\ 
   \label{A33d}{\bf Z}^{(3)}_4:&=&\lim_{\tau\rightarrow 0}\Bigg[\int^{0}_{-\infty}d\tau_1\frac{a^2(\tau_1)}{H^3(\tau_1)}\frac{{\cal G}_4(\tau_1)}{a^2(\tau_1)}\;\int^{0}_{-\infty}d\tau_2\;\frac{a^2(\tau_2)}{H^3(\tau_2)}\frac{{\cal G}_4(\tau_2)}{a^2(\tau_2)}\nonumber\\
  &&\quad\quad\quad\quad\times\int \frac{d^{3}{\bf k}_1}{(2\pi)^3} \int \frac{d^{3}{\bf k}_2}{(2\pi)^3} \int \frac{d^{3}{\bf k}_3}{(2\pi)^3} \int \frac{d^{3}{\bf k}_4}{(2\pi)^3} \int \frac{d^{3}{\bf k}_5}{(2\pi)^3} \int \frac{d^{3}{\bf k}_6}{(2\pi)^3}\nonumber\\
  &&\quad\quad\quad\quad\times \delta^3\bigg({\bf k}_1+{\bf k}_2+{\bf k}_3\bigg) \delta^3\bigg({\bf k}_4+{\bf k}_5+{\bf k}_6\bigg)\left({\bf k}_1\cdot{\bf k}_2\right)\left({\bf k}_4\cdot{\bf k}_5\right)|{\bf K}|^4\nonumber\\
  &&\quad\quad\quad\quad\times \langle \hat{\zeta}_{{\bf k}_1}(\tau_1)\hat{\zeta}_{{\bf k}_2}(\tau_1)\hat{\zeta}_{{\bf k}_3}(\tau_1)\hat{\zeta}_{\bf p}(\tau)\hat{\zeta}_{-{\bf p}}(\tau)\hat{\zeta}_{{\bf k}_4}(\tau_2)\hat{\zeta}_{{\bf k}_5}(\tau_2)\hat{\zeta}_{{\bf k}_6}(\tau_2)\rangle\Bigg],\end{eqnarray}
  where it is crucial to remember that:
\bea |{\bf K}|=|{\bf k}_1+{\bf k}_2+{\bf k}_3|=|{\bf k}_4+{\bf k}_5+{\bf k}_6|=\sqrt{k^2_1+k^2_2+k^3_3}=\sqrt{k^2_4+k^2_5+k^2_6}.\eea
The integral across conformal time and the momentum scales must also be addressed by splitting them into the three zones that were previously mentioned:
\bea {\bf Conformal\;time\;integral:}\quad\quad\quad\quad\lim_{\tau\rightarrow 0}\int^{\tau}_{-\infty}:\equiv \underbrace{\Bigg(\int^{\tau_s}_{-\infty}\Bigg)}_{\bf SRI}+\underbrace{\Bigg(\int^{\tau_e}_{\tau_s}\Bigg)}_{\bf USR}+\underbrace{\Bigg(\int^{\tau_{\rm end}\rightarrow 0}_{\tau_e}\Bigg)}_{\bf SRII},\eea
and 
\bea {\bf Momentum\;integral:}\quad\quad\quad\quad\int^{\infty}_{0}:\equiv \underbrace{\Bigg(\int^{k_s}_{k_*}\Bigg)}_{\bf SRI}+\underbrace{\Bigg(\int^{k_e}_{k_s}\Bigg)}_{\bf USR}+\underbrace{\Bigg(\int^{k_{\rm end}\rightarrow 0}_{k_e}\Bigg)}_{\bf SRII},\eea
where the IR and UV cut-offs, which are essential for extracting the finite contribution from the current calculation, fulfill the function of the finite limits of the integration in the description above. 

\section{Couplings and Coefficients in three phases including one-loop effects for Galileon inflation}
\label{A6a}

The appendix contains the equations for momentum-dependent functions and couplings that are required to characterize the one-loop contributions to the overall power spectrum.

\subsection{For the SRI phase}
The formula below represents the one-loop effect from the SRI phase that contributes to the overall power spectrum of scalar modes:
\bea
\Bigg[\Delta^{2}_{\zeta,\bf {One-Loop}}(k)\Bigg]_{\rm \textbf{SRI}} = \Bigg[\Delta^{2}_{\zeta,\bf {Tree}}(k)\Bigg]_{\rm \textbf{SRI}}\times \left( -\sum^{4}_{i=1}{\cal G}_{i,\mbf{SRI}}\mbf{F}_{i,\mbf{SRI}}(k_{s},k_{*})\right)
\eea
Using the CGT couplings, the values for ${\cal G}_{i,\textbf{SRI}}\;\forall i=1,2,3,4$ are defined:
\begin{eqnarray} {\cal G}_{1,{\bf SRI}}&=&{\cal G}^2_1(\tau_*),\\ 
    {\cal G}_{2,{\bf SRI}}&=&-{\cal G}^2_2(\tau_*)H^2(\tau_*)c^2_s,\\
    {\cal G}_{3,{\bf SRI}}&=&-\frac{{\cal G}^2_3(\tau_*)}{c^2_s},\\
    {\cal G}_{4,{\bf SRI}}&=&\frac{{\cal G}^2_4(\tau_*)H^2(\tau_*)}{c^6_s}.
    \end{eqnarray}
    and ${\cal K}_{*,s} \equiv k_{*}/k_{s}$ is used to create the momentum-dependent functions $\mbf{F}_{i,\textbf{SRI}}$.
\begin{eqnarray} {\bf F}_{1,{\bf SRI}}(k_s,k_*)&=&\frac{1}{2}\Bigg[{\cal K}_{*,s}^2 +3\Bigg],\\ 
         {\bf F}_{2,{\bf SRI}}(k_s,k_*)&=&\Bigg[\frac{17}{42}-\frac{2}{3}{\cal K}_{*,s}^6+\frac{24}{7}{\cal K}_{*,s}^7-\frac{9}{2}{\cal K}_{*,s}^8\Bigg],\\ 
         {\bf F}_{3,{\bf SRI}}(k_s,k_*)&=&\frac{2}{3}\Bigg[{\cal K}_{*,s}^6-1\Bigg],\\ 
         {\bf F}_{4,{\bf SRI}}(k_s,k_*)&=&\frac{1}{2}\Bigg[{\cal K}_{*,s}^2-1\Bigg].\end{eqnarray}

         \subsection{For the USR phase} The formula below represents the one-loop effect from the USR phase that contributes to the overall power spectrum of scalar modes:
\bea
\Bigg[\Delta^{2}_{\zeta,\bf {One-Loop}}(k)\Bigg]_{\rm \textbf{USR}} = \Bigg[\Delta^{2}_{\zeta,\bf {Tree}}(k)\Bigg]_{\rm \textbf{SRI}}\times \left( \sum^{4}_{i=1}{\cal G}_{i,\mbf{USR}}\mbf{F}_{i,\mbf{SRI}}(k_{e},k_{s})\right)
\eea
The CGT couplings are used to define the values for ${\cal G}_{i,\textbf{USR}}\;\forall i=1,2,3,4$:
\begin{eqnarray} {\cal G}_{1,{\bf USR}}&=&\Bigg(\frac{{\cal G}^2_1(\tau_e)}{c^3_s}{\cal K}_{e,s}^6-\frac{{\cal G}^2_1(\tau_s)}{c^3_s}\Bigg),\\
    {\cal G}_{2,{\bf USR}}&=&\Bigg(\frac{{\cal G}^2_2(\tau_e)H^2(\tau_e)}{c^5_s}{\cal K}_{e,s}^4-\frac{{\cal G}^2_2(\tau_s)H^2(\tau_s)}{c^5_s}\Bigg),\\
    {\cal G}_{3,{\bf USR}}&=&\Bigg(\frac{{\cal G}^2_3(\tau_e)}{c^6_s}{\cal K}_{e,s}^3-\frac{{\cal G}^2_3(\tau_s)}{c^6_s}\Bigg),\\
   {\cal G}_{4,{\bf USR}}&=&\Bigg(\frac{{\cal G}^2_2(\tau_e)H^2(\tau_e)}{c^7_s}{\cal K}_{e,s}^2-\frac{{\cal G}^2_2(\tau_s)H^2(\tau_s)}{c^7_s}\Bigg).
\end{eqnarray}
where ${\cal K}_{e,s} \equiv k_{e}/k_{s},\;\text{and}\;{\cal K}_{s,e} \equiv k_{s}/k_{e}$ has been used as the notation. We express the momentum-dependent functions $\mbf{F}_{i,\textbf{USR}}$ as follows using the same notations as before:
\bea {\bf F}_{1,{\bf USR}}(k_e,k_s)&=&\Bigg[\frac{1}{4}+\frac{9}{4}{\cal K}_{s,e}^2-\frac{1}{4}{\cal K}_{s,e}^4-9{\cal K}_{s,e}^4\ln{\cal K}_{s,e}-6\sin\left({\cal K}_{e,s}+1\right)\sin\left(1-{\cal K}_{e,s}\right)\nonumber\\
    &&\quad-6{\cal K}_{s,e}\sin\left(2{\cal K}_{e,s}\right)-\frac{9}{8}{\cal K}_{s,e}^2\Bigg\{2\left(\frac{k_e}{k_s}\right)\sin\left(2{\cal K}_{e,s}\right)-\cos\left(2{\cal K}_{e,s}\right)\Bigg\}\nonumber\\
    &&\quad-\frac{69}{16}\cos\left(2{\cal K}_{e,s}\right)+\frac{3}{4}{\cal K}_{s,e}^2\Bigg\{\sin\left(2{\cal K}_{e,s}\right)-2{\cal K}_{e,s}\cos\left(2{\cal K}_{e,s}\right)\Bigg\}\Bigg],\\
         {\bf F}_{2,{\bf USR}}(k_e,k_s)&=&\Bigg[\frac{1}{8}+\frac{9}{12}{\cal K}_{s,e}^2+\frac{9}{4}{\cal K}_{s,e}^4+\frac{9}{4}{\cal K}_{s,e}^6-\frac{43}{8}{\cal K}_{s,e}^8-\frac{9}{8}\cos\left(2{\cal K}_{e,s}\right)\Bigg],\\
    {\bf F}_{3,{\bf USR}}(k_e,k_s)&=&\Bigg[\frac{1}{8}+\frac{9}{12}{\cal K}_{s,e}^2+\frac{9}{4}{\cal K}_{s,e}^4+\frac{9}{4}{\cal K}_{s,e}^6-\frac{43}{8}{\cal K}_{s,e}^8-\frac{9}{8}\cos\left(2{\cal K}_{e,s}\right)\Bigg]\nonumber\\
         &=&{\bf F}_{2,{\bf USR}}(k_e,k_s),\\ 
         {\bf F}_{4,{\bf USR}}(k_e,k_s)&=&\Bigg[\frac{1}{12}+\frac{9}{20}{\cal K}_{s,e}^2+\frac{9}{8}{\cal K}_{s,e}^4+\frac{9}{12}{\cal K}_{s,e}^6-\frac{299}{230}{\cal K}_{s,e}^{12}\Bigg].\eea

       \subsection{For the SRII phase} The formula below represents the one-loop effect from the SRII phase that contributes to the overall power spectrum of scalar modes:
\bea
\Bigg[\Delta^{2}_{\zeta,\bf {One-Loop}}(k)\Bigg]_{\rm \textbf{SRII}} = \Bigg[\Delta^{2}_{\zeta,\bf {Tree}}(k)\Bigg]_{\rm \textbf{SRI}}\times \left( \sum^{4}_{i=1}{\cal G}_{i,\mbf{SRII}}\mbf{F}_{i,\mbf{SRII}}(k_{\rm end},k_{e})\right)
\eea
The CGT couplings are used to define the values for ${\cal G}_{i,\textbf{SRII}}\;\forall i=1,2,3,4$:
\begin{eqnarray} {\cal G}_{1,{\bf SRII}}&=&\Bigg(\frac{{\cal G}^2_1(\tau_{\rm end})}{c^3_s}{\cal K}_{e,s}^6-\frac{{\cal G}^2_1(\tau_e)}{c^3_s}\Bigg),\\
    {\cal G}_{2,{\bf SRII}}&=&\Bigg(\frac{{\cal G}^2_2(\tau_{\rm end})H^2(\tau_{\rm end})}{c^5_s}{\cal K}_{e,s}^4-\frac{{\cal G}^2_2(\tau_e)H^2(\tau_e)}{c^5_s}\Bigg),\\
   {\cal G}_{3,{\bf SRII}}&=&\Bigg(\frac{{\cal G}^2_3(\tau_{\rm end})}{c^6_s}{\cal K}_{e,s}^3-\frac{{\cal G}^2_3(\tau_e)}{c^6_s}\Bigg),\\
   {\cal G}_{4,{\bf SRII}}&=&\Bigg(\frac{{\cal G}^2_2(\tau_{\rm end})H^2(\tau_{\rm end})}{c^7_s}{\cal K}_{e,s}^2-\frac{{\cal G}^2_2(\tau_e)H^2(\tau_e)}{c^7_s}\Bigg). \end{eqnarray}
   where the preceding definitions of ${\cal K}_{e,s}\;\text{and}\;{\cal K}_{s,e}$ are from the USR phase. Here, the momentum-dependent functions $\mbf{F}_{i,\textbf{SRII}}$ are defined using only the leading order contributions removed with the suppressed quantities represented by $"\cdots"$. We decide to use the following notation in order to write the expressions more succinctly:
\bea
{\cal K}_{{\rm end},e} \equiv \frac{k_{\rm end}}{k_{e}},\quad\quad{\cal K}_{{\rm end},s} \equiv \frac{k_{\rm end}}{k_{s}},\quad\quad{\cal K}_{e,{\rm end}} \equiv \frac{k_{e}}{k_{\rm end}},\quad\quad{\cal K}_{s,{\rm end}} \equiv \frac{k_{s}}{k_{\rm end}}
\eea
The ultimate outcome is expressed as follows:
  \bea {\bf F}_{1,{\bf SRII}}(k_{\rm end},k_e)&=&\Bigg[\frac{81}{64}+\frac{81}{40}\left({\cal K}_{s,e}+1\right)-\frac{9}{8}\left({\cal K}_{s,e}^2 +1\right)+\frac{1}{6}\left(\left({\cal K}_{s,e}+1\right)^2+2{\cal K}_{s,e}\right)\nonumber\\&&+\frac{1}{8}{\cal K}_{s,e}^2+6
 +\frac{2}{7}{\cal K}_{s,e}\left(1+{\cal K}_{s,e}\right)\nonumber\\
 &&+\frac{27}{8}\left(1+{\cal K}_{s,e}^6\right)-\frac{3}{4}{\cal K}_{e,{\rm end}}\cos\left(2{\cal K}_{{\rm end},e}\right)+4{\cal K}_{e,{\rm end}}^2\ln\left({\cal K}_{e,{\rm end}}\right)\nonumber\\&&-\frac{3}{4}\sin\left(1+{\cal K}_{{\rm end},s}\right)\sin\left(1-{\cal K}_{{\rm end},s}\right)\nonumber\\
 &&-\frac{9}{4}\frac{\displaystyle{\cal K}_{s,e}}{\displaystyle\left(1-{\cal K}_{s,e}\right)^2}\Bigg\{\cos\left(2\left({\cal K}_{{\rm end},e}-{\cal K}_{{\rm end},s}\right)\right)\Bigg\}\nonumber\\&&-\frac{9}{4}\frac{\displaystyle 1}{\displaystyle\left(1-{\cal K}_{s,e}\right)}\Bigg\{{\cal K}_{s,{\rm end}}\sin\left(2\left({\cal K}_{{\rm end},e}-{\cal K}_{{\rm end},s}\right)\right)\Bigg\}\nonumber\\
    &&-9\frac{\displaystyle\left(1+{\cal K}_{s,e}\right)^2}{\displaystyle \left(1-{\cal K}_{s,e}\right)}\Bigg\{\sin\left(2\left({\cal K}_{{\rm end},e}-{\cal K}_{{\rm end},s}\right)\right)\Bigg\}+\frac{9}{8}\Bigg\{8-18{\cal K}_{s,e}^4\left({\cal K}_{s,e}+1\right)\Bigg\}\nonumber\\
    &&\Bigg\{\cos\left(2\left({\cal K}_{{\rm end},e}-{\cal K}_{{\rm end},s}\right)\right)-{\cal K}_{e,{\rm end}}\cos\left(2\left(1-{\cal K}_{e,s}\right)\right)\Bigg\}\nonumber\\&&+\frac{9}{7}{\cal K}_{s,e}^6+9{\cal K}_{s,e}^4\left(1+{\cal K}_{s,e}^2\right)-\frac{9}{16}{\cal K}_{s,e}^2\left(9+44{\cal K}^{2}_{s,e}\right)\ln{\cal K}_{e,{\rm end}}+22{\cal K}_{s,e}^2\left(1-{\cal K}_{e,{\rm end}}\right)\nonumber\\
    &&+\frac{27}{16}{\cal K}_{s,e}^4{\cal K}_{e,{\rm end}}\Bigg\{{\cal K}_{e,{\rm end}}\cos\left(2{\cal K}_{{\rm end},e}\right)-\sin\left(2{\cal K}_{{\rm end},e}\right)\Bigg\}\cdots\Bigg],\eea\bea
    {\bf F}_{2,{\bf SRII}}(k_{\rm end},k_e)&=&\Bigg[5{\cal K}_{s,e}^6\ln{\cal K}_{e,{\rm end}}+\frac{81}{8}\left(1-{\cal K}_{e,{\rm end}}\right)\left(1+{\cal K}_{s,e}\right)+\frac{81}{16}\ln{\cal K}_{e,{\rm end}}\nonumber\\
    &&+\frac{2}{3}{\cal K}_{s,e}\left(1-{\cal K}_{e,{\rm end}}^3\right)\left(1+{\cal K}_{s,e}\right)+\frac{1}{2}\left(2{\cal K}_{s,e}+\left(1+{\cal K}_{s,e}\right)^2\right)\nonumber\\&&+\frac{1}{4}+\frac{27}{64}+\frac{1}{4}{\cal K}_{s,e}^2+\frac{3}{8}\left(1+{\cal K}_{s,e}^2\right)\nonumber\\
    &&-\frac{27}{8}\left(1+{\cal K}_{s,e}^6\right)-\frac{9}{16}\frac{\displaystyle{\cal K}_{s,e}^3}{\displaystyle\left(1-{\cal K}_{s,e}\right)^2}\Bigg\{\cos\left(2\left({\cal K}_{{\rm end},e}-{\cal K}_{{\rm end},s}\right)\right)\Bigg\}\nonumber\\
    &&-\frac{27}{4}\frac{\displaystyle 1}{\displaystyle\left(1-{\cal K}_{s,e}\right)}\Bigg\{{\cal K}_{s,{\rm end}}\sin\left(2\left({\cal K}_{{\rm end},e}-{\cal K}_{{\rm end},s}\right)\right)\Bigg\}\nonumber\\&&+\frac{81}{16}\frac{\displaystyle\left(1+{\cal K}_{s,e}^2\right)}{\displaystyle \left(1-{\cal K}_{s,e}\right)}\Bigg\{\cos\left(2\left({\cal K}_{{\rm end},e}-{\cal K}_{{\rm end},s}\right)\right)\Bigg\}\nonumber\\
    &&-\frac{9}{16}\frac{\displaystyle{\cal K}_{s,e}^3}{\displaystyle\left(1-{\cal K}_{s,e}\right)}\Bigg\{\sin\left(2\left({\cal K}_{{\rm end},e}-{\cal K}_{{\rm end},s}\right)\right)\Bigg\}\nonumber\\&&-\frac{9}{16}\left(1+\left(\frac{k_e}{k_s}\right)\right)\Bigg\{{\cal K}_{e,{\rm end}}\cos\left(2\left({\cal K}_{{\rm end},e}-{\cal K}_{{\rm end},s}\right)\right)\Bigg\}\nonumber\\
    &&+\frac{27}{16}{\cal K}_{s,e}^6+\frac{81}{64}{\cal K}_{s,e}^6-\frac{81}{16}{\cal K}_{s,e}^4\left(1+{\cal K}_{s,e}^2\right)\nonumber\\&&-\frac{9}{32}{\cal K}_{s,e}^2\left(1-{\cal K}_{e,{\rm end}}\right)\left(9+8{\cal K}_{s,e}^2+36{\cal K}_{s,e}^4\right)\nonumber\\
    &&+\left(1+4{\cal K}_{s,e}^2+{\cal K}_{s,e}^4\right)\ln{\cal K}_{e,{\rm end}}-\frac{3}{4}{\cal K}_{e,{\rm end}}\cos\left(2{\cal K}_{{\rm end},e}\right)\cdots\Bigg]={\bf F}_{3,{\bf SRII}}(k_{\rm end},k_e),\eea\bea
         {\bf F}_{4,{\bf SRII}}(k_{\rm end},k_e)&=&\Bigg[\frac{81}{16}+\frac{81}{16}{\cal K}_{s,e}^6\ln{\cal K}_{e,{\rm end}}+\frac{81}{32}\left(2{\cal K}_{s,e}+\left(1+{\cal K}_{s,e}\right)^2\right)+\frac{27}{8}\left(1+{\cal K}_{s,e}\right)\nonumber\\
         &&+10{\cal K}_{s,e}\left(1+{\cal K}_{s,e}\right)+\frac{1}{12}+\frac{9}{40}\left(1+{\cal K}_{s,e}^2\right)\nonumber\\&&+\frac{27}{64}\left(1+{\cal K}_{s,e}^6\right)+\frac{81}{128}+6{\cal K}_{s,e}^2\ln{\cal K}_{e,{\rm end}}\nonumber\\
         &&-\frac{9}{16}\frac{\displaystyle\left(1+{\cal K}_{s,e}\right)^2}{\displaystyle\left(1-{\cal K}_{s,e}\right)^2}\Bigg\{\cos\left(2\left({\cal K}_{{\rm end},e}-{\cal K}_{{\rm end},s}\right)\right)\Bigg\}\nonumber\\&&-\frac{27}{8}\frac{\displaystyle \displaystyle\left(1+{\cal K}_{s,e}\right)^2}{\displaystyle\left(1-{\cal K}_{s,e}\right)^3}\Bigg\{\sin\left(2\left({\cal K}_{{\rm end},e}-{\cal K}_{{\rm end},s}\right)\right)\Bigg\}\nonumber\\
    &&-\frac{27}{4}\frac{\displaystyle \displaystyle\left(1+{\cal K}_{s,e}\right)^2}{\displaystyle\left(1-{\cal K}_{s,e}\right)}\Bigg\{{\cal K}_{e,{\rm end}}^2\sin\left(2\left({\cal K}_{{\rm end},e}-{\cal K}_{{\rm end},s}\right)\right)\Bigg\}\nonumber\\&&+\frac{27}{8}{\cal K}_{s,e}^4\left(1+{\cal K}_{s,e}^2\right)+\frac{9}{80}{\cal K}_{s,e}^2\left(9+44{\cal K}_{s,e}^2\right)\nonumber\\
    &&+\frac{81}{16}{\cal K}_{s,e}^6\left(1-{\cal K}_{e,{\rm end}}\right)+\frac{99}{56}+\frac{1}{4}{\cal K}_{e,{\rm end}}^2
   \nonumber\\&&-\frac{81}{32}{\cal K}_{s,e}^6{\cal K}_{e,{\rm end}}\Bigg\{{\cal K}_{e,{\rm end}}\cos\left(2{\cal K}_{{\rm end},e}\right)-\sin\left(2{\cal K}_{{\rm end},e}\right)\Bigg\}+\cdots\Bigg].\quad\quad
   \quad\eea

\section{Useful integrals in the Superhorizon regime}
\label{A7a}

The appendix discusses the limiting forms of the integrals found in equation (\ref{besselprod}). We analyze them in the super horizon regime using a product of Bessel functions. We adhere to the analysis that \cite{Domenech:2021ztg} presents. The variable $x=k\tau$ under this condition is expressed as $x \ll 1$. There are modes within this regime that, in the example considered, are super-Horizon with respect to the pivot scale. We feel it is necessary to give a quick explanation of the solutions under this regime, even if we have not taken into account the specifics of the super-Horizon kernel for our study. Another crucial premise is that the scale that contributes to the scalar power spectrum peak, or $k_{s} \gg k_{*}$, is supposed to enter far earlier than the pivot scale. Making this assumption lessens the likelihood that sourcing will cause notable changes before, during, or after the modes' re-entry at a pivot scale.

We may examine the variable $v \sim k_{s}/k \gg 1$ based on the aforementioned assumptions, which results in $u\sim v \gg 1$. In order to integrate under the assumptions $u \sim v$ and include these approximations into product of the Bessel functions, we must apply the following relations:
\bea
\int dx \;x\;J_{p}^{2}(ax) &=& \frac{x^{2}}{2}(J_{p}^{2}(ax)-J_{p-1}(ax)J_{p+1}(ax)),\\
\int dx\; x^{-2p+1}\;J_{p}^{2}(ax) &=& \frac{x^{-2p+2}}{2(1-2p)}(J^{2}_{p}(ax) + J^{2}_{p-1}(ax))
\eea
These relations allow us to reduce the integrals found in eqn.(\ref{besselprod}) to the following:
\bea
I_{J} &\approx & \frac{3+2b}{1+b}\frac{2^{-b-1/2}}{\Gamma(b+3/2)}\frac{x}{\pi v},\\
I_{Y}  &\approx & \frac{3+2b}{b(1+b)c_{s}\pi}\bigg(2^{b-1/2}\Gamma(b+1/2)\frac{x^{-2b}}{\pi v} - 2^{-b-3/2}\frac{c_{s}^{2b}}{\Gamma(b+3/2)}\frac{1+b+b^{2}}{1+b}\bigg).
\eea

\section{Useful integrals in the Subhorizon regime} 
\label{A8a}

We cover the various integrals using Bessel function products in this appendix, along with practical approximations under the sub-Horizon regime's limiting condition. In this regime, the variable $x=k\tau$ has the form $x \gg 1$. The integral in equation.(\ref{besselprod}), which includes a triple product of Bessel functions and, as we previously noted, is not analytically possible for generic $x$, is included in the kernel of equation.(\ref{simplekern}). Analytical findings exist for taking the limit $x \gg 1$; these integrals are mentioned here and may also be found in \cite{gervois1985integrals}. The findings are expressed in terms of several additional variables, which we designate as $p$, $q$, and $r$. The formula for the condition $|p-q| < r < p+q$ in eqn.(\ref{besselprod}) is as follows: 
\bea
I_{B}^{x \gg 1} &=& \int_{0}^{\infty}d\tilde{x}\;\tilde{x}^{1-a}\;B_{a}(r\tilde{x})J_{b}(p\tilde{x})J_{b}(q\tilde{x}), \nonumber\\
&=& \frac{\sqrt{2}}{\pi\sqrt{\pi}}\frac{(pq)^{a-1}}{r^{a}}(\sin{\theta})^{a-1/2}\times \left\{
	\begin{array}{ll}
		\displaystyle \frac{\pi}{2}P^{-a+1/2}_{b-1/2}  \quad\quad & \mbox{when}\quad  B_{a}(rx)=J_{a}(rx) \\ 
			\displaystyle 
			\displaystyle -Q^{-a+1/2}_{b-1/2} \quad\quad & \mbox{when }\quad  B_{a}(rx)=Y_{a}(rx)
	\end{array}
\right.
\eea
It has to do with the relationships: 
\bea
16\Delta^{2} \equiv (r^{2}-(p-q)^{2})((p+q)^{2}-r^{2}), \quad\quad \cos{\theta} = \frac{p^{2}+q^{2}-r^{2}}{2pq}, \quad\quad \sin{\theta} = \frac{2\Delta}{pq}.
\eea
In contrast, the following may be obtained from the integrals when $r > p+q$ is satisfied:
\bea
I_{B}^{x \gg 1} &=& \int_{0}^{\infty}d\tilde{x}\;\tilde{x}^{1-a}\;B_{a}(r\tilde{x})J_{b}(p\tilde{x})J_{b}(q\tilde{x}), \nonumber\\
&=& \frac{\sqrt{2}}{\pi\sqrt{\pi}}\frac{(pq)^{a-1}}{r^{a}}(\sinh{\theta})^{a-1/2}\Gamma(b-a+1){\cal Q}^{-a+1/2}_{b-1/2}(\cosh{\theta})\times \left\{
	\begin{array}{ll}
		\displaystyle -\sin{(b-a)\pi}  \quad\quad & \mbox{when}\quad  B_{a}(rx)=J_{a}(rx) \\ 
			\displaystyle 
			\displaystyle \cos{(b-a)\pi} \quad\quad & \mbox{when }\quad  B_{a}(rx)=Y_{a}(rx)
	\end{array}
\right.
\eea
This pertains to the new relationships: 
\bea
16\tilde{\Delta}^{2} \equiv (r^{2}-(p-q)^{2})(r^{2}-(p+q)^{2}), \quad\quad \cosh{\theta} = \frac{r^{2}-(p^{2}+q^{2})}{2pq}, \quad\quad \sinh{\theta} = \frac{2\tilde{\Delta}}{pq}.
\eea
The following variable renaming has to be completed in order for the integrals performed in this text to function properly:
\bea
r=1,\quad\quad p=c_{s}v, \quad\quad q=c_{s}u
\eea
Drawing from the aforementioned, the range $|1-v| < u < 1+v$ for the variables in eqn.(\ref{omegac}) may be divided into two categories: $c_{s}(u+v)< 1 ( p+q< r)$ and $1 < c_{s}(u+v) (r < p+q)$.

\section{Notes on Associate Legendre functions} 
\label{A9a}

We will now go over a few of the key equations related to Legendre's polynomials that we have utilized in our research. The formulation of the hypergeometric function which accepts a quadratic transformation is first presented:
\bea
\textbf{F}(l,m;n;x) = \frac{1}{\Gamma[n]}{\cal F}(l,m;n;x),
\eea
where ${\cal F}(l,m;n;x)$ denotes the Gauss's hypergeometric function and $l,m,n$ are rational parameters. Let's use this to write the functions of Ferrer and Olver's expression.
\begin{eqnarray}
P_{\nu}^{\mu}(x)&=& \bigg[\frac{1+x}{1-x}\bigg]^{\mu /2}\; \textbf{F}(\nu+1,-\nu;1-\mu;1/2-1/2x), \\
Q_{\nu}^{\mu}(x)&=&\frac{\pi}{2\sin{(\mu \pi)}}\bigg(\cos{(\mu \pi)}\bigg[\frac{1-x}{1+x}\bigg]^{\mu/2}\; \textbf{F}(\nu+1,-\nu;1-\mu;1/2-x/2) \nonumber \\ 
&& \quad \quad \quad \quad - \frac{\Gamma(\nu +\mu +1)}{\Gamma(\nu - \mu +1)}\bigg[\frac{1-x}{1+x}\bigg]^{\mu/2}\; \textbf{F}(\nu+1,-\nu;1+\mu;1/2-x/2)\bigg), \\
{\cal Q}_{\nu}^{\mu}(x) &=& \frac{\pi}{2\sin{(\mu \pi)} \Gamma(\nu+\mu+1)}\bigg(\bigg[\frac{x+1}{x-1}\bigg]^{\mu/2}\; \textbf{F}(\nu+1,-\nu;1-\mu;1/2-x/2) \nonumber \\
&& \quad \quad \quad \quad - \frac{\Gamma(\nu +\mu +1)}{\Gamma(\nu - \mu +1)}\bigg[\frac{x-1}{1+x}\bigg]^{\mu/2}\; \textbf{F}(\nu+1,-\nu;1+\mu;1/2-x/2)\bigg),
\end{eqnarray}
These are now Ferrer's functions, or Legendre's functions on the cut, denoted by $P_{\nu}^{\mu}(x)$ and $Q_{\nu}^{\mu}(x)$, defined exclusively for $|x|<1$. Nevertheless, ${\cal Q}_{\nu}^{\mu}(x)$, also referred to as Olver's functions, is defined exclusively for $\abs{x}>1$. Occasionally, working with a real-valued version of ${\cal Q}_{\nu}^{\mu}(x)$, as provided by:
\bea
{\cal Q}_{\nu}^{\mu}(x) \equiv \frac{{\cal Q}_{\nu}^{\mu}(x)}{\Gamma[\mu +\nu +1]}e^{-\mu \pi i}
\eea
The following are some helpful links between the first and second kind Legendre's functions:
\bea
{\cal Q}_{\nu}^{\mu}(x)&=& \sqrt{\frac{\pi}{2}}\frac{1}{(x^2-1)^{1/4}}P^{-\nu - 1/2}_{\mu- 1/2}\bigg(x(x^2 -1)^{-1/2}\bigg), \\
P_{\nu}^{\mu}(-x) &=& -(2/\pi)\sin{((\nu+\mu)\pi)}\;Q_{\nu}^{\mu}(x) +  \cos{((\nu + \mu)\pi)}\; P_{\nu}^{\mu}(x) \nonumber \\
Q_{\nu}^{\mu}(-x) &=& -(\pi/2) \sin{((\nu +\mu)\pi)}\; P_{\nu}^{\mu}(x) - \cos{((\nu+\mu)\pi)}\;Q_{\nu}^{\mu}(x)
\eea

\subsection{Asymptotic approximation}

The required asymptotic behavior of the corresponding Legendre functions around the singular points $x \sim 1$ and $x \to \infty$ is the reason for this part of the appendix. 
\begin{itemize}
    \item \underline{\textbf{Case I}} : Limit $x \to 1^{-}$ \\
Since $\abs{x}<1$ in this instance, only the Ferrer's functions are at issue. The first kind's Ferrer's function converges to:
\bea
\lim_{x\to 1^{-}} P_{\nu}^{\mu}(x) \sim \frac{1}{\Gamma(1-\mu)}\bigg(\frac{2}{1-x}\bigg)^{\mu /2}.
\eea
Conversely, Ferrer's second-kind function is assigned to:
\bea
\lim_{x \to 1^{-}}Q_{\nu}^{\mu}(x) &=& \frac{\Gamma(\mu)\Gamma(1+\nu - \mu)}{2 \Gamma (1+\nu+\mu)} \bigg(\frac{2}{1-x}\bigg)^{\mu /2} \; \; \text{for $\mu > 0$}\\
\lim_{x \to 1^{-}}Q_{\nu}^{\mu}(x) &=& \frac{\cos{(\mu \pi)}}{2} \Gamma(\mu) \bigg(\frac{2}{1-x}\bigg)^{\mu /2} \; \; \text{for $\mu < 0, \mu \ne 1/2$}
\eea
 \item \underline{\textbf{Case II}} : Limit $x \to 1^{+}$ \\
 Since $\abs{x} > 1$, Olver's functions are relevant in this scenario. This is where the limiting behavior appears:
\bea
\lim_{x \to 1^{+}}{\cal Q}_{\nu}^{\mu}(x) = \frac{\Gamma(\mu)}{2 \Gamma (\nu +\mu +1)}\bigg(\frac{2}{x-1}\bigg)^{\mu /2}
\eea 

    \item \underline{\textbf{Case III}} :Limit $x \to \infty$ \\
    The second kind of Olver's function is equally relevant to this scenario. This is the asymptotic limit:
\bea
\lim_{x \to \infty}{\cal Q}_{\nu}^{\mu}(x) = \frac{\sqrt{\pi}}{\Gamma(\nu +3/2)(2x)^{\nu+1}}
\eea
    \end{itemize}

    \subsection{Resonance approximation}
    As you can see from the preceding section, when $\mu \ne 0$, the integrals after taking their asymptotic limits show a divergent character, and their argument follows $x \sim 1$. The limiting form of the functions that occur in equation.(\ref{besselprod}) near this regime must be used when $x\sim 1$. The precise form of the these functions' argument, as previously mentioned in equation.(\ref{kernelavg}), and the unique nature of the associated Legendre functions discussed in the section before provide insight into the meaning of the term ``resonance'' for this limit. This argument is consistent with the behavior of the Heaviside Theta argument. Next, we will provide the necessary resonance approximations to evaluate the integrals near these resonant conditions.

Assuming that $y \rightarrow -1^{+}$ is the limit, we obtain:
\bea
P_{b}^{-b}(y) + \frac{b+2}{b+1}P_{b+2}^{-b}(y) \sim \frac{3+2b}{1+b}\frac{1}{\Gamma[1+b]}\bigg(\frac{2}{1+y}\bigg)^{-b/2}
\eea

\bea
Q_{b}^{-b}(y) + \frac{b+2}{b+1}Q_{b+2}^{-b}(y) \sim - \frac{3+2b}{1+b} \begin{cases}
(1+b+b^2)\displaystyle{\frac{\Gamma[b]}{\Gamma[2b+3]}\bigg(\frac{2}{1+y}\bigg)^{b/2}}, & \text{for } b>0  \\
\displaystyle{\frac{\cos{(b \pi)}}{2}}\Gamma[-b]\bigg(\frac{2}{1+y}\bigg)^{-b/2}, & \text{for } b<0
\end{cases}
\eea
In the instance when $y \rightarrow 1^{+}$, we obtain:
\bea
{\cal Q}_{b}^{-b}(y) + 2 \frac{b+2}{b+1}{\cal Q}_{b+2}^{-b}(y) \sim \frac{3+2b}{1+b} \begin{cases}
(1+b+b^2)\displaystyle{\frac{\Gamma[b]}{\Gamma[2b+3]}\bigg(\frac{2}{1-y}\bigg)^{b/2}}, & \text{for } b>0  \\
\displaystyle{\frac{\Gamma[-b]}{2}\bigg(\frac{2}{1-y}\bigg)^{-b/2}}, & \text{for } b<0
\end{cases}
\eea

\bibliography{Refss}
\bibliographystyle{utphys}

\end{document}